\documentclass[12pt]{iopart}

\usepackage{graphicx} 
\usepackage{tikz}
\usepackage{tikzscale}
\usetikzlibrary{decorations.pathmorphing}
\usetikzlibrary {shapes.geometric}
\usetikzlibrary{calc}
\usepackage{pgfplots}
\pgfplotsset{compat=1.9}
\usepackage{subcaption}
\usepackage{url}
\usepackage{hyperref}
\usepackage{booktabs}       
\usepackage{xspace}
\usepackage{xcolor}
\usepackage[section]{placeins}
\usepackage{bigdelim}
\usepackage{amssymb}
\usepackage{siunitx}
\usepackage{comment}
\usepackage{orcidlink}

\tikzset {
  half circle/.style= {
  semicircle,
  shape border rotate=0,
  anchor=chord center,
  minimum size=1
  }
}

\makeatletter
\AtBeginDocument{%
     \expandafter\renewcommand\expandafter\subsection\expandafter
       {\expandafter\@fb@subsecFB\subsection}%
     \newcommand\@fb@subsecFB{\FloatBarrier
     \gdef\@fb@afterHHook{\@fb@topbarrier \gdef\@fb@afterHHook{}}}
     \g@addto@macro\@afterheading{\@fb@afterHHook}
     \gdef\@fb@afterHHook{}
  }
\makeatother

\newcommand\geant{\textsc{Geant}4\xspace}
\newcommand{\LogPost}{LP}
\newcommand{\mean}[1]{\mathrm{mean}(#1)}
\newcommand{\PCC}[2]{\mathrm{PCC}(#1,#2)}
\newcommand{\mf}[1]{\mathrm{manifold}(#1)}
\newcommand{\sphere}[2]{\mathrm{B}(#1,#2)}
\newcommand{\nnd}[2]{\mathrm{NND}_{#2}(#1)}

\newcommand{\sg}{\mathrm{sg}\,}

\newcommand{\dsIph}{ds~1~--~$\gamma$\xspace}
\newcommand{\dsIpi}{ds~1~--~$\pi^+$\xspace}
\newcommand{\dsII}{ds~2\xspace}
\newcommand{\dsIII}{ds~3\xspace}

\newcommand{\cf}{\texttt{CaloFlow}\xspace}
\newcommand{\icalo}{i\texttt{CaloFlow}\xspace}

\newcommand{\submAmram}{\texttt{CaloDiffusion}\xspace}
\newcommand{\submAmramCite}{\submAmram~\cite{Amram:2023onf}\xspace}

\newcommand{\submBussMAF}{\texttt{L$2$LFlows-MAF}\xspace}
\newcommand{\submBussMAFCite}{\submBussMAF~\cite{Diefenbacher:2023vsw,Buss:2024orz}\xspace}
\newcommand{\submBuss}{\texttt{L$2$LFlows}\xspace} 

\newcommand{\submBussConv}{\texttt{conv.~L$2$LFlows}\xspace}
\newcommand{\submBussConvCite}{\submBussConv~\cite{Buss:2024orz}\xspace}

\newcommand{\submCardoso}{\texttt{CaloDiT}\xspace}
\newcommand{\submCardosoCite}{\submCardoso~\cite{CaloDIT_ACAT}\xspace}

\newcommand{\submCresswell}{\texttt{CaloForest}\xspace}
\newcommand{\submCresswellCite}{\submCresswell~\cite{Cresswell:2024fst}\xspace}

\newcommand{\submErnst}{\texttt{CaloVAE+INN}\xspace}
\newcommand{\submErnstCite}{\submErnst~\cite{Ernst:2023qvn}\xspace}

\newcommand{\submFavaro}{\texttt{CaloINN}\xspace}
\newcommand{\submFavaroCite}{\submFavaro~\cite{Ernst:2023qvn}\xspace}

\newcommand{\submKaech}{\texttt{MDMA}\xspace}
\newcommand{\submKaechCite}{\submKaech~\cite{Kach:2023rqw,Kach:2024yxi}\xspace}

\newcommand{\submKobylyansky}{\texttt{CaloGraph}\xspace}
\newcommand{\submKobylyanskyCite}{\submKobylyansky~\cite{kobylianskii2024calograph}\xspace}

\newcommand{\submKorol}{\texttt{CaloClouds}\xspace}
\newcommand{\submKorolCite}{\submKorol~\cite{Buhmann:2023bwk,Buhmann:2023kdg}\xspace}

\newcommand{\submLiu}{\texttt{Calo-VQ}\xspace}
\newcommand{\submLiuCite}{\submLiu~\cite{Liu:2024kvv}\xspace}

\newcommand{\submLiuNorm}{\texttt{Calo-VQ(norm)}\xspace}
\newcommand{\submLiuNormCite}{\submLiuNorm~\cite{Liu:2024kvv}\xspace}

\newcommand{\submMadula}{\texttt{CaloLatent}\xspace}
\newcommand{\submMadulaCite}{\submMadula~\cite{CaloLatent}\xspace}

\newcommand{\submMikuni}{\texttt{CaloScore}\xspace}
\newcommand{\submMikuniCite}{\submMikuni~\cite{Mikuni:2022xry,Mikuni:2023tqg}\xspace}

\newcommand{\submMikuniDist}{\texttt{CaloScore distilled}\xspace}
\newcommand{\submMikuniDistCite}{\submMikuniDist~\cite{Mikuni:2022xry,Mikuni:2023tqg}\xspace}

\newcommand{\submMikuniSingle}{\texttt{CaloScore single-shot}\xspace}
\newcommand{\submMikuniSingleCite}{\submMikuniSingle~\cite{Mikuni:2022xry,Mikuni:2023tqg}\xspace}

\newcommand{\submPangT}{\texttt{CaloFlow teacher}\xspace}
\newcommand{\submPangTCite}{\submPangT~\cite{Krause:2022jna}\xspace}

\newcommand{\submPangS}{\texttt{CaloFlow student}\xspace}
\newcommand{\submPangSCite}{\submPangS~\cite{Krause:2022jna}\xspace}

\newcommand{\submPangIT}{\texttt{iCaloFlow teacher}\xspace}
\newcommand{\submPangITCite}{\submPangIT~\cite{Buckley:2023daw}\xspace}

\newcommand{\submPangIS}{\texttt{iCaloFlow student}\xspace}
\newcommand{\submPangISCite}{\submPangIS~\cite{Buckley:2023daw}\xspace}

\newcommand{\submPangSuper}{\texttt{SuperCalo}\xspace}
\newcommand{\submPangSuperCite}{\submPangSuper~\cite{Pang:2023wfx}\xspace}

\newcommand{\submPalacios}{\texttt{CaloDREAM}\xspace}
\newcommand{\submPalaciosCite}{\submPalacios~\cite{Favaro:2024rle}\xspace}

\newcommand{\submReyes}{\texttt{CaloMan}\xspace}
\newcommand{\submReyesCite}{\submReyes~\cite{Cresswell:2022tof}\xspace}

\newcommand{\submRinaldi}{\texttt{BoloGAN}\xspace}
\newcommand{\submRinaldiCite}{\submRinaldi~\cite{ATL-SOFT-PUB-2020-006}\xspace}

\newcommand{\submSalamaniDNN}{\texttt{DNNCaloSim}\xspace}
\newcommand{\submSalamaniDNNCite}{\submSalamaniDNN~\cite{Salamani:2021zet,ATLAS:2022jhk}\xspace}

\newcommand{\submSalamaniTrans}{\texttt{Geant4-Transformer}\xspace}
\newcommand{\submSalamaniTransCite}{\submSalamaniTrans~\cite{G4TransDalila}\xspace}

\newcommand{\submScham}{\texttt{DeepTree}\xspace}
\newcommand{\submSchamCite}{\submScham~\cite{DeepTreeCHEP,DeepTreeNIPS}\xspace}

\newcommand{\submSchnake}{\texttt{CaloPointFlow}\xspace}
\newcommand{\submSchnakeCite}{\submSchnake~\cite{Schnake:2024mip}\xspace}

\newcommand{\submZhang}{\texttt{CaloShowerGAN}\xspace}
\newcommand{\submZhangCite}{\submZhang~\cite{FaucciGiannelli:2023fow}\xspace}

\newcommand{\submZhangTwo}{\texttt{CaloShower2GAN}\xspace}
\newcommand{\submZhangTwoCite}{\submZhangTwo~\cite{FaucciGiannelli:2023fow}\xspace}

\newcommand{\submZhangThree}{\texttt{CaloShower3GAN}\xspace}
\newcommand{\submZhangThreeCite}{\submZhangThree~\cite{FaucciGiannelli:2023fow}\xspace}

\newcommand{\submHeadlineSingle}[5]{
\subsection{#1}
\begin{center} By #2\label{#3}, with figures and tables referring to this approach as #4 and code being available at~\cite{#5}. \\ \rule{0.4\textwidth}{0.4pt}\end{center}
}
\newcommand{\submHeadlineMultiple}[5]{
\subsection{#1}
\begin{center} By #2\label{#3}, with figures and tables referring to these approaches as #4 and code being available at~\cite{#5}. \\ \rule{0.4\textwidth}{0.4pt}\end{center}
}

\newcommand*{\doi}[1]{%
   \href{https://doi.org/#1}{doi:~{\small\nolinkurl{#1}}}}

\definecolor{orange}{RGB}{255,163,0}

\begin{document}
\review[CaloChallenge Results]{CaloChallenge 2022: A Community Challenge for Fast Calorimeter Simulation}
\vspace{1cm}

\begin{center}
\includegraphics[width=0.5\textwidth]{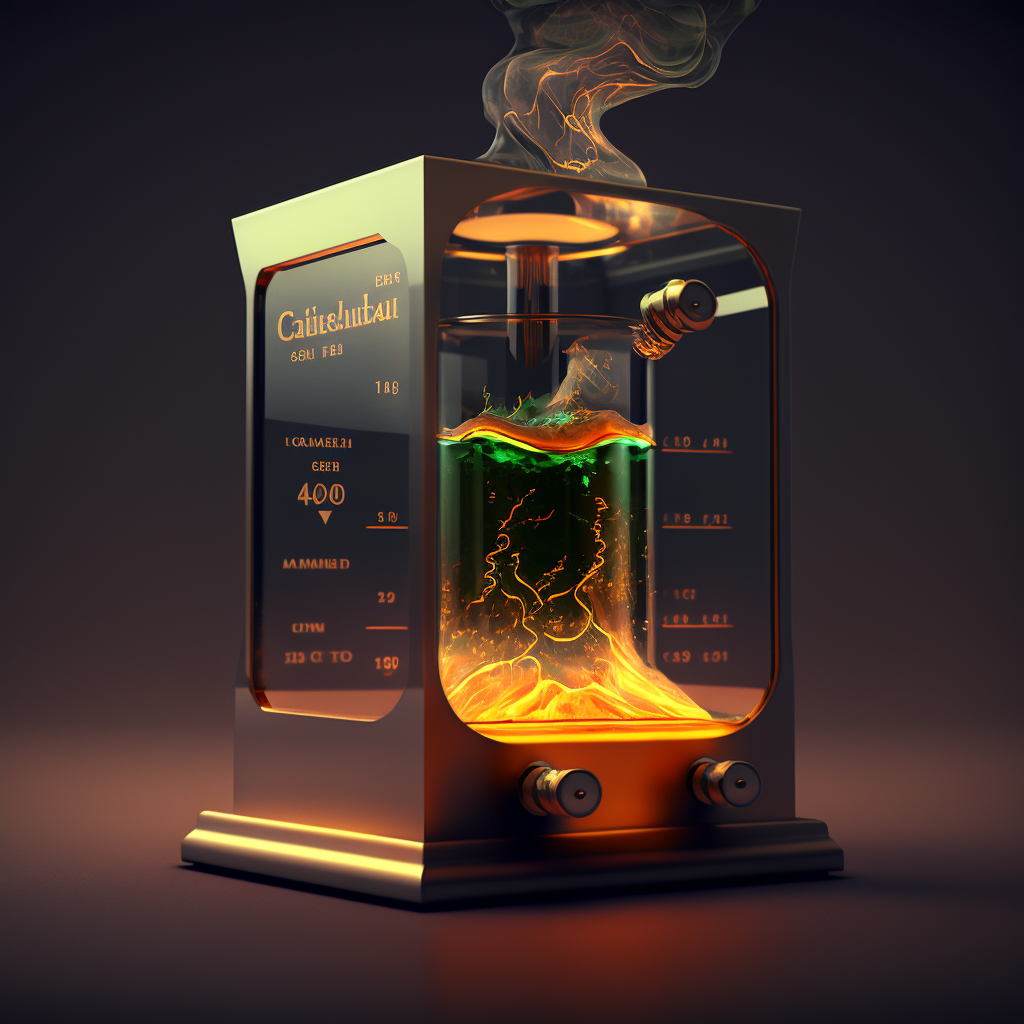}\\
\begin{small}``Calorimeter Simulation'', generated via \href{https://www.midjourney.com/home/}{midjourney, 2022}\end{small}
\end{center}

\author{
Claudius~Krause$^{1,2}$~(main editor)~\orcidlink{0000-0003-0924-3036}, 
Michele~Faucci~Giannelli$^{3,4}$~(editor)~\orcidlink{0000-0003-3731-820X}, 
Gregor~Kasieczka$^{5}$~(editor)~\orcidlink{0000-0003-3457-2755},
Benjamin~Nachman$^{6}$~(editor)~\orcidlink{0000-0003-1024-0932},
Dalila~Salamani$^{7}$~(editor)~\orcidlink{0000-0002-8780-5885}, 
David~Shih$^{8}$~(editor)~\orcidlink{0000-0003-3408-3871}, 
Anna~Zaborowska$^{7}$~(editor)~\orcidlink{0000-0001-6210-1921}, \\ $ $\\
Oz~Amram$^{9}$~\orcidlink{0000-0002-3765-3123},
Kerstin~Borras$^{10,11}$~\orcidlink{0000-0003-1111-249X},
Matthew~R.~Buckley$^{8}$~\orcidlink{0000-0003-1109-3460},
Erik~Buhmann$^{5}$ ~\orcidlink{0000-0002-4805-3721},
Thorsten~Buss$^{5,10}$~\orcidlink{0000-0002-1717-2138},
Renato~Paulo~Da~Costa~Cardoso$^{7}$,
Anthony~L.~Caterini$^{12}$~\orcidlink{0000-0002-0758-9562},
Nadezda~Chernyavskaya$^{7}$~\orcidlink{0000-0002-2264-2229},
Federico~A.G.~Corchia$^{13,14}$~\orcidlink{0000-0002-1788-3204},
Jesse~C.~Cresswell$^{12}$~\orcidlink{0000-0002-9284-8804},
Sascha~Diefenbacher$^{6}$~\orcidlink{0000-0003-4308-6804},
Etienne~Dreyer$^{15}$~\orcidlink{0000-0001-8955-9510},
Vijay~Ekambaram$^{16}$,
Engin~Eren$^{10}$~\orcidlink{0000-0002-6371-5252},
Florian~Ernst$^{2,7}$~\orcidlink{0009-0008-9363-6345},
Luigi~Favaro$^2$~\orcidlink{0000-0003-2421-7100},
Matteo~Franchini$^{13,14}$~\orcidlink{0000-0002-4554-252X},
Frank~Gaede$^{10}$~\orcidlink{0000-0002-7055-9200},
Eilam~Gross$^{15}$~\orcidlink{0000-0003-1244-9350},
Shih-Chieh~Hsu$^{17}$~\orcidlink{0000-0001-6214-8500},
Kristina~Jaruskova$^7$, 
Benno~K\"ach$^{5,10}$~\orcidlink{0000-0002-1194-2306},
Jayant~Kalagnanam$^{18}$,
Raghav~Kansal$^{9,19}$~\orcidlink{0000-0003-2445-1060},
Taewoo~Kim$^{12}$~\orcidlink{0009-0007-2691-4301},
Dmitrii~Kobylianskii$^{15}$~\orcidlink{0009-0002-0070-5900},
Anatolii~Korol$^{10}$~\orcidlink{0000-0002-2569-1771},
William~Korcari$^{5}$~\orcidlink{0000-0001-8017-5502},
Dirk~Kr\"ucker$^{10}$~\orcidlink{0000-0003-1610-8844},
Katja~Kr\"uger$^{10}$~\orcidlink{0000-0002-1956-6608},
Marco~Letizia$^{20,21}$~\orcidlink{0000-0001-9641-4352},
Shu~Li$^{22,23,24}$~\orcidlink{0000-0001-7879-3272},
Qibin~Liu$^{22,23,24}$~\orcidlink{0000-0001-5248-4391},
Xiulong~Liu$^{17}$~\orcidlink{0000-0001-8697-1489},
Gabriel~Loaiza-Ganem$^{12}$~\orcidlink{0009-0005-6767-2148},
Thandikire~Madula$^{25}$~\orcidlink{0000-0001-7689-8628},
Peter~McKeown$^{7,10}$~\orcidlink{0009-0006-9722-2233},
Isabell-A.~Melzer-Pellmann$^{10}$~\orcidlink{0000-0001-7707-919X},
Vinicius~Mikuni$^{6}$~\orcidlink{0000-0002-1579-2421},
Nam~Nguyen$^{18}$,
Ayodele~Ore$^2$~\orcidlink{0000-0001-6925-3565},
Sofia~Palacios~Schweitzer$^2$~\orcidlink{0009-0004-0296-7204},
Ian~Pang$^{8}$~\orcidlink{0000-0002-8225-7269},
Kevin~Pedro$^{9}$~\orcidlink{0000-0003-2260-9151},
Tilman~Plehn$^2$~\orcidlink{0000-0001-5660-7790},
Witold~Pokorski$^{7}$~\orcidlink{0009-0007-9910-414X},
Huilin~Qu$^{7}$~\orcidlink{0000-0002-0250-8655},
Piyush~Raikwar$^{7}$,
John~A.~Raine$^{26}$~\orcidlink{0000-0002-5987-4648},
Humberto~Reyes-Gonzalez$^{21,27,28}$~\orcidlink{0000-0003-3283-5208},
Lorenzo~Rinaldi$^{13,14}$~\orcidlink{0000-0001-9608-9940},
Brendan~Leigh~Ross$^{12}$~\orcidlink{0000-0003-0670-2225},
Moritz~A.W.~Scham$^{10,11,29}$~\orcidlink{0000-0001-9494-2151},
Simon~Schnake$^{10,11}$~\orcidlink{0000-0003-3409-6584},
Chase~Shimmin$^{30}$~\orcidlink{0000-0002-2228-2251},
Eli~Shlizerman$^{17}$~\orcidlink{0000-0002-3136-4531},
Nathalie~Soybelman$^{15}$~\orcidlink{0000-0003-0209-0858},
Mudhakar~Srivatsa$^{18}$,
Kalliopi~Tsolaki$^{7}$,
Sofia~Vallecorsa$^{7}$,
Kyongmin~Yeo$^{18}$,
Rui~Zhang$^{31,32}$~\orcidlink{0000-0002-8265-474X}
}

\address{$^1$ Institute of High Energy Physics (HEPHY), Austrian Academy of Sciences (OeAW), Dominikanerbastei 16, A-1010 Vienna, Austria}
\ead{Claudius.Krause@oeaw.ac.at}
\address{$^2$ Institut f\"ur Theoretische Physik, Universit\"at Heidelberg, Germany}
\address{$^3$ Istituto Nazionale di Fisica Nucleare (INFN), Sezione di Roma Tor Vergata, Roma, 00133, Italy}
\address{$^4$ Department of Microtechnology and Nanoscience, Chalmers University of Technology, 41296 Gothenburg, Sweden}
\address{$^5$ Institut f\"ur Experimentalphysik, Universit\"at Hamburg, Germany}
\address{$^6$ Lawrence Berkeley National Laboratory, Berkeley, CA 94720, USA}
\address{$^7$ CERN, Espl. des Particules 1, 1211 Meyrin, Switzerland}
\address{$^8$ NHETC, Department of Physics and Astronomy, Rutgers University, Piscataway, NJ 08854, USA}
\address{$^{9}$ Fermi National Accelerator Laboratory, Batavia, IL 60510, USA}
\address{$^{10}$ \href{https://ror.org/01js2sh04}{Deutsches Elektronen-Synchrotron DESY}, Hamburg, Germany}
\address{$^{11}$ III. Physikalisches Institut A, \href{https://ror.org/04xfq0f34}{RWTH Aachen University}, Germany}
\address{$^{12}$ Layer 6 AI, Toronto, Canada}
\address{$^{13}$ Department of Physics and Astronomy, Alma Mater Studiorum - University of Bologna, 6/2, Viale Carlo Berti Pichat, I-40127 Bologna, Italy}
\address{$^{14}$ Istituto Nazionale di Fisica Nucleare (INFN), Sezione di Bologna, 6/2, Viale Carlo Berti Pichat, I-40127 Bologna, Italy}
\address{$^{15}$ Weizmann Institute of Science, Rehovot, Israel}
\address{$^{16}$ IBM Research, India}
\address{$^{17}$ University of Washington, Seattle, WA 98195, USA}
\address{$^{18}$ IBM T. J. Watson Research Center, Yorktown Heights, NY USA}
\address{$^{19}$ California Institute of Technology, Pasadena, CA 91125, USA}
\address{$^{20}$ MaLGa--DIBRIS, University of Genova, Genova, Italy}
\address{$^{21}$ Istituto Nazionale di Fisica Nucleare (INFN), Sezione di Genova, Genova, Italy}
\address{$^{22}$ Tsung-Dao Lee Institute (TDLI), Shanghai Jiao Tong University, Shanghai 201210, China}
\address{$^{23}$ Key Laboratory for Particle Astrophysics and Cosmology (MOE), Shanghai Key Laboratory for Particle Physics and Cosmology (SKLPPC), Shanghai Jiao Tong University, Shanghai 200240, China}
\address{$^{24}$ Institute of Nuclear and Particle Physics, School of Physics and Astronomy, Shanghai Jiao Tong University, Shanghai 200240, China}
\address{$^{25}$ University College London (UCL), London,  WC1E 6BT, UK} 
\address{$^{26}$ Département de Physique Nucléaire et Corpusculaire, University of Geneva, 1211 Geneva, Switzerland}
\address{$^{27}$ Department of Physics, University of Genova, Genova, Italy}
\address{$^{28}$ Institut f\"{u}r Theoretische Teilchenphysik und Kosmologie, RWTH Aachen University, 52074 Aachen, Germany}
\address{$^{29}$ Institute for Advanced Simulation, \href{https://ror.org/02nv7yv05}{Forschungszentrum J\"ulich}, J\"ulich, Germany}
\address{$^{30}$ Yale University, New Haven, CT 06520, USA}
\address{$^{31}$ Department of Physics, Nanjing University, Nanjing 210093, China}
\address{$^{32}$ Department of Physics, University of Wisconsin Madison, Wisconsin 53706, USA}

\begin{abstract}
We present the results of the ``Fast Calorimeter Simulation Challenge 2022'' --- the CaloChallenge. We study state-of-the-art generative models on four calorimeter shower datasets of increasing dimensionality, ranging from a few hundred voxels to a few tens of thousand voxels. The 31 individual submissions span a wide range of current popular generative architectures, including Variational AutoEncoders (VAEs), Generative Adversarial Networks (GANs), Normalizing Flows, Diffusion models, and models based on Conditional Flow Matching. We compare all submissions in terms of quality of generated calorimeter showers, as well as shower generation time and model size. To assess the quality we use a broad range of different metrics including differences in 1-dimensional histograms of observables, KPD/FPD scores, AUCs of binary classifiers, and the log-posterior of a multiclass classifier. The results of the CaloChallenge provide the most complete and comprehensive survey of cutting-edge approaches to calorimeter fast simulation to date. In addition, our work provides a uniquely detailed perspective on the important problem of how to evaluate generative models. As such, the results presented here should be applicable for other domains that use generative AI and require fast and faithful generation of samples in a large phase space. 
\end{abstract}
\vspace{2pc}
\noindent{\it Report Numbers}: HEPHY-ML-24-05, FERMILAB-PUB-24-0728-CMS, TTK-24-43

\submitto{\RPP}

\maketitle
\newpage
\tableofcontents

\section{Introduction}
\label{sec:intro}
\markboth{\uppercase{Introduction}}{} 

\begin{figure}[b]
    \centering
    \includegraphics[height=11.5em]{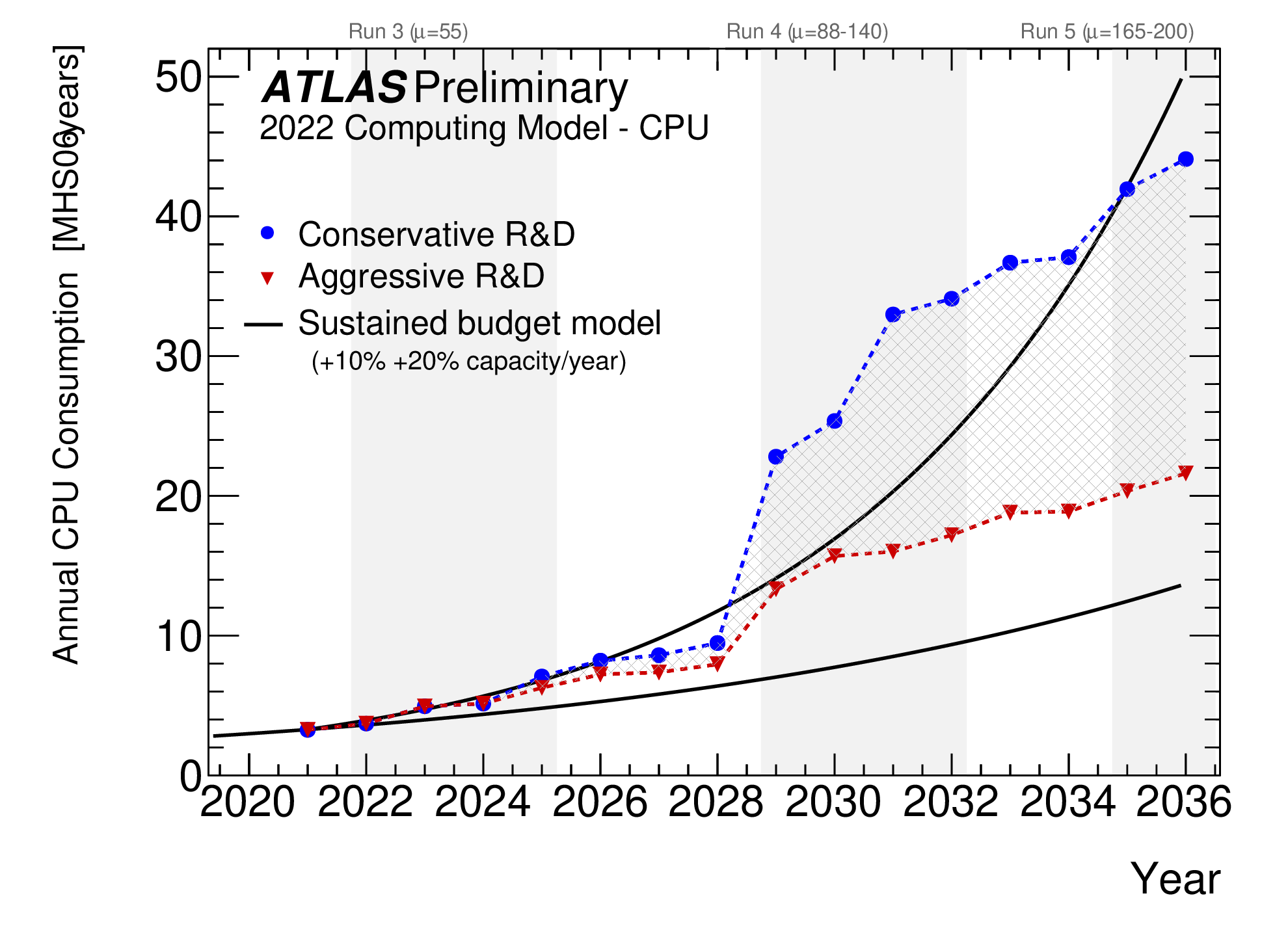}\hfill    \includegraphics[height=11.5em, trim= 20 30 10 30, clip]{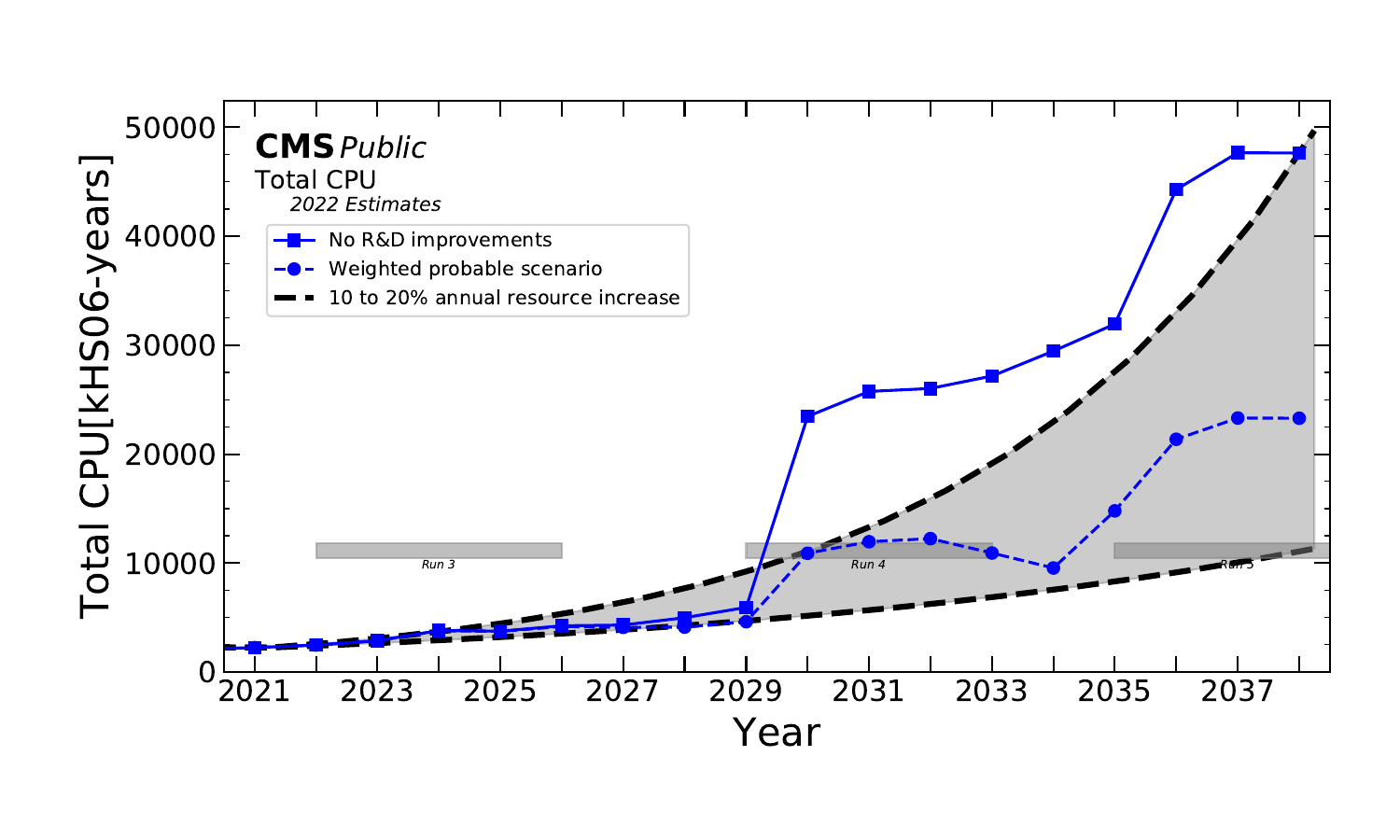}
    \caption{Projected CPU requirements. \textbf{Left:} For ATLAS~\cite{Collaboration:2802918}. \textbf{Right:} For CMS~\cite{Software:2815292}.}
    \label{fig:cpu_projection}
\end{figure}

At the Large Hadron Collider (LHC) and countless other particle or nuclear physics facilities, we aim to study Nature at the most fundamental level, searching for answers to questions such as the nature of dark matter and dark energy, the baryon-anti-baryon asymmetry in the universe, and the mass and hierarchy of neutrinos, which are all not explained in the Standard Model.
Simulations based on first principles provide a crucial bridge between theory and experiment and are at the core of the successful physics program of these facilities.  With the increasing amount of data that the LHC will generate in the upcoming runs, the amount of simulated events required for accurate and sensitive analyses will grow steadily, and with it the computational resources needed to generate them. In~\fref{fig:cpu_projection}, we see the projected CPU needs of the two general purpose experiments, ATLAS~\cite{Collaboration:2802918} and CMS~\cite{Software:2815292}, with similar challenges standing in front of other experiments, e.g. LHCb~\cite{Bozzi:2888939}. The largest fraction of the CPU consumption goes into simulation and within that, into the simulation of the detector responses and especially the calorimeters. These detectors are particularly challenging due to the need to track many secondary particles produced in extensive showers that result from particles stopping inside dense materials. State-of-the-art physics-based simulations use \geant~\cite{Agostinelli:2002hh,1610988,ALLISON2016186} and are a major computational bottleneck, forecast to overwhelm the computing budget of existing and future experiments. 

Without significant research and development of new simulation techniques and algorithms, the data collection will significantly outpace the Monte Carlo production capabilities of the experiments which in turn will limit the precision of many measurements as they will be limited by the statistics of the Monte Carlo simulation. Maintaining the current MC-to-data ratio is therefore a high priority for the LHC experiments. A possible mitigation can be achieved by replacing the expensive calorimeter simulations with faster alternatives.
Such faster calorimeter simulation techniques~\cite{ATL-PHYS-PUB-2010-013,ATL-SOFT-PUB-2014-001,SIMU-2018-04,Abdullin:2011zz,Hildreth:2297284}, which are usually called ``fast simulation'', typically rely on parametrized responses of the calorimeter, tailored to specific types of incoming particles.
By employing these parametrizations, effectively bypassing the intricate shower development process carried out by \geant, the simulation of an event is significantly sped up. However, these models usually lack the high fidelity that is required by the precision measurements carried out by the LHC experiments.

A possible alternative solution is provided by the immense progress in computer science, machine learning, and especially generative AI in the past two decades. Deep generative models (DGMs) learn, implicitly or explicitly, the distribution of (simulated) data from a given sample and then generate new data according to this distribution. Continuous research with impressive progress over nearly a decade~\cite{Paganini:2017hrr} has shown that these models have the potential to become fast and faithful alternatives for detector simulation, as was summarized in a recent review on DGMs for calorimeter simulation~\cite{Hashemi:2023rgo}. 
For that reason, such models also started to be included in the fast simulation packages of the experiments~\cite{SIMU-2018-04,Barbetti:2023bvi} in recent years. 

Motivated by the aim of spurring the further development and benchmarking of fast and high-fidelity calorimeter shower generation using deep learning methods, the Fast Calorimeter Simulation Challenge (``CaloChallenge'') was initiated in early 2022. It is modeled after two previous, highly successful data challenges in HEP --- the top tagging community challenge~\cite{Kasieczka:2019dbj} and the LHC Olympics 2020 anomaly detection challenge~\cite{Kasieczka:2021xcg}.

In the CaloChallenge, participants were tasked with training their favorite generative model on the provided calorimeter shower datasets, learning to sample from the conditional probability distribution $p(\mathcal{I}|E_{\rm inc})$, where $\mathcal{I}$ are the voxel energy deposits and $E_{\rm inc}$ is the incident energy. The particle and nuclear physics communities have been developing fast simulation methods for some time, and the goal of this challenge was to accelerate and expand on these efforts, while offering common benchmarks with which to assess the strengths and weaknesses of the new approaches, and a common evaluation pipeline for fair comparison. 

This is the community paper summarizing the outcome of the CaloChallenge. Over 60 participants 
contributed to the development of 31 different DGMs (some close variants or distillations of each other, 23 of them completely distinct) for fast calorimeter simulation, making use of cutting-edge techniques in generative modeling with deep learning, including GANs, VAEs, normalizing flows, diffusion models, and conditional flow-matching models. \Tref{tab:submissions} gives an overview of the presented models and links the corresponding code repositories. 

Many submissions were presented at ML4Jets 2022 at Rutgers~\cite{ML4Jets2022}, ML4Jets 2023 in Hamburg~\cite{ML4Jets2023}, and the CaloChallenge Workshop in Frascati~\cite{CaloChallengeWorkshop}. They have been published in separate research articles, either in peer-reviewed journals or in machine learning conferences. A small subset of the submissions have been compared previously in~\cite{Ahmad:2024dql}, independent of this study here and without submitted samples, but with retrained models based on code repositories instead. 


\begin{table}[bht]
    \caption{\label{tab:submissions}Models submitted to the CaloChallange. }
	\begin{tabular}{llcccccc}
	\toprule
	\multirow{2}{*}{Approach} & \multirow{2}{*}{Model} & \multirow{2}{*}{Code} &\multicolumn{4}{c}{Dataset} & \multirow{2}{*}{Section} \\
 & & & 1 -- $\gamma$ & 1 -- $\pi$ & \hphantom{\quad}2\hphantom{\quad} & \hphantom{\quad}3\hphantom{\quad} & \\
	\midrule
	\multirow{4}{*}{GAN} 
		  & \submZhangCite & \cite{submZhangGit} & $\checkmark$ & $\checkmark$ & & &\ref{subsec:CaloshowerGAN}\\
		& \submKaechCite & \cite{submKachGit} &  & & $\checkmark$ & $\checkmark$ &\ref{subsec:MDMA} \\
		& \submRinaldiCite & \cite{submRinaldiGit}& $\checkmark$ & $\checkmark$ & & &\ref{subsec:BoloGAN} \\
		& \submSchamCite & \cite{submSchamGit}& & & $\checkmark$ & & \ref{subsec:DeepTree}\\
	\midrule
	\multirow{5}{*}{NF} %
		& \submBuss~\cite{Diefenbacher:2023vsw,Buss:2024orz} & \cite{submBussGit} & & & $\checkmark$ & $\checkmark$ & \ref{subsec:L2LFlows}\\
		& \cf~\cite{Krause:2022jna,Buckley:2023daw} &\cite{submPang1Git,submPangIGit} & $\checkmark$ & $\checkmark$ & $\checkmark$ & $\checkmark$ & \ref{subsec:CaloFlow} \\
		& \submFavaroCite & \cite{submErnstGit} & $\checkmark$ & $\checkmark$ & $\checkmark$ & & \ref{subsec:caloinn}\\
        & \submPangSuperCite & \cite{submPangSuperGit}&  &  & $\checkmark$ & & \ref{subsec:supercalo}\\
        & \submSchnakeCite & \cite{submSchnakeGit}&  &  & $\checkmark$ & $\checkmark$ &\ref{subsec:calopointflow} \\
	\midrule
	\multirow{5}{*}{Diffusion} %
		& \submAmramCite & \cite{submAmramGit}& $\checkmark$ & $\checkmark$ & $\checkmark$ & $\checkmark$ & \ref{subsec:calodiffusion} \\
		& \submKorolCite & \cite{submKorolGit1,submKorolGit2}& & & & $\checkmark$  & \ref{subsec:caloclouds} \\
	    & \submMikuniCite  & \cite{submMikuniGit1,submMikuniGit2}& $\checkmark$ & & $\checkmark$ & $\checkmark$ & \ref{subsec:caloscore} \\
		&\submKobylyanskyCite & \cite{submKobylyanskyGit}& $\checkmark$ & $\checkmark$ & & & \ref{subsec:calograph}\\
		  & \submCardosoCite & \cite{submCardosoGit}& & & $\checkmark$ & & \ref{subsec:calodit}\\
	\midrule
	\multirow{6}{*}{VAE} %
		& \submLiuCite & \cite{submLiuGit}& $\checkmark$ & $\checkmark$ & $\checkmark$ & $\checkmark$ & \ref{subsec:vqvae}\\
		& \submReyesCite & \cite{submReyesGit} & $\checkmark$ & $\checkmark$ & & &\ref{subsec:caloman}\\
        & \submSalamaniDNNCite & \cite{submSalamaniDNNGit}&  & $\checkmark$ & & & \ref{subsec:dnncalosim}\\
		& \submSalamaniTransCite & \cite{submSalamaniTransGit} & & & & $\checkmark$ & \ref{subsec:g4trans} \\
		& \submErnstCite & \cite{submErnstGit}& $\checkmark$ & $\checkmark$ & $\checkmark$ & $\checkmark$ & \ref{subsec:calovaeinn} \\
		& \submMadulaCite & \cite{submMadulaGit}& & & $\checkmark$ & & \ref{subsec:calolatent} \\
	\midrule
	\multirow{2}{*}{CFM} %
		& \submPalaciosCite &\cite{submPalaciosGit} &  &  & $\checkmark$ & $\checkmark$ & \ref{subsec:calodream} \\
		& \submCresswellCite & \cite{submCresswellGit} & $\checkmark$ & $\checkmark$ &  &  & \ref{subsec:caloforest}\\
	\bottomrule
	\end{tabular}
\end{table}

The document is structured as follows:~in \Sref{sec:data}, we describe the calorimeter datasets that we provided. Then, we introduce the individual approaches, grouped by their main generative architecture: Generative Adversarial Networks (GANs) in \Sref{sec:gan}, Normalizing Flows (NFs) in \Sref{sec:flow}, Diffusion Models in \Sref{sec:diffusion}, Variational Autoencoders (VAEs) in \Sref{sec:vae}, and Conditional Flow Matching Models (CFMs) in \Sref{sec:CFM}. \Sref{sec:intro_metric} introduces the metrics which we employ to compare the submissions. We then show our results, where we first focus on the scores of the individual metrics in \Sref{sec:results} and then look at the correlations and Pareto fronts in \Sref{sec:correlations}. On the one hand, this sheds light on interesting trade-offs, while on the other hand, it tells us about the metrics themselves. We summarize and present an outlook in \Sref{sec:conclusions}. In the appendices, we collect additional reference plots as well as tables with the detailed numbers that are presented in the figures of Sections \ref{sec:results} and \ref{sec:correlations}. 
\FloatBarrier

\section{Datasets}
\label{sec:data}
\markboth{\uppercase{Datasets}}{} 
The challenge offers three datasets, ranging in difficulty from easy through medium to hard. The difficulty is set by the dimensionality of the calorimeter showers, \textit{i.e.}~the number layers and the number of voxels in each layer.

\begin{figure}[ht]
    \centering
    \includegraphics[width=0.85\textwidth]{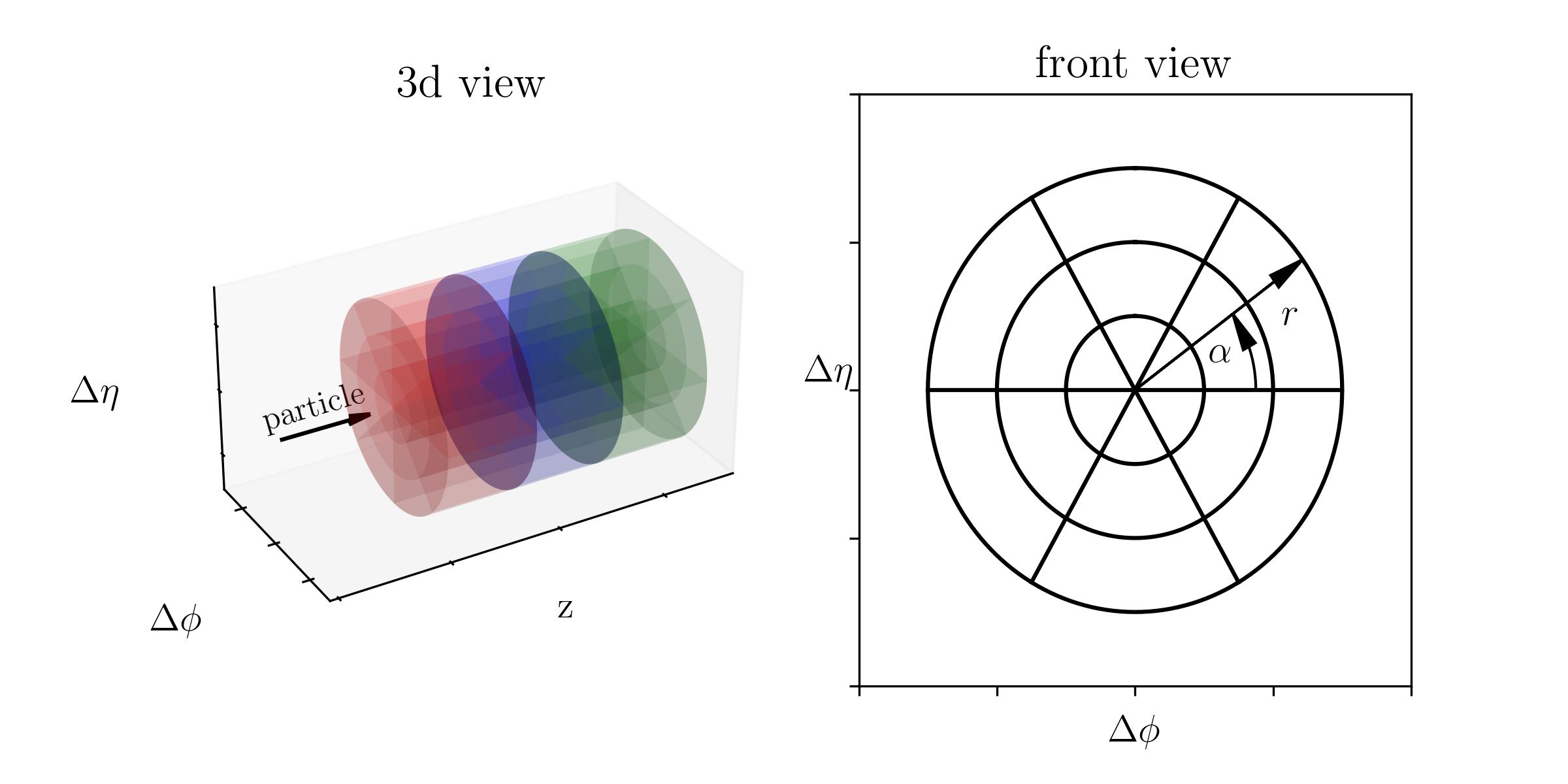}
    \caption{Schematic view of the voxelization in all datasets. Along the direction of the incoming particle ($z$), the volume is segmented in $N_z$ layers. Each layer has $N_r$ radial and $N_\alpha$ angular bins. \textbf{Left:} 3-dimensional view. \textbf{Right:} Front view. Images taken from~\cite{calochallenge}, separately published in~\cite{Krause:2022jna}.}
    \label{fig:coordsys}
\end{figure}

Each dataset has the same general format. The detector geometry consists of concentric cylinders with particles propagating along the $z$-axis. The detector is segmented along the $z$-axis into $N_z$ discrete layers. Each layer has $N_r$ bins along the radial direction and $N_\alpha$ bins in the angle $\alpha$. The number of layers and bins in $r$ and $\alpha$ is summarized in~\tref{tab:voxelization}. The coordinates $\Delta\phi$ and $\Delta\eta$ correspond to the $x$ and $y$ axis of the cylindrical coordinates. \Fref{fig:coordsys} shows a 3-dimensional view of a geometry with 3 layers, with each layer having 3 bins in the radial direction and 6 bins in the angular direction. The right image shows the front view of the geometry, as seen along the $z$ axis.

\begin{table}[th]
\caption{\label{tab:voxelization}Voxelization of layers in each dataset. We show $N_r \times N_\alpha$ and the total number of voxels, $N_i$, per layer. For datasets 1: a ``--'' indicates that this layer is not in the dataset, as the numbering is based on the ATLAS detector definitions~\cite{SIMU-2018-04}. }
\hspace*{-1cm}
\centering
\begin{small}
\begin{tabular}{@{}c|c|c|c|c|c|c|c|c|c|c||c}
\br
Layer & \multirow{2}{*}{0} & \multirow{2}{*}{1} & \multirow{2}{*}{2} & \multirow{2}{*}{3} &  \multirow{2}{*}{\dots} & \multirow{2}{*}{12} & \multirow{2}{*}{13} & \multirow{2}{*}{14} &  \multirow{2}{*}{\dots} & \multirow{2}{*}{44} & \multirow{2}{*}{total} \\
Number &  &  &  &  &   &  &  &  &  &  &  \\
\mr
\multirow{2}{*}{\dsIph} &$8\times 1$ &$16\times 10$ & $19\times 10$ &$5\times 1$ &\multirow{2}{*}{--}&$5\times 1$ &\multirow{2}{*}{--}&\multirow{2}{*}{--}&\multirow{2}{*}{--}&\multirow{2}{*}{--}& \multirow{2}{*}{368}\\
& $=8$ & $=160$ &$=190$ &$=5$ &&$=5$ &&&&&\\
\multirow{2}{*}{\dsIpi} &$8\times 1$ &$10\times 10$ & $10\times 10$ &$5\times 1$ &\multirow{2}{*}{--} &$15\times 10$ &$16\times 10$ &$10\times 1$ &\multirow{2}{*}{--}&\multirow{2}{*}{--}&\multirow{2}{*}{533}\\
& $=8$& $=100$ &$=100$ &$=5$ &&$=150$ &$=160$ &$=10$ &&&\\
\mr
\dsII & \multicolumn{10}{c||}{$9\times 16 =144$} &$6480$\\
\dsIII & \multicolumn{10}{c||}{$18\times 50 =900$} &$40\,500$\\
\br
\end{tabular}
\end{small}
\end{table}

\begin{table}[th]
\caption{\label{tab:ds1.sizes}Number of samples available per incident energy for each of the training and evaluation datasets for dataset 1 -- $\gamma$  and dataset 1 -- $\pi^+$.}
\hspace*{-1cm}
\centering
\begin{small}
\begin{tabular}{c|c|ccccc||c}
      \br
      $E_\mathrm{inc}$ & 256~MeV -- 131~GeV & 262~GeV   & 524~GeV  & 1.04~TeV  & 2.1~TeV   & 4.2~TeV  & total\\
      \mr
      \dsIph & 10\,000 per energy & 10\,000  & 5000  & 3000  & 2000  & 1000 & 121\,000\\
      \dsIpi & 10\,000 per energy & 9800   & 5000  & 3000  & 2000  & 1000 & 120\,800\\
      \br
\end{tabular}
\end{small}
\end{table}

Each CaloChallenge dataset comes as one or more \texttt{.hdf5} files that were written with python's \texttt{h5py} package~\cite{hdf5} using \texttt{gzip} compression. Within each file, there are two hdf5-datasets: \texttt{incident$\_$energies} has the shape \texttt{(num$\_$events, 1)} and contains the energy of the incoming particle in MeV;  \texttt{showers} has the shape \texttt{(num$\_$events, num$\_$voxels)} and stores the energy deposited in showers, where the energy depositions of each voxel (in MeV) are flattened. The mapping of array index to voxel location is done in the order (radial bins, angular bins, layer), so the first entries correspond to the radial bins of the first angular slice in the first layer. Then, the radial bins of the next angular slice of the first layer follow, and so on.

For every dataset in the CaloChallenge, there is one dataset file to be used for training the generative models and a second one for the evaluation (both by the individual collaborations and by us). For dataset 3, we split the training and evaluation data each into two separate files to have more manageable files sizes.

\subsection{Dataset 1 Photons and Pions} \label{subsec:ds1}

Dataset 1 can be downloaded from~\cite{michele_faucci_giannelli_2022_6368338,michele_faucci_giannelli_2023_8099322}. It is based on the ATLAS open datasets~\cite{ds1_atlas} and contains the simulation of single photons and single charged pions generated at the surface of the ATLAS calorimeter system and pointing back to the center of the detector. The interaction of the particles in the calorimeters was simulated with the official ATLAS software, which is based on~\geant, using a special configuration in which detailed hits were produced and noise from electronics and cross-talk was not included; this allows modeling perfect showers that can be injected in the simulation chain before these effects occur, making it more realistic. These samples were used to train the GANs presented in the AtlFast3 paper~\cite{SIMU-2018-04} and the FastCaloGAN note~\cite{ATL-SOFT-PUB-2020-006}.

Initially, only one dataset for the pion sample was available. Later, a second, independent dataset was provided by ATLAS, so we updated the Zenodo and all trainings and evaluations were done with two independent training and evaluation datasets~\cite{michele_faucci_giannelli_2023_8099322}. There are four datasets, two for photons and two for charged pions. Each dataset contains the voxelized shower information obtained from single particles in the range $0.2 < |\eta| < 0.25$; therefore the particles impact the detector with an angle. For each particle, there are 15 incident energies from 256 MeV up to 4 TeV produced in powers of two. 10\,000 events are available in each sample except for those at higher energies having lower statistics, see \tref{tab:ds1.sizes}.  The number of radial and angular bins varies from layer to layer and is also different for photons and pions, resulting in 368 voxels for photons (called ``\dsIph'' throughout) and 533 for pions (called ``\dsIpi'' throughout), see \tref{tab:voxelization}.

In the results section, a 1~MeV threshold is applied to all voxels to eliminate the low energy tail that affects some of the models but has no physics impact. This assessment is based on how the energy deposited in the calorimeter is transformed and calibrated into reconstructed objects (i.e. photons or jets) using clusters built from the calorimeters' cells~\cite{ATLAS:2016krp}. ATLAS calorimeters are segmented in cells to increase the granularity and improve the spatial reconstruction of showers, and this segmentation is reproduced in the simulation. The cells have rectangular shapes that are easier to construct, hence they do not match the cylindrical voxel geometry described above. This required an additional step in the AtlFast3 simulation in which the energy from the voxels was reassigned to the actual calorimeter cells.
In the offline reconstruction, ATLAS uses topological clusters that are started (seeded) from cells having at least 4 times the noise; they are subsequently grown to include neighboring cells with energy twice the noise level, and then they are finalized with any cell adjacent to the cluster that is above the noise threshold. The lowest cell noise in the layers considered in the datasets is about 10~MeV for layer 1 with other layers having up to 50~MeV. Therefore, a chosen 1~MeV threshold in the voxels' energy is reasonable even when taking into account the fact that multiple voxels could map to the same cell; this only occurs in the core of the shower where most of the energy is deposited and therefore the threshold cut will not take place, \textit{i.e.}~all masked voxels are peripheral voxels that actually map to multiple cells, further diluting the energy associated to each cell. 

\subsection{Datasets 2 and 3} 
\label{subsec:ds23}

\begin{figure}[ht]
\centering
\includegraphics[width=0.8\textwidth]{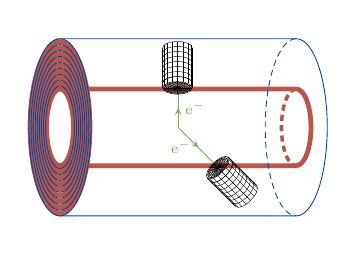}
  \caption{The Par04 detector~\cite{Par04} consists of concentric cylinders of absorber (red) and active material (blue). The energy deposited by incident particles is recorded in a cylindrical readout (black).}
  \label{fig:par04detector}
\end{figure}

Datasets 2 and 3 have been simulated with the Par04~\cite{Par04} example of \geant. The geometry used in the Par04 example is an idealised calorimeter, with concentric cylinders of alternating absorber and active materials. A draft of its layout is presented in~\fref{fig:par04detector}. Both datasets were simulated for the same detector which consists of 90 physical layers, with each layer composed of 1.4\,mm of tungsten (W) as an absorber and 0.3\,mm of silicon (Si) as active material. The inner radius of this calorimeter is 800\,mm and its depth is 153\,mm.

Particle showers are generated by electrons that enter the detector perpendicularly to the detector's cylinders' axis, as depicted in~\fref{fig:par04detector}'s upper electron. Datasets with different, varying incident angles, like the second electron in~\fref{fig:par04detector} pointing to the lower right, were also published but go beyond the scope of this challenge~\cite{dalila_salamani_2022_6082201}.

The Par04 example of \geant writes out only energy deposited in the active material, so it must be corrected for the deposits in the absorber. A simple scaling factor has been derived from the simulation, $f = 1/0.033$, uniform for all energies and cells of the detector. It means that on average $3.3\%$ of particle's energy is registered in the detector. The Par04 simulation has an energy threshold below which cell energy is not stored (to reduce the file size). It is chosen to a very low value of 0.5 keV, which translates to 15.15 keV after the energy scaling. We also apply this cutoff to all submissions before the final evaluation. 

The particle entrance position and direction determine the position (0,0,0) and orientation ($z$-axis) of the cylindrical readout, like the one shown in~\fref{fig:coordsys}. The size of each readout voxel is $\Delta r \times \Delta \varphi \times \Delta z$ and unlike for dataset 1, both datasets 2 and 3 have the same number of voxels in each of the $N_z$ layers. Also the number of voxels along $z$-axis is the same in datasets 2 and 3, but they differ in terms of segmentation in radius ($r$) and angle ($\alpha$). The size along $z$-axis is equal to $\Delta z=3.4$\,mm which corresponds to two physical layers (W-Si-W-Si). Taking into account only the absorber's value of radiation length ($X_0(\mathrm{W})=3.504$\,mm~\cite{pdg_W}) it
makes the size along $z$-axis approximately $\Delta z = 2\cdot1.4\mathrm{\,mm}/3.504\mathrm{\,mm} = 0.8 X_0$. In radius, the size of the voxels is $4.65$\,mm for dataset 2 and $2.325$\,mm for dataset 3, which in approximation, taking only the Moli\'ere radius of the absorber, is  $\Delta r = 4.65\mathrm{\,mm}/9.327\mathrm{\,mm} = 0.5 R_M$ for dataset 2 and $\Delta r = 2.325\mathrm{\,mm}/9.327\mathrm{\,mm} = 0.25 R_M$ for dataset 3.
The angular segmentation consists of 16 voxels for dataset 2 ($\Delta\varphi=2\pi/16\approx0.393\mathrm{\,rad}$) and 50 voxels for dataset 3 ($\Delta\varphi=2\pi/50\approx0.126\mathrm{\,rad}$).

The total number of voxels for dataset 2 is $N_z\times N_r \times N_\alpha=45\times 9\times 16 = 6480$ and for dataset 3 it is $N_z\times N_r \times N_\alpha=45\times 18\times 50 = 40500$, see \tref{tab:voxelization}.

 Files can be downloaded from \cite{faucci_giannelli_michele_2022_6366271} and \cite{faucci_giannelli_michele_2022_6366324} for dataset 2 and 3 respectively.
Dataset~2 consists of two files (one for training and one for evaluation) with 100\,000 showers of electrons each with energies sampled from a log-uniform distribution ranging from 1~GeV to 1~TeV.
Dataset~3 contains showers of electrons sampled from the same incident energy distribution. Due to the size, there are 4 files with 50\,000 showers each. One half of the available sample should be used in training, with the remaining half used as a reference file in evaluation. 
\FloatBarrier

\section{GAN-based Submissions}\label{sec:gan}
\markboth{\uppercase{GAN-based Submissions}}{}
Generative Adversarial Networks (GANs)~\cite{2014arXiv1406.2661G} are one of the earliest types of deep generative models and reached fame by being able to produce photorealistic images~\cite{karras2019stylebased}. A GAN consists of two networks, a generator and a critic\footnote{Also called discriminator, if the cross entropy loss is used.}. They are trained adversarially in a game where the generator produces fake samples that the discriminator tries to distinguish from real samples. 
On the upside, GANs are very flexible, as their only hard requirement is finding two networks that map to the correct space. Furthermore, GANs are typically very fast compared to other generative models and can produce samples with high fidelity. 
On the downside, their training is unstable, and they are difficult to optimize.
For this reason, several improvements were proposed, \textit{e.g.}~the Wasserstein GAN~\cite{2017arXiv170104862A,WGAN}. CaloGAN~\cite{Paganini:2017hrr,Paganini:2017dwg} was the first tool that demonstrated the feasibility of using a deep generative model to perform a fast calorimeter simulation. GANs are also the first model to be used in production, as FastCaloGAN~\cite{FastCaloGAN_code, ATL-SOFT-PUB-2020-006} was deployed as part of AtlFast3~\cite{SIMU-2018-04} and used by the ATLAS experiment to produce several billion events.

\FloatBarrier
\submHeadlineMultiple{CaloShowerGAN}{Michele Faucci Giannelli and Rui Zhang}{subsec:CaloshowerGAN}{\submZhangCite, \submZhangTwoCite, \submZhangThreeCite}{submZhangGit}
\paragraph{Introduction}
Building on the success of FastCaloGAN, \submZhangCite is designed to have a similar interface so that the ATLAS collaboration could easily integrate it.
However, \submZhang significantly diverges from FastCaloGAN in the internal structure of the tool and achieves a significant improvement in reproducing both photons and pions. This is realized through a new pre-processing of the training data by further optimizing the model architecture and hyperparameters.
\paragraph{Architecture}
For example, \submZhangThree employs three GANs for the parametrization of the photons in different energy ranges; this is motivated by how the energy is deposited in the different layers of the calorimeter as a function of the primary particle energy. The energy thresholds to define low, medium and high energy ranges are 4~GeV and 262~GeV, whereas \submZhangTwo merges the medium and high energy ranges.
Only one GAN is used for the pions in all three versions as the nature of the hadronic interaction allows even low-energy pions to interact in the deeper layers of the calorimeter. 

\begin{figure}[ht]
    \centering
    \includegraphics[width=0.6\textwidth]{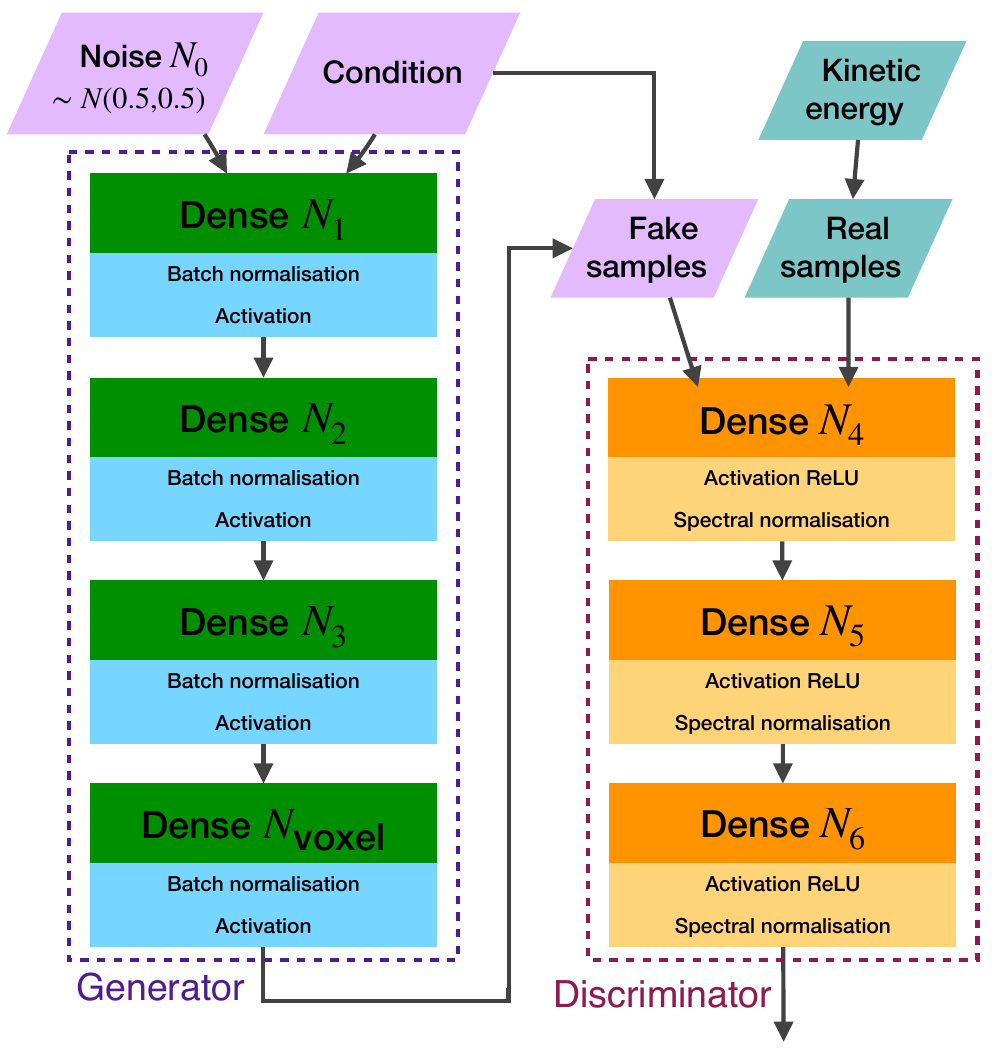}
    \caption{Architecture of \submZhang, taken from~\cite{FaucciGiannelli:2023fow}.} 
    \label{fig:caloshowergan}
\end{figure}

The GAN architecture (see \fref{fig:caloshowergan}) was significantly optimized for this challenge, details on the optimization process are described in \submZhangCite. 
The optimal hyperparameters used in the photon and pion GANs are shown in \tref{tab:hp}.

\paragraph{Preprocessing}
Several normalization steps are used to simplify the training of the GANs.
The first normalization is based on the kinetic energy of the particles as done in FastCaloGAN and other tools.
This normalization procedure allows standardizing all values within the input vector to a similar order of magnitude for all input momenta, eliminating the significant difference between the momenta of the samples.
In this way, the GAN can focus on reproducing the shape of the showers rather than its absolute value.

\submZhang employs additional normalization for layer-specific energy and total energy.
This information improves the training because the networks do not have to extract it from the data as it is explicitly provided; thus the GANs can focus on learning correlations and shapes improving the overall performance.
Details on the implementation of this normalization can be found in~\cite{FaucciGiannelli:2023fow}.

The condition label is also transformed to a normalized range of $[0, 1]$ using the following equation:
\begin{equation}
    \hat{E} = \frac{\log \frac{E_\mathrm{kin}}{E_\mathrm{min}}}
    {\log \frac{E_\mathrm{max}}{E_\mathrm{min}}}.
\end{equation}
Here $E_\mathrm{min}$ ($E_\mathrm{max}$) is the minimum (maximum) kinetic energy of the incoming particle in the training data.

\begin{table}[htb]
    \centering
    \caption{
        Optimal hyperparameter values for the photon and pion in \submZhang, taken from~\cite{FaucciGiannelli:2023fow}.
    }
    \label{tab:hp}
    \begin{tabular}{l cc}
        \toprule
        Hyperparameter                                  & Photon                & Pion \\
        \midrule
        Latent space size                         & $100$ & $200$ \\
        Generator size ($N_1, N_2, N_3$)  & $100$, $200$, $400$  & $200$, $400$, $800$ \\
        Discriminator size ($N_4$, $N_5$, $N_6$)  & $368$, $368$, $368$  & $800$, $400$, $200$ \\
        Generator optimizer                       & Adam~\cite{kingma2014adam}  & Adam~\cite{kingma2014adam} \\
        \hspace{1em} Learning rate                & \num{1e-4}      & \num{1e-4} \\
        \hspace{1em} $\beta_1$                    & \num{0.5}       & \num{0.5} \\
        \hspace{1em} $\beta_2$                    & \num{0.999}     & \num{0.999} \\
        Discriminator optimizer                   & Adam~\cite{kingma2014adam}  & Adam~\cite{kingma2014adam} \\
        \hspace{1em} Learning rate                & \num{1e-4}      & \num{1e-4} \\
        \hspace{1em} $\beta_1$                    & \num{0.5}       & \num{0.5} \\
        \hspace{1em} $\beta_2$                    & \num{0.999}     & \num{0.999} \\
        Batch size   & $1024$ & $1024$ \\
        D/G ratio   & $8$ & $5$ \\
        $\lambda$   & $3$ & $20$ \\
        Activation (generator) & Swish & ReLU \\
        Activation (discriminator) & ReLU & ReLU \\
        Neuron weight initialization (generator)        & Glorot Normal  & He Uniform \\
        Neuron weight initialization (discriminator)    & He Uniform     & He Uniform \\
        Trainable parameters (generator, discriminator) & $261$k, $408$k & $871$k, $829$k \\
        \bottomrule
    \end{tabular}
\end{table}

\paragraph{Training}
The batch size used for training the GANs is $1024$ and the training runs for a total of $10^{6}$ iterations.
Due to the adversarial nature of GAN training, the final iteration does not necessarily yield the best outcome, therefore the GANs are evaluated at intervals of $10^3$ iterations. This is a compromise between the time required for evaluation and the speed of learning of the GANs.

The evaluation is inspired by the methodology used in FastCaloGAN using the total energy distribution for all energies as a figure of merit.
The $\chi^2$ value for each GAN model is computed between the binned distributions of the \geant sample and generated sample by the model and then normalized by the number of degrees of freedom used in each distribution ($\chi^2/\mathrm{NDF}$). 
The model that gives the lowest $\chi^2/\mathrm{NDF}$ among the saved iterations is considered the best and is used in the challenge.

\begin{figure}[!th]
    \centering
    \includegraphics{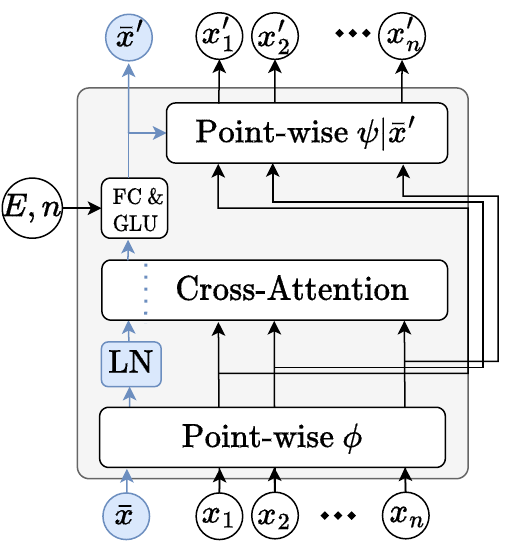}
    \caption{Main building block of the \submKaech architecture. The calorimeter is represented as a point cloud, where every point $\mathbf{x_i}$ is a hit in the detector and $\mathbf{\bar x}$ is the artificial mean-field. First, the points are mapped to a higher dimensional latent space, where after normalization cross-attention is calculated between the mean-field and the other points. Then the conditional information for the shower (\textit{i.e.}~the incident energy $E_{\rm inc}$ and the number of hits $n$) are introduced with a fully-connected (FC) layer and a gated linear unit (GLU). Finally, the updated mean-field is concatenated to the other points and a point-wise layer is used to update the points independently on each other. This architecture yields permutation-equivariance and scales linearly with the number of hits in the computational complexity. Figure adapted from~\cite{Kach:2023rqw,Kach:2024yxi}.} 
    \label{fig:mdma}
\end{figure}

\FloatBarrier
\submHeadlineSingle{Matching Deep Mean-field Attentive (MDMA) GAN}{Benno Käch, Dirk Krücker, Isabell Melzer-Pellmann, Moritz Scham, and Simon Schnake}{subsec:MDMA}{\submKaechCite}{submKachGit}
\paragraph{Introduction}
The \submKaechCite was first applied to the JetNet-150 \cite{kansal_raghav_2022_6975118} datasets, yielding state-of-the-art results not relying on kinematic inputs to the generative model. The model is designed to work on a point cloud (PC) representation of its input. As such, the calorimeter data is first preprocessed to convert the hits to a point cloud, where the coordinates of each point are given by $(E,z,\alpha,R)$ of every hit in the detector. This representation is especially efficient as the granularity of the detector grows and there are a large number of empty cells, thus only dataset 2 and 3 were considered for this model. 
\paragraph{Architecture}
The generator and critic consist of the same main building blocks, which use a cross-attention-based information aggregation mechanism. As there is a large number of hits on average (\textit{e.g.}$~\sim 1600$) the quadratic computational scaling of self-attention is not feasible. Therefore, a synthetic ``mean-field'' $\mathbf{\bar{x}}$ is introduced, initially set as the mean of all points in a cloud, acting as an intermediary for information exchange between points. First, the mean-field is updated via cross-attention (\textit{i.e.}~the Query $Q$ in the attention aggregation is the embedded mean-field $\mathbf{\bar{x}}$, whereas the Key $K$ and Value $V$ are an embedding of the hits in the detector). Then, the mean-field is further processed with a fully-connected layer, additionally using the number of hits as an input and a gated linear unit is applied to an embedding of the incoming energy and the mean-field. Finally, the mean-field is concatenated to every hit and a point-wise layer is applied. This aggregation is permutation-equivariant since cross-attention itself is permutation-equivariant and all the other aggregations are independent of the other points in the cloud. The difference between the generator and the critic is only in the final layer. For the critic a 2-layer deep neural network (DNN) is applied to obtain a score for every shower, whereas for the generator the output is mapped down to 4 dimensions, corresponding to the energy and index of the cell. A schematic for the minimal building block is shown in~\fref{fig:mdma}. The input for the generator is noise sampled from a normal distribution with dimensions four times the number of hits per shower. 
\paragraph{Training}
During training, showers of similar length are grouped together to form batches and padded to the same length. Padded points have no influence on the output. The model is trained as a Wasserstein GAN~\cite{WGAN} with gradient penalty~\cite{wgangp} to regularize. Additionally, weight normalized linear layers~\cite{weightnorm} are employed in the critic. To enhance the convergence of the generator, an L2-loss between the mean of the mean-field in the final layer of the critic for real and generated jets is calculated and minimized (hence the name mean-field matching). To enforce the conditioning with the incoming energy, the generator also minimizes an L2-loss between the detector responses for real and generated showers. During training, the condition of real showers are given to the generator allowing the matching for the L2-loss. The showers are post-processed by rotating  the shower by a random angle. This alleviates the suboptimal coordinate choice, which does not respect the periodicity in the angular coordinate. Note that since point clouds are generated, not only the incoming energy of the incoming particle is supplied as a condition, but also the number of hits in the shower needs to be supplied. For this study they were taken from the validation set --- in practice one would need another model to sample the probability mass function $p(n|E)$.

\begin{figure}[ht]
    \centering
    \includegraphics[width=\textwidth]{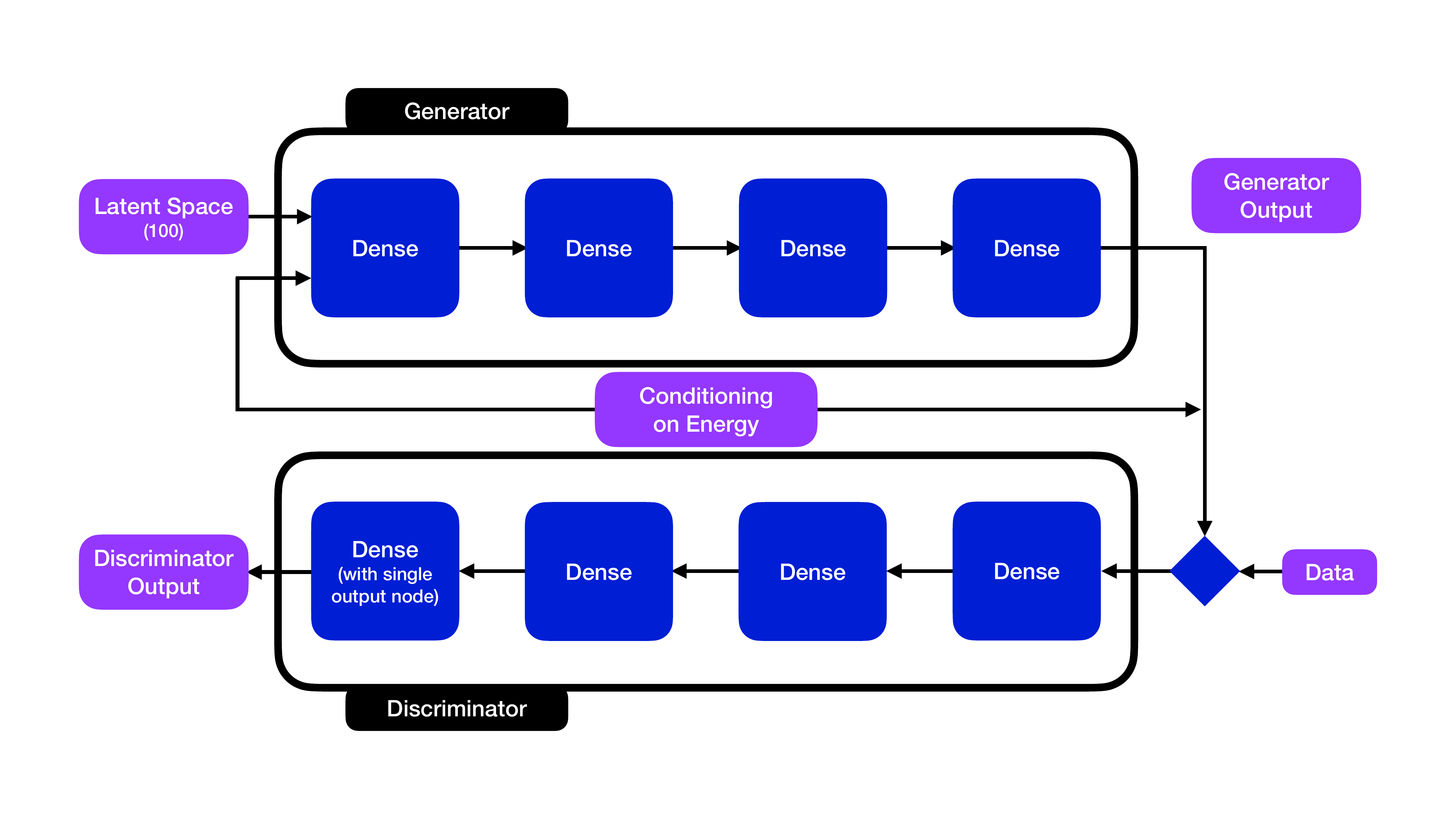}
    \caption{Architecture of \submRinaldi, adapted from~\cite{ATL-SOFT-PUB-2020-006}.} 
    \label{fig:bologan}
\end{figure}

\FloatBarrier
\submHeadlineSingle{BoloGAN}{Federico Andrea Guillaume Corchia, Matteo Franchini, and Lorenzo Rinaldi}{subsec:BoloGAN}{\submRinaldiCite}{submRinaldiGit}
\paragraph{Introduction}
\submRinaldiCite is a GAN-based calorimeter simulation tool derived and evolved from FastCaloGAN, a fast simulation tool developed in the ATLAS Collaboration at CERN~\cite{ATL-SOFT-PUB-2020-006, SIMU-2018-04}.
\paragraph{Architecture}
The tool uses the Wasserstein GAN~\cite{WGAN} with a gradient penalty (WGAN-GP) term~\cite{wgangp} in the loss function of the discriminator, providing good performance and training stability, and conditioning onto the kinetic energy of the particle (conditional WGAN-GP). The conditional WGAN-GP is implemented in TensorFlow 2.0 \cite{tensorflow2015-whitepaper}. The generator and the discriminator both employ three hidden layers, the generator being preceded by a latent space of 100 values and having the output layer with as size the number of voxels for the specific particle type and pseudorapidity interval; the last layer of the discriminator has one single output node. The general scheme is shown in~\fref{fig:bologan}. 
\submRinaldi WGAN-GP hyperparameters are set as shown in \tref{tab:BoloGAN_Hyperparams}, depending on particle type and on energy. These hyperparameters and the general architecture correspond to a trade-off between modeling performance and time required to train the GANs: the program is, in fact, intended to have the possibility to train multiple GANs at the same time, useful for modeling different particle types and pseudorapidity layers with accuracy. The GAN is trained first on a single energy point, then the other energy points are progressively added to training starting from the ones closest in energy to the initial sample. Conditioning is applied, as mentioned, onto the kinetic energy of the particle and the energy in each voxel is normalized by the true energy of the sample, so that all energy samples are scaled to the same values and training can focus on the shape of the total energy (which is the figure of merit, as shall be shown in the continuation). Truth energies used as labels for conditioning are also normalized to the highest energy, in this way all values are in the (0,1] range.

\begin{table}[ht]
    \centering
    \resizebox{\textwidth}{!}{
    \begin{tabular}{c|c|c|c}
        Parameter & Pions & Low En. Photons & High En. Photons \\
        \midrule
        Latent Space & 100 & 100 & 100 \\
        Generator Nodes Output Shape & 200, 400, 800, 533\textsuperscript{\textdagger} & 100, 200, 400, 368\textsuperscript{\textdagger} & 50, 100, 200, 368\textsuperscript{\textdagger} \\
        Discriminator Nodes Output Shape & 800, 400, 200, 1 & 368\textsuperscript{\textdagger}, 368\textsuperscript{\textdagger}, 368\textsuperscript{\textdagger}, 1 & 368\textsuperscript{\textdagger}, 368\textsuperscript{\textdagger}, 368\textsuperscript{\textdagger}, 1 \\
        Activation Function & ReLU & ReLU & Swish \\
        Optimizer & Adam~\cite{kingma2014adam} & Adam~\cite{kingma2014adam} & Adam~\cite{kingma2014adam} \\
        Learning Rate & $10^{-4}$ & $10^{-4}$ & $10^{-4}$ \\
        Discriminator/Generator Training Ratio & 5 & 8 & 8 \\
        Beta & 0.5 & 0.5 & 0.5 \\
        Lambda & 10 & 3 & 3 \\
        Batch Size & 512 & 1024 & 1024 \\
        Used Batch Normalization Layer & No & Yes & Yes \\
        
    \end{tabular}}
    \caption{\submRinaldi WGAN-GP hyperparameters. Low (high) energy photons are those up to (above) 4.096 GeV. Values marked with \textsuperscript{\textdagger} are equal to the number of voxels in the corresponding case.}
    \label{tab:BoloGAN_Hyperparams}
\end{table}
\paragraph{Training}
We performed training for 1 million epochs with a TensorFlow checkpoint saved every 1000 epochs. This granularity allows for monitoring improvement in training without having to save too many checkpoints, which would hamper speed and disk space. Because of the interplay between the generator and the discriminator, the final epoch is not necessarily the best one, also considering that there may be an unfavorable fluctuation in training. For this reason, a $\chi^2$ between the reference sample and the one simulated by the GAN, evaluated over the sum of the energy in all voxels (which corresponds to the total energy deposited into the calorimeter by the particle), is used to choose the best GAN iteration. The iteration with the lowest $\chi^2$ is considered the final choice to perform simulation activities. The total energy for each possible incident energy value was chosen as it is easy to define while it is difficult to reproduce. For every checkpoint, 10k events are generated per incident energy value and the $\chi^2$ between the reference sample and the GAN-simulated one is calculated; the total $\chi^2$ for a checkpoint is the sum of the $\chi^2$ for the individual incident energy values and the checkpoint with the lowest total $\chi^2$ is finally chosen as the best GAN iteration.

The program is currently able to simulate calorimeter showers for photons, electrons, and pions between 256 MeV and 4 TeV over the full detector acceptance. For the CaloChallenge, the tool was applied to Dataset 1 for both photons and pions. For pions one single GAN for all energy values has been trained, while for photons two GANs have been trained, one for low energies (\textit{i.e.}~up to 4.096 GeV) and the other one for high energies (above 4.096 GeV). 

\FloatBarrier
\submHeadlineSingle{DeepTree}{Moritz A.W.~Scham, Benno Käch, Simon Schnake, Dirk Krücker, and Kerstin Borras}{subsec:DeepTree}{\submSchamCite}{submSchamGit}
\paragraph{Introduction}
\submSchamCite is a point cloud (PC)-based GAN model, that uses a tree-like structure for upscaling PCs in the generator and for downscaling them in the critic. 
A calorimeter shower can be converted to a PC by taking the coordinates and the energy of the hits as points in an unordered set.
Representing calorimeter showers as PCs instead of voxels separates the hits from the detector geometry.
This offers multiple advantages: PCs are well-suited for handling sparsity in calorimeter data and they are very efficient if only a fraction of cells contain hits. Their adaptability to irregular calorimeter geometries makes PCs a versatile choice for various detector configurations. Lastly, the generator architecture developed for one calorimeter using PCs can be easily transferred to different calorimeter types. 
On the downside, this independence of the detector geometry also means that the model needs to learn the geometry of the detector from the dataset. 
In a postprocessing step, the generated points must be assigned to the individual cells of the calorimeter. Since the PC-based model does not know the detector geometry, several points may be generated and assigned to the same calorimeter cell. To obtain a unique output for each cell, these points must be combined in some way.
Designing a PC-based model that yields a varying number of points (cardinality) by itself is challenging. Here, the cardinality is sampled from the dataset and provided to the model.
Because dataset 1 (3) yields a too low (high) cardinality, this model targets dataset 2.

\begin{figure}[btp]
    \centering
    \includegraphics[width=0.85\textwidth]{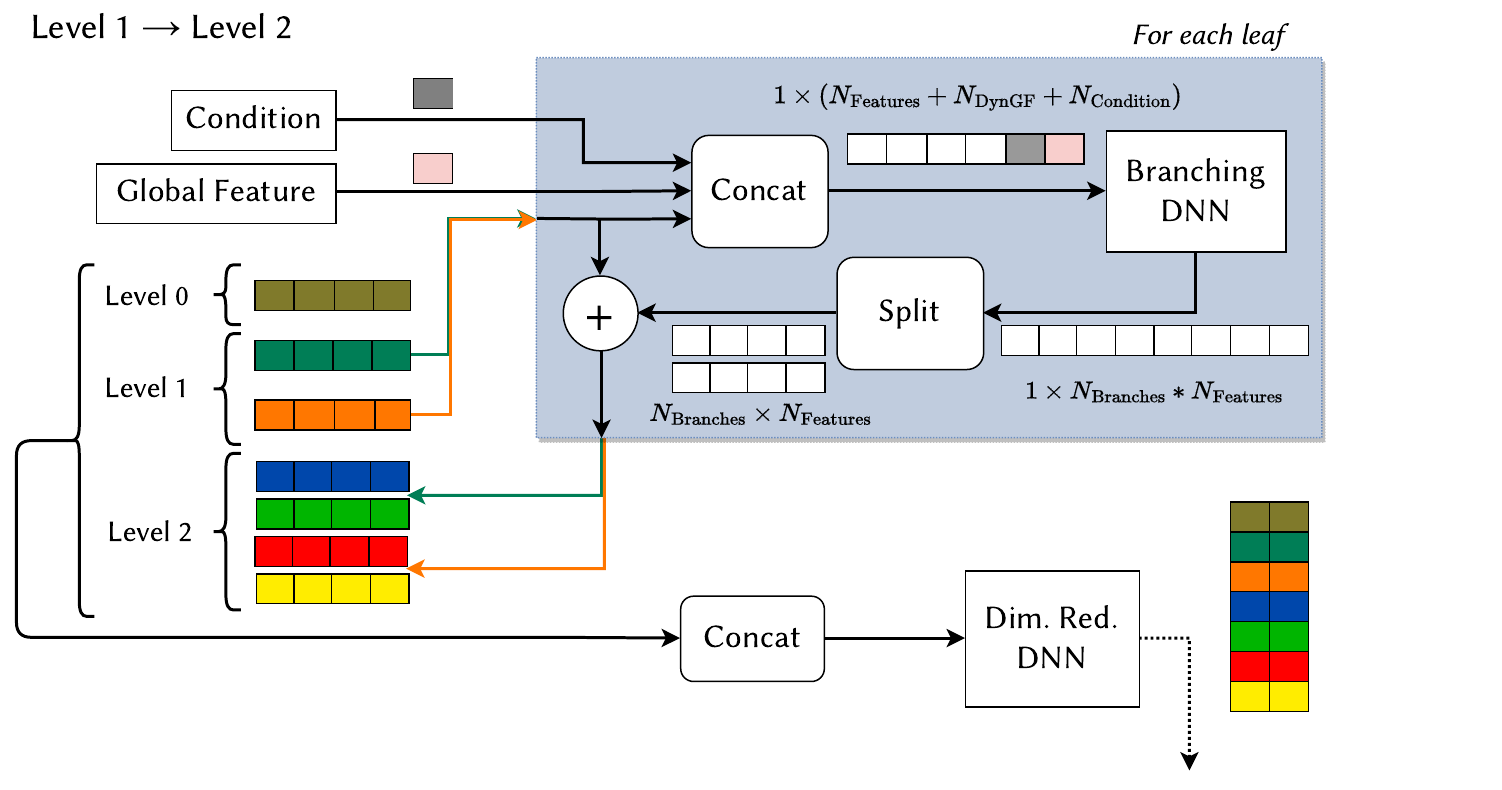}
    \caption{The Branching Layer of the \submScham generator, as described in \sref{subsec:DeepTree}. 
    In this example, the nodes of the 2nd level of the tree are produced and attached as the new leaves.
    With the nodes from level 1 (dark green / orange) as parents, the children in level 2 (blue+green / red+yellow) are generated independently: The condition and a vector representing the state of all points and are appended to the parent. The result is mapped by a branching DNN to the size of the parent times the number of branches. After splitting up the vectors into the new children, the parent is added to each of them. 
    With the new children added as leaves to the tree, all the levels of the tree are stacked up and passed through a dimensionality reduction DNN. Figure adapted from~\cite{DeepTreeCHEP}.}
    \label{fig:deeptreebranching}
\end{figure}

\paragraph{Generator}
The generator of this GAN constructs PCs by starting with a random vector as the root of a tree and then attaching one level of leaves after the other. The output of the generator is the last level of the tree. 
\subparagraph{Branching Layer} Starting with the root node, a \emph{Branching Layer} (\fref{fig:deeptreebranching}) takes the current leaves from the tree, maps each leaf to a given number of nodes $n_i$ and attaches these nodes as new leaves to the tree. Then further \emph{Branching Layers} are applied until the desired number $\prod_i \mathrm{n}_i$ of nodes is reached. For these projections, multiple deep feed-forward neural networks (DNNs) are used.
With each Layer, the number of nodes increases ($1\cdot2\cdot3\cdot4\cdot5\cdot5\cdot10=6000$) and the number of features decreases ($64,25,15,10,8,6,4$).
The cardinality $c$ is sampled from the evaluation set and only the first $c$ points produced by the generator are used.

\subparagraph{Ancestor MPL} In between the branching layers, a Message Passing Layer (MPL) is applied to this tree-structured graph. The edges in the graph are constructed so that each node receives messages from each of its ancestors as well as itself. As a message-passing algorithm, GINConv~\cite{xu2019powerful,pyg} is chosen. For GINConv, the messages are the features of the source node (here: the ancestor). These messages are aggregated by summing over all messages addressed to the target node. These aggregates are added to the  target node (scaled with a learnable weight) and passed through an DNN. The nodes are then updated with the output of the DNNs plus, as a residual connection, the nodes themselves.

In addition, the generator contains layers that condition the MPLs and the branching layers with a vector representing the current state of the leaves in the tree. For this, the leaves are passed separately through a first DNN, then summed up and passed through a second DNN. The DNNs of the generator consist of 3 hidden layers of 100 nodes without bias. The first two hidden layers are followed by a batch normalization~\cite{batchnorm} layer and LeakyReLU activation with a negative slope of 0.1.

\paragraph{Critic}
\begin{figure}[btp]
    \begin{center}
        \includegraphics[width=0.75\textwidth]{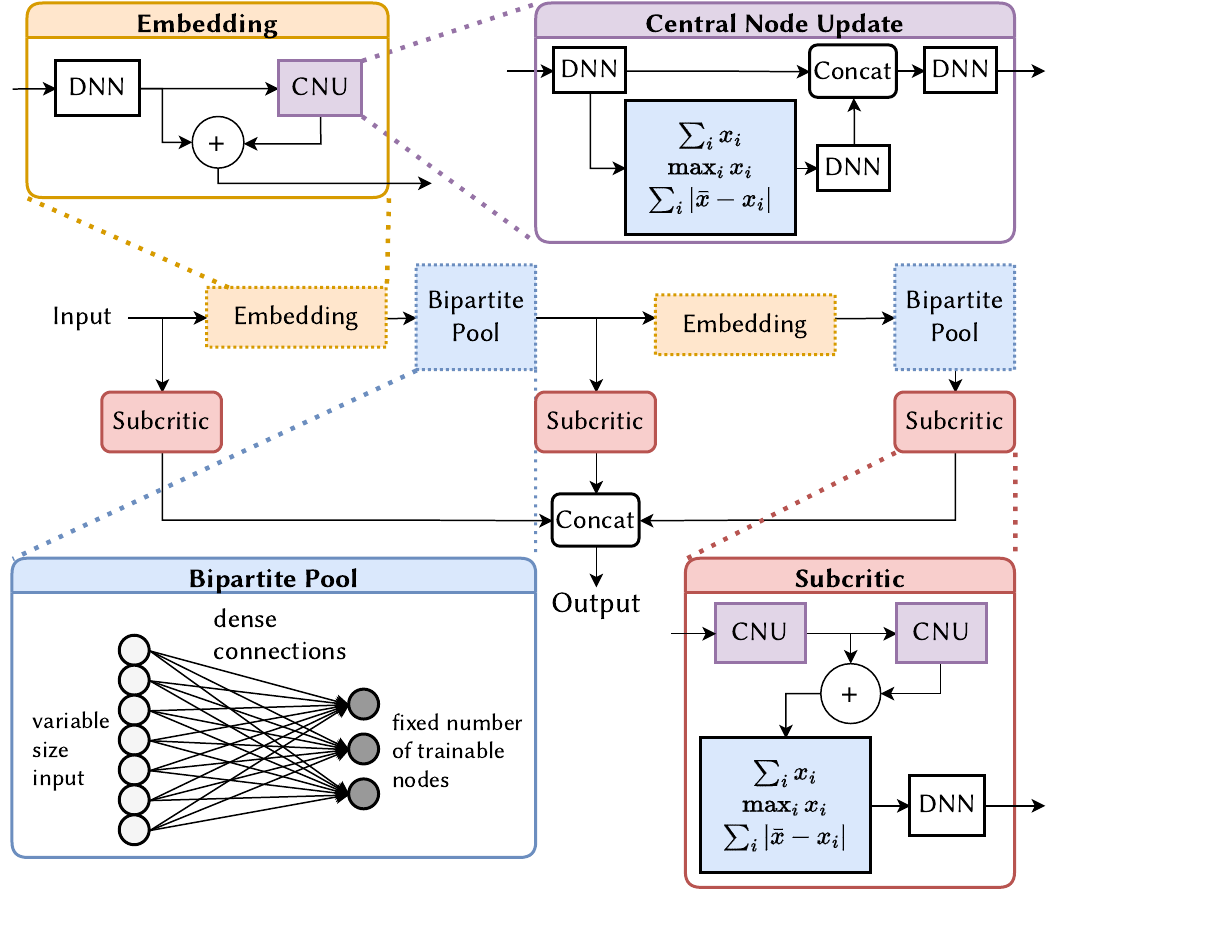}
    \end{center}
    \caption{The \submScham critic, as described in \sref{subsec:DeepTree}. Figure adapted from~\cite{DeepTreeNIPS}.}\label{fig:deeptreecritic}
\end{figure}

The critic, shown in~\fref{fig:deeptreecritic}, aims to reduce the size of the PC iteratively.
This is achieved by a pooling operation called Bipartite Pool. It constructs a bipartite graph that densely connects the input PC to a fixed number of trainable nodes and applies an MPL to this graph. As an MPL, Gatv2Conv~\cite{gatv2conv} is used with 16 attention heads.
Before each pooling, the points are processed by an embedding layer consisting of an DNN and a Central Node Update layer (CNU) with a residual connection.
The CNU transforms input points separately with an DNN, aggregates them with multiple methods (``multi-aggregation''), and maps the points back to their original dimension with another DNN.
This multi-aggregation is a concatenation of sum, maximum, cardinality, and width. The width is computed as mean absolute deviation from the mean $\frac{1}{n}\sum_i |x_i - \bar{x}|$.
Three ``subcritics'' are applied to different levels of pooling and the input PC. Each subcritic uses two CNUs with a residual connection, followed by the multi-aggregation. The aggregated vector is passed through an DNN to produce a single output.
Contrary to the generator, the DNNs of the critic use a dropout of 0.5 and spectral normalization, except for the DNN inside each embedding layer, which use batch normalization. All three subcritics are trained simultaneously, and their losses are added.

\paragraph{Preprocessing and Postprocessing}
The showers on the grid are converted to PCs by taking the $r, \alpha$ and $z$ indices, as well as the energy of each cell with an energy deposition. Uniform noise (0,1) is then added to these indices to make the distribution continuous (reversible by a floor operation). These values are then scaled to the interval $[0,1]$, transformed with a \emph{logit} function (inverse function: \emph{expit}) and finally normalized (mean $\rightarrow$ 0 and standard deviation $\rightarrow$ 1). The energy of the hits is scaled with a Box-Cox transformation with normal scaling (\texttt{PowerTransformer} in~\cite{scikit-learn}). For evaluation and generation, these transformations are inverted.\\
As conditional variables, the shower energy $E_\mathrm{gen}$, the average hit energy $\bar{E}$, and the cardinality $c$ are sampled from the evaluation set and provided to both the generator and the critic.
\footnote{This setup could introduce a bias in the evaluation. However, this does not impose a major restriction on the model, as producing two conditional variables is typically manageable and could be supported by an auxiliary NF in the future. In most practical cases, particularly when aiming to replace \geant, the energy would be provided as input, allowing for controlled generation of showers.}
 $E_\mathrm{gen}$ and $\bar{E}$ are first transformed with a Box-Cox transformation with normal scaling. The cardinality $c$ undergoes the same transformations as the cell indices, but instead of normal scaling, a quantile transformation (\texttt{QuantileTransformer} in~\cite{scikit-learn}) is applied.\\
As the generator is not directly aware of the calorimeter cells, it may produce multiple hits for a single calorimeter cell. This is especially true for events with a very high cardinality. Simply summing the hits in each cell would lead to a low cardinality and points containing very high energies. To mitigate this effect, an algorithm\footnote{Available on PyPI: \url{https://pypi.org/project/caloutils/}} is employed, that moves hits from ``overcrowded'' cells, to empty, neighboring cells (in $r / \alpha / z$).
Since hits of higher energy are more important, it tries to move the hits in the cells in order of energy, skipping the highest energy hit.
In case there are not enough empty neighboring cells are available, the remaining hits are summed up.
Due to this technique, the maximum cardinality that the generator produces goes from $\approx 3500$ to $\approx 4600$ (dataset: maximum $\approx 5300$ of $6000$ total cells).\\
As an additional postprocessing step, the generated PCs are shifted by a random value in $\alpha$, resulting in a uniform $\alpha$ distribution.

\FloatBarrier

\section{Normalizing Flow-based Submissions}\label{sec:flow}
\markboth{\uppercase{Normalizing Flow-based Submissions}}{}

A normalizing flow models a complex density by applying a sequence of transformations to a simpler base distribution, thereby constructing a flexible distribution over continuous random variables. The objective of the normalizing flow is to learn a bijective transformation $T$ between two spaces. Initially, a vector $x$ is sampled from an intricate and generally unknown probability density $p_x(x)$. We define $T$ as the transformation
$x = T(u)$, where $u \sim p_{u}(u)$ is a simple base distribution that is known and for which one can calculate the likelihood and sample from efficiently. Both $T$ and $p_{u}(u)$ can have parameters.

The transformation $T$ must be invertible, and both $T$ and $T^{-1}$ must be differentiable. Such transformations are categorized as diffeomorphisms. The density $p_x(x)$ is then well-defined and can be constructed by a change of variables 
\begin{equation}\label{eq:nf_cov}
p_{x}(x) = p_{u}(T^{-1}(x)) \left|\det J_{T}(u)\right|,
\end{equation}
where $J_T(u)$ is the Jacobian of $T$. Diffeomorphisms are notable for their composability, which allows us to construct $T$ from multiple smaller, invertible, and differentiable transformations $T = T_{K} \circ \dots \circ T_{1}$, where each $T_k$ maps $z_{k-1}$ to $z_k$. Assuming $z_0 =u$ and $z_K = x$, the transformations sequentially modify the distribution, illustrated in \fref{fig:normalizing_flow}.

\newcommand{\distcirc}[3]{
\node[draw, circle, minimum size=1.1cm] (#2) at (#1) {#3};
}

\newcommand{\dist}[3]{
        \begin{scope}[scale=1.5]
        \draw[domain=-1:1,samples=100] plot (\x,{#3});
        \draw (0,-0.1) node [half circle, draw=black, dashed, minimum size=1.55cm] {} node[below=0.1cm]{$#1$ #2};
        \end{scope}
        \distcirc{0, 2.5}{#1}{$#1$}
}

\begin{figure}
  \centering

\begin{tikzpicture}[scale=0.95]

\begin{scope}[shift={(0,0)}]
    \dist{x}{$ \sim p_x(x)$}{(1/(sqrt(2*pi)*0.25)*exp(-(\x)^2/(2*0.25^2))) + (0.8/(sqrt(2*pi)*0.3)*exp(-(\x-0.6)^2/(2*0.1^2))))/2 + (0.3/(sqrt(2*pi)*0.1)*exp(-(\x+0.3)^2/(2*0.1^2))))/2}
\end{scope}
\begin{scope}[shift={(5,0)}]
    \dist{z_k}{$\sim p_k(z_k)$}{(1/(sqrt(2*pi)*0.25)*exp(-(\x)^2/(2*0.25^2))) + (0.8/(sqrt(2*pi)*0.3)*exp(-(\x-0.5)^2/(2*0.1^2))))/2}
\end{scope}
\begin{scope}[shift={(14,0)}]]
    \dist{u}{$ \sim p_u(u)$}{0.5/(sqrt(2*pi)*0.25)*exp(-(\x)^2/(2*0.25^2))}
\end{scope}

\distcirc{$(z_k)!0.35!(u)$}{z_km1}{$z_{k-1}$}
\distcirc{$(z_k)!0.65!(u)$}{z_1}{$z_{1}$}

\node at ($(x)!0.5!(z_k)$) {\centering \LARGE ...};
\node at ($(z_k)!0.5!(u)$) {\centering \LARGE ...};

\draw[<-] ($(z_1.east)+(0,0.1)$) -- ($(u.west)+(0,0.1)$) node[midway, above] {\textbf{$T_1(u)$}};
\draw[->] ($(z_1.east)-(0,0.1)$) -- ($(u.west)-(0,0.1)$) node[midway, below] {\textbf{$T^{-1}_1(z_1)$}};
\draw[<-] ($(z_k.east)+(0,0.1)$) -- ($(z_km1.west)+(0,0.1)$) node[midway, above] {\textbf{$T_k(z_{k-1})$}};
\draw[->] ($(z_k.east)-(0,0.1)$) -- ($(z_km1.west)-(0,0.1)$) node[midway, below] {\textbf{$T^{-1}_k(z_{k})$}};

\end{tikzpicture}

  \caption{Visualization of a Normalizing Flow.}
  \label{fig:normalizing_flow}
\end{figure}

Normalizing flows offer two operational pathways: the inverse path, which is utilized for density estimation and transformation optimization, and the forward path, which functions as a generative model. In the inverse direction, samples from the complex distribution $p_x(x)$ are mapped to the base distribution $p_{u}(u)$, optimizing the process, typically by maximizing the likelihood (or minimizing the negative log-likelihood). Conversely, the forward path initiates with sampling from the base distribution $p_{u}(u)$ and maps these samples to the data space represented by $p_x(x)$. The designation of directions as inverse or forward is arbitrary.
The flow is easily extended to conditional distributions $p_x(x\mid c)$ by including the conditional information $c$ in each transformation in $T$.

The requirement of an efficient computation of the log-likelihood is addressed by specific design choices of the network architecture which make the Jacobian determinant in \eref{eq:nf_cov} tractable~\cite{Kobyzev:2019ydm,Papamakarios:2019fms,ross2021conformal}. Two common approaches are autoregressive flows~\cite{papamakarios2018masked, kingma2016improved} and coupling-based flows~\cite{Dinh:2016pgf,glow,ardizzone2019analyzinginverseproblemsinvertible}. Autoregressive flows have a fast and a slow direction of evaluation. When density estimation is fast, they are called ``Masked Autoregressive Flows'' (MAFs)~\cite{papamakarios2018masked}, when sampling is fast, they are called ``Inverse Autoregressive Flows'' (IAFs)~\cite{kingma2016improved}, with the autoregressive property being realized by masked neural networks called MADE (from ``Masked Autoencoder for Distribution Estimation'') blocks~\cite{germain2015made} in both cases. For further details and a review of common architectures for building these normalizing flows, please refer to the works by Kobyzev \textit{et al.}~\cite{Kobyzev:2019ydm} and Papamakarios \textit{et al.}~\cite{Papamakarios:2019fms}, from which the notation has been adapted.

\FloatBarrier
\submHeadlineMultiple{L2LFlows}{Thorsten Buss, Sascha Diefenbacher, Frank Gaede, Gregor Kasieczka, Claudius Krause, and David Shih}{subsec:L2LFlows}{\submBussMAFCite~and \submBussConvCite}{submBussGit}
\paragraph{Introduction}
Following~\cite{Krause:2021ilc,Diefenbacher:2023vsw,Buckley:2023daw}, we split the task of learning the distribution of showers into smaller pieces: A single Energy Distribution Flow and multiple Causal Flows.
The Energy Distribution Flow (EDF) learns the distribution of layer energies (\textit{i.e.}, the total energies deposited in a layer) conditioned on the incident energy. One of the Causal Flows (CFs) is trained for each of the 45 layers in the calorimeter, learning the shower shape conditioned on the incident energy, the layer energy in that particular layer, and the shower shapes of the previous layers. Conditioning on the output of previous flows is necessary to ensure consistency among the layers.
\begin{figure}[ht]
	\centering
	\includegraphics[width=0.6\textwidth]{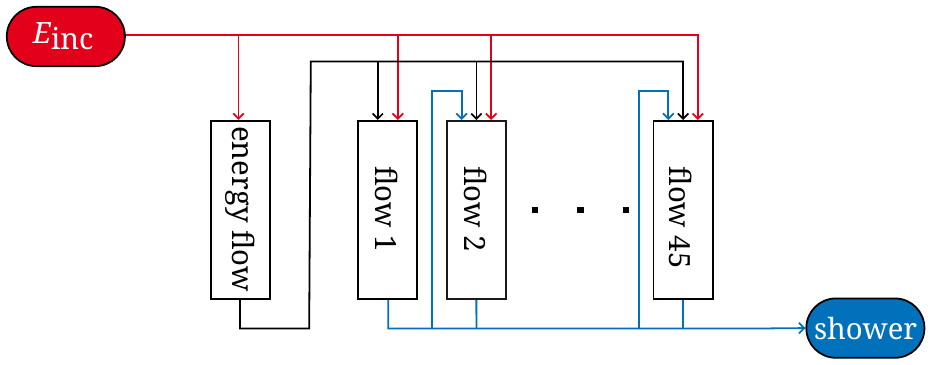}
	\caption{Diagram illustrating the overall architecture of \submBuss, taken from~\cite{Buss:2024orz}. Arrows directed at flows illustrate the conditional input of the flow. Arrows coming from flows illustrate what the flow generates. }
    \label{fig:l2lflows}
\end{figure}
Since we need the earlier flows' output as conditional input for the later flows, during generation, we first draw samples from the Energy Distribution Flow and then sequentially draw samples from the causal flows. In this sense, it is an auto-regressive model. \Fref{fig:l2lflows} illustrates the generation process starting from the incident energy $E_{\rm inc}$ and ending with a generated shower.

\paragraph{Energy Distribution Flow}
The task of the Energy Distribution Flow is to learn the distribution of layer energies conditioned on the incident energy
\begin{equation}
	p(E_1,\dots,E_N|E_{\rm inc})
\end{equation}
where $E_1$ to $E_N$ denote the layer energies. The architecture used is up to hyperparameters the one published in~\cite{Diefenbacher:2023vsw}. It is a MAF~\cite{papamakarios2018masked} consisting of 6 MADE blocks~\cite{germain2015made} with rational quadratic splines (RQS)~\cite{durkan2019neural_spline_flows}. We apply fixed permutations that are randomly initialized between these MADE blocks.

Similar to \cf, we use log and logit transformation as preprocessing. Logarithmic transformations help the network deal with inputs distributed over several orders of magnitude. Logit transformations help the network to generate only samples in an appropriate range. During inference time, the preprocessing is inverted.

Sometimes, the Energy Distribution Flow produces outliers with high energies. For that reason, we reject all sampled layer energies with an energy ratio of \mbox{$E_{\rm dep}/E_{\rm inc}>2.6$}, where $E_{\rm dep}$ is the total deposited energy.

\paragraph{Causal Flows}
\begin{figure}
	\centering
	\includegraphics[height=4.5cm]{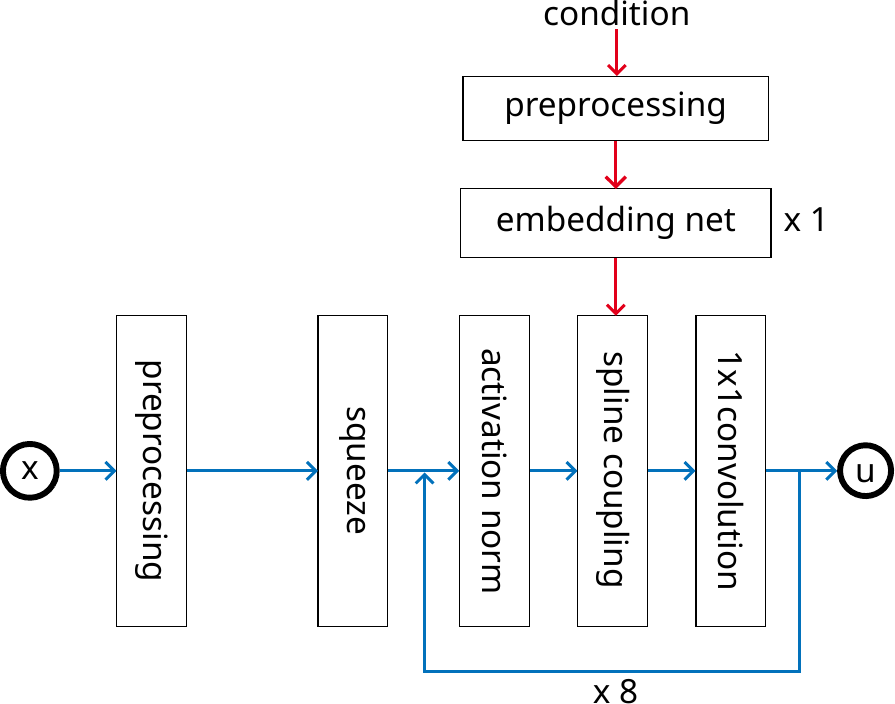}
	\hfill
	\includegraphics[height=4.5cm]{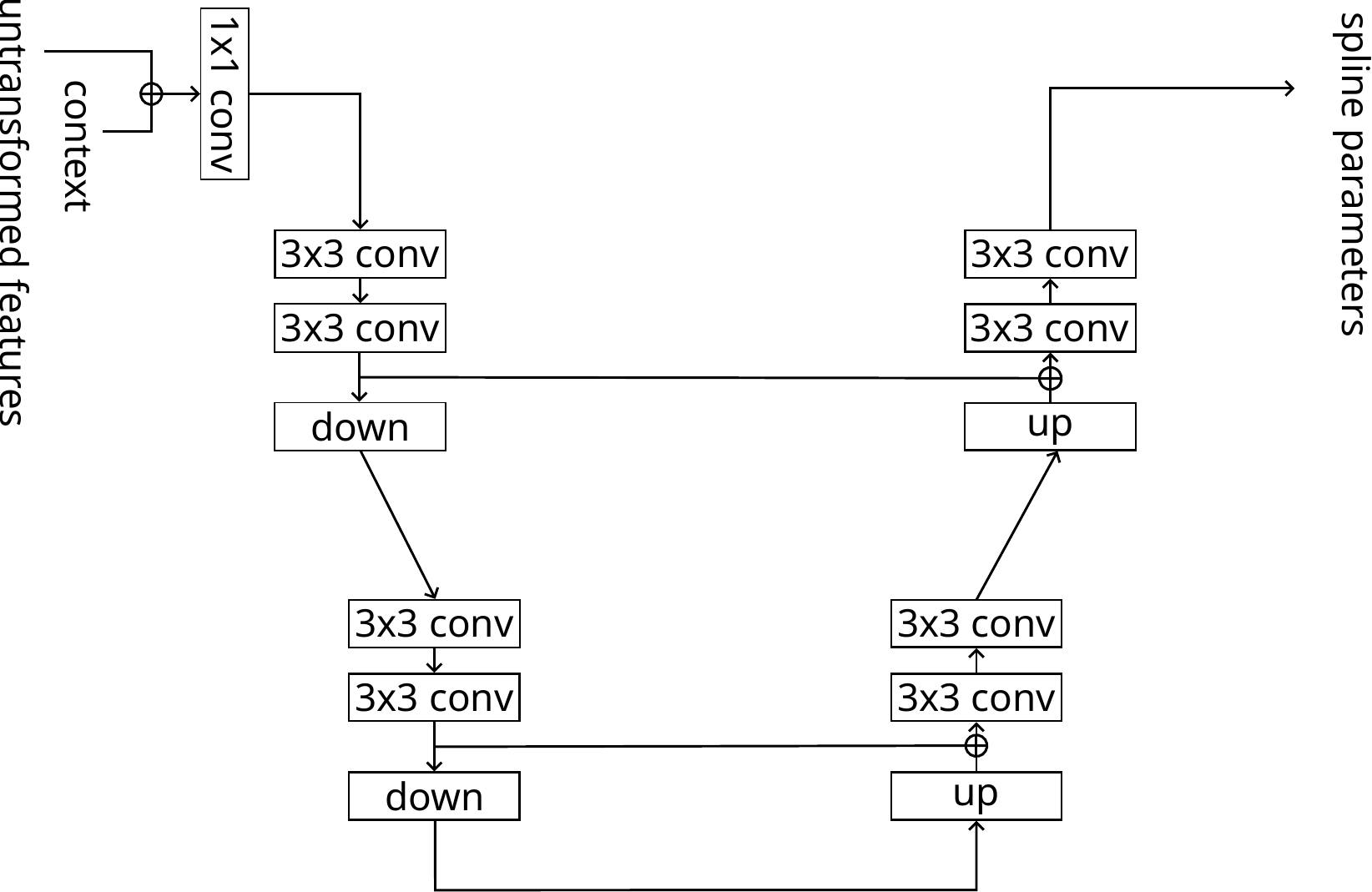}
	\caption{Diagrams illustrating the structure of our convolutional flows, taken from~\cite{Buss:2024orz}. \textbf{Left:} The overall structure of a single Causal Flow. \textbf{Right:} A U-net as it is used in the coupling blocks.}
    \label{fig:conv_flow}
\end{figure}

Each Causal Flow learns the probability distribution of shower shapes in one particular calorimeter layer. This distribution can be denoted as
\begin{equation}
	p(\mathcal{I}_i|\mathcal{I}_1,\dots,\mathcal{I}_{i-1},E_1,\dots,E_N,E_{\rm inc})
\end{equation}
where $\mathcal{I}_i\in\mathbb{R}^{n\times n}$ is the shower shape in layer $i$ given by the deposited energy in each calorimeter cell. We assume approximate locality and only pass up to five previous layers, the energy deposited in layer $i$, and the incident energy as conditional input to the flows. This helps the flow to focus on the most informative features.

We deploy two different network architectures to solve this task. The first one, called \submBussMAF, is a MAF like the Energy Distribution Flow. This architecture is similar to the one published in~\cite{Diefenbacher:2023vsw}. The second one, called \submBussConv, is a flow based on coupling blocks with convolutional U-nets~\cite{Unet} as sub-networks.

The MAF architecture consists of four MADE blocks alternated with randomly initialized but fixed permutations. To better deal with high-dimensional conditional inputs, a summary network is applied.  It receives all the conditional inputs and summarizes the information. Our summary network has 64 output knots.

In~\fref{fig:conv_flow}, the convolutional flow architecture is illustrated. On the right-hand side, we see how a data sample $x$ is transformed into a noise sample $u$. First, it is transformed using a preprocessing function. Next, a squeezing operation~\cite{Dinh:2016pgf} stacks pixels lying in small patches into different channels. This is necessary since, in the coupling blocks, we want to split the information along the channel dimension in order to preserve the spatial structure.

The heart of this architecture consists of eight GLOW blocks~\cite{glow}. They comprise an activation norm, a spline coupling block~\cite{Dinh2014NICE,durkan2019neural_spline_flows}, and an invertible $1\times 1$ convolution, where the activation norm is a normalization operation, and the invertible $1\times 1$ convolution replaces the random but fixed permutation in the MAF. The spline coupling block can learn correlations between pixels and transform inputs in a nonlinear way. 

We used U-nets to learn features on different scales. This is in contrast to RealNVP~\cite{Dinh:2016pgf} and GLOW\cite{glow}, which used a so-called multi-scale architecture. The U-nets are employed in the coupling blocks as sub-networks. We found this setup to be more flexible and to result in higher fidelity than the multi-scale architecture from RealNVP. The U-net architecture is illustrated on the left-hand side of~\fref{fig:conv_flow}.

Since convolutional architectures scale much better with input dimension, the main task of the embedding network is not to reduce the number of inputs but rather to bring the input in a shape the flow can handle.

\paragraph{Training}
We train the \submBussConv on datasets 2 and 3 for 800 epochs. The MAF version, \submBussMAF, is only trained on dataset 3 for 1000 epochs. In both cases, Adam~\cite{kingma2014adam} is used as an optimizer. An exponential decay learning rate scheduler is used in the MAF version, while a one-cycle learning rate scheduler~\cite{smith2018superconvergence} is used in the convolutional version. To ensure a stable training behavior, L2 regularization~\cite{2017arXiv171105101L} and L2 gradient clipping are applied. 

To mitigate the challenges arising from data sparsity, we fill zero voxels with log Gaussian distributed values. Since this noise is below the energy threshold, it will be cut away after generation. Furthermore, we add noise between zero and $1$~keV to all voxels. We rotate showers by random angles during training as data augmentation.

The log-likelihood is additive under joining distributions. Therefore, training each flow individually is equivalent to training all flows jointly. This allows for straightforward parallelization on different compute nodes.

We implement our models using PyTorch~\cite{pytorch} and NFlows~\cite{nflows}. Using single floating point precision is sufficient since we fixed numerical instabilities in NFlows.

\FloatBarrier

\begin{figure}
    \centering
    \includegraphics[width=0.6\textwidth]{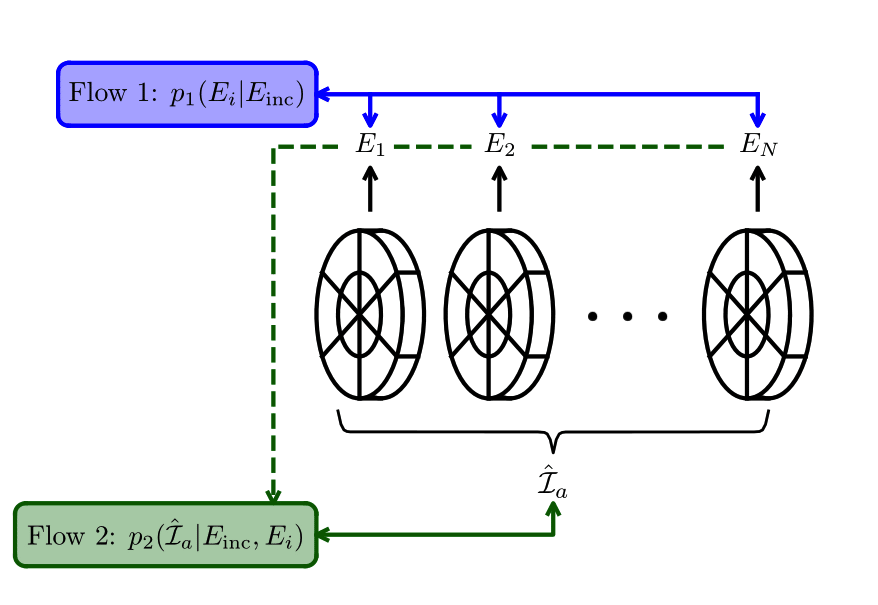}
    \caption{Schematic of the flows used in the \cf~approach. Solid lines are bidirectional --- the direction into each flow denotes the density estimation step and the direction out of the flow denotes the sample generation step. Dashed lines indicate the conditional input to the respective flows.}
    \label{fig:caloflow}
\end{figure}

\submHeadlineMultiple{(inductive) CaloFlow}{Matthew R. Buckley, Claudius Krause, Ian Pang, and David Shih}{subsec:CaloFlow}{\submPangTCite, \submPangSCite, \submPangITCite, and \submPangISCite}{submPang1Git,submPangIGit}
\paragraph{Introduction}
Following the excellent performance of \cf~\cite{Krause:2021ilc,Krause:2021wez} on a simplified calorimeter setup, we adapt \cf\ to the more realistic setup in dataset 1. This corresponds to \submPangTCite and \submPangSCite submissions.
\paragraph{Architecture and Preprocessing}
In \cf, we implement a two-flow method that learns the normalized voxel-level shower energies $\hat\mathcal{I}_a$ conditioned on the corresponding incident energies of the showers $E_{\rm inc}$ denoted by $p(\hat\mathcal{I}_a|E_{\rm inc})$. Here $a$ is the voxel index and the normalization is performed for each layer such that the normalized voxel energies in each layer sum to unity. In the original \cf~studies~\cite{Krause:2021ilc,Krause:2021wez}, it was found that training a single flow to obtain $p(\hat\mathcal{I}_a|E_{\rm inc})$ resulted in problems related to energy conservation. Hence, \cf~makes use of a two-flow (flow-1 and flow-2) setup. Flow-1 is constructed to learn the probability density of calorimeter layer energies\footnote{The layer energy of a given calorimeter layer is the sum of all the voxel energies in that layer.} conditioned on incident energy $p_1(E_i|E_{\rm inc})$, while flow-2 is designed to learn the probability density of the voxel level shower energies conditioned on incident energy and calorimeter layer energies $p_2(\hat\mathcal{I}_a|E_{\rm inc}, E_i)$. When sampling from \cf, the layer energies are first sampled using flow-1 given an input shower incident energy $E_{\rm inc}$. Next, the layer energies $E_i$ from flow-1 and the incident energy $E_{\rm inc}$ are used as conditional inputs for flow-2 which outputs the shower distribution $\hat\mathcal{I}_a$ of the event. Both flow-1 and flow-2 are chosen to be MAFs~\cite{papamakarios2018masked}. In particular, their transformation functions are compositions of rational quadratic splines (RQS)~\cite{durkan2019neural_spline_flows}. A class of neural networks known as MADE blocks~\cite{germain2015made} are used to define the parameters $\vec \kappa$ of the RQS transformations. \Fref{fig:caloflow} shows a schematic of the \cf~approach. Separate flows were trained for the photon and pion datasets.

\paragraph{Training}
Since MAFs are fast in performing density estimation but slow in generation, we opt to train a corresponding IAFs~\cite{kingma2016improved} that is fast in generation instead. We only trained an IAF for flow-2 as flow-1 has a lower dimensional output and is relatively fast to sample from. Since training the IAF with the negative log-likelihood of the data is prohibitively slow, the IAF is trained by fitting it to a pre-trained MAF using the Probability Density Distillation (PDD) method~\cite{pmlr-v80-oord18a} that was also applied in \cite{Krause:2021wez}. This method is also known as teacher-student training where the MAF (IAF) is referred to as the teacher (student). The objective of this training is to enable the IAFs to learn $f_{\rm IAF} = f_{\rm MAF}$, or equivalently, to ensure that $f_{\rm MAF}$ and $f^{-1}_{\rm IAF}$ serve as inverse functions of each other. This equivalence is crucial, as only the fast passes through the flows can be used meaningfully for optimization. In practice, the fitting is implemented based on two training loss terms that we refer to as $z$ and $x$-losses. To compute the $z$-loss, we begin with a sample $z$ which is then passed through the student IAF to obtain a sample $x^{\prime}$ in data space and the corresponding likelihood $s(x^{\prime})$. The data sample $x^{\prime}$ is then mapped via the teacher MAF back to the latent space which obtains the likelihood $t(x^{\prime})$. Similarly, to compute the $x$-loss, one can start with a data sample $x$ which maps to latent space $z^{\prime}$ via the teacher, and then map back to data space via the student. In the original PDD study~\cite{pmlr-v80-oord18a} study, the KL divergence of $s(x^{\prime})$ and $t(x^{\prime})$ was initially used as the training loss. However, the authors noted that it does not converge well. Hence, as in \cite{Krause:2021wez}, we used a training loss function that is based on a mean square error that compares relevant values\footnote{For the student model for \dsIph, these consist of coordinates before and after passing them through the flows and RQS parameters from individual MADE blocks within the bijectors. For the student model for \dsIpi, we did not enforce agreement with the teacher at the level of individual MADE blocks, but only at the endpoints of the flows.} at each equivalent stage of the teacher and student passes.

In \cf~, each flow-1 MAF consists of six MADE blocks, while each flow-2 MAF/IAF consists of eight MADE blocks. We used two hidden layers for each flow-1 MADE block and a single layer for each flow-2 MADE block. The hidden layers in flow-1 each have 64 nodes. We experimented with different flow-2 MADE block hidden layers sizes and settled for the following choice: 378 for $\gamma$ teacher; 736 for $\gamma$ student; 533 for $\pi^+$ teacher; 500 for $\pi^+$ student.

\paragraph{Introduction}
Applying \cf~to the higher dimensional voxelization in datasets 2 and 3 is extremely memory intensive. The number of splines and therefore RQS parameters grows linearly with data dimension $d$, making the number of parameters in the output layer grow as $\mathcal{O}(d^2)$, easily dominating over the number of parameters in the hidden layers. Hence, we proposed a new method, that we dub inductive \cf~or \icalo, to overcome this obstacle. This method corresponds to the \submPangITCite and \submPangISCite submissions.

\setcounter{footnote}{0}

\begin{figure}
\includegraphics[width=\textwidth]{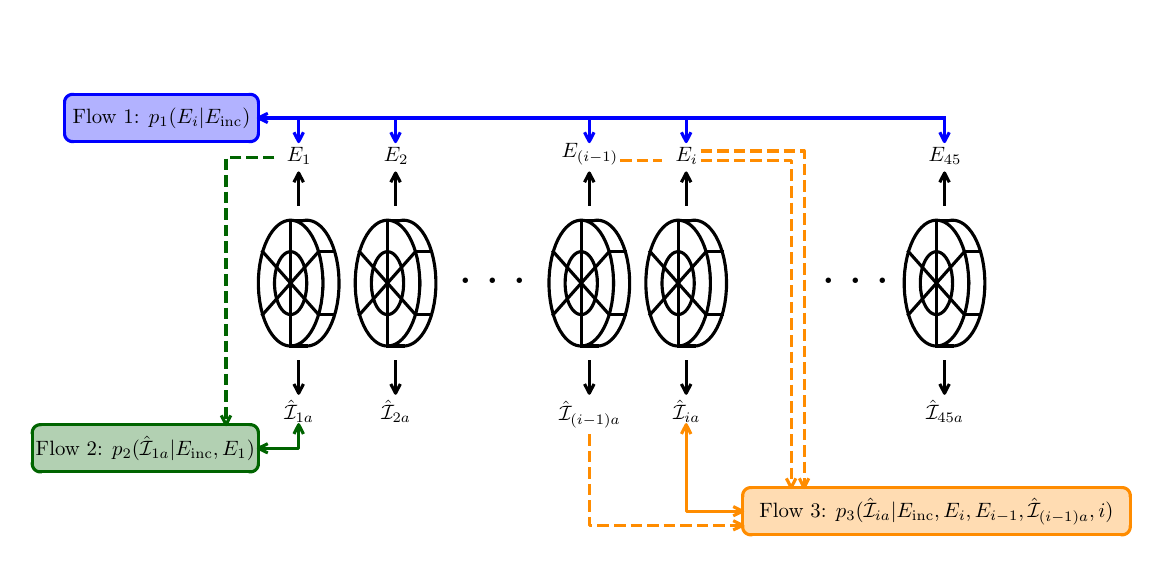}
\caption{Schematic of the three \icalo{} flows, taken from~\cite{Buckley:2023daw}. Solid lines are bidirectional --- the direction into each flow denotes the density estimation step and the direction out of the flow denotes the sample generation step. Dashed lines indicate the conditional input to the respective flows. Flow-3 is used iteratively on subsequent layers.}
\label{fig:icaloflow}
\end{figure}

\paragraph{Architecture and Preprocessing}
Our \icalo~method uses three normalizing flows to learn and generate calorimeter showers. Flow-1 learns the joint probability distribution of total energy deposited in each layer $E_i$, conditioned on the incident energy of the event $E_{\rm inc}$: $p_1(E_i|E_{\rm inc})$. It is necessary to learn this probability distribution as $E_i$ is a conditional input for flow-2 and flow-3 in the generation step. Flow-2 learns the probability distribution of the unit-normalized voxel energies in the first layer of the calorimeter, $\hat{\mathcal{I}}_{1a} \equiv \mathcal{I}_{1a}/\sum_b \mathcal{I}_{1b}$, conditioned on $E_{\rm inc}$ and the energy deposited in the first layer, $E_1$: $p_2\left(\hat{\mathcal{I}}_{1a}|E_{\rm inc},E_1\right)$. Here $a$ is the voxel index. Finally, flow-3 learns the probability distribution of unit-normalized voxel energies in every layer after the first, $\hat{\mathcal{I}}_{ia} \equiv \mathcal{I}_{ia}/\sum_b \mathcal{I}_{ib}$ for $i\in[2,45]$, where the $i^{\rm th}$ layer is conditioned on the energy deposited in the layers $i$ and $i-1$ ($E_i$ and $E_{i-1}$), the incident energy $E_{\rm inc}$, the unit-normalized voxel energies in the $(i-1)^{\rm th}$ layer $\hat{\mathcal{I}}_{(i-1)a}$, and the one-hot\footnote{One-hot encoding is used for layer numbers instead of ordinal encoding using the layer number directly, because other than the location in the detector, there is no information in the layer number, \textit{i.e.}, layer 30 is not 15 times more important than layer 2.} encoded layer number $i$: $p_3\left(\hat{\mathcal{I}}_{ia}|E_{\rm inc},E_i,E_{i-1}, \hat{\mathcal{I}}_{(i-1)a}, i\right)$. \Fref{fig:icaloflow} shows a schematic of the \icalo~approach. Like in \cf, we used MAFs with RQS transformations for flow-1, and MAF-IAF pairs for flow-2 and flow-3. For dataset 2, flows-2 and -3 in \icalo consist of eight MADE blocks with two hidden layers of size 256\footnote{With the exception of the flow-3 IAF, which has hidden layer sizes of 384.}. For dataset 3, the configuration is similar, only flow-3 uses just one hidden layer for both MAF and IAF setups. Flow-1 always has just one hidden layer. 

The number of trainable parameters for the \cf~models are included in Tables~\ref{tab:ds1-photons.numparam} and~\ref{tab:ds1-pions.numparam}. For the teacher models, the total parameter count matches that of sample generation, which is the sum of parameters in flow-1 and flow-2 (teacher). As for the student models, the parameter count during sample generation is the sum of parameters in flow-1 and flow-2 (student). Given the necessity of a pre-trained teacher model for each student model, the total parameter count encompasses parameters from flow-1, flow-2 (teacher), and flow-2 (student).

\FloatBarrier

\submHeadlineSingle{CaloINN}{Luigi Favaro, Florian Ernst, Claudius Krause, Tilman Plehn, and David Shih}{subsec:caloinn}{\submFavaroCite}{submErnstGit}
\paragraph{Introduction}
In \submFavaroCite we train a normalizing flow for the generation of showers in dataset 1 and
dataset 2. We use the INN variant of FrEIA~\cite{freia} with coupling layers, which 
unlike autoregressive methods provides fast evaluation in the
forward and backward directions. This is achieved by transforming only a subset of the 
input features  with a reversible transformation. 
The parameters of the transformation are predicted by a network conditioned on the remaining features and the incident energy. 
The \submFavaro architecture allows for a generation step of $\mathcal{O}(1)$ ms per shower 
on a single GPU without the necessity of a second distillation process.
\paragraph{Architecture and Preprocessing}
The architecture takes voxels normalized by the layer energy as input.
The information of the energy per layer is encoded in extra energy dimensions,
similarly preprocessed as in \submPangT, as shown in \eref{eq:us}. To explore the expressive power
of a single flow network, we simply append the energy ratio variables to the feature vector.
We do not explore a separate training for the energy and the voxel dimensions which would
simplify and improve the learning process of the energy dimensions.
\begin{equation}
  u_0 = \frac{\sum_i E_i}{E_\mathrm{inc}}
  \qquad\quad \mathrm{and} \qquad\quad 
  u_i = \frac{E_i}{\sum_{j\ge i} E_j} \; ,
\label{eq:us}
\end{equation}
After creating the final feature vector, we add uniform noise sampled from $\mathcal{U}(0, b)$, where $b=5\cdot10^{-6}$ for dataset 1 and $b=10^{-6}$ for dataset 2. Then, we apply a regularized log transformation with parameter $\alpha=10^{-8}$ according to
\begin{equation}
    x = \log (x_0 + \alpha) \, .
\end{equation}
We show the schematic of a coupling block in~\fref{fig:caloinn_schematic}.
In each coupling block, the input vector is split in two halves, $x_t$ and $x_c$,
of equal size. The block only transforms half of the input features, which are selected randomly
during initialization, defining the transformation
\begin{equation}
   T(x_t, x_c; E_{\mathrm{inc}}) = \left\{
    \begin{array}{ll}
        y_t = f(x_t; x_c, E_{\mathrm{inc}}) \\
        y_c = x_c &  \, .
    \end{array} 
    \right.
\end{equation}
The splitting between $x_t$ and $x_c$ is fundamental in coupling layer-based normalizing flows.
This choice provides a triangular Jacobian matrix which is tractable and can be promptly evaluated both during training and generation.
The transformation $f$ applied to the features is a RQS~\cite{durkan2019neural_spline_flows} for
dataset 1 and a cubic spline for dataset 2.
The prediction of the spline parameters is obtained with a sub-network consisting of a sequence of dense layers with
256 nodes for each hidden layer. The number of hidden layers is four for dataset 1 and
three for dataset 2. After permuting the order of the features, we normalize the output to mean zero
and unit standard deviation with an ActNorm~\cite{glow} layer. This allowed us to improve the stability of the training and utilize a deeper model.
\begin{figure}
    \centering
    \scalebox{.5}{\tikzstyle{crc} = [circle, rounded corners=0.3ex, minimum width=1.5cm, minimum height=1cm, text centered, align=center, inner sep=0, fill=white, font=\LARGE, draw]

\begin{tikzpicture}[node distance=1cm, line width=2pt, font=\LARGE]

\coordinate (A);
\coordinate (B) at ([xshift=+7cm] A);
\coordinate (C) at ([yshift=+10mm] B);
\coordinate (D) at ([yshift=+10mm] A);

\draw[fill={rgb,1:red,0.902;green,0.902;blue,0.902}] (A) -- (B) -- (C) -- (D) -- cycle;

\node at ($(A)!0.5!(B) + (0, 0.4cm)$) {$x$, $u$};

\coordinate (A'') at ([xshift = 2.cm, yshift=-2cm] A);
\coordinate (B'') at ([xshift=+3.0cm] A'');
\coordinate (C'') at ([yshift=-2cm] B'');
\coordinate (D'') at ([yshift=-2cm] A'');

\draw[fill={rgb,1:red,0.902;green,0.902;blue,0.902}] (A'') -- (B'') -- (C'') -- (D'') -- cycle;

\coordinate (A') at ([yshift=-13cm] A);
\coordinate (B') at ([yshift=-13cm] B);
\coordinate (C') at ([yshift=-13cm] C);
\coordinate (D') at ([yshift=-13cm] D);

\draw[fill={rgb,1:red,0.902;green,0.902;blue,0.902}] (A') -- (B') -- (C') -- (D') -- cycle;

\node at ($(A')!0.5!(B') + (0, 0.4cm)$) {$\hat{x}$, $\hat{u}$};

\draw[fill={rgb,1:red,0.815;green,0.909;blue,0.949}] ([yshift=-15mm]A) -- ([yshift=-15mm]B) -- ([yshift=-15mm]C) -- ([yshift=-15mm]D) -- cycle;

\draw[fill={rgb,1:red,0.972;green,0.807;blue,0.8}] ([yshift=15mm]A') -- ([yshift=15mm]B') -- ([yshift=15mm]C') -- ([yshift=15mm]D') -- cycle;

\draw[fill={rgb,1:red,0.815;green,0.909;blue,0.949}] ([yshift=30mm]A') -- ([yshift=30mm]B') -- ([yshift=30mm]C') -- ([yshift=30mm]D') -- cycle;

\draw[->] ([xshift =5mm, yshift=-18mm]A) -- ([xshift=5mm, yshift=-8.8cm]A);
\draw[->] ([xshift =-4mm, yshift=-18mm]B) -- ([xshift=-4mm, yshift=-8.8cm]B);

\node (trafo) [crc, below of = C, yshift=-7.3cm, xshift=-4mm, font=\Large, fill={rgb,1:red,0.835;green,0.909;blue,0.831}]{$T$};

\draw[->] ([xshift =5mm, yshift=-5cm]A) -- (trafo);

\draw[fill={rgb,1:red,0.972;green,0.807;blue,0.8}] ([yshift=-1.2cm]D'') -- ([yshift=-1.2cm]C'') -- ([yshift=-3.3cm]C'') -- ([yshift=-3.3cm]D'') -- cycle;

\draw[->] ([xshift =1.5cm, yshift=-0.1cm]D'') -- ([xshift=1.5cm, yshift=-1.1cm]D'');

\node at ($(A'')!0.5!(B'') - (0, 1.0cm)$) {$\log E_{\text{inc}}$};

\node[text width=1.9cm] at ($(A'')!0.5!(B'') - (0, 4.3cm)$) {Spline params.};

\node at ($(A)!0.5!(B) - (0, 1.0cm)$) {Split};

\node at ($(A')!0.5!(B') + (0, 2.0cm)$) {ActNorm};

\node at ($(A')!0.5!(B') + (0, 3.5cm)$) {Permute};

\end{tikzpicture}}
\caption{Schematic representation of the \submFavaro coupling block, taken from~\cite{Ernst:2023qvn}.}
\label{fig:caloinn_schematic}
\end{figure}
In the large-scale architecture, we stack twelve blocks for dataset 1 and fourteen blocks
for dataset 2 to construct the full flow. 
\FloatBarrier

\submHeadlineSingle{SuperCalo}{Ian Pang, John Andrew Raine, and David Shih}{subsec:supercalo}{\submPangSuperCite}{submPangSuperGit}
\paragraph{Introduction}
Our approach, which we dub as \submPangSuperCite, presents a way to generate high-dimensional calorimeter showers by super-resolving low-resolution calorimeter showers. The showers used in the CaloChallenge datasets are represented as 3-dimensional images that are binned into voxels in position space. We will refer to these voxels as \textit{fine voxels}. A coarse-grained representation of each shower can be obtained by grouping together neighboring fine voxels to make \textit{coarse voxels}. In this approach, we split the task of learning $p(\vec E_{\rm fine}|E_{\rm inc})$ into two parts. Here $\vec E_{\rm fine}$ is the energy deposited in the fine voxels and $E_{\rm inc}$ is the incident energy of the particle. First, we learn to sample from $p(\vec E_{\rm coarse}|E_{\rm inc})$, where $\vec E_{\rm coarse}$ is the energy deposited in the coarse voxels. Next, we learn to super-resolve the coarse voxels to obtain the fine voxels, which is equivalent to sampling from $p(\vec E_{\rm fine}|\vec E_{\rm coarse})$.

\paragraph{Architecture and Preprocessing}
However, trying to learn $p(\vec E_{\rm fine}|\vec E_{\rm coarse})$ with a single model would be no better in terms of model size than the original problem of learning $p(\vec E_{\rm fine}|E_{\rm inc})$. As a result, we rewrite the distribution according to the following ansatz:

\begin{equation}
p(\vec E_{\rm fine}|\vec E_{\rm coarse}) = \prod_{i=1}^{N_{\rm coarse}}p(\vec e_{{\rm fine},i}|E_{{\rm coarse},i},\dots).
\end{equation}

In other words, each coarse voxel, with deposited energy $E_{{\rm coarse}, i}$, is upsampled to its fine voxels, with deposited energies $\vec e_{{\rm fine},i}$, using a universal super-resolution model that may be conditioned on some coarse shower information. Here is the list of the conditional inputs that we used in the super-resolution model:

\begin{itemize}
    \item Incident energy of the incoming particle, $E_{\rm inc}$
    \item Deposited energy in coarse voxel $i$, $E_{{\rm coarse},i}$
    \item Fine layer energies of layers spanned by coarse voxel $i$
    \item Deposited energy in neighboring coarse voxels in $\alpha$, $r$ and $z$ directions\footnote{There is maximum of 6 neighboring coarse voxels for each coarse voxel. For coarse voxels with fewer than 6 adjacent coarse voxels, the missing neighboring coarse voxel energies are padded with zeros.}
    \item One-hot encoded coarse layer number 
    \item One-hot encoded coarse radial bin 
\end{itemize} 

Since it is not obvious which choice of coarse shower representation would result in the highest fidelity high-resolution showers (fine voxels), we experimented with a few choices and picked the one that gave the best results. In particular, we grouped the fine voxels such that 1 coarse voxel = 1 $r$ bin $\times$ 2 $\alpha$ bins $\times$ 5 $z$ bins. This choice results in a 648 dimensional coarse shower.

Similar to \submPangTCite and \submPangSCite, we used a flow-1 + flow-2 setup to learn the distribution of energy deposited in coarse voxels in each shower conditioned on the incident energy of the particle $p(\vec E_{\rm coarse}|E_{\rm inc})$. Next, we train our super-resolution flow to learn $p(\vec e_{{\rm fine},i}|E_{{\rm coarse},i},\dots)$. Then, generating showers with the full model chain involves sampling sequentially from flow-1, flow-2 and the super-resolution flow. All the flows used in this work are MAFs~\cite{papamakarios2018masked} with RQS~\cite{durkan2019neural_spline_flows} transformations. Like in \cf, the parameters of each RQS transformation are computed using a MADE block~\cite{germain2015made}. Each flow consists of eight MADE blocks. Flow-1 (2) has MADE blocks with a single hidden layer with 256 (648) nodes. The super-resolution flow has MADE blocks with two hidden layers, each with size 128.

\FloatBarrier

\submHeadlineSingle{CaloPointFlow}{Simon Schnake, Benno Käch, Moritz Scham, Dirk Krücker, and Kerstin Borras}{subsec:calopointflow}{\submSchnakeCite}{submSchnakeGit}
\paragraph{Introduction}
The requirement for fast simulation of calorimeter showers has led to a growing interest in using machine learning models for their efficient and high-fidelity generation. Calorimeter showers are generally sparse, with a majority of calorimeter cells being empty, necessitating a representation that is efficient and effective at capturing the essential features of the data. Point clouds offer an apt solution for representing sparse data structures due to their innate efficiency. Our modified model builds upon the original PointFlow~\cite{yang2019pointflow} model known for its exceptional ability to produce high-quality point clouds.  The \submSchnake model leverages PointFlow's advantages while making specific adjustments to specialize in generating calorimeter data. 

The model consists of four sub-models, as shown in~\fref{fig:cpf-arch}. The initial sub-model, CondFlow, is responsible for generating the number of hits, referred to as $n_\textrm{hits}$, and the total energy, $E_\textrm{sum}$, in the calorimeter cells by a normalizing flow. The second stage comprises the permutation invariant encoder, which transforms the entire point cloud $X$ into a latent representation $z$. This transformation is based on the DeepSets \cite{NIPS2017_f22e4747} architecture. Subsequently, the LatentFlow, the third sub-model, produces the latent representation $z$, which is conditioned on the values of $E_\textrm{sum}$ and $n_\textrm{hits}$. The final component, the PointFlow, is a permutation equivariant normalizing flow that performs pointwise transformations. PointFlow is conditioned on $z$, as well as $E_\textrm{sum}$ and $n_\textrm{hits}$. 

\begin{figure}[ht]
    \centering
    \begin{subfigure}{.5\textwidth}
    \centering
    \includegraphics[width=\textwidth]{figures/calopointflow/training.tikz}
    \caption{Training}
    \end{subfigure}%
    \begin{subfigure}{.5\textwidth}
    \centering
    \includegraphics[width=\textwidth * 1.1]{figures/calopointflow/generation.tikz}
    \caption{Sampling}
    \end{subfigure}%
    \caption{Architecture of \submSchnake, taken from~\cite{Schnake:2024mip}. \textbf{Left:} In training. \textbf{Right:} In sampling.}
    \label{fig:cpf-arch}
\end{figure}

\paragraph{Preprocessing}
Concerning the conditional variables learned through CondFlow, the number of hits, $n_{\textrm{hits}}$, is processed by adding uniform noise ranging from 0 to 1 and then dividing by the square root of the input energy, $E_{\textrm{in}}$. The total energy of the shower, $E_{\textrm{sum}}$, is normalized by dividing it by $E_{\textrm{in}}$. We then log-transform and normalize the resulting values.

\begin{figure}
    \centering
    \includegraphics[width=\textwidth]{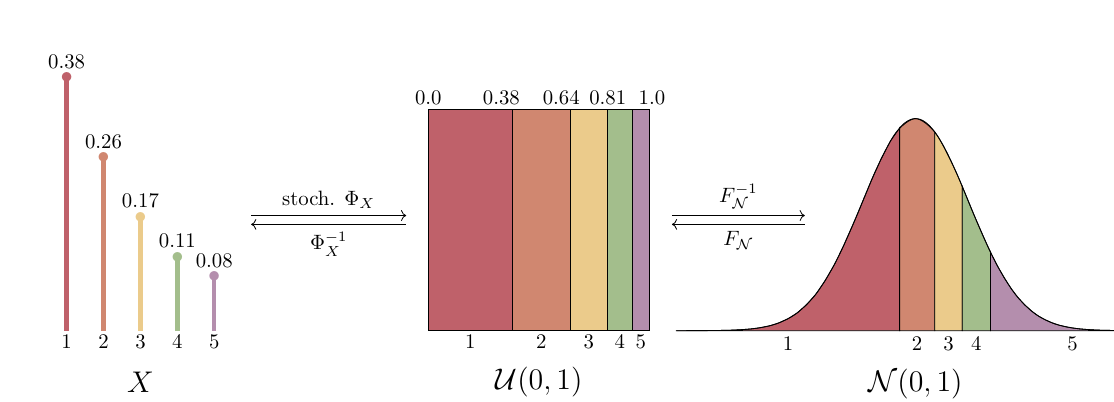}
    \caption{The CDFDequantization, adapted from~\cite{Schnake:2024mip}.}
    \label{fig:cpf-cdfdequantization}
\end{figure}

Discrete distributions in continuous space resemble delta distributions and pose modeling challenges using a normalizing flow. To overcome this, dequantization techniques transform discrete distributions into continuous ones. This usually requires adding uniform noise to fill the space between 0 and 1, followed by a logit transformation~\cite{uria2013rnade, Dinh:2016pgf}. We introduce a new dequantization strategy, CDFDequantization, displayed in~\fref{fig:cpf-cdfdequantization}. This approach utilizes the quantile function and the cumulative distribution function (CDF), which establish invertible mappings between their distribution and the standard uniform distribution. This principle is inherent in inverse sampling, where the uniform distribution is segmented into parts corresponding to the probabilities of certain discrete values. Although the CDF mapping is not directly invertible, a stochastic inverse can be developed by transforming each discrete value $ d $ into $ CDF(d) + PMF(d) \cdot u $, where $ u $ is randomly drawn from $ \mathcal{U}(0,1) $. This process effectively raises our discrete distribution to $ \mathcal{U}(0,1) $. To map the uniform distribution to the entire real space, and convert $ \mathcal{U}(0, 1) $ to a logistic distribution, a logit transform is typically applied in standard dequantization. However, we choose to employ the quantile function of the standard normal distribution to map our values onto the normal distribution. This design ensures the marginal distributions of our model are intrinsically normal, with correlations being the only remaining aspect to be modeled.

\paragraph{Architecture}
All three flows --- CondFlow, LatentFlow, and PointFlow --- are coupling layer-based flows~\cite{Dinh:2016pgf} utilizing RQS~\cite{durkan2019neural_spline_flows}. In the previous iteration of the \submSchnake model, a coupling-layer-based normalizing flow was applied on a point-by-point basis. This involved dividing point features into two equal segments using a system of coupling blocks, with one segment being transformed based on the other. The final result is a permutation invariant system capable of managing varying numbers of points. However, a significant challenge remained in this model: it lacked a mechanism to facilitate the exchange of information between points, making it prohibitive to model inter-point correlations.

In order to solve this problem, we have implemented a modification that makes use of Deep Sets~\cite{NIPS2017_f22e4747}. Following our coupling strategy, we opted to aggregate half of the point features objectively. We maintain transformation permutation invariance by initially mapping each point to a latent space that has high dimensionality. Afterwards, we perform a pooling operation such as max, sum, or mean to merge the information. We then transformed the aggregated latent information and incorporated it as an additional conditional variable. This process enables our point-wise normalizing flows to transfer inter-point information and therefore enhances the model's ability to capture correlations.

A major concern in point cloud-based models for simulating calorimeters is the presence of multiple hits per cell. This occurs when points, produced in continuous space, are mapped onto the discrete space of calorimeter cells, which may result in multiple points being assigned to a single cell. This inconsistency contradicts the real data, where each cell can have only one energy value. To address this matter, the model must accurately determine the direct distances between points as well as the occupancy status of each cell, a task that is notably complex. The \submSchnake model, for instance, could not carry out this function.

Our strategy for mitigating the issue leverages the rotational invariance characteristic of the detector. The principle of rotational invariance implies that the marginal energy distribution in the angular coordinate $\alpha$ should exhibit a uniform, or flat distribution. To utilize this characteristic, we confine the generation of particle showers to the $(z, r)$ plane. Afterward, we randomly assign the angular position $\alpha$. This method effectively loosens the constraint of limiting the number of showers per cell. In dataset 2, this relaxation allows up to 16 showers per cell, and for dataset 3, the number rises to up to 50 showers per cell.

However, there may be instances where the number of hits surpasses the specified limits. In these cases, the extra hits are randomly allocated to already occupied $\alpha$ regions. Although this approach does not offer a complete resolution to the problem of multiple hits, it has been demonstrated to considerably improve the experimental outcomes in practical settings. The efficacy of the mentioned approach in augmenting the quality of data highlights its potential as a significant temporary remedy, while we attempt to devise more resilient techniques to tackle the intricacies linked with the occurrence of multiple hits in particle detectors.


\FloatBarrier

\section{Diffusion-based Submissions}\label{sec:diffusion}
\markboth{\uppercase{Diffusion-based Submissions}}{}
Diffusion models are a class of generative models based on applying a chosen perturbation to the data and then training a model to invert that perturbation.  
The model is defined in terms of a forward process, in which perturbations are gradually applied to the data sample eventually reaching a known end-point distribution (such as Gaussian noise). 
A model is then trained to learn the reverse process, which inverts the perturbations to recover the original data sample. 
Once trained, new samples can be generated by sampling from the end-point distribution and iteratively applying the reverse process model.
Diffusion models have been defined under two different formalisms, a score-based formulation and a denoising formalism.  

In score-based models, the forward process is defined by a stochastic differential equation (SDE):
\begin{equation}
dx = f(x,t)dt + g(t)dW,
\label{eq:score_diffu}
\end{equation}
where $f(x,t)$ and $g(t)$ are user-specified diffusion and drift functions, respectively.
$W$ is a Wiener process (Brownian motion), indexed by a time parameter $t \in [0,1]$.
The reverse process can then be solved by the following SDE:
\begin{equation}
dx = [f(x,t) - g(t)^2 \nabla_x \mathrm{log } p_t(x) ] dt + g(t) dW,
\label{eq:score_diffu_reverse}
\end{equation}
where the $\nabla_x \mathrm{log } p_t(x)$ is the `score' of the data, or the gradient of the log probability. A denoising score-matching strategy is used to learn the score function \cite{denoising_score_matching} and then used in \eref{eq:score_diffu_reverse} to generate samples. 

In the denoising diffusion formalism, as formulated in the original DDPM paper \cite{ho2020denoising}, the forward process is defined by the repeated addition of Gaussian noise to the original data in $t$ steps:
\begin{eqnarray}
        x_t  &=& \sqrt{1 - \beta_t} x_{t-1} + \beta_t \epsilon, \\ 
        q(x_t \mid x_{t-1}) &=& \mathcal{N}(x_t \mid \sqrt{1 - \beta_t}x_{t-1}, \beta_t),
\end{eqnarray}
where $\mathcal{N}(x \mid \mu, \sigma)$ is a Gaussian likelihood and $\beta_t$ is a 
user-chosen `noise schedule' that specifies how much noise is added at each step.
For a sufficiently large $T$ (the total number of diffusion steps), the Gaussian noise will overwhelm the original data and $x_T$ will follow a multivariate Gaussian distribution.
The reverse process is also assumed to follow a Gaussian likelihood:
\begin{equation}
    p(x_{t-1} \mid x_t) = \mathcal{N}(x_{t-1} \mid \mu_{\theta}(x_t,t,z), \beta_t \mathcal{I}),
\end{equation}
with an unknown mean $\mu_{\theta}$ that is learned by a neural network during training.

Though appearing conceptually different, these two formalisms have been shown to be mathematically equivalent \cite{luo2022understanding} for a particular choice of the drift and diffusion functions (the `variance-preserving' choice): the optimal model trained under one formalism is optimal for the other as well.
Though in practice, because optimality is never reached, the two formalisms may offer different practical advantages and disadvantages.
The diffusion literature is rapidly evolving and newer models generally alter these formalisms slightly, but the key conceptual ideas remain.

\FloatBarrier

\submHeadlineSingle{CaloDiffusion with GLaM}{Oz Amram and Kevin Pedro}{subsec:calodiffusion}{\submAmramCite}{submAmramGit}
\paragraph{Introduction}
\submAmramCite is based on denoising diffusion models \cite{ho2020denoising}, in which the perturbation applied to the image is an addition of Gaussian noise. We use the cosine noise schedule proposed in~\cite{improved_diffu} with 400 diffusion steps for all datasets. Shower preprocessing is done similarly to other approaches, where the voxel energies are divided by the incident particle energy and logit transformed. 

\paragraph{Architecture}
The denoising model follows a U-net architecture \cite{Unet}, with 3 sets of ResNet \cite{RESNET} blocks with linear attention \cite{linear_attention}. The input is compressed by a factor of two in each dimension after each of the first two ResNet blocks. The architecture is then mirrored, with 3 ResNet blocks with 2 upsampling layers to return to the original data shape. Skip connections are used to ensure no information bottleneck. Conditioning variables --- the diffusion noise level and the incident particle energy --- are processed by a DNN and then added to the model in the middle of each ResNet block. 

We make several optimizations focused on the cylindrical geometry of shower datasets. This includes cylindrical convolutions that the respect the periodic nature of the angular dimension and a novel method to condition the convolutions on the layer and radial bin values.

To handle the irregular geometry of dataset 1, we introduce a new approach: Geometry Latent Mapping (GLaM). GLaM learns an embedding from the dataset 1 geometry to a perfectly regular cylindrical geometry. Unlike an autoencoder, the embedding space is larger than the input space, so that no information is lost. The mapping is learned separately for each layer and initialized based on the geometric overlap of the input cells with a perfect cylinder. The cylindrical data is then processed using the cylindrical convolutions and then a reverse embedding is learned to restore the original shape.

\FloatBarrier

\submHeadlineSingle{CaloClouds: Fast Geometry-Independent Highly-Granular Calorimeter Simulation}{Erik Buhmann, Sascha Diefenbacher, Engin Eren, Frank Gaede, Gregor Kasieczka, Anatolii Korol, William Korcari, Katja Krüger, and Peter McKeown}{subsec:caloclouds}{\submKorolCite}{submKorolGit1,submKorolGit2}
\begin{figure}[ht]
    \begin{subfigure}[htbp]{0.40\textwidth}
         \centering
         \includegraphics[width=\textwidth]{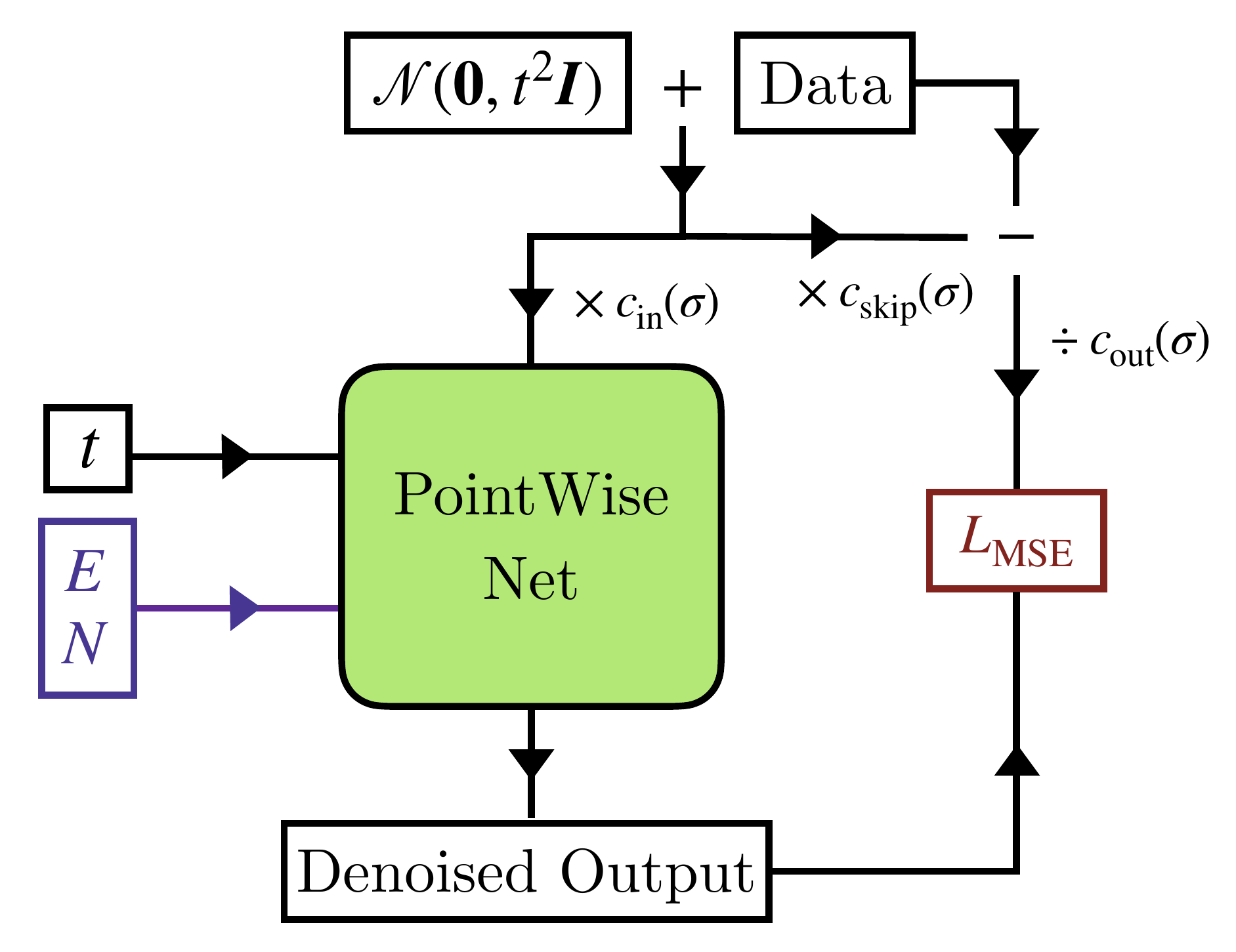}
         \caption{Training}
         \label{fig:training_CC2}
    \end{subfigure}
    \hspace{0.1pt}
    \begin{subfigure}[htbp]{0.55\textwidth}
         \centering
         \includegraphics[width=\textwidth]{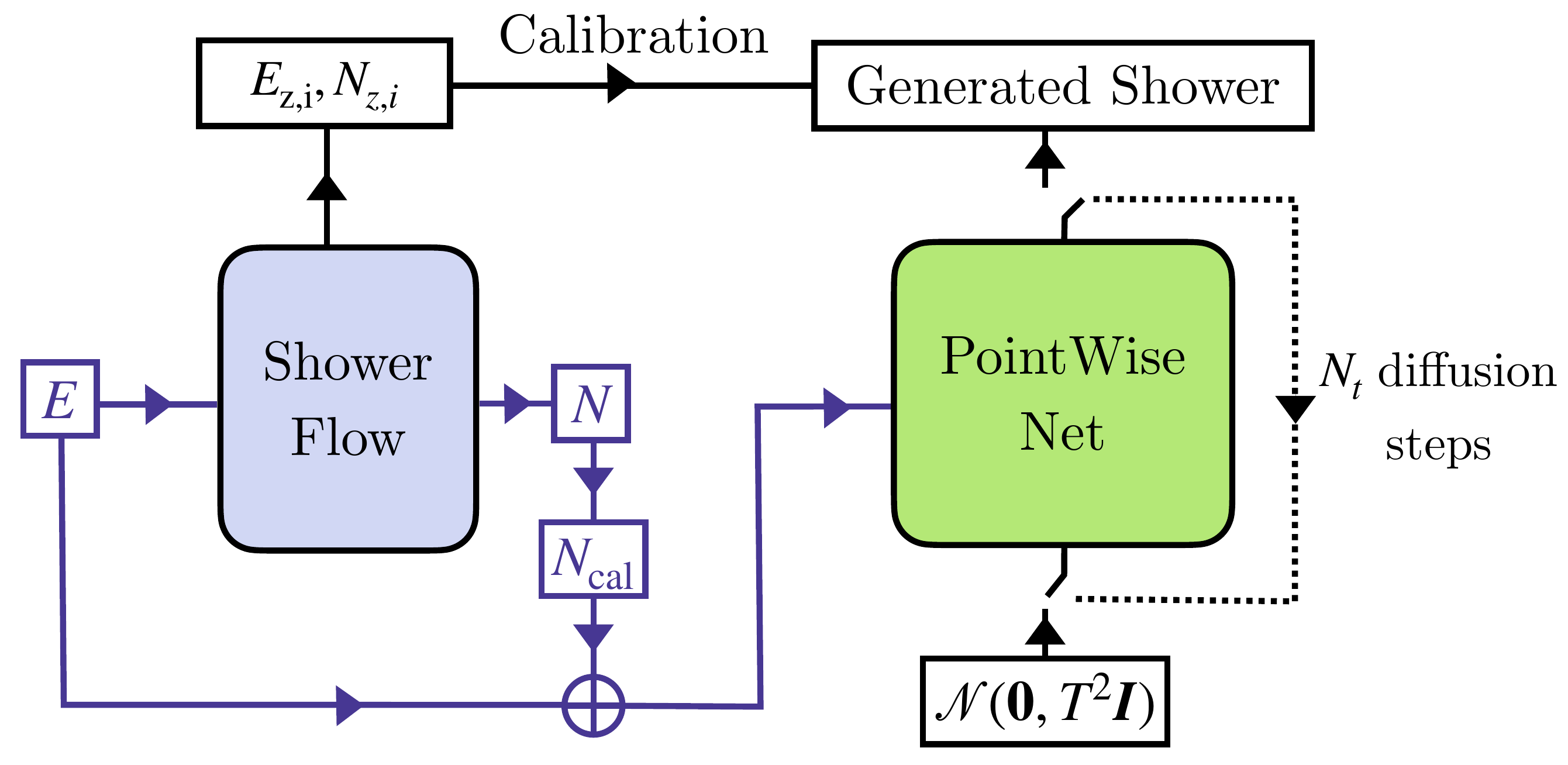}
         \caption{Sampling}
         \label{fig:sampling_CC2}
    \end{subfigure}
\caption{
\submKorol training and sampling pipeline, taken from~\cite{Buhmann:2023kdg}.  \textbf{Left:} Training of the PointWise Net. The training of the Shower Flow is not shown. \textbf{Right:} Sampling of \submKorol.
}
\label{fig:CaloCloud_model}
\end{figure}

\paragraph{Introduction}
The \submKorol model family was introduced in~\cite{Buhmann:2023bwk}. The improved version, \texttt{CaloClouds~II}~\cite{Buhmann:2023kdg} with consistency distillation~\cite{song2023consistency}, is here adapted to dataset~3.
As a point cloud generative model, \submKorol consists of two sub-models: A normalizing flow model dubbed \textit{Shower Flow} and a diffusion model named \textit{PointWise Net}.
An overview of the training and sampling pipeline is shown in~\fref{fig:CaloCloud_model}.

\paragraph{Architecture}
As the names suggest, the Shower Flow generates several global calorimeter shower observables, \textit{i.e.}~the layer-wise visible energy, the layer-wise number of hits, as well as the center of gravity (center of energy) in $x$- and $y$-direction.
The Shower Flow is conditioned on the incident particle's energy. 
We implemented it with a single bijector consisting of ten normalizing flow blocks, each containing seven coupling layers. Out of those, six are affine transformations~\cite{Dinh:2016pgf} and one a RQS~\cite{durkan2019neural_spline_flows}. 
The generated observables are used for a post-diffusion calibration of said shower features and the total number of hits is used for the conditioning of the diffusion model.

The PointWise Net diffusion model is conditioned on the incident energy as well as the number of hits. 
The diffusion model is based on the implementation in~\cite{karras2022elucidating_EDM_diffusion} and the network architecture is adapted from~\cite{luo2021diffusion}.
As the name suggests, it generates each calorimeter hit i.i.d. (independent and identically distributed). 
From a shower physics perspective, this i.i.d. assumption is inaccurate, however it yields decent performance and allows for a fast sampling necessary for large cardinality calorimeter showers such as the ones studied in~\cite{Buhmann:2023bwk, Buhmann:2023kdg} and in dataset~3. 
Layer structures taking into account inter-point correlations could be considered for smaller point clouds such as dataset~1 and are implemented, \textit{i.e.}~in \submKobylyanskyCite. 

\paragraph{Training}
Both the Shower Flow and the PointWise Net are trained separately and used sequentially during sampling of each batch. 
For the generation with the PointWise Net, we apply 13 diffusion steps with the Heun ODE solver, resulting in 25 function evaluations per batch. 
To apply the \submKorol model to dataset 3, we transformed each shower into a point cloud with four features: hit energy and the Cartesian coordinates in 3-dimensional space. We normalized the 3-dimensional coordinates to the range $\{x,y,z\} \in [-1,1]$.
This results in calorimeter shower point clouds with a cardinality of up to 20\,000~hits. 
To compare the \submKorol model to other models, we projected the calorimeter point clouds back to the (voxelized) fixed grid. 
As this leads to a clustering of a few generated point into a single voxel, we apply a calibration step between the predicted number of hits $N$ and the (larger) calibrated generated number of hits $N_\mathrm{cal}$ used for conditioning of PointWise Net. 

In~\cite{Buhmann:2023bwk, Buhmann:2023kdg}, the \submKorol models are used to generate point clouds with each point clouds representing clustered \geant steps (simulated energy depositions) with a resolution $36$ times higher than the actual resolution of the simulated electromagnetic calorimeter. A subsequent projection of there ultra-high granular calorimeter point clouds (up to 6000 points per shower) into the regular high-granularity calorimeter cells (up to 1500 calorimeter hits) results in a clustering of points leading to a precise estimation of the number of (regular cell) hits. 
As dataset 3 only contains regular cell hits, we expect the \submKorol model performance to improve when using a ultra-high granularity point cloud dataset.
Nonetheless, even when generating showers with regular cell hits we observe a decent performance.
Further speed-ups of the diffusion process could be achieved by applying consistency distillation~\cite{song2023consistency}, which allows for single shot generation without significant loss in fidelity. 

\FloatBarrier

\submHeadlineMultiple{Score-based Generative Models for Calorimeter Shower Simulation}{Vinicius Mikuni and Ben Nachman}{subsec:caloscore}{\submMikuniCite, \submMikuniDistCite, and \submMikuniSingleCite}{submMikuniGit1,submMikuniGit2}
\paragraph{Introduction}
Continuous diffusion generative models, or score-based models aim to approximate the score function of the data $\nabla\log(p(x))$ for data described by the probability density $p(x)$. The advantage of this approach is that both stochastic and deterministic solvers can be used for the generation of new observations, often leading to faster sampling times. The first diffusion generative model applied to collider physics problems was introduced in~\cite{Mikuni:2022xry} and later updated to improve the generation quality and speed in~\cite{Mikuni:2023tqg}. In the updated version, a neural network output $v_{\theta}(x_t,t)$ is used to calculate the loss function by minimizing the quantity
\begin{equation}
    \mathcal{L} = \textbf{E}_{x_t,t} \left\| v_t - v_{\theta}(x_t,t)\right\|^2.
    \label{eq:loss_v2}
\end{equation}
The velocity term $v_t \equiv \alpha_t\epsilon-\sigma_t x$ is calculated based on data $x_t$ that has been perturbed by a time-dependent Gaussian perturbation $q(x_t|x) = \mathcal{N}(x_t,\alpha_t x,\sigma_t^2 I)$. The velocity parametrization is observed to lead to a lower variance loss, improving the quality of the generated samples. The approximation to the score function $s_\theta(x_t,t)$ is identified as 
\begin{equation}
    s_\theta(x_t,t) =  x_t  - \frac{\alpha_t}{\sigma_t}v_\theta(x_t,t).
\end{equation}
New samples are generated from the trained model by solving the following ordinary differential equation,
\begin{equation}
    \frac{\mathrm{d}x_t}{\mathrm{d}t} = f(x,t)-\frac{1}{2}g(t)^2\nabla_x\log q(x_t)\,,
    \label{eq:ode}
\end{equation}
with the DDIM solver~\cite{DBLP:journals/corr/abs-2010-02502} with update rule then specified by
\begin{equation}
    x_s = \alpha_s x_\theta(x_t,t)  + \sigma_s\frac{x_t -\alpha_t x_\theta(x_t,t)}{\sigma_t},
\end{equation}
for time $s<t$ and position prediction $x_\theta(x_t,t) = \alpha_t x_t - \sigma_t v_\theta(x_t,t)$. While the solver still require a large number of function evaluations ($\mathcal{O}(100)$), we are able to reduce this number trough a distillation procedure~\cite{salimans2022progressive}, resulting in faster generation times requiring even a single step for the generation. 
\paragraph{Architecture}
The neural network architecture used for the training is similar to the one used in the initial 
\submMikuni paper, based on the U-net~\cite{Unet} architecture with additional attention layers. More specifically, datasets 2 and 3 have the number of spatial components in each dimension reduced by a factor 2 every other convolutional layer (resulting in a factor $2\times2\times2=8$ reduction) with fixed kernel size set to 3. This process is repeated 3 times, with lowest dimensional representation reduced by a factor 512 compared to the initial number of voxels. The 3-dimensional convolution operations used for datasets 2 and 3 use 32, 64, and 96 hidden nodes  with swish~\cite{ramachandran2017searching} activation function. The attention layer is only used at the lowest dimensional representation, with data patches determined by the flattened array describing the data at the lowest dimensionality. The upsampling section of the architecture is a mirrored version, with dimensions increased by a factor 8 every other layer. Skip connections between the downsampling and upsampling sides of the architecture are combined with a concatenation operation, completing the architecture. Conditional information consisting of the time information, incident particle energy, and deposited energy per layer (in case of the diffusion model trained to generate normalized voxels), are included through an addition operation after every convolutional layer. A trainable embedding of the conditional features is created by a fully connected layer over the conditional inputs. The output size is fixed to match the expected output size of the convolutional layers. For dataset 1, the strategy is similar. The number of voxels to be simulated is reduced by a factor 2 every other layer, with this process repeated 4 times and overall reduction of factor 16 compared to the initial size. The number of hidden nodes for the 1-dimensional convolutional layers is then chosen to be 16, 32, 64, and 96 for each fixed dimensionality. Since this dataset is smaller compared to datasets 2 and 3, attention layers are used in all lower dimensional representations of the initial data. 
A second diffusion model is introduced to learn only the energy deposition per layer, similar to the approach used in the original \cf\ paper~\cite{Krause:2021ilc}. The model used to train the diffusion model is based on the ResNet~\cite{RESNET} architecture, consisting of multiple fully connected layers with additional skip connections. The number of ResNet layers is set to 3 in all datasets, with 128 hidden nodes in dataset 1 and 1024 in datasets 2 and 3. Additional models are trained to reduce the sampling time of the baseline \submMikuni model. The model architecture of the distilled version is the same as the baseline model, also using the initial baseline weights as the starting point to accelerate the training procedure. 
\FloatBarrier

\submHeadlineSingle{CaloGraph}{Dmitrii Kobylianskii, Nathalie Soybelman, Etienne Dreyer, and Eilam Gross}{subsec:calograph}{\submKobylyanskyCite}{submKobylyanskyGit}
\begin{figure}[ht]
    \centering
    \includegraphics[trim={0 15cm 0 10cm},width=0.7\textwidth]{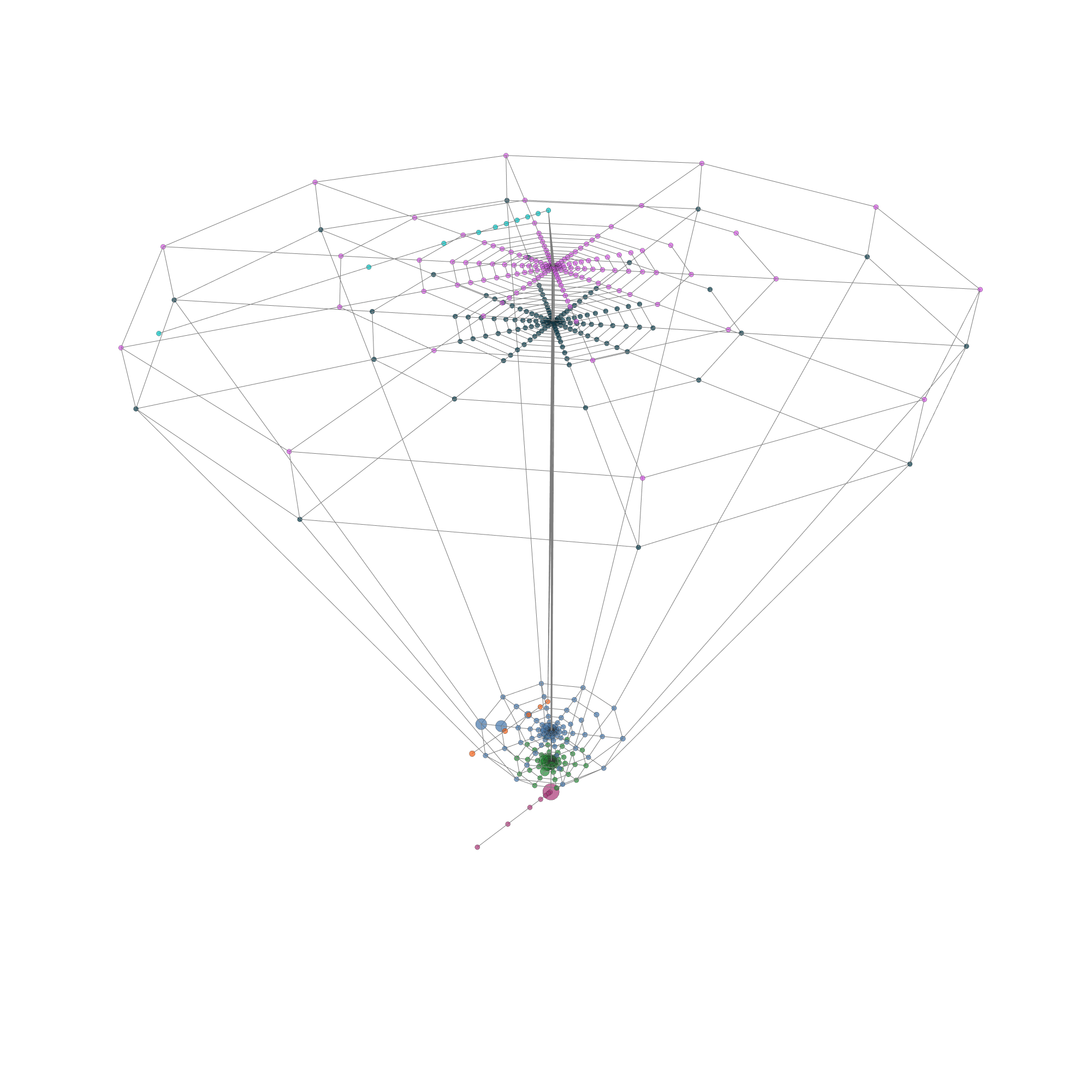}
    \caption{Event display from the pion dataset in graph form, taken from~\cite{kobylianskii2024calograph}. The nodes represent the center of the calorimeter cells, and their size relates \iffalse linearly \fi to the cell energy.}
    \label{fig:pion-graph}
\end{figure}

\paragraph{Introduction}
\submKobylyanskyCite stands out as a diffusion model based on graphs specifically designed for low-granularity calorimeters with irregular geometries, such as those found in ATLAS. Unlike image-based methods that necessitate unique mappings for non-regular geometries and point cloud generation techniques that predict point positions requiring specific grid summation, a graph representation requires no pre- or postprocessing, except for the initial one-time graph construction. Calorimeter cells are nodes in the graph with fixed positions, and edges connect nearest neighbors within the given layer and the layers below and above. An example from the pion dataset 1 is shown in~\fref{fig:pion-graph}.
\begin{figure}[b]
    \centering
    \includegraphics[width=\textwidth]{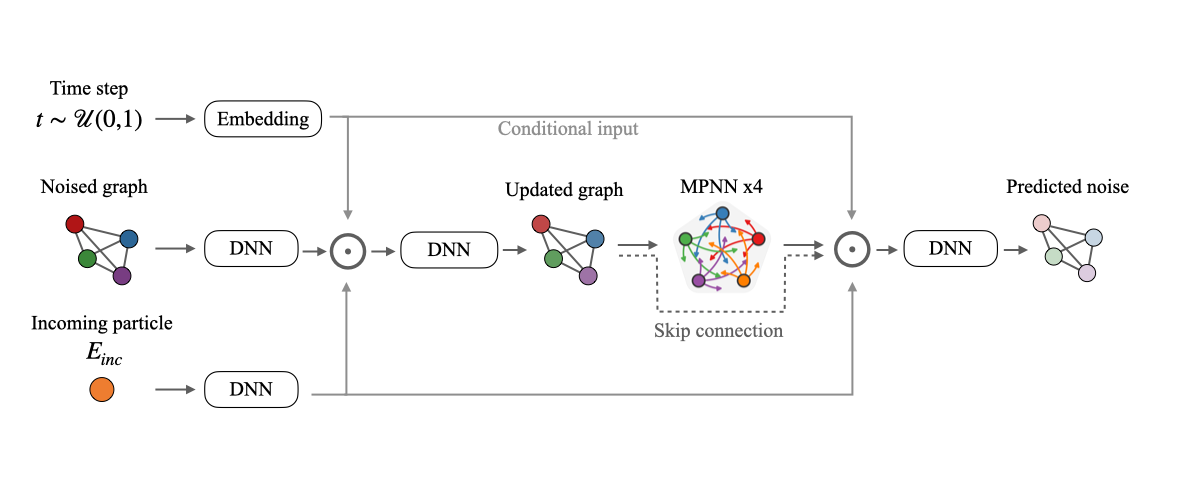}
    \hspace{-30pt}
    \caption{Architecture of \submKobylyansky, adapted from~\cite{kobylianskii2024calograph}.}
    \label{fig:calograph}
\end{figure}
However, managing a large number of edges will result in high memory needs, making this approach mostly suited for low-granularity detectors. Therefore, we present results only for dataset 1. 

\paragraph{Architecture}
Our approach relies on a denoising diffusion model. We use a cosine noise schedule from~\cite{improved_diffu}. During the backward process, we use the ``pseudo numerical methods for diffusion models'' (PNDM) from~\cite{liu2022pseudo} to solve the diffusion ODE. The architecture of the neural network is presented in~\fref{fig:calograph}.

The input to the network is the noised target graph: constructed as described above, the node features consist of the cell position ($\eta$, $\phi$, layer) and the noised cell energy.
It is passed through the initial DNN and then combined with the conditional input consisting of the embedded, uniformly sampled time step as well as the incoming particle energy. The combined input is passed through another DNN, resulting in the updated graph. Subsequently, four rounds of message passing are applied before predicting the noise through a final DNN. The network has a total of 0.8 million parameters, see \tref{tab:ds1-photons.numparam} and \tref{tab:ds1-pions.numparam}.

\FloatBarrier

\submHeadlineSingle{Diffusion Transformer}{Renato Paulo Da Costa Cardoso, Piyush Raikwar, Anna Zaborowska, Dalila Salamani, Kristina Jaruskova, Sofia Vallecorsa, Kyongmin Yeo, Vijay Ekambaram, Nam Nguyen, Jayant Kalagnanam, and Mudhakar Srivatsa}{subsec:calodit}{\submCardosoCite}{submCardosoGit}
\paragraph{Introduction}
Currently, the state-of-the-art approach for image generation is diffusion, while the state-of-the-art architecture for almost any data modality is based on transformers~\cite{vaswani2017attention}. We combine both methodologies for a transformer-based diffusion model.

The use of transformer models for the image generation task is not new, with approaches such as Vision transformers (ViT)~\cite{dosovitskiy2021image}, Swin transformers~\cite{liu2021swin}, etc. getting good results for image classification tasks. When paired with the diffusion process, we have impressive generative models like OpenAI Sora~\cite{videoworldsimulators2024}. In our case, the model architecture is based on Diffusion with transformers (DiT)~\cite{peebles2023scalable}. As for the diffusion process, we go with denoising diffusion probabilistic models (DDPM)~\cite{ho2020denoising}, with the modification of using a cosine scheduler as described in~\cite{improved_diffu}. We present the results of \submCardoso for dataset 2.

\paragraph{Preprocessing}
We preprocess the input data to ease the diffusion process. The preprocessing is done by scaling the shower energies in the range of -1 and 1, followed by applying a logit function and normalization of those values. The energy condition is also preprocessed to be in the 0 to 1 range via scaling it by the max energy of 1000 GeV.

\begin{figure}
    \centering
    \includegraphics[width=\textwidth]{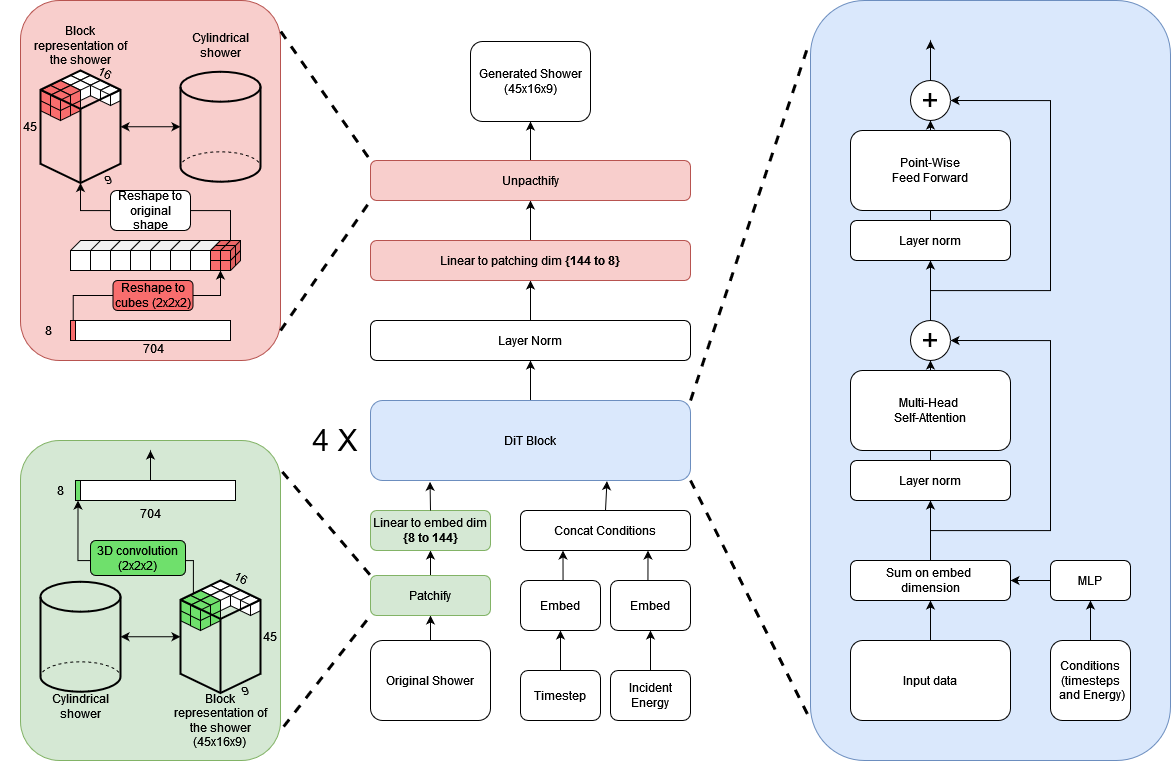}
    \caption{Architecture of \submCardoso.}
    \label{fig:caloditmodel}
\end{figure}

\paragraph{Architecture}
We define the architecture of the \submCardoso in figure~\ref{fig:caloditmodel}. We use a stack of 4 DiT blocks, which are ViT-like transformer blocks with a modified conditioning unit to accommodate diffusion timesteps. In our case, we also concatenate the energy condition along with the timestep, which is then passed to each DiT block. These conditions along with a noisy shower are passed to the model to get a denoised shower as an output. During inference, this process is repeated 400 times to generate a shower, where 400 is the number of diffusion steps our model uses.

As with any other transformer-based model, we need to represent the input as some form of sequence. While ViT and DiT are used for 2-dimensional images, the shower dataset is 3-dimensional. Thus, we split the 3-dimensional shower into multiple smaller 3-dimensional patches (patchify), 704 patches to be precise. These patches are linearly projected to a higher dimension of 144 and then passed to the model. We do the opposite for the output to combine smaller 3-dimensional patches into the 3-dimensional shower (unpatchify). Note that, we use a $2\times 2\times 2$ convolution layer for ``patchification'' to better extract the representations, but ``unpatchification'' is a simple reshape operation. Within each DiT block, the conditions are first passed through a DNN of 2 layers and then summed up with patch embeddings of size 144. After the final DiT block, we have layer normalization and a linear projection to match the shower dimensions, followed by unpatchification.

Since our sequence is 3-dimensional, we also adapt the sinusoidal positional embeddings~\cite{vaswani2017attention} from 2 dimensions to 3 dimensions to represent the patches in the 3-dimensional space. This is done by allocating space for an extra dimension in the positional embedding vector. These positional embeddings are added to the patches after their linear projection before the first DiT block.

\FloatBarrier

\section{VAE-based Submissions}\label{sec:vae}
\markboth{\uppercase{VAE-based Submissions}}{}
Variational Autoencoders (VAEs)~\cite{kingma2013auto,rezende2014stochastic} are a class of generative models which combine deep learning with probabilistic methods. A VAE is composed of two stacked neural networks acting as encoder and decoder. The encoder learns a mapping from the input space to a latent space in which a meaningful representation of the data is learned. The decoder learns the inverse mapping by reconstructing the original input from the latent representation. The VAE is designed with a prior on the representation space, hence once the model is trained to reconstruct the input, the decoder can be used independently as a generator of new data by sampling from the prior.  

Moreover, improved accuracy can be achieved by performing density estimation on the latent space. The VAE thus serves as a dimensionality reduction method by learning a low-dimensional manifold of the ambient space. The approach is explored with different techniques and density estimators in this section.

The key idea behind training VAEs lies in variational inference (VI). VI approximates probability densities based on the optimization of the Kullback-Leibler (KL) divergence~\cite{jordan1999introduction}. VI uses a family of densities and finds the closest member of that family to the target density using the KL divergence. The KL divergence is a fundamental quantity in information theory to measure the difference between two probability distributions. If a probability distribution $q$ is used to approximate $p$ then the KL divergence measures the loss in information using the approximation. 

\FloatBarrier

\begin{figure}[bh]
    \centering
    \includegraphics[width=.8\linewidth]{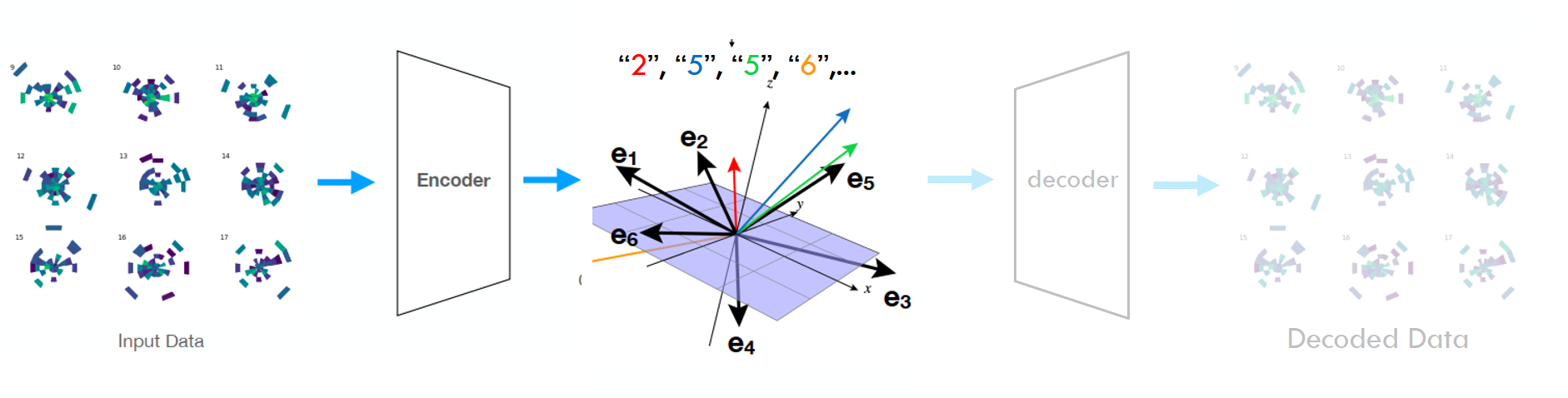}
    \includegraphics[width=.8\linewidth]{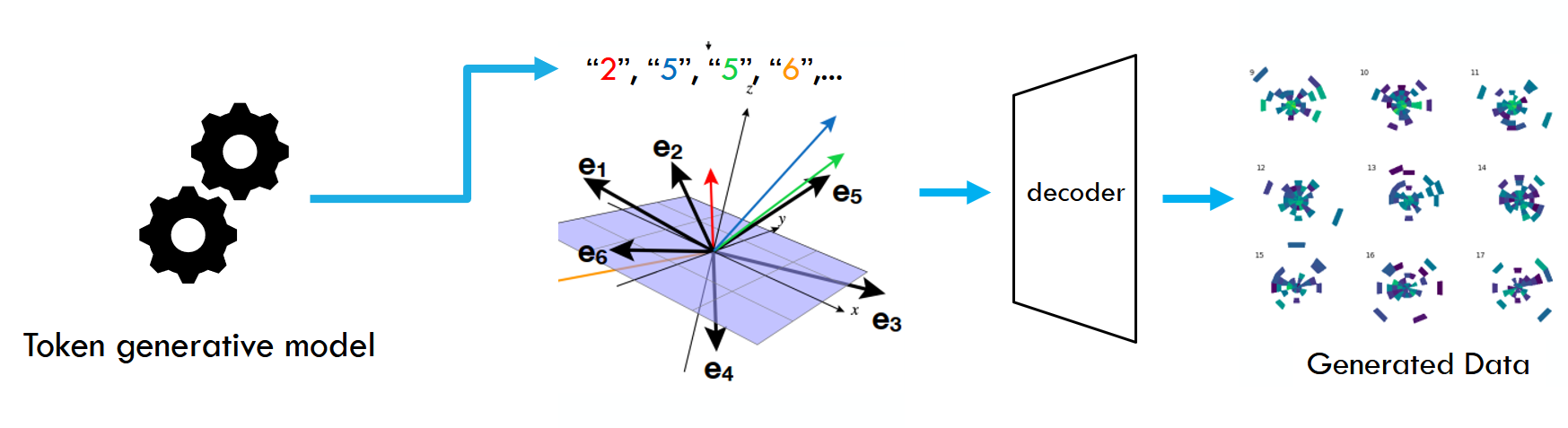}
    \caption{Demonstration of the \submLiu architecture, taken from~\cite{Liu:2024kvv}. The upper and lower parts show the two stages of the model, respectively.}
    \label{fig:caloVQ_demo}
\end{figure}

\submHeadlineMultiple{Latent Generative Models for Calo Simulation with VQ-VAE}{Qibin Liu, Chase Shimmin, Xiulong Liu, Eli Shlizerman, Shu Li, and Shih-Chieh Hsu}{subsec:vqvae}{\submLiuCite~and~\submLiuNormCite}{submLiuGit}
\paragraph{Introduction}
Calorimeters with high granularity often feature a large number of voxels, reaching up to tens of thousands. Directly sampling of such high-dimensional and highly sparse data is usually challenging and inefficient. To address this, a two-stage method, as illustrated in~\fref{fig:caloVQ_demo}, is proposed. This method is based on a vector-quantized variational auto-encoder (VQVAE) \cite{van2017neural} and a transformer-based token generative model \cite{vaswani2017attention,minGPT}. In the following, we describe the implementation of this model, including processing of calorimeter data, representation, architecture and training procedure.

\paragraph{Preprocessing}
To handle the large dynamic range of calorimeter energy, the data is normalized per detector layer (for \dsIph and \dsIpi) or the entire detector (for \dsII and \dsIII, except the model implicitly marked with ``norm'') for each sample. It is then transformed using a scaled and shifted logarithm. The pre-processing is described in~\eref{eq:caloVQ_preprocessing}.

\begin{equation}
x_i=\frac{1}{c}\log \left(a + b \frac{\mathcal{I}_{i}}{E_{\rm sum}}\right )
\label{eq:caloVQ_preprocessing}
\end{equation}

Here, $i$ is the index of the voxel, $\mathcal{I}_i$ is the original value of each voxel, and $E_{\rm sum}$ is the sum of energies within the same layer or the entire calorimeter, depending on the dataset. The hyperparameters $a$, $b$, and $c$ are tuned according to the input range.

\paragraph{Architecture}
The first stage of the model aims to reduce the dimensionality of the input. The encoder transforms the input into the representation in the latent space, followed by the decoder, which reconstructs the input. To achieve high compression ratio and effective usage of latent space, a vector quantization technique \cite{van2017neural} is implemented. This technique labels each latent vector with one index of a fixed set of representative ``code vectors''. The code vectors are updated during training, minimizing the quantization loss and commitment loss, as shown in the following,

\begin{equation}
L_{VQ} = ||\sg[q(z|x)] - e_k||^2 + ||q(z|x) - \sg[e_k]||^2
\end{equation}

Here $\sg[X]$ denotes the stop-gradient operator which will not take the gradient of $X$ into calculation. The first term denotes the quantization loss which moves the codebook ($e_k$) to better represent the latent space ($z$). The second term is called commitment loss which limits the arbitrarily growth of embedding space and makes the encoded vector commit to one of the codes.

The L2 distance between input and decoded output, following the vanilla VAE architecture, is used as one loss term. Additionally, the discriminator loss~\cite{esser2021taming} and physics-aware losses are added to improve the quality of reconstruction, particularly for the detailed feature. 

\begin{figure}[ht]
    \centering
    \includegraphics[width=.6\linewidth]{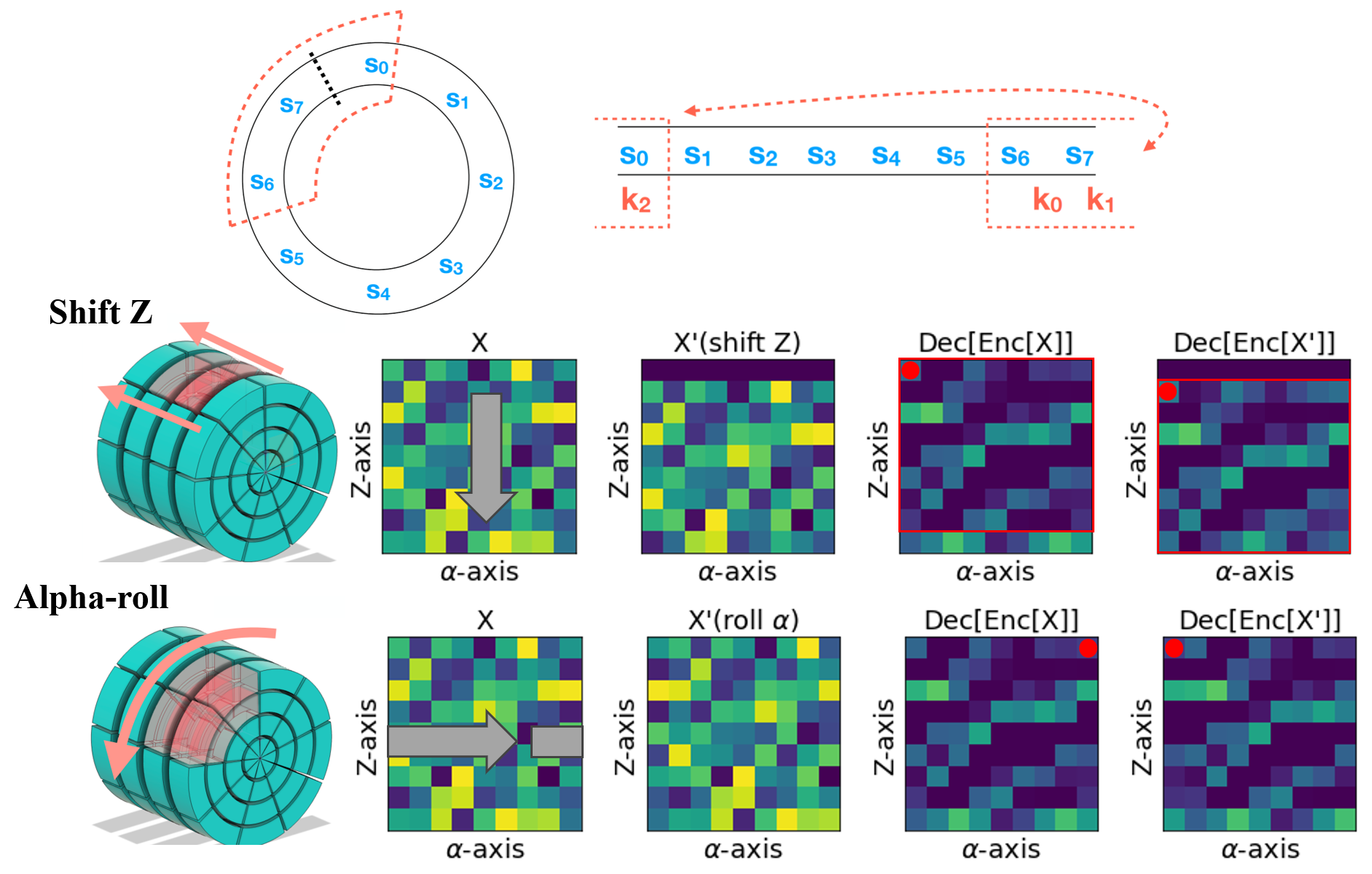}
    \caption{Cylindrical convolution operator, taken from~\cite{Liu:2024kvv}. The bottom plots show the equivariant property is well kept. }
    \label{fig:caloVQ_cyconv}
\end{figure}

The encoder and decoder consist of convolutional and dense layers. For photon and pion datasets 1, 1-dimensional convolution and fully connected layers are combined to better process the irregular geometry. For datasets 2 and 3, since the transitional symmetry only exists in the $z$ (depth) and $\alpha$ (angular) direction, 2-dimensional convolutions are used, treating the 3-dimensional data as a multi-channel 2-dimensional image defined on the $(Z,\alpha)$ coordinates. The cylindrical convolution operation is shown in~\fref{fig:caloVQ_cyconv}. It maintains the equivariant property of the calorimeter data mapped in cylindrical coordinates. The radial direction is therefore treated as the channel of this image-like representation.

\begin{figure}[ht]
    \centering
    \includegraphics[width=.5\linewidth]{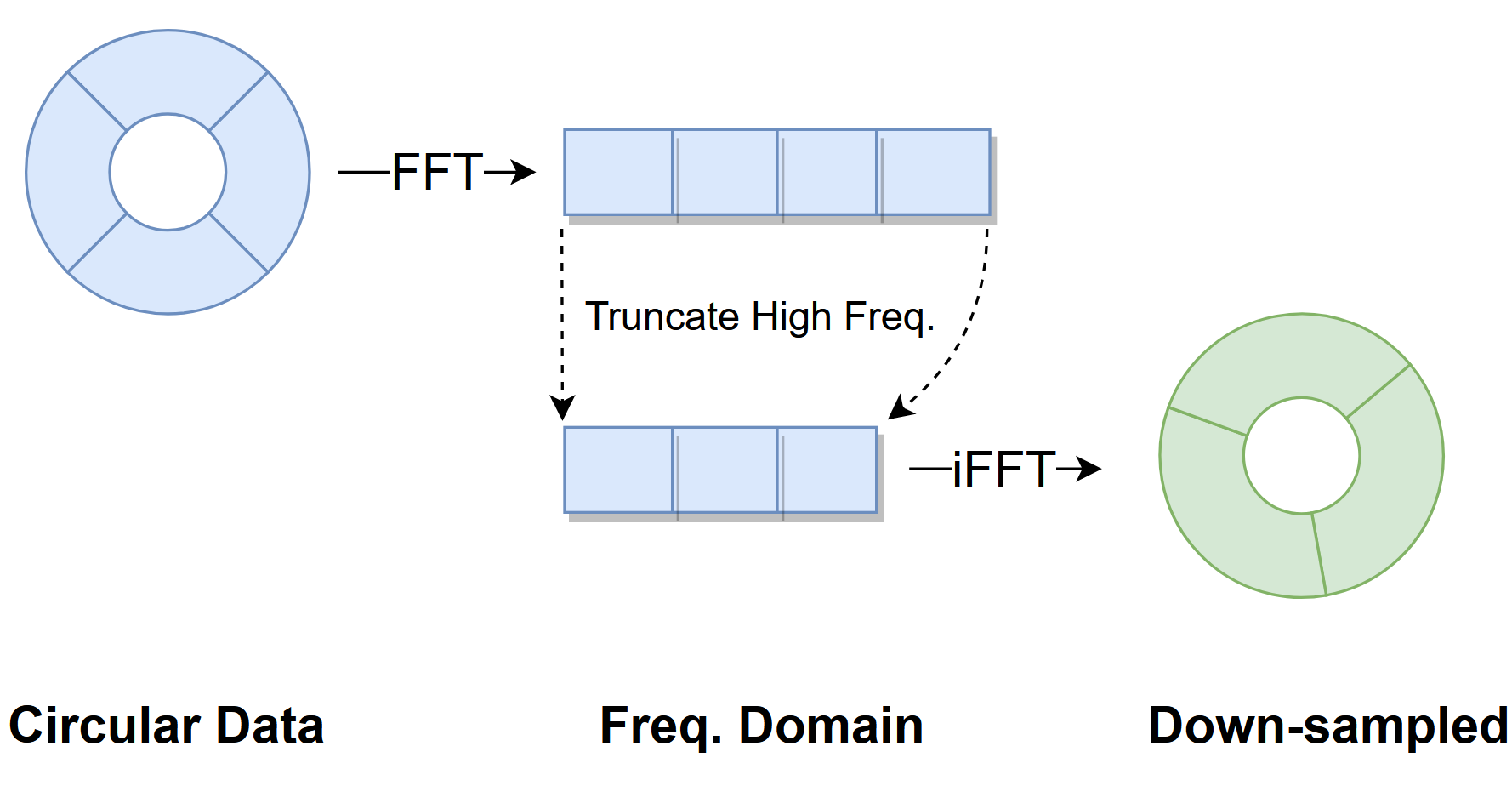}
    \caption{Illustration of the FFT down-sampling, taken from~\cite{Liu:2024kvv}. }
    \label{fig:caloVQ_FFT}
\end{figure}

Arbitrary up-/down-sampling on the angular direction with circular (periodical) boundaries is achieved using the FFT-resampling method, as illustrated in~\fref{fig:caloVQ_FFT}. The data along the angular direction is first transformed with discrete FFT into frequency space, then truncated to the desired dimension and transformed back with inverse operation.

The softmax activation is used in the last layer to ensure the correct normalization as mentioned in~\eref{eq:caloVQ_preprocessing}. By definition, the output of softmax satisfies the required normalization in log scale with proper shift and scaling. 

The second stage model focuses on learning and sampling of the probabilistic distribution, in a highly reduced and regularized latent space, characterized by a fixed-length sequence of discrete codes (tokens). The transformer-based model minGPT~\cite{minGPT} is adapted to sample the latent codes, conditioned on the incident energy. As discussed in the previous section, the normalization factor(s) $E_{\mathrm{sum}}$, which control the energy response of the entire calorimeter or each calorimeter layer, is digitized to discrete codes and sampled together with the latent codes of the first stage. Two codes are used to digitize one floating number with 20bit accuracy assuming 1024 choices for each code are utilized.

The final sequence to learn for the second stage is
\begin{equation}
\underbrace{A_1 A_2 ...}_{E_{\rm sum}~codes}\overbrace{B_1 B_2 B_3 B_4 ...}^{latent~codes}
\end{equation}

The parameters of the second stage model are tuned towards less transformer heads, layers and embedding while retaining the same quality of generated data (such as smaller error on $E_{\rm sum}$). 

\paragraph{Training}
The first stage model is trained adversarially, updating the encoder/decoder and the discriminator alternately. Learning rate is constant during the training and the model with best reconstruction loss on validation dataset is selected after training for a fixed number of epochs. 

Then the quantized latent space, as discrete numbers (tokens), is used to train the second stage model. The training objective is to minimize the cross-entropy between the predicted token based on previous ones and the truth token. A constant learning rate is used and the model with best validation loss is chosen. 

The main hyper-parameters are summarized in~\tref{tab:caloVQ_parameters}.

\begin{table}[ht]
\centering
\begin{tabular}{cccccc}
\toprule
                              & \dsIph & \dsIpi & \dsII  & \dsIII   &\dsIII(norm)   \\
\midrule
Pre-processing~\eref{eq:caloVQ_preprocessing}: & & & & & \\
a              & 1        & 1        & 1    & 1       & 1     \\
b              & 8000     & 8000     & 3000 & 40\,000   & 40\,000 \\
c              & 10       & 10       & 7    & 10      & 10   \\
\midrule
Stages-1: & & & & & \\
Hidden layer                 & 5        & 5        & 7    & 10      & 10   \\
VQ dim                        & 256      & 256      & 256  & 192     & 256  \\
Condition dim                     & 3        & 3        & 1    & 1       & 1   \\
Codebook size                 & 1024     & 1024     & 1024 & 1024    & 1024 \\
R code                       & 10       & 14       & 2    & 2       & 90   \\
Shower code                  & 32       & 32       & 68   & 182     & 624  \\
\#pars / M               & 3.8      & 4.1      & 3.1  & 2.1     & 2.2 \\
\#pars (gen) / M           & 1.9      & 2.0      & 1.0  & 0.8     & 0.9 \\
\midrule
Stages-2: & & & & & \\
Layer                 & 2        & 2        & 2    & 1       & 1   \\
Head                  & 2        & 2        & 2    & 1       & 1   \\
Embed                 & 64       & 64       & 64   & 16      & 128 \\
\#pars / K               & 231      & 231      & 235  & 38      & 551 \\
\bottomrule
\end{tabular}
\caption{Setup of hyper-parameter and number of trainable parameters of preprocessing, first stage model and second stage model of \submLiu, taken from~\cite{Liu:2024kvv}. The numbers for ``hidden\_layers'' are halved since symmetric encoder and decoder. Only the decoder and quantization module in stage 1 are used in generation mode and the number of parameters are denoted with ``(gen)'' in the table. }
\label{tab:caloVQ_parameters}
\end{table}
\FloatBarrier

\submHeadlineSingle{CaloMan: Fast Generation of Calorimeter Showers with Density Estimation on Learned Manifolds}{Jesse C.~Cresswell, Brendan Leigh Ross, Gabriel Loaiza-Ganem, Humberto Reyes-Gonzalez, Marco Letizia, and Anthony L.~Caterini}{subsec:caloman}{\submReyesCite}{submReyesGit}
\paragraph{Introduction}
As surveyed in the present work, many types of generative models have been used to model calorimeter showers, and particular emphasis has been given to normalizing flows~\cite{Krause:2021ilc, Krause:2021wez}. Despite their expressivity, NFs suffer from the fact that they model a density that has the same dimensionality as the input data. For high-dimensional data, this would mean dealing with very large NF models that compromise training and prediction speed. For calorimeter showers, size and speed can quickly become major problems as the dimensionality of raw shower representations surpasses $10^{4}$. However, we expect that shower generation is governed by simple underlying physical processes, and thus can be represented in a much lower dimensional space. In the context of machine learning, this is an example of the \textit{manifold hypothesis}, which states that high-dimensional natural data actually lies on a low-dimensional embedded sub-manifold in the ambient space \cite{bengio2013representation,pope2021,brown2022union}.

\begin{figure}
\begin{center}
\centerline{
\includegraphics[width=0.3\textwidth,trim=20 125 80 140, clip]{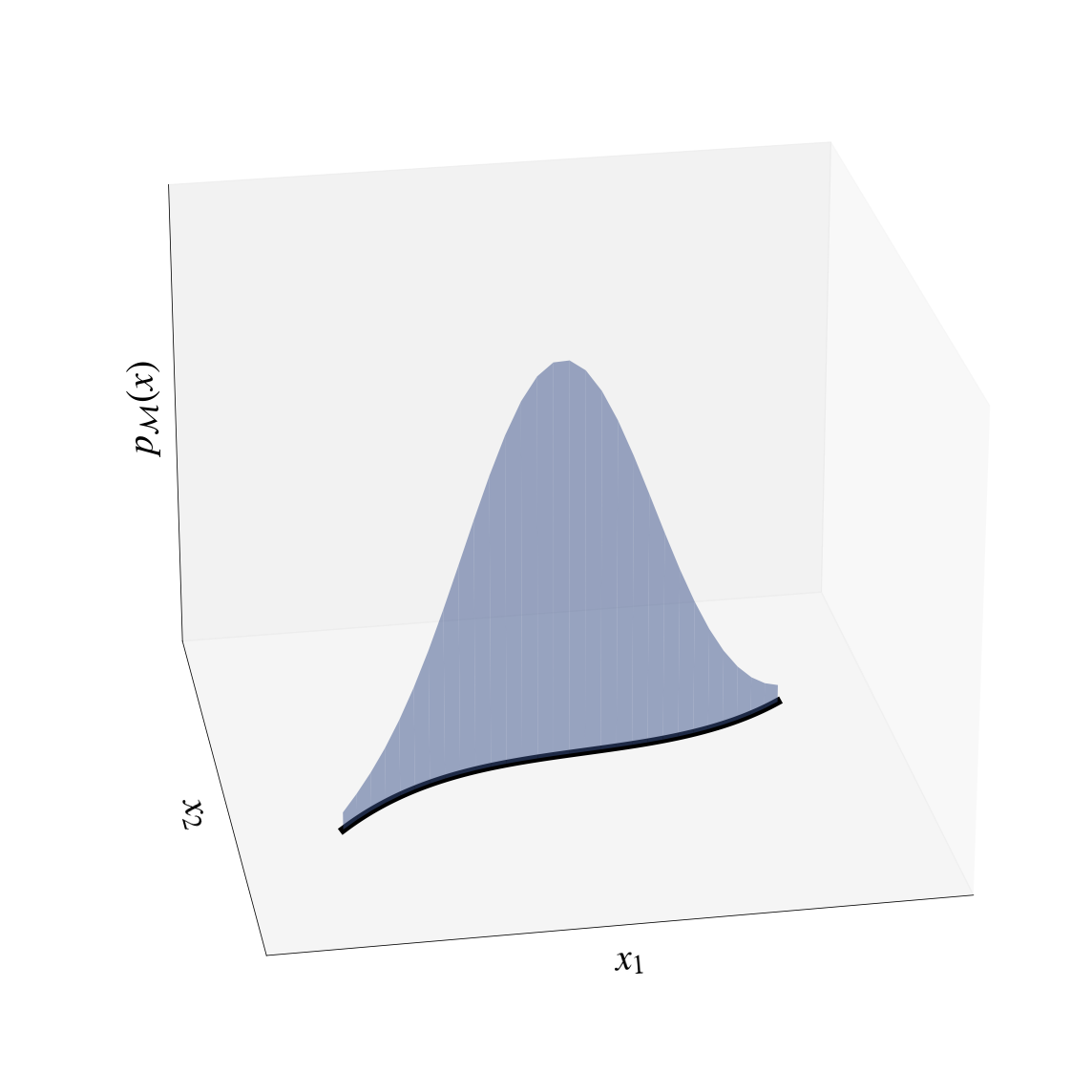}
\includegraphics[width=0.3\textwidth,trim=80 125 20 140, clip]{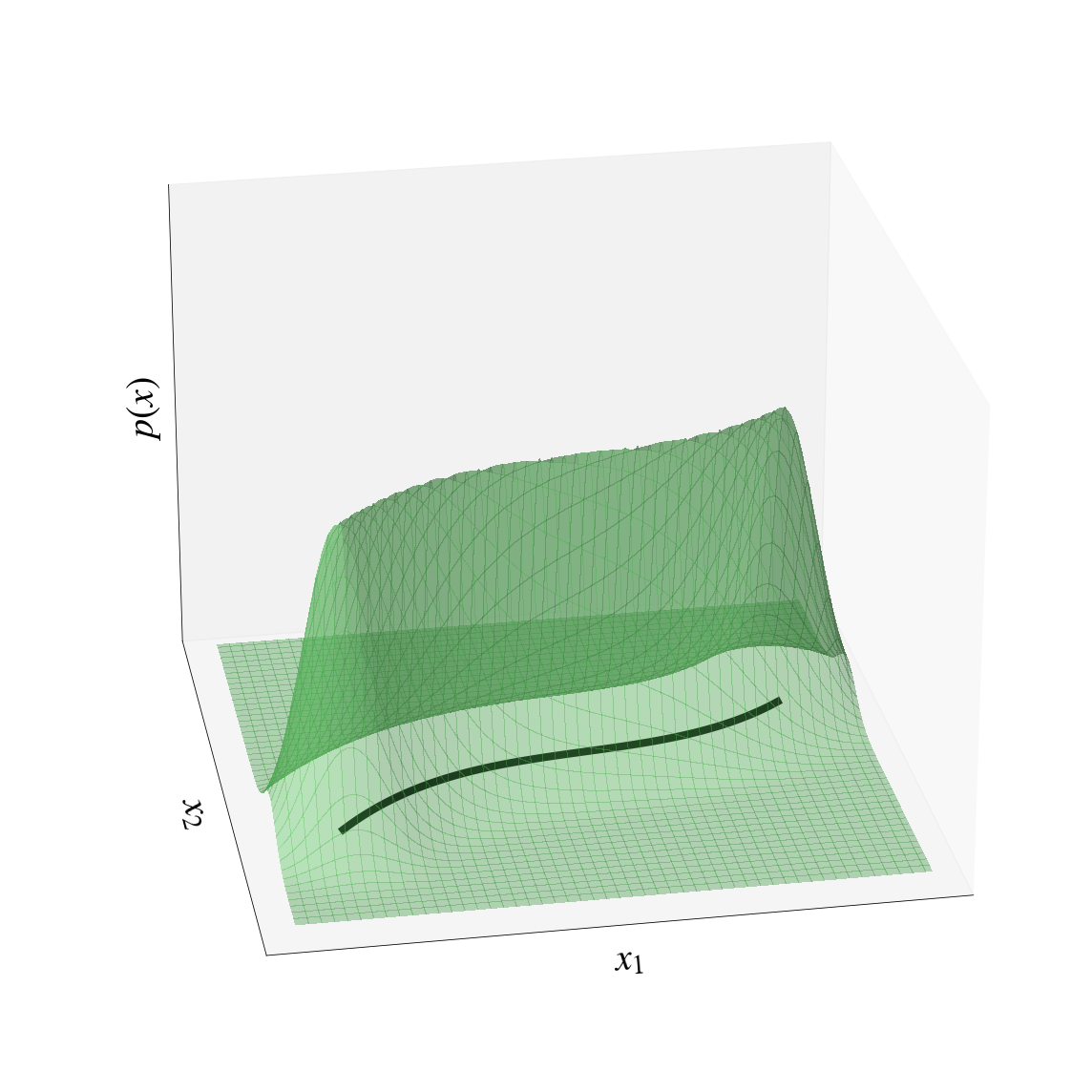}
}
\caption{A low-dimensional density on a manifold (left), and a full-dimensional density model undergoing manifold overfitting (right). Although the full-dimensional model concentrates around the manifold, it distributes the density incorrectly along the manifold. Figure taken from \cite{Cresswell:2022tof}.}
\label{fig:manifold_overfitting}
\end{center}
\end{figure}

Moreover, maximum-likelihood methods, including NFs, rely on the assumption that the underlying distribution possesses a full-dimensional probability density $p(x)$ in the ambient space. This may not always be the case: if the data is confined to a low-dimensional manifold, the data manifold is a subset of measure zero, over which no continuous density can be integrated to obtain non-zero probabilities. In this situation, training a likelihood-based model typically leads to densities that spike to infinity around the manifold, but not in accordance with the data distribution. This phenomenon, illustrated in~\fref{fig:manifold_overfitting}, is known as \textit{manifold overfitting}~\cite{loaiza2022diagnosing, loaiza2024survey}.

To avoid manifold overfitting, while also delivering fast, light-weight models, we propose \submReyes, which follows the two-step procedure outlined by \cite{loaiza2022diagnosing} to build our calorimeter shower simulators. The first step of the approach is to learn a lower dimensional manifold using a \textit{generalized autoencoder}. This can be any ML model capable of learning a latent space, and transforming it back to the ambient space. Examples include autoencoders \cite{rumelhart1985learning}, variational autoencoders  \cite{kingma2013auto}, Wasserstein autoencoders \cite{tolstikhin2017wasserstein}, bidirectional GANs \cite{ donahue2016adversarial, dumoulin2016adversarially}, and adversarial variational Bayes \cite{mescheder2017adversarial}. The second step is to perform \textit{density estimation} on the learned manifold. Any explicit likelihood estimator can be used. This includes NFs, energy based models \cite{du2019implicit}, auto-regressive models \cite{uria2013rnade}, score-based models \cite{song2019generative}, and diffusion models \cite{ho2020denoising}.

\paragraph{Intrinsic dimensionality of CaloChallenge datasets}
Most methods for learning low-dimensional manifold structure require the dimensionality $d$ to be provided as an input. Hence, we applied methods for intrinsic dimension estimation \cite{campadelli2015}, which also shed light on the fundamental nature of the calorimeter shower data.

\begin{figure}[t]
\centering
  \includegraphics[width=0.4\textwidth, trim={0 0 0 0}, clip]{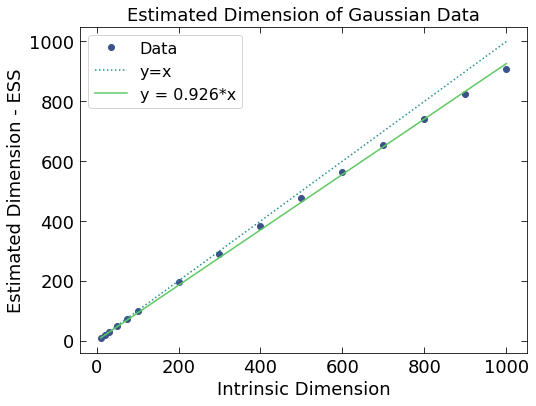}
  \quad
  \includegraphics[width=0.56\textwidth, trim={0 0 0 0}, clip]{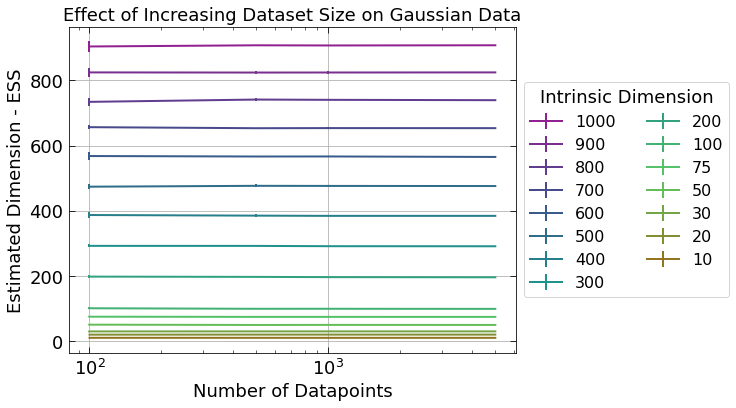}
    \caption{Benchmarking the ESS estimator. \textbf{Left:} As the dimension of random Gaussian data is increased, the ESS estimate of intrinsic dimension linearly increases ($n=5000$). Standard deviations are smaller than the size of the dots. \textbf{Right:} ESS provides consistent estimates regardless of the number of datapoints used. Notably, accurate estimates can be obtained even when the number of datapoints is less than the dimensionality. The error bars show standard deviations over 10 seeds.}
    \label{fig:ess}
\end{figure}

Here we use the Expected Simplex Skewness (ESS) estimator \cite{johnsson2015} which is based on angular information between $k$-nearest neighbor points in a dataset. Most literature on estimating the intrinsic dimension of datasets focuses on relatively low dimensions (\textit{i.e.}~$d\leq 20$) and few datapoints $n$. Hence, we first benchmark ESS on synthetic datasets of known dimensionality $d$ more comparable to CaloChallenge data, using the implementation from \cite{bac2021} with default hyperparameters. We randomly generate $n$ datapoints from $d$-dimensional Gaussian distributions, with $n\in\{100,\ 500,\ 1000,\ 5000\}$, and $d$ spanning values from 10 to 1000, and apply the ESS estimator to each dataset. We repeat the test for 10 random seeds on each setting. \Fref{fig:ess} shows that ESS has a consistent linear behavior as $d$ is increased, and that it is insensitive to $n$, even when $n<d$. Noticing a slight negative bias, we fit a scale factor to the data for $n=5000$, and use its inverse to calibrate the ESS estimates on shower data. We also repeat the experiment for data drawn from a uniform distribution on a hypercube with extremely similar results. The calibration scaling factors are $1.0795$ for Gaussian data, and $1.0793$ for Uniform data\footnote{We acknowledge that the synthetic data we used does not itself have low-dimensional structure, however similar benchmarking has been done for such data with consistent results \cite{tempczyk2022}. More recent work shows some limitations of ESS \cite{kamkari2024geometric}.}.

With evidence that ESS can scale to high-dimensional datasets, we applied the ESS estimator to the four CaloChallenge training datasets without any preprocessing of the data, and correct the results with our calibration factors. The intrinsic dimension estimates are given in~\tref{tbl:estimates}, and we based our latent space dimensions on these values in the following. The results provide evidence for low-dimensional structure in calorimeter data, and emphasize the potential efficiency gains over approaches that model a full-dimensional latent space.

\begin{table}[t]
\caption{Estimated Dimensionality $d$ with Gaussian (G) and Uniform (U) Calibration}\label{tbl:estimates}
\centering
\begin{tabular}{ l r | r r r}
 Dataset & Features &  ESS & ESS (G) & ESS (U)\\ 
 \toprule
 Photons & 368 & 28.6 & 30.9 & 30.9 \\  
 Pions & 533 & 17.2 & 18.5 & 18.5 \\
 Electrons 2 & 6480 & 209.0 & 225.6 & 225.5\\  
 Electrons 3 & 40\,500 & 749.2 & 808.7 & 808.6
\end{tabular}
\end{table}

\paragraph{Architecture and Preprocessing} 
Similarly to \cite{Krause:2021ilc}, for \submReyes we separated the training procedure into two stages. In the first stage, using a NF, we learn the distributions of energies per layer, $E_{\ell}$, conditioned on the incident energies $E_{\mathrm{inc}}$. In the second stage, we model the voxel energies $\mathcal{I}$, conditioned on  $E_{\mathrm{inc}}$ and $E_{\ell}$ by first learning the manifold with a generalized autoencoder, then estimating a density on the manifold as described above.

\subparagraph{Preprocessing} Following \cite{Krause:2022jna}, to ensure energy conservation, the $E_{\ell}$ are first transformed as  
\begin{equation}\label{u_transf}
u_{0}=\frac{\sum^{n}_{\ell=0} E_{\ell}}{E_{\mathrm{inc}}}, \ \ u_{1}=\frac{E^{0}_{\ell}}{\sum^{n}_{\ell=0} E_{\ell}}, u_{2}=\frac{E^{1}_{\ell}}{E^{2}_{\mathrm{rem}}}, \ ... \ ,u_{n}=\frac{E^{n-1}_{\ell}}{E^{n}_{\mathrm{rem}}},
\end{equation}
where $n$ is the total number of layers,  and $E^{i}_{\mathrm{rem}}=\sum^{n}_{\ell=0} E_{\ell}-\sum^{i}_{\ell=0} E_{\ell}$.
The resulting variables are then transformed into logit space as
\begin{equation}\label{eq:logit}
u_i' = \log \frac{x_i}{1-x_i} ; 
\end{equation}
where $x_i=\alpha+(1-2\alpha)u_i$ and $ \alpha=10^{-6}$. The incident energy $E_{\mathrm{inc}}$ is preprocessed as:
\begin{equation}\label{eq:einc}
E_{{\mathrm{inc}}}\leftarrow\log_{10}\left(\frac{E_{\mathrm{inc}}}{33.3\mathrm{GeV}}\right).
\end{equation}

For the second stage, $E_{\mathrm{inc}}$ and $E_{\ell}$ are transformed as
\begin{equation}
E_{\mathrm{inc}}\leftarrow\log_{10}\left(E_{\mathrm{inc}}+1\mathrm{keV}\right),\quad E_{\ell}\leftarrow \log_{10}\left(\frac{E_{\ell}+1\mathrm{keV}}{100\mathrm{GeV}}\right) - 1.
\end{equation}
Finally, the voxel energy $\mathcal{I}$ is normalized so that the energies per layer sums up to one. The conditioning vector for the second stage is given by the concatenation of the incident energy $E_{\mathrm{inc}}$ and the energies per layer $E_{\ell}$.

It is worth pointing out that, according to our tests, the most important modeling aspects are the separation of the pipeline in two stages and the inclusion of the energy per layer in the conditioning vector for the second stage. For simplicity, we adopted a preprocessing strategy that closely follows \cite{Krause:2022jna} but we found that other reasonable preprocessing choices at each step do not significantly impact the performance of the model.

\subparagraph{Architecture} For all the experiments, we used a NF for the first stage and a two-step model for the second stage with a VAE as the generalized autoencoder and a NF of the same type as the density estimator.

For stage one and the photon dataset we used a 8 layer $\times$ 384 units coupling rational-quadratic neural spline flow \cite{durkan2019neural_spline_flows} with a 6-block residual network \cite{RESNET} in each layer. For the pion dataset, we used the same model with 8 layer $\times$ 512 units and 3-block residual networks. The output of each residual block was combined with the conditioning input, namely the incident energy, using a gated linear unit. The NF’s prior distribution was a unit variance diagonal Gaussian.

For stage two, the VAE’s encoder and decoder were both DNNs with three hidden
layers of 512 units each, and ReLU activations. The encoder output the means and variances for a diagonal Gaussian over the latent dimensions. The decoder output was also treated as a diagonal Gaussian with means for each data dimension but only a single variance shared across all dimensions.
The prior distribution was a unit variance diagonal Gaussian over the latent space. The latent dimension was 35 for photons and 20 for pions, values which are slightly higher than the estimates reported in~\tref{tbl:estimates}. The NF was of the same type as in stage one with 4 layer $\times$ 128 units and 3-block residual networks. The output of each residual block was combined with the conditioning input, now the incident energy concatenated with the energies per layer, using a gated linear unit. The NF’s prior distribution was a unit variance diagonal Gaussian.

\paragraph{Training} All models were trained with batch sizes of 512 and the Adam optimizer \cite{kingma2014adam} with a learning rate of $10^{-4}$ in stage one and $10^{-3}$ in stage two. We also applied gradient clipping with a max gradient norm of 10. The models were trained for a maximum of 200 epochs each with early stopping after 20 epochs of no validation improvement on a 20\% hold-out set.

For stage two, the VAE and NF were trained sequentially. Once the VAE was trained, its parameters were frozen and the training dataset was encoded deterministically with the VAE encoder means. The encoded data were then passed as inputs to the NF. The validation and early stopping metric was the average $\chi^2$ separation power over all high level features for stage one, the reconstruction error for the VAE and the negative log-likelihood for the NF in stage two. Both the VAE and the NF were conditioned on the incident energies and the energies per layer during training. The models with the best validation metrics were used for evaluation.

\FloatBarrier

\submHeadlineSingle{DNN CaloSim}{Dalila Salamani}{subsec:dnncalosim}{\submSalamaniDNNCite}{submSalamaniDNNGit}
\paragraph{Introduction}
The VAE model explored for \dsIpi is inspired by the VAE model developed in the context of shower simulation in the ATLAS experiment~\cite{Salamani:2021zet,ATLAS:2022jhk}. 
\paragraph{Architecture}
It comprises four dense layers for the encoder with 1500, 1000, 500, and 100 nodes for each layer respectively. The decoder has also four dense layers with a reversed order of the number of nodes. The latent dimension is 50. A batch normalization layer is used after each dense layer. The encoder and decoder networks are jointly trained to maximize the variational lower bound on the marginal log-likelihood for the data, approximated with the reconstruction loss (binary cross-entropy) and the KL divergence. 
\paragraph{Preprocessing}
The model is not trained on the absolute voxel energies but rather on the voxel energy ratios, where per shower, the energies of the voxels within each layer are normalized to the total energy deposited in that layer. This reparametrization of the input allows the model to better preserve the correlations of the energies across layers. In order to re-scale back the energies after generation, the VAE model learns, in addition, the energy per layer and the total energy of the shower deposited in the calorimeter. These quantities are encoded as additional $N$+1 nodes in the input and output of the VAE model, where $N$ represents the number of calorimeter layers (in dataset 1, for pions $N$=7). As the voxel energies are encoded as ratios the additional $N$+1 nodes are also encoded as ratios, where for each layer its energy is divided by the total energy and the total energy is divided by the incident energy. In total, the number of nodes in the input and output layers is 541 = 533+7+1. 
The model is conditioned on the incident energy of the incoming particle.   

\begin{figure}
    \centering
    \includegraphics[width=\linewidth]{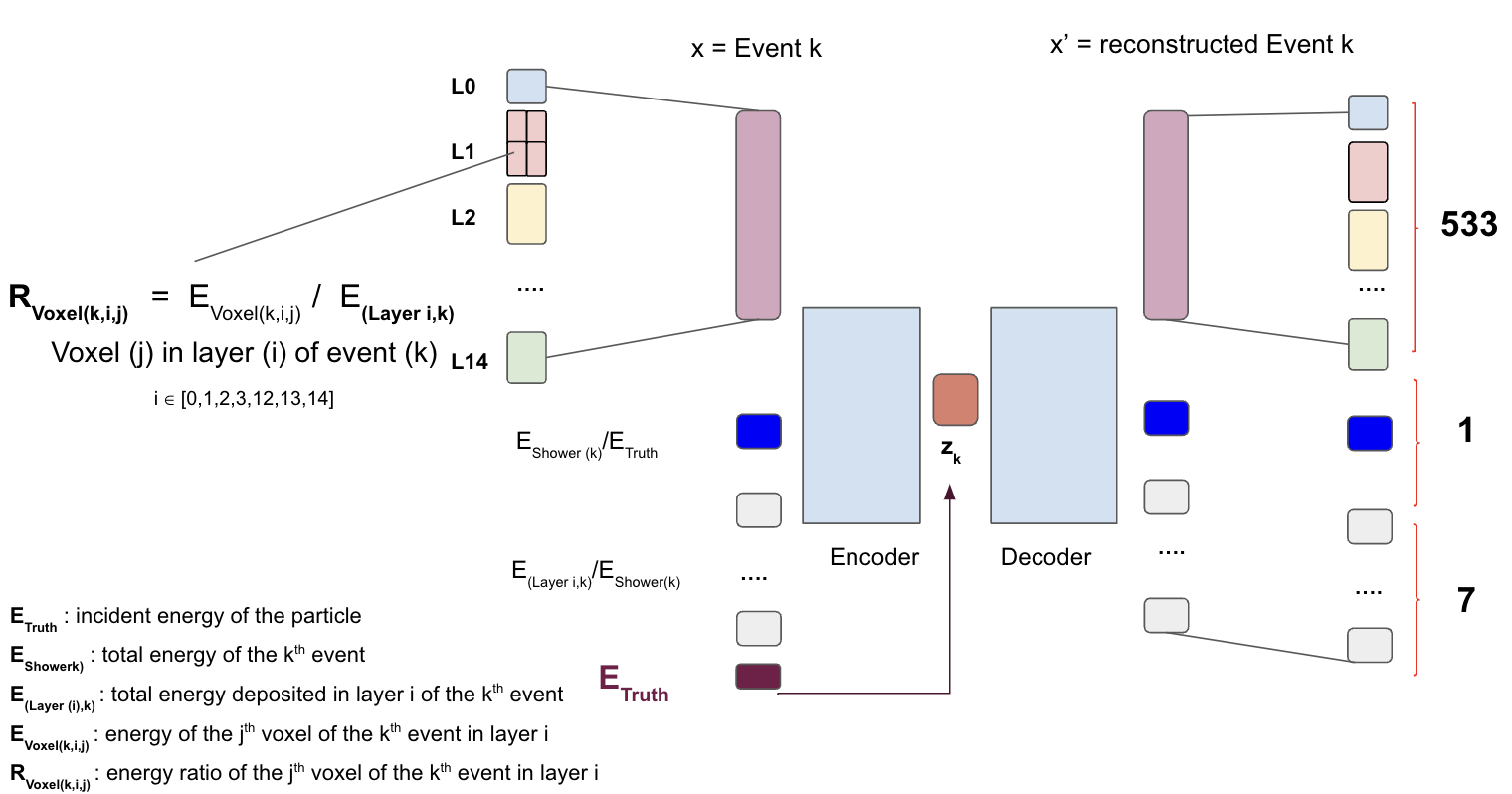}
    \caption{For the \submSalamaniDNN model, the VAE is trained to reconstruct the voxel energy ratios. The 533 inputs/outputs (the large pink boxes) represent the voxel energies for layers 0, 1, 2, 3, 12, 13, and 14. The additional input/output (dark blue) represents the normalized energy of the shower to the energy of the particle ($E_{\rm inc}$, called $E_{\rm Truth}$ here) and the 7 following inputs/outputs (in gray) are the ratios of the energy of the layer to the energy of the shower. The model is conditioned on the energy of the incoming particle, this is added as one additional input to the latent space (brown box).  }
    \label{fig:dnncalosim}
\vspace{-0.5cm}
\end{figure}

From prior knowledge on the deposited energy, ratios of voxel values and all total energies per layer should sum up to one. This can be ensured in the output layer of the VAE by applying a softmax activation function. By using the softmax function, the values are converted into probabilities that sum to one and automatically values fall in the range of [0,1]. The softmax is applied for the voxels of each calorimeter layer and for the ratios of the energy per layer. \Fref{fig:dnncalosim} shows a schematic representation of the VAE model.  

\paragraph{Training}
One effective trick for improving the training of the VAE model is using an iterative approach, where the model is trained on varying batch sizes and learning rates. By cycling through different combinations, the model is exposed to a variety of training conditions, which helps to avoid local minima and leads easier to potentially generalization. In total 8 iterations are used with different values of batch sizes and learning rates.

\FloatBarrier

\submHeadlineSingle{Geant4 Transformer}{Piyush Raikwar, Renato Paulo Da Costa Cardoso, Nadezda Chernyavskaya, Kristina Jaruskova, Witold Pokorski, Dalila Salamani, Mudhakar Srivatsa, Kalliopi Tsolaki, Sofia Vallecorsa, and Anna Zaborowska}{subsec:g4trans}{\submSalamaniTransCite}{submSalamaniTransGit}
\paragraph{Introduction}
Given the recent success of transformer-based models in various tasks, \textit{i.e.}, from image classification (ViT)~\cite{dosovitskiy2021image} to text generation (GPT-3)~\cite{NEURIPS2020_1457c0d6}, we explore the applicability of transformers for the task of generating non-trivially structured particle showers, specifically for dataset 3. The presence of an attention mechanism and the lack of a strong inductive bias in the architecture should help in better modeling of energy distributions in a highly granular mesh, given enough data.  

\paragraph{Preprocessing}
For the preprocessing, we divide the voxel energies (in MeV) by a scalar of 4300 to bring all entries between 0 and 1. Also, to use the incident energies of the particles as the condition, we divide the incident energies (in GeV) by 1024 so that their distribution is between 0 and 1.  

Since transformers are permutation-invariant sequence-to-sequence models, a shower needs to be in the form of a sequence. A naive way would be to treat each voxel as an element of the sequence, but that would be computationally expensive. Therefore, inspired by ViT, we create non-overlapping 3-dimensional patches of the shower to feed it to our model. To be specific, the dimensions of the patch are $\Delta r \times \Delta \phi \times \Delta z = 18 \times 5 \times 3$, \textit{i.e.}, 150 patches per shower. In addition to having a sequence, the transformer also needs to know the position of an element in the sequence, which we feed by using sinusoidal positional embeddings.

\paragraph{Architecture}

\begin{figure}[hbt]
    \centering
    \includegraphics[width=\linewidth]{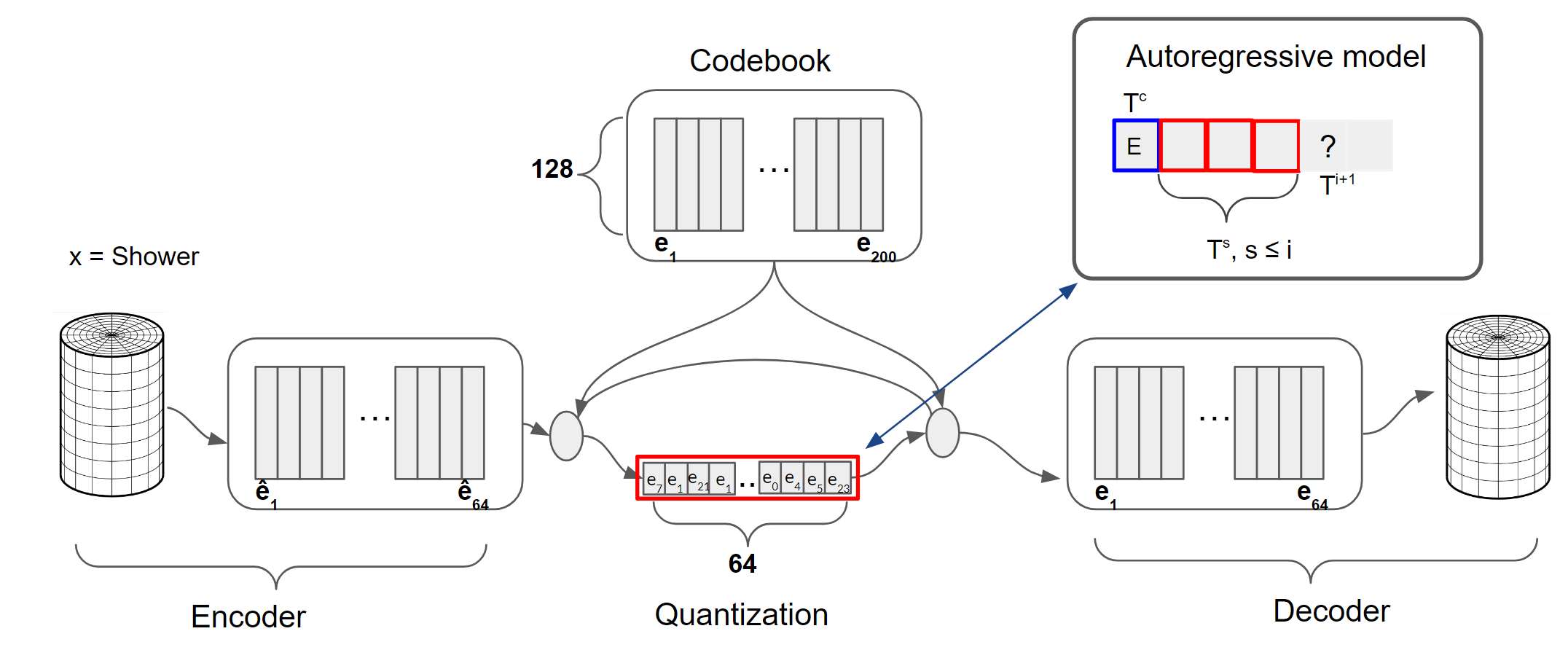}
    \caption{The figure shows the components of the VQ-VAE and the AR (top-right). The shower is tokenized using the VQ-VAE's encoder and by referring to the codebook. The tokenized representation of the shower is then fed to VQ-VAE's decoder to get back the shower. The AR models the tokenized shower's space (red tokens) separately in an autoregressive manner conditioned on $E_{\rm inc}$ (blue token). During inference, the VQ-VAE encoder is not used. Figure taken from~\cite{G4TransDalila}.}
    \label{fig:ar}
\vspace{-0.5cm}
\end{figure}

Our model is a two-stage model inspired by Dall-E~\cite{pmlr-v139-ramesh21a}, where the first stage is a Vector Quantized Variational Autoencoder (VQ-VAE)~\cite{van2017neural}, followed by an Autoregressive model (AR) (see \fref{fig:ar}). The VQ-VAE models the low-level features by tokenizing the shower, and the AR models the high-level features by learning the distribution of these tokens. These models are trained separately.

The VQ-VAE, as already explained for \submLiuCite in \sref{subsec:vqvae}, is similar to a traditional VAE, except that its latent space is discrete and can be represented by a set of codebook vectors (or tokens, $T^i$) chosen from a codebook. The codebook is a learnable entity that can be trained in parallel with the encoder and decoder to represent the latent space optimally. The output from the encoder is quantized using a nearest-neighbor search on the codebook to obtain a discrete latent space. So, the VQ-VAE is trained to reconstruct the showers where the latent space is a sequence of tokens. The distribution of these tokens is unknown, hence we cannot generate new showers. Therefore, the task of AR is to model the distribution of the tokens generated by the VQ-VAE, given the initial conditions, \textit{i.e.}, incident energy of the particle in our case. AR learns the latent space of VQ-VAE by autoregressively predicting these tokens, \textit{i.e.}, learning the probability of the next token given all the previous tokens. The process of sampling a new token from a multinomial distribution makes the AR a generative model. Note that VQ-VAE is not conditioned on the incident energy of the particle.  

For the generation of new showers, we start by creating a condition token ($T^c$). Given the condition token, we sample the next token and this continues till we have all the required tokens. All tokens except the condition token are then passed to the VQ-VAE decoder to get the final shower.  


Both VQ-VAE and AR have a ViT-like uniform architecture. Each of them is described in detail as follows.  

In the case of the VQ-VAE, the patches form the sequence. These patches are linearly projected to match the projection dimension of the VQ-VAE, which is 256. To this, 3-dimensional positional embeddings of 256 dimensions are added to inject the position information. These patches are then fed to the VQ-VAE. The encoder and decoder of the VQ-VAE consist of 4 encoder-only~\cite{vaswani2017attention} transformer blocks each. Each block consists of 16 attention heads and 512 nodes in the DNN sub-block. The patches after the last transformer block in the encoder are concatenated and projected to the desired dimensions of the latent space. That is, the number of patches is independent of the number of tokens in the latent space. Thus, a token can represent information from any of the patches. In our case, the latent space consists of 64 tokens of 128 dimensions. The codebook however consists of 200 tokens, out of which a combination of 64 tokens is used to represent a shower. The opposite is done to project the latent space back to the patches, which are then fed to the decoder. The activation function at the end of the decoder is a sigmoid, and binary cross-entropy is used as the loss function.  

For the AR, the tokens form the sequence. The tokens are projected linearly to the projection dimension of the AR, which is 128. Here, we use 1-dimensional positional embeddings to denote a token's position. These tokens are then passed to the AR. AR consists of 4 decoder-only~\cite{NEURIPS2020_1457c0d6} transformer blocks having 8 attention heads and 256 nodes in the DNN. The condition token is created by linearly projecting our conditions to match the projection dimension of the AR. The activation function at the end of the last block is a softmax, and the model is trained with categorical cross-entropy loss, where the target tokens are obtained from the VQ-VAE encoder.

\FloatBarrier

\submHeadlineSingle{CaloVAE+INN}{Florian Ernst, Luigi Favaro, Claudius Krause, Tilman Plehn, and David Shih}{subsec:calovaeinn}{\submErnstCite}{submErnstGit}
\paragraph{Introduction}
In this section we describe our effort to improve the scaling of the \submFavaro model. The idea is similar to the previously described VAE-Flow models. We train a VAE on the individual CaloChallenge datasets in a first step and a down-scaled \submFavaro in the corresponding latent space. The advantages of this approach were already discussed in~\sref{subsec:caloman} in the context of \submReyes. Namely, the enhanced topological properties and the smaller input dimensionality for the normalizing flow.\\
However, we were approaching with a different perspective. Our main goal is not to find the true manifold dimensionality, but we consider the VAE as a pure compression tool. Using this point of view it is of utmost importance to get good reconstructions first, before applying the INN in the latent space.\\

\paragraph{Loss function}
For the INN part we are employing the same loss function that was introduced in \submFavaro (\sref{subsec:caloinn}).\\
For the VAE we are using the $\beta$-VAE \cite{higgins2017betavae, Burgess:2018hqi} ELBO loss with a Gaussian encoder and a Bernoulli decoder. For the latent space we chose a standard normal distribution resulting in the following loss function:

\begin{eqnarray}
    \label{eq:BetaVAELoss}
        \mathcal{L}_\text{BVAE} 
        &=& \sum_{x \in TS}\sum_{z \sim E(z|x)} \left[x \log\left(\lambda(z)\right) + (1-x)\log\left(1-\lambda(z)\right)\right]\\ \nonumber
        &+& \beta \cdot\sum_{x \in TS}\left[1+\log\left(\sigma_{E}(x)^2\right)
        -\mu_{E}(x)^2
        -\sigma_{E}(x)^2\right].
\end{eqnarray}

Here, $\lambda$ is the parameter of the decoder Bernoulli distribution; $\mu_E$ and $\sigma_E$ are the parameters of the Gaussian encoder, as predicted by a neural network; $E(z|x)$ is the encoder; and $TS$ indicates the training dataset. We decided to explicitly not add a physics loss term since we did not want to bias the VAE to improve its reconstruction of the relevant physics variables, as we use these variables as performance metrics. 

\begin{figure}[ht!]
    \centering
    \includegraphics[width=0.7\textwidth]{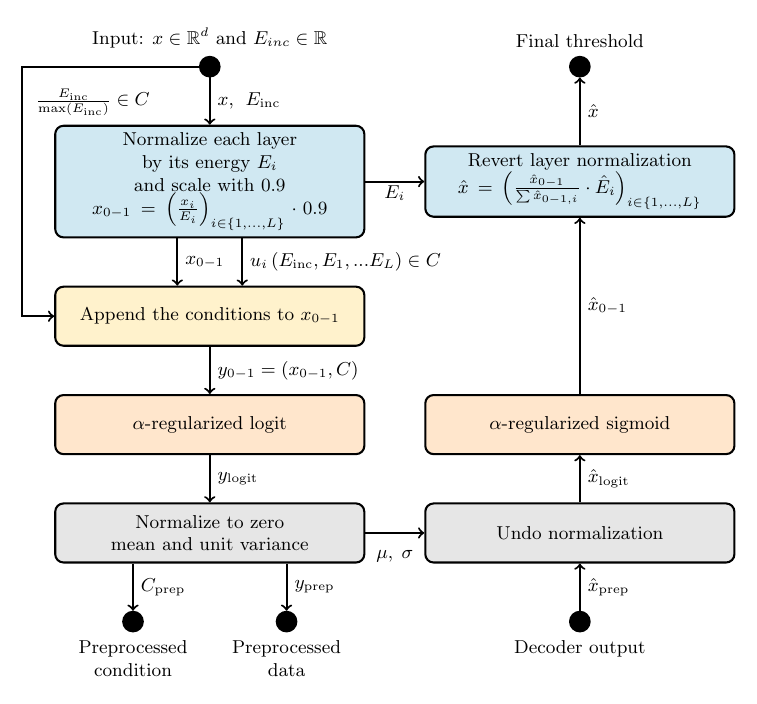}
    \raisebox{2.2cm}{\includegraphics[width=0.28\textwidth]{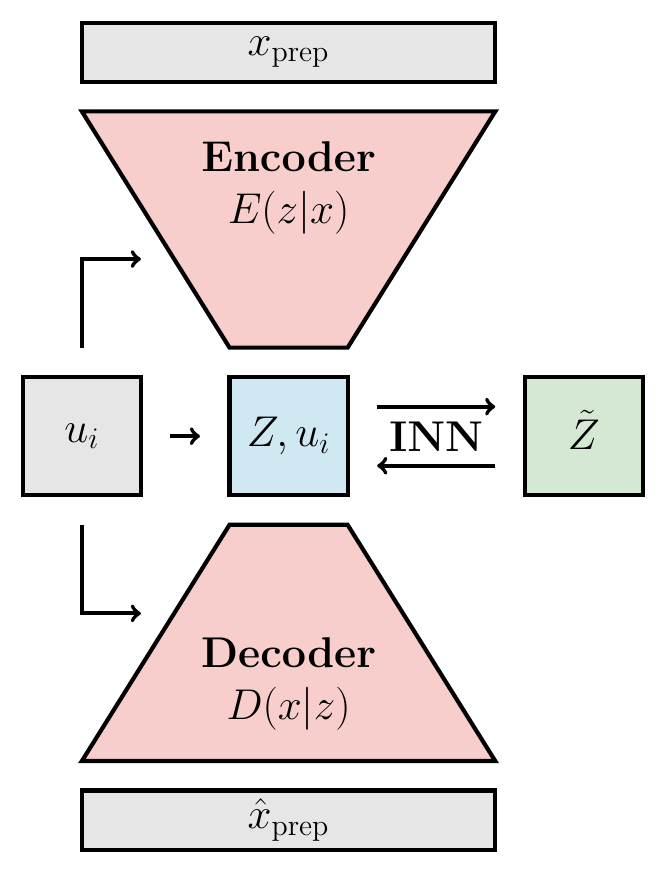}}
\caption{\textbf{Left:} Visualization of the preprocessing steps before the VAE-compression. \textbf{Right:} Schematic visualization of the VAE-INN combination of \submErnst, taken from~\cite{Ernst:2023qvn}.}
\label{fig:VAE+INN-prepocessing}
\end{figure}

\paragraph{Preprocessing}
Our preprocessing consists of four steps and is similar to the \submFavaro preprocessing. We keep the initial calorimeter layer normalization and extract our ``extra dimensions'' just like before. However, we scale them with a factor of 0.9 to prevent float precision problems. We did not add noise as we found it to be not helpful during the reconstruction process. We replaced the logarithm with an $\alpha$-regularized logit as in \eref{eq:logit} and added a final standardization layer. For the datasets 2 and 3 we used a learnable affine transformation, for dataset 1, we simply normalized to zero mean and unit variance. The entire preprocessing is illustrated in~\fref{fig:VAE+INN-prepocessing} (left).

The biggest difference to the \submFavaro preprocessing is the fact that the extra dimensions are not learned explicitly but used as additional conditions for the VAE. Afterwards, they are learned directly by the INN.

\paragraph{Architecture}

For the practical implementation we chose a fully connected encoder and decoder for dataset 1 and a kernel VAE (KVAE) for dataset 2 and 3. The KVAE is an architecture that is a compromise between a fully connected network and and a convolutional network. It tries to find a optimum between the reconstruction quality of the DNN and the scaling properties of the convolutional architecture. The idea is to use a two-step encoding, where neighboring detector layers are jointly encoded into a sub-latent space. Afterwards these sub-latent spaces are concatenated and encoded for a second time. This architecture emphasizes the stronger correlations between neighboring layers. We call the number of jointly encoded detector layers the ``kernel size'' and the distance between the two first layers of neighboring encoding ``blocks'' the ``kernel stride''\\
The decoder architecture is a copy of the encoder with inverted order of the size of the hidden layers.\\

The INN (a NF) is trained in the latent space of the second encoding step before the sampling happens, as seen in~\fref{fig:VAE+INN-prepocessing} (right). This means, the INN is trained to predict the actual $\sigma_E$ and $\mu_E$ values, effectively doubling the size of the input space for the INN. We found that this procedure improved the sampling quality of the resulting combined architecture significantly. The reason is probably that our $\beta$-parameter in the ELBO-loss is so small that VAE learned to store non-trivial information in the $\sigma_E$-parameters. Therefore, the input space of the INN consists of the encoder means $\mu_E$, the encoder widths $\sigma_E$ and the extra dimensions $u_i$.\\

Our final hyperparameter configurations for the three datasets can be seen in \tref{tab:VAE+INN-hyp}. We used $\beta = 10^{-9}$ as the KL term is just a regularization in this setup. The latent space is not required to be actually Gaussian, that is the task of the latent INN. However, the latent space must be ``well enough'' behaved for the INN to be trained in it.

\begin{table}[h!t]
	\centering    
    \begin{tabular}{l@{\hskip 10pt}|@{\hskip 10pt}c@{\hskip 5pt}l}
        \toprule
		Parameter          & VAE  \\
		\hline
		lr scheduler       &  Constant LR  & \hspace{0em}\rdelim\}{10}{*}[Inner VAE]\\
		lr                 &  $10^{-4}$ \\
		hidden dimension   & 5000, 1000, 500 (ds~1) \\
                           & 1500, 1000, 500 (ds~2) \\
                           & 2000, 1000, 500 (ds~3) \\
        latent dimension   & 50 (ds~1,2) / 300 (ds~3) \\
        \# of epochs       & 1000 \\
		batch size         & 256 \\ 
        $\beta$            & $ 10^{-9}$\\
        threshold $t$ [keV]& 2 (ds~1) / 15.15 (ds~2,3)\\
        \hline
		hidden dimension   & 1500, 800, 300  & \hspace{0em}\rdelim\}{3}{*}[Kernel]\\
        kernel size        & 7 \\
        kernel stride      & 3 (ds~2), 5 (ds~3)\\
		\bottomrule  
	\end{tabular}
 
    \vspace{20 pt}
    
    \begin{tabular}{@{\hskip 10pt}c@{\hskip 5pt}|@{\hskip 10pt}c}
		\toprule
		Parameter          & INN (after VAE)  \\
		\hline
		coupling blocks    & RQS \\
		\# layers          & 3 \\
		hidden dimension   & 32 \\
		\# of bins         & 10 \\
		\# of blocks       & 18 \\
		\# of epochs       & 200 \\
		batch size         &  256 \\ 
		lr scheduler       &  ``one cycle'' \\
		max. lr            & $10^{-4}$ \\
		$\beta_{1,2}$ (Adam)      & $(0.9, 0.999)$ \\
		$\alpha$           & $10^{-6}$ \\
		\bottomrule
    \end{tabular}
 
	\caption{Network and training parameters for the \submErnst.}
	\label{tab:VAE+INN-hyp}
\end{table}

\FloatBarrier

\submHeadlineSingle{CaloLatent: Score-based Generative Modeling in the Latent Space for Calorimeter Shower Generation}{Thandikire Madula and Vinicius M. Mikuni}{subsec:calolatent}{\submMadulaCite}{submMadulaGit}
\paragraph{Introduction}
In our work, we introduce \submMadulaCite, a latent diffusion inspired surrogate model. The main idea in latent diffusion is to map the data into a compressed latent representation using a VAE. Once the latent representation has been obtained, a diffusion model can be deployed to learn the distribution of the latent space. The motivations behind this approach are similar to those outlined by \submReyes and \submErnst.

In our approach we use the VAE backbone primarily for compression, therefore we prioritize the VAE’s reconstruction ability over its generation ability. To this end, we utilize the $\beta$-VAE~\cite{higgins2017betavae, Burgess:2018hqi} formulation where the KL divergence is weighted by a factor of $\beta$. We chose $\beta = 10^{-6}$. The diffusion model used to learn the latent distribution is a score-based diffusion model, the intricacies of which have been outlined by \submMikuni in \sref{subsec:caloscore}. 

\paragraph{Preprocessing} To evaluate the performance of our model we focused on dataset 2 of the challenge. We processed the data for training using the following steps. First, we normalized the voxel energy using \eref{eq:normvoxel}.
\begin{equation}
    x'_i = \frac{\mathcal{I}_i}{E_i}
    \label{eq:normvoxel}
\end{equation}
 where $\mathcal{I}_i$ are the voxels in layer $i$ of the detector and $E_i$ is the total energy of that layer. Secondly, we apply minmax scaling to the data which is defined by \eref{eq:minmax}.
 \begin{equation}
    x' = \frac{x - x_{\rm min}}{x_{\rm min} - x_{\rm max}}
     \label{eq:minmax}
 \end{equation}

 Where $x_{\rm min}$ and $x_{\rm max}$ are the minimum and maximum voxel values respectively. The resulting data is then transformed using the logit function outlined in \eref{eq:logit}. Finally, we take the values in the logit space and apply a standardization as given by \eref{eq:standard}.

\begin{equation}
    x' = \frac{x - \mu}{\sigma}
    \label{eq:standard}
\end{equation}

Where $\mu$ and $\sigma$ are the means and standard deviations, repectively.

\begin{figure}[ht!]
    \centering
    \includegraphics[width=0.8\textwidth]{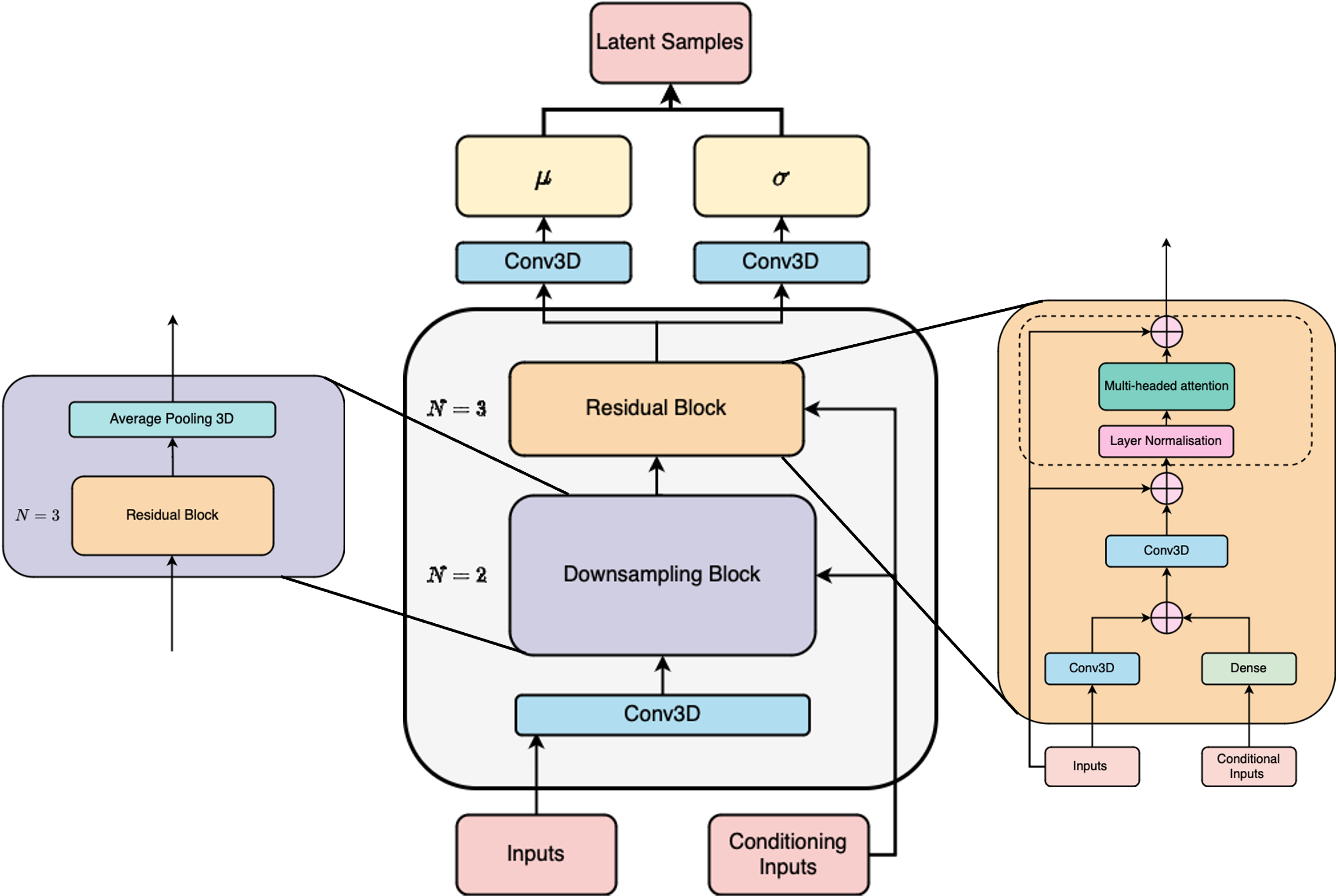}
    \caption{Schematic diagram of the \submMadula VAE encoder.}
    \label{fig:CaloLatentEncoder}
\end{figure}

\paragraph{Architecture} \submMadula is comprised of three networks. First, a scored-based diffusion model that is used to learn the distribution of the layer energy. The layer score model is a simple ResNet~\cite{RESNET} consisting of 3 layers each with 512 nodes. Second, we have the VAE backbone. The encoder and decoder of the VAE are 3-dimensional convolutional neural networks also inspired by the ResNet architecture. \Fref{fig:CaloLatentEncoder} shows a schematic diagram of the VAE encoder used for \submMadula. The decoder of the VAE is a mirror image of the encoder; however, it employs up-sampling blocks in place of the encoder down-sampling blocks. The VAE reduces the data dimensionality from 6408 to 1008. The third and final network in \submMadula is the score-based diffusion model used to learn the latent space, the architecture of this model is identical to that of the layer model.
\paragraph{Training} All three networks are trained independently. The layer score model and the VAE are trained for 500 epochs using a cosine decaying learning rate with an initial learning rate of $4\cdot 10^{-4}$. The latent score model is trained for 250 epochs using the same learning schedule as the other models.
\FloatBarrier

\section{Conditional Flow Matching-based Submissions}\label{sec:CFM}
\markboth{\uppercase{Conditional Flow Matching-based Submissions}}{}
The \emph{conditional flow matching} (CFM) algorithm was proposed in various forms simultaneously by several groups \cite{liu2023flow, albergo2023building, lipman2023flow}, and developed in \cite{tong2023improving} which we follow here. Like continuous normalizing flow models \cite{chen2018neuralode}, flow matching learns to interpolate probability densities $p_t$ between the data $p_1$, and a simple prior $p_0=\mathcal{N}( x_0 \mid 0, I)$. The interpolation is determined by a vector field at each time $\mu_t(x)$, which transports datapoints $x$ via the ODE
\begin{equation}\label{eq:fm_ode}
    dx = \mu_t(x)dt.
\end{equation}
When $p_t$ and $\mu_t(x)$ jointly satisfy the continuity equation for conservation of probability
\begin{equation}
    \frac{d}{dt} p_t + \nabla_x \cdot (p_t \mu_t) = 0,
\end{equation}
$p_t$ will be a properly normalized density at each $t$. Hypothetically, one could train a model $\nu_\theta(t, x)$ of the vector field $\mu_t(x)$ by direct regression,
\begin{equation}\label{eq:fm_loss}
    L_{\mathrm{FM}} = \mathbb{E}_{t\sim \mathcal{U}(0,1), \, x\sim p_t}\Vert \nu_\theta(t, x) - \mu_t(x)\Vert^2_2,
\end{equation}
however in practice neither $p_t$ nor $\mu_t(x)$ is uniquely determined, we can only sample from $p_t$ for $t=1$ (data) and $0$ (prior), and we do not have access to $\mu_t(x)$ for evaluation. As a workaround, CFM proposes to use conditional densities $p_t(x\mid (x_1, x_0))$ and vector fields $\mu_t(x\mid (x_1, x_0))$, where $x_1\sim p_1$ is a training datapoint and $x_0\sim p_0$ is noise, such that both are tractable. For example, when
\begin{eqnarray}\label{eq:cfm_prob}
    p_t(x\mid (x_1, x_0)) &=& \mathcal{N}( x \mid tx_1 - (1-t)x_0, \sigma^2),\\
    \mu_t(x\mid (x_1, x_0)) &=& x_1 - x_0,
\end{eqnarray}
the CFM loss
\begin{equation}\label{eq:cfm_loss}
    L_{\mathrm{CFM}} = \mathbb{E}_{t\sim \mathcal{U}(0,1), \, x_1\sim p_1, \, x_0\sim p_0,  \, x\sim p_t(\cdot\mid (x_1, x_0)),}\Vert \nu_\theta(t, x) - \mu_t(x\mid (x_1, x_0))\Vert^2_2,
\end{equation}
has the same gradients as~\eref{eq:fm_loss}, and therefore will lead to the same model $\nu_\theta(t, x)$, but now $p_t(x\mid (x_1, x_0))$ and $\mu_t(x\mid (x_1, x_0))$ are tractable for all $t$. Finally, new datapoints are generated by solving the ODE in~\eref{eq:fm_ode} starting from $x_0\sim p_0$ but using the learned vector field $\nu_\theta(t, x)$.

\FloatBarrier

\submHeadlineSingle{CaloDREAM: Vision Transformer CFM}{Luigi Favaro, Ayodele Ore, Sofia Palacios Schweitzer, and Tilman Plehn}{subsec:calodream}{\submPalaciosCite}{submPalaciosGit}
\paragraph{Introduction}
The \submPalaciosCite architecture consists of two continuous normalizing flows trained with the CFM objective given in \eref{eq:cfm_loss}\;\footnote{Here the resolution parameter $\sigma$ from \eref{eq:cfm_prob} is taken to be zero.}. The first is an \emph{energy} model, which is responsible for generating the total energy deposited in each calorimeter layer. The model uses the energy ratio variables $u$, defined in \eref{eq:us}, as a basis for the layer energies. \submPalacios then employs a \emph{shape} model to generate voxel values, given the $u$ ratios as conditions. In order to enforce energy conservation, the shape model is trained on voxels normalized by their layer energy. In the following, the unique aspects of the two models comprising \submPalacios are detailed.

\begin{figure}
    \centering
    \includegraphics[scale=0.6]{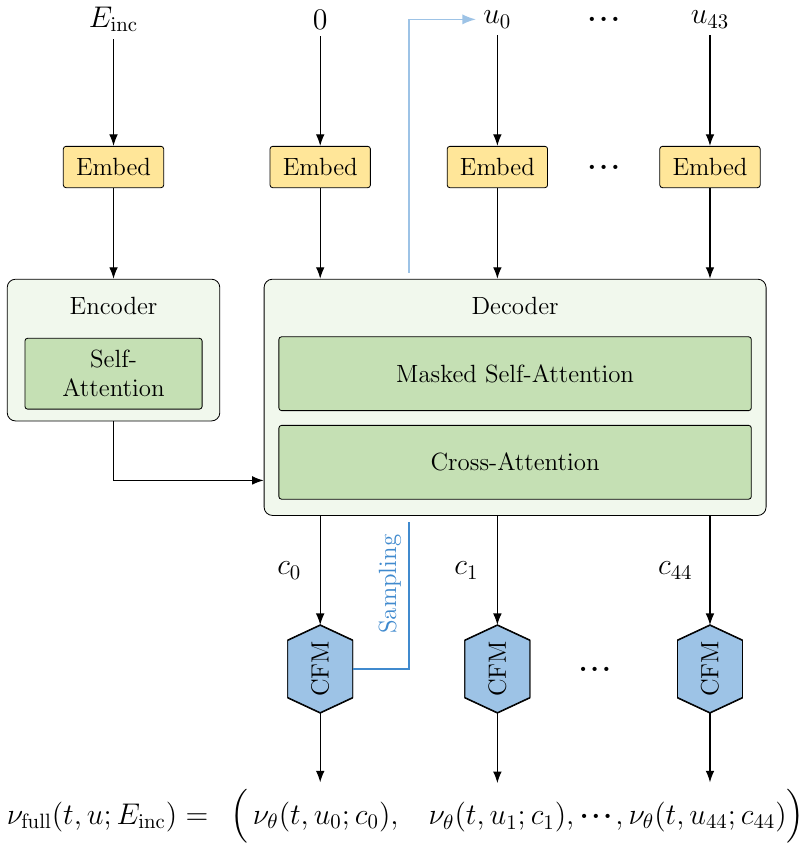}
    \hspace{1.3cm}
    \raisebox{0.1cm}{\includegraphics[scale=0.6]{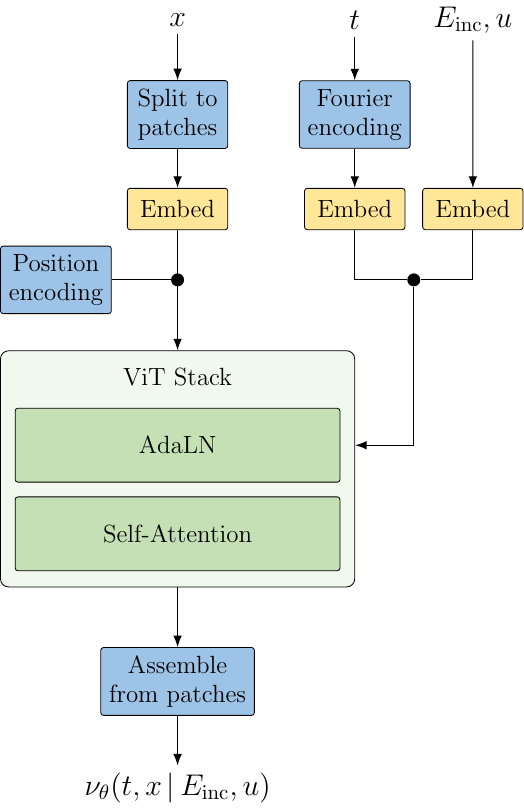}}
    \caption{\textbf{Left:} Schematic diagram of the \submPalacios energy network. \textbf{Right:} Schematic diagram of the shape network. Figure adapted from~\cite{Favaro:2024rle}.}
    \label{fig:calodream_arches}
\end{figure}
\paragraph{Architecture --- Energy model}
As discussed above, the heart of a CFM model is a learnable vector field $\nu_\theta$. Although it is typical to use a single neural network as a direct parameterization, other choices are also viable. In particular, the causal nature of energy flow through a calorimeter inspires an autoregressive construction of the full vector field.

\submPalacios adopts an autoregressive CFM architecture introduced in~\cite{Heimel:2023mvw} to learn the distribution of energy ratios given an incident energy, $p(u|E_\mathrm{inc})$. The model structure is illustrated in~\fref{fig:calodream_arches} (left). Instead of directly learning a 45-dimensional vector field, a single network is trained to solve 45 CFM tasks --- one for each layer. In order to distinguish these tasks, the network is conditioned on previous layer energy ratios. The full vector field can be written as
\begin{equation}
    \nu_{\mathrm{full}}(t, u\,|\, E_{\mathrm{inc}}) = \left( \nu_\theta(t, u_0\,|\, c_0),\;\dots\;, \nu_\theta(t, u_{44}\,|\, c_{44}) \right),
\end{equation}
where $\nu_\theta$ is a neural network and each $c_i$ is a condition that encodes the incident energy and the sequence of previous $u$'s
\begin{equation}
    \label{eq:causal_conds}
   c_i = \left\{
    \begin{array}{ll}
        c_i(u_0, \dots, u_{i-1}, E_{\mathrm{inc}}) &  i>0\\
        c_i(E_{\mathrm{inc}}) & i=0
    \end{array} 
    \right. \; .
\end{equation}
In practice, an encoder-decoder transformer is used to learn the conditions. Both the encoder and decoder contain four attention blocks with four heads and an embedding dimension of 64. By applying a triangular mask to the relevant attention matrices, the structure prescribed in \eref{eq:causal_conds} is respected. For $\nu_\theta$, a DNN with eight layers and width of 512 is used.

When training the energy model, all components of the full vector field can be evaluated in parallel. During inference, on the other hand, the CFM network must be sampled for each component in sequence. Specifically, $u_0$ is first sampled by integrating $\nu_\theta(t, u_0\,|\, c_0)$ and can then be used to compute $c_1$. This condition is in turn required to generate $u_1$ and so forth.

\paragraph{Architecture --- Shape model}
\submPalacios also uses CFM for the shape model, which is responsible for learning to sample showers $\mathcal{I}$ from $p(\mathcal{I}\,|\,E_\mathrm{inc}, u)$. Unlike the energy model, here the vector field is parameterized directly with a neural network $\nu_\psi(t,\mathcal{I}\,|\,E_\mathrm{inc}, u)$. In order to obey energy conservation, the shape network is trained on layer-normalized voxels. The network is a vision transformer (ViT) similar to~\cite{peebles2023scalable} and illustrated in \fref{fig:calodream_arches} (right).
As a first step, the network divides the shower into non-overlapping rectangular blocks in the three-dimensional calorimeter space $(z, \alpha, r)$. Each of these regions defines a \emph{patch}, which is embedded with a shared linear layer.
The embeddings are supplemented with learnable position encodings which break the permutation invariance among patches.
The network uses a joint embedding for the conditional inputs, $t,E_\mathrm{inc}$ and $u$. The time embedding is first transformed to Fourier space and embedded with a two-layer dense network.
The energy conditions are instead directly embedded with a separate dense network using the same architecture. The final operation sums the two embedded conditions into a single vector.

After applying these initial transformations, the patches and the conditions are passed to a stack of ViT blocks.
Each block contains a self-attention over patches followed by a dense network transformation.
The conditional information is introduced via affine transformations with shift and scale $a,b\in\mathbb{R}$ and an additional rescaling factor $\gamma\in\mathbb{R}$ learned by a dense network, referred to as AdaLN~\cite{peebles2023scalable,2017arXiv170907871P} in \fref{fig:calodream_arches}.
Concretely, the operation inside the ViT block is summarized by
\begin{eqnarray}
 x_{\mathrm{h}} &=& x + \gamma_{\mathrm{h}} g_{\mathrm{h}} (a_{\mathrm{h}}x + b_{\mathrm{h}}),\\
 x_{\mathrm{l}} &=& x_{\mathrm{h}} + \gamma_{\mathrm{l}} g_{\mathrm{l}} (a_{\mathrm{l}}x_{\mathrm{h}} + b_{\mathrm{l}}),
\end{eqnarray}
where $g_{\mathrm{h}}$ is the multi-head self-attention step and $g_{\mathrm{l}}$ is the fully connected transformation.
After the stack of ViT blocks, the same modulation is applied to the final patch features before projecting back to the original size. The patches can then be assembled into the full calorimeter shape.

\paragraph{Training}
The training is carried out using the AdamW optimizer with an initial learning rate of $10^{-3}$ and a cosine learning rate scheduler. We train the model for 800 and 600 epochs for dataset 2 and dataset 3 respectively. The patch sizes used in each dataset are $(3,16,1)$ and $(3,5,2)$. In both cases there are six self-attention blocks with six heads each. The \submPalacios samples evaluated in the results section are obtained by solving the energy and shape model ODEs with the Runge-Kutta 4 solver with step size 0.02.

\FloatBarrier

\submHeadlineSingle{CaloForest}{Jesse C.~Cresswell and Taewoo Kim}{subsec:caloforest}{\submCresswellCite}{submCresswellGit}
\paragraph{Introduction} 
The methods in the above sections differ in their learning tasks and architectures, but all use neural networks as function approximators. Neural network architectures are often carefully designed to have inductive biases that are beneficial for a specific data modality. A case in point are datasets 2 and 3 of~\sref{subsec:ds23} which have an image-like structure; each layer of the calorimeter corresponds to a channel, and the voxels of each layer are arrayed in a consistent manner giving a familiar $c\times h\times w$ format. Convolutional neural networks are well-adapted for this structure, achieving efficiency through parameter sharing by applying the same kernel across the image.

However, tabular datasets (like dataset 1 in~\sref{subsec:ds1}) have minimal structure that neural networks can take advantage of. Researchers still often resort to basic DNN architectures with little-to-no useful inductive bias. Historically, tree-based algorithms have outperformed on \emph{discriminative} tasks for tabular data at scale \cite{grinsztajn2022, mcelfresh2023neural}, with only very recent DNN-based foundation models starting to consistently improve upon tree-based methods~\cite{ma2024tabdpt}. There are additional advantages of tree-based models: they usually do not require any data pre-processing (whereas neural networks are highly sensitive to input data scale and distribution); they can operate on data that contains null values (whereas neural networks require null values to be dropped or imputed); they can be trained efficiently on GPU or CPU (whereas large neural networks usually necessitate GPU training, but most high performance computing clusters are still CPU based); and they have improved explainability (for example, Shapley values \cite{shapley1951} are generally intractable to compute for large neural networks, but the TreeSHAP algorithm makes them workable for trees \cite{lundberg2017, lundberg2018}). Yet, tree-based learning is not common for \emph{generative} tasks, even on tabular data.

We propose training generative models for the tabular dataset 1 using a tree-based function approximator, namely XGBoost \cite{chen2016xgb}. Just like a DNN, XGBoost is a universal function approximator, meaning that with large enough number of parameters (\textit{i.e.}~tree depth) and training datapoints, it can in principle fit any function \cite{friedman2001greedy}. This begs the question of why tree-based models are still rarely used for generative tasks \cite{nock22, jolicouer2023}. In principle, XGBoost could be used as a replacement for the neural network function approximator in any generative modeling algorithm, such as the ones used throughout Sections \ref{sec:gan}--\ref{sec:vae}. In practice, the mechanics of training trees deviates significantly from how neural networks are trained, requiring non-trivial reengineering of algorithms.

\paragraph{Generative Modeling with Trees} The difficulties of replacing neural network function approximators with XGBoost (in this case for $\nu_\theta(t, x)$) are well-illustrated by the work \cite{jolicouer2023} which provides an implementation of CFM backed by XGBoost. First, notice that when using neural networks, \eref{eq:cfm_loss} would ordinarily be optimized by sampling a minibatch of data $x_1$ from $p_1$, sampling a $t$, sampling $x_0$ from the prior $p_0$ and then generating $x$ from $p_t(x\mid (x_1, x_0))$ in \eref{eq:cfm_prob}. In particular, the noise vector $x_0$ would be sampled anew every batch, eventually leading to good coverage of the distribution in the expectation. XGBoost is not trained with minibatches; for regression tasks it requires an entire dataset to be fed in and then minimizes the squared error loss overall. Therefore, the noise $x_0$ associated to each training point $x_1$ would only be sampled once. For better coverage of the noise distribution, \cite{jolicouer2023} proposes to duplicate each of the $n$ training datapoints $K$ times, and for each copy of $x_1$ generate different noise $x_0$. 

Second, whereas with a neural network the time step $t$ could be fed in as an additional input to the network during training and generation, simply adding $t$ as a feature to XGBoost is unlikely to give sufficient emphasis to it, because only a single feature is used in each split of each tree in the ensemble. Instead, \cite{jolicouer2023} proposes to discretize $t$ into $n_t$ uniform steps and train a different XGBoost ensemble to represent $\nu_\theta(t_j, x)$ for each timestep $t_j$. The expectation over $t$ is removed in the loss function \eref{eq:cfm_loss}, and it is instead treated as a constant for each of $n_t$ separate loss functions.

Third, whereas a neural network can easily be designed with a number of outputs equal to the number of features $p$ in $x$ (the same size as the regression target $\mu_t(x\mid(x_1, x_0))$ for a given $t$), XGBoost only outputs a scalar. A brute-force workaround is to simply train a different ensemble to predict each element of the vector field $\mu_t(x\mid(x_1, x_0))$.

Fourth, when conditional generation on a class label $y$ is required, a neural network can accept $y$ as an input during training and generation to adapt its behavior while sharing parameters. Like conditioning on $t$, conditioning on $y$ is better done by training a separate XGBoost ensemble for each of the $n_y$ classes. This is only suited to conditioning on categorical labels, not continuous ones, so it does not allow for sampling at arbitrary incident energies. 

Despite these challenges, the promises of tree-based generative modeling are seductive: better performance on tabular regression tasks may translate to better tabular generation; lack of need for preprocessing; native handling of missing values; efficient training on CPU; and improved explainability. As a proof of concept for tree-based generative modeling of calorimeter showers, we applied CFM backed by XGBoost to the tabular dataset 1.

\paragraph{Modeling Dataset 1} In total, for a tabular dataset of size $[n,\ p]$, the method described by \cite{jolicouer2023} requires training $n_t\times p \times n_y$ XGBoost ensembles, each on a dataset of size $[n_i\times K,\ p]$, where $n_i$ is the number of points belonging to class $i$ (with $\sum_{i}^{n_y} n_i = n$). This poses a practical challenge. The largest training datasets benchmarked in \cite{jolicouer2023} had sizes $[16\,512, \ 9]$ (largest $n$), $[288,\ 90]$ (largest $p$), and $[10\,888, \ 16]$ (largest product $np$), while dataset 1 -- $\pi^+$ is 370$\times$ larger at $[120\,800, \ 533]$.

Unfortunately, the implementation of CFM with XGBoost published by \cite{jolicouer2023} does not scale to problems of this magnitude.\footnote{We accessed this code repository \url{https://github.com/SamsungSAILMontreal/ForestDiffusion}, commit hash \texttt{855281b} dated Nov. 2, 2023.} Noting that \cite{jolicouer2023} recommend $n_t=50$ and $K=100$, dataset 1 -- $\pi^+$ (with $n_y=15$) would necessitate training 399\,750 XGBoost ensembles, most of which would use a dataset of size $[1\,000\,000, \ 533]$. From the original $[n, \ p]$ dataset, the implementation attempts to create a duplicated version of size $[n_t, K\times n, p]$ as a \texttt{numpy} array in memory all at once, which for the pions training dataset requires 2.34 TiB of CPU memory. Training thousands of XGBoost ensembles on slices of the data in parallel further exacerbates the memory burden, since many copies of the data array are created and persisted in RAM or RAM disk. We estimate that a full training run using the default hyperparameters on dataset 1 -- $\pi^+$ would require more than 1.2 PiB of CPU memory.

However, this memory burden is not a fundamental limitation of the proposed method, but rather a lack of optimization of the original implementation. For our proof-of-concept, we reimplemented CFM with XGBoost solving many engineering challenges around memory efficiency and parallelization in python. Our implementation runs with a peak CPU memory burden of 78 GiB on dataset 1 -- $\pi^+$, or roughly 16\,000 times less.

In addition to improving the memory efficiency and runtime, we also mention methods to increase model performance. From hyperparameter ablation we found that $n_t$ has the largest effect on model fit and should be increased as high as feasible, noting that this comes with a linear increase in the training time and number of model parameters. We found that for datasets with larger $n$ a more conservative value of $K$ was sufficient compared to the recommendation in \cite{jolicouer2023}. Larger tabular datasets tend to have more redundant rows built in, and  different noise is added to these rows giving sufficient coverage with lower $K$. We observe that the model consistently underfits the dataset for all sizes we tested, in agreement with \cite{jolicouer2023}, but underfitting can be mitigated by increasing the learning rate substantially. Although XGBoost typically does not require data preprocessing, the CFM algorithm adds noise which must be commensurate to the data's typical scale. Hence, it is important to at least scale the data to a finite domain similar to the standard deviation of the added noise. The original implementation uses MinMax scaling over the entire dataset. However, when the data has distinct classes with different properties, as is the case for the incident energy levels of Dataset 1, we find that it is more beneficial to scale each class separately since XGBoost models are trained separately for each class.

In summary, we trained models for the photons and pions datasets using a single desktop workstation with 250 GiB RAM and 40 CPUs (Intel Xeon Silver 4114T). We discretized time into $n_t=100$ steps, and duplicated each datapoint $K=20$ times. Each XGBoost ensemble had 20 trees of maximum depth 7, a learning rate of 1.5, and all other hyperparameters left as defaults.\footnote{We used XGBoost version 2.0.0.} We trained up to 20 XGBoost ensembles in parallel at a time, each with 2 CPUs. In total, for the photons model 552\,000 XGBoost ensembles were trained in 135 hours with a peak memory burden of 54 GiB, while the pions model used 799\,500 ensembles, completed in 281 hours, and required 78 GiB of memory.

This proof of concept shows that tabular generative modeling with tree-based function approximators trained on CPU is feasible for calorimeter shower simulation. We have worked through an example of how to convert from neural networks to XGBoost using a modern generative framework. However, our trained models have clear room for improvement on several fronts. First, performance is not yet competitive with highly-tuned neural network approaches. Second, our models are massively overparameterized (although we do not observe overfitting), with the number of XGBoost ensembles several times larger than the number of datapoints they were trained on (each ensemble having thousands of parameters). Third, the sheer number of trees trained contributes to slow training and large model size on disk. We believe these are solvable issues. Performance could be improved by replacing the simple Euler ODE solver with more advanced methods, though we point out that the learned vector field $\nu_\theta(t, x)$ only allows sampling at discrete values of $t$. We anticipate model size could still be reduced with additional hyperparameter optimization, or by moving to multi-output trees. Training time could be slashed by parallelizing training steps across a cluster of CPUs, which is straightforward for this method.

\FloatBarrier

\section{Introduction to metrics}
\label{sec:intro_metric}
\markboth{\uppercase{Introduction to metrics}}{}

The evaluation of DGMs is a challenging task that has seen significant research in itself in the past years~\cite{Kansal:2022spb,Das:2023ktd,Ahmad:2024dql}. For the application of DGMs as part of the detector simulation, we are interested in surrogate models that are faithful (\textit{i.e.}~reproduce the showers of \geant as close as possible), light-weight (\textit{i.e.}~do not require much space to store and are fast to load), and fast in generation. Each of these aspects by itself is hard to capture with a single number, so we will report a set of different metrics to give a more complete picture. It is expected that there will not necessarily be a single clear winner, and different submissions will have their advantages and disadvantages.

\subsection{High-level Features (Histograms)}
\label{sec:results_hlf_intro}
We begin the evaluation by looking at high-level features, \textit{i.e.}~physical observables that are derived from the energy depositions in the calorimeter. We focus on the following set:

\begin{itemize}
    \item The energy deposition in each voxel: $\mathcal{I}_{ia}$.
    \item The energy depositions in each layer of the calorimeter, as the sum over all voxels in that layer: $E_i = \sum_a \mathcal{I}_{ia}$.
    \item The total energy deposition in the shower, as sum over all voxels, normalized to the incident energy: $E_{dep} / E_{inc} = \sum_{a,i} \mathcal{I}_{ia} / E_{inc}$. 
    \item The centers of energy in $\eta$, $\phi$, and $r$ direction, defined via $\sum_{a} l_a \mathcal{I}_{ia} / \sum_{a} \mathcal{I}_{ia}$. The locations $l_a$ are either $\phi_a = r_a \sin{\alpha_a}$, $\eta_a = r_a \cos{\alpha_a}$ or $r_a$, where $r_a$ and $\alpha_a$ are the centers of the voxels in $\alpha$ and $r$. These are taken as the mean of the voxel boundary values defined in the \texttt{binning.xml} files. The sum goes over all voxels $a$ in a given layer. 
    \item The width of the center of energy distributions in $\eta$, $\phi$, $r$ direction: $\sqrt{\frac{\sum_{a} l_a^2 \mathcal{I}_{ia}}{\sum_{a} \mathcal{I}_{ia}}-\left(\frac{\sum_{a} l_a \mathcal{I}_{ia} }{\sum_{a} \mathcal{I}_{ia}}\right)^2}$
    \item The sparsity, defined as 1 minus the activity, with the activity being the fraction of voxels per layer with an energy deposition above threshold (threshold is defined per dataset in section~\ref{sec:data}).
\end{itemize}

For each of these observables, we compute the \emph{separation power} between the submissions and the held-out test set. We use the same binning as shown in appendix~\ref{app:histograms} in the reference histograms for the two \geant datasets. The separation power between two histograms is defined as~\cite{Diefenbacher:2020rna} 
\begin{equation}
    \label{eq:sep.power}
    S(h_1, h_2) = \frac{1}{2}\sum_i\frac{(h_{1,i}-h_{2,i})^2}{h_{1,i}+h_{2,i}}, 
\end{equation}
where $h_{j,i}$ is count of the $i$th bin of histogram $j$. The histogram counts are expected to be normalized: $\sum_i h_{j,i} = 1$. With these definitions, we have $S=0$ if and only if $h_1 = h_2$ and $S=1$ if the two distributions have no overlap. 

The separation power is closely related to the $\chi^2$ homogeneity test~\cite{Pearson:1900otctag,cramer1946mathematical,Gagunashvili:2006nva}. The difference is that the $\chi^2$ test statistic does not include normalization of the histogram counts.

To get a better feeling for the natural statistical spread of the separation power between different \geant datasets, we show a gray band in all figures of separation powers, indicating the minimal and maximal separation power we found comparing ten different pairs of \geant datasets. For dataset 1, we constructed these pairs by joining, shuffling, and then splitting the events from the two given datasets (\textit{i.e.}~drawing without replacement from the joined dataset), ensuring that the $E_{inc}$ distribution is always the same. 
For datasets 2 and 3, we generated 9 additional datasets with \geant, 100\,000 showers each, such that we get ten sets of pairs. 

\subsection{Correlations}
\label{sec:results_corr_intro}
The energies deposited in subsequent layers are correlated with each other due to the size of the particle shower in $z$ direction. One measure to study if these correlations are learned correctly is given by Pearson correlation coefficient (PCC) between the layer-wise energy depositions~\cite{Ahmad:2024dql}. For two sets of layer energies $\{E_i\}$ and $\{E_j\}$ of the same size, the PCC is given by
\begin{equation}
    \label{eq:PCC}
    \PCC{E_i}{E_j} = \frac{\sum_k\left(E_{i,k} - \mean{E_i}\right)\left(E_{j,k} - \mean{E_j}\right)}{\sqrt{\sum_k \left(E_{i,k} - \mean{E_i}\right)^2}\sqrt{\sum_k \left(E_{j,k} - \mean{E_j}\right)^2}},
\end{equation}
where $k$ runs over all samples in the set, and $i$ and $j$ are layer numbers.

\subsection{Classifier-based Metrics.}
\label{sec:results_cls_intro}
Classifiers offer a way to perform a two-sample test~\cite{2016arXiv161006545L} that is sensitive to the full distribution, including correlations between features. In the context of generative models for calorimeter simulation, they have been proposed as metric in~\cite{Krause:2021ilc} and were further discussed in~\cite{Das:2023ktd}, where it was also shown that they can give valuable insights to what failure mode the generative model has. 

Here we focus on two different classifier tests. The first one, a binary classification task, compares each submission with the \geant test set. The second one, a multiclass classification task, compares all submissions with each other. For each, we consider two different neural network architectures. 

\paragraph{Binary classification}
The binary classification test evaluates how well the underlying distribution was learned and therefore how close the generated distribution is to the reference. It relies on the Neyman-Pearson Lemma~\cite{Neyman:1933wgr}, stating that the most powerful classifier to distinguish two samples is their likelihood ratio. If a well-trained classifier is unable to distinguish submitted samples from the \geant test set, we conclude that the submission replicates the \geant distribution well~\cite{Krause:2021ilc,Das:2023ktd}. The result of this test, however, depends on the preprocessing that was applied to the data. Using the calorimeter showers in the physical space lets the classifier focus on the brightest voxels only, since energy depositions in them are orders of magnitude above the low-energy depositions. Applying a logarithm or logit transformation, enhances the sensitivity to mismodeling in them. While this gives a better understanding on whether or not the entire distribution was learned well, it might be that the difference is only in features and correlations that are irrelevant for the down-stream physics analysis. For that reason, we consider two different sets of input features. The first one are the energy depositions in the voxels (called ``low-level'' observables), the second one are the observables we introduced in \Sref{sec:results_hlf_intro} (called ``high-level'' observables). 

The figure of merit in this setup is the AUC, the area under the receiver operating characteristic (ROC) curve. The ROC curve shows the true positive rate (TPR) as a function of the false positive rate (FPR). In a random classifier, the TPR will grow linearly with the FPR giving a AUC of $0.5$. In a classifier that can separate the two datasets perfectly well, the ROC curve will become a step function, so the AUC becomes 1. We train ten classifiers with different random initialization and average the AUCs when reporting the results. 

In training, we split the submission and \geant dataset each into training, testing, and evaluation sets first, before merging them with the corresponding labels. This ensures having always a balanced setup. The number of events in each set is shown in~\tref{tab:ttv}. We select the model state with the highest accuracy on the test set for the final evaluation. Before evaluating the AUC on the evaluation dataset, we calibrate the classifier with isotonic regression~\cite{2017arXiv170604599G} on the test set. 

\begin{table}
\caption{\label{tab:ttv}Number of samples in training, testing and evaluation datasets in the binary classification setup. }
\begin{indented}
\item[]
\centering
\begin{tabular}{@{}c|ccc}
\br
dataset & training & testing & evaluation \\
\mr
1 - photon & 80\,000 & 20\,000 & 21\,000 \\
1 - pion & 80\,000 & 20\,000& 20\,800 \\
2 & 60\,000 & 20\,000 & 20\,000\\
3 & 60\,000 & 20\,000 & 20\,000\\
\br
\end{tabular}
\end{indented}
\end{table}

However, since a different neural classifier is trained for each submission, a comparison between submissions on equal conditions is harder to make. Therefore, we consider a second classifier test based on a multiclass classification setup below.

\paragraph{Multiclass classification}
With the multiclass classification setup, we try to assess which of the submissions is closest to \geant. The method was introduced in~\cite{Lim:2022nft} in the context of comparing hydrodynamical galaxy simulations, and subsequently applied to high-energy physics scenarios in~\cite{Diefenbacher:2023vsw,Golling:2023yjq}. It relies on training a single classifier with cross entropy loss on the task ``submission~1 vs.~submission~2 vs.~\dots~vs.~submission~$n$''. When evaluating the trained classifier on a \geant-based test set, we can read off which submission the \geant sample is closest to. 

As figure of merit, we consider the average of the log posterior~\cite{Lim:2022nft}. It is defined as
\begin{equation}
    \label{eq:log.post}
    \LogPost(\mathrm{model~}i| \mathrm{samples~}j) = \frac{1}{N} \sum_{x_k \in j} \log{p_{\mathrm{model~}i}(x_k)},
\end{equation}
the average logarithm of the probability that samples $j$ come from the model (submission) $i$. Here, the index $k$ goes over all $N$ samples in the set $j$. 
As a cross check of the quality of the trained multiclass classifier, we look at its performance in identifying the held-out test sets of each submission. A well-trained classifier will be able to distinguish the individual submissions from each other, so
\begin{equation}
    \label{eq:log.post.crosscheck}
    \LogPost(\mathrm{model~}i| \mathrm{samples~}j=i) > \LogPost(\mathrm{model~}i| \mathrm{samples~}j\neq i). 
\end{equation}
We check that this holds for each trained multiclass classifier before using it for final evaluation. We train ten classifiers with different random initialization and average the mean log posteriors of the ten runs. The results of the cross check can be found in appendix~\ref{app:crosscheck}. 

The submissions are split in training, testing, and evaluation sets as shown in~\tref{tab:ttv}, before they are merged and shuffled into single training, testing and evaluation sets. In training (both the DNN and the CNN ResNet architecture), the best model state based on the validation loss is used for the final evaluations.

\paragraph{DNN} We consider a regular DNN for the binary classification on low- and high-level observables, and the multiclass classification setup. The DNN of the binary classification consists of an input layer, two hidden layers of 2048 neurons each, and an output layer. We use leaky ReLU activations (with negative slope $0.01$) in all layers except the last one, where we use a sigmoid activation. We do not use dropout or batch normalization. The network is optimized with the Adam optimizer~\cite{kingma2014adam}, a learning rate of $2\cdot 10^{-4}$, and in batches of 1000 samples for 50 epochs. 

The DNN of the multiclass classification test consists of an input layer, one hidden layer with 4096 neurons, and an output layer. We use leaky ReLU activations (with negative slope $0.01$) in all layers except the last one, where we use a softmax activation. No dropout or batch normalization is used. We optimize the network with a schedule-free AdamW optimizer~\cite{2024arXiv240515682D} and an initial learning rate of $1\cdot 10^{-3}$ in batches of 2000 samples for 25 epochs (or fewer, if the validation loss already increases).  

When classifying ``low-level'' observables, we use the voxel energies normalized to the incident energy and the decadic logarithm of the incident energy as input features. ``High-level'' observables are given by the observables we introduced in~\sref{sec:results_hlf_intro}. 

\paragraph{CNN ResNet} An alternative architecture based on 3D CNNs is considered for the binary and multiclass classification on low-level features. Compared to a fully-connected DNN, a CNN is more capable of exploiting the spatial structures of particle showers, therefore allowing it to provide stronger separation between different models. We adapt a 3D CNN implementation \cite{hara3dcnns} based on the ResNet architecture~\cite{RESNET} to process the particle showers. Each shower is treated as a 3-dimensional image where the intensity of each pixel is the energy deposition in the corresponding voxel. This leads to images of a shape $(45, 16, 9)$ for dataset 2, and $(45, 50, 18)$ for dataset 3. The shower image is first processed by a 3D convolution with a kernel size of 7 and a stride of 2, followed by a max pooling layer with a kernel size of 3 and a stride of 2, for downsampling. Then, an 18-layer ResNet is applied to the downsampled image. A kernel size of 3 is used in all the convolutions in the ResNet, and the number of output channels in each convolution ranges between 32 to 128. A global average pooling is used to aggregate the output to a 1D feature vector summarizing the full image. This feature vector is then concatenated with the incidence energy, normalized with a batch normalization layer \cite{batchnorm}, before being processed by a final fully-connected layer for the classification. 

For dataset 2, we optimize the network for 48 epochs using the AdamW optimizer~\cite{2017arXiv171105101L} with learning rate $2.5\cdot 10^{-5}$ and otherwise default settings. For dataset 3, it was sufficient to use the same optimizer setup, but with learning rate $5\cdot 10^{-5}$ for 12 epochs.

\subsection{Computer Science-inspired Metrics}
\label{sec:results_kpd_intro}

A standard quantitative benchmark for state-of-the-art generative models in computer vision is the Fr\'{e}chet Inception distance (FID) \cite{NIPS2017_8a1d6947}.
The idea behind FID is to extract salient high-level features of real and generated images via the activations of the penultimate layer of a high-performing inception classifier, and then compare them using the Fr\'{e}chet, or 2-Wasserstein, distance between Gaussian fits to the two sets of features.
This metric has been shown to be highly sensitive to the quality and diversity of generated images and has been extended as well to evaluate jet simulations using the ParticleNet classifier \cite{Kansal:2021cqp}.
Recently, however, \cite{Kansal:2022spb} studied a physics-informed alternative to this method, referred to as the Fr\'{e}chet physics distance (FPD) based on high-level \textit{physical} features of the samples, rather than DNN classifier activations, which proved to be highly performant.
The complementary kernel physics distance (KPD) metric was proposed as well, similarly inspired by the popular kernel Inception distance (KID) \cite{2018arXiv180101401B}, which calculates a kernel-based estimate of the maximum mean discrepancy between the two sets of features. 
In this work, we apply FPD and KPD to evaluate the various surrogate models by using the meaningful high-level features of calorimeter simulations outlined in \sref{sec:results_hlf_intro}.
We also importantly derive baseline scores and errors with which to compare the submissions for the different datasets using the procedures described in \cite{Kansal:2022spb}. We use the implementation of \cite{Kansal_JetNet_2023} with a minimum sample size of 10\,000 in the computation of FPD and a batch size of 10\,000 for the KPD.

\subsection{Manifold-based Metrics}
\label{sec:results_coverage_intro}

Manifold-based metrics construct a proxy for the generated and reference data manifold and provide a computationally straightforward way to asses the diversity of the submitted samples. The diversity measures how well the generated samples populate the entire data manifold. There is a trade-off between realism and diversity~\cite{2024arXiv240610429A} observed for natural images, which immediately triggers the question if such a trade-off also exists for calorimeter showers. Here, we study four different metrics: Precision, Recall, Density, and Coverage, which are defined in the following~\cite{2018arXiv180600035S,2019arXiv190406991K,2020arXiv200209797F}. While Precision and Density are more a measure of shower quality, Recall and Coverage measure diversity. We report results on the former two as well because all four metrics are closely related to each other and correlations between them provide additional insights. 

Precision and Recall first construct a manifold of ``real'', \textit{i.e.} reference, and ``fake'', \textit{i.e.} generated samples. These are defined as the union of all $d$-dimensional spheres around the points $x_i$, with the radii chosen such that the $k$ nearest samples are inside the sphere, 

\begin{equation}
    \label{eq:manifold}
    \mf{x_1, \dots, x_n} = \bigcup_{i=1}^{n} \sphere{x_i}{\nnd{x_i}{k}}.
\end{equation}

Here, $\sphere{x}{r}$ defines a sphere around $x$ with radius $r$ and $\nnd{x}{k}$ denotes the distance of $x$ to its $k$th nearest neighbor. 

\begin{itemize}
    \item \textit{Precision}. Following the definition of the improved precision of \cite{2019arXiv190406991K}, it counts the binary decision of whether the generated data $y_j$ is contained in any neighborhood sphere of reference samples $x_i$. It is bounded by 1.
    \begin{equation}
        \label{eq:precision}
        \textrm{precision} = \frac{1}{m}\sum_{j=1}^{m} \mathbf{1}_{y_j \in \mf{x_1, \dots, x_n}}
    \end{equation}
    Here, $\mathbf{1}$ is the indicator function and $n(m)$ is the number of reference (generated) samples. 
    \item \textit{Density} improves the Precision metric by taking into account that the manifold around reference outliers is overestimating the manifold~\cite{2020arXiv200209797F}. It counts how many reference-sample neighborhood spheres contain $y_j$. The manifold is now defined as the superposition of spheres instead of the union, and models that place samples in regions where the reference samples are densely packed are getting a higher score. However, it is not bounded by 1 anymore. 
    \begin{equation}
        \label{eq:density}
        \textrm{density} = \frac{1}{km}\sum_{j=1}^{m}\sum_{i=1}^{n} \mathbf{1}_{y_j \in \sphere{x_i}{\nnd{x_i}{k}}}
    \end{equation}    
    \item \textit{Recall}. Following the definition of the improved recall of \cite{2019arXiv190406991K}, it is symmetrically with respect to precision. It counts the binary decision of whether the reference data $x_i$ is contained in any neighborhood sphere of generated samples $y_j$. It is also bounded by 1.
    \begin{equation}
        \label{eq:recall}
        \textrm{recall} = \frac{1}{n}\sum_{i=1}^{n} \mathbf{1}_{x_i \in \mf{y_1, \dots, y_m}}
    \end{equation}    
   \item \textit{Coverage} measures the fraction of reference samples whose neighborhoods contain at least one generated sample. It is bounded between 0 and 1.
    \begin{equation}
        \label{eq:coverage}
        \textrm{coverage} = \frac{1}{n} \sum_{i=1}^{n} \mathbf{1}_{\exists j~\textrm{s.t.}~y_j \in \sphere{x_i}{\nnd{x_i}{k}}}
    \end{equation}
\end{itemize}
In our analysis, we chose $k=5$~\cite{2020arXiv200209797F} and preprocess all voxels by $\log_{10}{\mathcal{I}}$. Voxels without energy deposition, \textit{i.e.}~below threshold, are set to $\log_{10}{0.1~\textrm{MeV}} = -1$. 

\subsection{Generation Timings}
\label{sec:results_timing_intro}
To properly compare the generation times of all submissions, each submitting group created a singularity container~\cite{10.1371/journal.pone.0177459} of the necessary software environment.  We transferred them to the \texttt{clip} cluster~\cite{clip} and measured the time it takes to load the container, load the model (weights and biases), if applicable move it on the GPU, generate the samples, and save them as \texttt{.hdf5}~\cite{hdf5} to disk. While this contains an additional overhead, we think that it is more realistic, closer to the real-life application. There is some scatter from run to run, but that comes from the execution on a cluster and most likely is also present in a full simulation chain. We therefore repeat these steps ten times and show the mean and standard deviation of the run times.  

Current fast simulation frameworks, with or without deep generative models, simulate with batch size of 1, since this is how simulation is handled in \geant, with the parallelization applied commonly on the event, and not particle, level (different events are simulated simultaneously in different threads). Most of the DGM architectures, however, benefit from larger batch sizes. We therefore study batch sizes of 1, 100, and 10\,000 to show how the models behave under different use cases. 

Unless explicitly noted otherwise, we generate as many samples as were in the training set. However, for some models and smaller batch sizes we had to reduce the overall number of generated events. In this cases, the overhead of loading the model will have a higher share in the overall generation time compared to the sample generation. 

DGMs usually run a lot faster on graphics processing units (GPU), since these are optimized for matrix-vector multiplications. Yet, these are not as widely available on HPC clusters, where the majority of nodes have only CPUs. We therefore run the timing evaluations on both types of hardware and report the results. We run the CPU timings on an Intel\textsuperscript{\tiny\textregistered} Xeon\textsuperscript{\tiny\textregistered} Gold 6138 CPU @2.00GHz with 170 GB RAM. While this is more on the slow end, we used this node because of the larger RAM requirements of some models. The GPU timings were done with a NVIDIA\textsuperscript{\tiny\textregistered} A100-SXM4 with 40GB Graphics RAM, 360GB RAM, and Intel\textsuperscript{\tiny\textregistered} Xeon\textsuperscript{\tiny\textregistered} Gold 6226R CPU @2.90GHz. These are the C2 and G4 partitions of the~\texttt{clip}~cluster~\cite{clip}, respectively. 

\subsection{Memory Requirements}
\label{sec:results_memory_intro}
As a proxy of the memory requirements to store each model on disk, we report the number of trainable parameters that each model requires. In particular, we report two numbers. One refers to how many trainable parameters are involved in training the generative model. The other one refers to how many trainable parameters are required for generation, \textit{i.e.}~how many need to be loaded in production. These numbers can differ for example in GANs, where only the generator network is needed in production, or in cases where the generative network is a distilled version of another model. We know that the actual memory requirements depend on the floating point representation used for the parameters and on the number of additional, non-trainable parameters that are required to load and run the model. Techniques like node pruning and weight quantization can reduce the number of parameters and the memory footprint significantly, sometimes without loss in sample quality. Nevertheless, we decided to not focus on these aspects and just work with the number of trainable parameters. 
\FloatBarrier

\section{Results: Individual Metrics}
\label{sec:results}
\markboth{\uppercase{Results: Individual Metrics}}{}

\begin{figure}[th]
    \centering
\includegraphics[width=0.75\textwidth]{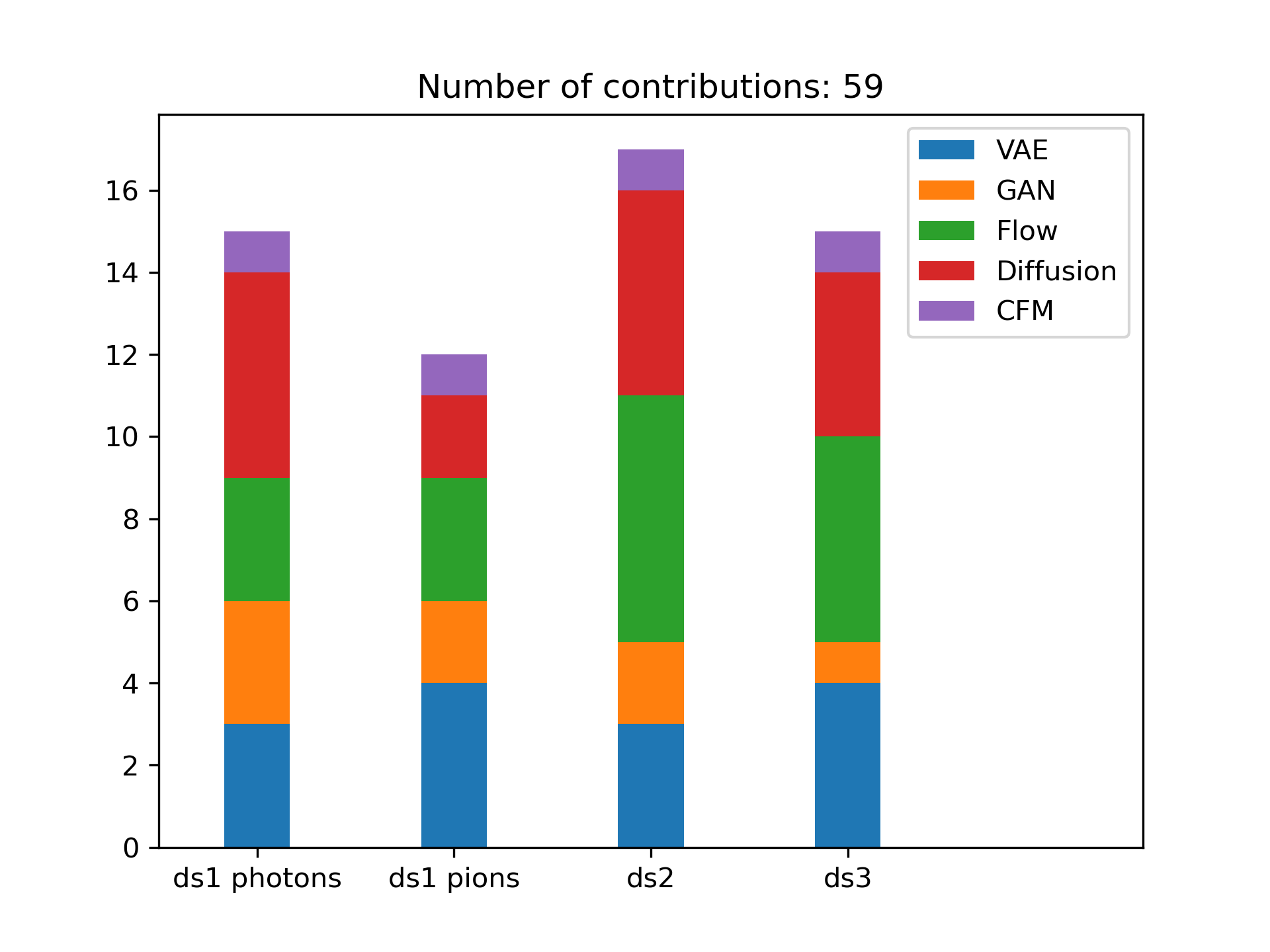}
    \caption{Number of submissions per dataset and DGM architecture.}
    \label{fig:num_contrib}
\end{figure}

In total, we present here the results of 59 submissions. These are evenly spread across the different datasets and generative model architectures as can be seen in~\fref{fig:num_contrib} and \tref{tab:submissions}. 

\subsection{Preprocessing}

All submitted files were checked for \texttt{NaN} entries, if the $E_{\textrm{inc}}$ distribution matches the expectation, and if the correct number of samples were submitted. Then they were saved as \texttt{np.float32} numbers in a \texttt{hdf5} file~\cite{hdf5} with gzip compression. A threshold cut was applied to all voxels before they were used in evaluation. The \geant reference was treated the same way, and all results below use the second \geant dataset that was provided at~\cite{michele_faucci_giannelli_2023_8099322,faucci_giannelli_michele_2022_6366271,faucci_giannelli_michele_2022_6366324}.

\subsection{\texorpdfstring{Dataset 1, Photons (\dsIph)}{Dataset 1, Photons}}
\label{sec:results_ds1-photons}
\begin{figure}
\centering
\includegraphics[width=\textwidth]{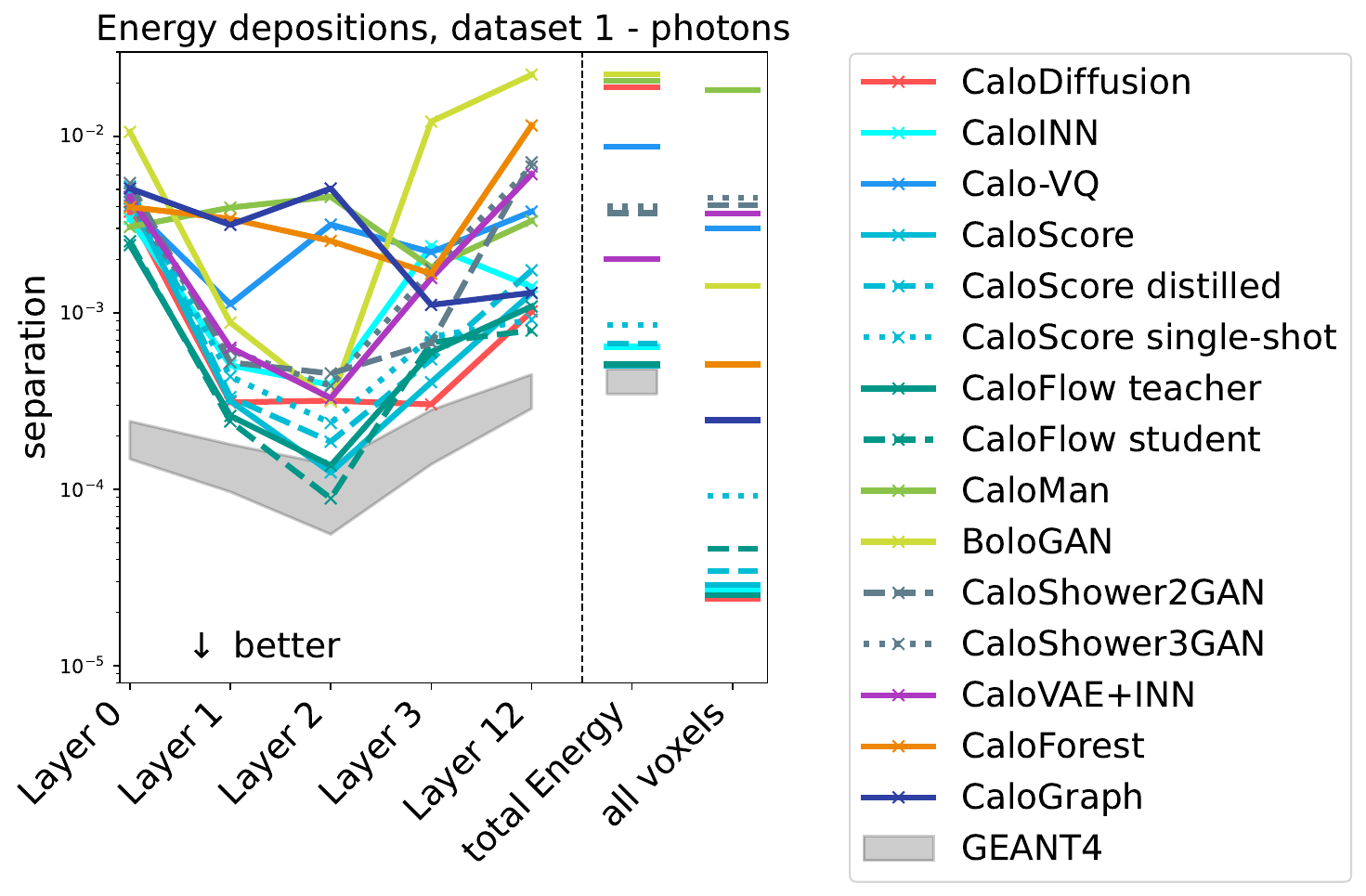}
\caption{Separation power of energy depositions with threshold at 1~MeV.}
\label{fig:ds1-photons-1-depositions}
\end{figure}

\begin{figure}
\centering
\includegraphics[width=\textwidth]{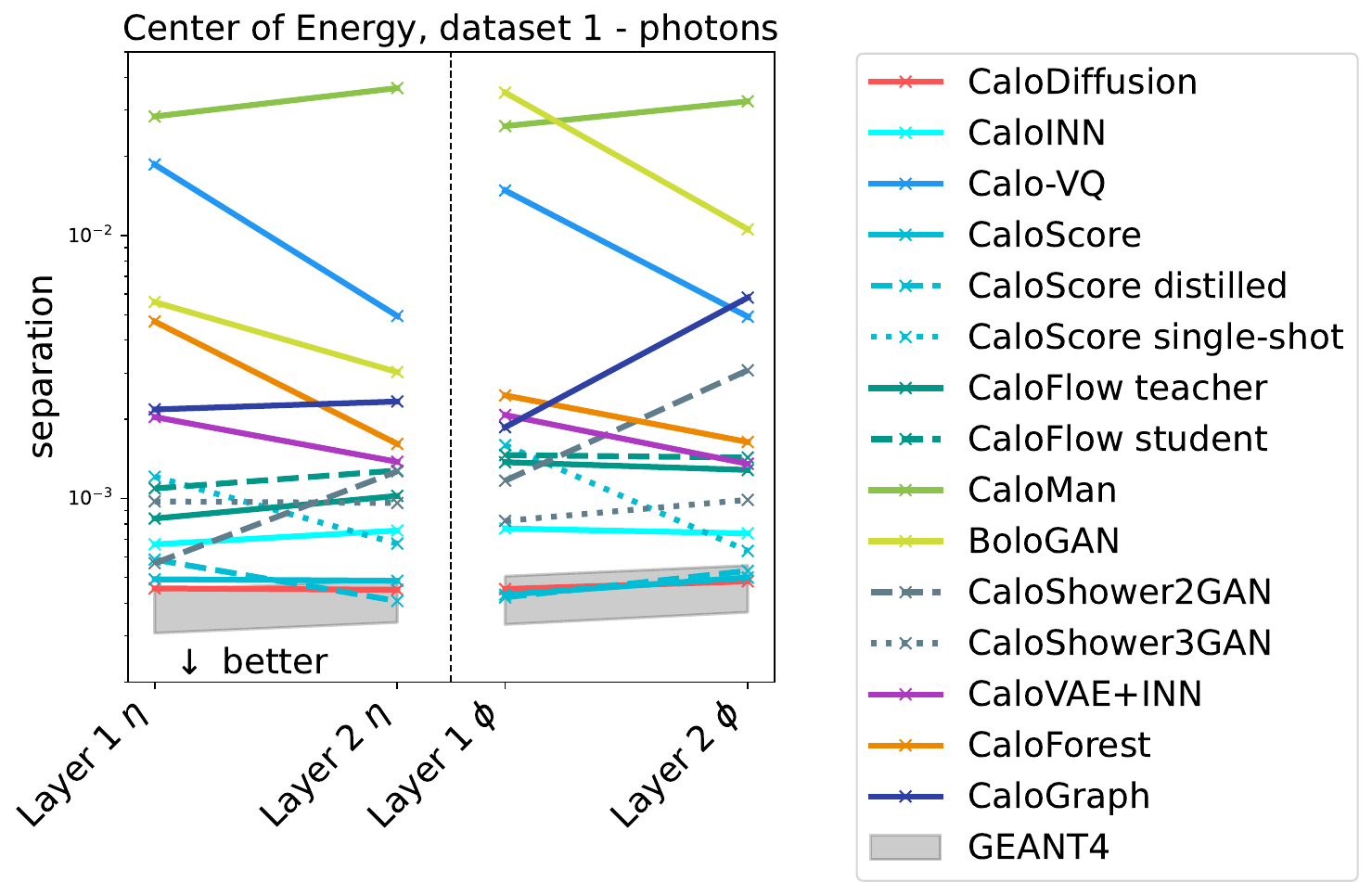}
\caption{Separation power of centers of energy with threshold at 1~MeV.}
\label{fig:ds1-photons-1-CE}
\end{figure}

\begin{figure}
\centering
\includegraphics[width=\textwidth]{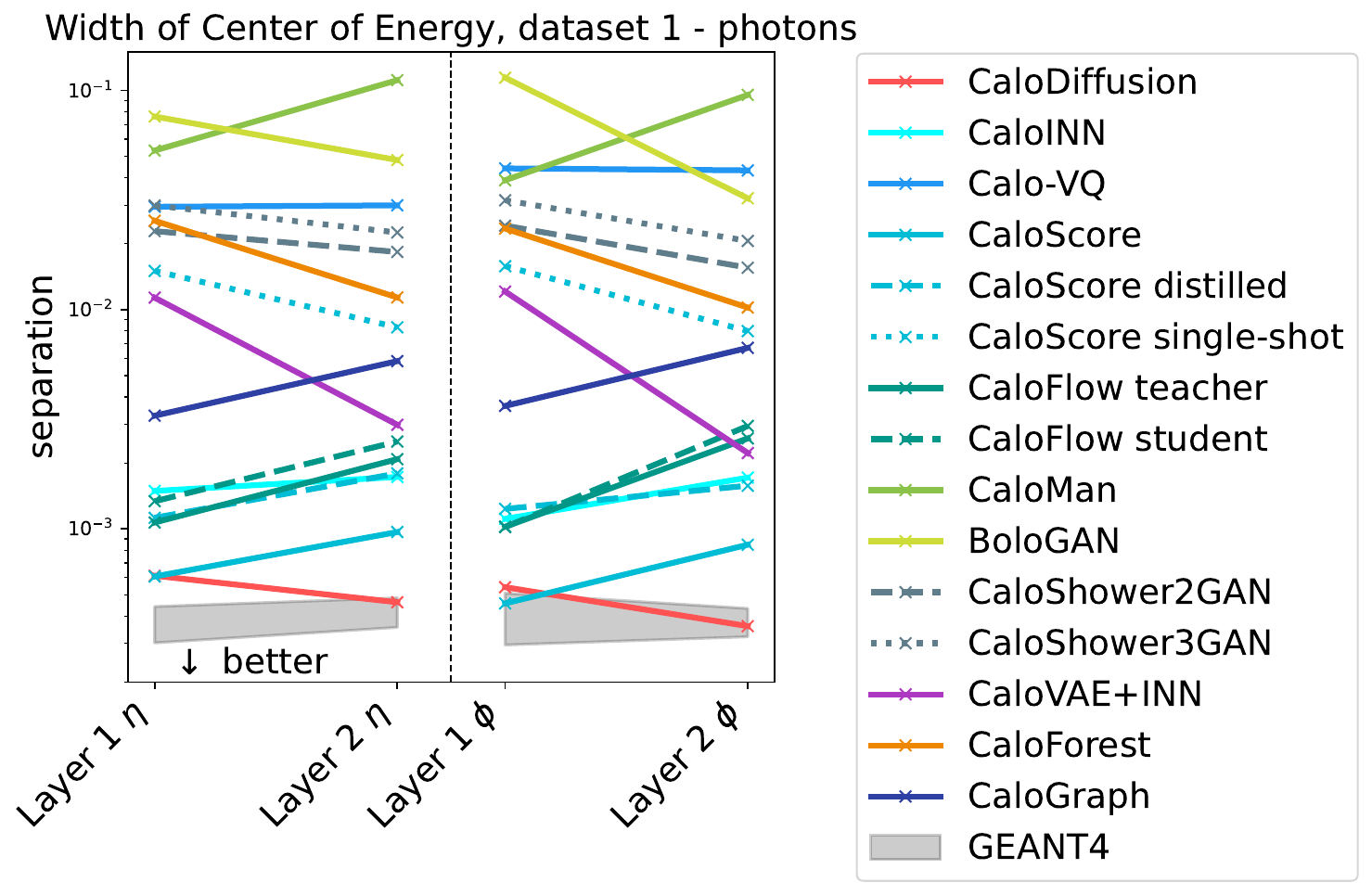}
\caption{Separation power of widths of centers of energy with threshold at 1~MeV.}
\label{fig:ds1-photons-1-width-CE}
\end{figure}

\begin{figure}
\centering
\includegraphics[width=\textwidth]{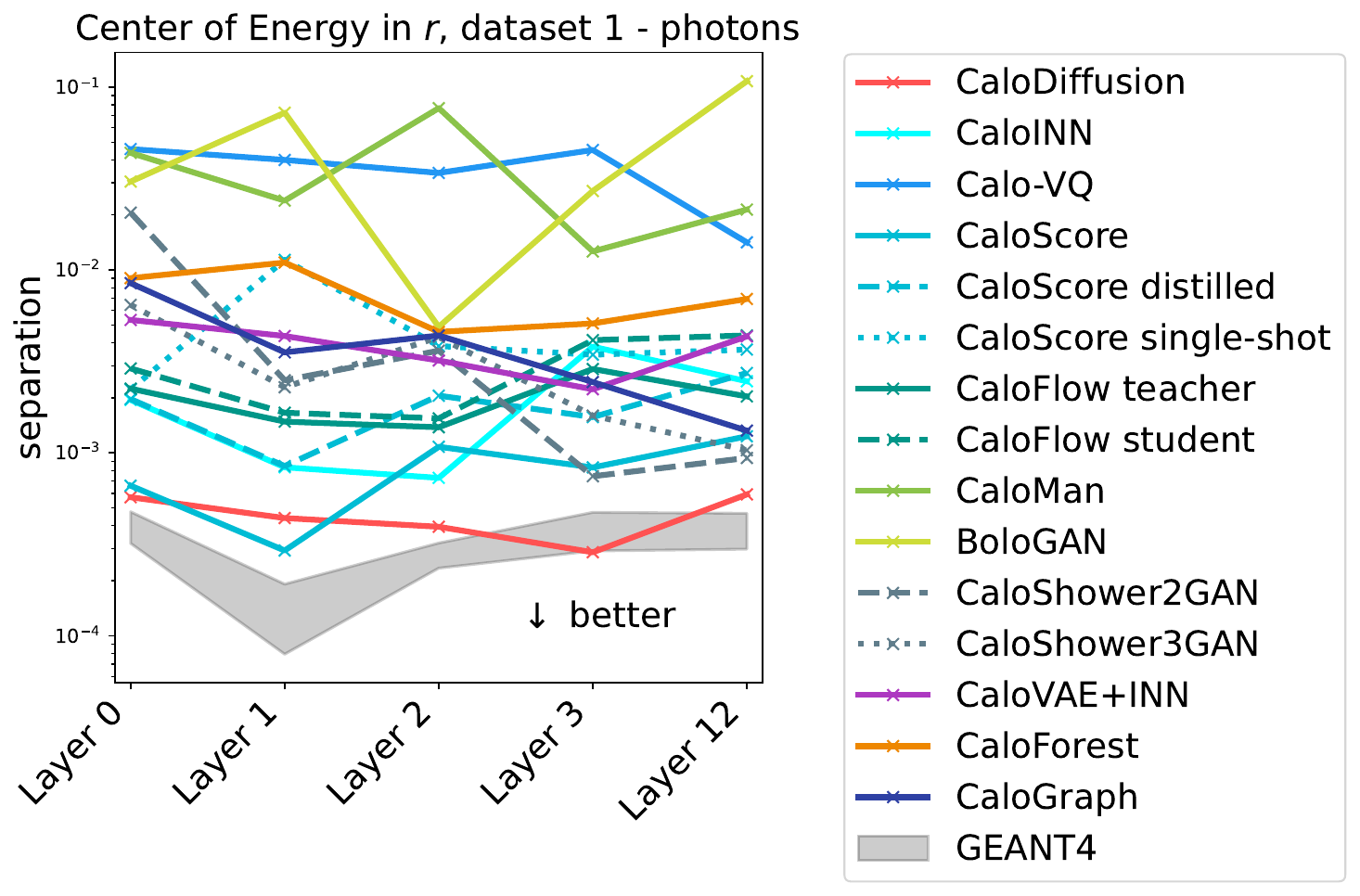}
\caption{Separation power of centers of energy with threshold at 1~MeV.}
\label{fig:ds1-photons-1-CE-r}
\end{figure}

\begin{figure}
\centering
\includegraphics[width=\textwidth]{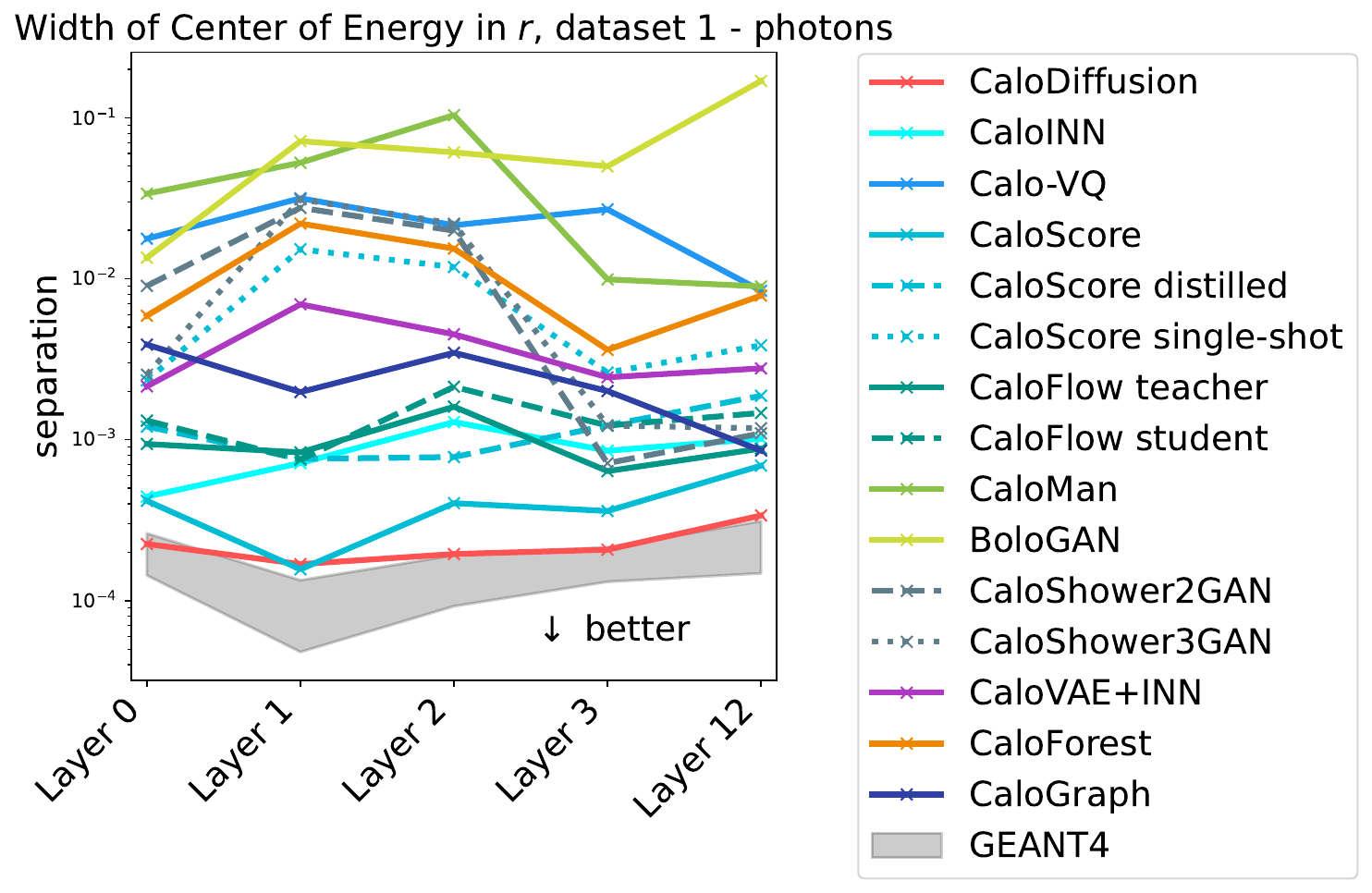}
\caption{Separation power of widths of centers of energy with threshold at 1~MeV.}
\label{fig:ds1-photons-1-width-CE-r}
\end{figure}

\begin{figure}
\centering
\includegraphics[width=\textwidth]{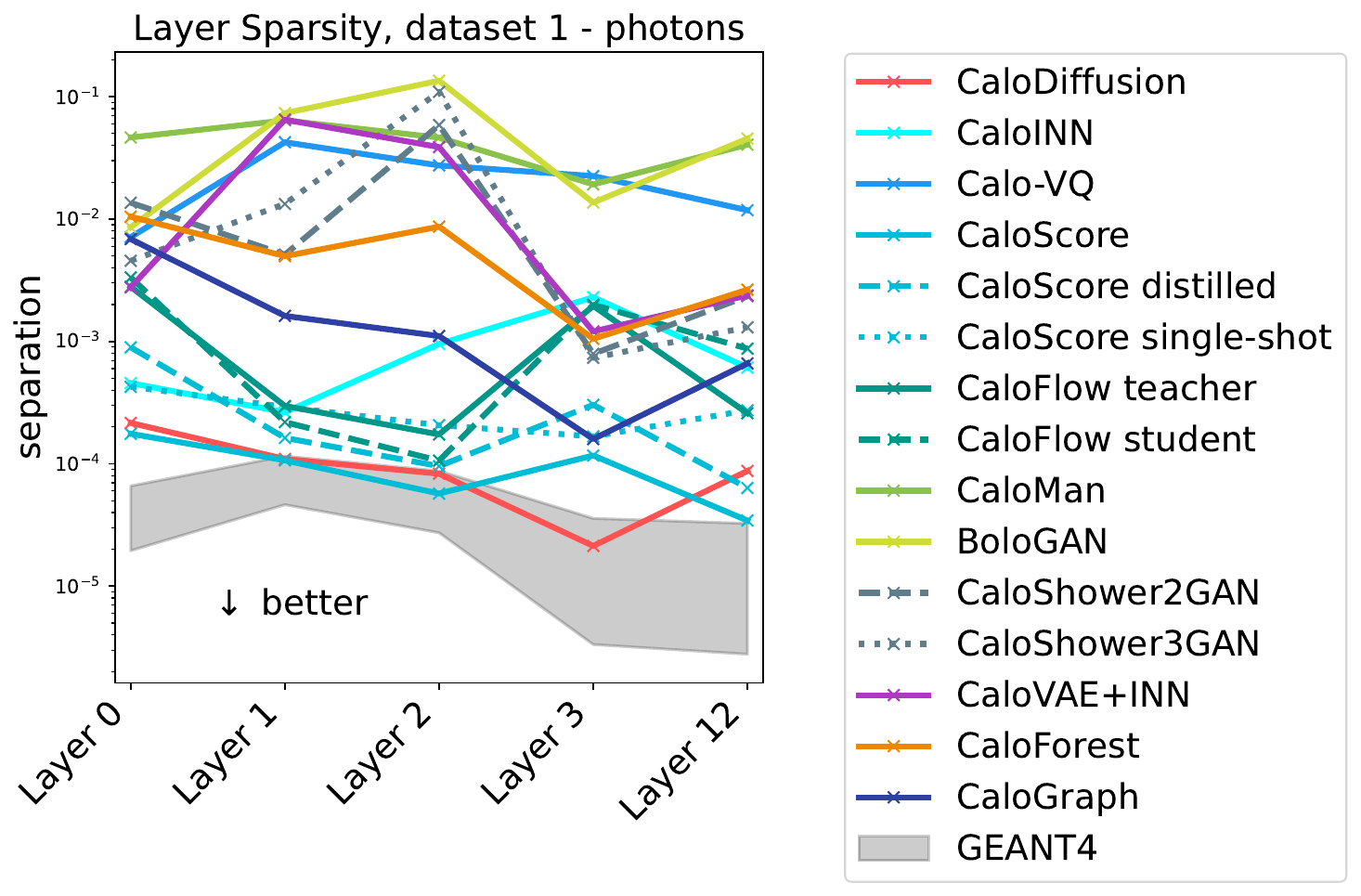}
\caption{Separation power of the sparsity with threshold at 1~MeV.}
\label{fig:ds1-photons-1-sparsity}
\end{figure}

\begin{figure}
\centering
\hfill\includegraphics[width=0.4\textwidth]{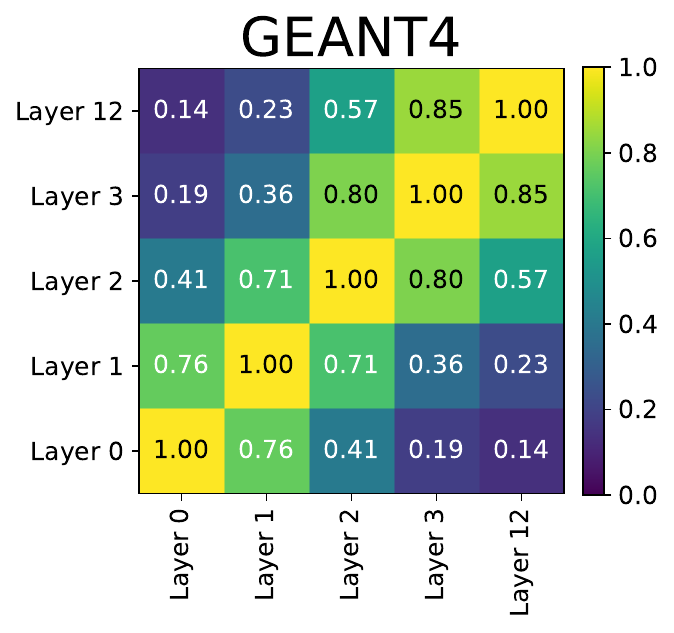}\hfill $ $
\\
\includegraphics[width=0.2\textwidth]{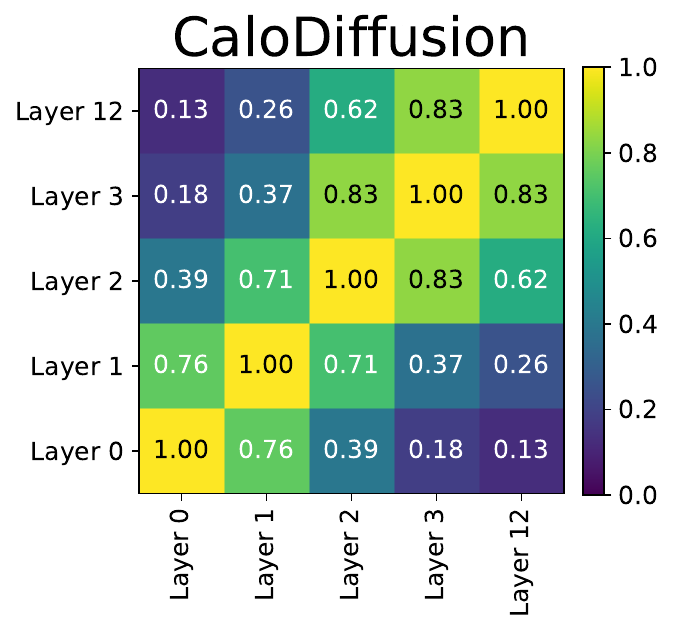}\hfill
\includegraphics[width=0.2\textwidth]{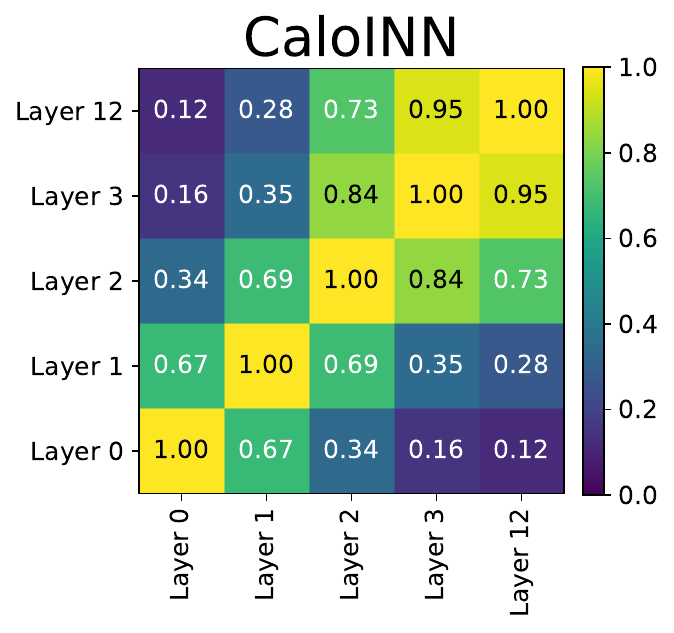}\hfill
\includegraphics[width=0.2\textwidth]{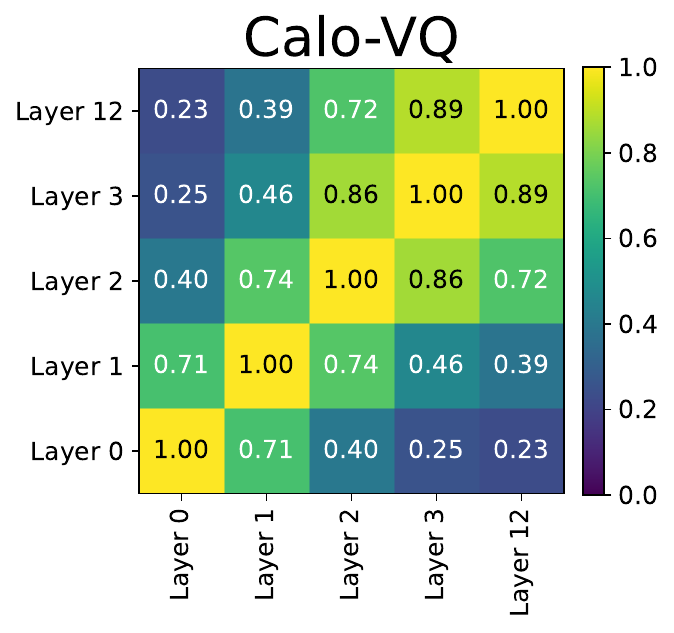}\hfill
\includegraphics[width=0.2\textwidth]{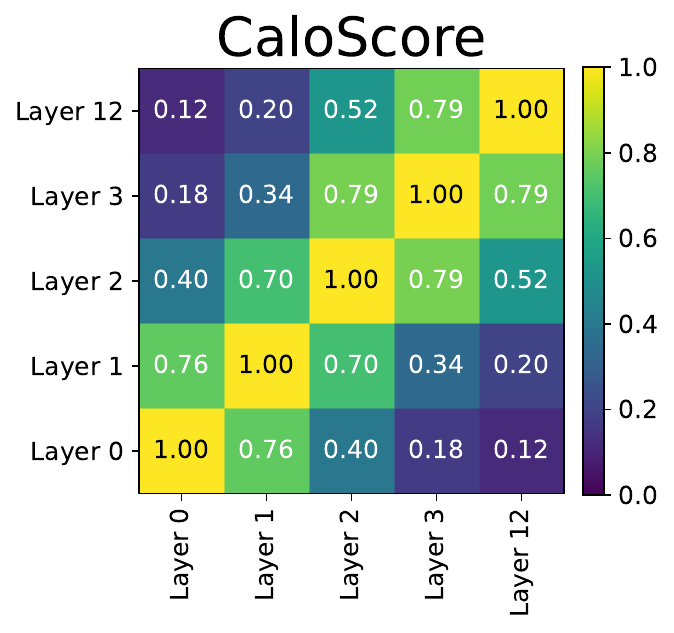}\hfill
\includegraphics[width=0.2\textwidth]{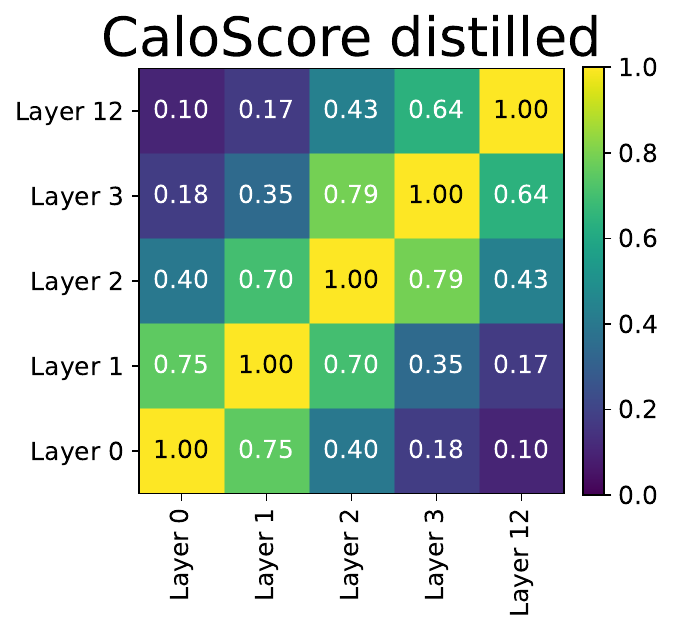}\\
\includegraphics[width=0.2\textwidth]{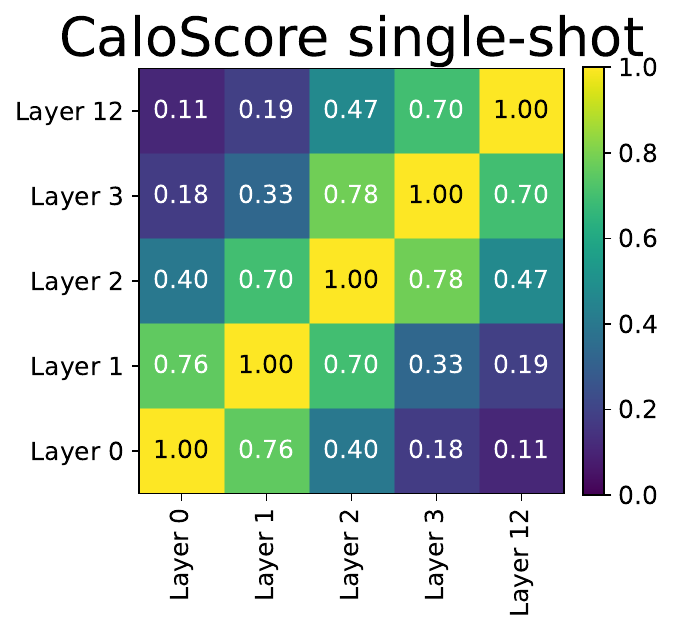}\hfill
\includegraphics[width=0.2\textwidth]{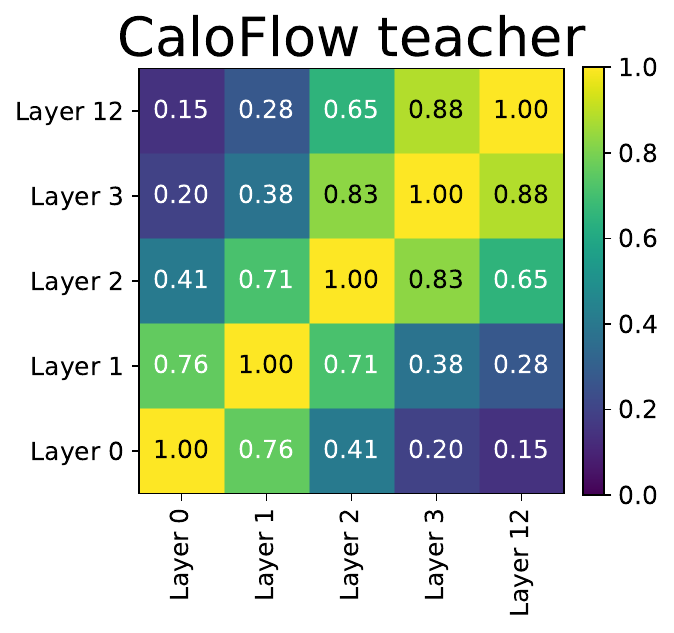}\hfill
\includegraphics[width=0.2\textwidth]{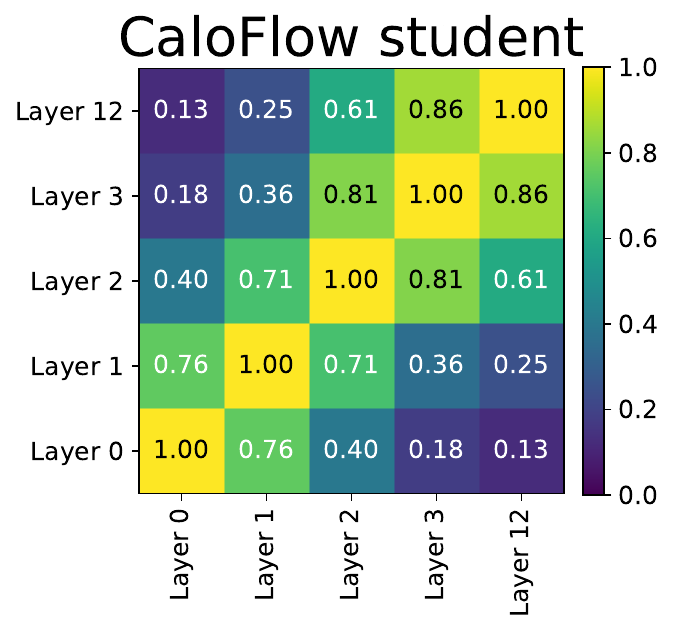}\hfill
\includegraphics[width=0.2\textwidth]{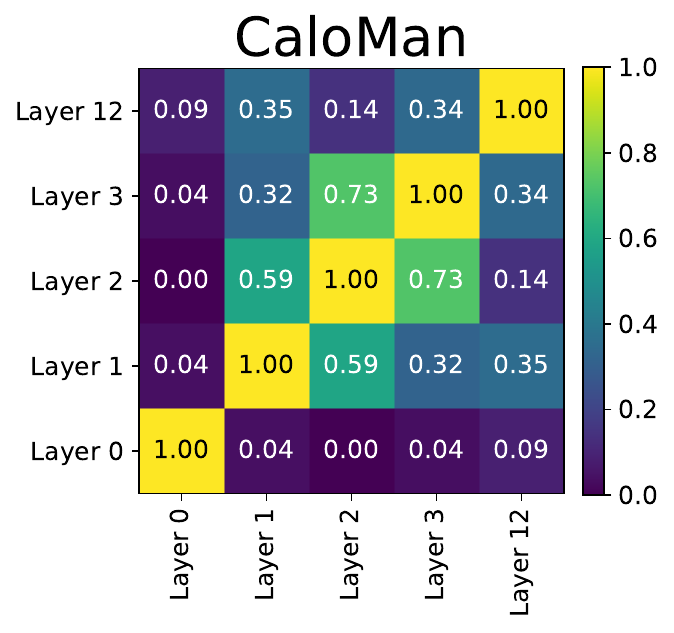}\hfill
\includegraphics[width=0.2\textwidth]{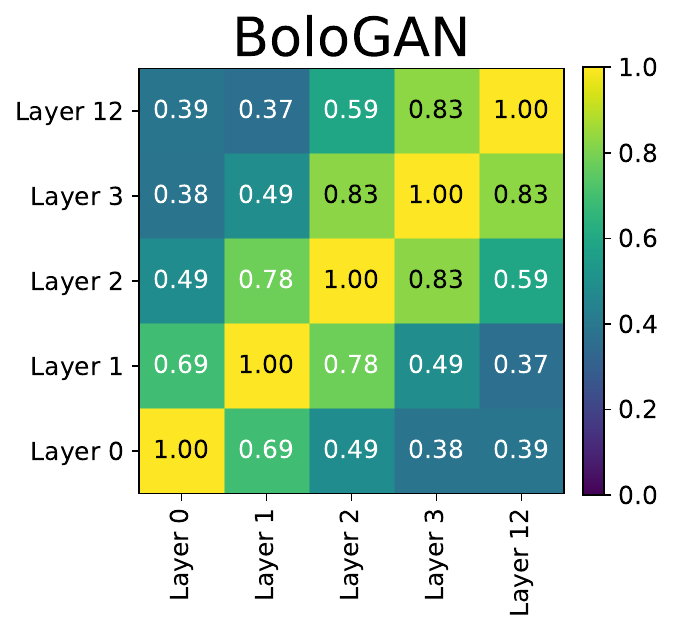}\\
\includegraphics[width=0.2\textwidth]{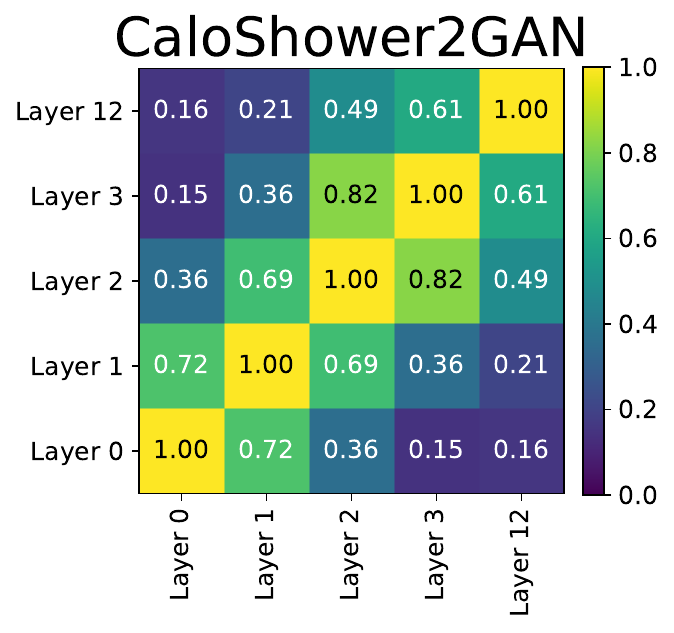}\hfill
\includegraphics[width=0.2\textwidth]{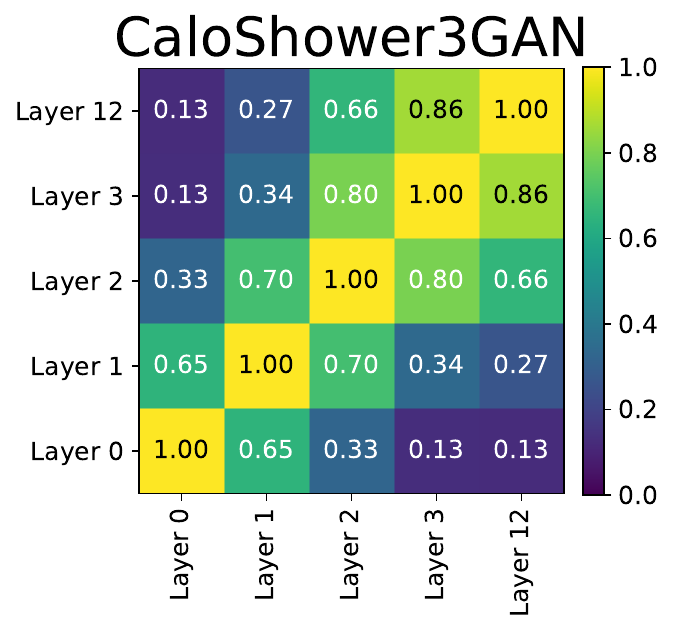}\hfill
\includegraphics[width=0.2\textwidth]{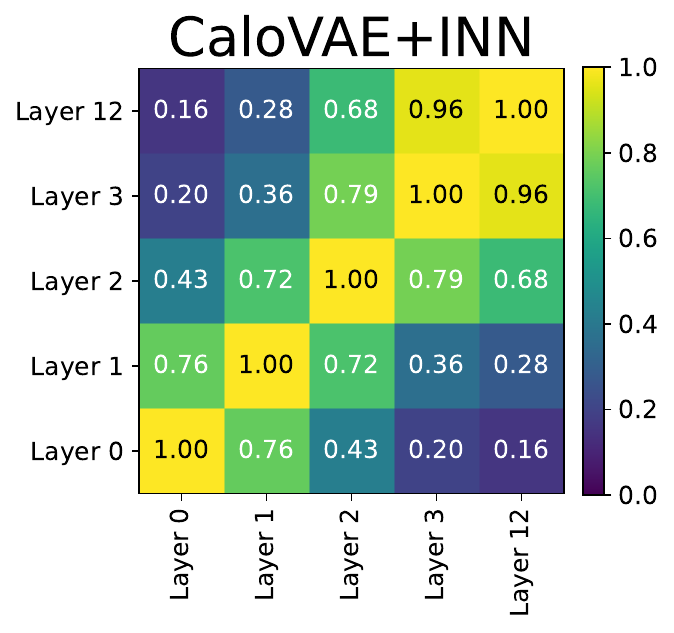}\hfill
\includegraphics[width=0.2\textwidth]{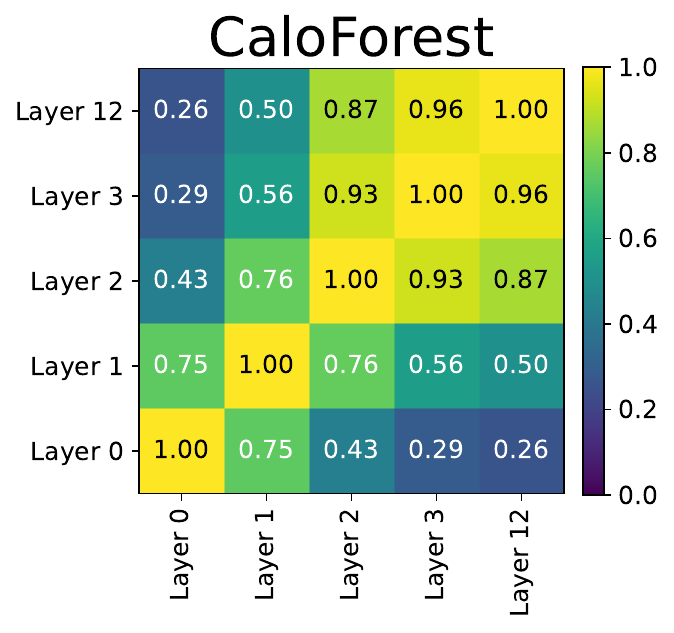}\hfill
\includegraphics[width=0.2\textwidth]{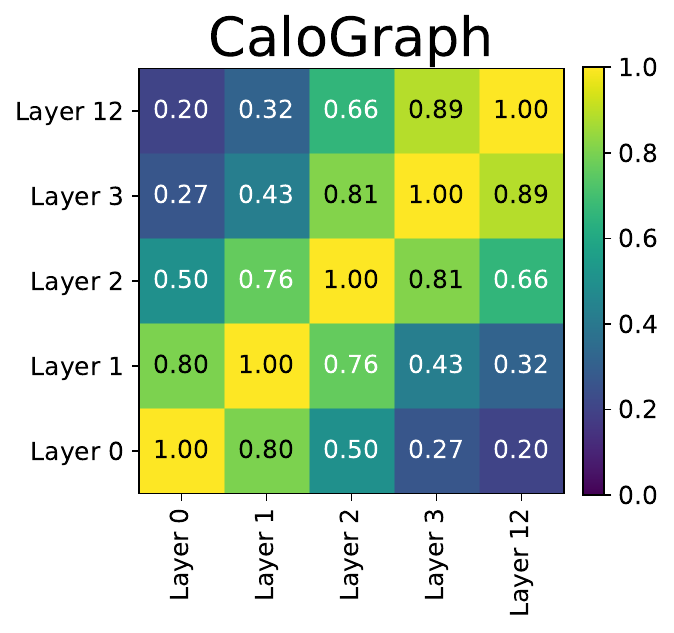}\\
\caption{Pearson correlation coefficients of layer energies in \dsIph, with threshold at 1~MeV.}
\label{fig:ds1-photons-1-corr}
\end{figure}

We begin the evaluation with the high-level features, and especially the energy depositions in~\fref{fig:ds1-photons-1-depositions}. The separation power of the submissions vary roughly within 2 orders of magnitude and they stay almost everywhere just at the upper limit of the \geant reference. It is interesting to note that almost all submissions show a better performance, \textit{i.e.}~a smaller separation power, in layers with an angular segmentation (1 and 2, see \tref{tab:voxelization}). Having more voxels per layer seems therefore beneficial for modeling the layer energies. The best performance is given by normalizing flow (\cf) and diffusion model (\submMikuni) based submissions. We should note that the otherwise well-performing \submAmram has a bad separation power in the total energy deposition, one of the crucial observables. 

The centers of energy in $\eta$ and $\phi$ direction are summarized in ~\fref{fig:ds1-photons-1-CE}. Here we see the diffusion model based submissions \submAmram and \submMikuni with the best performance, at the level of the \geant reference. In general, we observe all models performing equally well in $\eta$ and $\phi$ direction. 

The widths of centers of energy in $\eta$ and $\phi$ in~\fref{fig:ds1-photons-1-width-CE} tell a similar story. Again, the model performance in $\eta$ and $\phi$ directions is about the same, and the diffusion models \submAmram and \submMikuni have the best separation power, just slightly worse than the \geant reference. However, the distilled versions \submMikuniDist and \submMikuniSingle are worse now, having a larger separation power than the first normalizing flow models of \cf. 

The centers of energy in $r$ in~\fref{fig:ds1-photons-1-CE-r} and its width in~\fref{fig:ds1-photons-1-width-CE-r} show separation powers that are more or less constant from layer to layer, stemming from the fact that $N_r$ roughly stays within one order of magnitude. A few submissions show worse performance in the width for layers 1 and 2, where the angular segmentation is present. The ordering of the different DGMs is about the same as for the $\eta$ and $\phi$ directions, as these are correlated. \submAmram is at the upper level of the \geant reference and \submMikuni is next, with the distilled versions a little worse and at the level of the normalizing flow submissions \submFavaro and \cf. 

The last observables we compare with separation powers are the sparsities shown in~\fref{fig:ds1-photons-1-sparsity}. These show the largest spread among the considered observables, spanning four orders of magnitude between the \geant reference and the worst performing submission. The ordering of the models, however, is similar to all the other considered observables. The diffusion models are at the upper end of the reference, with distilled versions in between normalizing flow based models. \submKobylyansky, which was just slightly worse than these in all other observables too, is now at the same level. 

We now move on to investigate the correlations between the energies deposited in the layers in~\fref{fig:ds1-photons-1-corr}. Overall, most of the submissions reproduce the pattern induced by \geant well, but there is a noticeable tendency of models to overestimate the correlation between layers 3 and 12, as seen in the top right corners. Some models based on GANs and VAEs, which had higher separation powers, also seem to have a harder time reproducing these correlations. 

\begin{figure}
\centering
\includegraphics[width=\textwidth]{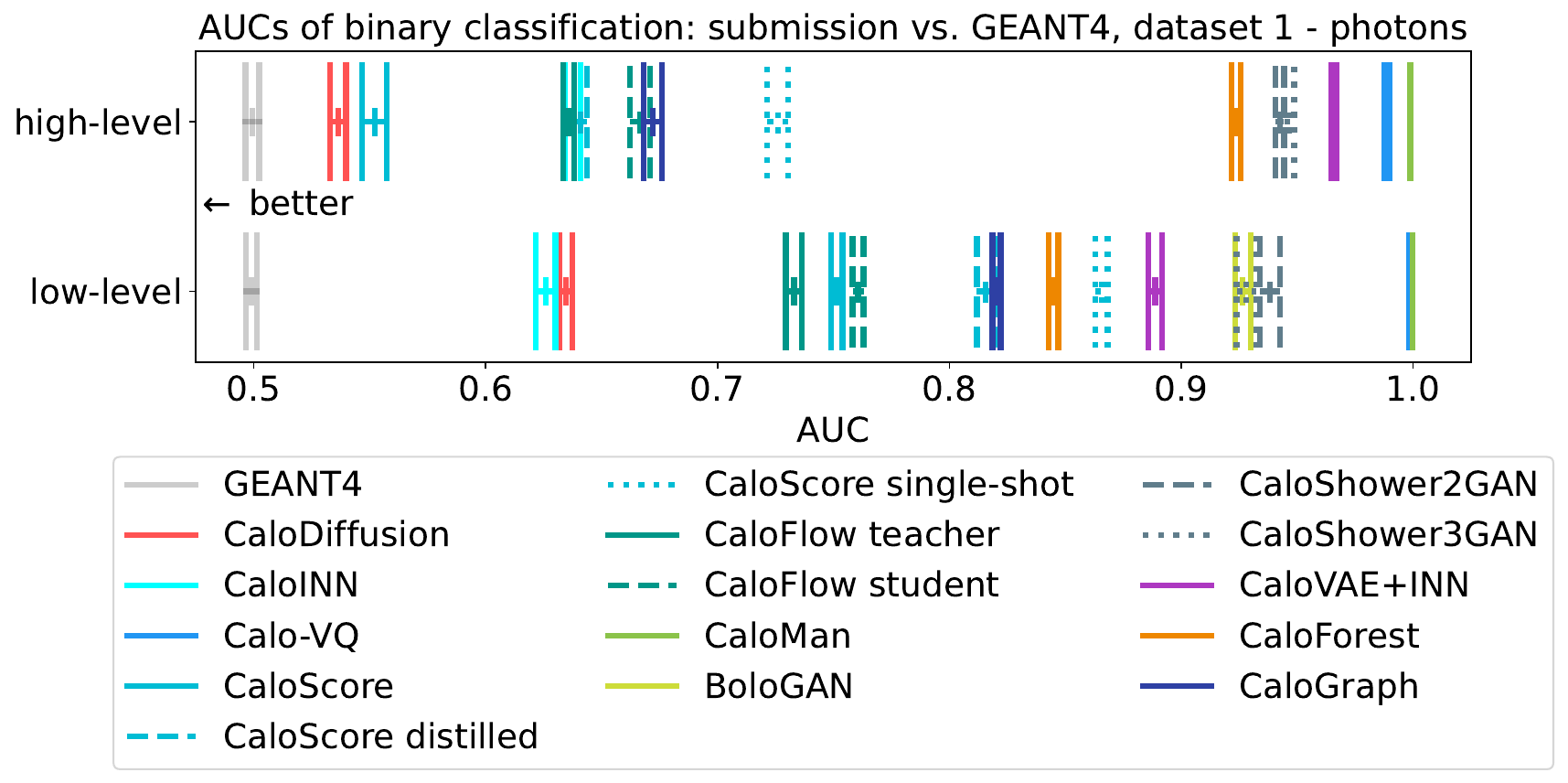}
\caption{Low-level and high-level AUCs for evaluating \geant\ vs.~submission of \dsIph, averaged over 10 independent evaluation runs. For the precise numbers, see \Tref{tab:ds1-photons.aucs}.}
\label{fig:ds1-photons.aucs}
\end{figure}

\begin{figure}
\centering
\includegraphics[width=\textwidth]{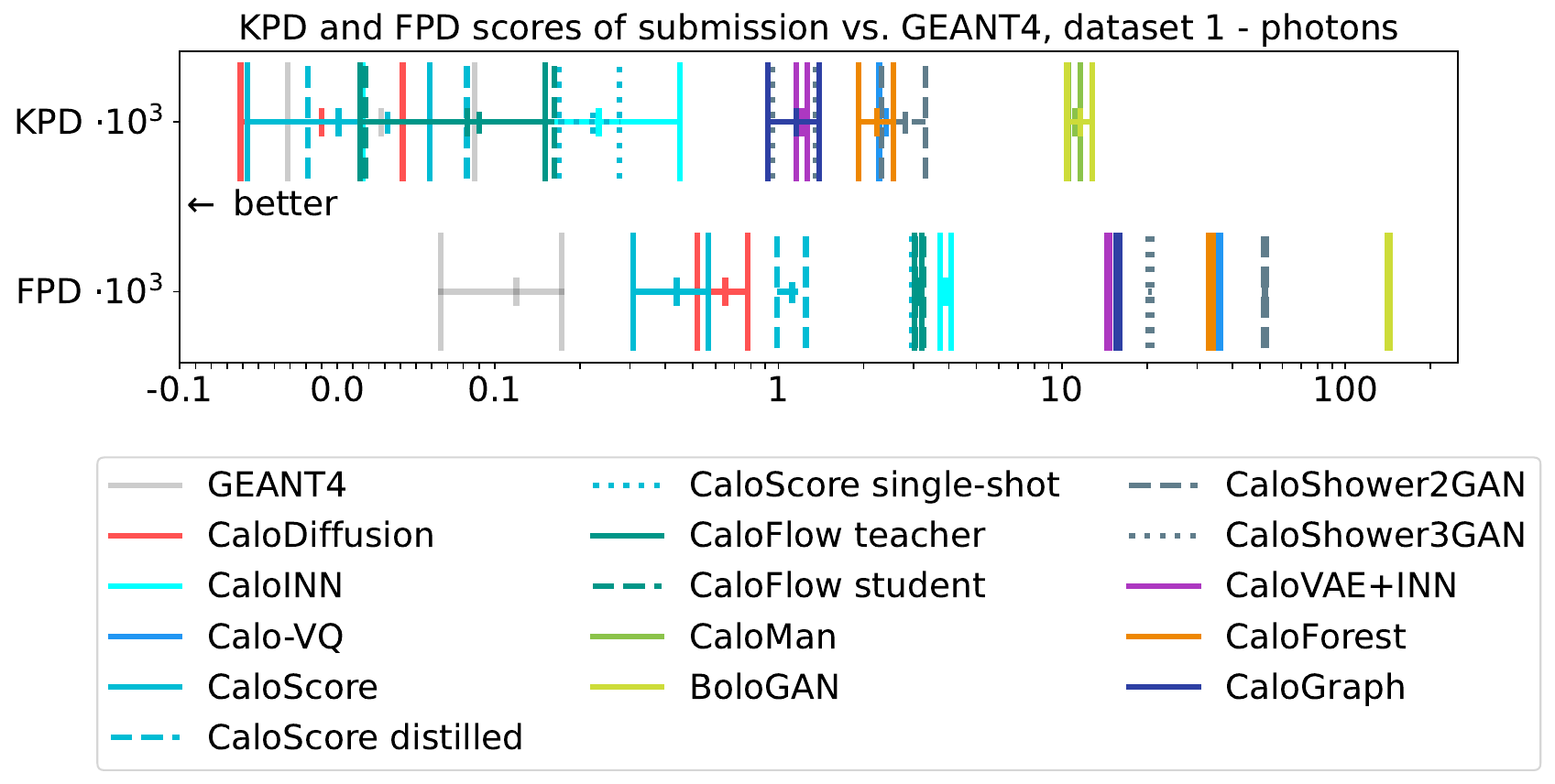}
\caption{KPD and FPD for evaluating \geant\ vs.~submission of \dsIph. For the precise numbers, see \Tref{tab:ds1-photons.kpd}.}
\label{fig:ds1-photons.kpd}
\end{figure}

Another way to look at the correlations between all observables is given by the classifier metric. \Fref{fig:ds1-photons.aucs} (and \tref{tab:ds1-photons.aucs}) shows the AUCs of classifying low-level and high-level observables of the submission against the \geant reference. In general, we observe a good consistency between the two sets of observables and a small spread of AUCs for reruns with different initialization. Submissions that have a high (low) score in the low-level observables also have a high (low) score when high-level observables are used as an input. The difference between the AUCs of the same submission is below 0.2. These results are also consistent to what we have seen in the separation power. DGMs based on diffusion models or normalizing flows achieve the best results, with AUCs of $\mathcal{O}(0.6)$. We also observe that distilled versions tend to perform worse compared to their base model. This is more prominent for \submMikuni distilled to \submMikuniDist and \submMikuniSingle than for \submPangT distilled to \submPangS.  

When judged by KPD and FPD in \fref{fig:ds1-photons.kpd} (see also \tref{tab:ds1-photons.kpd}), the relative performance of the submissions is confirmed by this metric, too. We do see, however, that these metrics (especially the KPD) scatter a bit more, so the flow-based and diffusion model-based submission's scores now almost all agree with each other within the uncertainties. This larger scatter of the KPD would also result in concluding that some submissions are indistinguishable from the reference data, since the KPD is consistent with 0. This, however, cannot be confirmed by the FPD and the AUCs of \fref{fig:ds1-photons.aucs}, which see the best scores of \submAmram still significantly away from the baseline scores obtained with the \geant reference. 

\begin{figure}
\centering
\includegraphics[width=\textwidth]{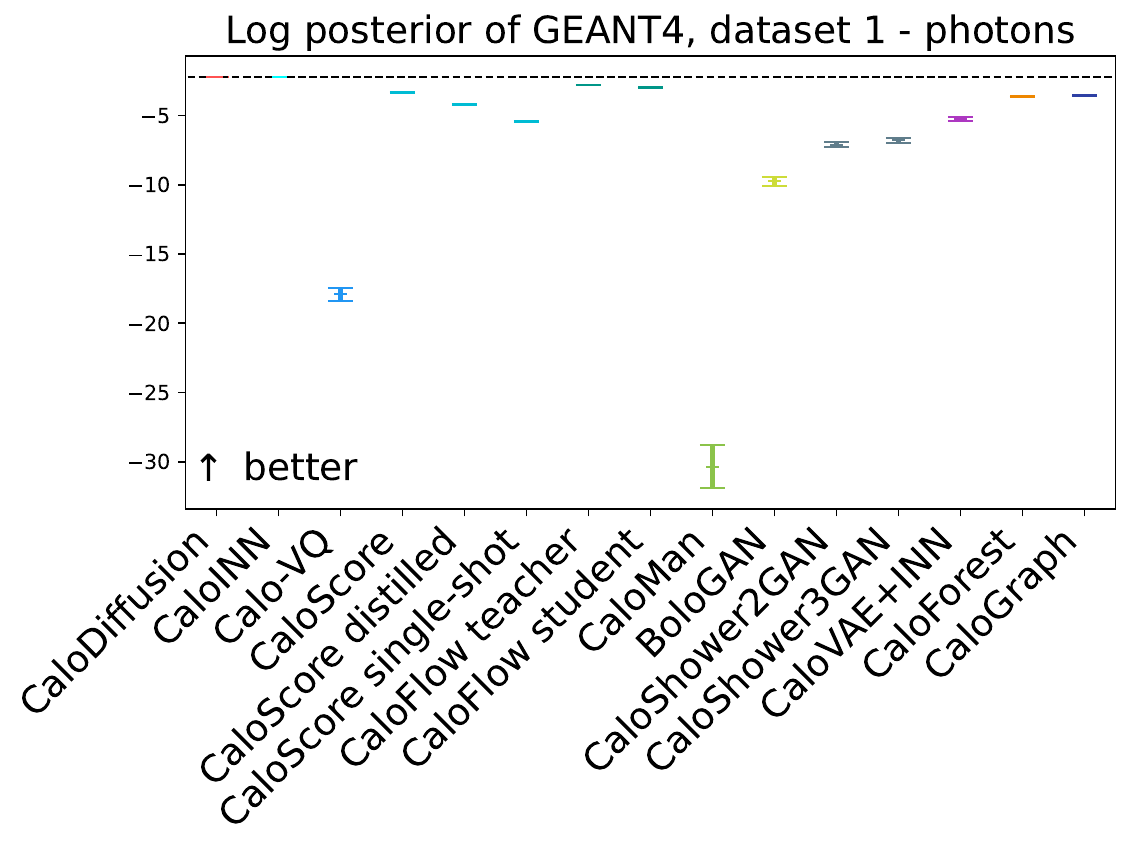}
\caption{Log-posterior scores for \dsIph \geant test data, averaged over 10 independent classifier trainings. For the precise numbers, see \Tref{tab:ds1-photons.multi}.}
\label{fig:ds1-photons.multi}
\end{figure}

Now we move on to the multiclass classifier. The crosscheck of the well-trained classifier can be found in \fref{fig:consistency.ds1photons}. \Fref{fig:ds1-photons.multi} shows the main results, the mean log-posterior for the \geant test set (and the same results are reported in~\tref{tab:ds1-photons.multi}). These results are consistent with the other classifier test, with \submAmram and \submFavaro in the lead. Interestingly, \submMikuni, which was having good results in terms of the separation power of the high-level observables, was overtaken by the classifier metrics by normalizing flow-based submissions like \cf and \submFavaro. 

Overall, in terms of shower quality of \dsIph, we observe that some models approach the \geant reference, telling us that the comparatively easy and low-dimensional distribution of photon showers can indeed be learned by DGMs. In particular, we see that diffusion model and normalizing flow-based submissions get consistently better scores than GAN and VAE-based submissions. 

\begin{figure}
\centering
\includegraphics[width=\textwidth]{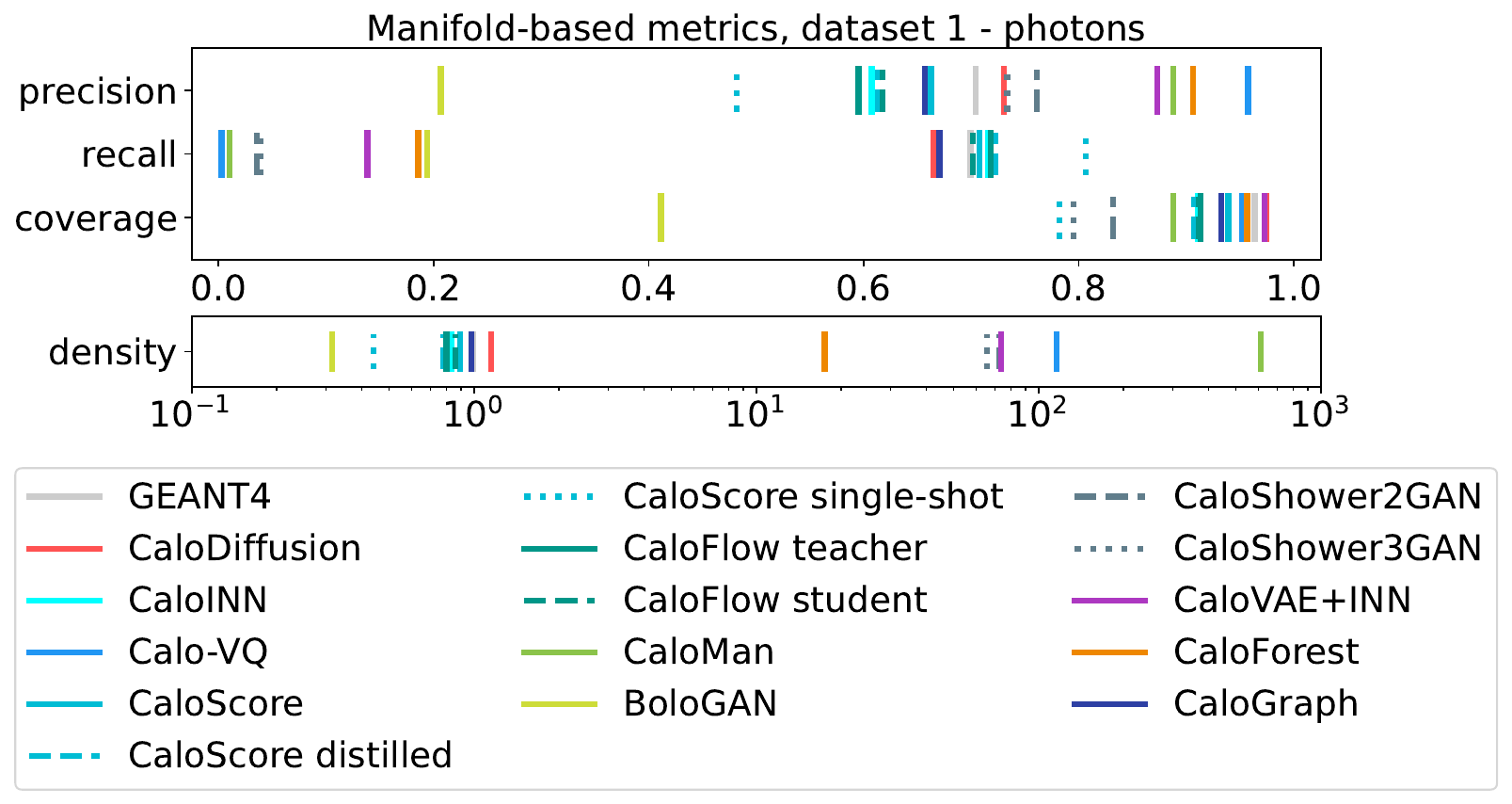}
\caption{Precision, density, recall, and coverage for \dsIph submissions. For the precise numbers, see \Tref{tab:ds1-photons.prdc}.}
\label{fig:ds1-photons.prdc}
\end{figure}

In \fref{fig:ds1-photons.prdc} and \tref{tab:ds1-photons.prdc} we show precision, density, recall, and coverage of the \dsIph submissions. We observe different classes of results. The first one shows values for all 4 metrics that are of the same order as the scores for the \geant\ reference, indicating a diverse and realistic dataset. \submAmram, \submFavaro, \submMikuni, and \submKobylyansky fall in this class. 

The second prominent pattern we observe shows values of precision and coverage that are of the same size as for \geant, but a much larger density and a much smaller recall. Most of the GAN and VAE-based models like \submLiu, \submReyes, \submZhang, and \submErnst, but also \submCresswell fall in this class. The high density suggests that the generated samples all fall close to the bulk of the reference data, but the low recall indicates that the relative distance between the generated samples is fairly small, so not many of the reference samples lie on the generated manifold. Overall, these generative models seem to focus on generating samples in the bulk that are similar to each other. 

The third class have good scores for recall and coverage, but a small precision, with the density being at the order of \geant or smaller. In \tref{tab:ds1-photons.prdc}, we see \submMikuniSingle and \cf in this category. They have a good distribution of samples close to the reference manifold, but a noticeable subset of them falls outside the manifold. When the density is low, it also indicates that the bulk is not as densely populated. 

The last pattern we observe has all four metrics below the \geant reference, as seen for \submRinaldi. 

\begin{figure}
\centering
\includegraphics[width=\textwidth]{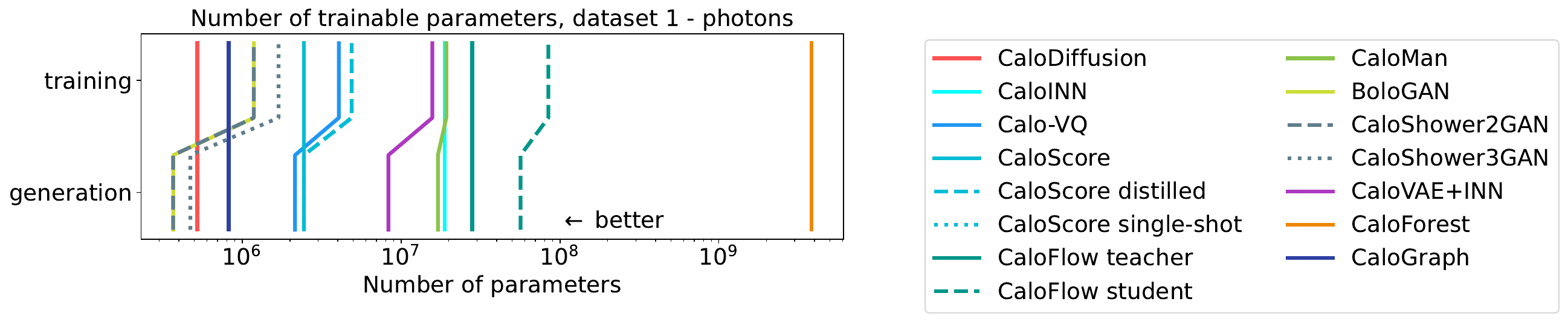}
\caption{Number of trainable parameters for training and generation of \dsIph submissions. For the precise numbers, see \Tref{tab:ds1-photons.numparam}.}
\label{fig:ds1-photons.numparam}
\end{figure}

In terms of the the requirements of resources, the situation is different. \Fref{fig:ds1-photons.numparam} shows the number of trainable parameters of each submission, with the precise numbers in \tref{tab:ds1-photons.numparam}. Normalizing Flow-based models are now at the back of the list, as they usually require larger models. GANs and VAEs are much more lightweight, as can be seen by \submZhang and \submRinaldi, which need the fewest parameters. Given the rather small dimensionality of \dsIph, the diffusion model of \submAmram also only needs a comparatively small number of parameters. 

\begin{figure}
\centering
\includegraphics[width=\textwidth]{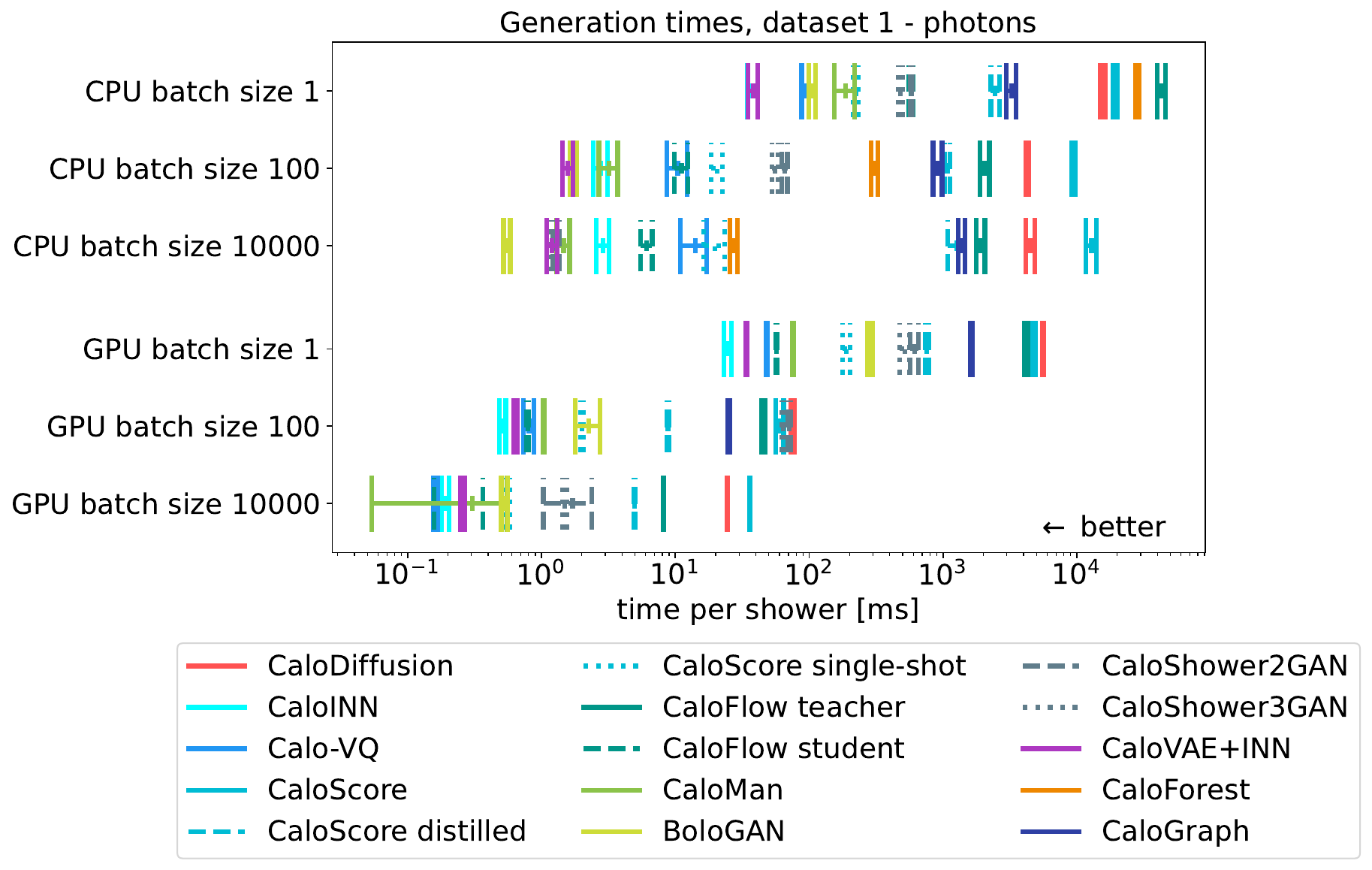}
\caption{Timing of \dsIph submissions on CPU and GPU architectures. Not all submissions are shown everywhere due to memory and other constraints. More details are in \tref{tab:ds1-photons.timing.CPU} and \tref{tab:ds1-photons.timing.GPU}.}
\label{fig:ds1-photons.timing}
\end{figure}

Which model is the fastest really depends on specific setup of the evaluation. We see the generation times per shower of the submissions in~\fref{fig:ds1-photons.timing} (with details in~\tref{tab:ds1-photons.timing.CPU} and~\tref{tab:ds1-photons.timing.GPU}). On the CPU in~\fref{fig:ds1-photons.timing} (and \tref{tab:ds1-photons.timing.CPU}), we observe a reduction in generation time when moving from batch size 1 to batch size 100 for all submissions. Further increasing the batch size to 10000 does not decrease the generation time further, indicating that now the algorithms are not dominated by the \texttt{for} loop over all batches anymore. The fastest models, \submRinaldi and \submErnst reach generation times of about one millisecond per shower for batch size 100, and even below for larger batch size. On the GPU in~\fref{fig:ds1-photons.timing} (and \tref{tab:ds1-photons.timing.GPU}), generation times are usually smaller than on the CPU, but different models gained differently under the changing hardware. For batch size 100, we now have five submissions at or below one millisecond generation time. For batch size 10000, only \submAmram, \submMikuni, and \submPangT are well above the one millisecond mark. The fastest models are now GAN-based (like \submRinaldi) or VAE-based (like \submLiu, \submReyes, and \submErnst). We now also observe improvements when increasing the batch size to 10000, even though the advantage in going from 100 to 10000 is not as big as the one going from 1 to 100. Rather surprisingly, we observe a larger generation time for the GAN-based models \submZhangTwo and \submZhangThree. We suspect that this is a remnant of these being part of the larger ATLAS software pipeline that was not fully optimized for the challenge submission.  

\FloatBarrier

\subsection{\texorpdfstring{Dataset 1, Pions (\dsIpi)}{Dataset 1, Pions}}
\label{sec:results_ds1-pions}

\begin{figure}
\centering
\includegraphics[width=\textwidth]{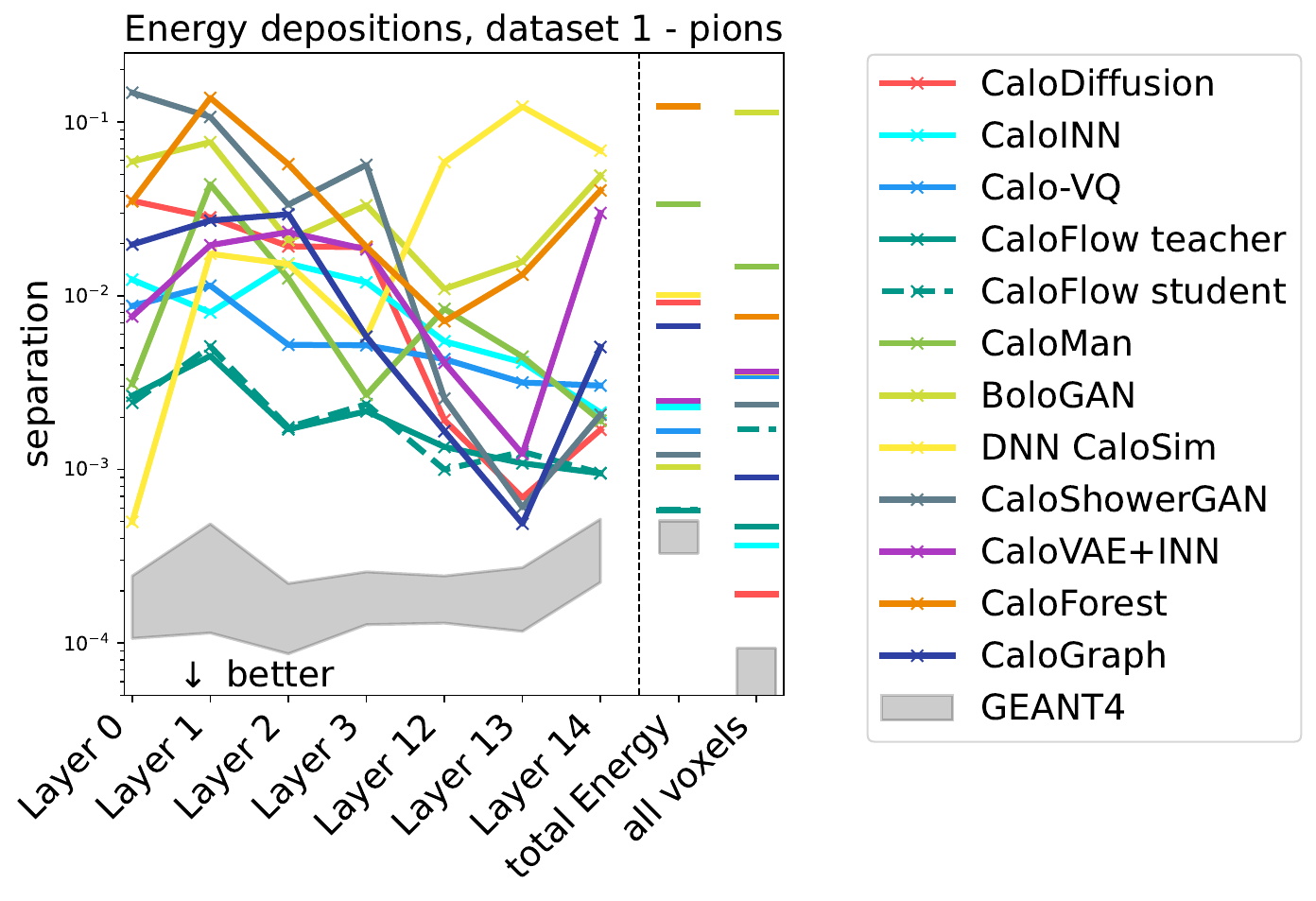}
\caption{Separation power of energy depositions with threshold at 1~MeV.}
\label{fig:ds1-pions-1-depositions}
\end{figure}

\begin{figure}
\centering
\includegraphics[width=\textwidth]{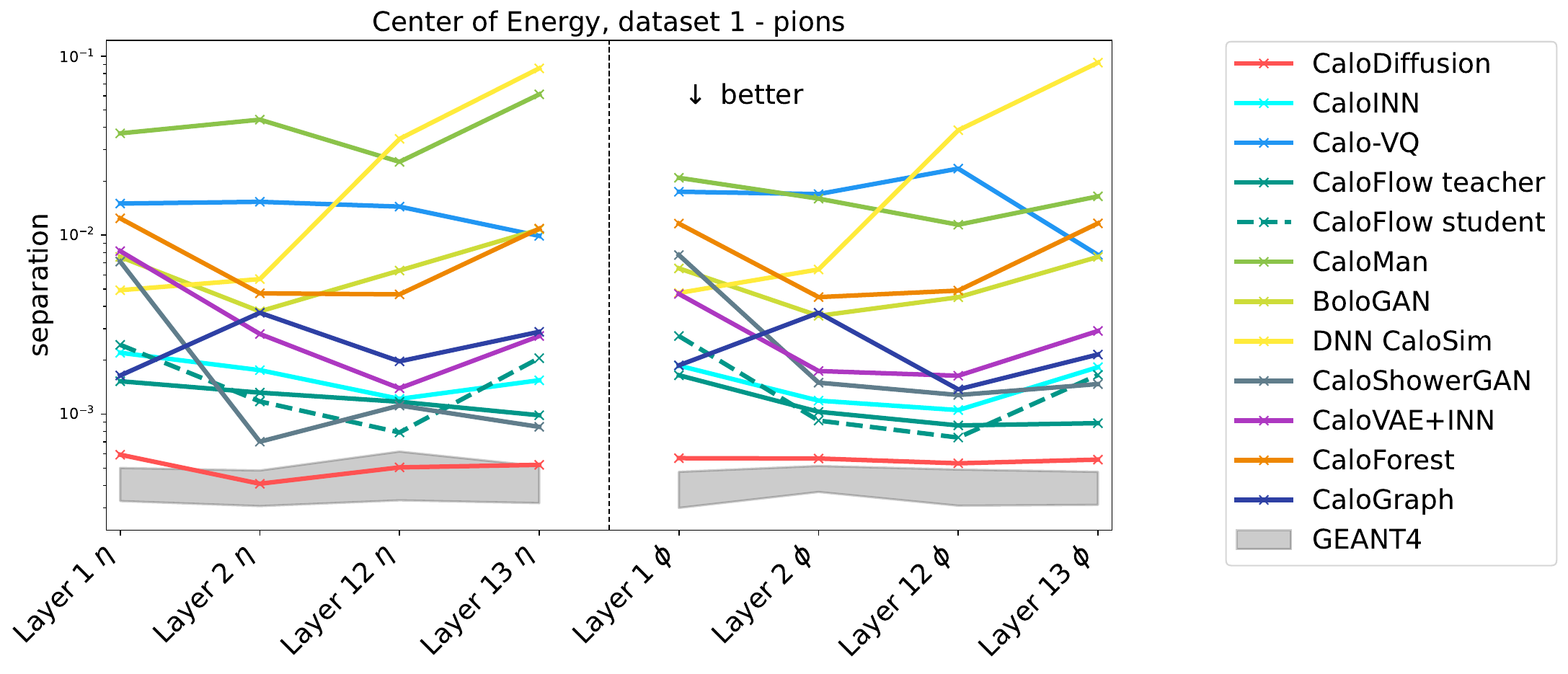}
\caption{Separation power of centers of energy with threshold at 1~MeV.}
\label{fig:ds1-pions-1-CE}
\end{figure}

\begin{figure}
\centering
\includegraphics[width=\textwidth]{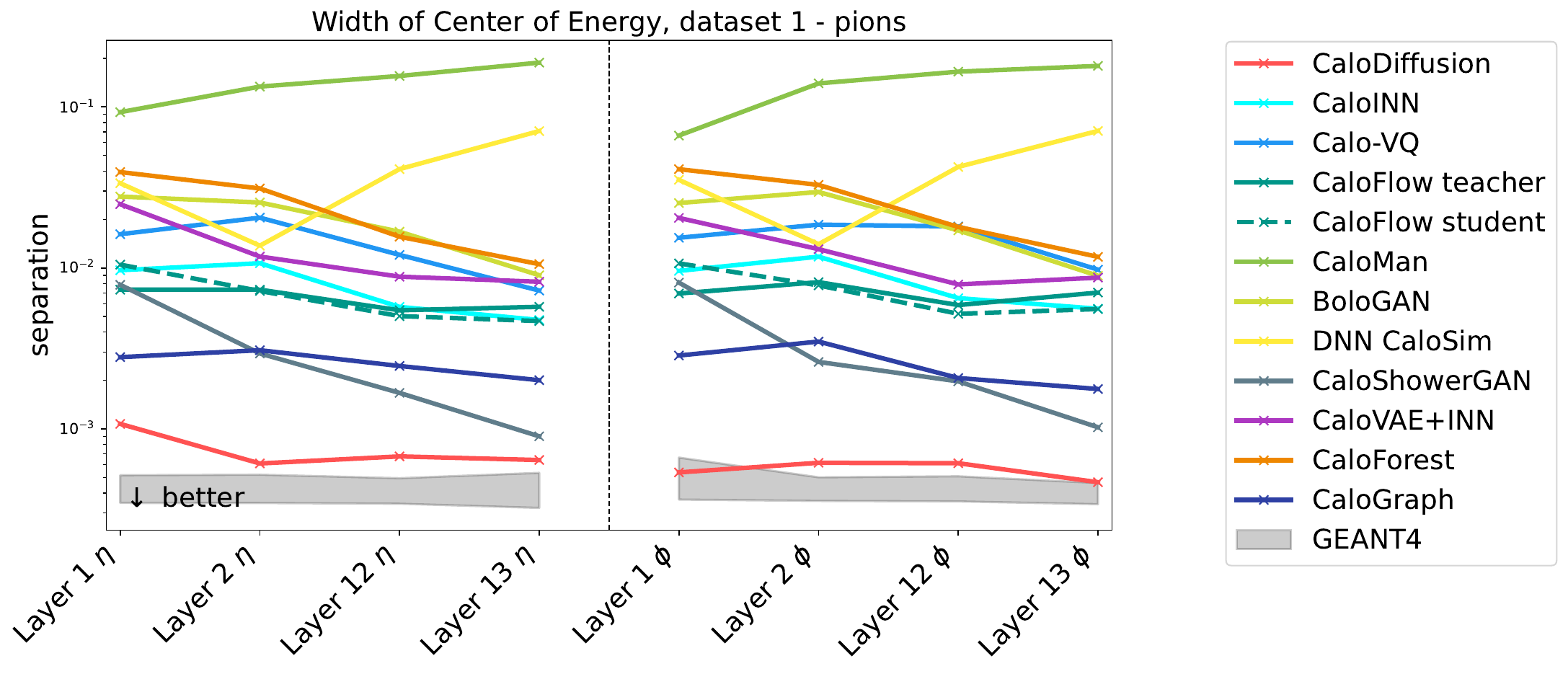}
\caption{Separation power of widths of centers of energy with threshold at 1~MeV.}
\label{fig:ds1-pions-1-width-CE}
\end{figure}

\begin{figure}
\centering
\includegraphics[width=\textwidth]{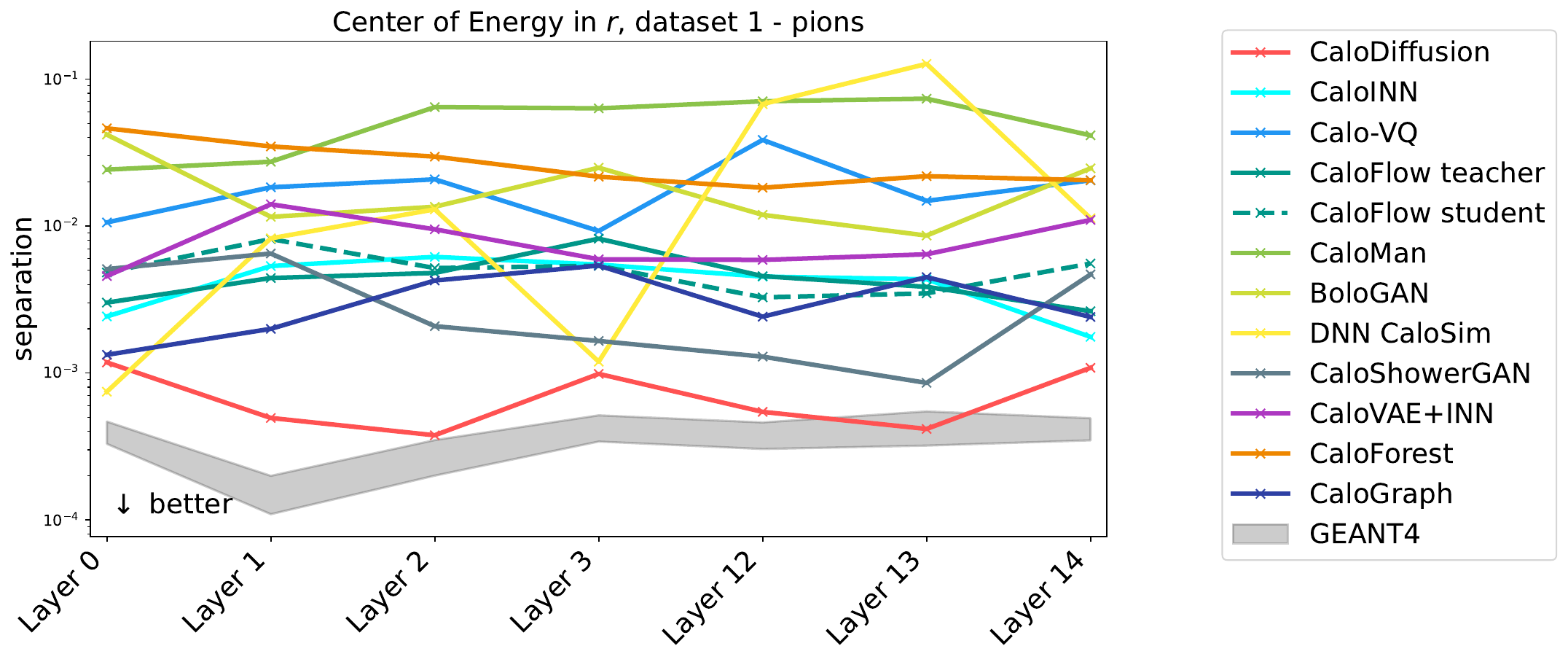}
\caption{Separation power of centers of energy with threshold at 1~MeV.}
\label{fig:ds1-pions-1-CE-r}
\end{figure}

\begin{figure}
\centering
\includegraphics[width=\textwidth]{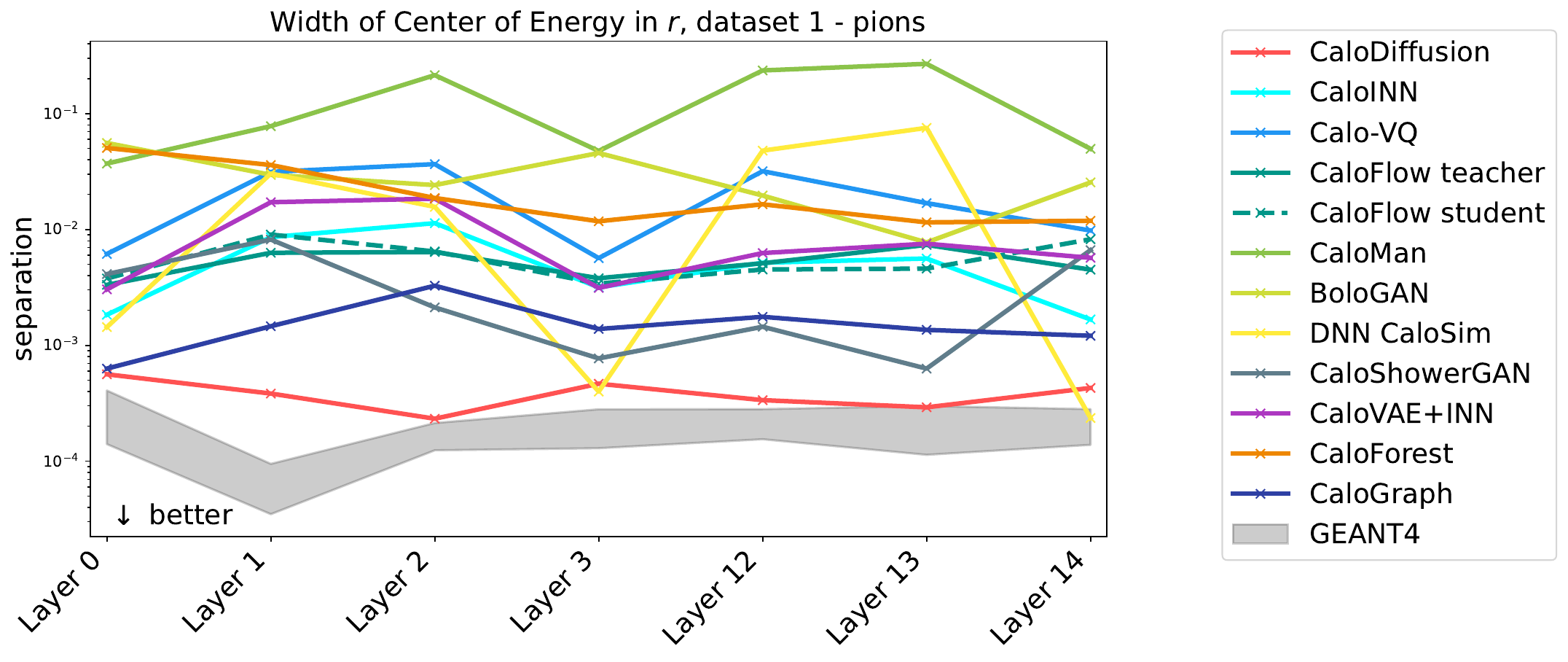}
\caption{Separation power of widths of centers of energy with threshold at 1~MeV.}
\label{fig:ds1-pions-1-width-CE-r}
\end{figure}

\begin{figure}
\centering
\includegraphics[width=\textwidth]{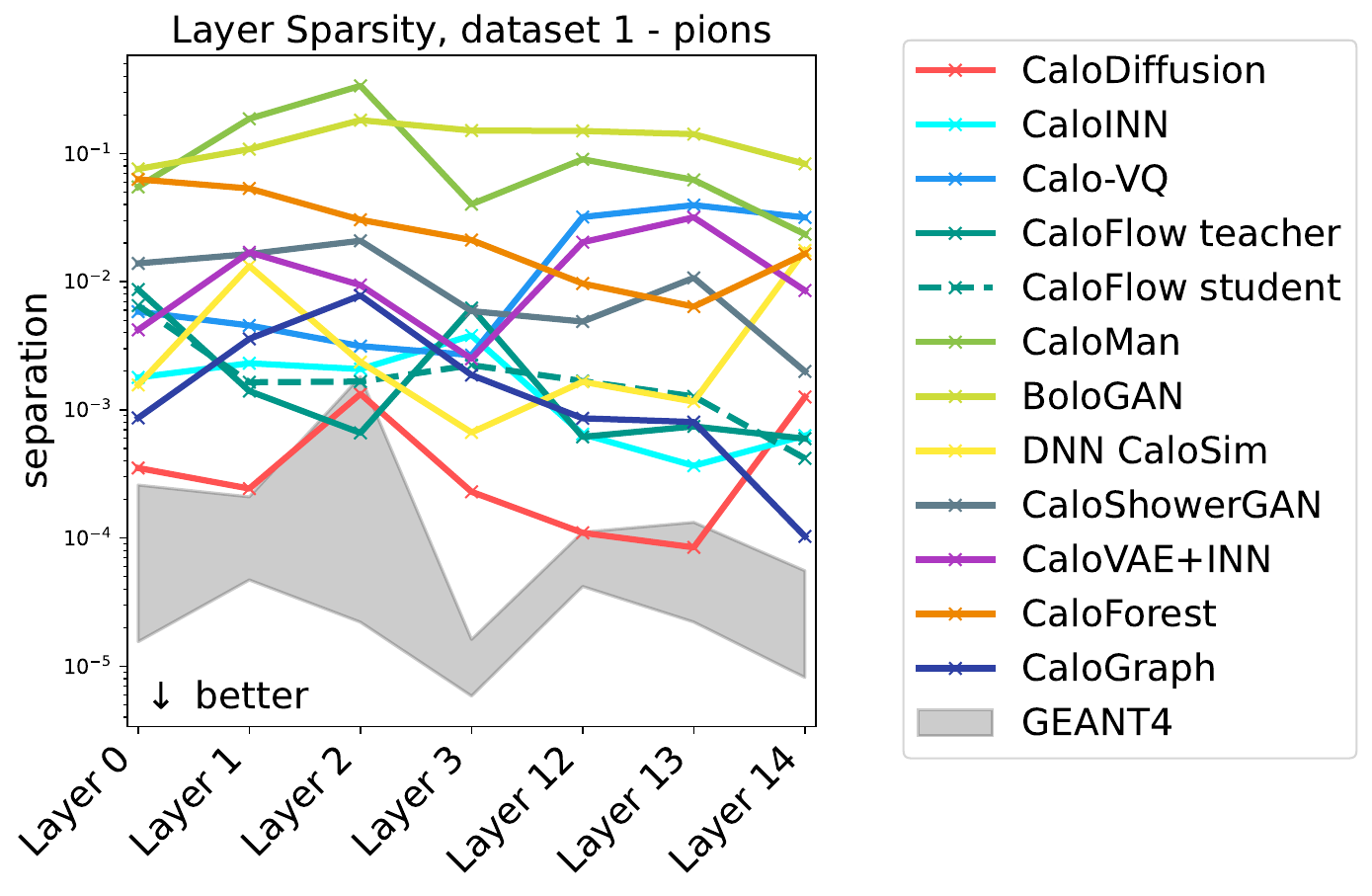}
\caption{Separation power of the sparsity with threshold at 1~MeV.}
\label{fig:ds1-pions-1-sparsity}
\end{figure}

\begin{figure}
\centering
\hfill\includegraphics[width=0.4\textwidth]{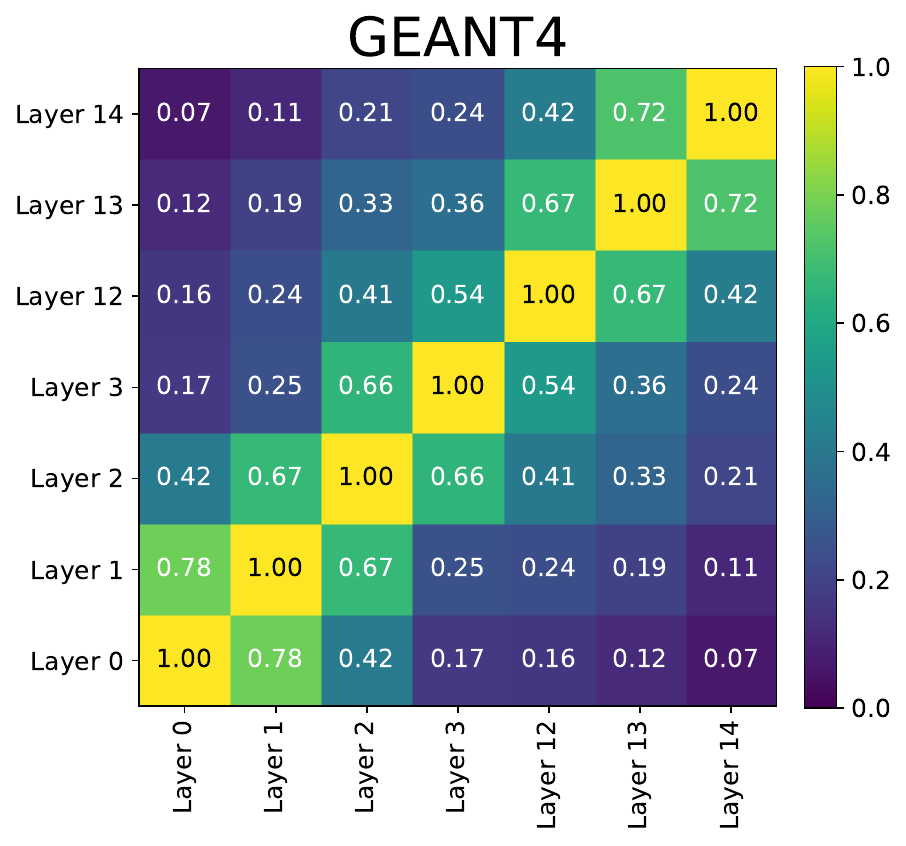}\hfill $ $
\\
\hfill\includegraphics[width=0.2\textwidth]{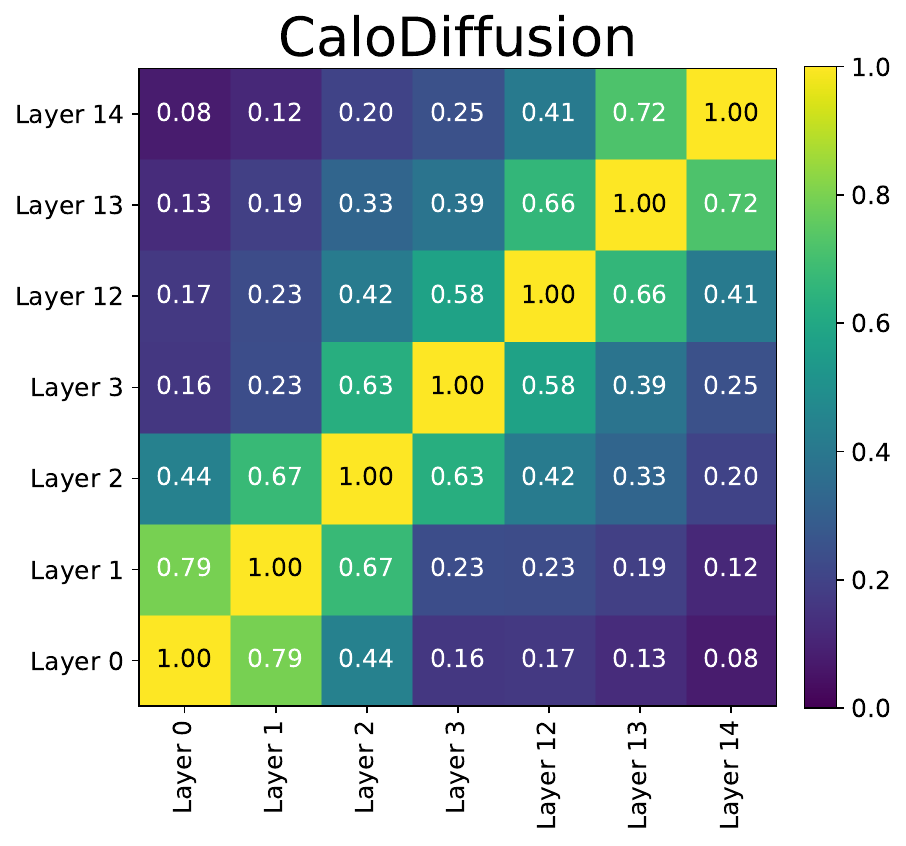}\hfill
\includegraphics[width=0.2\textwidth]{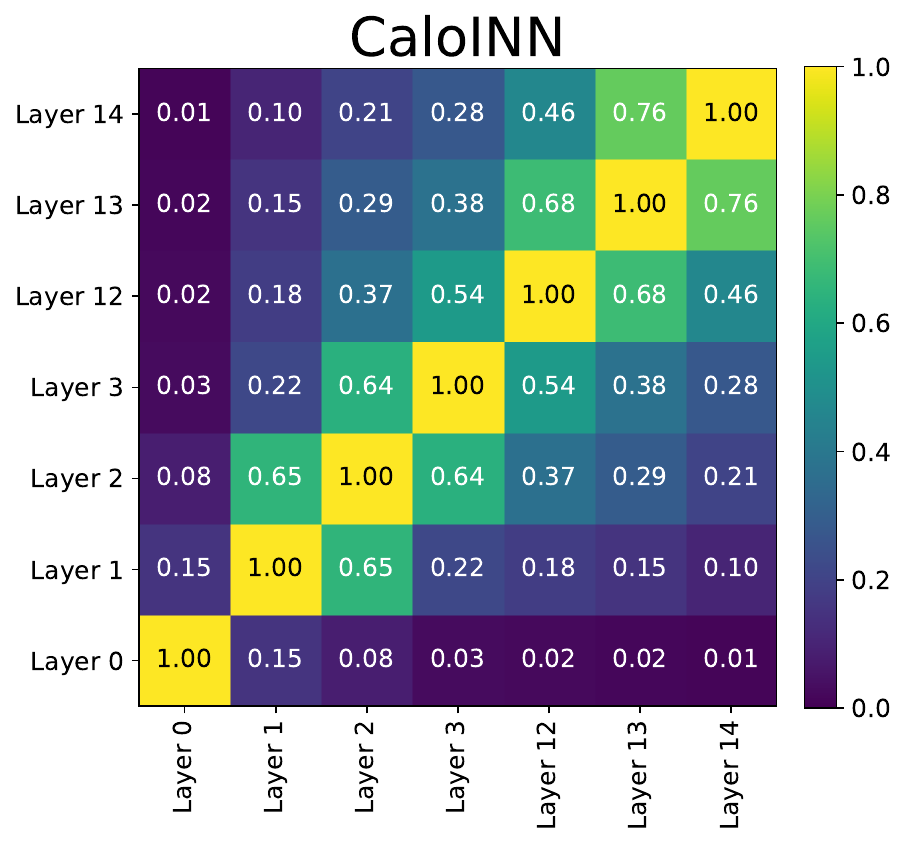}\hfill
\includegraphics[width=0.2\textwidth]{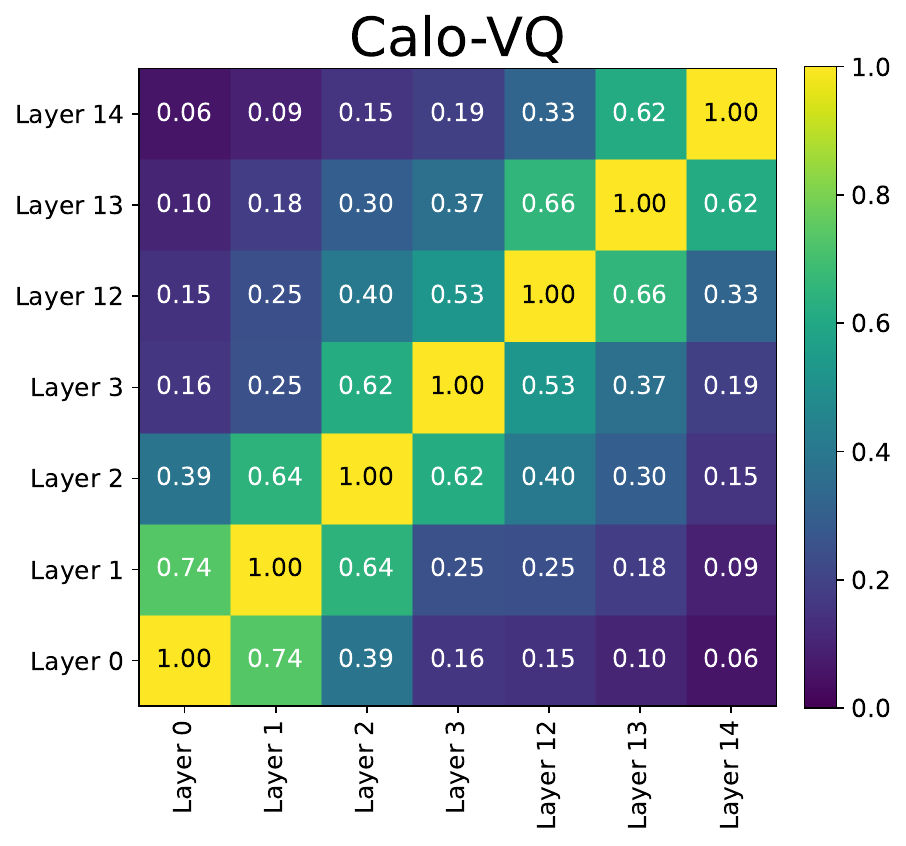}\hfill
\includegraphics[width=0.2\textwidth]{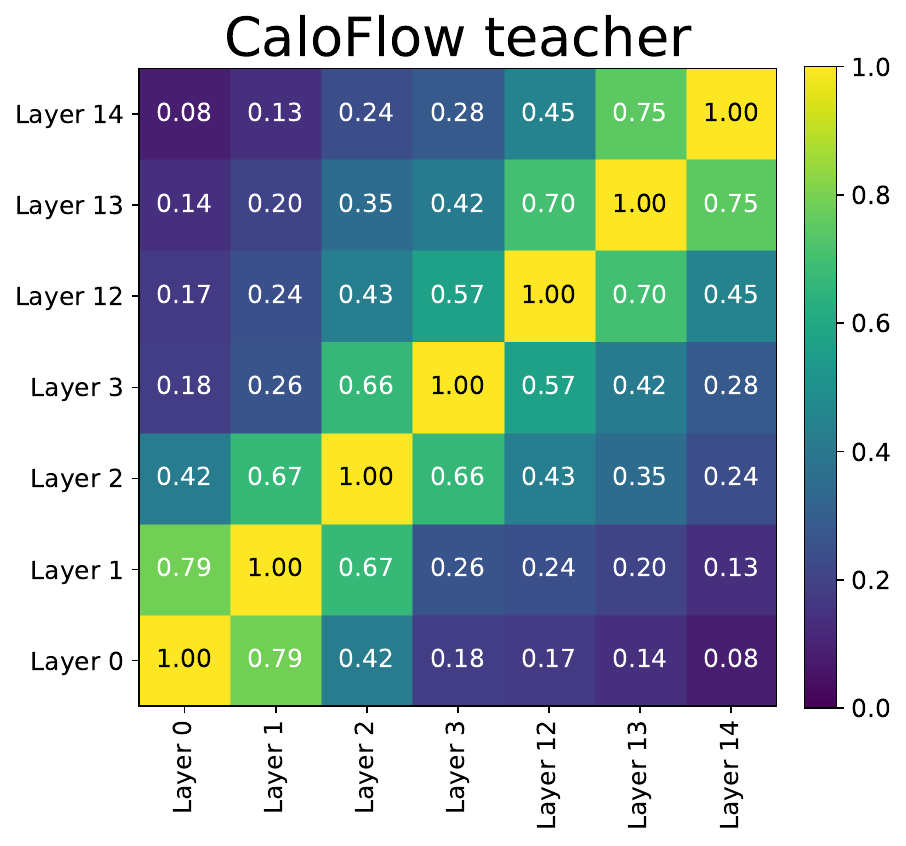}\hfill $ $\\
\hfill\includegraphics[width=0.2\textwidth]{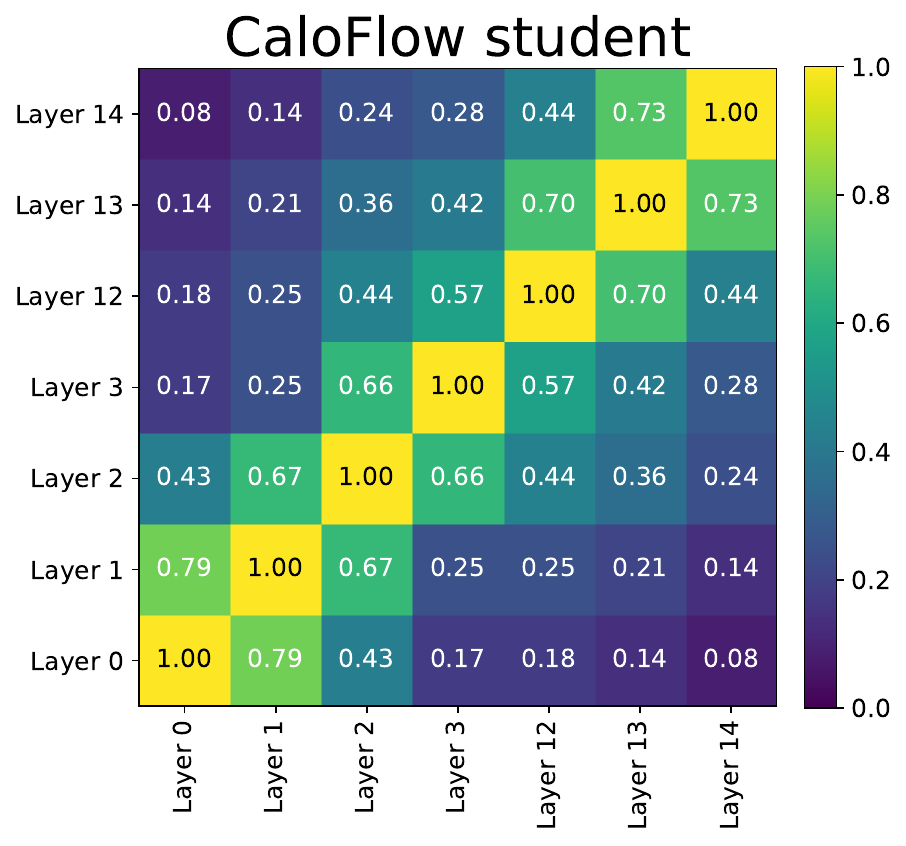}\hfill
\includegraphics[width=0.2\textwidth]{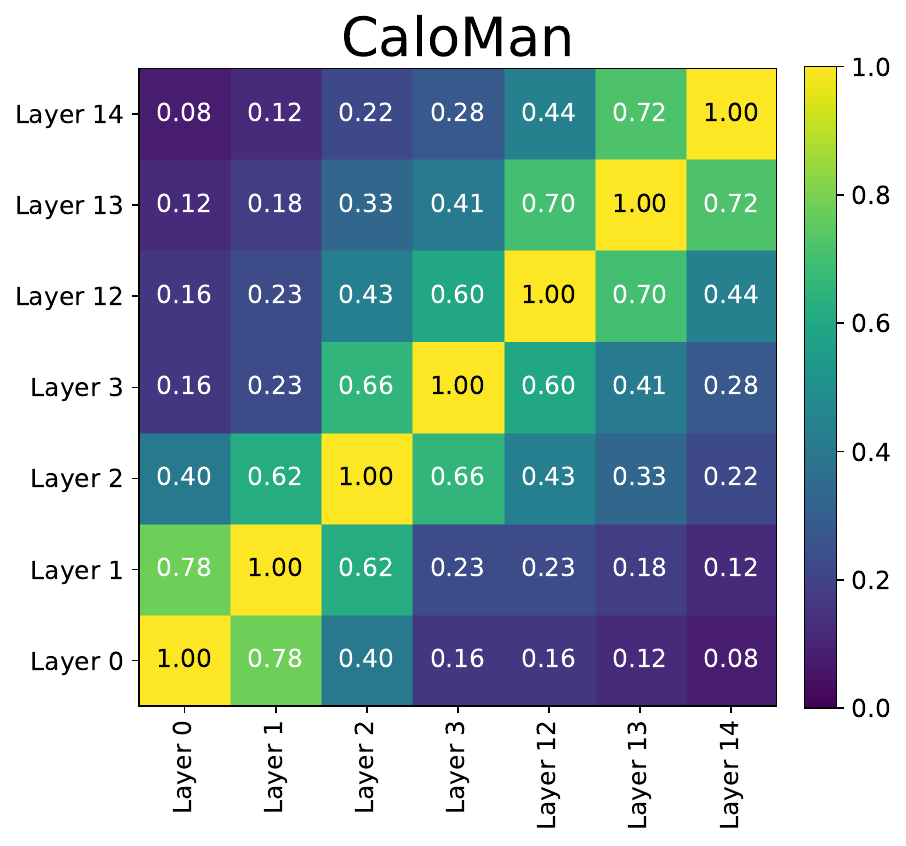}\hfill
\includegraphics[width=0.2\textwidth]{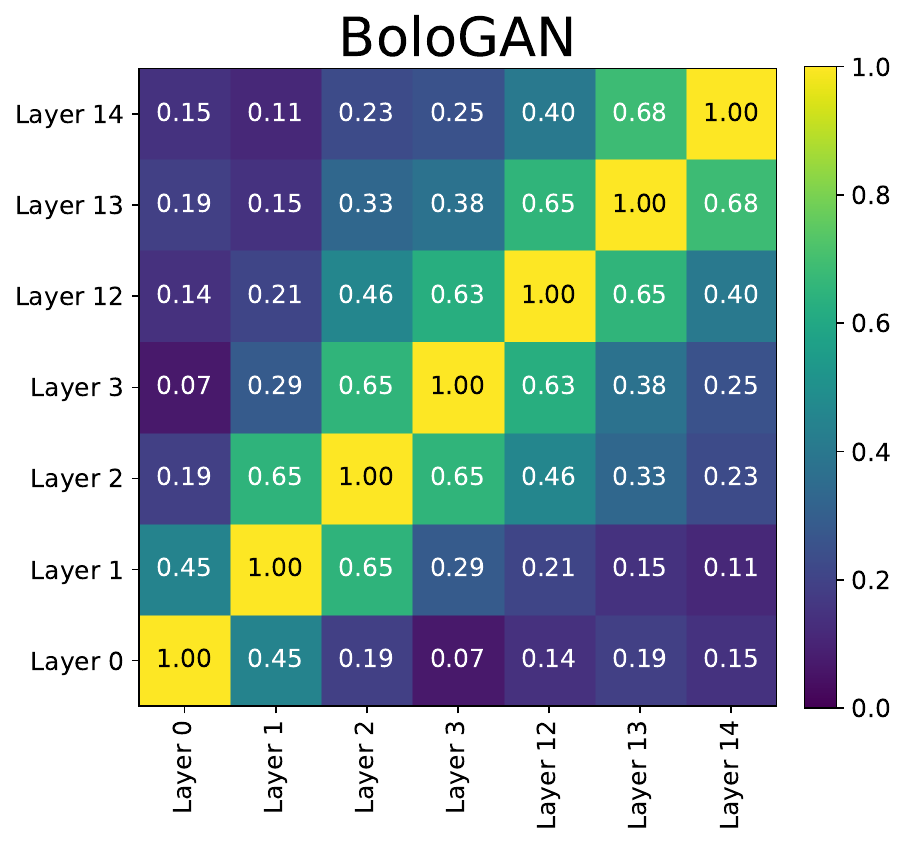}\hfill
\includegraphics[width=0.2\textwidth]{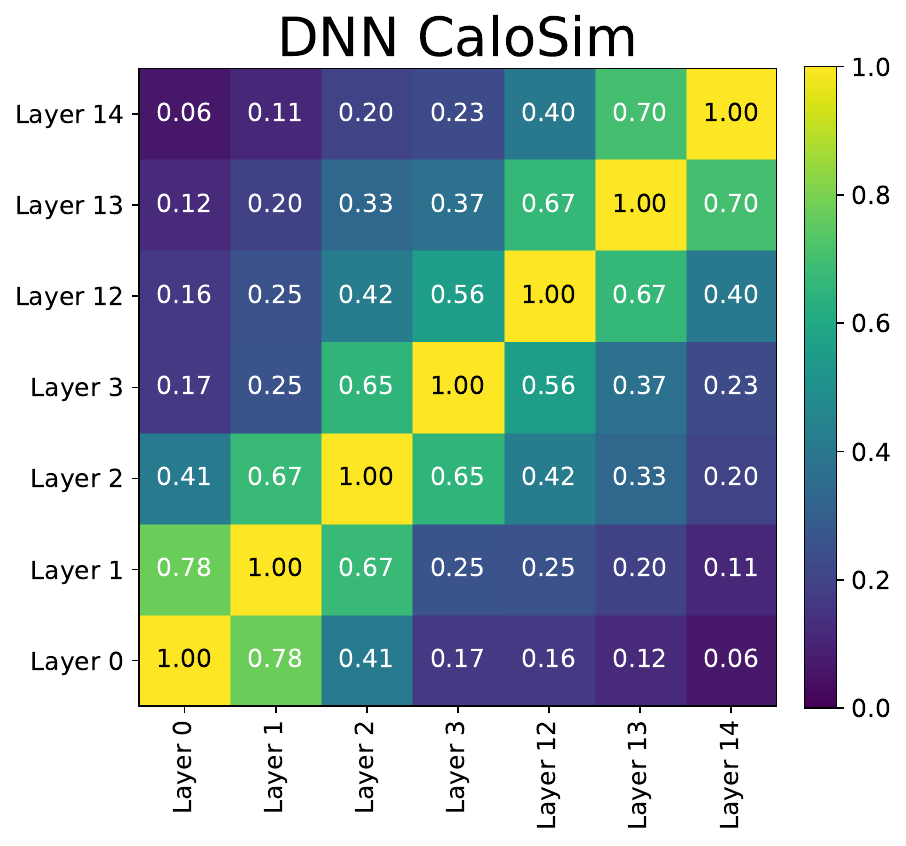}\hfill $ $\\
\hfill\includegraphics[width=0.2\textwidth]{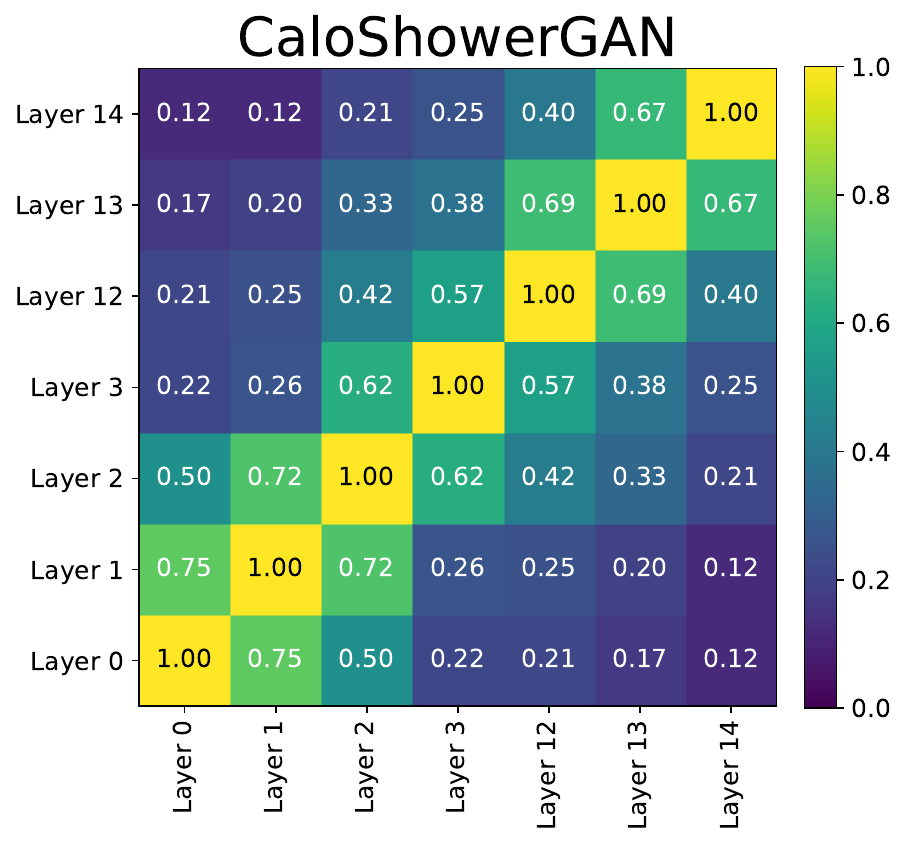}\hfill
\includegraphics[width=0.2\textwidth]{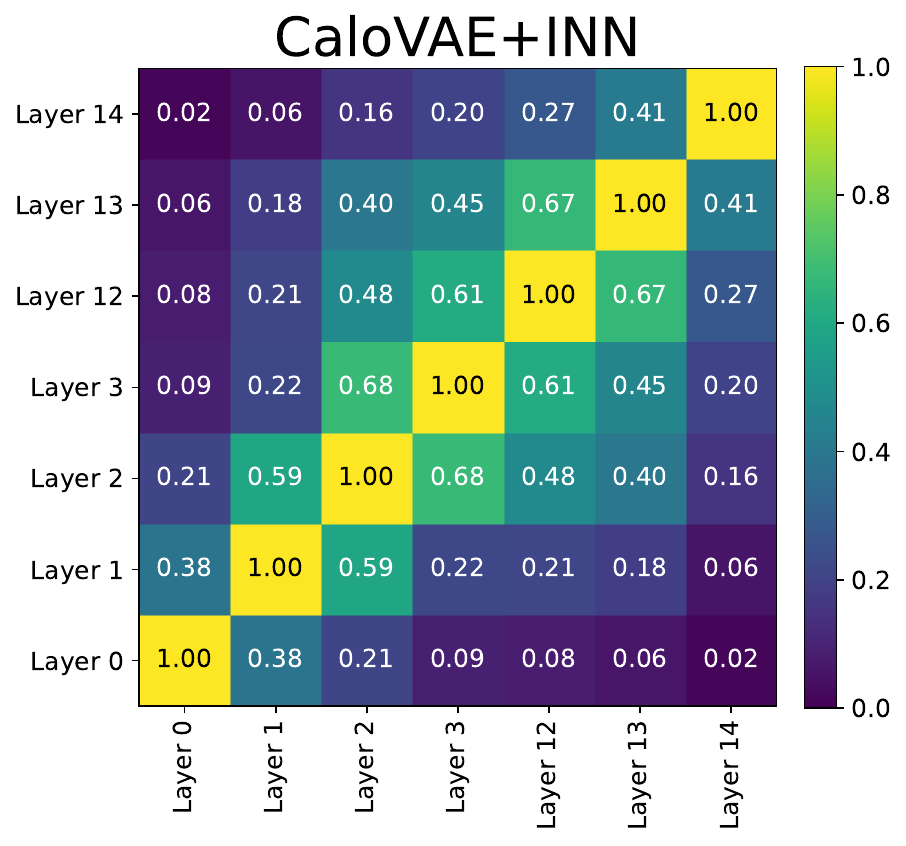}\hfill
\includegraphics[width=0.2\textwidth]{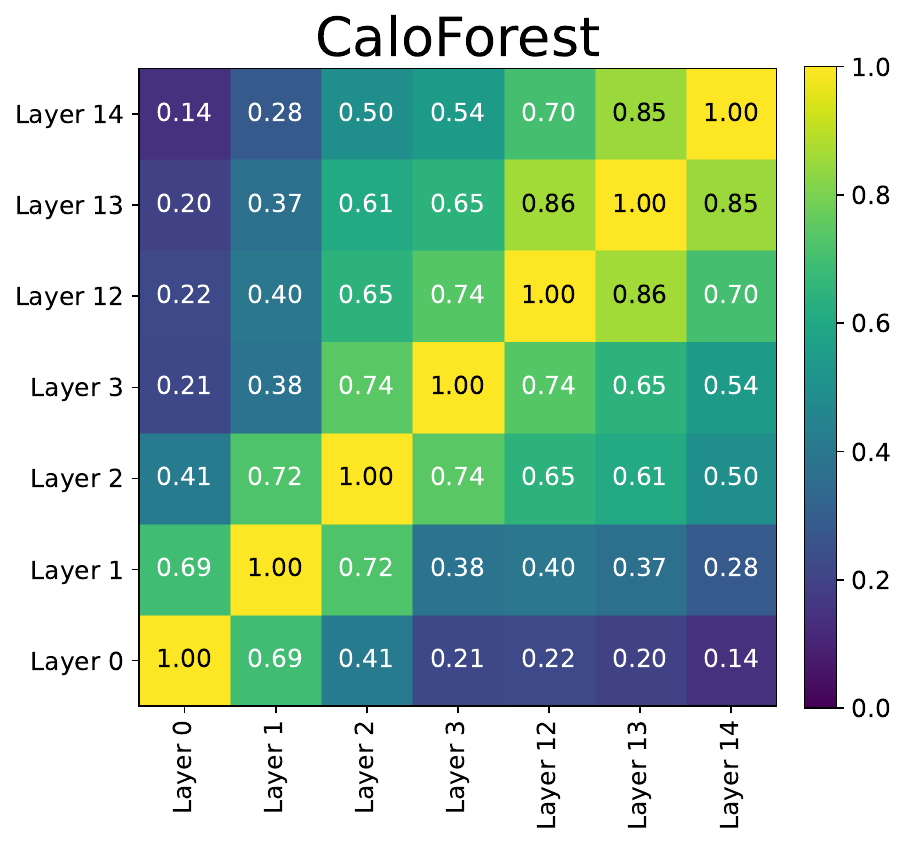}\hfill
\includegraphics[width=0.2\textwidth]{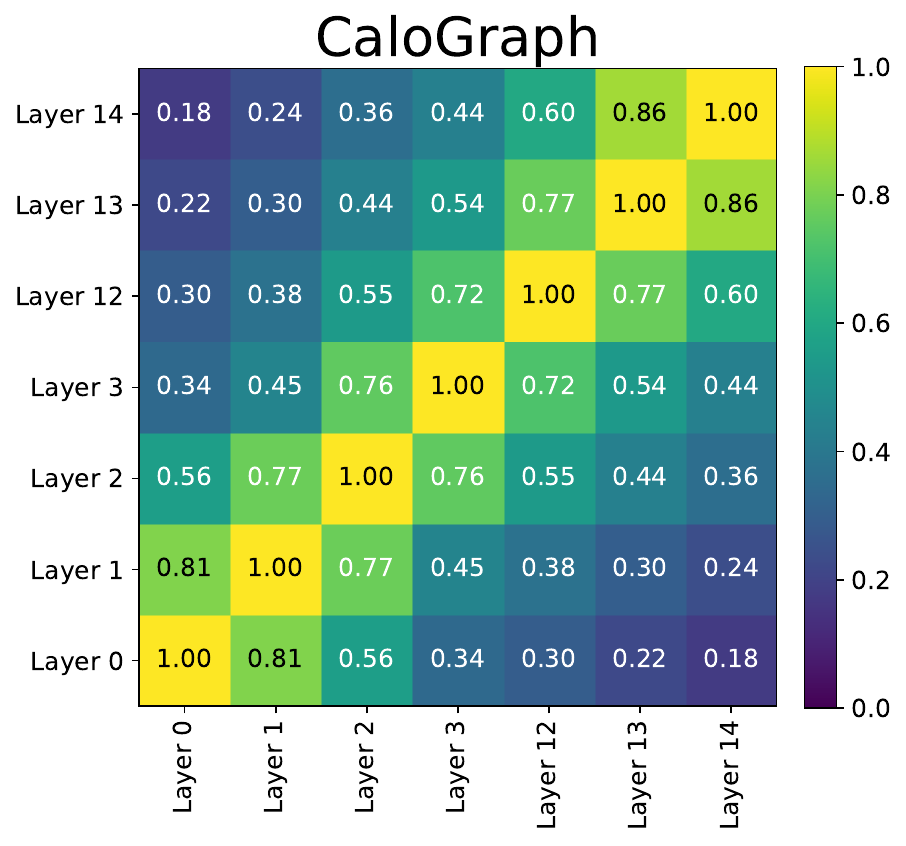}\hfill $ $\\
\caption{Pearson correlation coefficients of layer energies in \dsIpi, with threshold at 1~MeV.}
\label{fig:ds1-pions-1-corr}
\end{figure}

Starting again with high-level features, we first look at the energy depositions in~\fref{fig:ds1-pions-1-depositions}. The separation power of the submissions vary roughly within 2 orders of magnitude and they stay about one order of magnitude worse than the \geant reference. \cf shows the best performance overall, but occasionally another model is better in modeling a single layer. Diffusion models are not as good as for \dsIph, now VAE-based models like \submSalamaniDNN, \submLiu, or \submReyes are better, especially for earlier layers. Again, many models show better performance in layers 12 and 13, which have a higher segmentation in angular direction. 

The centers of energy, shown in~\fref{fig:ds1-pions-1-CE}, show a consistent picture in both directions $\eta$ and $\phi$. The separation power again spans about two orders of magnitude with \submAmram just at the \geant reference, followed by \cf, \submFavaro, and \submZhang. Interestingly, \submSalamaniDNN shows larger separation powers even though its performance in other metrics indicates otherwise, as we will see below.  

The widths of these center of energy distributions are compared to each other in~\fref{fig:ds1-pions-1-width-CE}. We again observe a very good performance of \submAmram, but now \submKobylyansky and \submZhang come in second before the flow-based models. 

When turning to the radial direction, the centers of energy in~\fref{fig:ds1-pions-1-CE-r} and its width in~\fref{fig:ds1-pions-1-width-CE-r} show again results consistent with the evaluation along $\eta$ and $\phi$: \submAmram with the smallest separation powers, followed by \submKobylyansky and \submZhang. While most submissions show separation powers of the same size for each layer, \submSalamaniDNN does a lot better in layers 0, 3, and 14 than in layers 1, 2, 12, and 13. 

For the sparsities in~\fref{fig:ds1-pions-1-sparsity}, we see a lot more variation from layer to layer in each of the submission. Even the separation power of the \geant reference varies almost two orders of magnitude between layers 2 and 3. The best performing submission is still \submAmram, but the gap to the other submissions is smaller. 

\Fref{fig:ds1-pions-1-corr} shows the correlation in layer energies for the submissions. The submissions \submAmram, \cf, \submReyes, and \submSalamaniDNN reproduce the pattern of \geant well. Other submissions, such as \submFavaro, \submLiu, \submRinaldi, \submErnst, and \submZhang, have some problems with correlations of layers 0 and/or 14, which are the first and last layers. \submCresswell finds a smaller correlation between the first layers and a too large correlation for the rest while \submKobylyansky has too large correlations everywhere.  

\begin{figure}
\centering
\includegraphics[width=\textwidth]{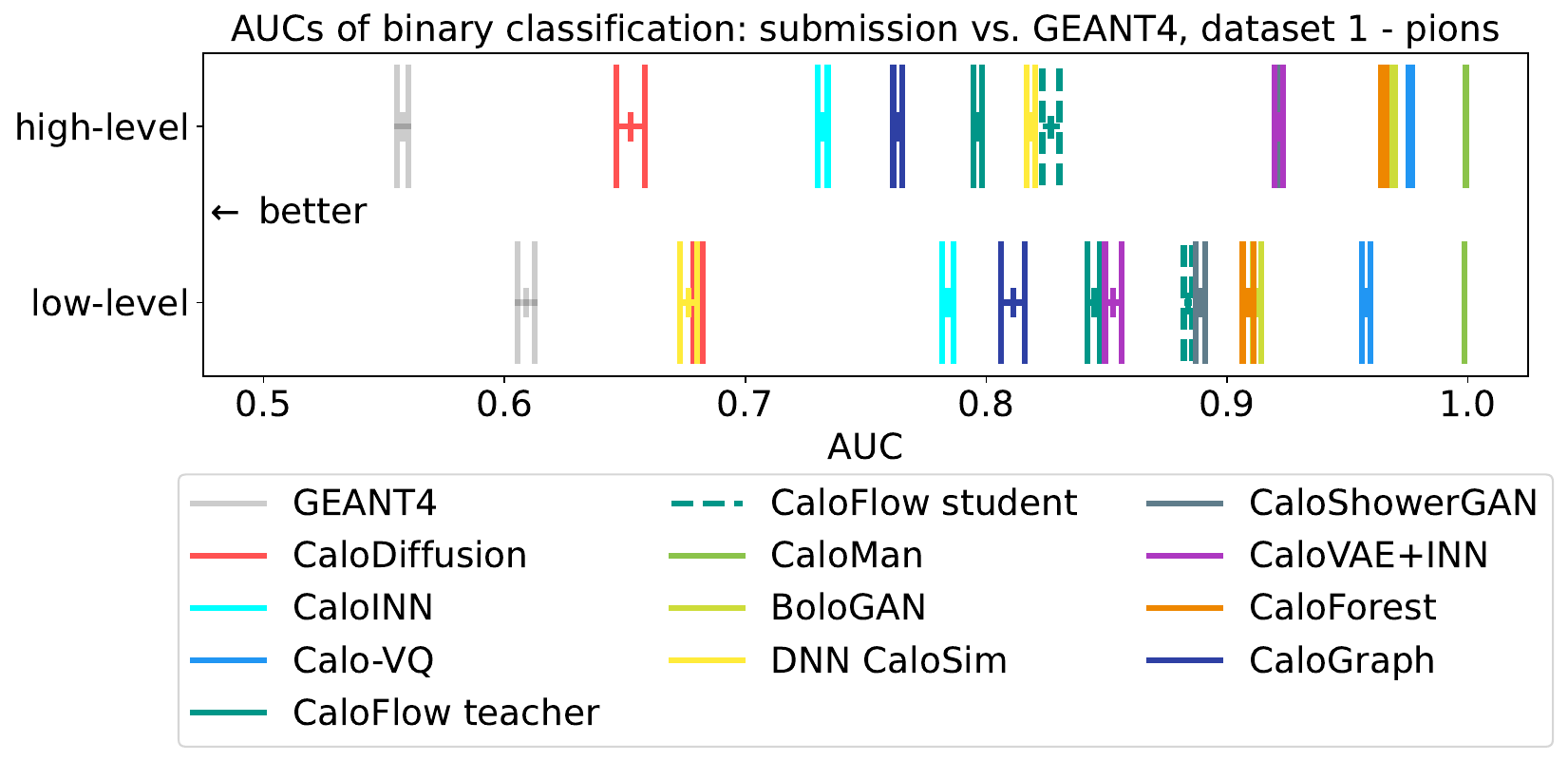}
\caption{Low-level and high-level AUCs for evaluating \geant\ vs.~submission of \dsIpi, averaged over 10 independent evaluation runs. For the precise numbers, see \Tref{tab:ds1-pions.aucs}.}
\label{fig:ds1-pions.aucs}
\end{figure}

\begin{figure}
\centering
\includegraphics[width=\textwidth]{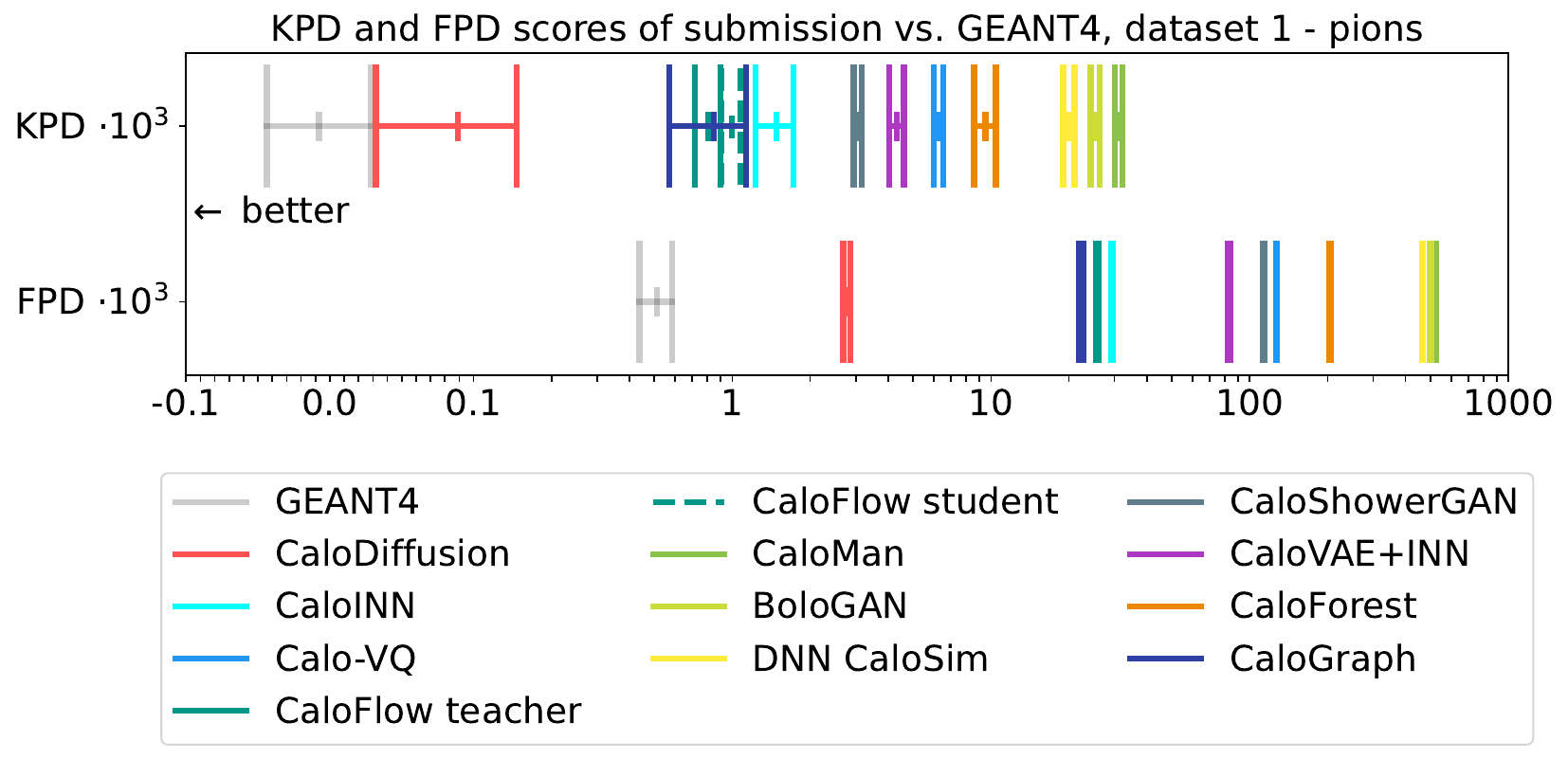}
\caption{KPD and FPD for evaluating \geant\ vs.~submission of \dsIpi. For the precise numbers, see \Tref{tab:ds1-pions.kpd}.}
\label{fig:ds1-pions.kpd}
\end{figure}

Moving on to classifier-based metrics, we find the AUCs of high- and low-level observables in \fref{fig:ds1-pions.aucs} (and \tref{tab:ds1-pions.aucs}). Here we observe several things. First, the AUC for separating the training and test \geant samples is larger than the expected value of $0.5$. This is due to the fact that two slightly different versions of the ATLAS software were used due to technical problems in generating high statistics with the old version used for the ATLAS internal training. The differences were expected and deemed small enough to be irrelevant for physics applications. Detailed comparison between the two samples that justify this statement are provided in~\ref{app:histograms.pions}. The AUC from the generative models will have this value as the maximum achievable separation instead of the usual $0.5$. Second, we see very low AUCs for \submAmram, which was already indicated by the separation powers of the obervables before. Third, we see a low AUC for \submSalamaniDNN in the low-level observables which is, however, not reflected in the AUC of the high-level observables. This latter fact also correlates with the separation powers seen before. Other than that, we see overall good scores from diffusion and normalizing flow-based models, whereas GAN and VAE-based models show AUCs worse than $0.9$. 

The same is true for KPD and FPD metrics shown in \fref{fig:ds1-pions.kpd} (and also in \tref{tab:ds1-pions.kpd}). The best scores are attained for \submAmram, followed by \submKobylyansky and \cf. The submission of \submSalamaniDNN is not among the top contestants here. 

\begin{figure}
\centering
\includegraphics[width=\textwidth]{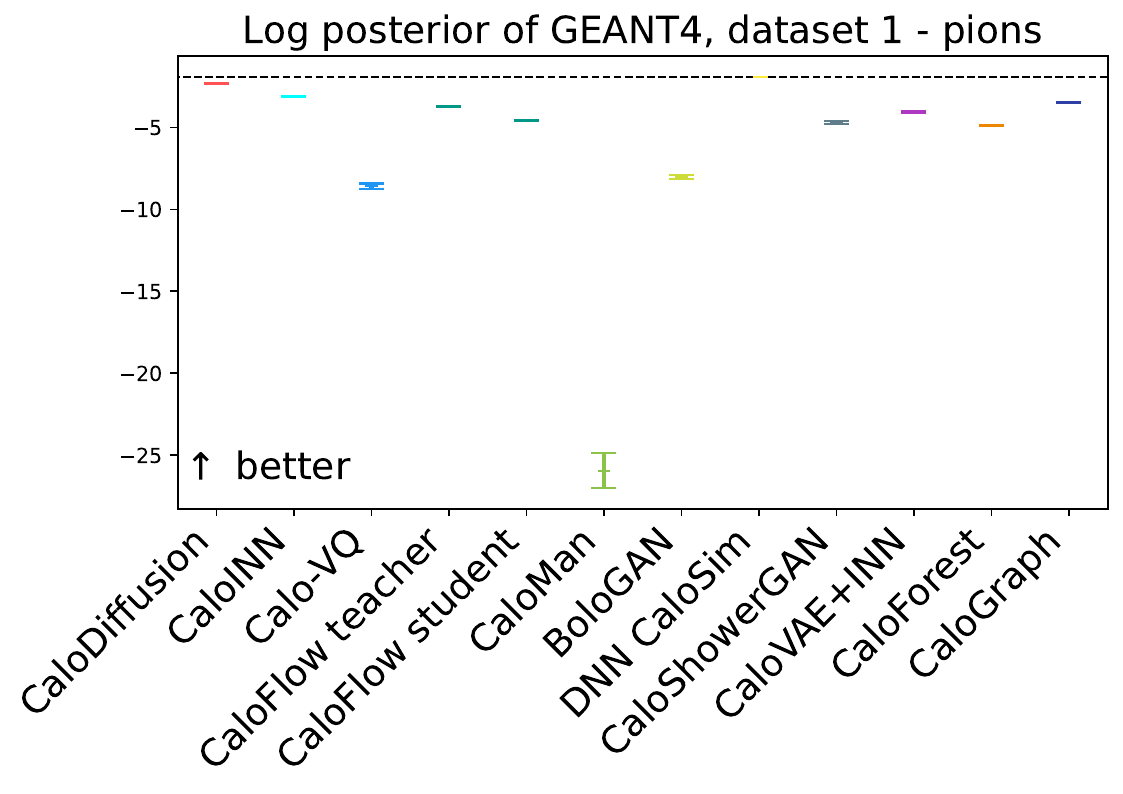}
\caption{Log-posterior scores for \dsIpi \geant test data, averaged over 10 independent classifier trainings. For the precise numbers, see \Tref{tab:ds1-pions.multi}.}
\label{fig:ds1-pions.multi}
\end{figure}

When looking at the results of the multiclass classification in \fref{fig:ds1-pions.multi}, the situation is slightly different. \submAmram, \submFavaro, and \submKobylyansky show again good scores, but \submSalamaniDNN is outperforming them. Since the multiclass classification is also based on low-level observables, this observation confirms the low-level AUC of \tref{tab:ds1-pions.aucs}. The consistency check of the multiclass classifier can be seen at \fref{fig:consistency.ds1pions}. 

\begin{figure}
\centering
\includegraphics[width=\textwidth]{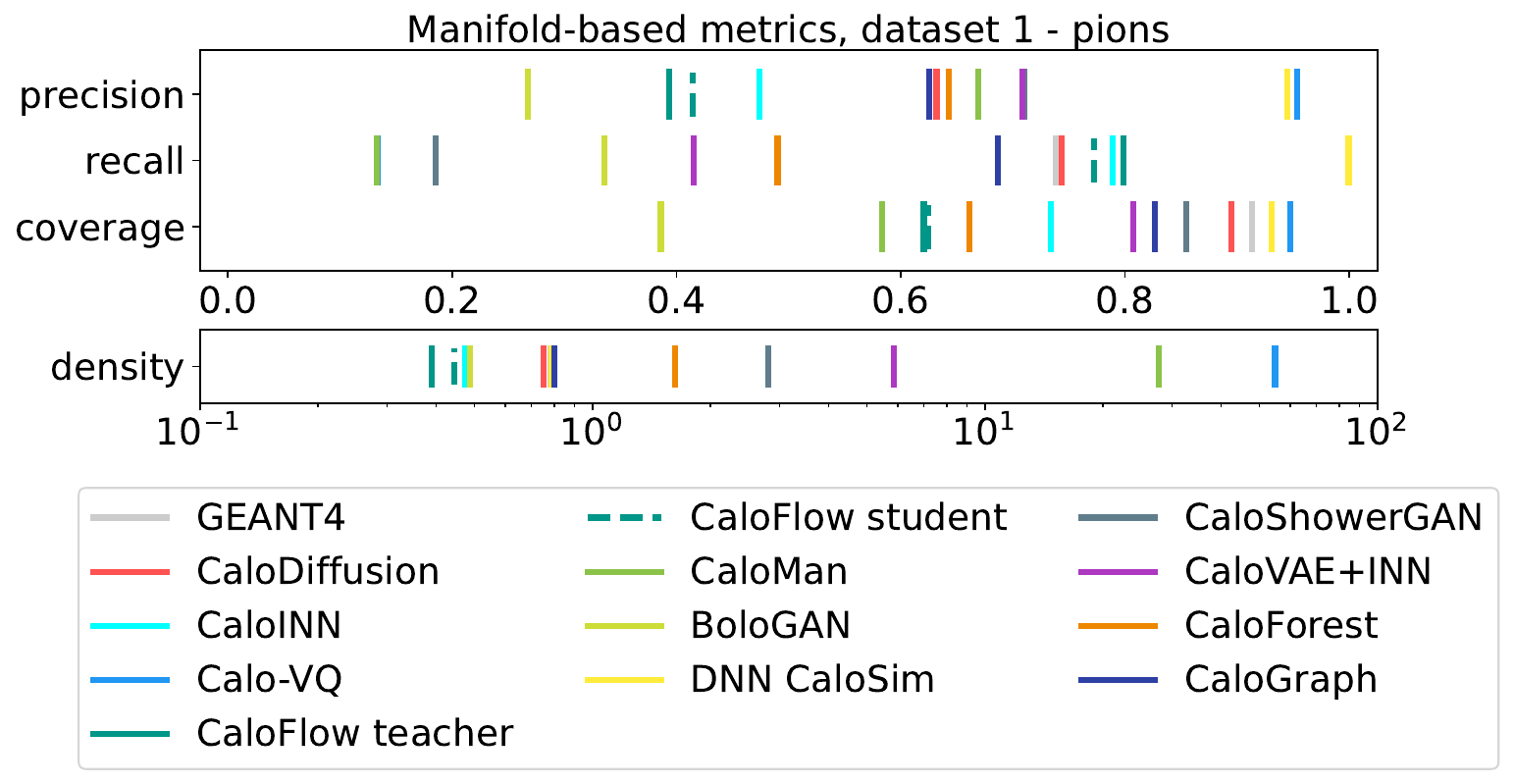}
\caption{Precision, density, recall, and coverage for \dsIpi submissions. For the precise numbers, see \Tref{tab:ds1-pions.prdc}.}
\label{fig:ds1-pions.prdc}
\end{figure}

In \fref{fig:ds1-pions.prdc} and \tref{tab:ds1-pions.prdc} we show precision, density, recall, and coverage of the \dsIpi submissions. We again observe similar patterns as in the \dsIph case. The submissions \submAmram, \submSalamaniDNN, and \submKobylyansky have their scores around the scores of the \geant reference, indicating a good fit to the underlying distribution. The normalizing flow-based submissions \submFavaro and \cf have good recall and coverage, but a relatively small precision and density, indicating that a large enough subset of samples were generated away from the reference manifold. VAE-based submissions \submLiu, \submReyes, \submErnst, and to some extend also \submZhang show a large density paired with a very small recall. As for \dsIph, we interpret this as generative models that seem to focus on generating samples in the bulk of the data, with all samples being fairly similar to each other. 

\begin{figure}
\centering
\includegraphics[width=\textwidth]{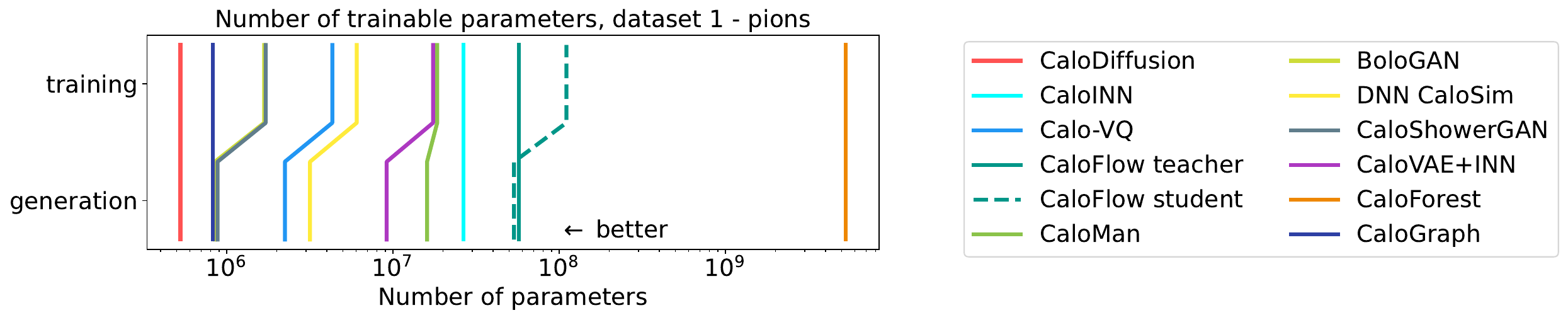}
\caption{Number of trainable parameters for training and generation of \dsIpi submissions. For the precise numbers, see \Tref{tab:ds1-pions.numparam}.}
\label{fig:ds1-pions.numparam}
\end{figure}

\Fref{fig:ds1-pions.numparam} compares the sizes of the submissions, with \tref{tab:ds1-pions.numparam} giving the precise numbers. Most models require (at least in training) more than $10^6$ trainable parameters, only \submAmram and \submKobylyansky stay below that. Overall, we observe normalizing-flow based models to be much larger than diffusion and GAN-based models. The BDT-based \submCresswell stands out because of the many parameters that are required to define all trees. 

\begin{figure}
\centering
\includegraphics[width=\textwidth]{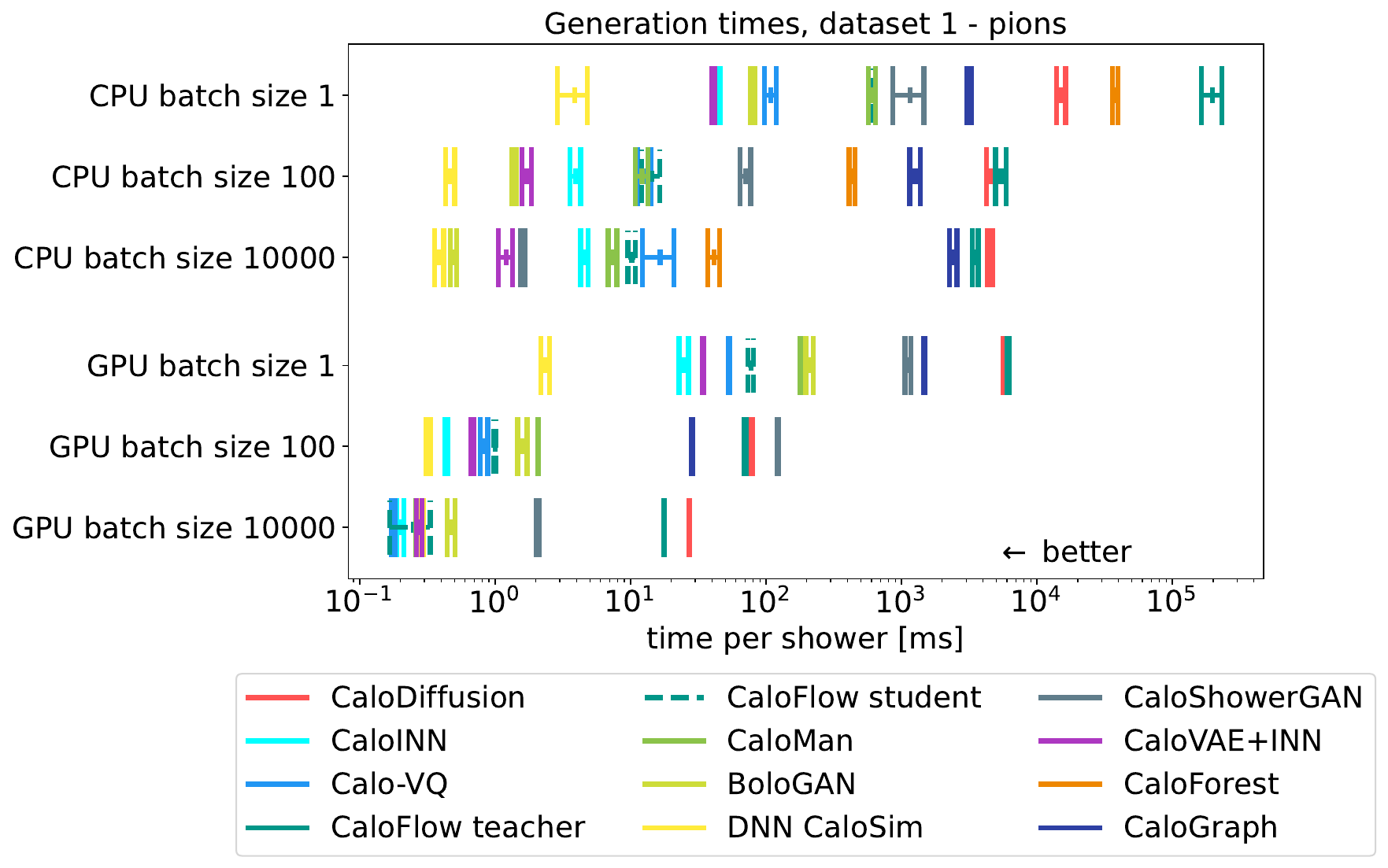}
\caption{Timing of \dsIpi submissions on CPU and GPU architectures. Not all submissions are shown everywhere due to memory and other constraints. More details are in \tref{tab:ds1-pions.timing.CPU} and \tref{tab:ds1-pions.timing.GPU}.}
\label{fig:ds1-pions.timing}
\end{figure}

\Fref{fig:ds1-pions.timing} (with details in \tref{tab:ds1-pions.timing.CPU} and~\tref{tab:ds1-pions.timing.GPU}) show the generation times per particle shower of the submissions. Across all batch sizes and architectures, we see \submSalamaniDNN as being the fastest model, only beaten for very large batch sizes on a GPU but not by a large margin. This model only needs a few milliseconds (for batch size 1) to a fraction of a millisecond (for lager batch sizes) to generate a single shower. Other GAN-based and VAE-based models like \submRinaldi and \submErnst also show fast shower generation times. Normalizing-flow-based submissions, however, show a strong dependence on the implemented algorithm. The coupling-layer based implementation of \submFavaro is much faster than the MAF/IAF-based implementations of \cf, with the MAF being much slower than the IAF, as expected~\cite{Krause:2021wez}. \submCresswell does not have timings on a GPU, as the tree-based algorithm only runs on a CPU. Also here, we observe a larger generation time for the GAN-based model \submZhang. Again, we suspect that this is a remnant of \submZhang being part of the larger ATLAS software pipeline that was not fully optimized for the challenge submission.  

\FloatBarrier

\subsection{\texorpdfstring{Dataset 2, Electrons (\dsII)}{Dataset 2, Electrons}}
\label{sec:results_ds2}
As explained in \sref{subsec:ds23}, the minimal energy that can be read out is given by 15.15~keV and we apply a threshold cut to all submissions before evaluation. 

\begin{figure}
\centering
\includegraphics[width=\textwidth]{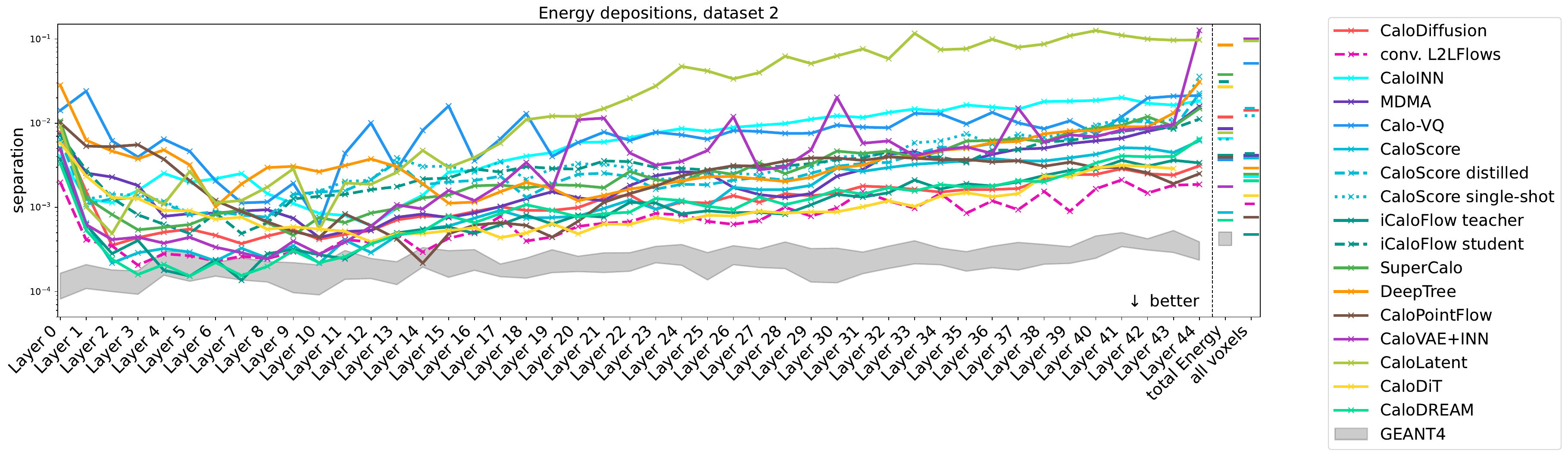}
\caption{Separation power of energy depositions.}
\label{fig:ds2-depositions}
\end{figure}

\begin{figure}
\centering
\includegraphics[width=\textwidth]{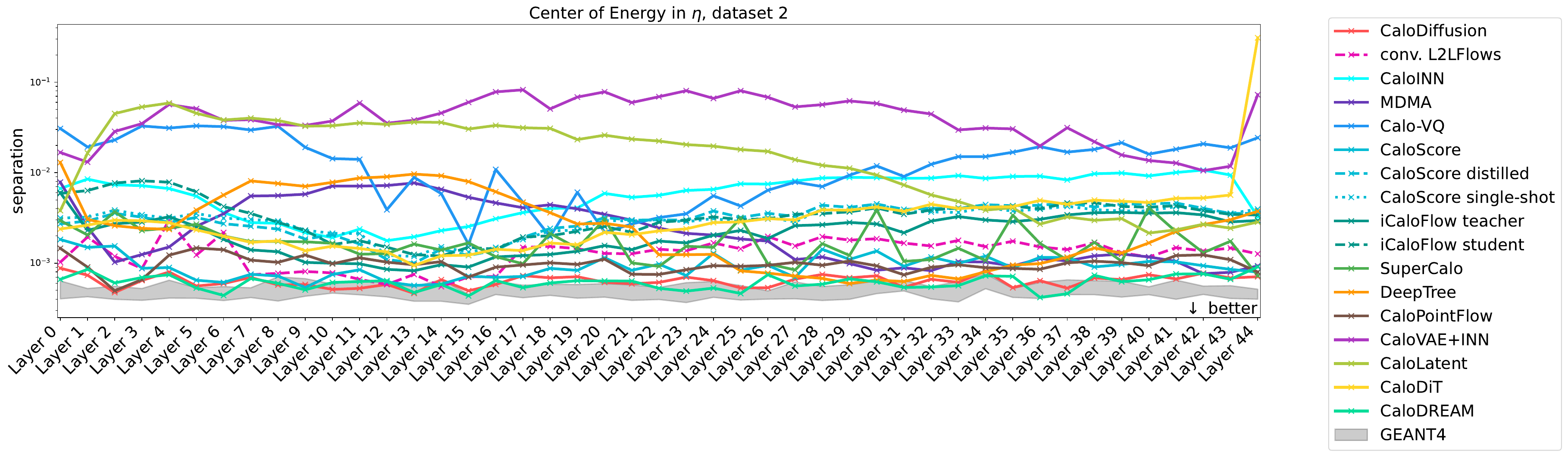}
\caption{Separation power of centers of energy in $\eta$ direction.}
\label{fig:ds2-CE-eta}
\end{figure}

\begin{figure}
\centering
\includegraphics[width=\textwidth]{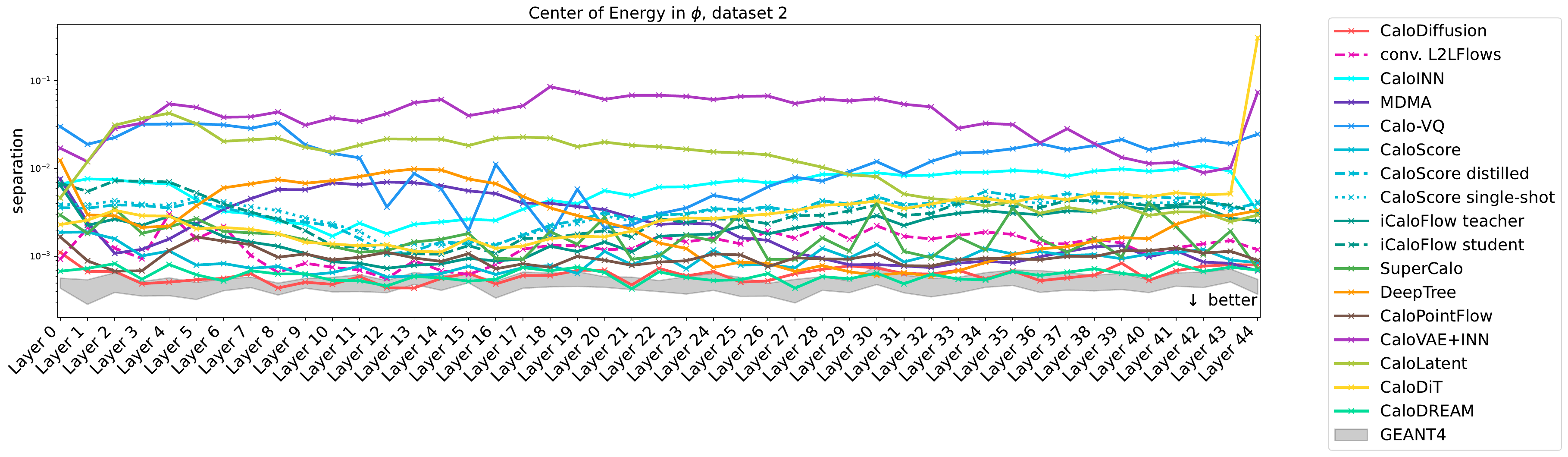}
\caption{Separation power of centers of energy in $\phi$ direction.}
\label{fig:ds2-CE-phi}
\end{figure}

\begin{figure}
\centering
\includegraphics[width=\textwidth]{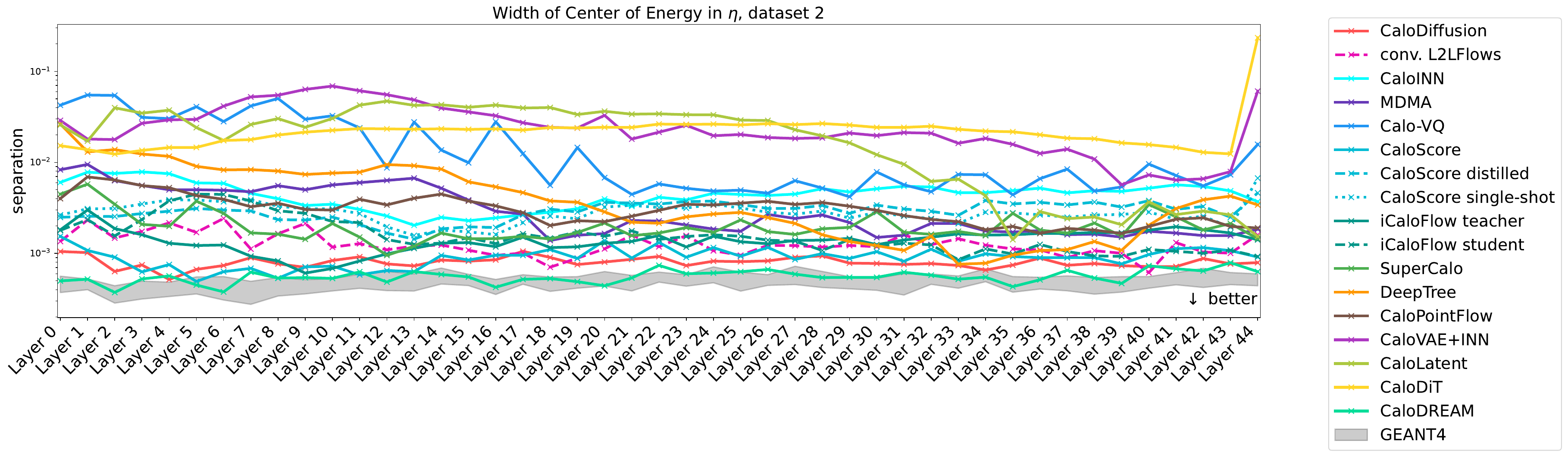}
\caption{Separation power of widths of centers of energy in $\eta$ direction.}
\label{fig:ds2-width-CE-eta}
\end{figure}

\begin{figure}
\centering
\includegraphics[width=\textwidth]{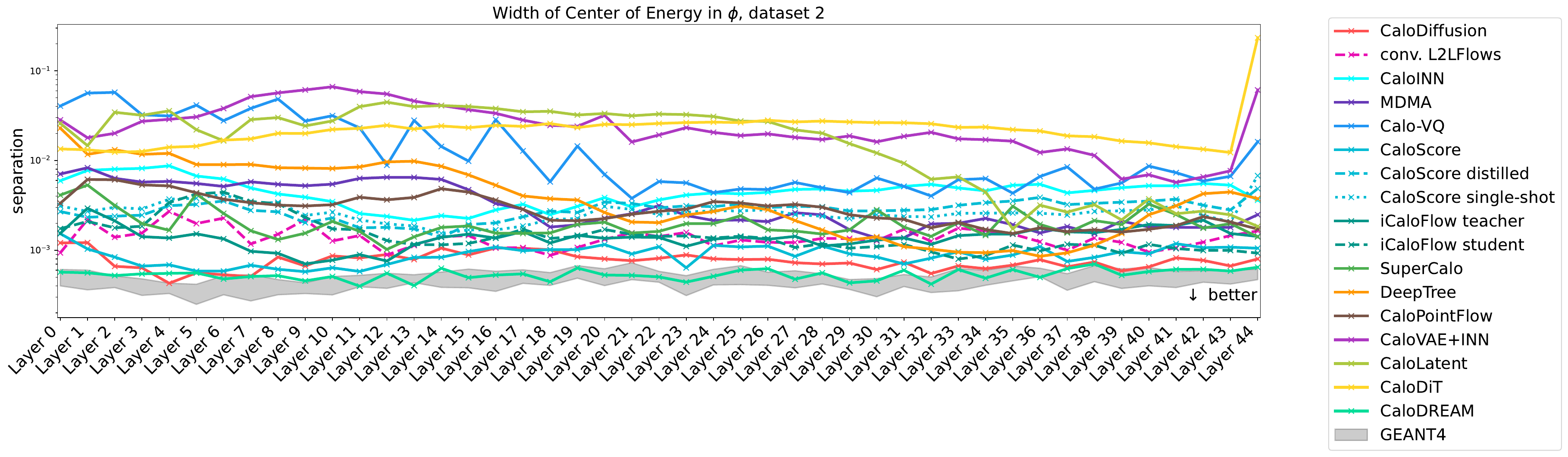}
\caption{Separation power of widths of centers of energy in $\phi$ direction.}
\label{fig:ds2-width-CE-phi}
\end{figure}

\begin{figure}
\centering
\includegraphics[width=\textwidth]{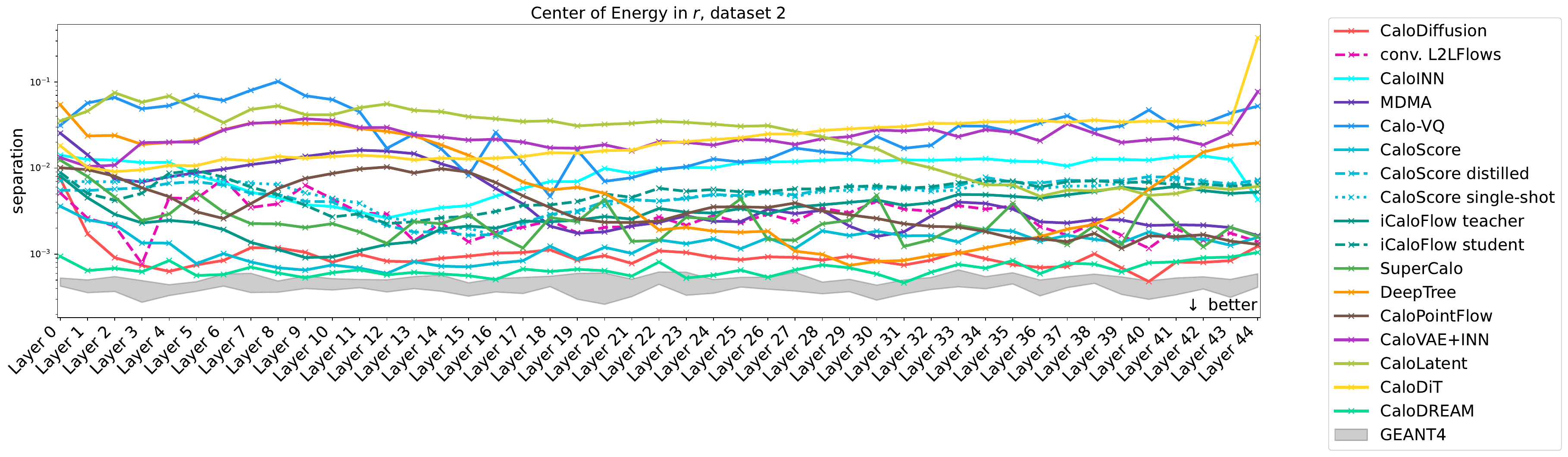}
\caption{Separation power of centers of energy in radial direction.}
\label{fig:ds2-CE-r}
\end{figure}

\begin{figure}
\centering
\includegraphics[width=\textwidth]{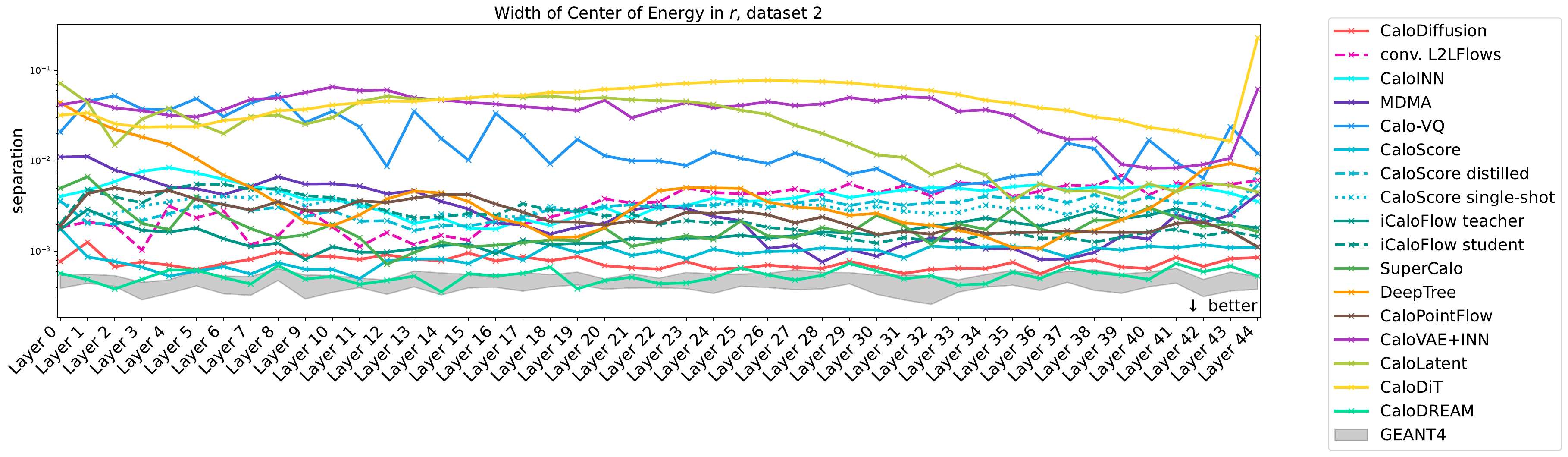}
\caption{Separation power of widths of centers of energy in radial direction.}
\label{fig:ds2-width-CE-r}
\end{figure}

\begin{figure}
\centering
\includegraphics[width=\textwidth]{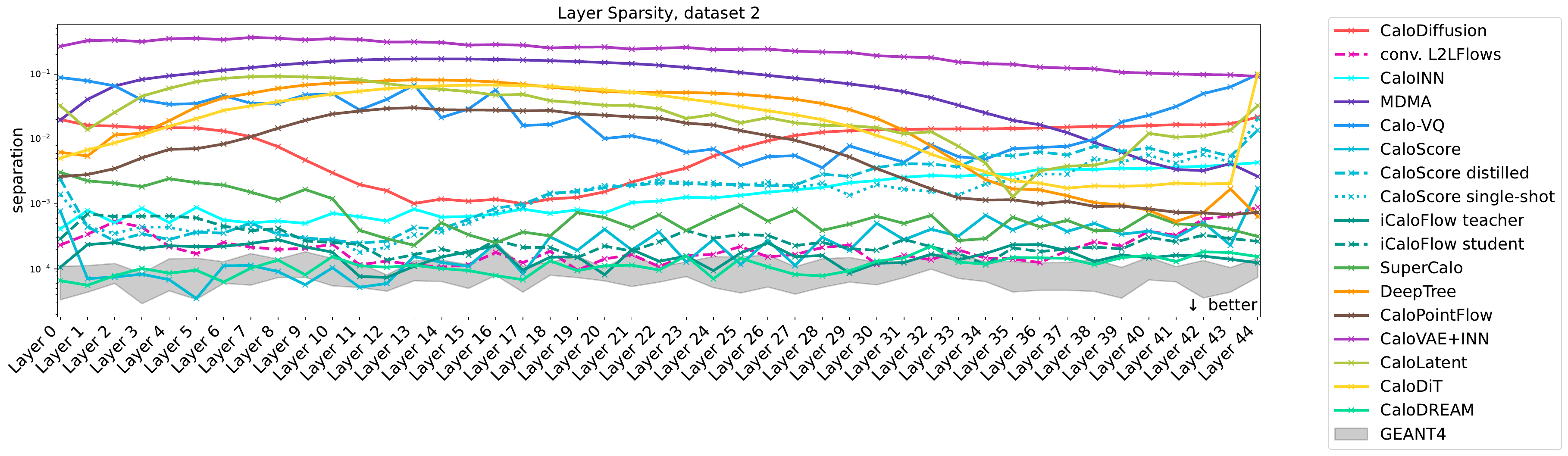}
\caption{Separation power of the sparsity.}
\label{fig:ds2-sparsity}
\end{figure}

\begin{figure}
\centering
\hfill\includegraphics[width=0.4\textwidth]{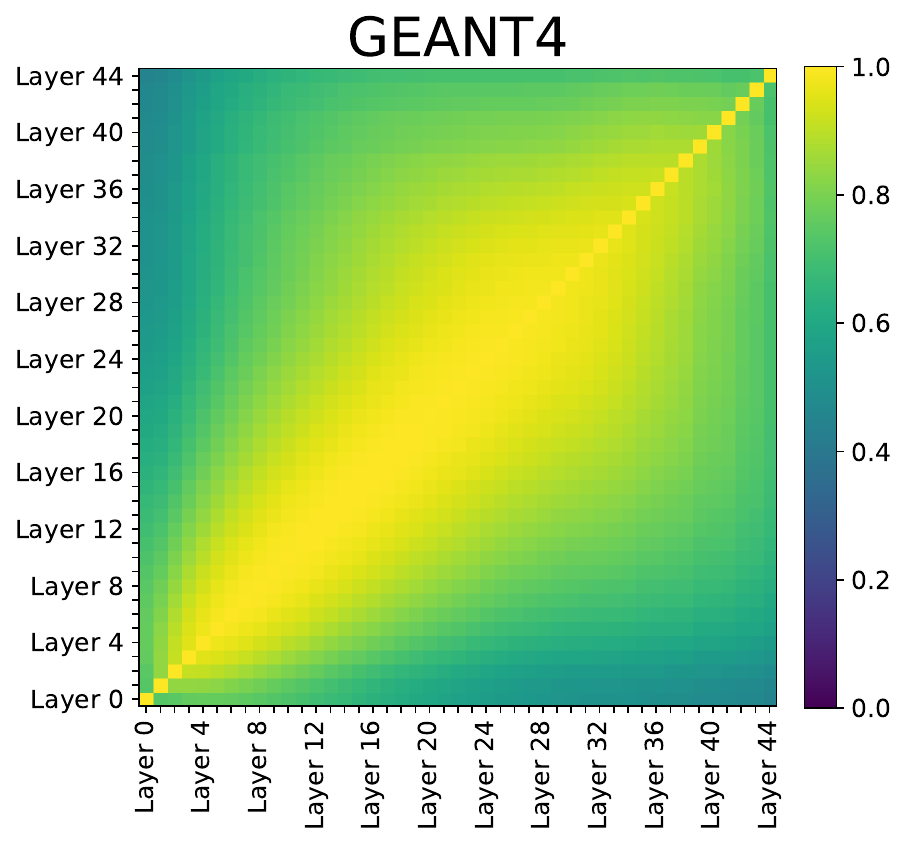}\hfill $ $
\\
\includegraphics[width=0.2\textwidth]{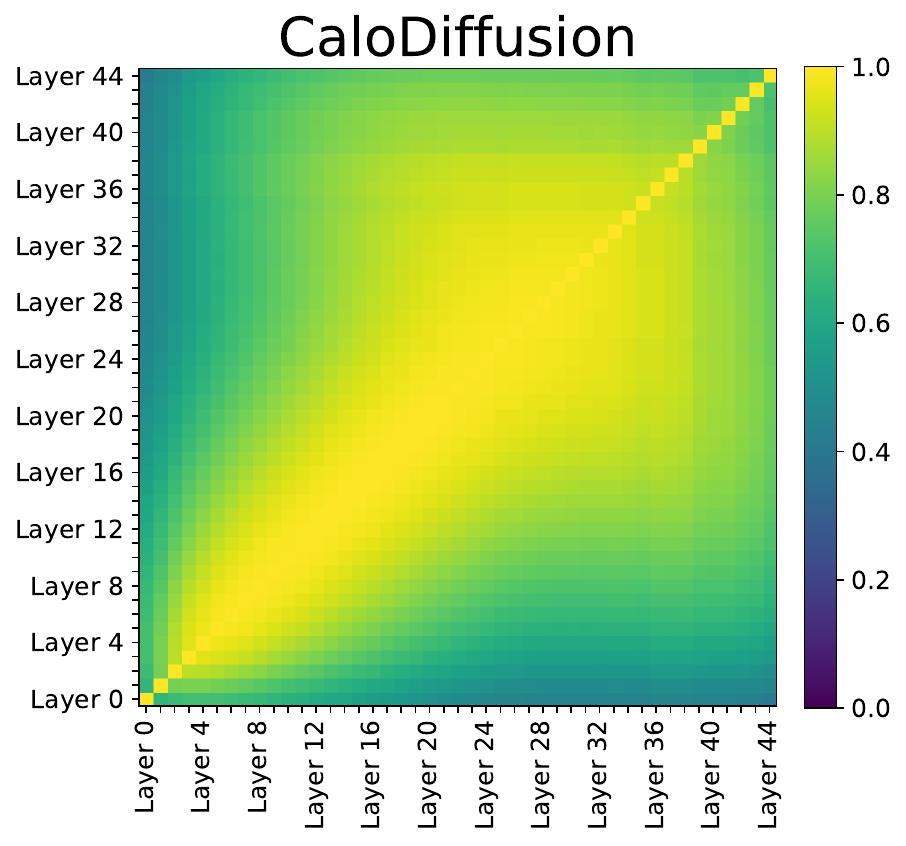}\hfill
\includegraphics[width=0.2\textwidth]{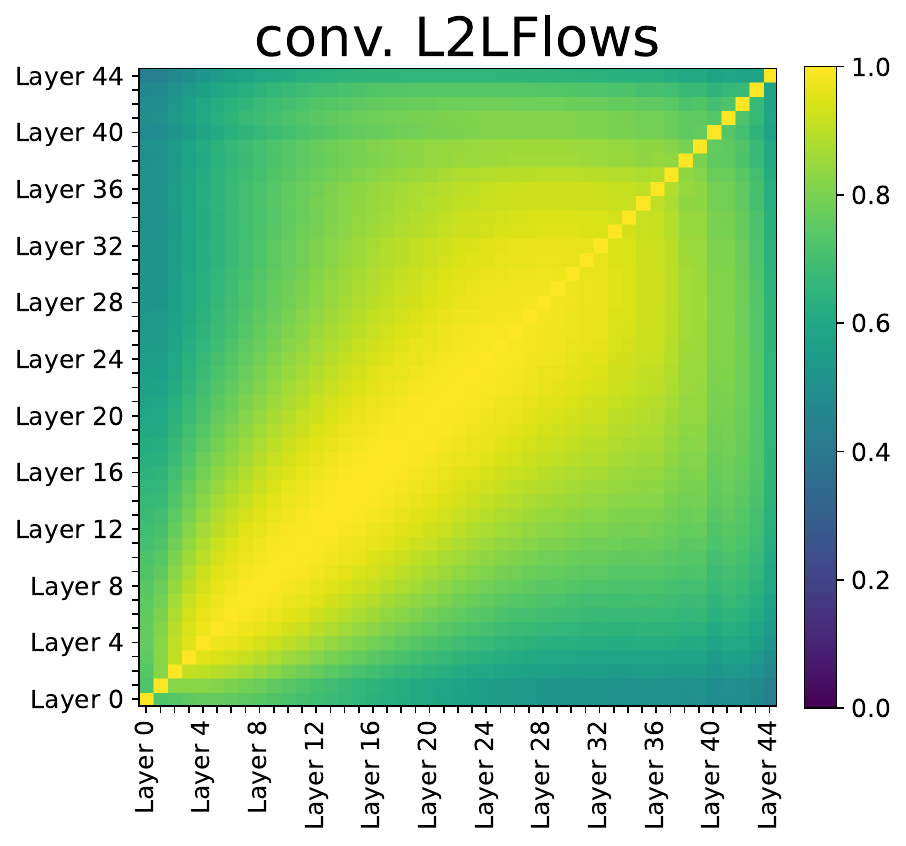}\hfill
\includegraphics[width=0.2\textwidth]{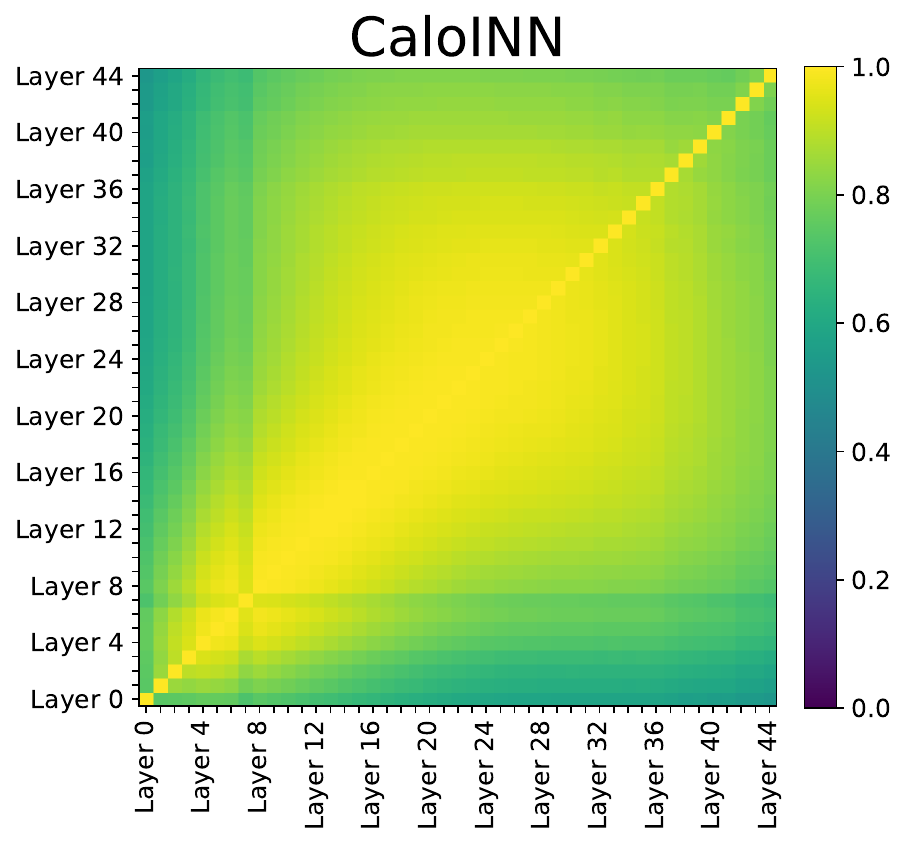}\hfill
\includegraphics[width=0.2\textwidth]{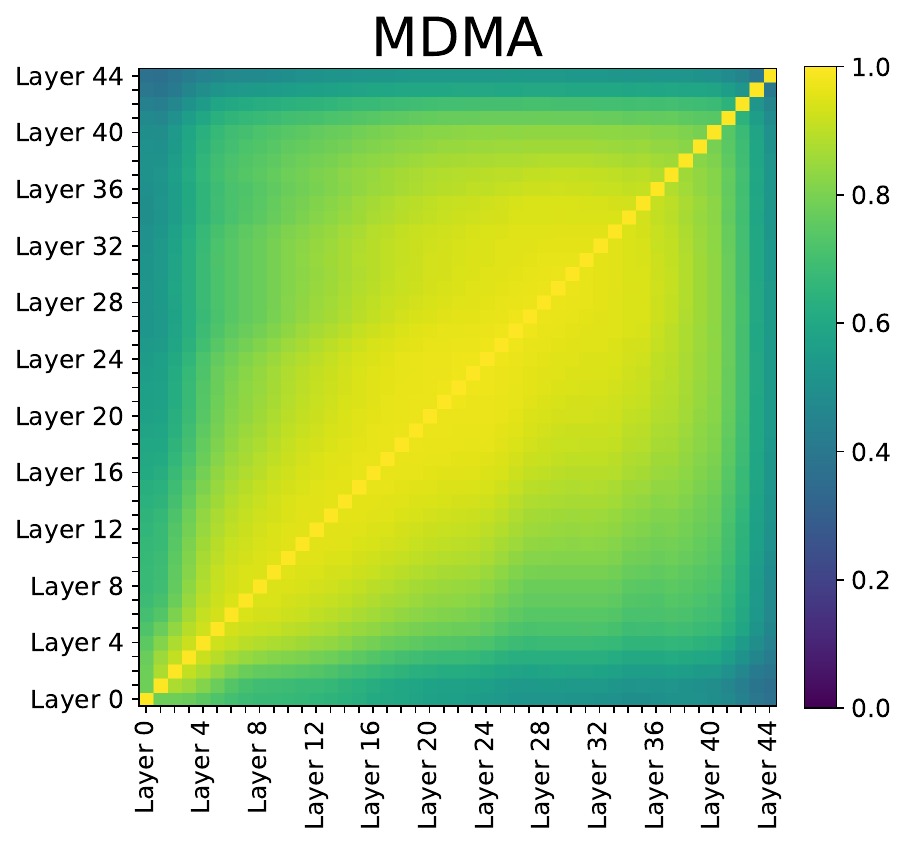}\hfill
\includegraphics[width=0.2\textwidth]{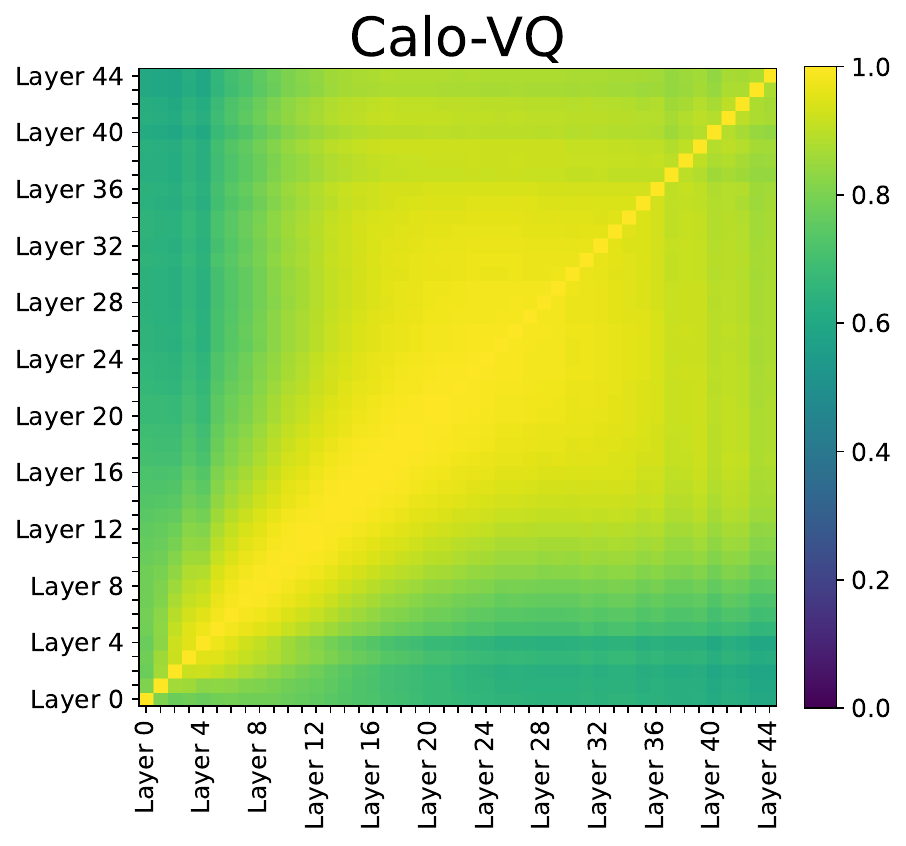}\\
\includegraphics[width=0.2\textwidth]{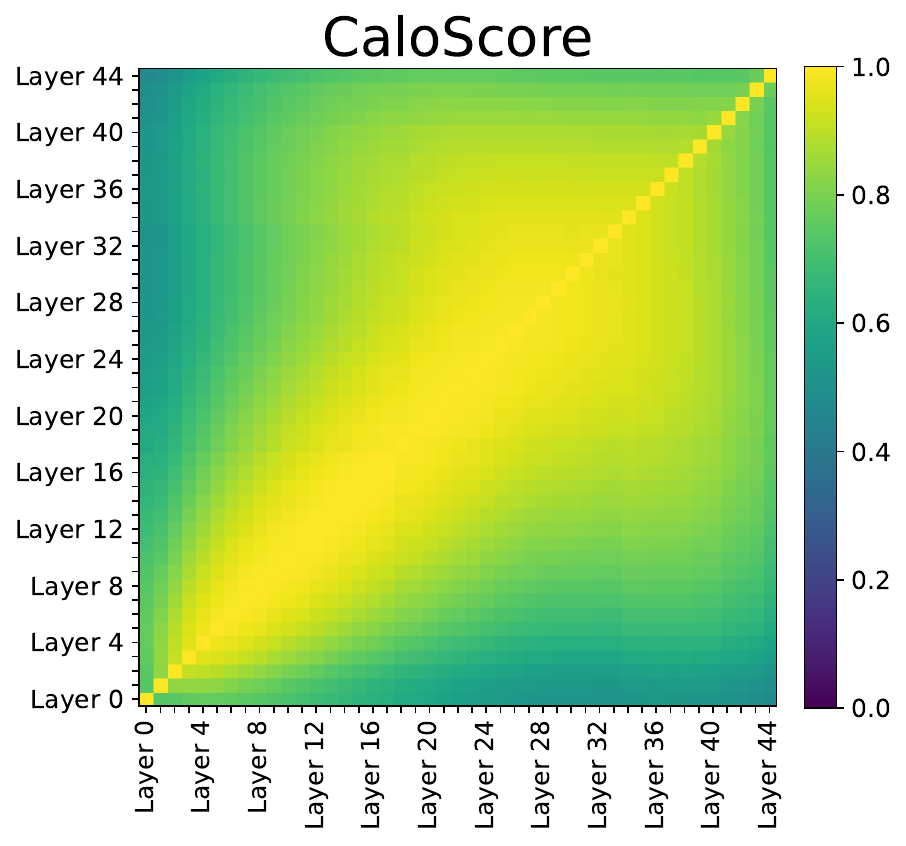}\hfill
\includegraphics[width=0.2\textwidth]{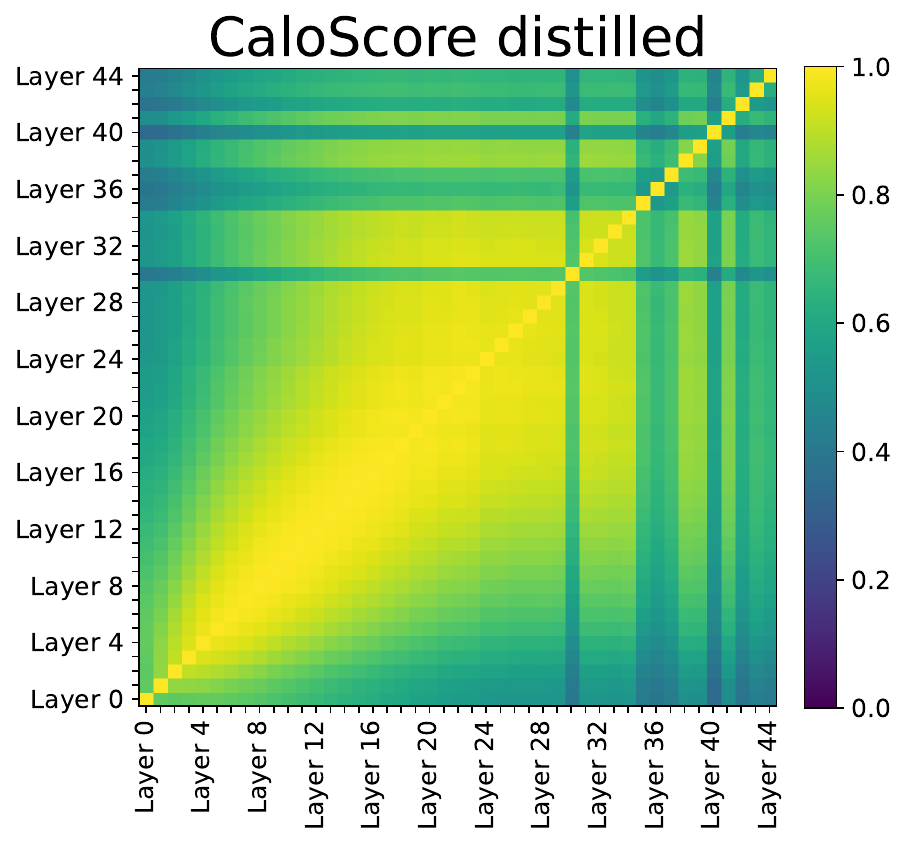}\hfill
\includegraphics[width=0.2\textwidth]{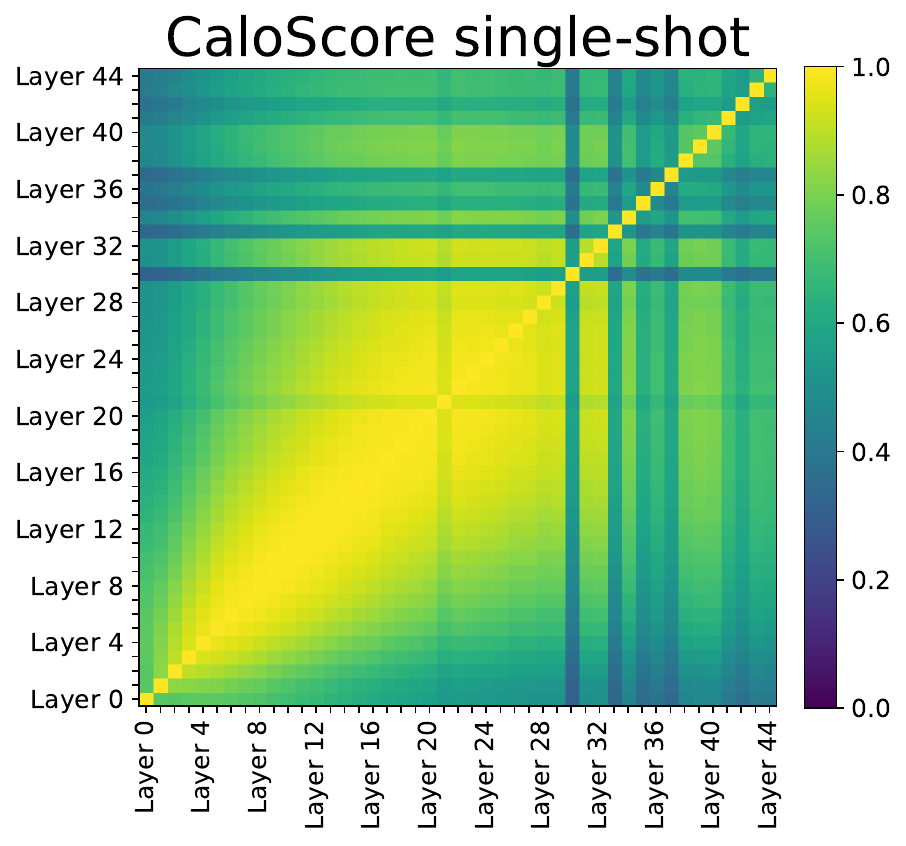}\hfill
\includegraphics[width=0.2\textwidth]{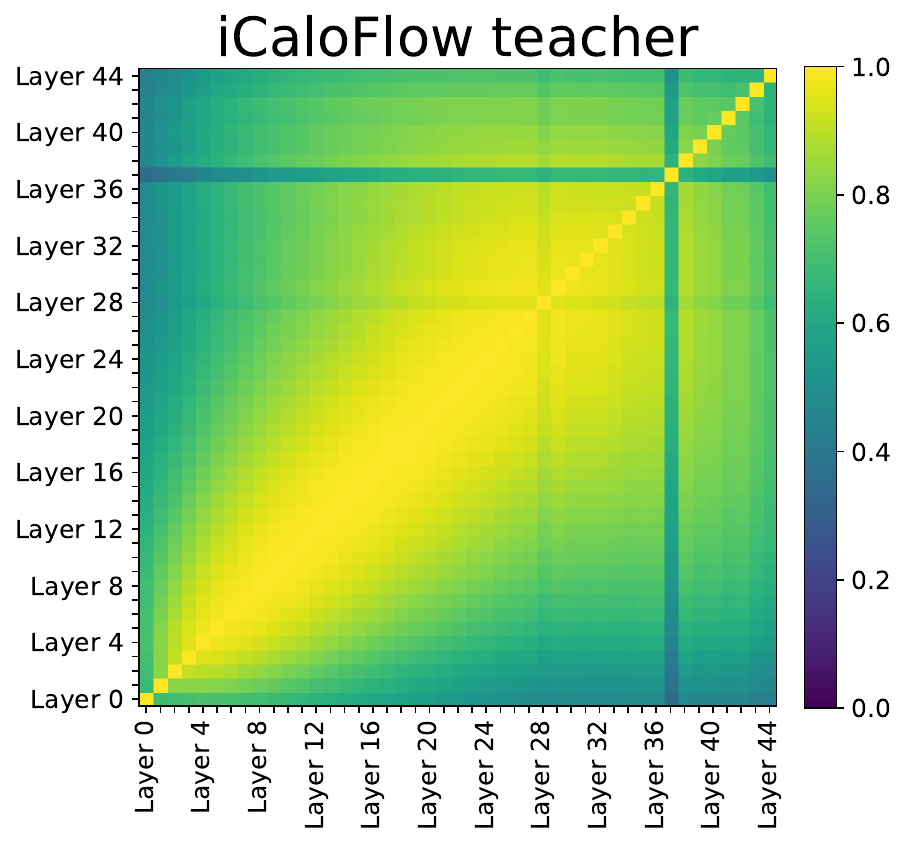}\hfill
\includegraphics[width=0.2\textwidth]{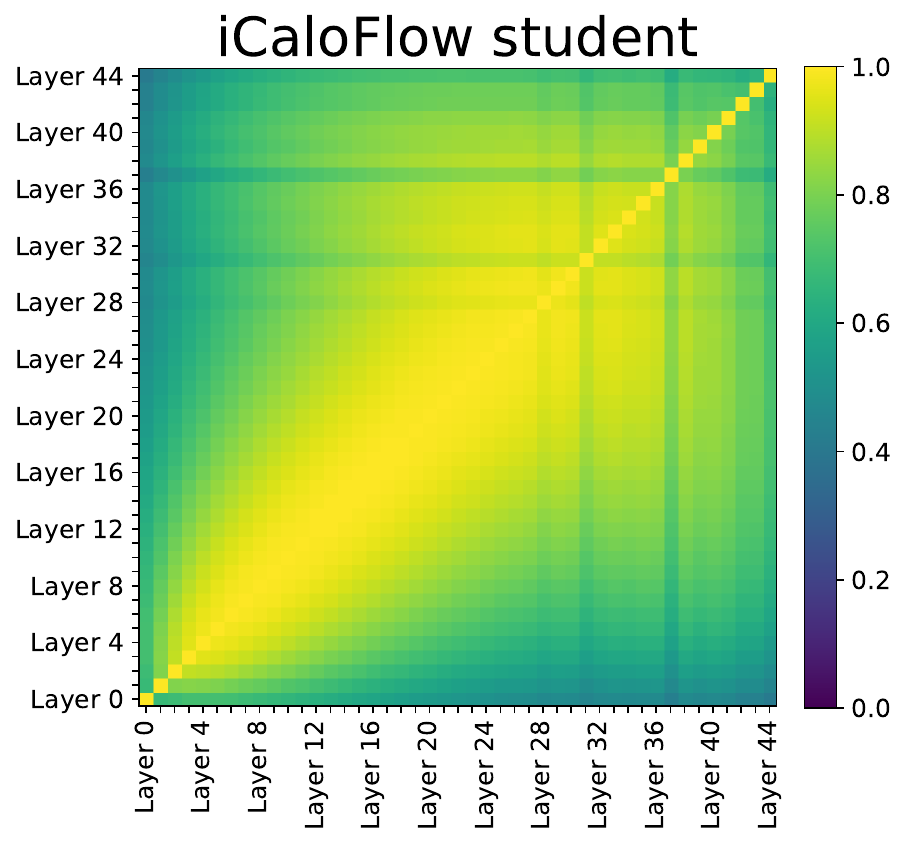}\\
\hfill \includegraphics[width=0.2\textwidth]{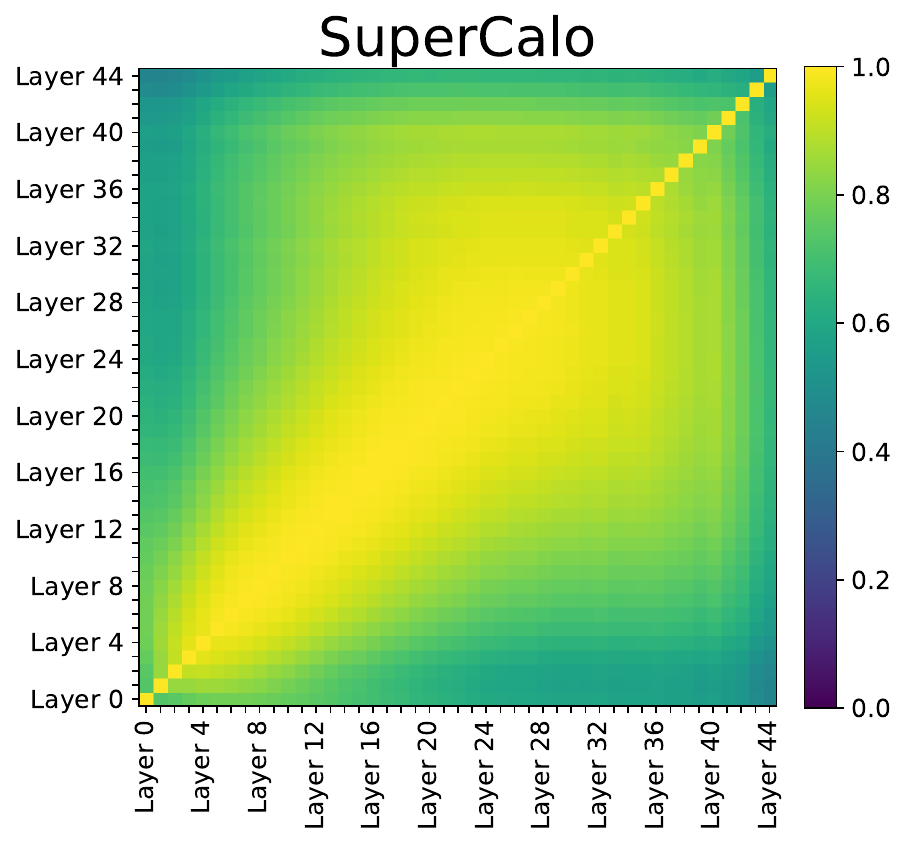}\hfill
\includegraphics[width=0.2\textwidth]{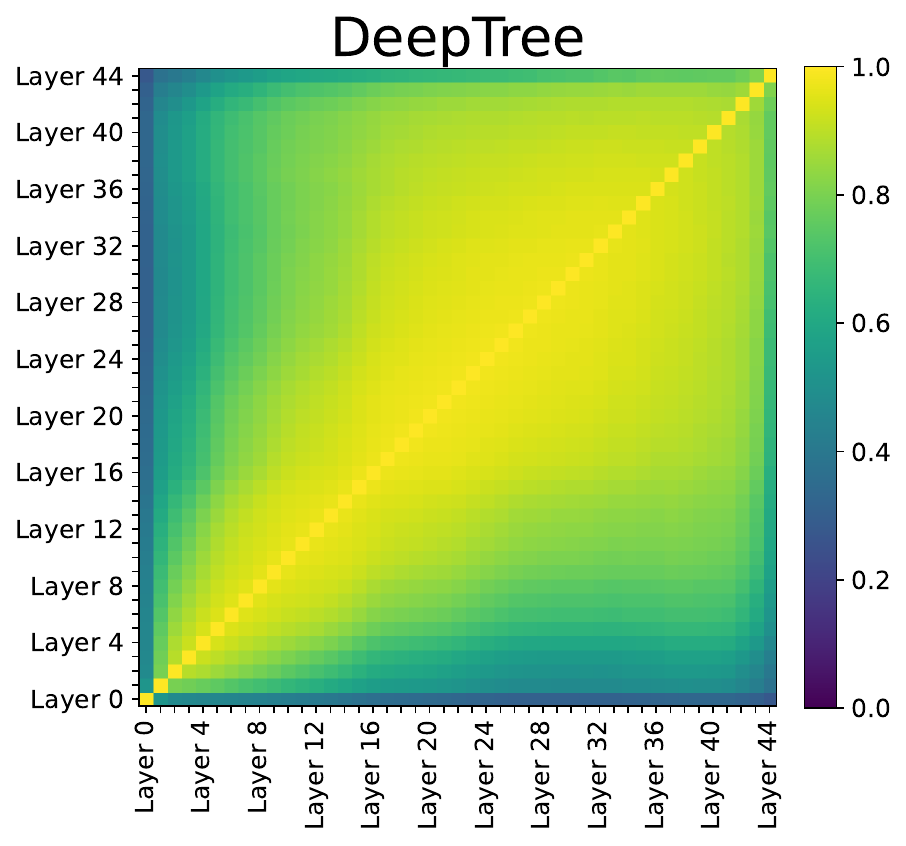}\hfill
\includegraphics[width=0.2\textwidth]{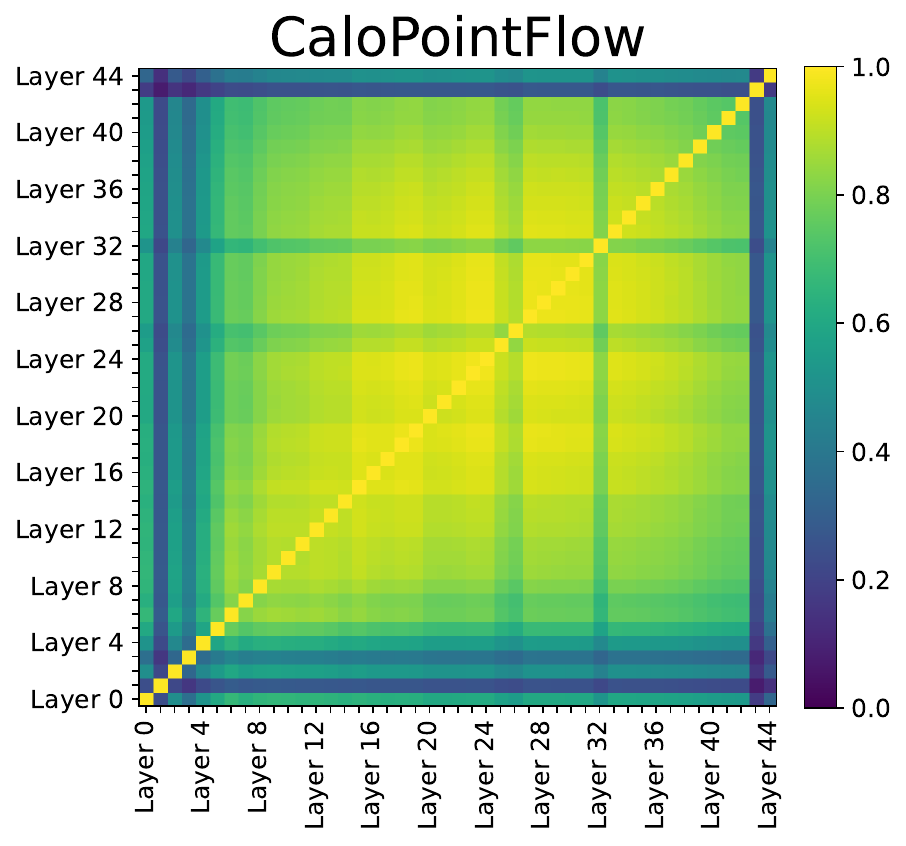}\hfill
\includegraphics[width=0.2\textwidth]{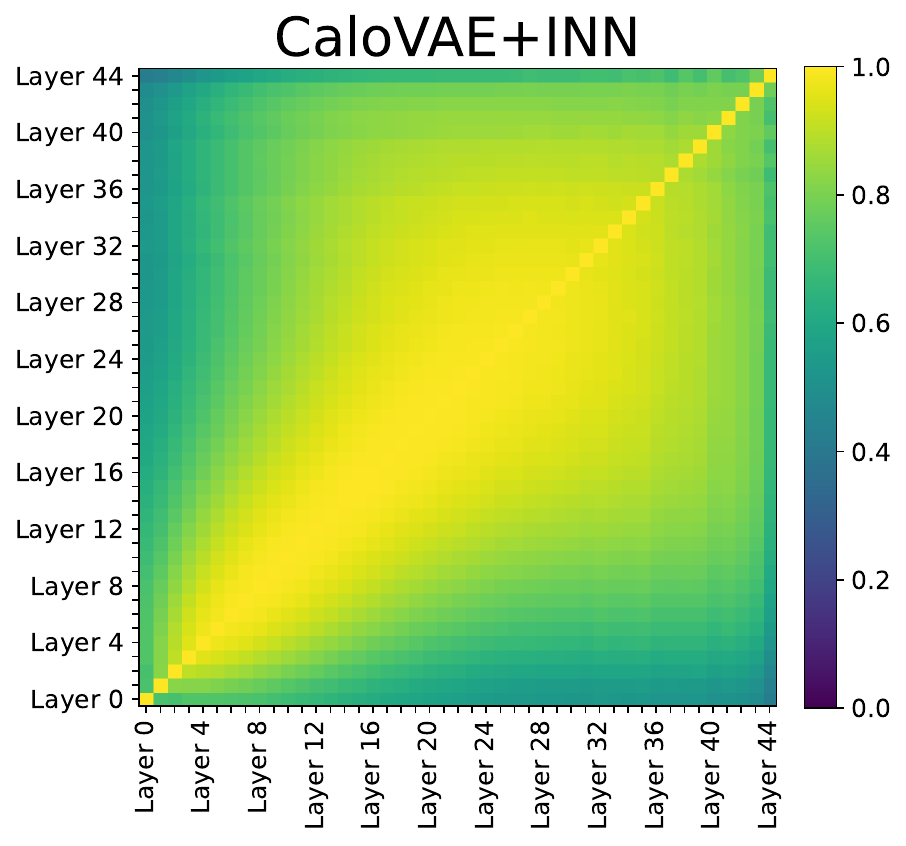}\hfill $ $\\
\hfill \includegraphics[width=0.2\textwidth]{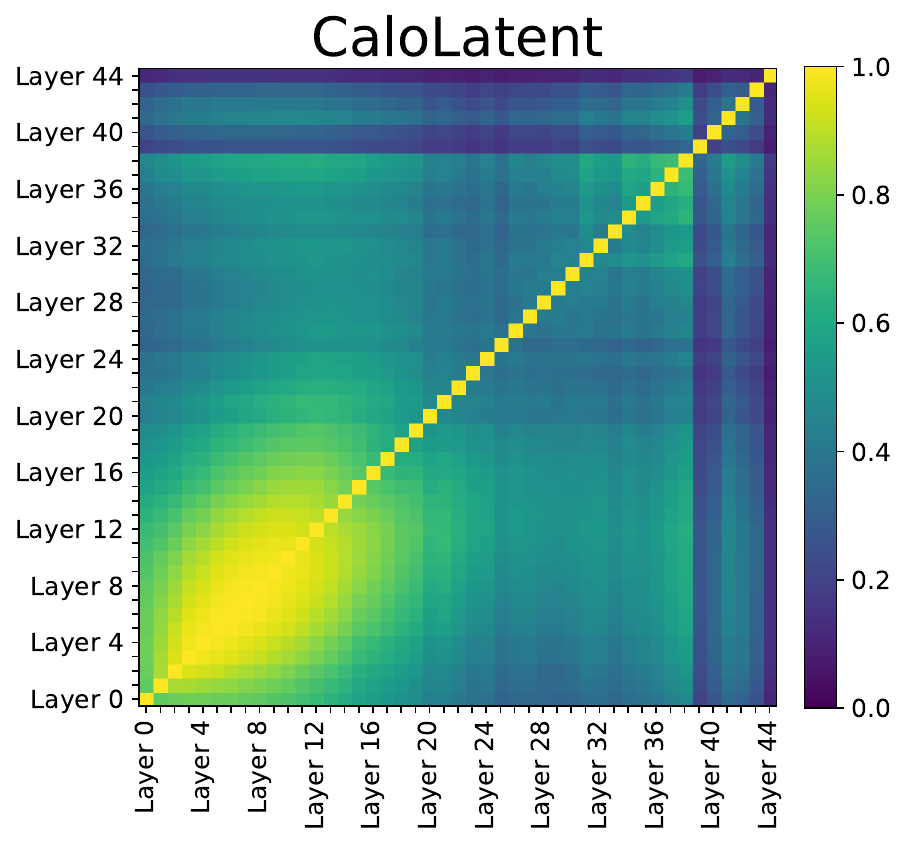}\hfill
\includegraphics[width=0.2\textwidth]{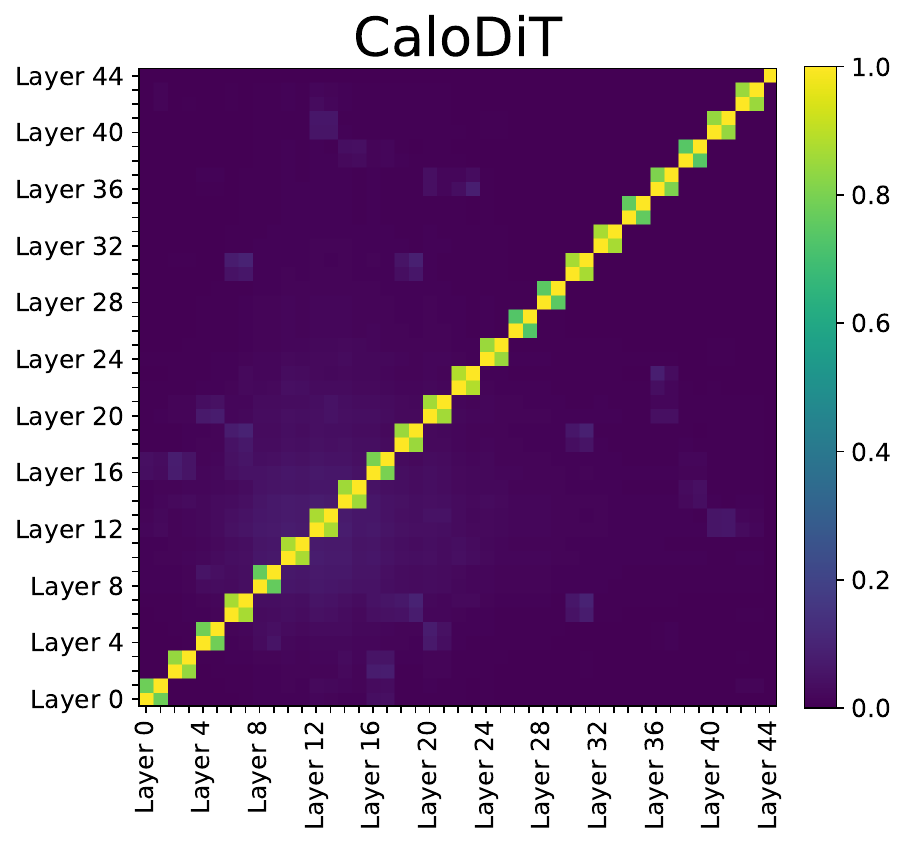}\hfill
\includegraphics[width=0.2\textwidth]{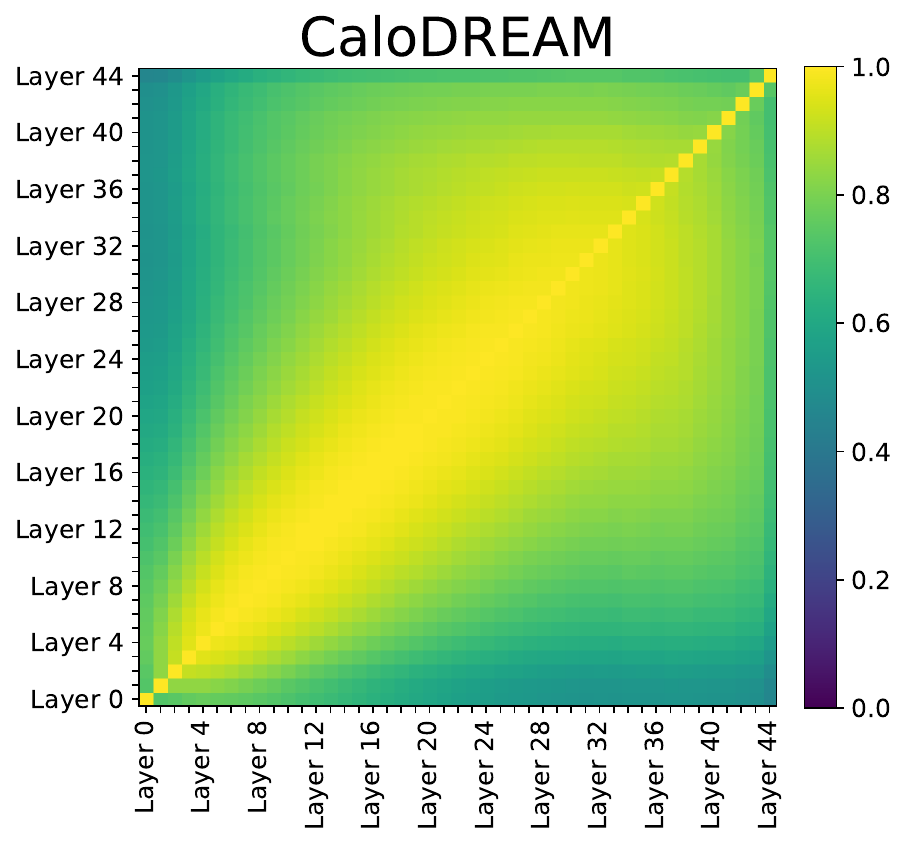}\hfill $ $
\caption{Pearson correlation coefficients of layer energies in \dsII.}
\label{fig:ds2-corr}
\end{figure}

We again start the evaluation with the separation power of the energy depositions in all layers, all voxels, and the total deposited energy in~\fref{fig:ds2-depositions}. The values for the submissions span roughly 2 orders of magnitude and only for early layers they reach down to the reference values given by \geant. In general, we observe all submissions getting worse towards the end of the detector, \textit{i.e.} for a larger layer number. While some submissions show a smooth change of separation powers from layer to layer, some others oscillate with a period of a few layers. 

The centers of energy in $\eta$ and $\phi$ are shown in~\fref{fig:ds2-CE-eta} and~\fref{fig:ds2-CE-phi}. There is a rotational symmetry in the data, so the distributions in $\eta$ and $\phi$ look very similar to each other (see \fref{fig:app_ref.ds2.3} and \fref{fig:app_ref.ds2.5}). Judging by the separation powers, the models learn the distributions in these two variables equally well, reflecting this symmetry. In detail, we see \submAmram and \submPalacios having the smallest separation powers, just at the upper bound of the \geant reference band. VAE-based submissions like \submErnst, \submReyes, or \submLiu again have the the largest separation powers. Looking at the change from layer to layer, we now see a different pattern compared to the energy distributions in~\fref{fig:ds2-depositions}. Now, only some submissions show an increasing separation power for an increasing layer number. Others are either rather constant or have a large separation power for small layer numbers, show better results in the central part of the detector and then increase again towards the end. We also notice some models having a rather steep increase only in the last layer.     

The separation powers of the widths of centers of energy in $\eta$ (\fref{fig:ds2-width-CE-eta}) and $\phi$ (\fref{fig:ds2-width-CE-phi}) are very similar to the separation powers of the centers of energies themselves. Both directions, $\eta$ and $\phi$, show almost identical results. Now \submPalacios is having the best score, at the level of the \geant reference band. Other submissions show again their best performance in the central region of the calorimeter segment, before the separation power rises again at larger layer numbers.  

Given the rotational symmetry in $\eta$ and $\phi$, the separation powers in centers of energy and its width in radial direction resemble the ones in $\eta$ and $\phi$ strongly, as can be seen in \fref{fig:ds2-CE-r} and \fref{fig:ds2-width-CE-r}.

The last set of separation powers we look at are from the sparsities in~\fref{fig:ds2-sparsity}. Here, the spread between different models is larger, ranging more than three orders of magnitude. Interestingly, \submPalacios still shows very good results, at the level of the \geant reference band. \submAmram on the other hand does not reproduce the sparsities well, with \submMikuni, \submPangSuper, \submFavaro, and \icalo outperforming it in all layers.  

We show the Pearson correlation coefficients in layer energies in~\fref{fig:ds2-corr}. Interestingly, we do not reproduce all findings of~\cite{Ahmad:2024dql}, which trained a few models from scratch, indicating that some of the observed patterns fluctuate from training to training. In general, we observe two different failure modes in these figures: One group (most prominently \submMadula and \submCardoso) do not reproduce the correlations in a large region. A second group (consisting of \submMikuniDist, \icalo, and \submSchnake) show problems in single layers, indicated by streaks in~\fref{fig:ds2-corr}. Also, the distillation procedure worsened the pattern for \submMikuniSingle, but slightly improved it for \submPangIS. \submCardoso shows only correlation to one of the nearest neighbor layers, nothing beyond that, which is consistent with the larger separation powers we saw before. 

\begin{figure}
\centering
\includegraphics[width=\textwidth]{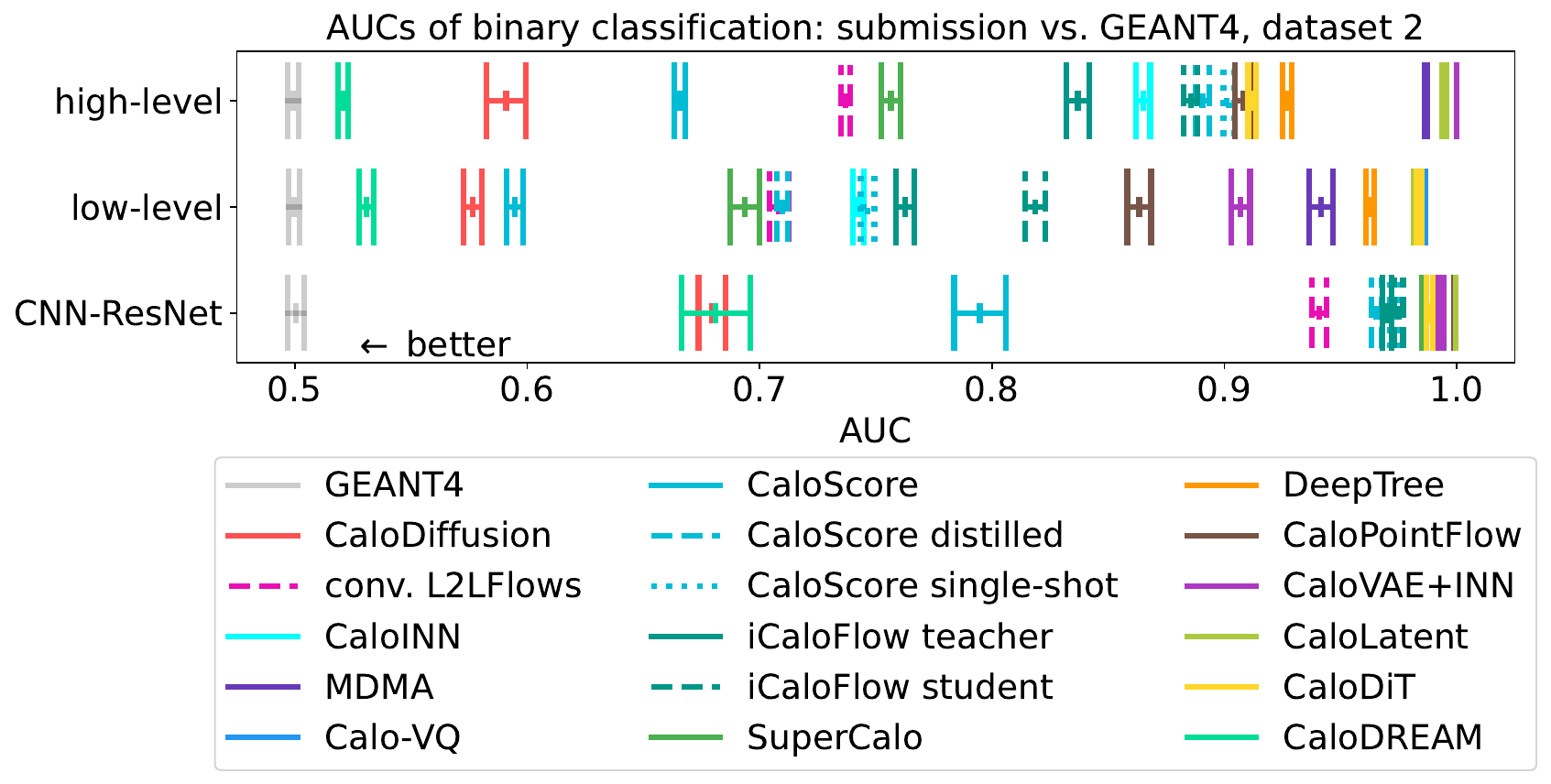}
\caption{Low-level and high-level AUCs for evaluating \geant\ vs.~submission of \dsII, averaged over 10 independent evaluation runs. For the precise numbers, see \Tref{tab:ds2.aucs}.}
\label{fig:ds2.aucs}
\end{figure}

\begin{figure}
\centering
\includegraphics[width=\textwidth]{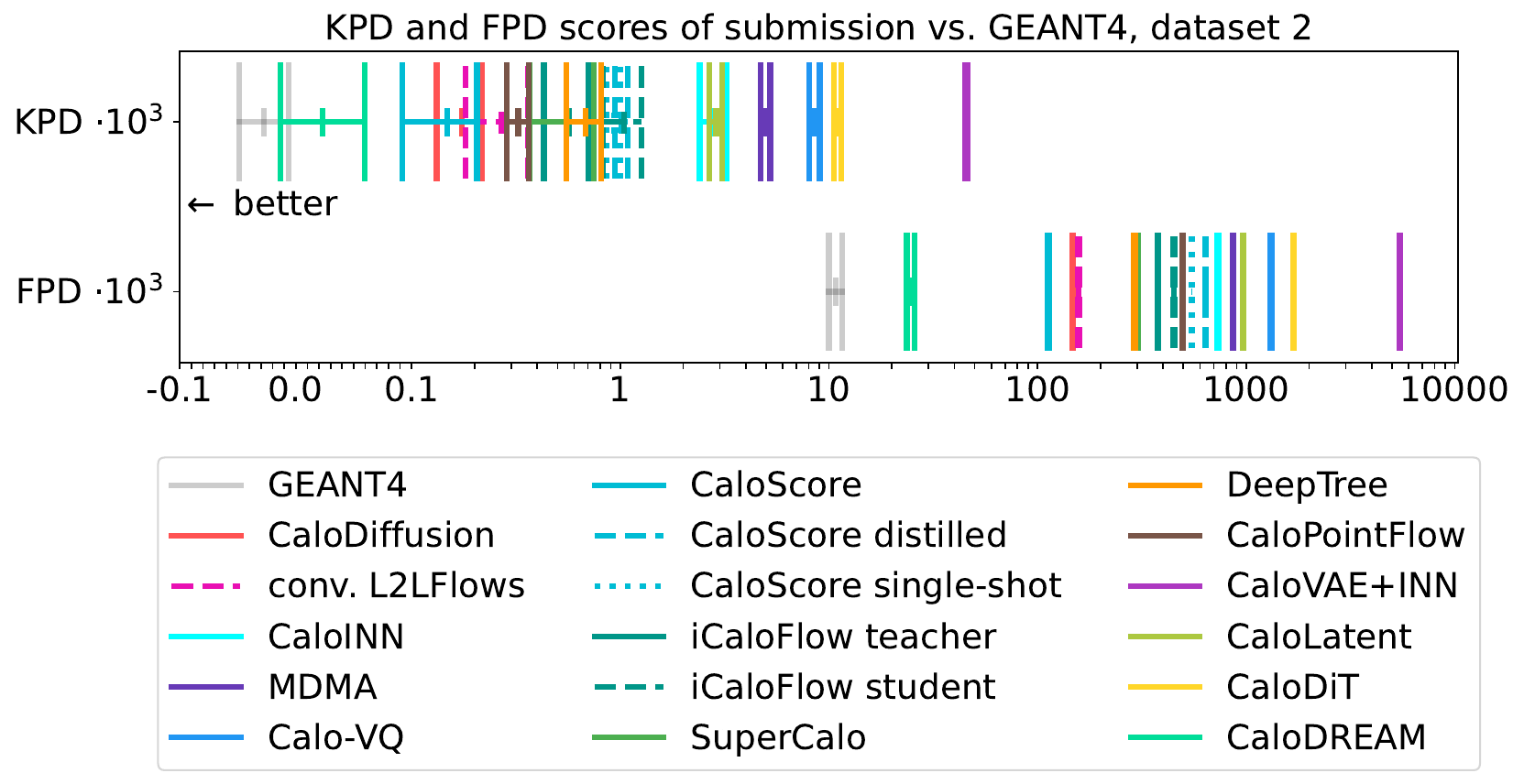}
\caption{KPD and FPD for evaluating \geant\ vs.~submission of \dsII. For the precise numbers, see \Tref{tab:ds2.kpd}.}
\label{fig:ds2.kpd}
\end{figure}

We now turn to classifier-based metrics and start with the AUC of the binary classifiers in~\fref{fig:ds2.aucs} (and \tref{tab:ds2.aucs}). In addition to the DNN architecture that we also used for dataset 1, we now have an additional, CNN-ResNet-based architecture that we use for the evaluation. This CNN-ResNet architecture is much more sensitive to differences in the distributions and it moves the AUC of many submissions close to 1.0. While \submPalacios has the best scores in the DNN-based classifiers, it is tied with \submAmram in the stronger CNN-ResNet classifier. However, as before, the submissions \submPalacios, \submAmram, and \submMikuni show in general the best (lowest) binary AUC scores, independent of the classifier architecture used. Flow-based models follow, while VAE and GAN-based submissions have the highest AUC.   

The computer science-inspired metrics KPD and FPD in \fref{fig:ds2.kpd} (with details in \tref{tab:ds2.kpd}) show results consistent with the classifier AUCs. \submPalacios, \submAmram, and \submMikuni again have the best (lowest) scores, but now \submMikuni is slightly better than \submAmram, which is in fact overlapping with \submBussConv now. At the other end of the spectrum we again see submissions based on GANs and VAEs.

\begin{figure}
\centering
\includegraphics[width=\textwidth]{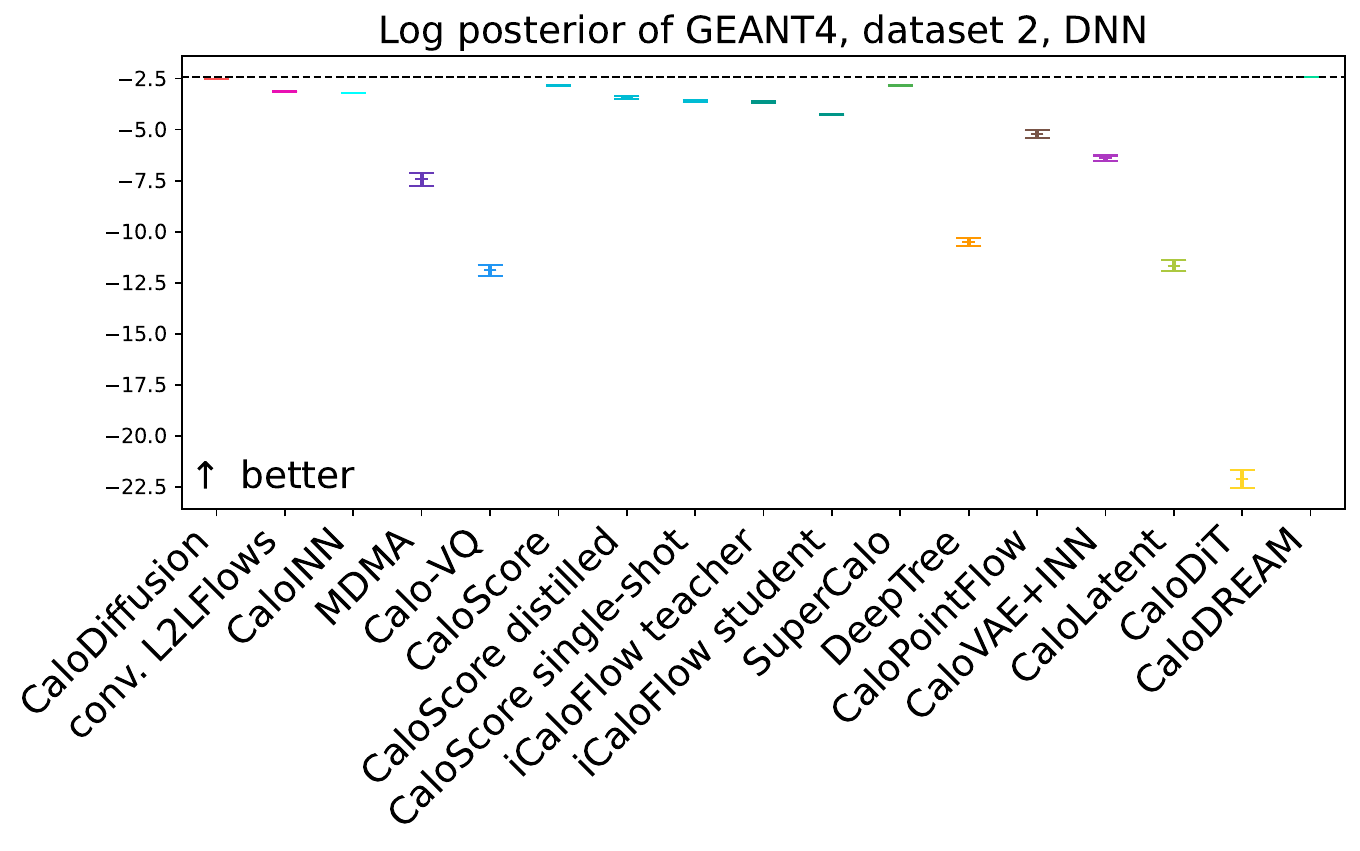}
\caption{Log-posterior scores for \dsII \geant test data, averaged over 10 independent DNN classifier trainings. For the precise numbers, see \Tref{tab:ds2.multi.dnn}.}
\label{fig:ds2.multi.dnn}
\end{figure}

\begin{figure}
\centering
\includegraphics[width=\textwidth]{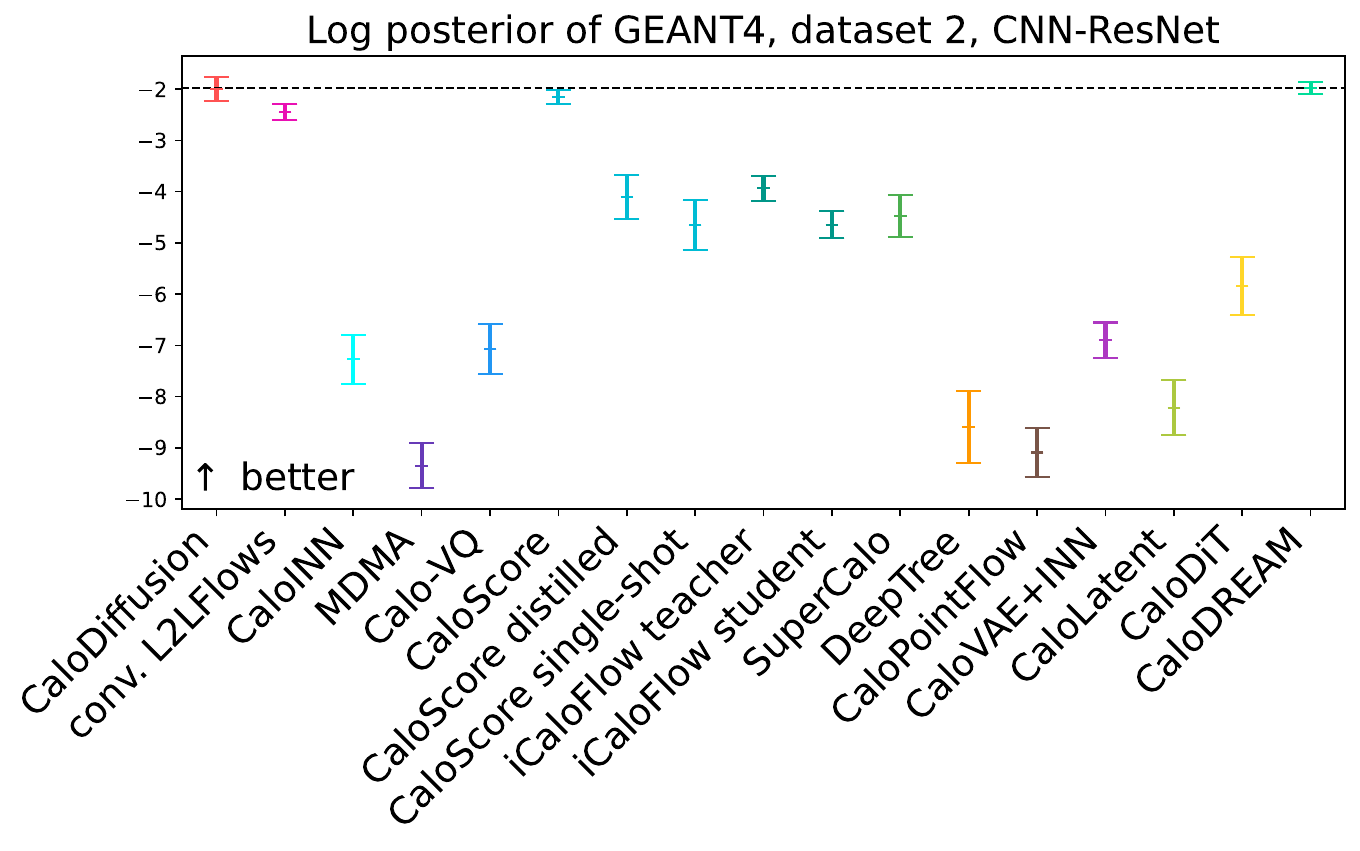}
\caption{Log-posterior scores for \dsII \geant test data, averaged over 10 independent CNN ResNet classifier trainings. For the precise numbers, see \Tref{tab:ds2.multi.resnet}.}
\label{fig:ds2.multi.resnet}
\end{figure}

For the multiclass classification we also employ a DNN and a CNN-ResNet architecture. Both of these have well-trained classifiers, as can be seen in \fref{fig:consistency.ds2} and \fref{fig:consistency.ds2.ResNet}. In \fref{fig:ds2.multi.dnn} (as well as \tref{tab:ds2.multi.dnn}), we see the results for the DNN architecture. \submPalacios is again leading with \submAmram at a very close second place. \submMikuni with its distilled versions and the flow-based submissions of \submPangSuper, \submBussConv, \submFavaro, and \icalo follow with very small differences. Distilled submissions of \submMikuni and \icalo perform in general slightly worse than the original versions that they have been distilled from. Turning to the CNN-ResNet architecture in \fref{fig:ds2.multi.resnet} (and \tref{tab:ds2.multi.resnet}), the story is roughly the same as for the DNN before. Overall, we observe the errorbars becoming larger, indicating a larger spread of result in different trainings. However, the spread in the log posterior from the best to the worst model decreased by a factor two from about 20 to about 10. The three submissions \submAmram, \submMikuni, and \submPalacios are on top and have comparable scores within their error bars. \submBussConv follows closely and has a small gap to the midfield, which is composed of \icalo, the distilled versions of \submMikuni, and \submPangSuper. 

\begin{figure}
\centering
\includegraphics[width=\textwidth]{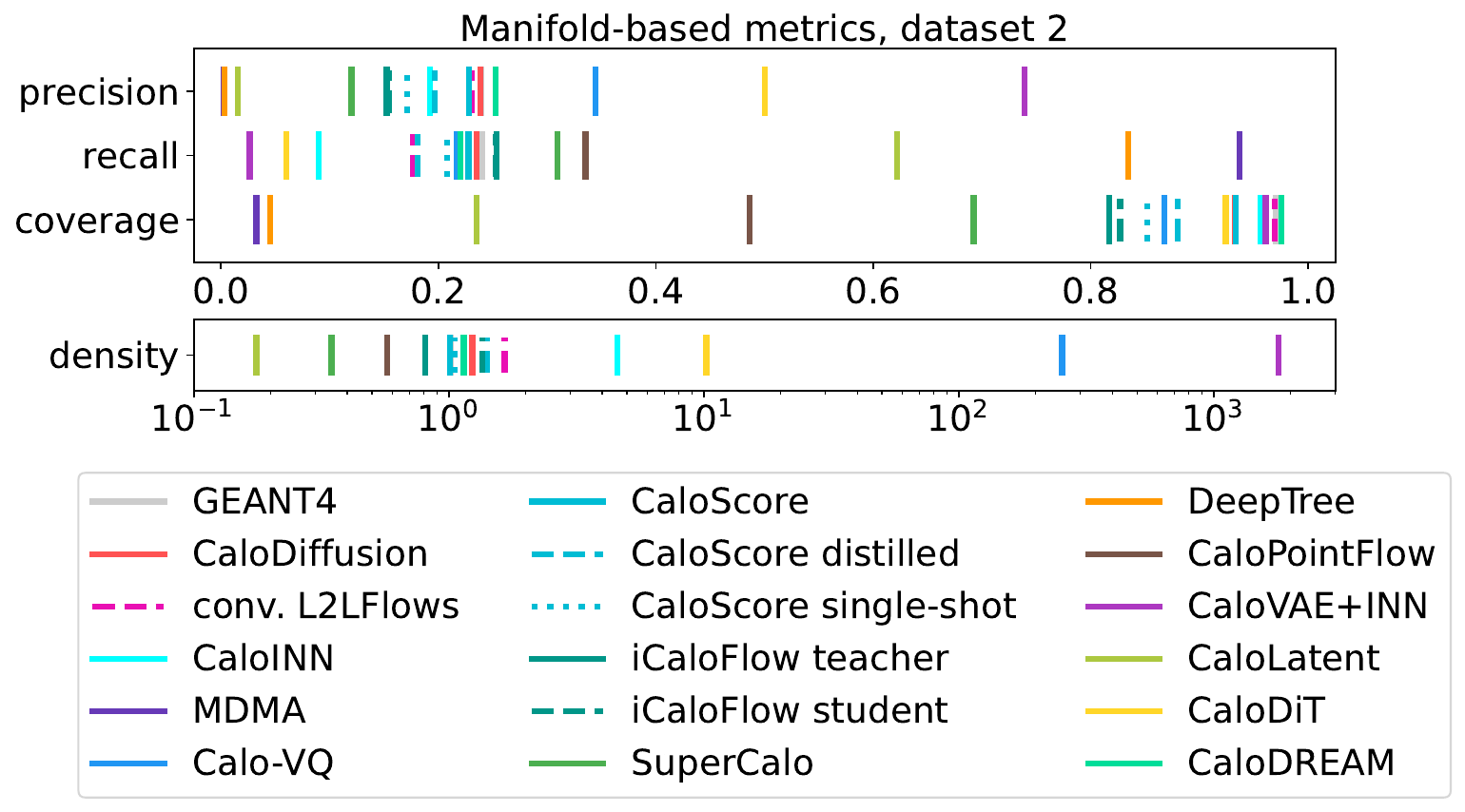}
\caption{Precision, density, recall, and coverage for \dsII submissions. For the precise numbers, see \Tref{tab:ds2.prdc}.}
\label{fig:ds2.prdc}
\end{figure}

In \fref{fig:ds2.prdc} (with details in \tref{tab:ds2.prdc}) we show precision, density, recall, and coverage of the \dsII submissions. We first notice that there is a group of submissions --- consisting of \submAmram, \submBussConv, \submMikuni and its distillations, \icalo, and \submPalacios~--- that gets all four metrics close to the \geant reference. This is another indication that these models generate high-quality showers. Another group stands out with a large density value. These are \submFavaro, \submLiu, \submErnst, and \submCardoso. The large density, together with the small recall that most of the submissions in this group have, suggests again that samples are generated very similar to each other, but not diverse enough. The GAN submissions \submKaech and \submScham stand out in a third group, with low precision, density and coverage, but large recall. We interpret this pattern as generating samples that are fairly spread out, but not really close to the reference samples. That way, precision and density are low, but the large distance between the submitted samples makes the recall manifold large enough to contain the references. The remaining submissions, \submPangSuper, \submSchnake, and \submMadula do not really fit in these groups, but are somehow similar to the GAN submissions with smaller precision, density, and coverage, but larger recall than the \geant reference. However, the gap is smaller in these cases. 

To summarize the shower quality, we see a similar pattern than already for dataset 1: The diffusion and conditional flow matching models have the best quality, followed then by Normalizing Flows and GAN and VAE-based models at the end. 

\begin{figure}
\centering
\includegraphics[width=\textwidth]{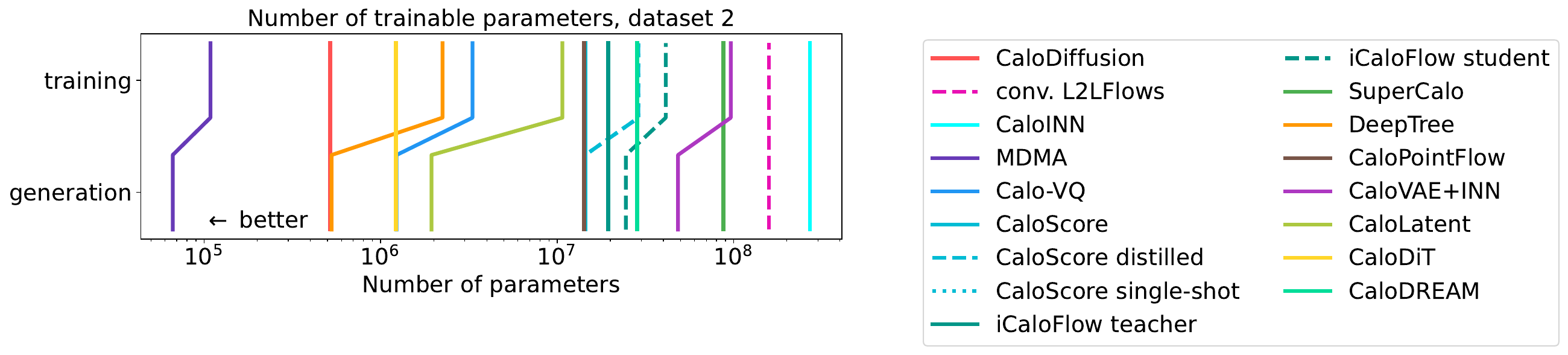}
\caption{Number of trainable parameters for training and generation of \dsII submissions. For the precise numbers, see \Tref{tab:ds2.numparam}.}
\label{fig:ds2.numparam}
\end{figure}

\Fref{fig:ds2.numparam} compares the sizes of the submissions, with \tref{tab:ds2.numparam} giving the precise numbers. The by-far smallest model is \submKaech, with about an order of magnitude fewer parameters than the next submissions, \submAmram and \submScham. Normalizing-flow-based architectures like \submBussConv and \submFavaro have the most parameters, so the bijective transformation in this 6480-dimensional space takes it toll on the required number of parameters.  

\begin{figure}
\centering
\includegraphics[width=\textwidth]{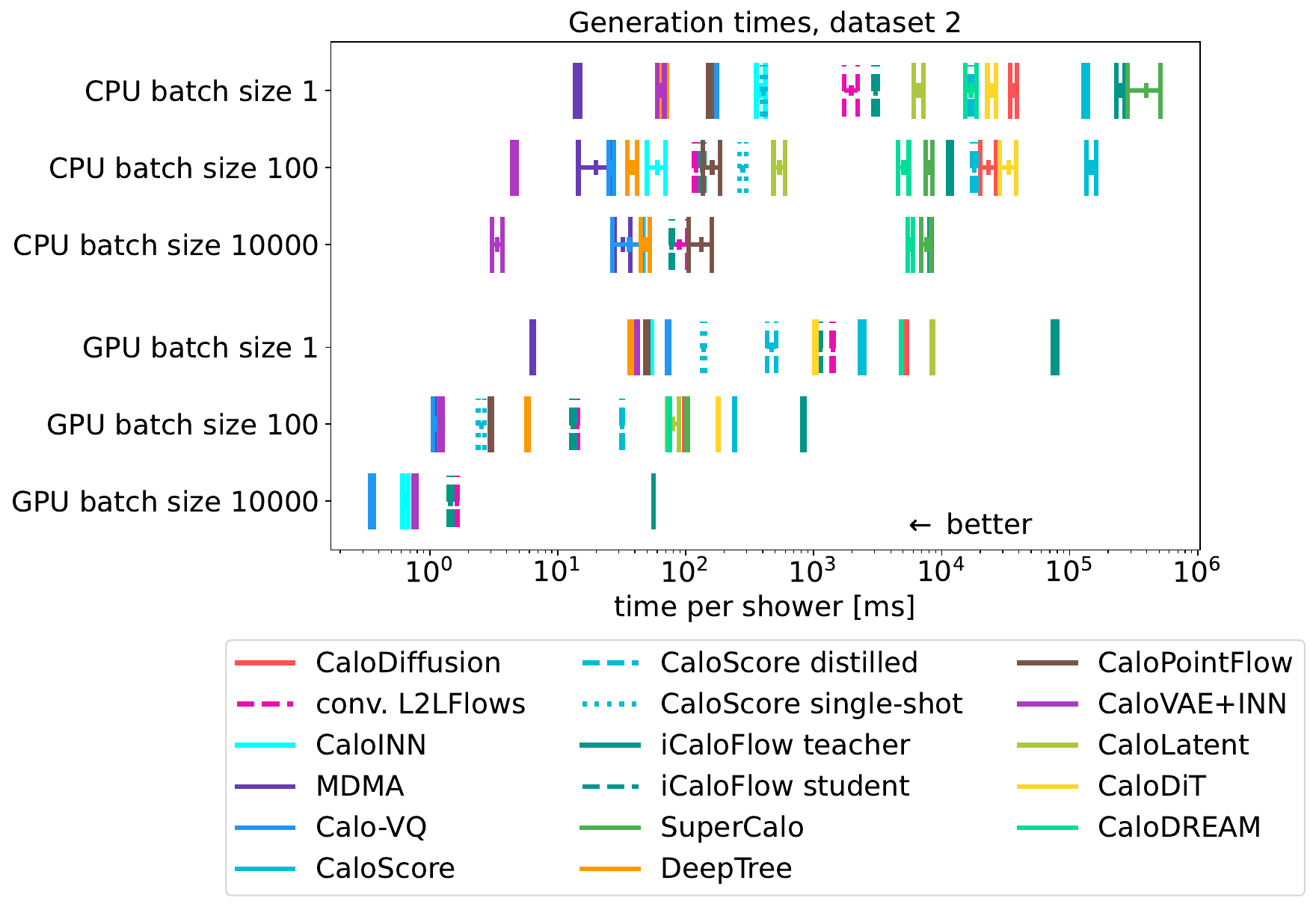}
\caption{Timing of \dsII submissions on CPU and GPU architectures. Not all submissions are shown everywhere due to memory and other constraints. More details are in \tref{tab:ds2.timing.CPU} and \tref{tab:ds2.timing.GPU}.}
\label{fig:ds2.timing}
\end{figure}

\Fref{fig:ds2.timing} (with detailed numbers in~\tref{tab:ds2.timing.CPU} and~\tref{tab:ds2.timing.GPU}) show the generation times of the submissions normalized to generating a single shower. Overall, they span several orders of magnitude, even when only looking at one of the two architectures alone. This spread depends also on the batch size, with smaller batch sizes having a larger spread between slowest and fastest submission. For example, for a batch size of 1, we see four to five orders of magnitude difference. On the CPU, sample generation is in general slower and more spread out between slowest and fastest submission than on the GPU. Batching helps to speed up generation, but some of the models run into memory problems at very large batch sizes, even more on a GPU with limited VRAM. As for the DGM types, we see VAE and GAN-based models in the lead, with \submKaech, \submErnst, and \submLiu being the fastest. The symmetric flow architecture of \submFavaro is also fast in generation, but only on a GPU and for larger batch sizes. The diffusion models and MAF-based normalizing flows are the slowest submissions. Distillation of models clearly improves the generation speed in all cases, as expected. Generating showers in batches improves the generation speed in all cases, but also leads to out of memory errors in 11 out of 17 cases on the GPU. Given that some of the generation times (especially for smaller batch sizes) get considerably large, we restricted the number of samples used to time the models to fewer than 100\,000 events, see the details in~\tref{tab:ds2.timing.CPU} and~\tref{tab:ds2.timing.GPU}.  

\FloatBarrier

\subsection{\texorpdfstring{Dataset 3, Electrons (\dsIII)}{Dataset 3, Electrons}}
\label{sec:results_ds3}
\begin{figure}
\centering
\includegraphics[width=\textwidth]{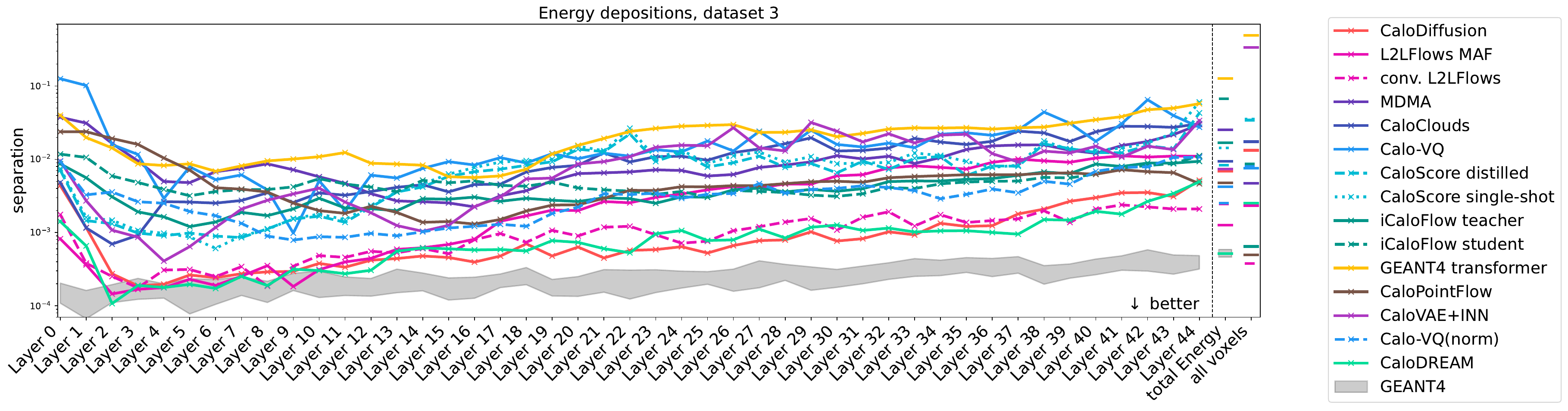}
\caption{Separation power of energy depositions.}
\label{fig:ds3-depositions}
\end{figure}

Also for dataset 3, the minimal energy that can be read out is given by 15.15 keV. We again start our evaluation with the separation power of high-level observables, in particular with the energy depositions per layer and in total in~\fref{fig:ds3-depositions}. We notice that many models show the best performance around layers 3 -- 10, and separation powers then grow towards the end of the detector. \submPalacios, \submAmram, \submBussConv, and \submBussMAF even reach the \geant reference band in this region. Further, \submPalacios matches the total energy deposition very well. 

\begin{figure}
\centering
\includegraphics[width=\textwidth]{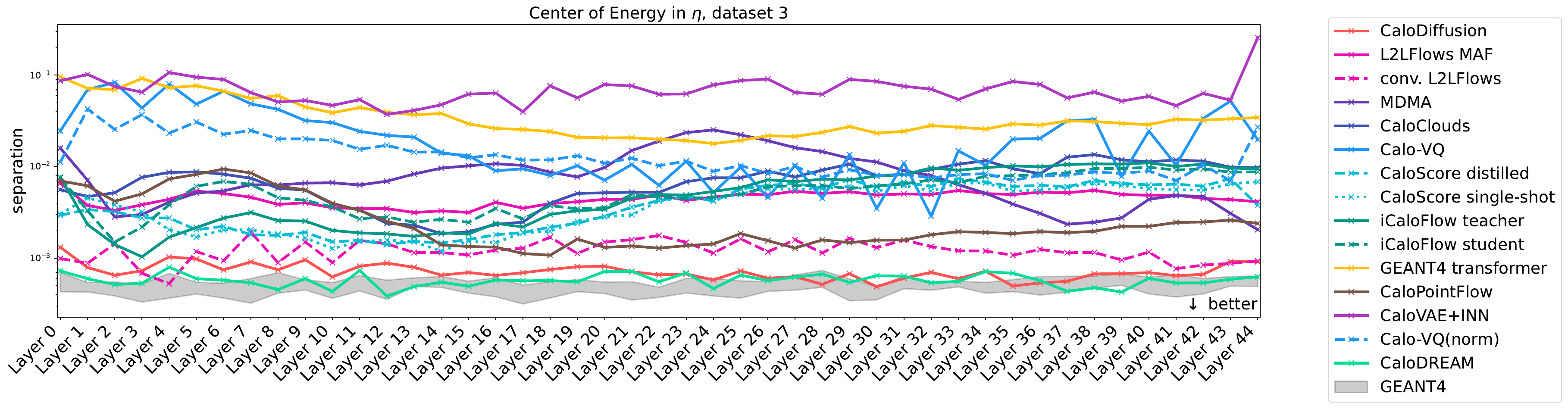}
\caption{Separation power of centers of energy in $\eta$ direction.}
\label{fig:ds3-CE-eta}
\end{figure}

\begin{figure}
\centering
\includegraphics[width=\textwidth]{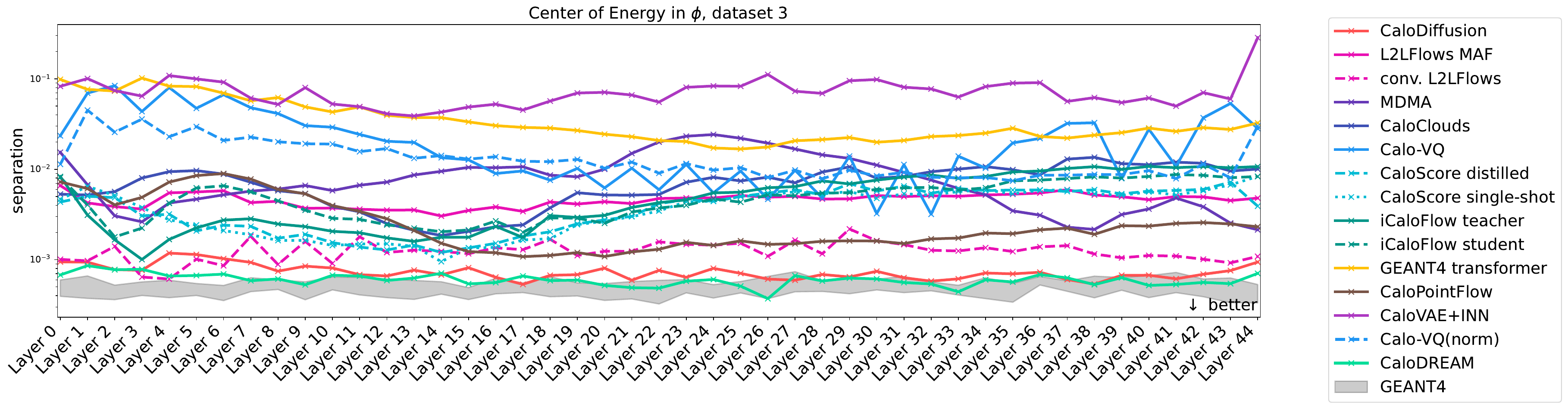}
\caption{Separation power of centers of energy in $\phi$ direction.}
\label{fig:ds3-CE-phi}
\end{figure}

\begin{figure}
\centering
\includegraphics[width=\textwidth]{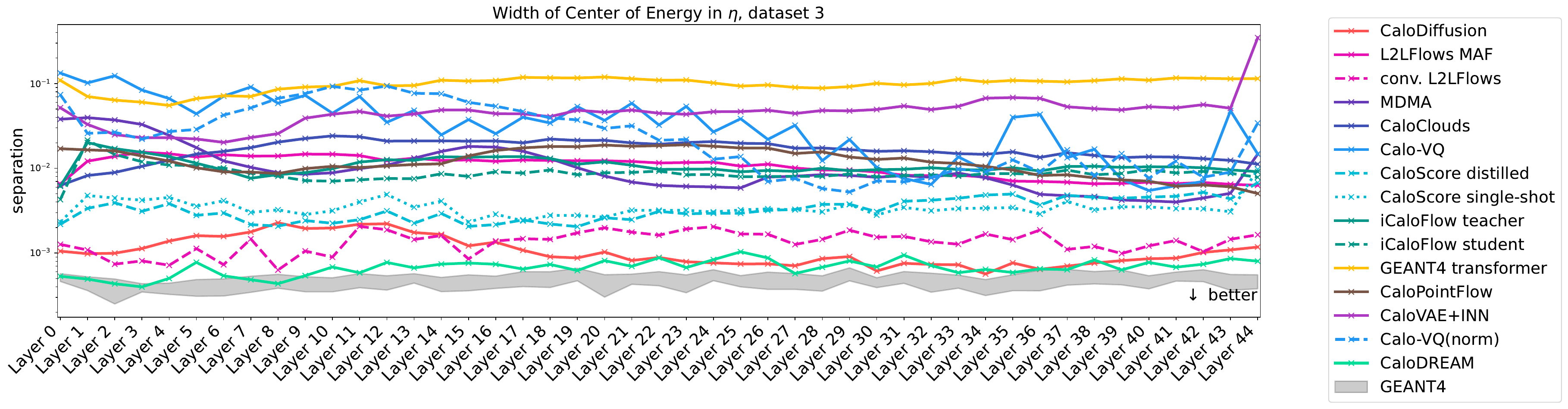}
\caption{Separation power of widths of centers of energy in $\eta$ direction.}
\label{fig:ds3-width-CE-eta}
\end{figure}

\begin{figure}
\centering
\includegraphics[width=\textwidth]{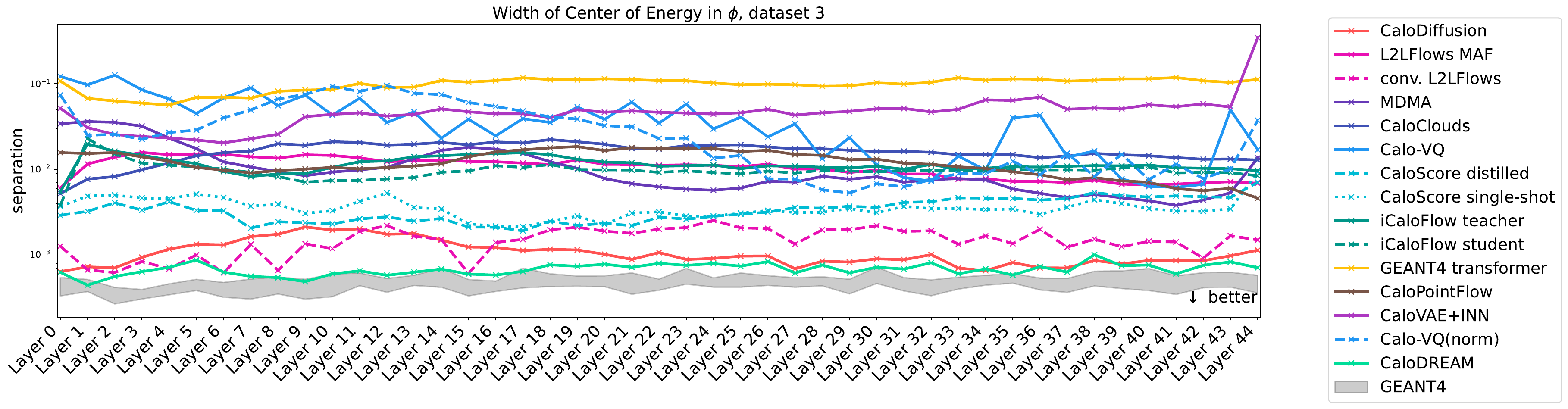}
\caption{Separation power of widths of centers of energy in $\phi$ direction.}
\label{fig:ds3-width-CE-phi}
\end{figure}

\begin{figure}
\centering
\includegraphics[width=\textwidth]{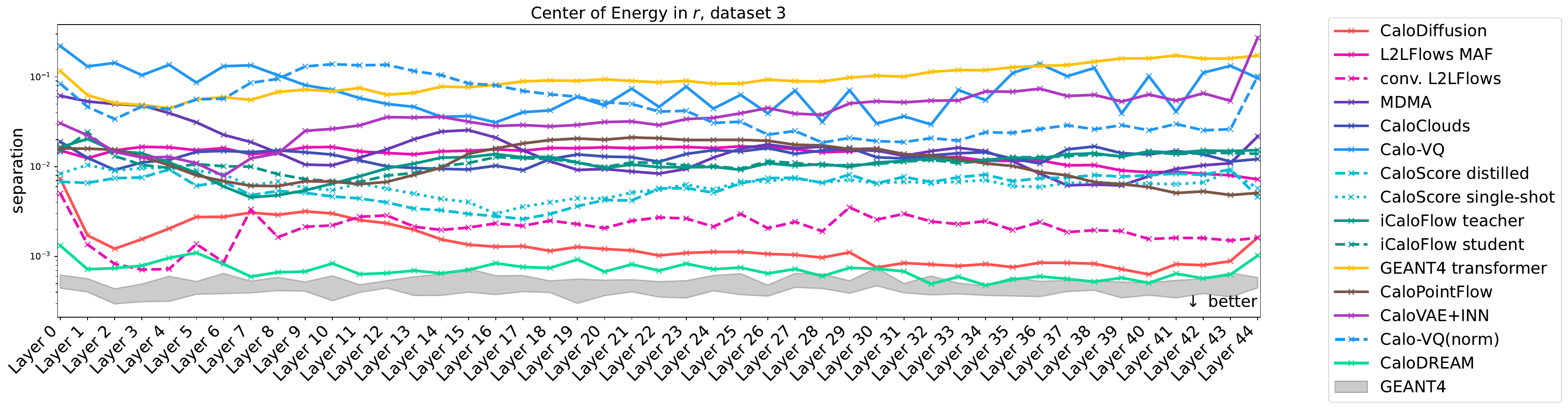}
\caption{Separation power of centers of energy in radial direction.}
\label{fig:ds3-CE-r}
\end{figure}

\begin{figure}
\centering
\includegraphics[width=\textwidth]{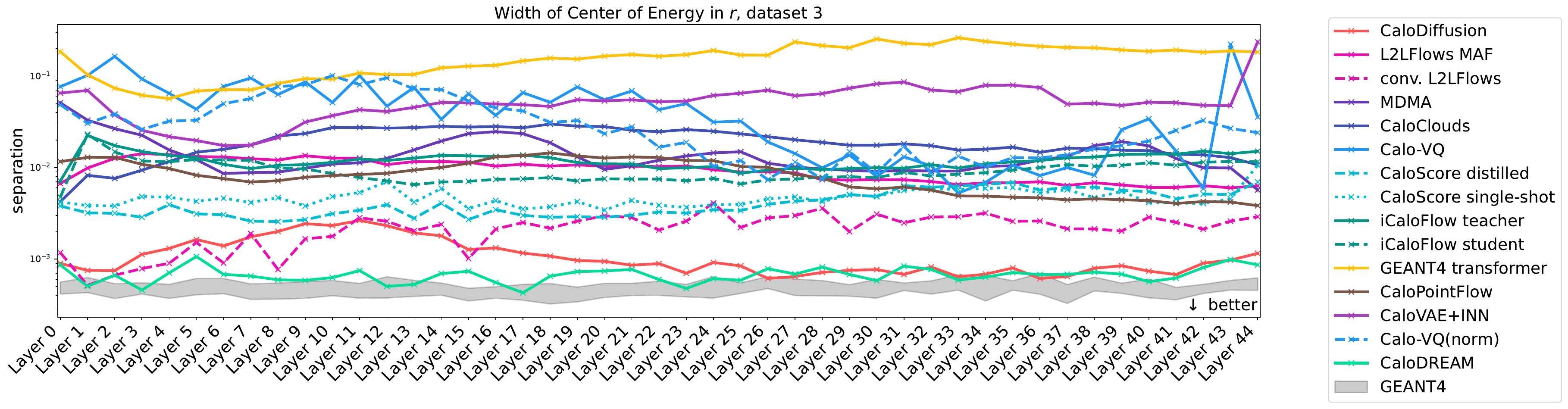}
\caption{Separation power of widths of centers of energy in radial direction.}
\label{fig:ds3-width-CE-r}
\end{figure}

\begin{figure}
\centering
\includegraphics[width=\textwidth]{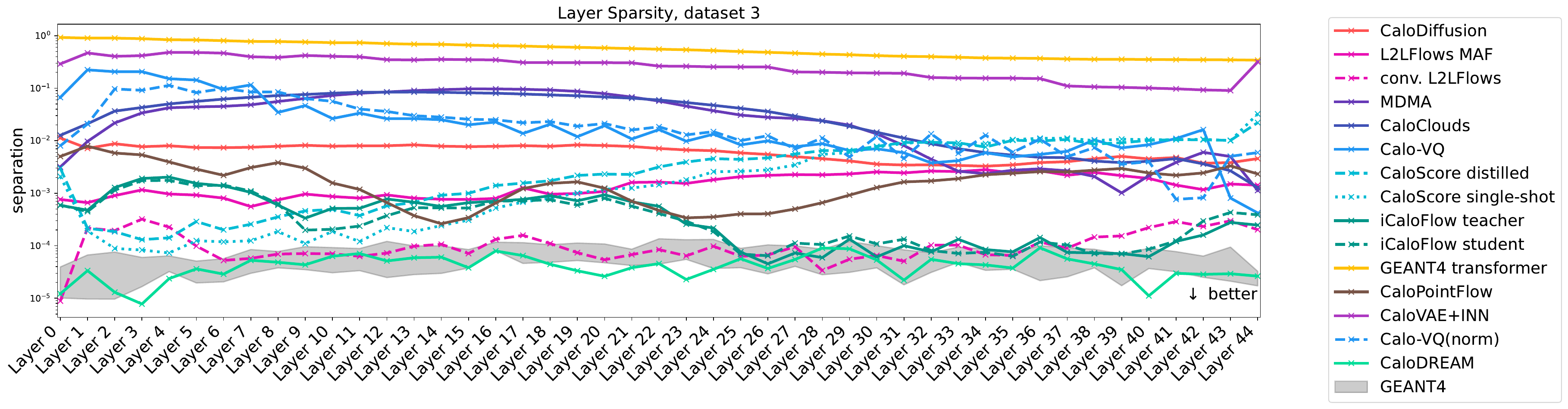}
\caption{Separation power of the sparsity.}
\label{fig:ds3-sparsity}
\end{figure}

\begin{figure}
\centering
\hfill\includegraphics[width=0.4\textwidth]{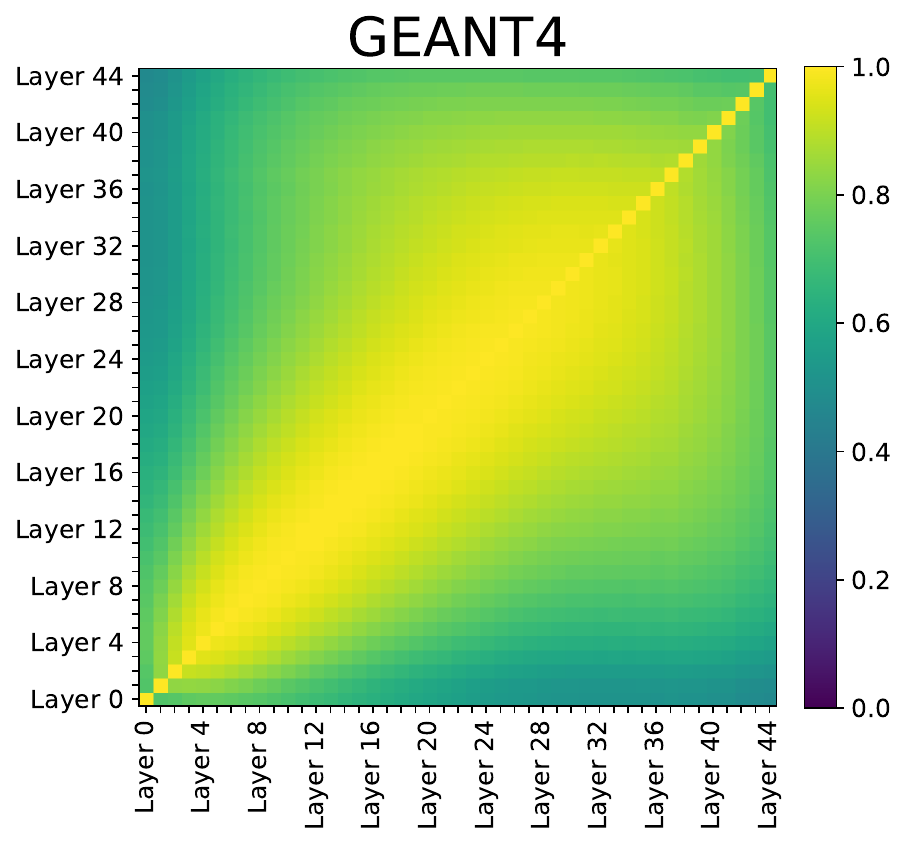}\hfill $ $
\\
\includegraphics[width=0.2\textwidth]{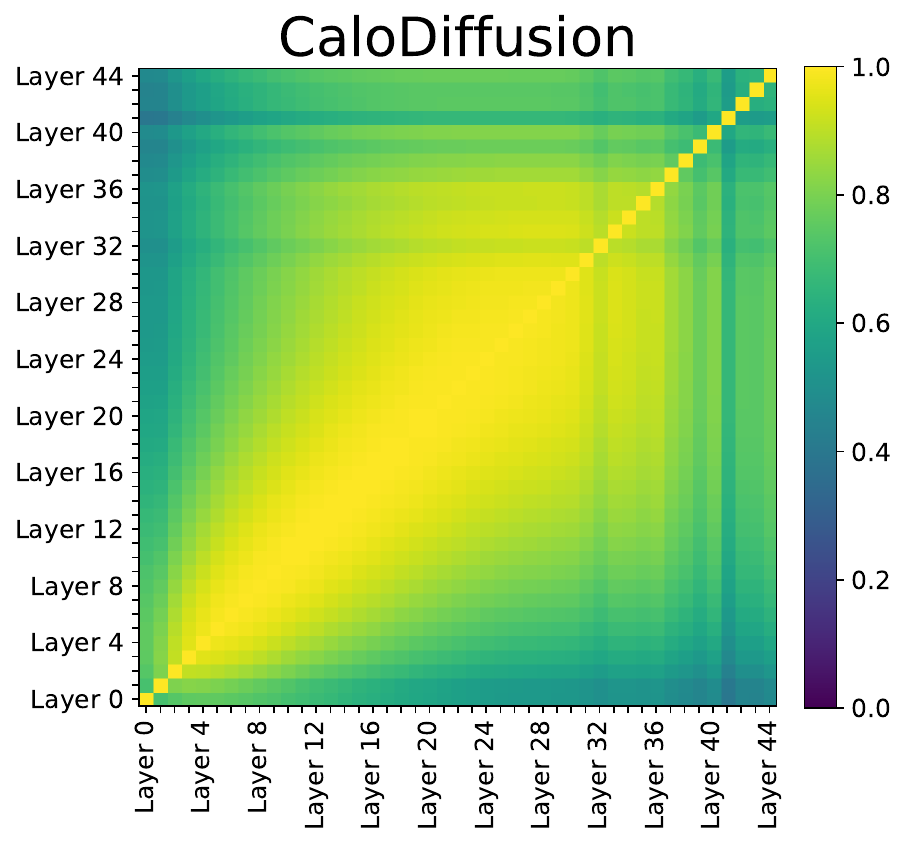}\hfill
\includegraphics[width=0.2\textwidth]{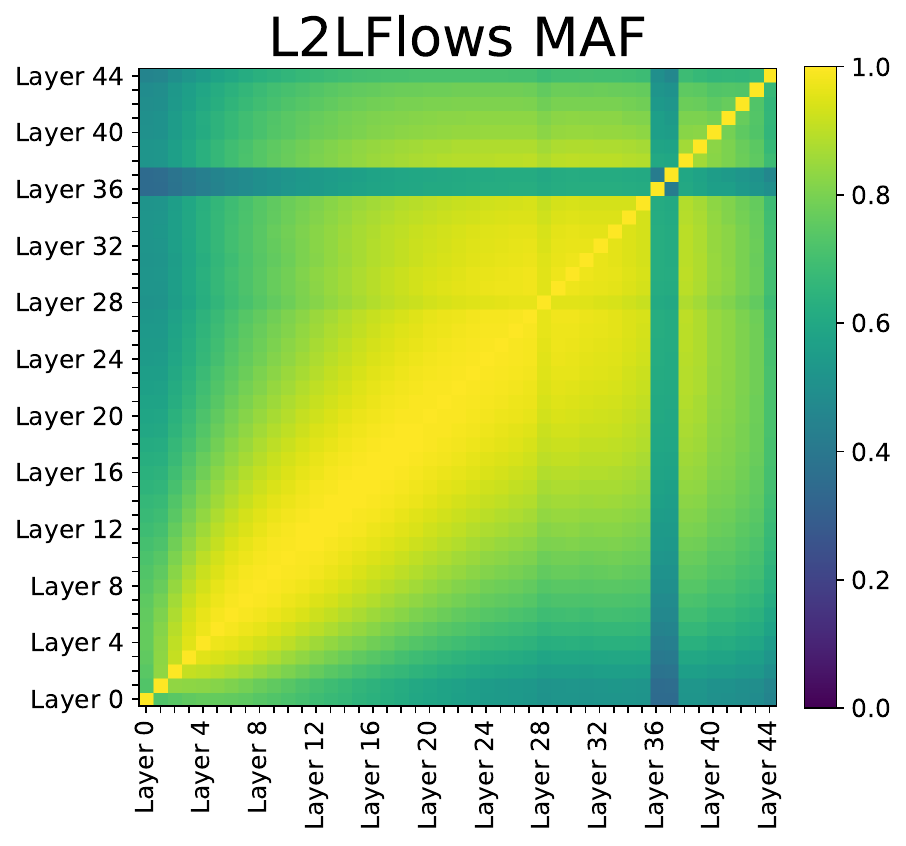}\hfill
\includegraphics[width=0.2\textwidth]{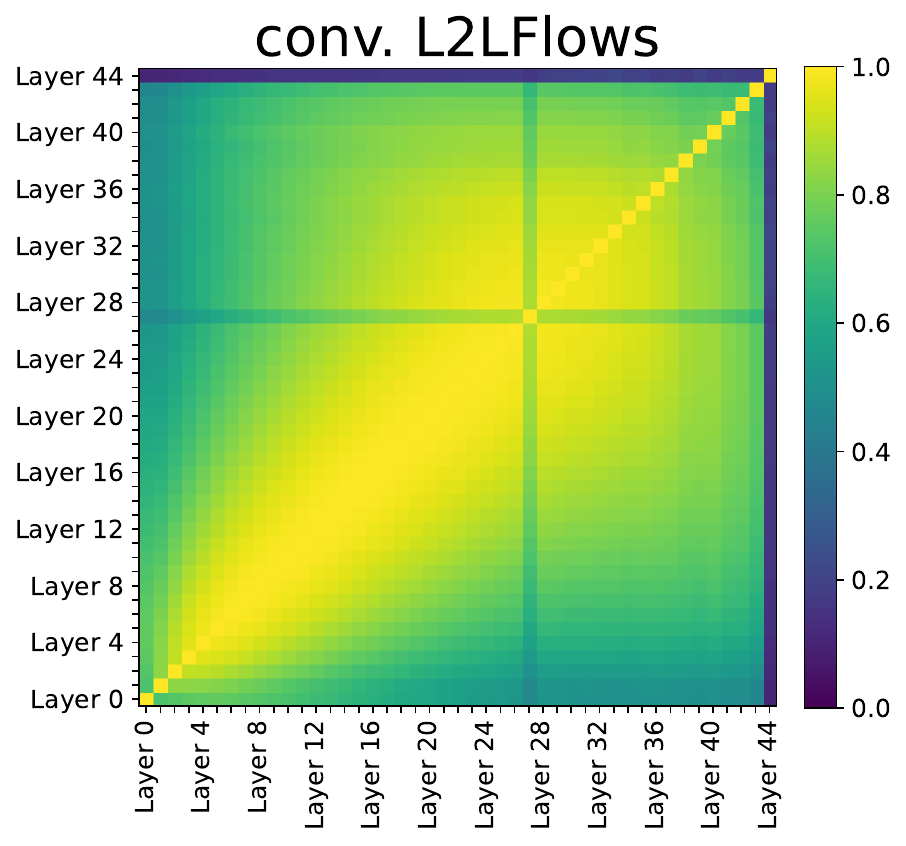}\hfill
\includegraphics[width=0.2\textwidth]{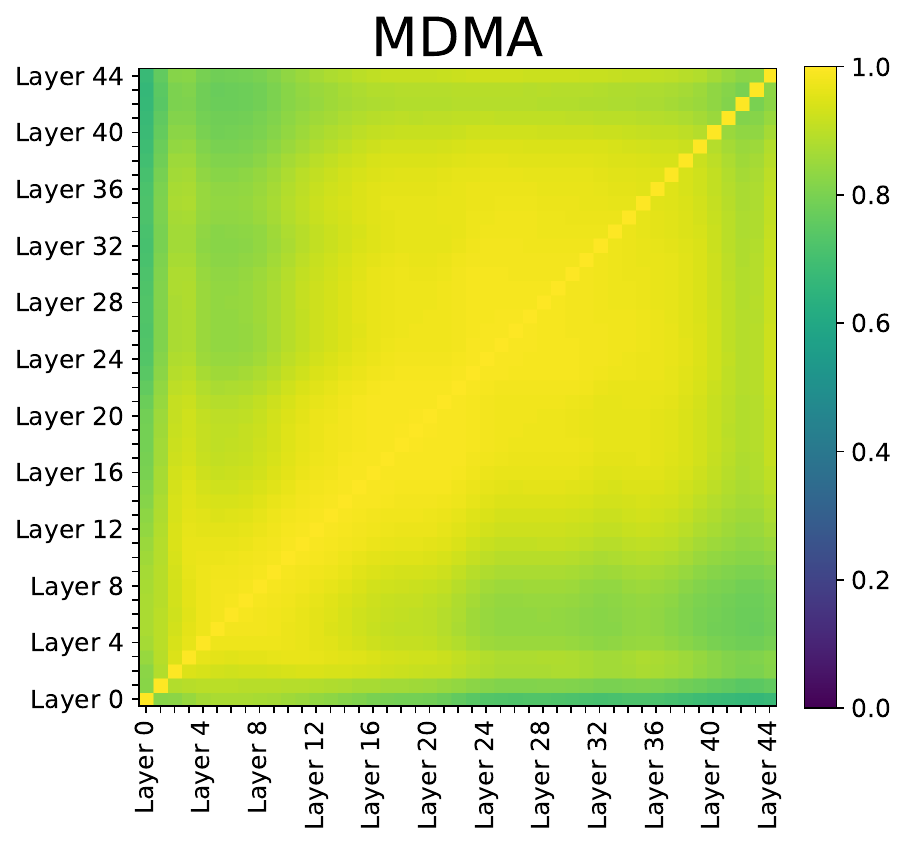}\hfill
\includegraphics[width=0.2\textwidth]{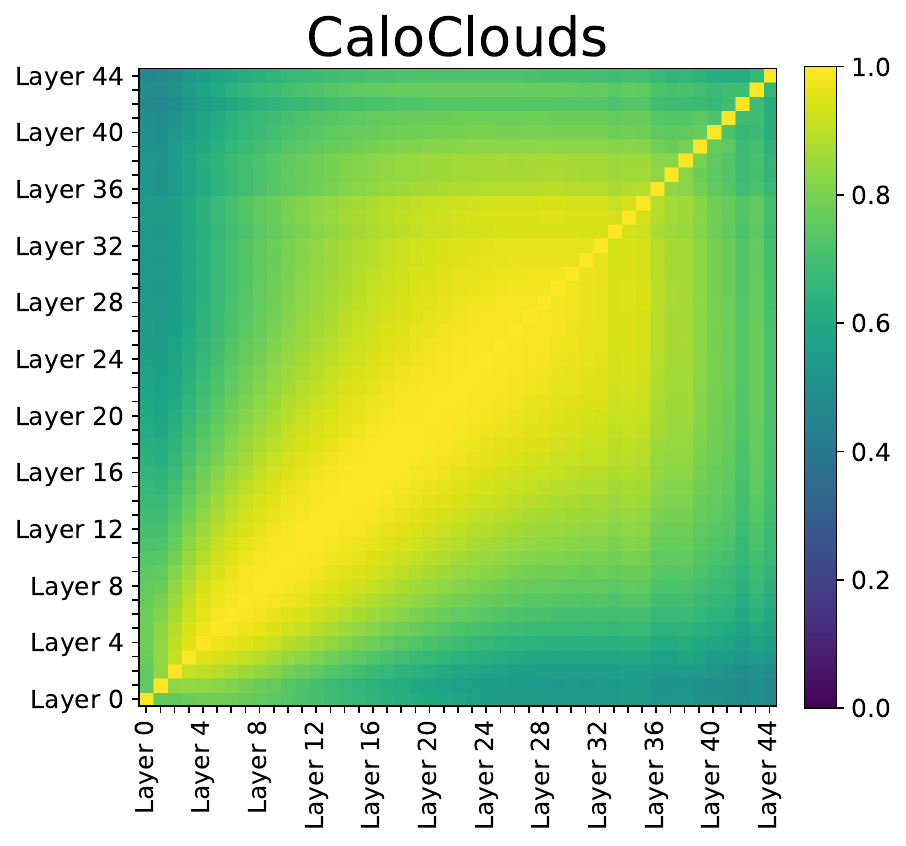}\\
\includegraphics[width=0.2\textwidth]{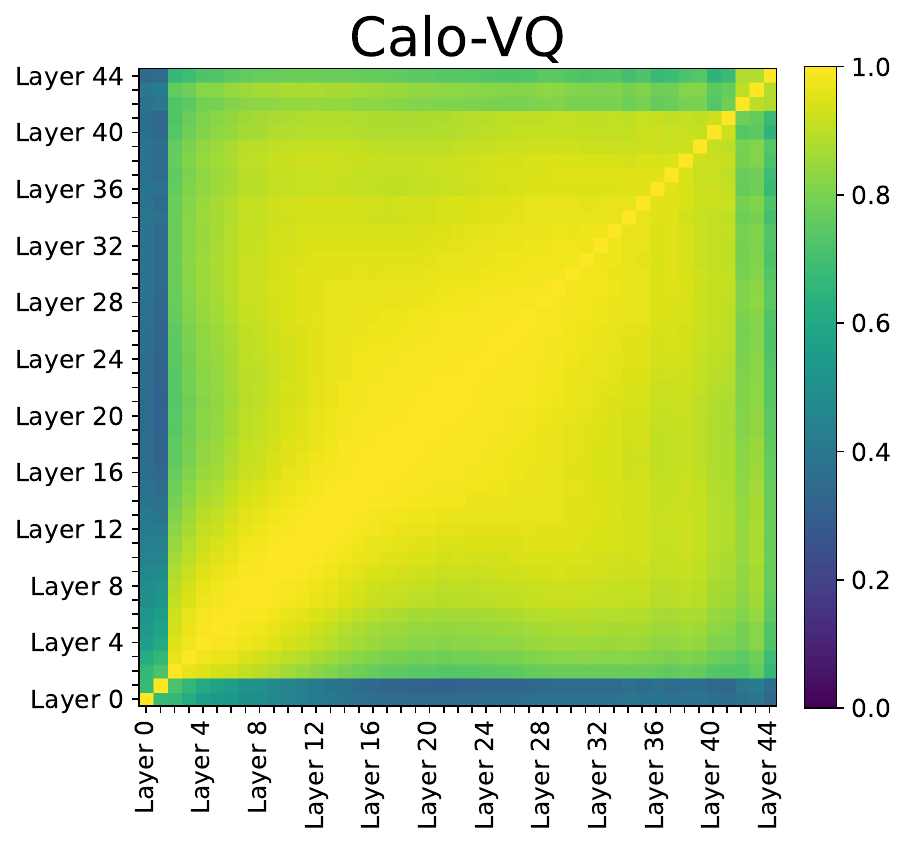}\hfill
\includegraphics[width=0.2\textwidth]{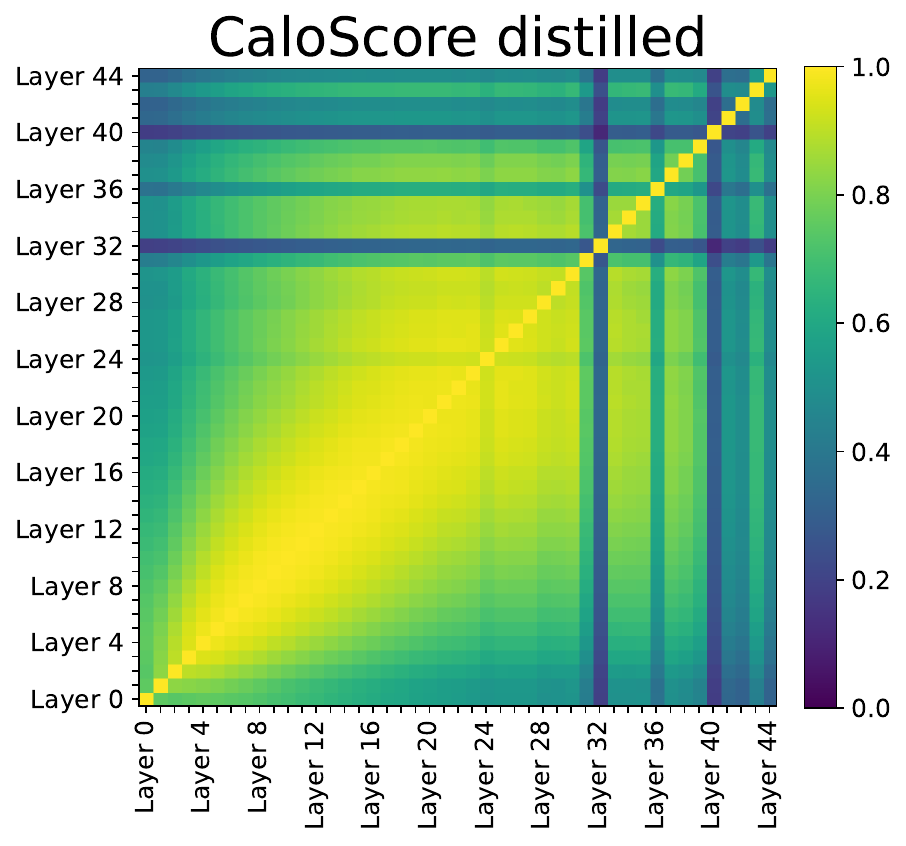}\hfill
\includegraphics[width=0.2\textwidth]{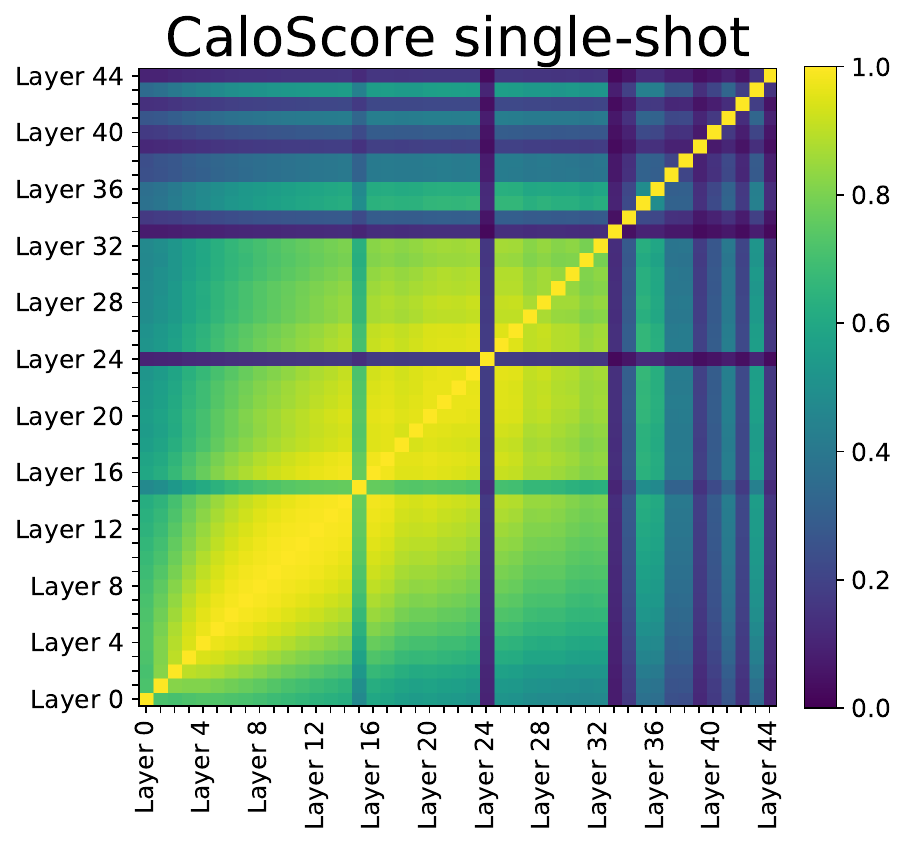}\hfill
\includegraphics[width=0.2\textwidth]{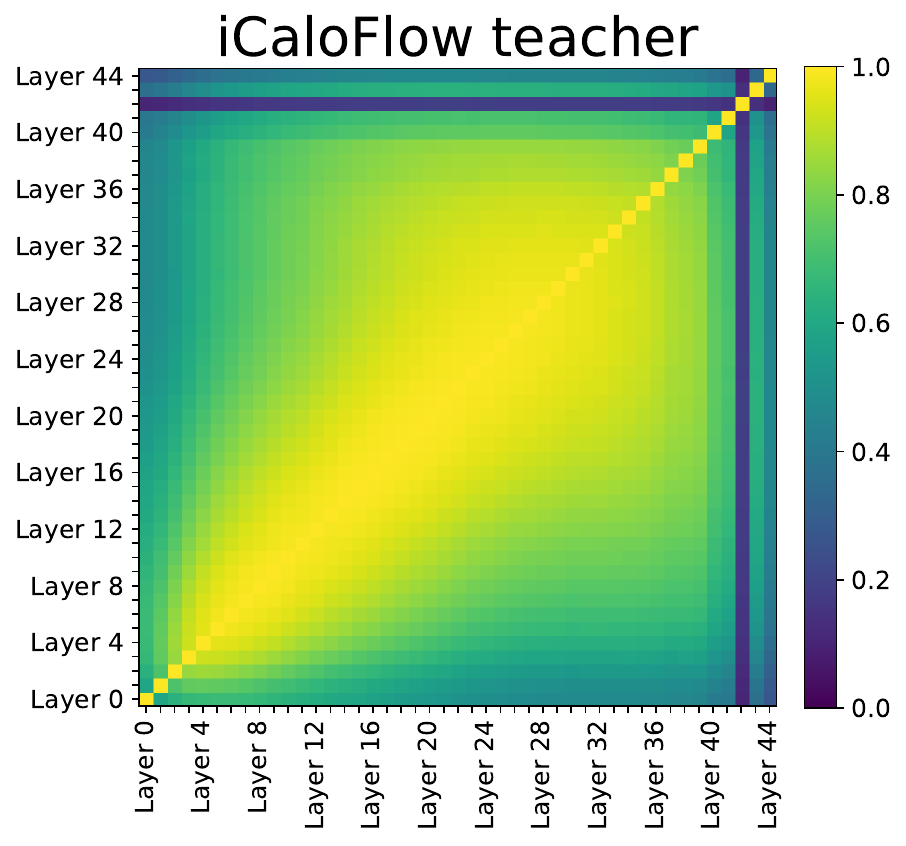}\hfill
\includegraphics[width=0.2\textwidth]{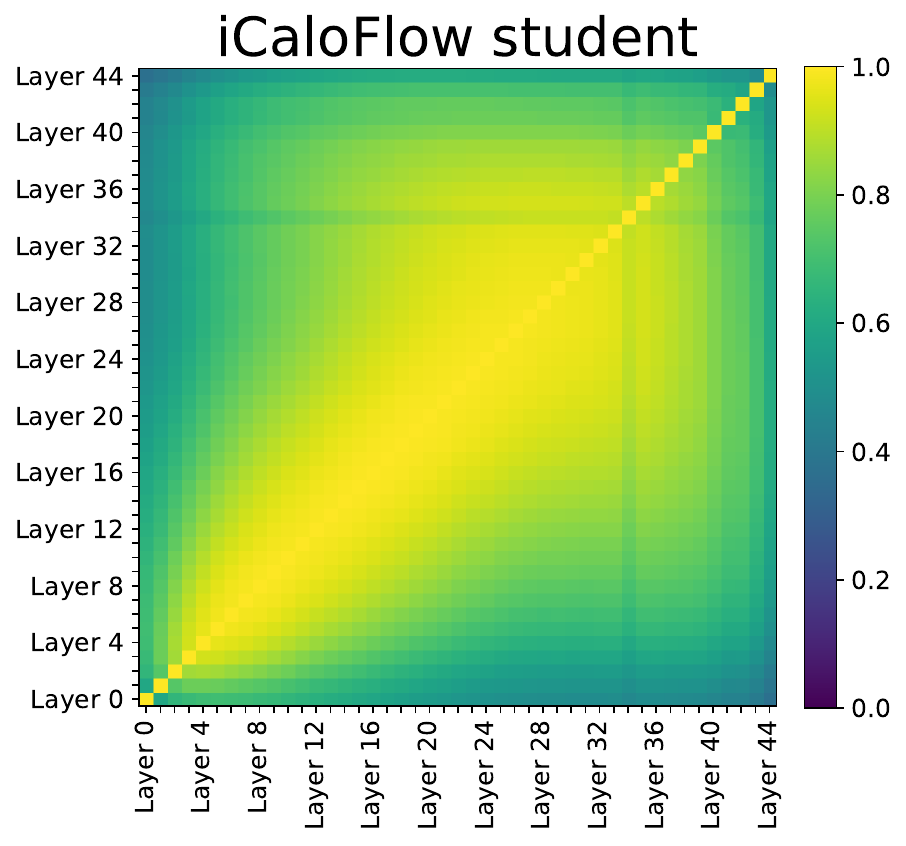}\\
\includegraphics[width=0.2\textwidth]{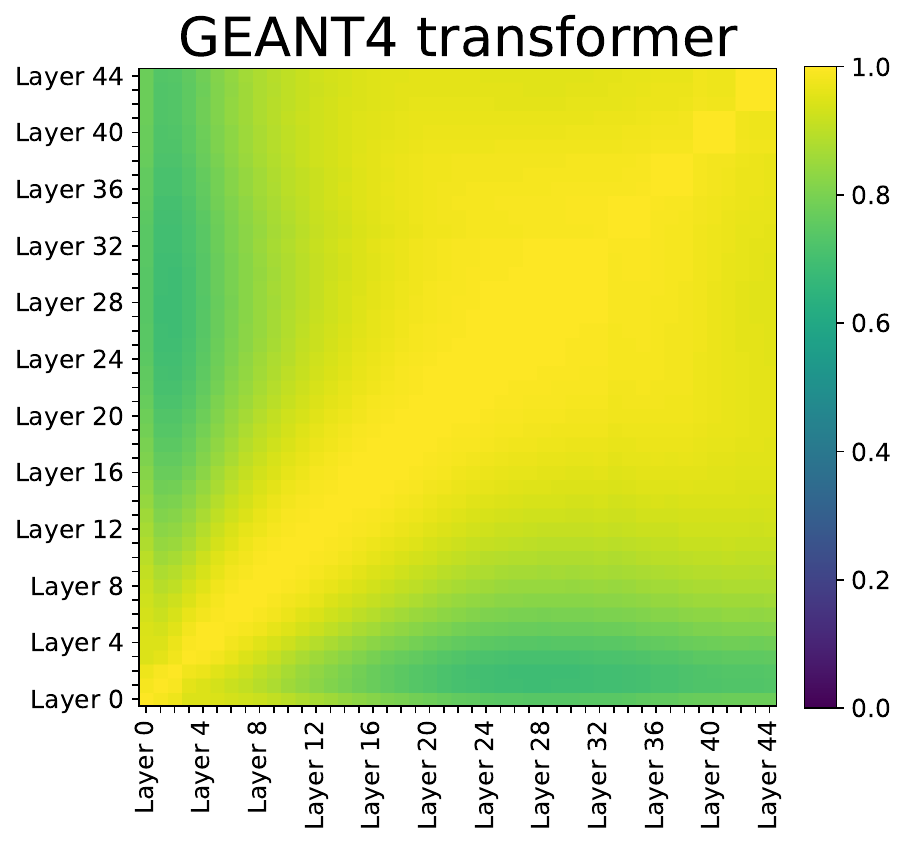}\hfill
\includegraphics[width=0.2\textwidth]{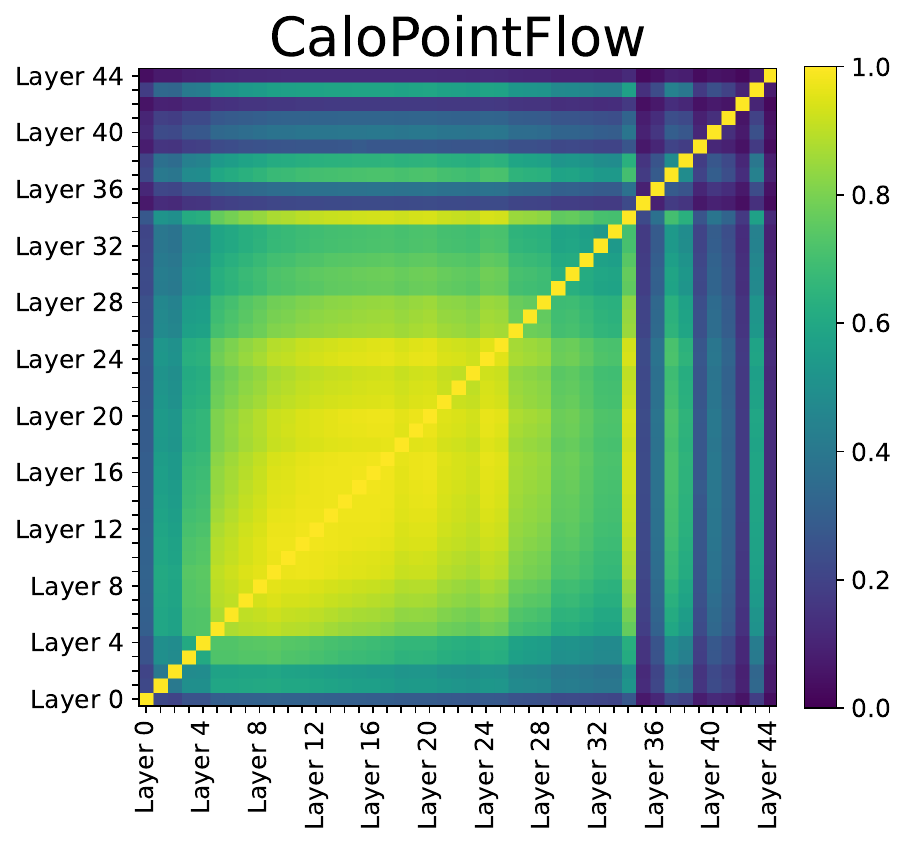}\hfill
\includegraphics[width=0.2\textwidth]{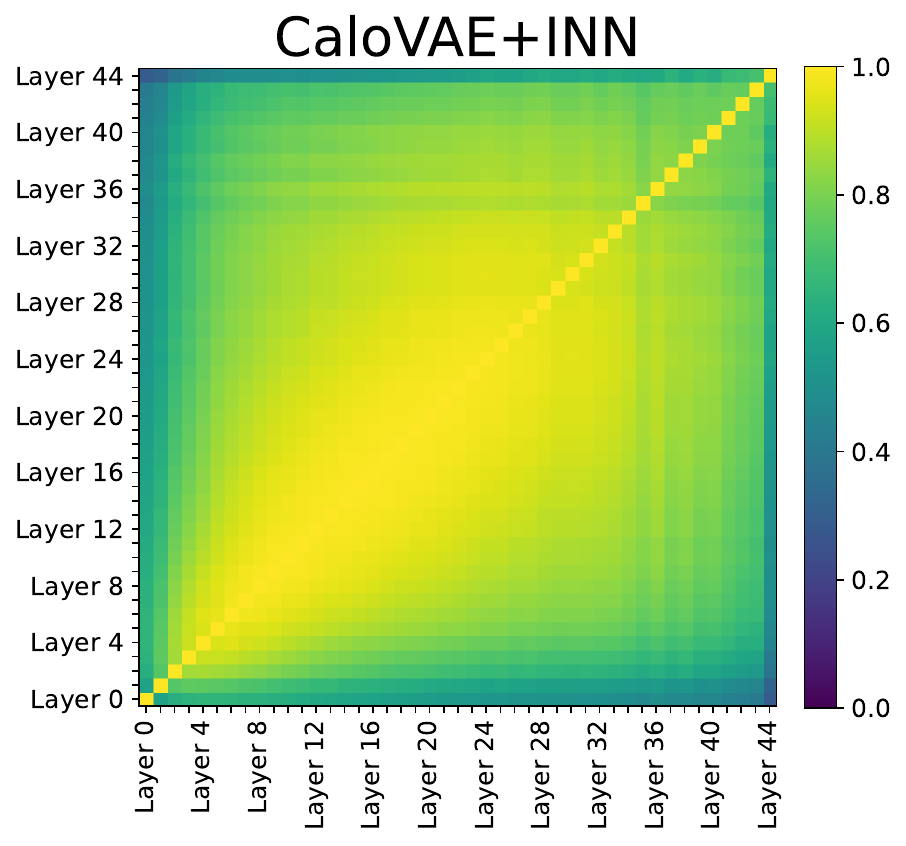}\hfill
\includegraphics[width=0.2\textwidth]{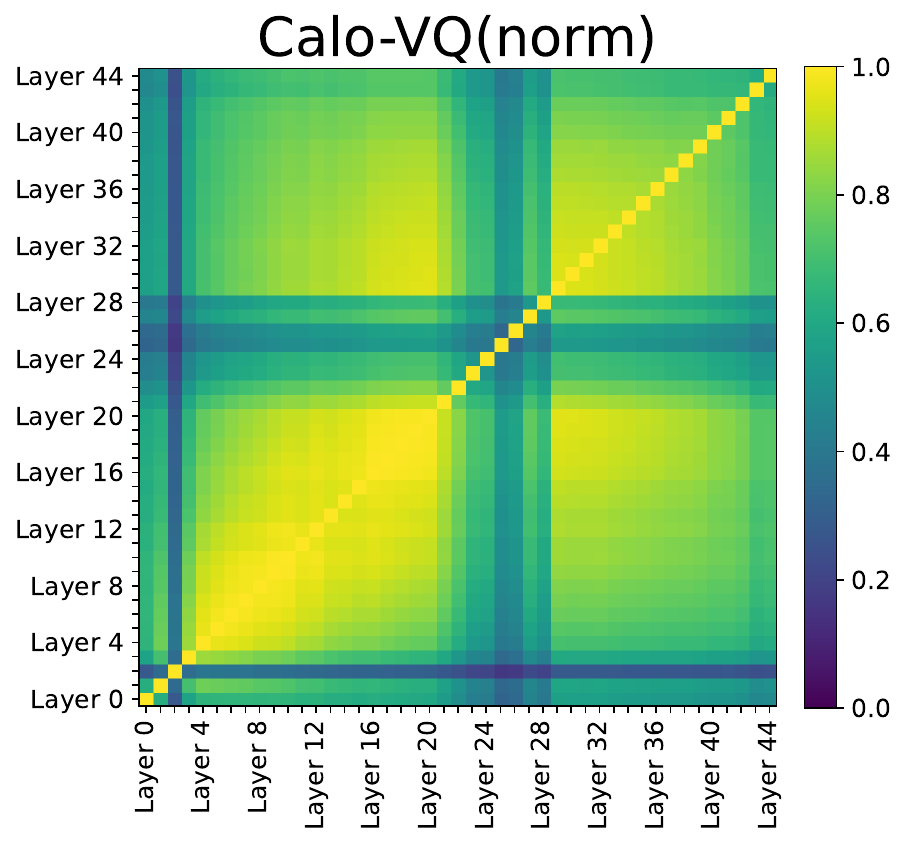}\hfill
\includegraphics[width=0.2\textwidth]{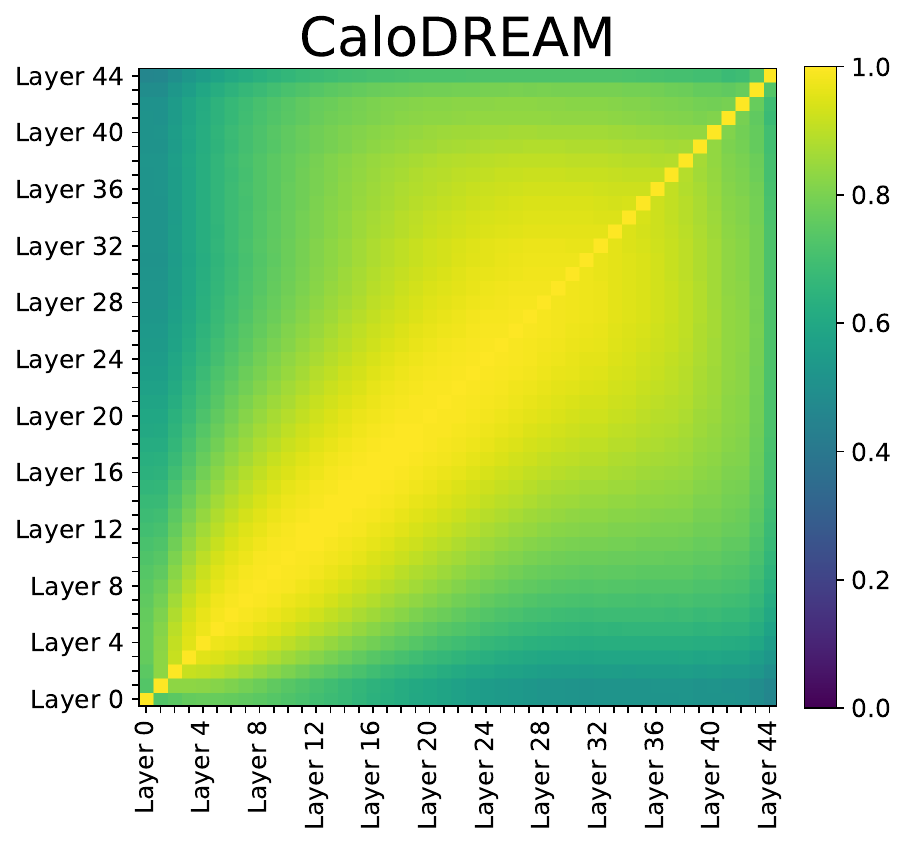}\\
\caption{Pearson correlation coefficients of layer energies in \dsIII.}
\label{fig:ds3-corr}
\end{figure}

Moving on to centers of energy in $\eta$ and $\phi$ in~\fref{fig:ds3-CE-eta} and~\fref{fig:ds3-CE-phi}, we see again two very similar distributions in both of these directions, indicating that the rotational symmetry was learned well by all submissions. At the level of the \geant reference, we see \submAmram and \submPalacios, both with rather smooth separation powers from one calorimeter layer to the next. Slightly worse, we have \submSchnake, which was also smooth, and \submBussConv, which shows some ups and downs from layer to layer. Among the rest, we notice a similar, but stronger up and down pattern for \submLiu (which is however not present in the improved version \submLiuNorm). \submKaech shows the largest spread between the lowest and largest separation power. The VAE-based submissions \submErnst and \submSalamaniTrans have the most problems reproducing the \geant data. 

Most of the statements of the centers of energy also apply to their widths in~\fref{fig:ds3-width-CE-eta} and~\fref{fig:ds3-width-CE-phi}. \submPalacios has the smallest separation powers, close to \geant. \submAmram comes second, but with a larger gap for earlier layers, where \submBussConv shows a better match to the reference. In between these submissions and the bulk, we see \submMikuniDist and \submMikuniSingle. \submLiu again has an oscillating behavior over the entire size of the detector, and \submSalamaniTrans and \submErnst have the largest separation powers. The order of submissions is also preserved when looking at the centers of energy in $r$ in~\fref{fig:ds3-CE-r} and their widths in~\fref{fig:ds3-width-CE-r}. 

Only for the sparsities in~\fref{fig:ds3-sparsity} we see a difference. \submPalacios still shows the best performance, again at the level of \geant, but \submAmram has a much harder time reproducing the correct distribution. Instead, \submMikuniDist and \submMikuniSingle (for early layers), \icalo (for later layers), and \submBussConv have small separation powers and get close to \submPalacios. Also in this observable, the VAE-based submissions \submSalamaniTrans and \submErnst show the largest separation powers. In fact, overall we see the separation powers ranging over five orders of magnitude between best and worst submission. 

In~\fref{fig:ds3-corr} we look at the Pearson correlation coefficients of the layer energies. Also in this case we do not reproduce all findings of~\cite{Ahmad:2024dql}, again indicating that some of the observed patterns might fluctuate from training to training. Similar to what we have observed for the other datasets, we see three different groups of correlation patterns. The first one reproduces the \geant shape quite well and consists of \submKorol, \submPangIS, and \submPalacios. The submissions \submKaech, \submSalamaniTrans in the second group are also very smooth and only have small regions that appear slightly brighter than the reference. The third group consists of submissions that have single layers that do not have the correct correlation, indicated by stripes in the figures. While some are very faint and just a few (like for \submAmram, \submLiu, \submErnst, or \submPangIT), others have more and a stronger contrast (like \submMikuniDist, \submMikuniSingle, \submSchnake, or \submLiuNorm). We also find again the intriguing pattern that the distillation of \submMikuni made the correlations worse, but the distillation of \icalo improved the correlations.  

\begin{figure}
\centering
\includegraphics[width=\textwidth]{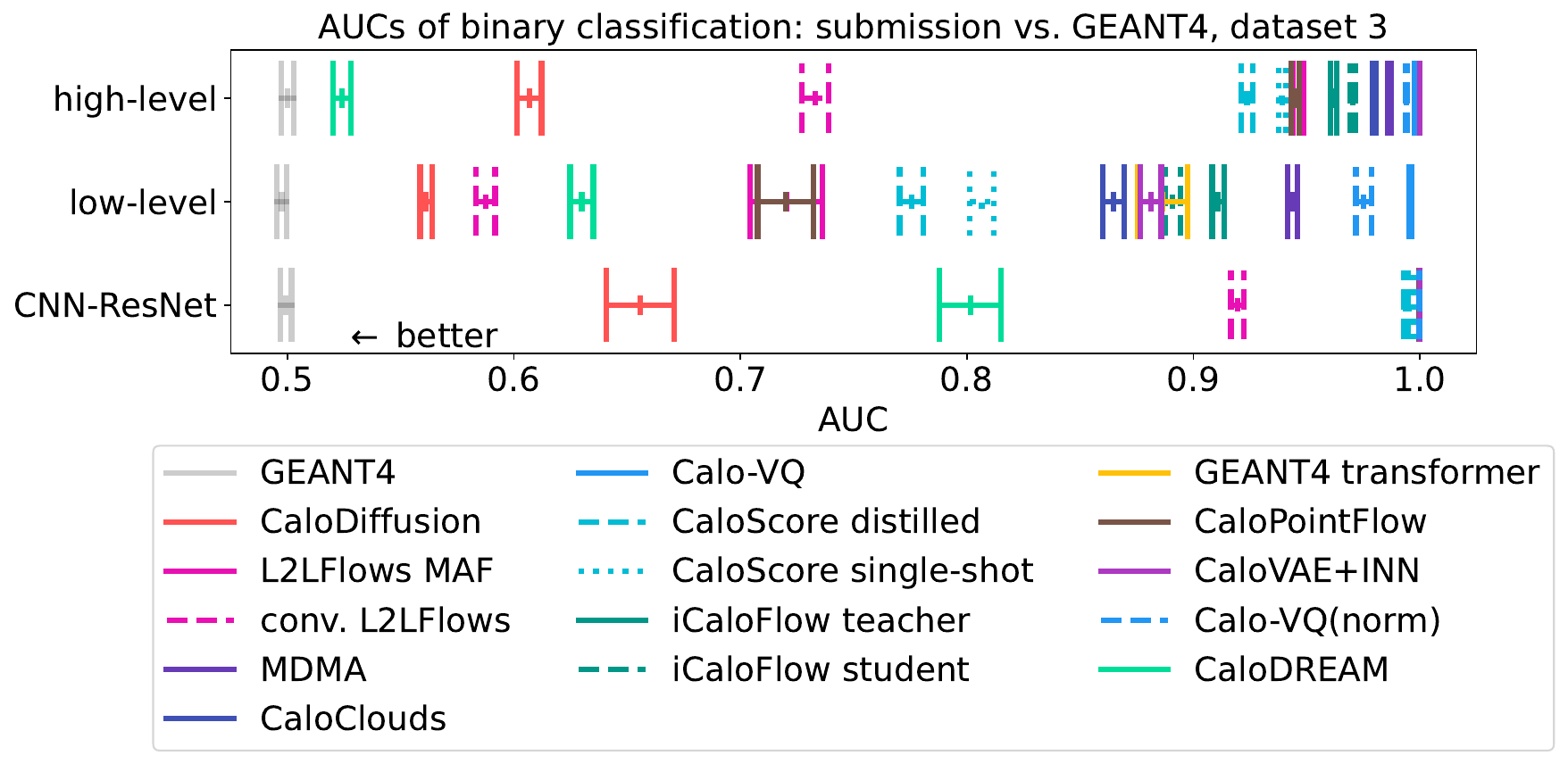}
\caption{Low-level and high-level AUCs for evaluating \geant\ vs.~submission of \dsIII, averaged over 10 independent evaluation runs. For the precise numbers, see \Tref{tab:ds3.aucs}.}
\label{fig:ds3.aucs}
\end{figure}

\begin{figure}
\centering
\includegraphics[width=\textwidth]{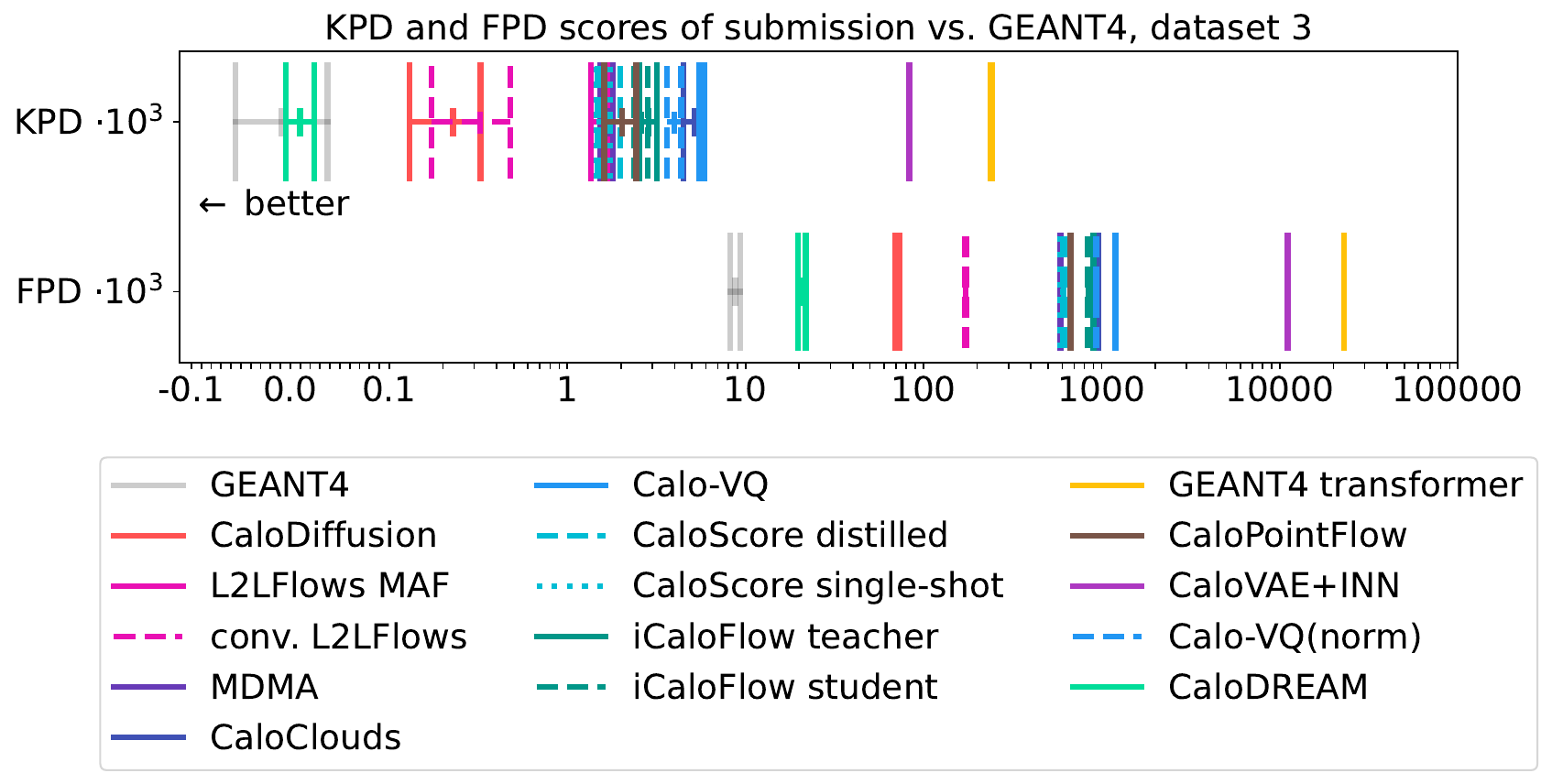}
\caption{KPD and FPD for evaluating \geant\ vs.~submission of \dsIII. For the precise numbers, see \Tref{tab:ds3.kpd}.}
\label{fig:ds3.kpd}
\end{figure}
The AUCs of the binary classifiers in~\fref{fig:ds3.aucs} (and \tref{tab:ds3.aucs}) corroborate the results of the separation power. For the high-level features, the best three models --- \submPalacios, \submAmram, and \submBussConv --- are clearly separated from the other submissions. For low-level features, these three submissions still have the best performance, independent of the classifier architecture, but the ordering changed with \submAmram having the best AUC. While the DNN indicates differences between submissions, yielding a spread between all the AUCs, the CNN-ResNet architecture essentially identifies the three best submissions --- \submAmram, \submPalacios, and \submBussConv --- and gives all other submissions an AUC of 1. 

A similar ordering, at least in terms of the top three models, is also seen in the KPD and FPD scores in \fref{fig:ds3.kpd} (with details in \tref{tab:ds3.kpd}). Now \submPalacios is closest to the \geant reference. Also for these scores (especially for the KPD), the bulk of all other submissions is very close to each other with scores overlapping within uncertainties.  

\begin{figure}
\centering
\includegraphics[width=\textwidth]{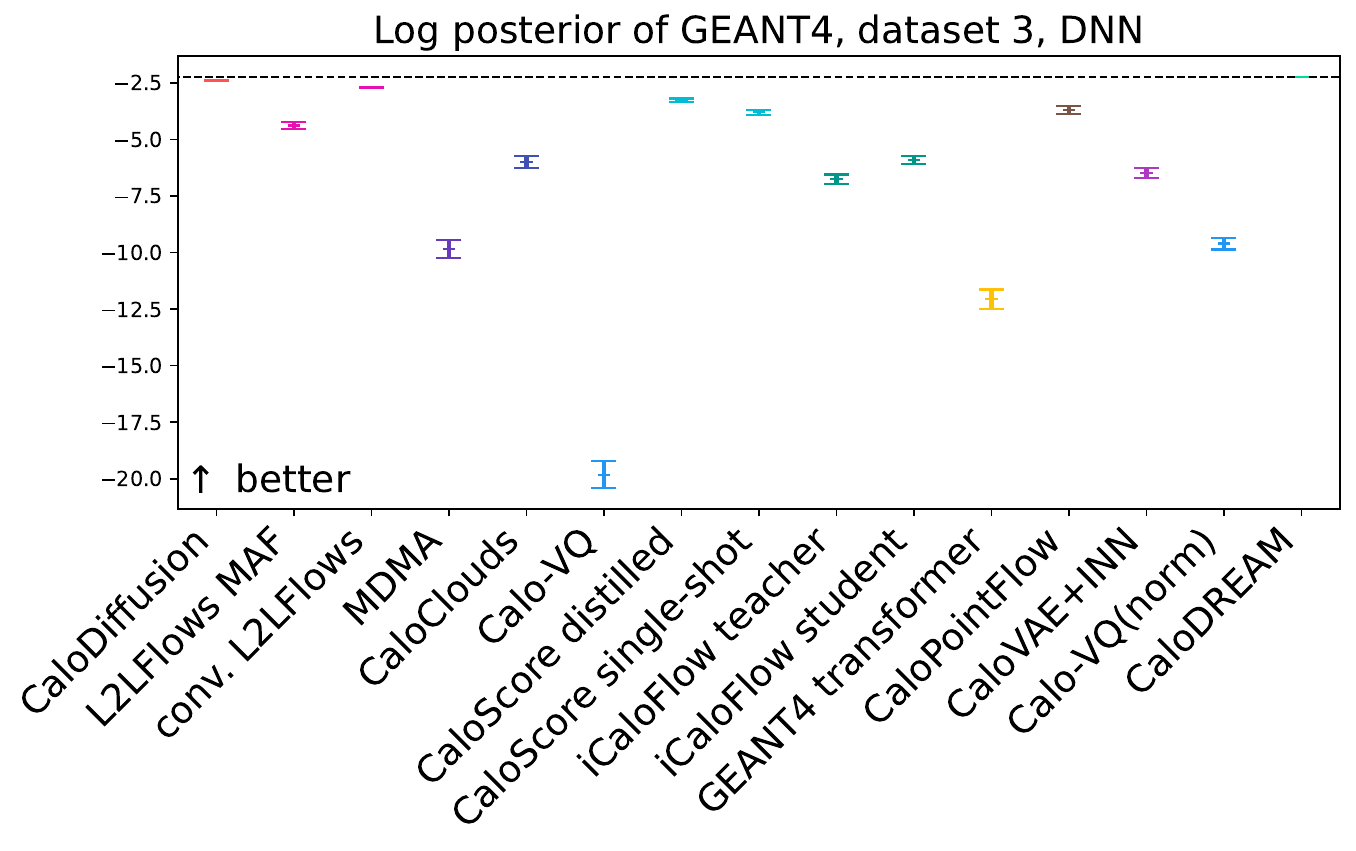}
\caption{Log-posterior scores for \dsIII \geant test data, averaged over 10 independent DNN classifier trainings. For the precise numbers, see \Tref{tab:ds3.multi.dnn}.}
\label{fig:ds3.multi.dnn}
\end{figure}

\begin{figure}
\centering
\includegraphics[width=\textwidth]{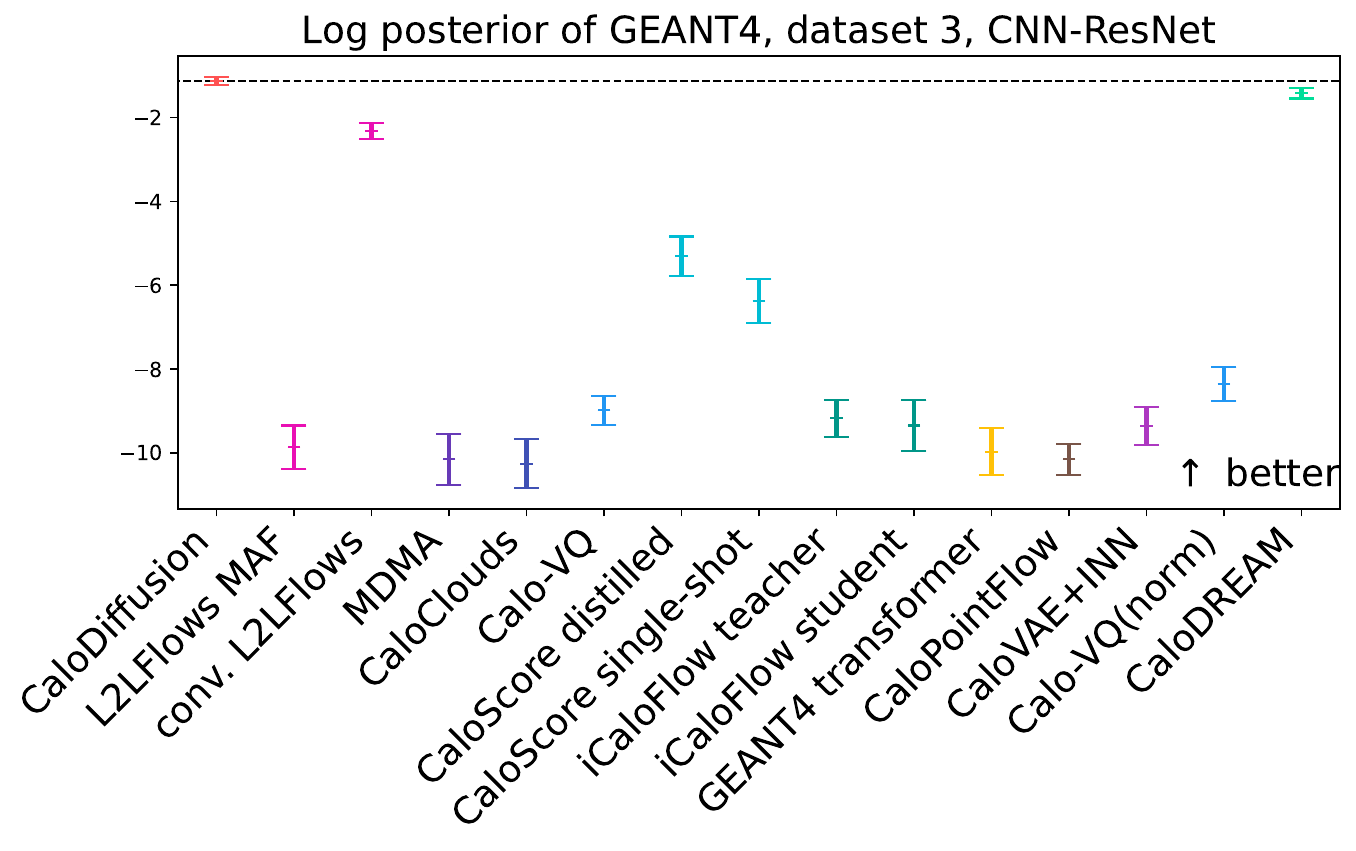}
\caption{Log-posterior scores for \dsIII \geant test data, averaged over 10 independent CNN ResNet classifier trainings. For the precise numbers, see \Tref{tab:ds3.multi.resnet}.}
\label{fig:ds3.multi.resnet}
\end{figure}

The multiclass classifier metric, shown in \fref{fig:ds3.multi.dnn}, \fref{fig:ds3.multi.resnet}, \tref{tab:ds3.multi.dnn}, and \tref{tab:ds3.multi.resnet} is consistent with the binary AUCs shown before. \submPalacios and \submAmram have the highest log-posterior, and \submBussConv comes in third before there is a gap to the remaining submissions. Again, we see the CNN-ResNet being more powerful, giving low scores to almost all submissions when compared to \geant. As with \dsII, we also observe here that the spread in log-posterior between the best and worst model is smaller in the CNN-ResNet compared to the DNN architecture. However, both of the considered architectures have well-trained classifiers, as can be seen in \fref{fig:consistency.ds3} and \fref{fig:consistency.ds3.ResNet}. The size of the error bars, coming from ten independent retrainings of the classifier, seems to be correlated with the central value of the log-posterior, with smaller (worse) log-posterior scores having larger error bars. 

\begin{figure}
\centering
\includegraphics[width=\textwidth]{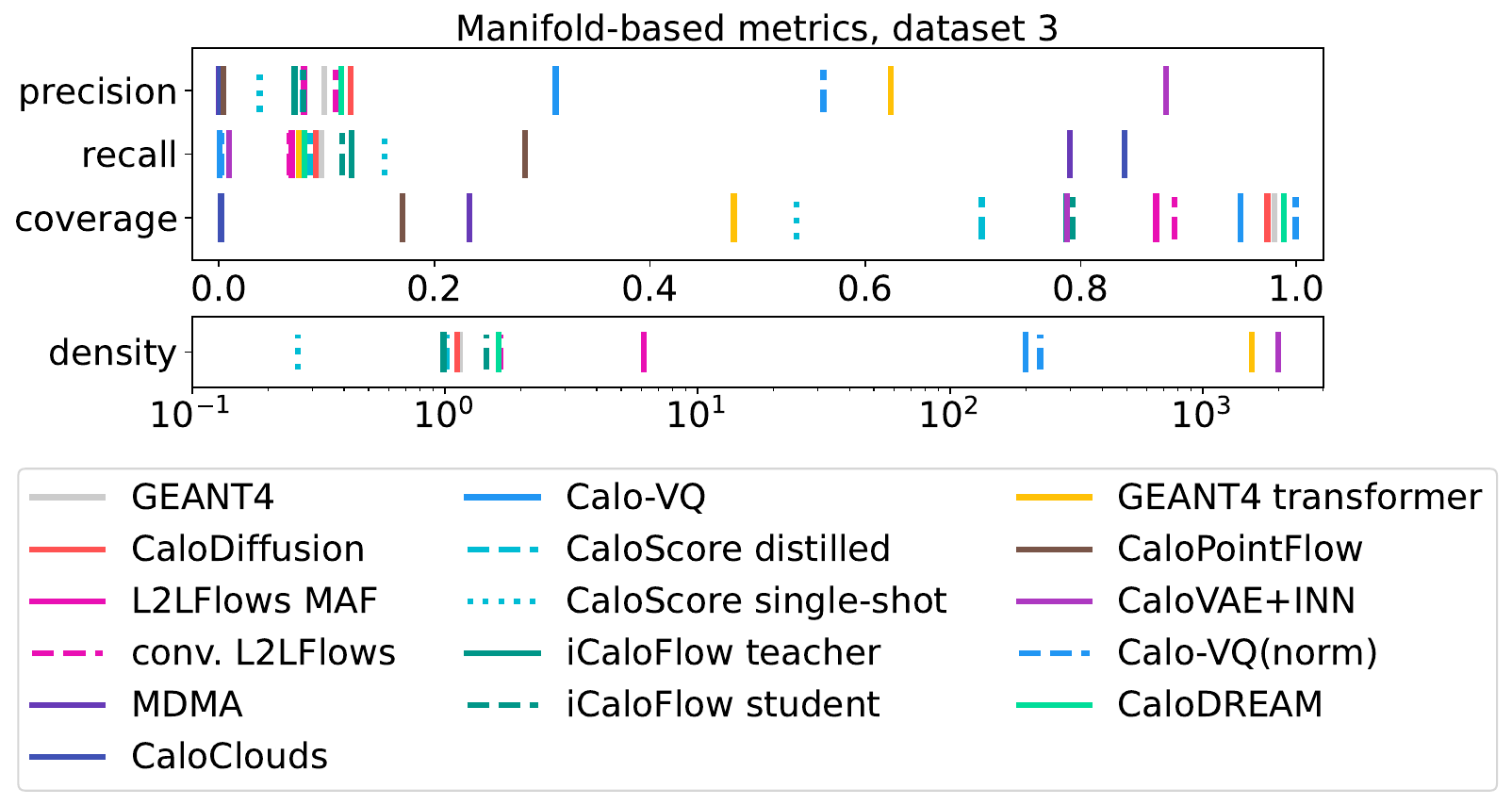}
\caption{Precision, density, recall, and coverage for \dsIII submissions. For the precise numbers, see \Tref{tab:ds3.prdc}.}
\label{fig:ds3.prdc}
\end{figure}

In \fref{fig:ds3.prdc} and \tref{tab:ds3.prdc} we show precision, density, recall, and coverage of the \dsIII submissions. The first thing we notice are the \geant scores, which now have much smaller precision and recall compared to \dsII in \fref{fig:ds2.prdc}, maybe a sign for the much higher-dimensional dataset. When looking at the submissions, we observe groups with similar patterns as for the other datasets before. The first one of these are \submAmram, \submMikuniDist, and \submPangIT, which have scores comparable to the \geant reference. Similar to these, but with a slightly larger density are \submPalacios, \submBussConv and \submPangIS. These groups overlap to a large extend with the ``winners'' of the classifier-based metrics, but have with \icalo also new members. With increasing density, we but otherwise similar scores is \submBussMAF. These all indicate samples that are distributed similarly close to the validation data like the training data. Another group of submissions, consisting of \submKaech, \submKorol, \submMikuniSingle, and \submSchnake, have precision, density and coverage below the \geant scores, and at the same time a very large recall. As already for \dsII, we interpret such a pattern as samples being generated fairly spread out, but not really close to the reference samples. Also these observations are consistent with what we saw for other metrics before. The last group, with a very large density, a larger precision and a low recall was also present in \dsII. In this group we have \submLiu, \submLiuNorm, \submSalamaniTrans, and \submErnst.

\begin{figure}
\centering
\includegraphics[width=\textwidth]{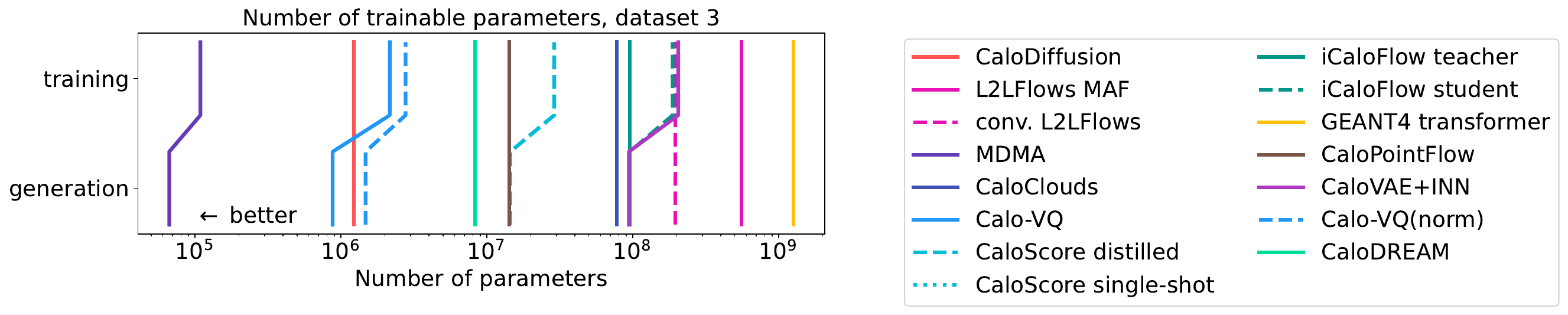}
\caption{Number of trainable parameters for training and generation of \dsIII submissions. For the precise numbers, see \Tref{tab:ds3.numparam}.}
\label{fig:ds3.numparam}
\end{figure}

\Fref{fig:ds3.numparam} compares the sizes of the submissions, with \tref{tab:ds3.numparam} giving the precise numbers. Overall, the entire span in number of parameters is more than four orders of magnitude. Similar to \dsII, \submKaech has by far the fewest number of trainable parameters, making it a very economic submission. Following behind are with \submLiu and \submAmram a VAE and a diffusion model, showing that these architectures can generate high-dimensional data much more economically than normalizing flows.

\begin{figure}
\centering
\includegraphics[width=\textwidth]{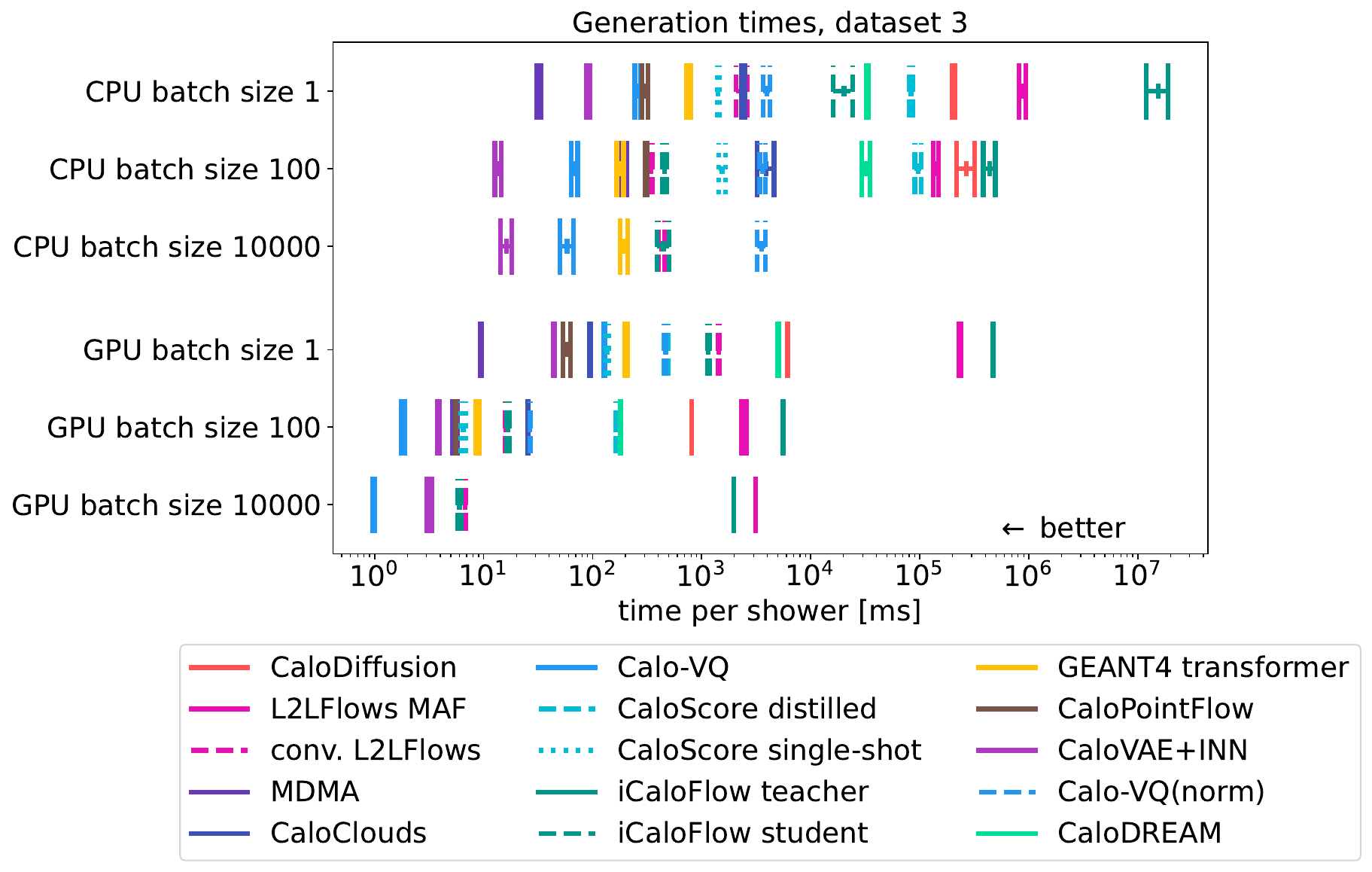}
\caption{Timing of \dsIII submissions on CPU and GPU architectures. Not all submissions are shown everywhere due to memory and other constraints. More details are in \tref{tab:ds3.timing.CPU} and \tref{tab:ds3.timing.GPU}.}
\label{fig:ds3.timing}
\end{figure}

Lastly, we look at the generation time per shower in \fref{fig:ds3.timing} (see~\tref{tab:ds3.timing.CPU} for CPU and~\tref{tab:ds3.timing.GPU} for GPU details). Overall, we see the same pattern as for all datasets before. Increasing the batch size and moving from a CPU to a GPU architecture speeds up generation. Depending on the architecture, sometimes by several orders of magnitude. However, the high dimensionality of \dsIII makes generation with large batch sizes sometimes impossible due to memory constraints. For example at batch size 10000, nine out of 15 submissions run into CUDA out of memory errors on the GPU. The large spread in generation times also required us to restrict the number of samples used to time the models to fewer than 100\,000 events, especially for smaller batch sizes. For the largest batch size of 10\,000, we had three cases on the CPU in which generation of a single batch took longer than two days. Details for this are given in~\tref{tab:ds3.timing.CPU} and~\tref{tab:ds3.timing.GPU}. The fastest models are again GAN-based submissions like \submKaech and VAE-based submissions like \submLiu or \submErnst. As for the previous datasets, we observe that distillation worked and speeds up generation in all cases. 

\FloatBarrier

\section{Results: Correlations Between Metrics}
\markboth{\uppercase{Results: Correlations Between Metrics}}{} 
\label{sec:correlations}

In this section we study how the scores in different metrics are related to each other. The goal of that is two-fold: First, in \sref{sec:metric.comparison}, we study how various different metrics that all measure the same property correlate with each other. In the case of the sample quality, this will shed light on various aspects regarding the evaluation of generative models, a result of great importance beyond detector fast simulation. Second, in \sref{sec:pareto.fronts}, we are interested in the Pareto fronts in the ``quality \textit{vs.}~speed \textit{vs.}~resource consumption'' space, as these will be the ultimate results of the CaloChallenge. The observations made in the first part, \textit{i.e.}~how which quality metrics correlate with each other, will be especially important for the choice of metrics shown in the final evaluation Pareto Fronts.  

\subsection{Metric Comparison}
\label{sec:metric.comparison}
As a nice side result of the challenge, we can evaluate how different metrics that measure the  quality of the showers correlate with each other. These tests also justify that the Pareto fronts we will show below are representative. 

\begin{figure}[ht]
\centering
\includegraphics[width=\textwidth]{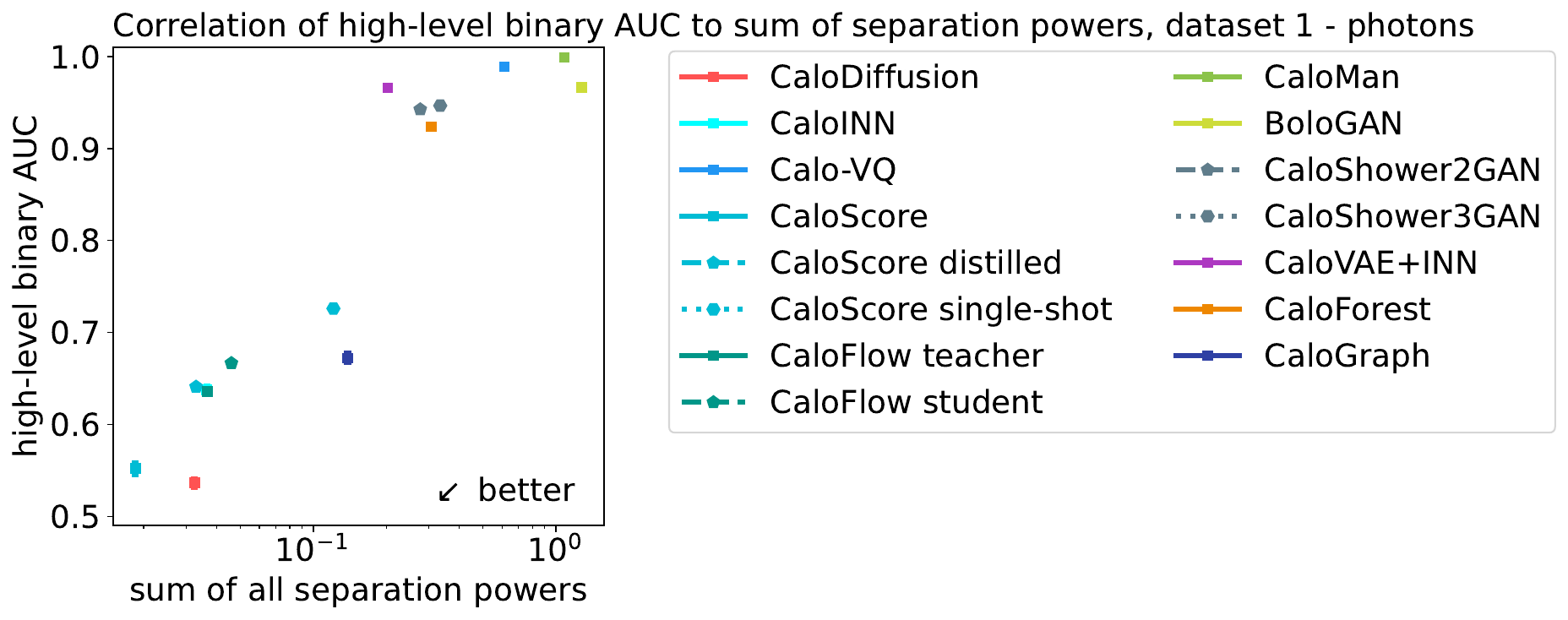}
\caption{Correlation of two metrics based on high-level observables: the sum of all the separation powers (\fref{fig:ds1-photons-1-depositions}--\fref{fig:ds1-photons-1-sparsity}) vs. the binary AUC (\fref{fig:ds1-photons.aucs} and \tref{tab:ds1-photons.aucs}).}
\label{fig:ds1-photons.scatter.high}
\end{figure}

\begin{figure}[ht]
\centering
\includegraphics[width=\textwidth]{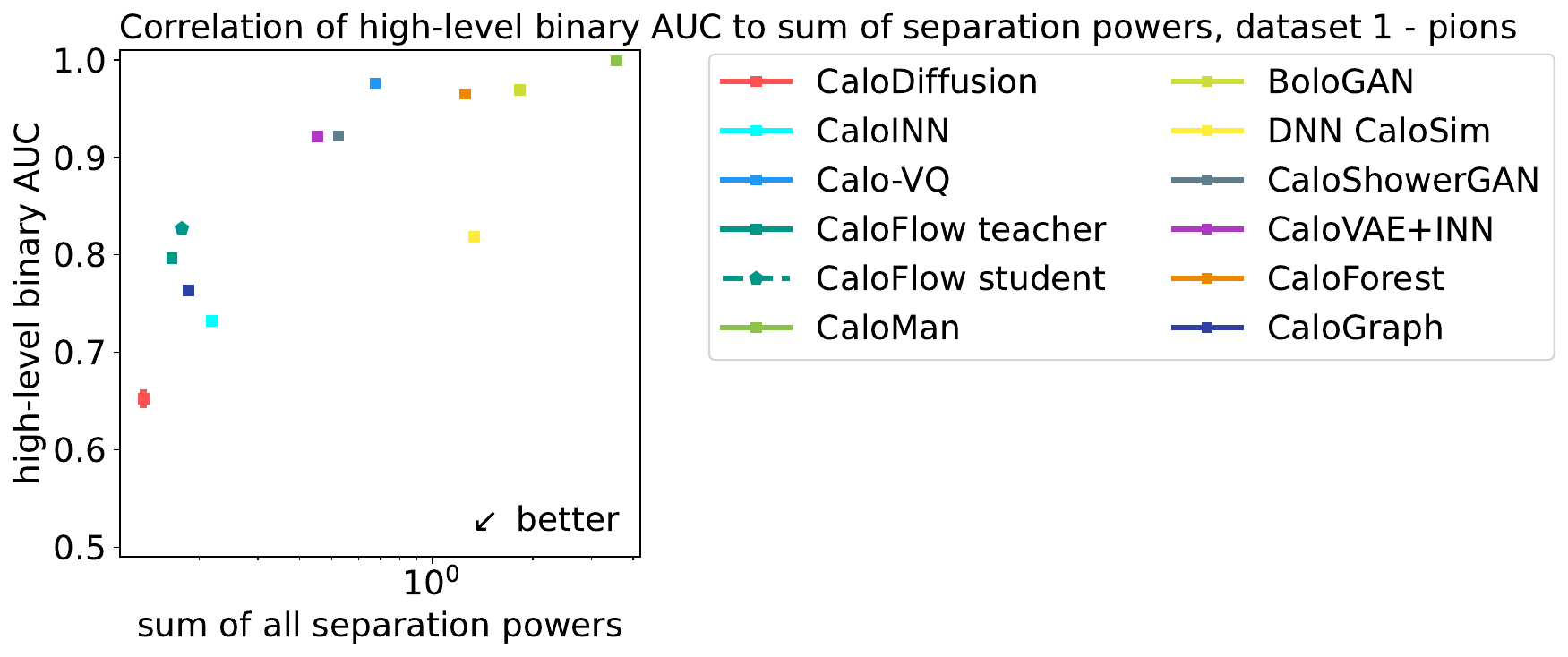}
\caption{Correlation of two metrics based on high-level observables: the sum of all the separation powers (\fref{fig:ds1-pions-1-depositions}--\fref{fig:ds1-pions-1-sparsity}) vs. the binary AUC (\fref{fig:ds1-pions.aucs} and \tref{tab:ds1-pions.aucs}).}
\label{fig:ds1-pions.scatter.high}
\end{figure}

\begin{figure}[ht]
\centering
\includegraphics[width=\textwidth]{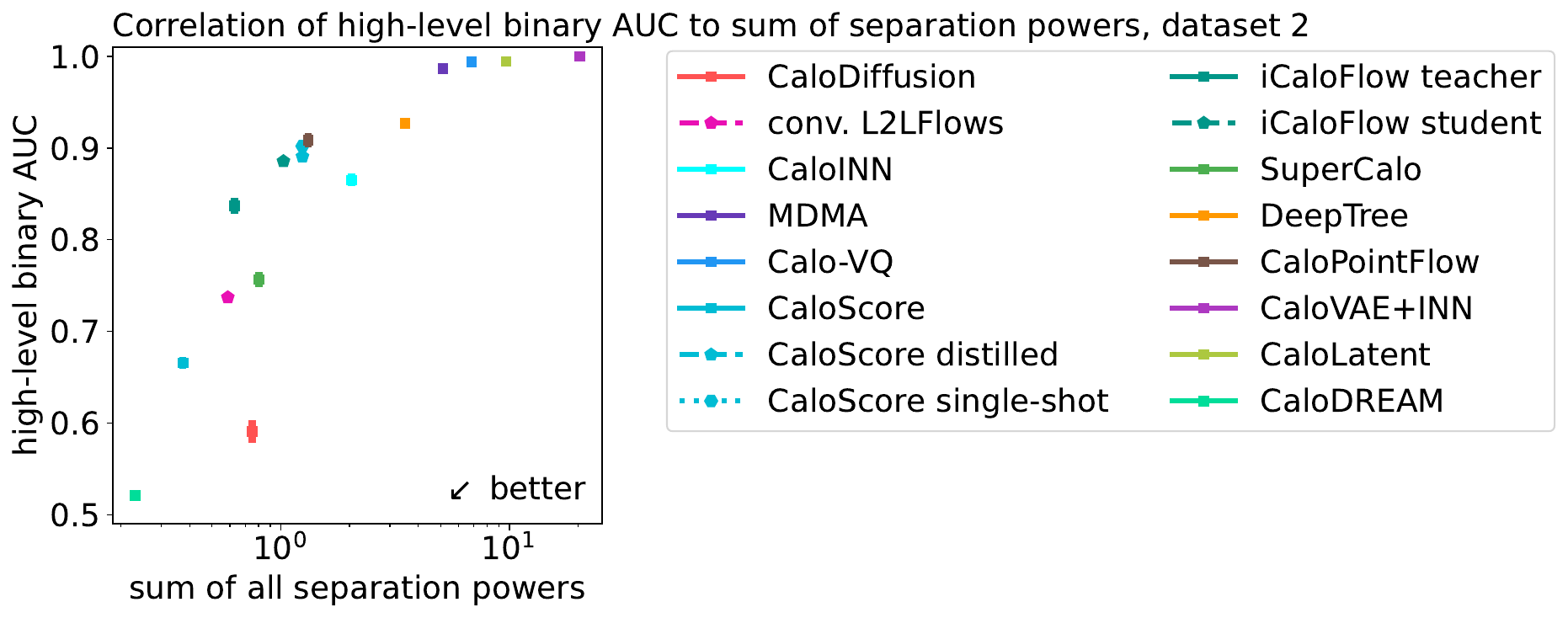}
\caption{Correlation of two metrics based on high-level observables: the sum of all the separation powers (\fref{fig:ds2-depositions}--\fref{fig:ds2-sparsity}) vs. the binary AUC (\fref{fig:ds2.aucs} and \tref{tab:ds2.aucs}).}
\label{fig:ds2.scatter.high}
\end{figure}

\begin{figure}[ht]
\centering
\includegraphics[width=\textwidth]{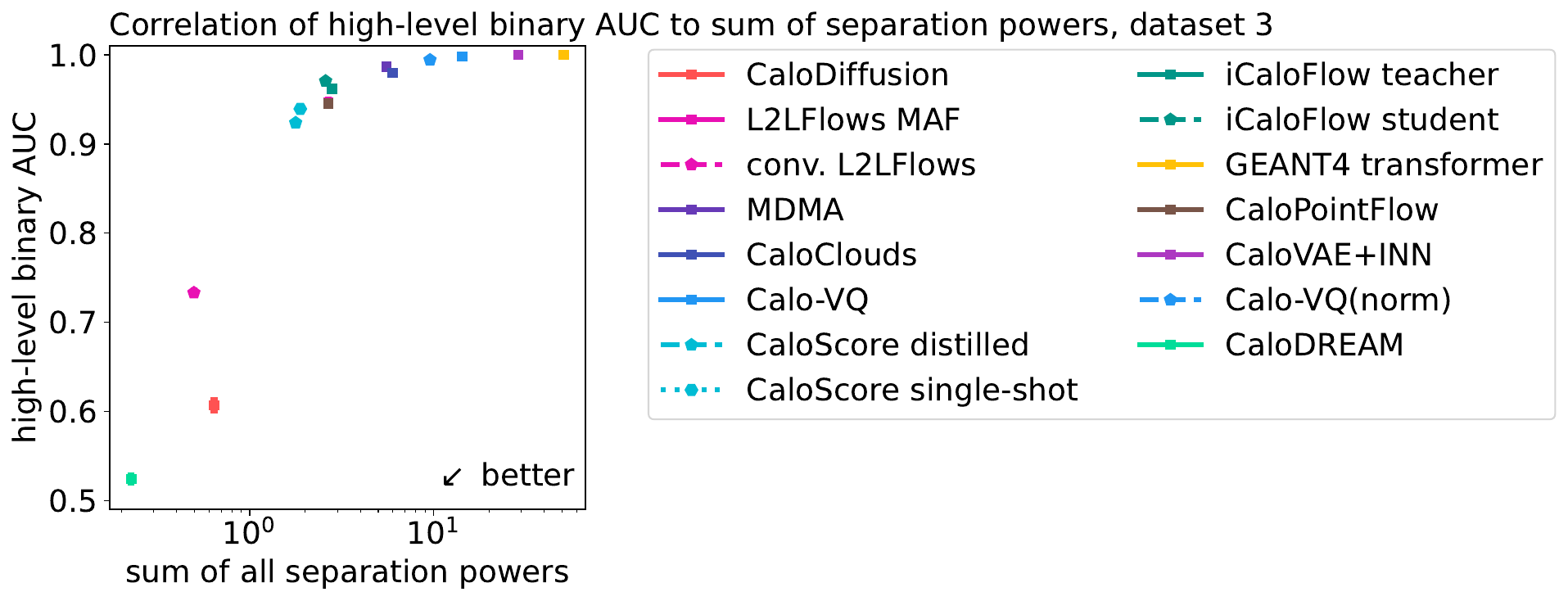}
\caption{Correlation of two metrics based on high-level observables: the sum of all the separation powers (\fref{fig:ds3-depositions}--\fref{fig:ds3-sparsity}) vs. the binary AUC (\fref{fig:ds3.aucs} and \tref{tab:ds3.aucs}).}
\label{fig:ds3.scatter.high}
\end{figure}
The first of these tests looks at the high-level observables that were defined in \sref{sec:results_hlf_intro} and compares the sum of all separation powers to the AUC of the binary classifier. While the former is only sensitive to the distribution of the individual observables, the latter also captures correlations between them. We see in \fref{fig:ds1-photons.scatter.high} that the results for \dsIph show a clear correlation. Submissions with a higher AUC also have a larger sum of their separation powers. The situation is similar for \dsIpi in \fref{fig:ds1-pions.scatter.high}, but there the submissions are a little more spread out, indicating that some models struggled a bit more to capture all correlations between the observables. Also datasets 2 and 3 in \fref{fig:ds2.scatter.high} and \fref{fig:ds3.scatter.high} show a clear correlation of the two metrics, confirming that they both capture the essential features of the high-level observables. 

\begin{figure}[ht]
\centering
\includegraphics[width=\textwidth]{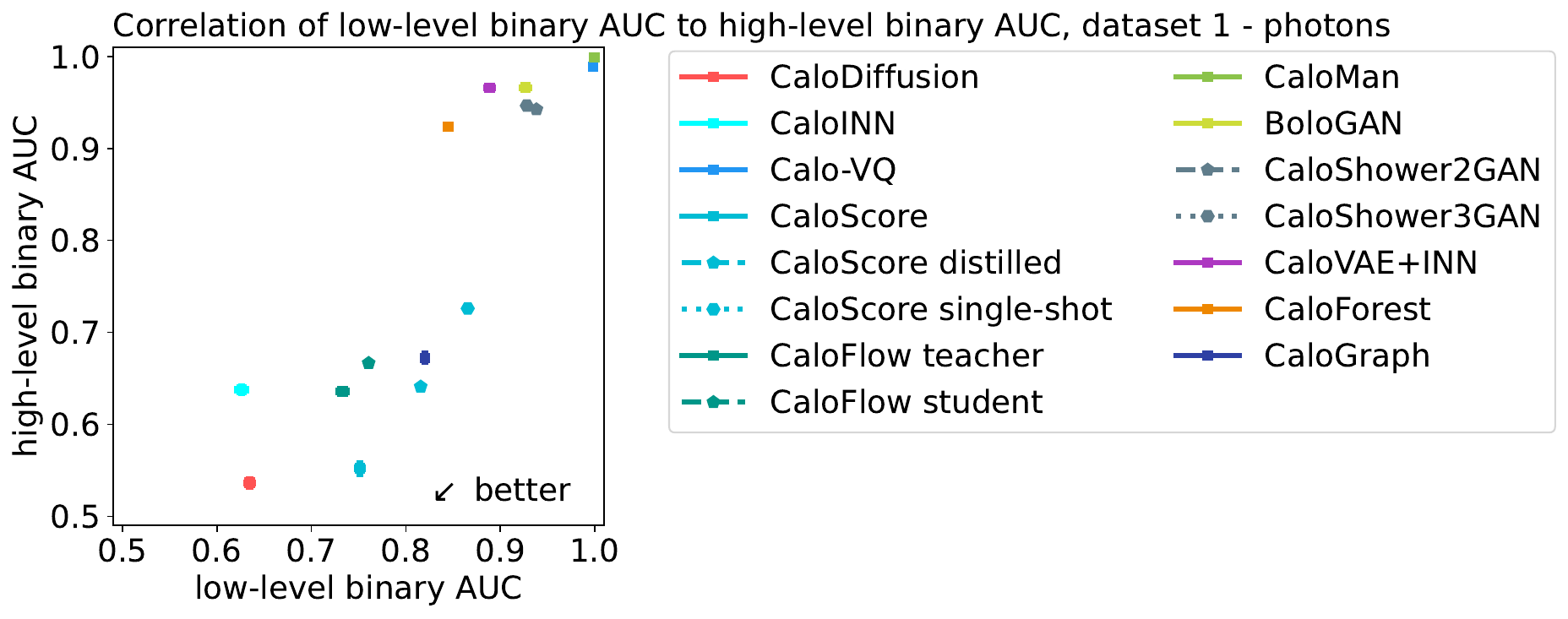}
\caption{Correlation of two metrics based on binary classifiers (\fref{fig:ds1-photons.aucs} and \tref{tab:ds1-photons.aucs}): the AUC based on low-level observables vs. the AUC based on high-level observables.}
\label{fig:ds1-photons.scatter.auc}
\end{figure}

\begin{figure}[ht]
\centering
\includegraphics[width=\textwidth]{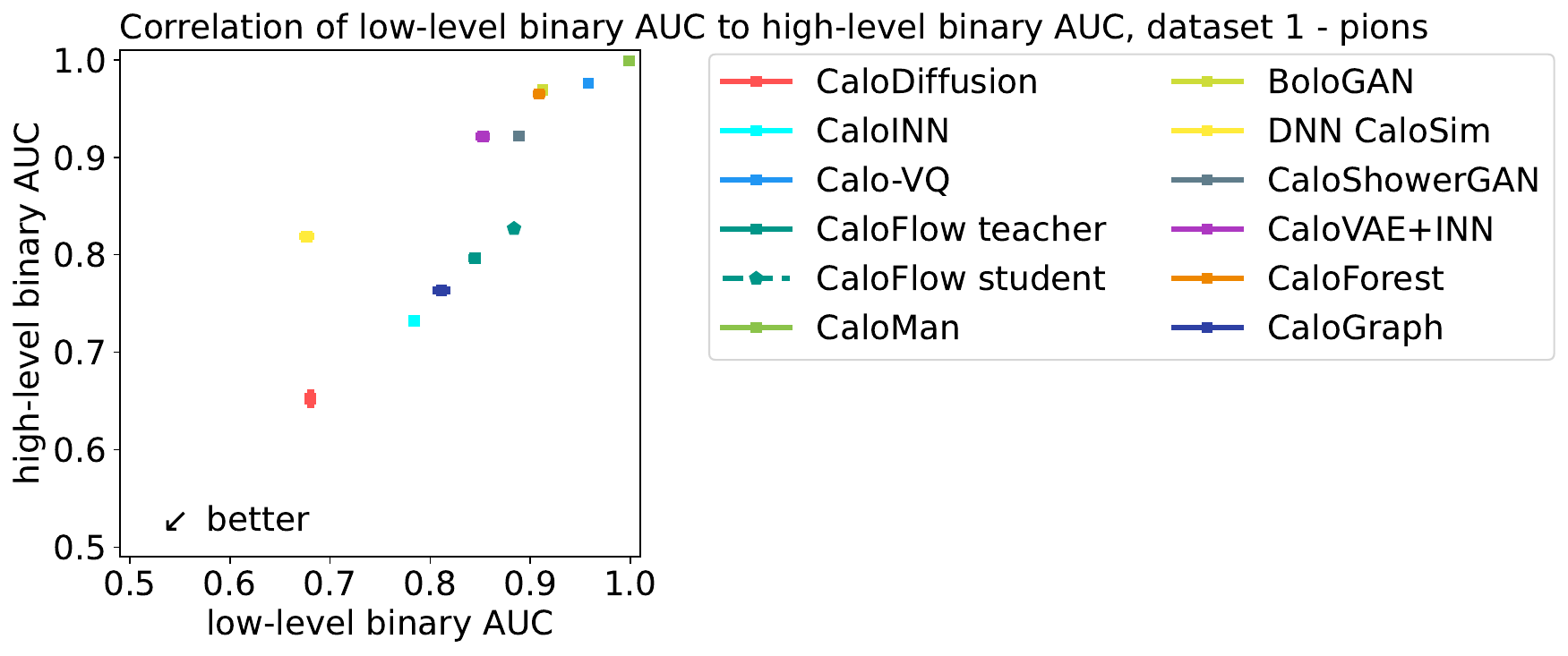}
\caption{Correlation of two metrics based on binary classifiers (\fref{fig:ds1-pions.aucs} and \tref{tab:ds1-pions.aucs}): the AUC based on low-level observables vs. the AUC based on high-level observables.}
\label{fig:ds1-pions.scatter.auc}
\end{figure}

\begin{figure}[ht]
\centering
\includegraphics[width=\textwidth]{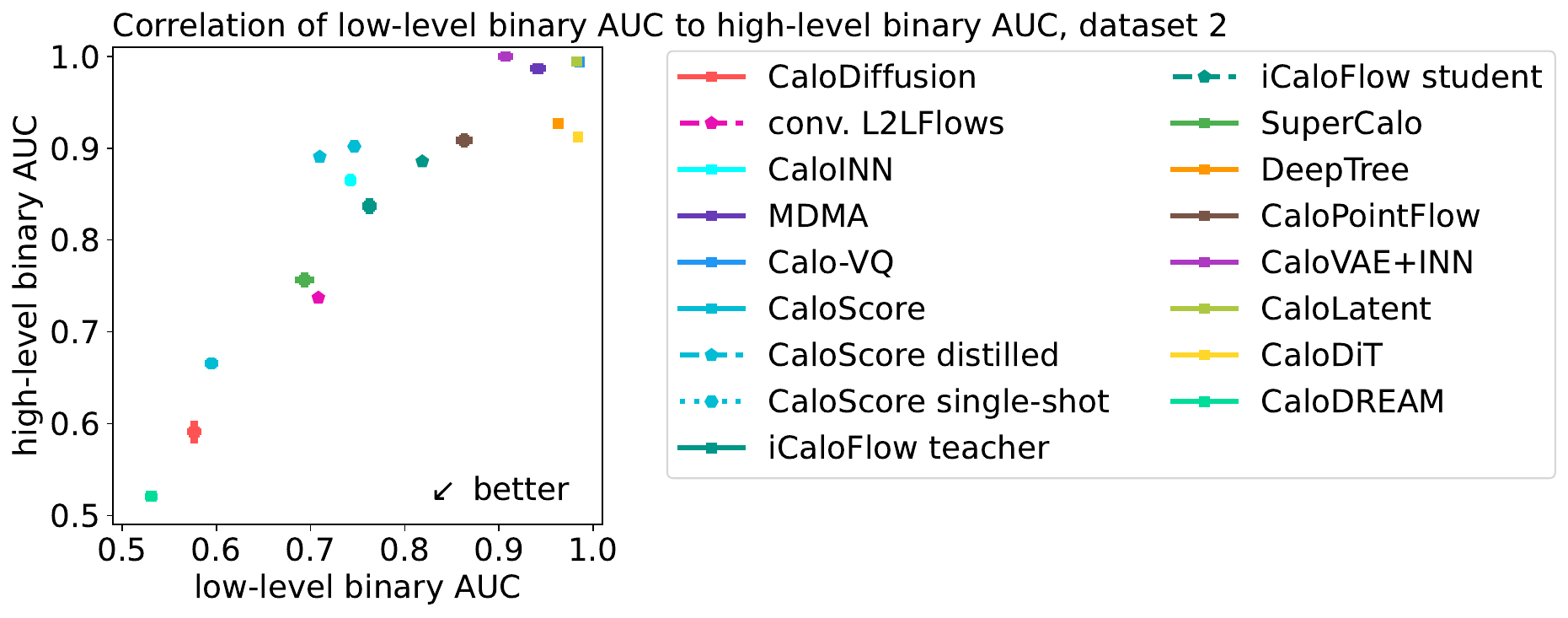}
\caption{Correlation of two metrics based on binary classifiers (\fref{fig:ds2.aucs} and \tref{tab:ds2.aucs}): the AUC based on low-level observables vs. the AUC based on high-level observables.}
\label{fig:ds2.scatter.auc}
\end{figure}

\begin{figure}[ht]
\centering
\includegraphics[width=\textwidth]{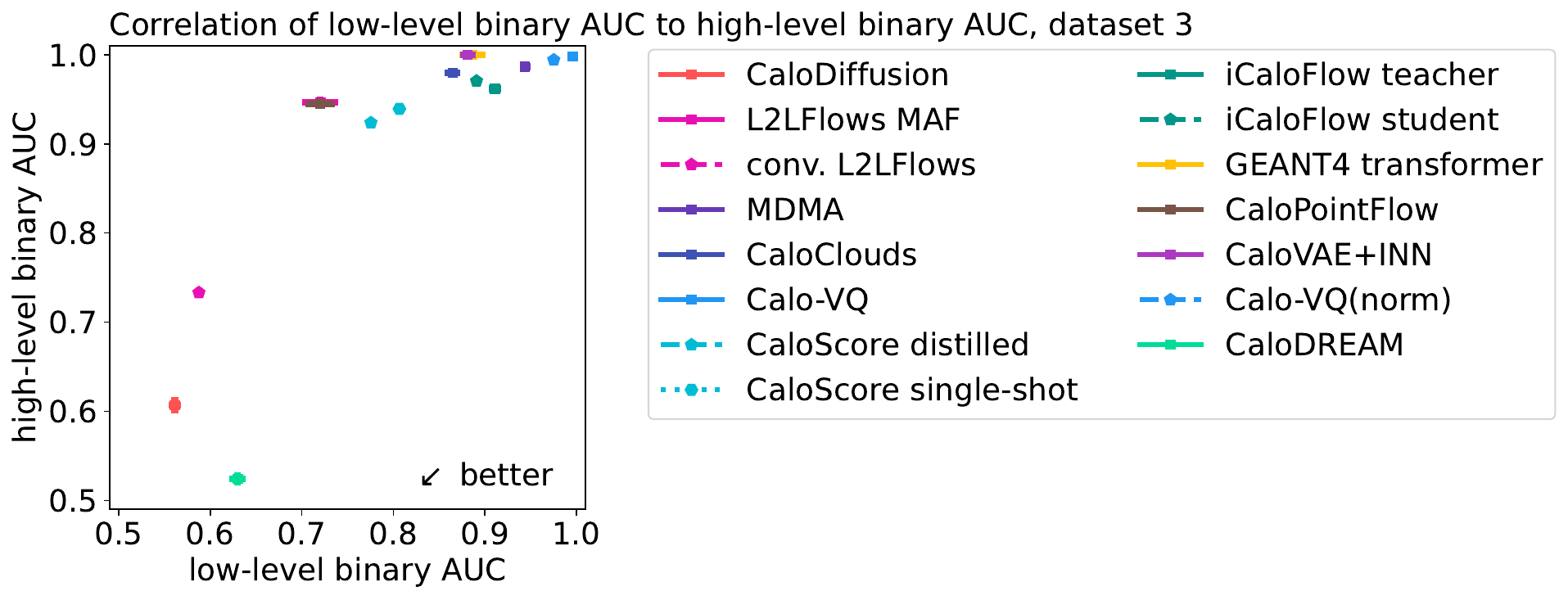}
\caption{Correlation of two metrics based on binary classifiers (\fref{fig:ds3.aucs} and \tref{tab:ds3.aucs}): the AUC based on low-level observables vs. the AUC based on high-level observables.}
\label{fig:ds3.scatter.auc}
\end{figure}

\begin{figure}[ht]
\centering
\includegraphics[width=\textwidth]{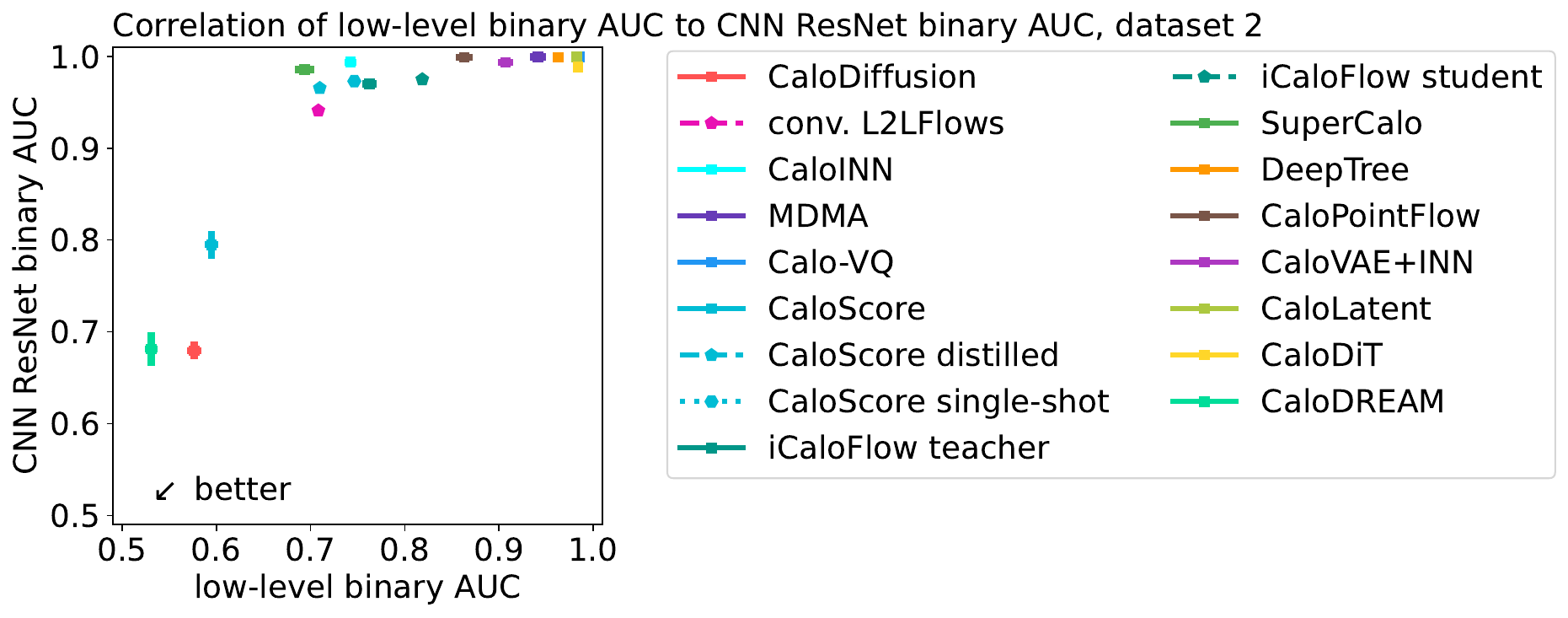}
\caption{Correlation of two metrics based on binary classifiers and low-level observables (\fref{fig:ds2.aucs} and \tref{tab:ds2.aucs}): the AUC based on a DNN classifier vs. the AUC based on CNN ResNet classifier.}
\label{fig:ds2.scatter.cnn}
\end{figure}

\begin{figure}[ht]
\centering
\includegraphics[width=\textwidth]{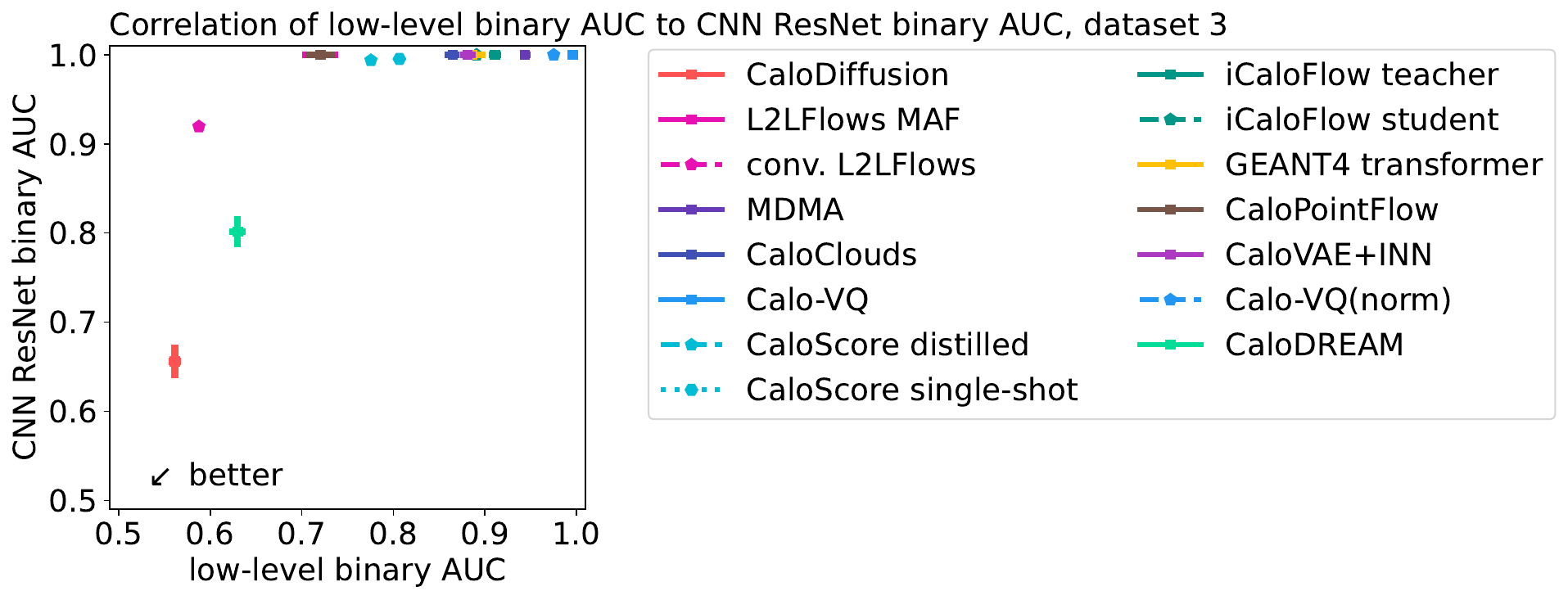}
\caption{Correlation of two metrics based on binary classifiers and low-level observables (\fref{fig:ds3.aucs} and \tref{tab:ds3.aucs}): the AUC based on a DNN classifier vs. the AUC based on CNN ResNet classifier.}
\label{fig:ds3.scatter.cnn}
\end{figure}

Next we investigate how the choice of the input representation to the binary classifier influences the AUC. In particular, we look at the correlation of the AUC of the binary classifier with low-level inputs \textit{vs.}~the AUC of the binary classifier with high-level inputs. \Fref{fig:ds1-photons.scatter.auc} shows the result for \dsIph. While there is a clear correlation between the two metrics visible, there is also a noticeable spread between submissions, for example when comparing \submFavaro to \submMikuni. The situation is more clear for \dsIpi in \fref{fig:ds1-pions.scatter.auc}. Here, we see two different lines forming. One with \submAmram, \submFavaro, \submKobylyansky, and \cf, where the low-level AUC is slightly worse than the corresponding high-level AUC. The other one with \submSalamaniDNN, \submErnst, \submZhang, \submCresswell, \submLiu, and \submReyes, where the high-level AUC is larger than the low-level AUC. Interestingly, the division in these two sets aligns with the underlying architectures, with the diffusion models and normalizing flows in the first group and the VAEs and GANs in the second group. We interpret these differences as follows: the first group (diffusion and Normalizing Flow-based) generates showers which better capture the correlations between voxels that form the high-level observables and the remaining mismodeling between the submissions and \geant is in the lower-energetic, subleading voxels. The second group (VAE and GAN-based), however, already mismodels the correlations that form the high-level observables leading to a larger AUC for this classifier. The strong correlation between the AUCs is also present for dataset 2 in \fref{fig:ds2.scatter.auc}. For dataset 3 in \fref{fig:ds3.scatter.auc} it is not as pronounced, but that is mostly due to the high-level AUCs being close to 1 for many submissions. The correlations between high and low-level AUCs also tell us something about the classifier metric itself. Since all the high-level observables are derived from the low-level ones, there cannot be any additional information in the high-level observables. The AUC based on low-level inputs should therefore be strictly larger, \textit{i.e.}~indicating a better classifier than the AUC based on high-level features alone. The fact that we do not see this here indicates that the DNN classifier used in this study is not at the Neyman-Pearson limit and additional studies based on the high-level observables are indeed necessary to get a better understanding on the quality of the generated samples. 

For datasets 2 and 3 we can also compare the AUCs obtained by the two different architectures used for the binary classification: the DNN and the CNN ResNet. The results are shown in \fref{fig:ds2.scatter.cnn} and \fref{fig:ds3.scatter.cnn}. In both cases we see a correlation, but we also see many submissions having a CNN ResNet-based AUC close to 1, making it hard to order them by this metric. This is especially true for dataset 3. We therefore use the DNN architecture for the Pareto fronts below.  

\begin{figure}[ht]
\centering
\includegraphics[width=\textwidth]{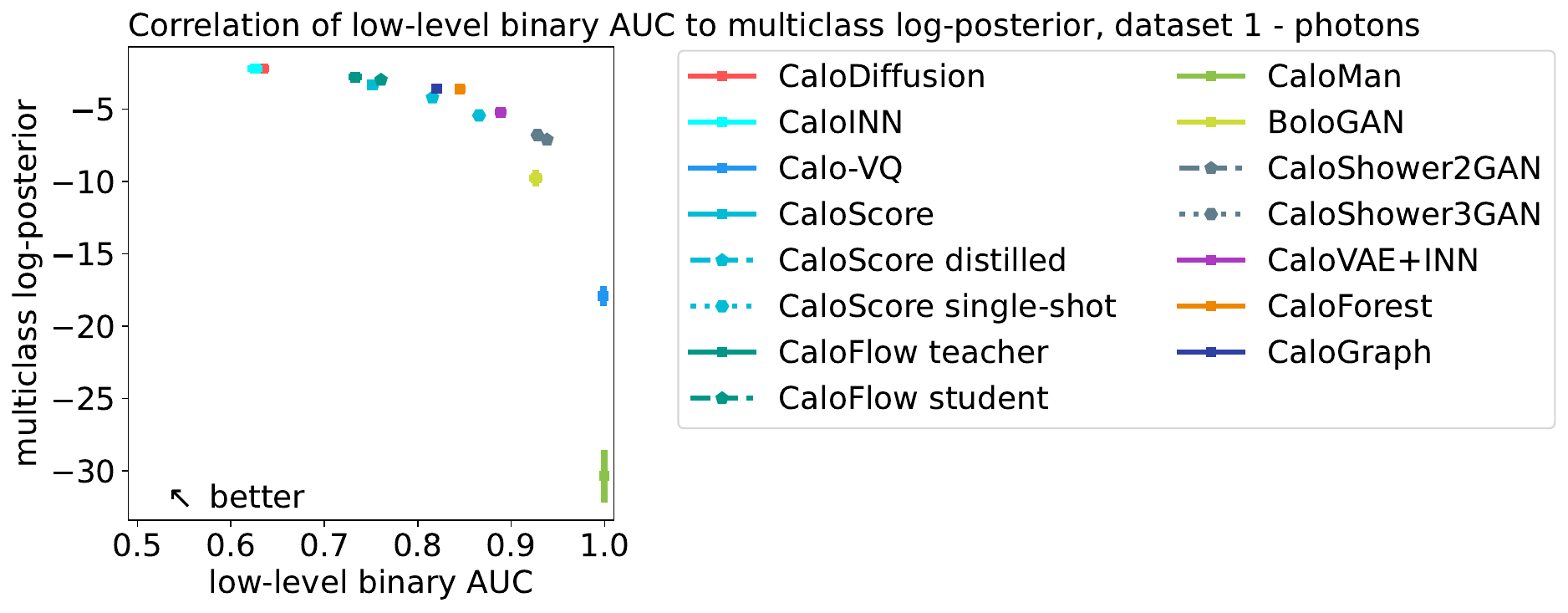}
\caption{Correlation of two metrics based on classifiers : the log posterior (\fref{fig:ds1-photons.multi} and \tref{tab:ds1-photons.multi}) of the multiclass classification vs. the AUC of the binary classification (\fref{fig:ds1-photons.aucs} and \tref{tab:ds1-photons.aucs}).}
\label{fig:ds1-photons.scatter.cls}
\end{figure}

\begin{figure}[ht]
\centering
\includegraphics[width=\textwidth]{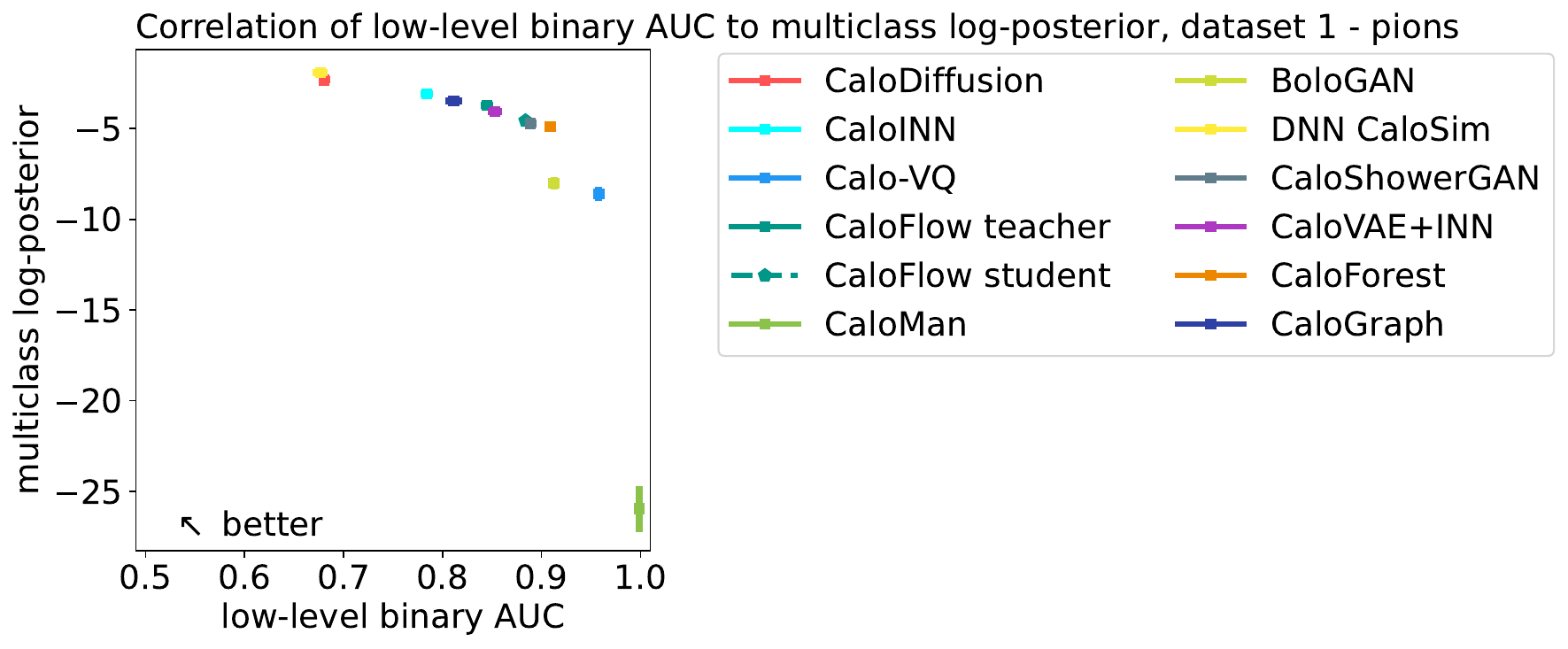}
\caption{Correlation of two metrics based on classifiers : the log posterior (\fref{fig:ds1-pions.multi} and \tref{tab:ds1-pions.multi}) of the multiclass classification vs. the AUC of the binary classification (\fref{fig:ds1-pions.aucs} and \tref{tab:ds1-pions.aucs}).}
\label{fig:ds1-pions.scatter.cls}
\end{figure}

\begin{figure}[ht]
\centering
\includegraphics[width=\textwidth]{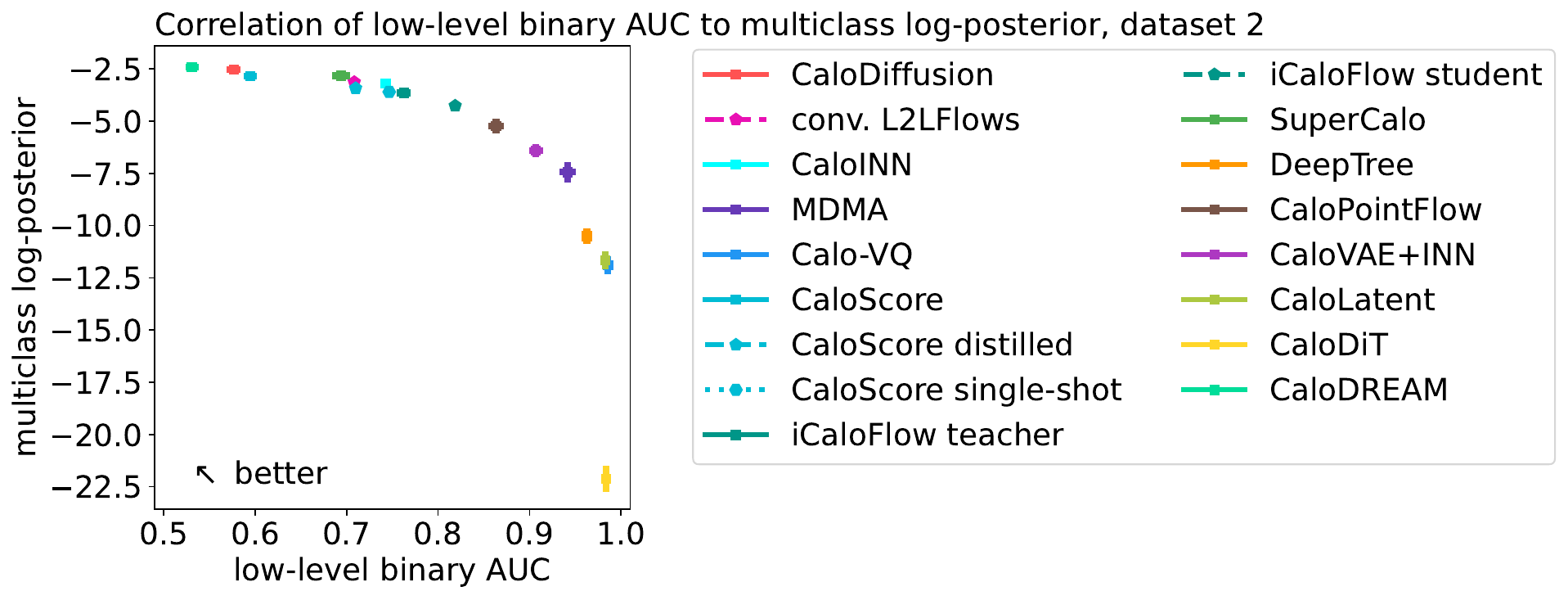}
\caption{Correlation of two metrics based on classifiers : the log posterior (\fref{fig:ds2.multi.dnn} and \tref{tab:ds2.multi.dnn}) of the multiclass classification vs. the AUC of the binary classification (\fref{fig:ds2.aucs} and \tref{tab:ds2.aucs}).}
\label{fig:ds2.scatter.cls}
\end{figure}

\begin{figure}[ht]
\centering
\includegraphics[width=\textwidth]{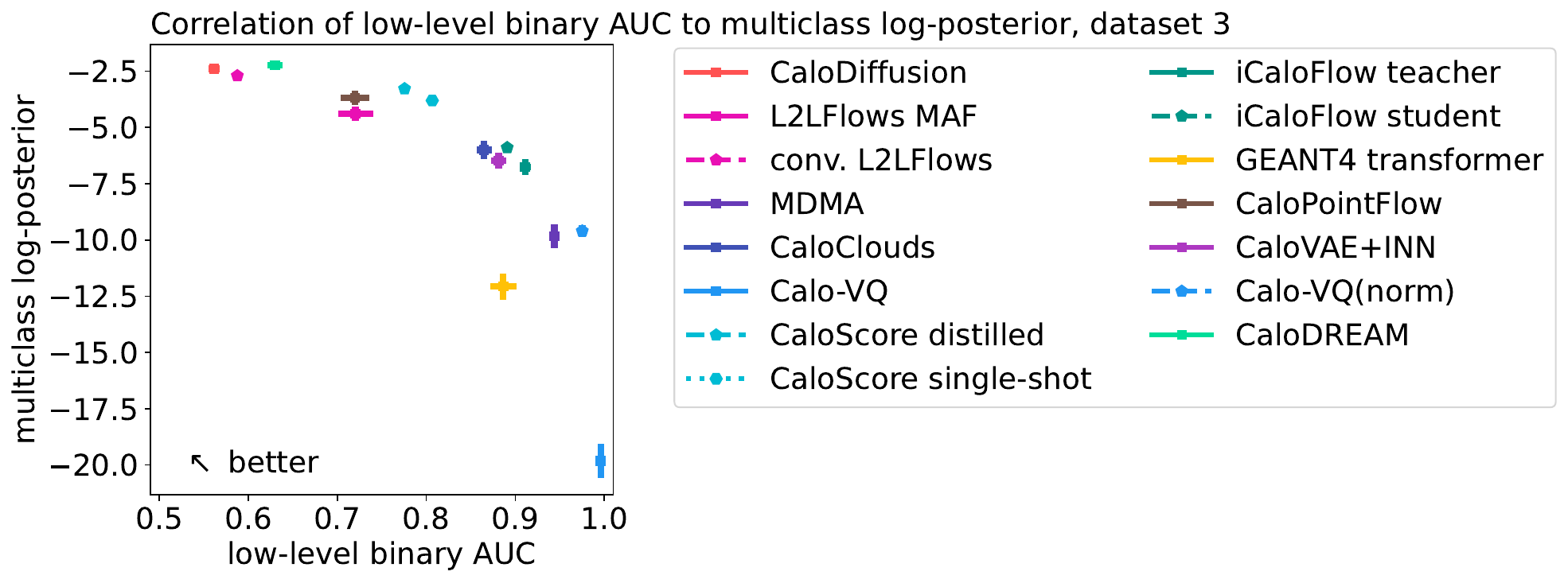}
\caption{Correlation of two metrics based on classifiers : the log posterior (\fref{fig:ds3.multi.dnn} and \tref{tab:ds3.multi.dnn}) of the multiclass classification vs. the AUC of the binary classification (\fref{fig:ds3.aucs} and \tref{tab:ds3.aucs}).}
\label{fig:ds3.scatter.cls}
\end{figure}
Lastly, we compare the results of the binary classification to the results of the multiclass classification. Also in these cases (\dsIph in \fref{fig:ds1-photons.scatter.cls}, \dsIpi in \fref{fig:ds1-pions.scatter.cls}, \dsII in \fref{fig:ds2.scatter.cls}, and \dsIII in \fref{fig:ds3.scatter.cls}), we observe a clear correlation: submissions performing well in one metric also perform well in the other metric, indicating that both the binary and multiclass classification capture the main differences between the submissions. The spread for \dsIII in \fref{fig:ds3.scatter.cls} is larger than for \dsII in \fref{fig:ds2.scatter.cls}, which is maybe due to the rather small sample size compared to the high-dimensionality of \dsIII. Overall, this implies that the binary classification can be used for further model development and it is not required to have all other submitted samples at hand to perform a multiclass classification for model evaluation. 

\begin{figure}[ht]
\centering
\includegraphics[width=\textwidth]{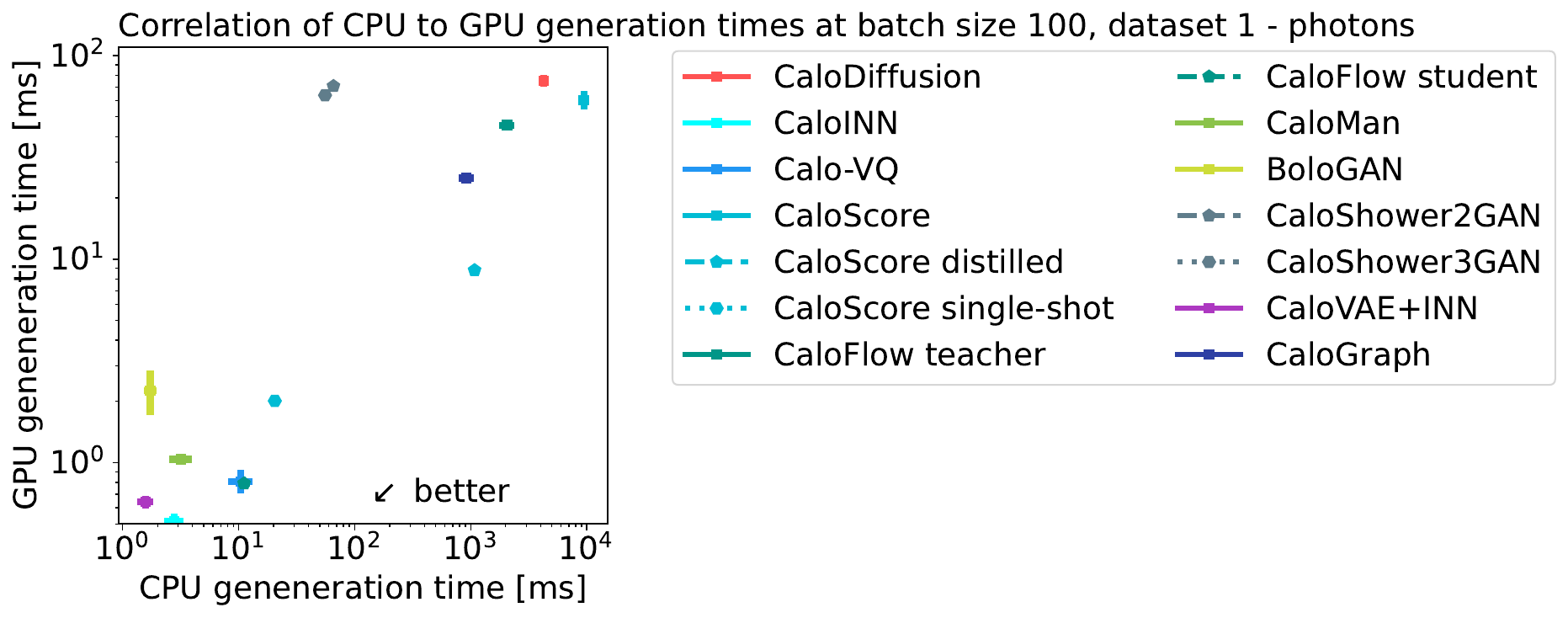}
\caption{Correlation of generation times on CPU and GPU (see \fref{fig:ds1-photons.timing}, \tref{tab:ds1-photons.timing.CPU} and \tref{tab:ds1-photons.timing.GPU}).}
\label{fig:ds1-photons.scatter.time}
\end{figure}

\begin{figure}[ht]
\centering
\includegraphics[width=\textwidth]{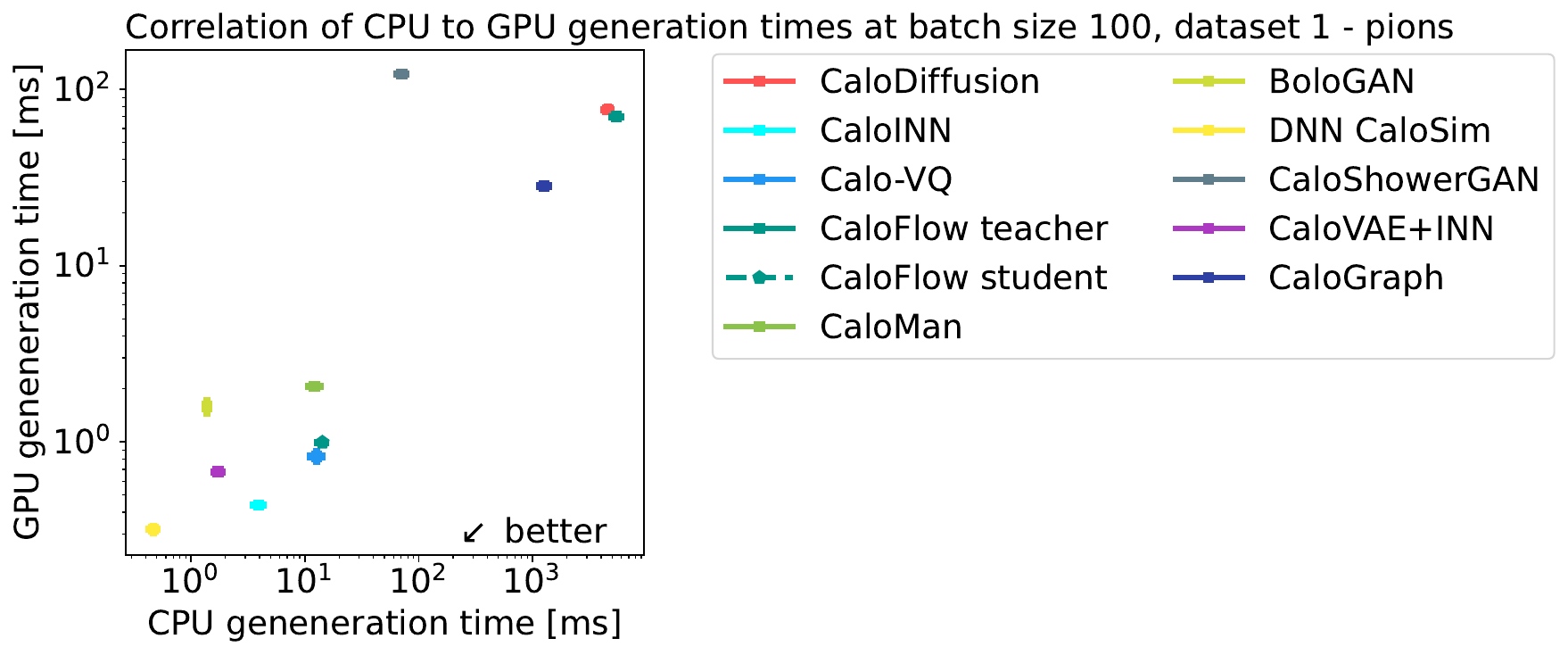}
\caption{Correlation of generation times on CPU and GPU (see \fref{fig:ds1-pions.timing}, \tref{tab:ds1-pions.timing.CPU} and \tref{tab:ds1-pions.timing.GPU}).}
\label{fig:ds1-pions.scatter.time}
\end{figure}

\begin{figure}[ht]
\centering
\includegraphics[width=\textwidth]{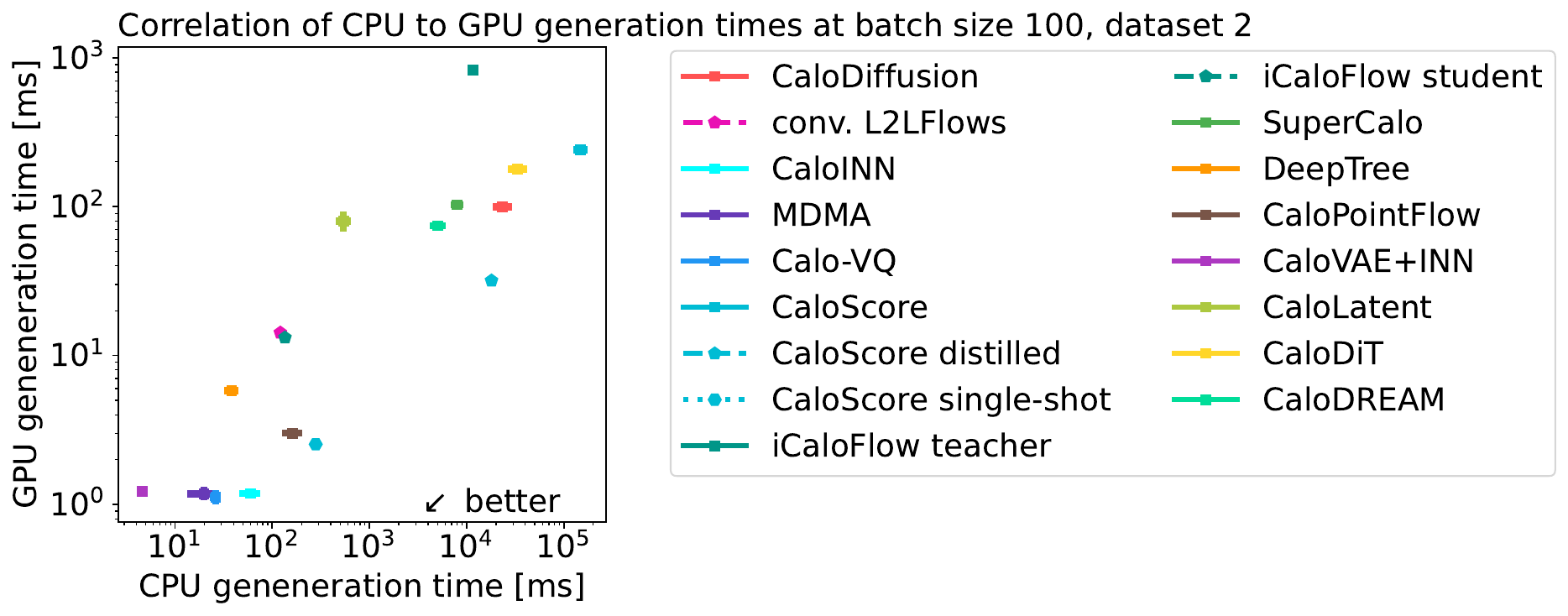}
\caption{Correlation of generation times on CPU and GPU (see \fref{fig:ds2.timing}, \tref{tab:ds2.timing.CPU} and \tref{tab:ds2.timing.GPU}).}
\label{fig:ds2.scatter.time}
\end{figure}

\begin{figure}[ht]
\centering
\includegraphics[width=\textwidth]{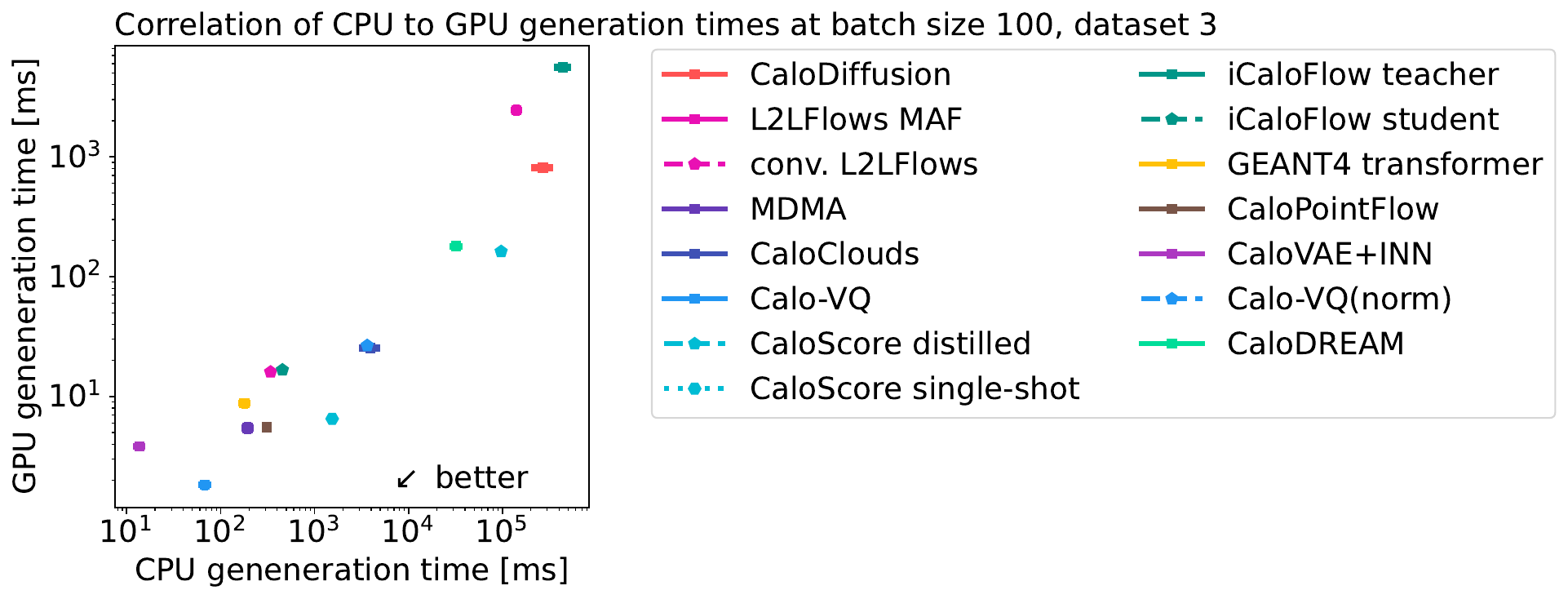}
\caption{Correlation of generation times on CPU and GPU (see \fref{fig:ds3.timing}, \tref{tab:ds3.timing.CPU} and \tref{tab:ds3.timing.GPU}).}
\label{fig:ds3.scatter.time}
\end{figure}

In addition to the quality metrics, we also look at the correlation between the generation times per shower on CPU and GPU architectures. In particular, we consider generation batch sizes of 100 in \fref{fig:ds1-photons.scatter.time} for \dsIph, \fref{fig:ds1-pions.scatter.time} for \dsIpi, \fref{fig:ds2.scatter.time} for \dsII, and \fref{fig:ds3.scatter.time} for \dsIII. In all cases, we see the scatter between fastest and slowest model to be much larger on the CPU than on the GPU. On top of the usual speed-up on the GPU, we observe the actual speed-up factor vary from model to model, depending on the specific building blocks of the models.  

\FloatBarrier

\subsection{Pareto Fronts}
\label{sec:pareto.fronts}

This section compiles the main results of the ``Fast Calorimeter Simulation Challenge 2022''. We show the performance of the submissions in the abstract ``quality \textit{vs.}~speed \textit{vs.}~resource consumption'' space. We are interested in submissions which are lightweight (\textit{i.e.}~have few parameters), are fast in generation, and have good sample quality. In particular, we focus on two planes in which there is a trade-off between two properties: quality \textit{vs.}~resource consumption and quality \textit{vs.}~speed. The third option, speed \textit{vs.}~resource consumption, does not show a real trade-off, so we collect the figures in appendix~\ref{app:time.num_param}. 

\begin{figure}[ht]
\centering
\includegraphics[width=\textwidth]{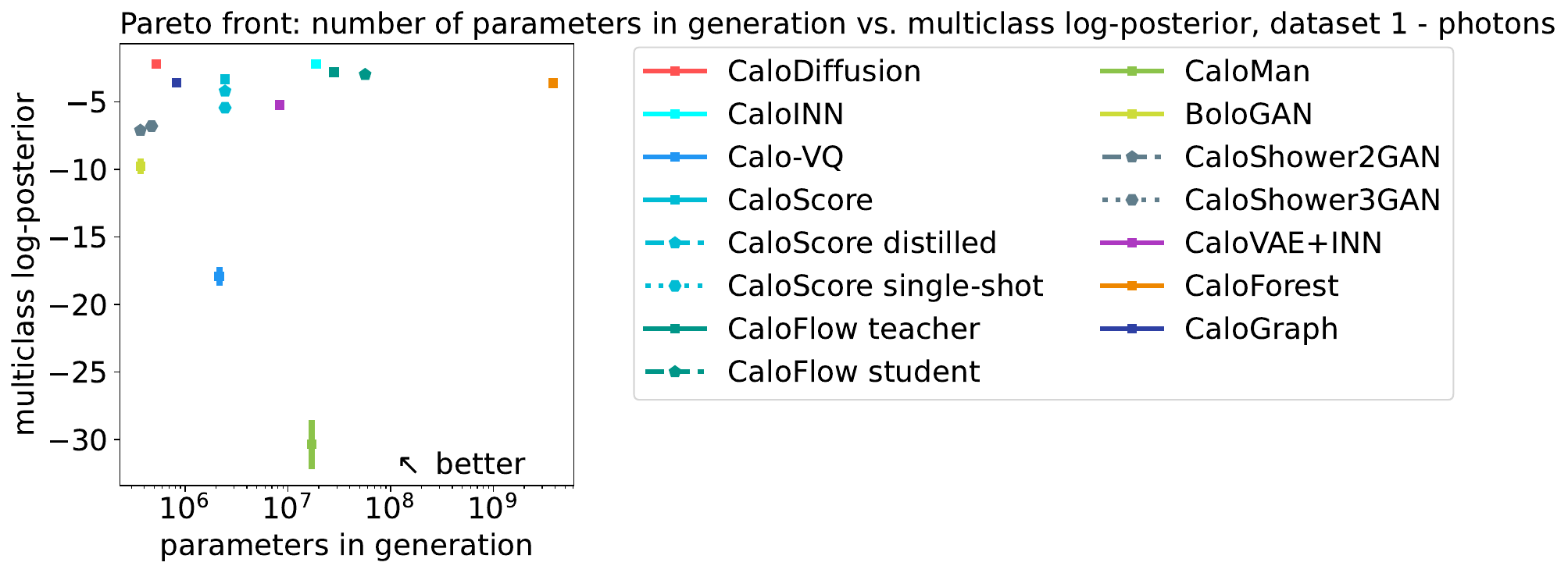}
\caption{Pareto front in sample quality (from \fref{fig:ds1-photons.multi} and \tref{tab:ds1-photons.multi}) and number of parameters in generation (from \fref{fig:ds1-photons.numparam} and \tref{tab:ds1-photons.numparam}).}
\label{fig:ds1-photons.pareto.quality.size}
\end{figure}

\begin{figure}[ht]
\centering
\includegraphics[width=\textwidth]{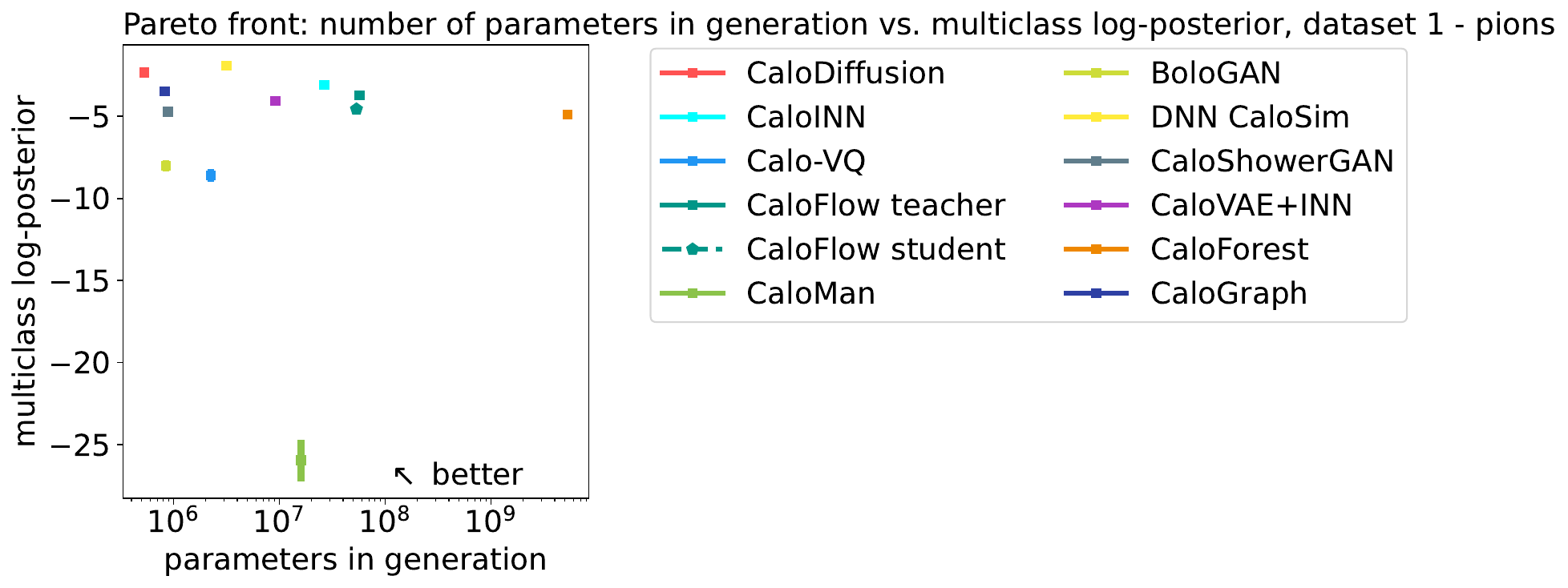}
\caption{Pareto front in sample quality (from \fref{fig:ds1-pions.multi} and \tref{tab:ds1-pions.multi}) and number of parameters in generation (from \fref{fig:ds1-pions.numparam} and \tref{tab:ds1-pions.numparam}).}
\label{fig:ds1-pions.pareto.quality.size}
\end{figure}

\begin{figure}[ht]
\centering
\includegraphics[width=\textwidth]{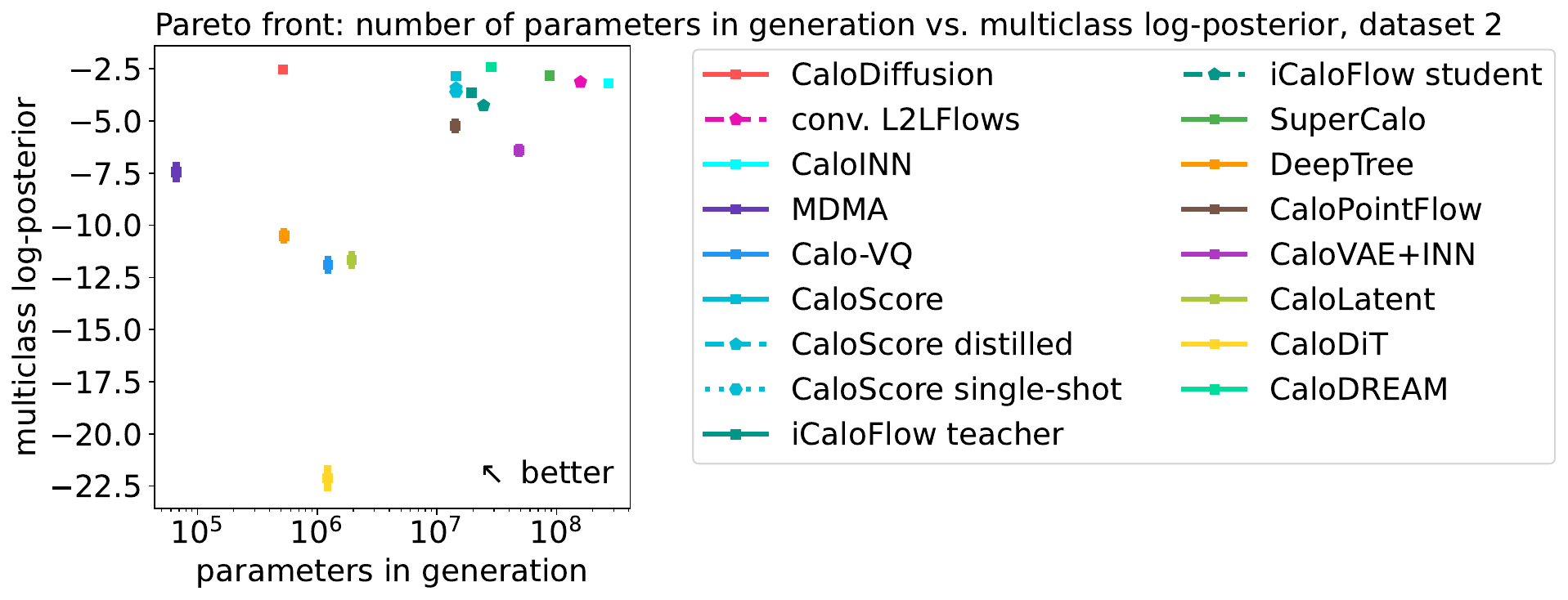}
\caption{Pareto front in sample quality (from \fref{fig:ds2.multi.dnn} and \tref{tab:ds2.multi.dnn}) and number of parameters in generation (from \fref{fig:ds2.numparam} and \tref{tab:ds2.numparam}).}
\label{fig:ds2.pareto.quality.size}
\end{figure}

\begin{figure}[ht]
\centering
\includegraphics[width=\textwidth]{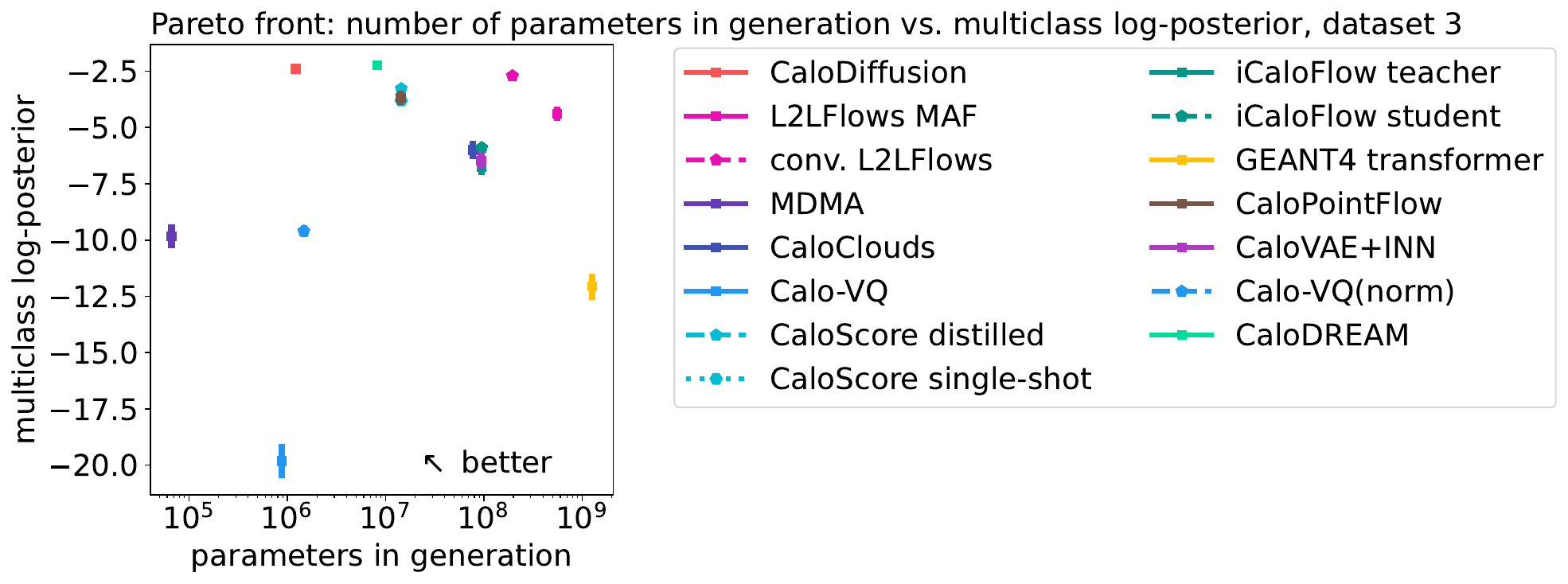}
\caption{Pareto front in sample quality (from \fref{fig:ds3.multi.dnn} and \tref{tab:ds3.multi.dnn}) and number of parameters in generation (from \fref{fig:ds3.numparam} and \tref{tab:ds3.numparam}).}
\label{fig:ds3.pareto.quality.size}
\end{figure}
We start by comparing sample quality to model size by plotting the DNN multiclass log-posterior with respect to the number of trainable parameters in generation. While we expect models with more parameters to better learn the underlying probability distribution that generates the showers, the use of the generative model inside a fast simulation framework prefers models that require less memory and are faster to load, \textit{i.e.}~have fewer parameters in generation. The figures for both particles in dataset 1 (photons in~\fref{fig:ds1-photons.pareto.quality.size} and pions in~\fref{fig:ds1-pions.pareto.quality.size}) are very similar. In both cases we see \submAmram in the top left corner, indicating that this diffusion model can generate high-quality showers with a comparatively small number of parameters. For dataset 2 in~\fref{fig:ds2.pareto.quality.size}, we do not have a clear winner in the corner. Instead, we observe a cluster of various submissions (including \submMikuni, its distillations, \icalo, and \submPangSuper) at good scores, but relatively large number of parameters. \submAmram is part of the Pareto front, with similar or better quality than submissions of said cluster, but more than an order of magnitude fewer parameters. Sacrificing some quality moves the Pareto front to even fewer parameters with the submission \submKaech. Dataset 3 in \fref{fig:ds3.pareto.quality.size} shows a similar trade-off between \submAmram and \submKaech around the top-left corner, but not such a large cluster of submission in the top-right.

\begin{figure}[ht]
\centering
\includegraphics[width=\textwidth]{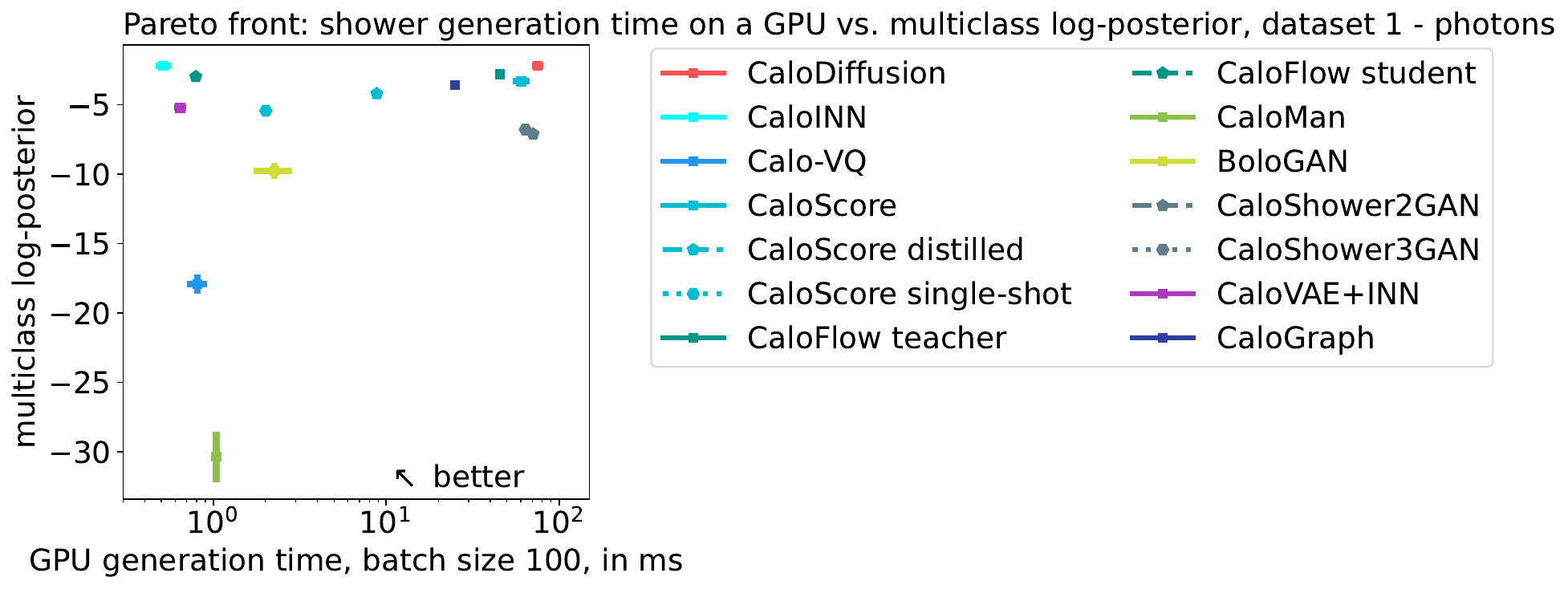}
\caption{Pareto front in sample quality (from \fref{fig:ds1-photons.multi} and \tref{tab:ds1-photons.multi}) and generation speed (from \fref{fig:ds1-photons.timing} and \tref{tab:ds1-photons.timing.GPU}).}
\label{fig:ds1-photons.pareto.quality.speed}
\end{figure}

\begin{figure}[ht]
\centering
\includegraphics[width=\textwidth]{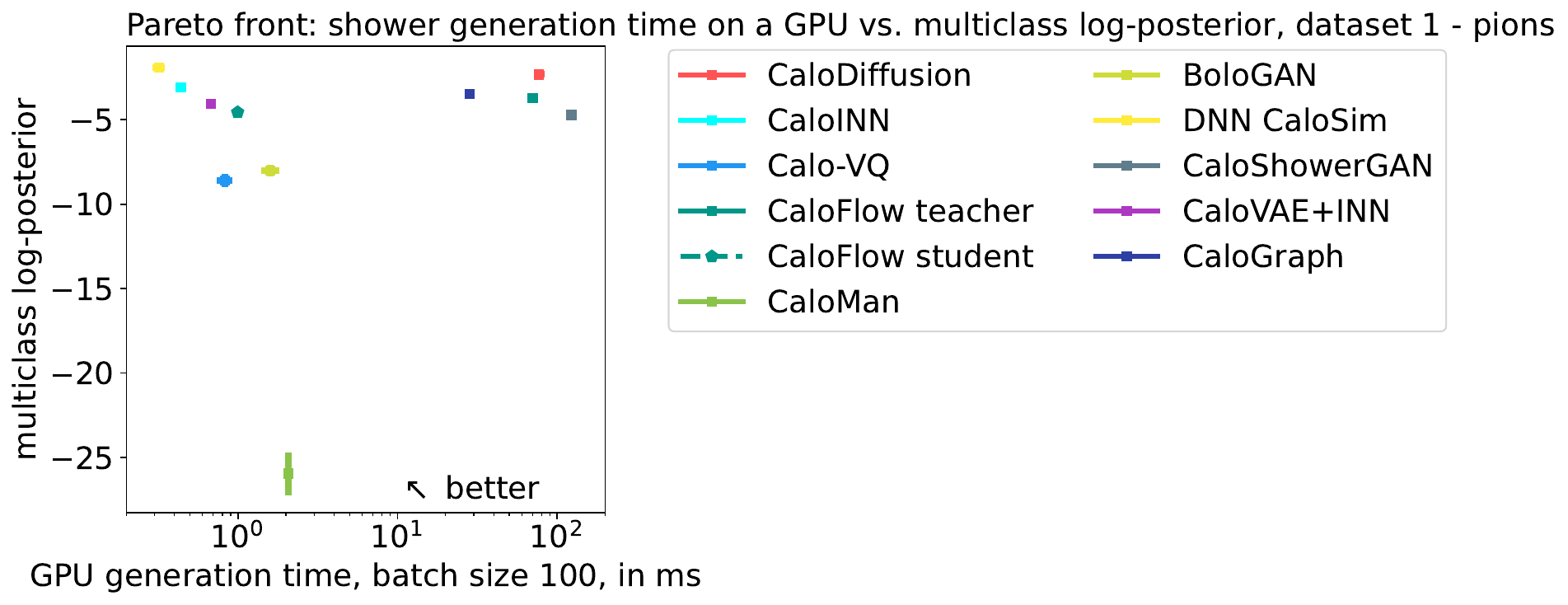}
\caption{Pareto front in sample quality (from \fref{fig:ds1-pions.multi} and \tref{tab:ds1-pions.multi}) and generation speed (from \fref{fig:ds1-pions.timing} and \tref{tab:ds1-pions.timing.GPU}).}
\label{fig:ds1-pions.pareto.quality.speed}
\end{figure}

\begin{figure}[ht]
\centering
\includegraphics[width=\textwidth]{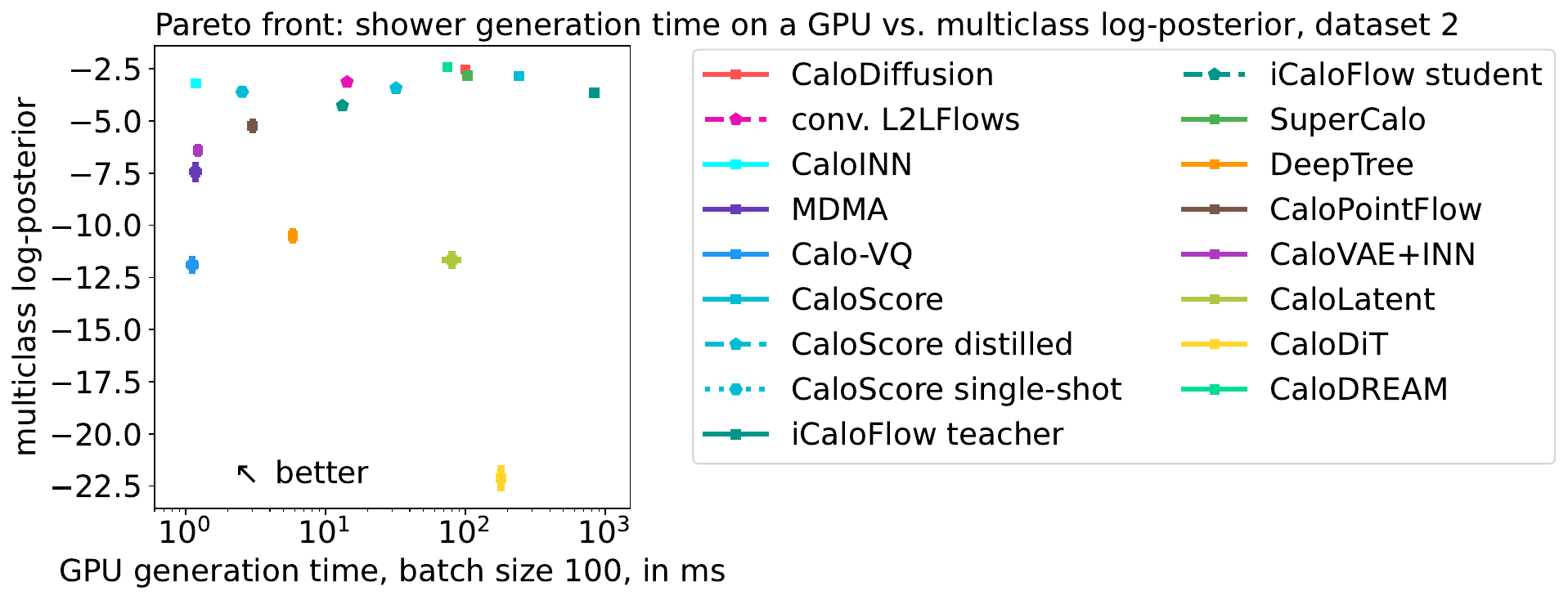}
\caption{Pareto front in sample quality (from \fref{fig:ds2.multi.dnn} and \tref{tab:ds2.multi.dnn}) and generation speed (from \fref{fig:ds2.timing} and \tref{tab:ds2.timing.GPU}).}
\label{fig:ds2.pareto.quality.speed}
\end{figure}

\begin{figure}[ht]
\centering
\includegraphics[width=\textwidth]{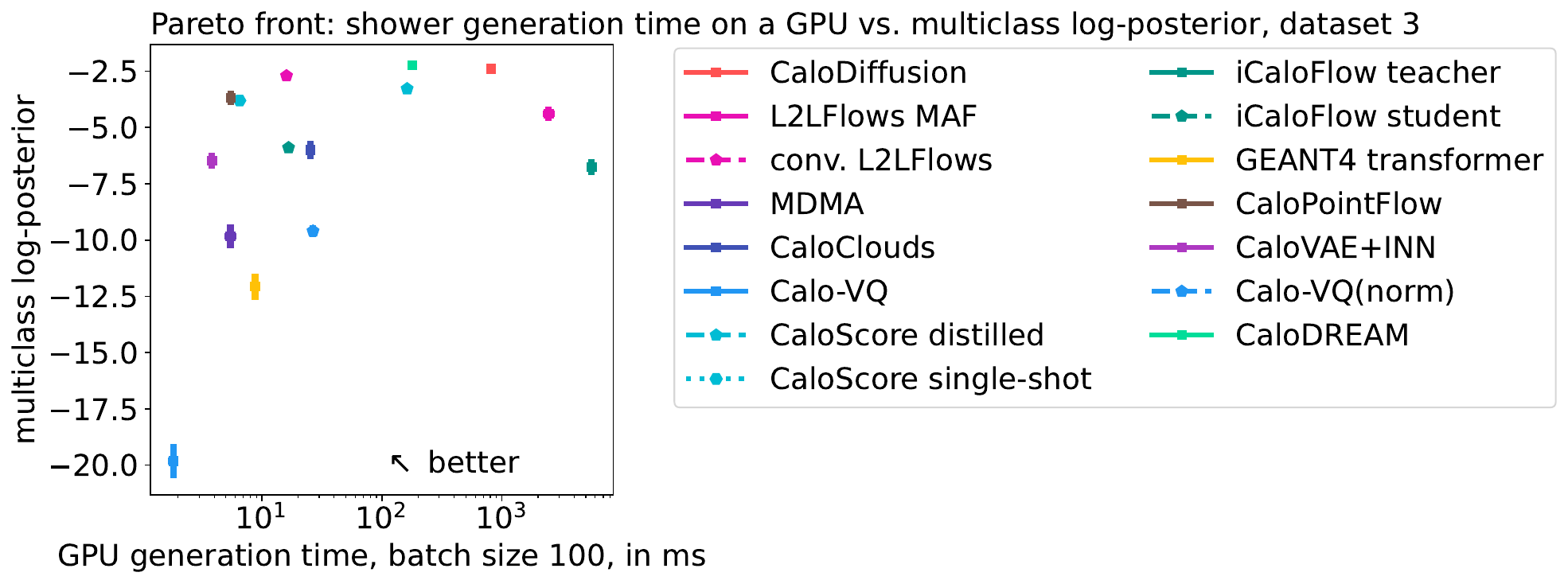}
\caption{Pareto front in sample quality (from \fref{fig:ds3.multi.dnn} and \tref{tab:ds3.multi.dnn}) and generation speed (from \fref{fig:ds3.timing} and \tref{tab:ds3.timing.GPU}).}
\label{fig:ds3.pareto.quality.speed}
\end{figure}
Next, we show the money plots in \fref{fig:ds1-photons.pareto.quality.speed}, \fref{fig:ds1-pions.pareto.quality.speed}, \fref{fig:ds2.pareto.quality.speed}, and \fref{fig:ds3.pareto.quality.speed}. Here, we compare the sample quality, measured by the DNN multiclass log-posterior, to the generation time, measured by the per-shower-time it takes to generate the entire dataset in batches of 100 on a GPU. 

For \dsIph, we truly see a trade-off between the two metrics in \fref{fig:ds1-photons.pareto.quality.speed}. On the one side, we have submissions with good sample quality, \textit{i.e.}~a high log-posterior and large generation time in the top-right corner. The submissions \submAmram, \submMikuni, \submKobylyansky, and \submPangT belong to this group. The distillations \submMikuniDist and \submMikuniSingle for a line to smaller generation times at the expense of a little shower quality, as we had seen in the individual metrics before. On the other side, we have submissions with lower log-posterior score, but a much faster generation time per shower. The VAE-based submissions \submReyes, \submLiu, and \submErnst belong to this group. In the corner with the best scores of both dimensions, we have the Normalizing Flow-based submission \submFavaro. Also \submPangS is close, indicating that the low-dimensional data of dataset 1 -- photons can be described well with normalizing flows and that a good choice for the architecture can also make the generation fast. 

With \dsIpi in \fref{fig:ds1-pions.pareto.quality.speed}, the situation is similar than with \dsIph, given that it has comparable dimensionality. We again observe a cluster of submissions in the top-right at good shower quality and large generation time. Again, these are the diffusion models \submAmram, \submKobylyansky, the MAF-based normalizing flow of \submPangT, and the GAN of \submZhang. Much faster, but also worse in quality we again see VAE and GAN-based models of \submLiu, \submRinaldi, and \submReyes. In the corner of fast generation of good showers, we see four submissions that actually form a line, making it easy to give them an order towards better showers at shorter times: From worst to best, these are \submPangS, \submErnst, \submFavaro, and \submSalamaniDNN. Out of these, we have two normalizing flow-based submissions with \submFavaro and \submPangS --- as we had before for the photon dataset. However this time, we also have two VAE-based submissions at the Pareto front: \submSalamaniDNN and \submErnst. This indicates that the larger shower-to-shower variability of pion showers is better captured by VAEs as the rather uniform photon showers we had before. 

Dataset 2 now increases the dimensionality of the samples by an order of magnitude. The Pareto front in \fref{fig:ds2.pareto.quality.speed}, however, does show similar features as we have seen for dataset 1 before. There is a group of diffusion and normalizing flow-based submissions in the top right with very high log-posterior scores, but also rather big generation times per shower. In this group, we have \submPalacios, \submAmram, \submMikuni, \icalo, \submPangSuper, and \submBussConv. At the other end of the spectrum, we have again fast submissions with worse log-posterior scores. In this group we have GAN-based submissions \submKaech and \submScham, and VAE-based submissions \submLiu and \submErnst. In the corner with both scores being good, we have three submissions: \submSchnake, \submMikuniSingle, and \submFavaro from ``worst'' to best. So also for this dataset, the normalizing flow-based submissions have the best trade-off between shower quality and generation speed. 

In dataset 3, the Pareto front in \fref{fig:ds3.pareto.quality.speed} is a little more diffuse, with the individual groups more spread out and no single submission in the best corner. Nevertheless, the similar general trends than before also apply. Diffusion models like \submAmram and \submMikuniDist, the CFM model \submPalacios, and normalizing flow-based submissions \submBussMAF and \submPangIT have good shower quality, but need longer to generate the showers. VAE and GAN-based submissions \submLiu, \submLiuNorm, \submSalamaniTrans and \submKaech are much faster in generation, but at the expense of shower quality. In the top-left corner, we see the remaining submissions. While \submKorol and \submPangIS are outperformed by \submErnst, \submMikuniSingle, \submSchnake and \submBussConv they still show a decent trade-off of quality and generation speed. The latter group now forms the Pareto front. The fastest among them is \submErnst. With a better shower quality, but at slightly bigger generation time, we have \submSchnake and \submMikuniSingle almost at the exact same spot, just slightly slower. Slowest of these four, but best in quality, is \submBussConv. For this high-dimensional dataset finding the optimal point really influences the choice of generative architecture, since this group consists of normalizing flows, a diffusion model, and a VAE.  

\FloatBarrier

\section{Conclusions and Outlook}
\label{sec:conclusions}
\markboth{\uppercase{Conclusions and Outlook}}{} 

In this document, we summarize the results of the Fast Calorimeter Simulation Challenge 2022. We present a broad survey of state-of-the-art generative AI architectures on four different calorimeter shower datasets with dimensionalities ranging from a few hundred to a few tens of thousand voxels. The data has a few physics-specific characteristics, like a high degree of sparsity, energy depositions in voxels spanning several orders of magnitude, and correlations between voxels across several layers, that are not present in natural images and other datasets. With about 15 submissions per dataset, and at least one submission for each type of generative architecture (GAN, VAE, Normalizing Flow, Diffusion, and Conditional Flow Matching) per dataset, this document provides the most detailed and complete survey of generative AI for high-energy physics. 

First announced in February 2022, the challenge quickly motivated the first publications using the dataset with \submMikuniCite, \cf~\cite{Krause:2022jna}, and \submReyesCite. More followed and were presented at the ML4Jets conference in November 2022 at Rutgers~\cite{ML4Jets2022}. While we first planned to close the challenge with the dedicated meeting in Frascati~\cite{CaloChallengeWorkshop} in May 2023, we saw a constant interest in the challenge with new submissions being presented at ML4Jets in Hamburg in November 2023~\cite{ML4Jets2023}. In total, we have received 59 submissions, sampled from 31 models, from 23 collaborations consisting of researchers from the theory and the experimental communities, as well as from outside academia. By now most of the submissions have been published by physics journals or ML conferences, highlighting the high quality of the individual works. 

While the main focus of this challenge was on generative models for calorimeter showers, with the requirements of the (HL)-LHC and future colliders in mind, many of the results will likely translate to other domains in high-energy physics in which generative AI is used as well, such as generative unfolding~\cite{Datta:2018mwd,Bellagente:2019uyp,Bellagente:2020piv,Vandegar:2020yvw,Howard:2021pos,Leigh:2022lpn,Backes:2022sph,Shmakov:2023kjj,Ackerschott:2023nax,Diefenbacher:2023wec,Butter:2023ira,Shmakov:2024gkd,Huetsch:2024quz}, modeling of hadronization effects~\cite{Ilten:2022jfm,Ghosh:2022zdz,Chan:2023ume,Bierlich:2023zzd,Chan:2023icm}, end-to-end simulations like flashsim~\cite{Vaselli:2024hml}, or anomaly detection with generative aspects~\cite{Hallin:2021wme,Raine:2022hht,Golling:2022nkl}.

\subsection{Overall Physics Results}
Since fast simulation frameworks are in the ideal case faithful, fast, and light-weight, it was expected that with such ambitious objectives there will be no clear winner of the CaloChallenge. Instead, our goal was to create a survey of different generative architectures, their advantages and disadvantages, and especially their scaling behavior when increasing the dimensionality of the dataset. Ultimately, the objectives of experiments will differ, with some in need of high fidelity simulation, others prioritizing the speed, trading off physics accuracy to a certain extent.

For low-dimensional datasets, \textit{i.e.}~dataset 1 -- photons with 368 voxels, we saw that diffusion models like \submAmramCite and Normalizing Flow-based models like \submFavaroCite have the best quality, meaning they reproduce the \geant distribution most faithfully. The diffusion model has a rather small number of trainable parameters, so it's also more lightweight than the normalizing flow, but since generation requires multiple steps and calls to the neural network, the diffusion model is much slower in generation. The invertible architecture of \submFavaro does not require multiple calls to the same neural network, it also avoids a more resource-consuming distillation step like \cf, making \submFavaro an optimal submission for \dsIph, see~\fref{fig:ds1-photons.pareto.quality.speed}. GAN and VAE-based architectures are in general also fast in generation, but do not produce high-quality showers, making them less favorable if high fidelity is top priority.

For dataset 1 -- pions, the situation is very similar. With 533 voxels, the dataset is still relatively low-dimensional, so normalizing flows and diffusion models, namely \submFavaroCite and \submAmramCite show again a good performance. In addition, the VAE-based submission \submSalamaniDNNCite showed great performance in many of the quality metrics. The statement regarding model size and generation speed of dataset \dsIph also applies here. The diffusion model does not need a lot of trainable parameters, which makes it very lightweight. Generation speed, however, is lower due to the subsequent denoising steps in generation. The normalizing flow, on the other hand, is growing at least linearly in size with the dimensionality of the dataset, so it is now already about 1.4 times bigger than for the photons. Nevertheless, it is still very fast in generation, and, at least on the GPU, only marginally behind the well-performing submission \submSalamaniDNN. This VAE-based model is the fastest in generation on the CPU for all batch sizes and is only beaten marginally on the GPU for very large batch sizes. It is midrange in terms of model size, with some GANs having fewer parameters and most normalizing flows having more. It is also interesting to note that \submSalamaniDNN had the best scores in the low-level binary AUC and the multiclass log-posterior, but had worse scores for KPD and FPD, as well as the separation powers we looked at. In these latter cases, \submAmram, \submPangSCite, and \submKobylyanskyCite performed better than \submSalamaniDNN. 

Dataset 2 now increases the dimensionality by an order of magnitude to 6480. This also increases the number of parameters in the so-far well-performing normalizing flow of \submFavaroCite by an order of magnitude to about $2.7 \cdot 10^8$, making it the largest submission for dataset 2. Nevertheless, it still gives the best trade-off in quality and generation speed in~\fref{fig:ds2.pareto.quality.speed}. In terms of quality alone, the diffusion models \submAmramCite and \submMikuniCite as well as the conditional flow matching model \submPalaciosCite have better multiclass log-posteriors, KPD/FPD, and binary AUCs. However, in generation all of these require multiple steps and hence they are slower than \submFavaro. With the use of distillation, \submMikuni was able to speed up generation generation times by an order of magnitude to \submMikuniDist and another order of magnitude to \submMikuniSingleCite at the expense of a little shower quality. Similar techniques can also be applied to \submAmram and \submPalacios, which would bring them closer to \submFavaro in~\fref{fig:ds2.pareto.quality.speed}. In terms of model size, \submKaechCite needed by far the fewest parameters, making it also the fastest in generation, especially for small batch sizes.

Dataset 3 increases the complexity of the showers by another order of magnitude, to 40500 voxels. This was too big for the bijector in \submFavaroCite, so it was not submitted to this dataset. Diffusion and conditional flow matching models \submAmramCite, \submMikuniDistCite, and \submPalaciosCite show again the best shower quality, but not the fastest generation. Splitting the entire shower into individual calorimeter layers makes the problem again manageable for a normalizing flow, as can be seen by the good shower quality of \submBussConvCite. In terms of model sizes, \submKaechCite again is the smallest submission, followed by the VAE-based model \submLiuCite and its variant \submLiuNormCite and then \submAmram and \submPalacios. In terms of generation speed, GAN-based submission \submKaech and VAE-based submissions \submLiu and \submErnstCite are the fastest, which is correlated to the model sizes. When looking at the trade-off between quality and speed in~\fref{fig:ds3.pareto.quality.speed}, we see four submissions competing with each other. Fastest, but worst in quality of those four is \submErnst. In the center, we have \submSchnakeCite and \submMikuniSingleCite, and slowest, but best in quality, is \submBussConv. The potential speed-up of \submAmram and \submPalacios with model distillation, as discussed at the end of the dataset 2 paragraph, also applies here.   

Summarizing, there is no single submission that excels in all three types of metrics: speed, quality, and size. Normalizing Flows show the best trade-off in sample quality and generation speed, but since they have to train a bijective mapping they do not scale well to higher dimensional datasets. Diffusion and conditional flow matching models have the highest sample quality, but suffer from a slow generation process. GAN and VAE-based submissions have fewer trainable parameters and are usually very fast in generation, but that comes at the expense of shower quality. Even the best performing model is not perfect for the high-dimensional datasets 2 and 3, so there is still a lot of room for improvement in generative architectures to be even more faithful and resource-efficient in the future. 

Model distillation improves speed at expense of quality and we have seen some submissions that use this technique already, while it could be applied to others, too. Techniques like weight quantization or node pruning can have a large effect on the resource requirements with some or little effect on the sample quality. This has not been studied here and should be investigated more in the future.

\subsection{Take-aways of the CaloChallenge beyond Detector Simulation }
This challenge triggered the development and adaptation of a lot of generative architectures to high-dimensional calorimeter shower data, leading to more than 20 publications in physics and ML journals or conferences, as well as talks at the central machine learning conference in particle physics, ML4Jets~\cite{ML4Jets2022,ML4Jets2023} and other specialized workshops~\cite{CaloChallengeWorkshop}. This collaborative effort was done by experimentalists, theorists, and scientists working outside academia in industry alike, but mostly outside of the big collaborations ATLAS and CMS. We hope that the presented results are useful for the experiment-specific development of fast simulation frameworks in the future. 

The challenge also provided four datasets that will now serve as benchmarks for future generative models. Despite the large body of results we reported, there are a few questions that this challenge cannot answer. For example, some submissions (\submSchnakeCite, \submKaechCite, \submSchamCite, and \submKorolCite) worked with point clouds instead of with the voxelized data that we provided. Since they did not have access to the hits that \geant simulated before we voxelized the data, they had to rely on suboptimal methods to create the point clouds. Further studies that directly use the point clouds coming from \geant are needed to understand if that had an effect on shower quality. 

Something else we noticed but were unable to disentangle and study in detail was the effect of model distillation. \submMikuniDist and \submMikuniSingle were distilled from \submMikuni and \submPangS and \submPangIS were distilled from \submPangT and \submPangIT respectively. Since both, samples from the original model and samples from the distilled version of the same original model were submitted, there is a correlation between the scores of these submissions. This is most visible in the multiclass classification metric, where original and distilled model were sometimes confused with each other (see for example \fref{fig:consistency.ds2}). We also see it in the Pearson correlation coefficients of layer energies in \fref{fig:ds2-corr} and \fref{fig:ds3-corr}, where the distinct pattern of \submMikuni got worse with distillation. The situation is, however, different for \icalo, where the pattern got fainter with distillation. Other metrics were also sometimes better, sometimes worse in distilled versions, as previously seen also in~\cite{Krause:2021wez}. We suspect that a smoothing that takes place in distillation can improve an incorrectly learned feature. One way of disentangling such effects would be to train multiple instances of the original model and use one for sample generation of the submission and the other one for training the distilled model. 

The irregular geometry of datasets 1 posed a special challenge, in particular for models that were using 3-dimensional convolutions. While for example \submBussConv decided not to work on \dsIph and \dsIpi for that reason, \submAmram came up with a special solution to the problem. It also triggered some dedicated approaches for irregular geometries, like for example \submKobylyansky. 

In addition, we also gained more insights in the evaluation of generative models for physics applications. We studied how different quality metrics, motivated by physics or coming from computer science, correlate with each other. 

\subsection{Outlook to the Future}
\label{sec:outlook}

To serve as a benchmark for future developments in calorimeter simulation, we collected the raw data that went into all the figures of \sref{sec:results} and tables in \sref{app:tables} in a pandas~\cite{jeff_reback_2021_5774815} dataframe that we publish together with the jupyter notebook~\cite{Kluyver2016jupyter} required to reproduce the figures on the GitHub page of the CaloChallenge~\cite{calochallenge}. The 59 submitted samples are available for download at~\cite{faucci_giannelli_2025_15961728,faucci_giannelli_2025_15961924,faucci_giannelli_2025_15962050,faucci_giannelli_2025_15962527}. 

For a better understanding on the resource requirements and best working point in the shower quality \textit{vs.}~generation speed trade-off, a full end-to-end implementation in fast simulation frameworks of experiments is needed. Since the generation times improve a lot for generating showers in batches, this should also be taken into account properly. In that sense, the results presented here focus only on one single step of the full fast simulation chain. The produced showers still need to be projected back into the detector geometry and the generative model needs to be embedded in the appropriate software framework. These additional constraints go beyond the scope of this challenge, but they are required to get a full picture on the impact of generative AI in making the simulation faster. It could therefore very well be that sacrificing a little performance for a better speed, or sacrificing some speed advantage for a better performance is more beneficial when looking at the end-to-end performance. Also, conditioning on more initial conditions, like for example the incident angle \textit{vs.}~training more individual models can only be evaluated in a more complete framework. Another practical question that arises is the computing architecture which will run the final fast simulation. While inference clearly benefits GPU utilization and large batch sizes, this must be well incorporated in experiments' computing workflows so that speed-up factors can be maximized. In any case, further studies in all of these directions are therefore needed, and the corresponding results are applicable well beyond (HL)-LHC. 

It is also important to stress that while many figures of merit are presented in this work, any experiment should not simply pick a technology but should carry out a careful evaluation of several models. The granularity of the calorimeters, the geometry of the cells in the sub-systems and the overall detector geometry will impose constraints on which models can be used. For example, ATLAS trains and runs 100 models (one per $\eta$ slice) but has a relatively low detector granularity while the CMS high-granularity calorimeter (HGCAL) covers only a small region of the detector but has a much higher granularity. Therefore, ATLAS may struggle to handle hundreds of models with many parameters but will be less affected by poor modeling of the shape, making some of the less performing models better candidates. For the HGCAL the opposite is true, although in this case the complex geometry of the calorimeter may require additional studies for the voxelisation strategy. What needs to be mentioned here is an important work of the LHCb on the implementation of the workflow presented in the Par04 example of \geant into their simulation framework, featuring a Par04-inspired VAE model~\cite{Gaussino_ACAT}. This allows them to test any of the models submitted to the CaloChallenge, looking not only at the simulation level observables, but at the broader spectrum of important variables that are typically a part of validation chain. 

To summarize, we are very excited to have received so many different submissions to the CaloChallenge. We now have a full toolbox with publicly available models as well as a detailed set of comparisons of several different approaches for experiments and other interested users to try out. It will be highly exciting to see how these methods evolve in the future and how they are deployed in experiments, expanding our understanding of Nature by improved simulation techniques!

\ack
We thank James Tuan for the IT support at Rutgers University, where the intermediate evaluations and preparations were run. The main computational results presented were obtained using the CLIP cluster (\url{https://clip.science}).

Oz Amram is supported by the U.S. CMS Software and Computing Operations Program under the U.S. CMS HL-LHC R\&D Initiative.
Erik Buhmann, Thorsten Buss, Gregor Kasieczka, and William Korcari are supported by the Deutsche Forschungsgemeinschaft under Germany’s Excellence Strategy — EXC 2121 Quantum Universe — 390833306 and via the KISS consortium (05D23GU4, 13D22CH5) funded by the German Federal Ministry of Education and Research BMBF in the ErUM-Data action plan.
Erik Buhmann further acknowledges funding through a scholarship by the Friedrich Naumann Foundation for Freedom.
The work of Florian Ernst has been sponsored by the Wolfgang Gentner Programme of the German Federal Ministry of Education and Research (grant no. 13E18CHA).
Michele Faucci Giannelli received funding from the European Union’s Horizon 2020 research and innovation programme under the Marie Skłodowska-Curie grant agreement No 754496.
Luigi Favaro and Ayodele Ore are supported by the Deutsche Forschungsgemeinschaft (DFG, German Research Foundation) under grant 396021762 -- TRR~257: \textsl{Particle Physics Phenomenology after the Higgs Discovery}.
Benno Käch is funded by Helmholtz Association's Initiative and Networking Fund through Helmholtz AI (grant number: ZT-I-PF-5-64).
Dmitrii Kobylianskii, Nathalie Soybelman, Etienne Dreyer, and Eilam Gross are supported by the Israel Science Foundation (ISF), Grant No. 2871/19 Centers of Excellence and BSF-NSF Grant No. 2020780. They are also grateful for the support provided by the collaborative Weizmann Institute and Mohamed bin Zayed University of Artificial Intelligence (MBZUAI) research grant, as well as the Benoziyo Center for High Energy Physics.
Marco Letizia acknowledges the financial support of the European Research Council (grant SLING 819789).
Qibin Liu and Shu Li acknowledge the support from National Key R$\&$D Program of China (Grant No.: 2023YFA1606904 and 2023YFA1606900), National Natural Science Foundation of China (Grant No.: 12150006), and Shanghai Pilot Program for Basic Research—Shanghai Jiao Tong University (Grant No.: 21TQ1400209).
Benjamin Nachman is supported by the U.S. Department of Energy (DOE), Office of Science under contract DE-AC02-05CH11231.
Sofia Palacios Schweitzer is supported by the BMBF Junior Group Generative Precision Networks for Particle Physics (DLR 01IS22079).
Matthew Buckley, Ian Pang, and David Shih are supported by the U.S. Department of Energy (DOE), Office of Science grant DOE-SC0010008.
Kevin Pedro and Oz Amram are supported by Fermi Research Alliance, LLC under Contract No. DE-AC02-07CH11359 with the U.S. Department of Energy, Office of Science, Office of High Energy Physics.
Piyush Raikwar, Dalila Salamani, and Anna Zaborowska were supported by the CERN Strategic R$\&$D Programme on Technologies for Future Experiments and have received funding from the European Union’s Horizon 2020 Research and Innovation programme under Grant Agreement No. 101004761.
Humberto Reyes-Gonzalez is supported by the Deutsche Forschungsgemeinschaft (DFG, German Research Foundation) under grant 396021762 -- TRR 257: Particle Physics Phenomenology after the Higgs Discovery. He also acknowledges the support from the Italian PRIN grant 20172LNEEZ.
Moritz Scham is funded by Helmholtz Association's Initiative and Networking Fund through Helmholtz AI (grant number: ZT-I-PF-5-3). 
Rui Zhang is supported by US High-Luminosity Upgrade of the Large Hadron Collider (HL-LHC) under Work Authorization No.\ KA2102021.

\appendix
\markboth{\uppercase{Appendix}}{}
\section{Histograms of high-level features}
\label{app:histograms}
Here we show the histograms of the high-level features that were used to compute the separation powers in section~\ref{sec:results_hlf_intro}. In particular, we show the distributions of the \geant training and evaluation datasets. The exact same binning was chosen to compute the separation powers with~\eref{eq:sep.power}. 
\subsection{\texorpdfstring{Dataset 1, Photons (\dsIph)}{Dataset 1, Photons}}
\begin{figure}[ht]
    \centering
    \includegraphics[height=0.2\textheight]{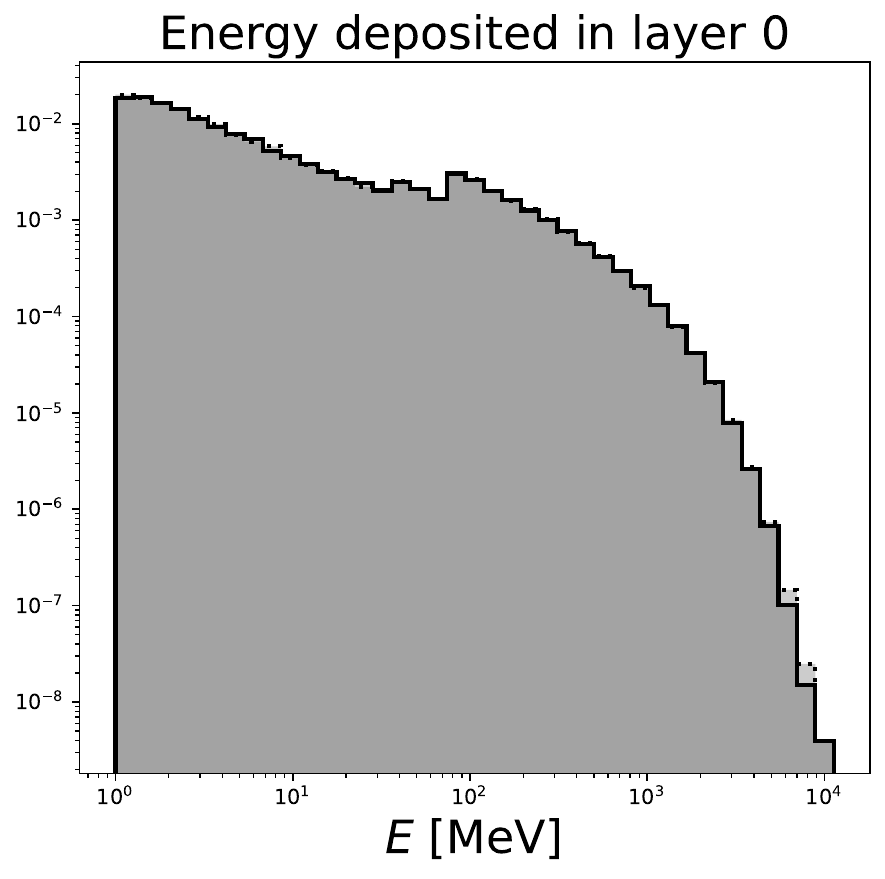} \hfill     \includegraphics[height=0.2\textheight]{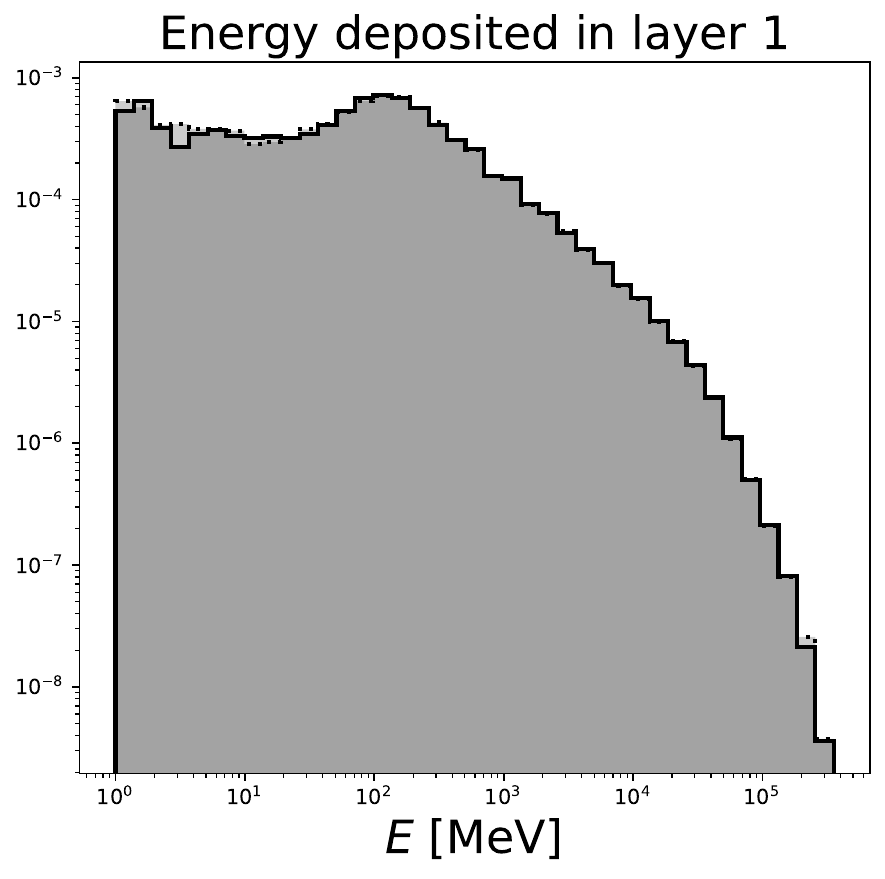} \hfill     \includegraphics[height=0.2\textheight]{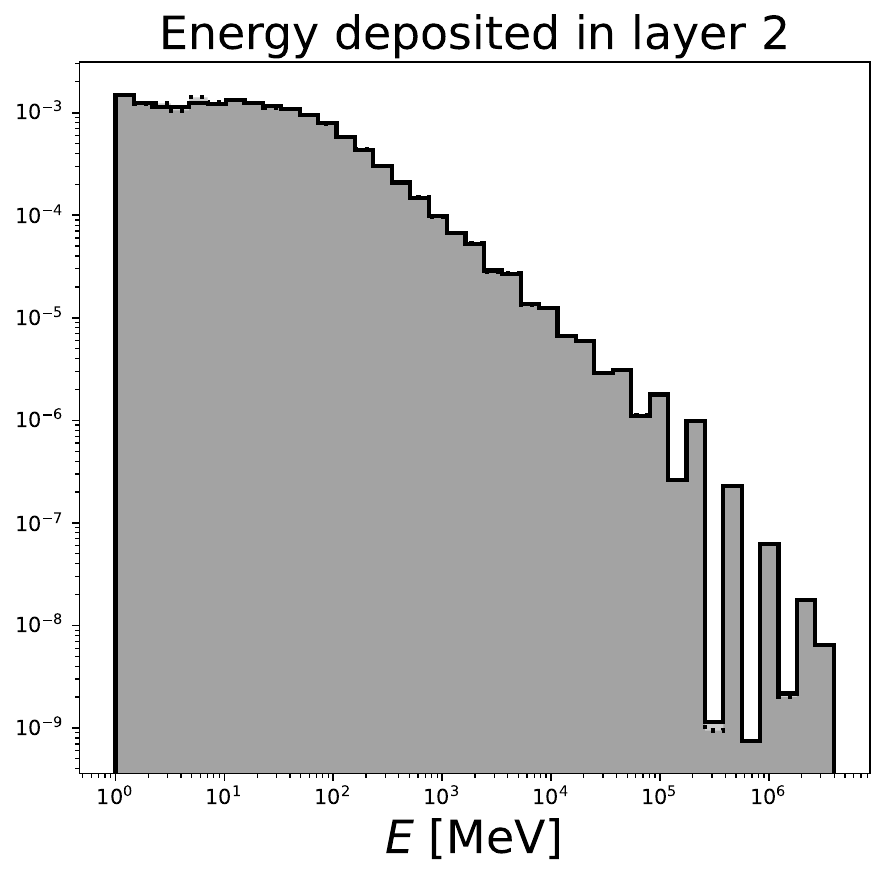} \\
    \includegraphics[height=0.2\textheight]{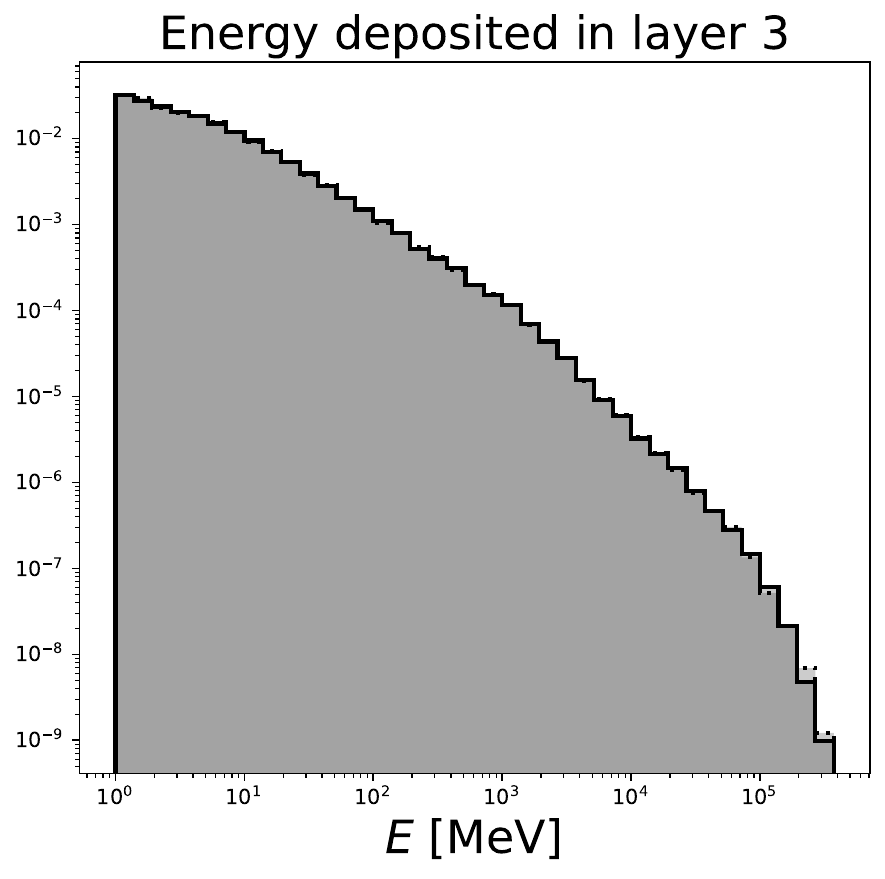} \hfill     \includegraphics[height=0.2\textheight]{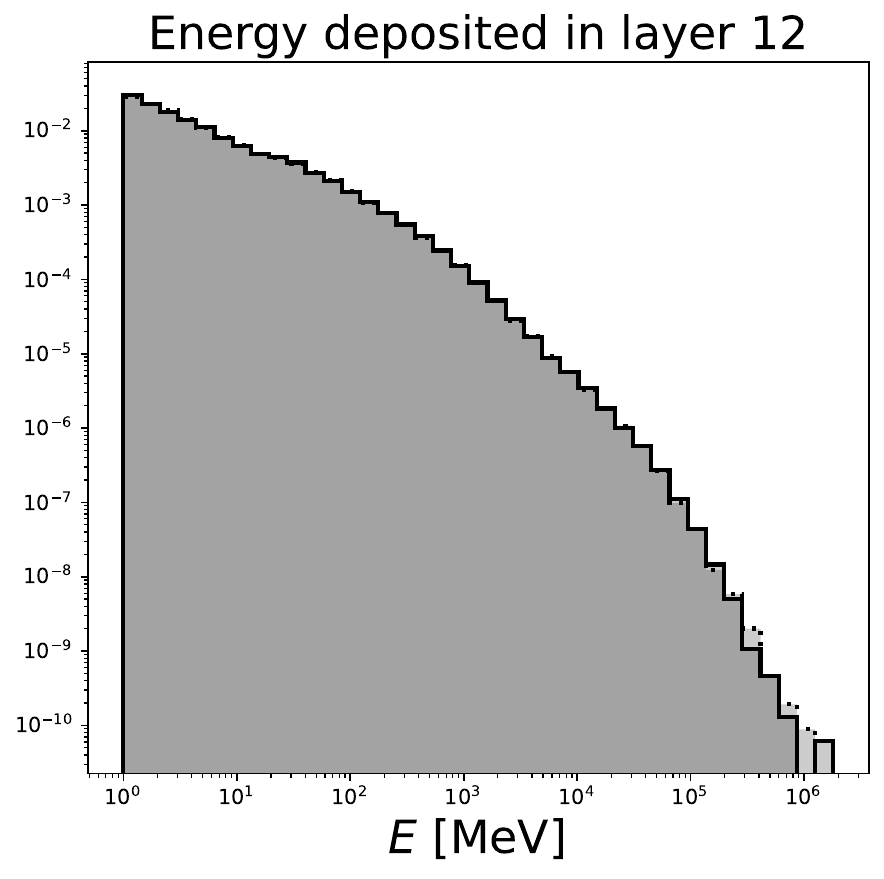} \hfill     \includegraphics[height=0.2\textheight]{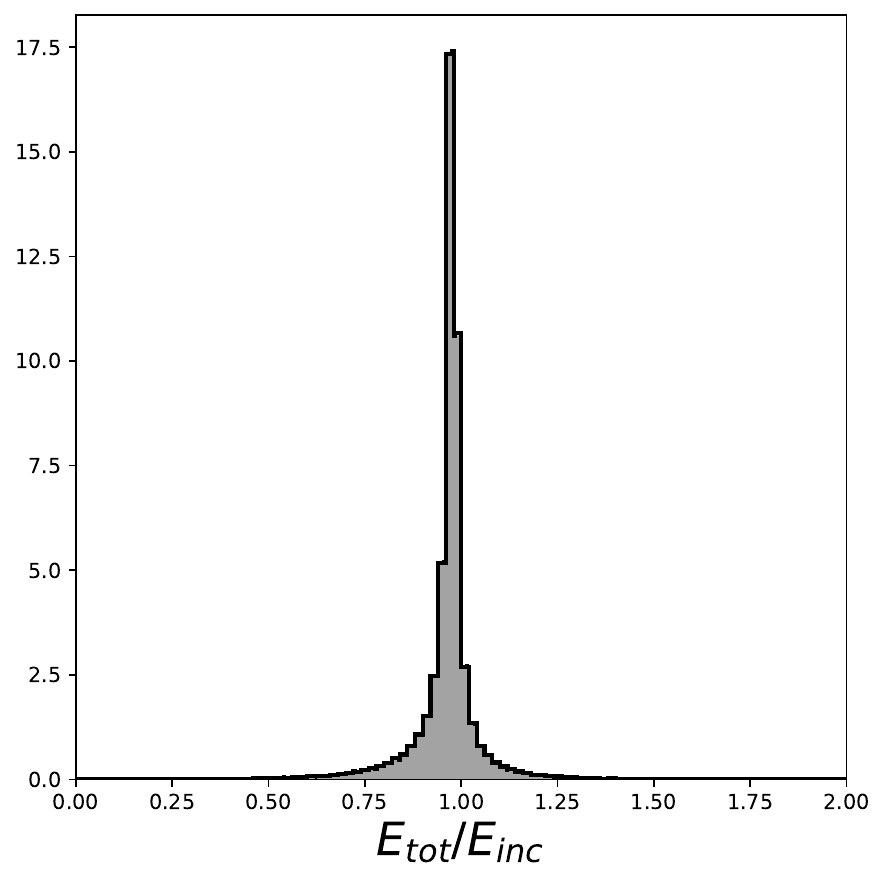}\\
    \includegraphics[height=0.2\textheight]{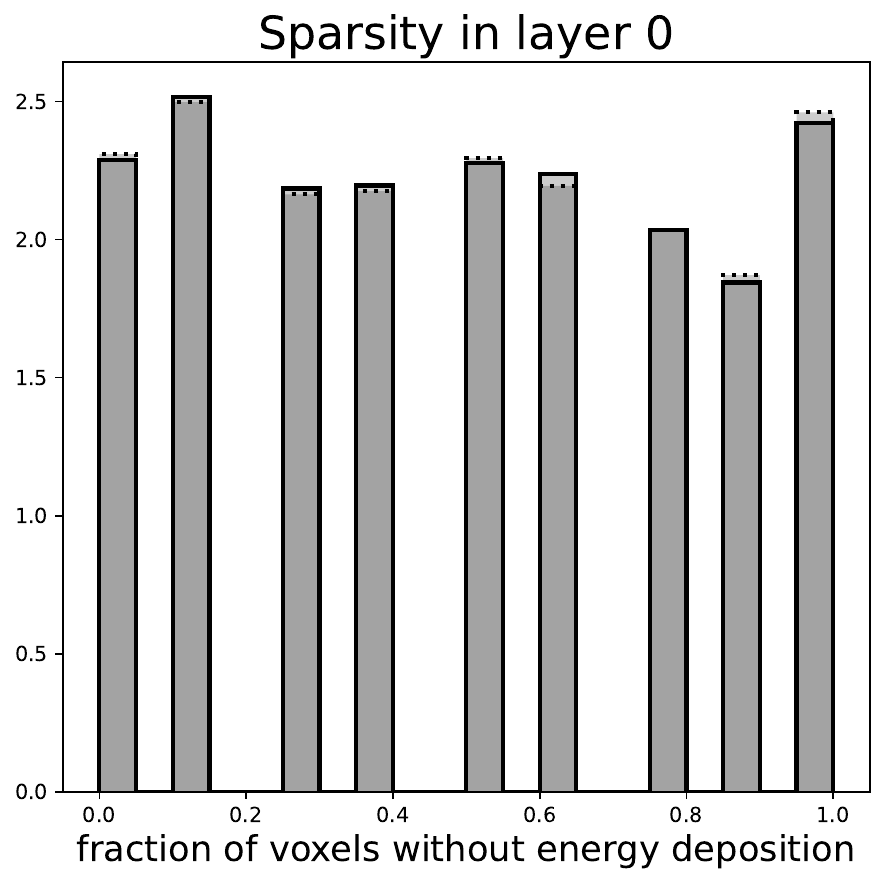} \hfill     \includegraphics[height=0.2\textheight]{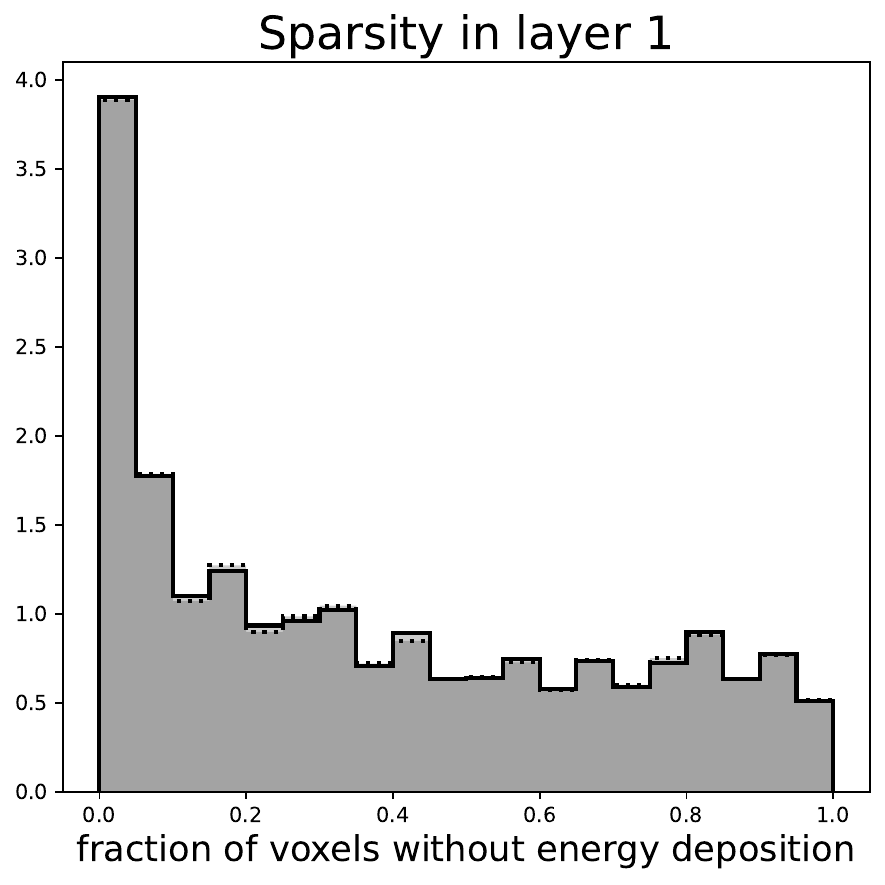} \hfill     \includegraphics[height=0.2\textheight]{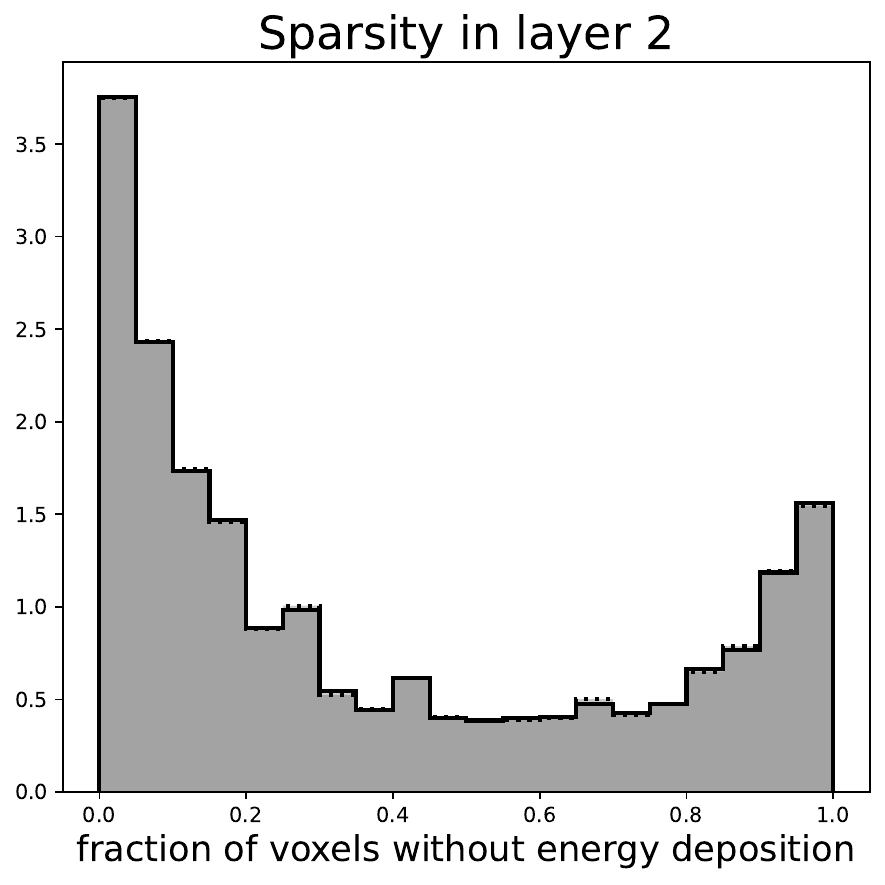} \\
    \includegraphics[height=0.2\textheight]{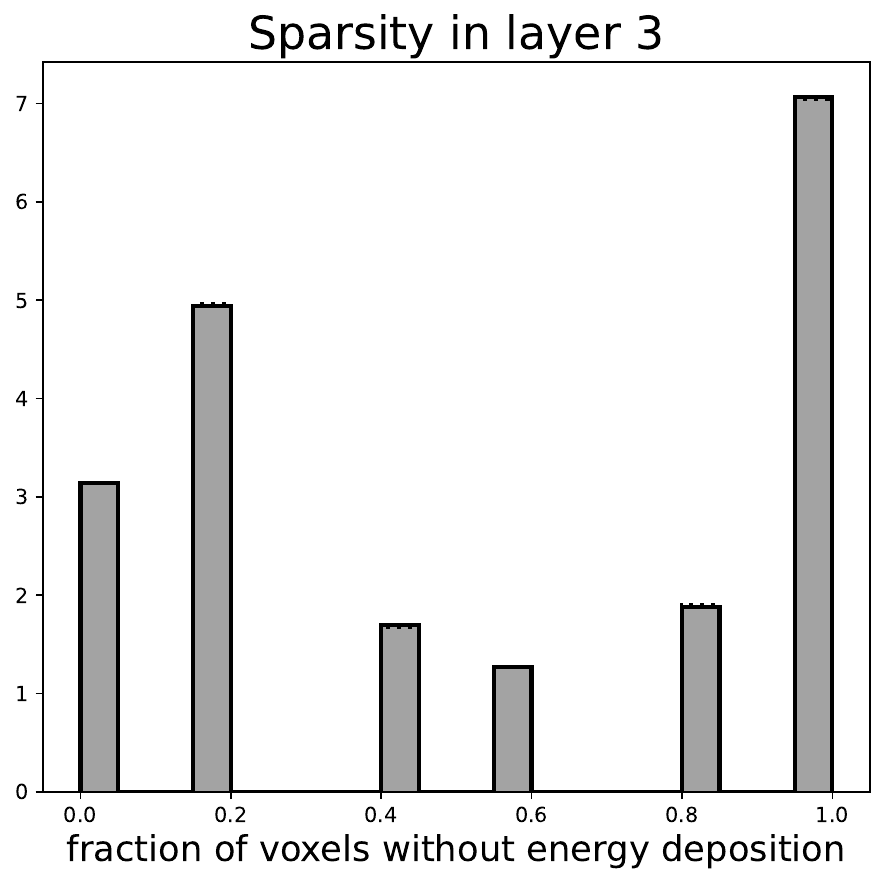} \hfill     \includegraphics[height=0.2\textheight]{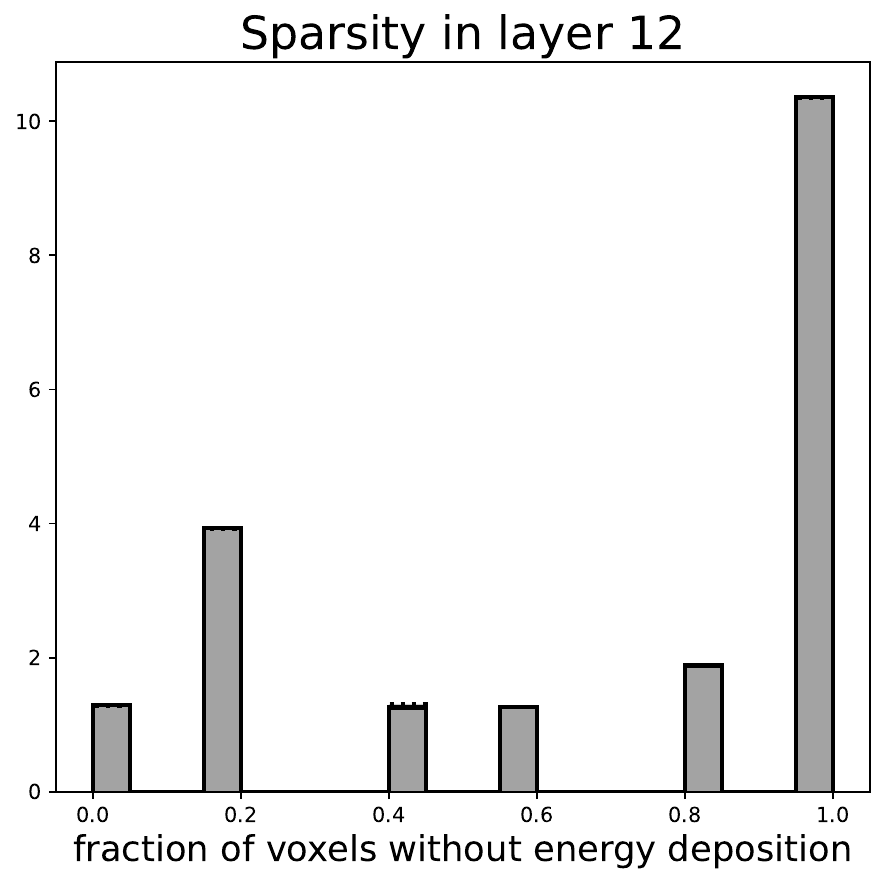} \hfill     \includegraphics[height=0.2\textheight]{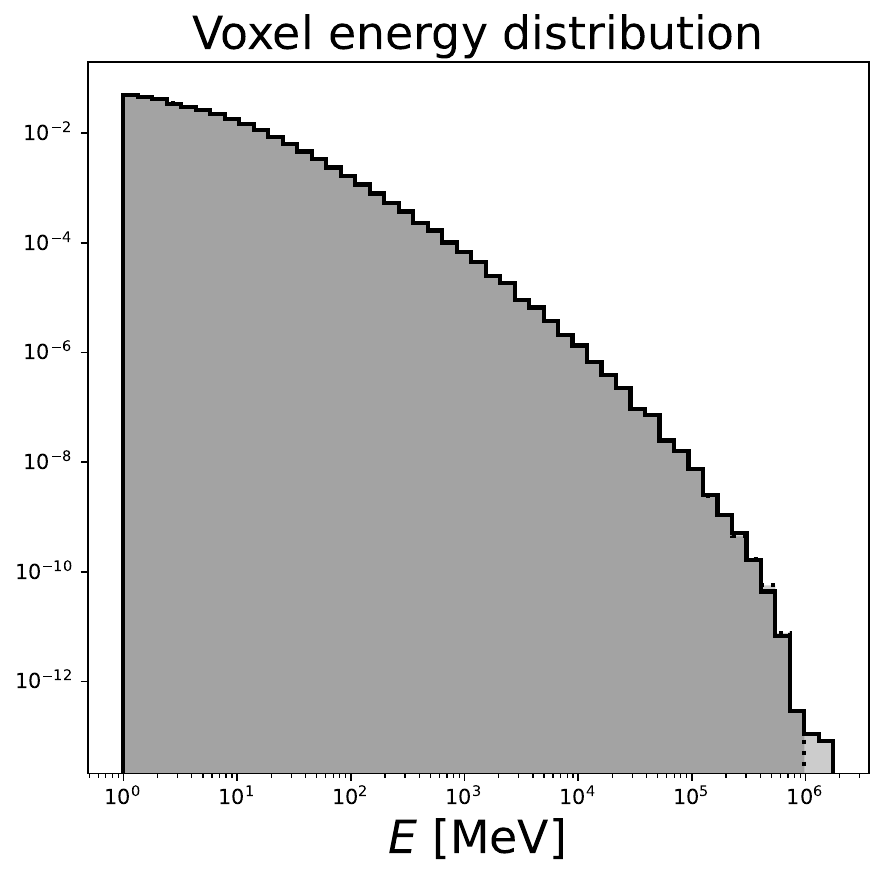}\\
    \includegraphics[width=0.5\textwidth]{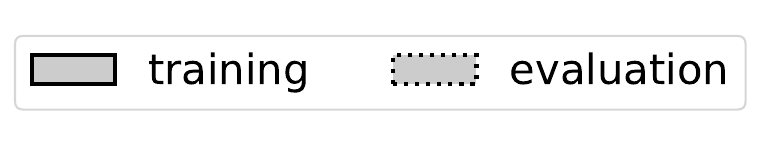}
    \caption{Distribution of \geant training and evaluation data in layer energies $E_i$, ratio of total deposited energy to incident energy, sparsity, and energy per voxel for ds1 --- photons.}
    \label{fig:app_ref.ds1-photons.1}
\end{figure}

\begin{figure}[ht]
    \centering
    \hfill \includegraphics[height=0.2\textheight]{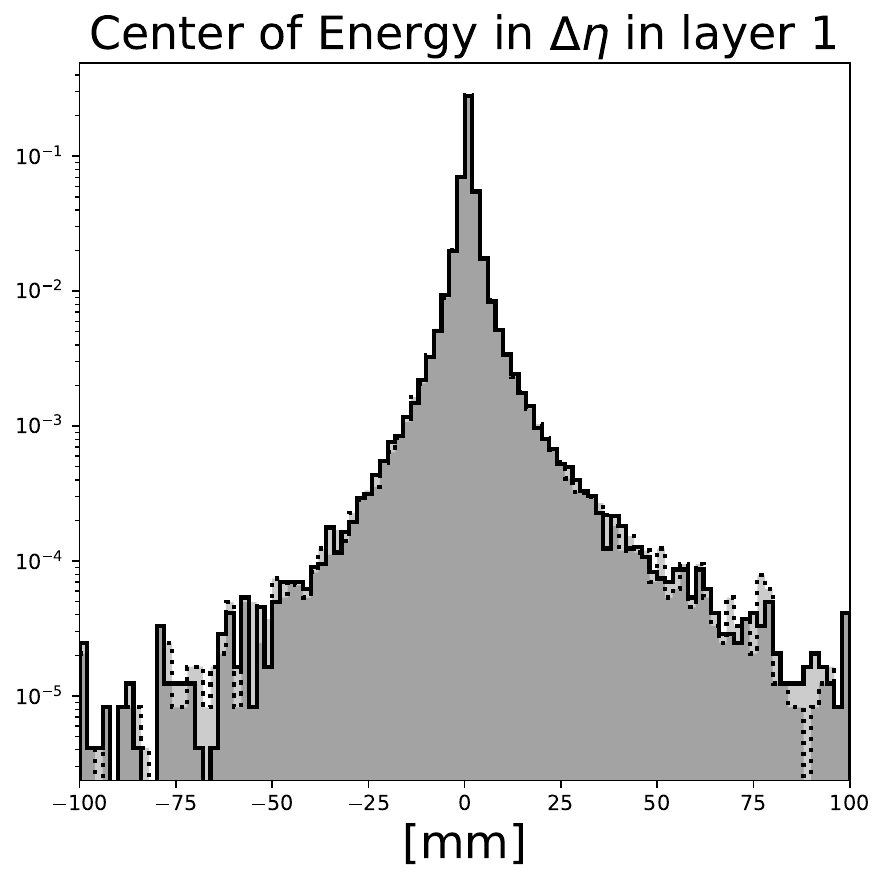} \hfill     \includegraphics[height=0.2\textheight]{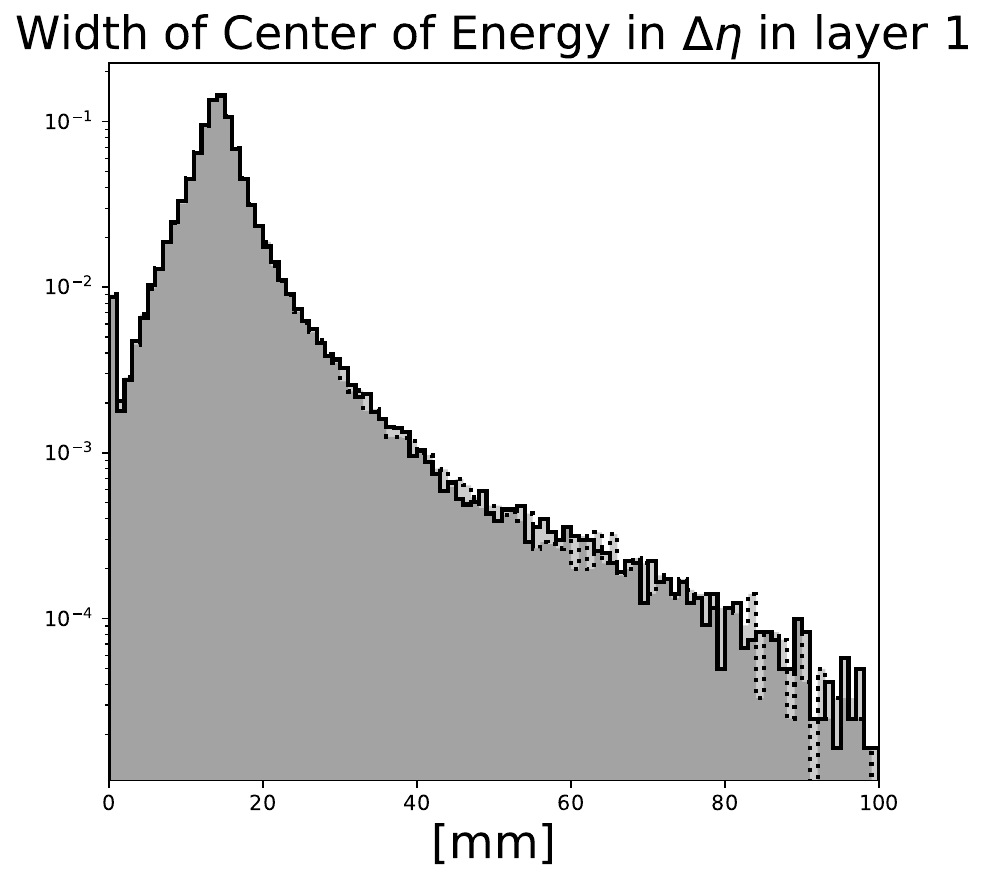} \hfill $ $\\
    \hfill \includegraphics[height=0.2\textheight]{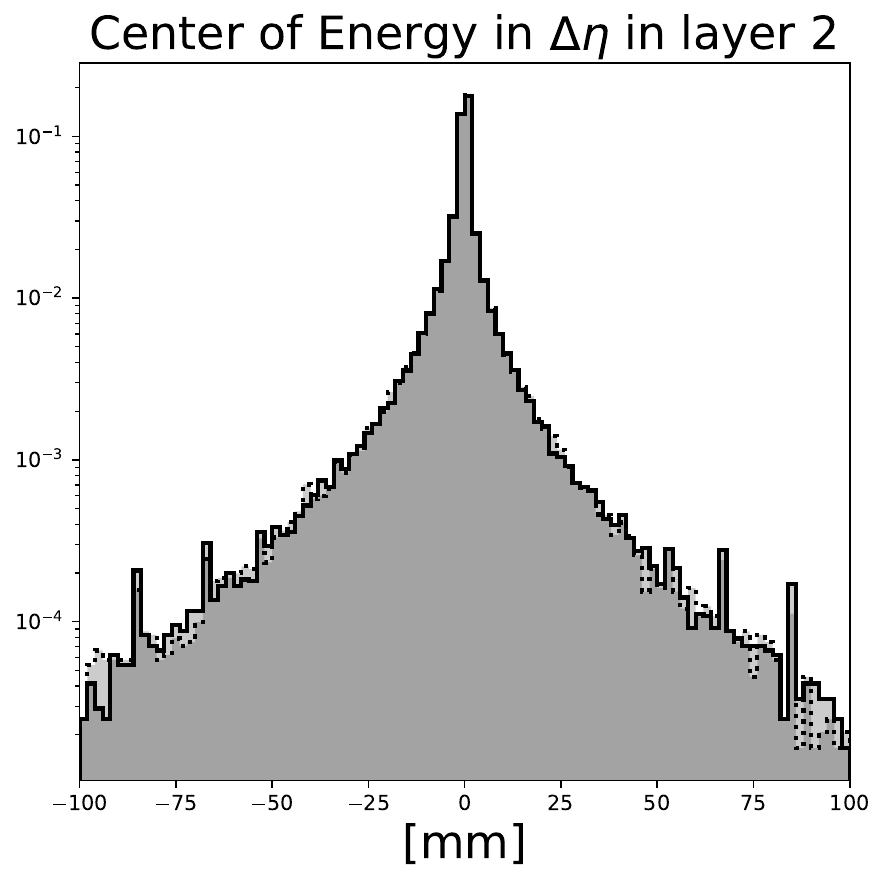} \hfill     \includegraphics[height=0.2\textheight]{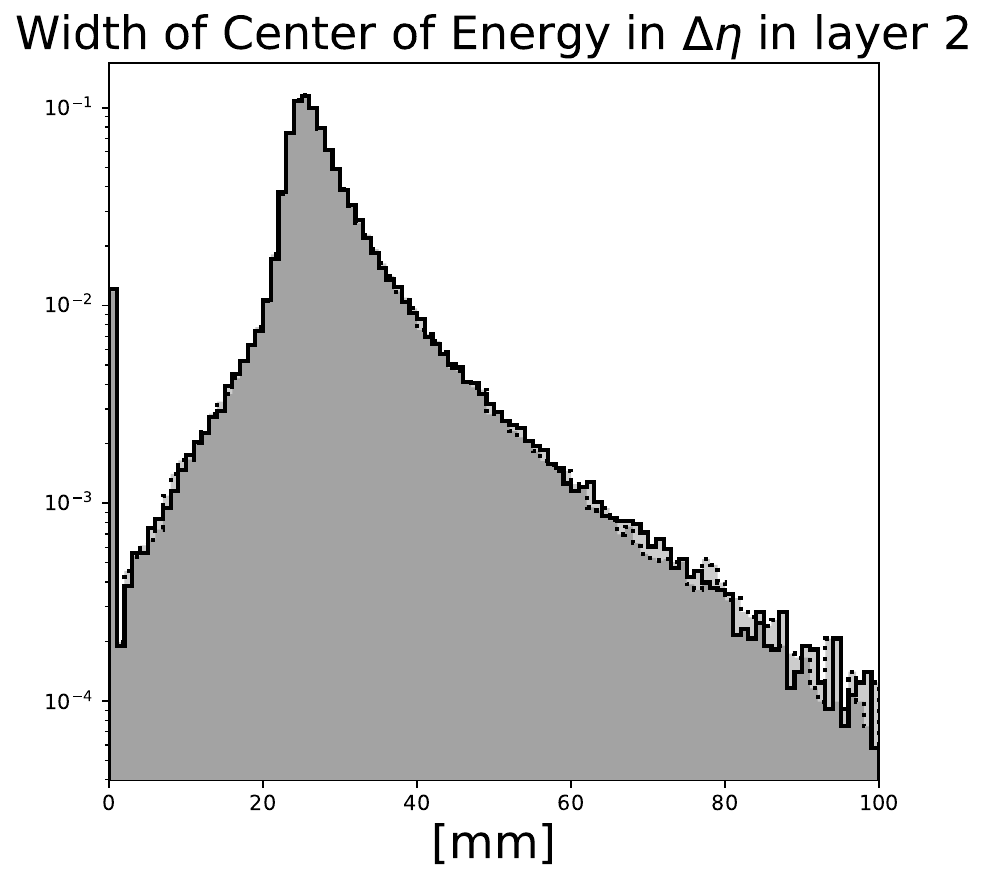} \hfill $ $\\
    \hfill\includegraphics[height=0.2\textheight]{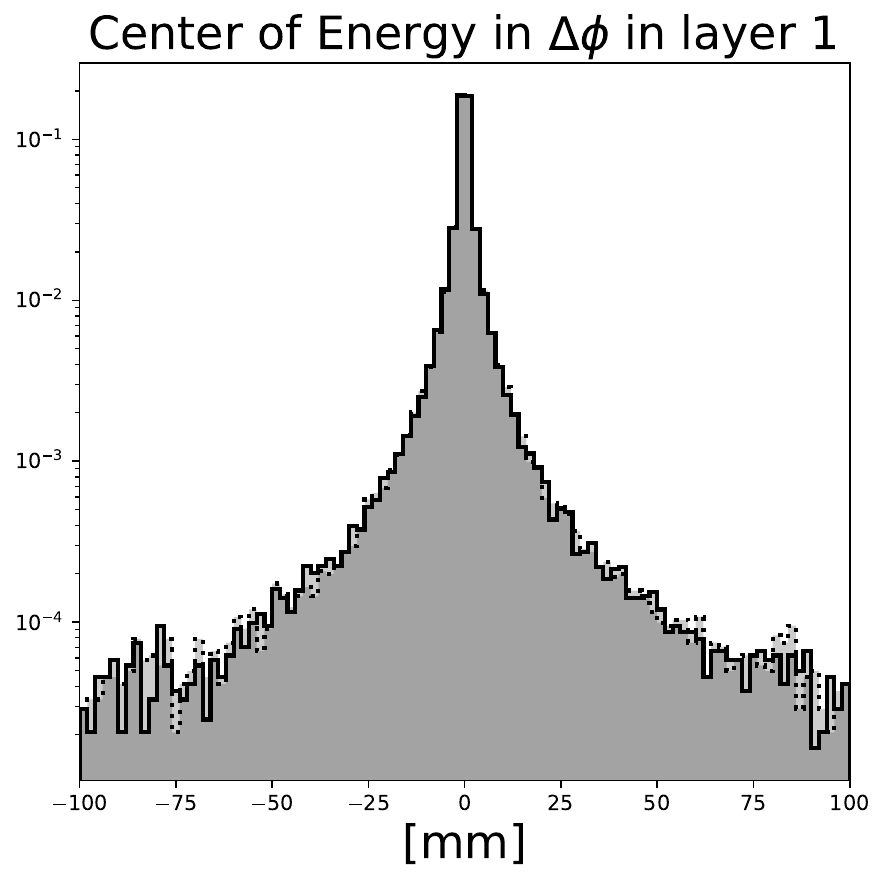} \hfill     \includegraphics[height=0.2\textheight]{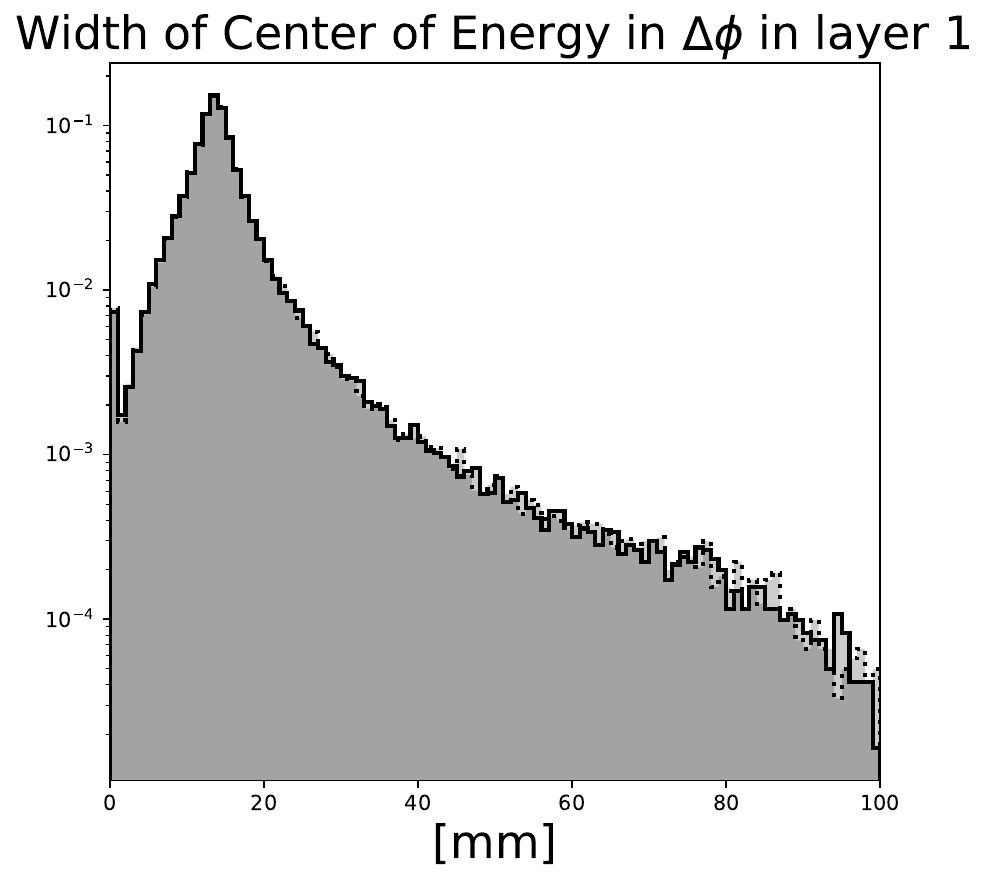}\hfill $ $\\
    \hfill\includegraphics[height=0.2\textheight]{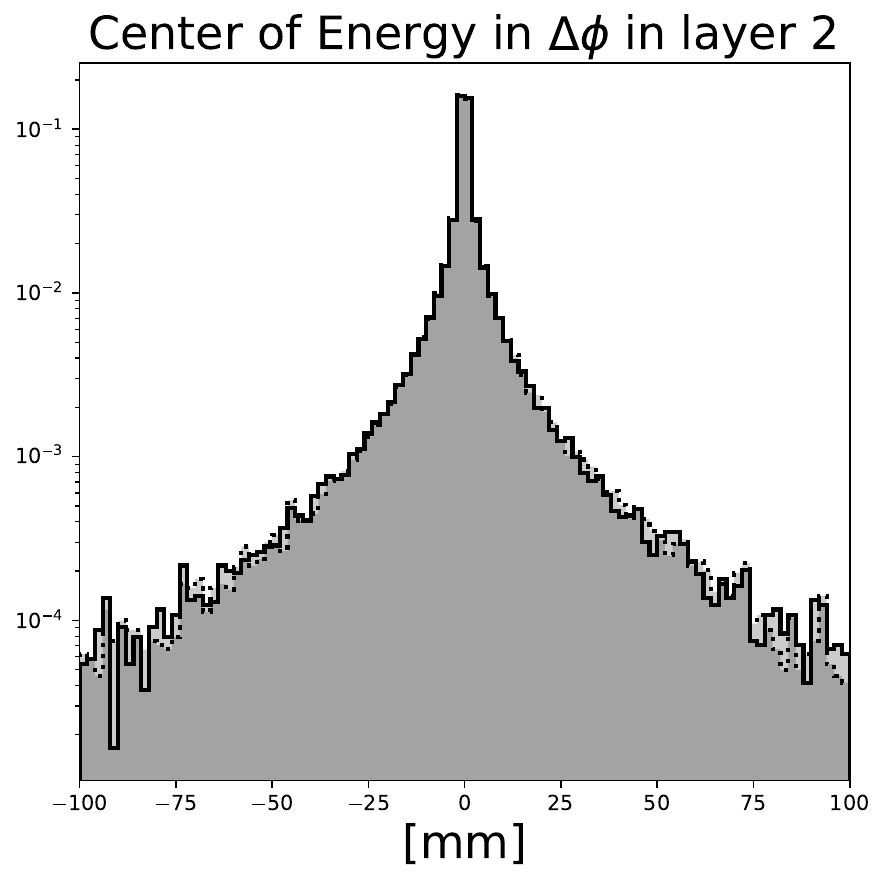} \hfill     \includegraphics[height=0.2\textheight]{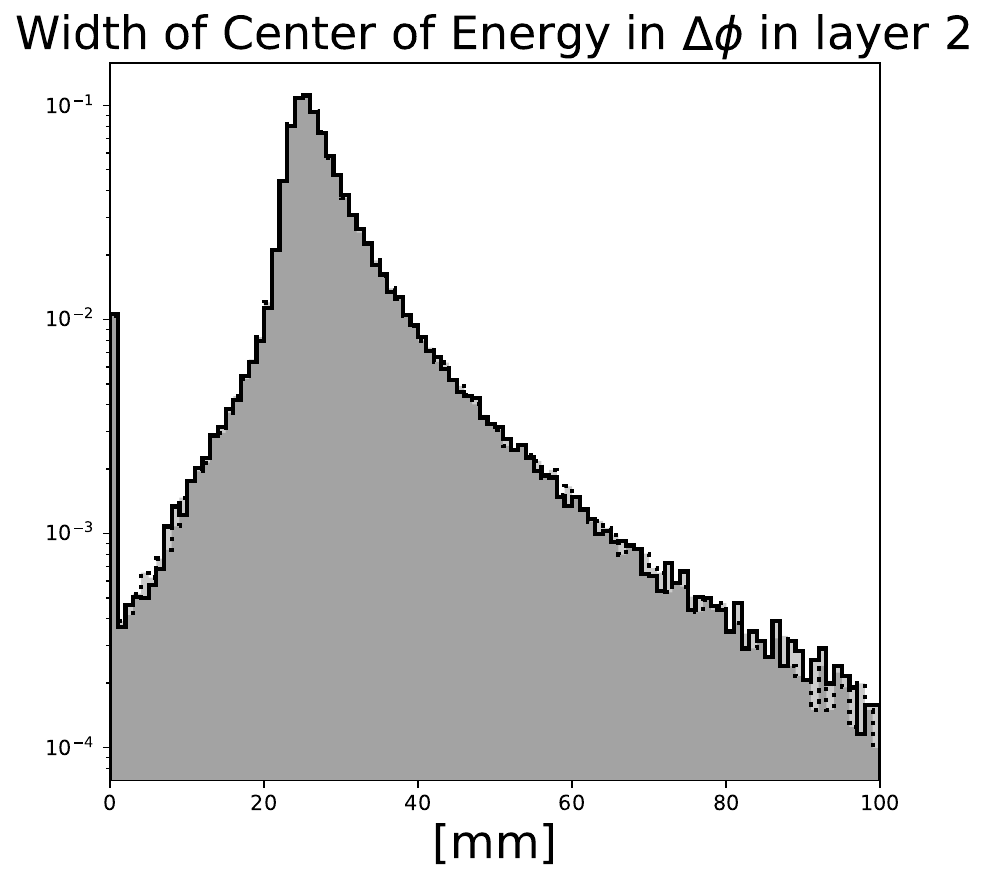}\hfill $ $\\
    \includegraphics[width=0.5\textwidth]{figures/Appendix_reference/legend.pdf}
    \caption{Distribution of \geant training and evaluation data in centers of energy along $\eta$ and $\phi$, as well as the widths of these distributions for ds1 --- photons. }
    \label{fig:app_ref.ds1-photons.2}
\end{figure}

\begin{figure}[ht]
    \centering
    \includegraphics[height=0.2\textheight]{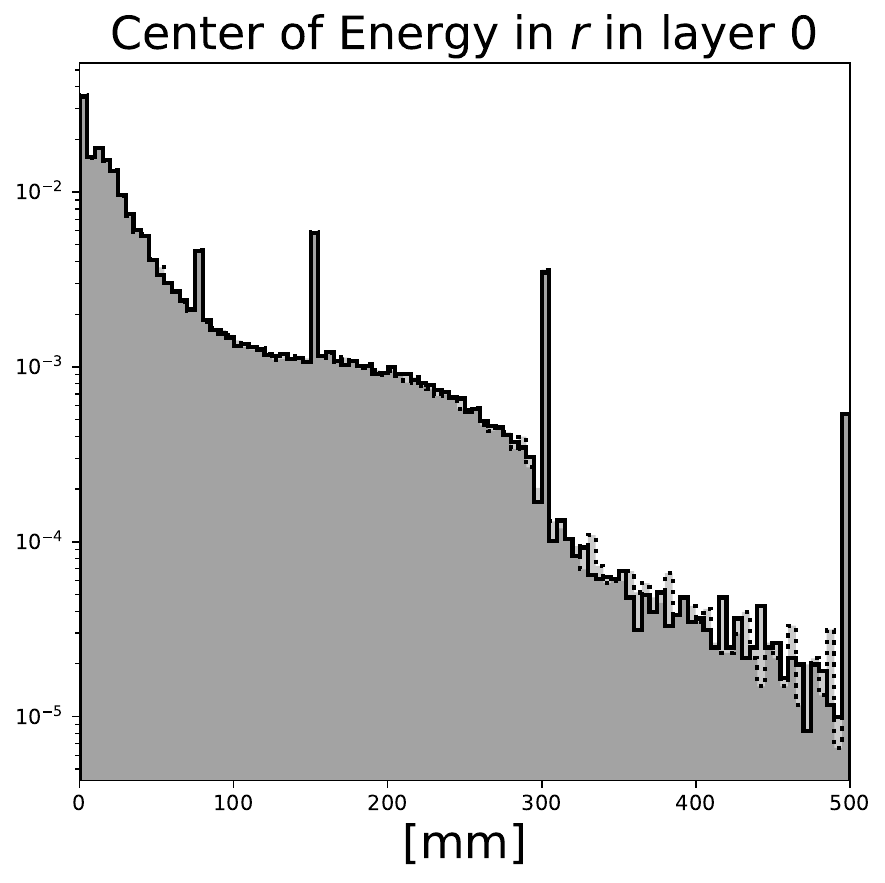} \hfill     \includegraphics[height=0.2\textheight]{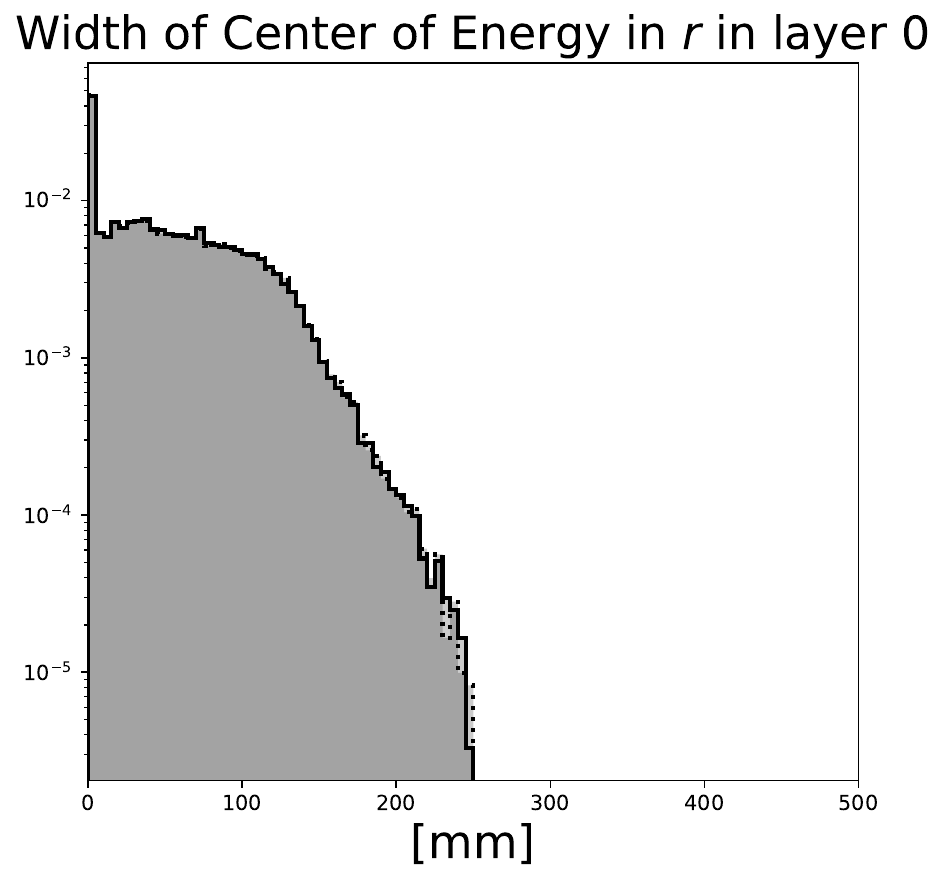} \hfill     \includegraphics[height=0.2\textheight]{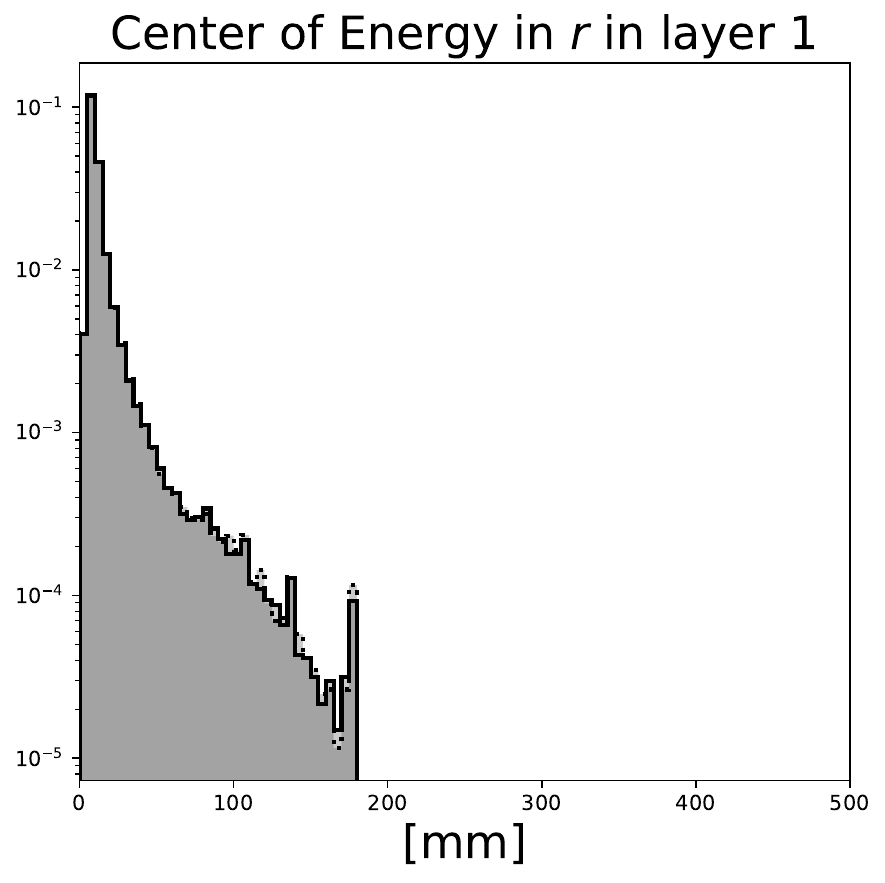} \\
    \includegraphics[height=0.2\textheight]{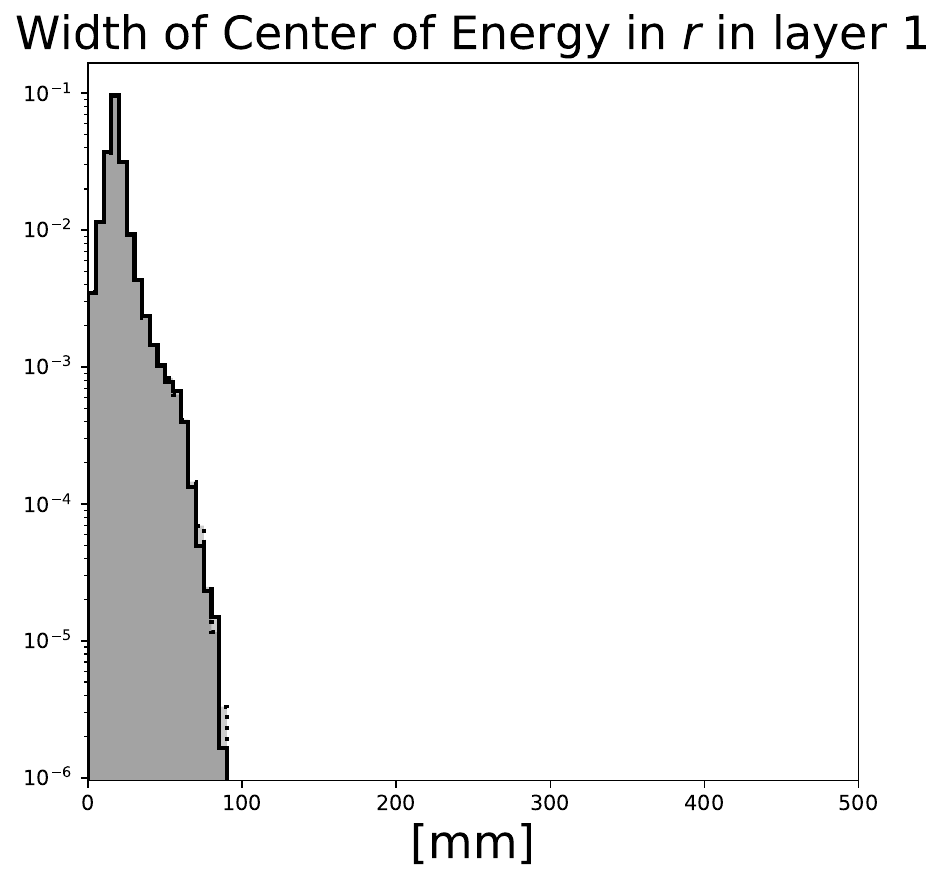} \hfill     \includegraphics[height=0.2\textheight]{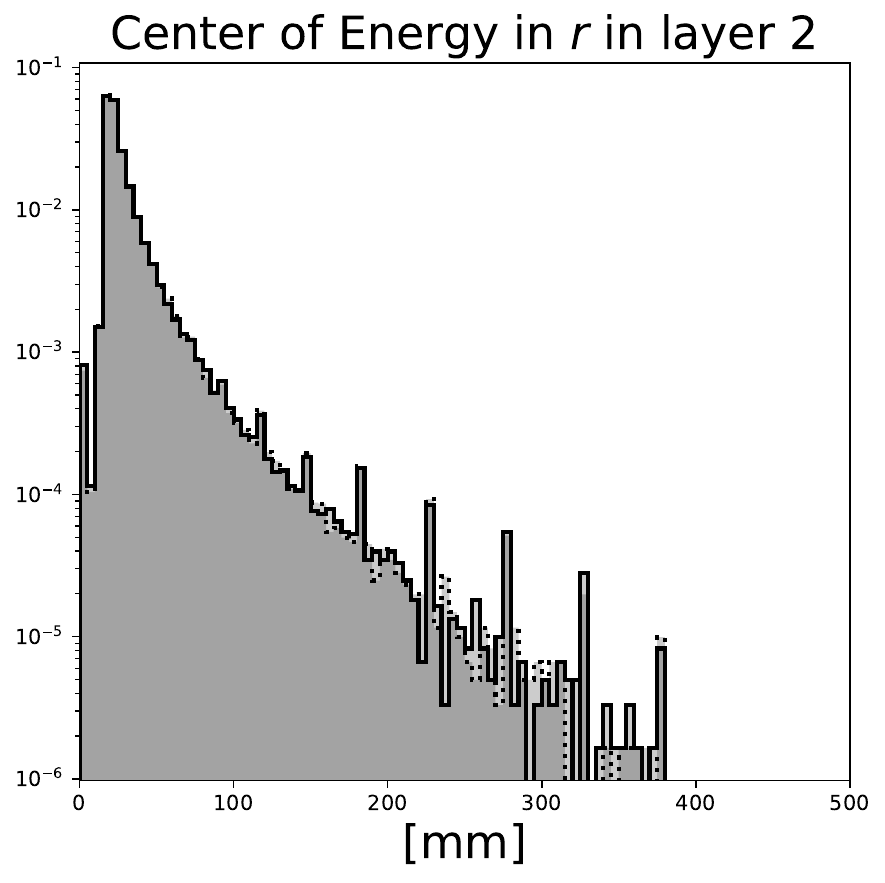} \hfill     \includegraphics[height=0.2\textheight]{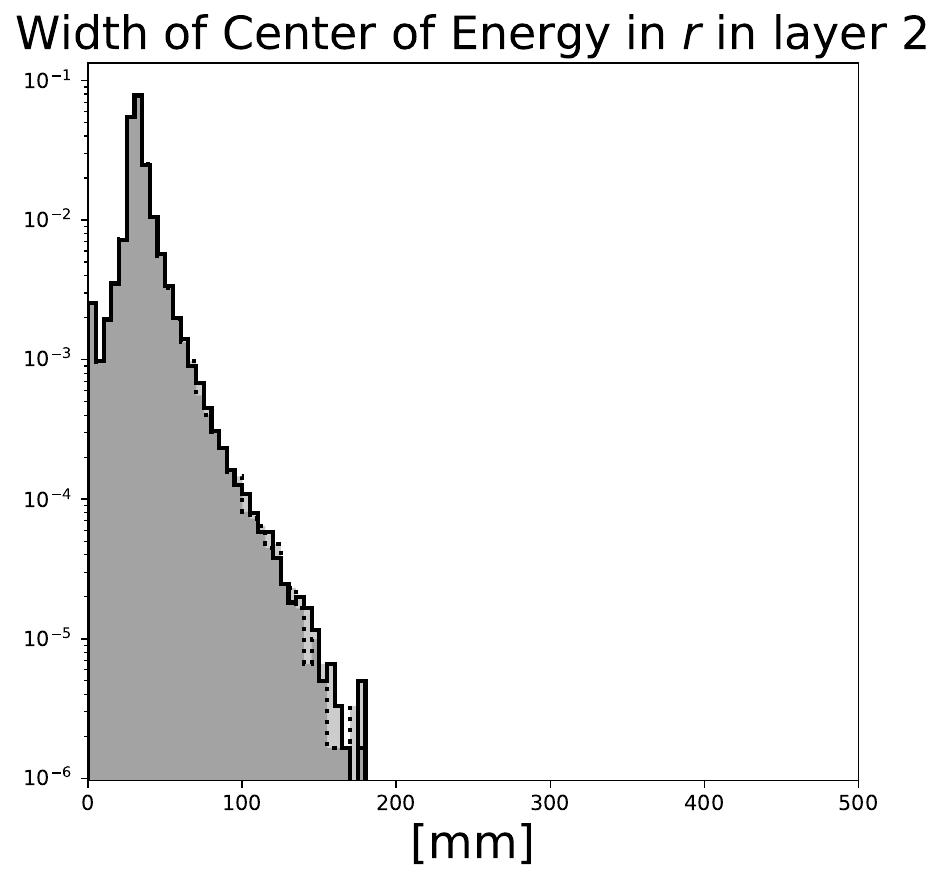}\\
    \includegraphics[height=0.2\textheight]{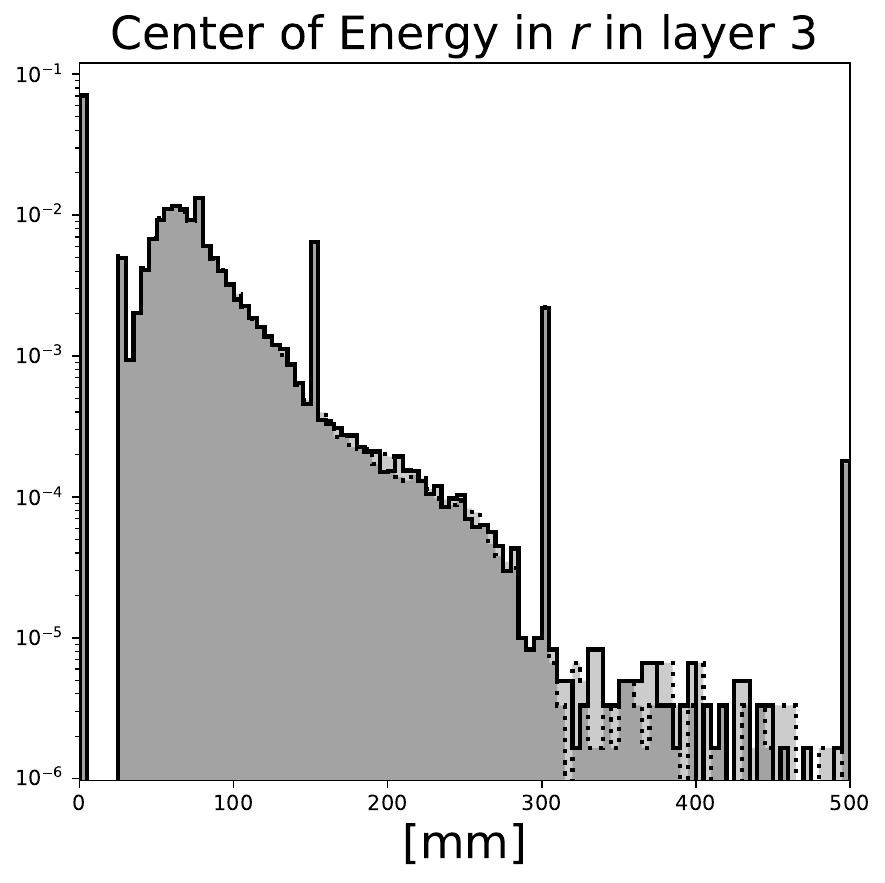} \hfill     \includegraphics[height=0.2\textheight]{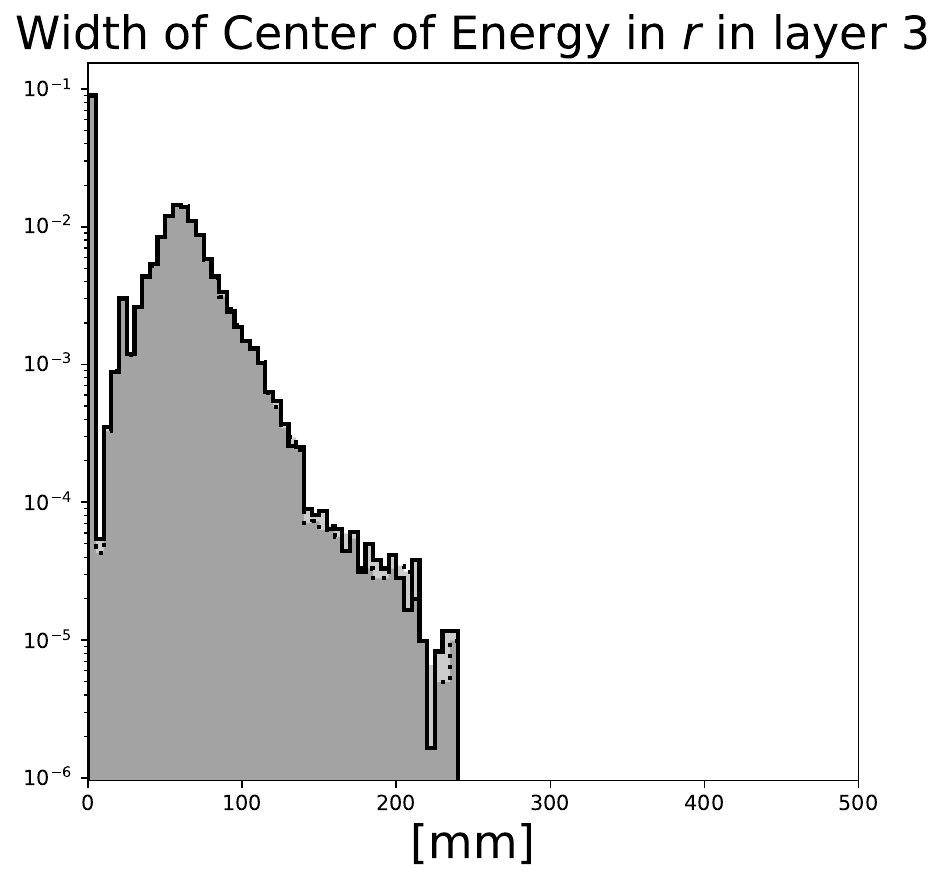} \hfill     \includegraphics[height=0.2\textheight]{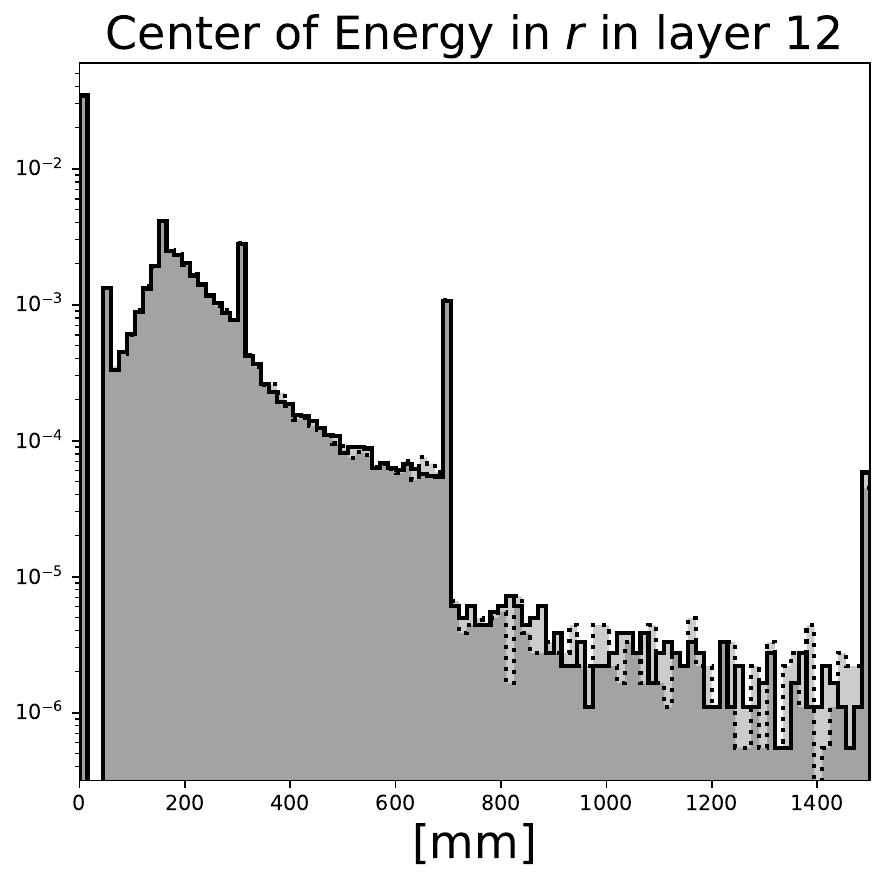} \\
    \includegraphics[height=0.2\textheight]{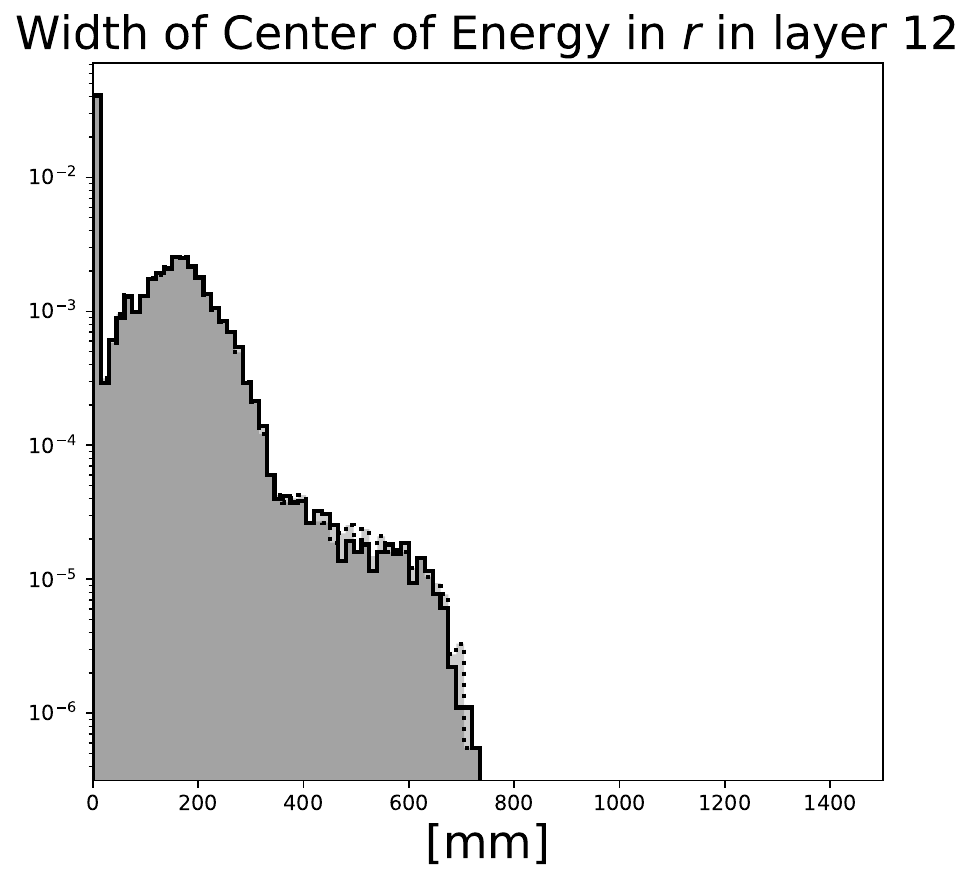} \\
    \includegraphics[width=0.5\textwidth]{figures/Appendix_reference/legend.pdf}
    \caption{Distribution of \geant training and evaluation data in centers of energy along the radial direction, as well as their widths for ds1 --- photons. }
    \label{fig:app_ref.ds1-photons.3}
\end{figure}
\subsection{\texorpdfstring{Dataset 1, Pions (\dsIpi)}{Dataset 1, Pions}}
\label{app:histograms.pions}
\begin{figure}[ht]
    \centering
    \hfill \includegraphics[height=0.2\textheight]{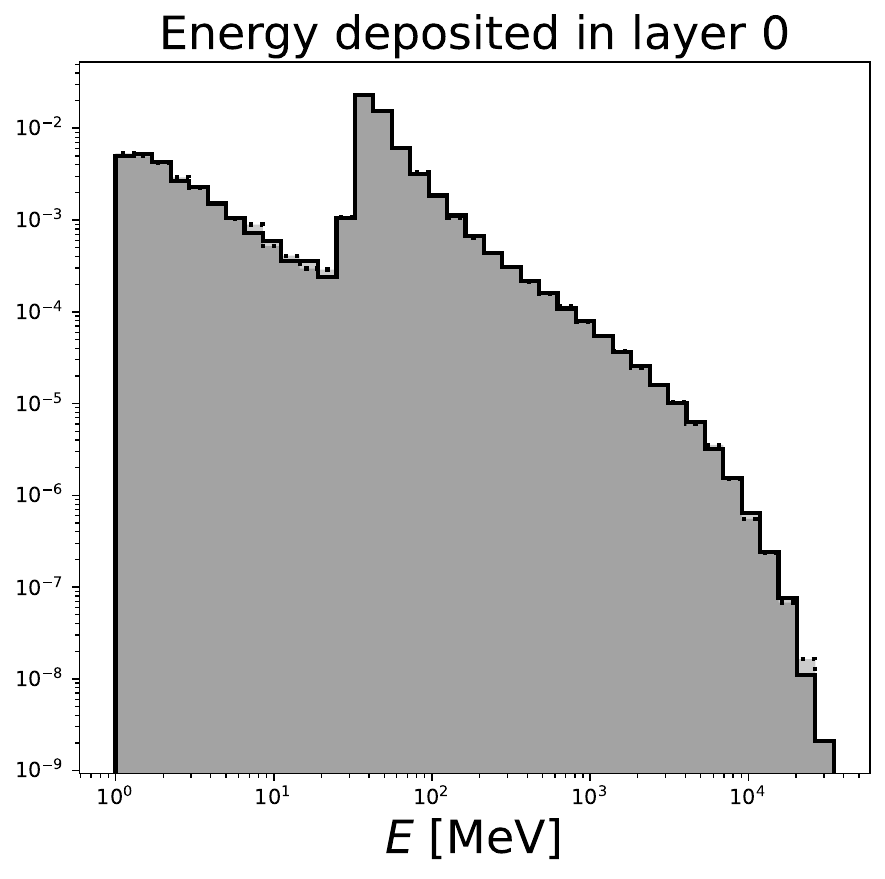} \hfill     \includegraphics[height=0.2\textheight]{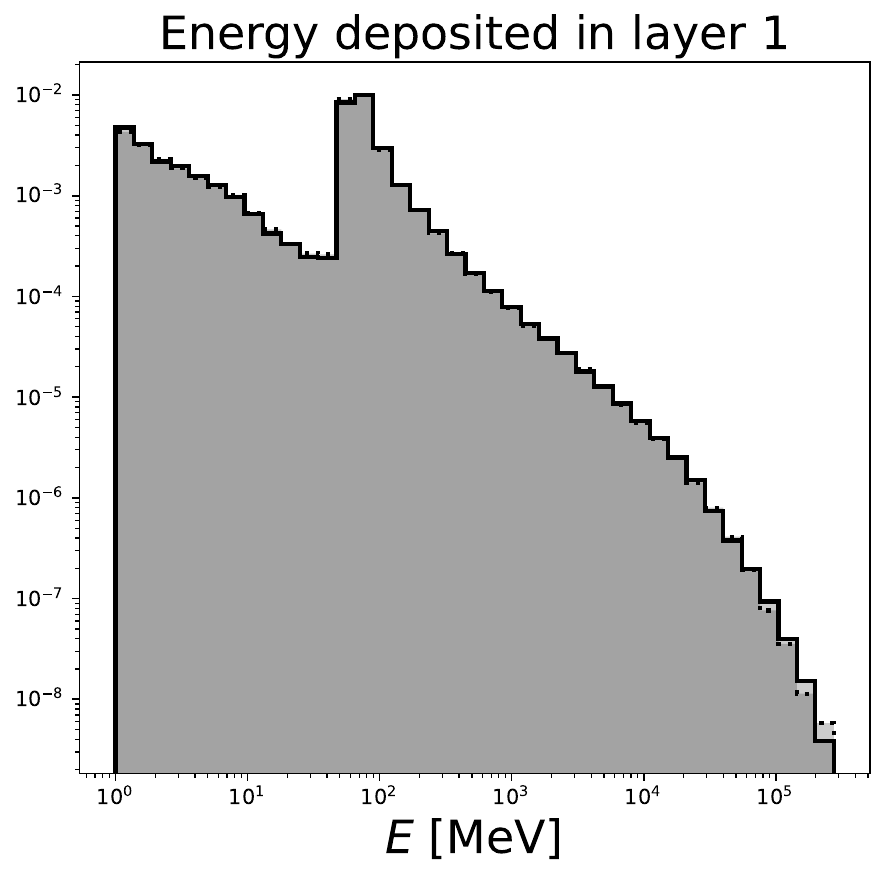} \hfill $ $\\
    \hfill \includegraphics[height=0.2\textheight]{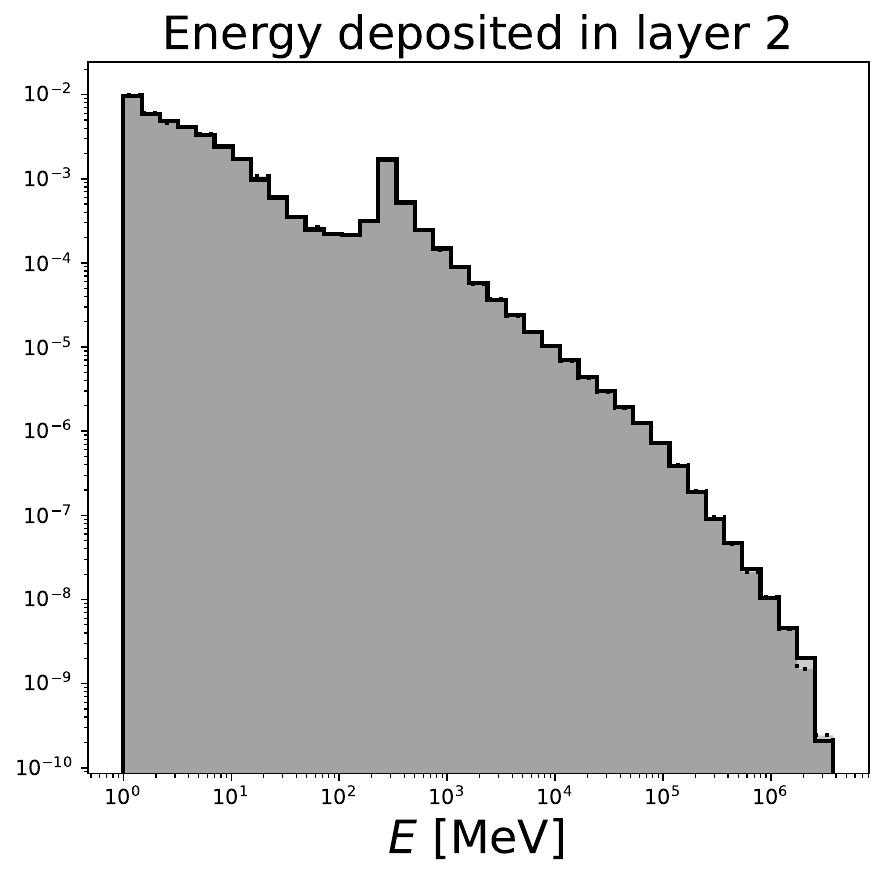} \hfill     \includegraphics[height=0.2\textheight]{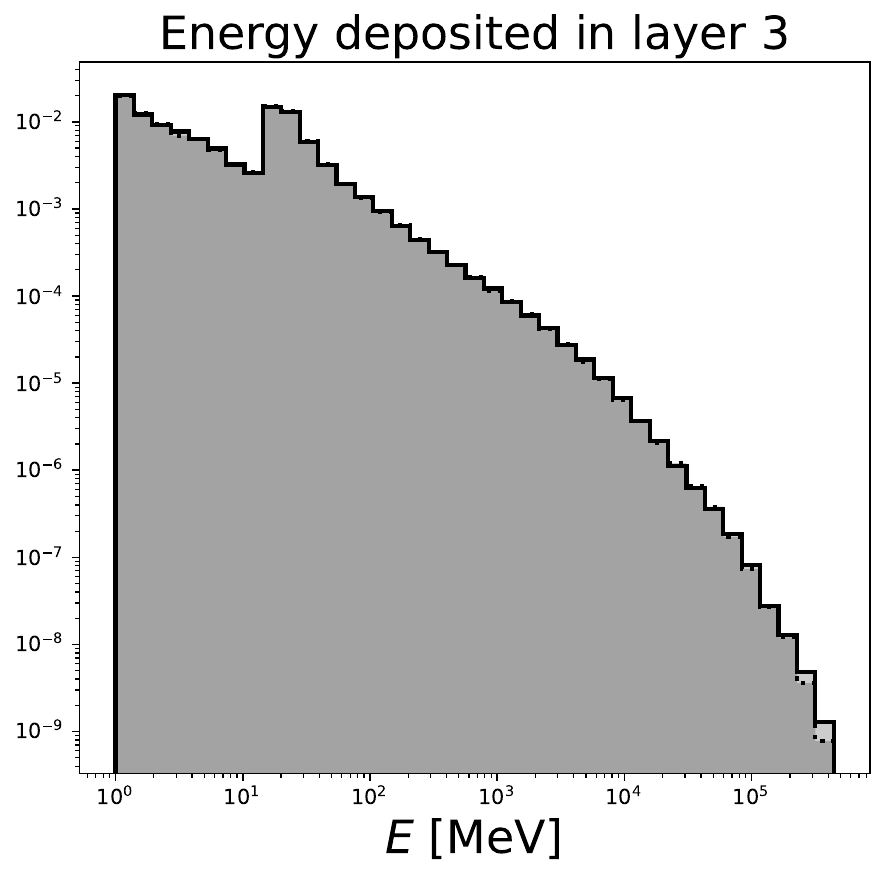} \hfill $ $\\
    \hfill\includegraphics[height=0.2\textheight]{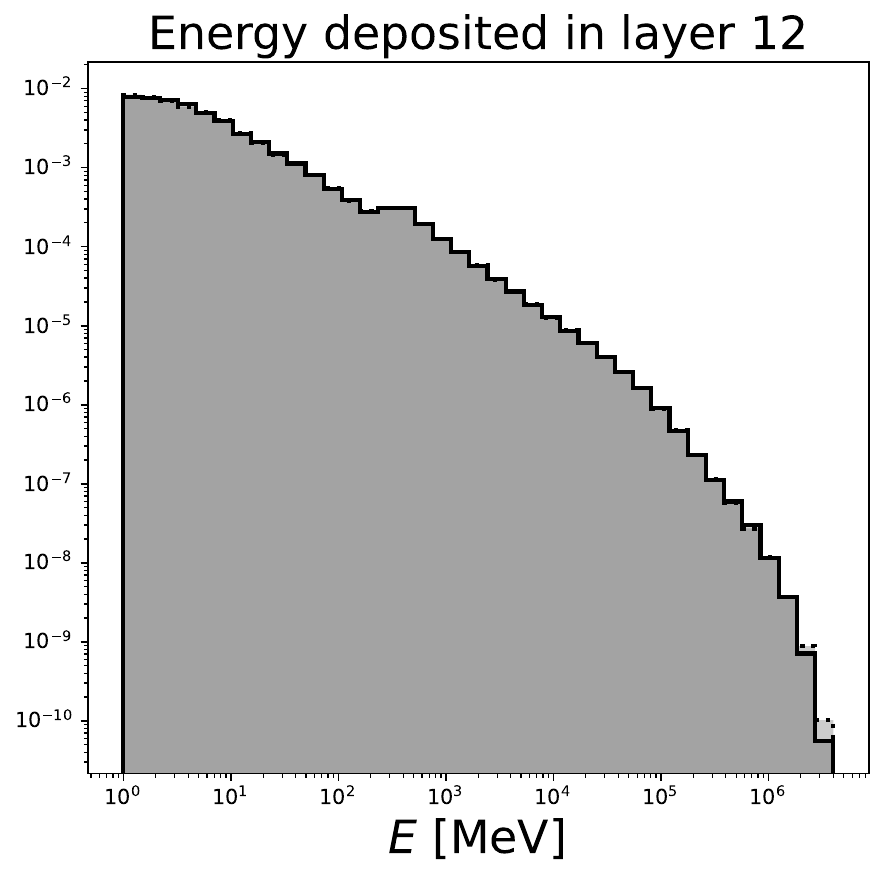} \hfill     \includegraphics[height=0.2\textheight]{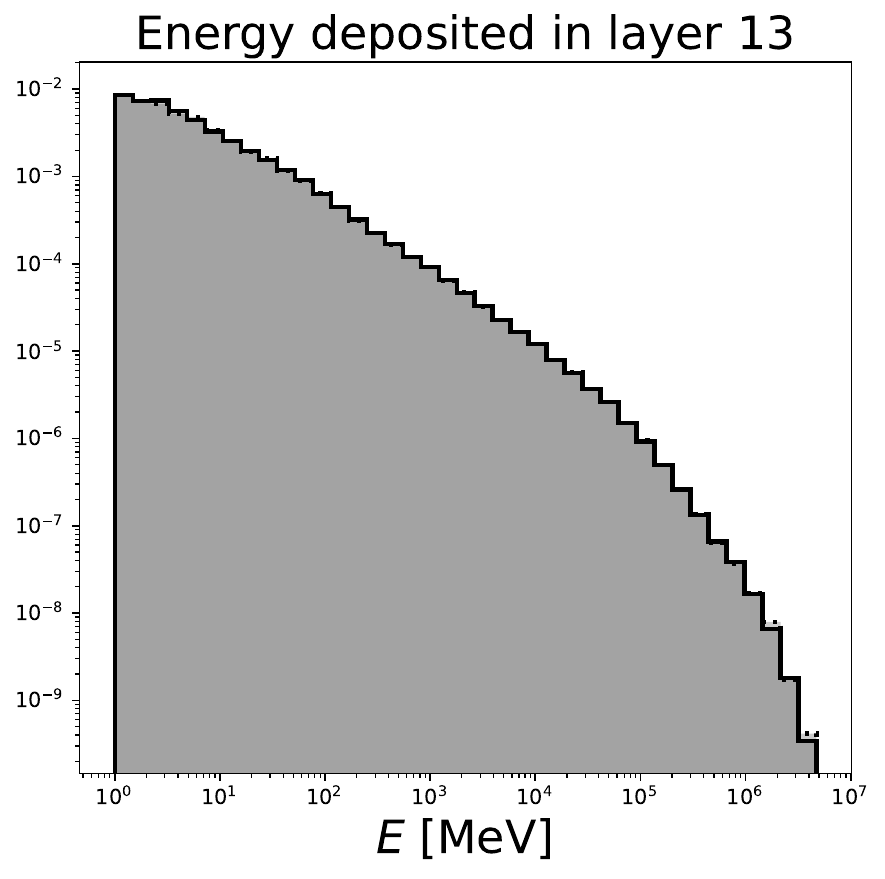}\hfill $ $\\
    \hfill\includegraphics[height=0.2\textheight]{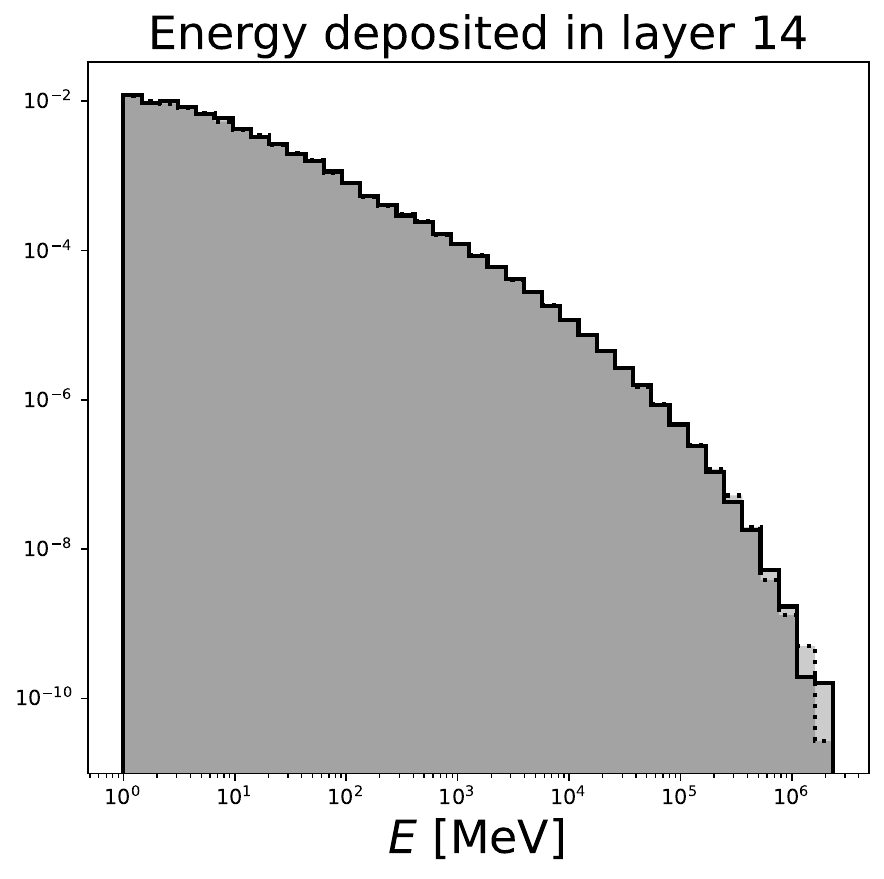} \hfill     \includegraphics[height=0.2\textheight]{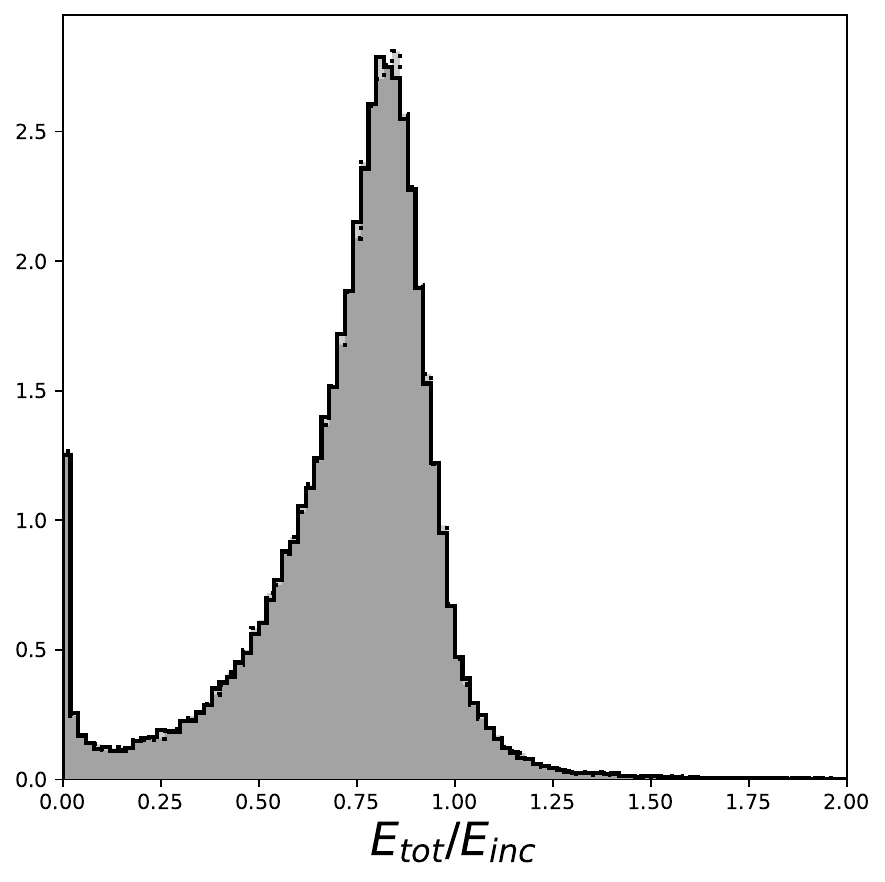}\hfill $ $\\
    \includegraphics[width=0.5\textwidth]{figures/Appendix_reference/legend.pdf}
    \caption{Distribution of \geant training and evaluation data in layer energies $E_i$, and ratio of total deposited energy to incident energy for ds1 --- pions. }
    \label{fig:app_ref.ds1-pions.1}
\end{figure}

\begin{figure}[ht]
    \centering
    \hfill \includegraphics[height=0.2\textheight]{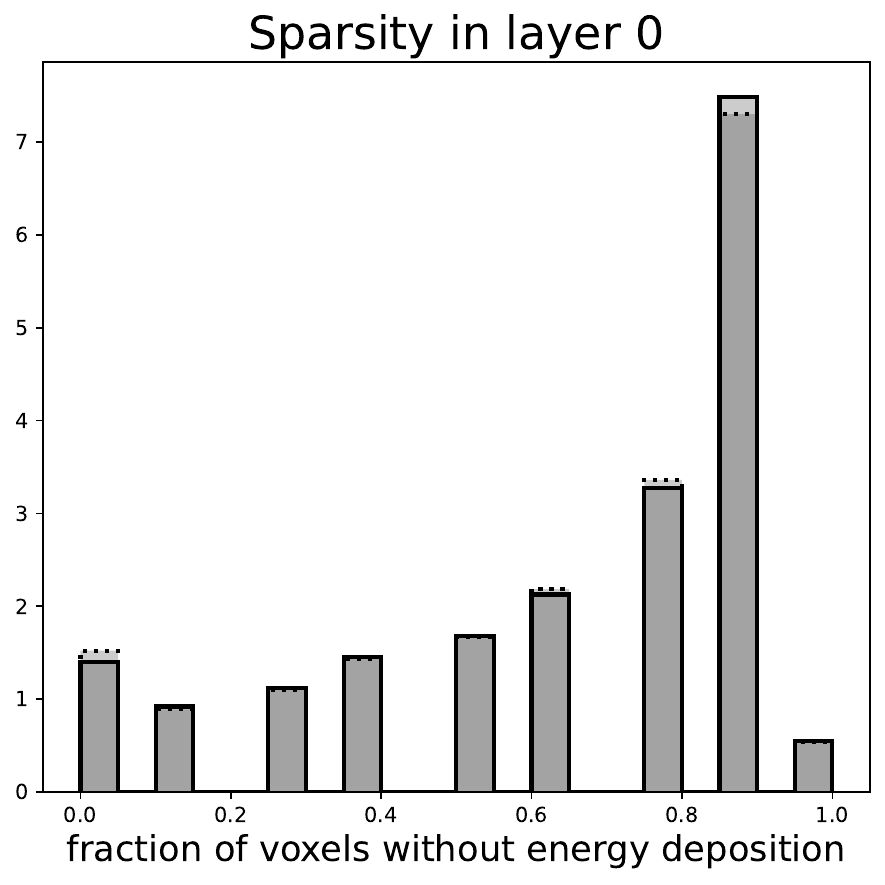} \hfill     \includegraphics[height=0.2\textheight]{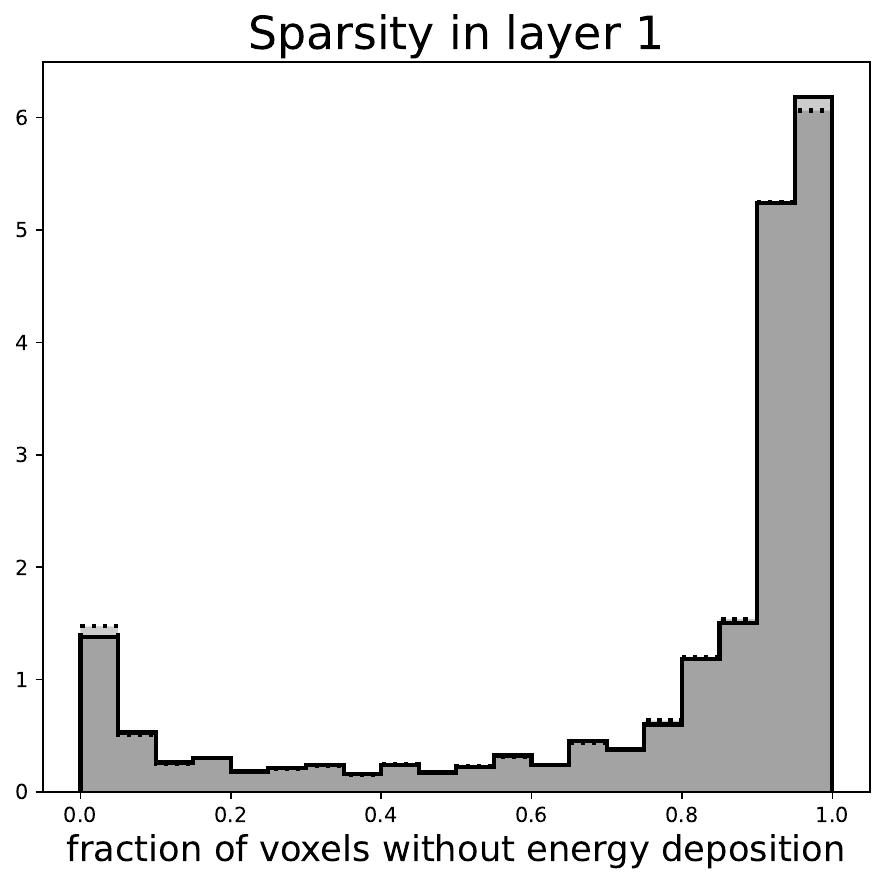} \hfill $ $\\
    \hfill \includegraphics[height=0.2\textheight]{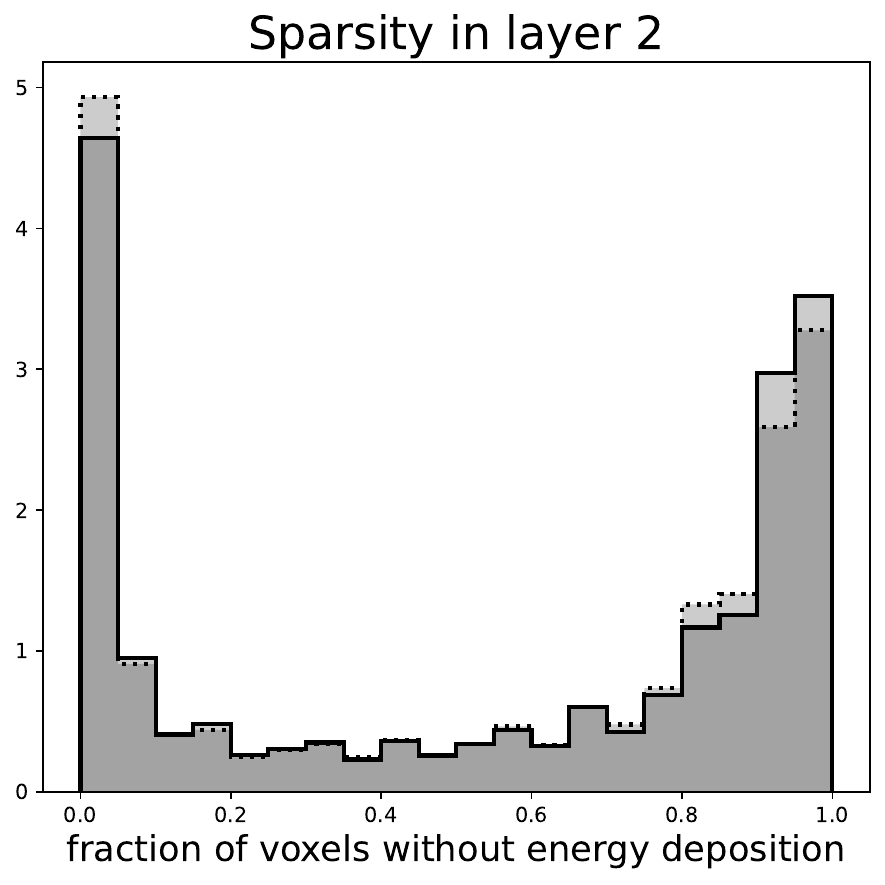} \hfill     \includegraphics[height=0.2\textheight]{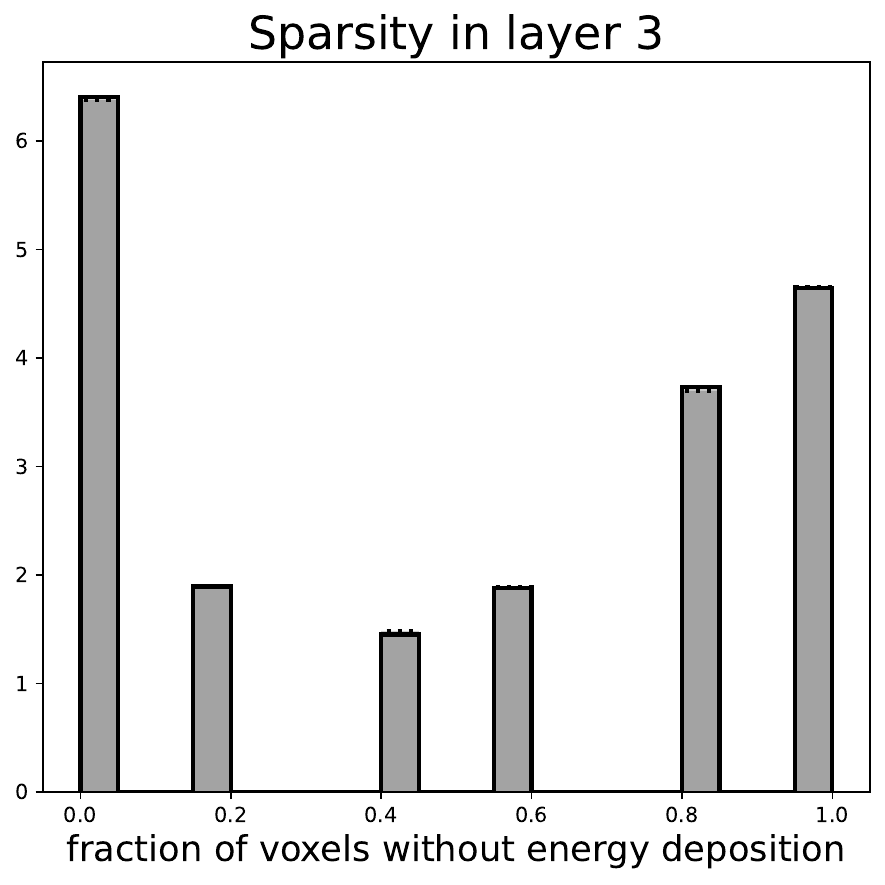} \hfill $ $\\
    \hfill\includegraphics[height=0.2\textheight]{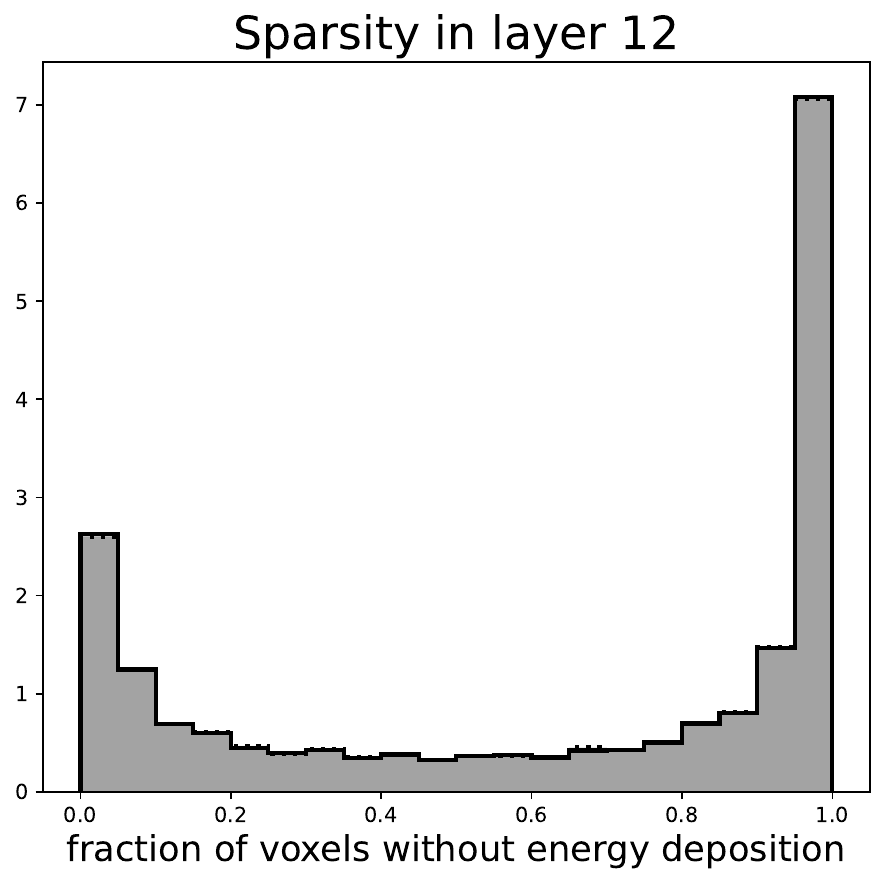} \hfill     \includegraphics[height=0.2\textheight]{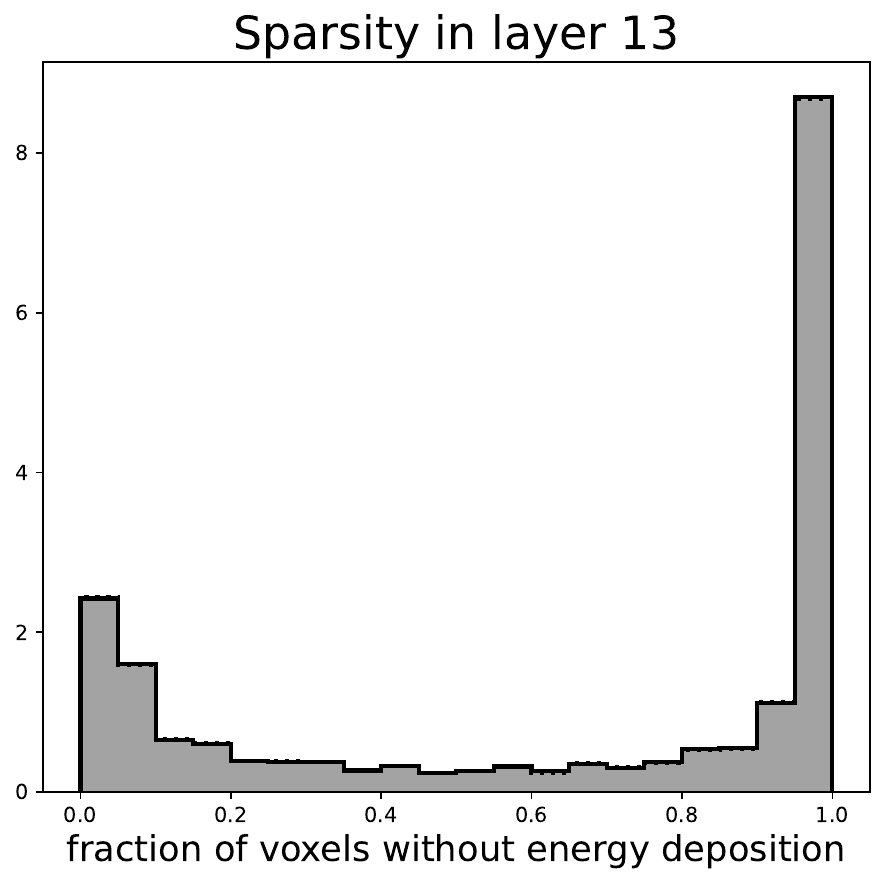}\hfill $ $\\
    \hfill\includegraphics[height=0.2\textheight]{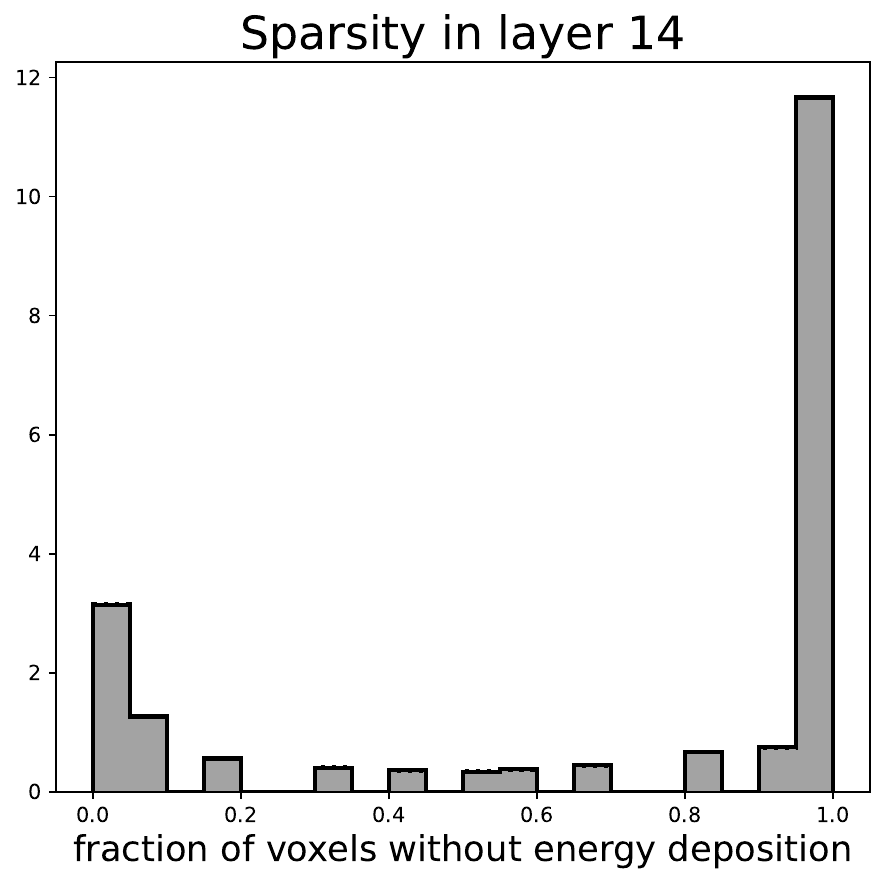} \hfill     \includegraphics[height=0.2\textheight]{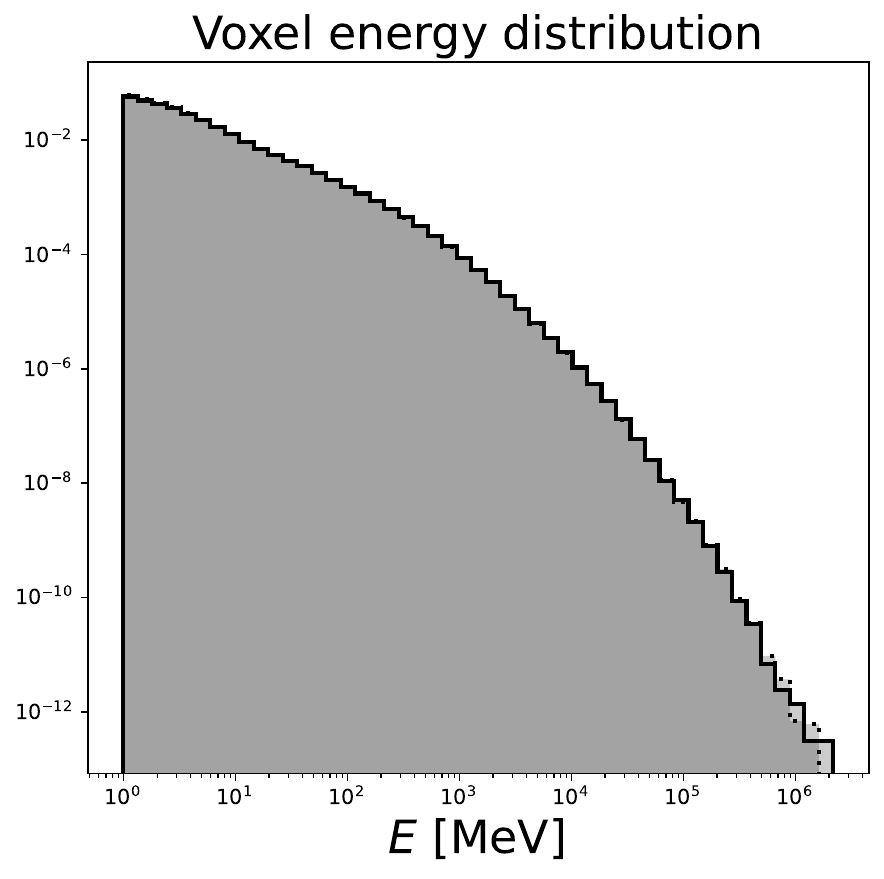}\hfill $ $\\
    \includegraphics[width=0.5\textwidth]{figures/Appendix_reference/legend.pdf}
    \caption{Distribution of \geant training and evaluation data in sparsity and energy per voxel for ds1 --- pions. }
    \label{fig:app_ref.ds1-pions.2}
\end{figure}

\begin{figure}[ht]
    \centering
    \includegraphics[height=0.15\textheight]{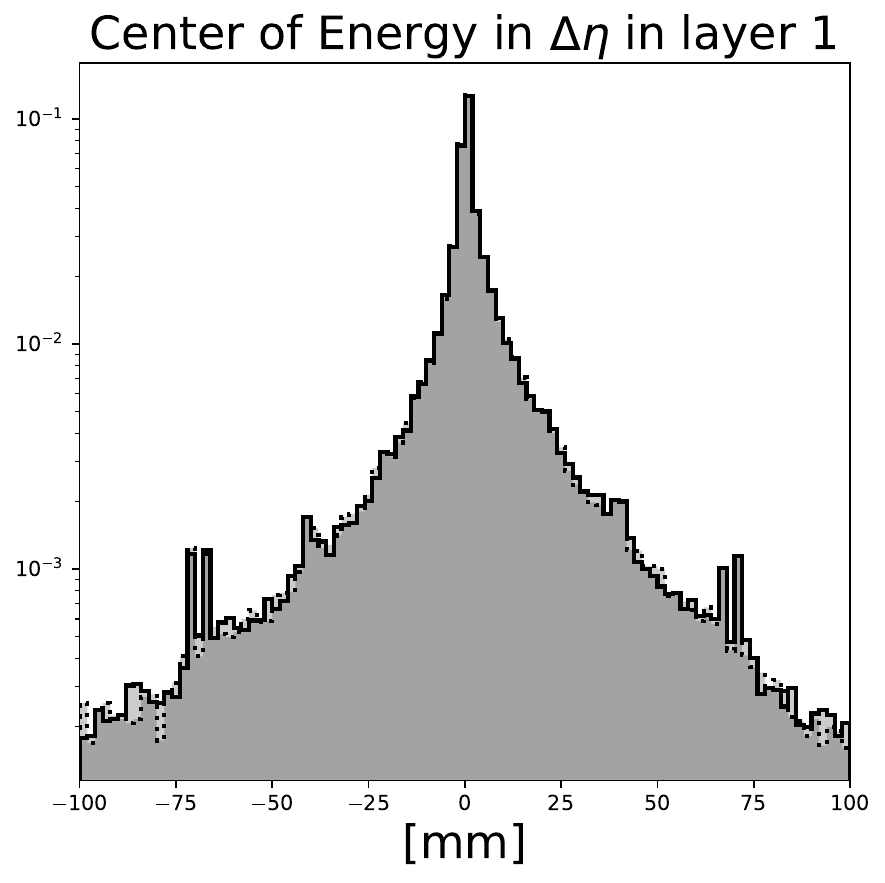} \hfill \includegraphics[height=0.15\textheight]{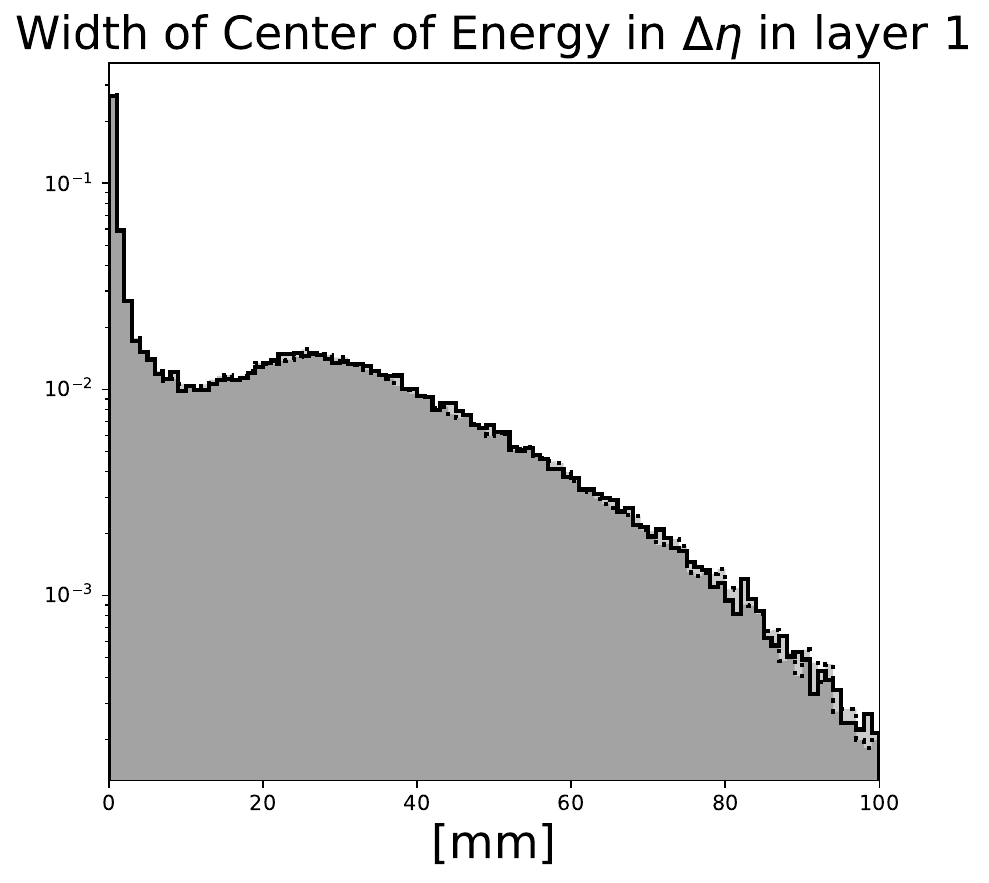} \hfill \includegraphics[height=0.15\textheight]{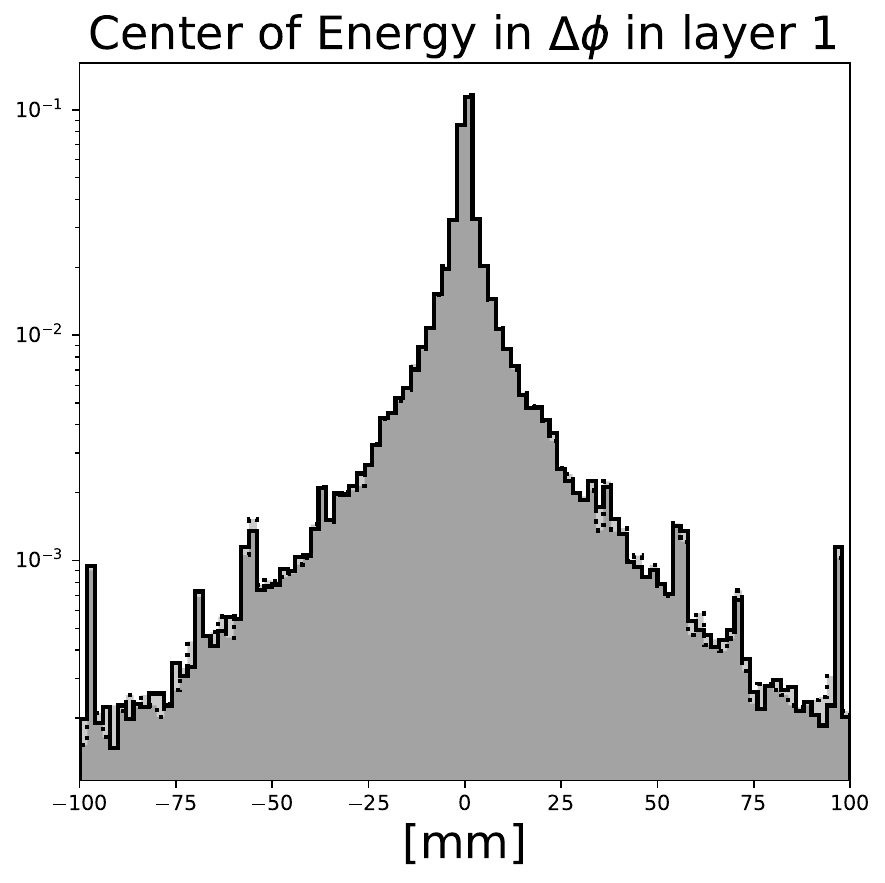} \hfill \includegraphics[height=0.15\textheight]{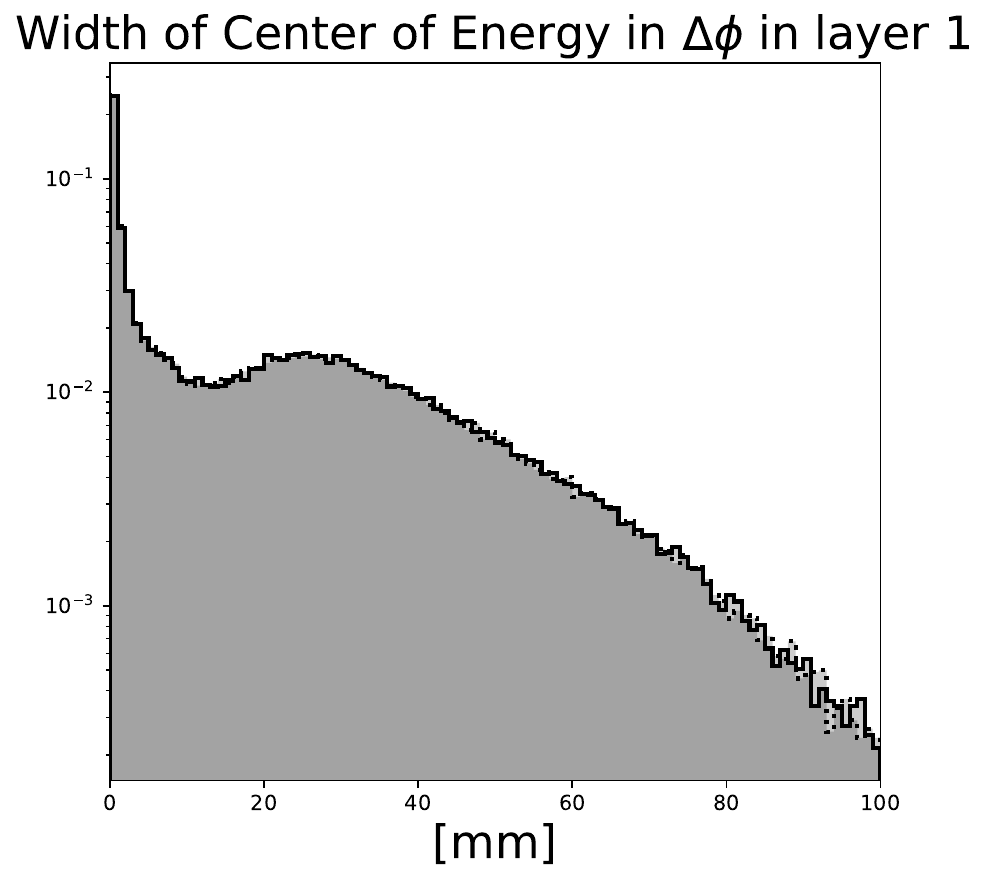}\\
    \includegraphics[height=0.15\textheight]{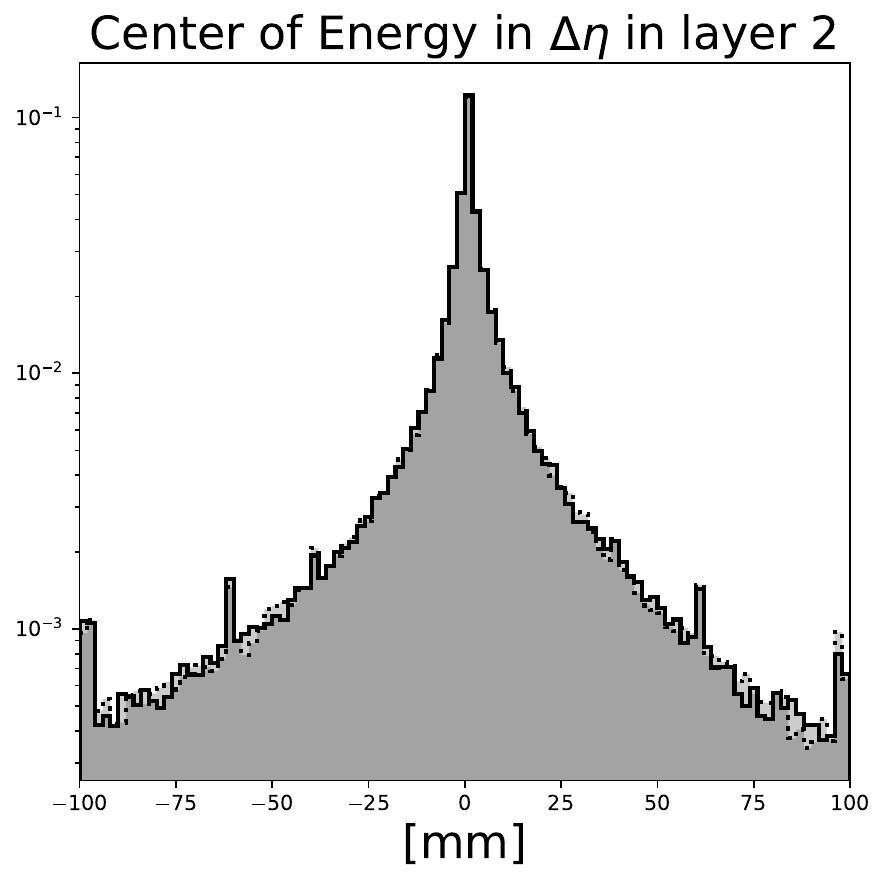} \hfill \includegraphics[height=0.15\textheight]{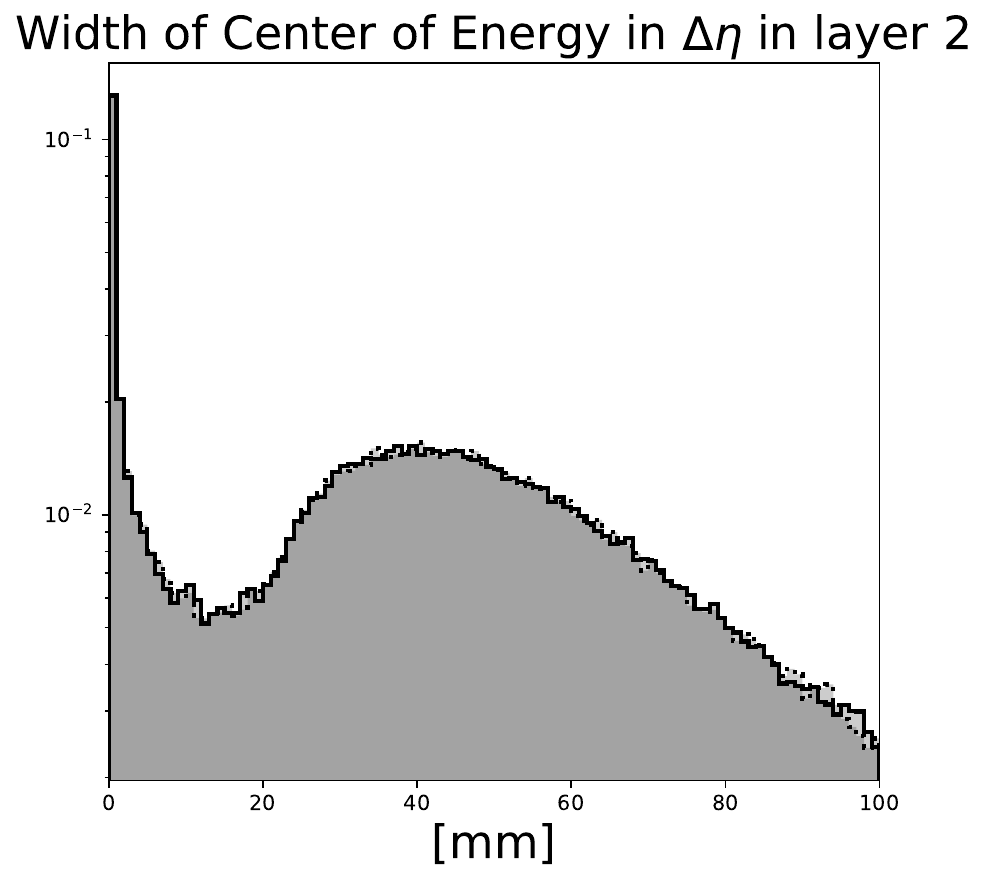} \hfill \includegraphics[height=0.15\textheight]{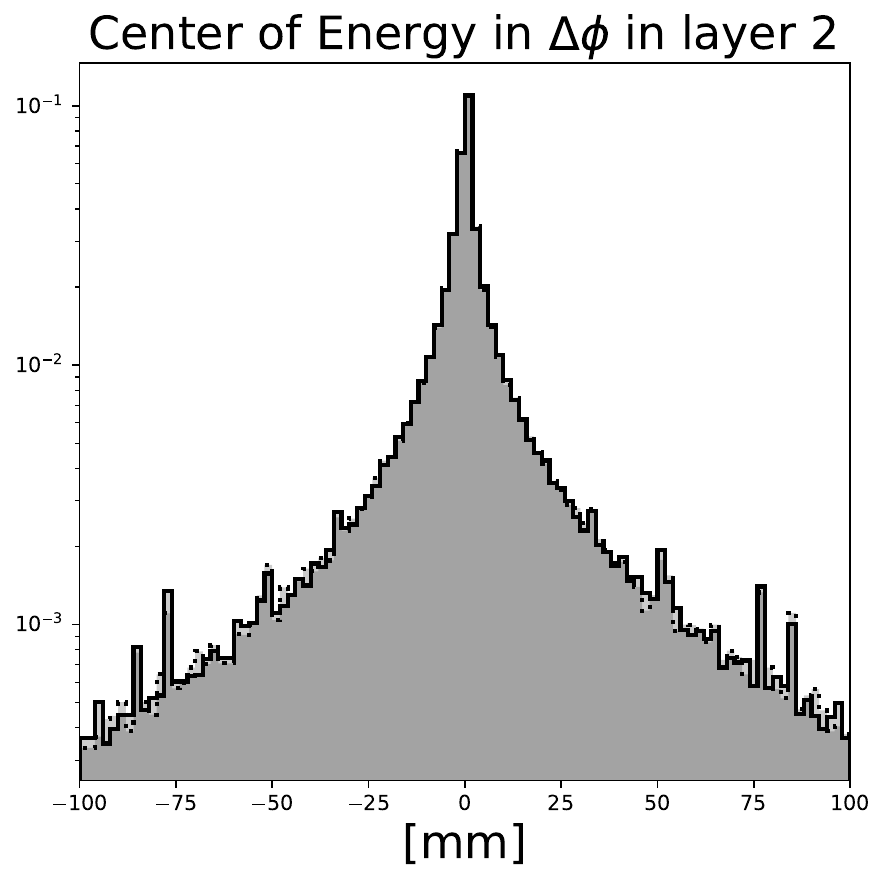} \hfill \includegraphics[height=0.15\textheight]{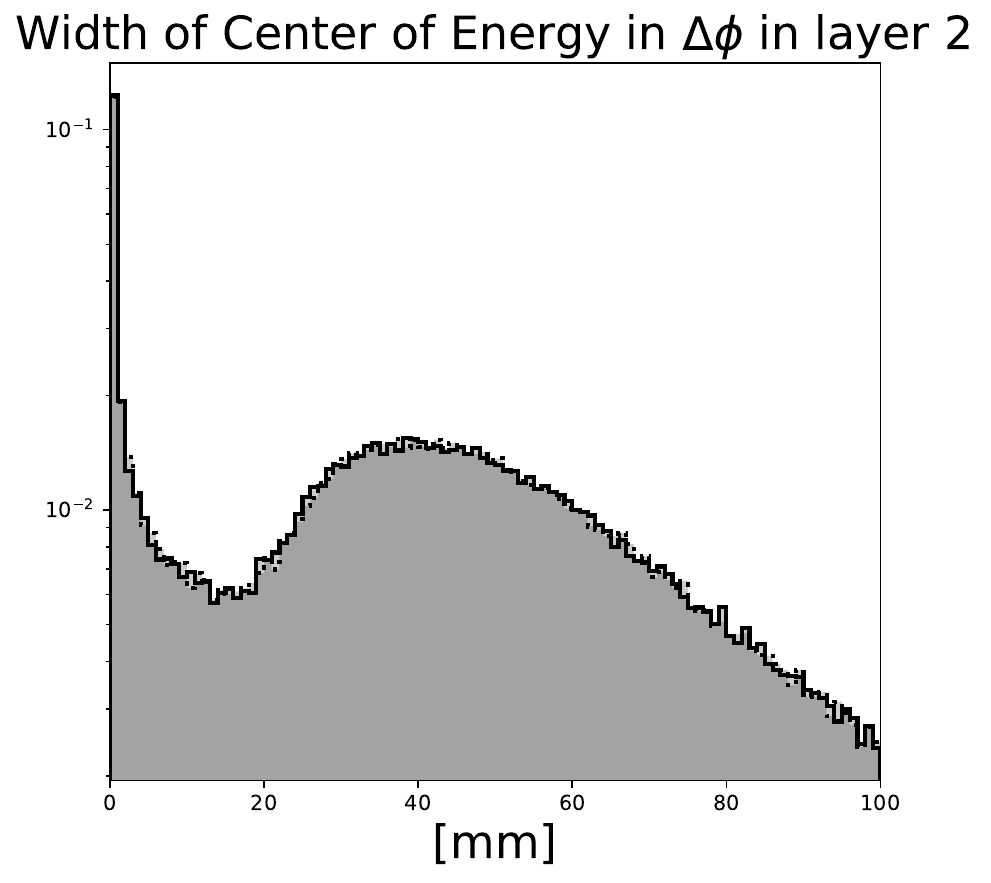}\\
    \includegraphics[height=0.15\textheight]{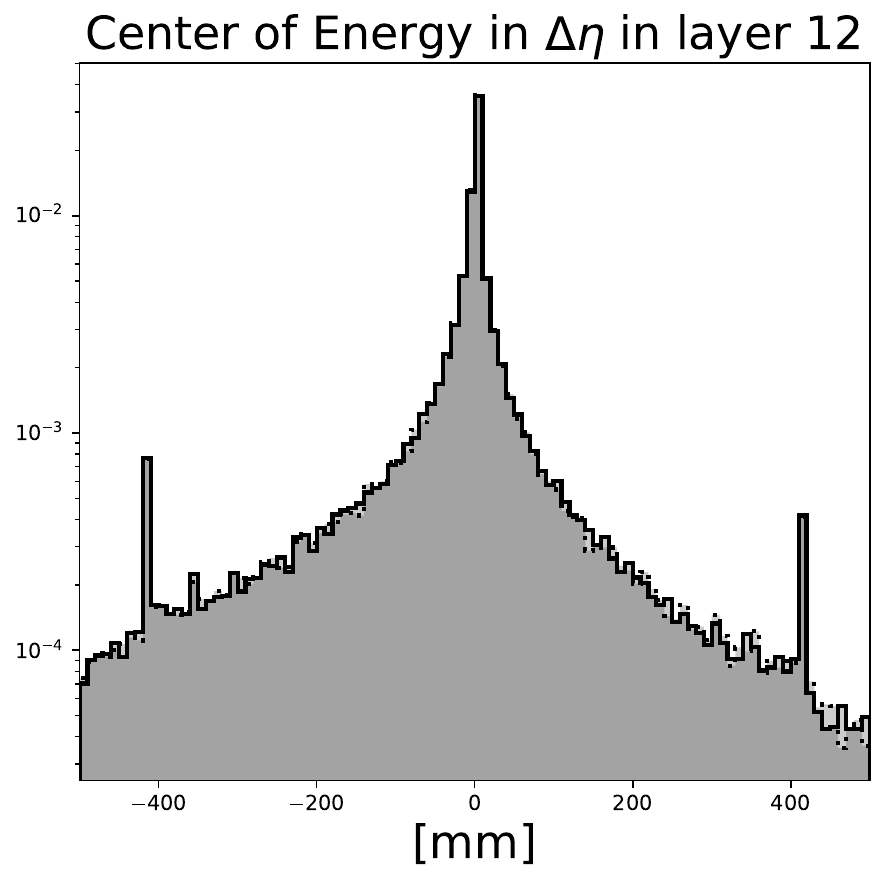} \hfill \includegraphics[height=0.15\textheight]{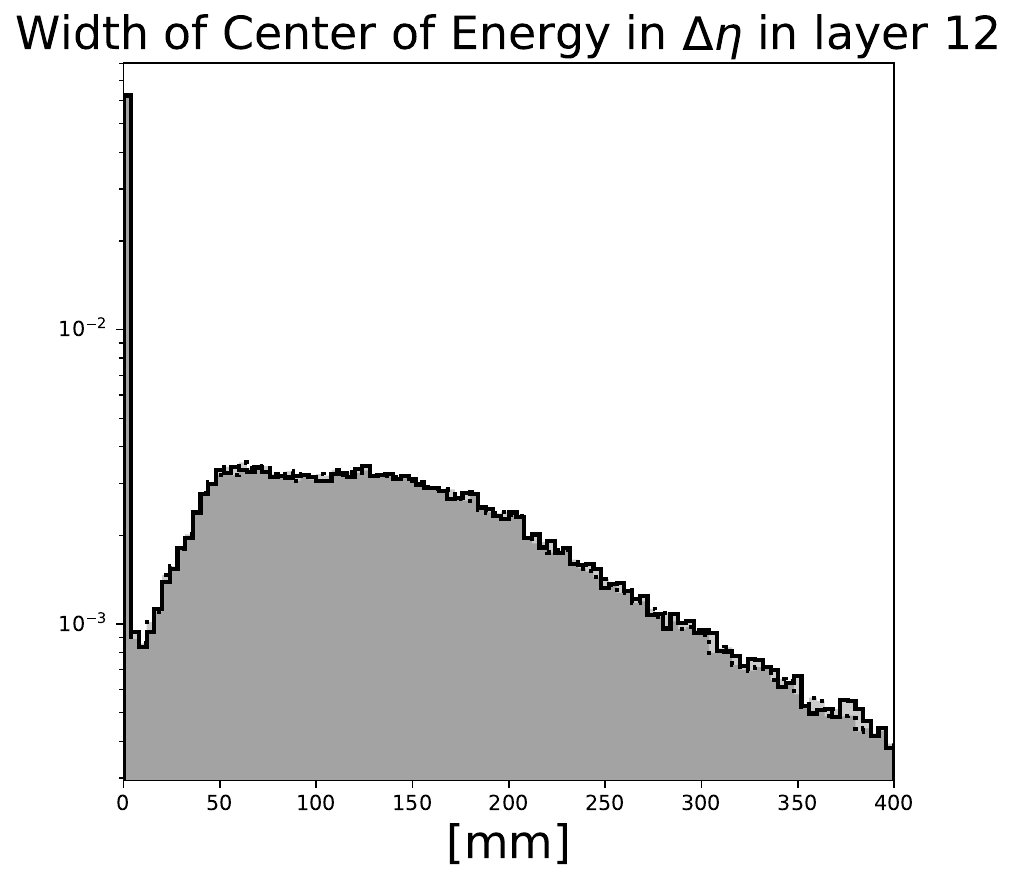} \hfill \includegraphics[height=0.15\textheight]{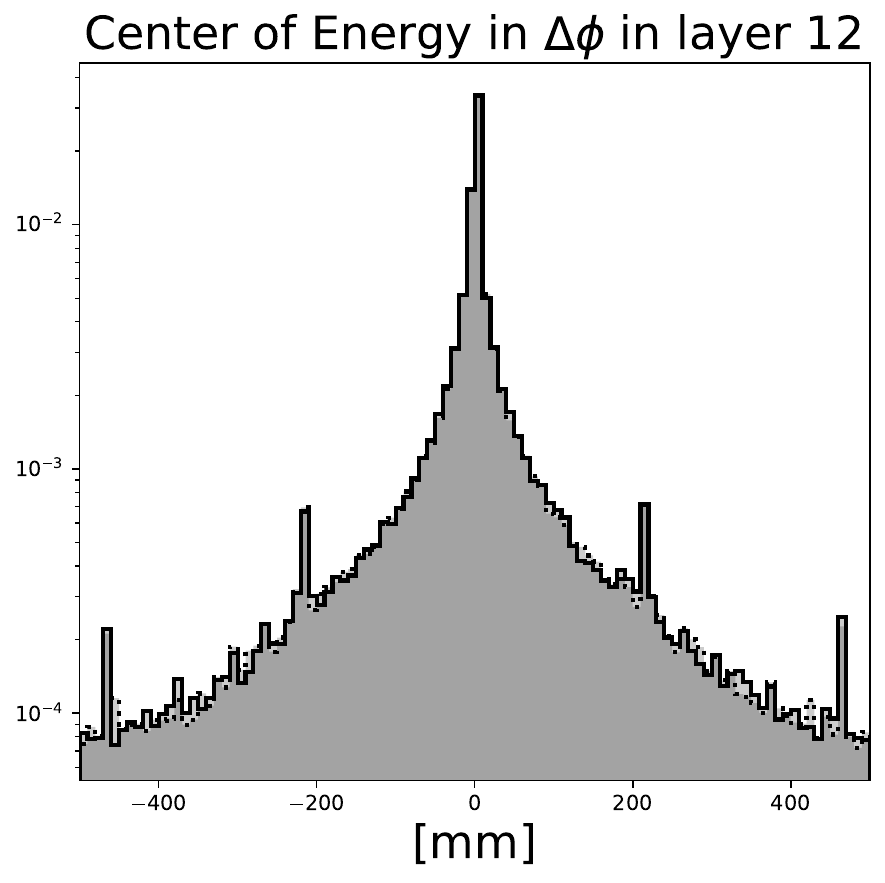} \hfill \includegraphics[height=0.15\textheight]{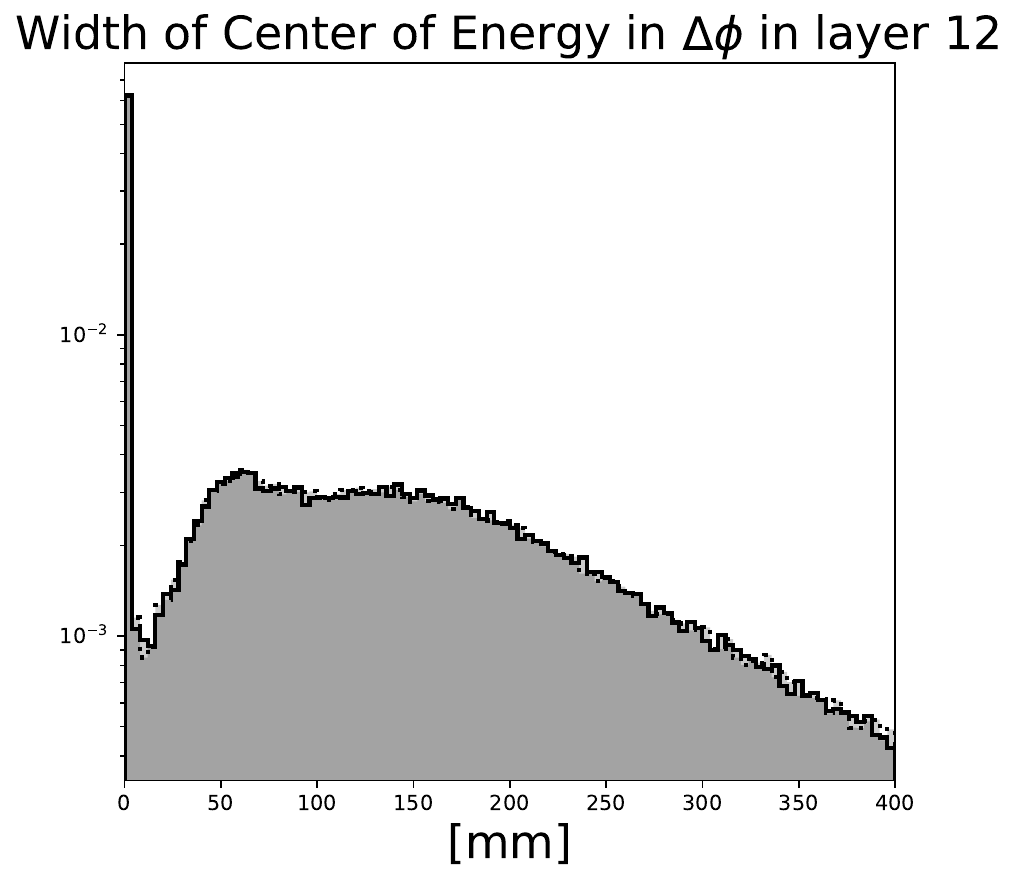}\\
    \includegraphics[height=0.15\textheight]{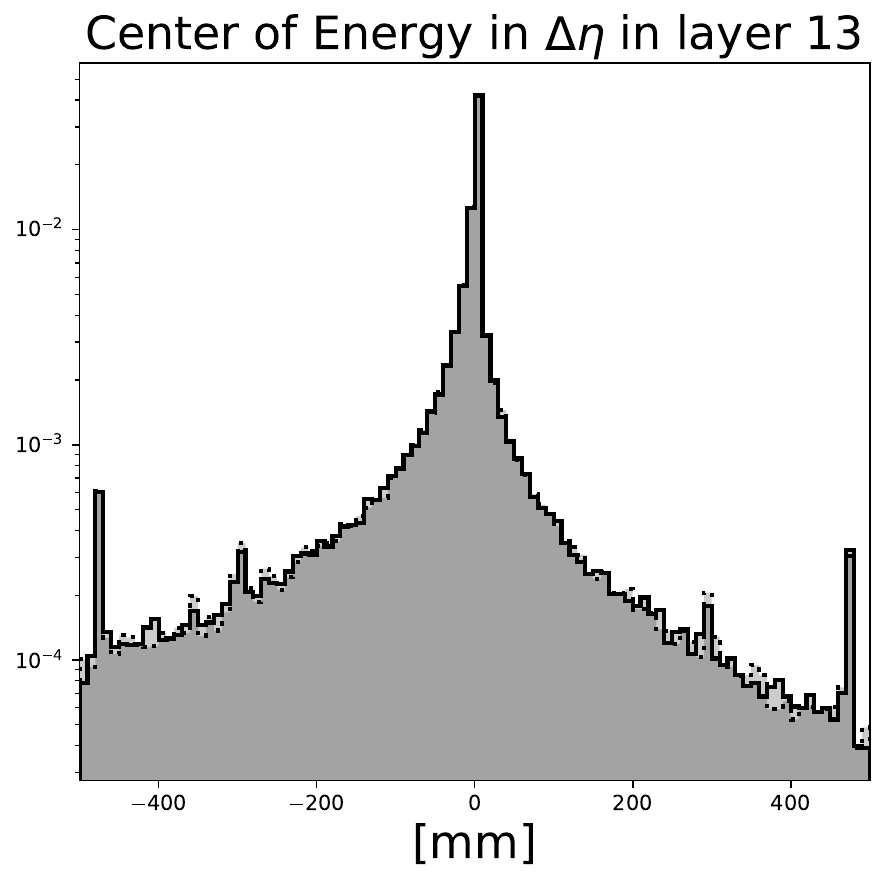} \hfill \includegraphics[height=0.15\textheight]{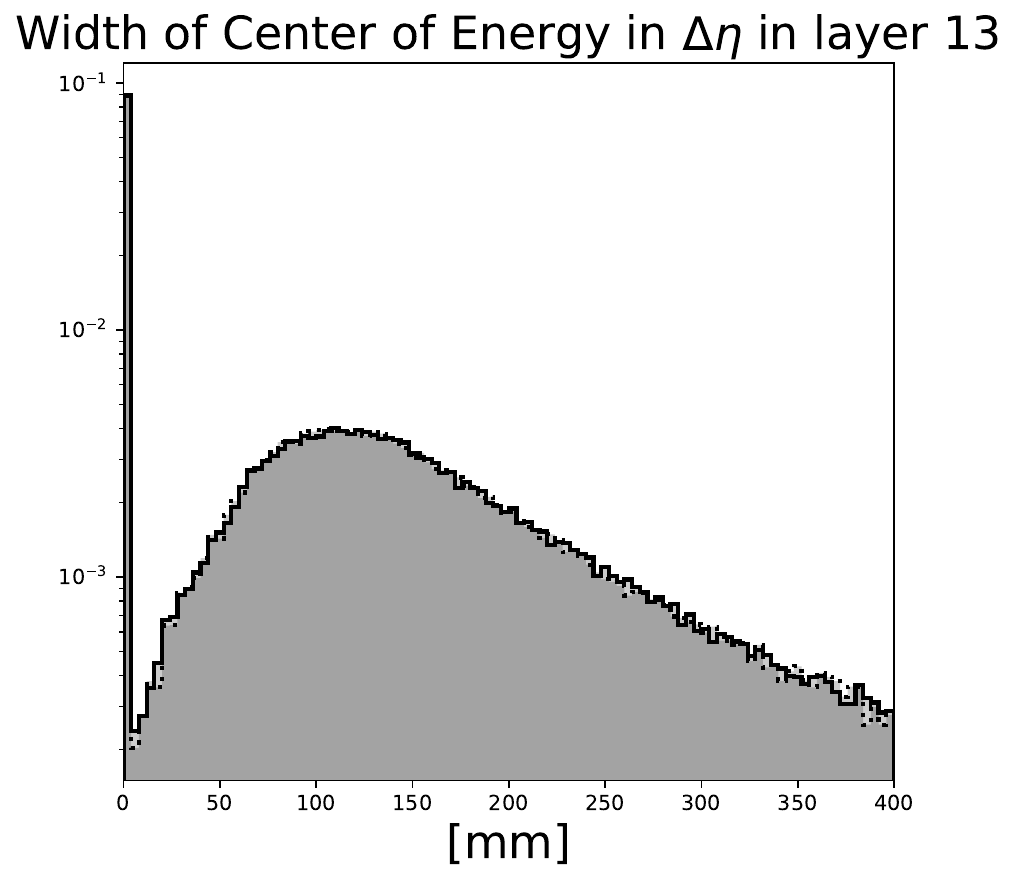} \hfill \includegraphics[height=0.15\textheight]{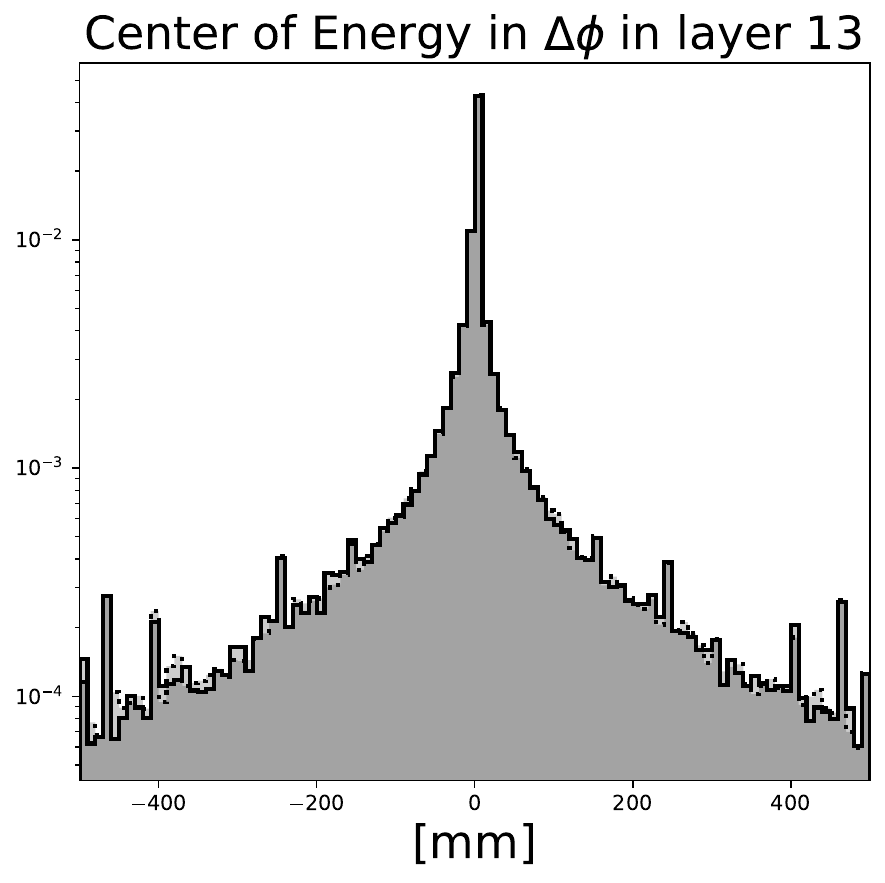} \hfill \includegraphics[height=0.15\textheight]{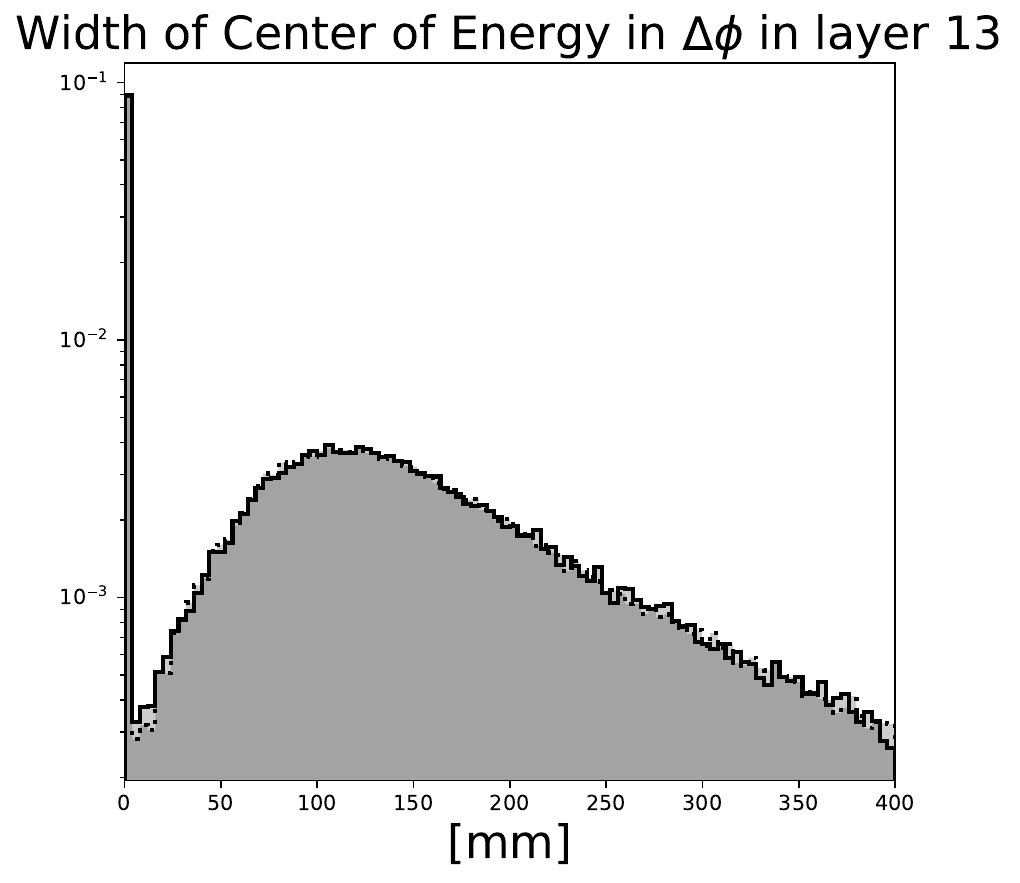}\\
    \includegraphics[width=0.5\textwidth]{figures/Appendix_reference/legend.pdf}
    \caption{Distribution of \geant training and evaluation data in centers of energy along $\eta$ and $\phi$, as well as their widths for ds1 --- pions. }
    \label{fig:app_ref.ds1-pions.3}
\end{figure}

\begin{figure}[ht]
    \centering
    \includegraphics[height=0.15\textheight]{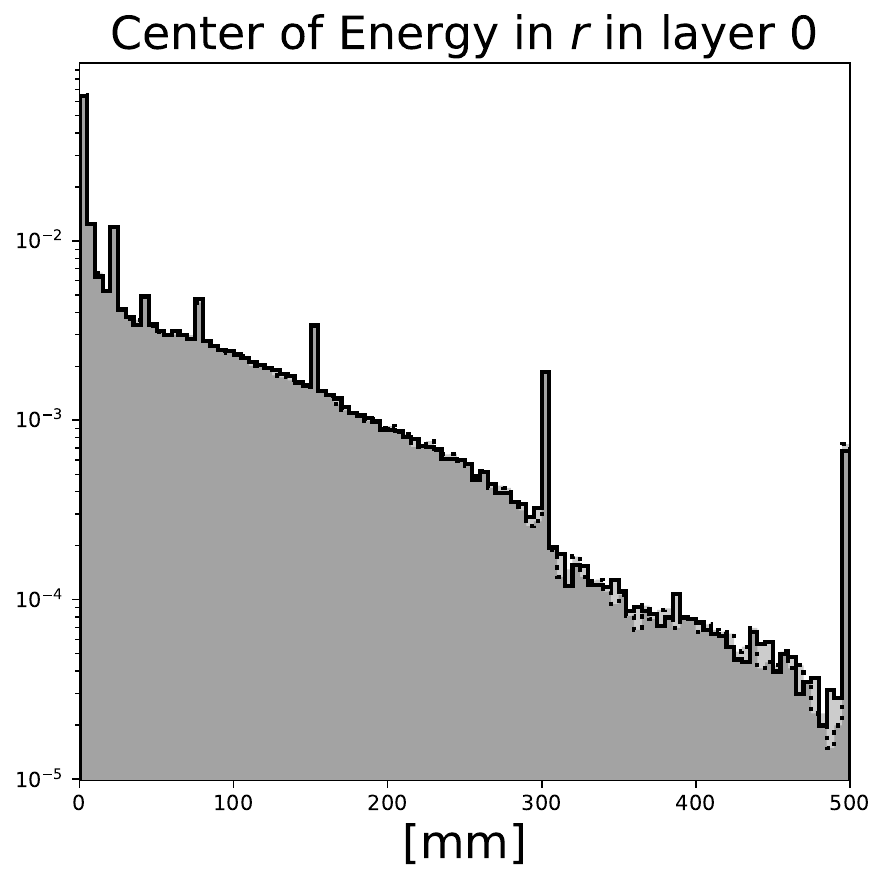} \hfill     \includegraphics[height=0.15\textheight]{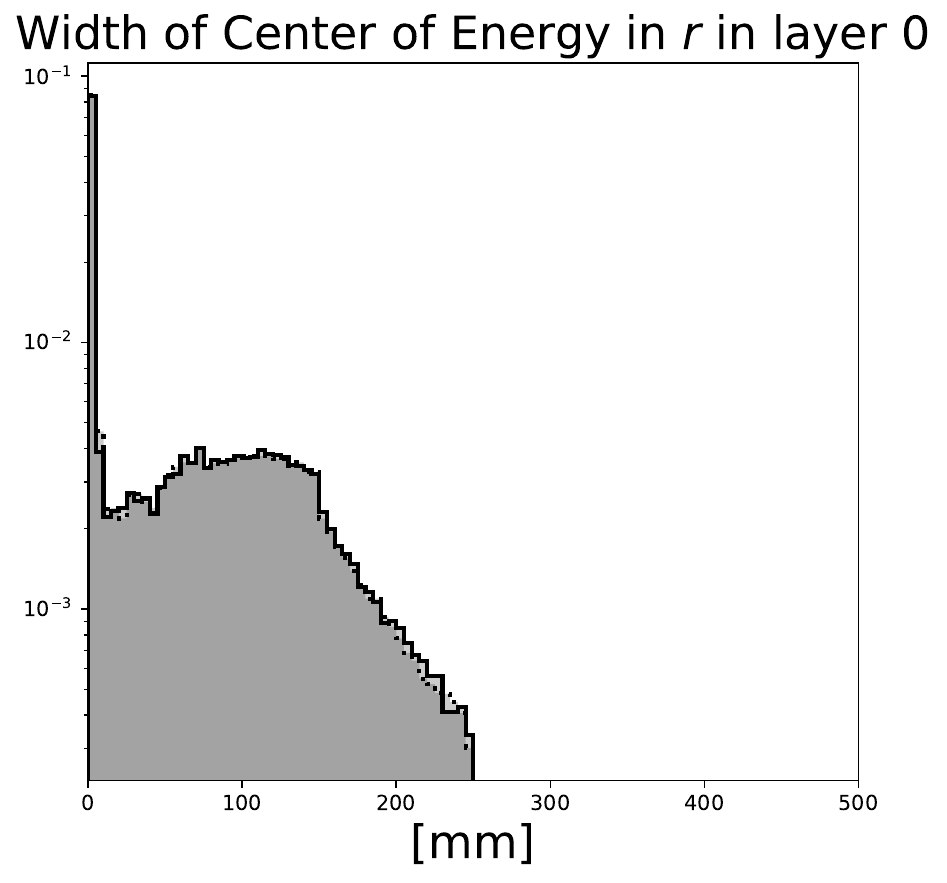} \hfill     \includegraphics[height=0.15\textheight]{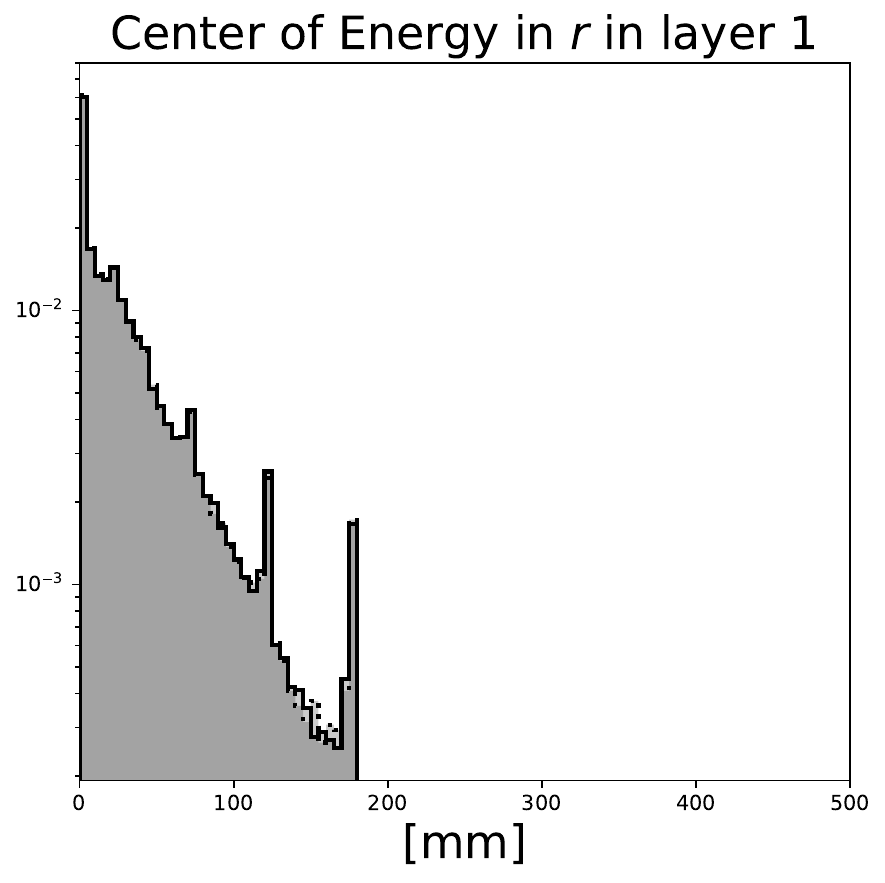} \hfill \includegraphics[height=0.15\textheight]{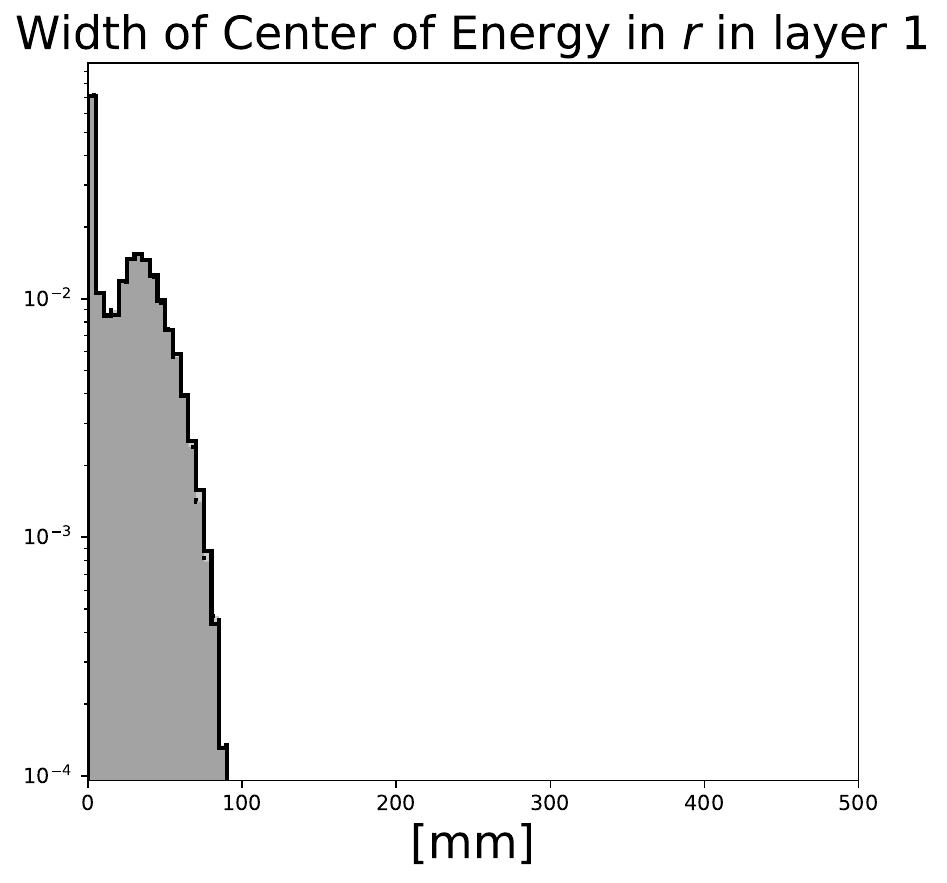}\\
    \includegraphics[height=0.15\textheight]{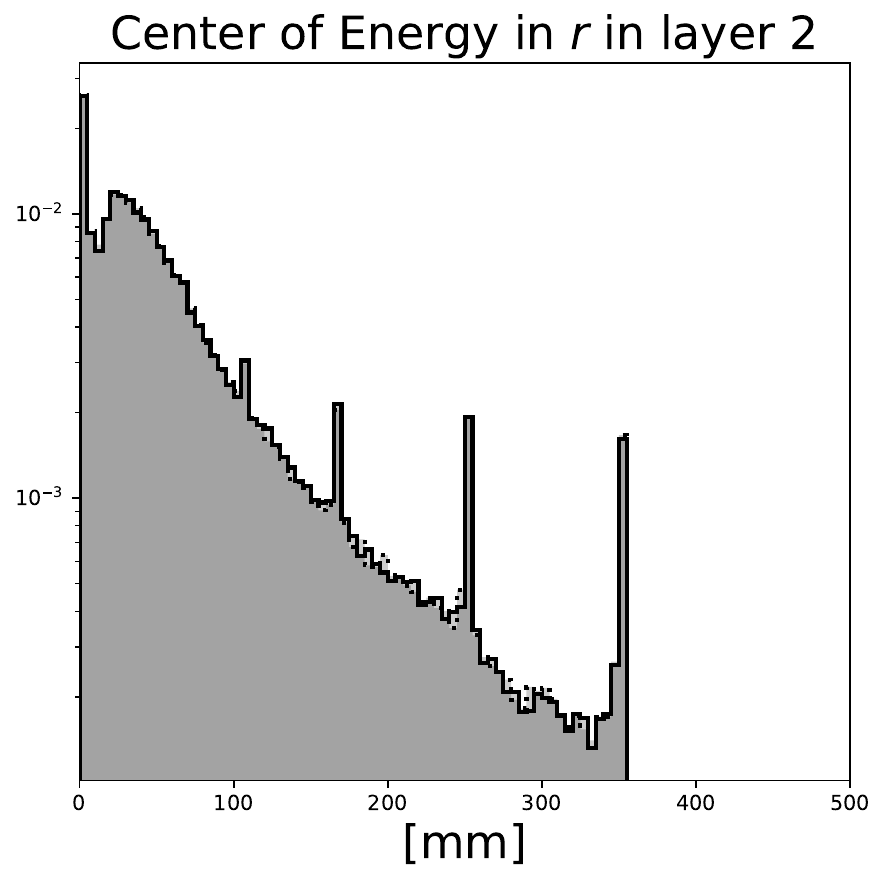} \hfill     \includegraphics[height=0.15\textheight]{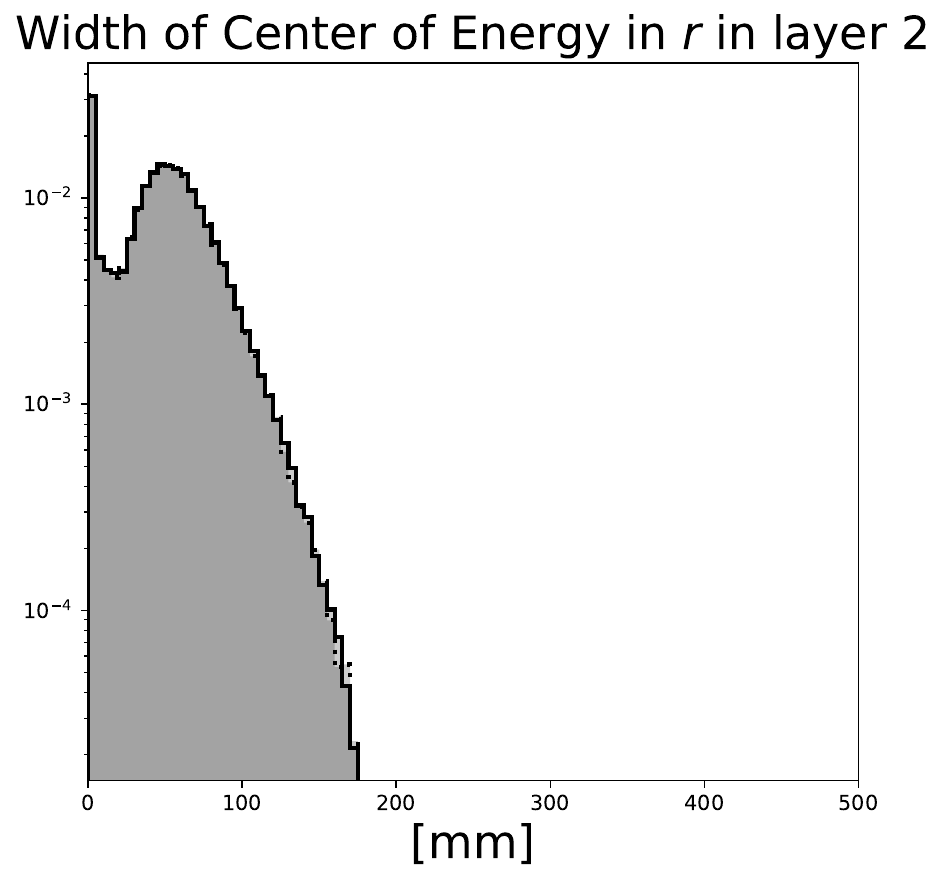}\hfill \includegraphics[height=0.15\textheight]{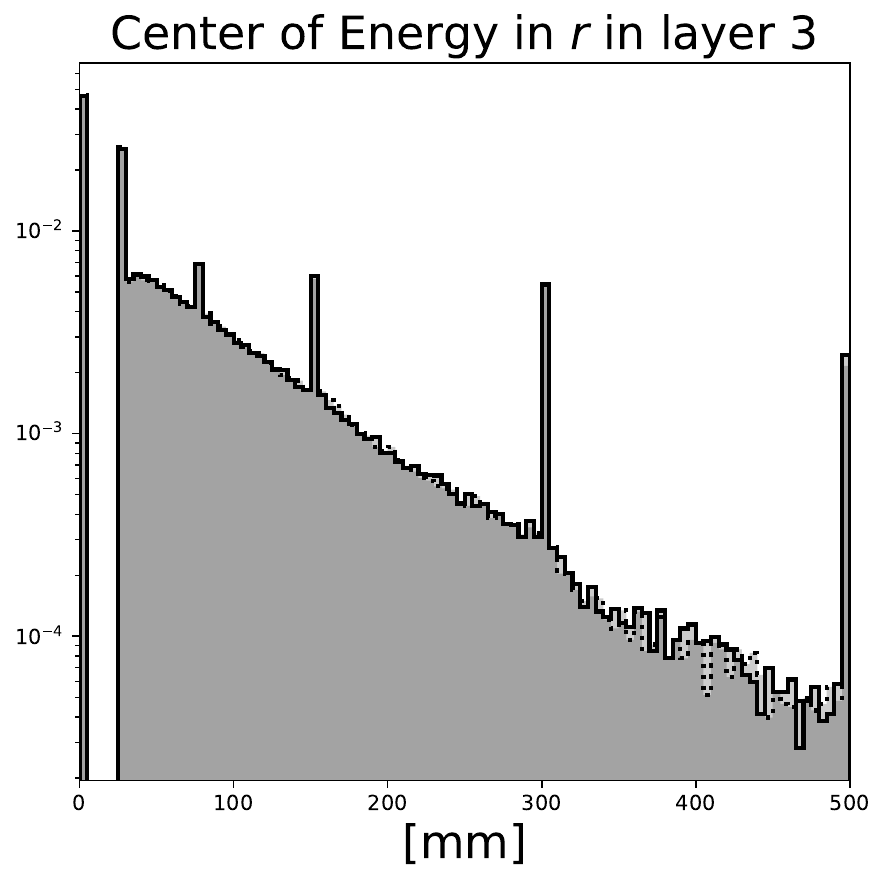} \hfill     \includegraphics[height=0.15\textheight]{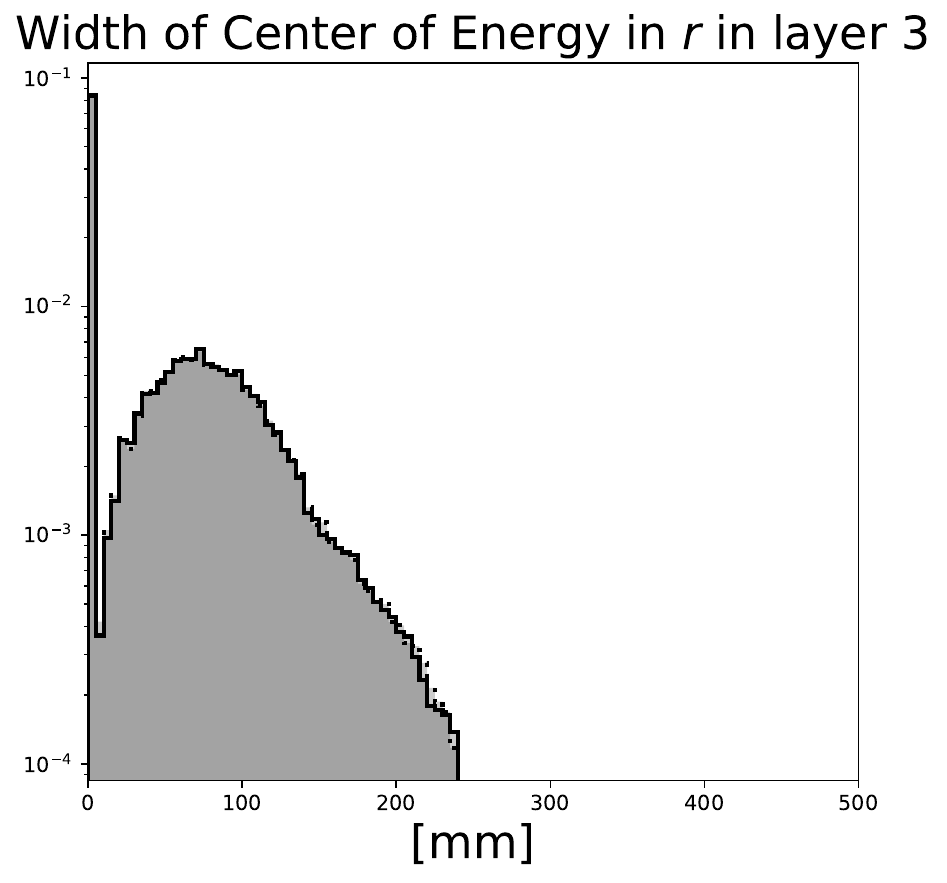} \\
    \includegraphics[height=0.15\textheight]{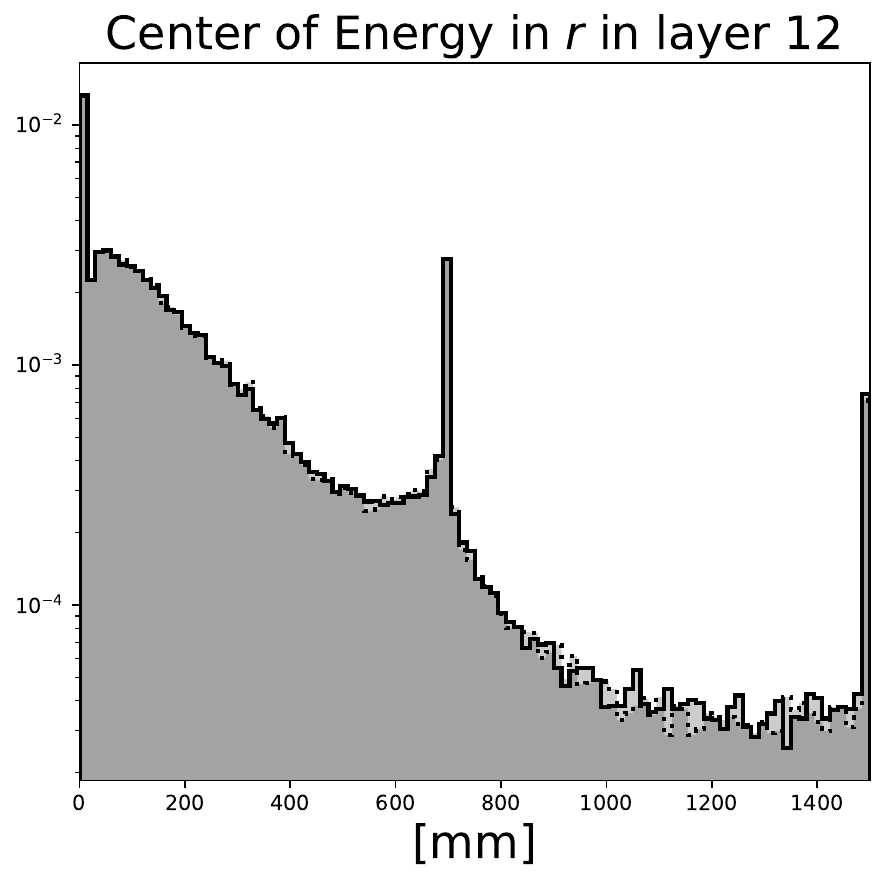} \hfill \includegraphics[height=0.15\textheight]{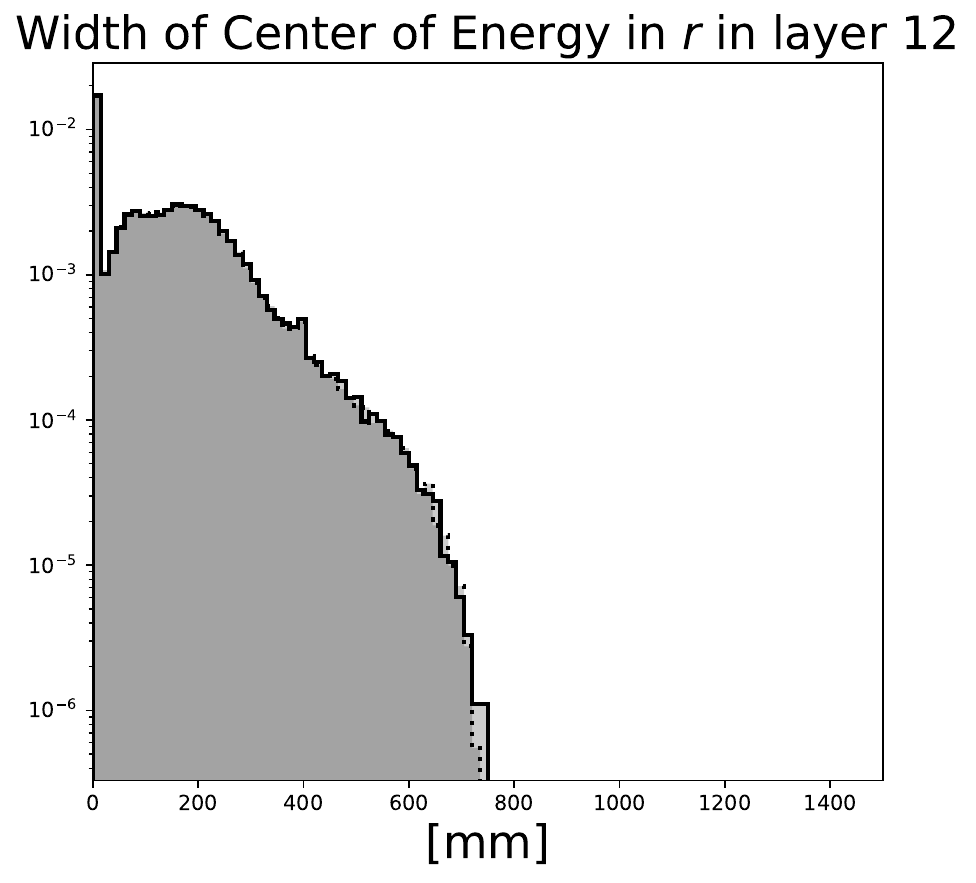} \hfill     \includegraphics[height=0.15\textheight]{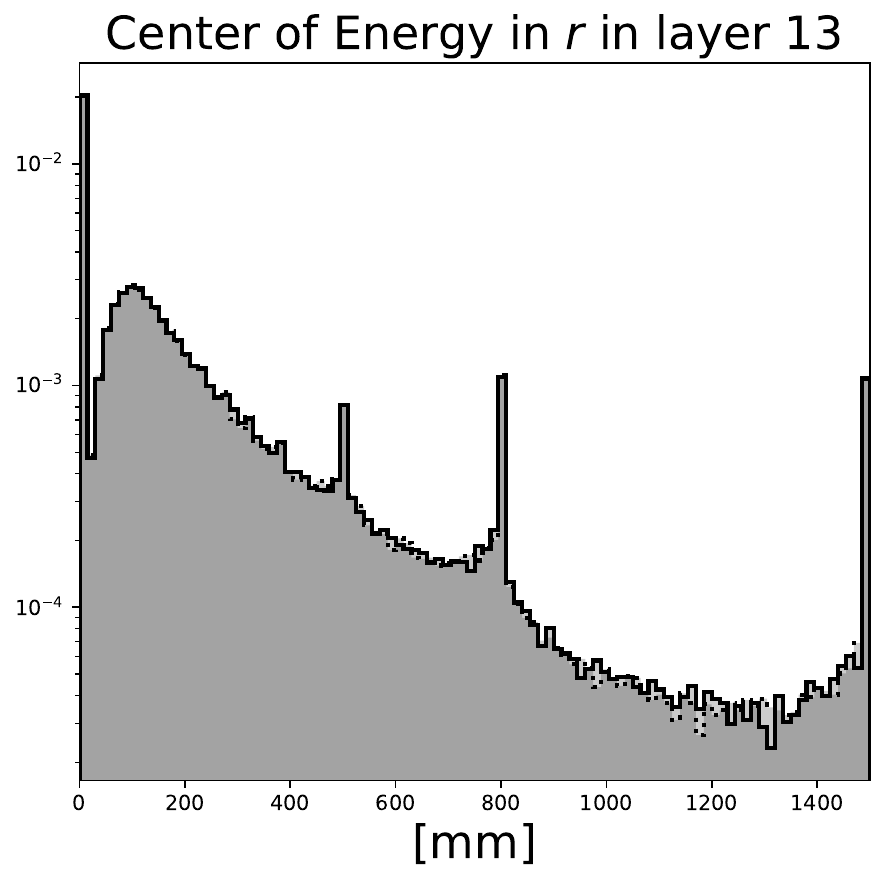} \hfill     \includegraphics[height=0.15\textheight]{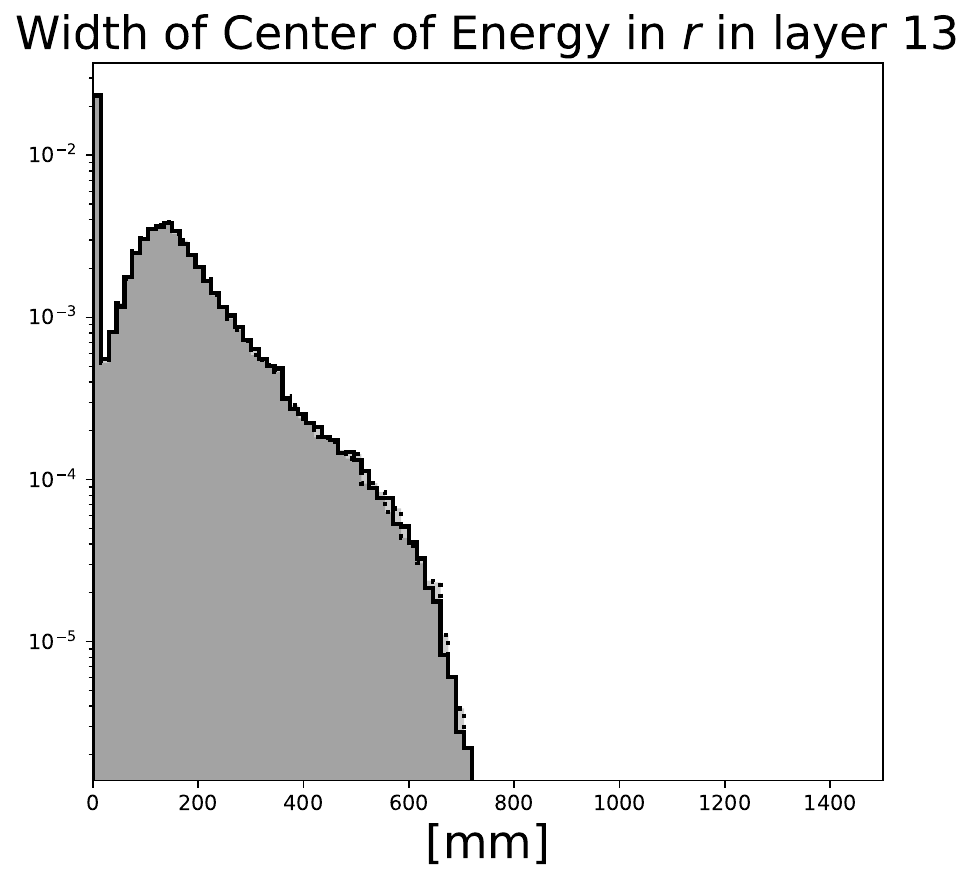}\\
    \hfill \includegraphics[height=0.15\textheight]{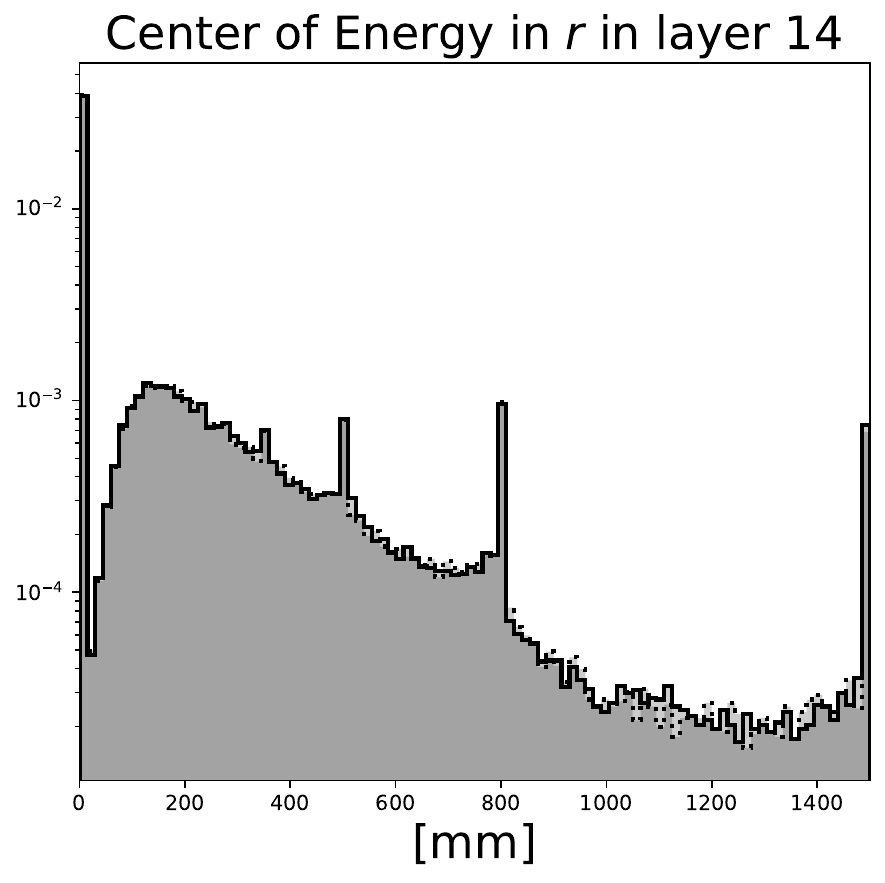} \hfill     \includegraphics[height=0.15\textheight]{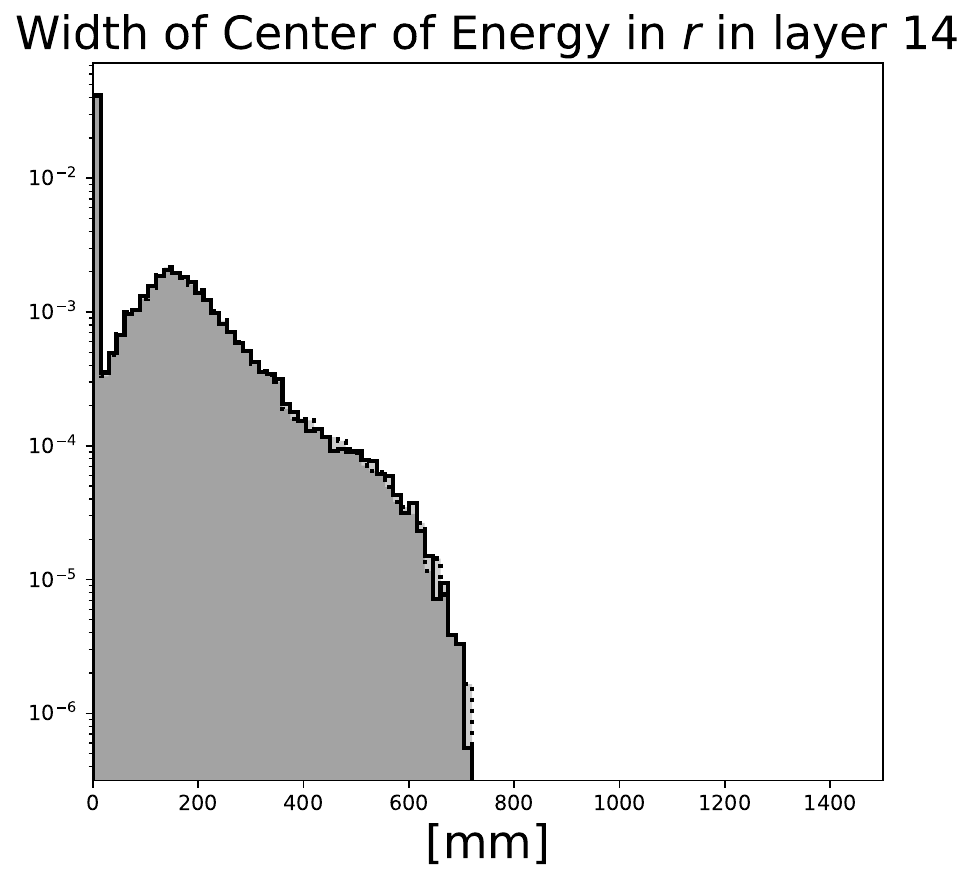} \hfill $ $\\
    \includegraphics[width=0.5\textwidth]{figures/Appendix_reference/legend.pdf}
    \caption{Distribution of \geant training and evaluation data in centers of energy along the radial direction, as well as their widths for ds1 --- pions.}
    \label{fig:app_ref.ds1-pions.4}
\end{figure}

\subsection{\texorpdfstring{Dataset 2, Electrons (\dsII)}{Dataset 2, Electrons}}
\begin{figure}[ht]
    \centering
    \hfill \includegraphics[height=0.2\textheight]{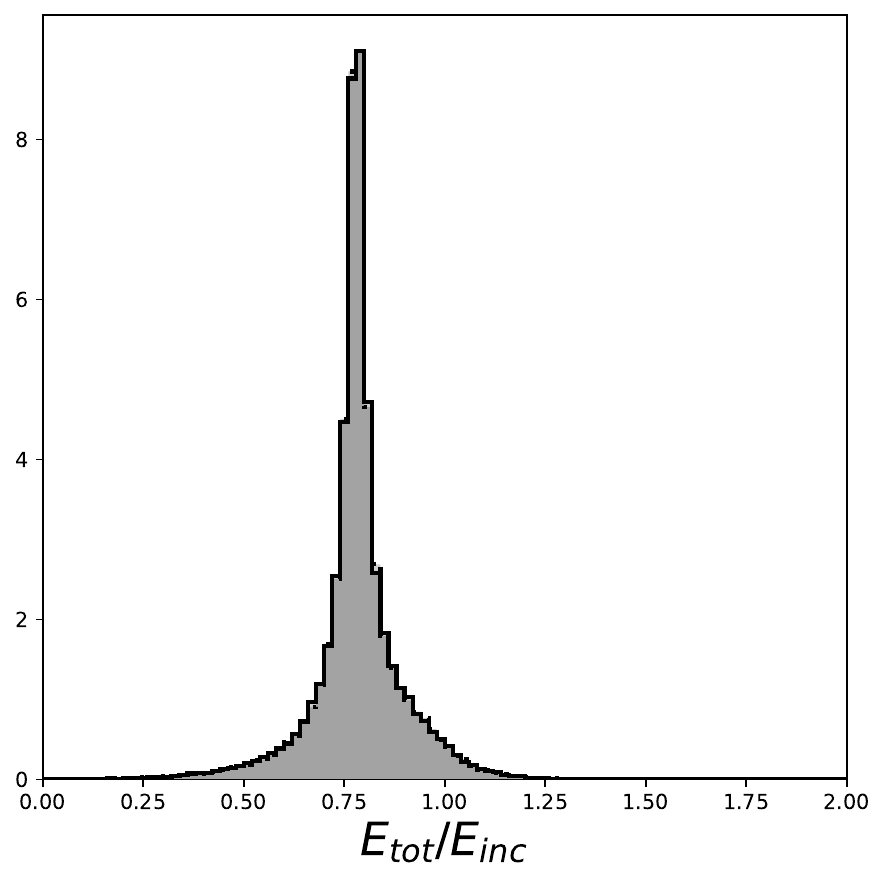} \hfill     \includegraphics[height=0.2\textheight]{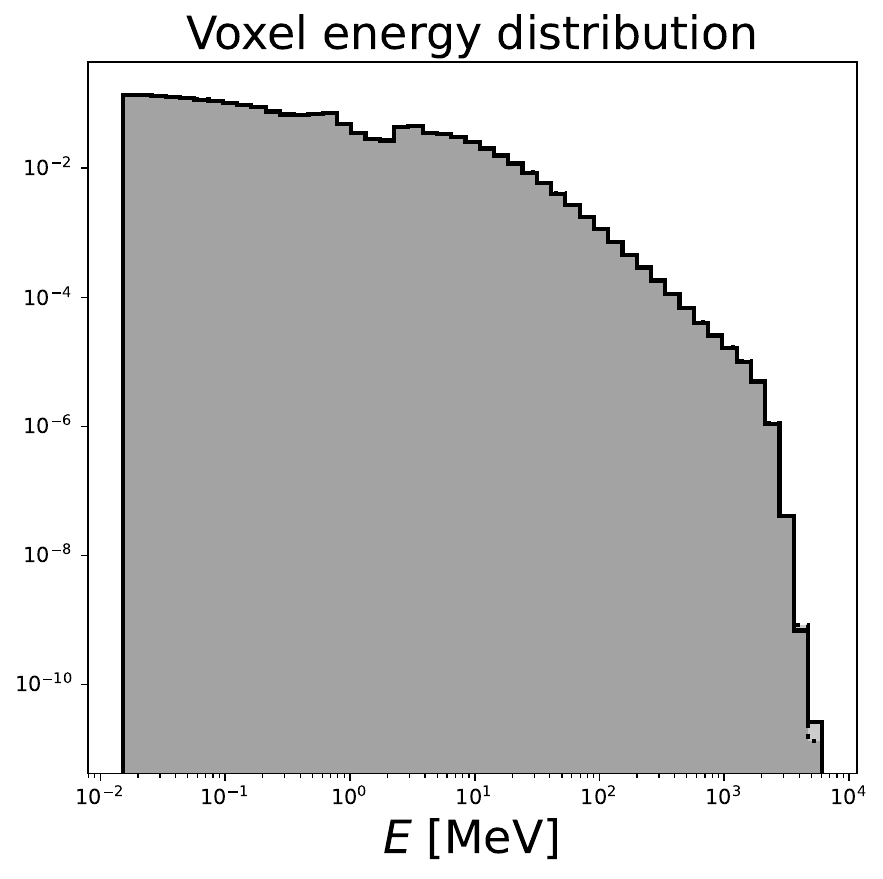} \hfill $ $\\
    \includegraphics[width=0.5\textwidth]{figures/Appendix_reference/legend.pdf}
    \caption{Distribution of \geant training and evaluation data in ratio of total deposited energy to incident energy and energy per voxel for ds2. }
    \label{fig:app_ref.ds2.1}
\end{figure}

\begin{figure}[ht]
    \centering
    \includegraphics[height=0.1\textheight]{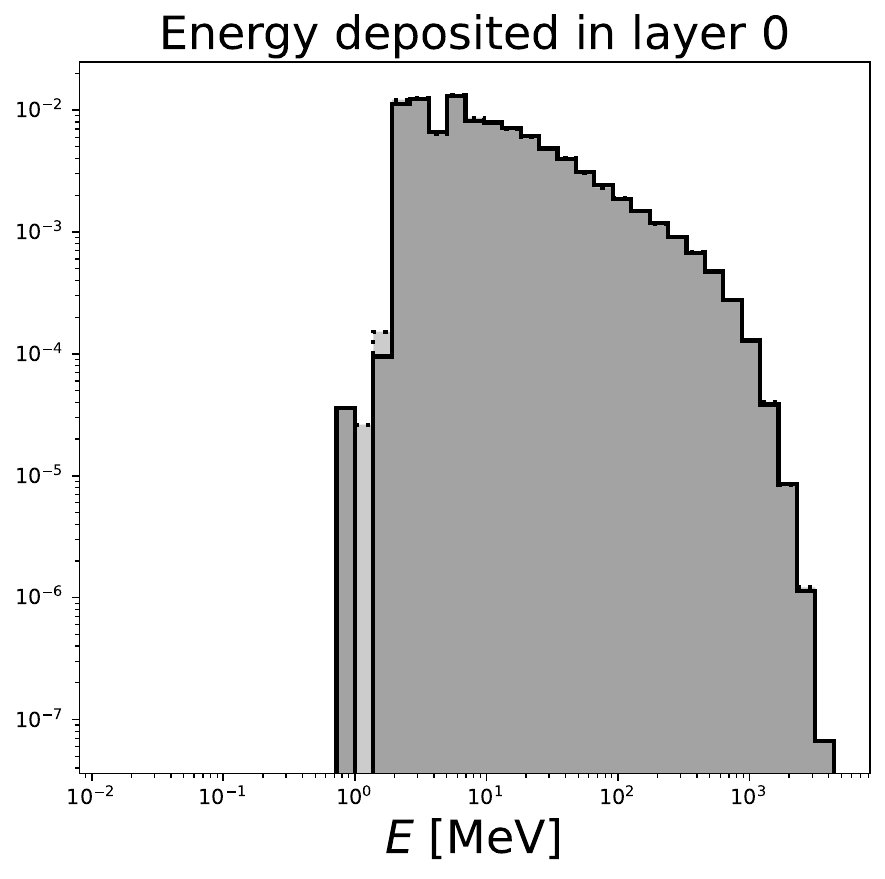} \hfill \includegraphics[height=0.1\textheight]{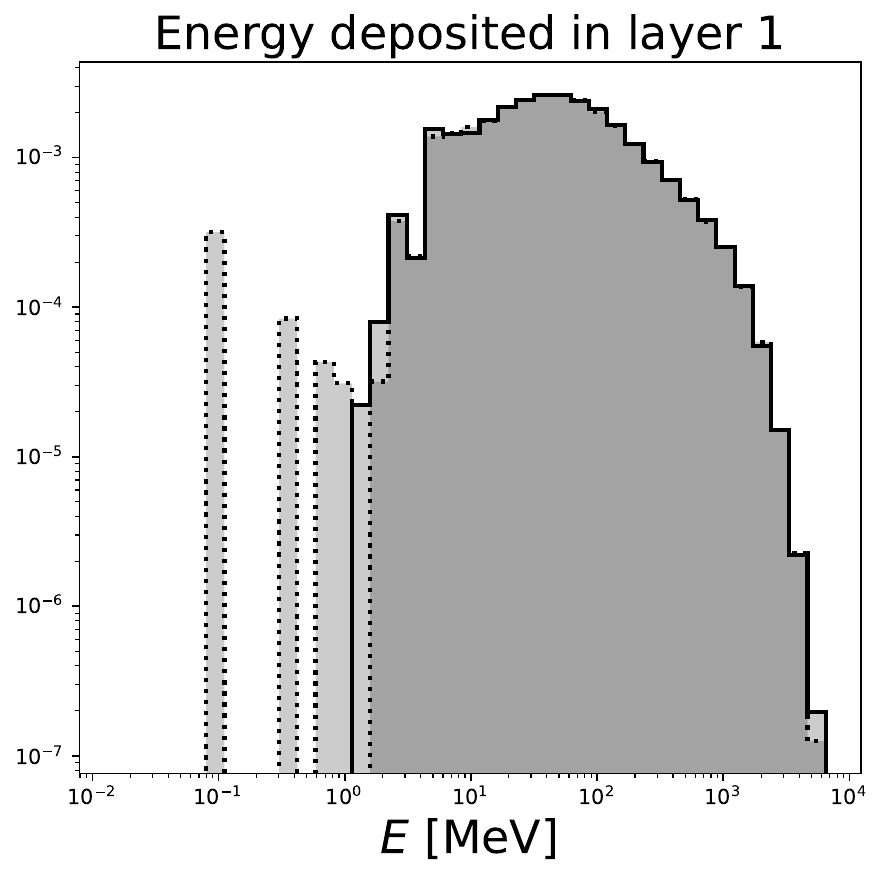} \hfill \includegraphics[height=0.1\textheight]{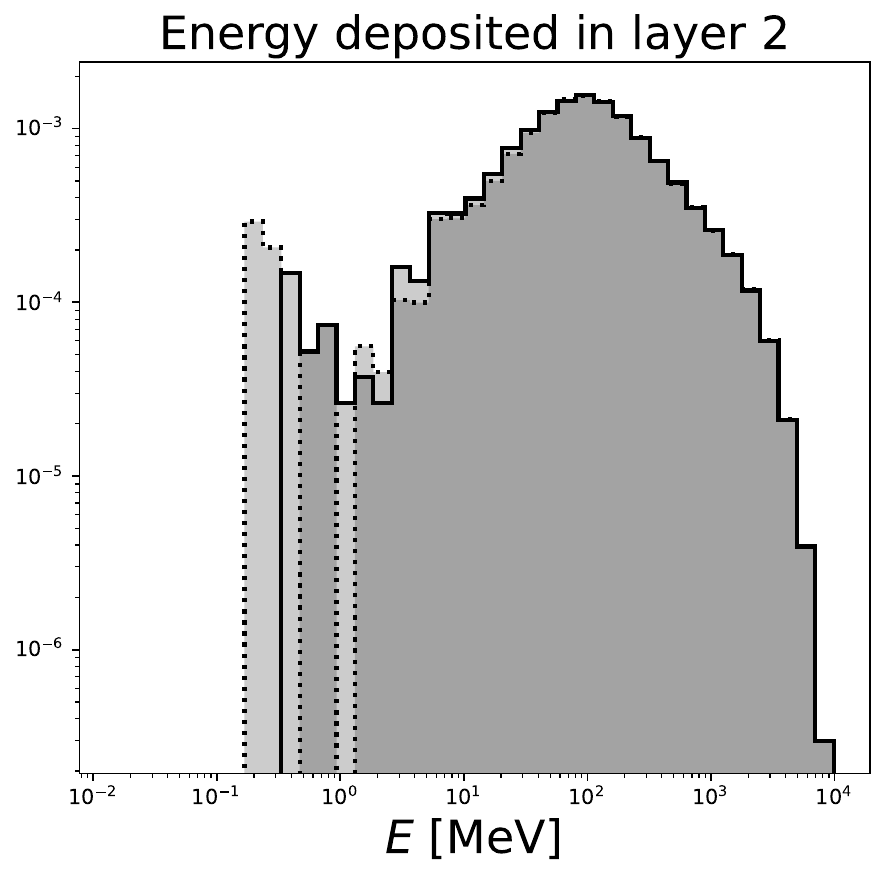} \hfill \includegraphics[height=0.1\textheight]{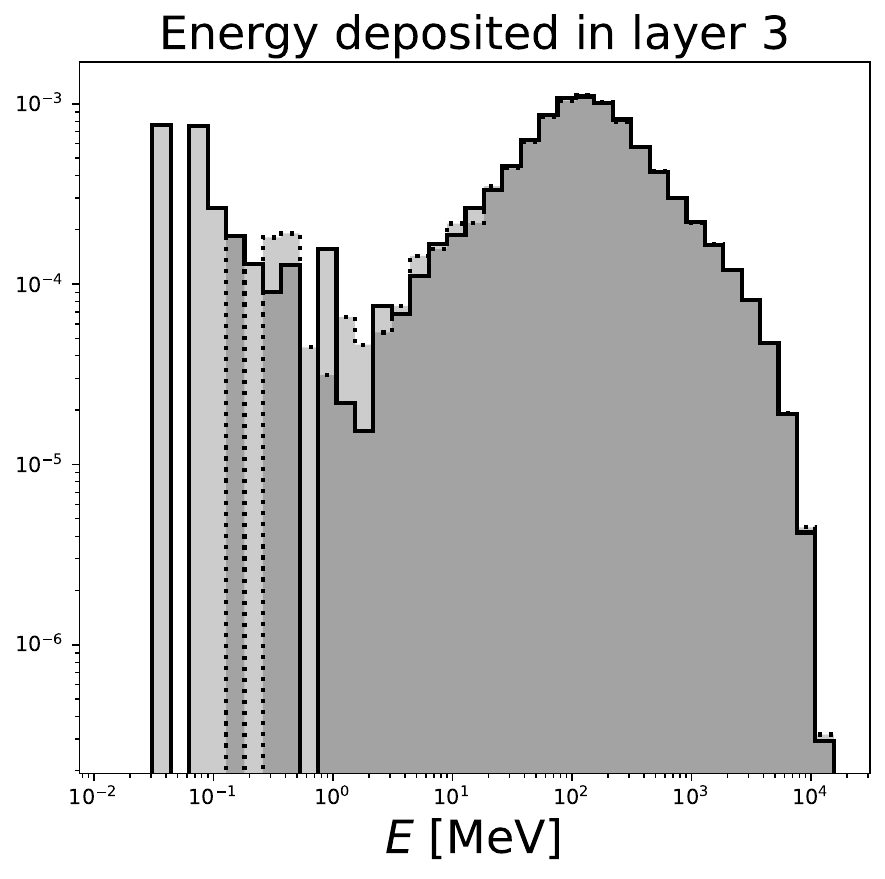} \hfill \includegraphics[height=0.1\textheight]{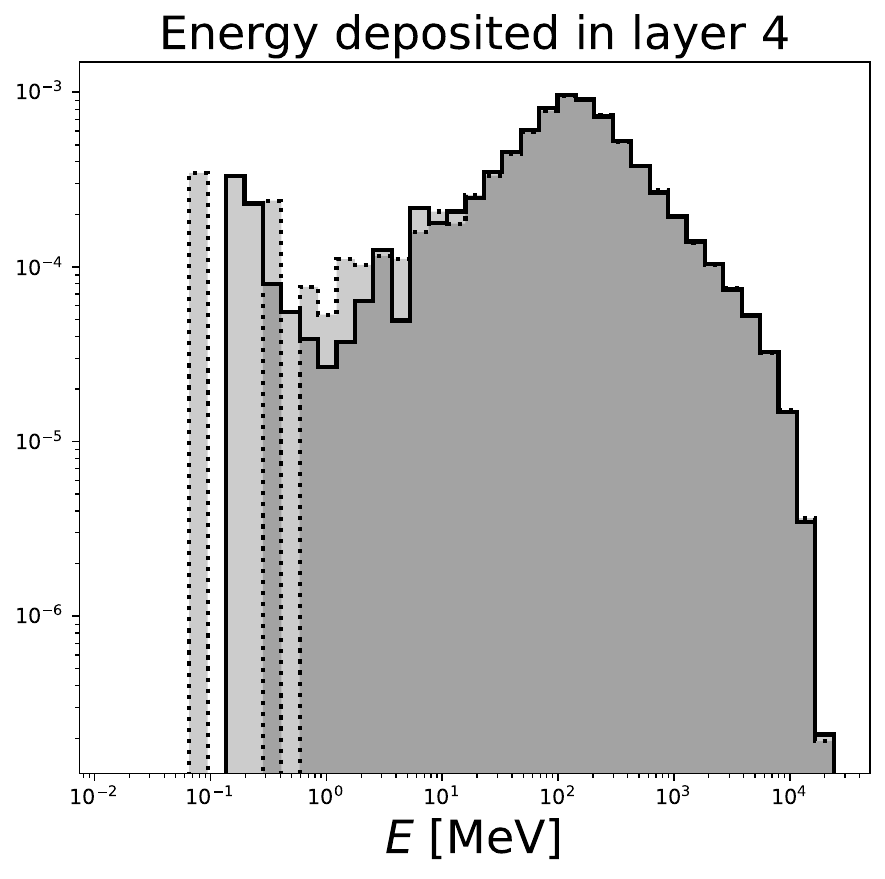}\\
    \includegraphics[height=0.1\textheight]{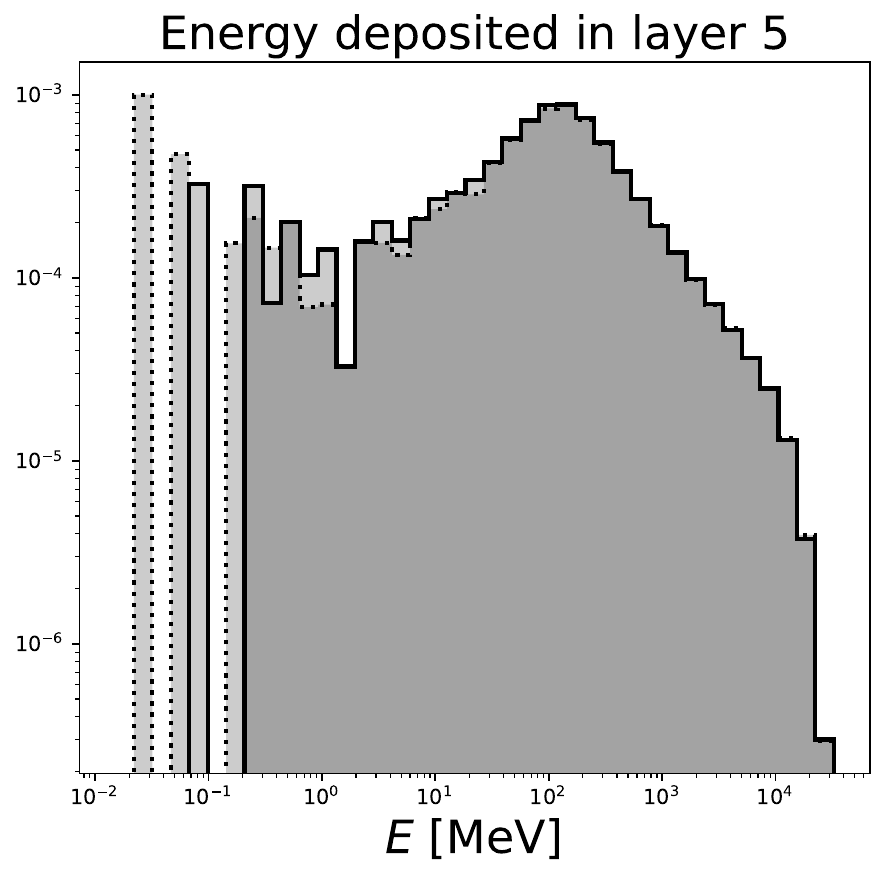} \hfill \includegraphics[height=0.1\textheight]{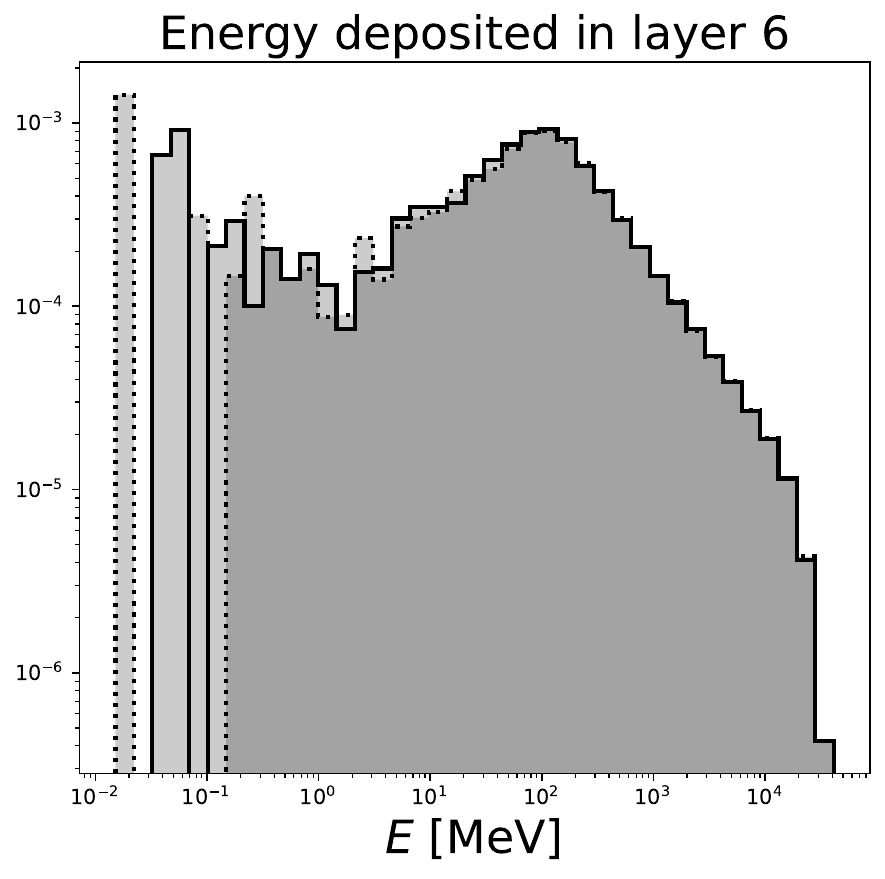} \hfill \includegraphics[height=0.1\textheight]{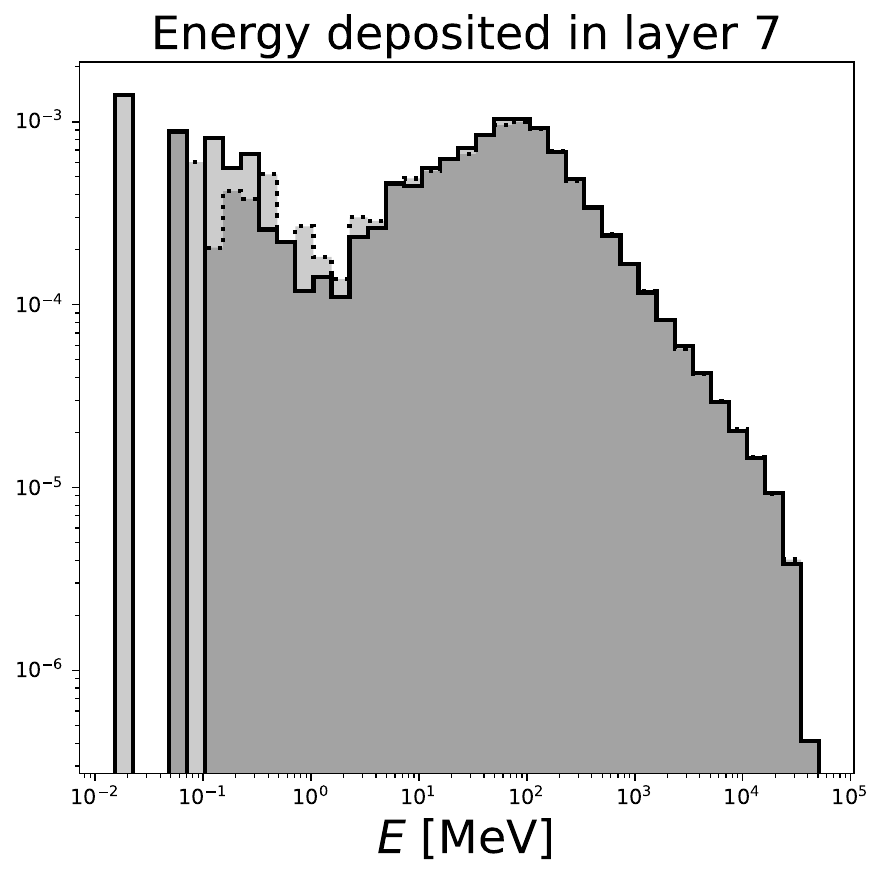} \hfill \includegraphics[height=0.1\textheight]{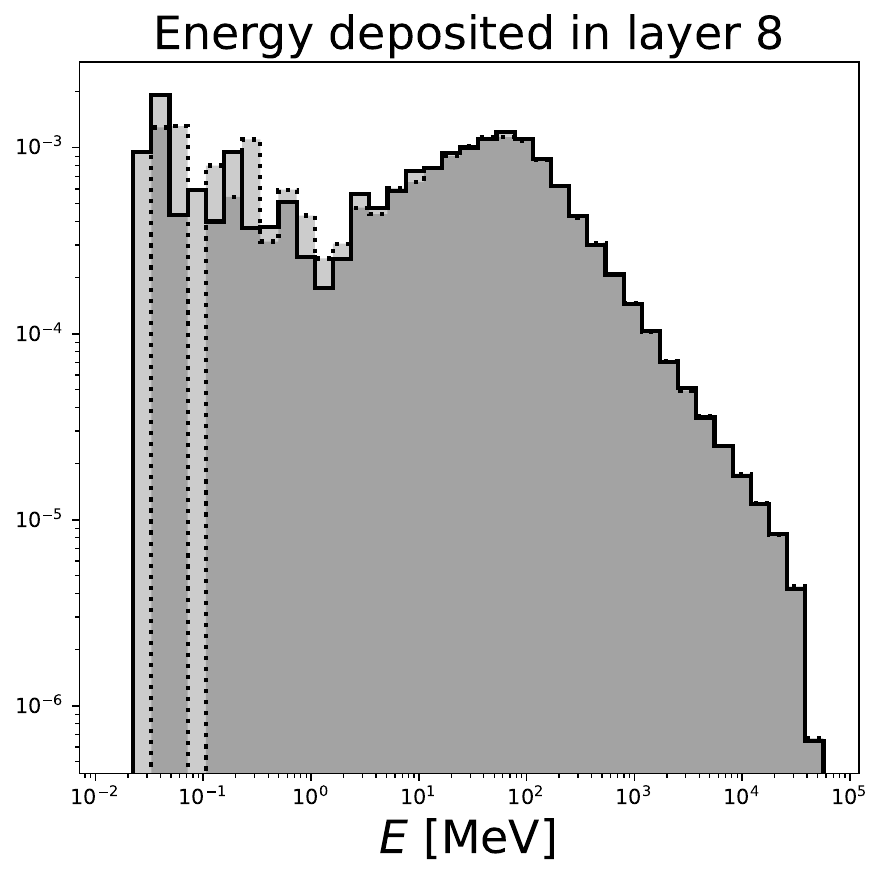} \hfill \includegraphics[height=0.1\textheight]{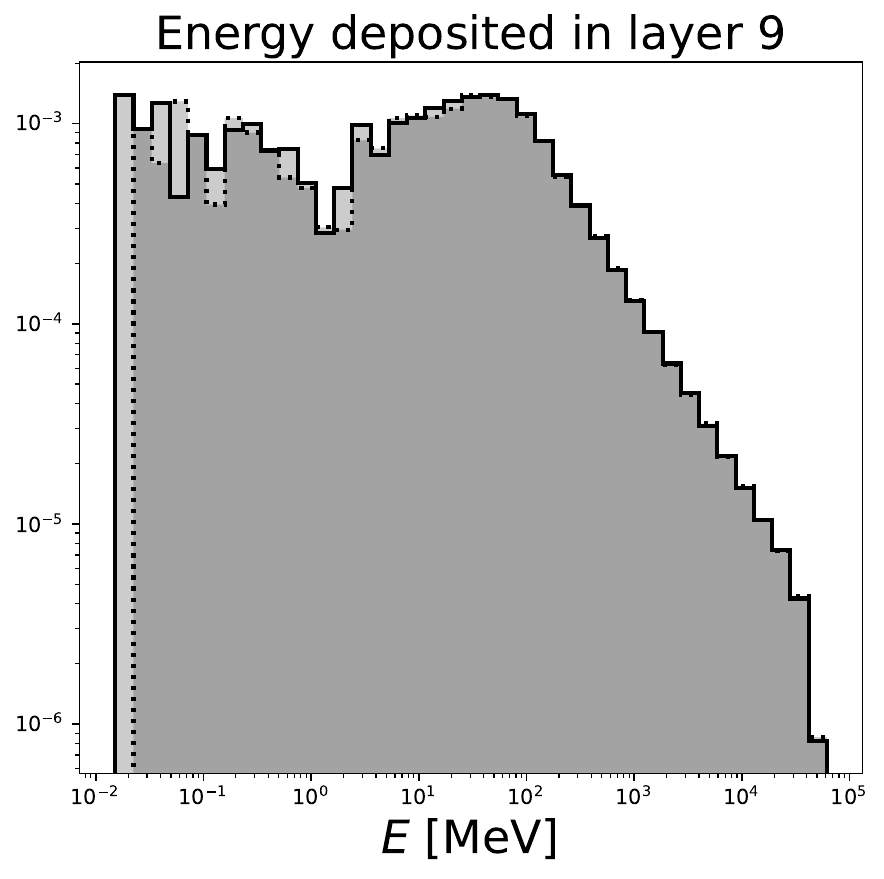}\\
    \includegraphics[height=0.1\textheight]{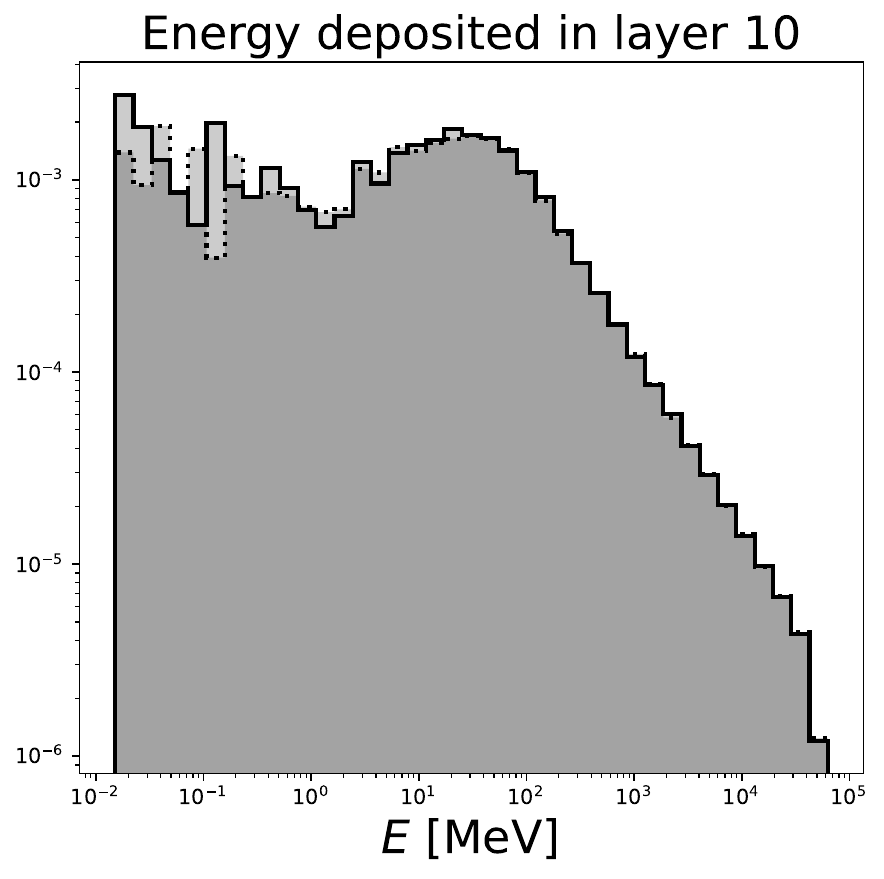} \hfill \includegraphics[height=0.1\textheight]{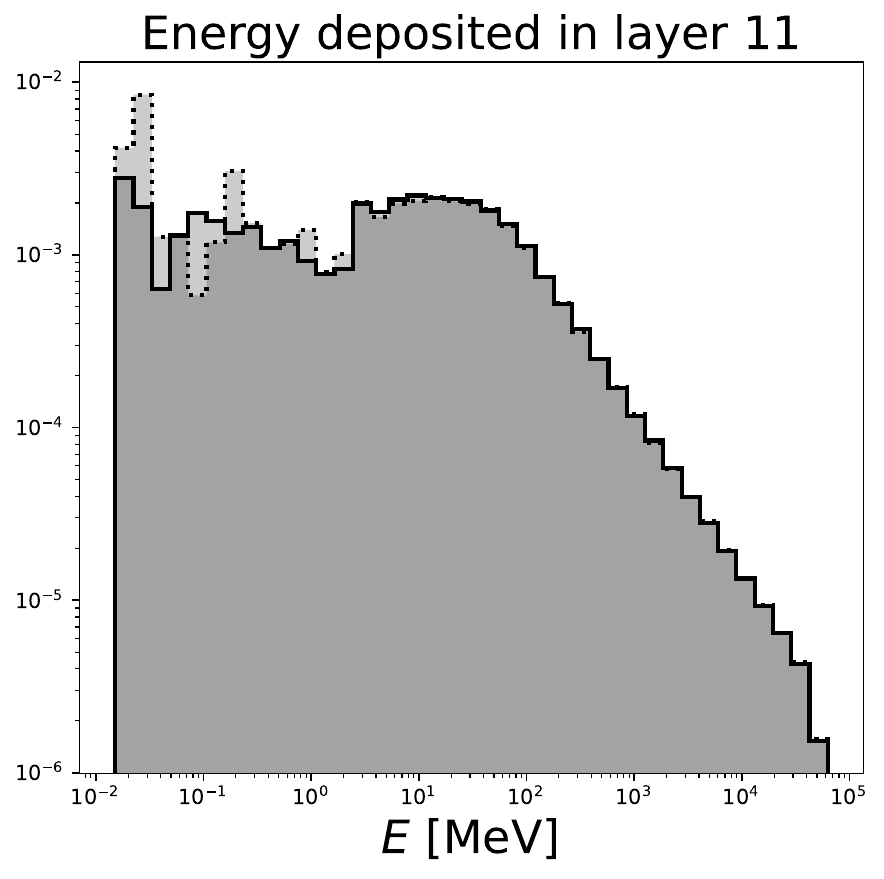} \hfill \includegraphics[height=0.1\textheight]{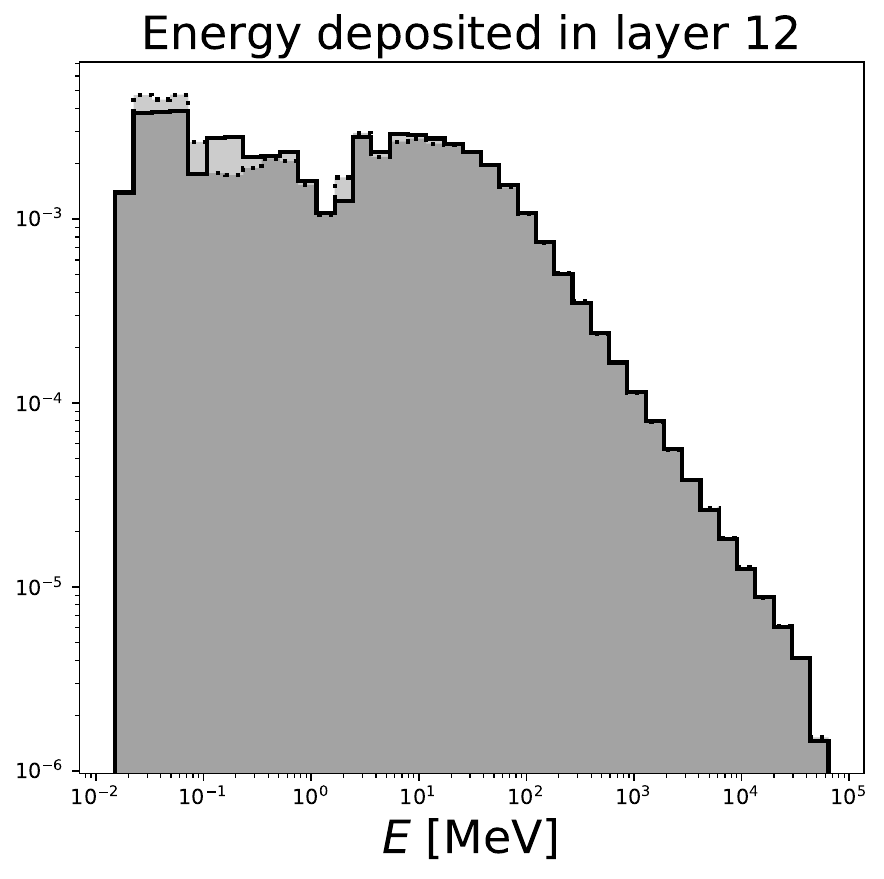} \hfill \includegraphics[height=0.1\textheight]{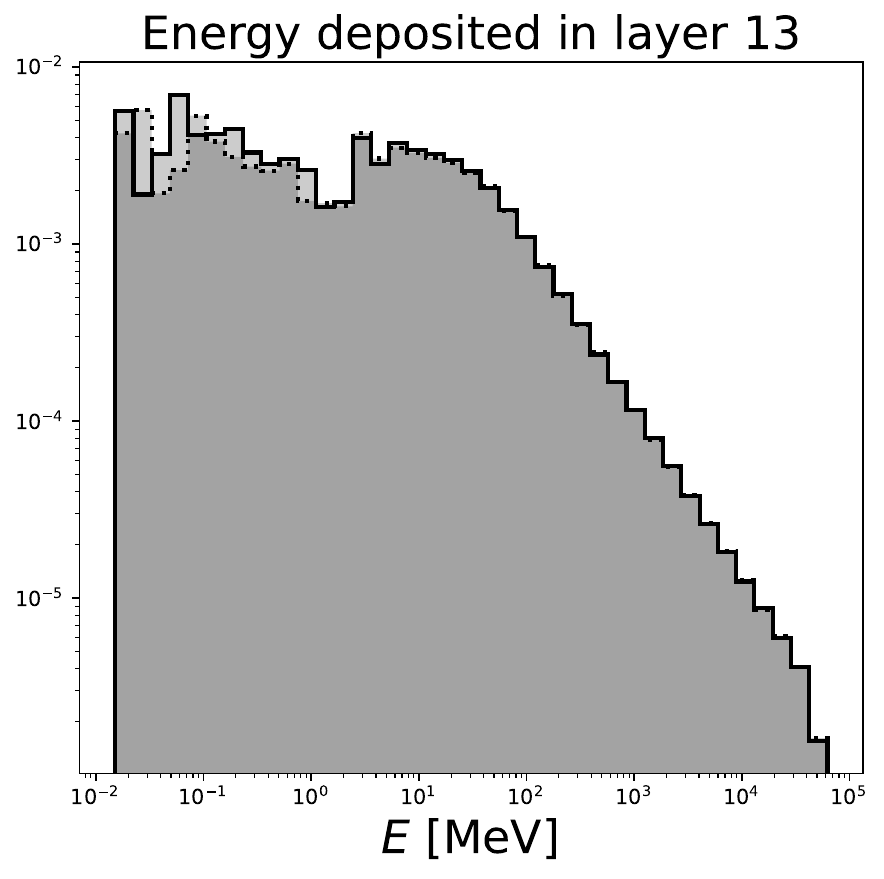} \hfill \includegraphics[height=0.1\textheight]{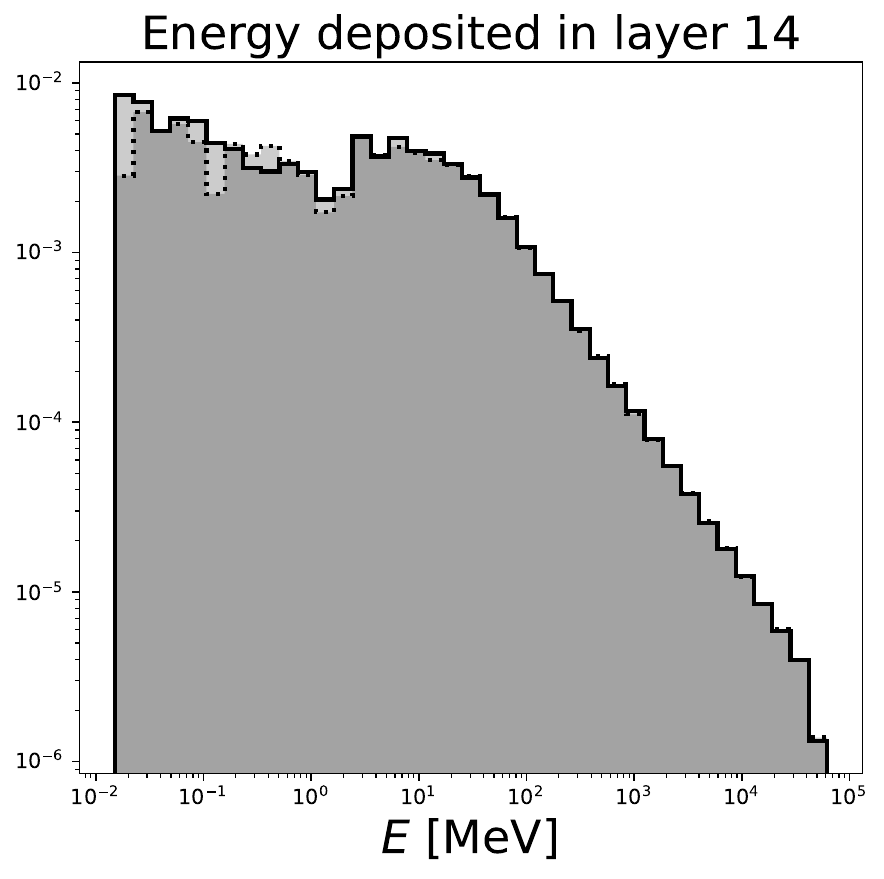}\\
    \includegraphics[height=0.1\textheight]{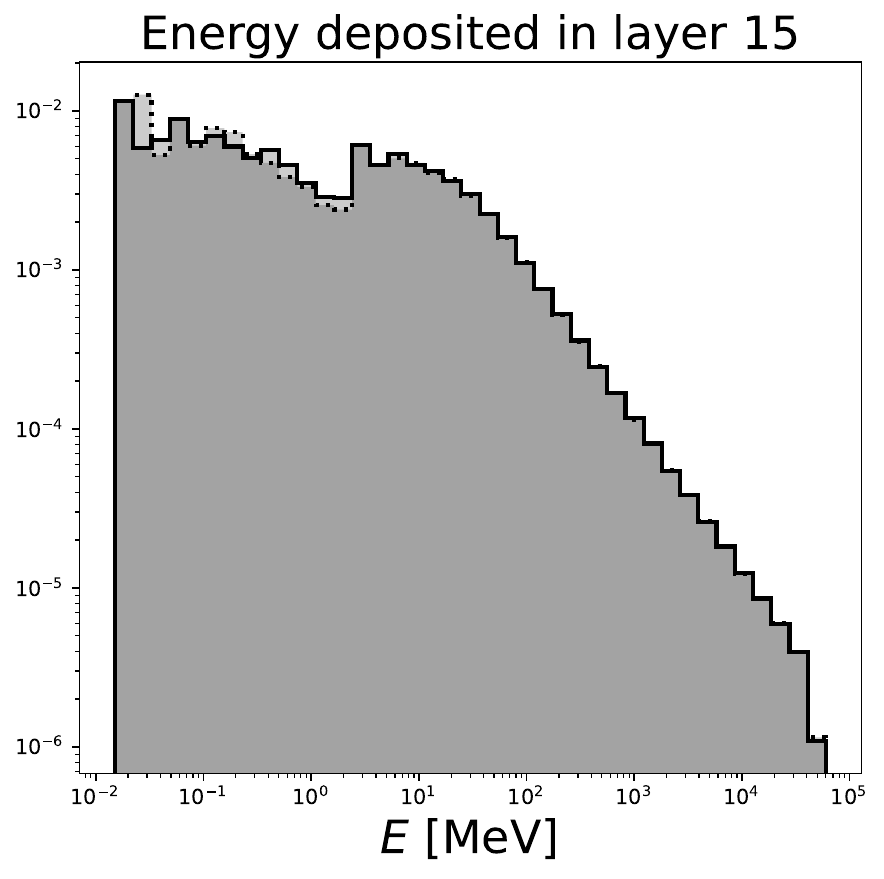} \hfill \includegraphics[height=0.1\textheight]{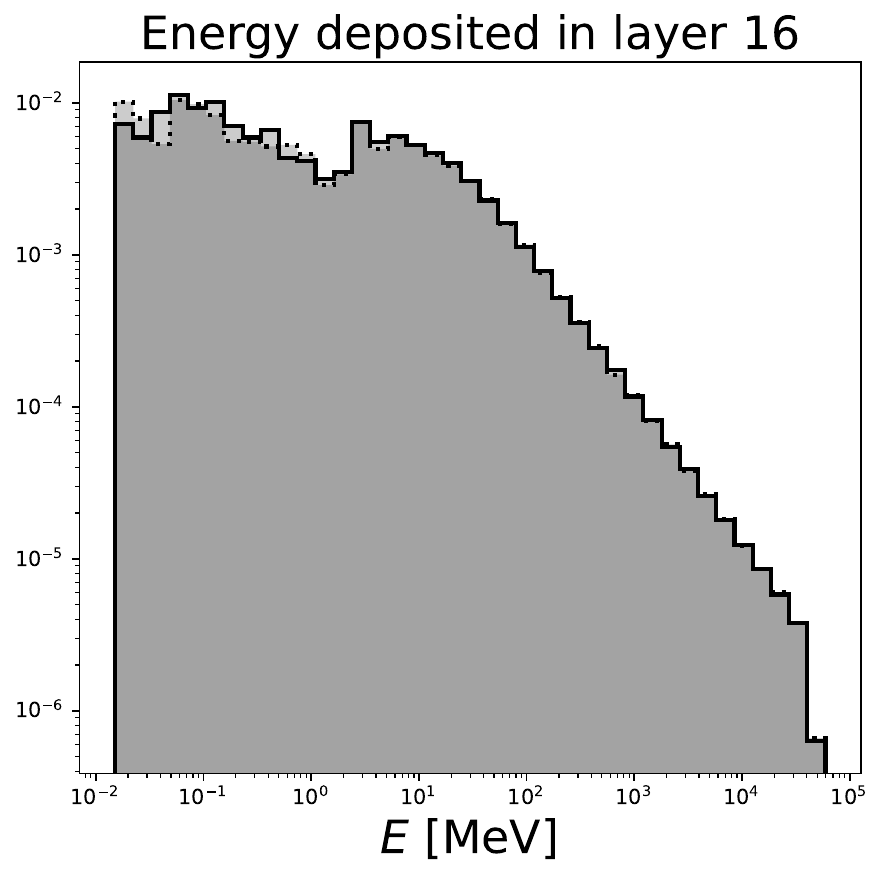} \hfill \includegraphics[height=0.1\textheight]{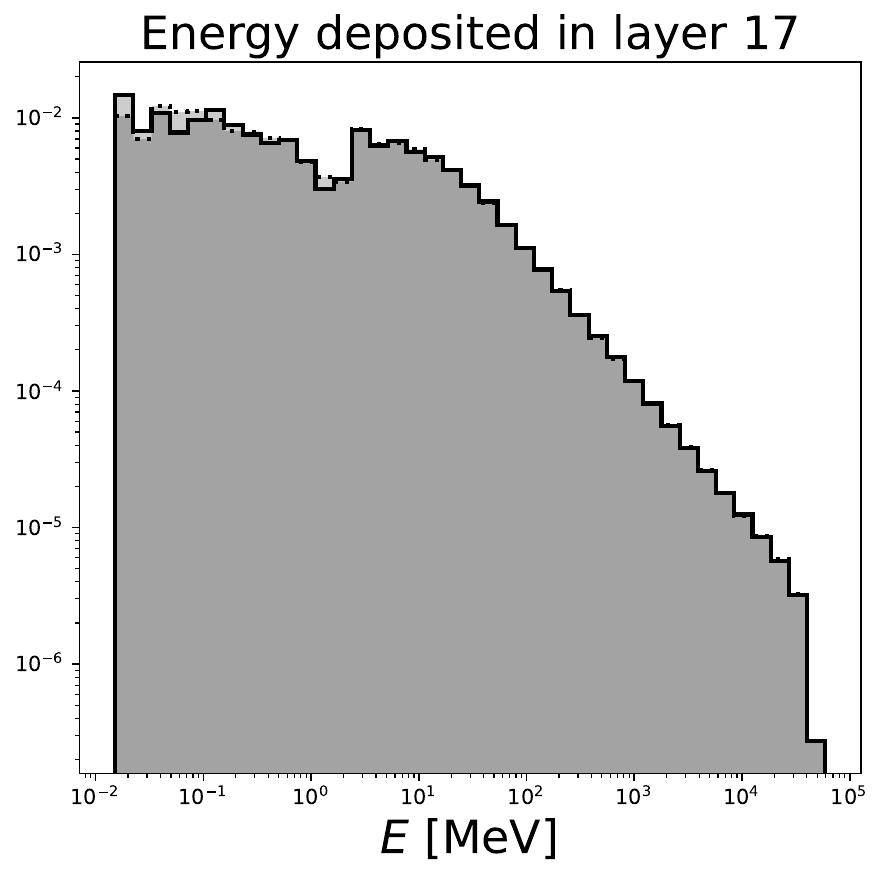} \hfill \includegraphics[height=0.1\textheight]{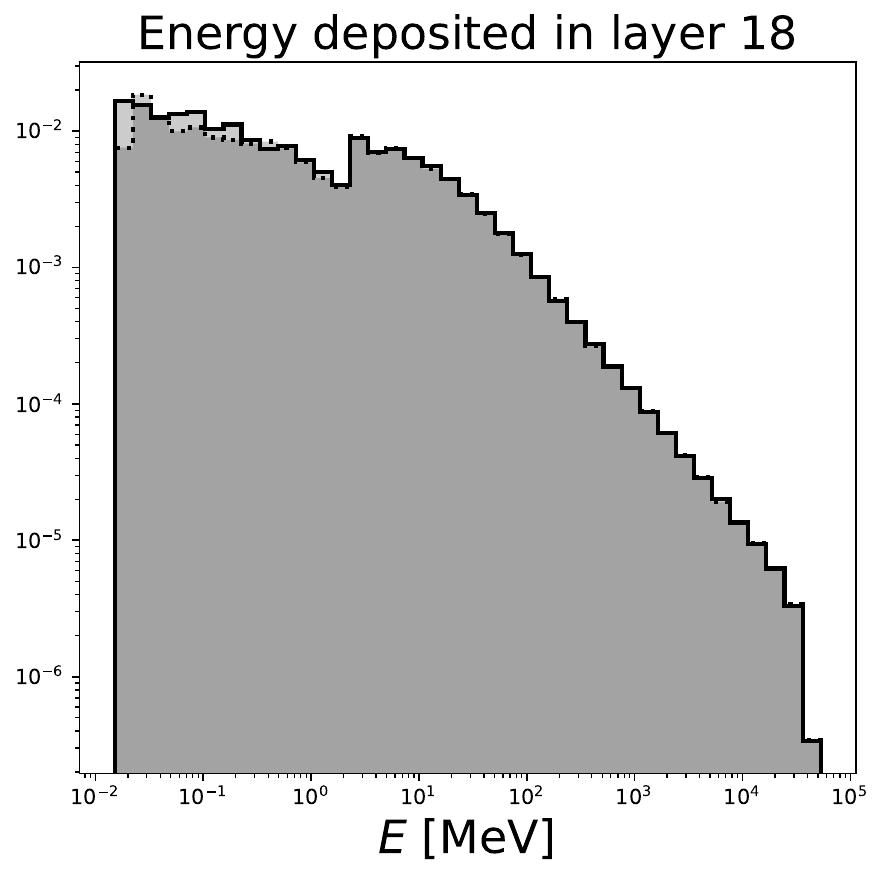} \hfill \includegraphics[height=0.1\textheight]{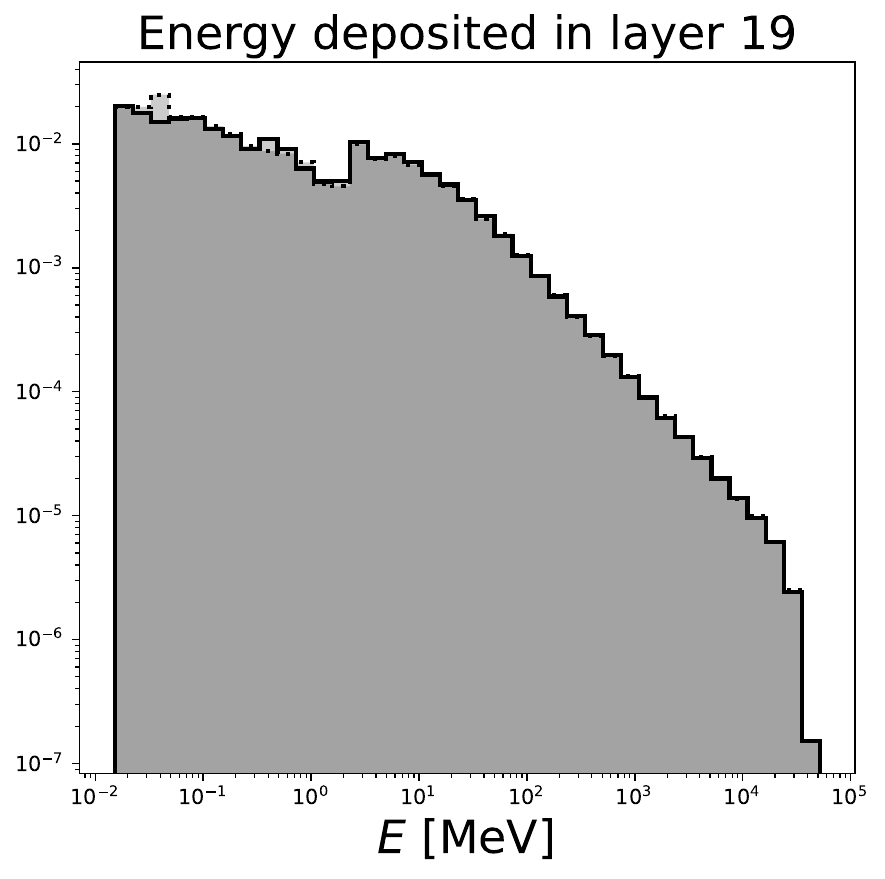}\\
    \includegraphics[height=0.1\textheight]{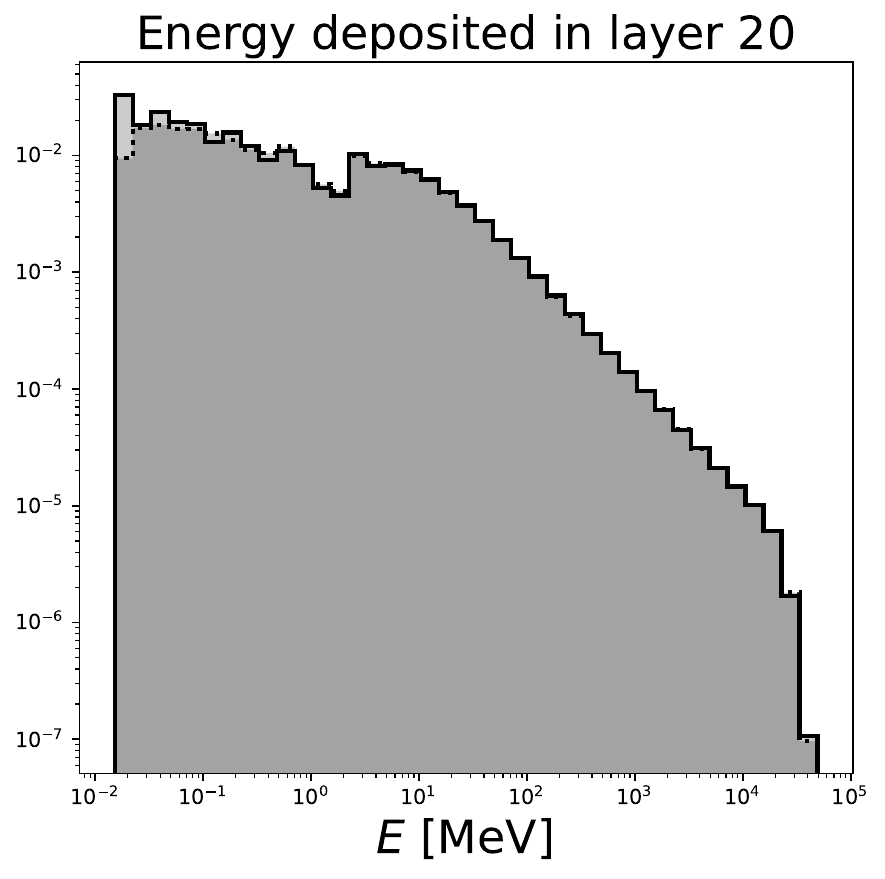} \hfill \includegraphics[height=0.1\textheight]{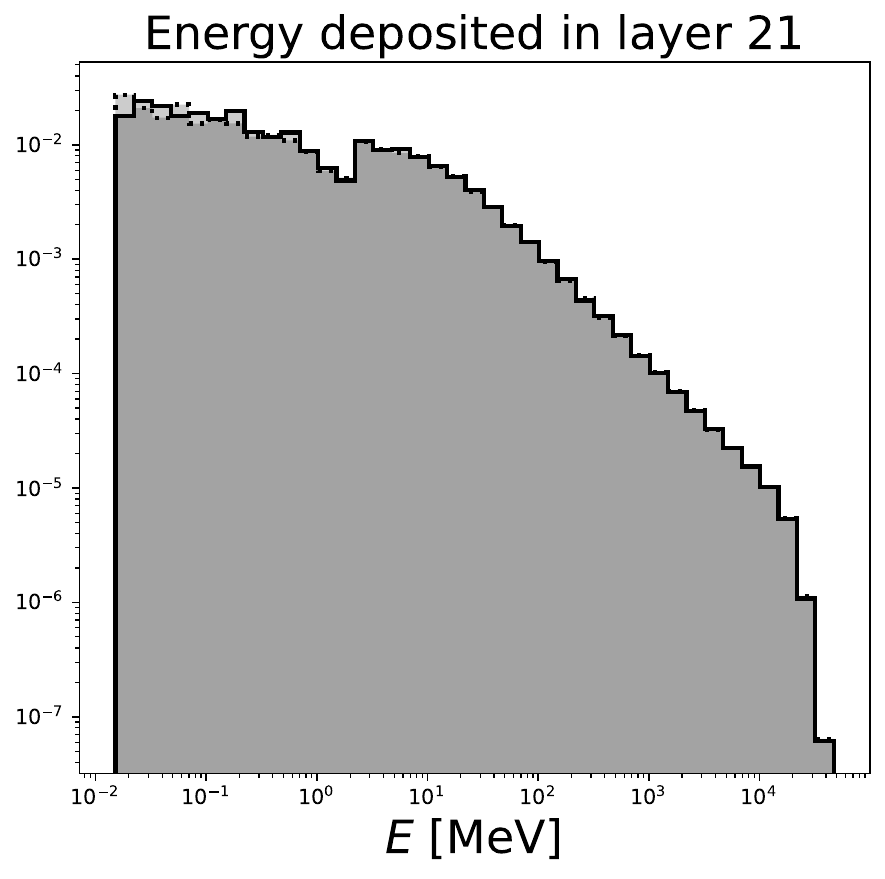} \hfill \includegraphics[height=0.1\textheight]{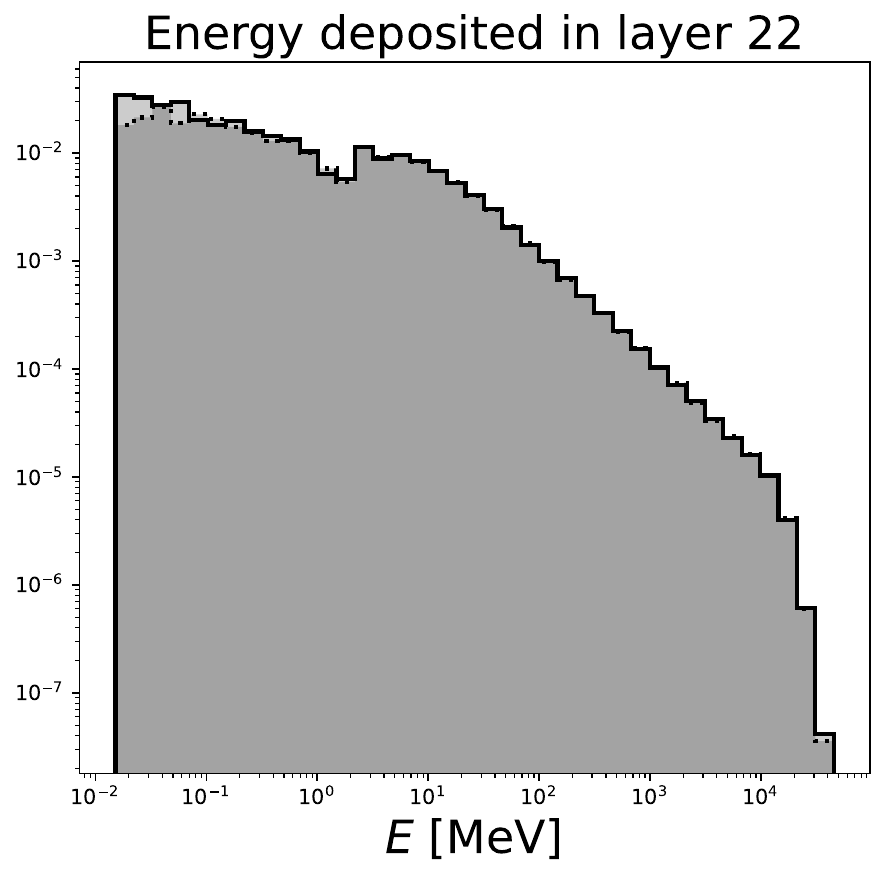} \hfill \includegraphics[height=0.1\textheight]{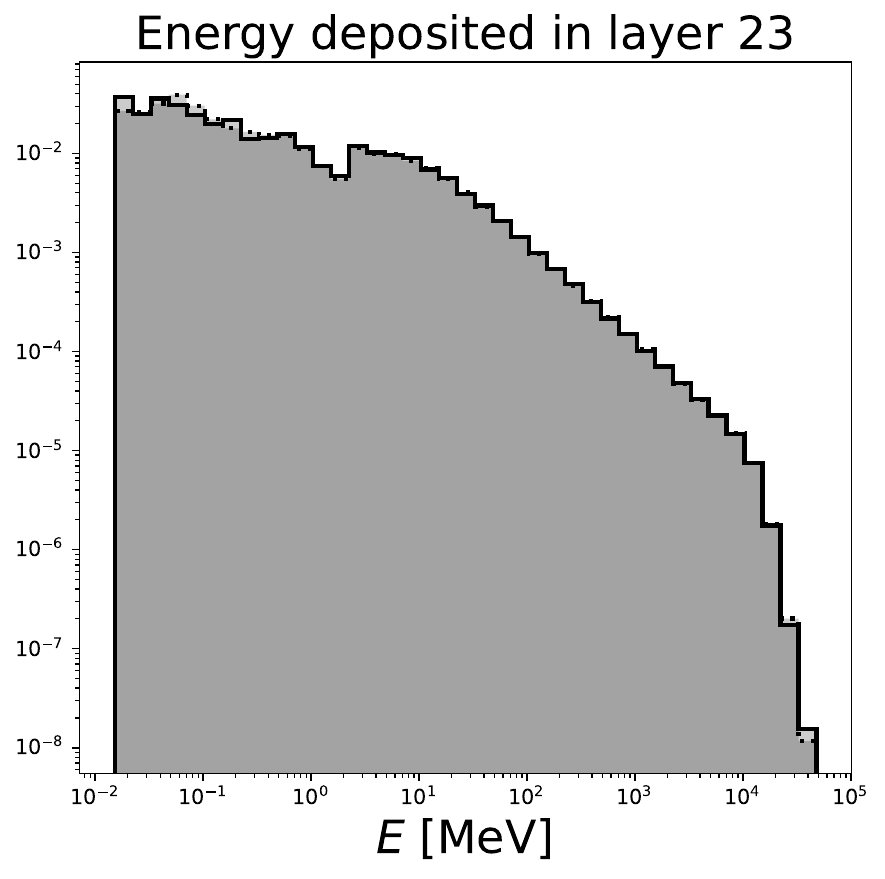} \hfill \includegraphics[height=0.1\textheight]{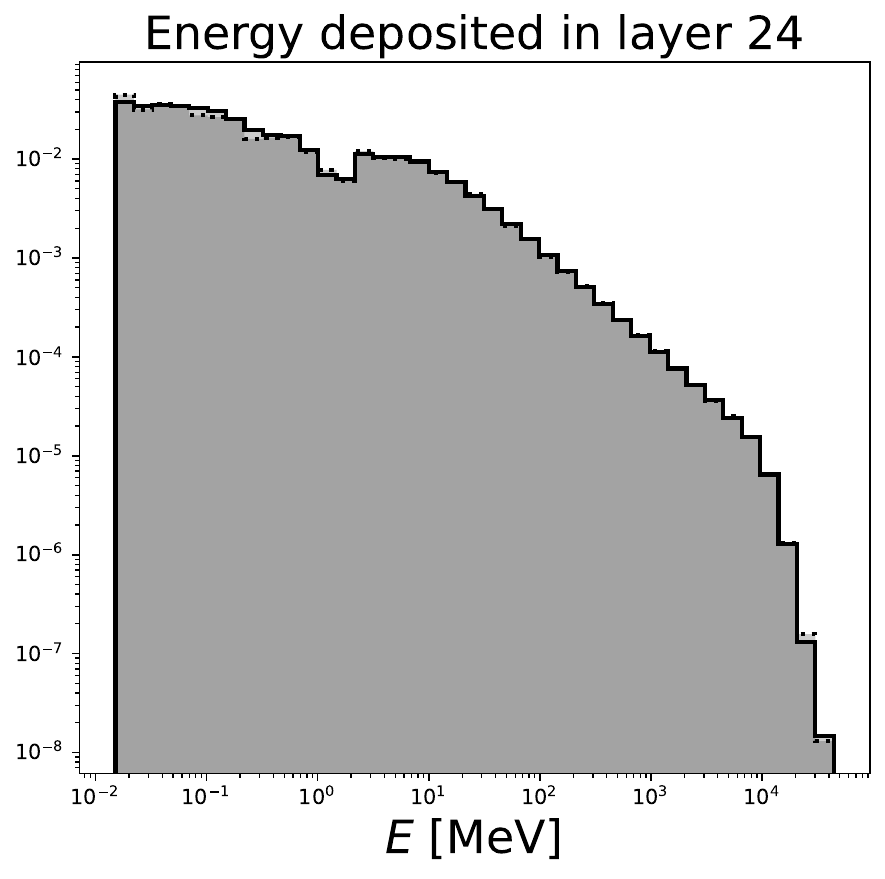}\\
    \includegraphics[height=0.1\textheight]{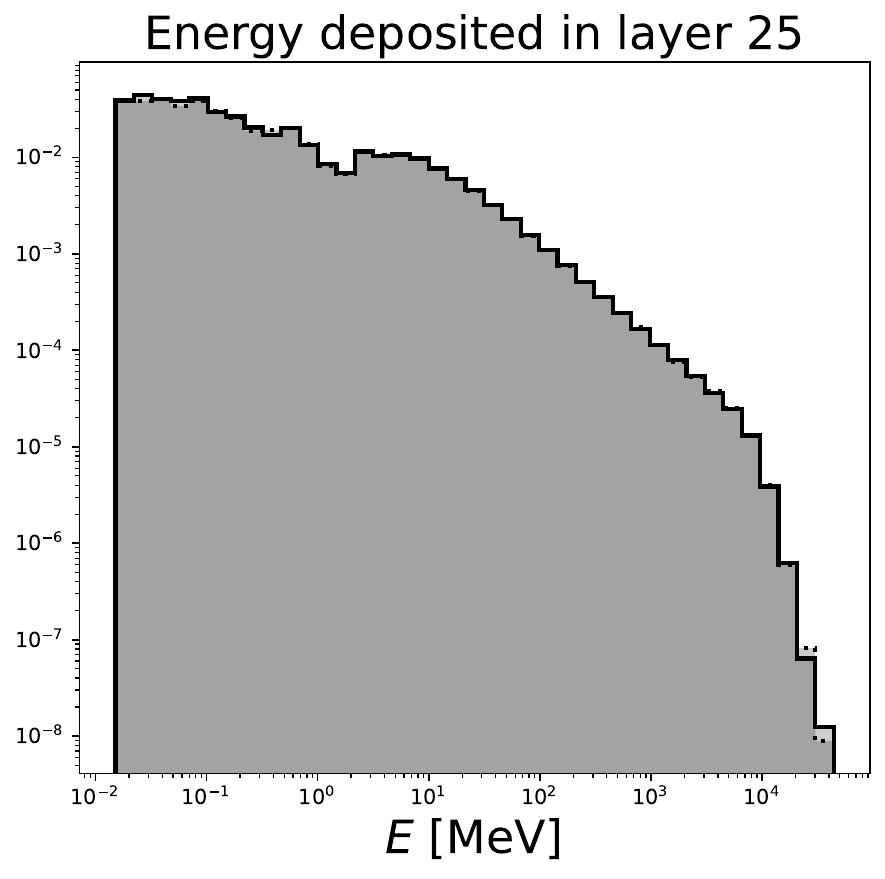} \hfill \includegraphics[height=0.1\textheight]{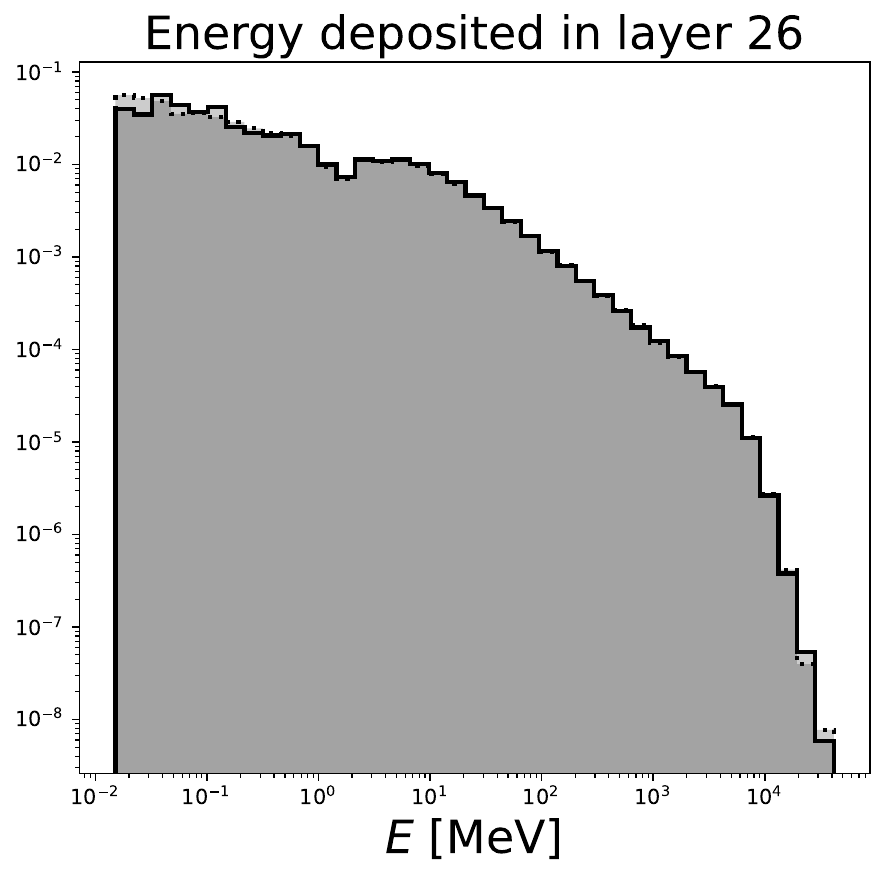} \hfill \includegraphics[height=0.1\textheight]{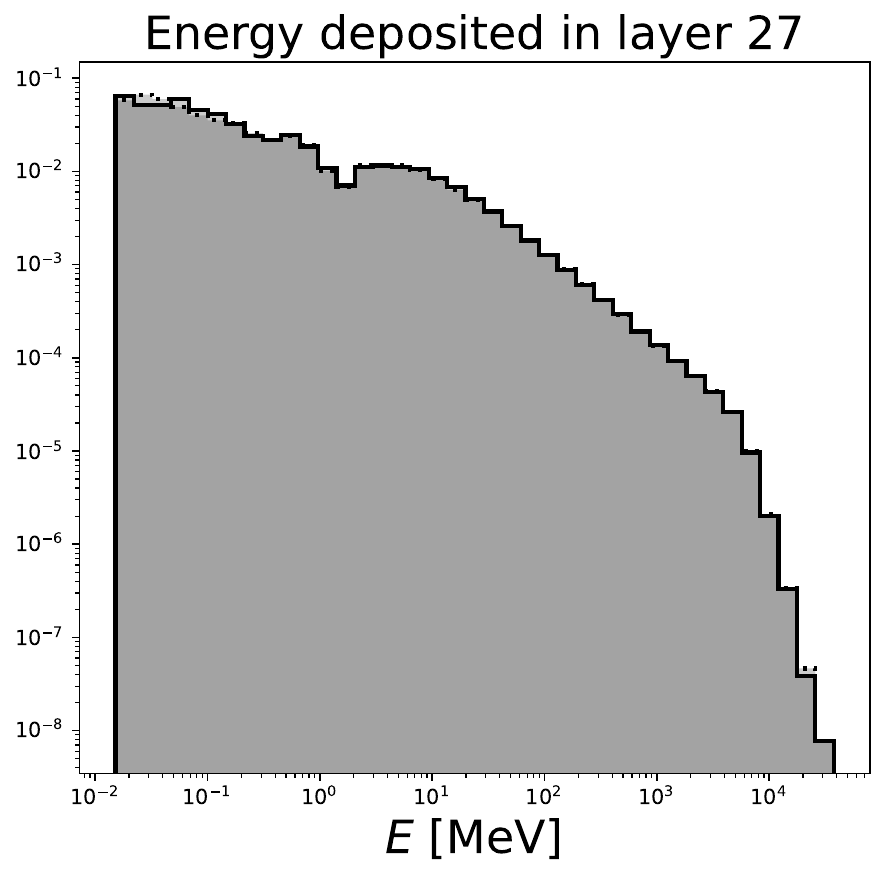} \hfill \includegraphics[height=0.1\textheight]{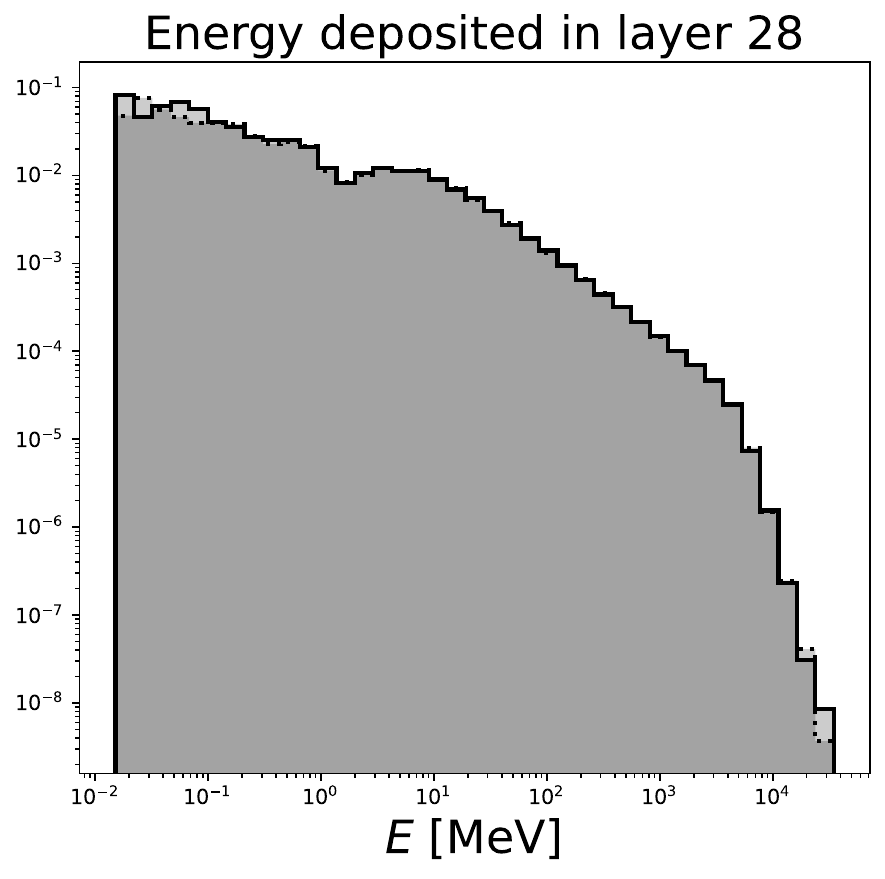} \hfill \includegraphics[height=0.1\textheight]{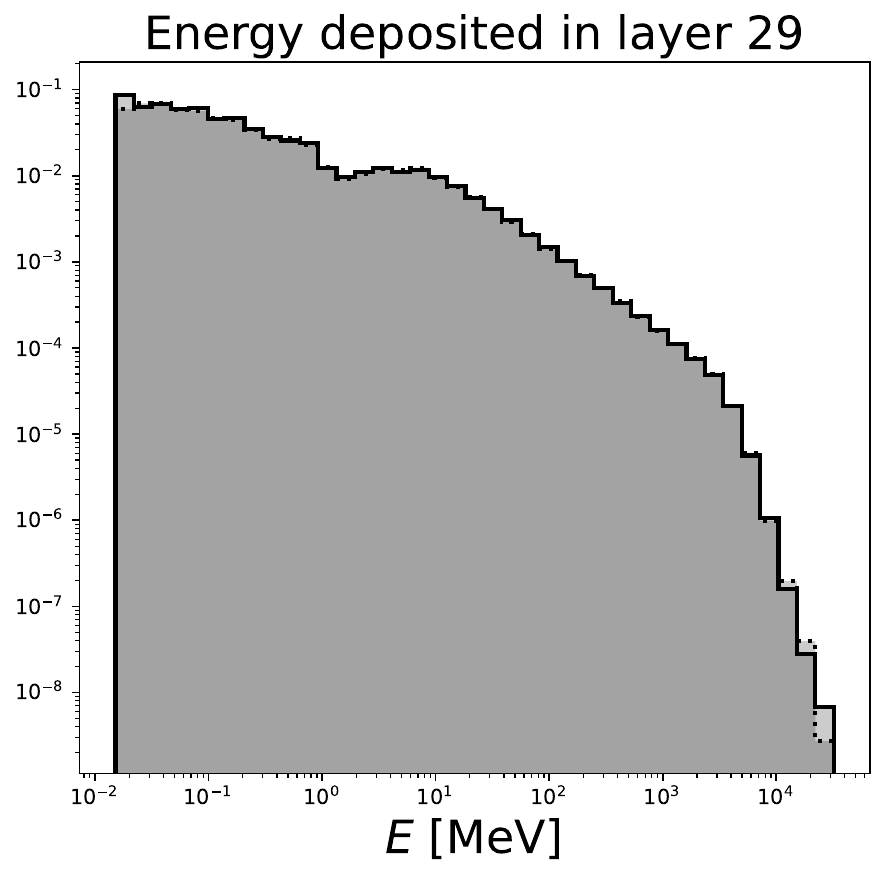}\\
    \includegraphics[height=0.1\textheight]{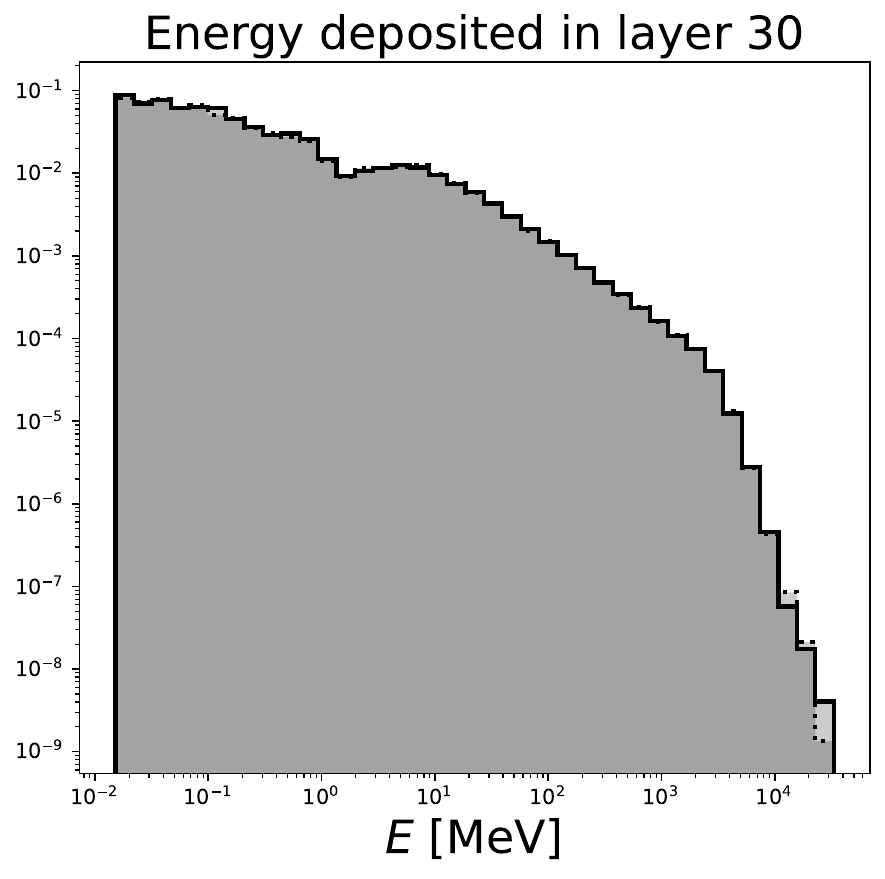} \hfill \includegraphics[height=0.1\textheight]{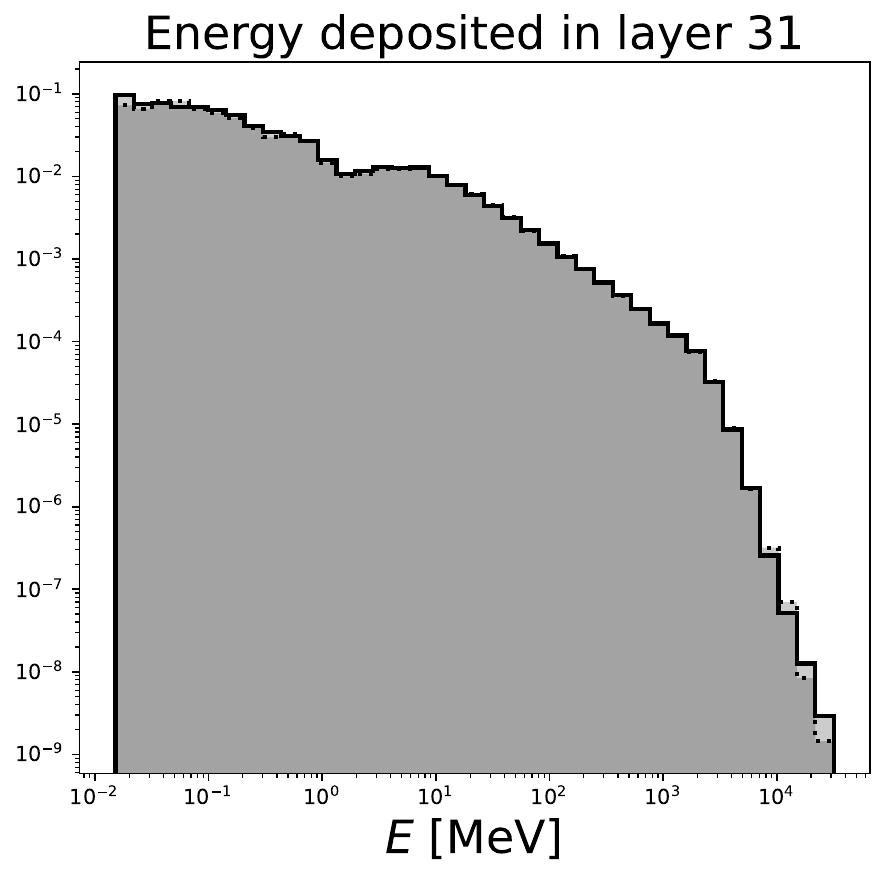} \hfill \includegraphics[height=0.1\textheight]{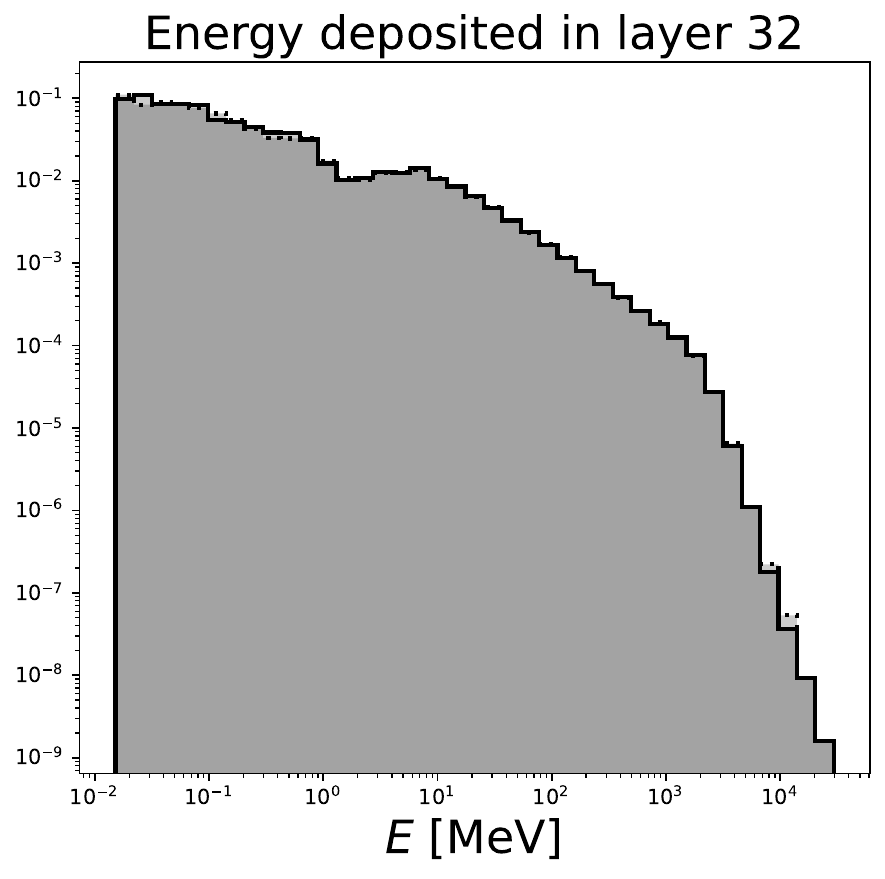} \hfill \includegraphics[height=0.1\textheight]{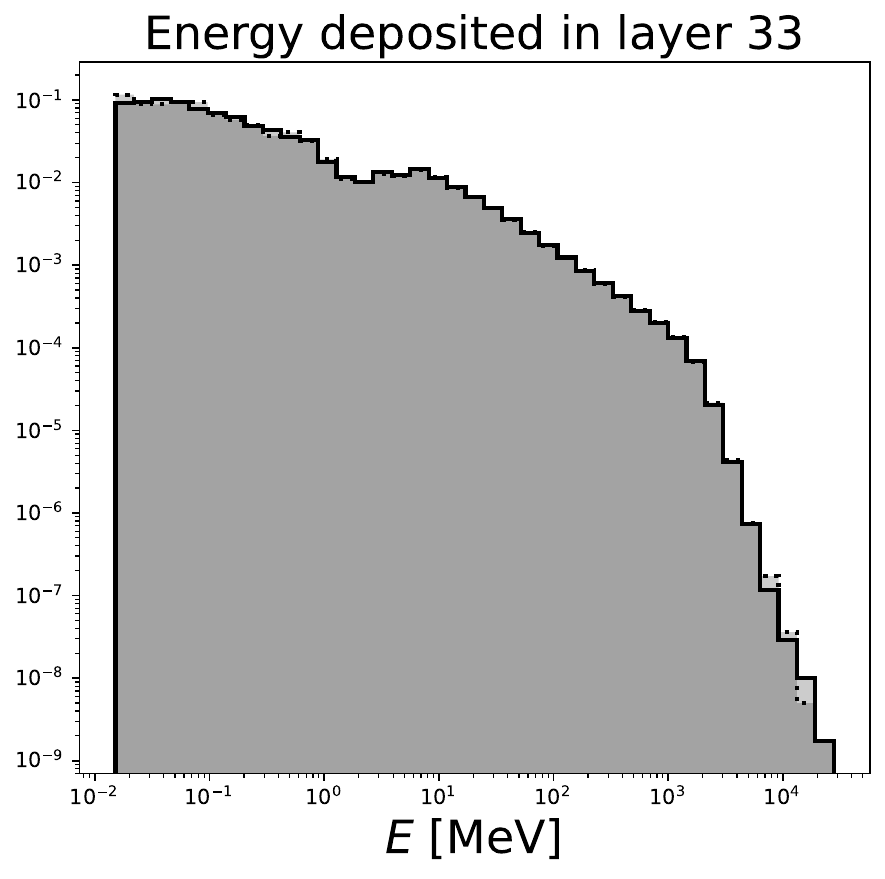} \hfill \includegraphics[height=0.1\textheight]{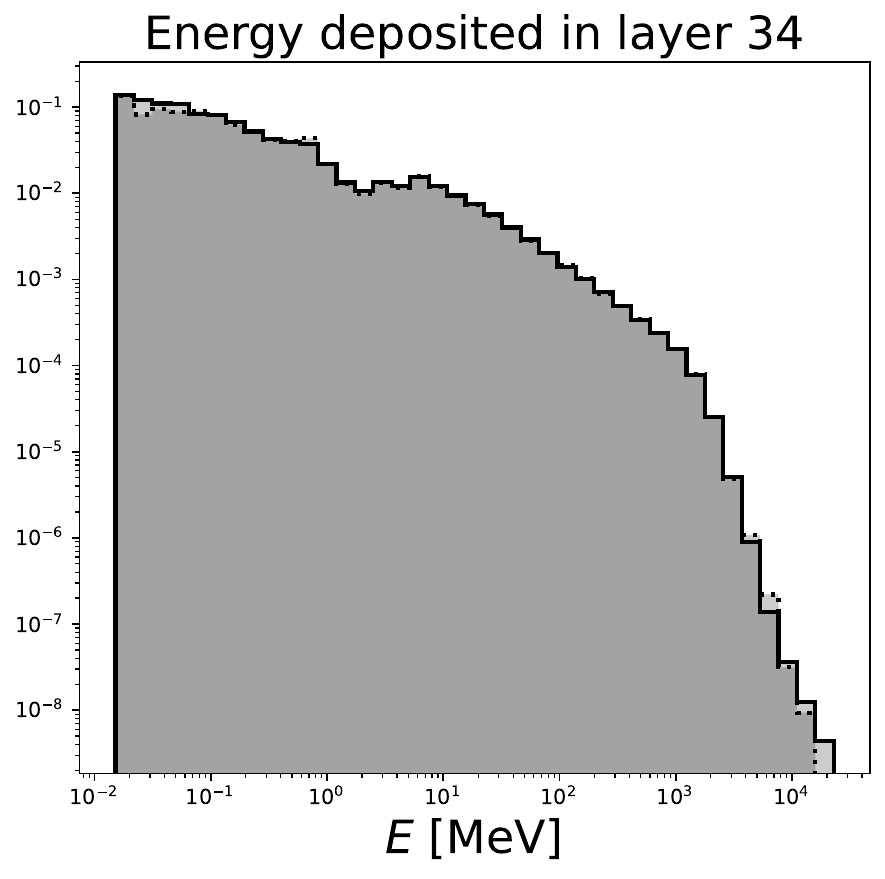}\\
    \includegraphics[height=0.1\textheight]{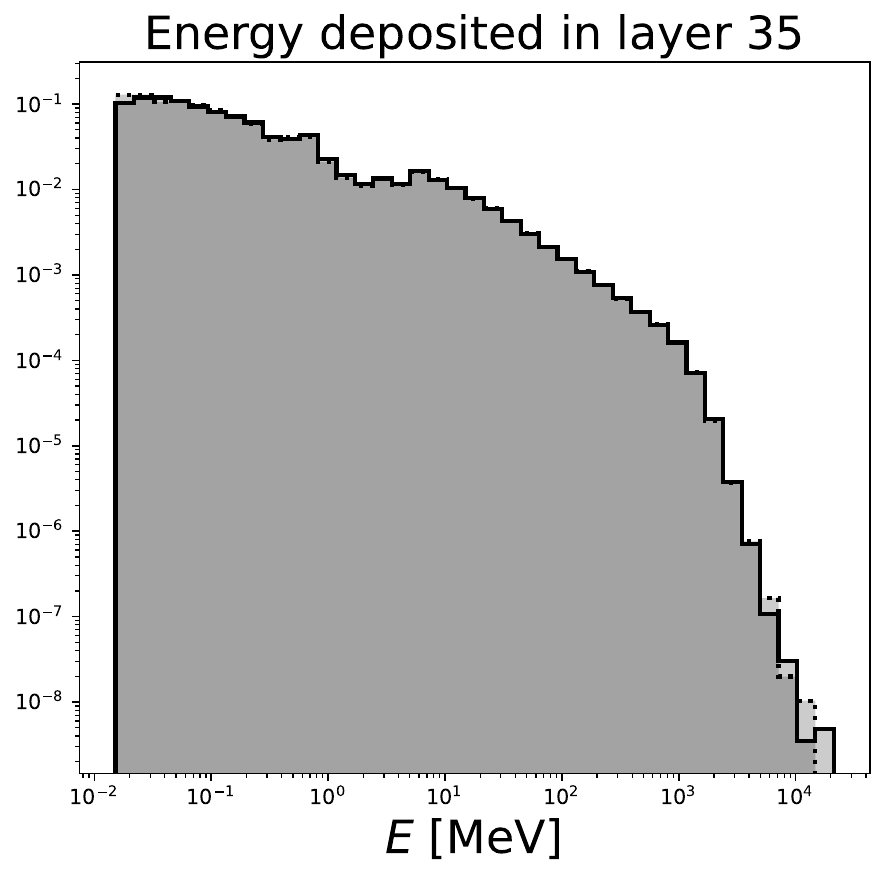} \hfill \includegraphics[height=0.1\textheight]{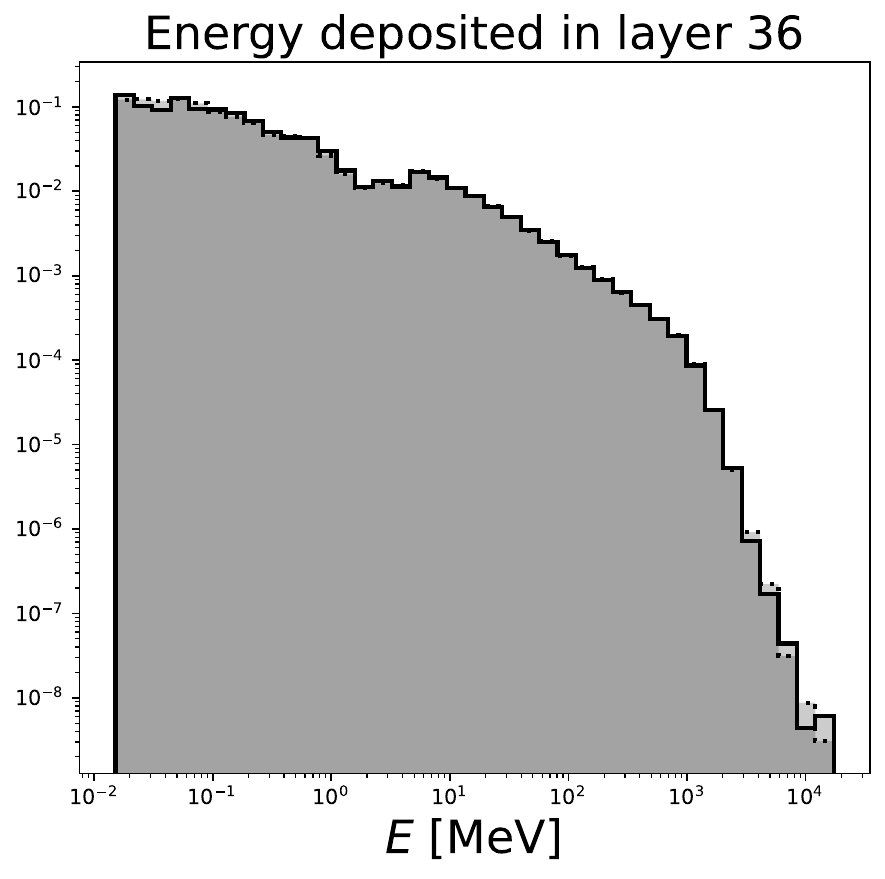} \hfill \includegraphics[height=0.1\textheight]{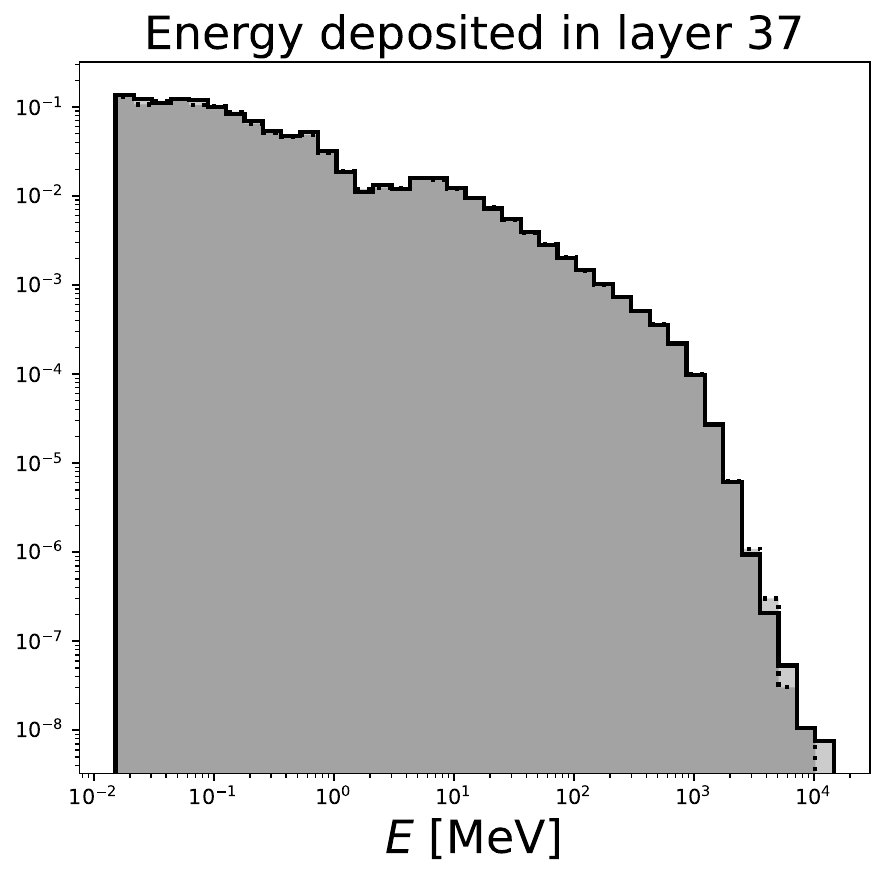} \hfill \includegraphics[height=0.1\textheight]{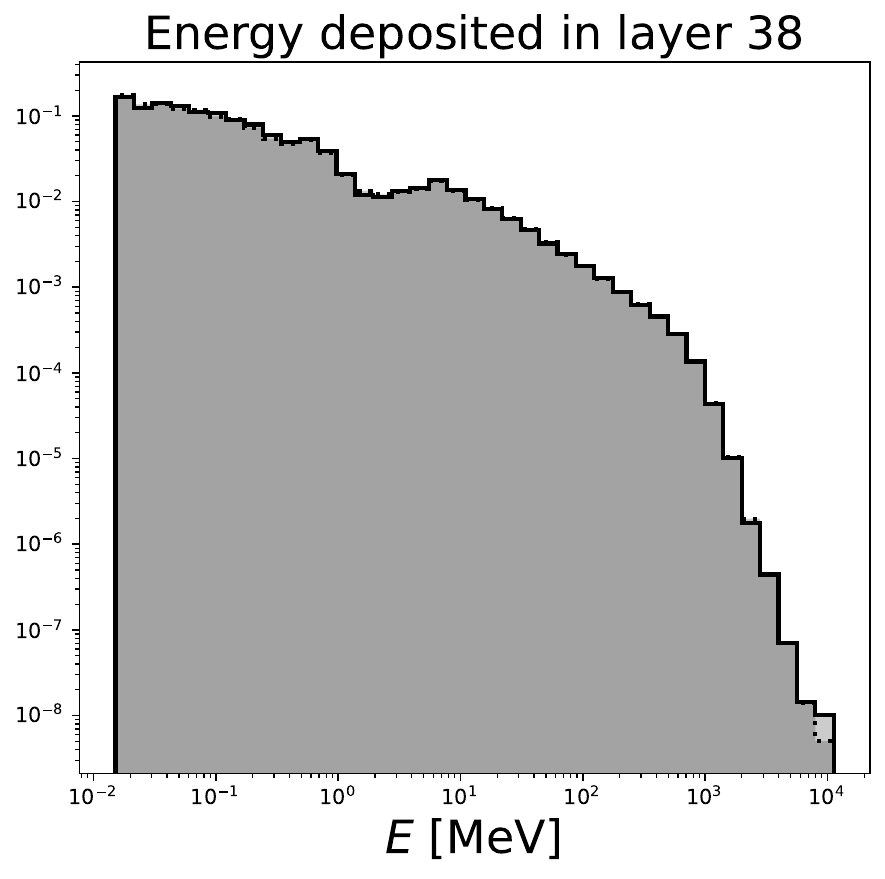} \hfill \includegraphics[height=0.1\textheight]{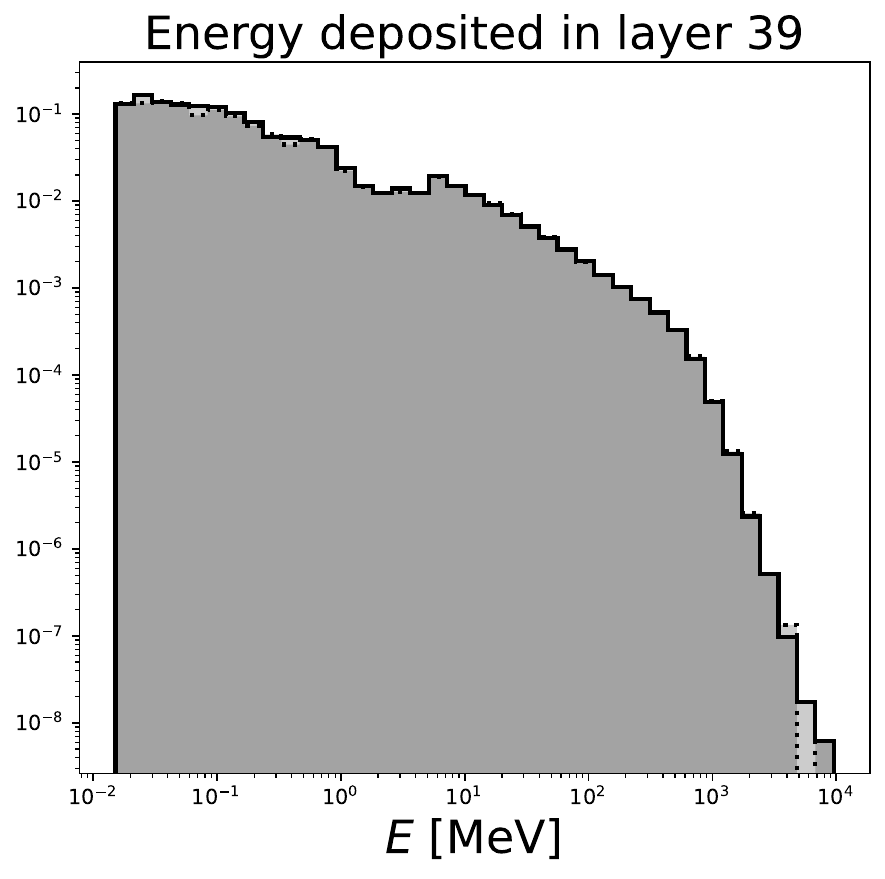}\\
    \includegraphics[height=0.1\textheight]{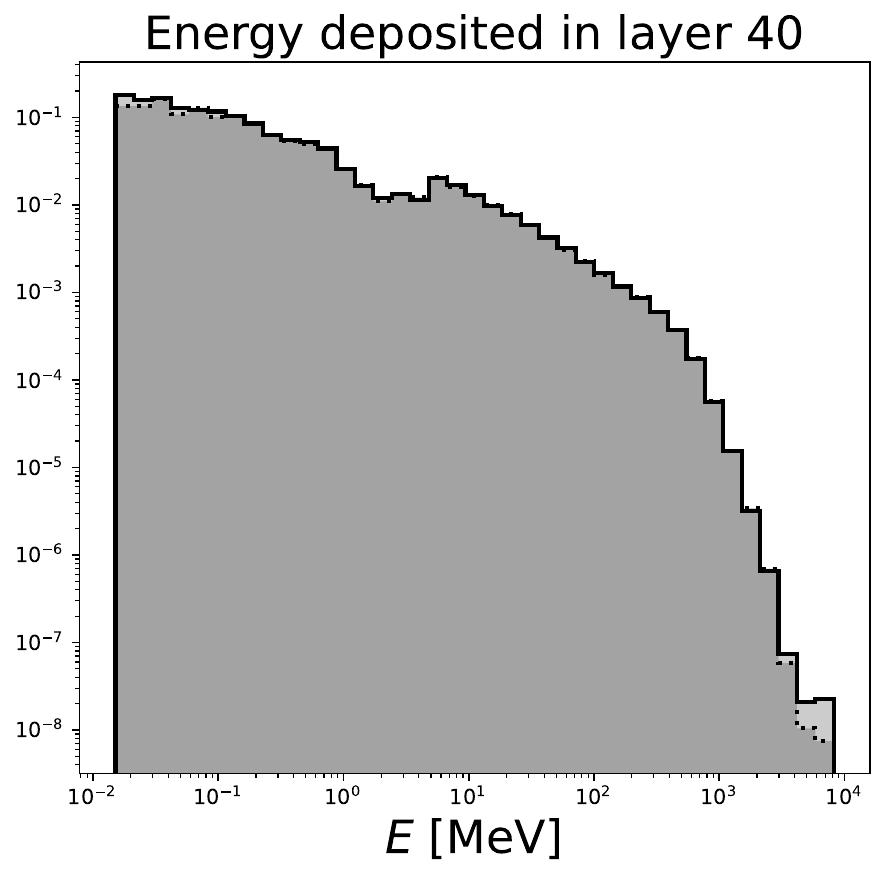} \hfill \includegraphics[height=0.1\textheight]{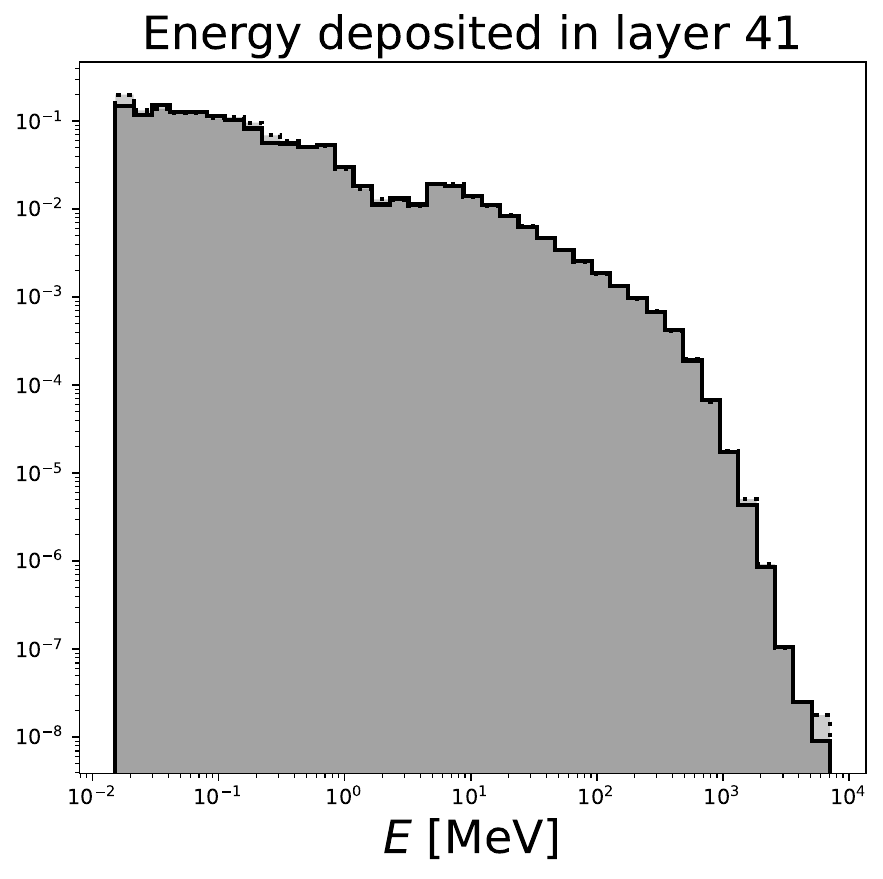} \hfill \includegraphics[height=0.1\textheight]{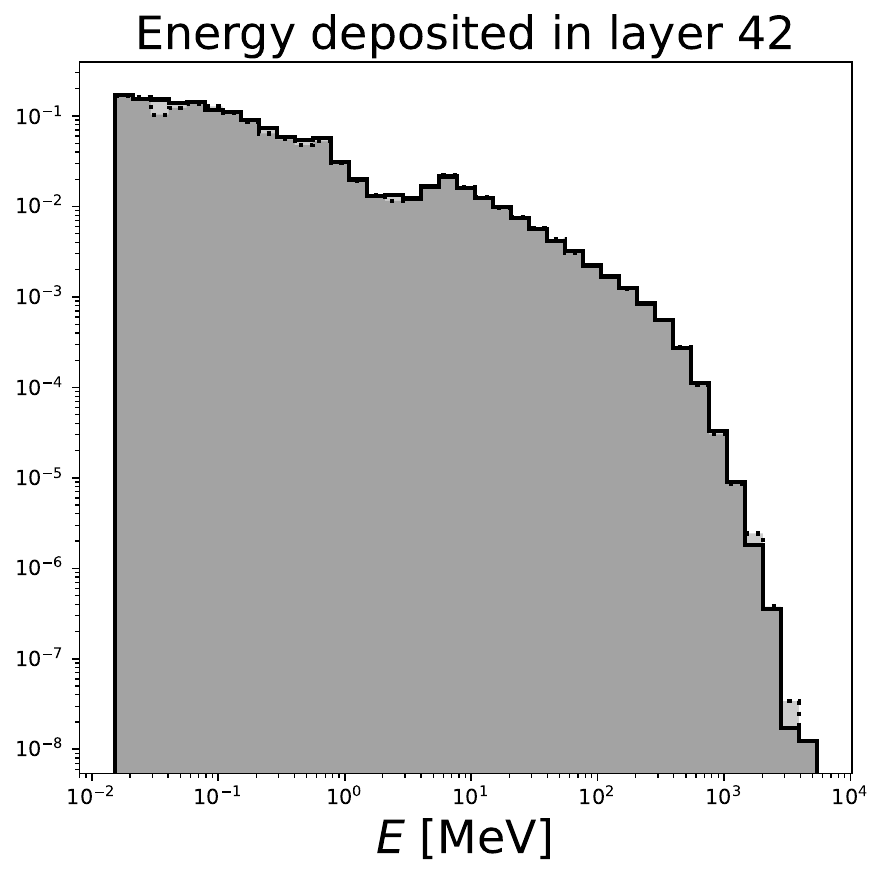} \hfill \includegraphics[height=0.1\textheight]{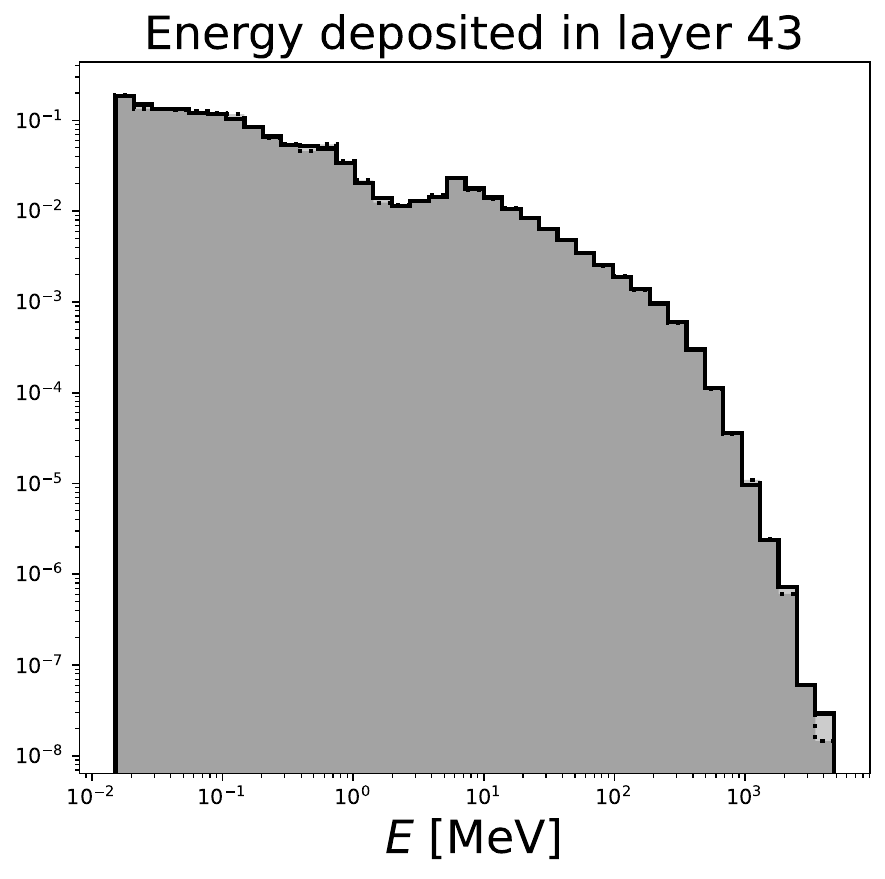} \hfill \includegraphics[height=0.1\textheight]{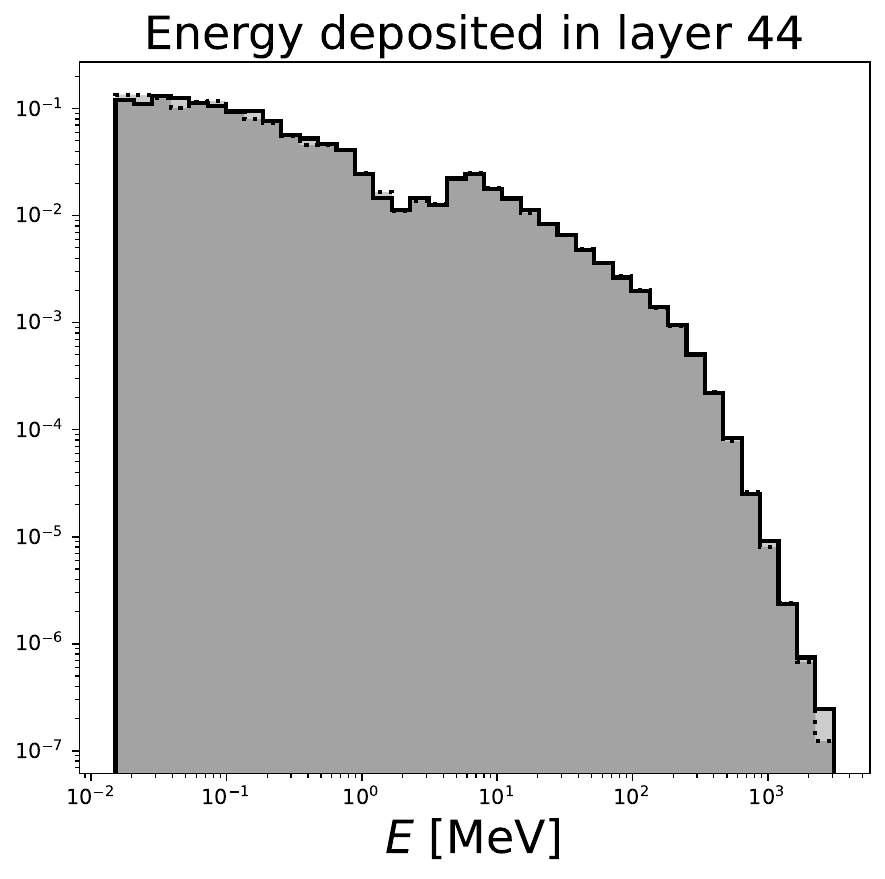}\\
    \includegraphics[width=0.5\textwidth]{figures/Appendix_reference/legend.pdf}
    \caption{Distribution of \geant training and evaluation data in layer energies $E_i$ for ds2. }
    \label{fig:app_ref.ds2.2}
\end{figure}

\begin{figure}[ht]
    \centering
    \includegraphics[height=0.1\textheight]{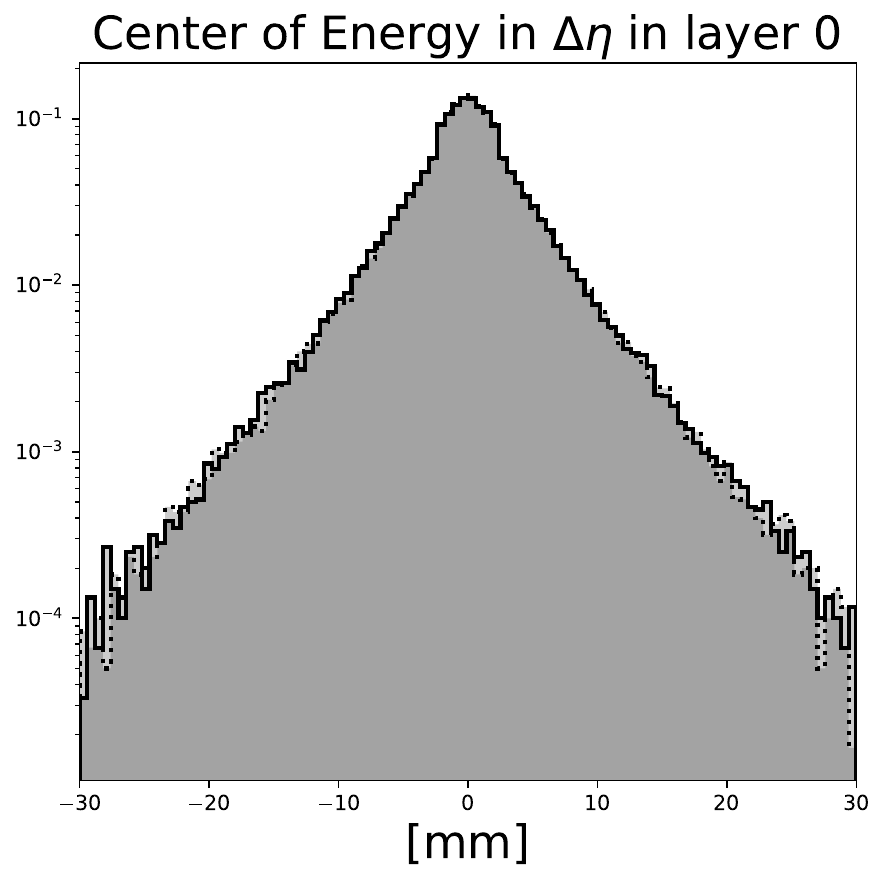} \hfill \includegraphics[height=0.1\textheight]{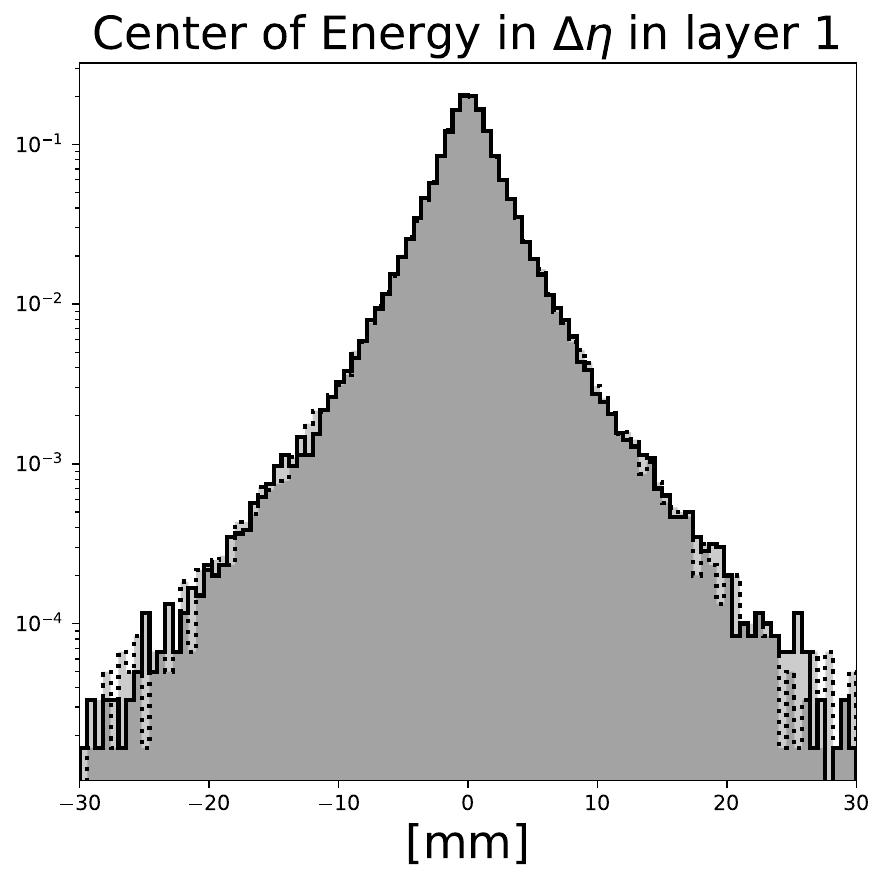} \hfill \includegraphics[height=0.1\textheight]{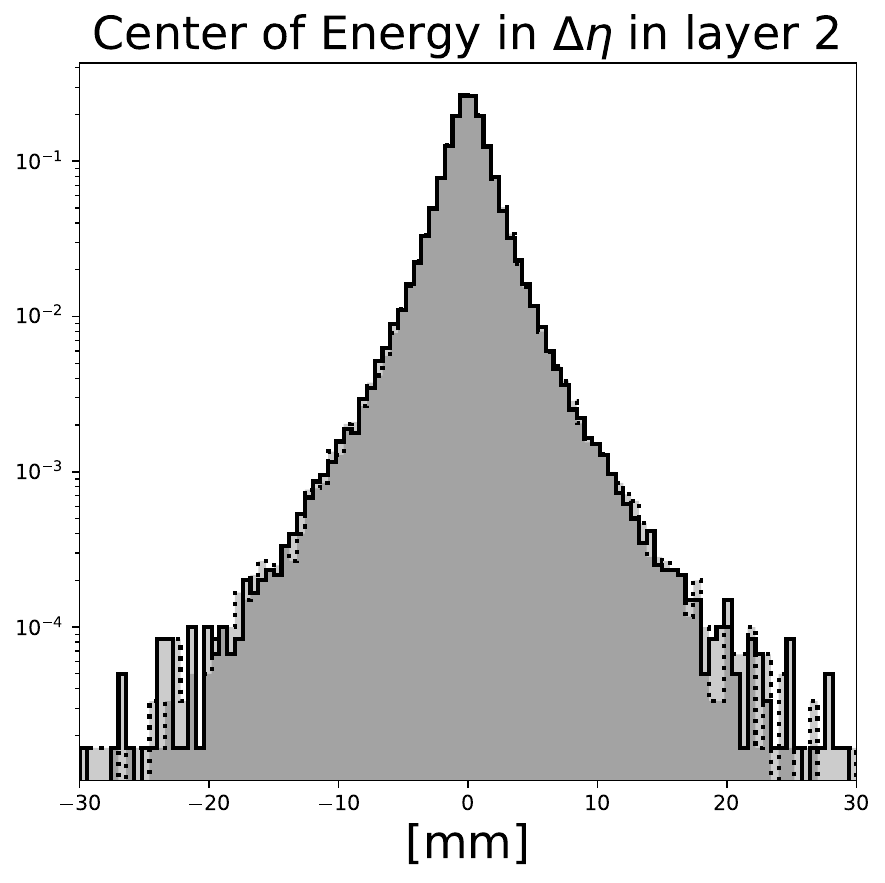} \hfill \includegraphics[height=0.1\textheight]{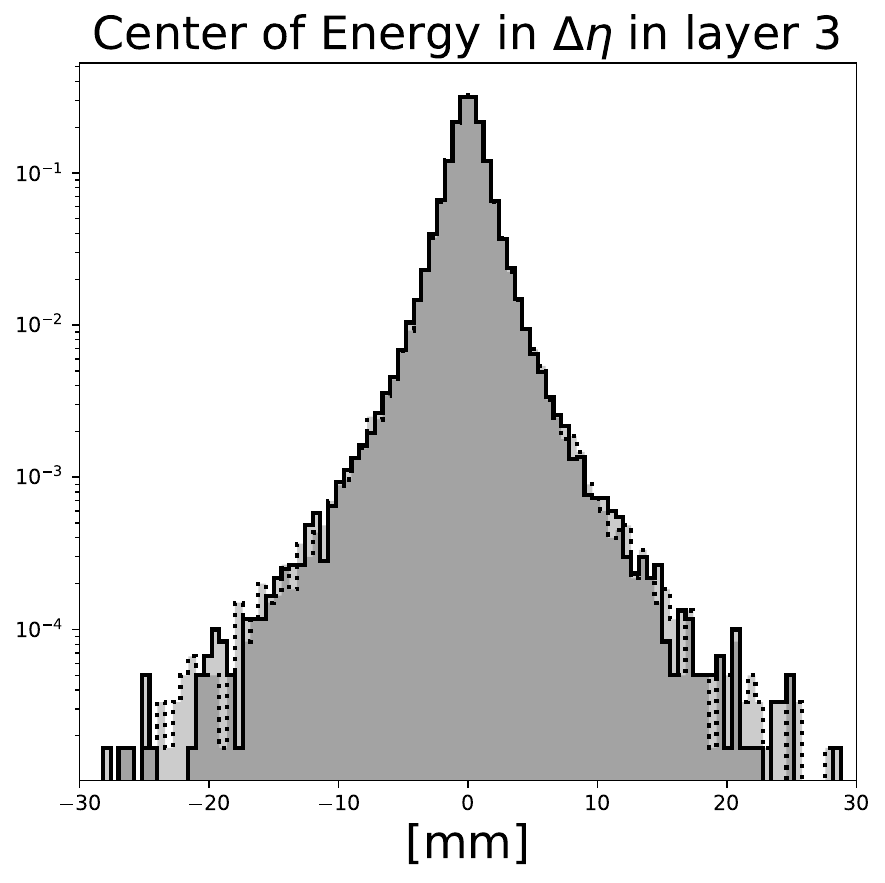} \hfill \includegraphics[height=0.1\textheight]{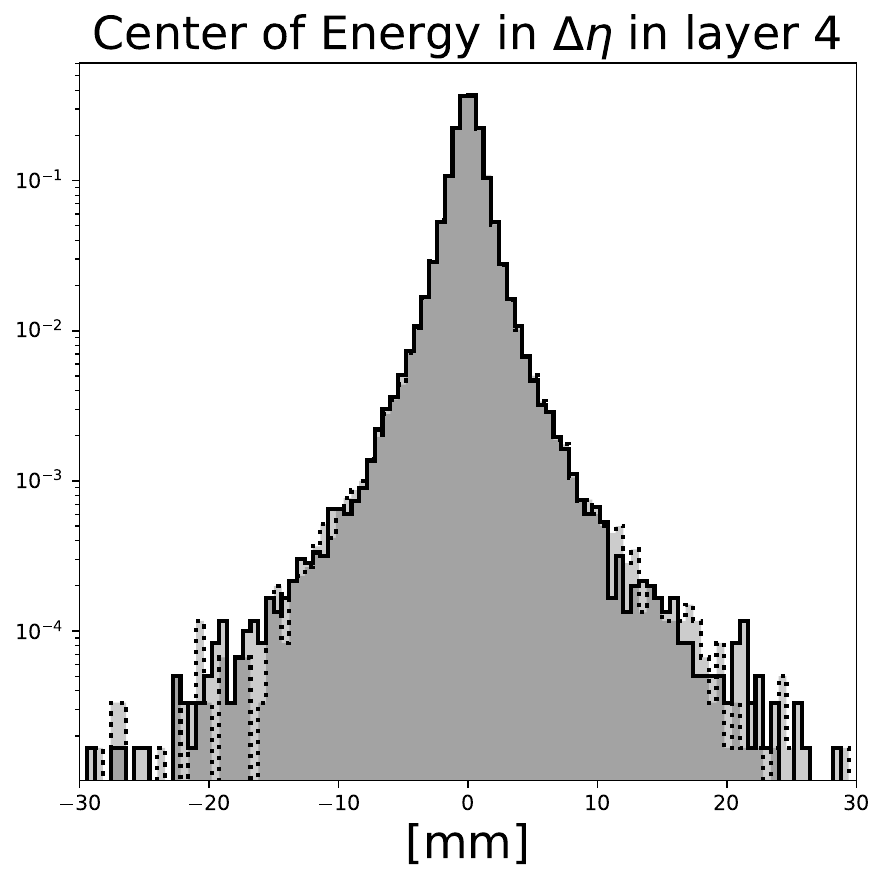}\\
    \includegraphics[height=0.1\textheight]{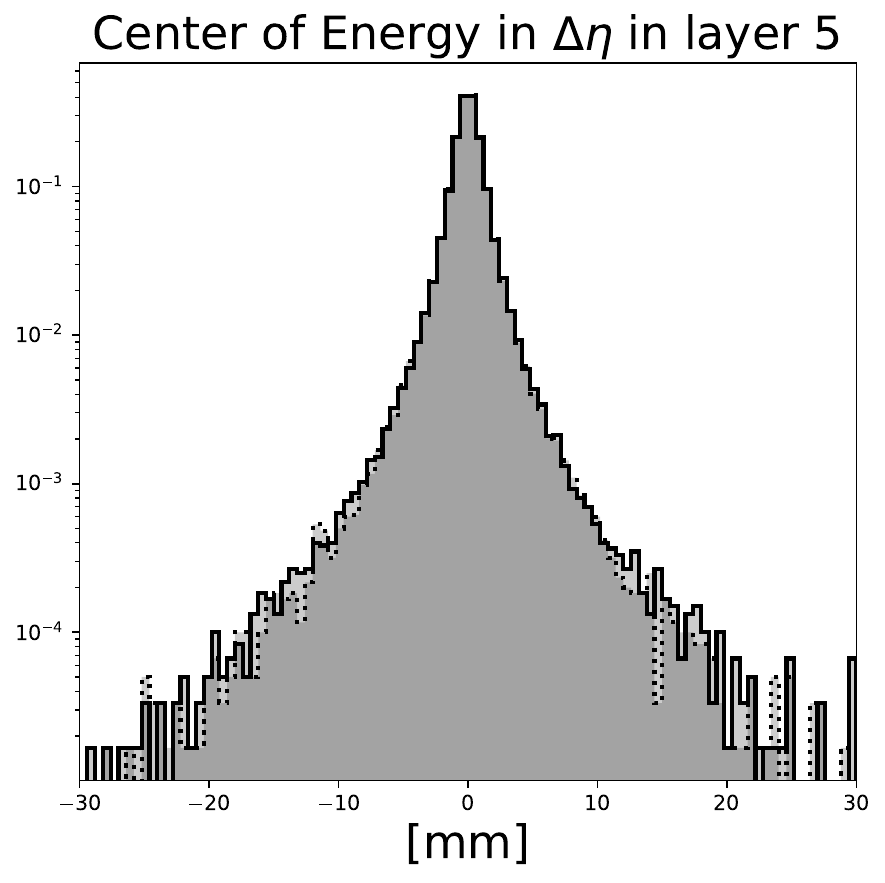} \hfill \includegraphics[height=0.1\textheight]{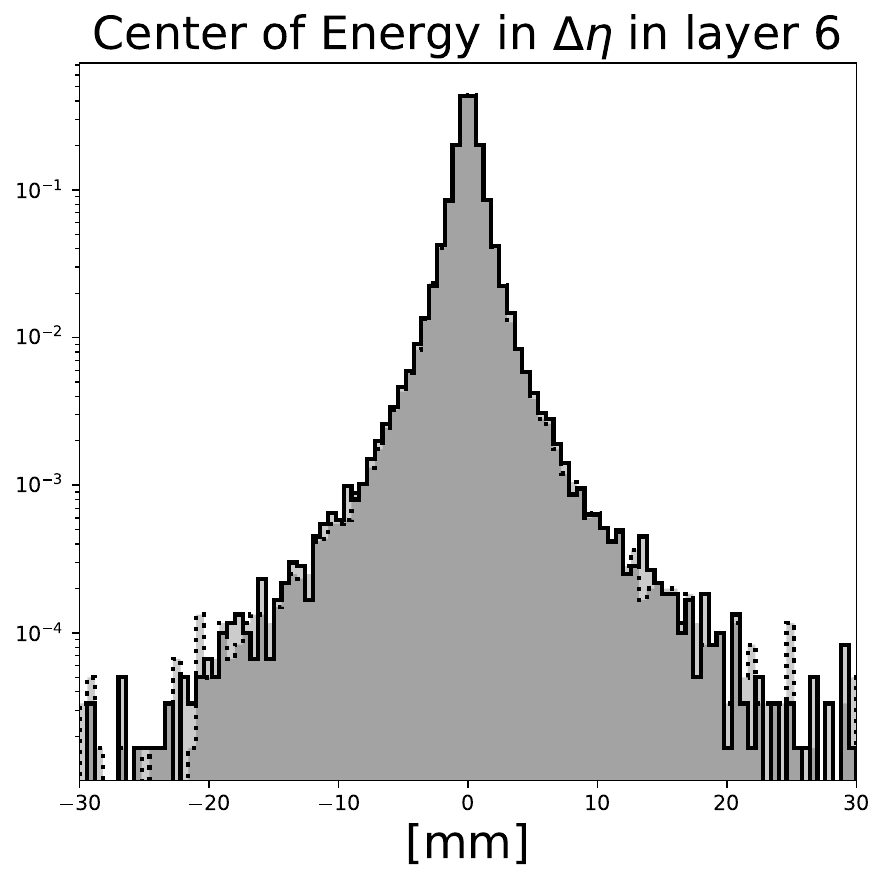} \hfill \includegraphics[height=0.1\textheight]{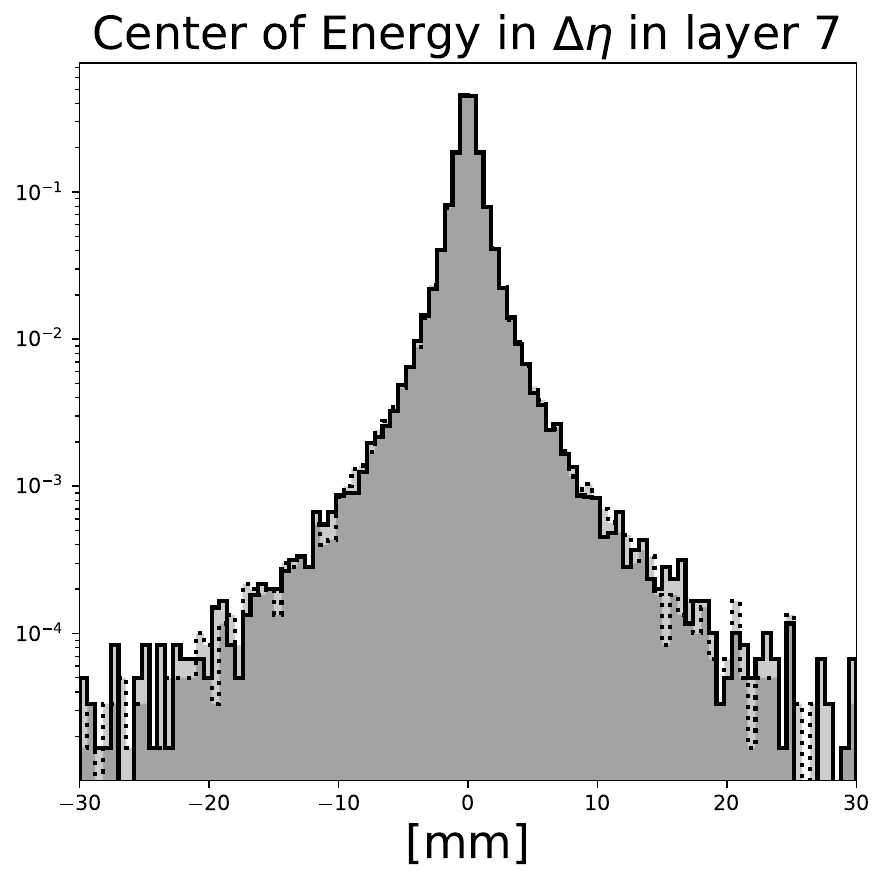} \hfill \includegraphics[height=0.1\textheight]{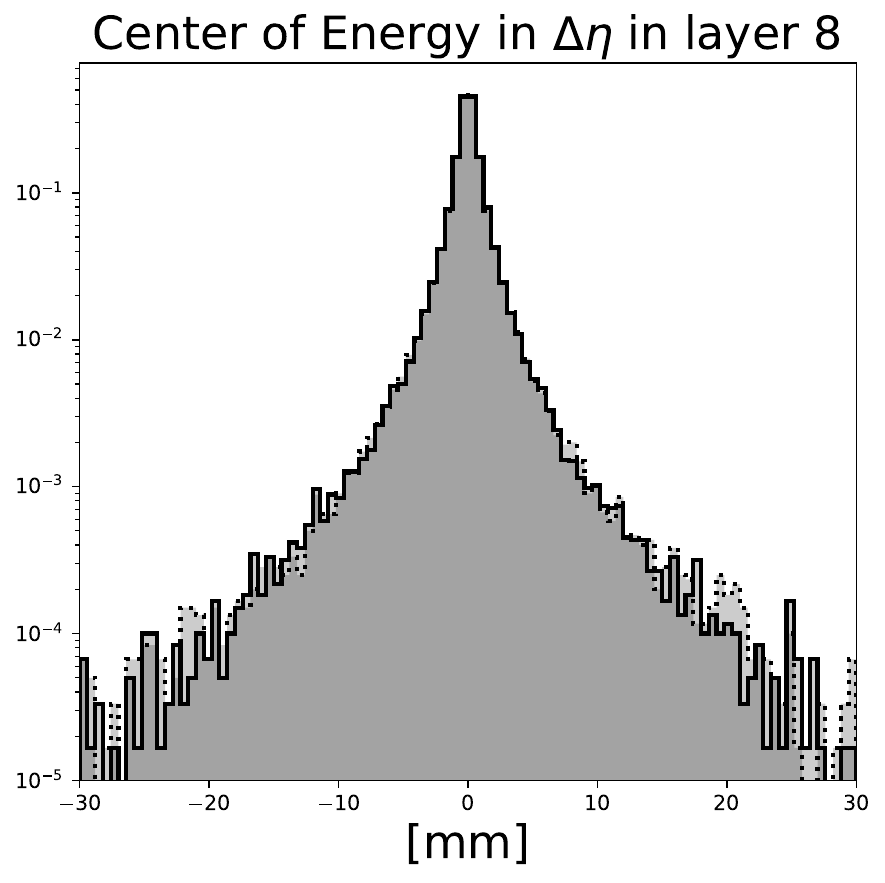} \hfill \includegraphics[height=0.1\textheight]{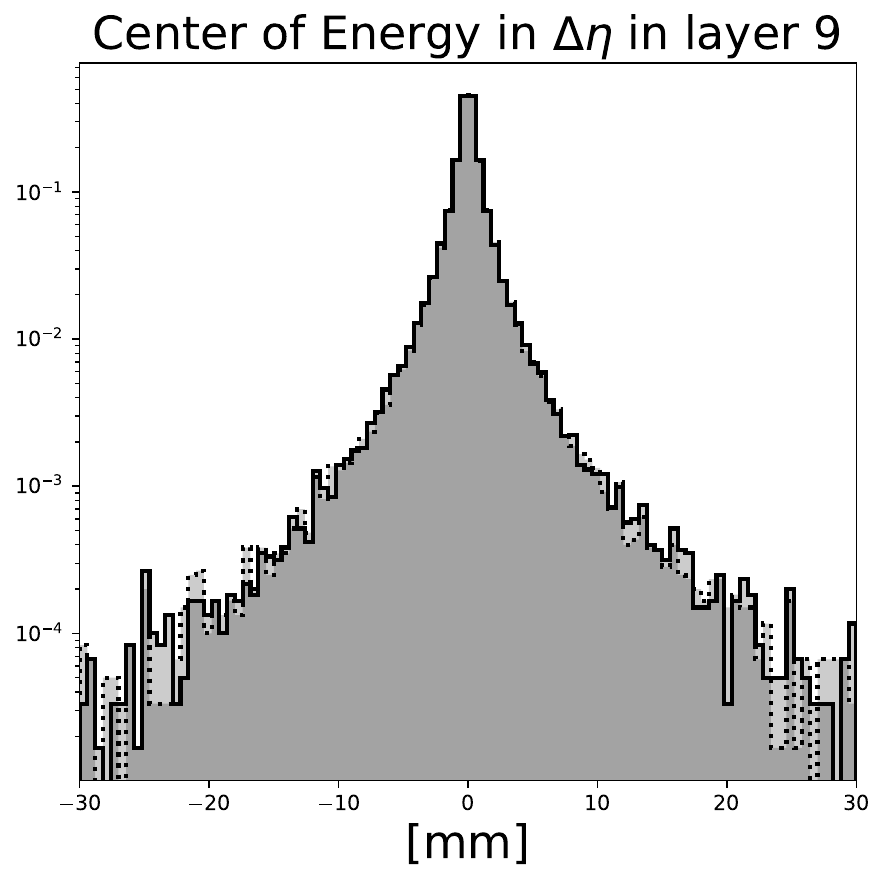}\\
    \includegraphics[height=0.1\textheight]{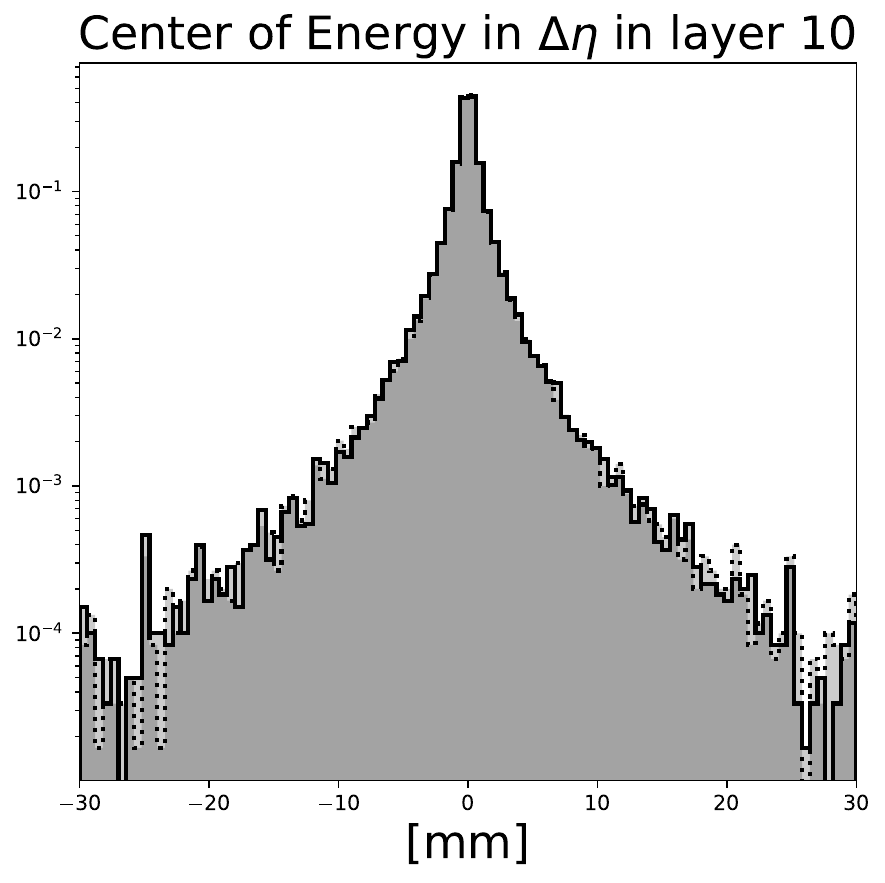} \hfill \includegraphics[height=0.1\textheight]{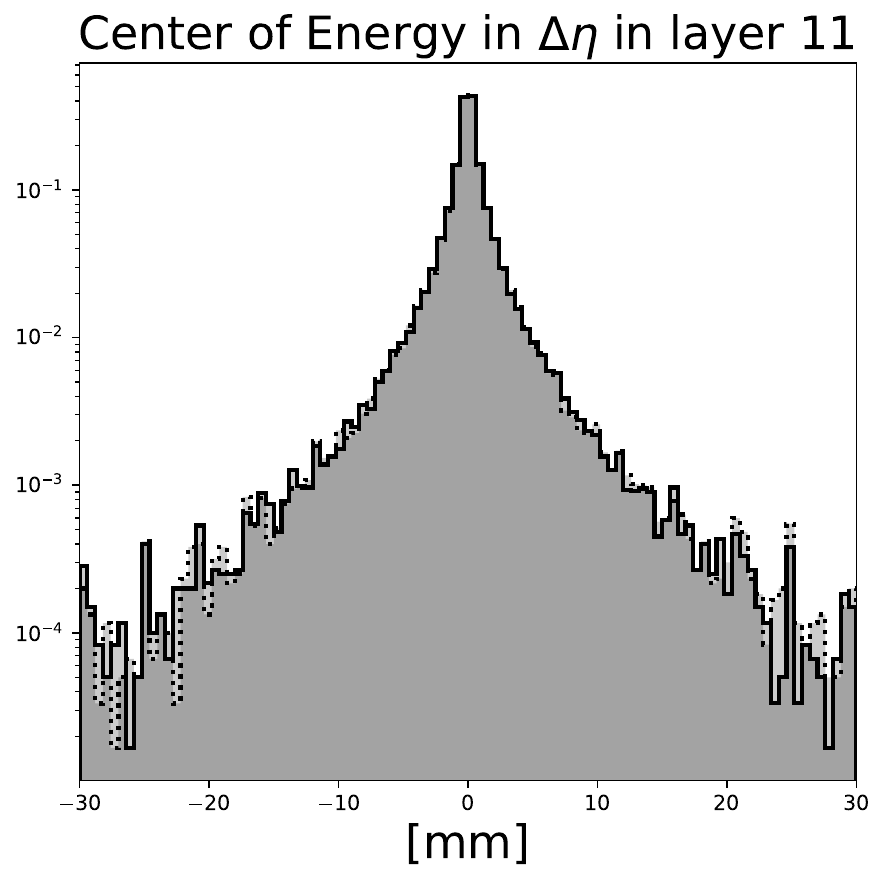} \hfill \includegraphics[height=0.1\textheight]{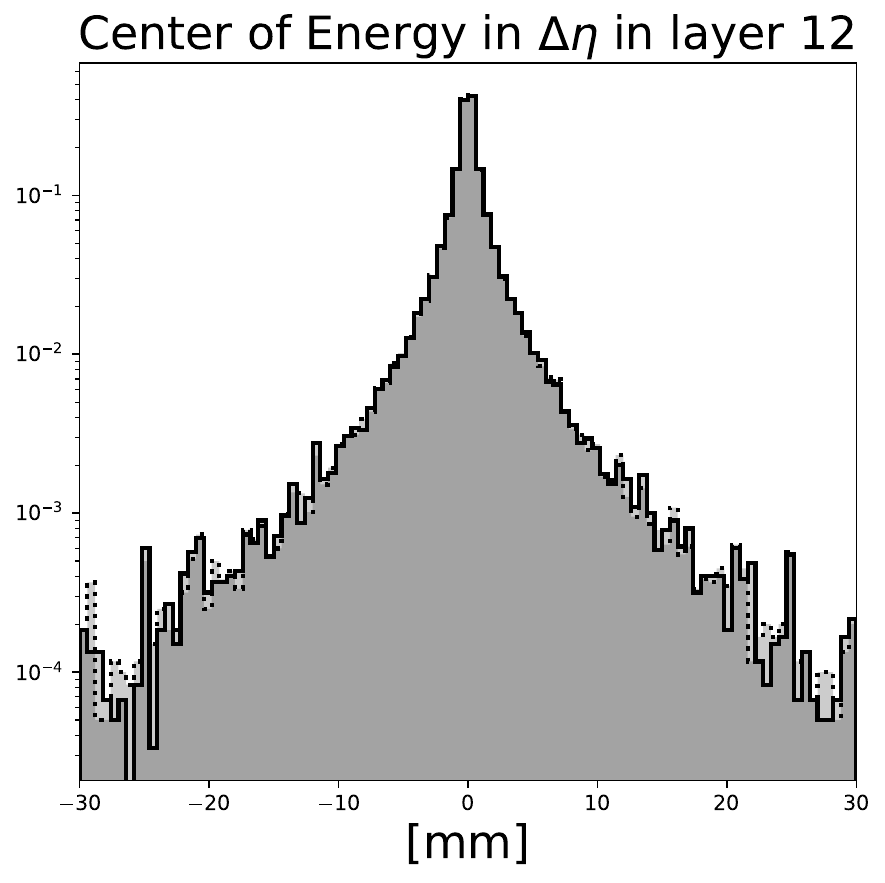} \hfill \includegraphics[height=0.1\textheight]{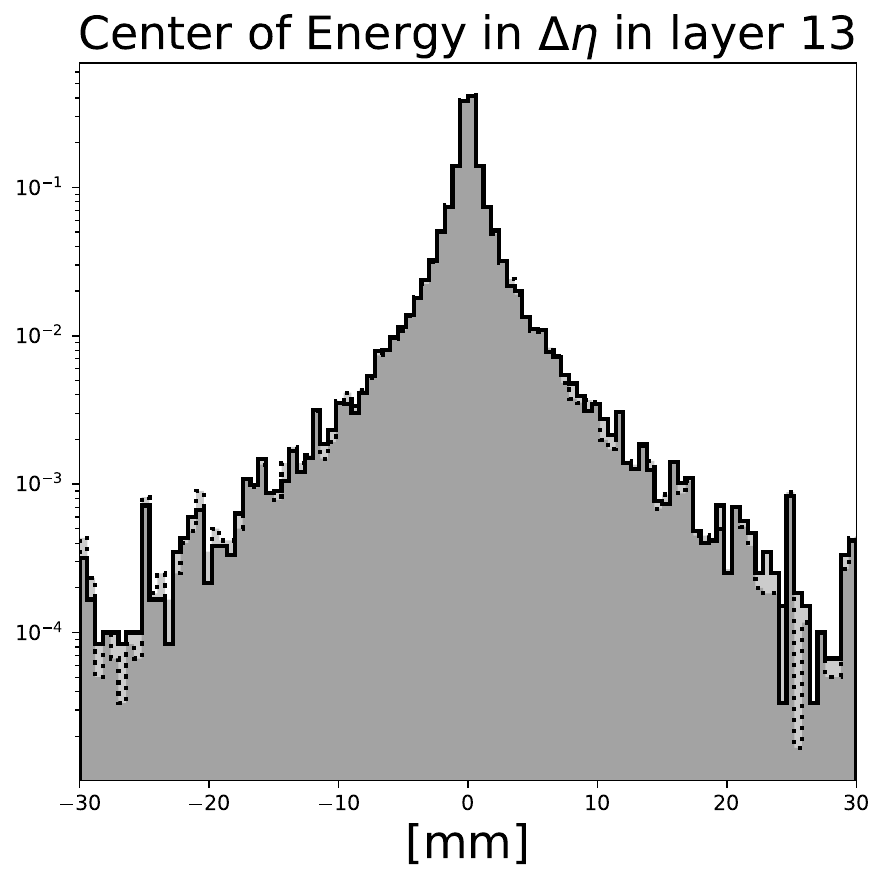} \hfill \includegraphics[height=0.1\textheight]{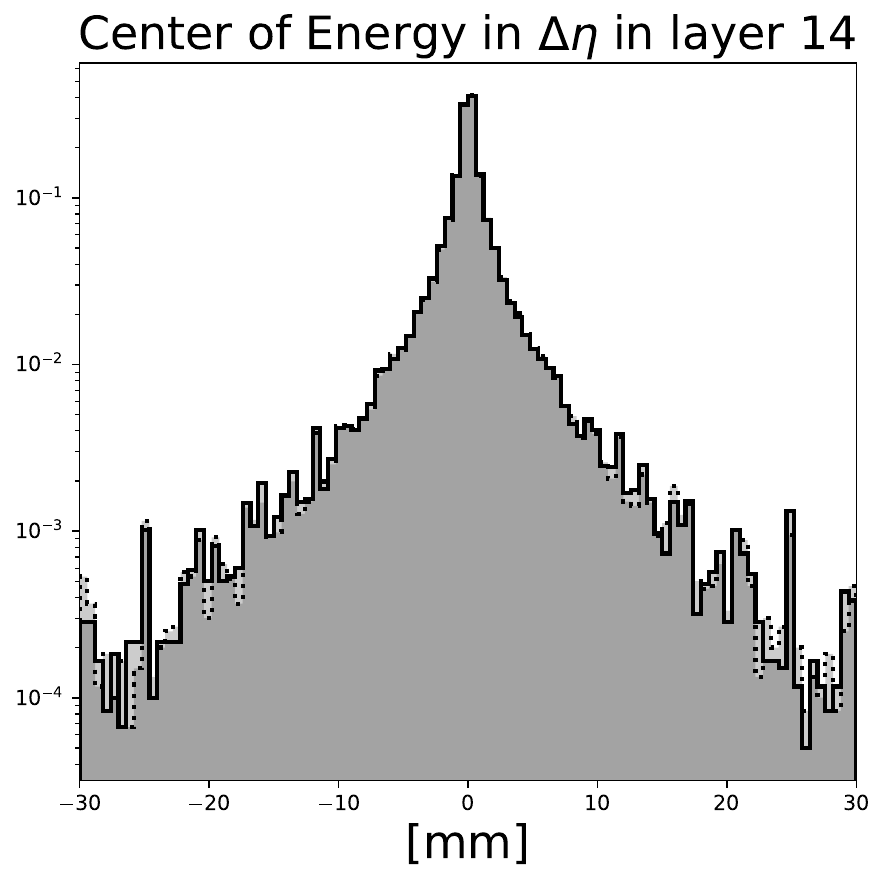}\\
    \includegraphics[height=0.1\textheight]{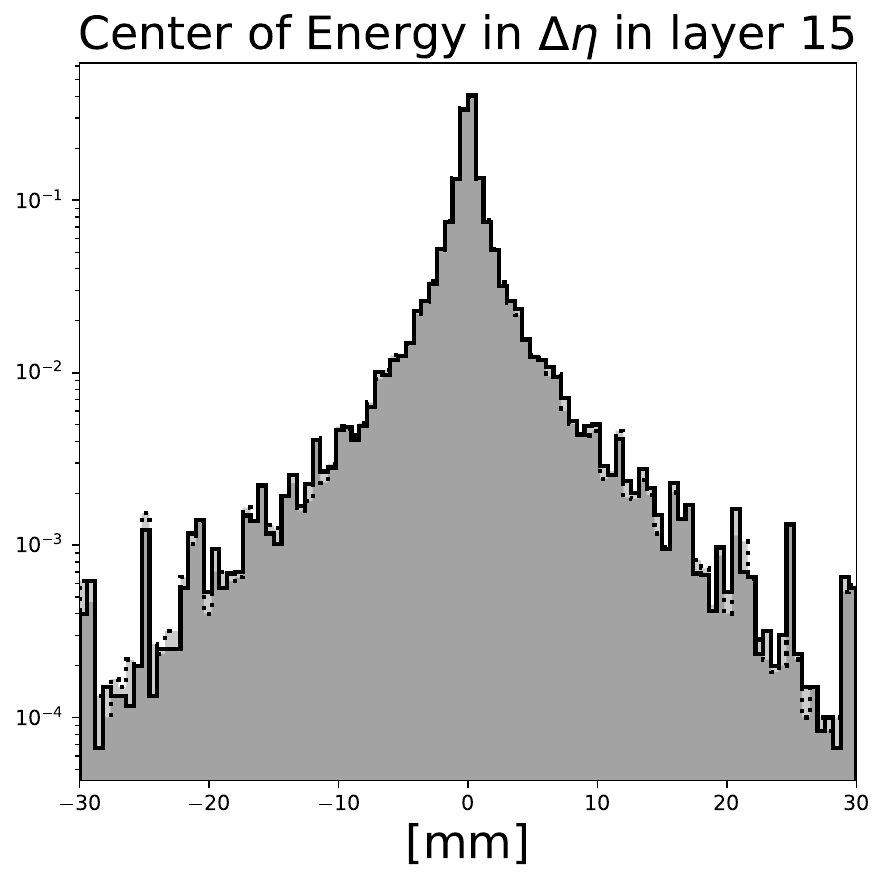} \hfill \includegraphics[height=0.1\textheight]{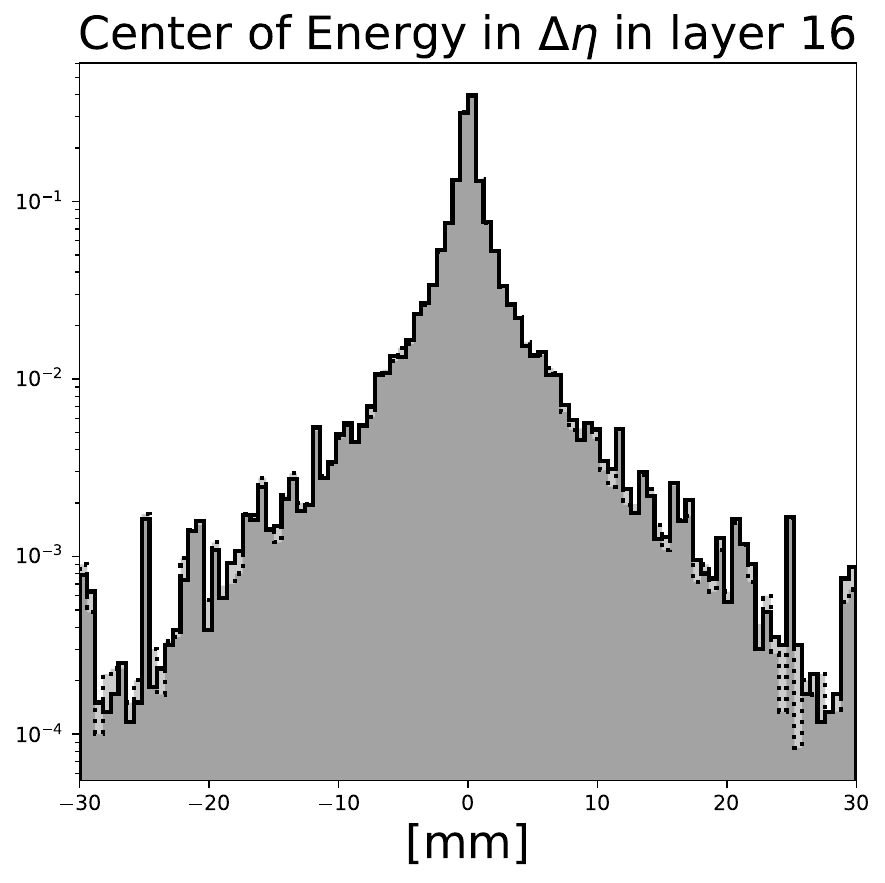} \hfill \includegraphics[height=0.1\textheight]{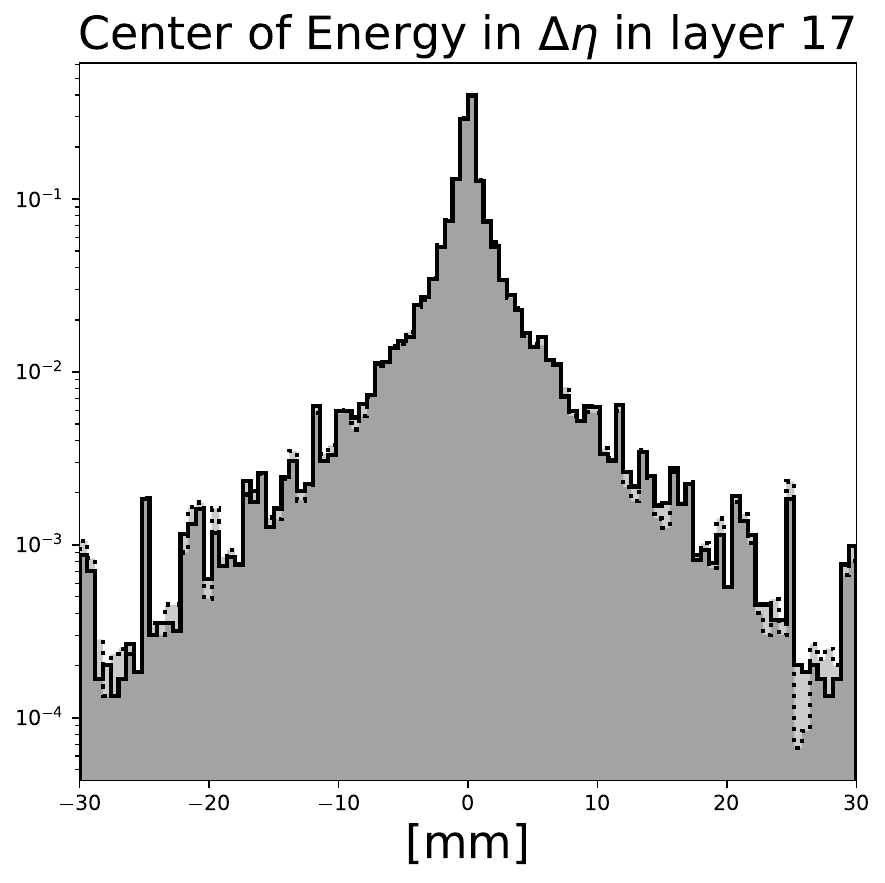} \hfill \includegraphics[height=0.1\textheight]{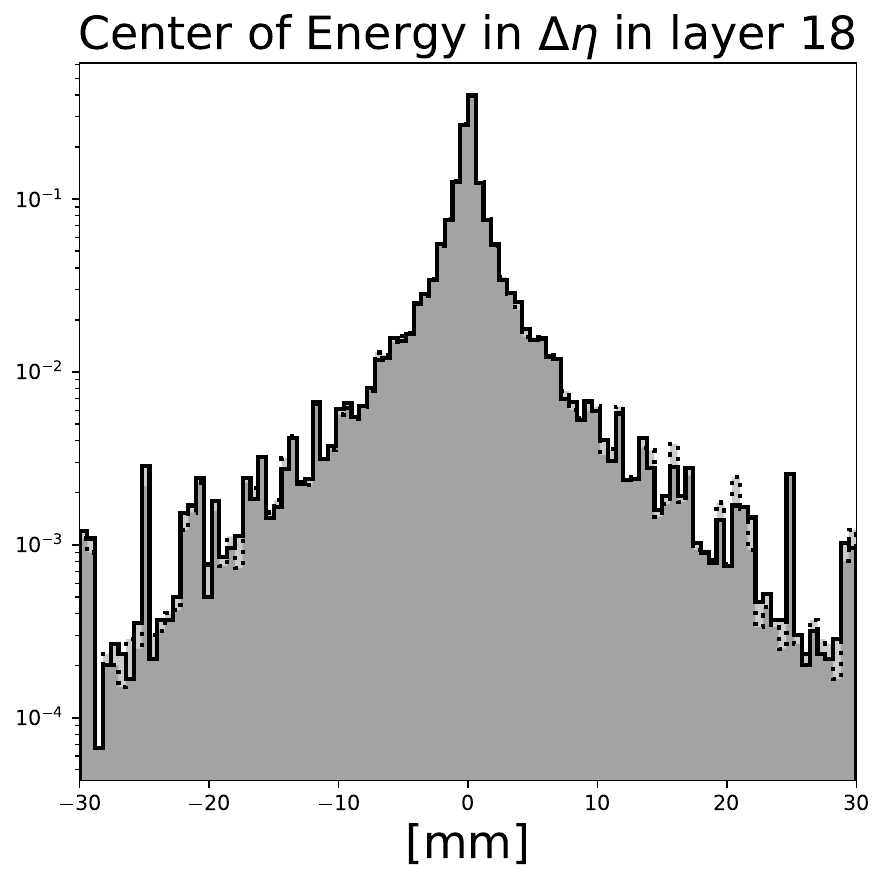} \hfill \includegraphics[height=0.1\textheight]{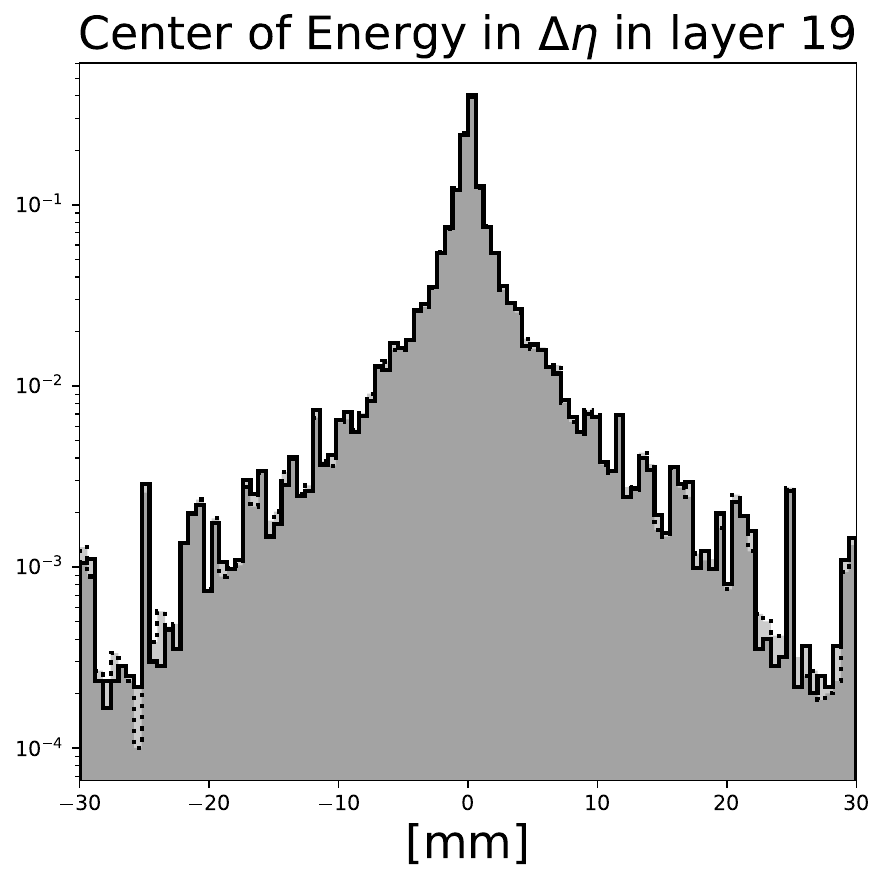}\\
    \includegraphics[height=0.1\textheight]{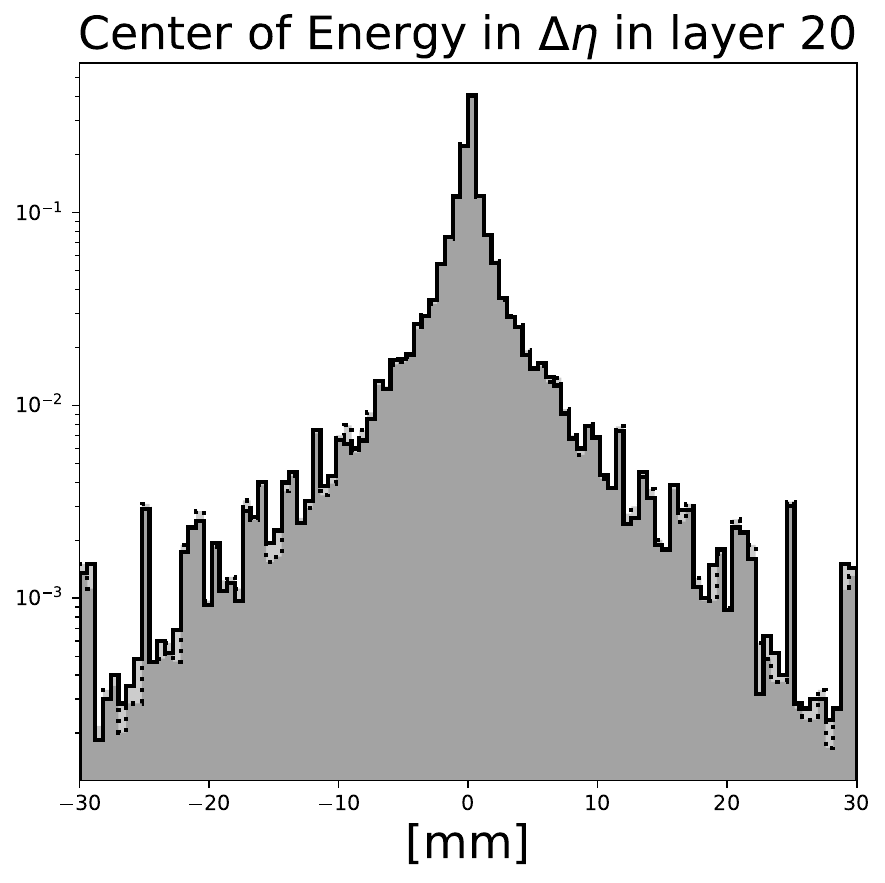} \hfill \includegraphics[height=0.1\textheight]{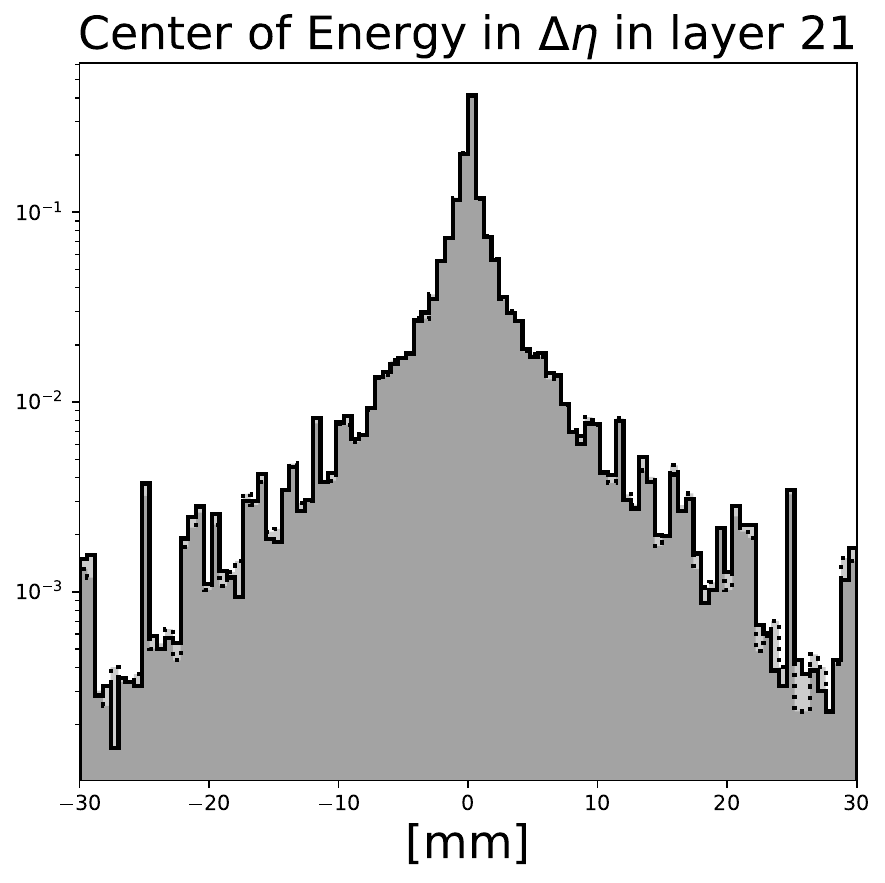} \hfill \includegraphics[height=0.1\textheight]{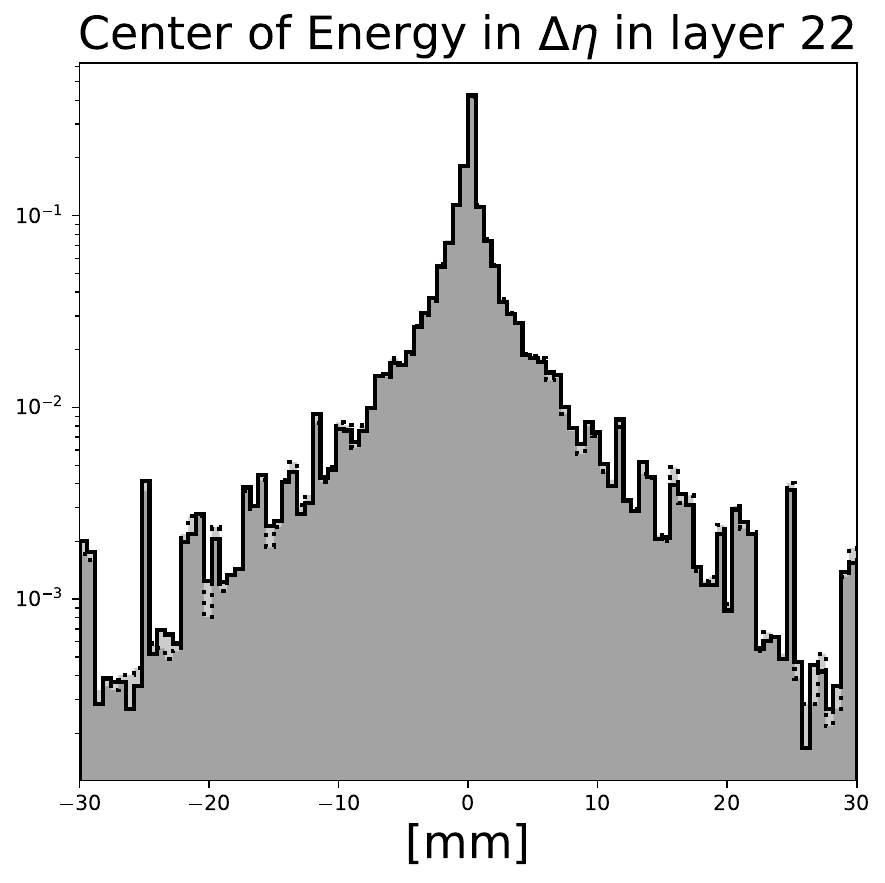} \hfill \includegraphics[height=0.1\textheight]{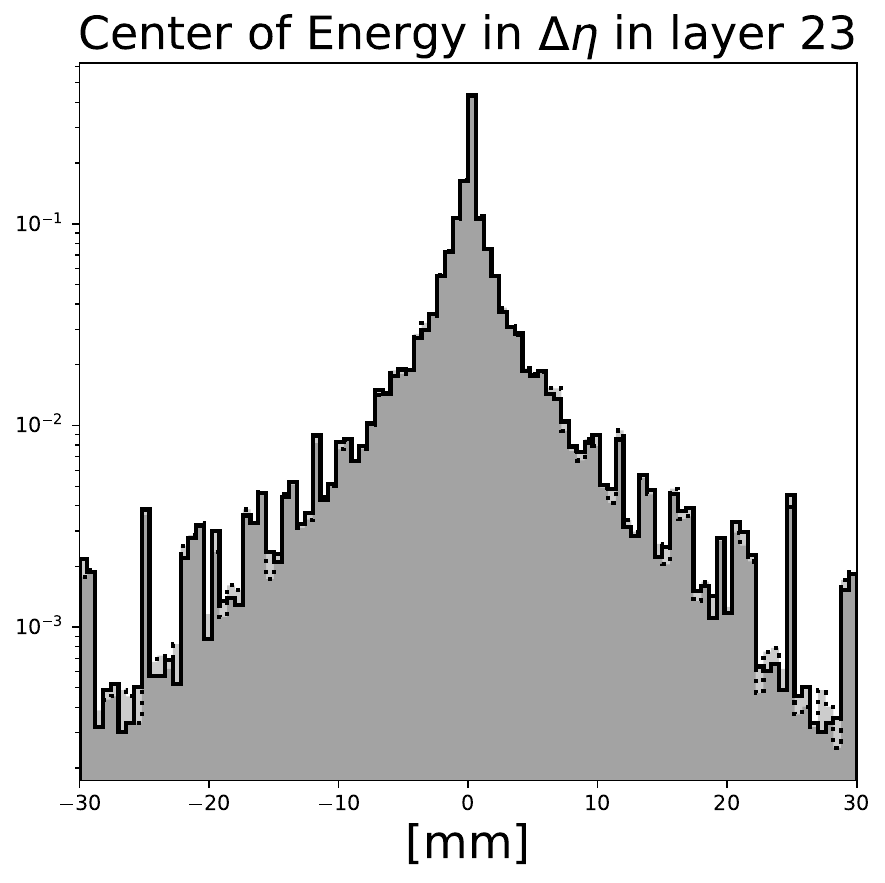} \hfill \includegraphics[height=0.1\textheight]{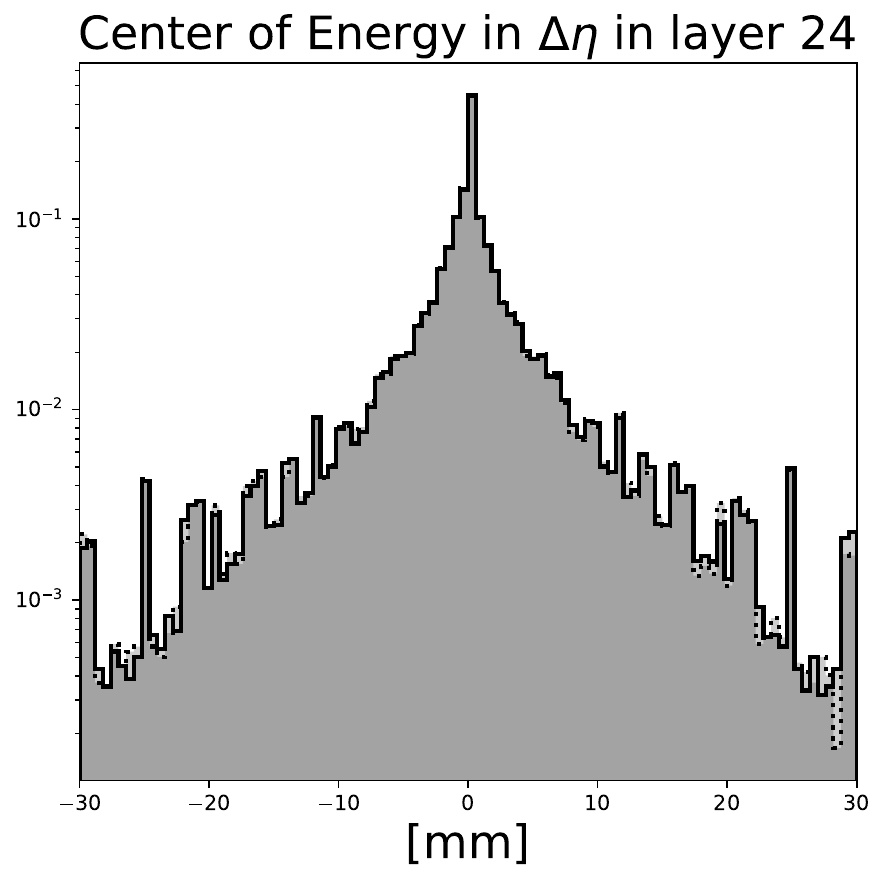}\\
    \includegraphics[height=0.1\textheight]{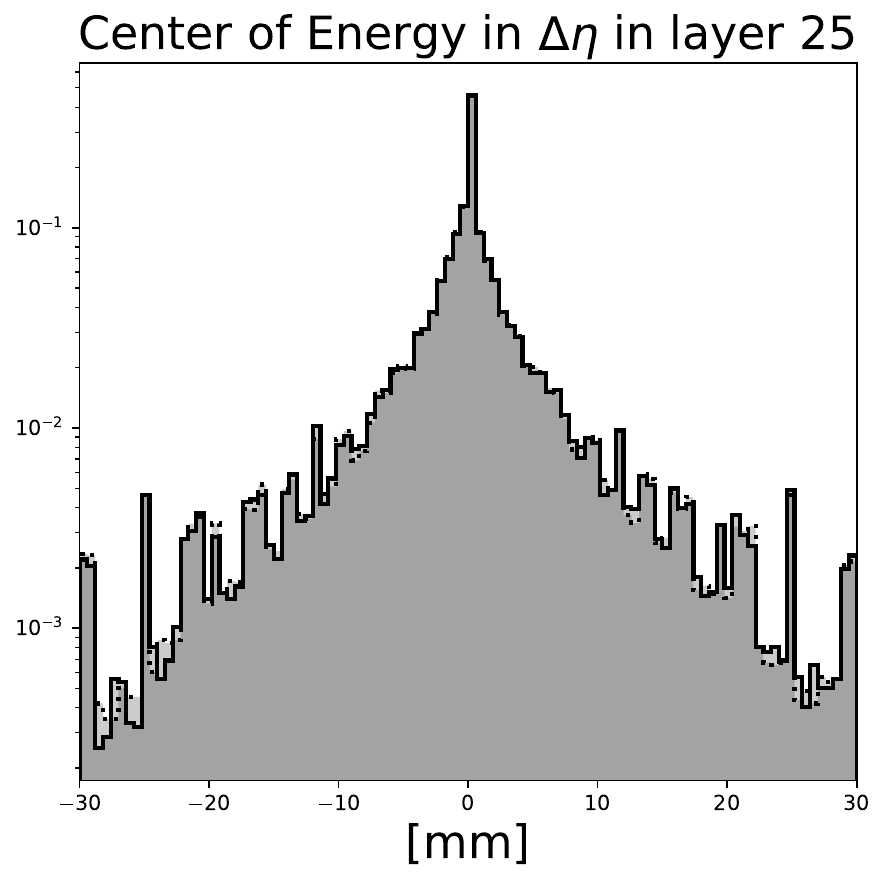} \hfill \includegraphics[height=0.1\textheight]{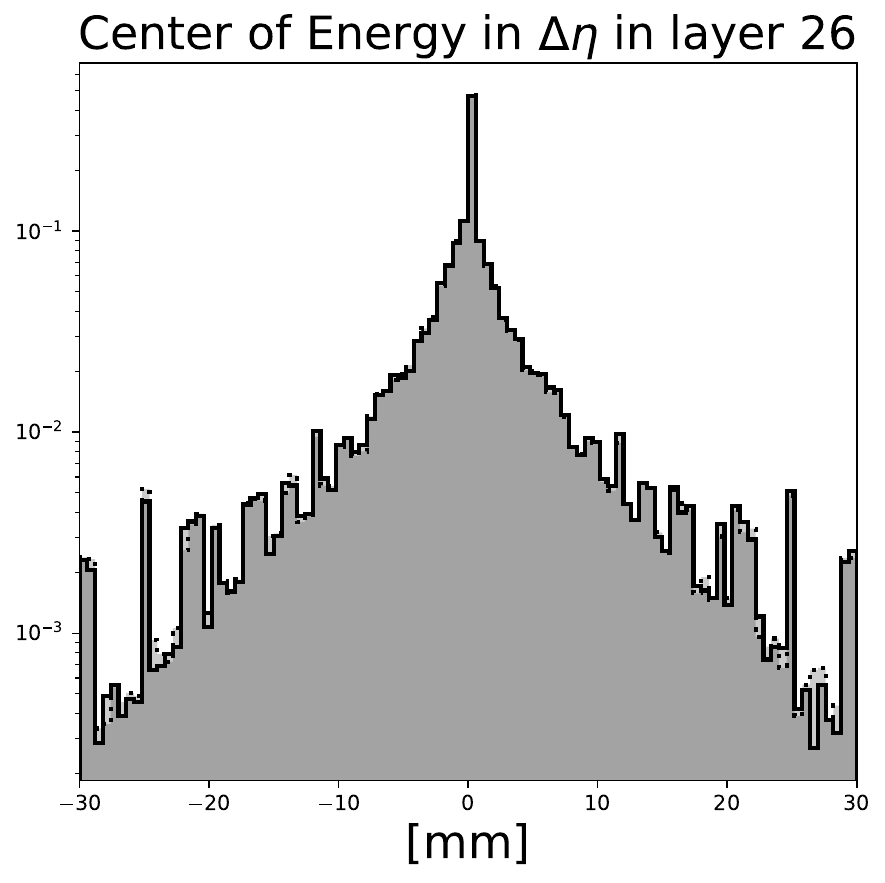} \hfill \includegraphics[height=0.1\textheight]{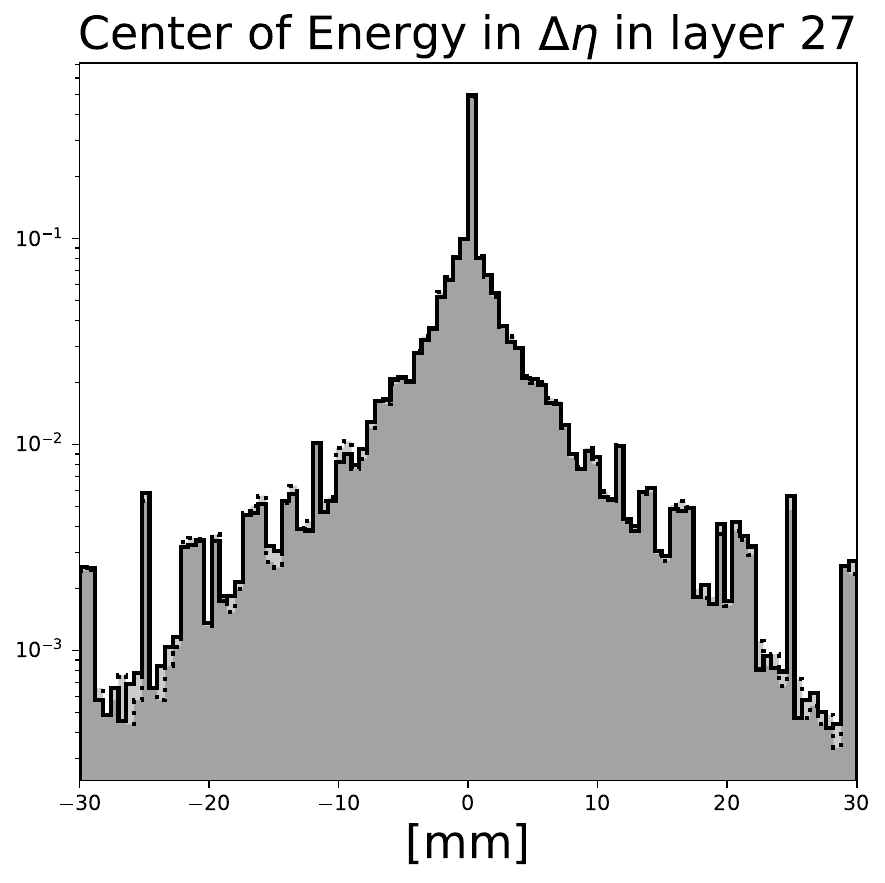} \hfill \includegraphics[height=0.1\textheight]{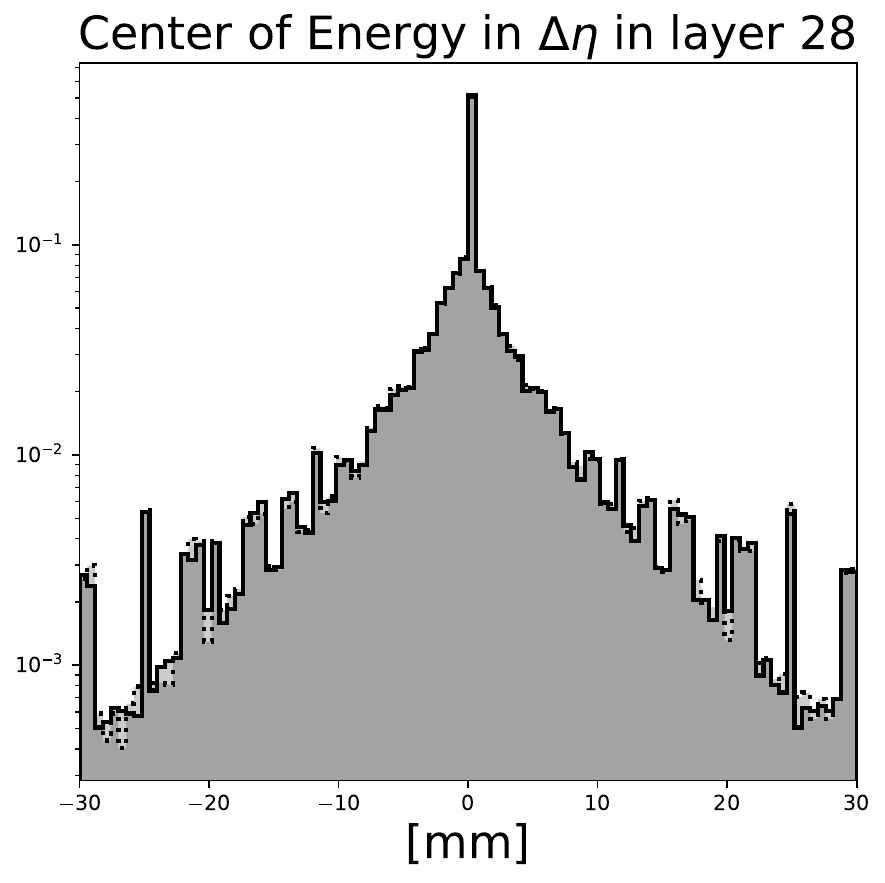} \hfill \includegraphics[height=0.1\textheight]{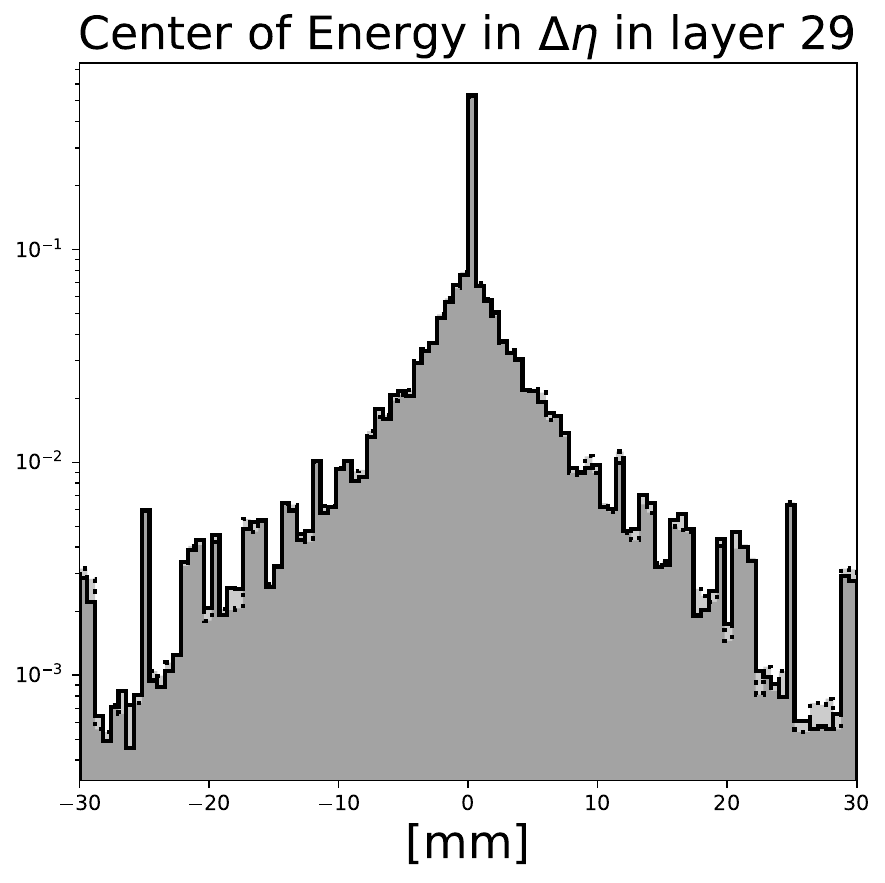}\\
    \includegraphics[height=0.1\textheight]{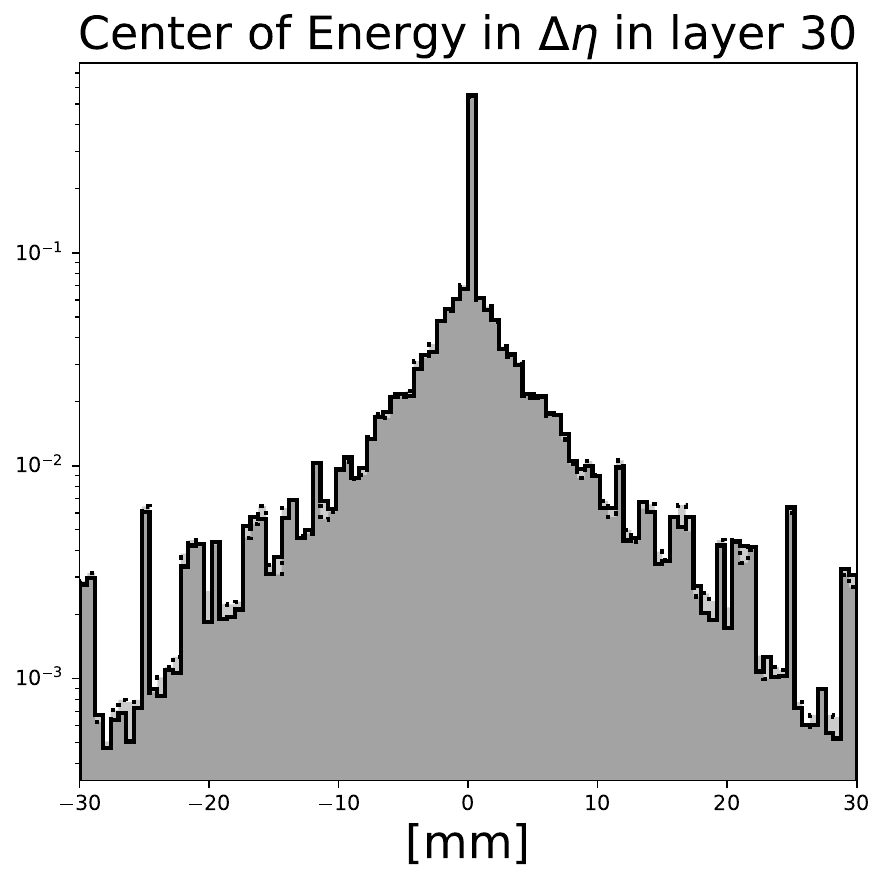} \hfill \includegraphics[height=0.1\textheight]{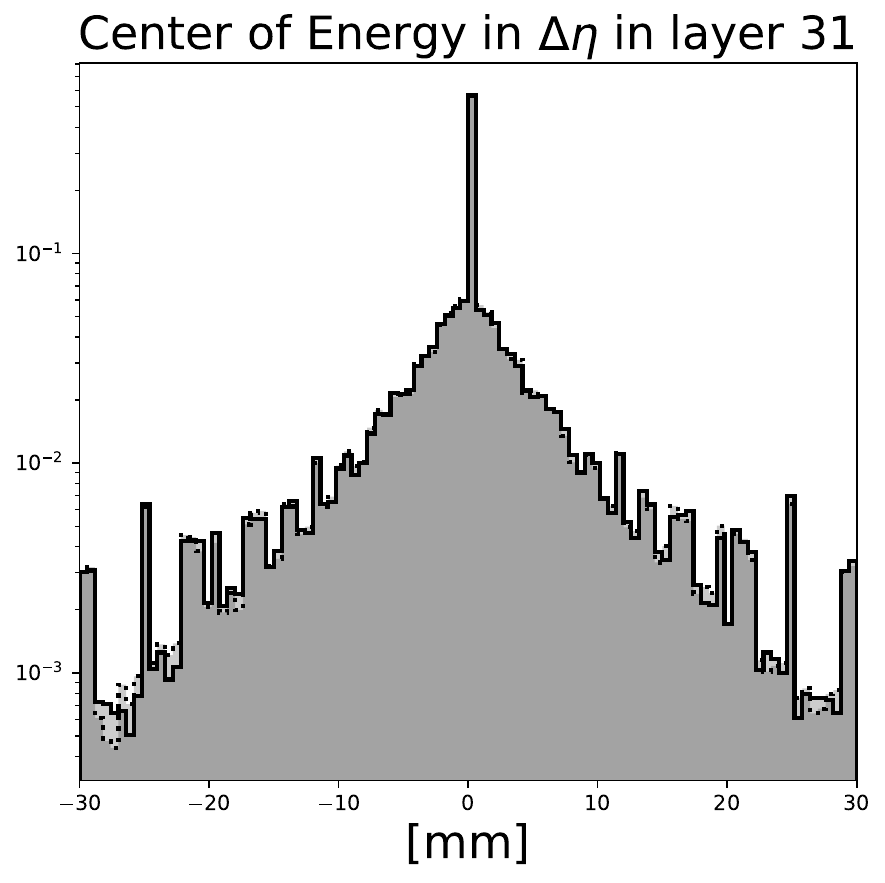} \hfill \includegraphics[height=0.1\textheight]{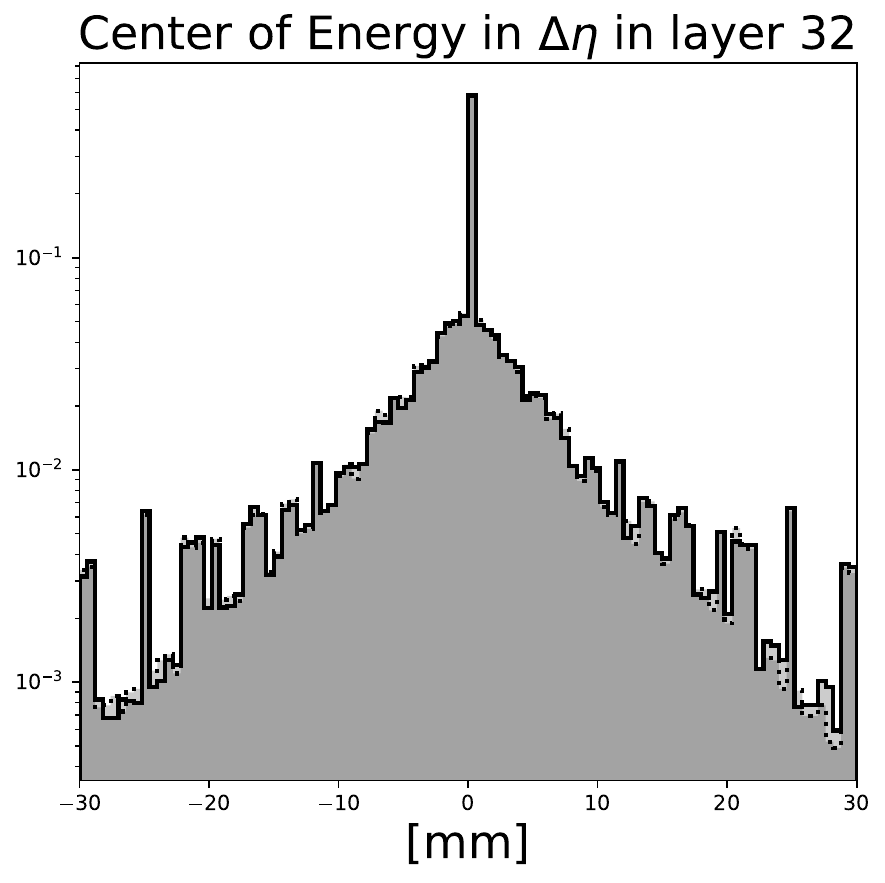} \hfill \includegraphics[height=0.1\textheight]{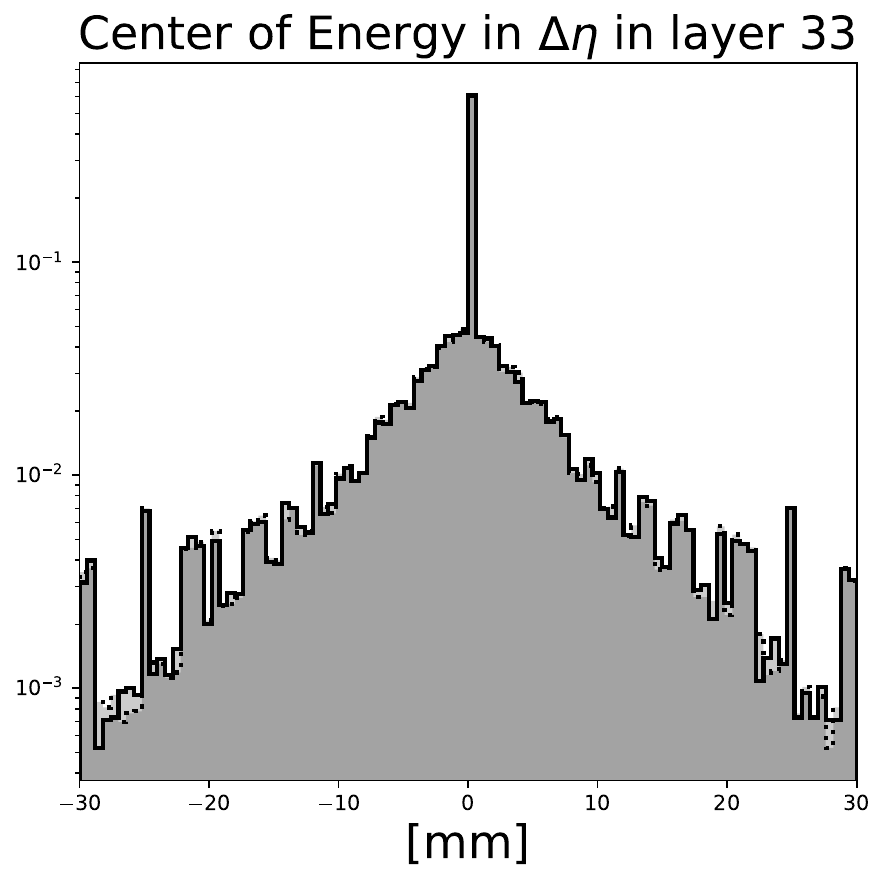} \hfill \includegraphics[height=0.1\textheight]{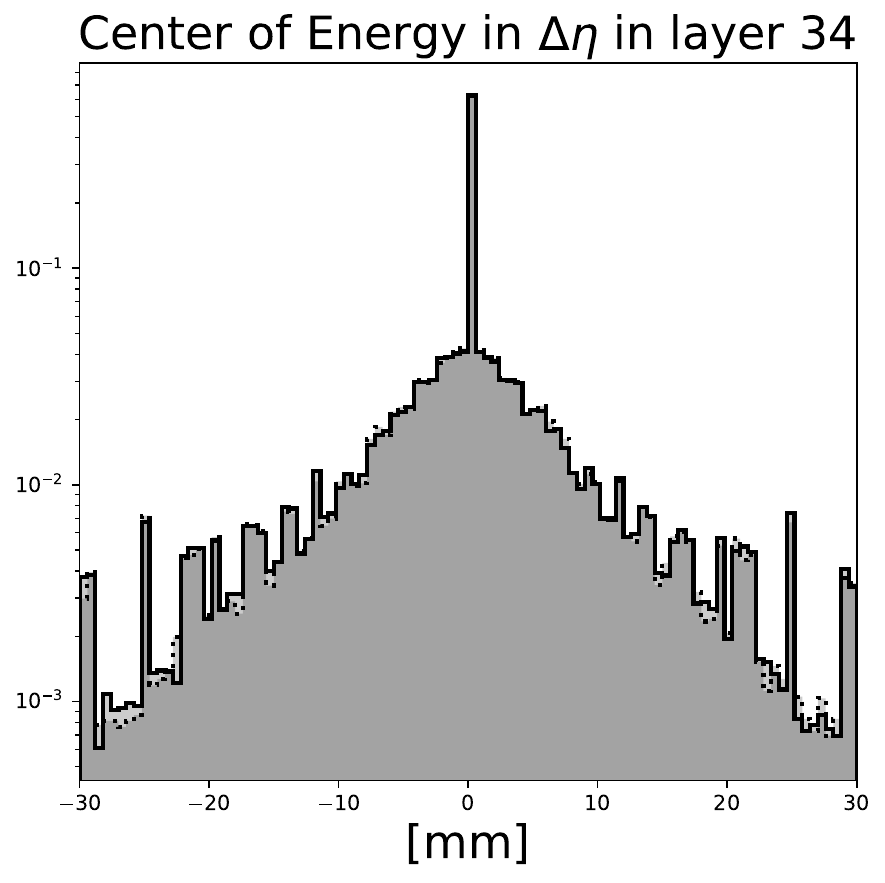}\\
    \includegraphics[height=0.1\textheight]{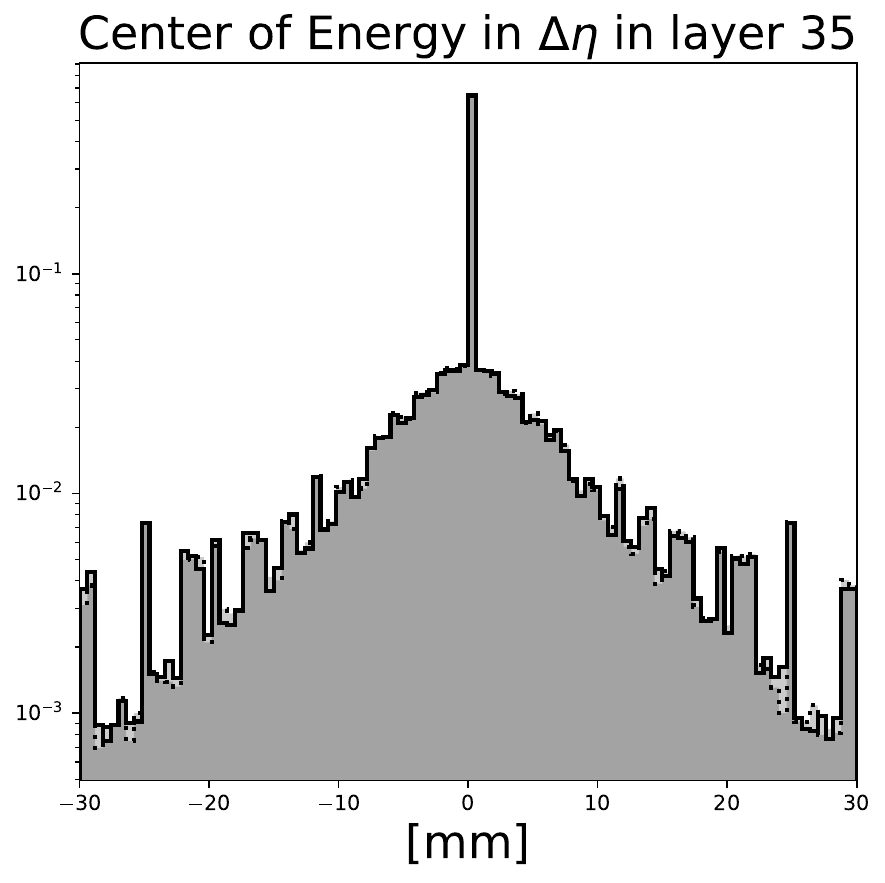} \hfill \includegraphics[height=0.1\textheight]{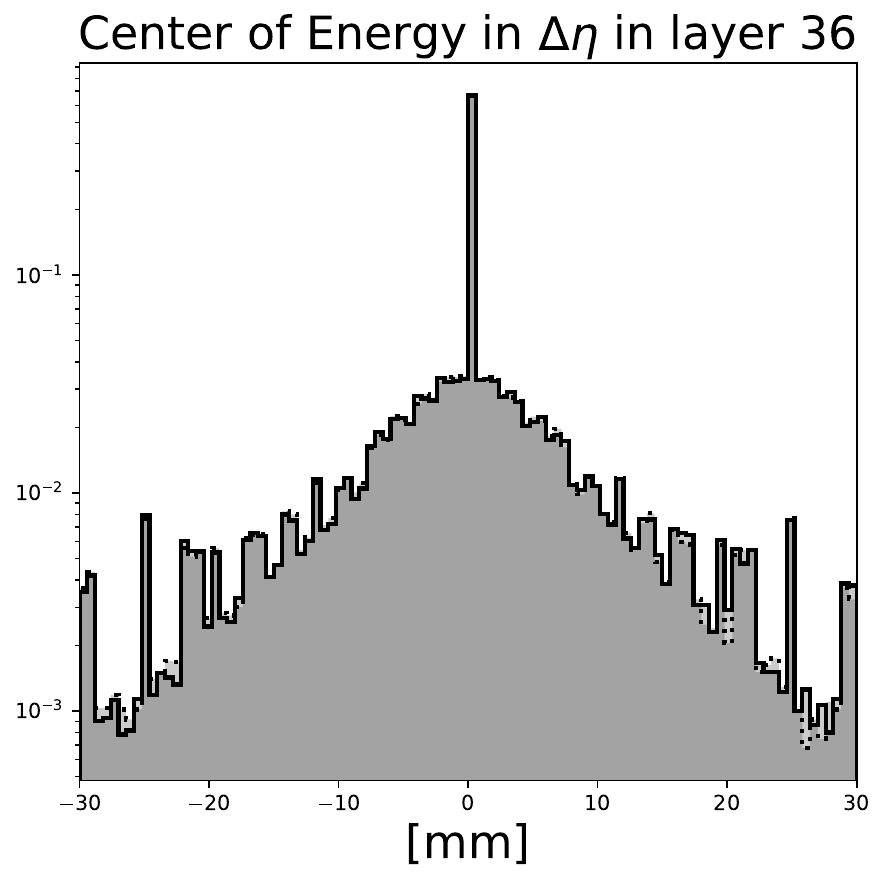} \hfill \includegraphics[height=0.1\textheight]{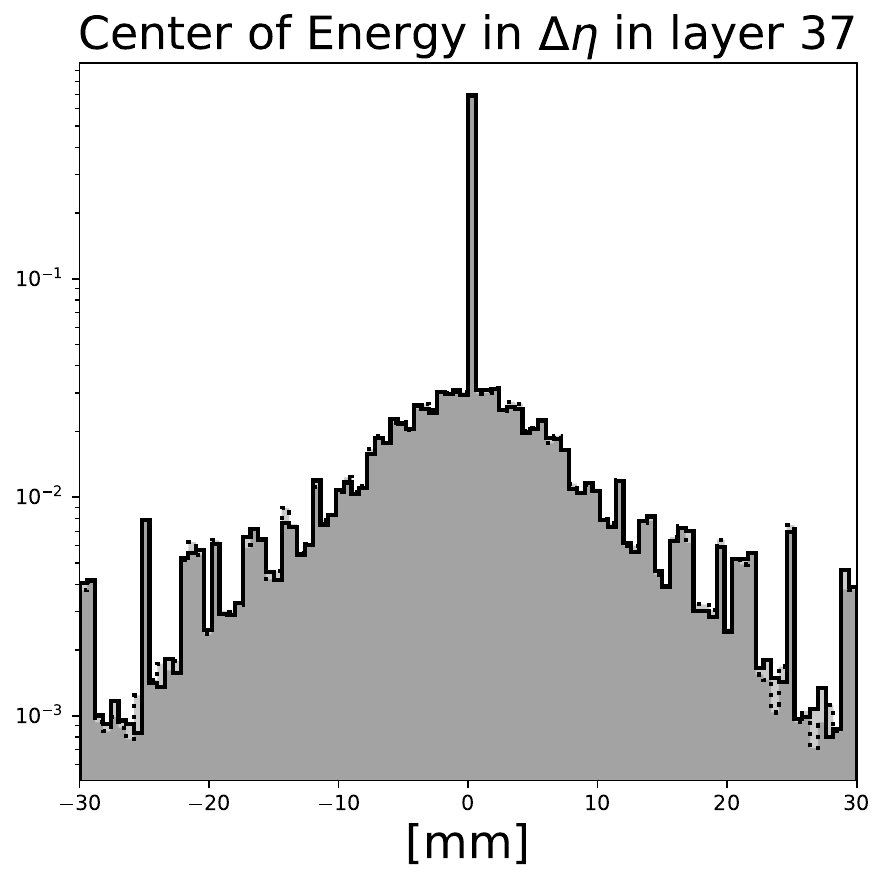} \hfill \includegraphics[height=0.1\textheight]{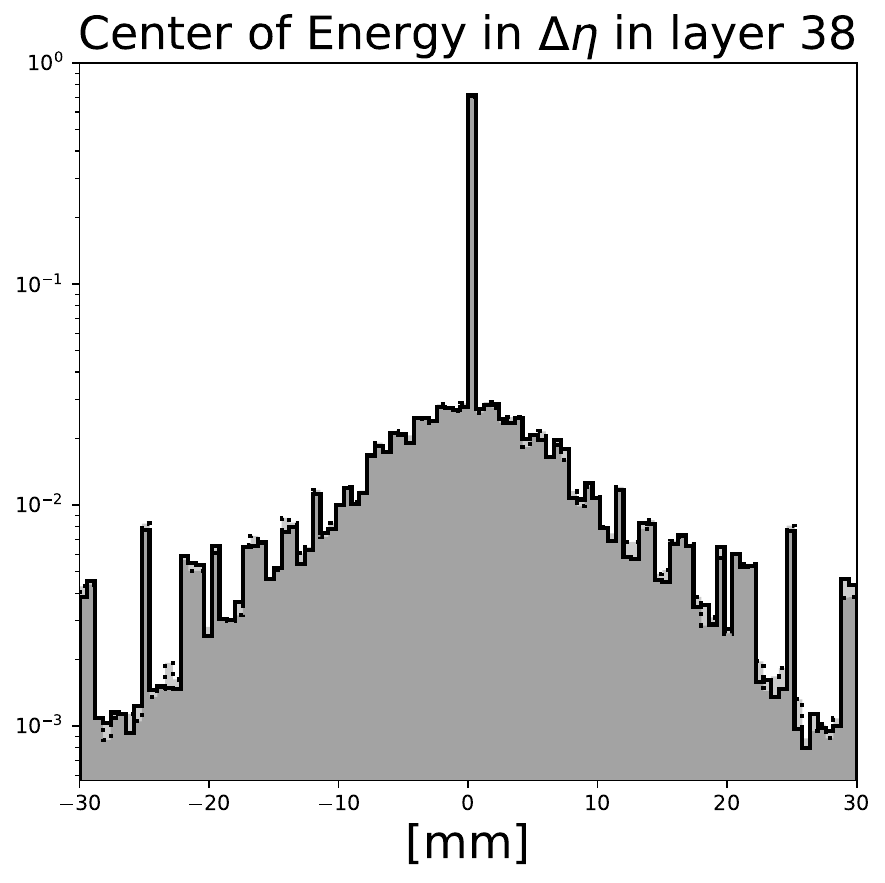} \hfill \includegraphics[height=0.1\textheight]{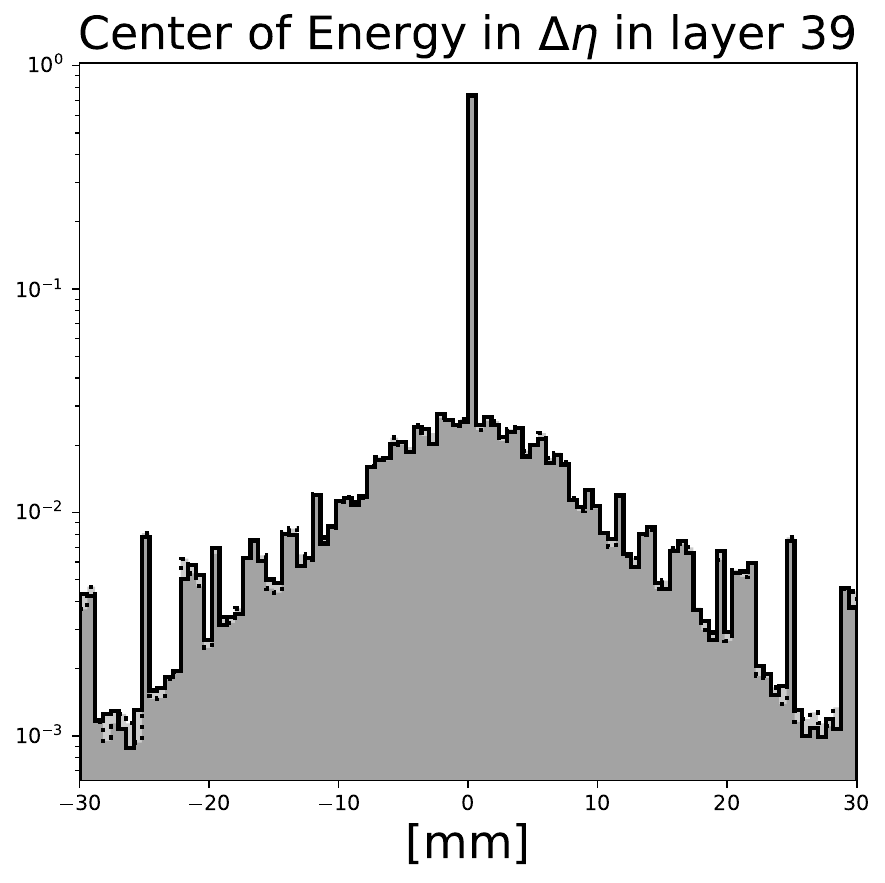}\\
    \includegraphics[height=0.1\textheight]{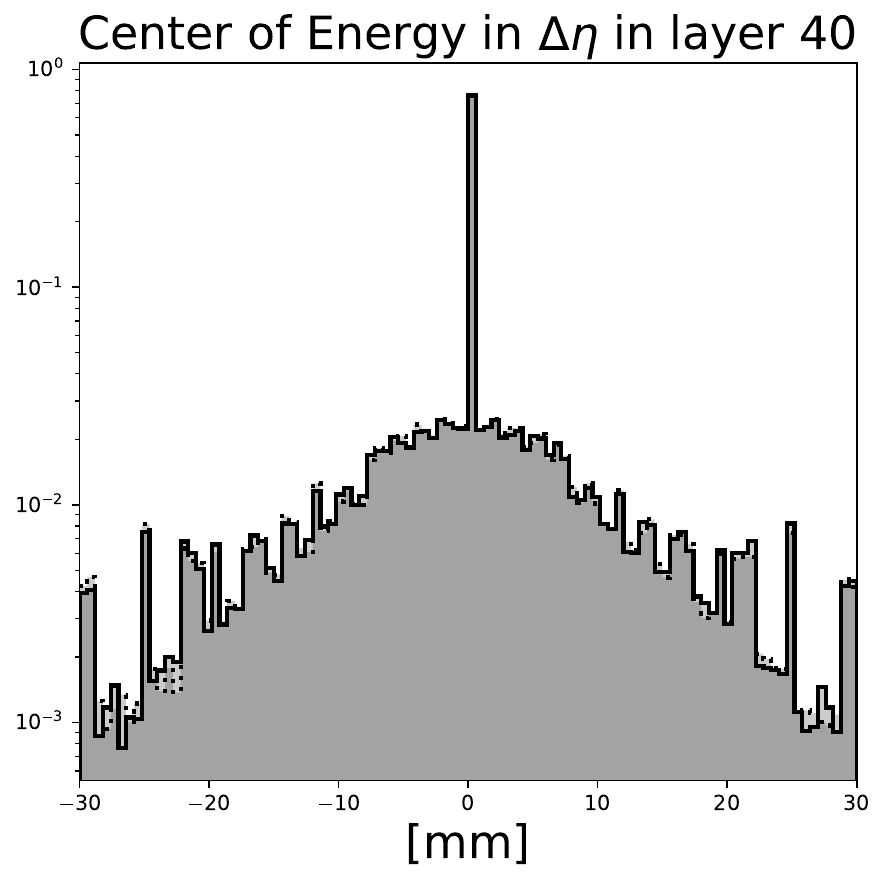} \hfill \includegraphics[height=0.1\textheight]{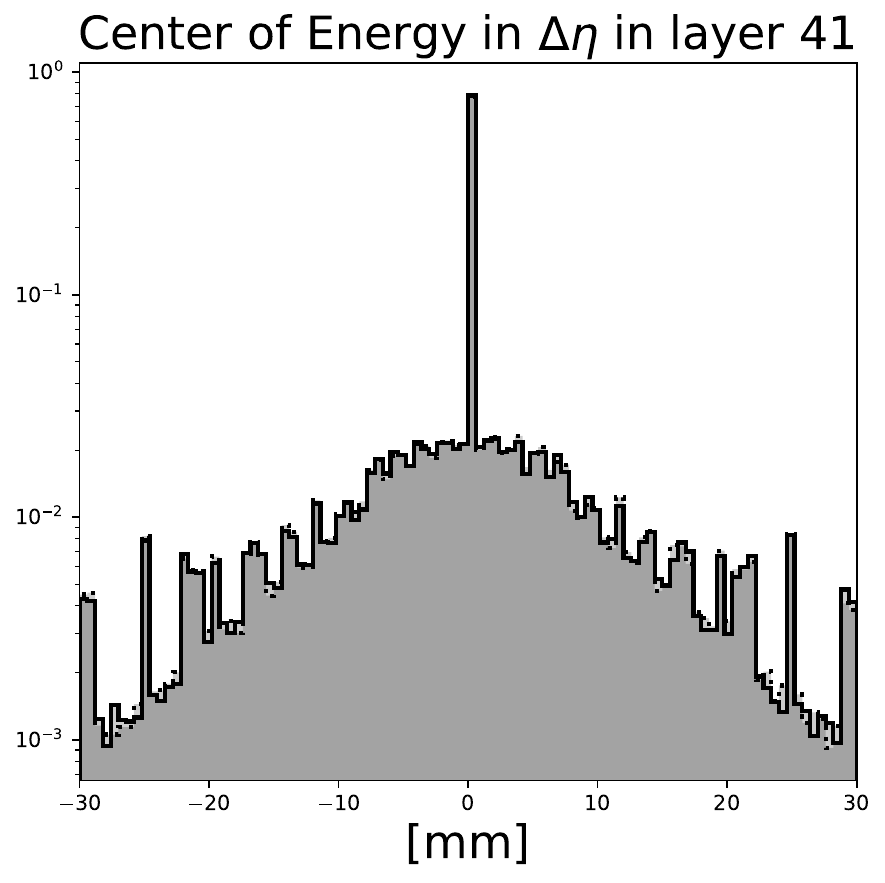} \hfill \includegraphics[height=0.1\textheight]{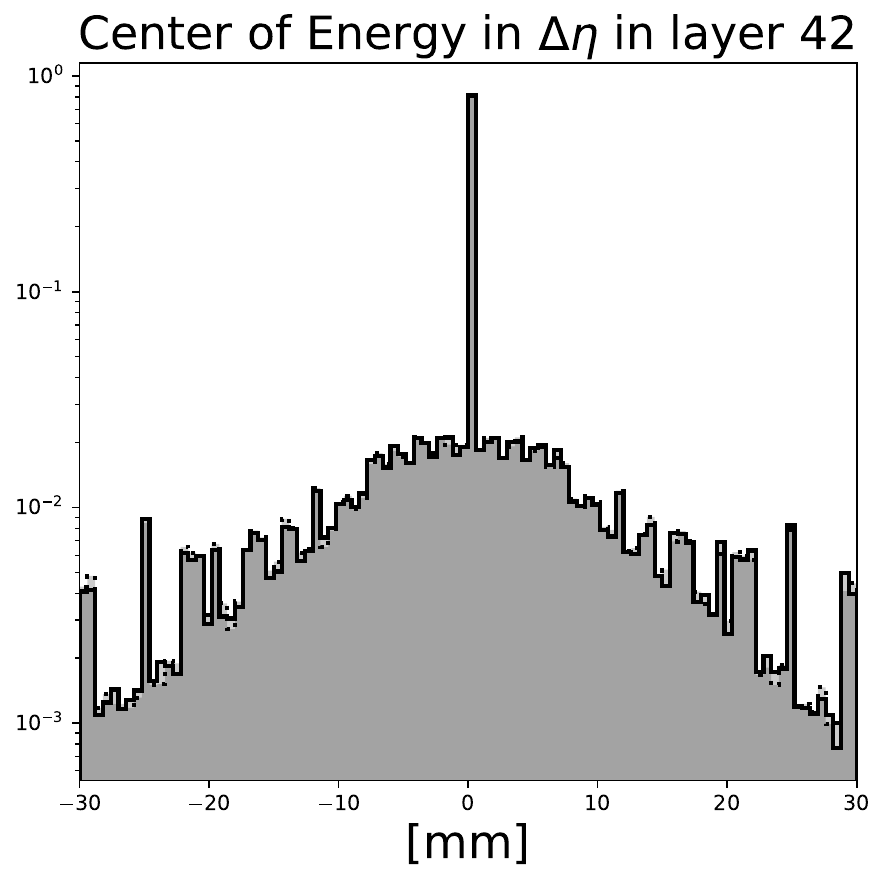} \hfill \includegraphics[height=0.1\textheight]{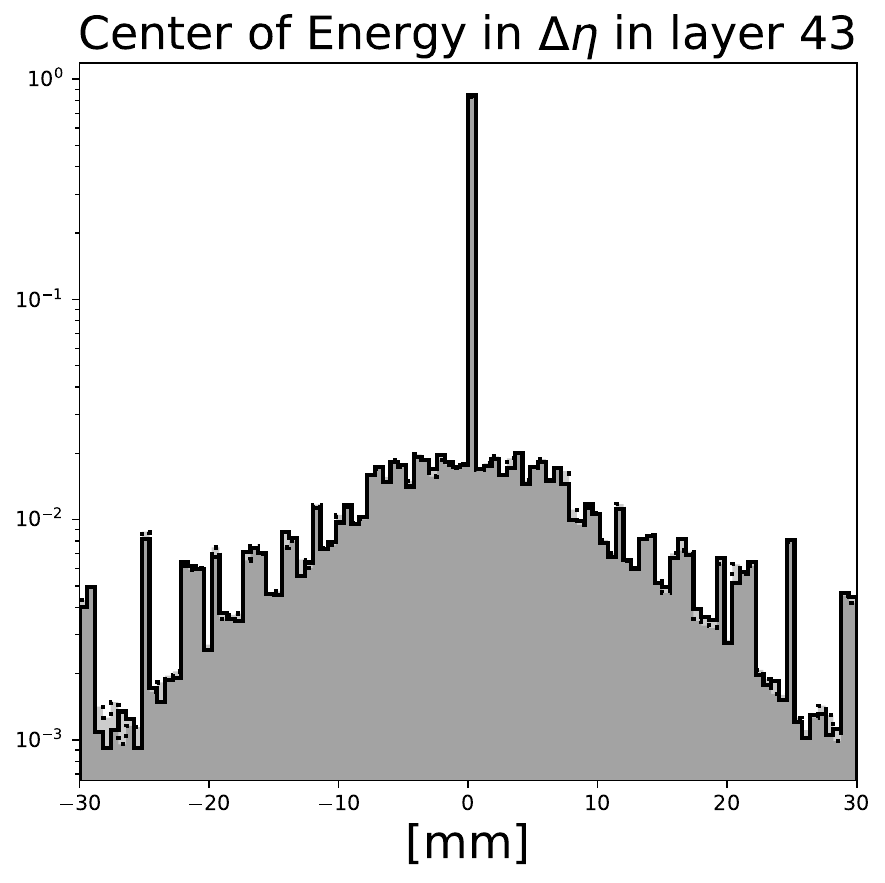} \hfill \includegraphics[height=0.1\textheight]{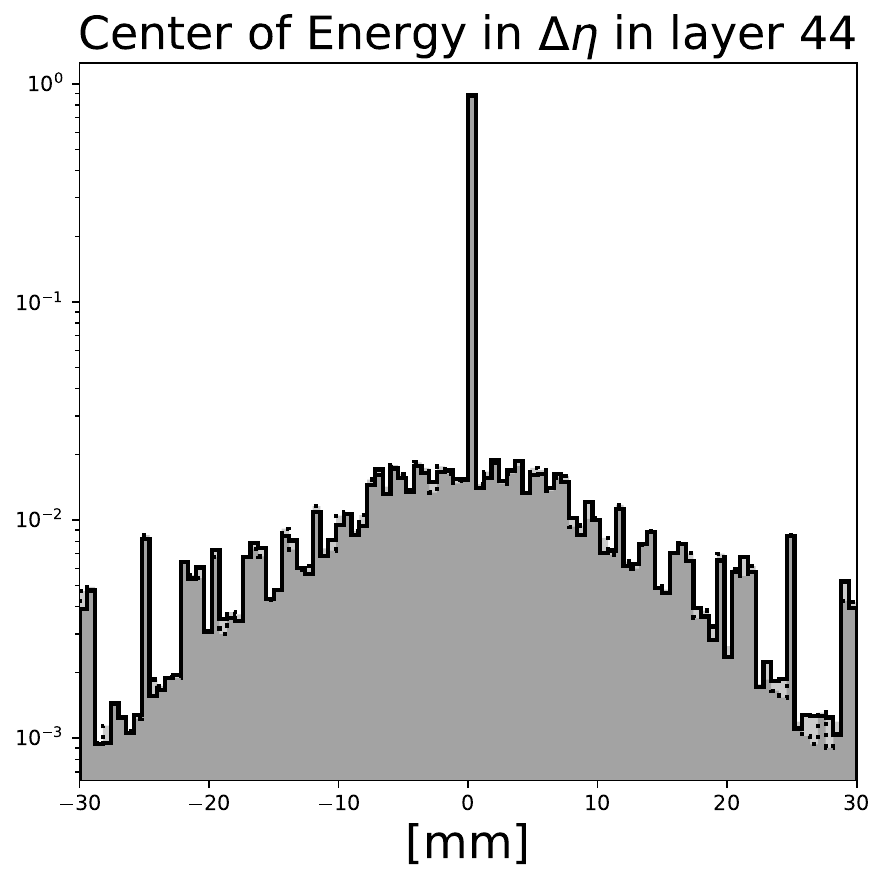}\\
    \includegraphics[width=0.5\textwidth]{figures/Appendix_reference/legend.pdf}
    \caption{Distribution of \geant training and evaluation data in centers of energy in $\eta$ direction for ds2. }
    \label{fig:app_ref.ds2.3}
\end{figure}

\begin{figure}[ht]
    \centering
    \includegraphics[height=0.1\textheight]{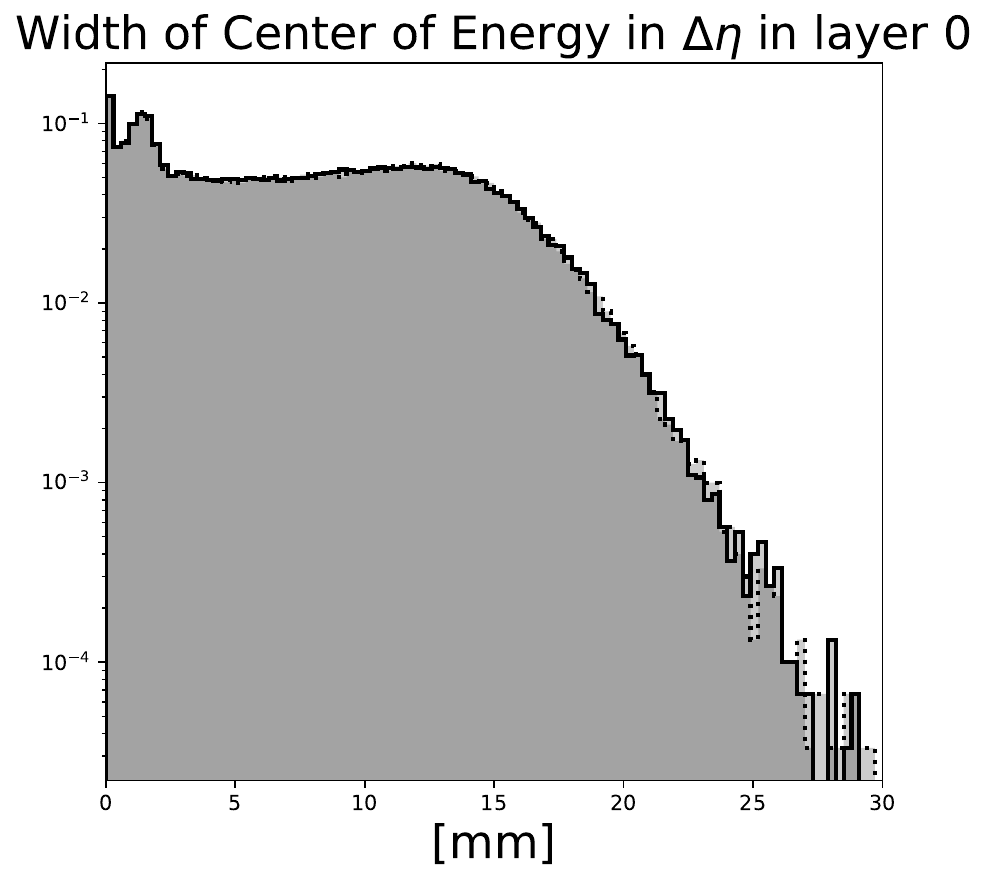} \hfill \includegraphics[height=0.1\textheight]{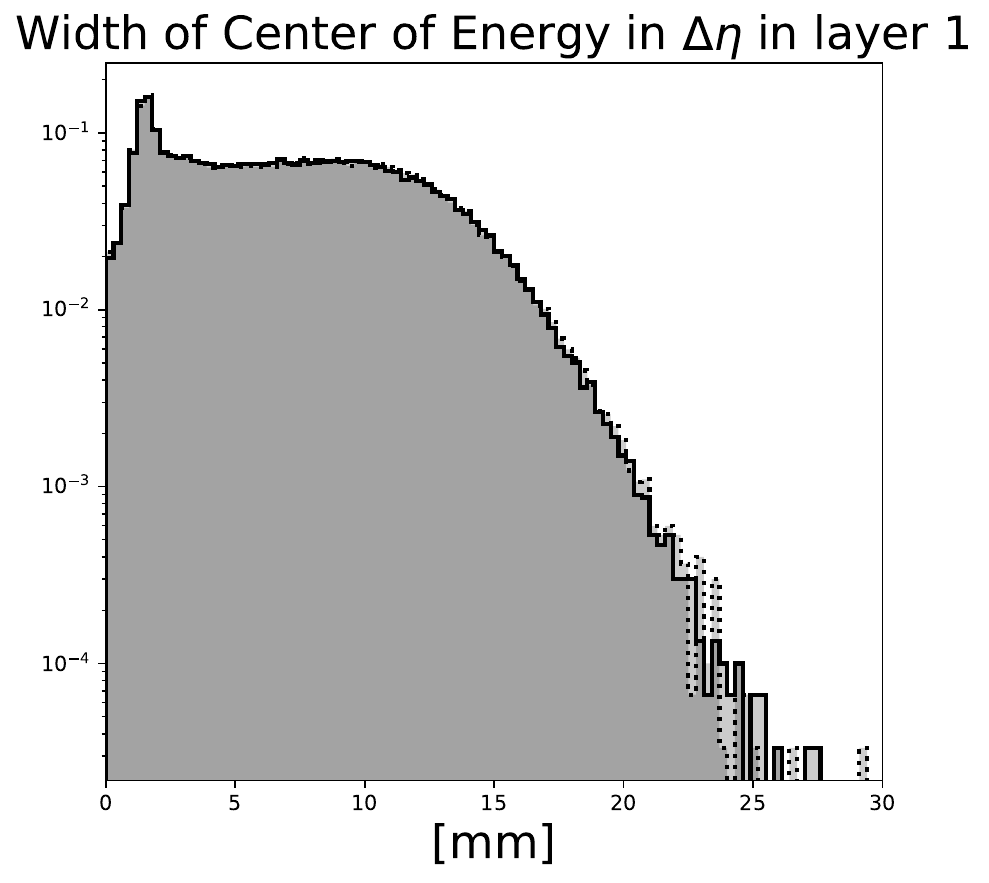} \hfill \includegraphics[height=0.1\textheight]{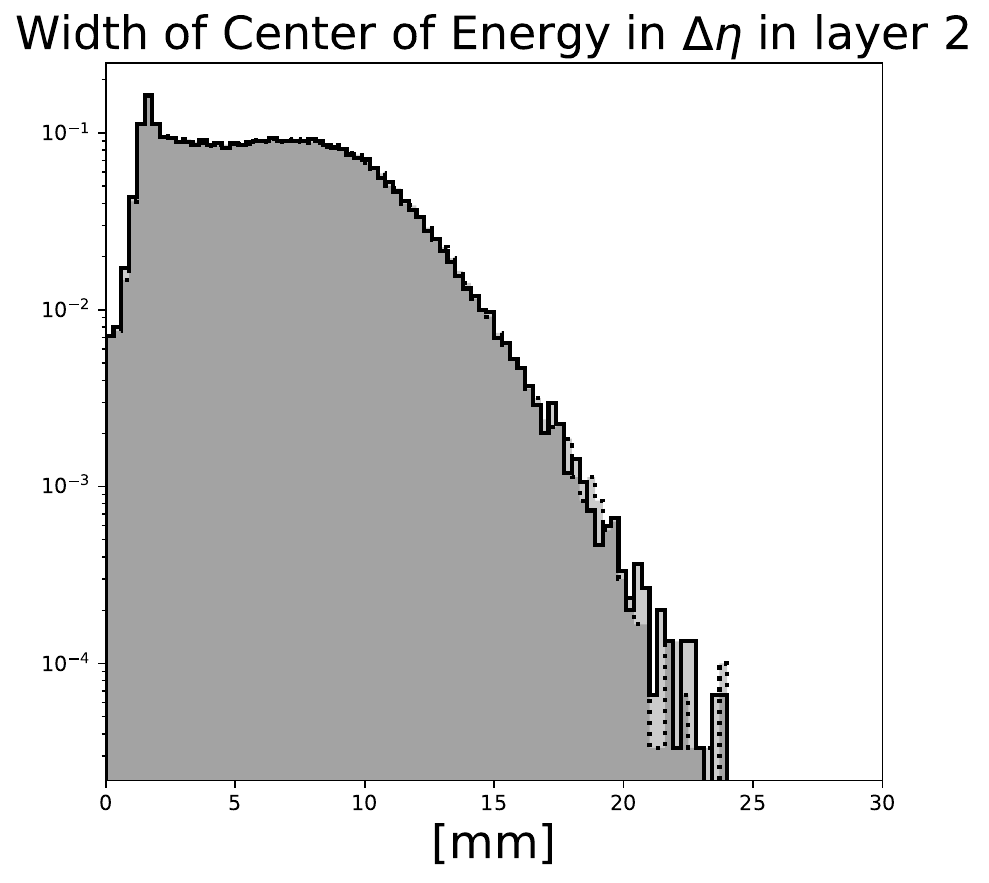} \hfill \includegraphics[height=0.1\textheight]{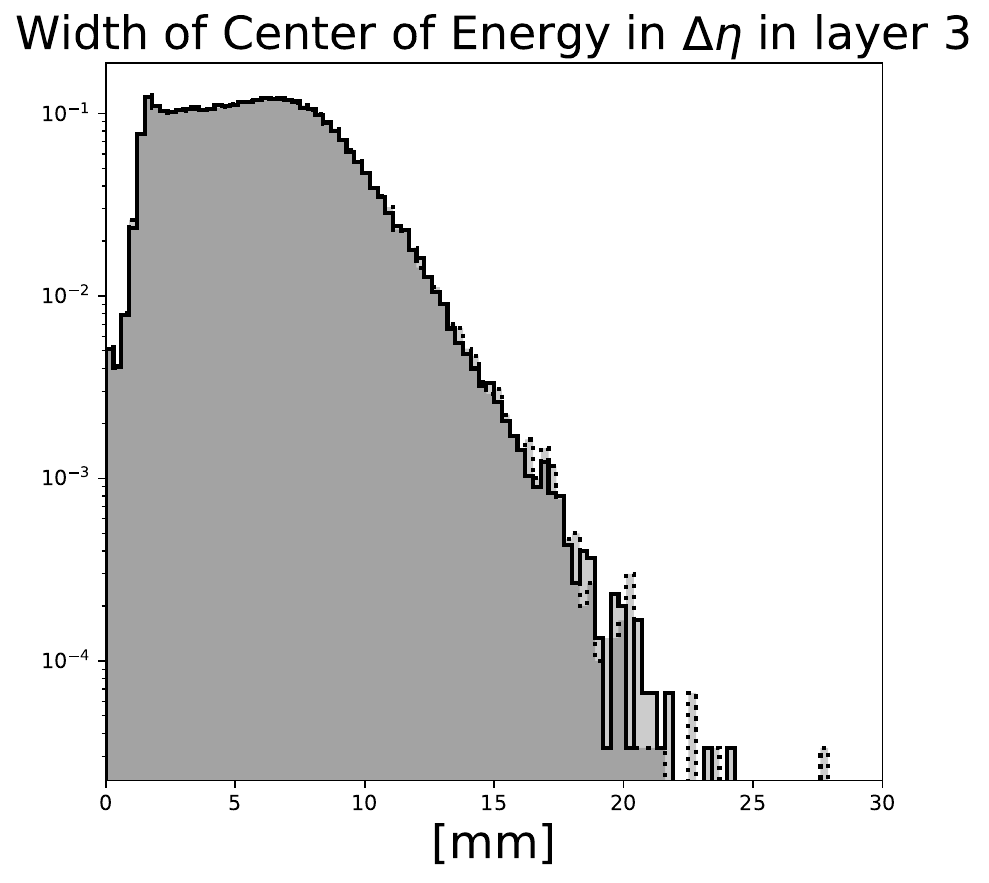} \hfill \includegraphics[height=0.1\textheight]{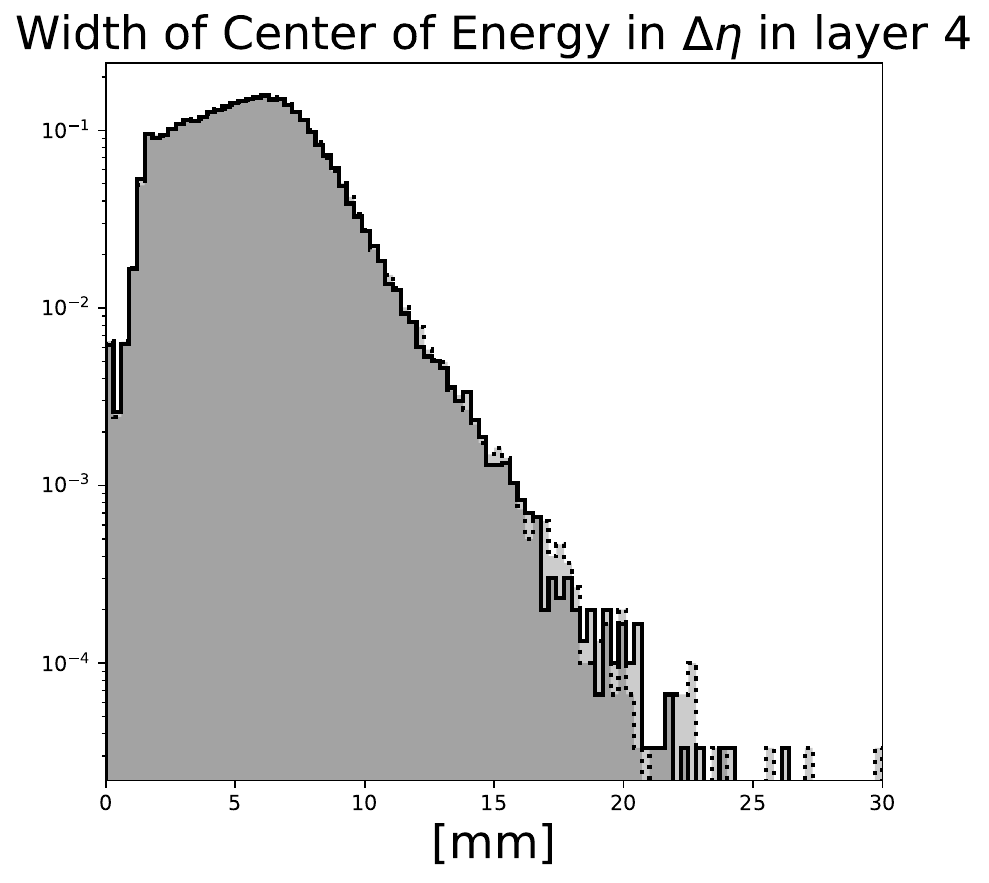}\\
    \includegraphics[height=0.1\textheight]{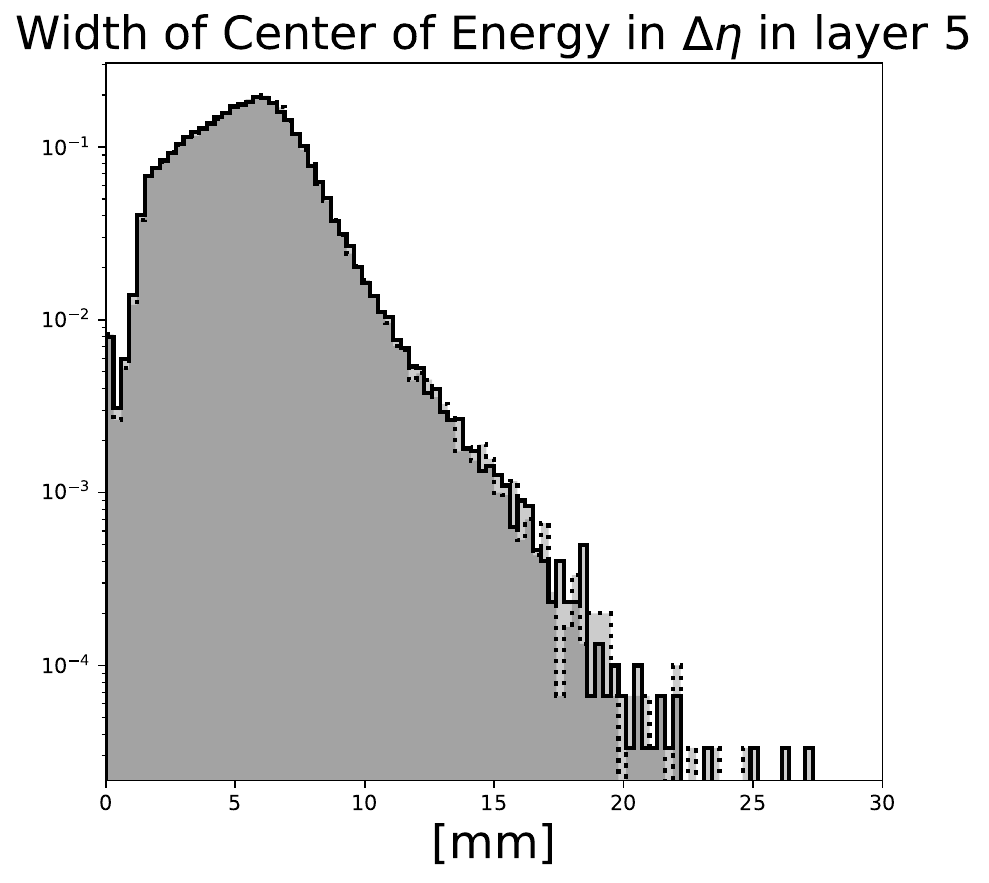} \hfill \includegraphics[height=0.1\textheight]{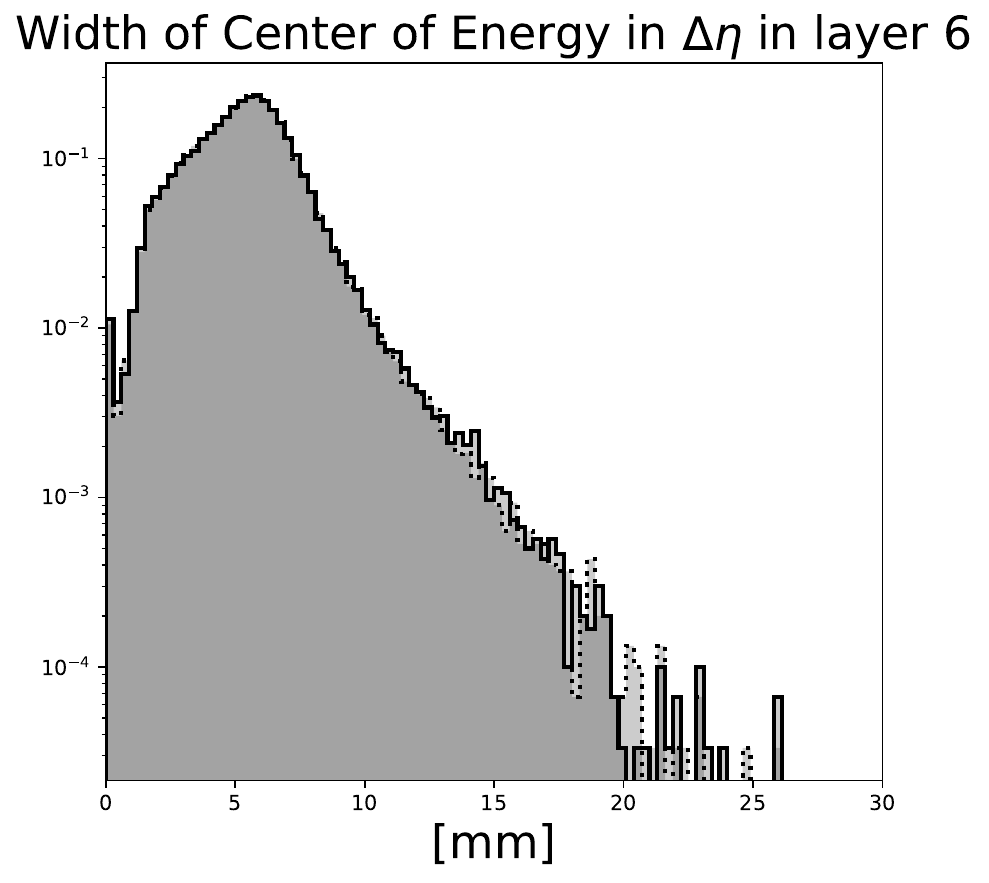} \hfill \includegraphics[height=0.1\textheight]{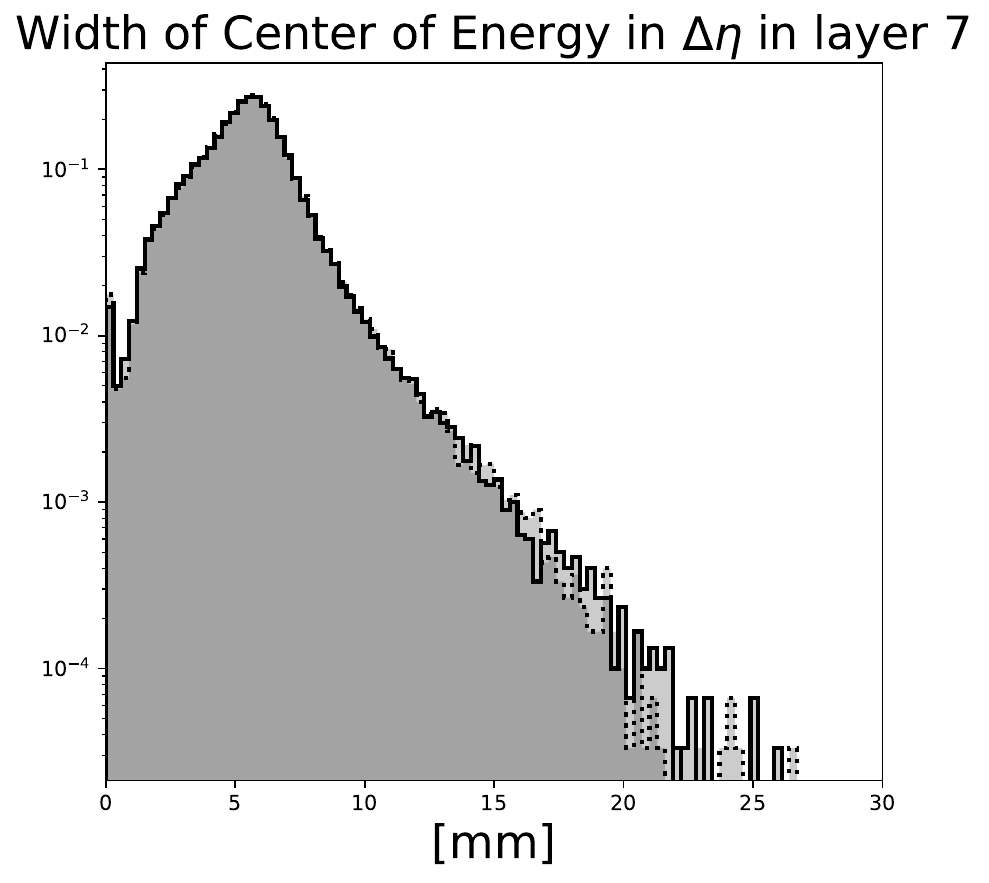} \hfill \includegraphics[height=0.1\textheight]{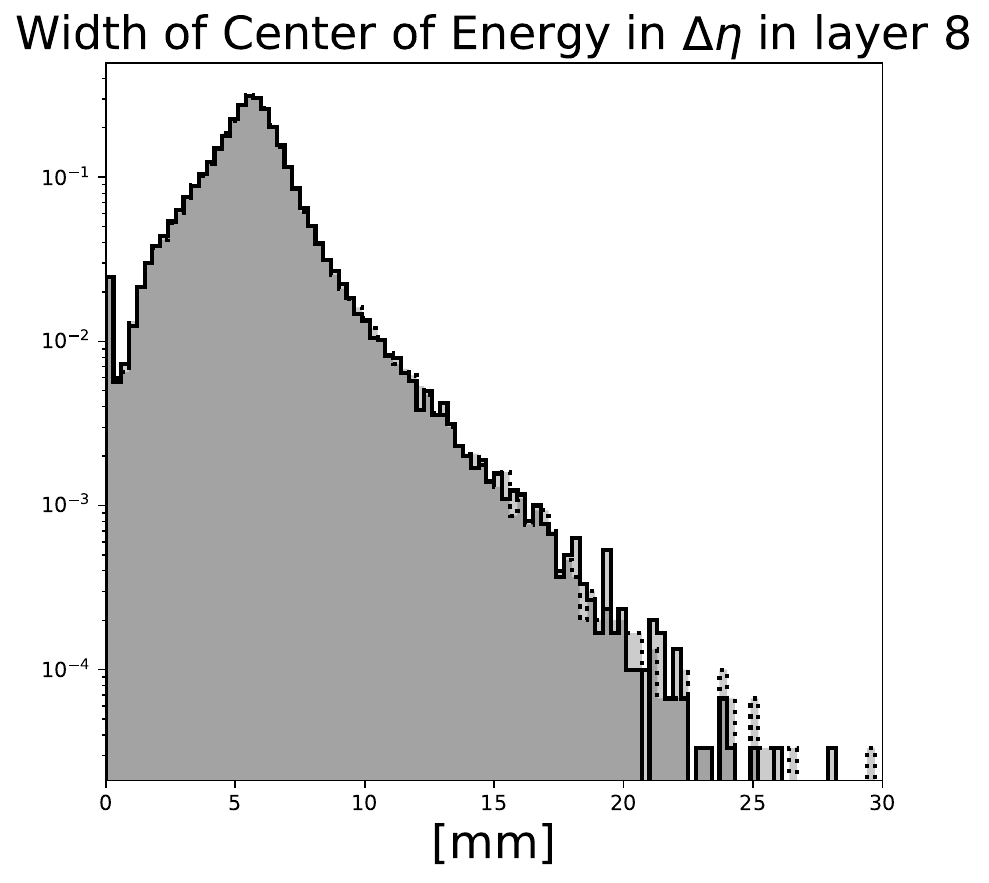} \hfill \includegraphics[height=0.1\textheight]{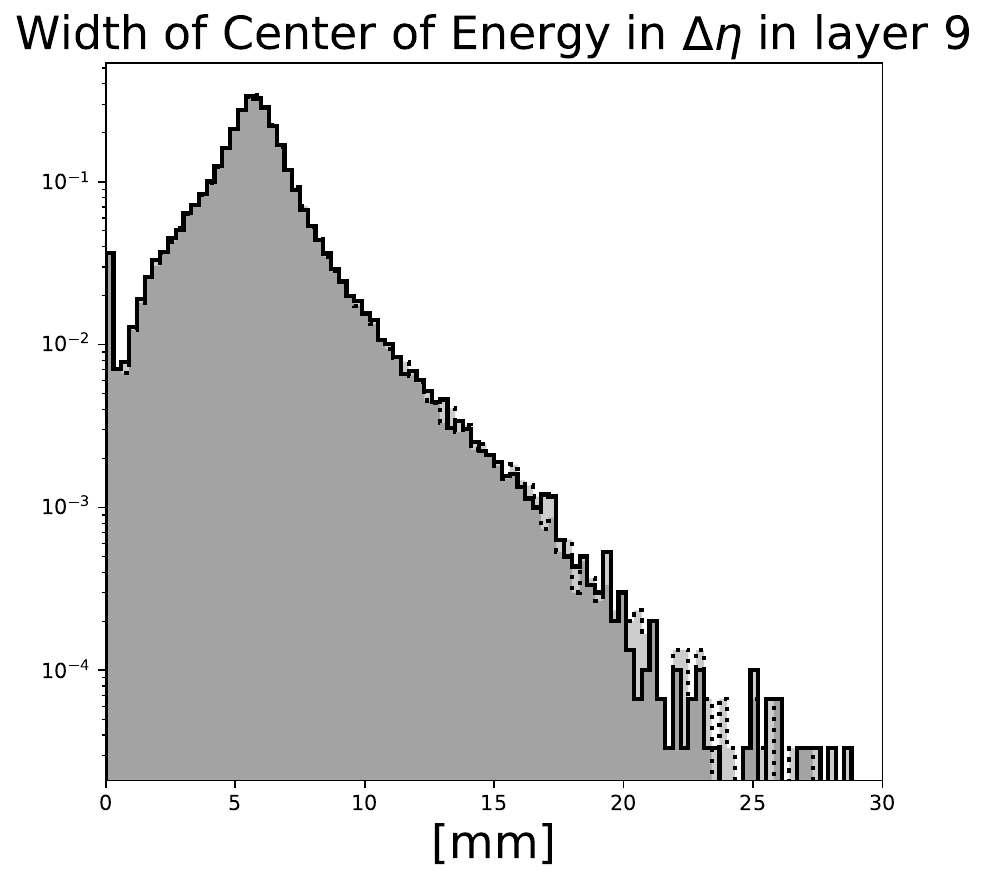}\\
    \includegraphics[height=0.1\textheight]{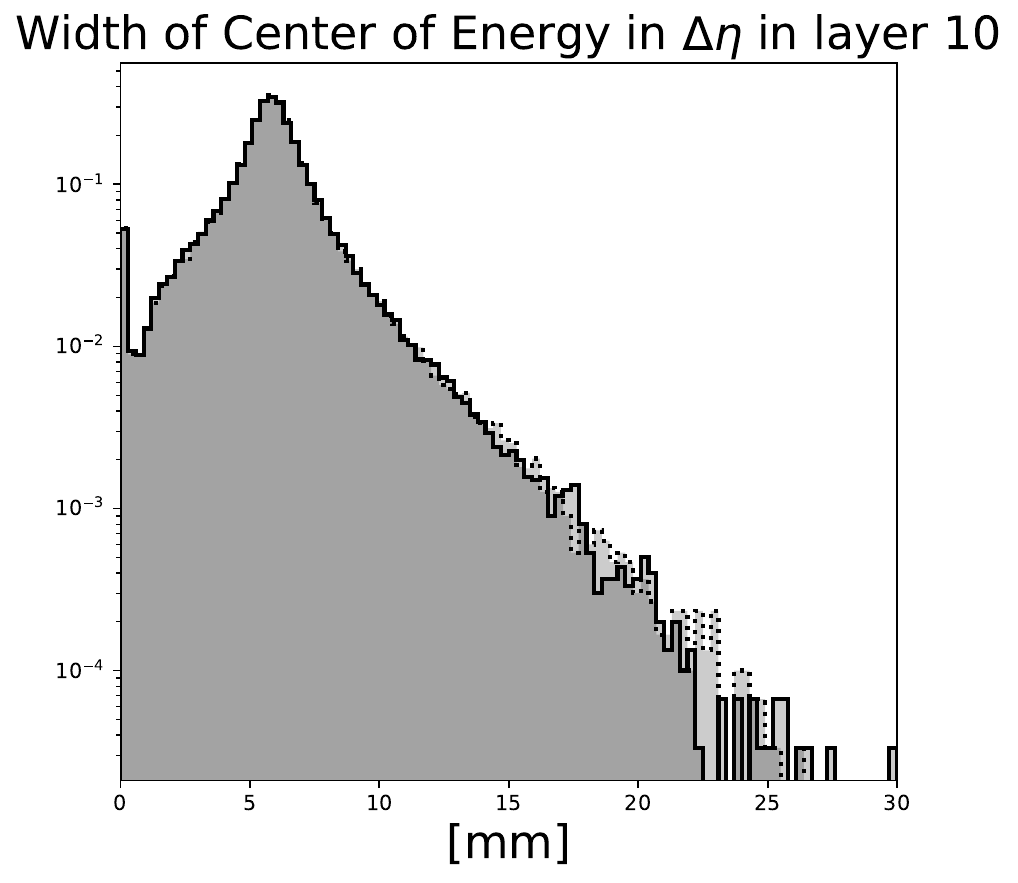} \hfill \includegraphics[height=0.1\textheight]{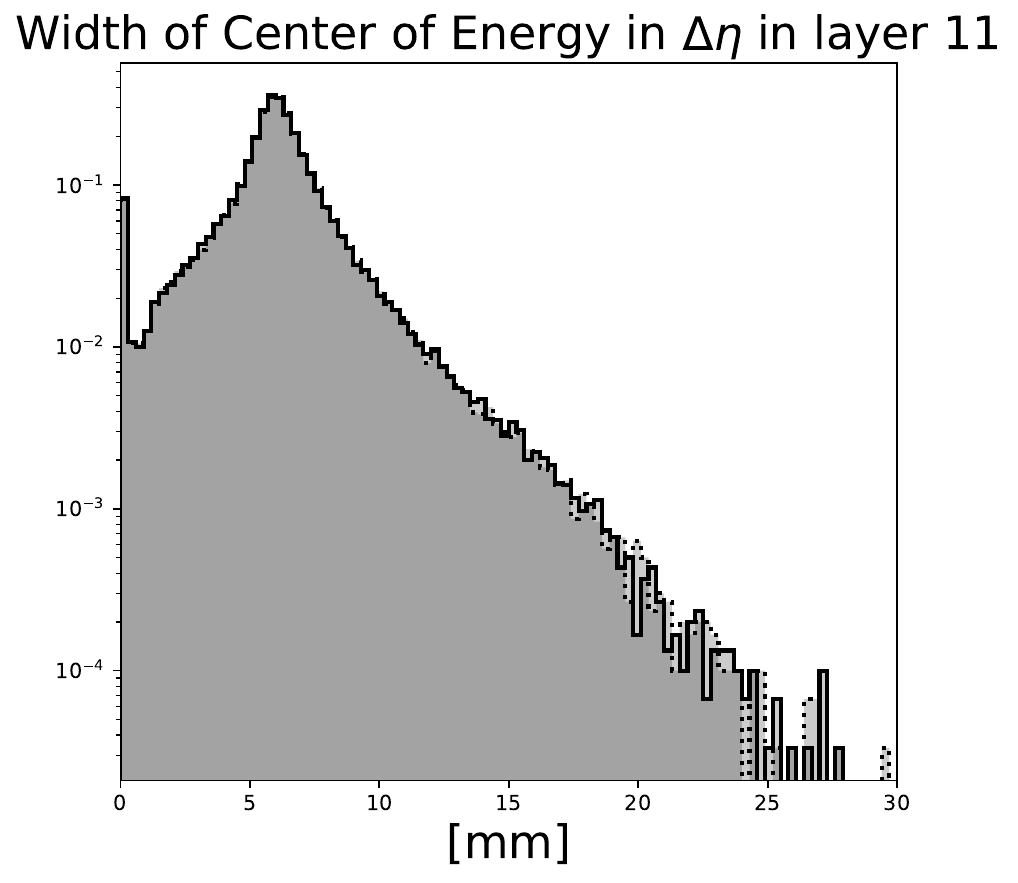} \hfill \includegraphics[height=0.1\textheight]{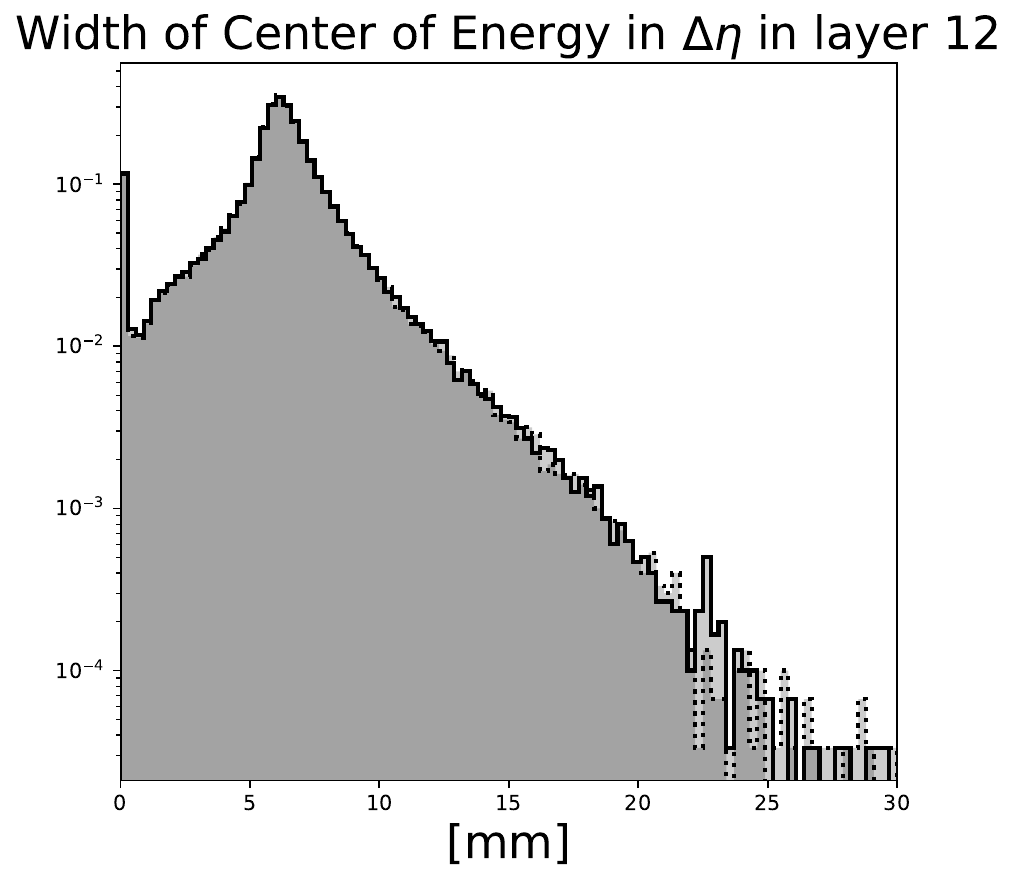} \hfill \includegraphics[height=0.1\textheight]{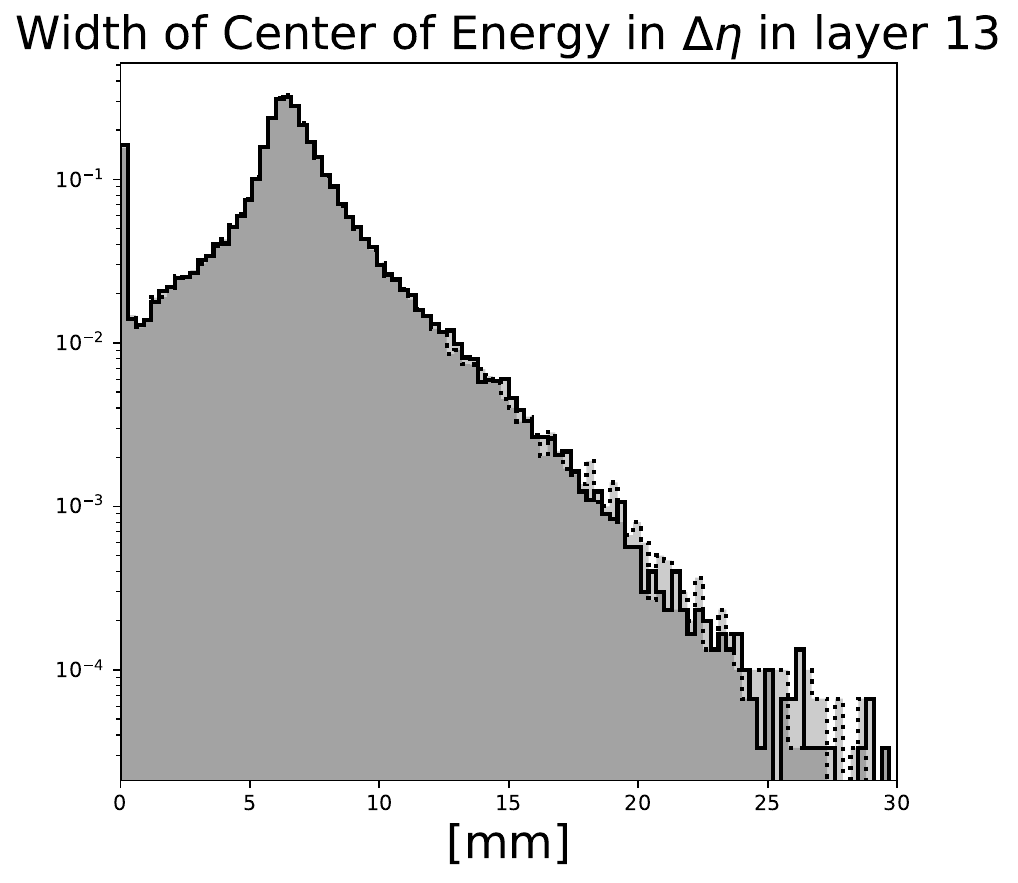} \hfill \includegraphics[height=0.1\textheight]{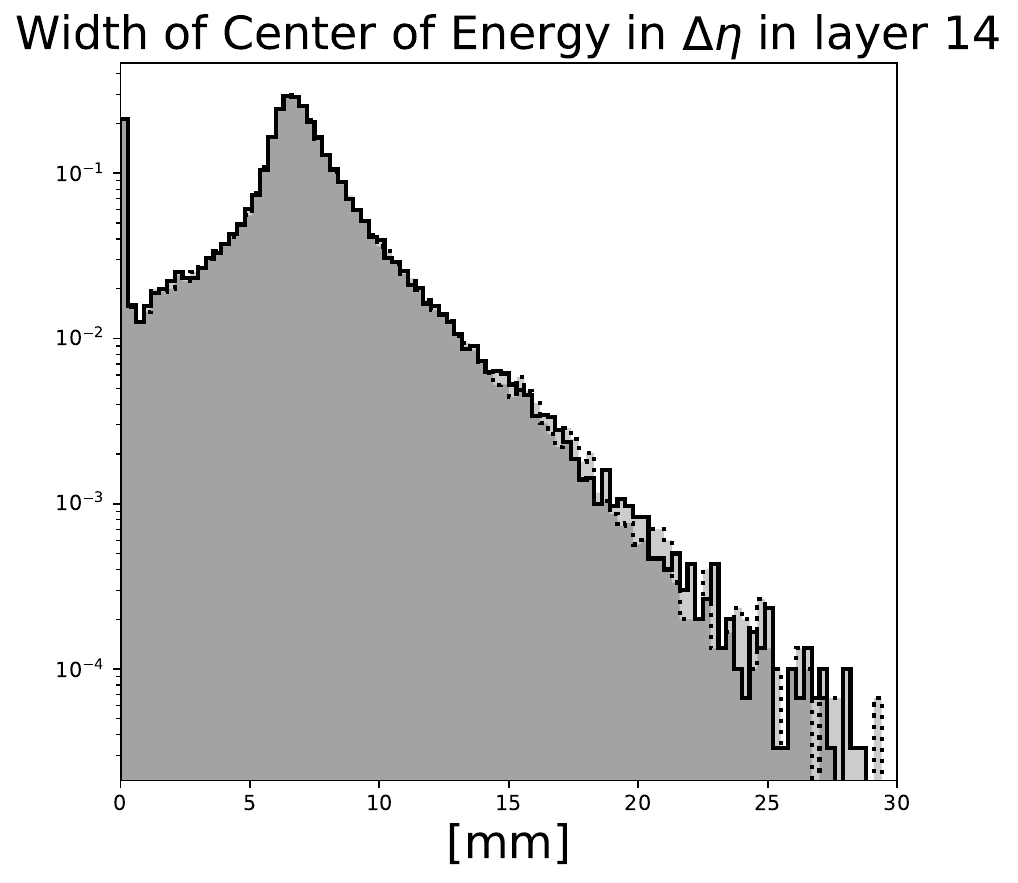}\\
    \includegraphics[height=0.1\textheight]{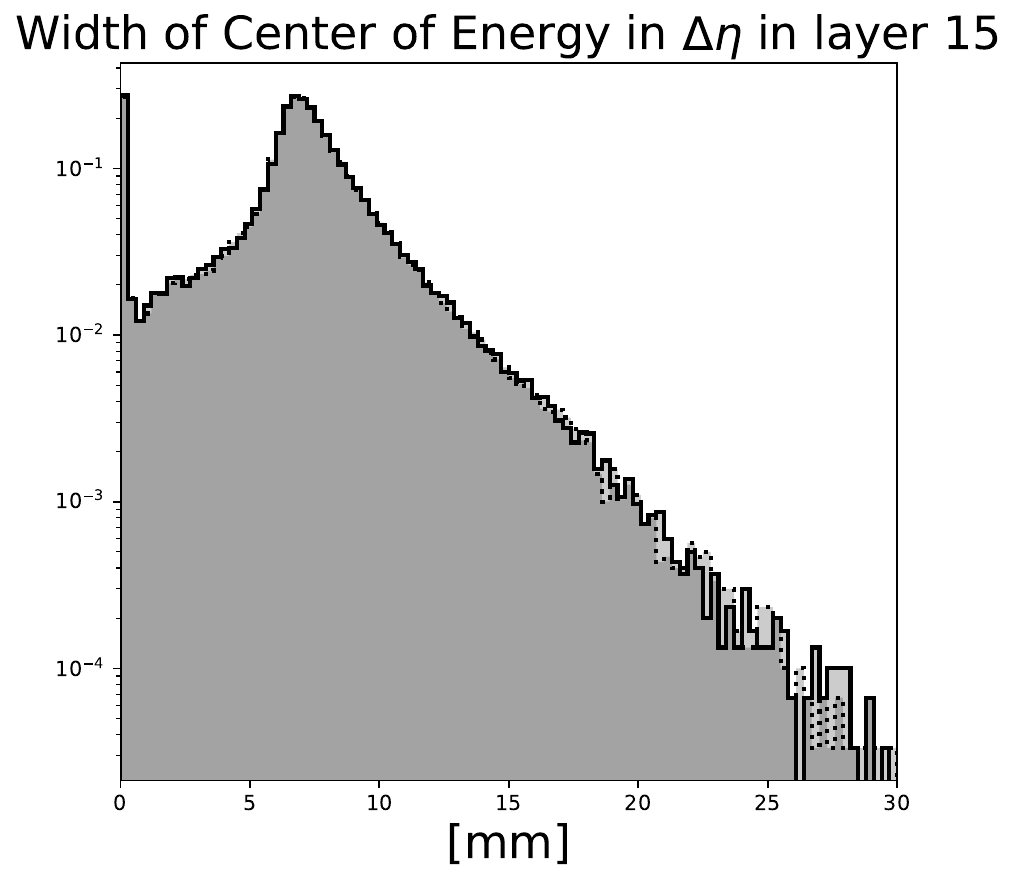} \hfill \includegraphics[height=0.1\textheight]{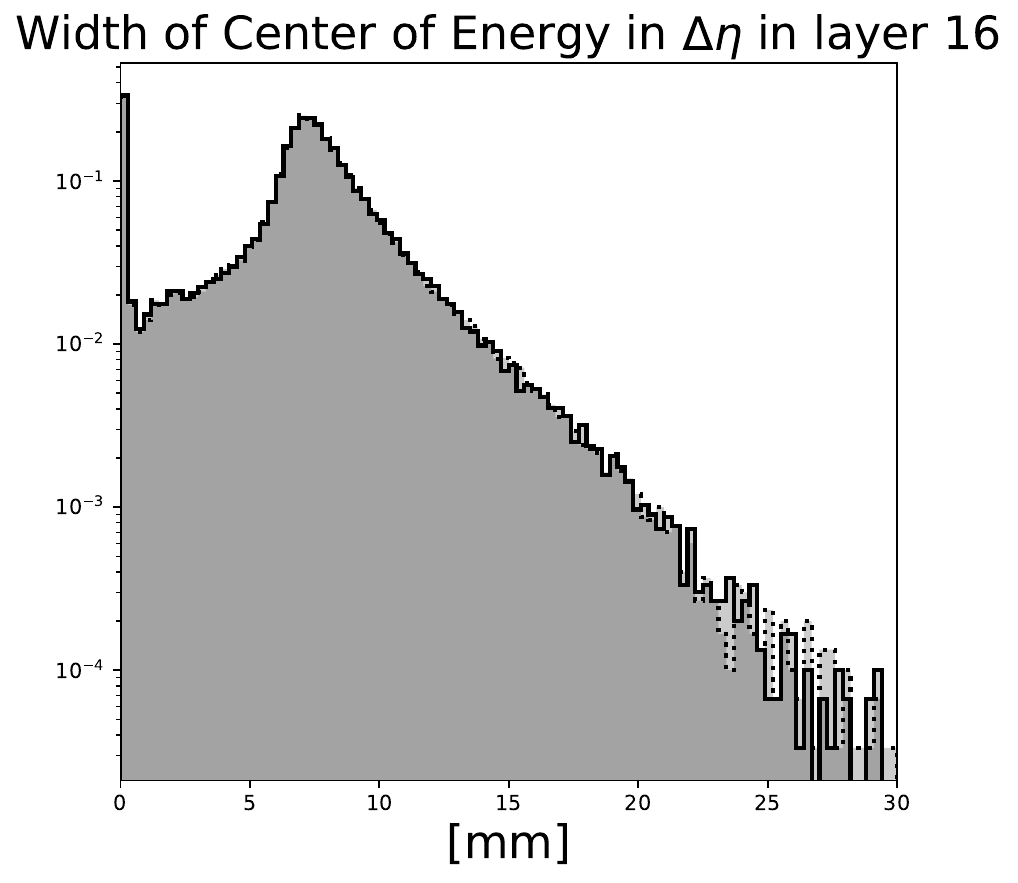} \hfill \includegraphics[height=0.1\textheight]{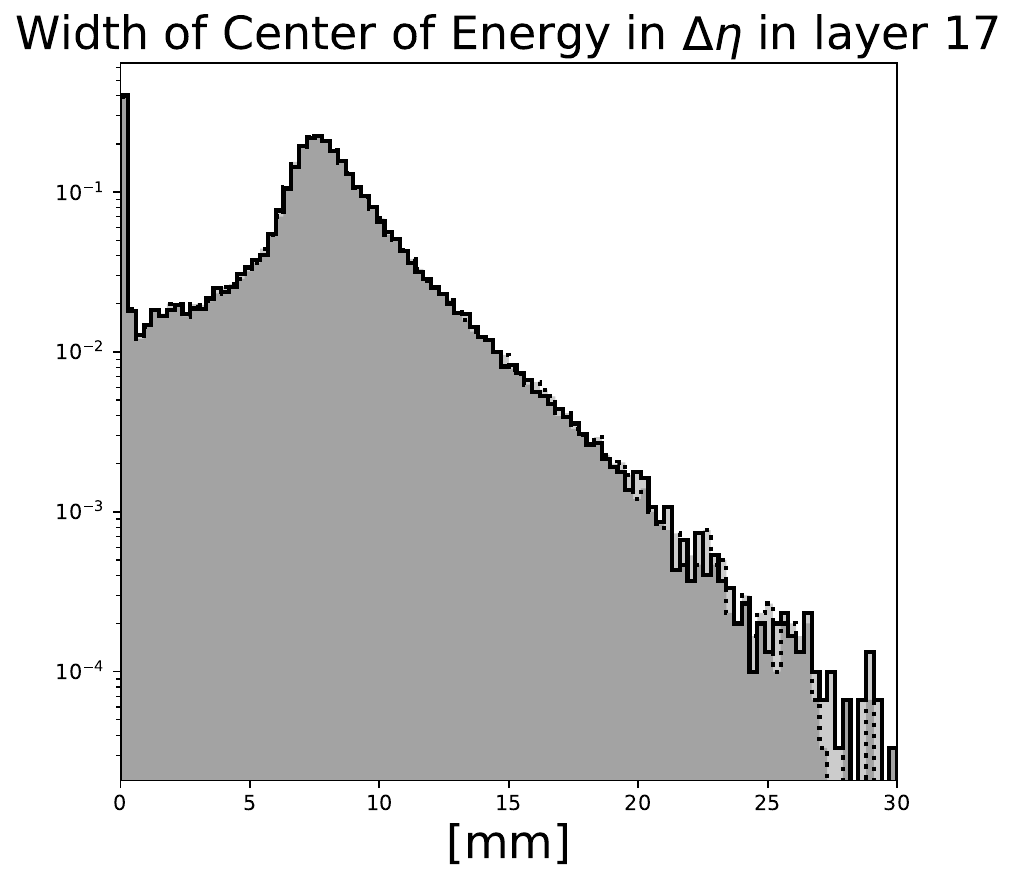} \hfill \includegraphics[height=0.1\textheight]{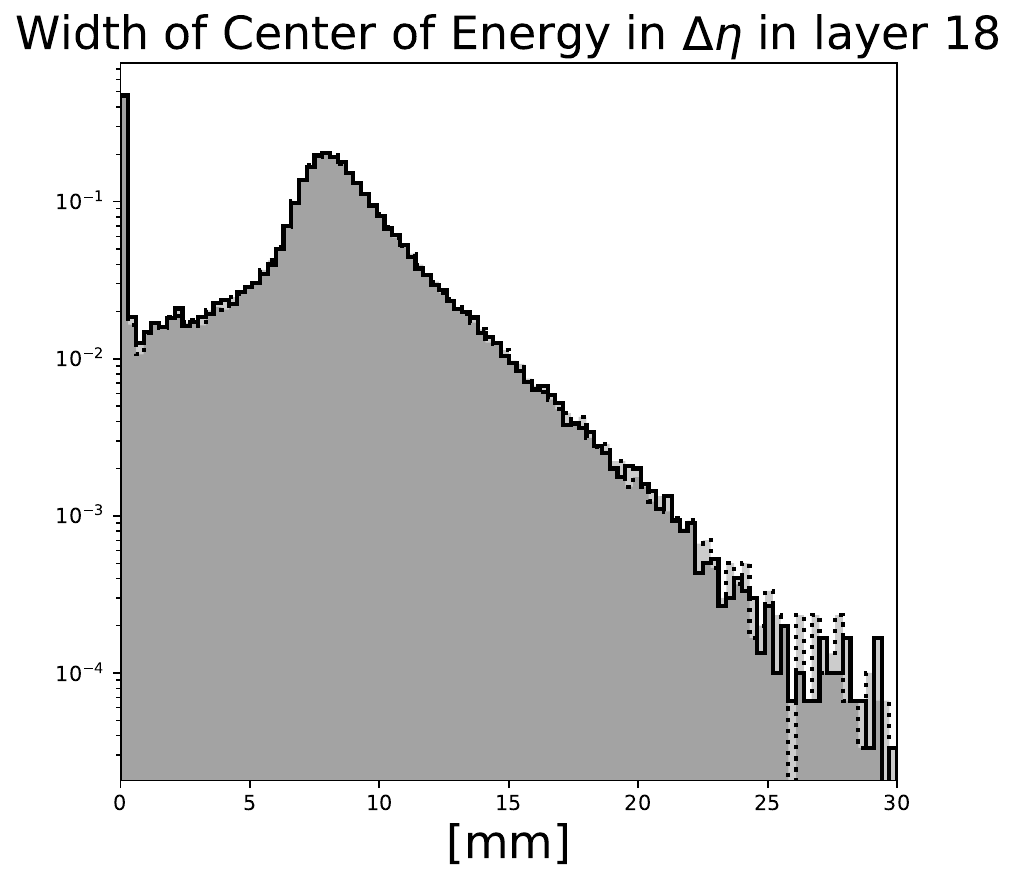} \hfill \includegraphics[height=0.1\textheight]{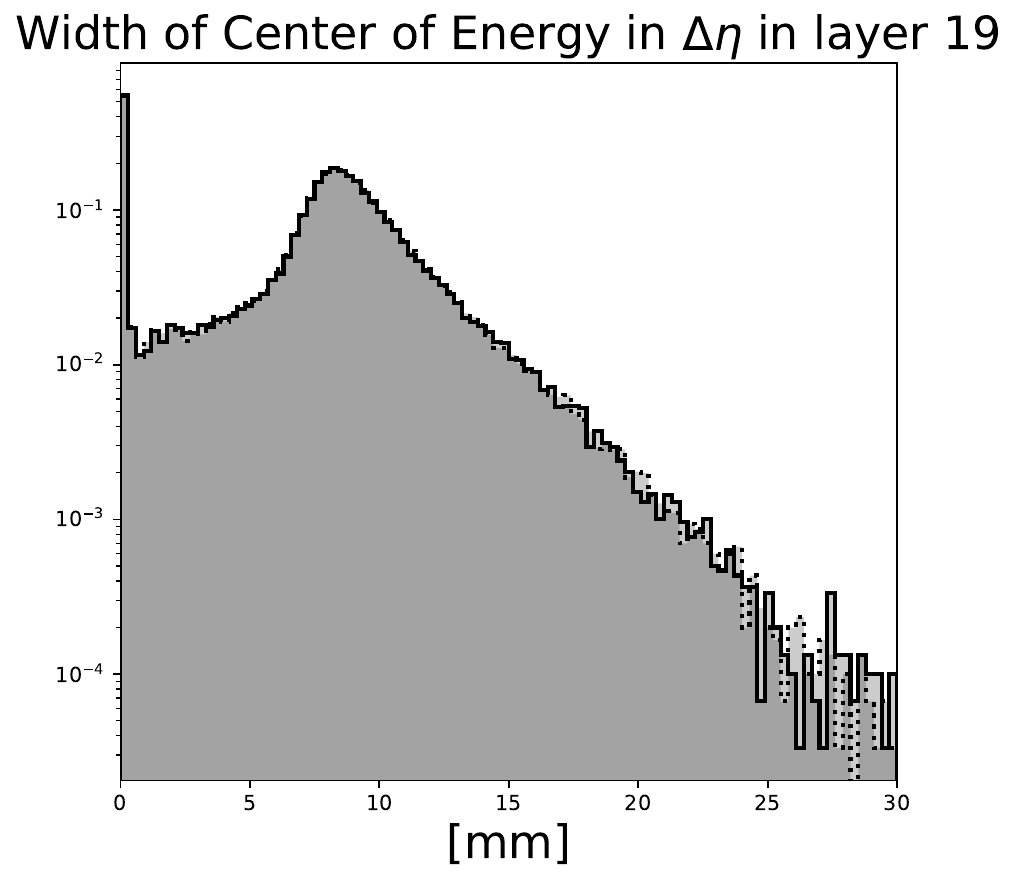}\\
    \includegraphics[height=0.1\textheight]{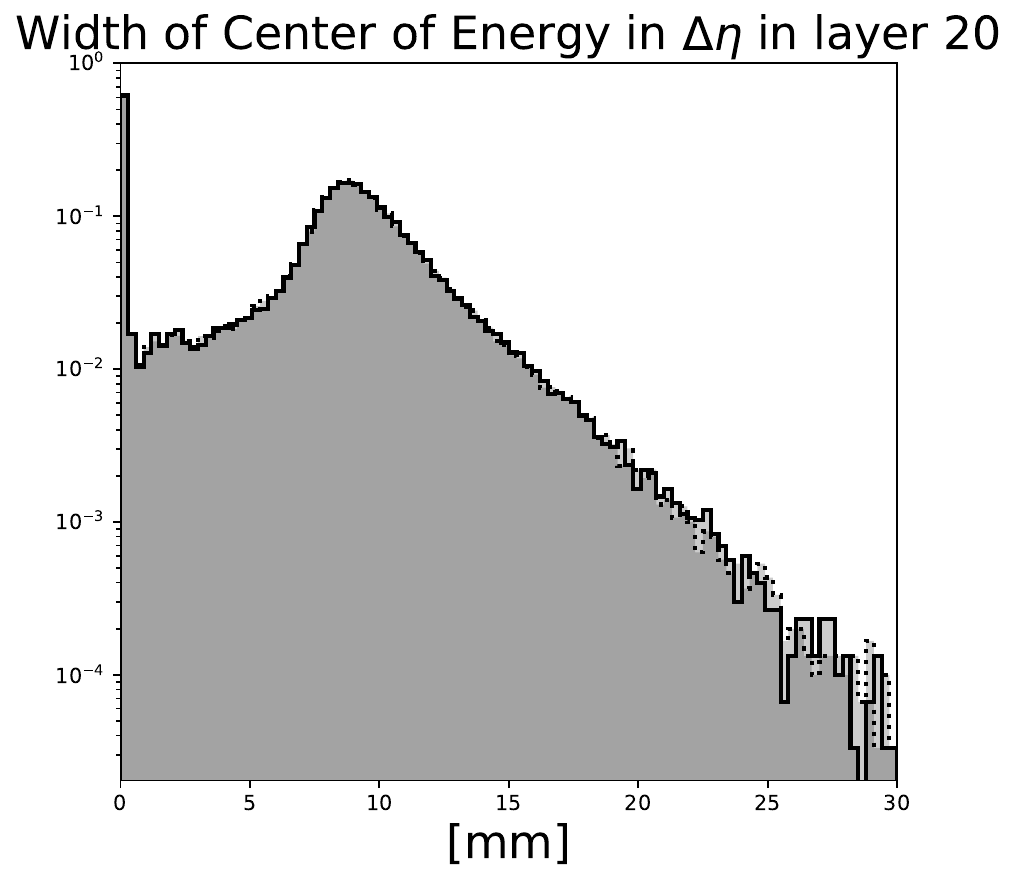} \hfill \includegraphics[height=0.1\textheight]{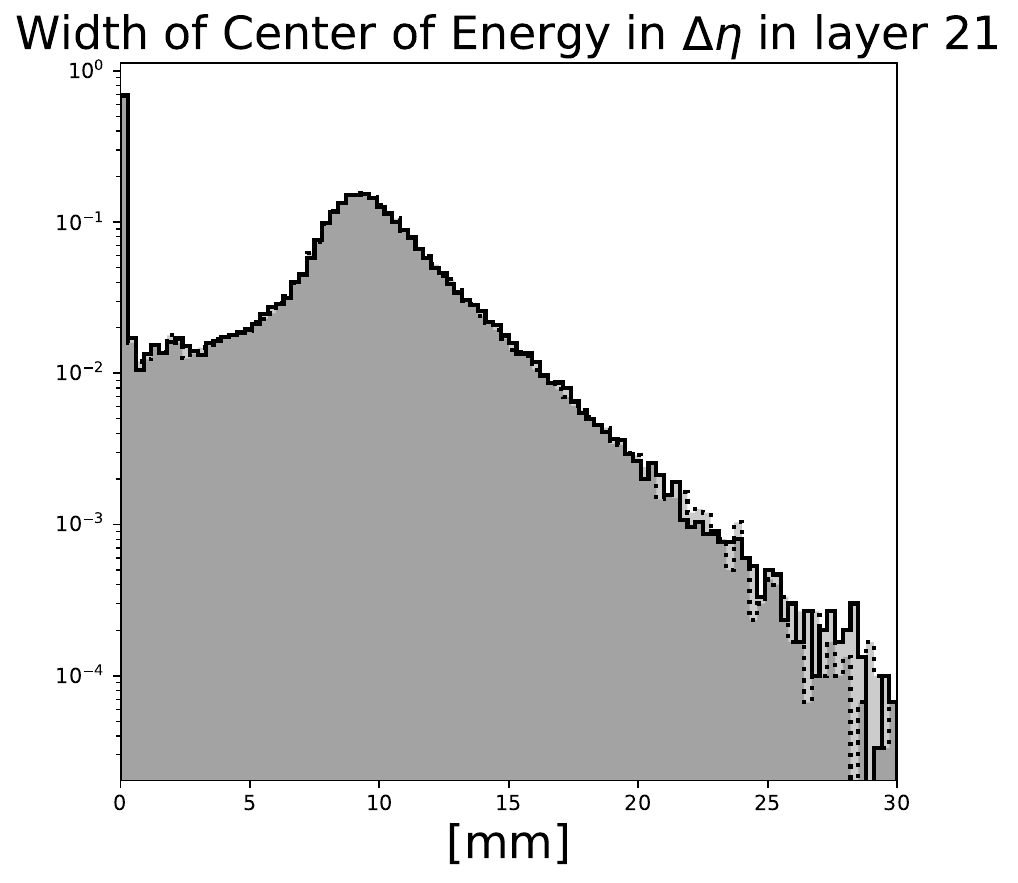} \hfill \includegraphics[height=0.1\textheight]{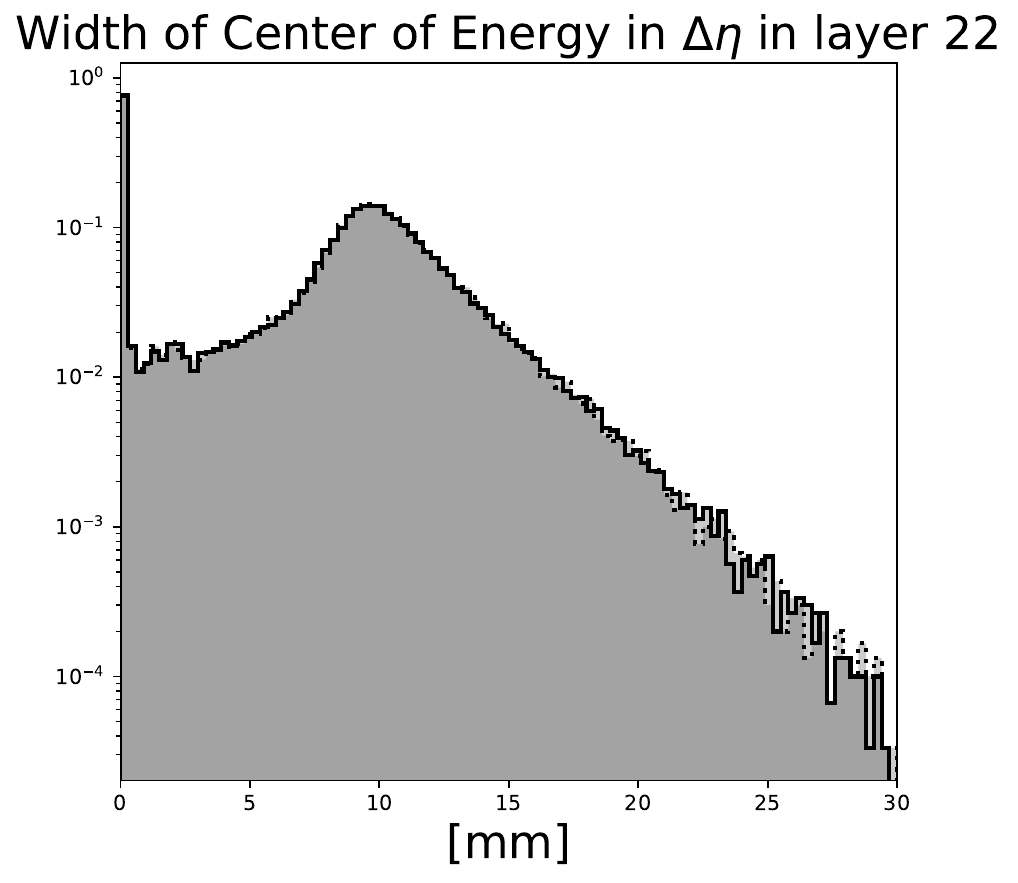} \hfill \includegraphics[height=0.1\textheight]{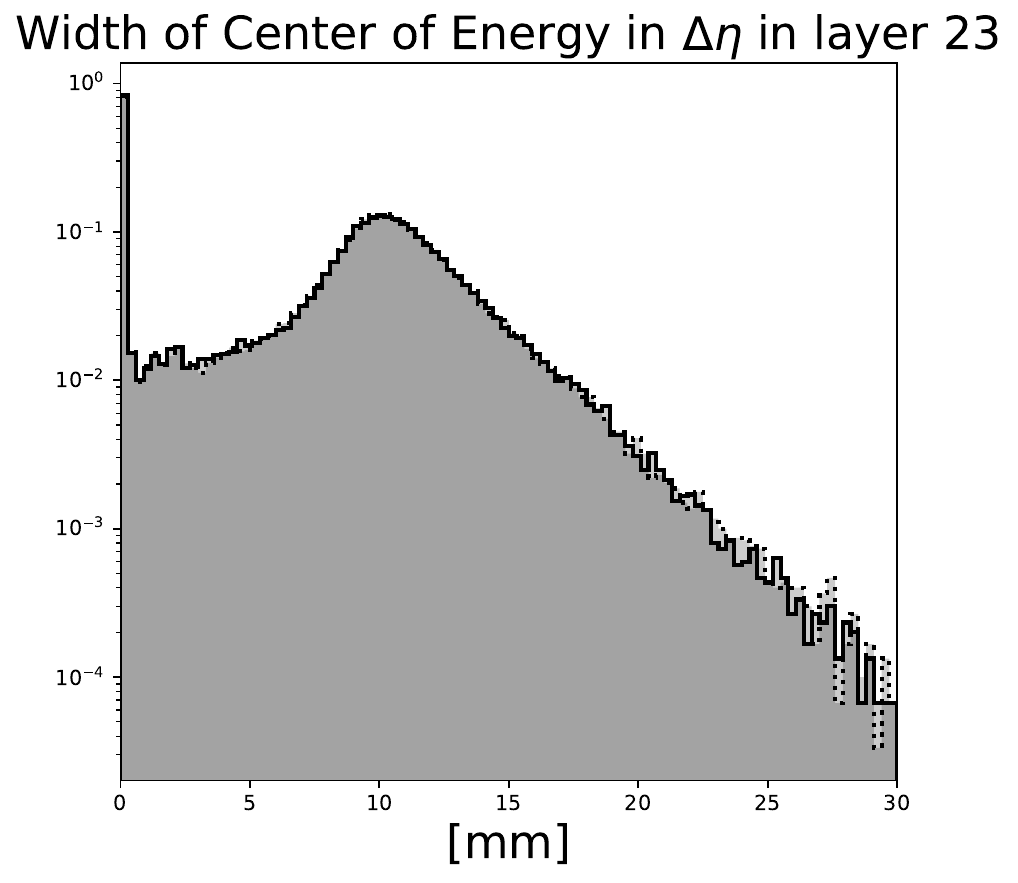} \hfill \includegraphics[height=0.1\textheight]{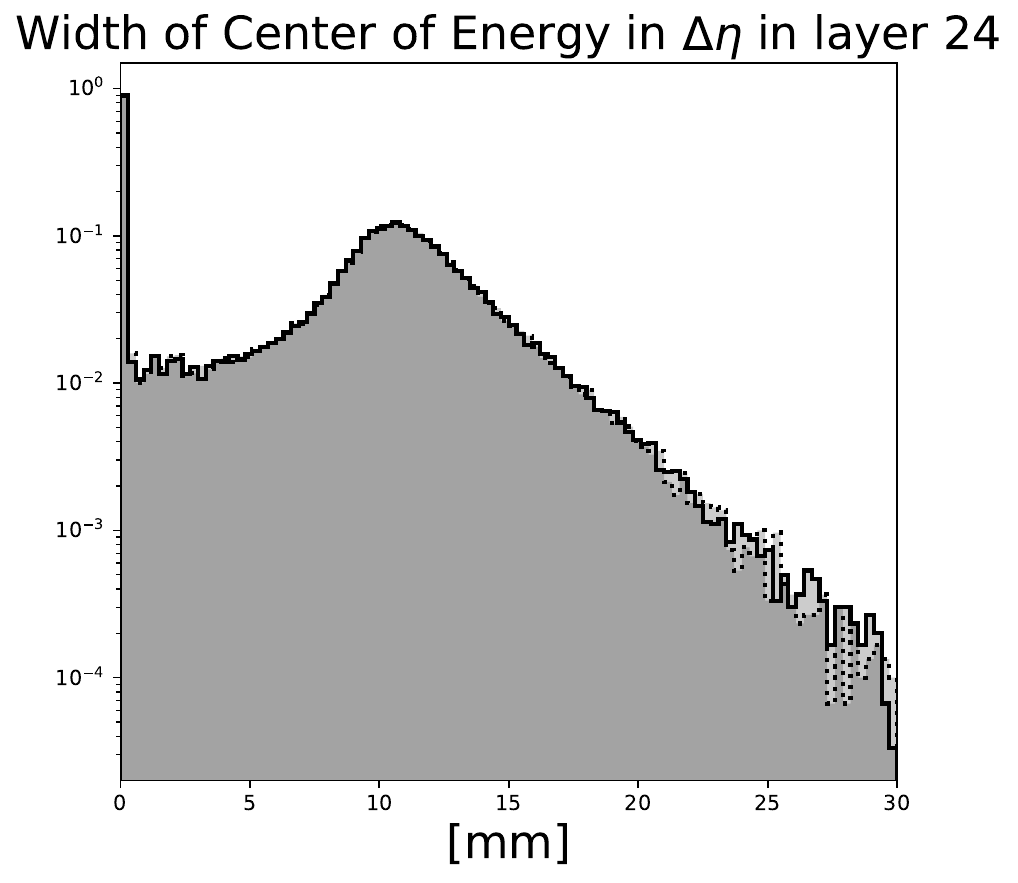}\\
    \includegraphics[height=0.1\textheight]{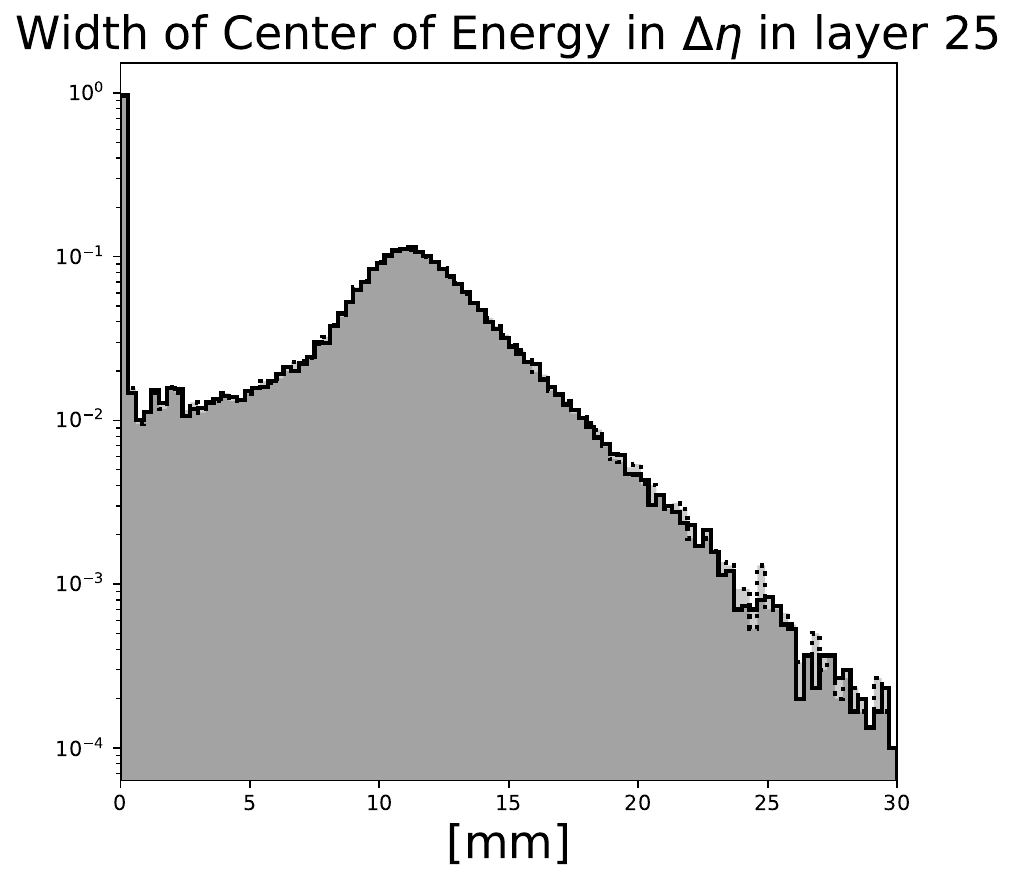} \hfill \includegraphics[height=0.1\textheight]{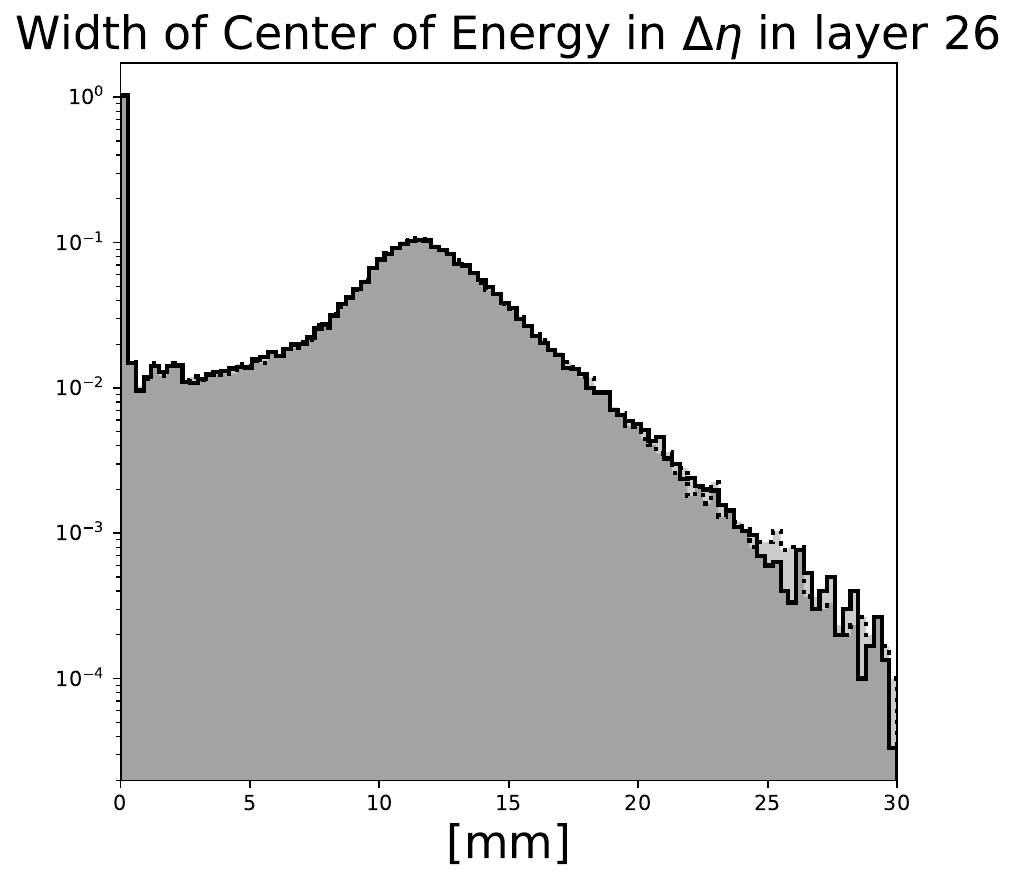} \hfill \includegraphics[height=0.1\textheight]{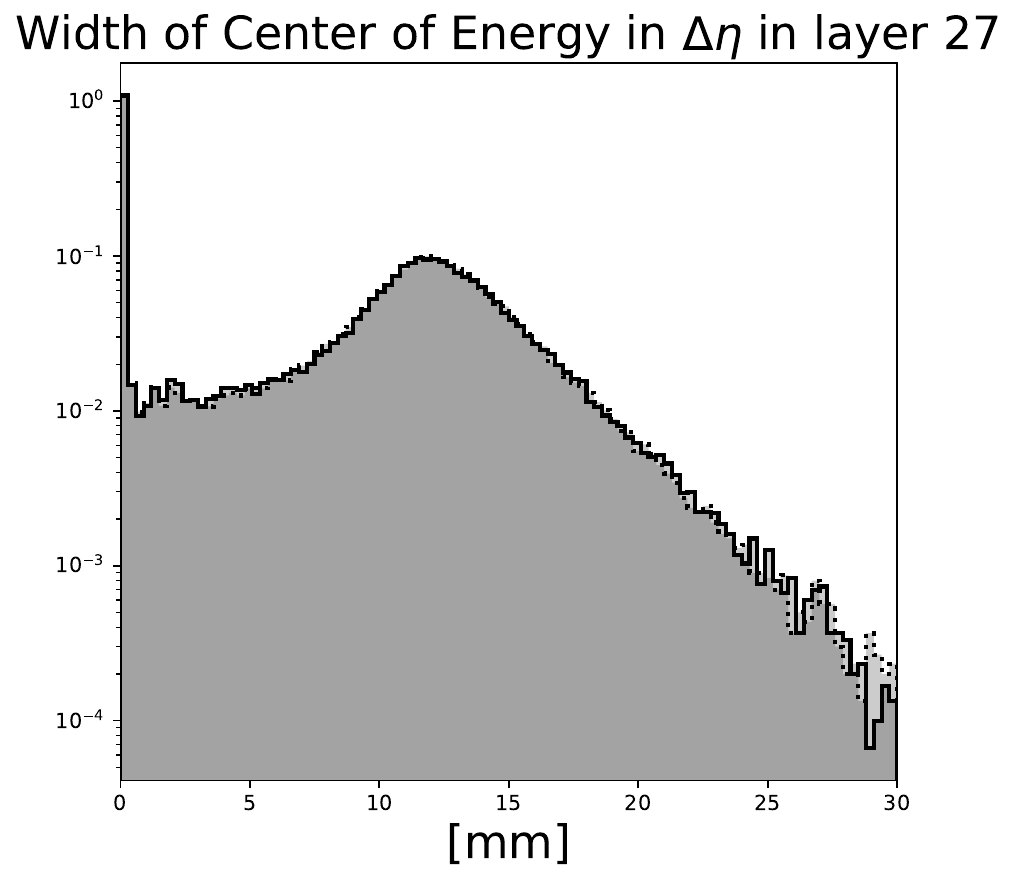} \hfill \includegraphics[height=0.1\textheight]{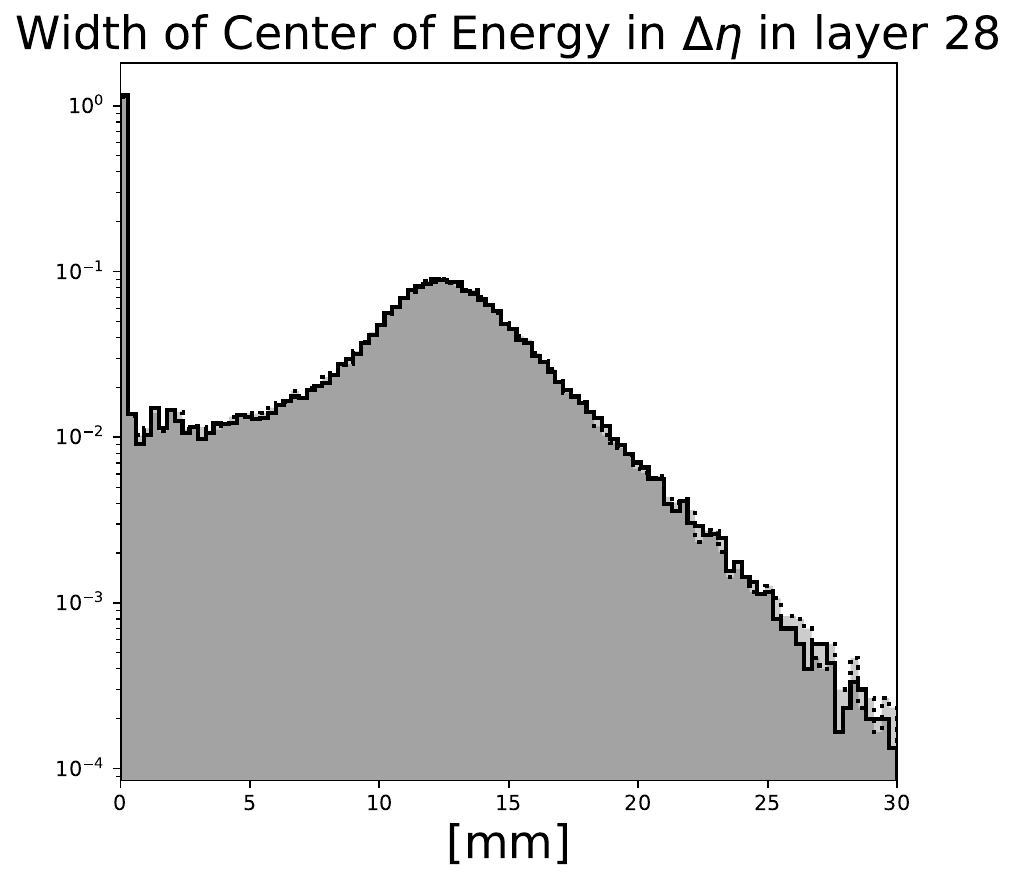} \hfill \includegraphics[height=0.1\textheight]{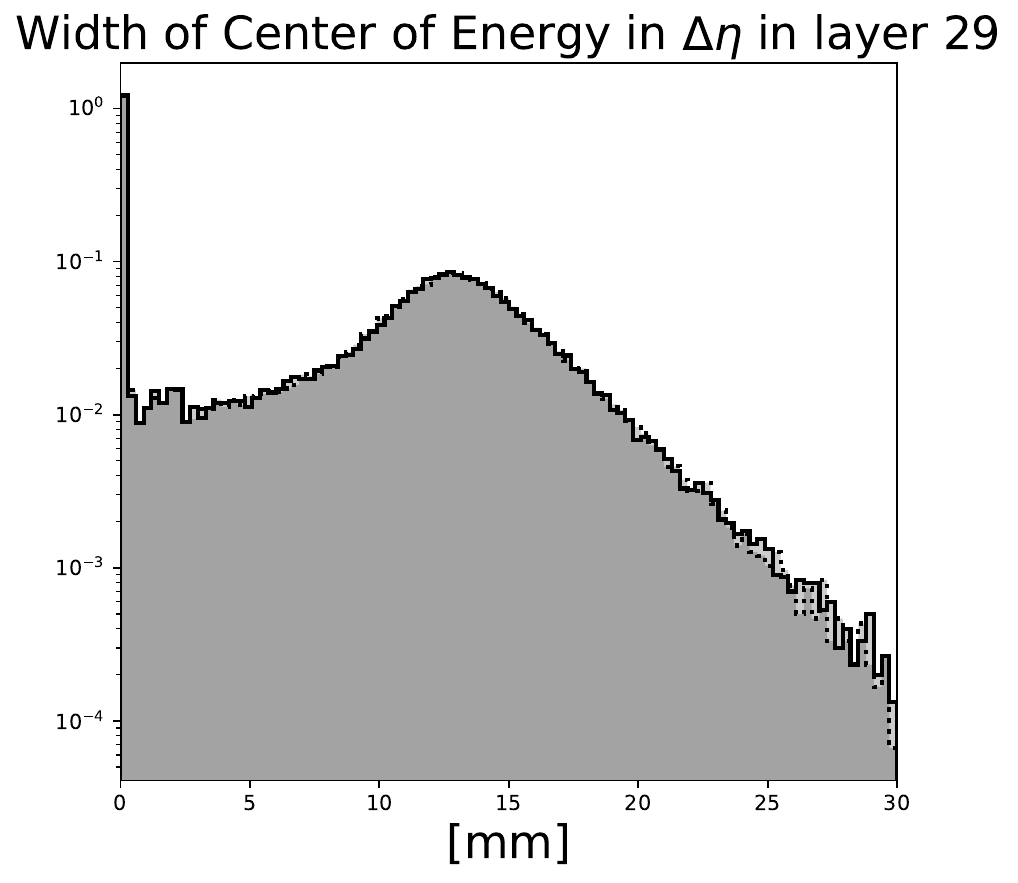}\\
    \includegraphics[height=0.1\textheight]{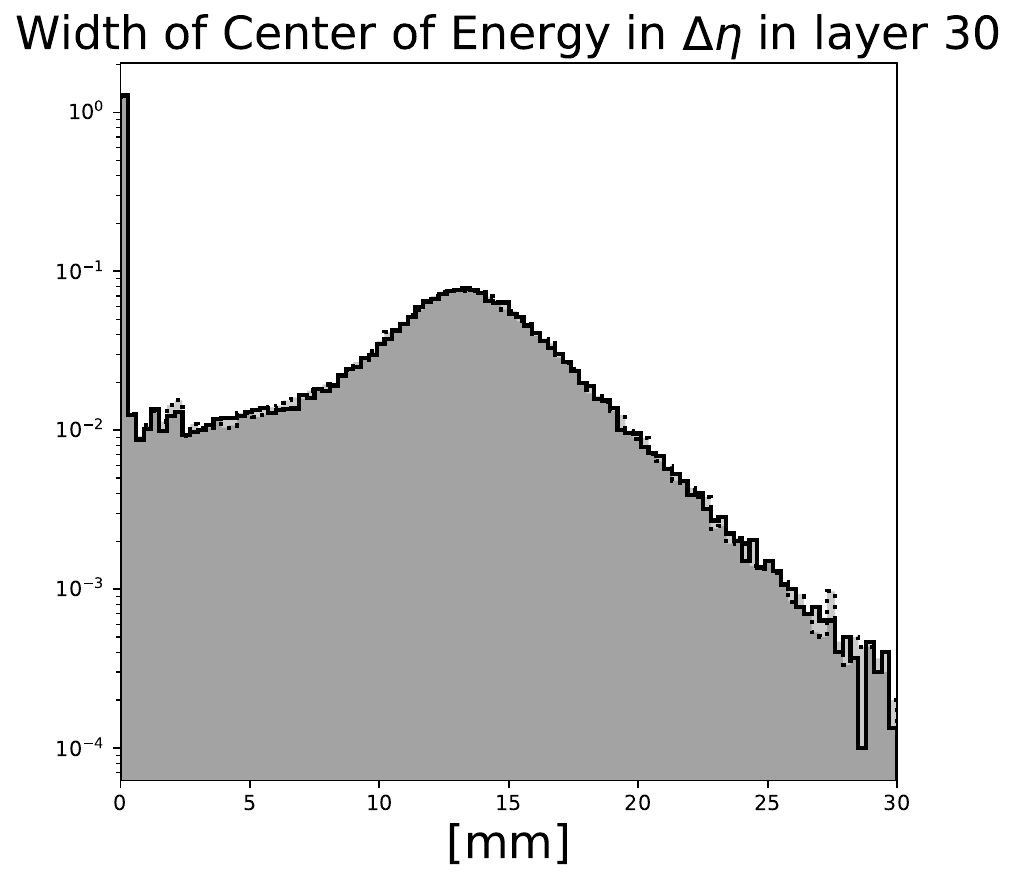} \hfill \includegraphics[height=0.1\textheight]{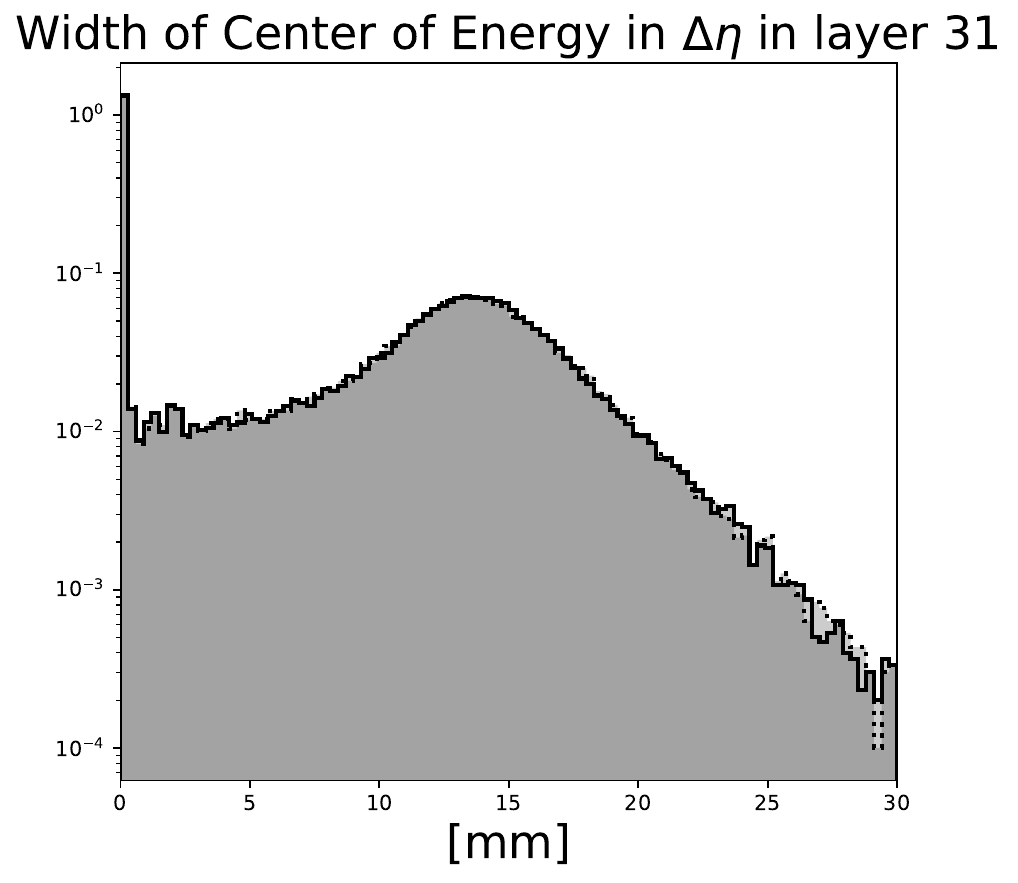} \hfill \includegraphics[height=0.1\textheight]{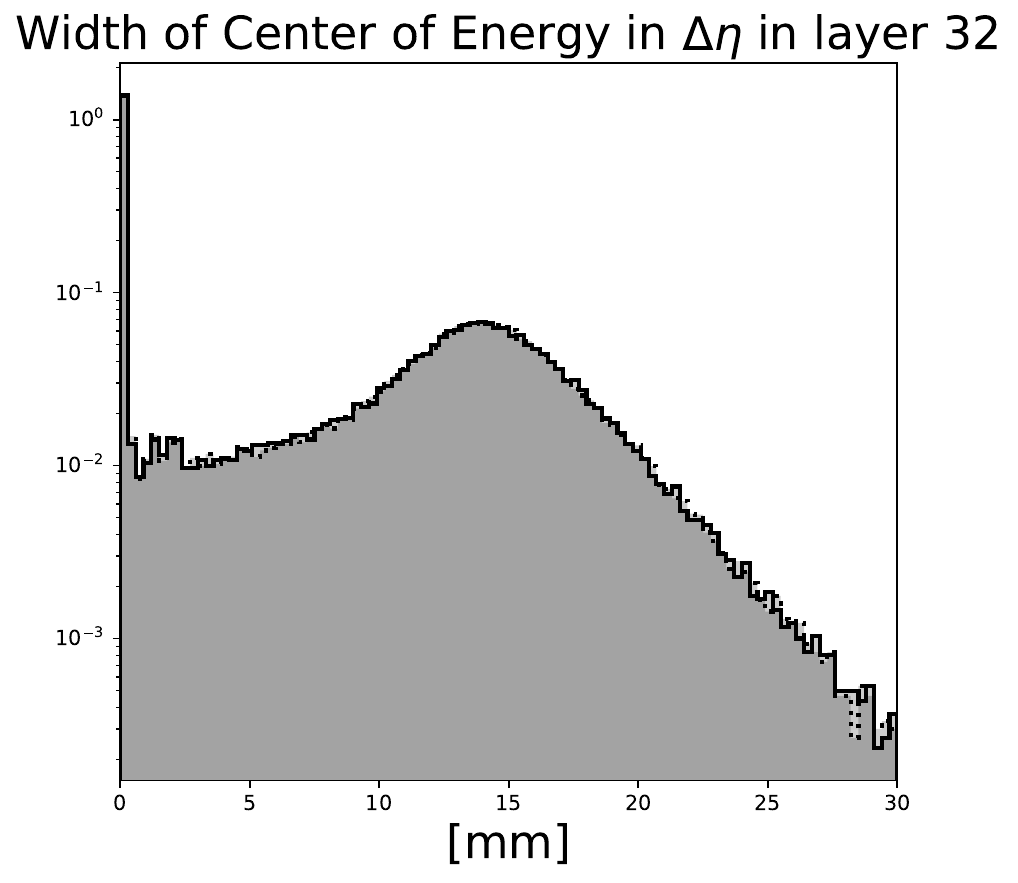} \hfill \includegraphics[height=0.1\textheight]{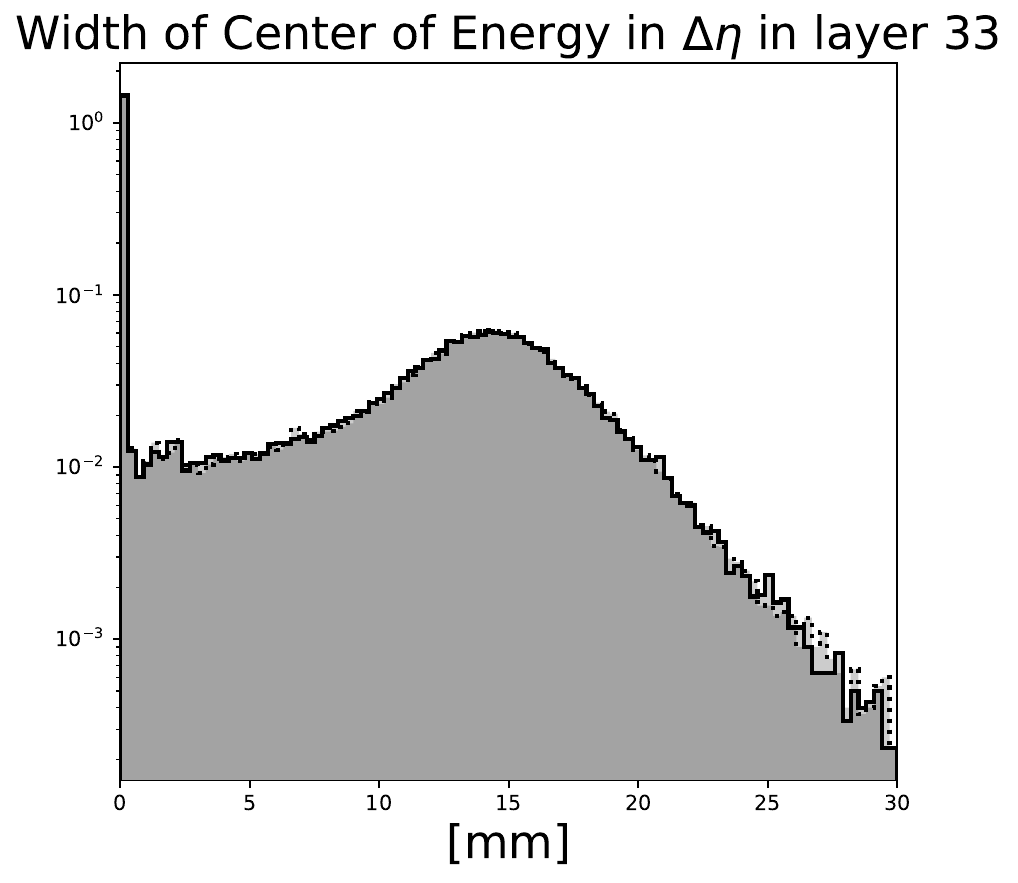} \hfill \includegraphics[height=0.1\textheight]{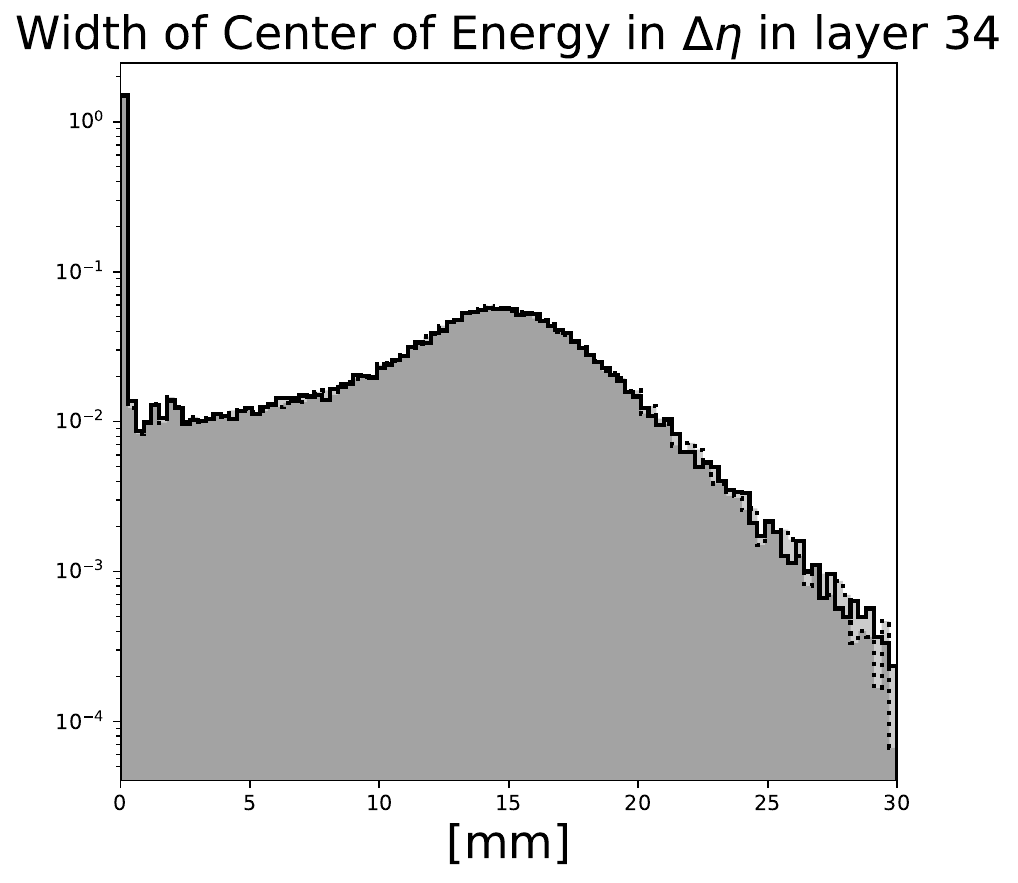}\\
    \includegraphics[height=0.1\textheight]{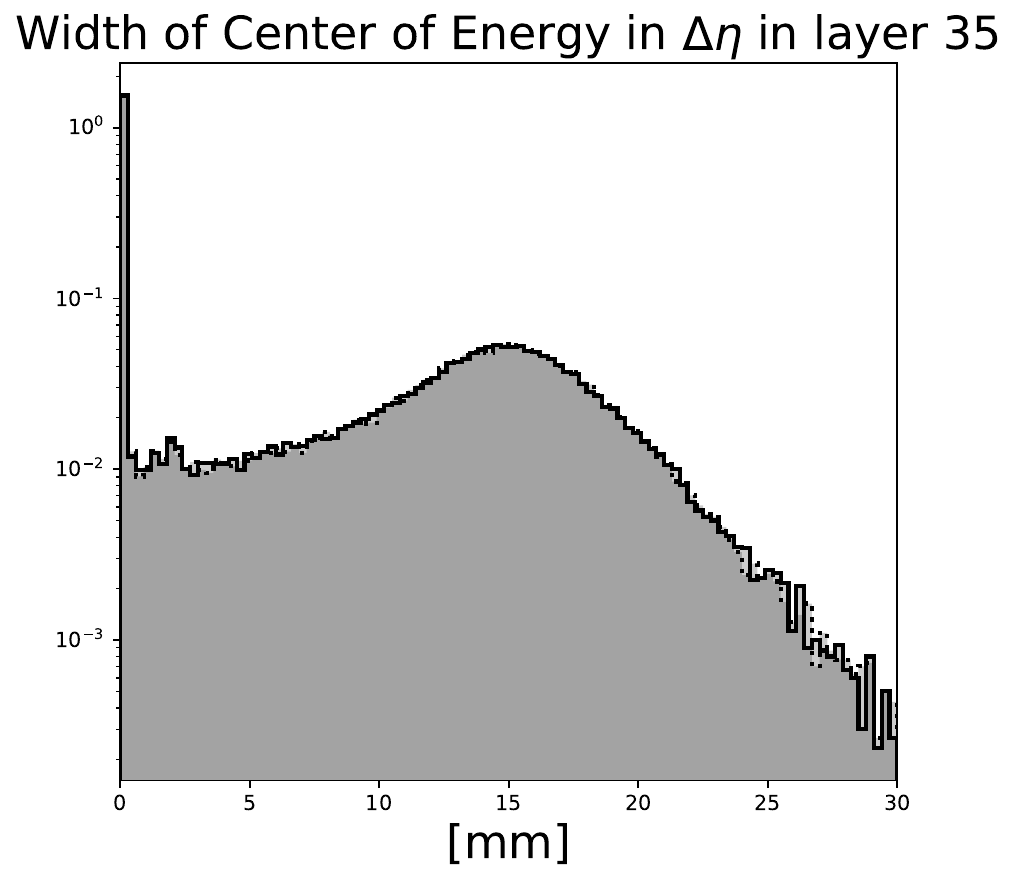} \hfill \includegraphics[height=0.1\textheight]{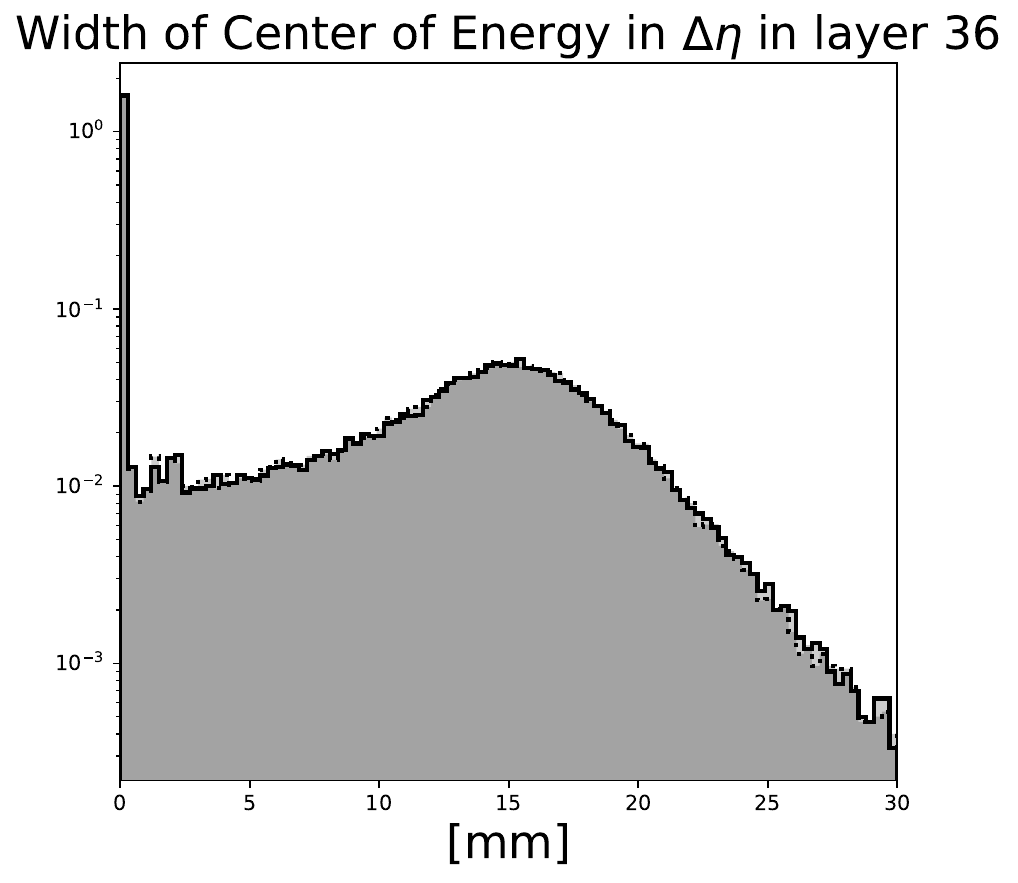} \hfill \includegraphics[height=0.1\textheight]{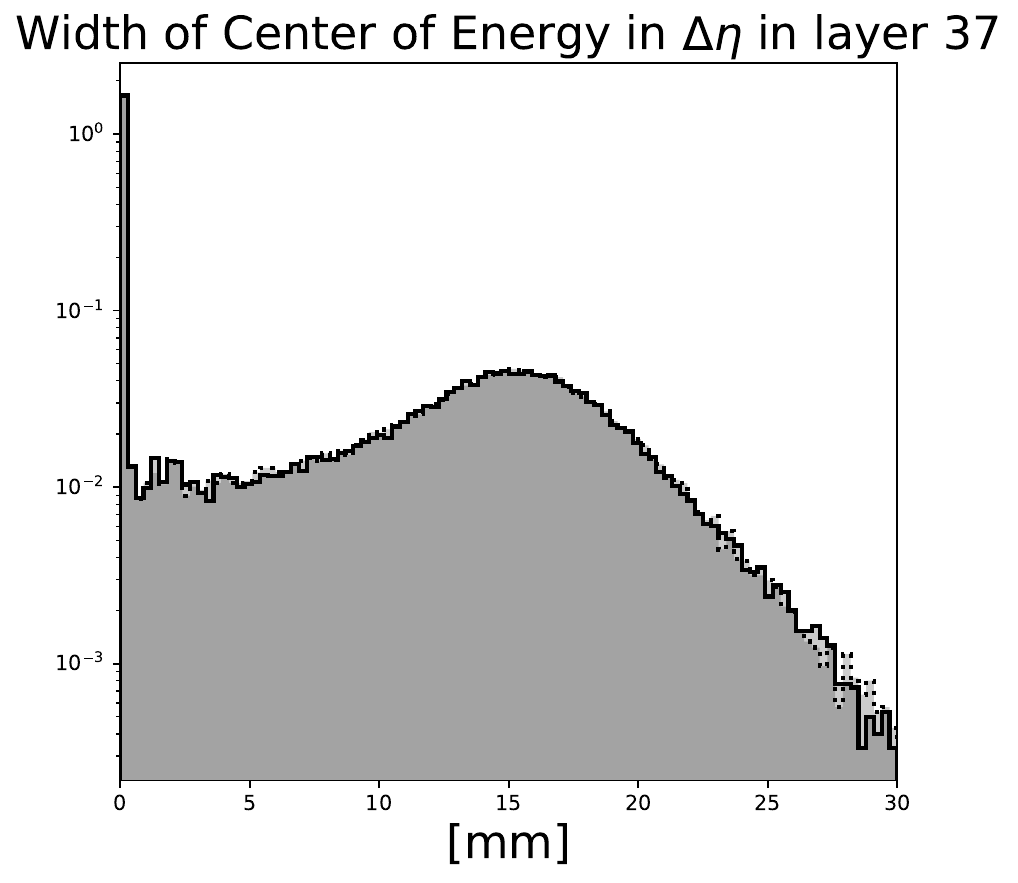} \hfill \includegraphics[height=0.1\textheight]{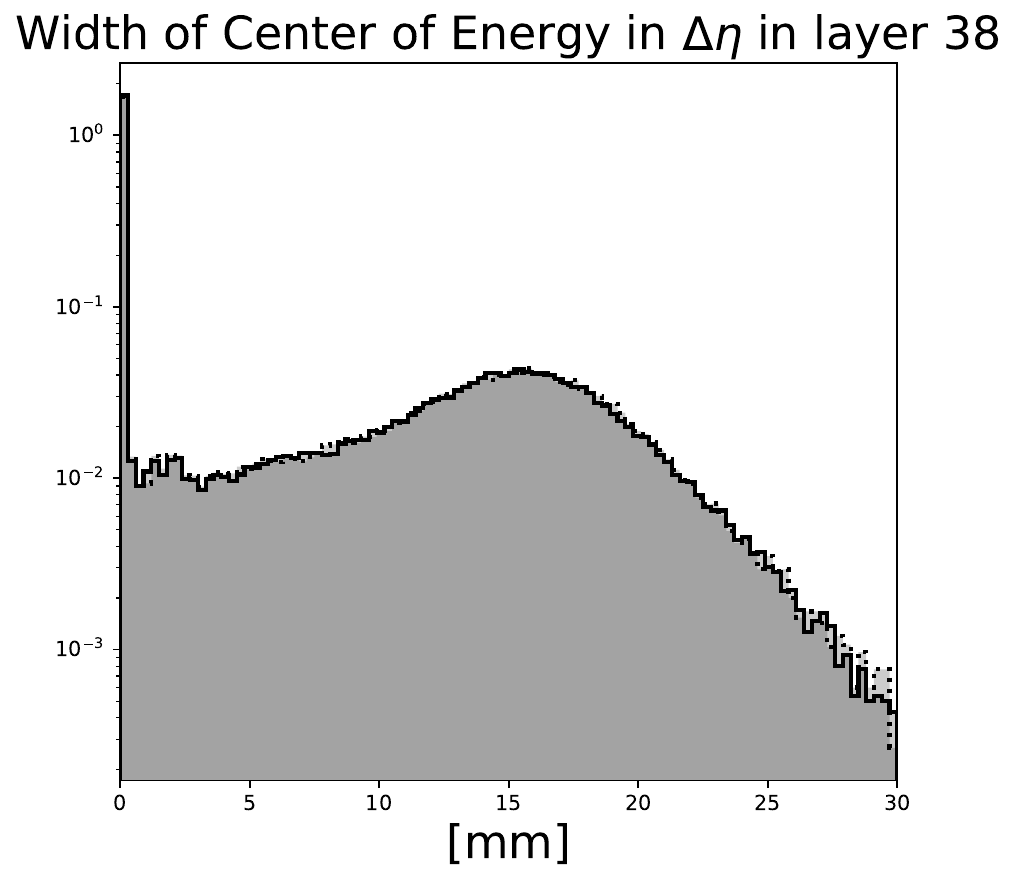} \hfill \includegraphics[height=0.1\textheight]{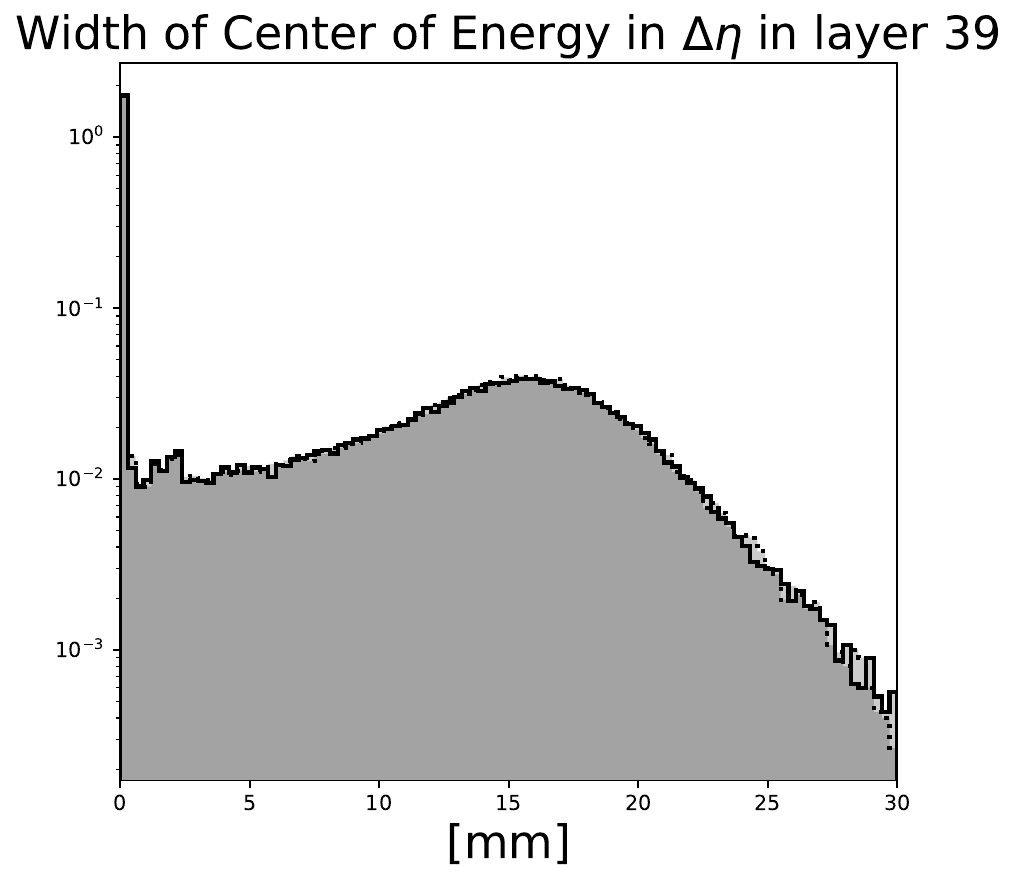}\\
    \includegraphics[height=0.1\textheight]{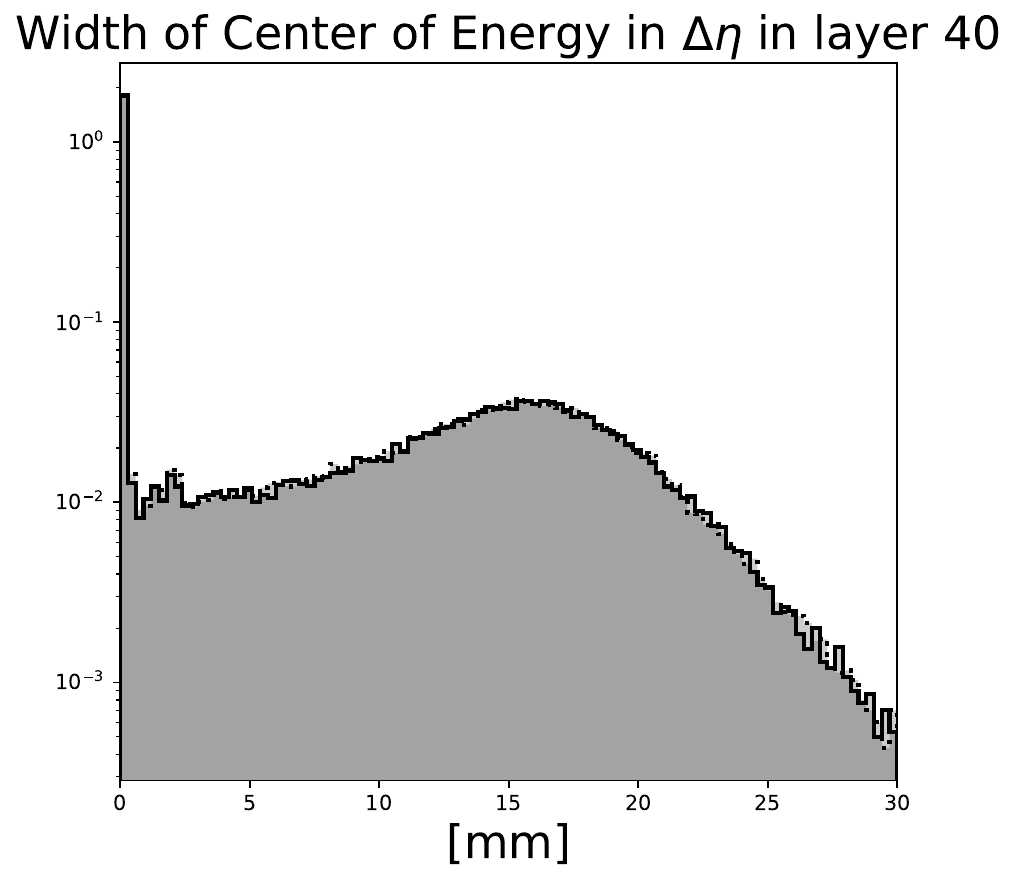} \hfill \includegraphics[height=0.1\textheight]{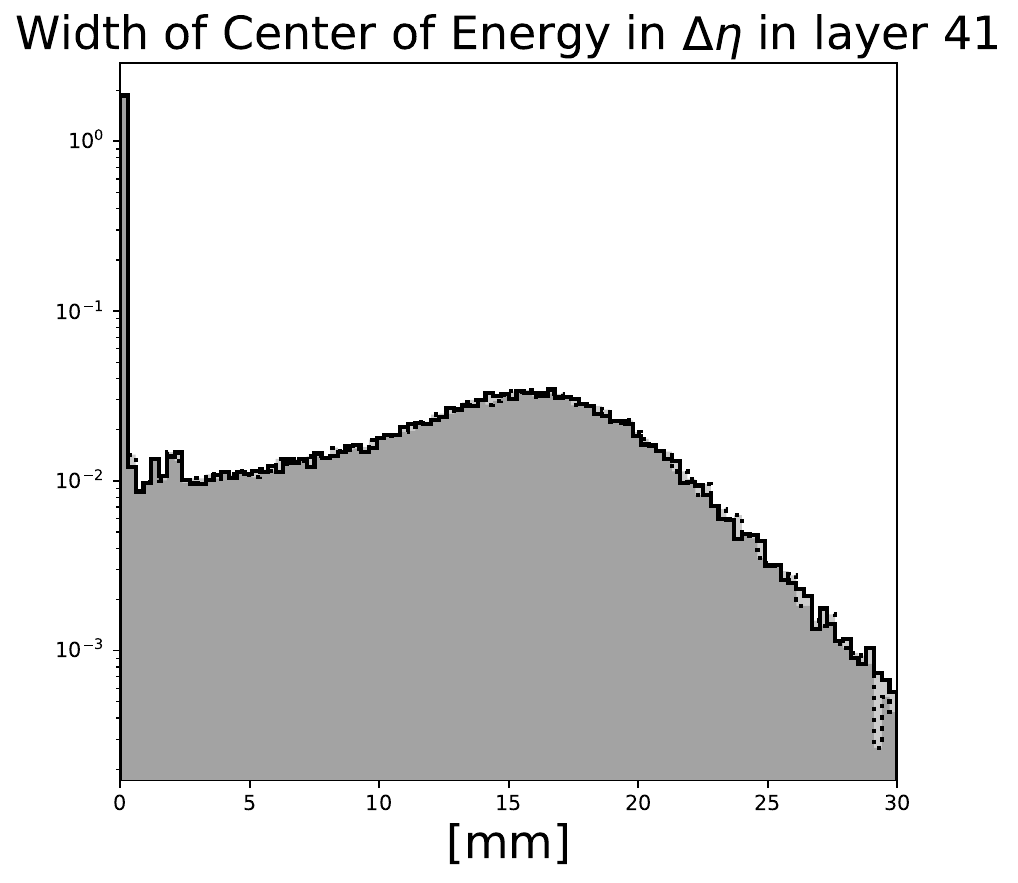} \hfill \includegraphics[height=0.1\textheight]{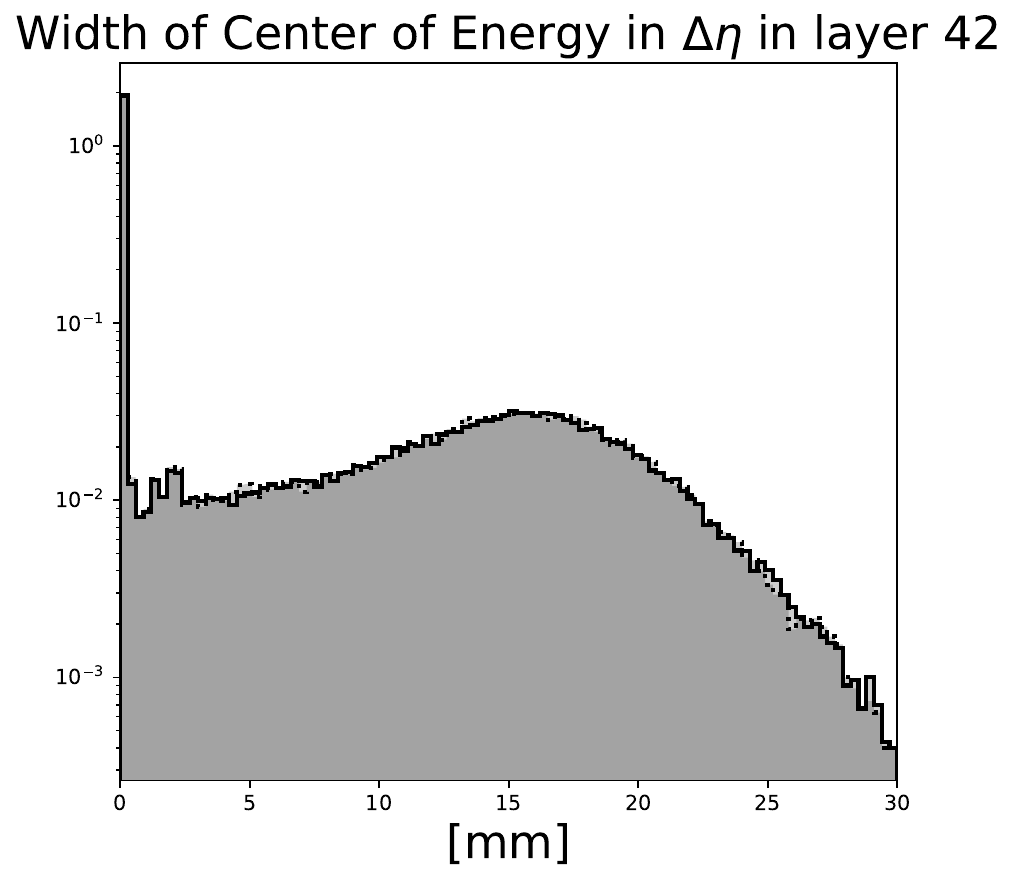} \hfill \includegraphics[height=0.1\textheight]{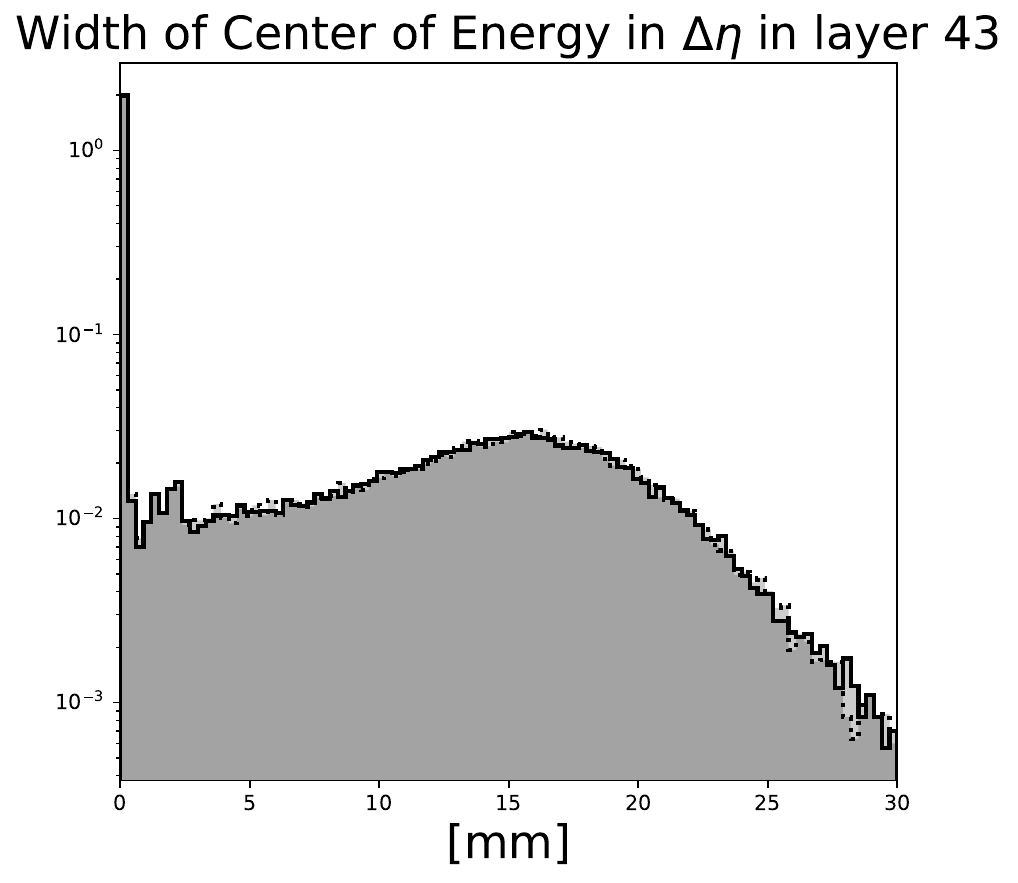} \hfill \includegraphics[height=0.1\textheight]{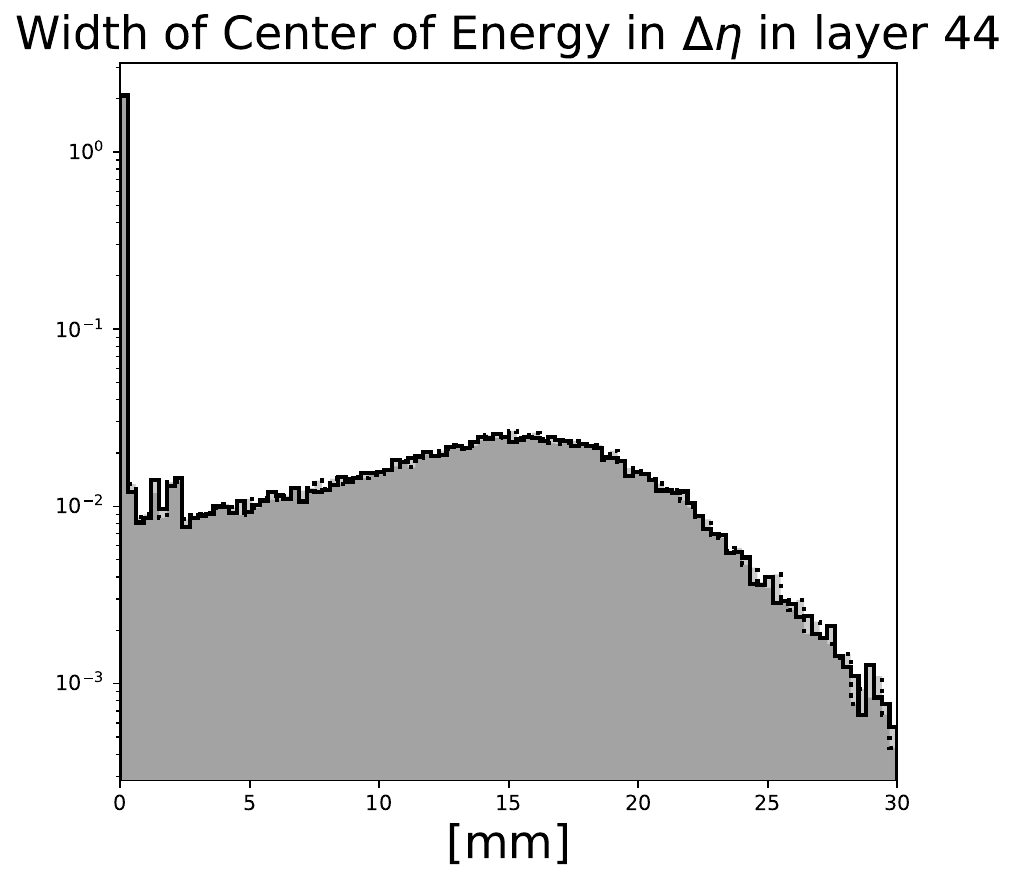}\\
    \includegraphics[width=0.5\textwidth]{figures/Appendix_reference/legend.pdf}
    \caption{Distribution of \geant training and evaluation data in width of the centers of energy in $\eta$ direction for ds2. }
    \label{fig:app_ref.ds2.4}
\end{figure}

\begin{figure}[ht]
    \centering
    \includegraphics[height=0.1\textheight]{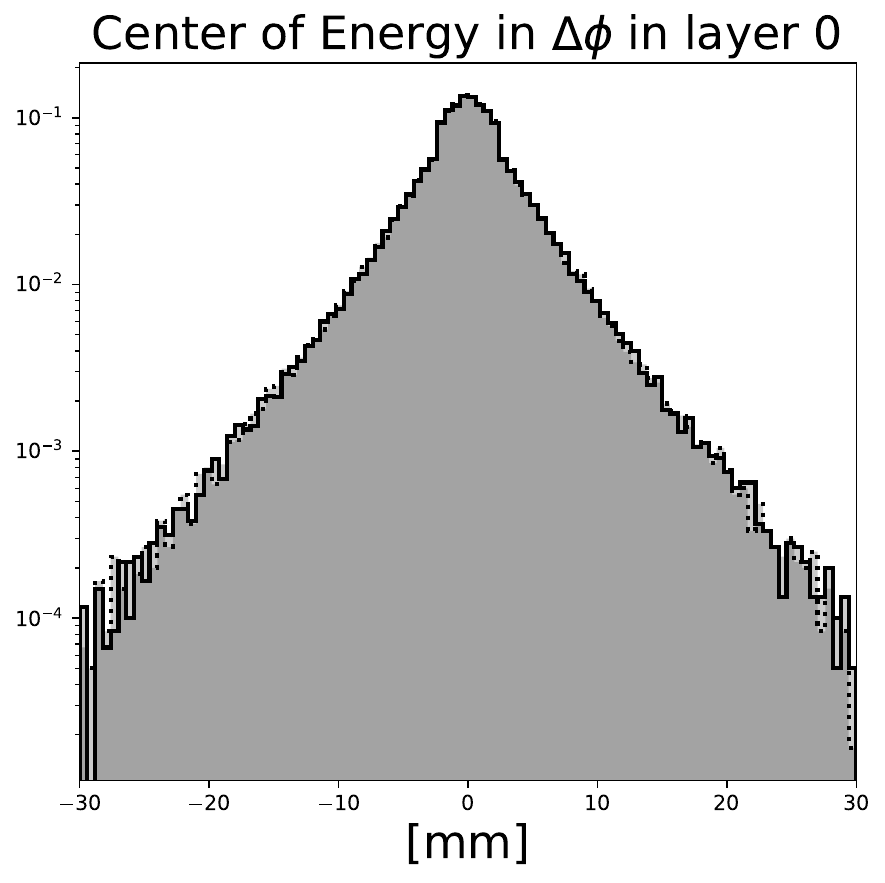} \hfill \includegraphics[height=0.1\textheight]{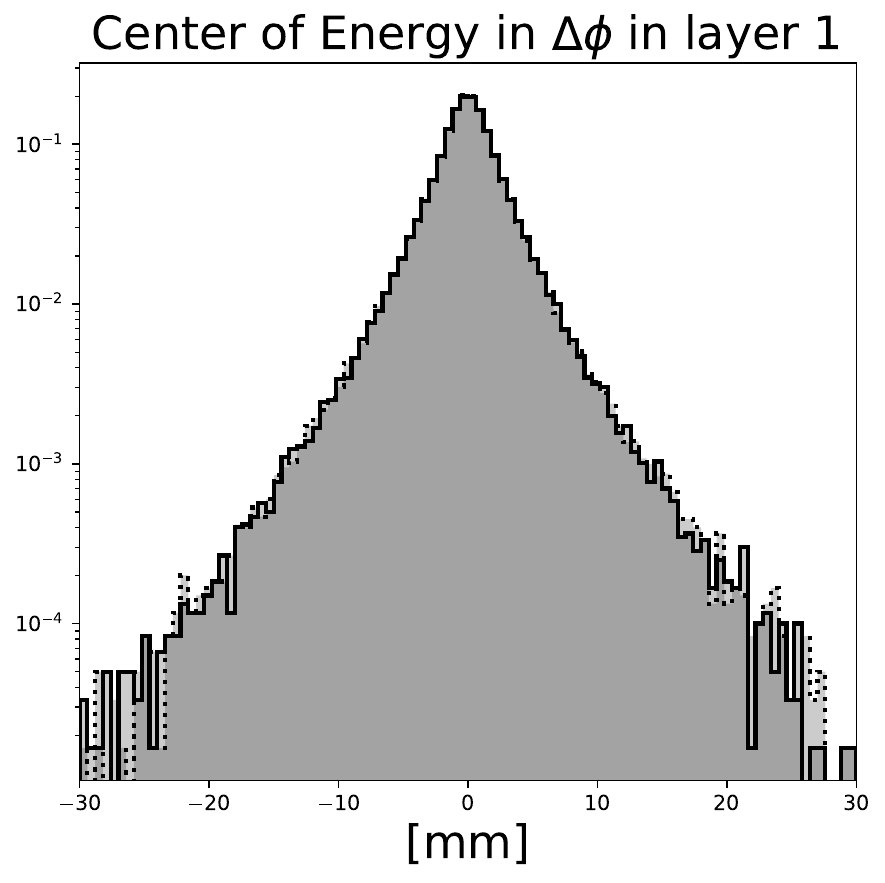} \hfill \includegraphics[height=0.1\textheight]{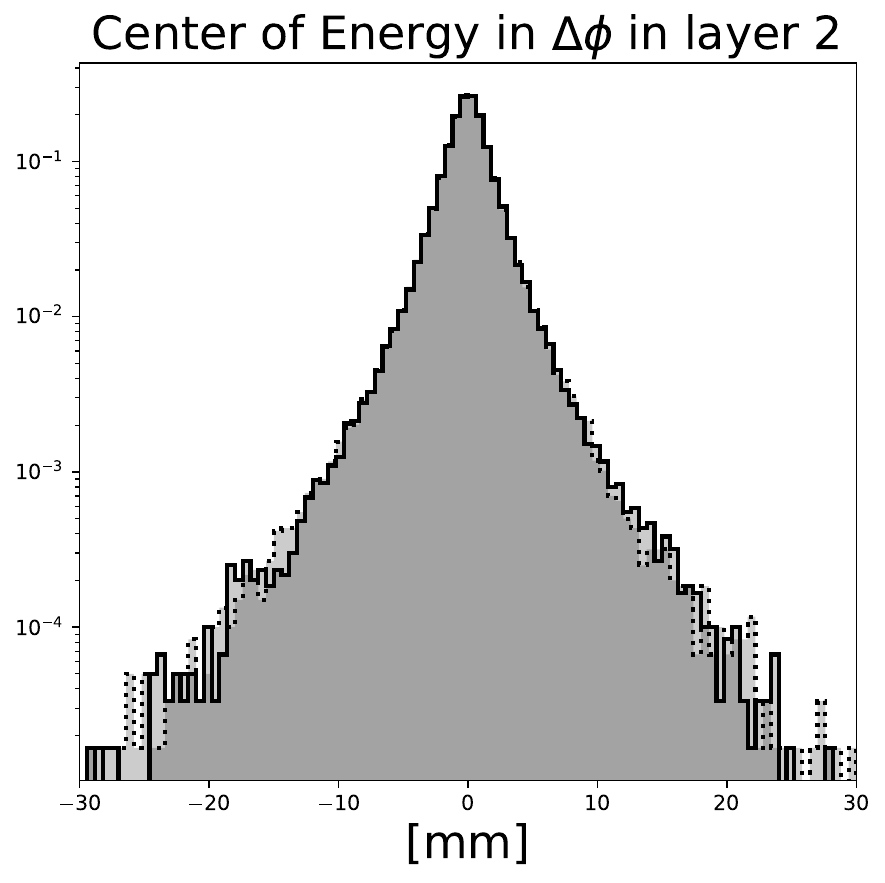} \hfill \includegraphics[height=0.1\textheight]{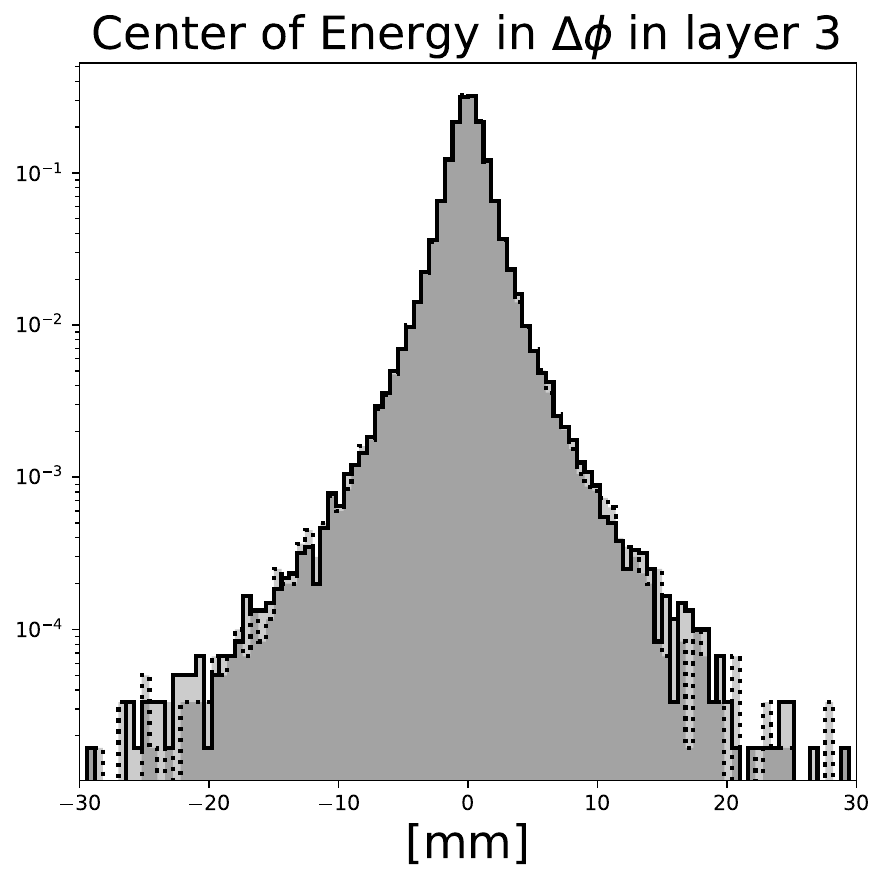} \hfill \includegraphics[height=0.1\textheight]{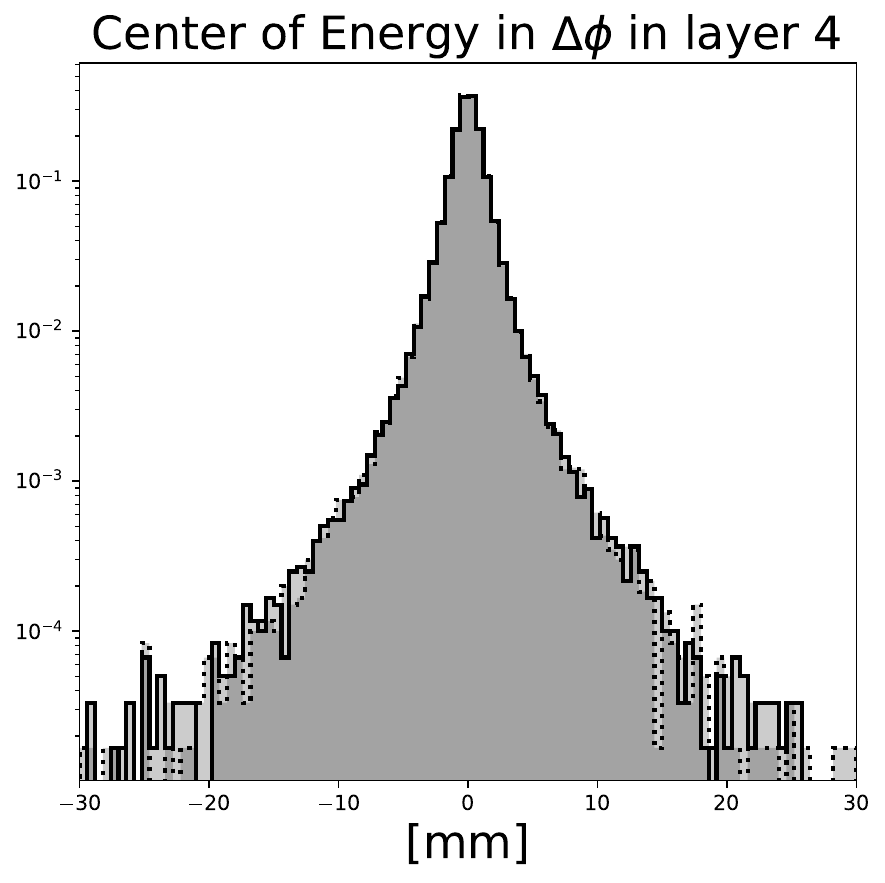}\\
    \includegraphics[height=0.1\textheight]{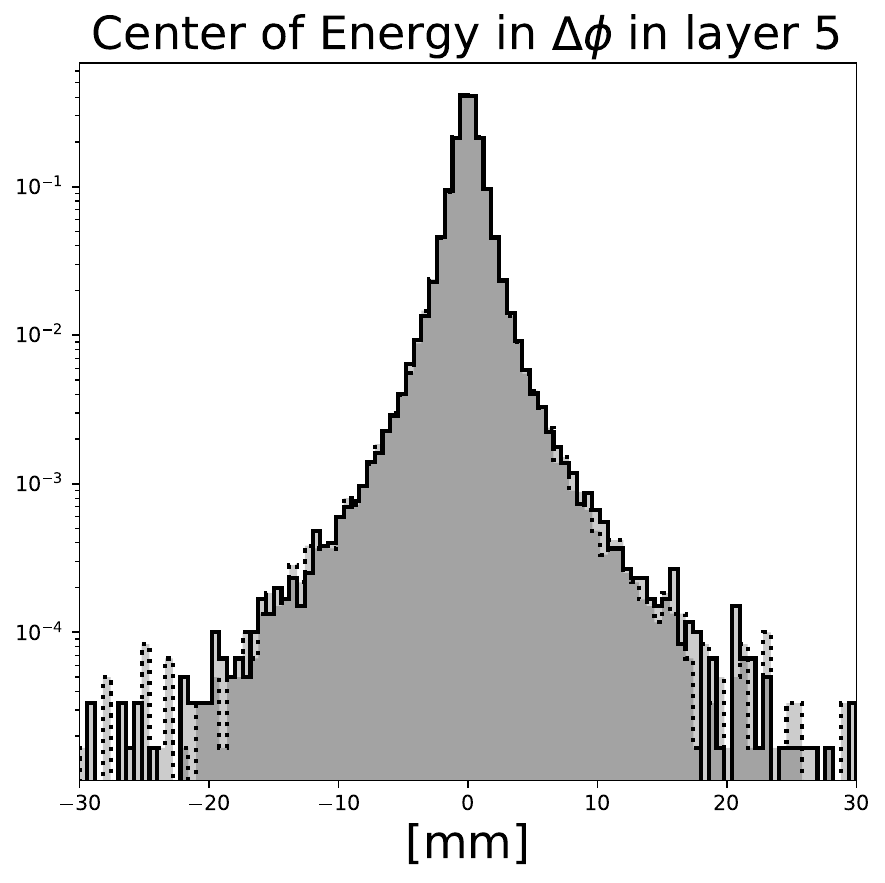} \hfill \includegraphics[height=0.1\textheight]{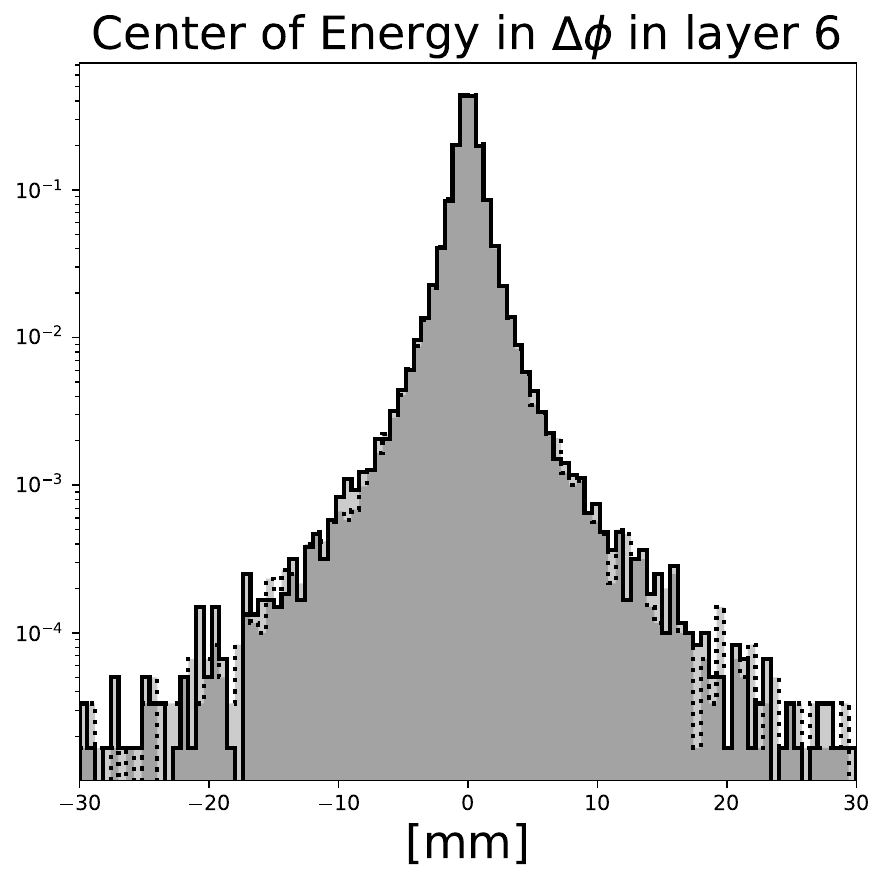} \hfill \includegraphics[height=0.1\textheight]{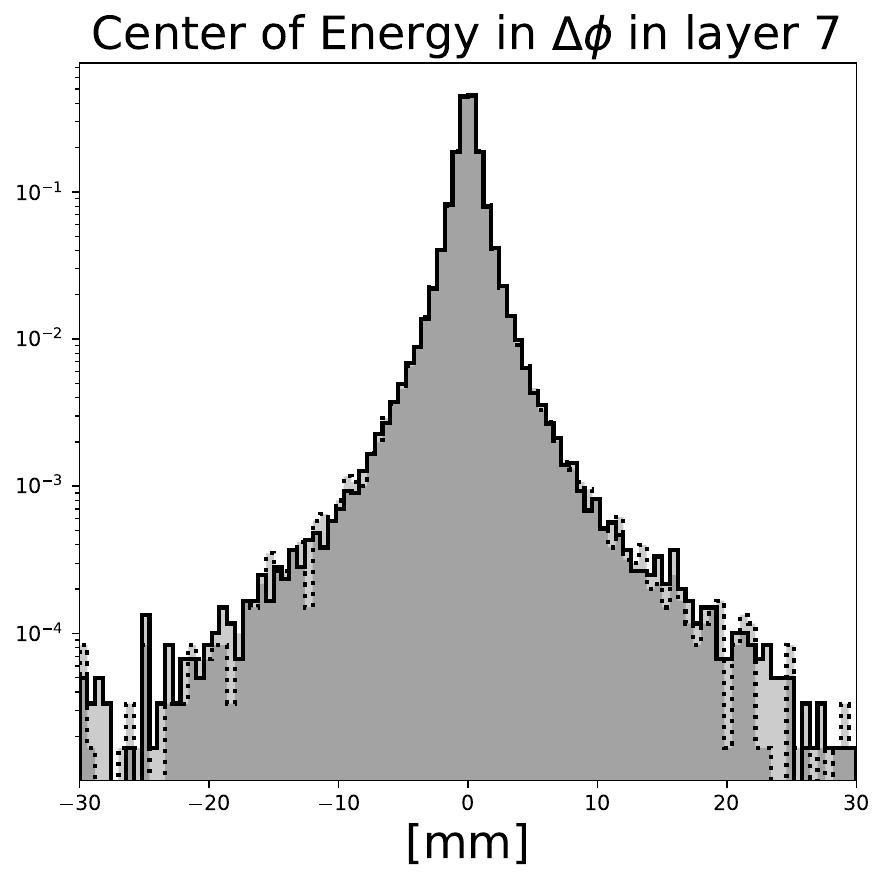} \hfill \includegraphics[height=0.1\textheight]{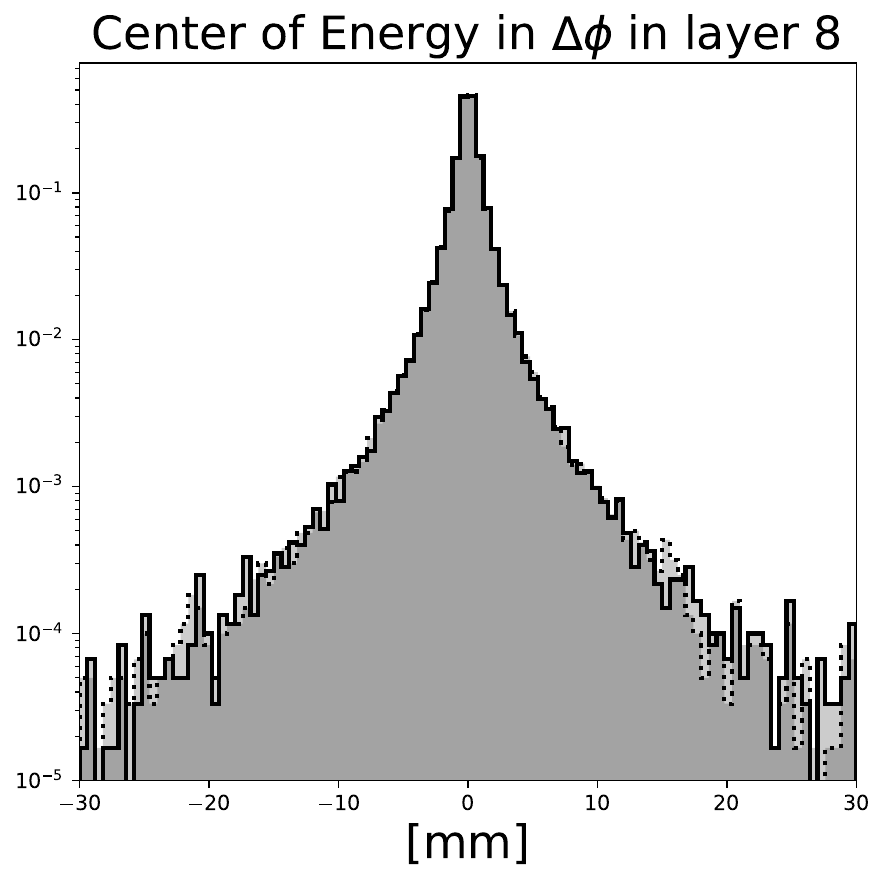} \hfill \includegraphics[height=0.1\textheight]{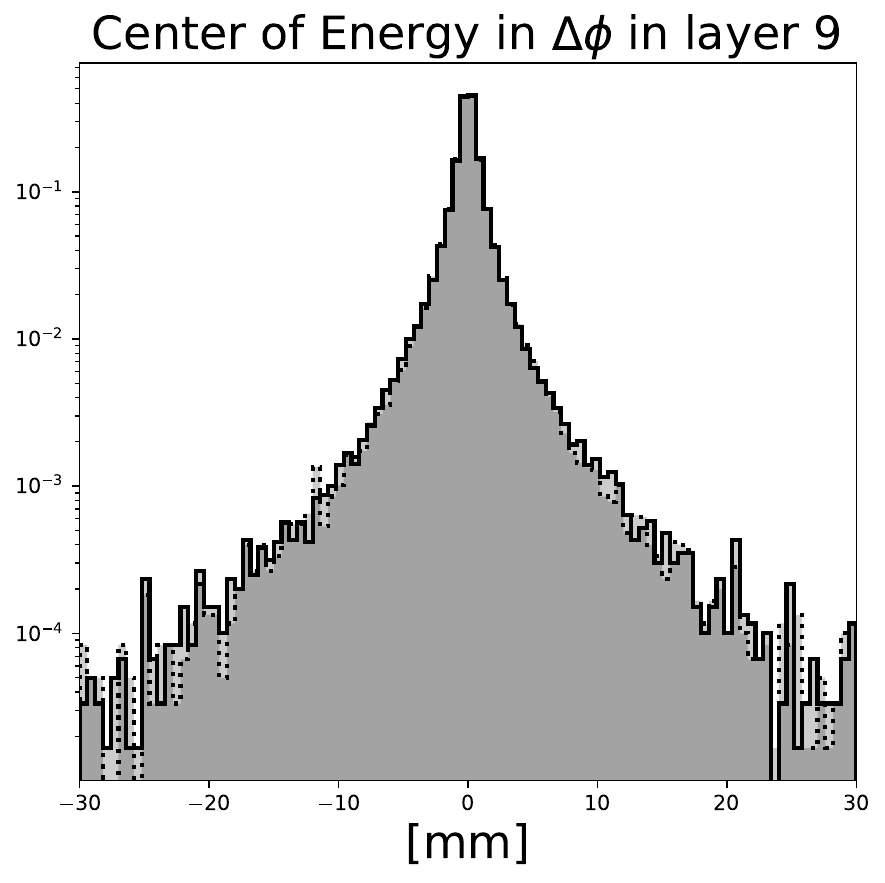}\\
    \includegraphics[height=0.1\textheight]{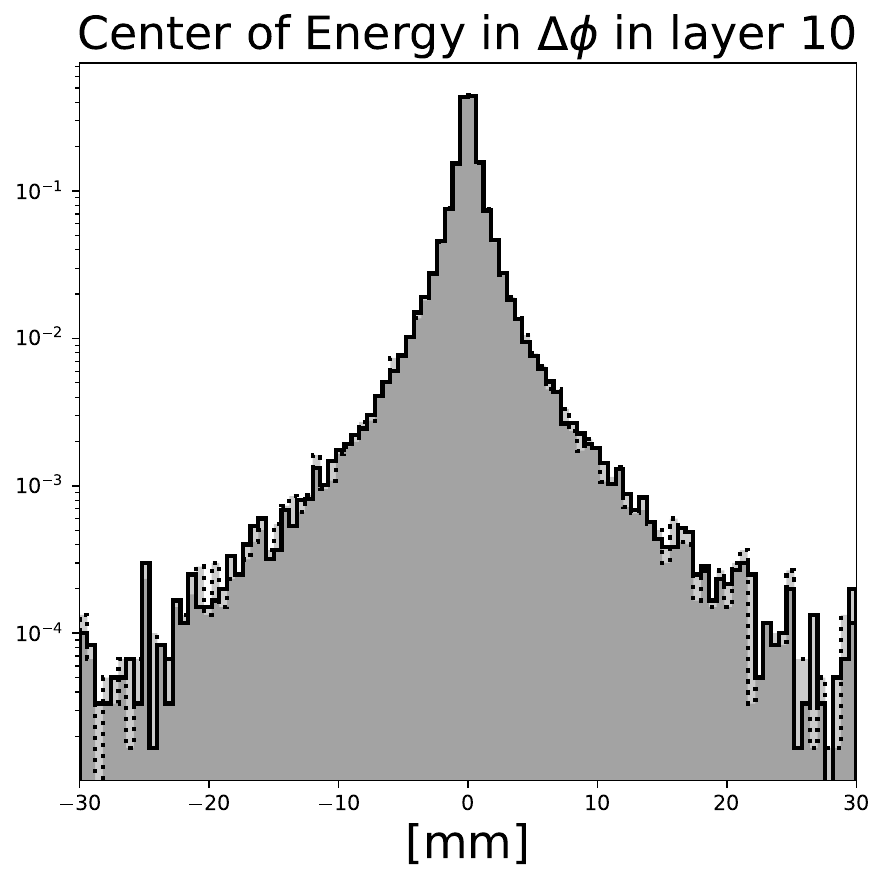} \hfill \includegraphics[height=0.1\textheight]{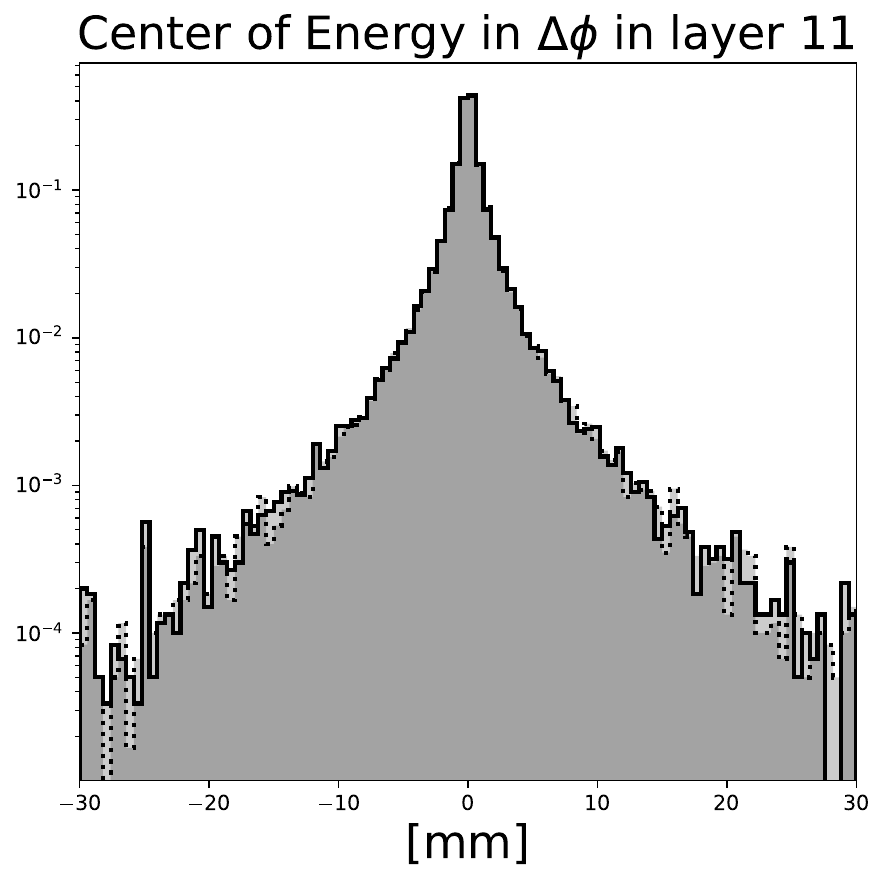} \hfill \includegraphics[height=0.1\textheight]{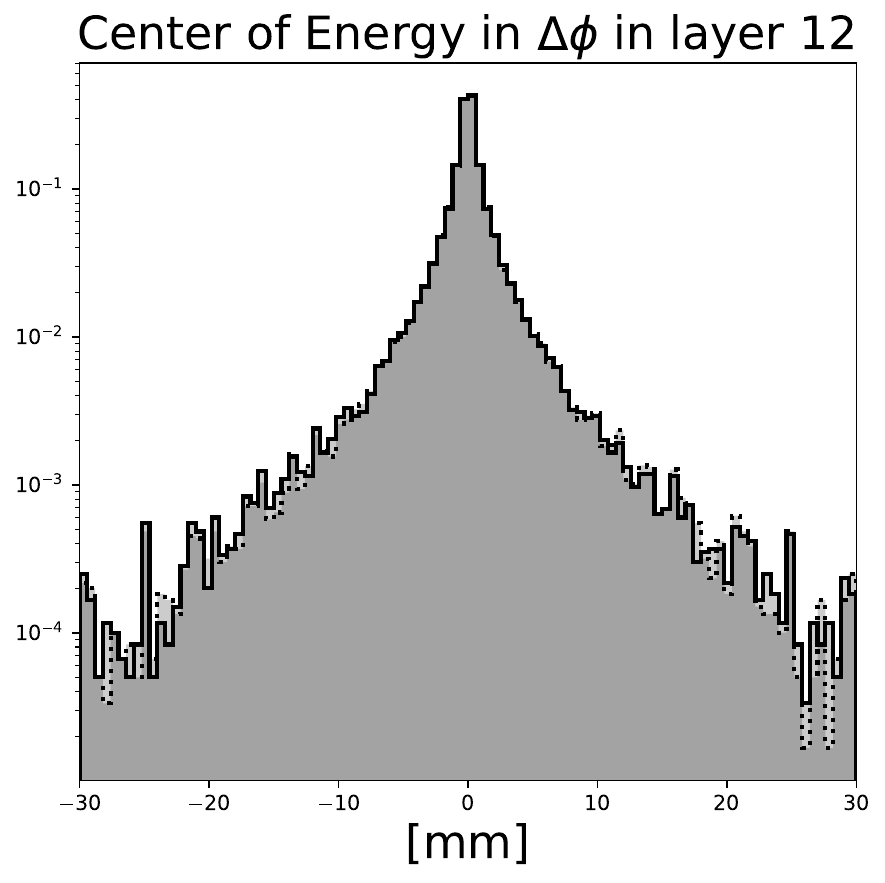} \hfill \includegraphics[height=0.1\textheight]{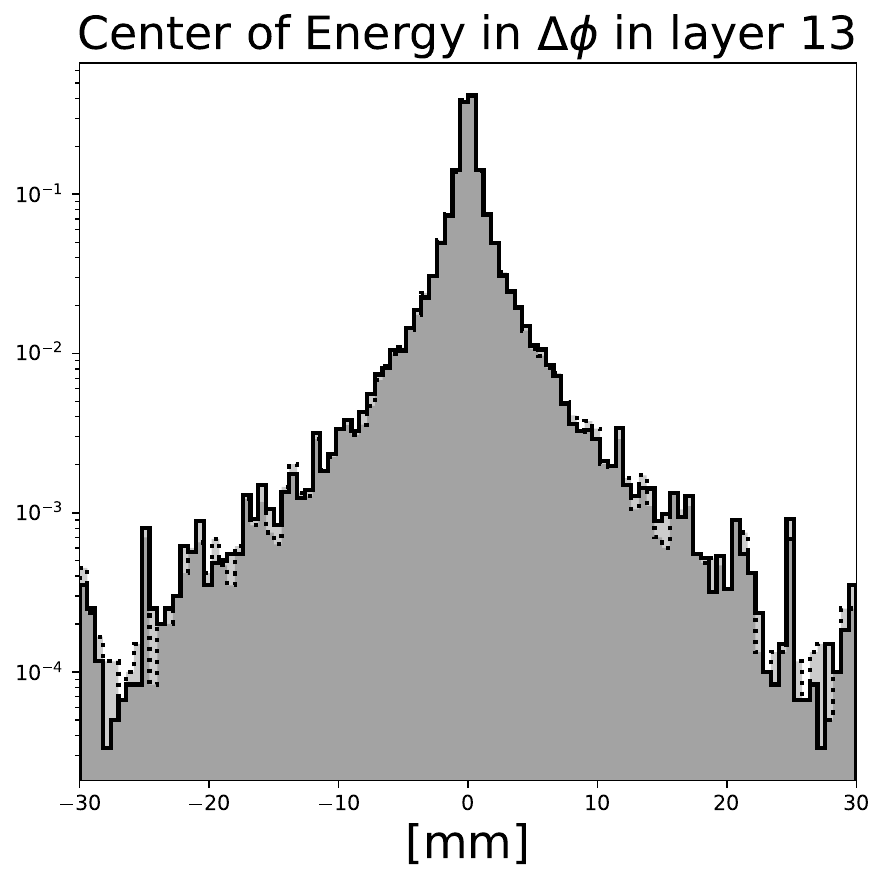} \hfill \includegraphics[height=0.1\textheight]{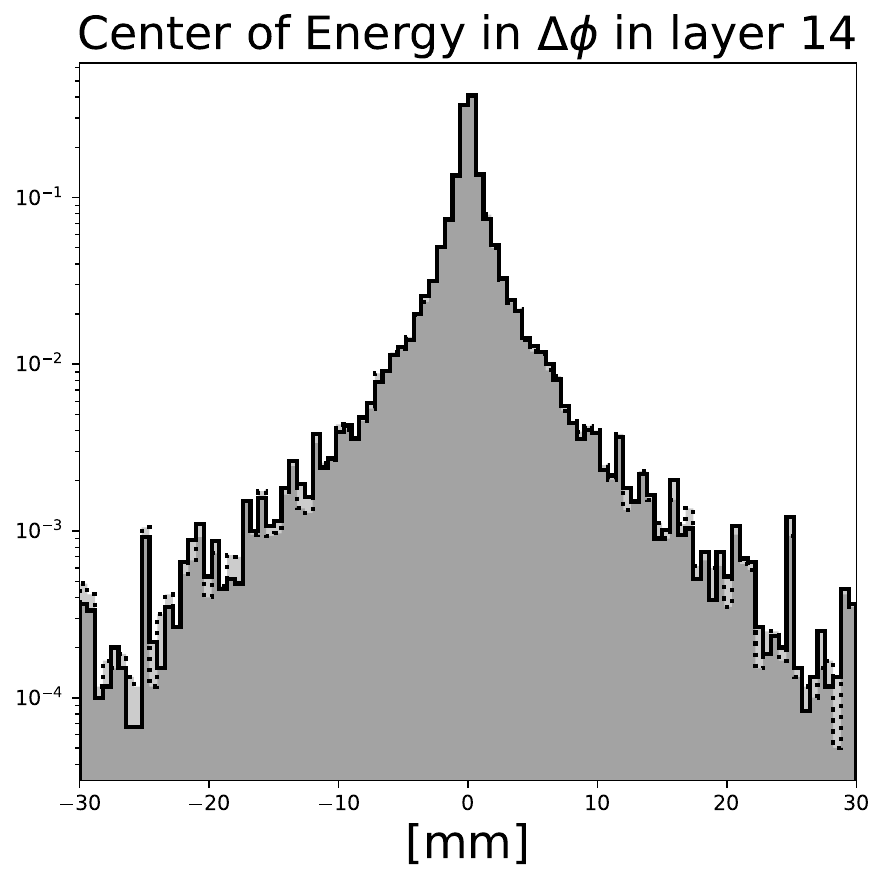}\\
    \includegraphics[height=0.1\textheight]{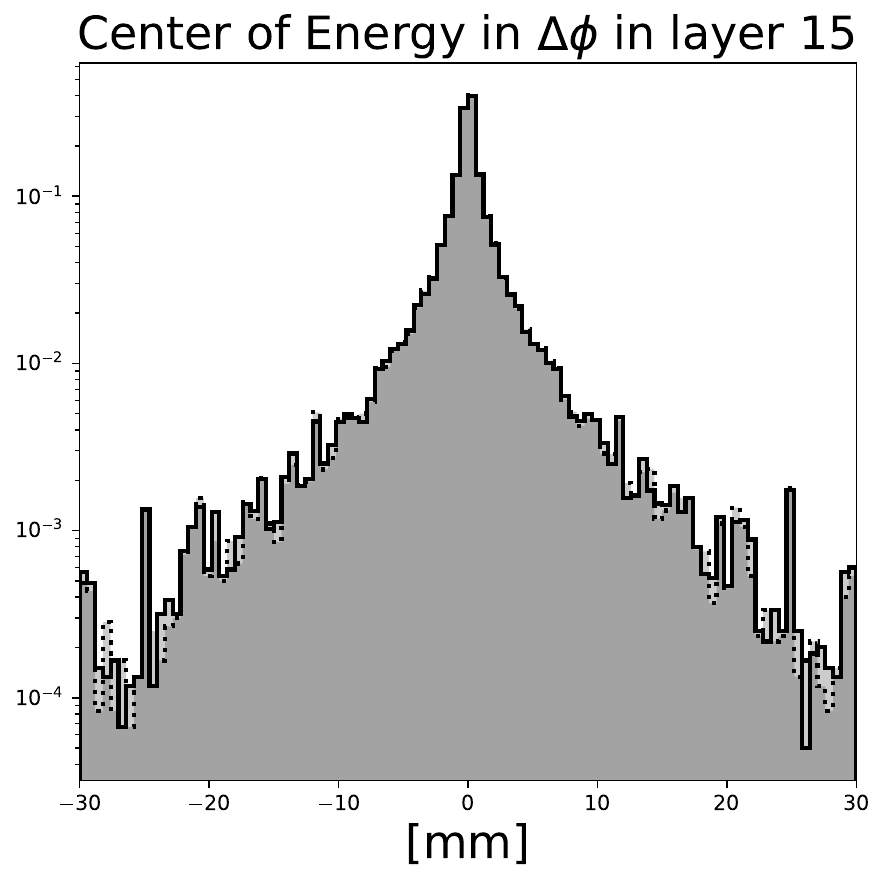} \hfill \includegraphics[height=0.1\textheight]{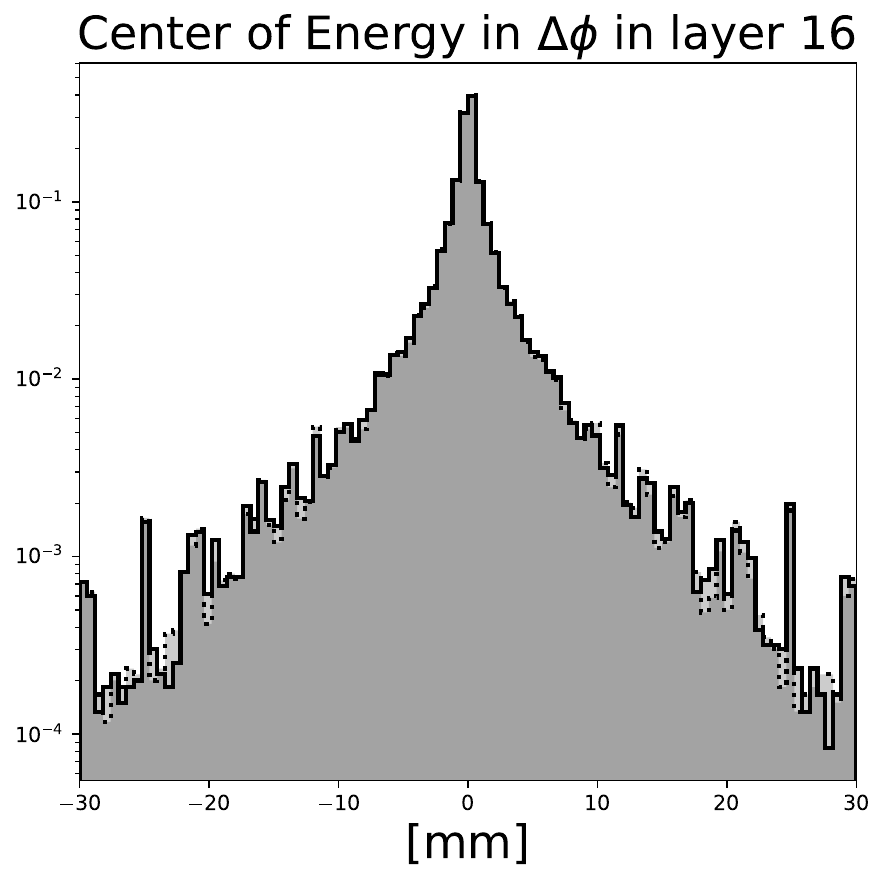} \hfill \includegraphics[height=0.1\textheight]{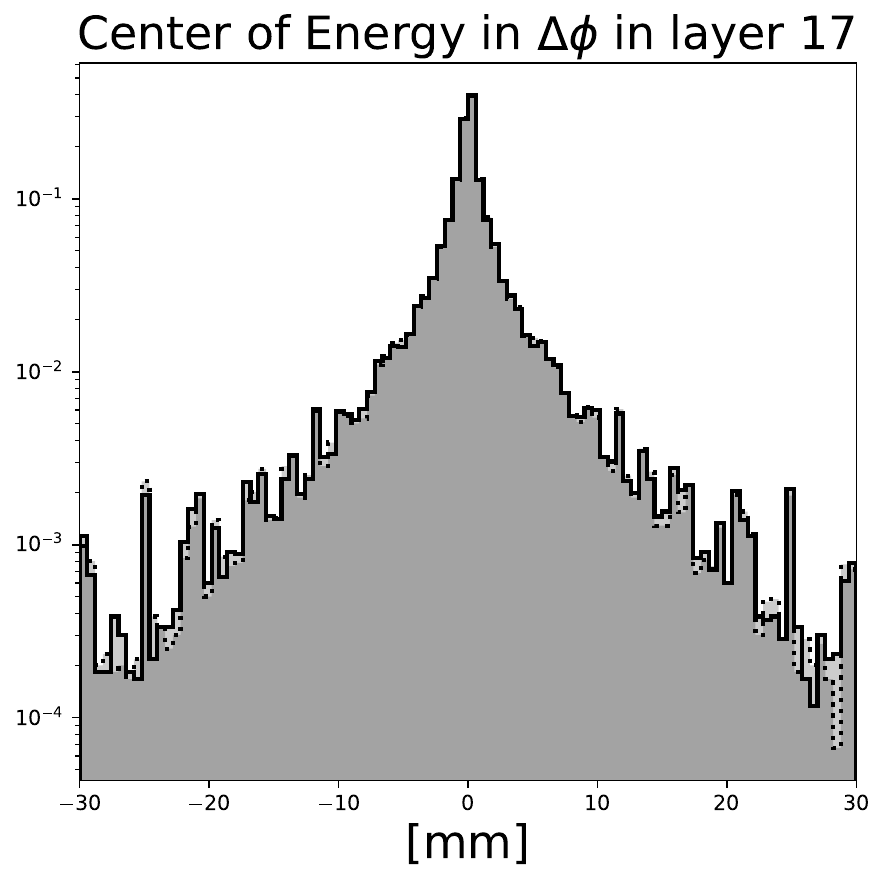} \hfill \includegraphics[height=0.1\textheight]{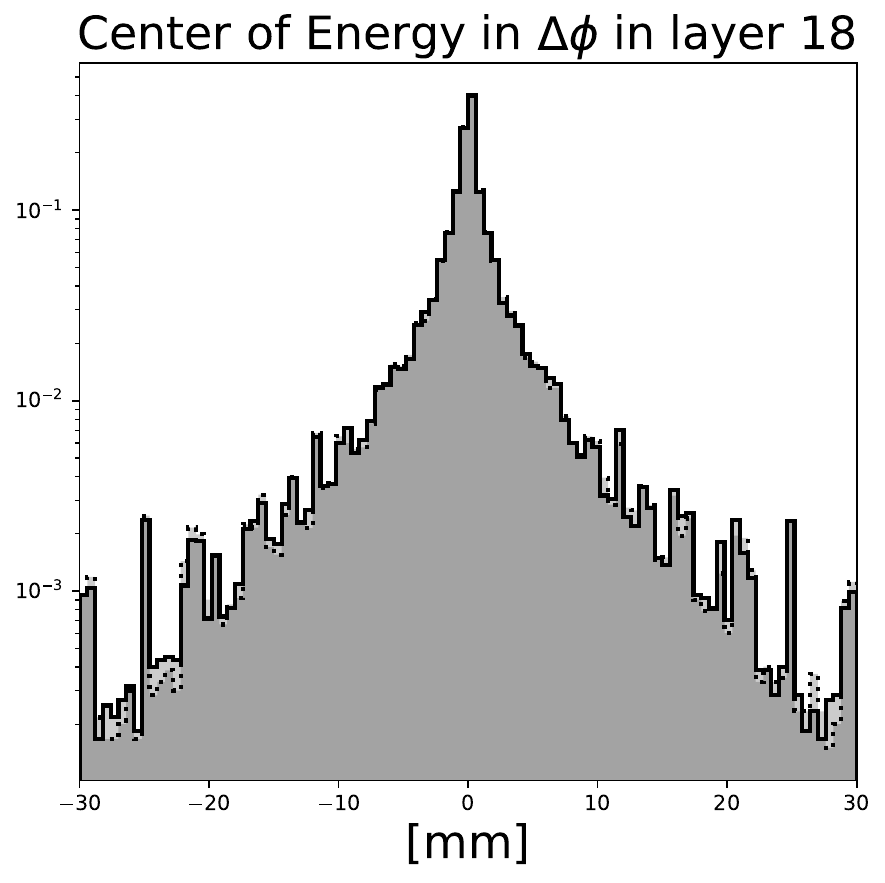} \hfill \includegraphics[height=0.1\textheight]{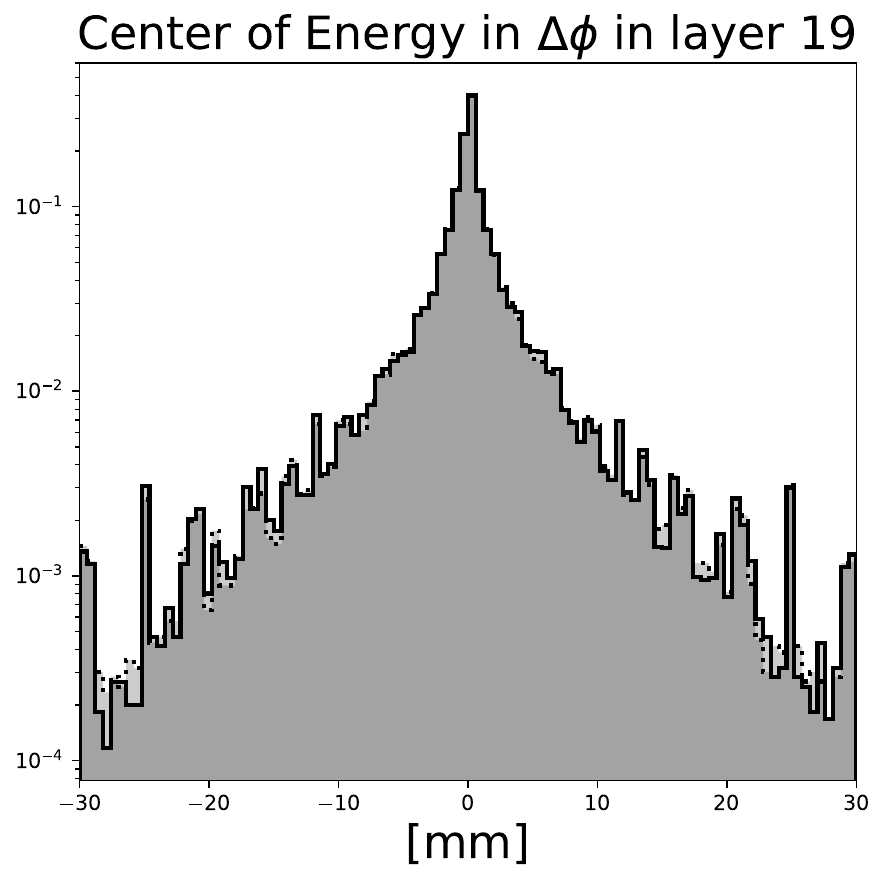}\\
    \includegraphics[height=0.1\textheight]{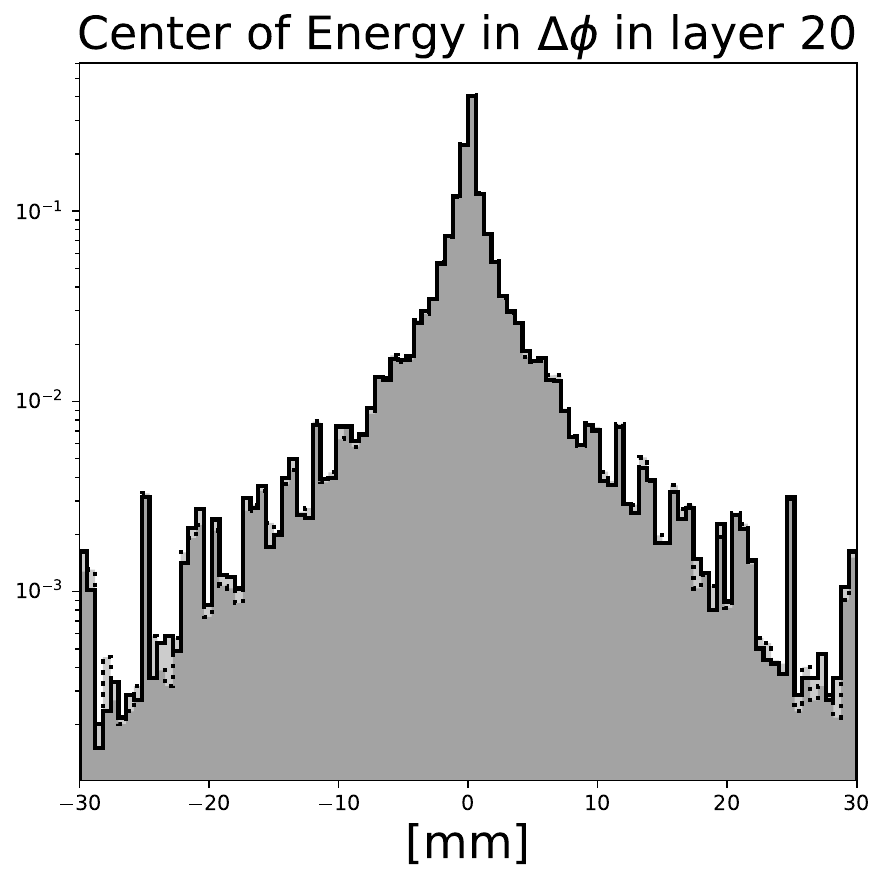} \hfill \includegraphics[height=0.1\textheight]{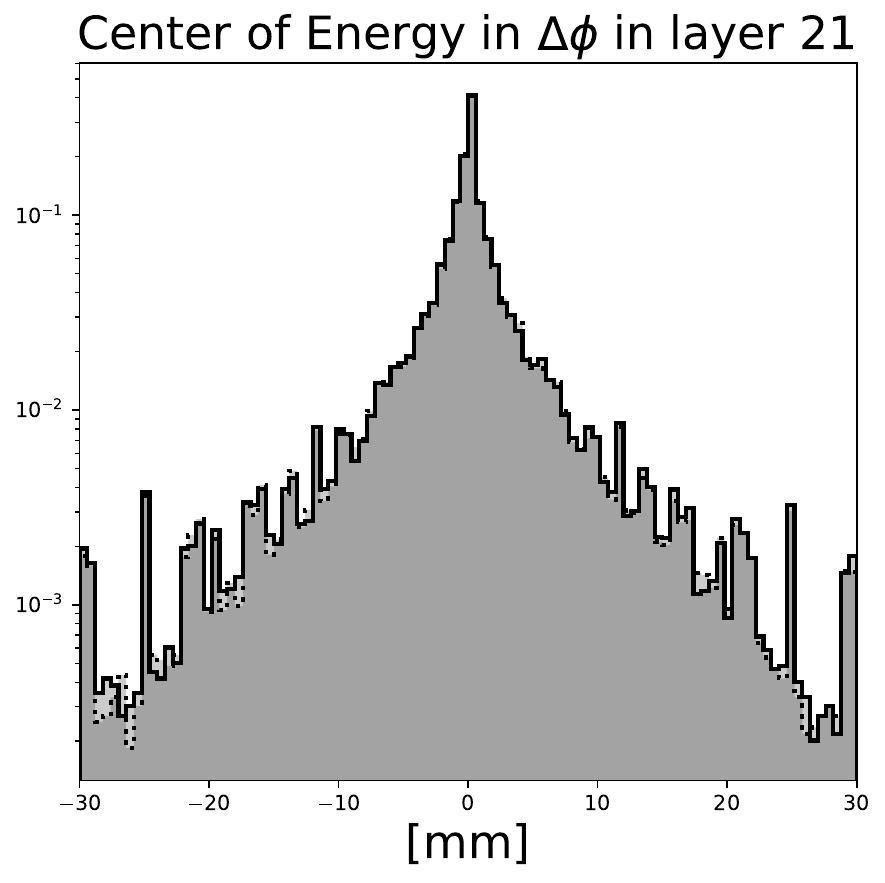} \hfill \includegraphics[height=0.1\textheight]{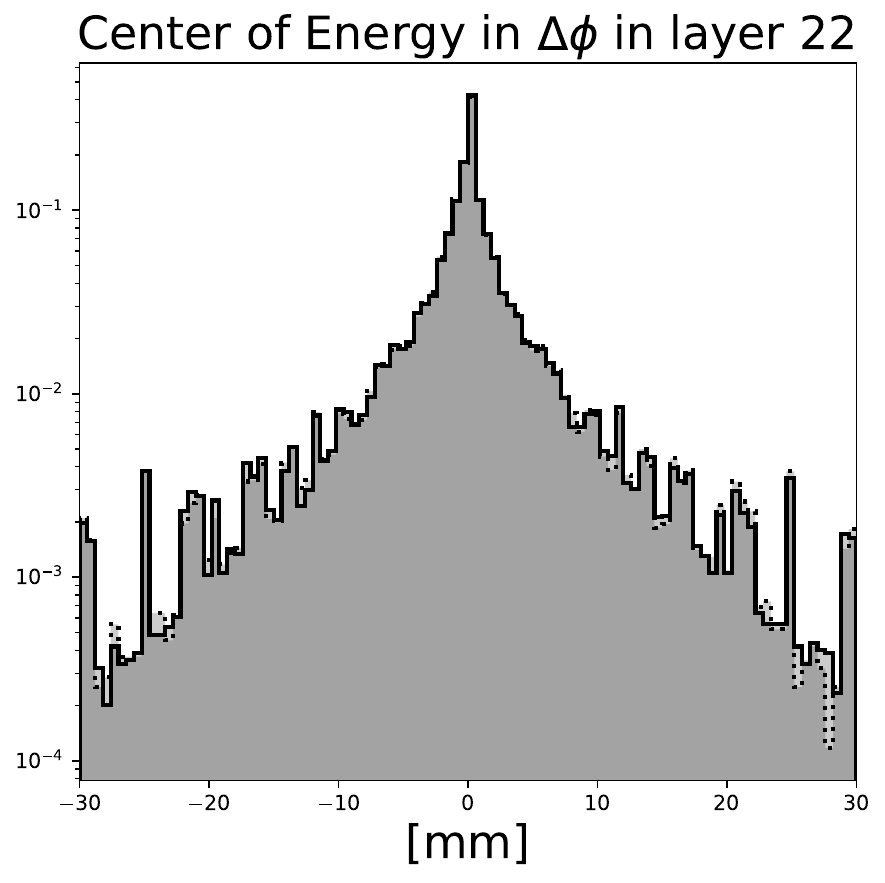} \hfill \includegraphics[height=0.1\textheight]{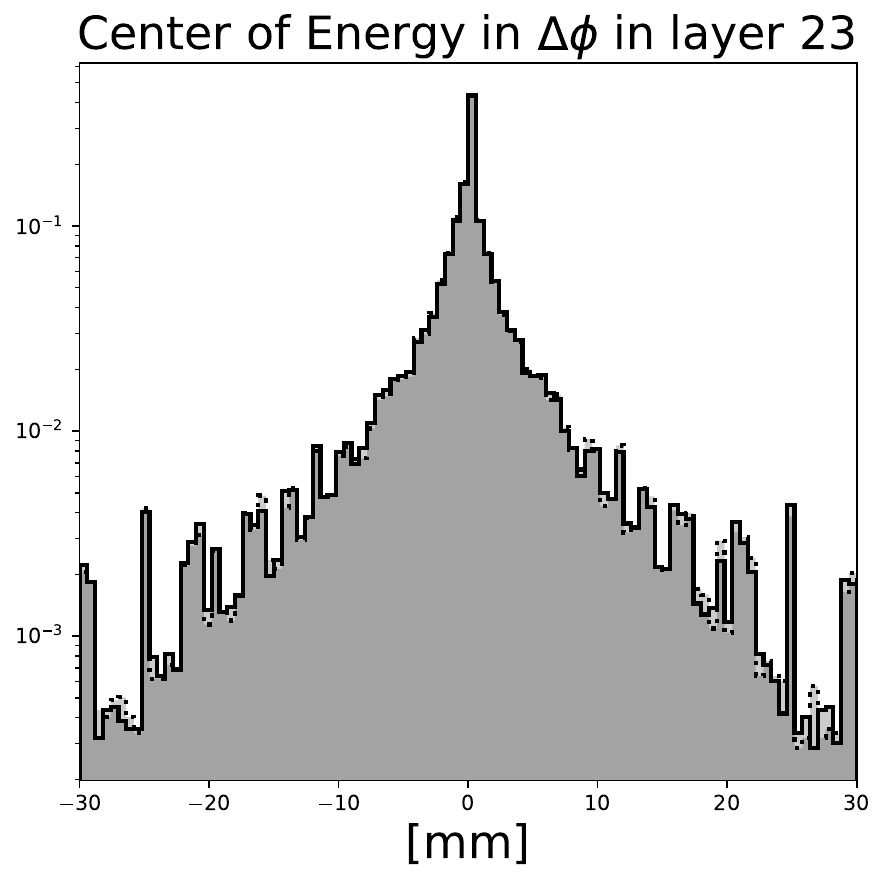} \hfill \includegraphics[height=0.1\textheight]{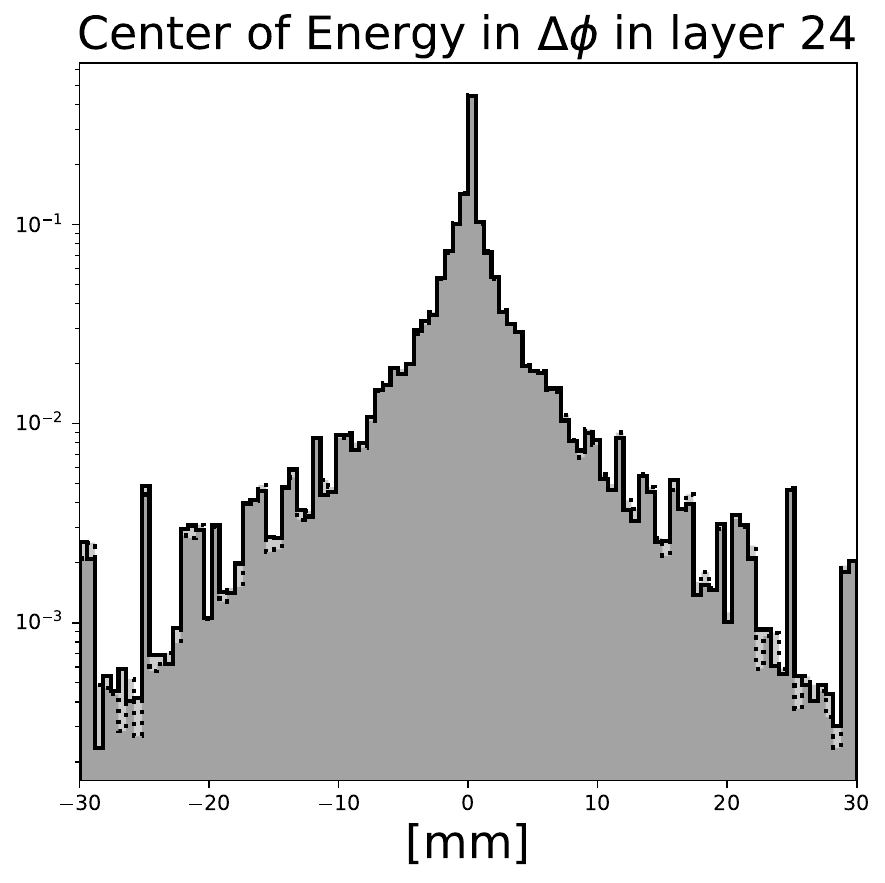}\\
    \includegraphics[height=0.1\textheight]{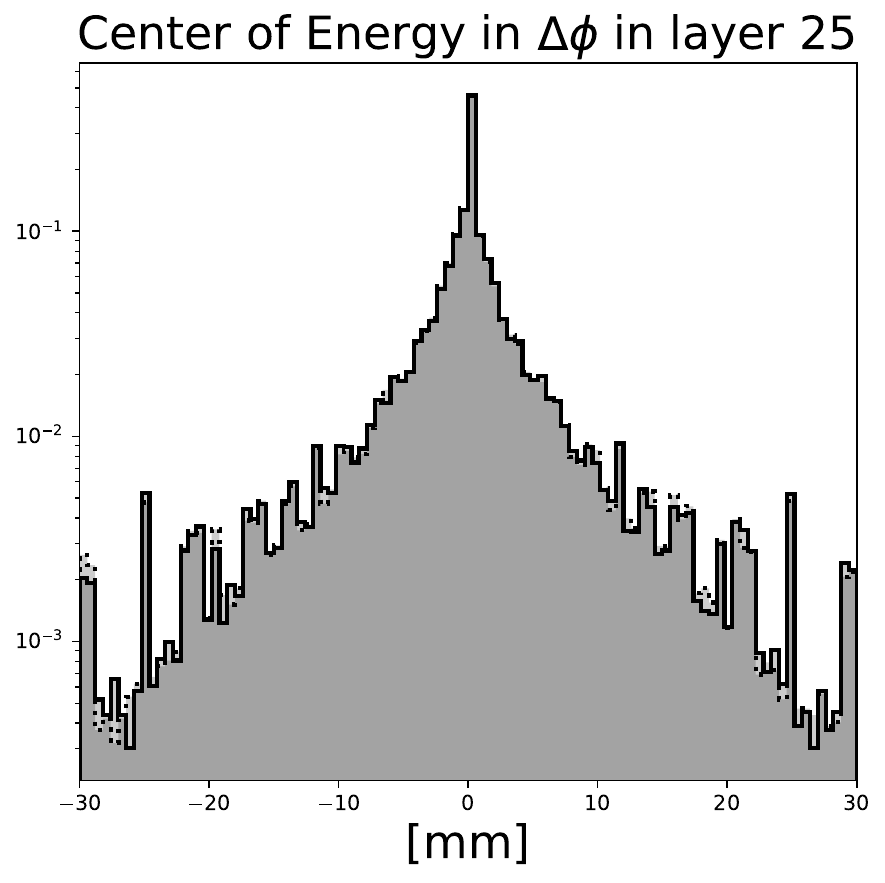} \hfill \includegraphics[height=0.1\textheight]{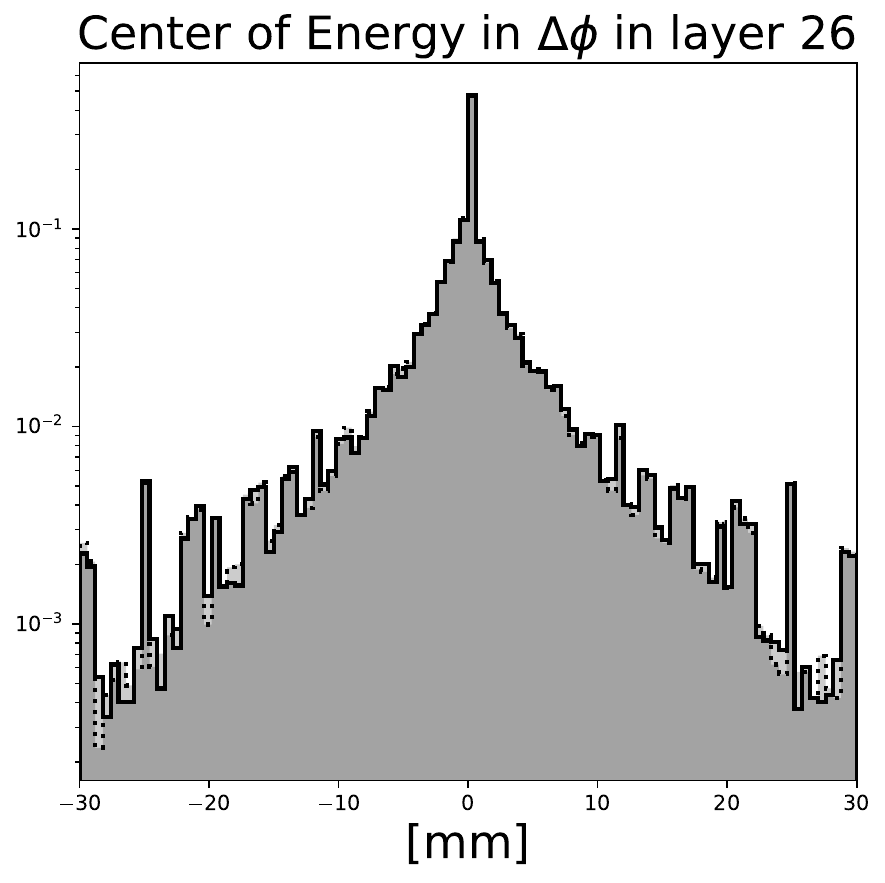} \hfill \includegraphics[height=0.1\textheight]{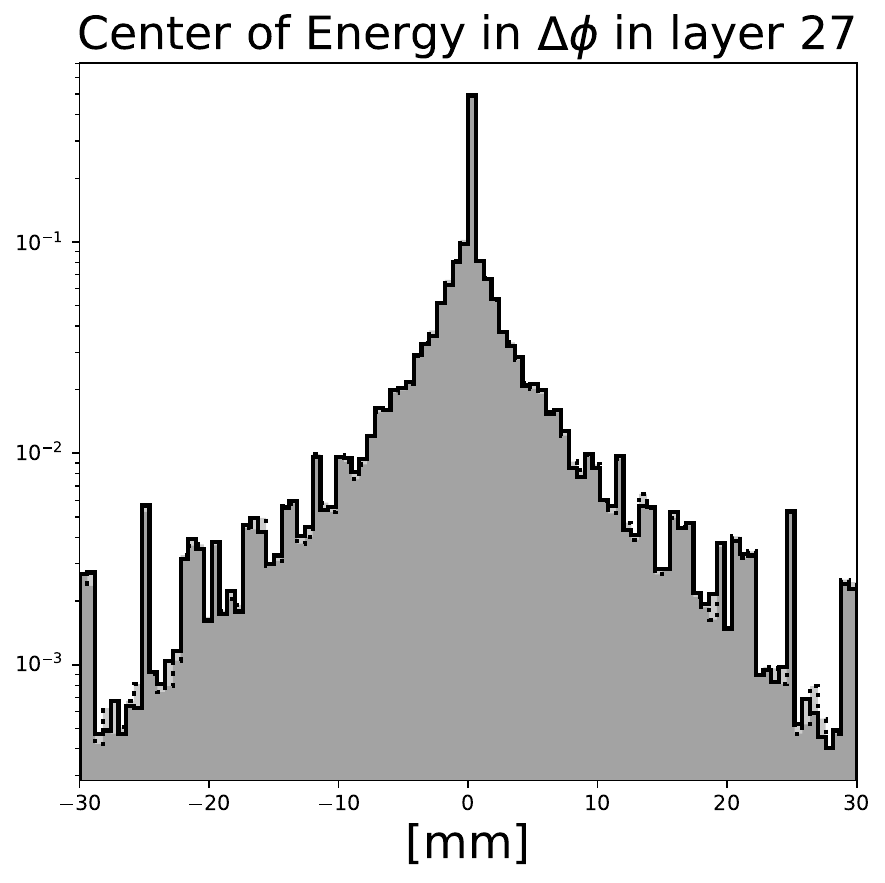} \hfill \includegraphics[height=0.1\textheight]{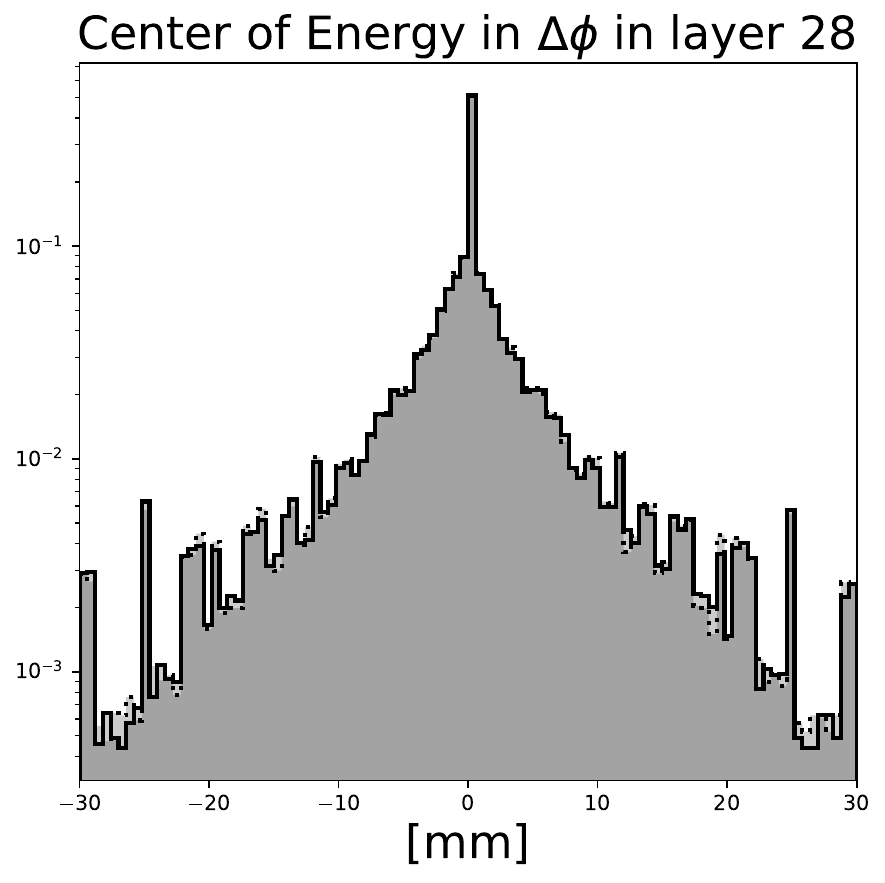} \hfill \includegraphics[height=0.1\textheight]{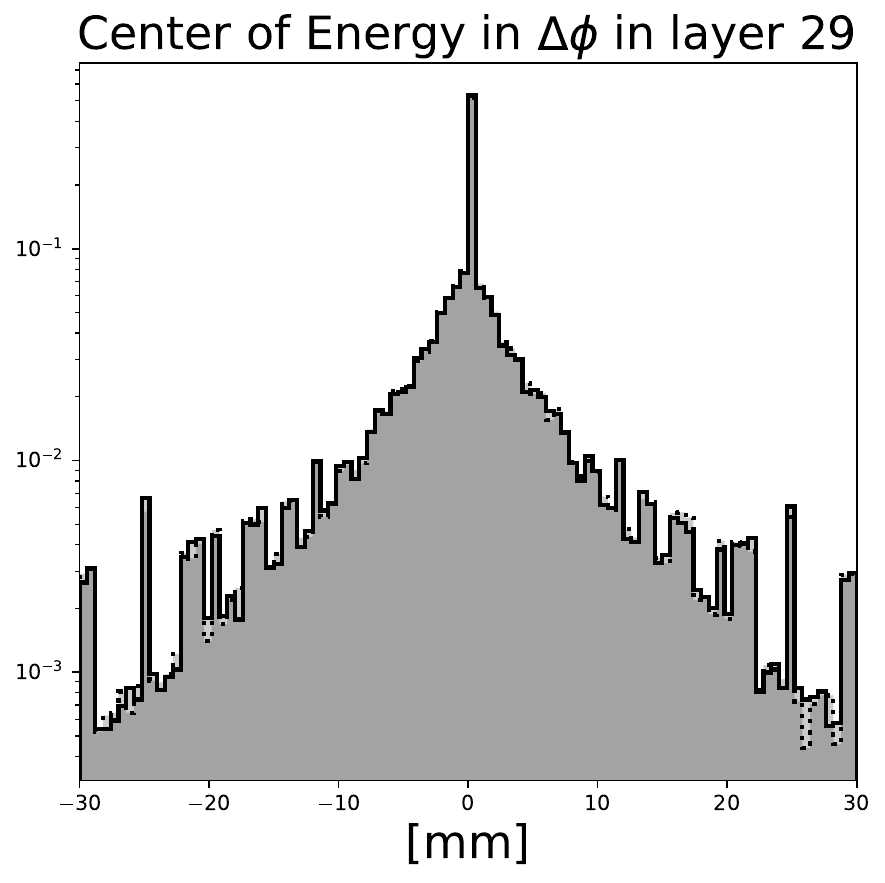}\\
    \includegraphics[height=0.1\textheight]{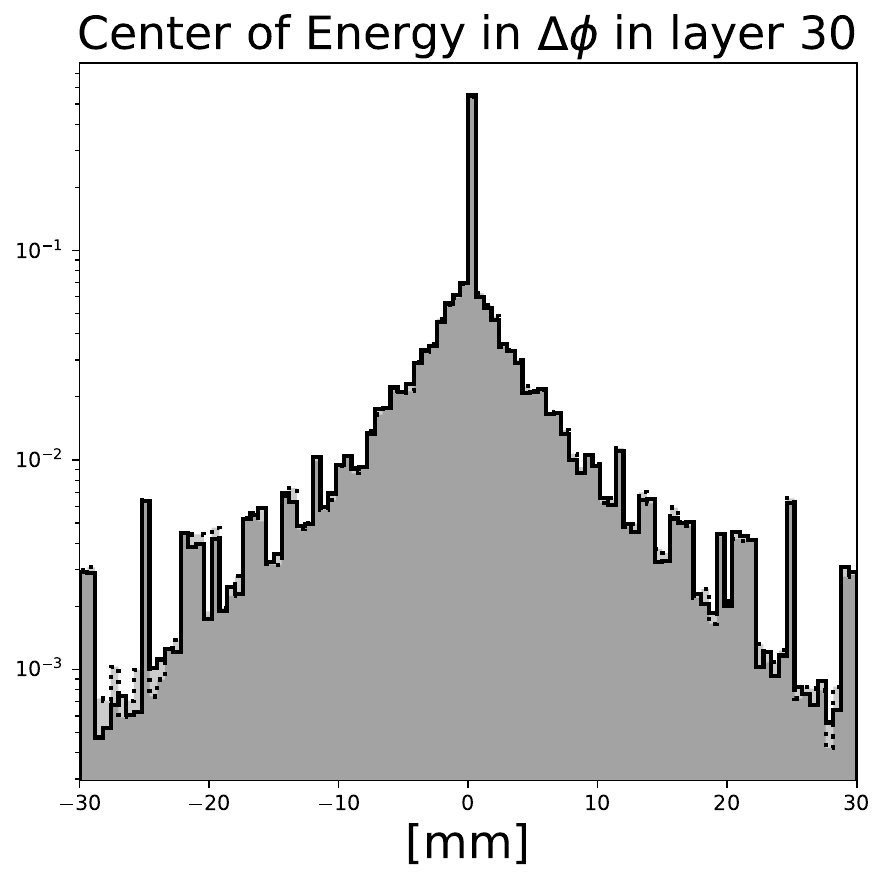} \hfill \includegraphics[height=0.1\textheight]{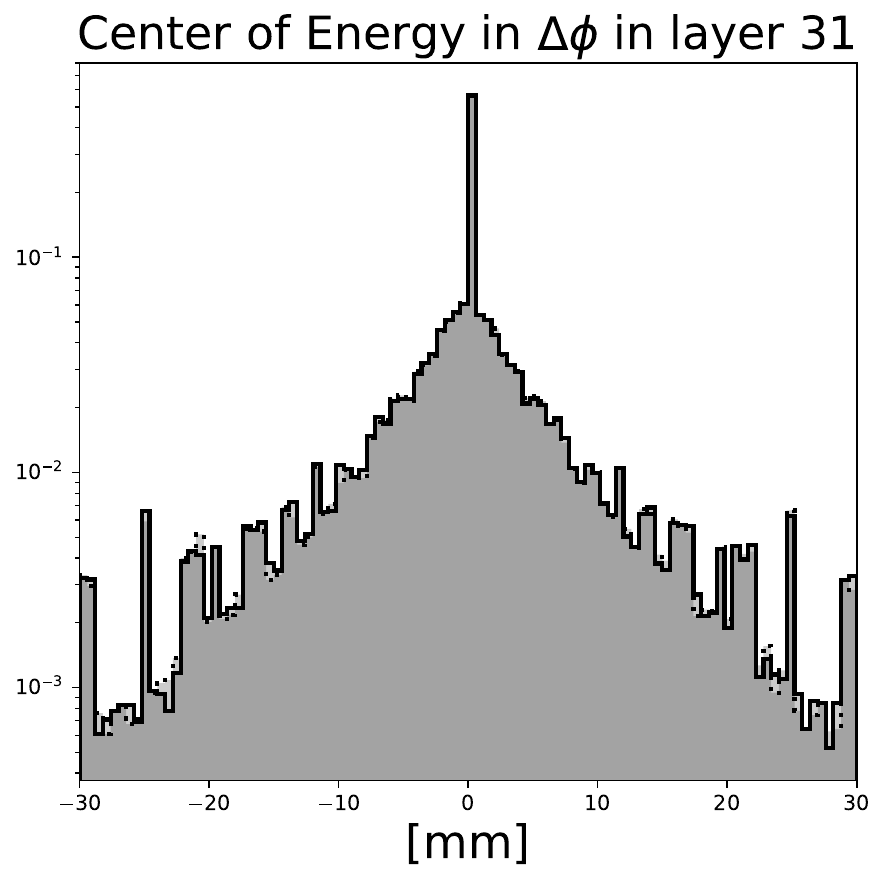} \hfill \includegraphics[height=0.1\textheight]{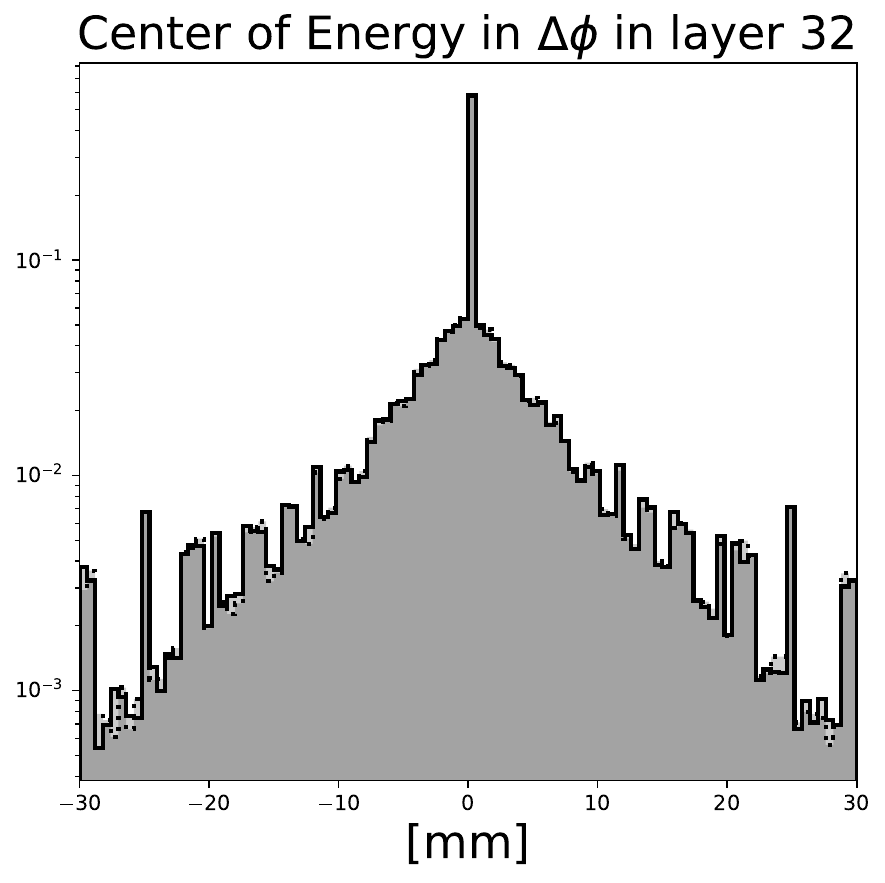} \hfill \includegraphics[height=0.1\textheight]{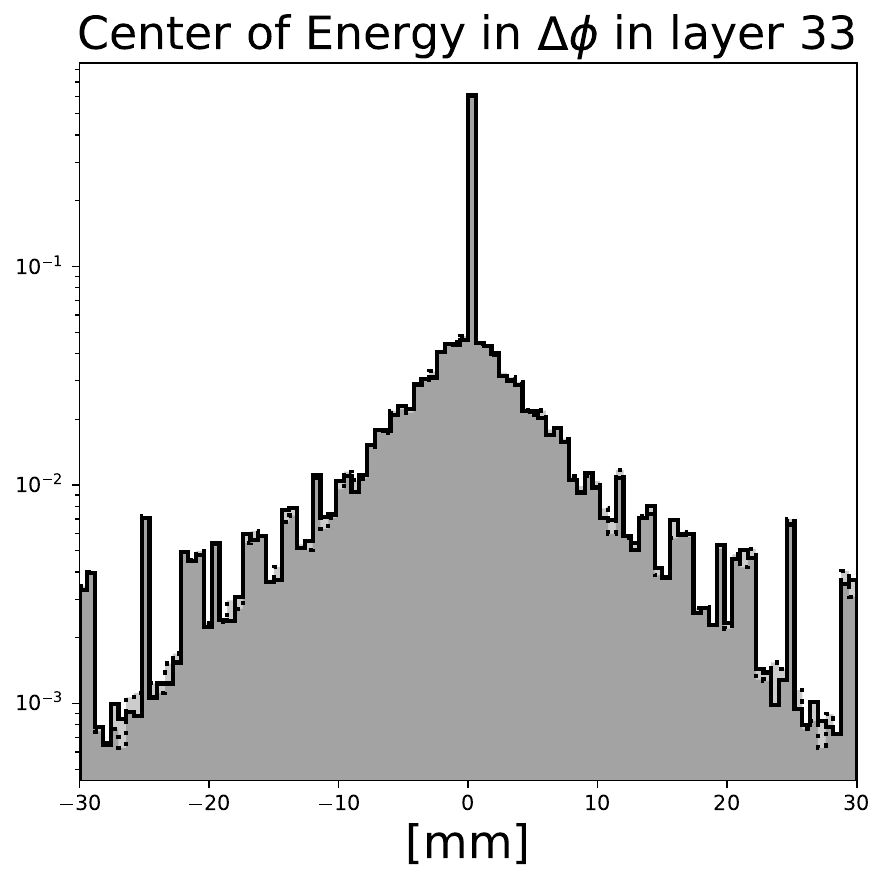} \hfill \includegraphics[height=0.1\textheight]{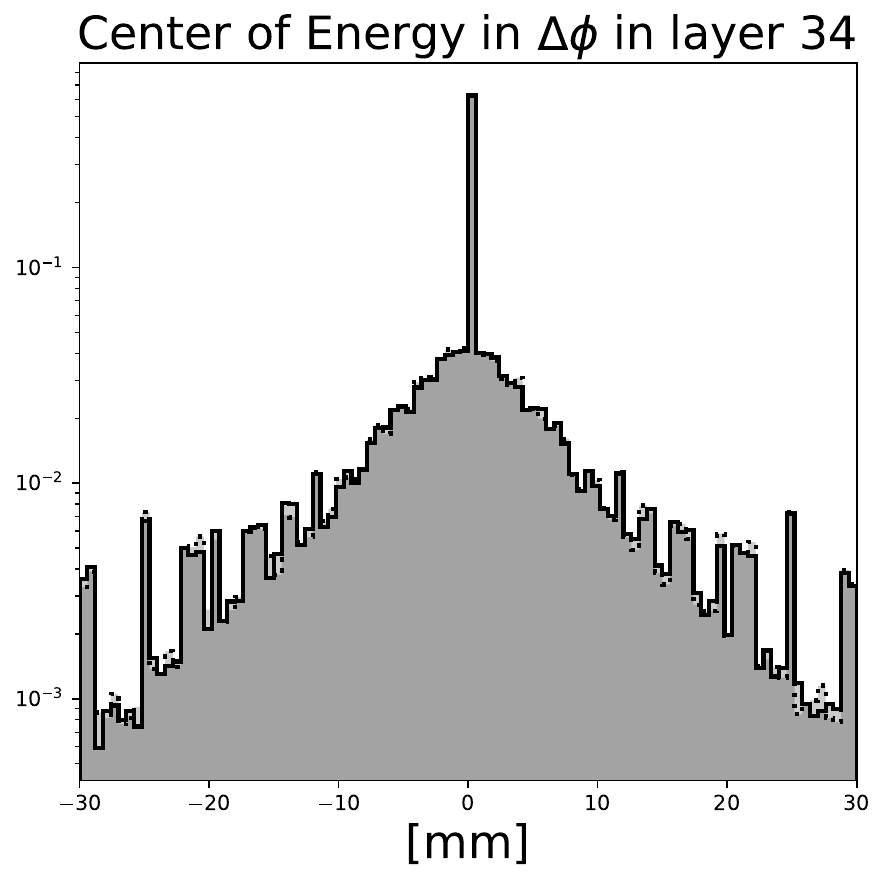}\\
    \includegraphics[height=0.1\textheight]{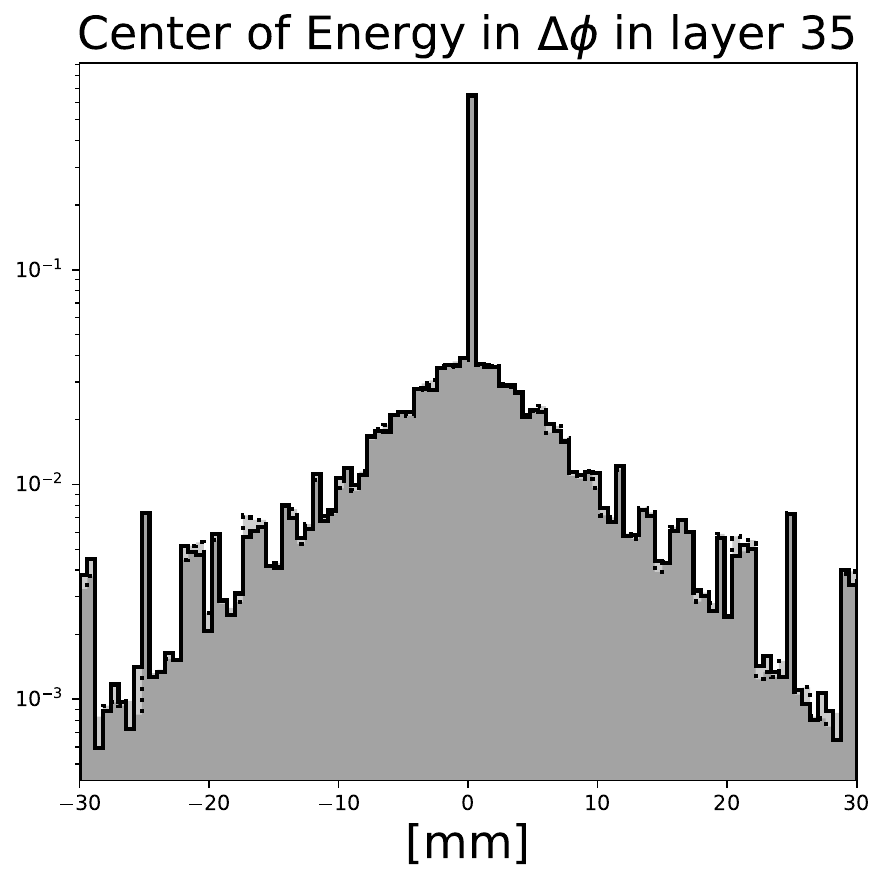} \hfill \includegraphics[height=0.1\textheight]{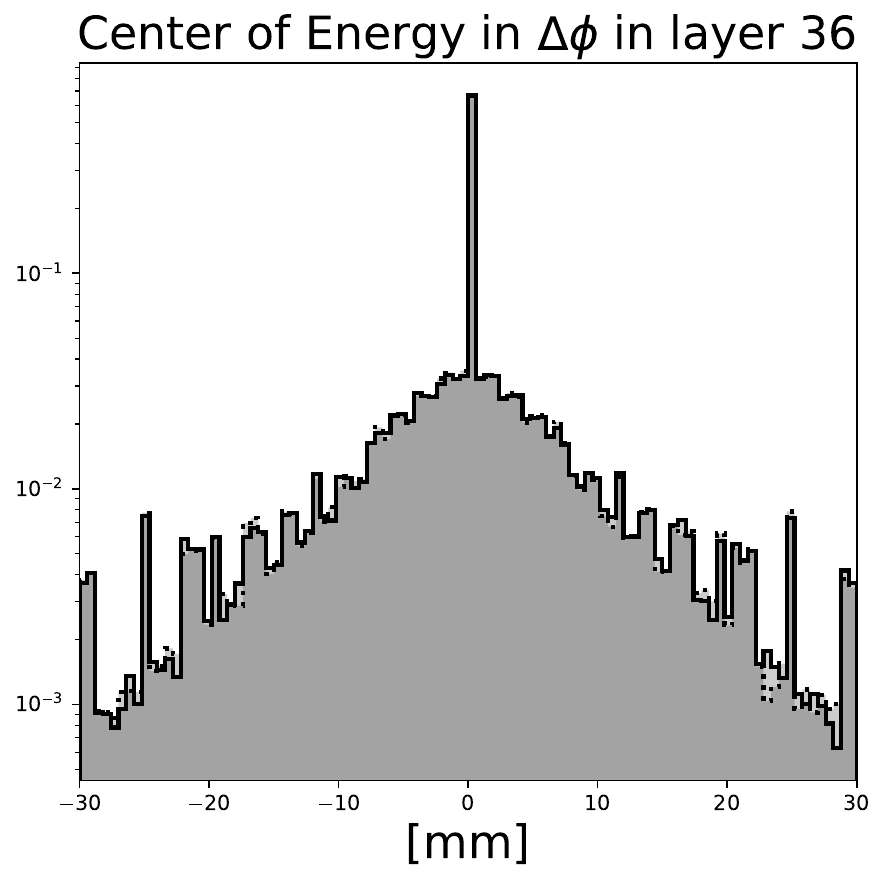} \hfill \includegraphics[height=0.1\textheight]{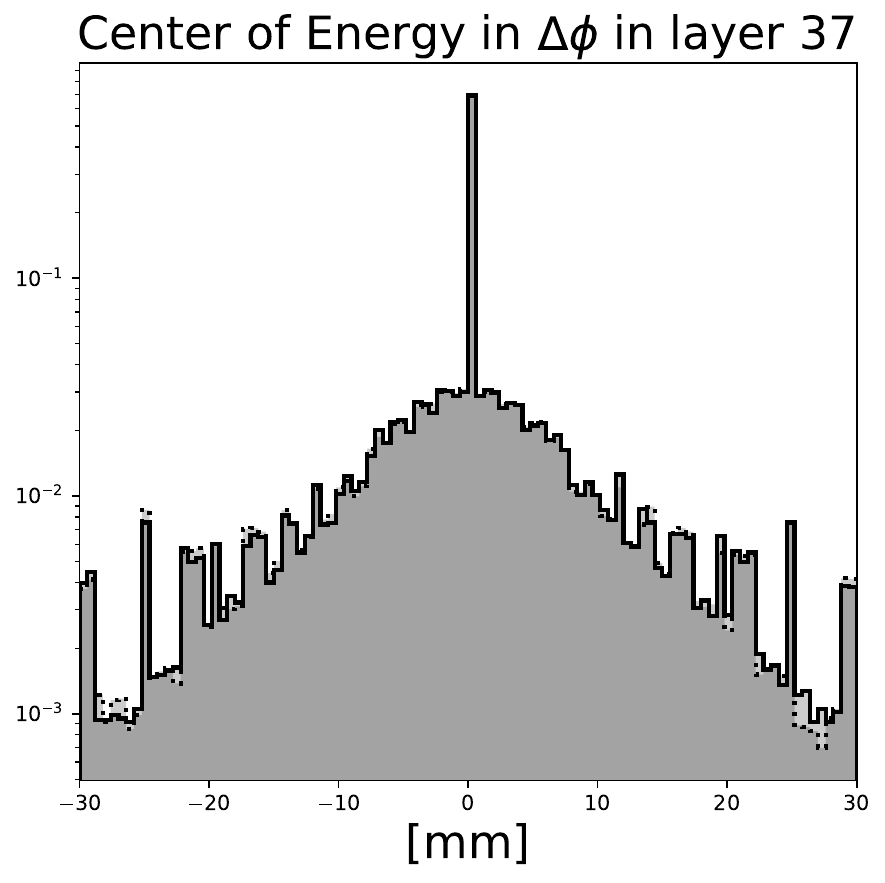} \hfill \includegraphics[height=0.1\textheight]{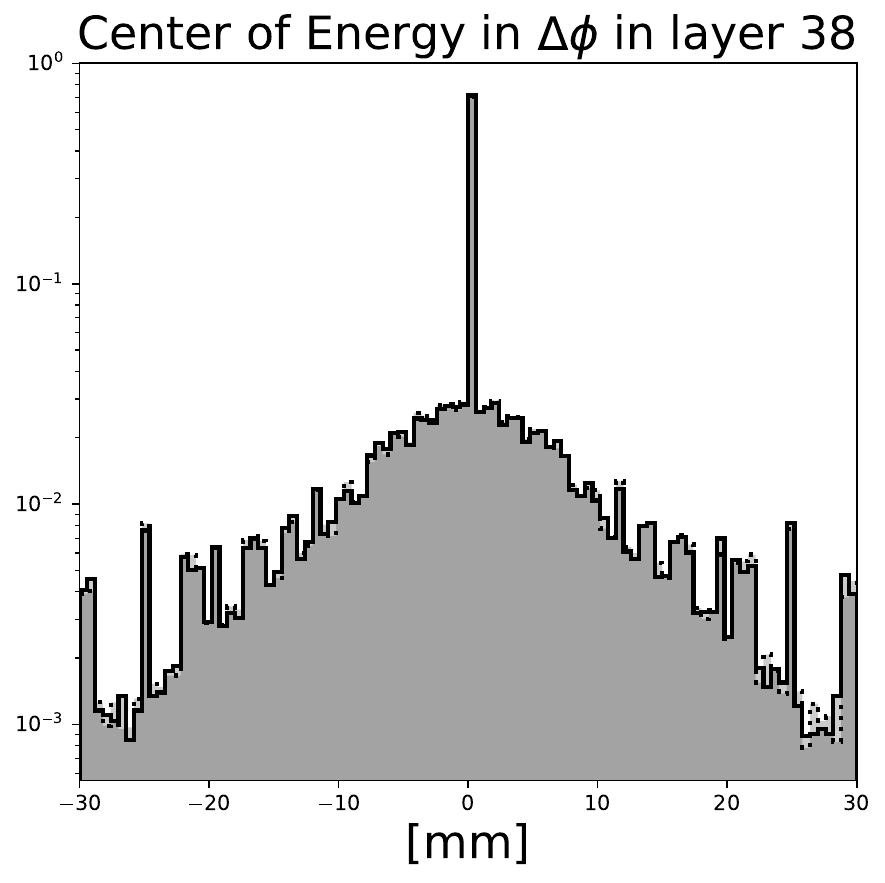} \hfill \includegraphics[height=0.1\textheight]{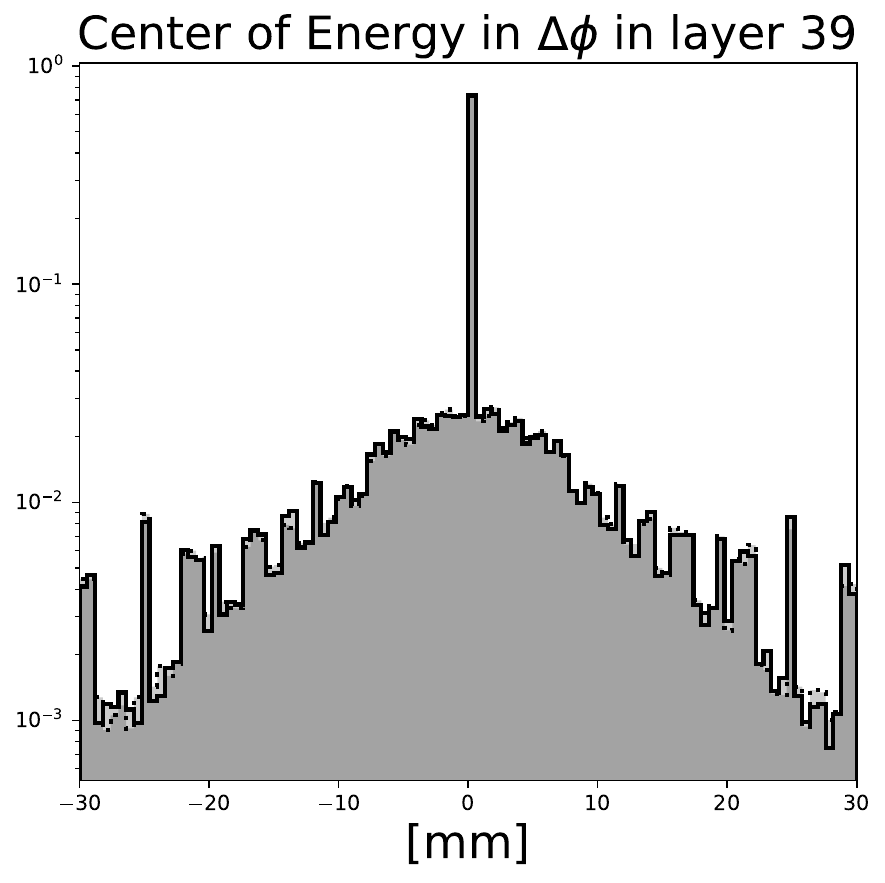}\\
    \includegraphics[height=0.1\textheight]{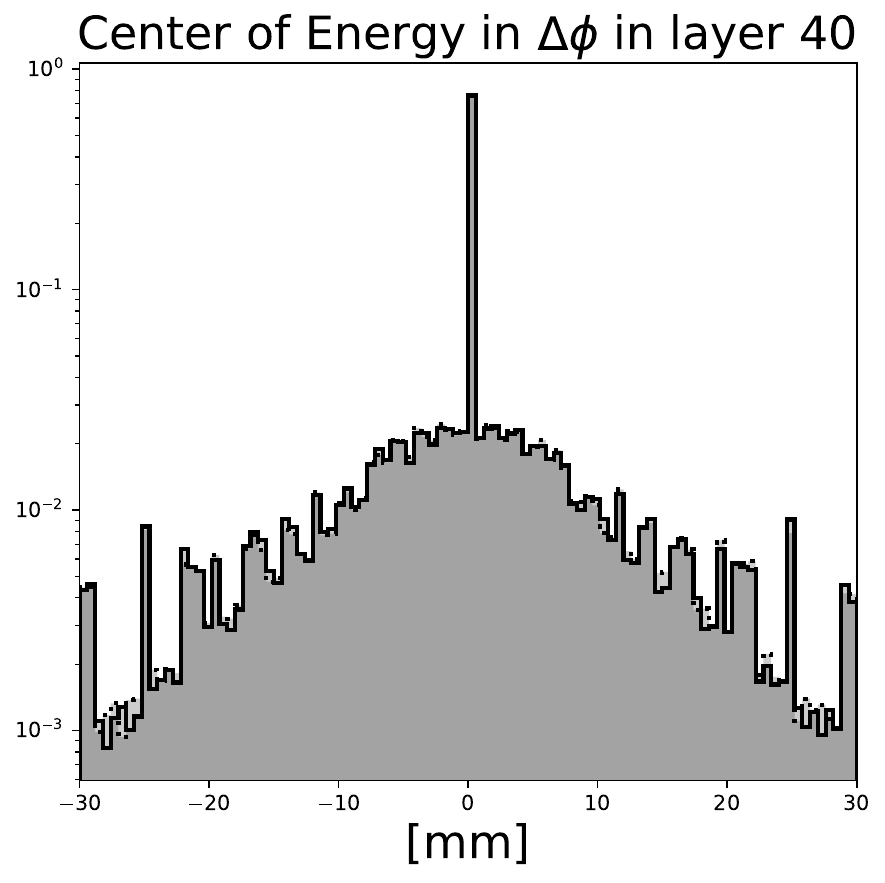} \hfill \includegraphics[height=0.1\textheight]{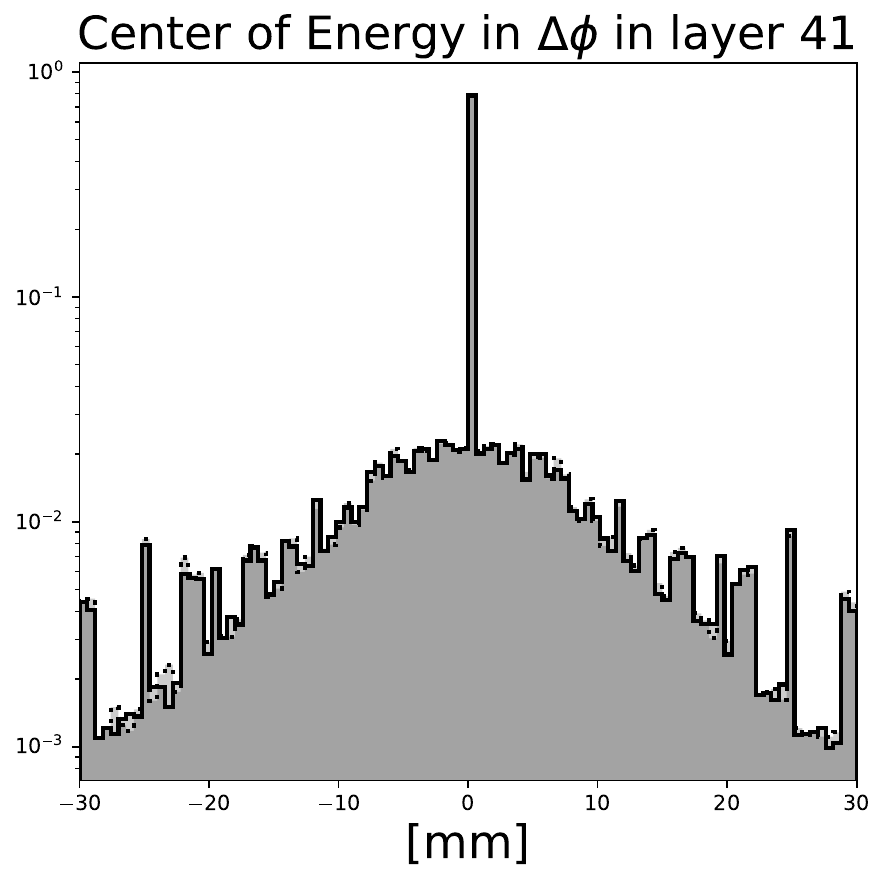} \hfill \includegraphics[height=0.1\textheight]{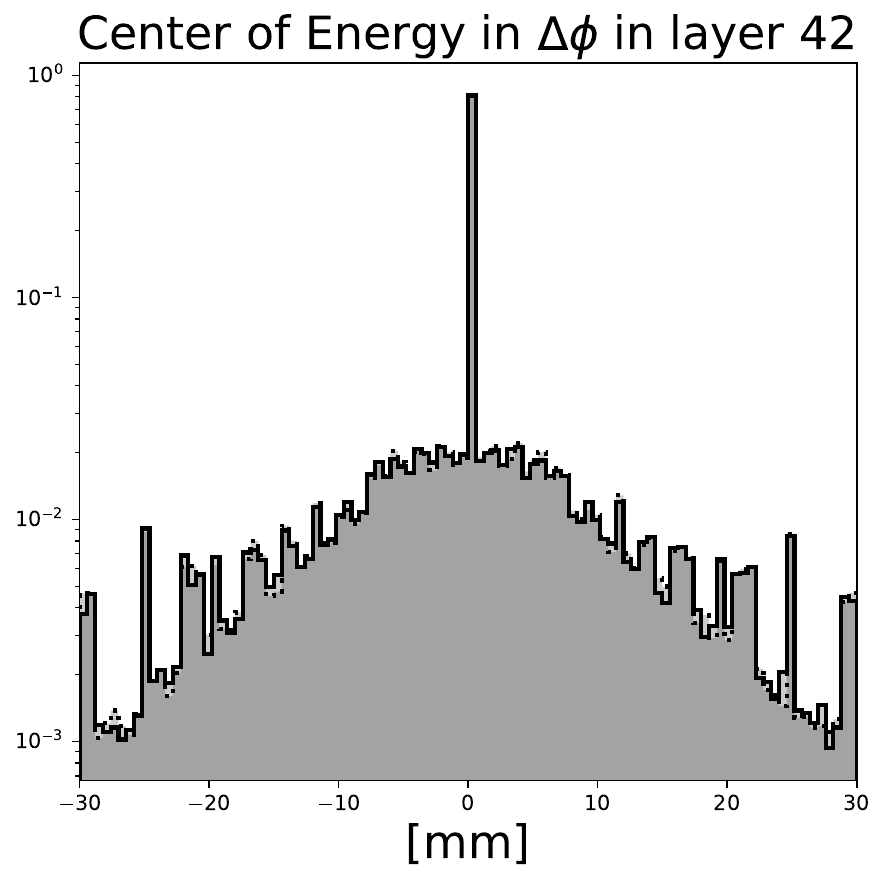} \hfill \includegraphics[height=0.1\textheight]{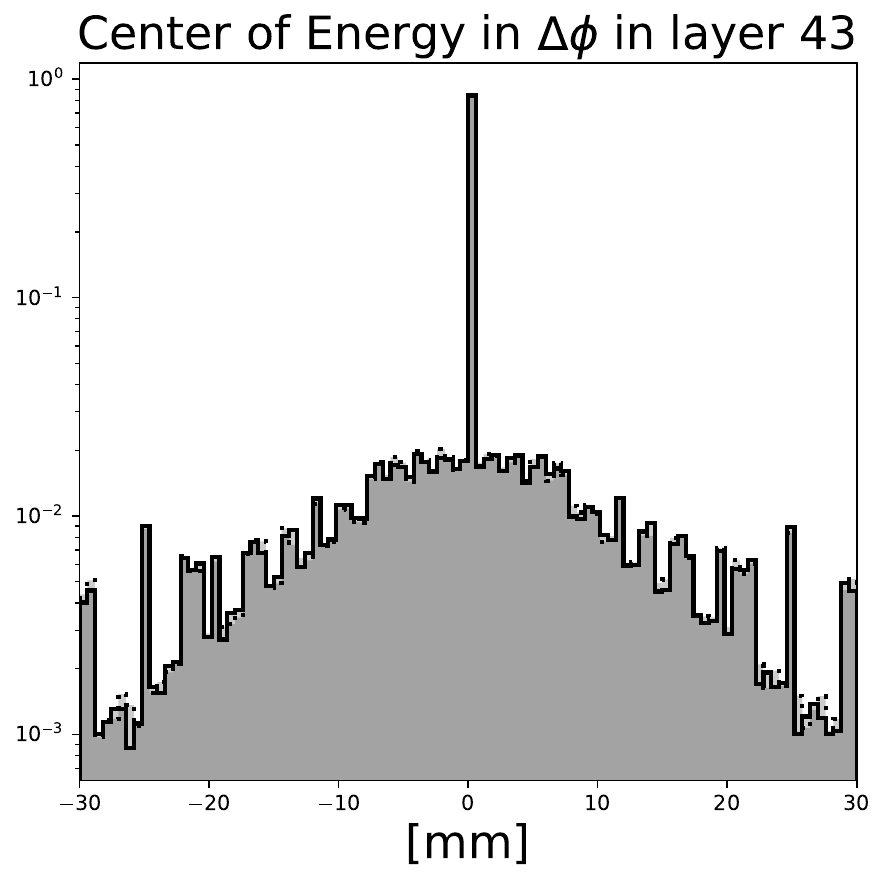} \hfill \includegraphics[height=0.1\textheight]{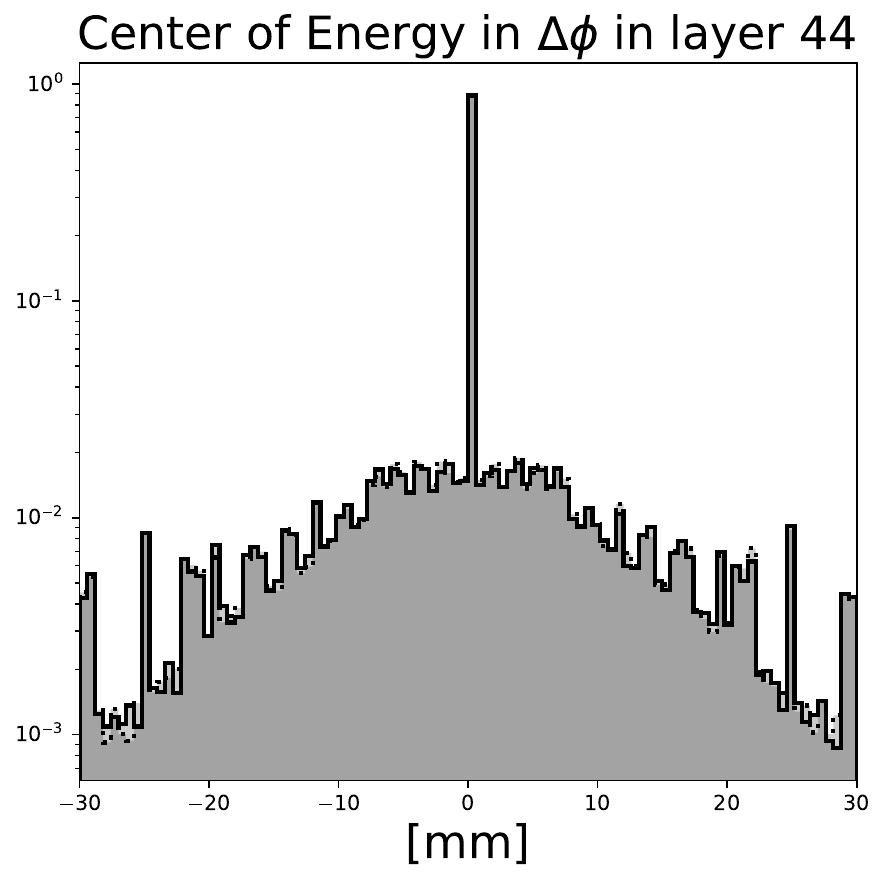}\\
    \includegraphics[width=0.5\textwidth]{figures/Appendix_reference/legend.pdf}
    \caption{Distribution of \geant training and evaluation data in centers of energy in $\phi$ direction for ds2. }
    \label{fig:app_ref.ds2.5}
\end{figure}

\begin{figure}[ht]
    \centering
    \includegraphics[height=0.1\textheight]{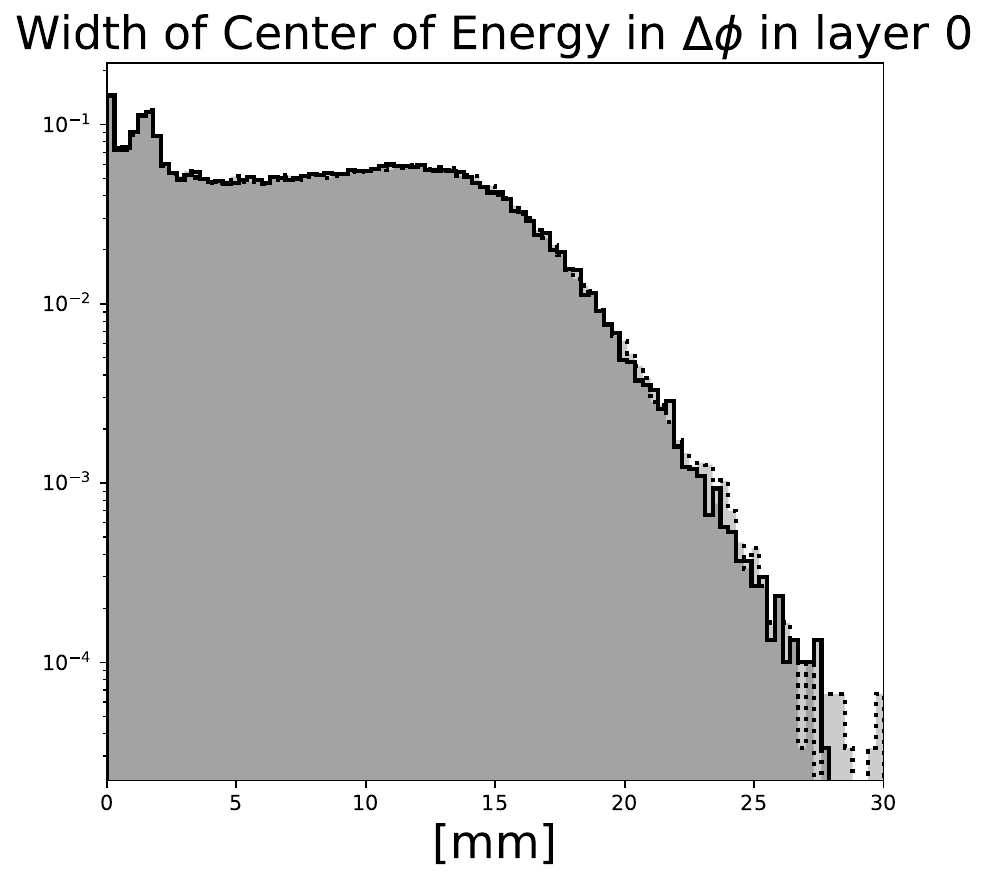} \hfill \includegraphics[height=0.1\textheight]{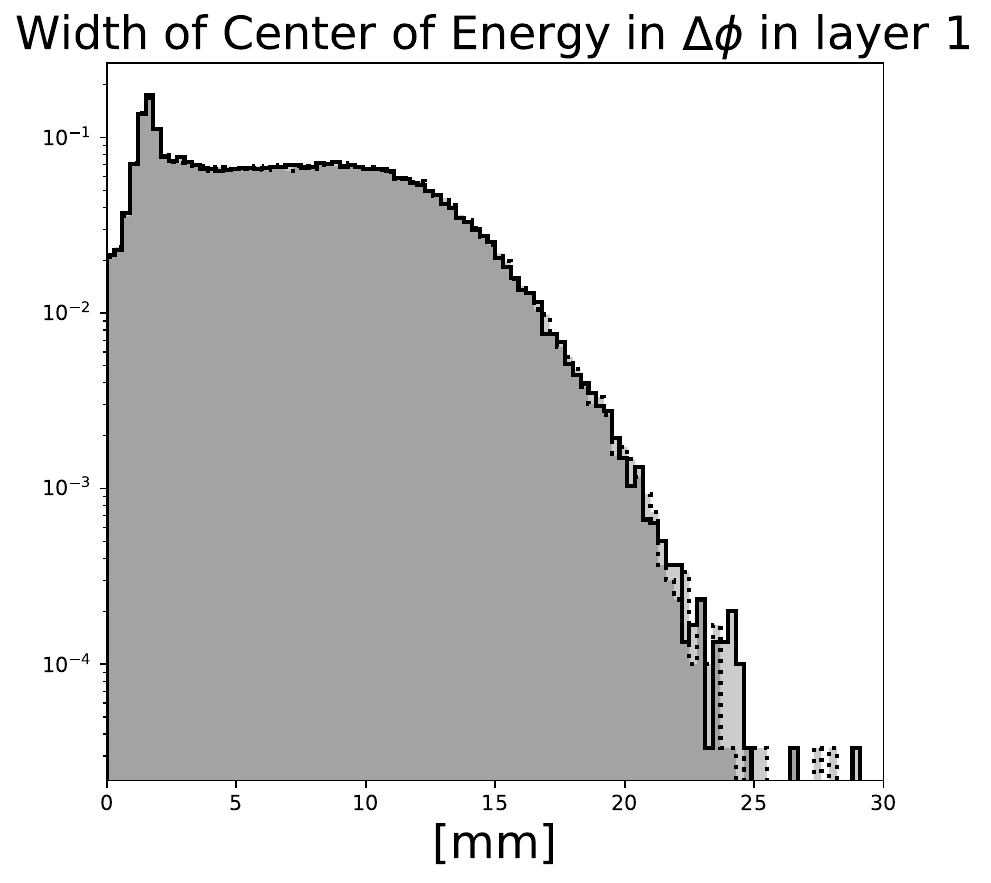} \hfill \includegraphics[height=0.1\textheight]{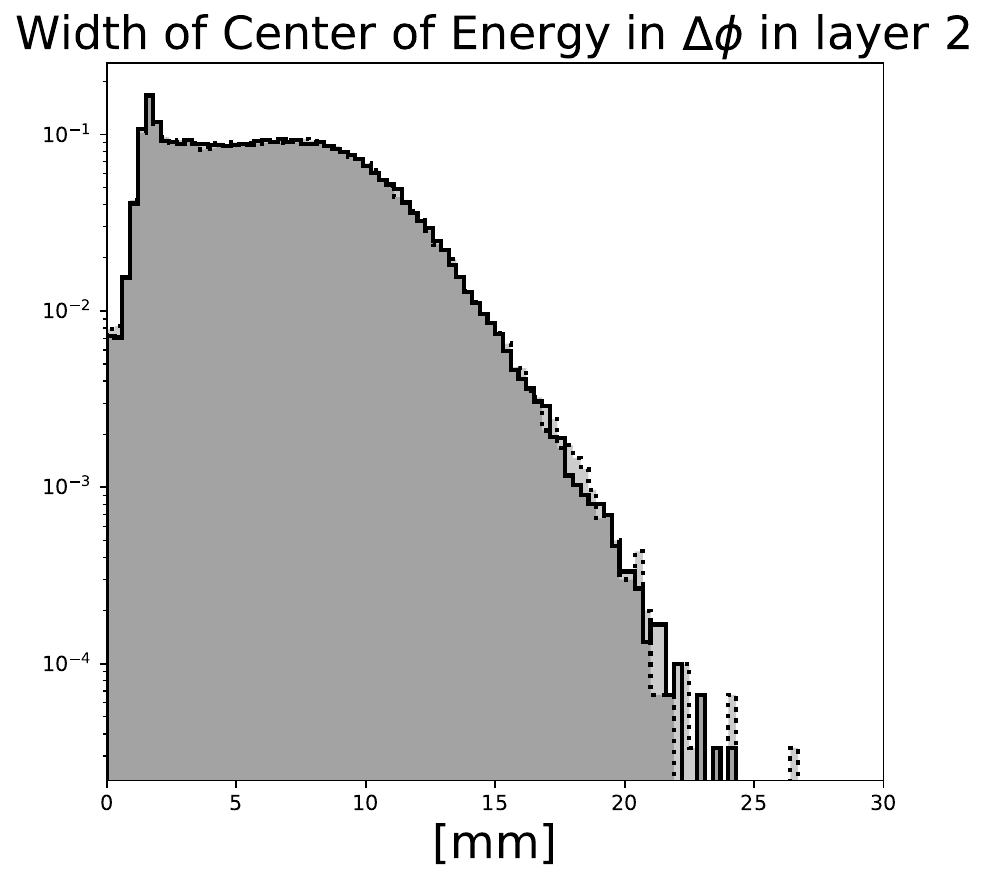} \hfill \includegraphics[height=0.1\textheight]{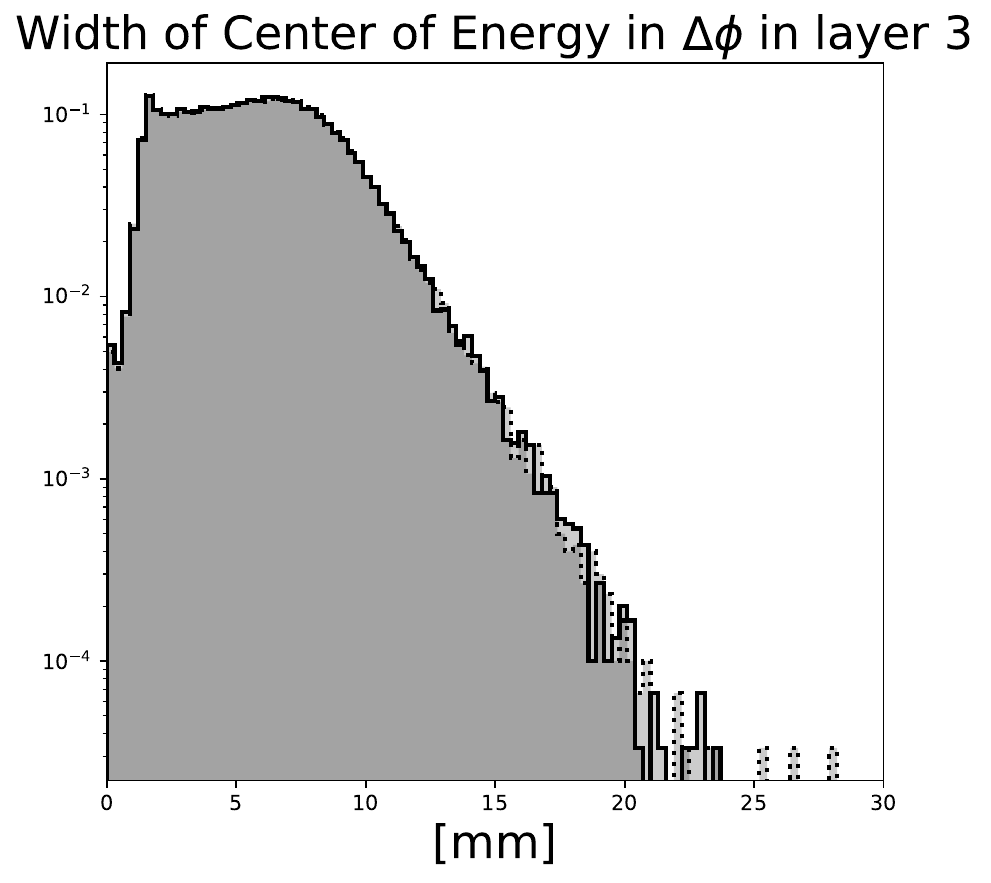} \hfill \includegraphics[height=0.1\textheight]{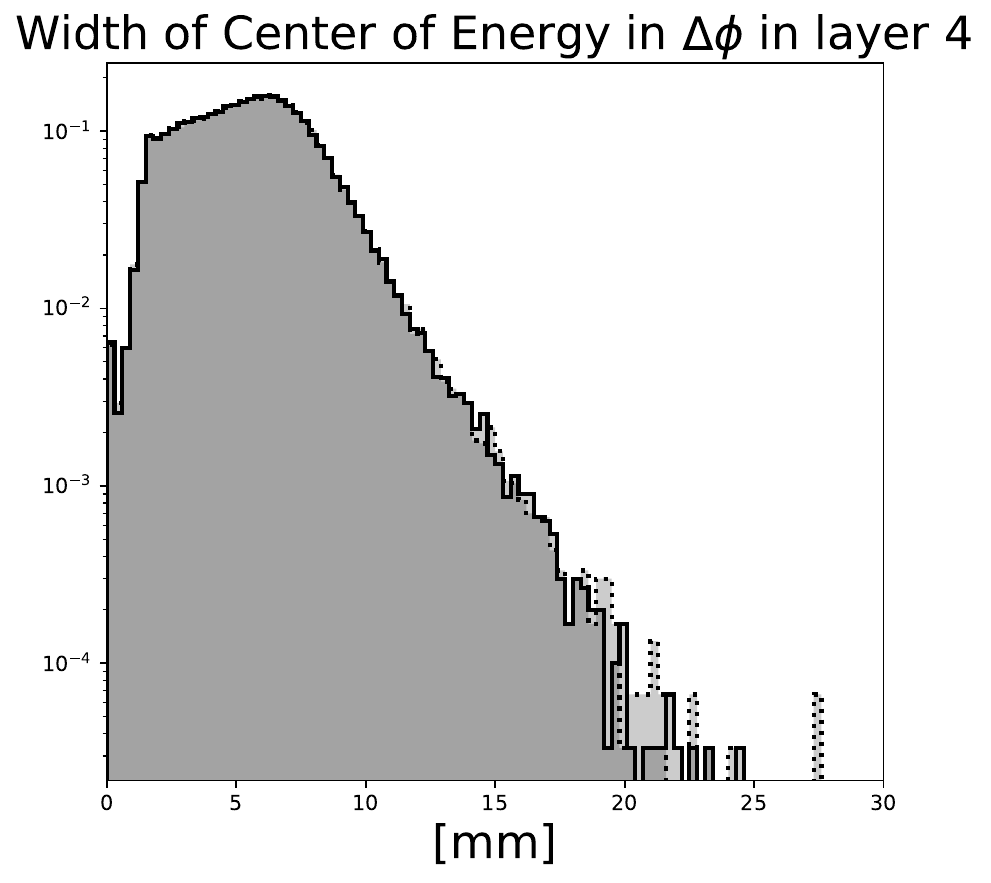}\\
    \includegraphics[height=0.1\textheight]{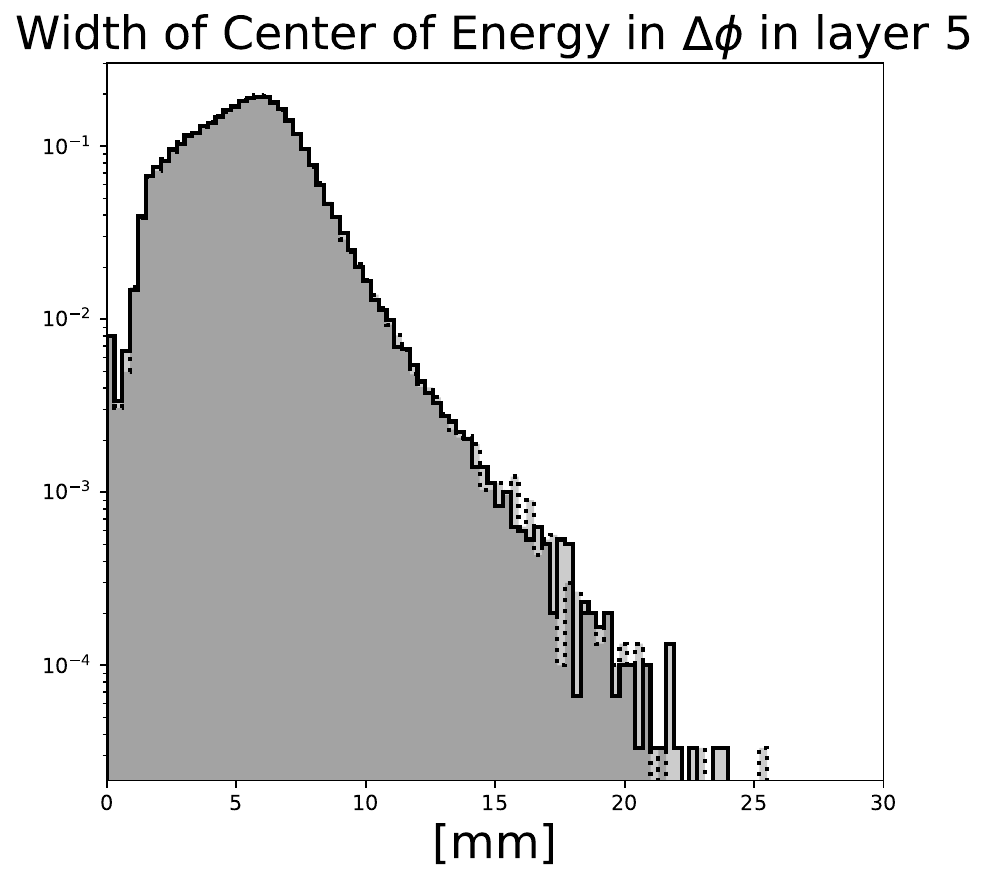} \hfill \includegraphics[height=0.1\textheight]{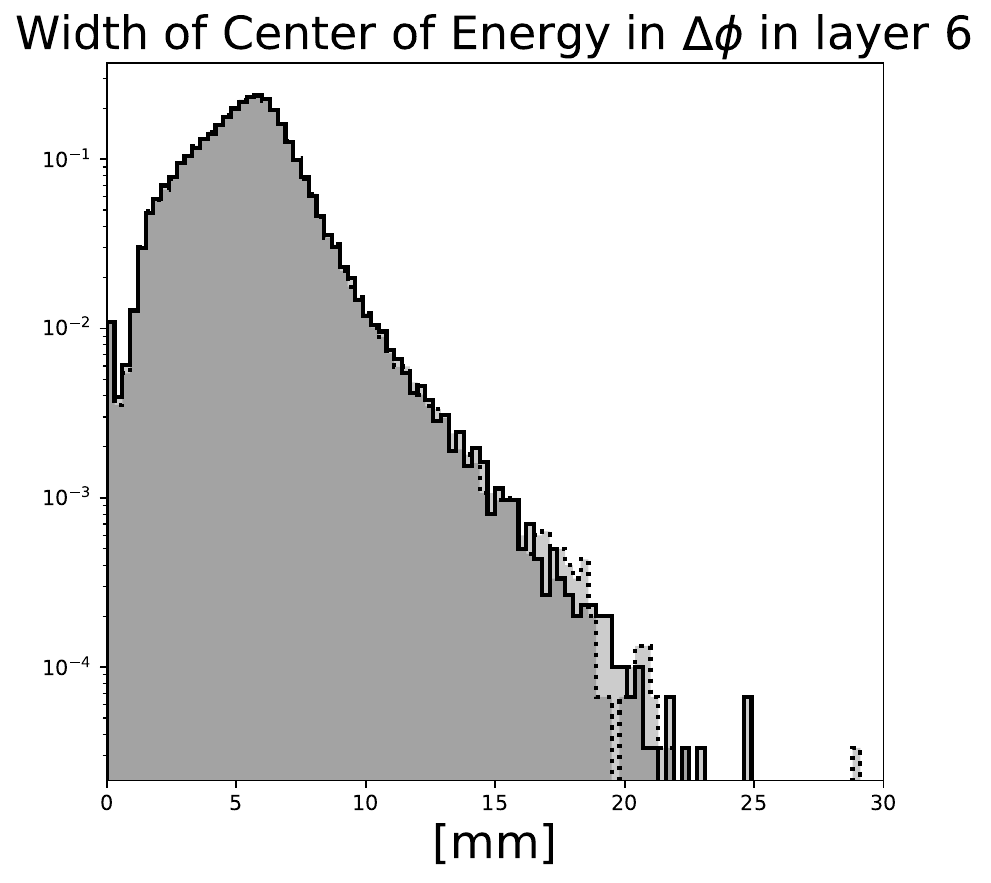} \hfill \includegraphics[height=0.1\textheight]{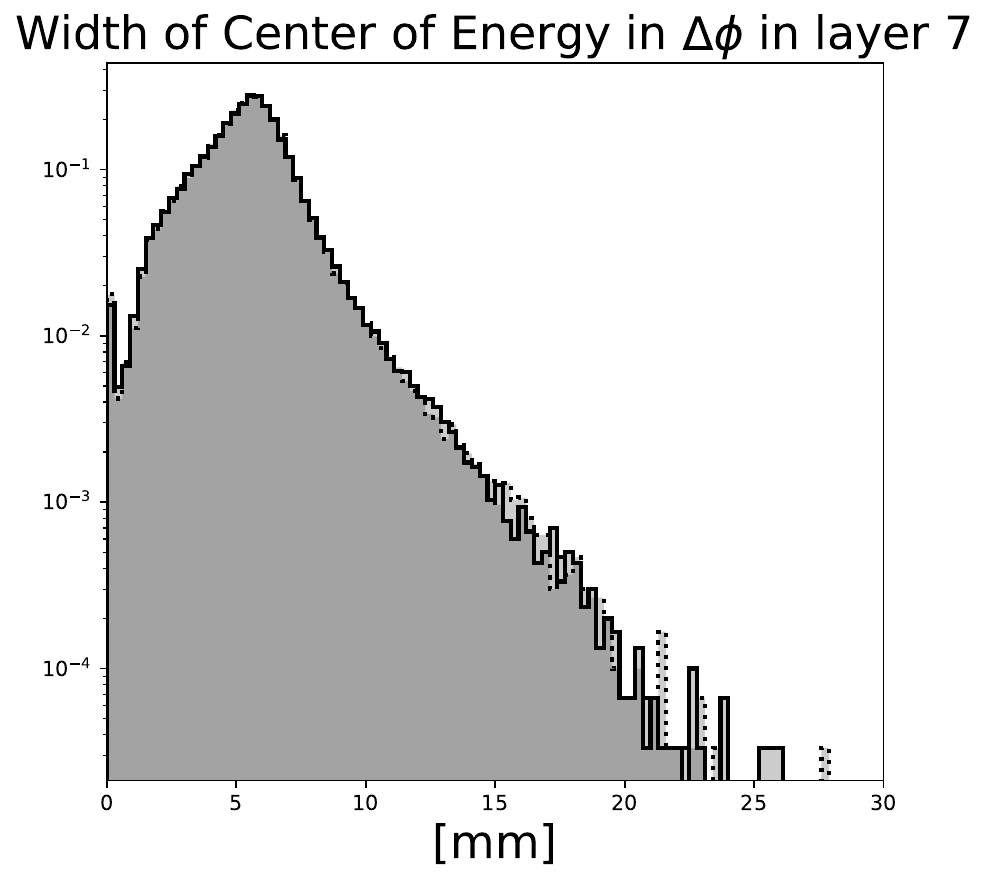} \hfill \includegraphics[height=0.1\textheight]{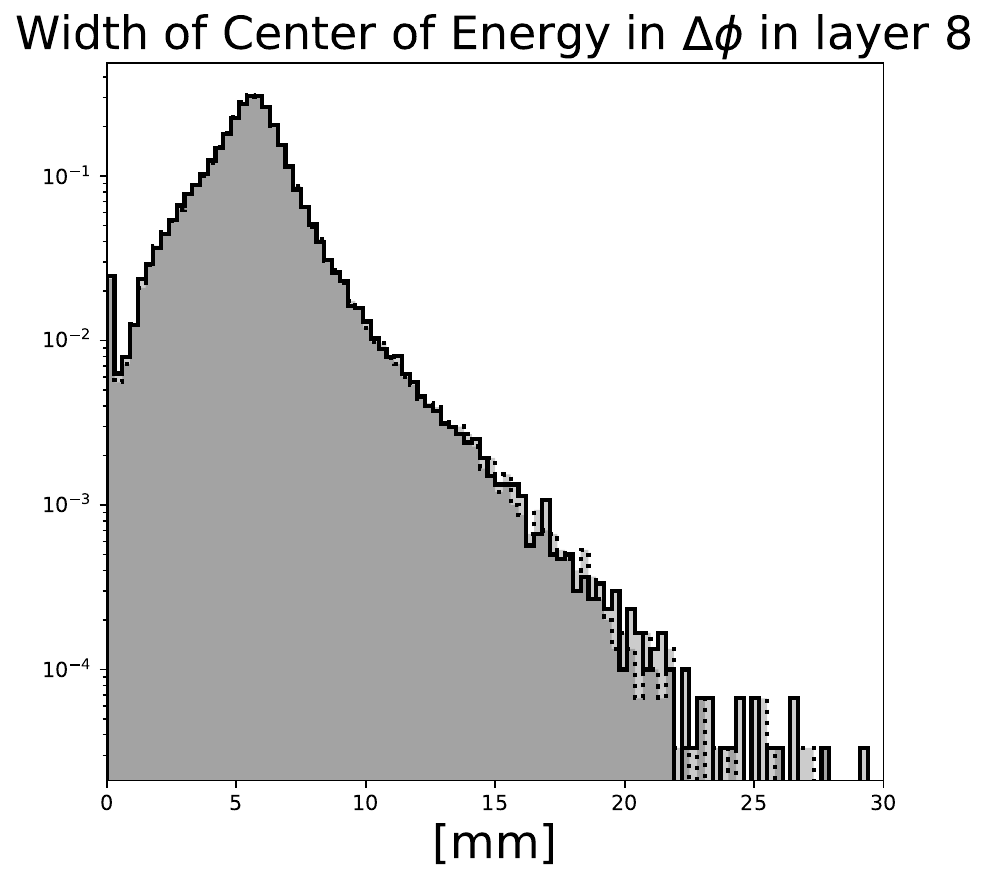} \hfill \includegraphics[height=0.1\textheight]{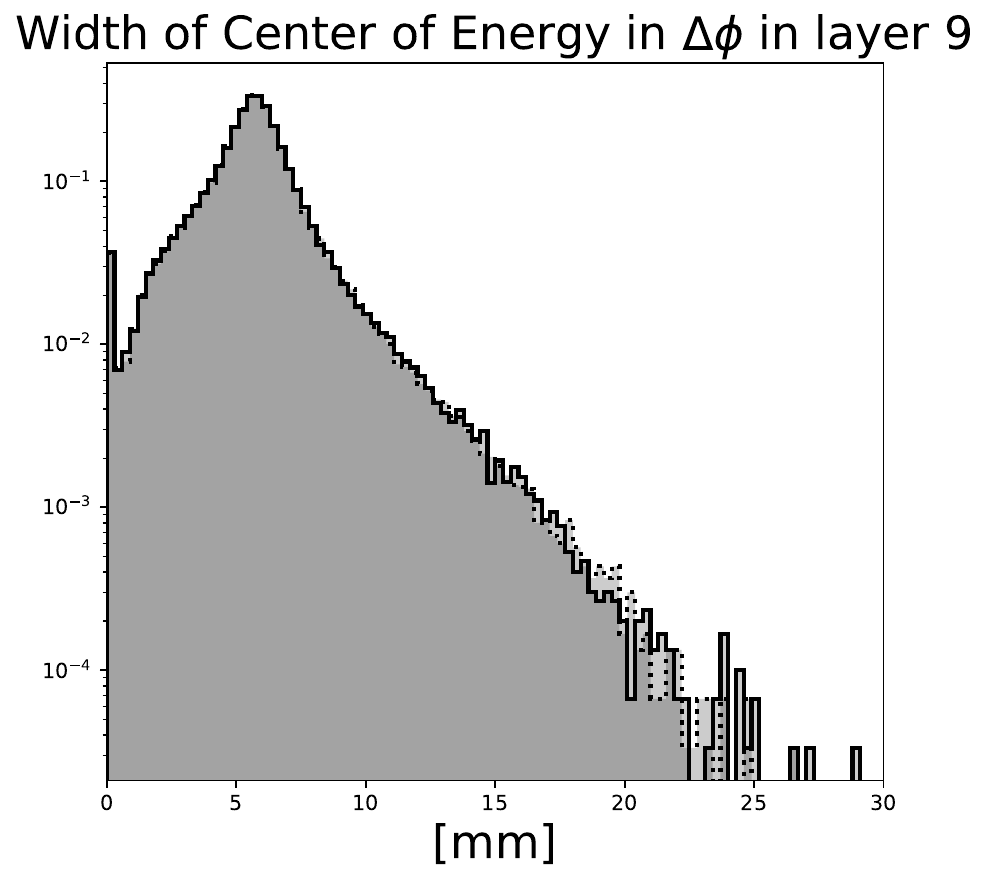}\\
    \includegraphics[height=0.1\textheight]{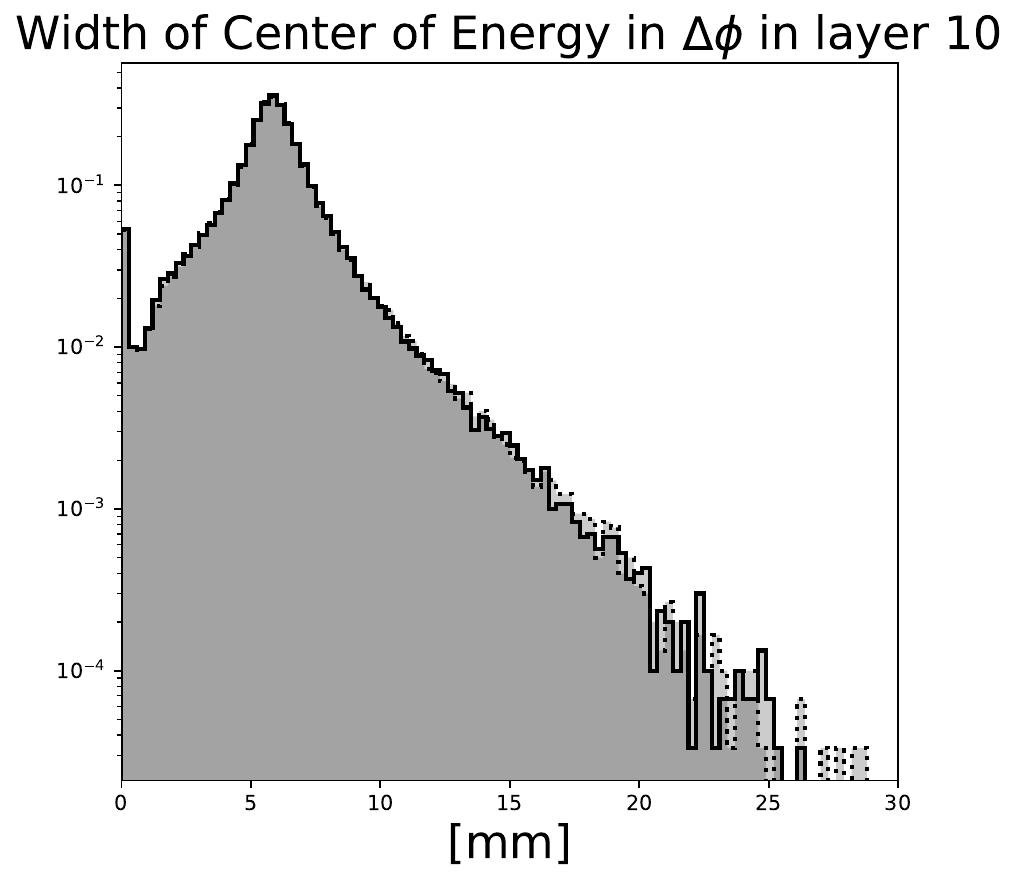} \hfill \includegraphics[height=0.1\textheight]{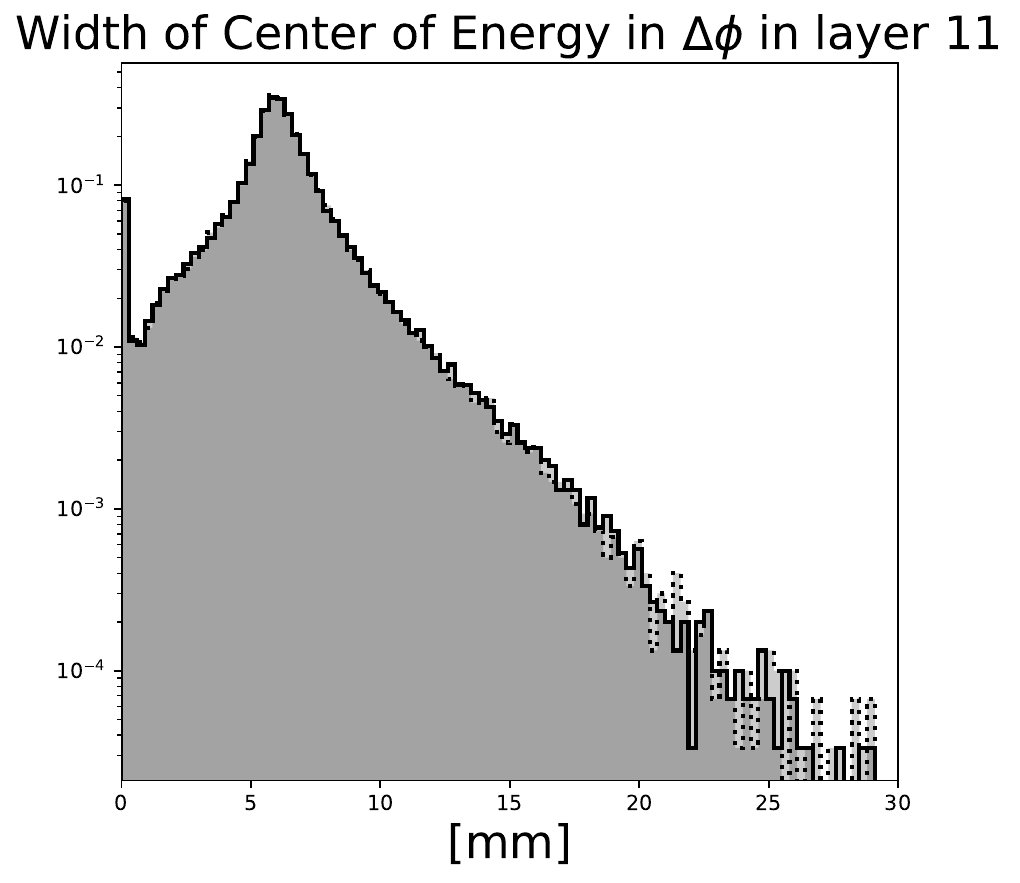} \hfill \includegraphics[height=0.1\textheight]{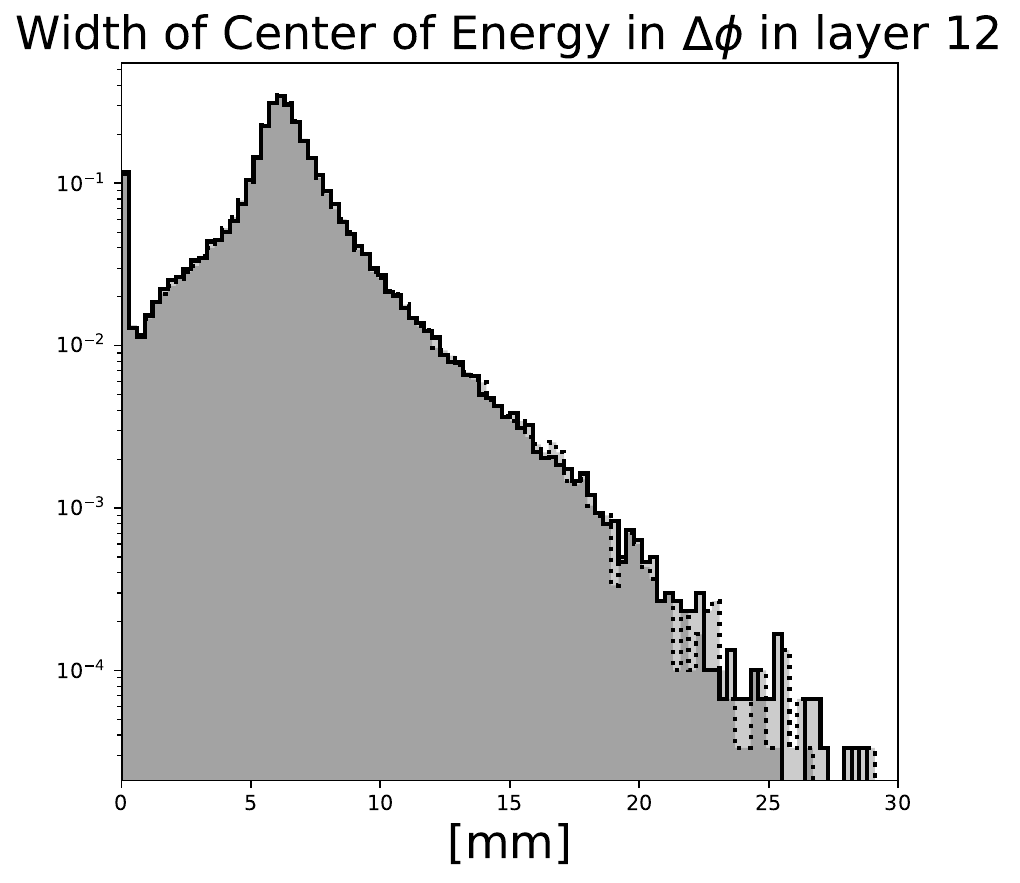} \hfill \includegraphics[height=0.1\textheight]{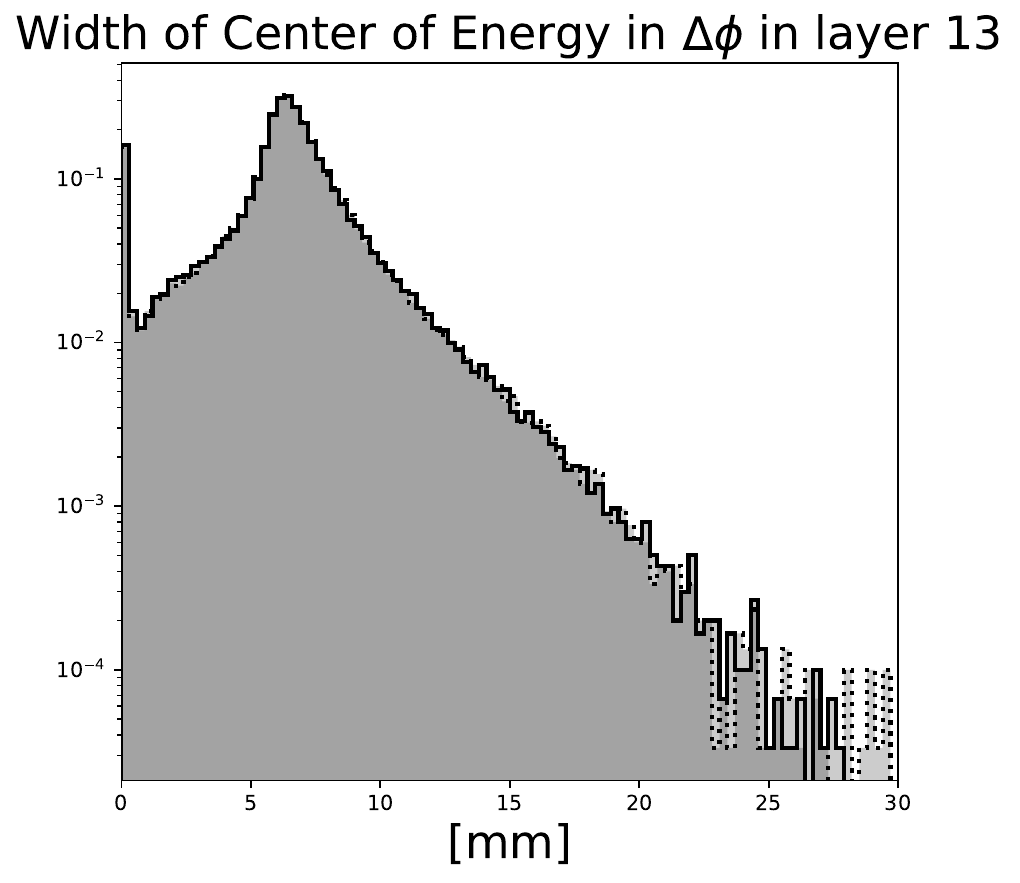} \hfill \includegraphics[height=0.1\textheight]{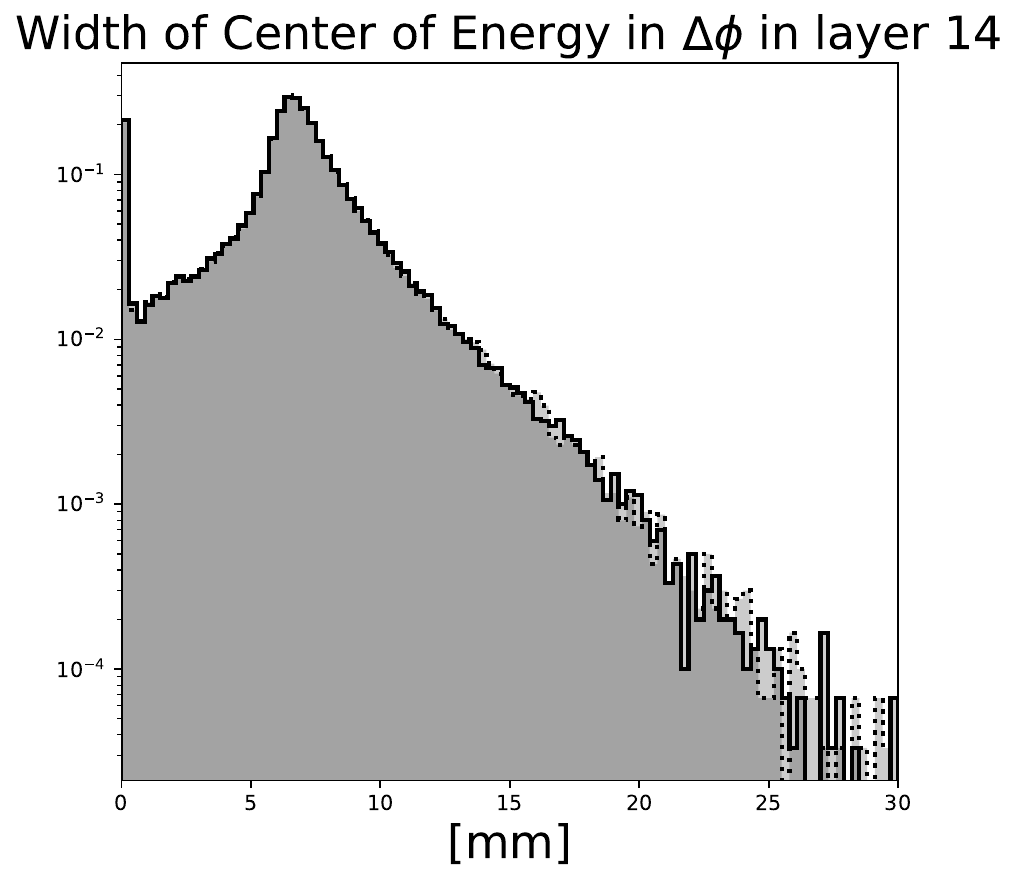}\\
    \includegraphics[height=0.1\textheight]{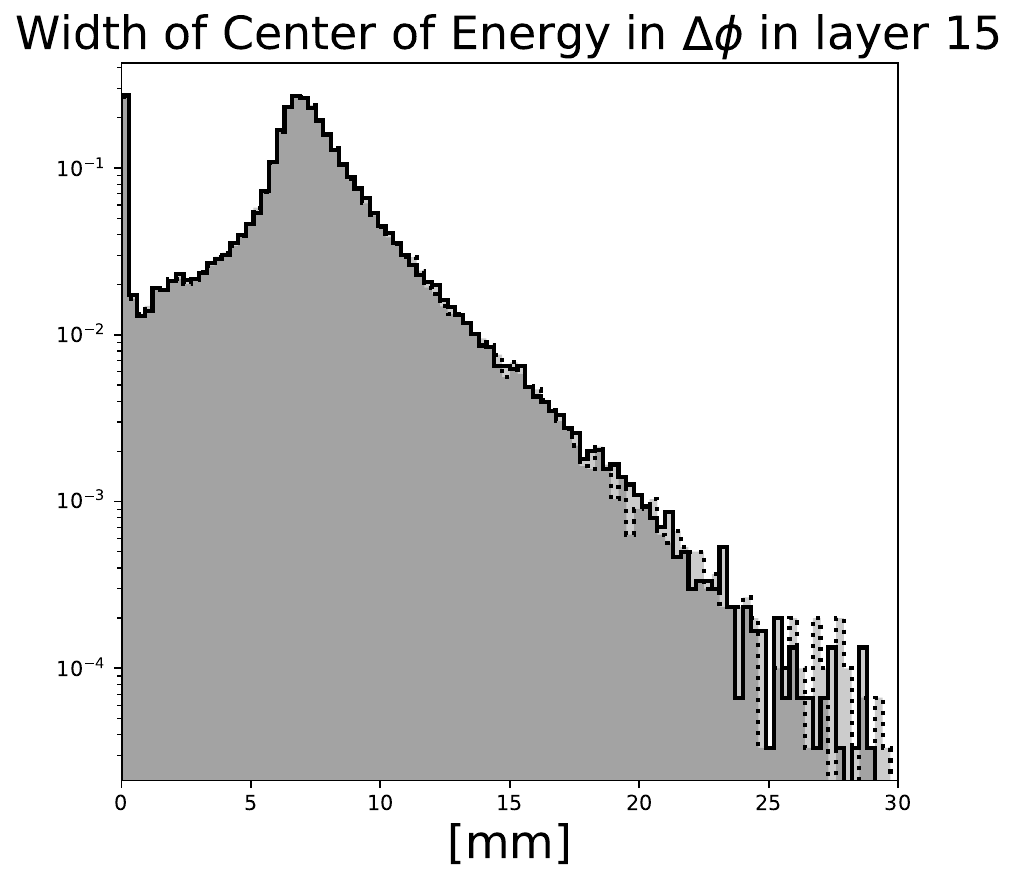} \hfill \includegraphics[height=0.1\textheight]{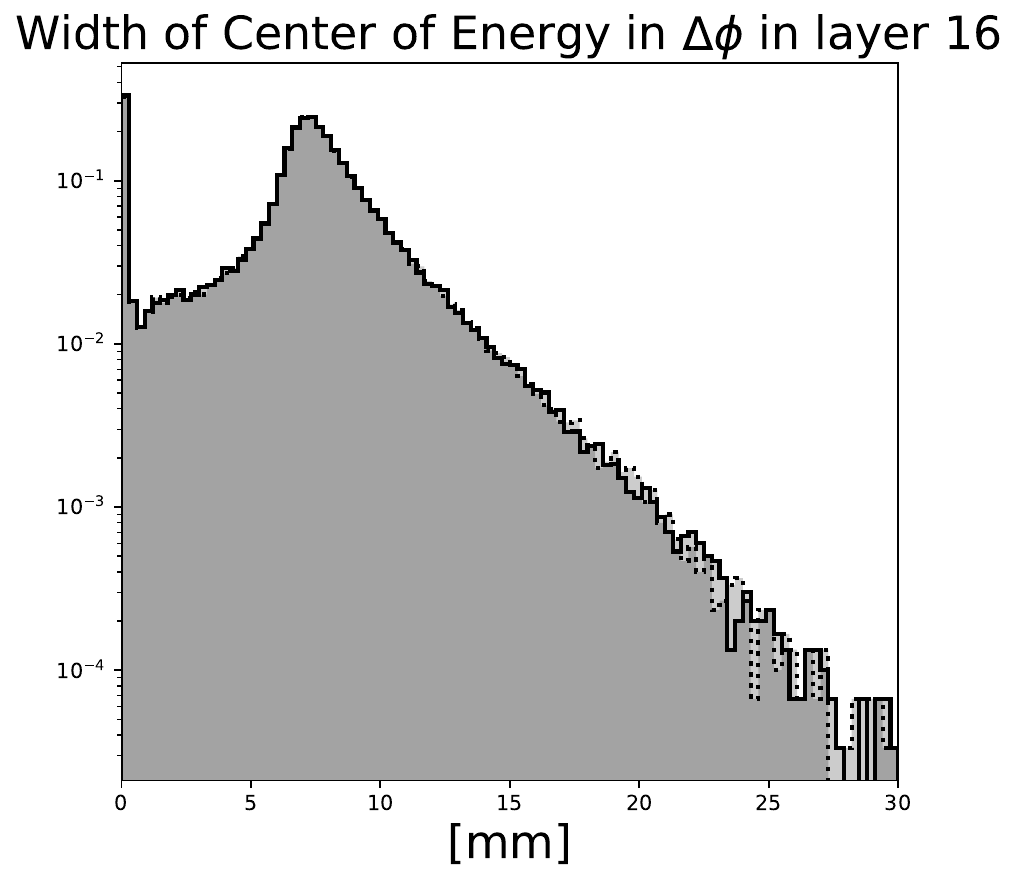} \hfill \includegraphics[height=0.1\textheight]{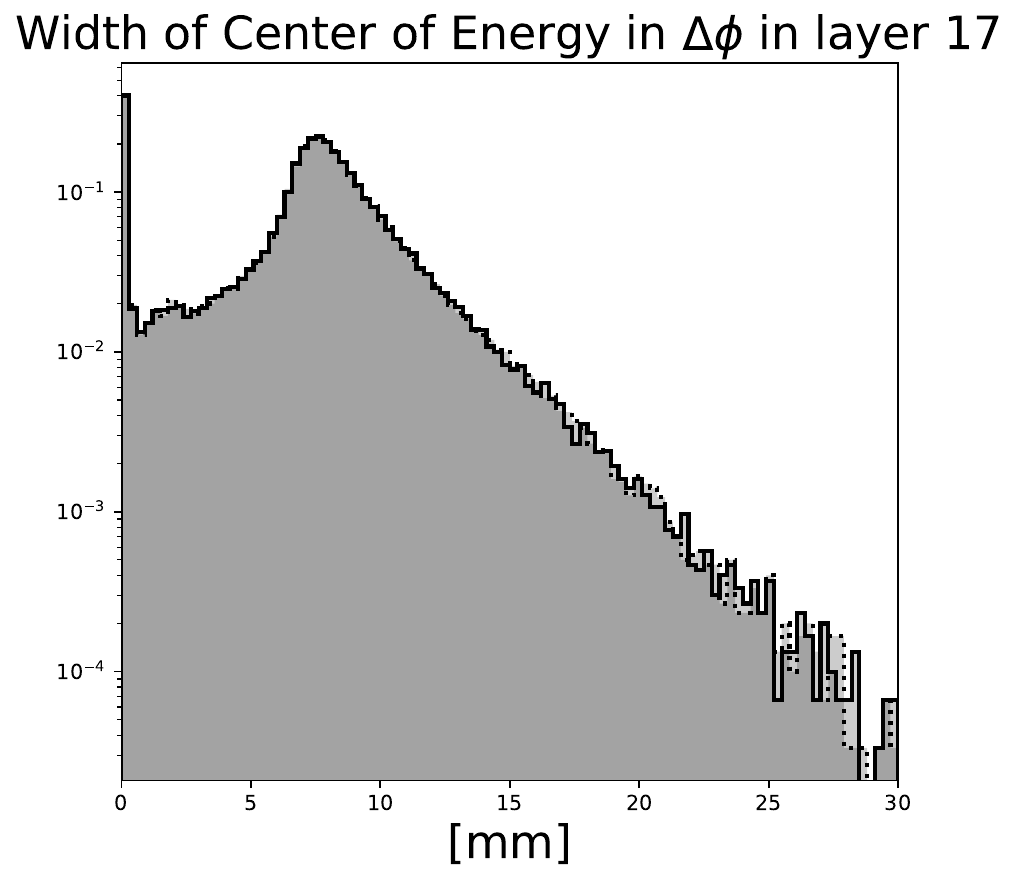} \hfill \includegraphics[height=0.1\textheight]{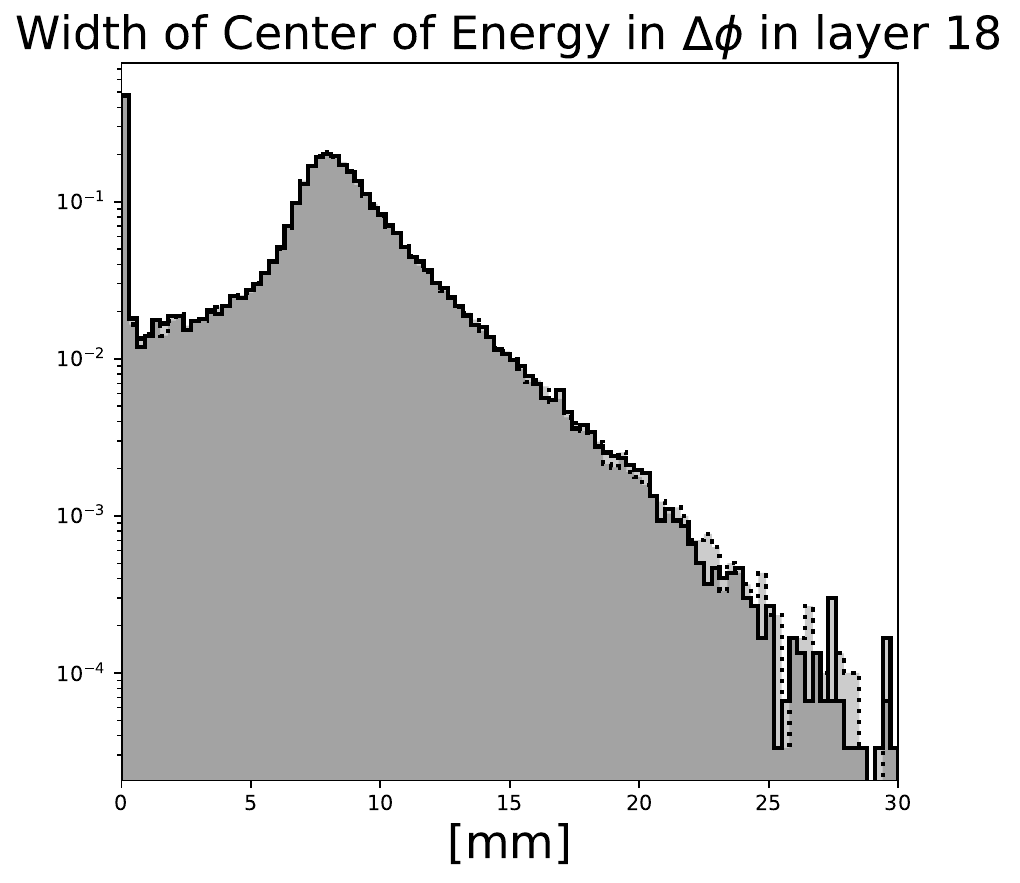} \hfill \includegraphics[height=0.1\textheight]{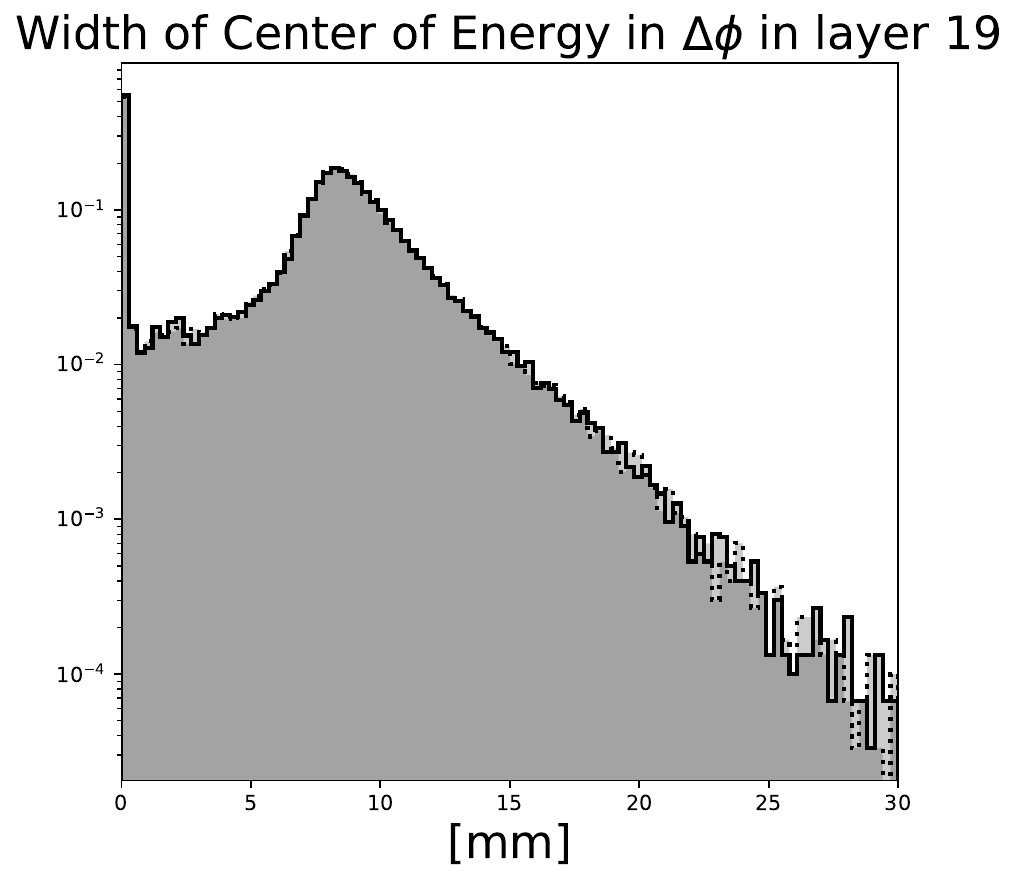}\\
    \includegraphics[height=0.1\textheight]{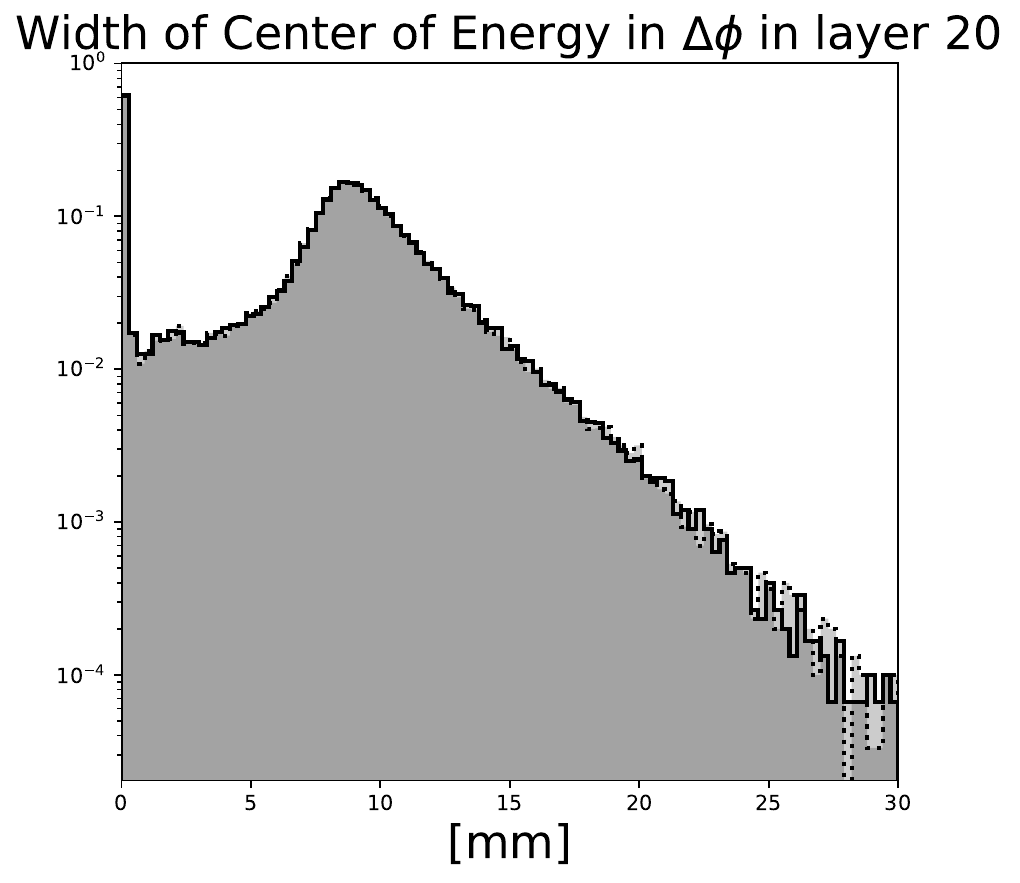} \hfill \includegraphics[height=0.1\textheight]{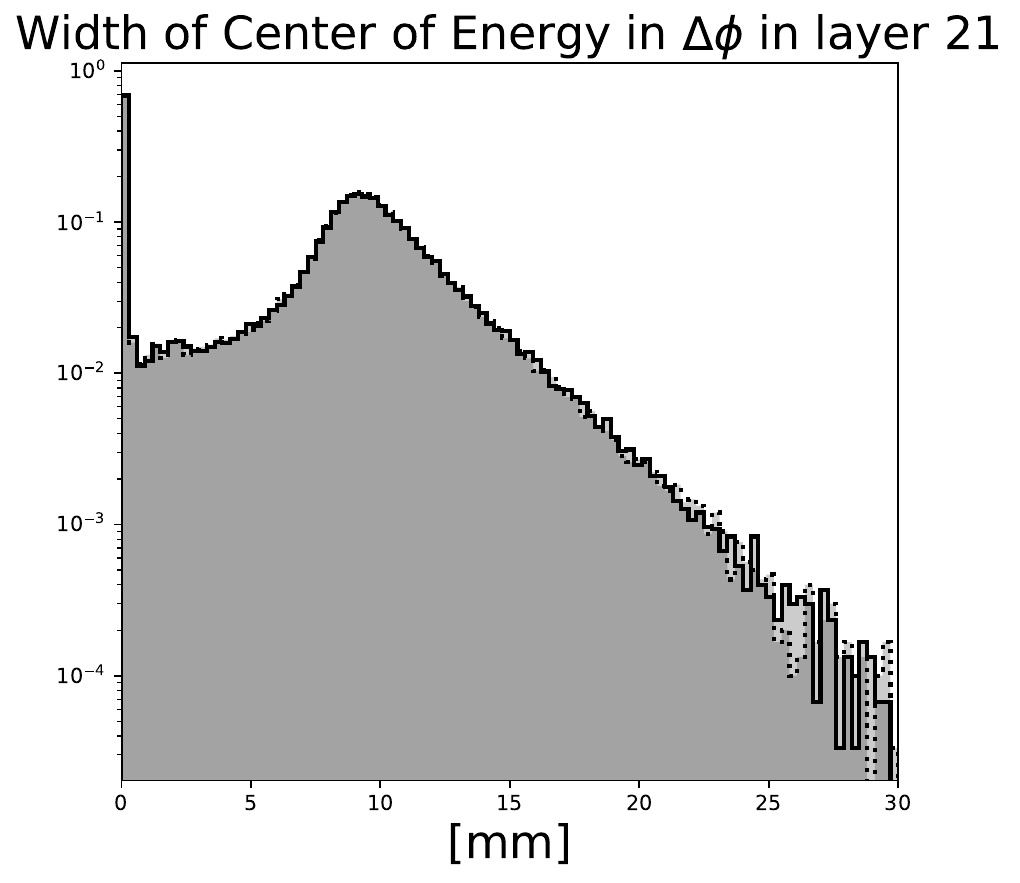} \hfill \includegraphics[height=0.1\textheight]{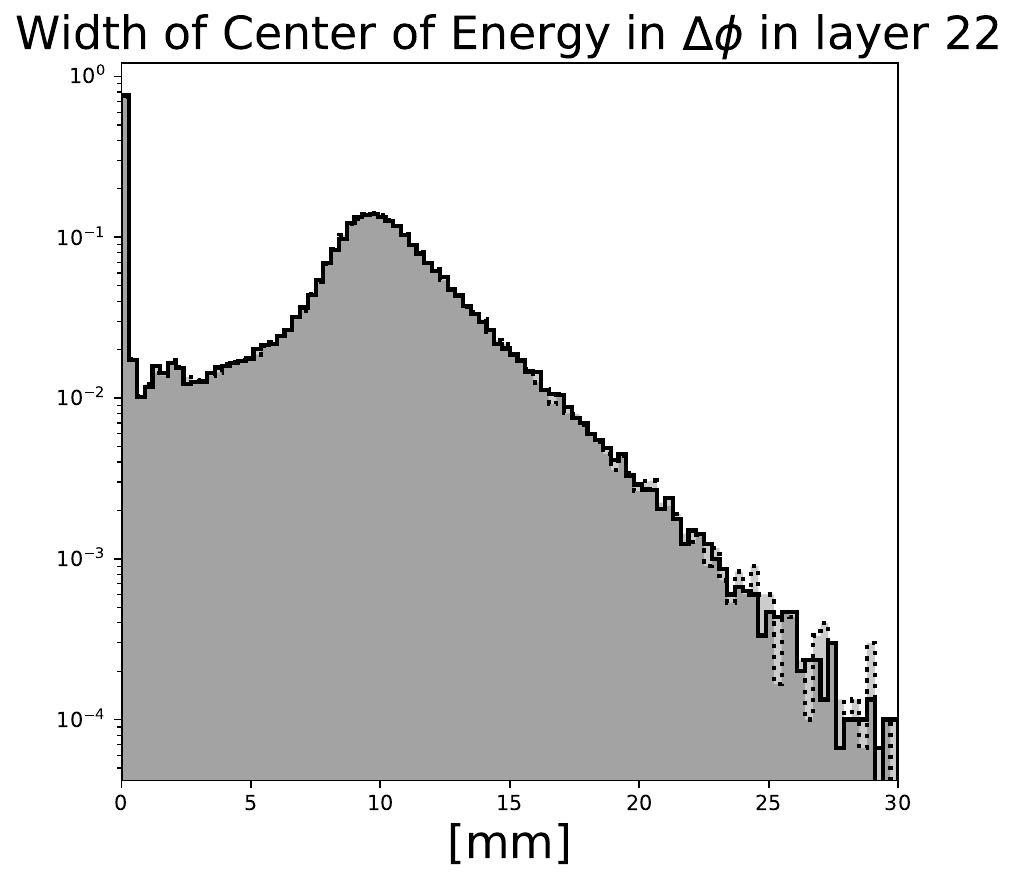} \hfill \includegraphics[height=0.1\textheight]{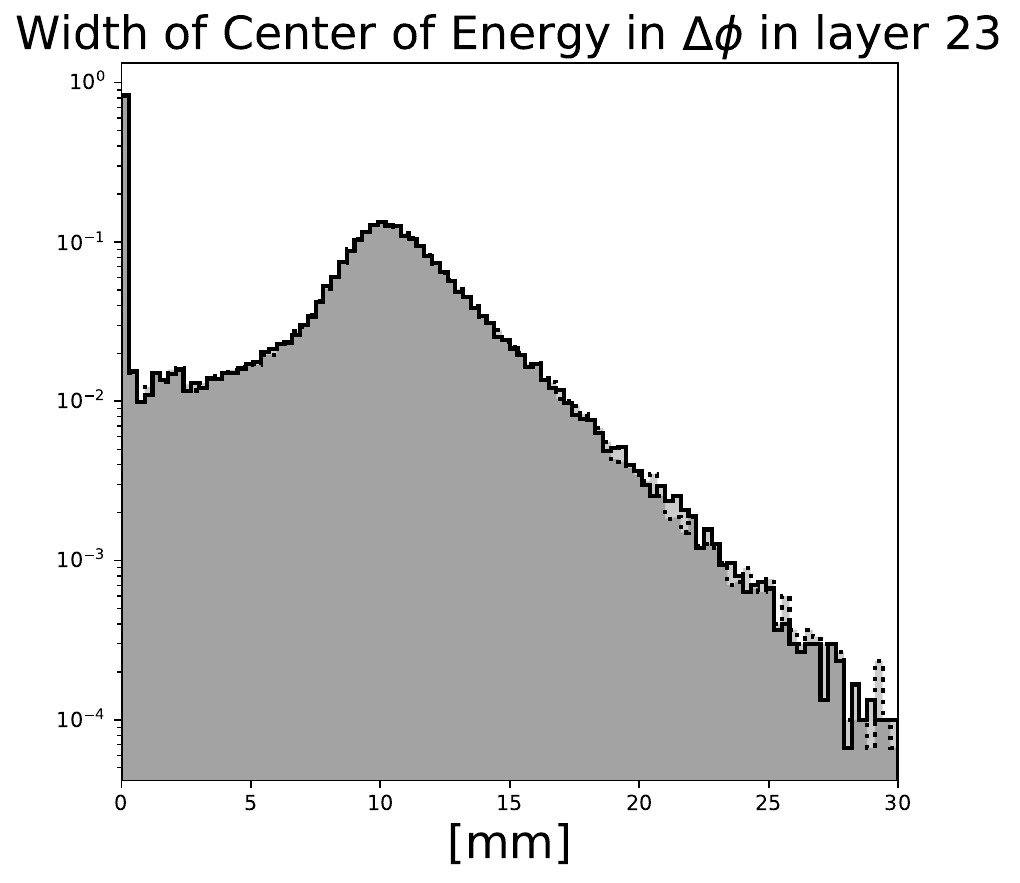} \hfill \includegraphics[height=0.1\textheight]{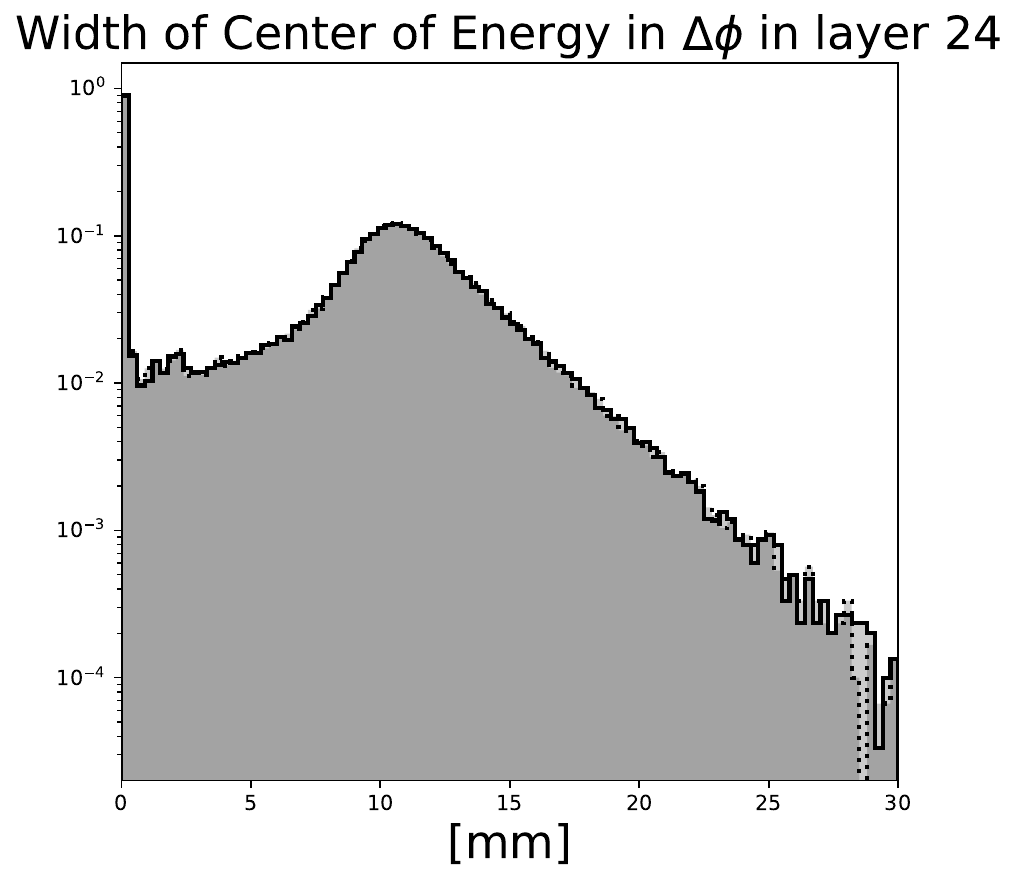}\\
    \includegraphics[height=0.1\textheight]{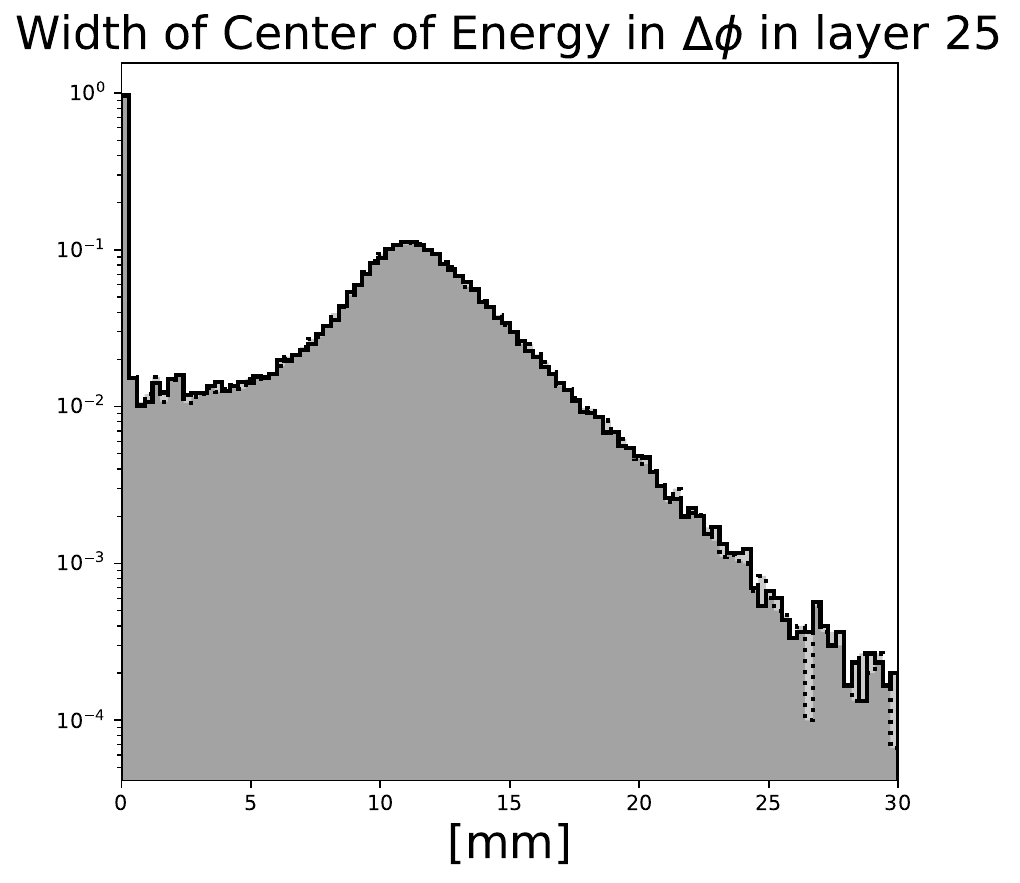} \hfill \includegraphics[height=0.1\textheight]{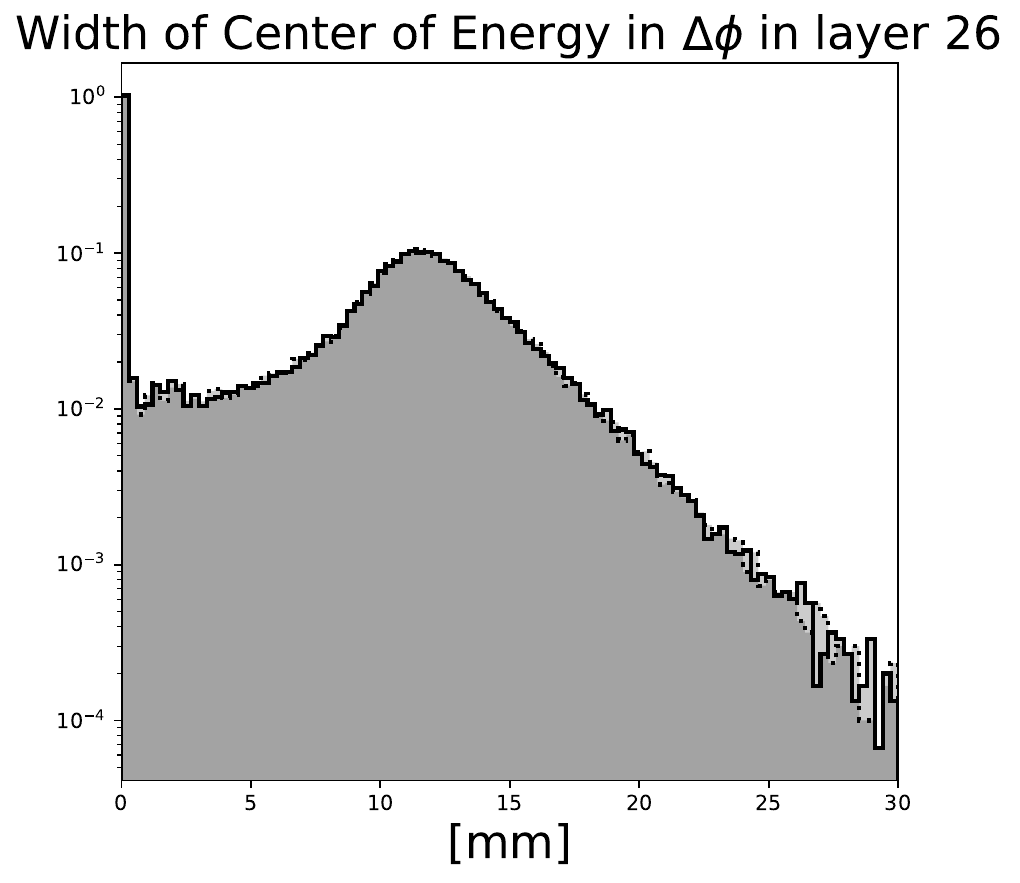} \hfill \includegraphics[height=0.1\textheight]{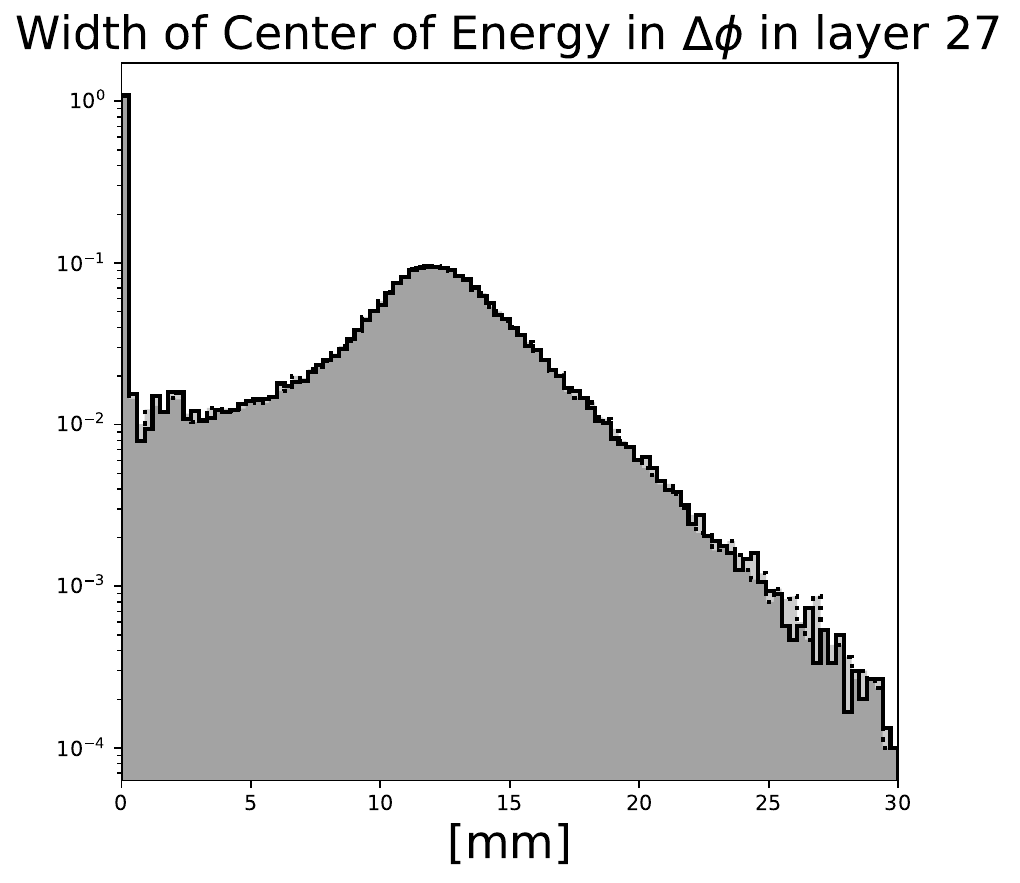} \hfill \includegraphics[height=0.1\textheight]{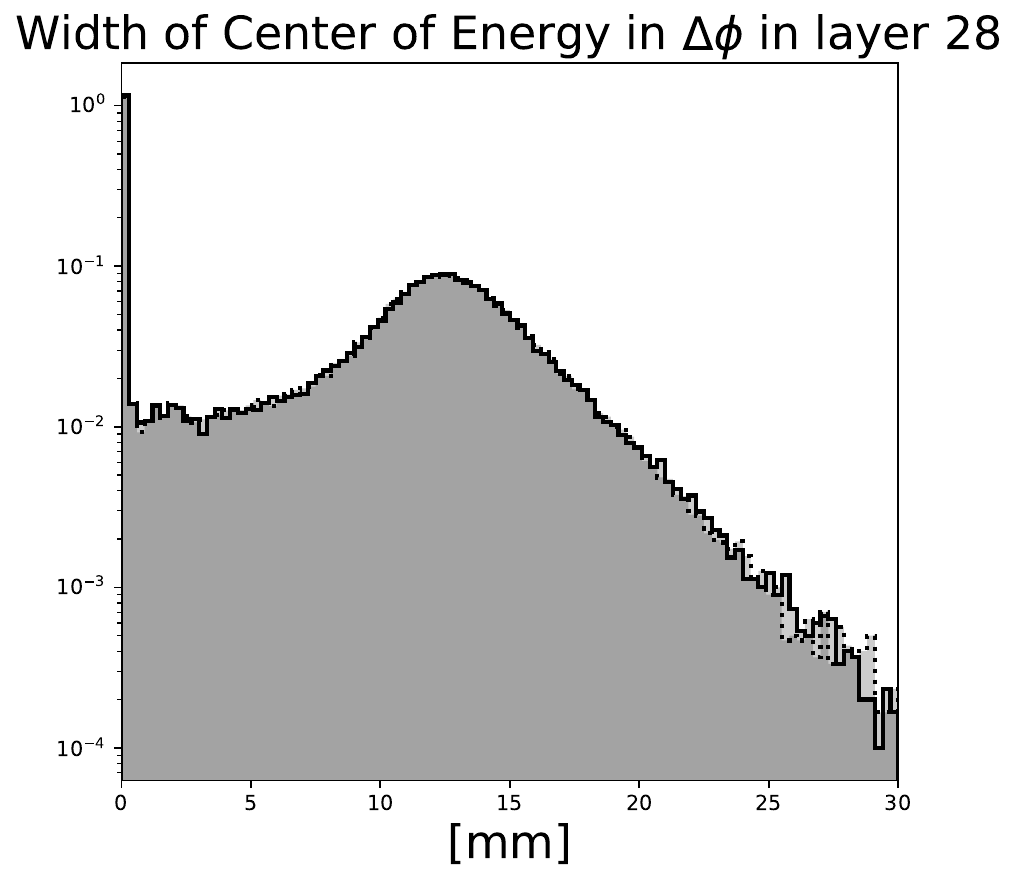} \hfill \includegraphics[height=0.1\textheight]{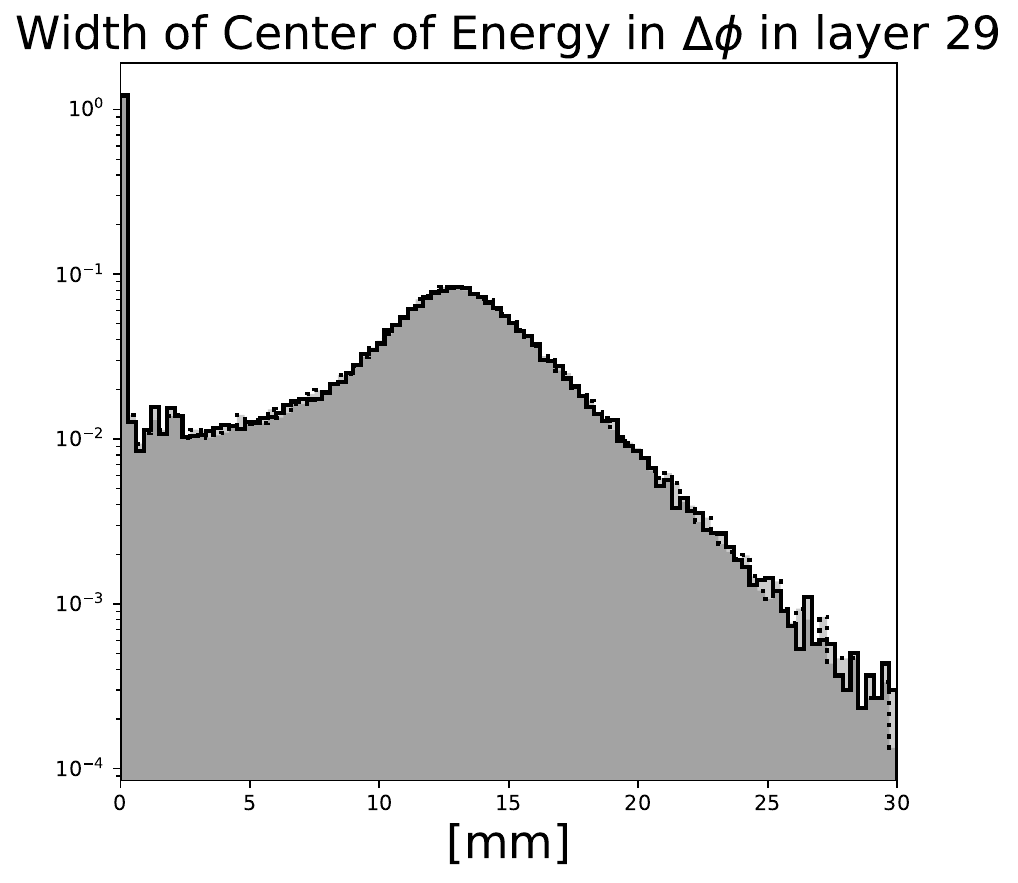}\\
    \includegraphics[height=0.1\textheight]{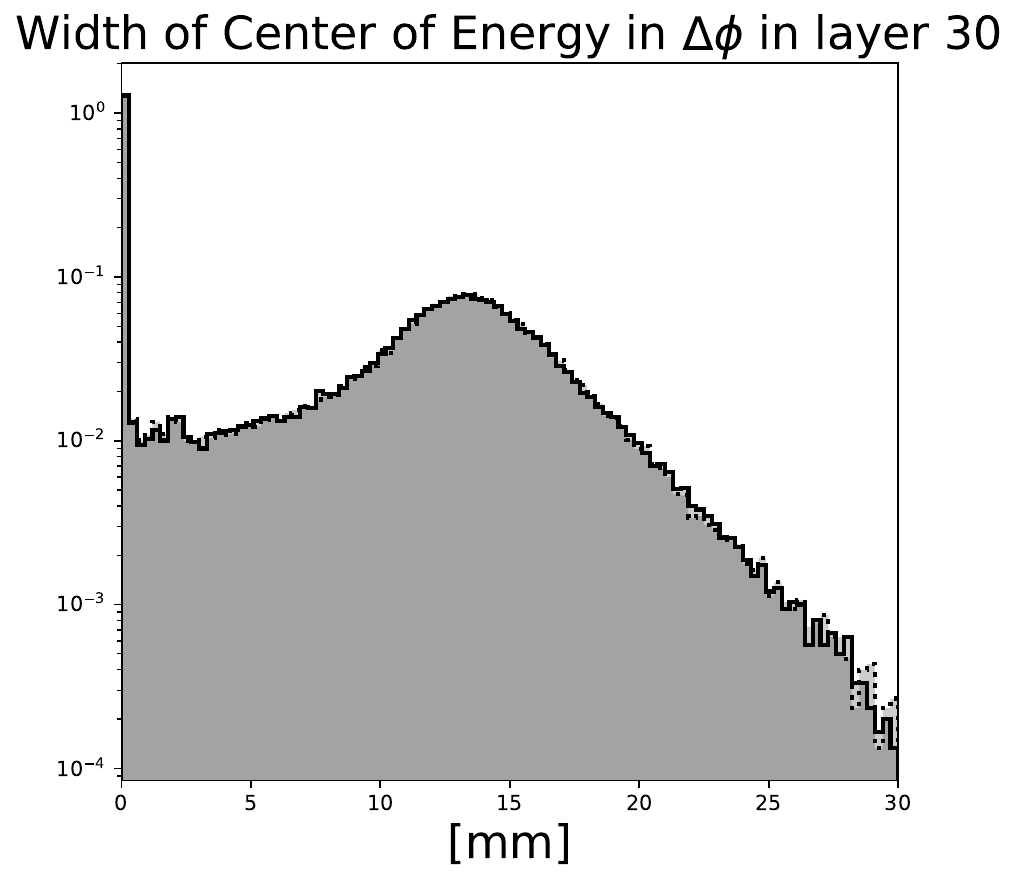} \hfill \includegraphics[height=0.1\textheight]{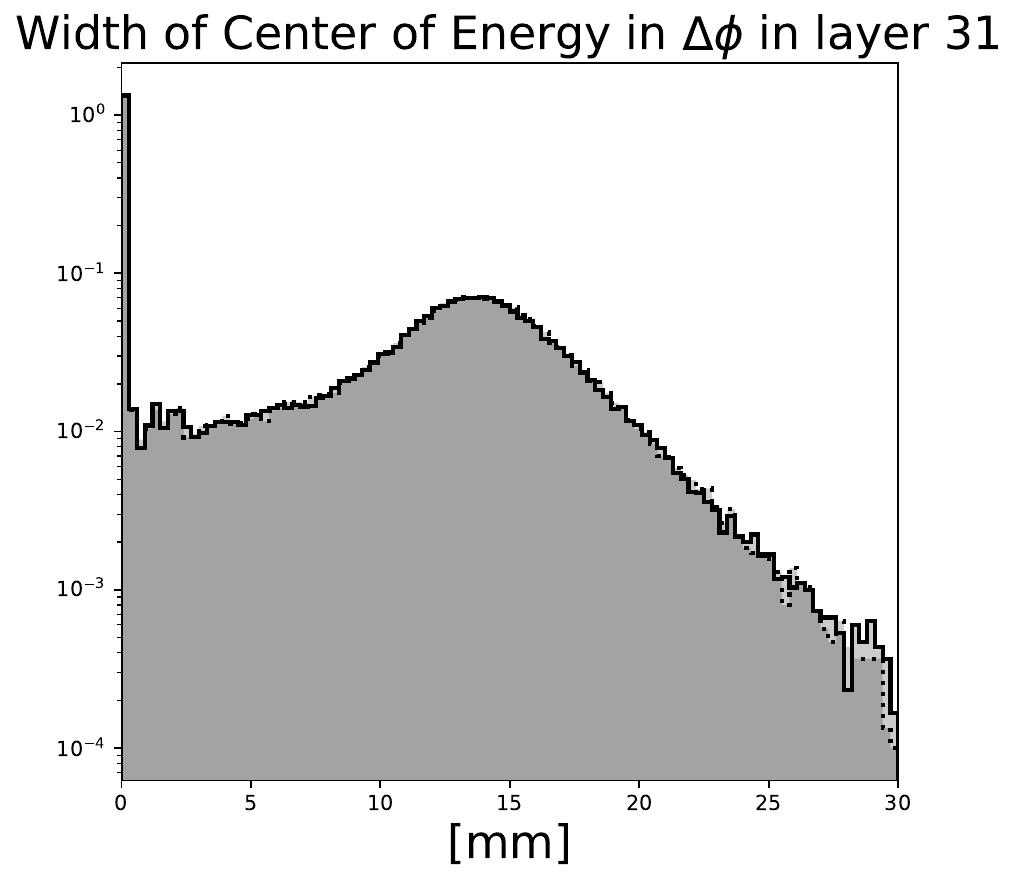} \hfill \includegraphics[height=0.1\textheight]{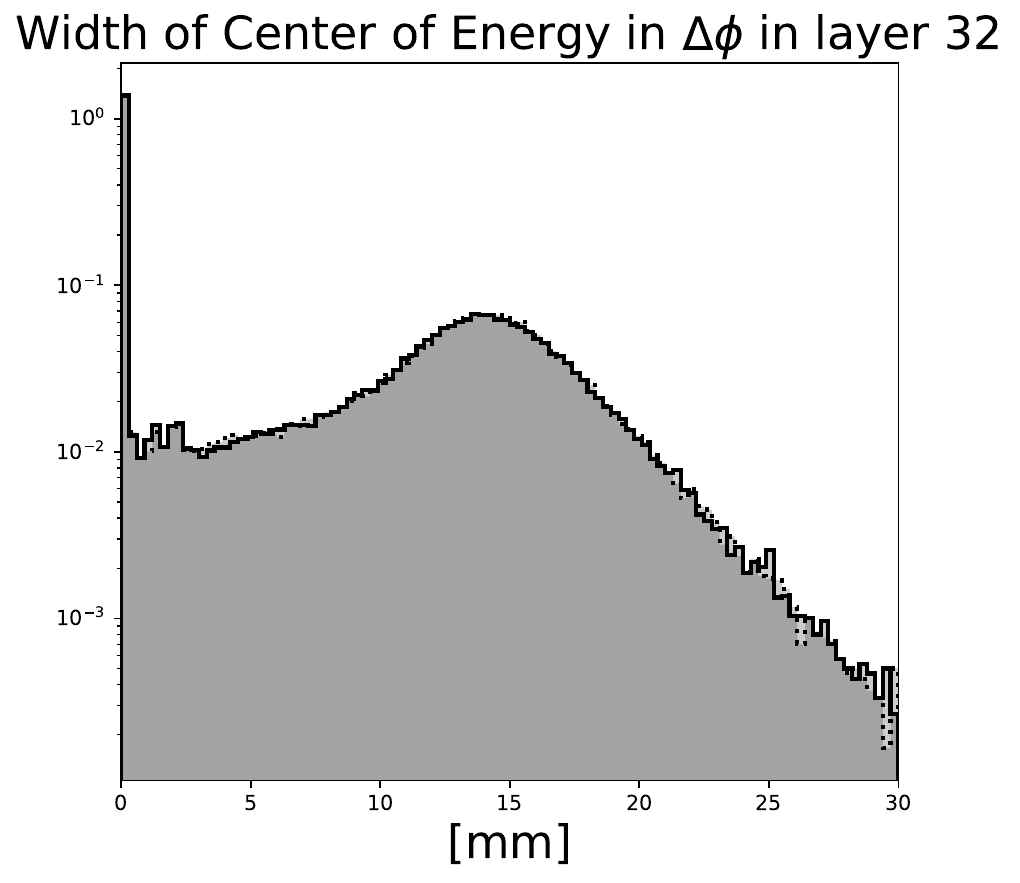} \hfill \includegraphics[height=0.1\textheight]{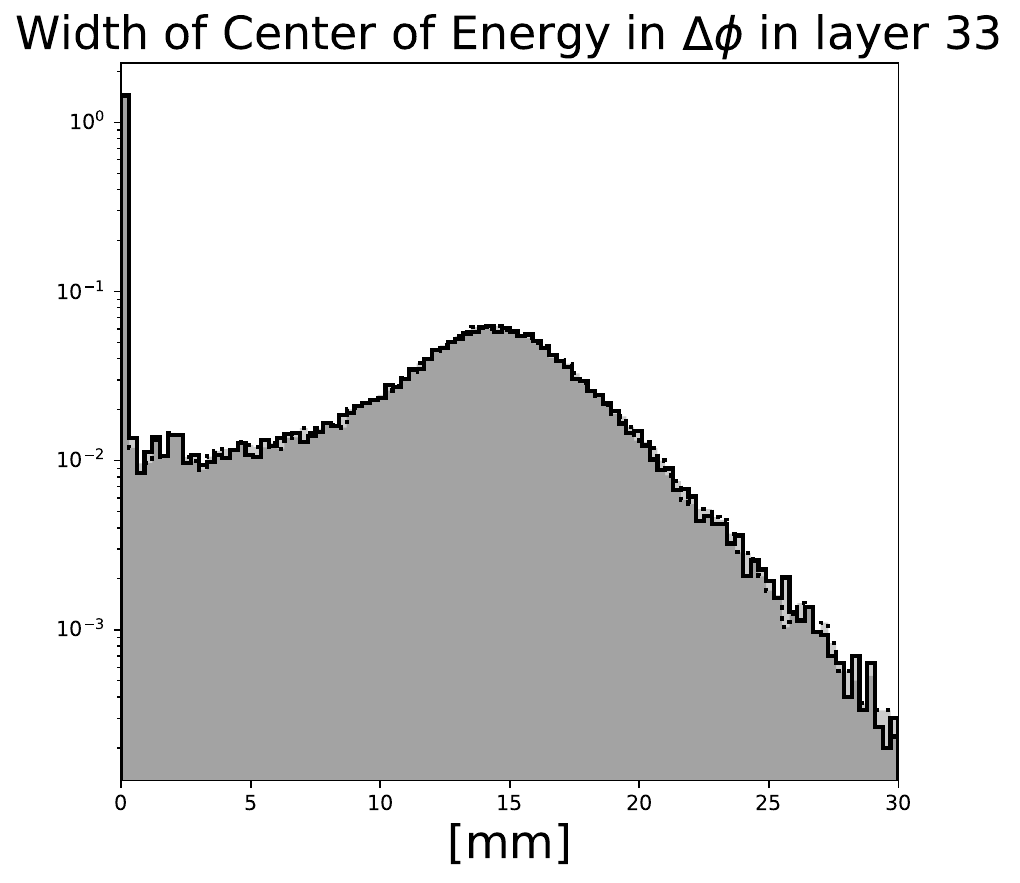} \hfill \includegraphics[height=0.1\textheight]{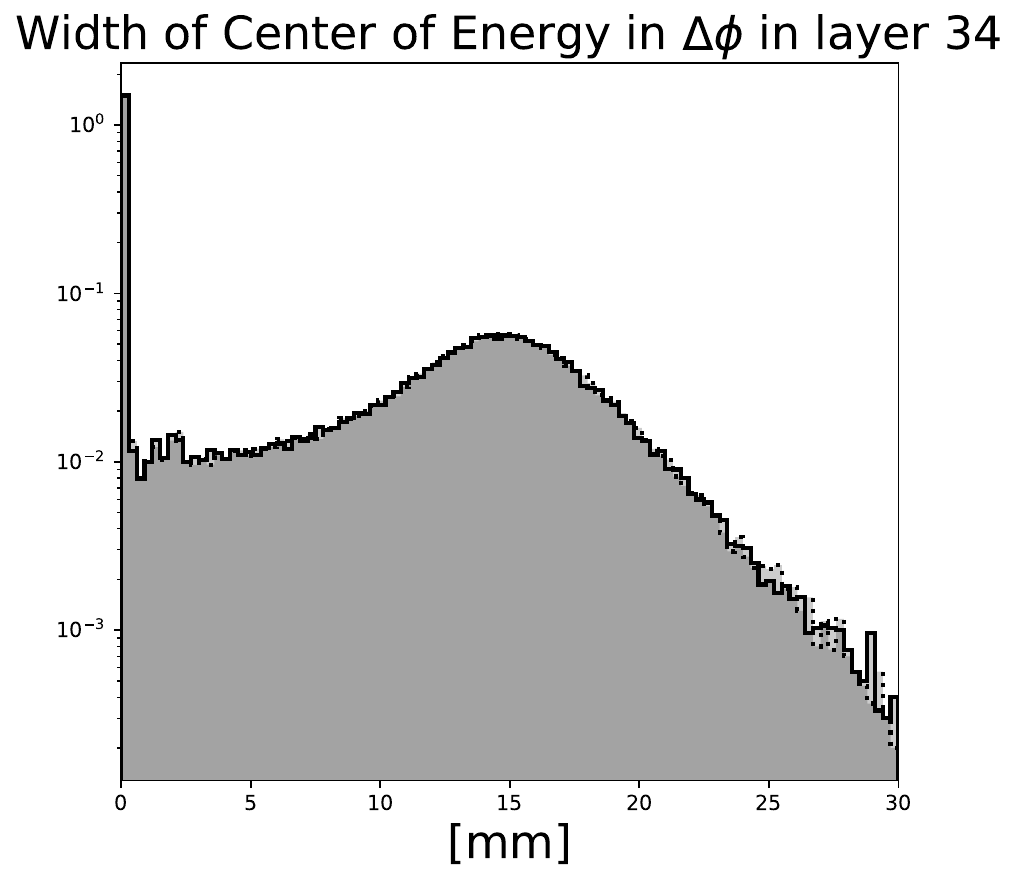}\\
    \includegraphics[height=0.1\textheight]{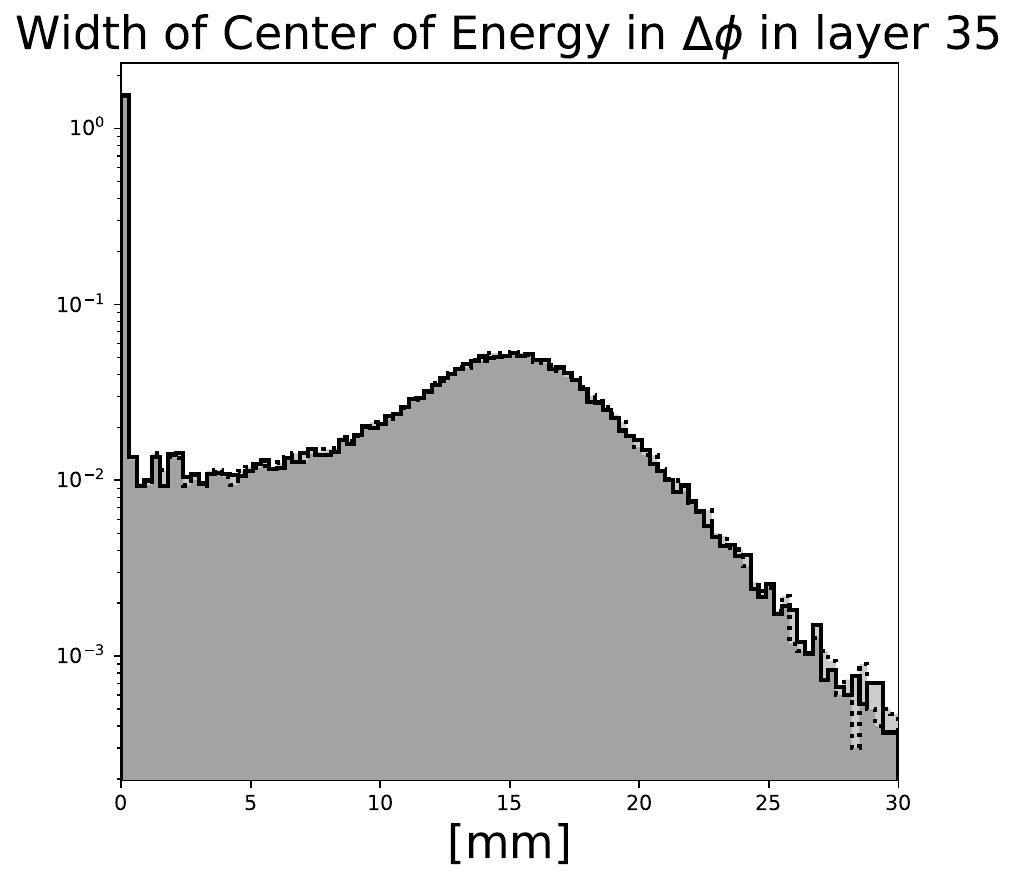} \hfill \includegraphics[height=0.1\textheight]{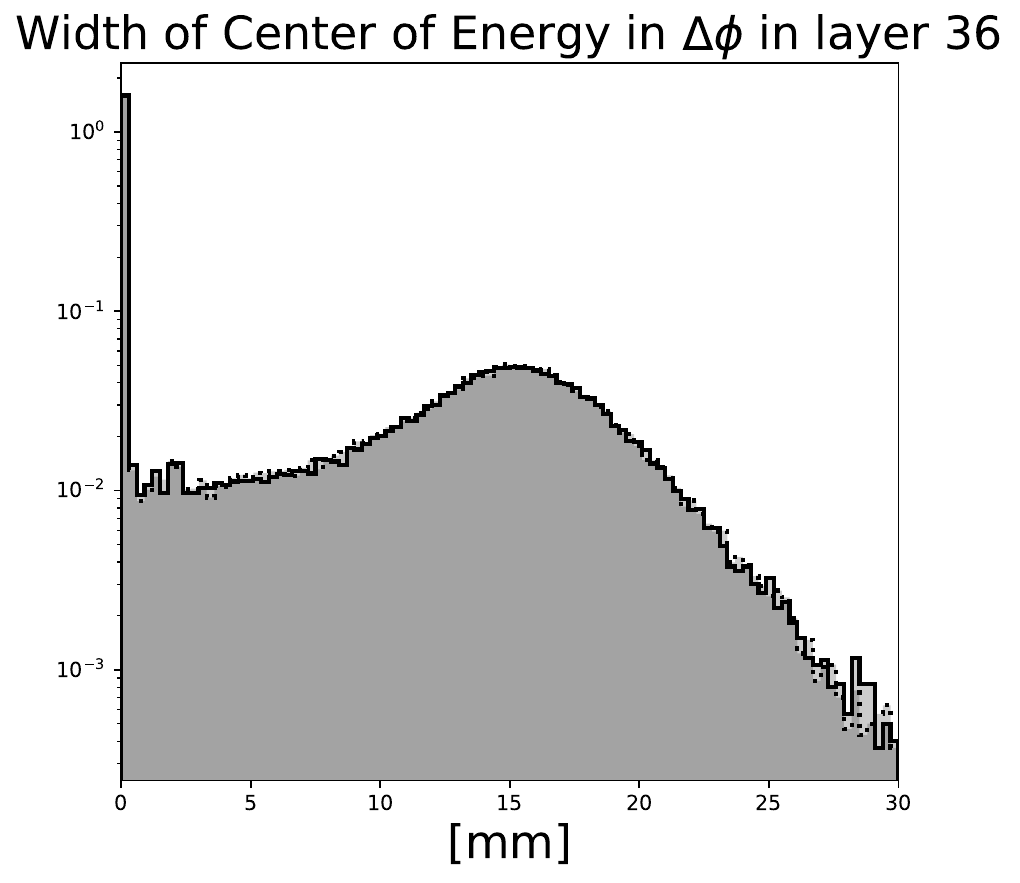} \hfill \includegraphics[height=0.1\textheight]{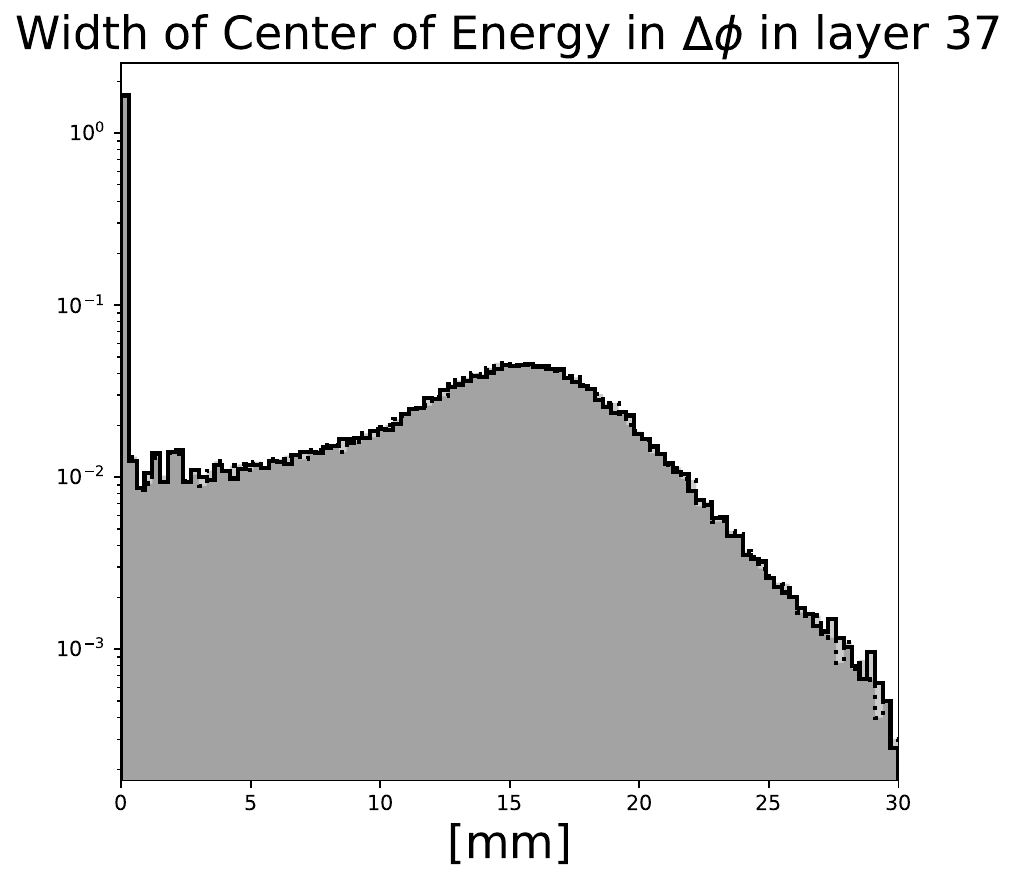} \hfill \includegraphics[height=0.1\textheight]{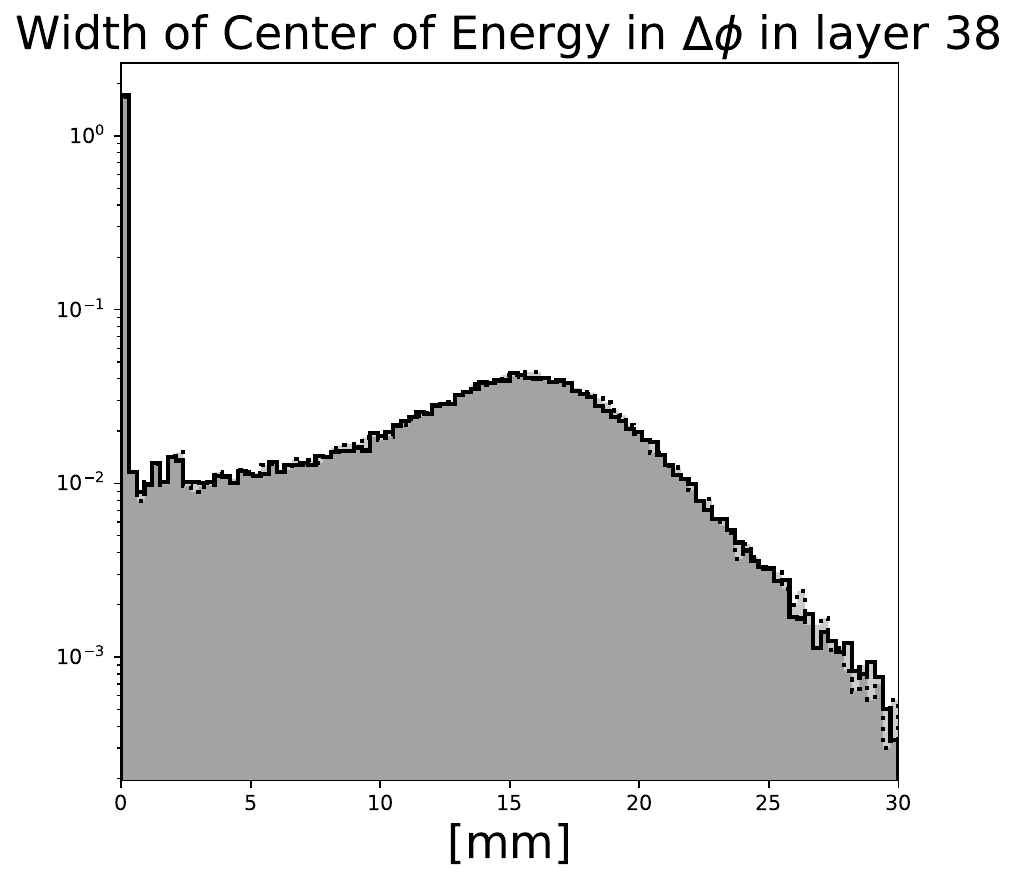} \hfill \includegraphics[height=0.1\textheight]{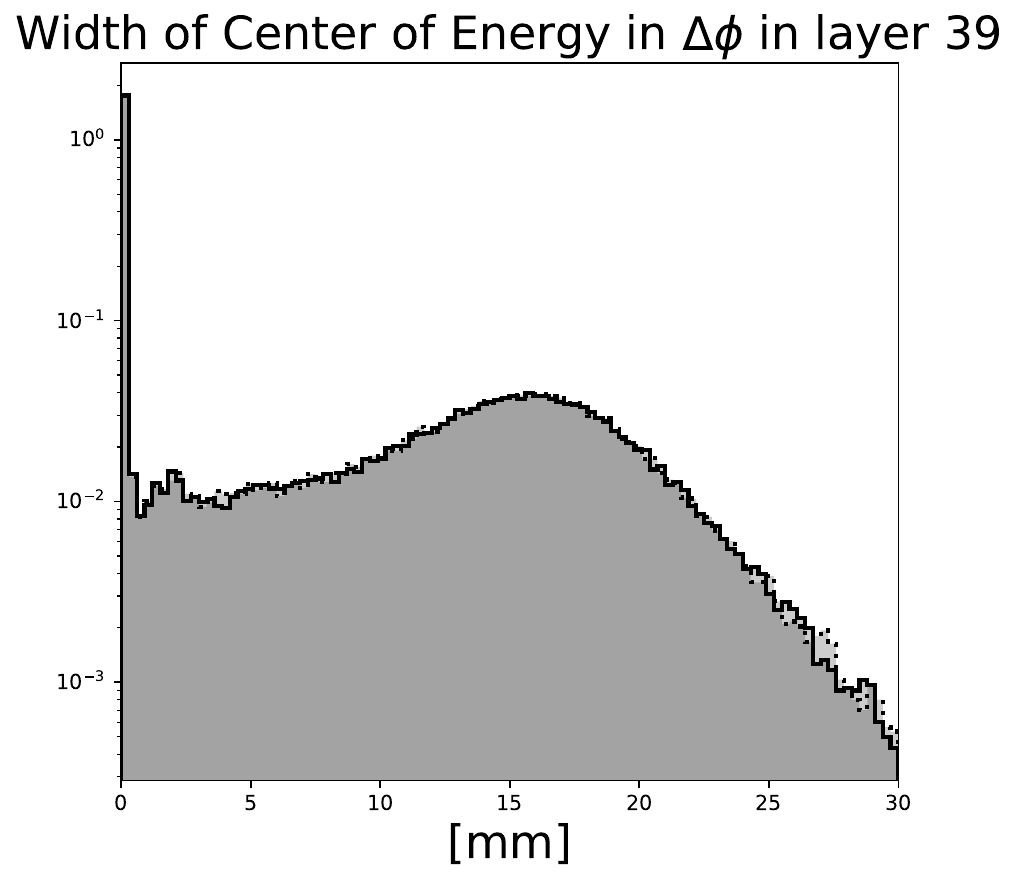}\\
    \includegraphics[height=0.1\textheight]{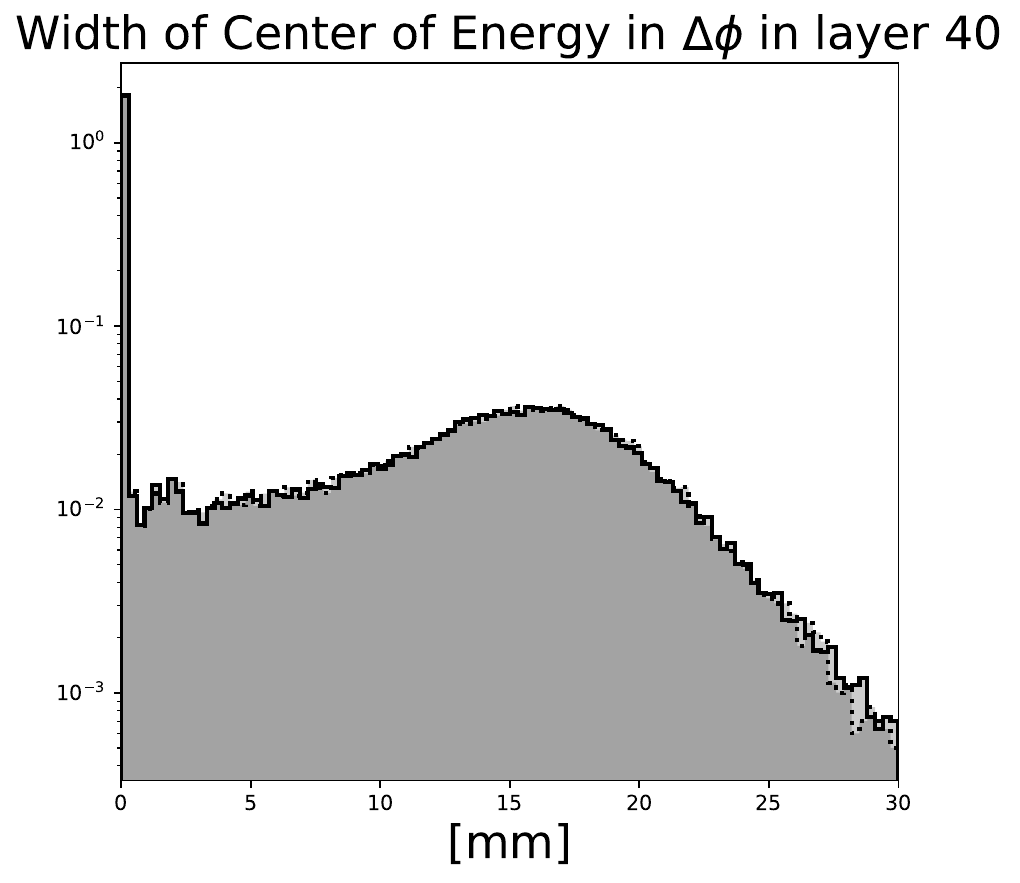} \hfill \includegraphics[height=0.1\textheight]{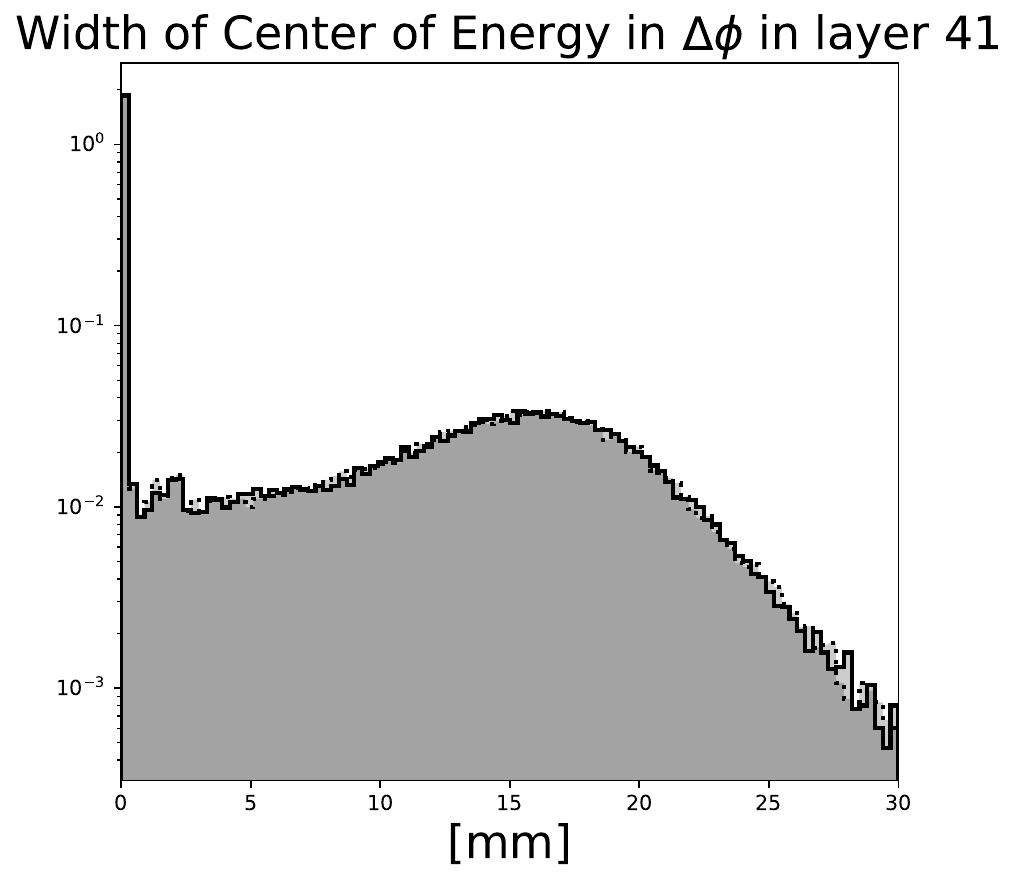} \hfill \includegraphics[height=0.1\textheight]{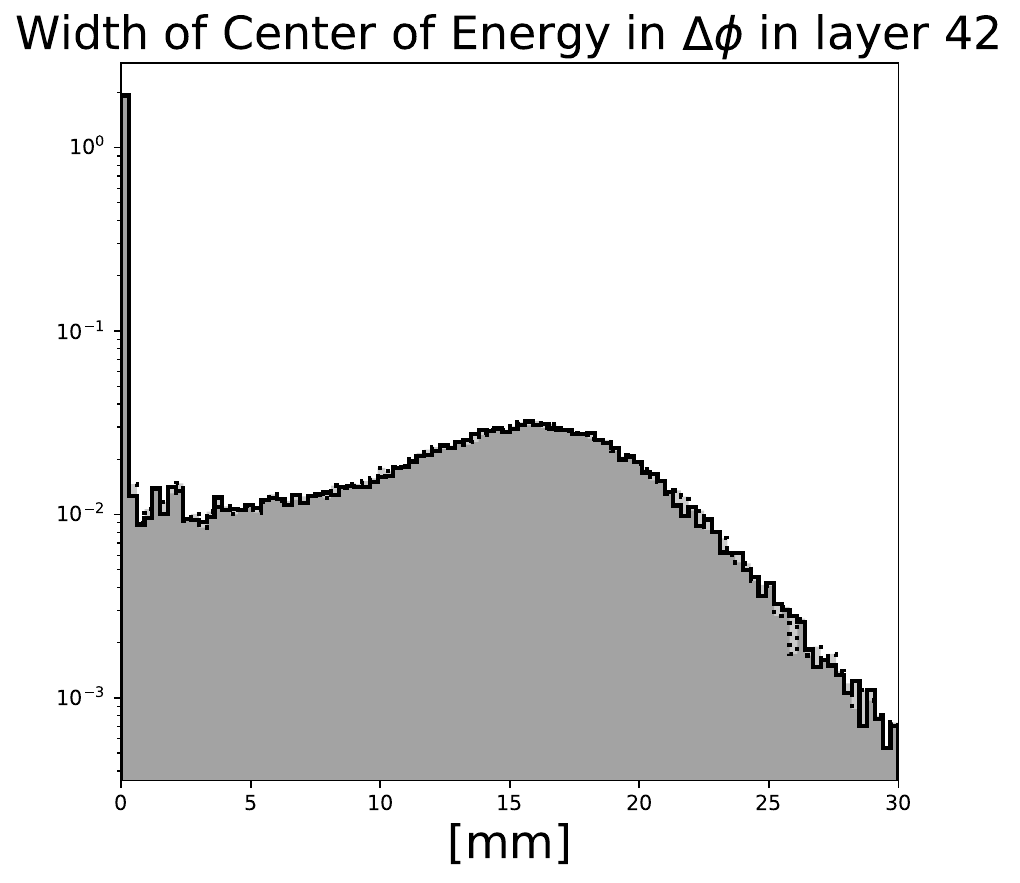} \hfill \includegraphics[height=0.1\textheight]{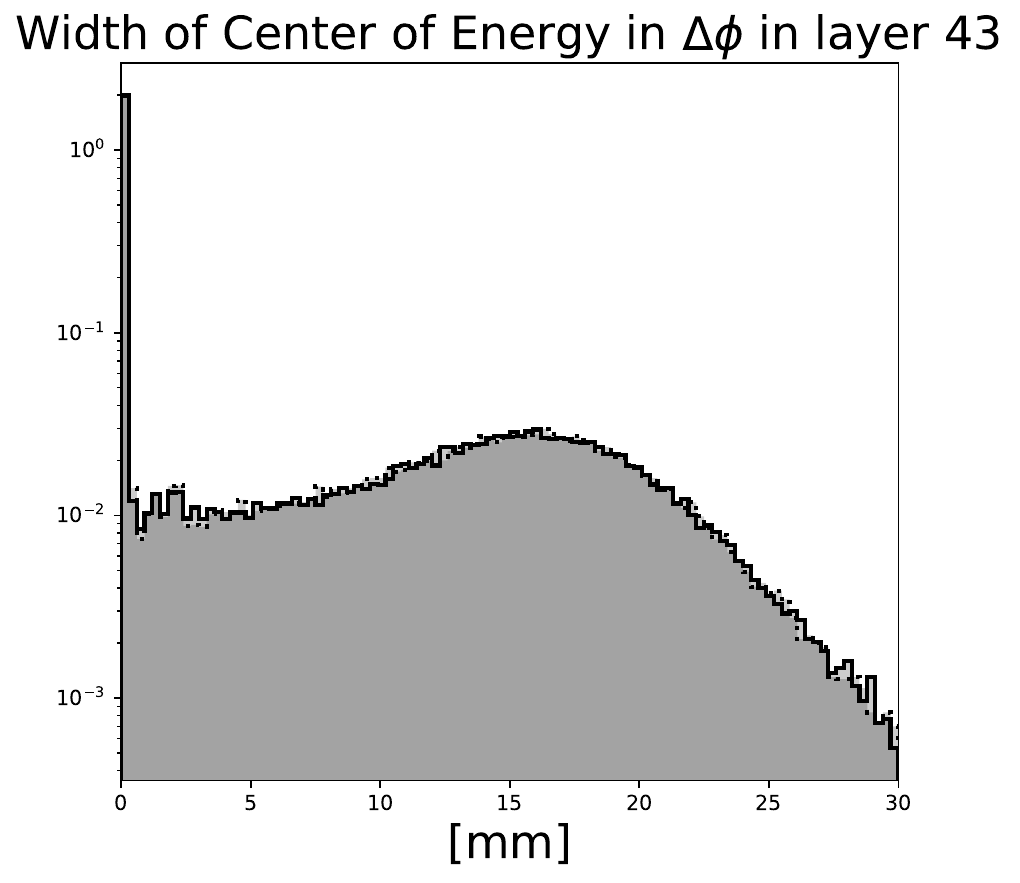} \hfill \includegraphics[height=0.1\textheight]{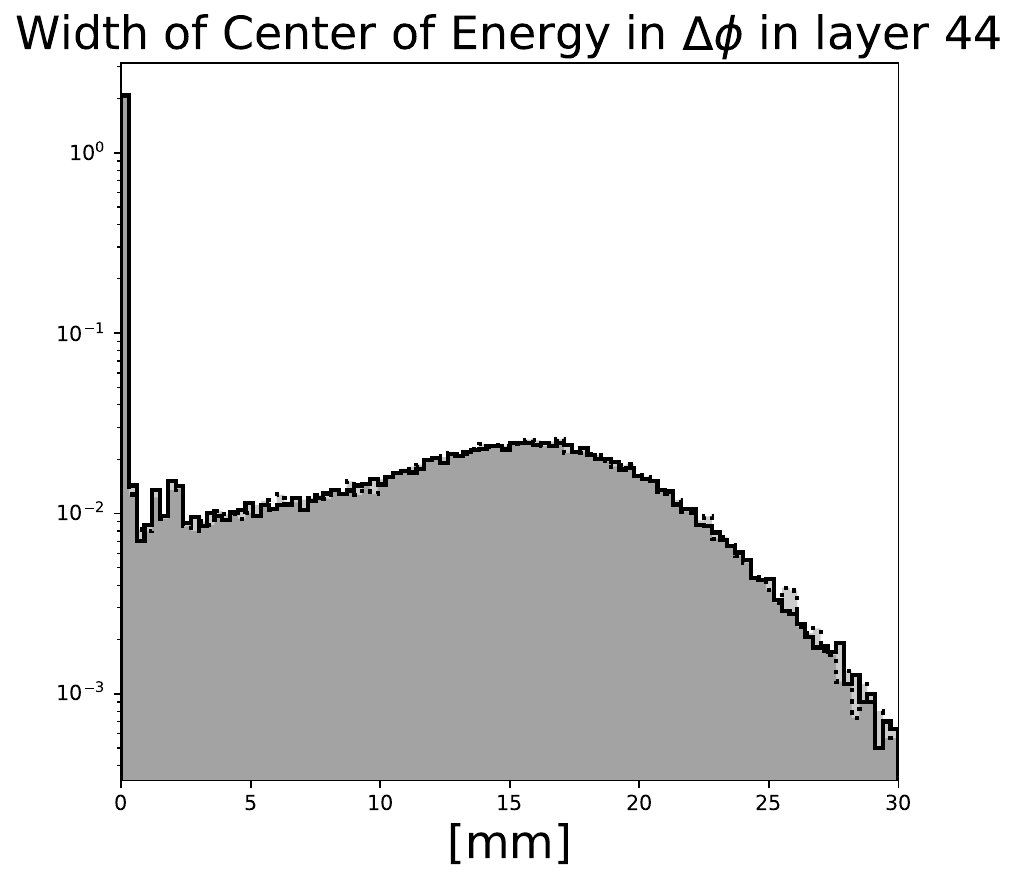}\\
    \includegraphics[width=0.5\textwidth]{figures/Appendix_reference/legend.pdf}
    \caption{Distribution of \geant training and evaluation data in width of the centers of energy in $\phi$ direction for ds2. }
    \label{fig:app_ref.ds2.6}
\end{figure}

\begin{figure}[ht]
    \centering
    \includegraphics[height=0.1\textheight]{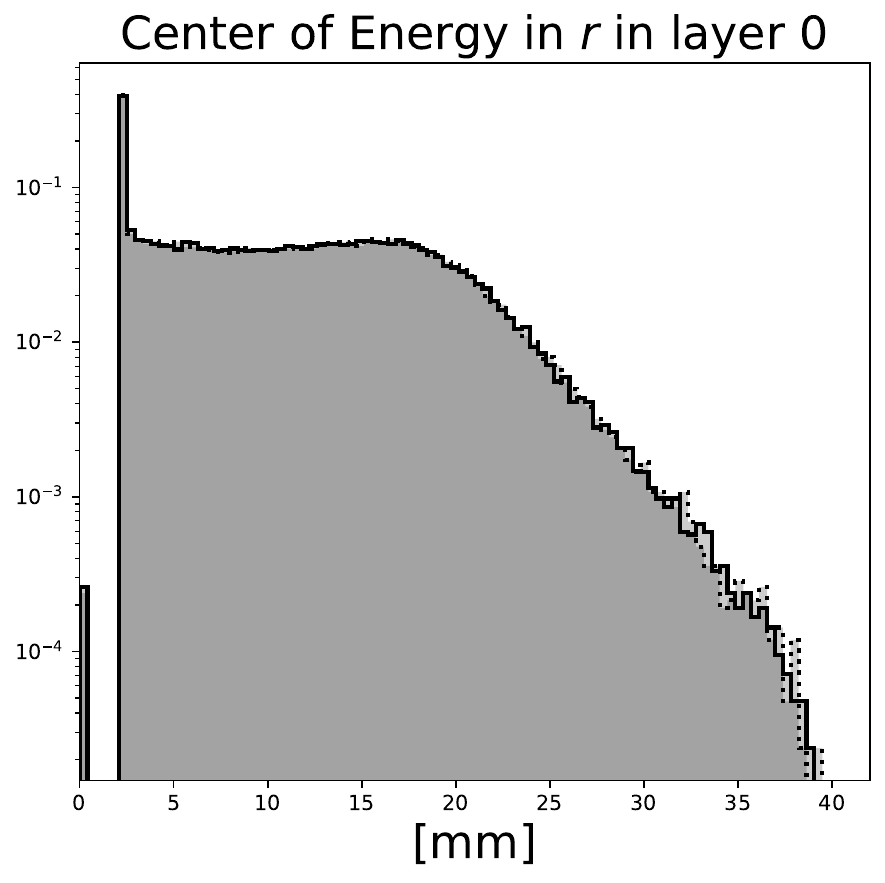} \hfill \includegraphics[height=0.1\textheight]{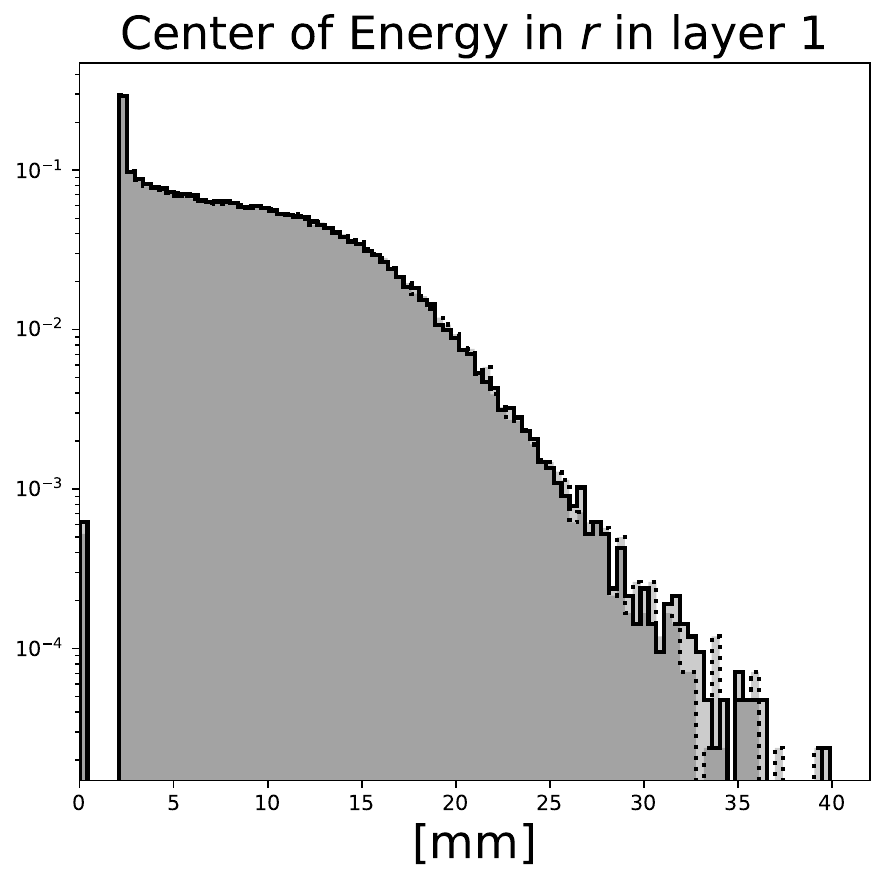} \hfill \includegraphics[height=0.1\textheight]{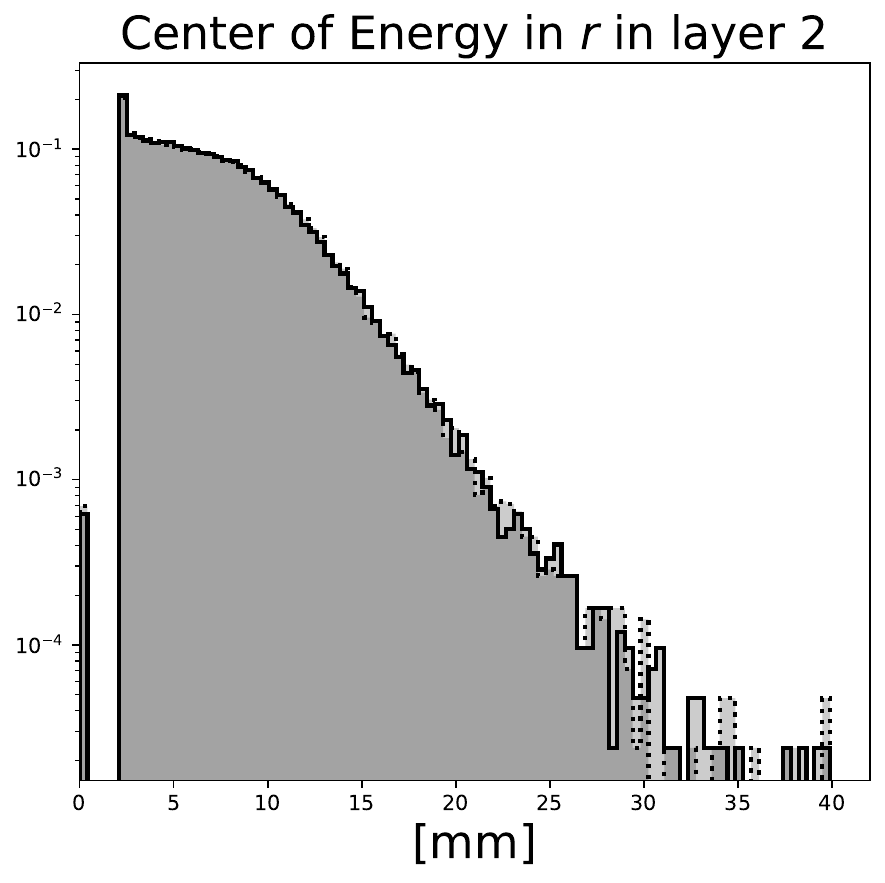} \hfill \includegraphics[height=0.1\textheight]{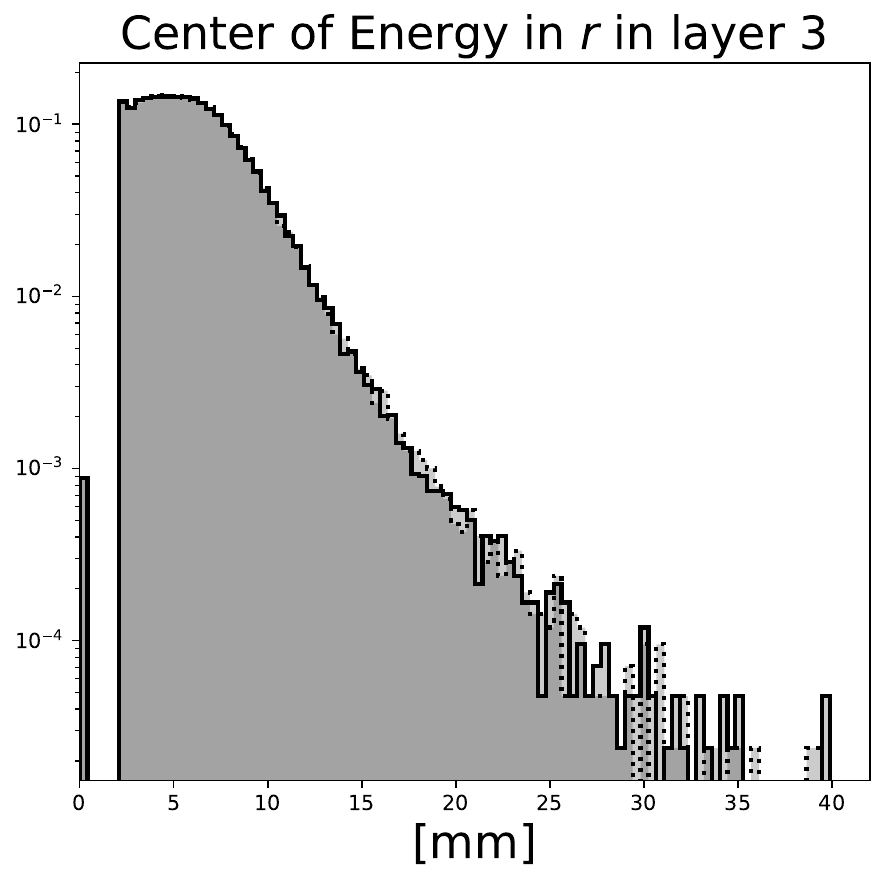} \hfill \includegraphics[height=0.1\textheight]{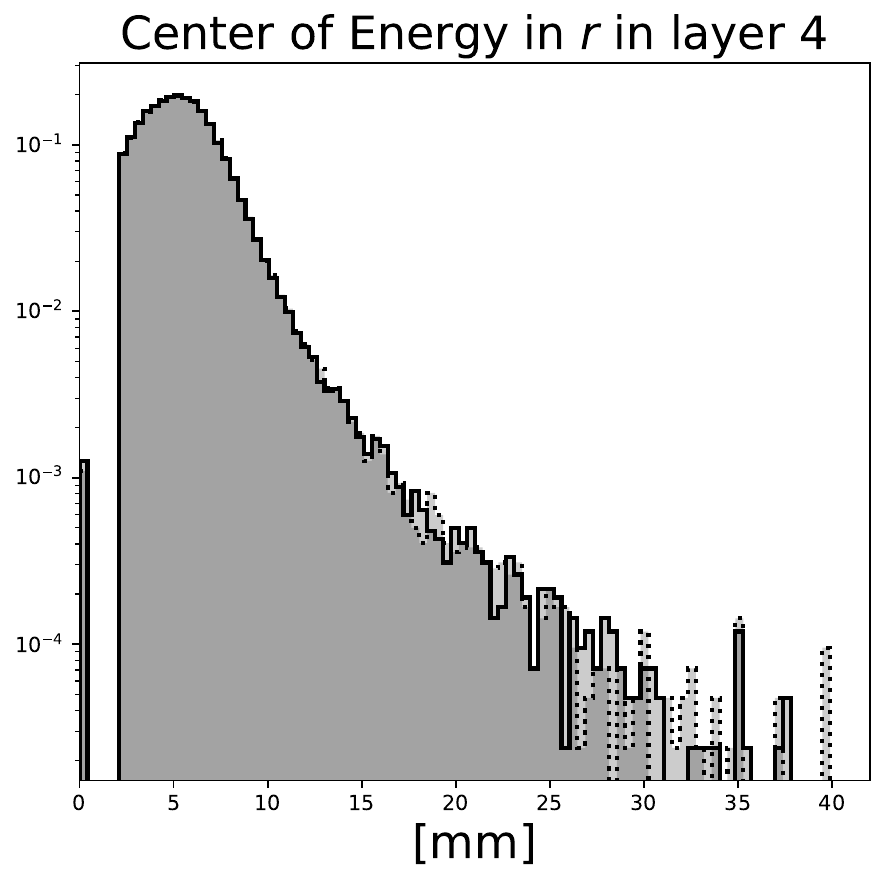}\\
    \includegraphics[height=0.1\textheight]{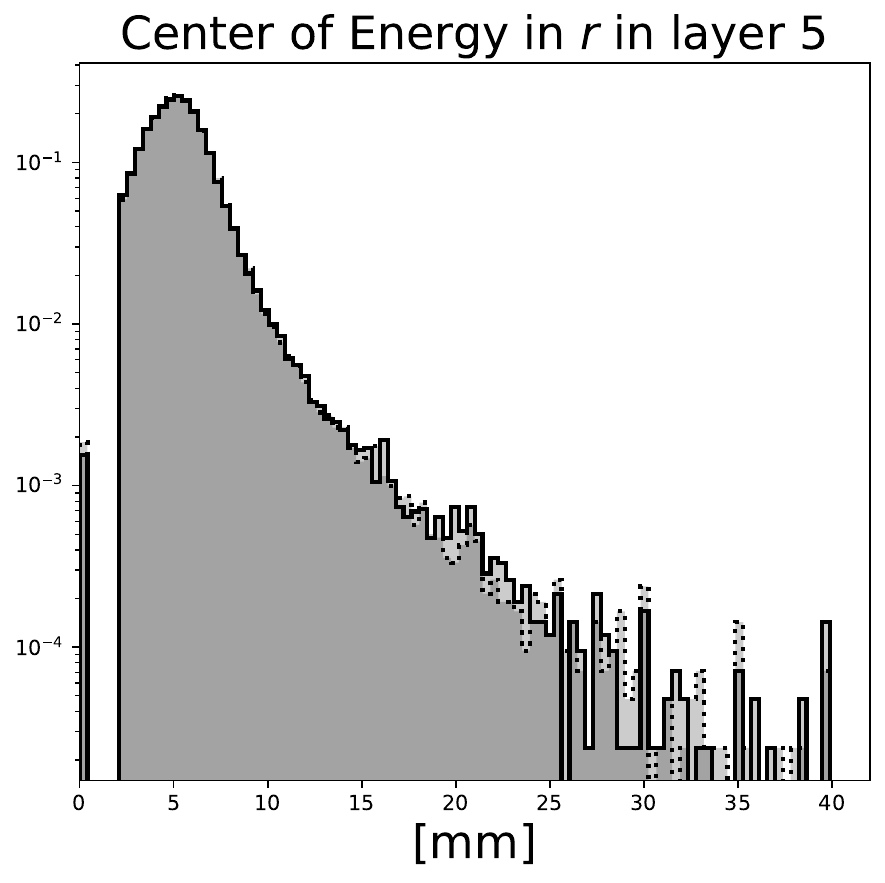} \hfill \includegraphics[height=0.1\textheight]{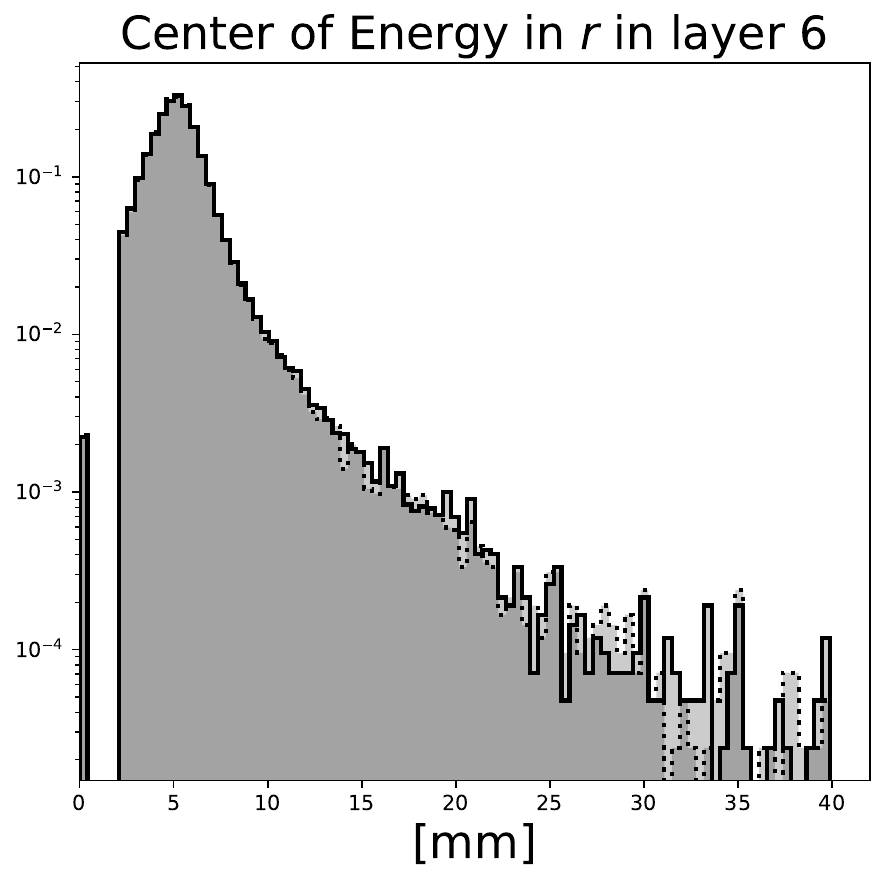} \hfill \includegraphics[height=0.1\textheight]{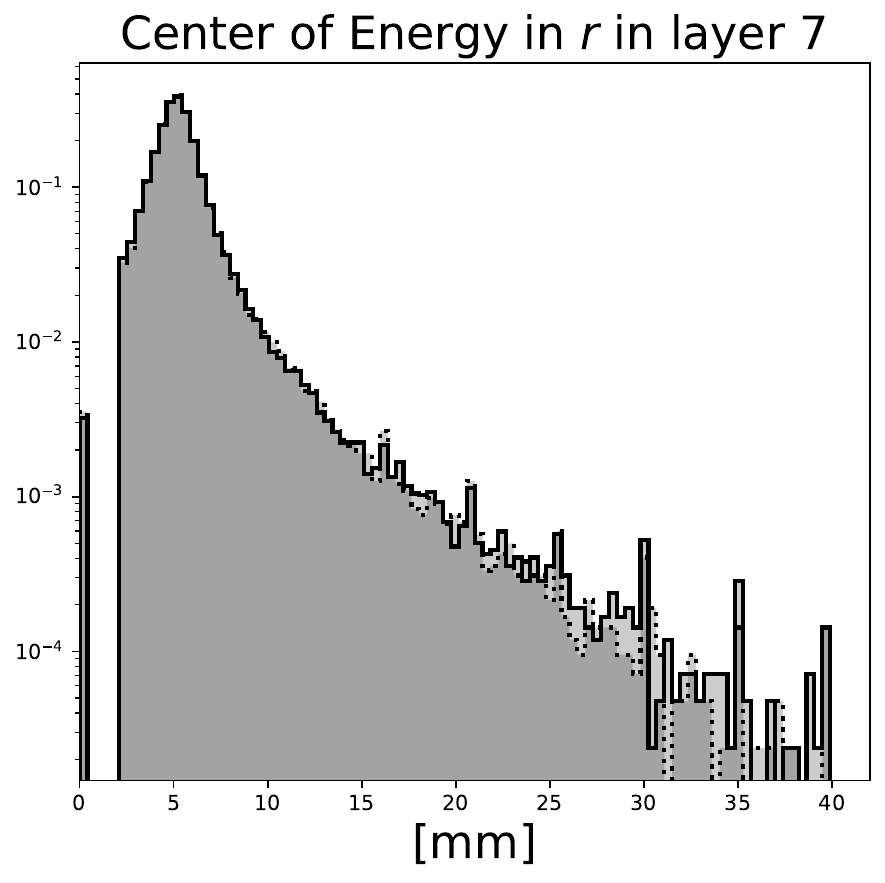} \hfill \includegraphics[height=0.1\textheight]{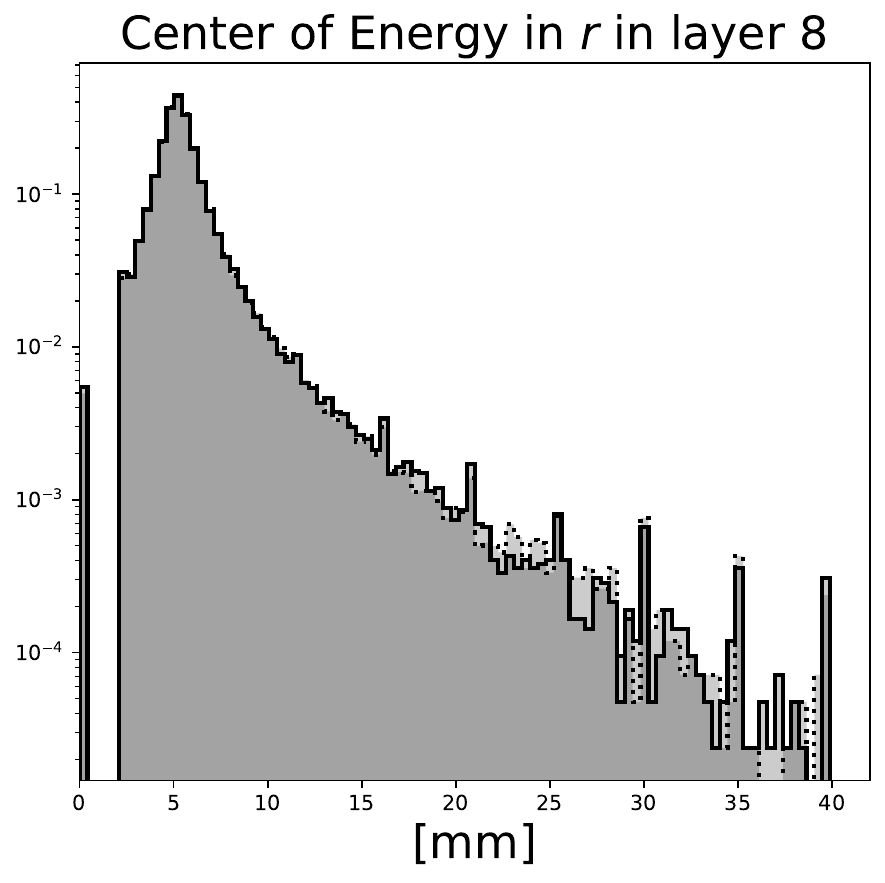} \hfill \includegraphics[height=0.1\textheight]{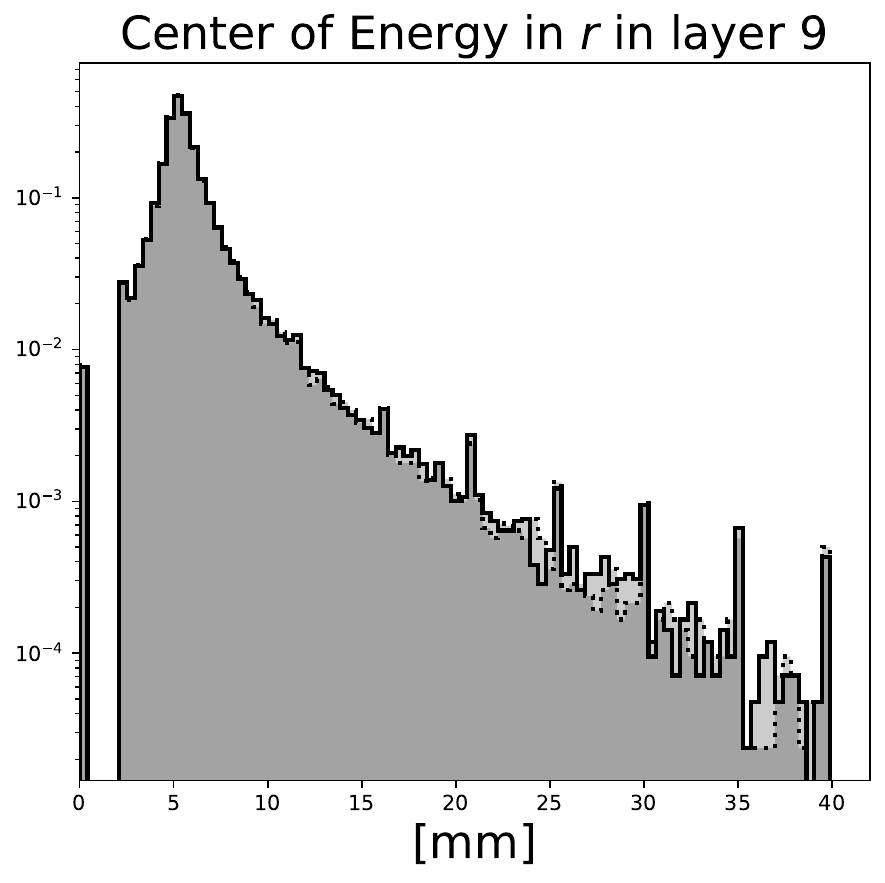}\\
    \includegraphics[height=0.1\textheight]{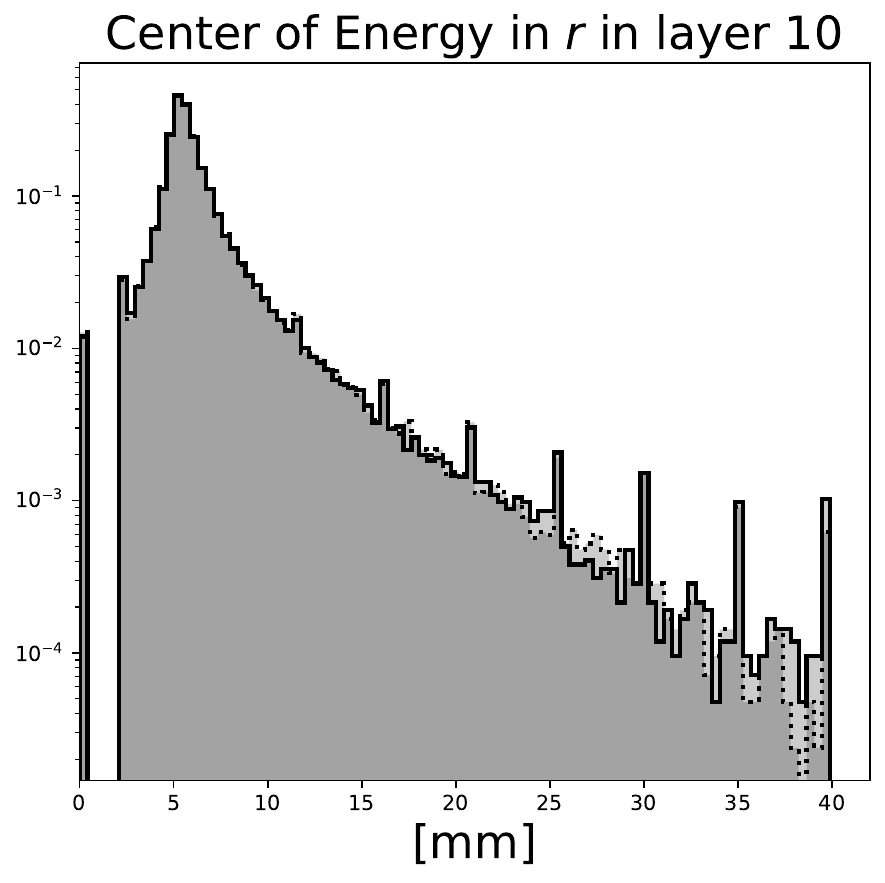} \hfill \includegraphics[height=0.1\textheight]{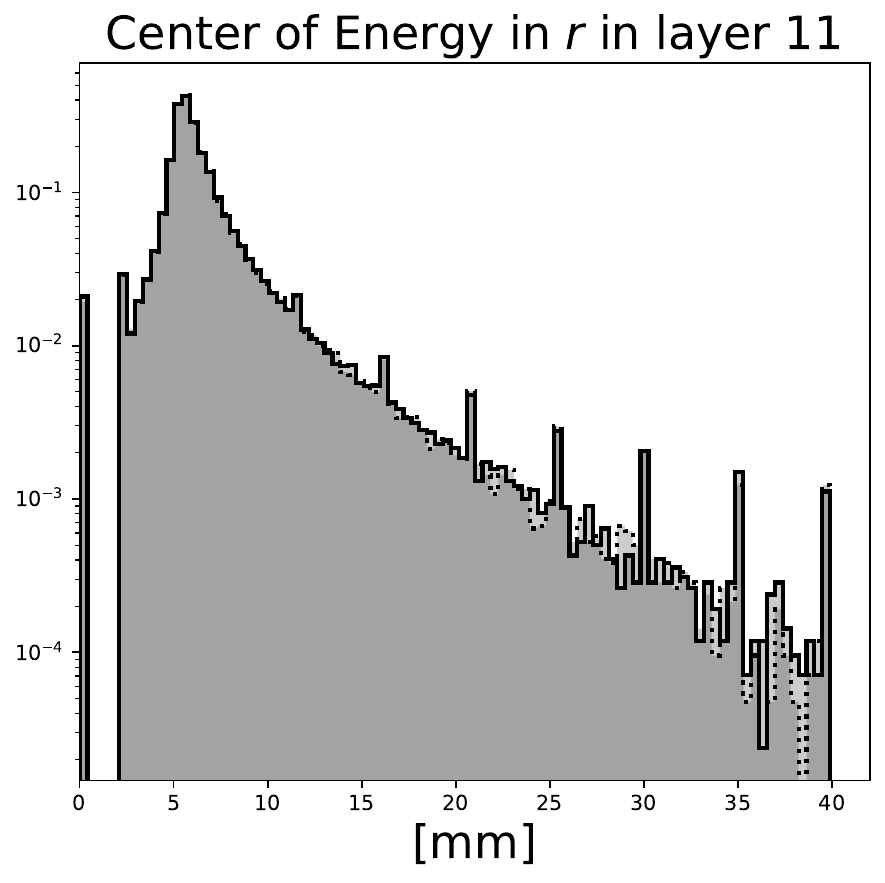} \hfill \includegraphics[height=0.1\textheight]{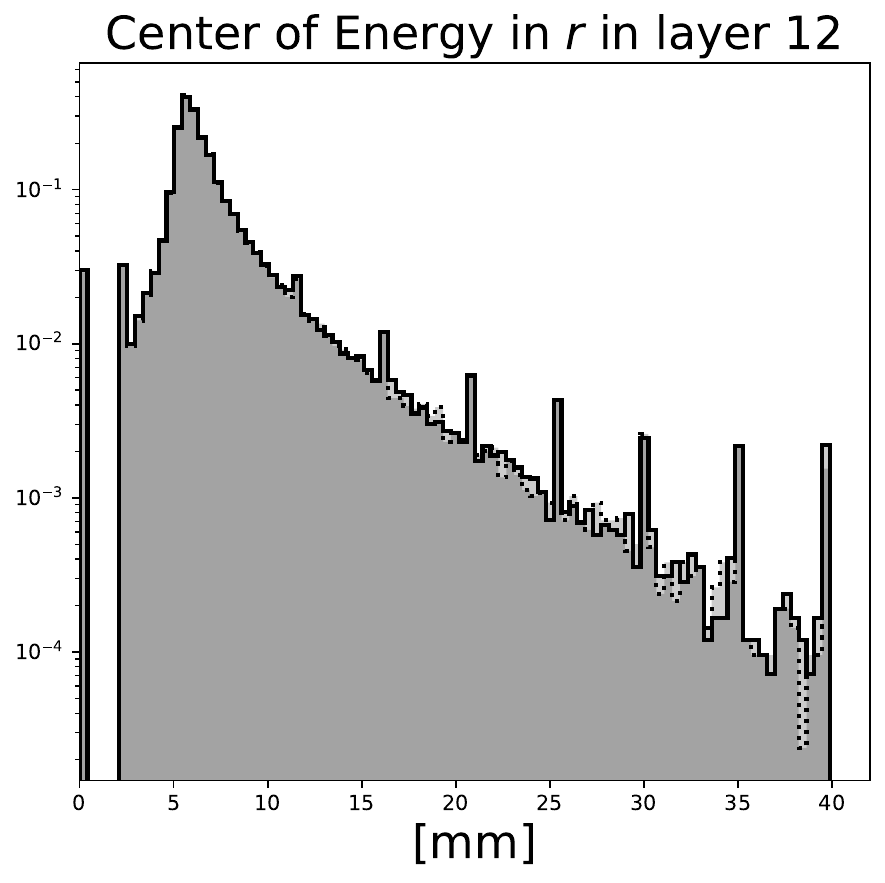} \hfill \includegraphics[height=0.1\textheight]{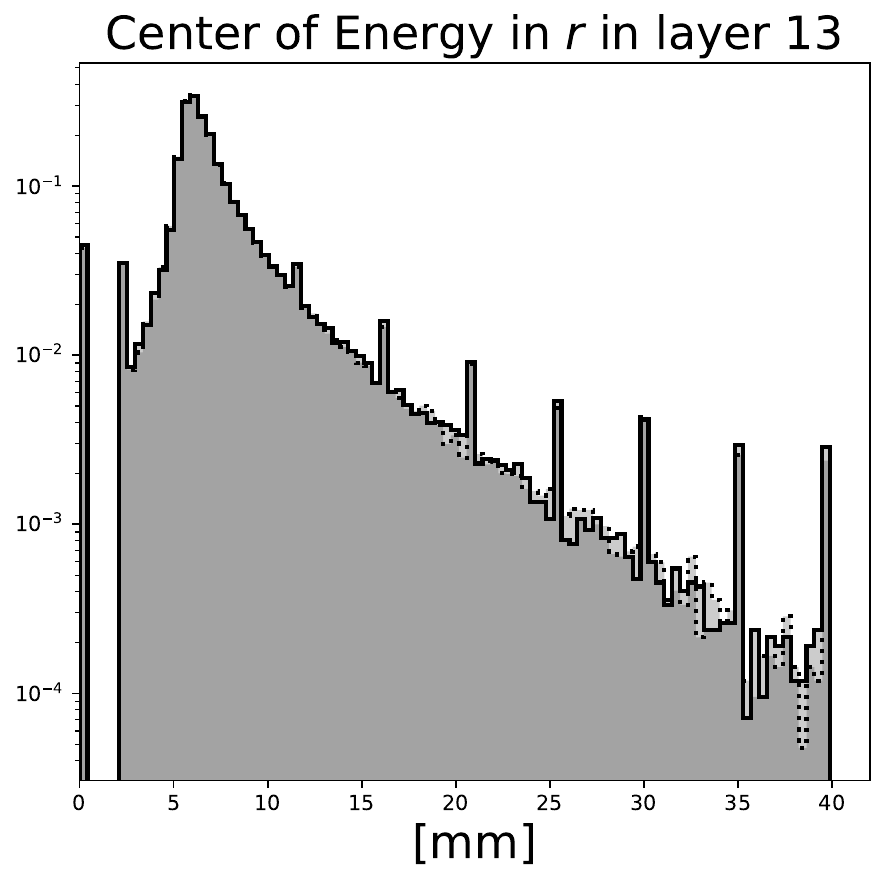} \hfill \includegraphics[height=0.1\textheight]{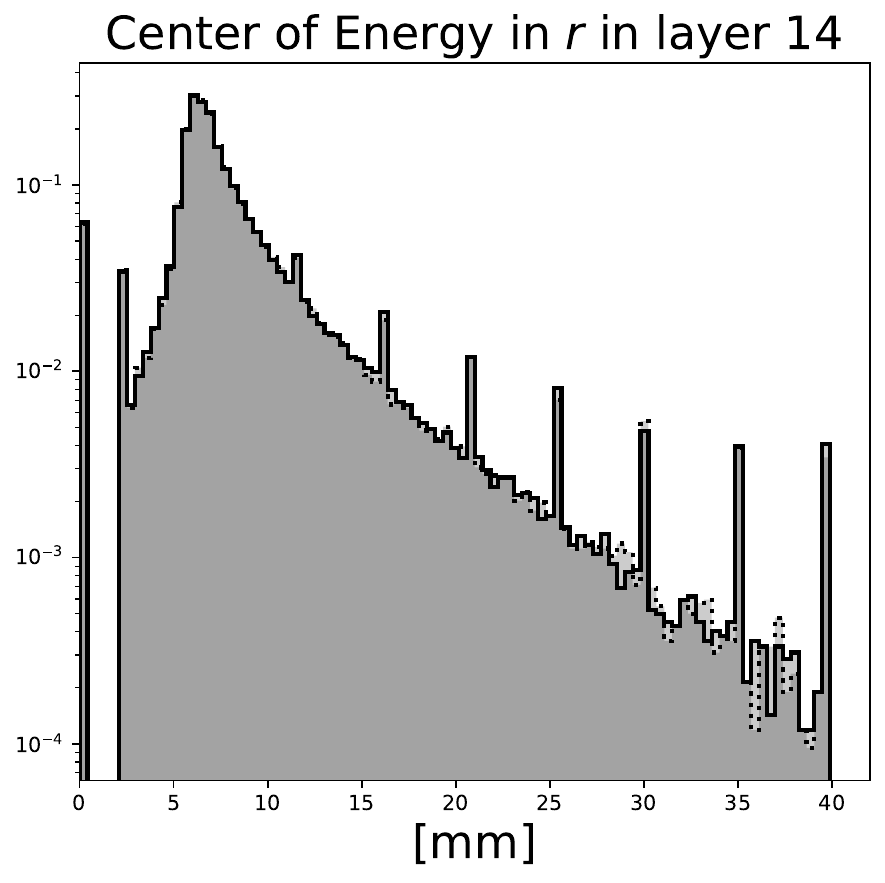}\\
    \includegraphics[height=0.1\textheight]{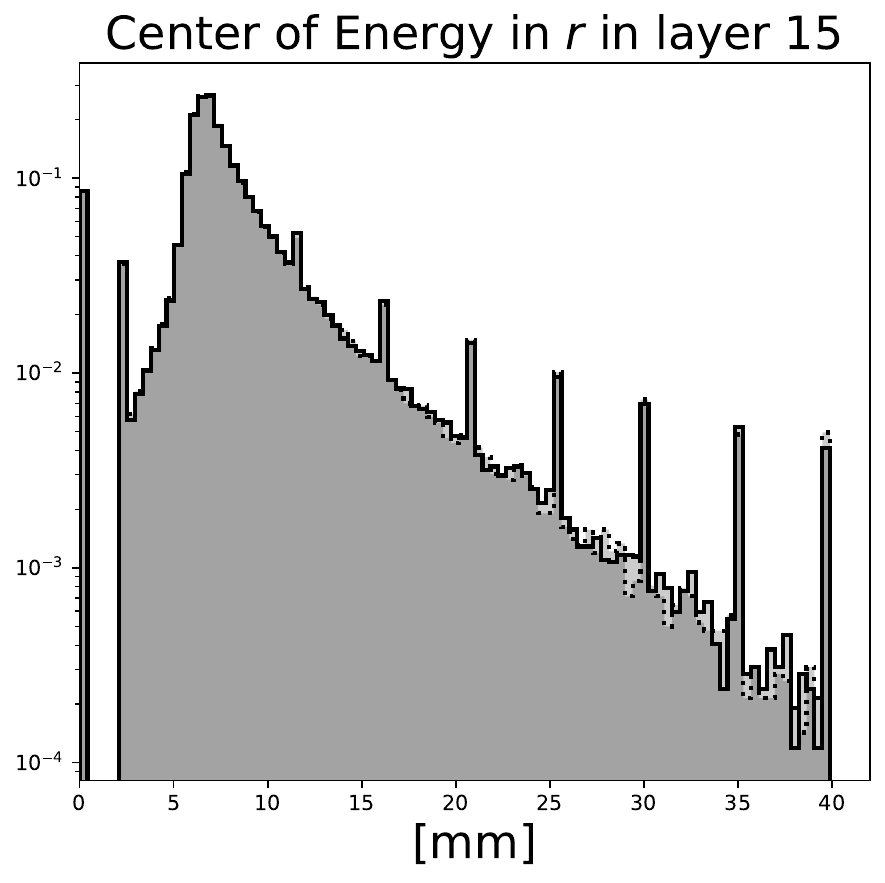} \hfill \includegraphics[height=0.1\textheight]{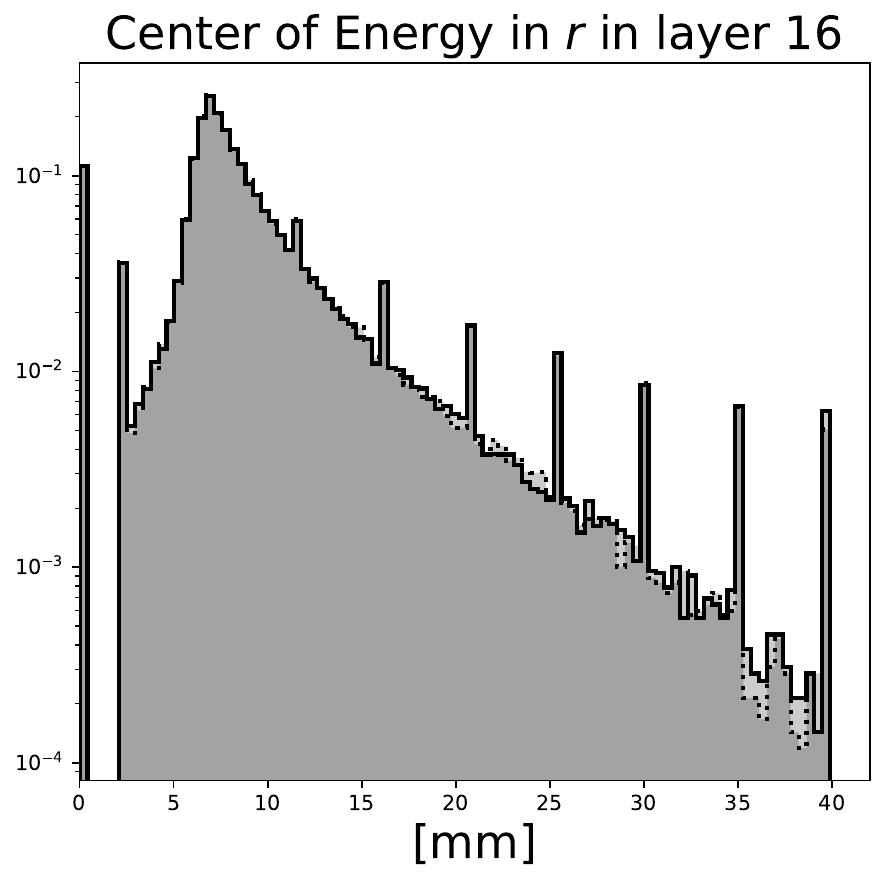} \hfill \includegraphics[height=0.1\textheight]{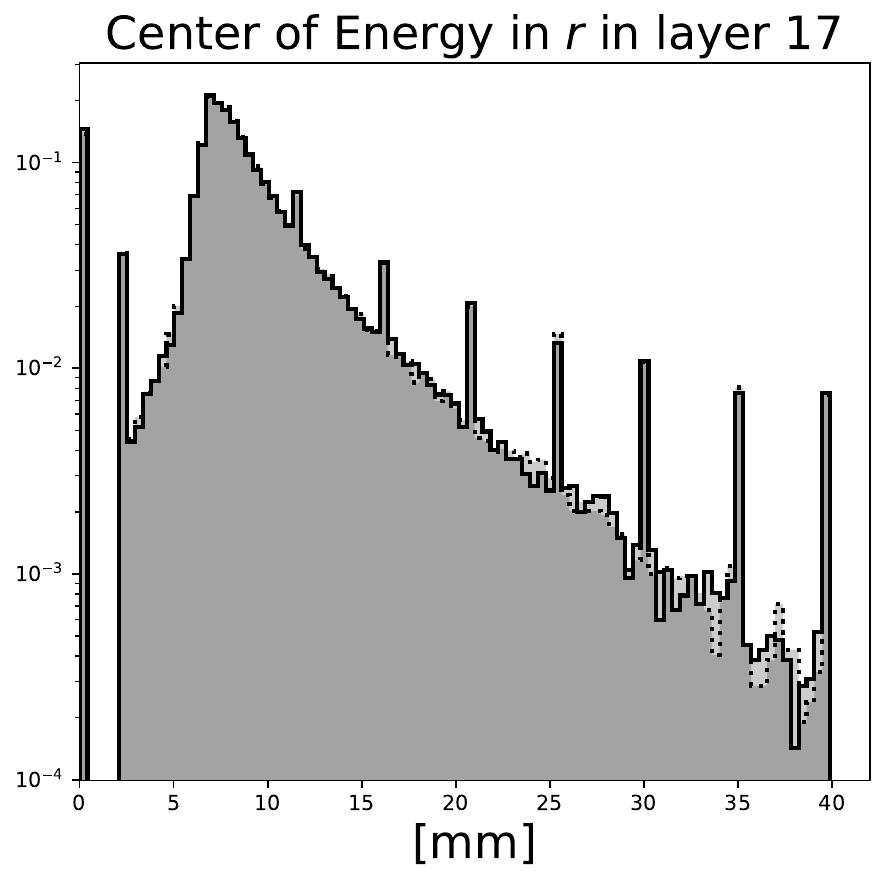} \hfill \includegraphics[height=0.1\textheight]{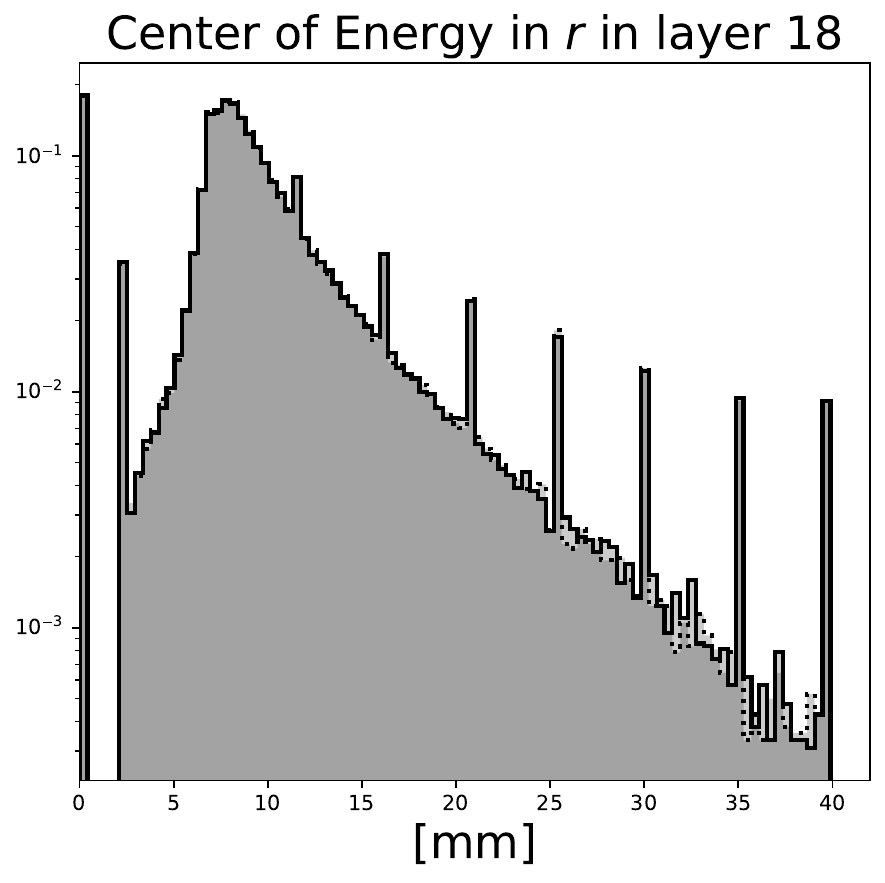} \hfill \includegraphics[height=0.1\textheight]{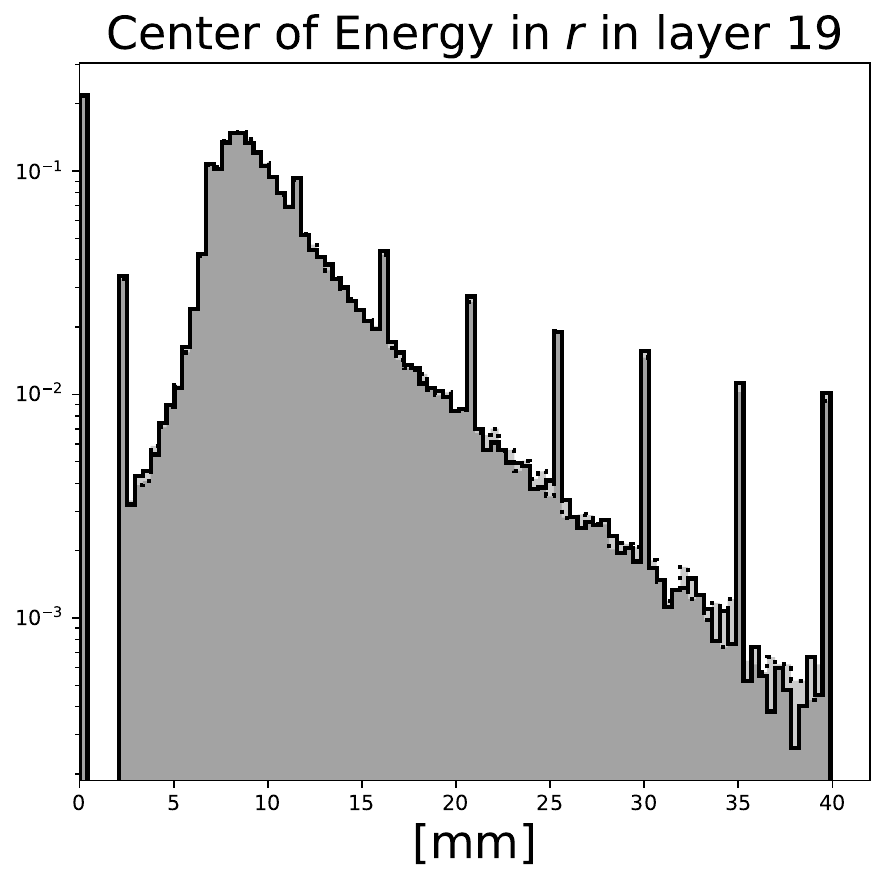}\\
    \includegraphics[height=0.1\textheight]{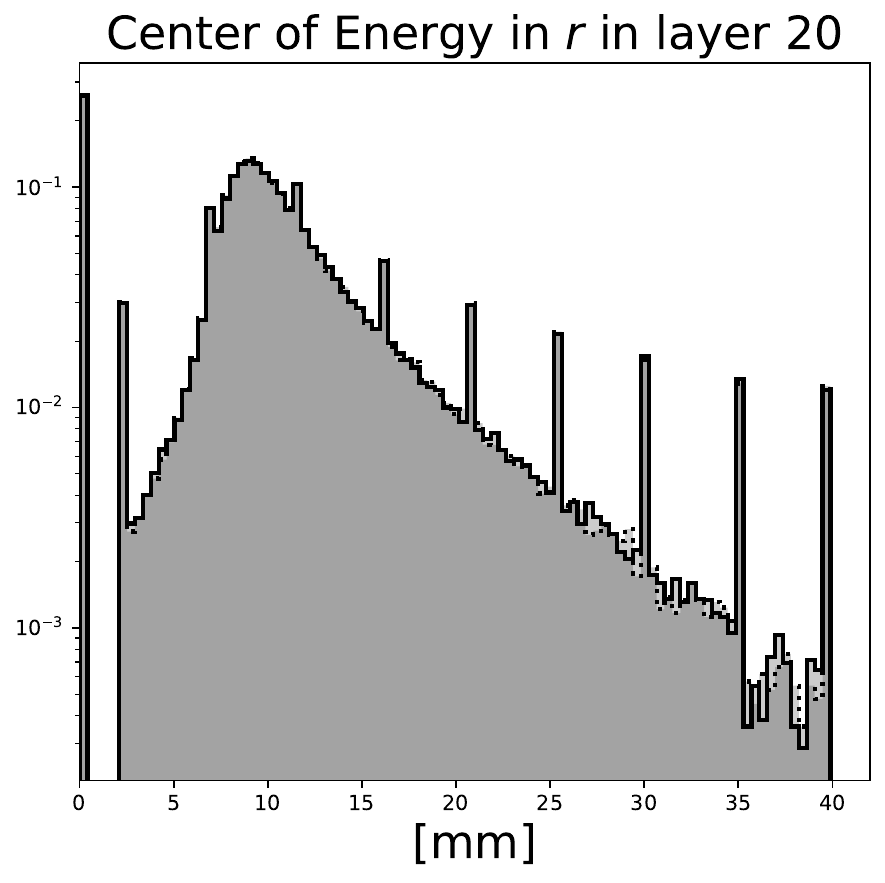} \hfill \includegraphics[height=0.1\textheight]{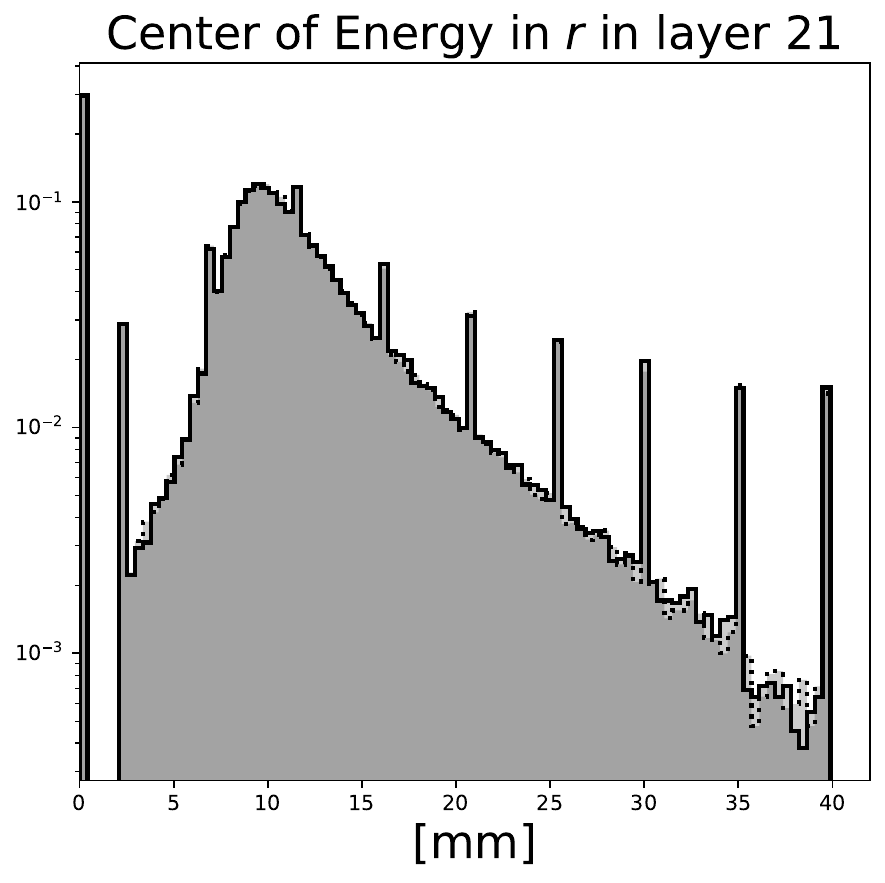} \hfill \includegraphics[height=0.1\textheight]{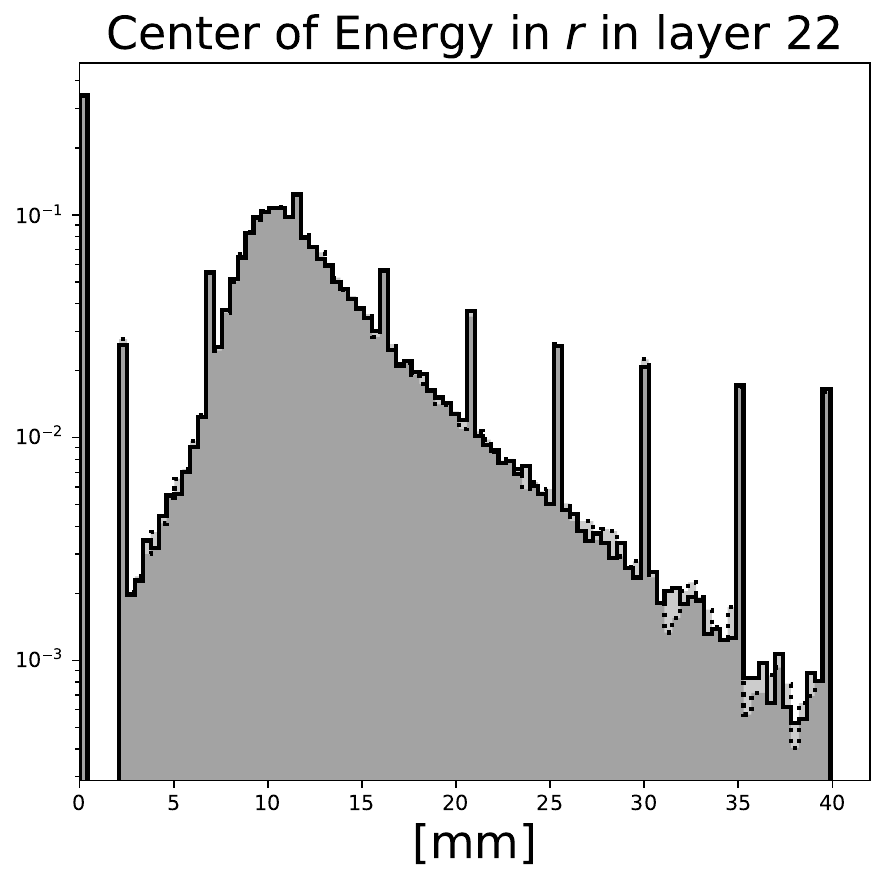} \hfill \includegraphics[height=0.1\textheight]{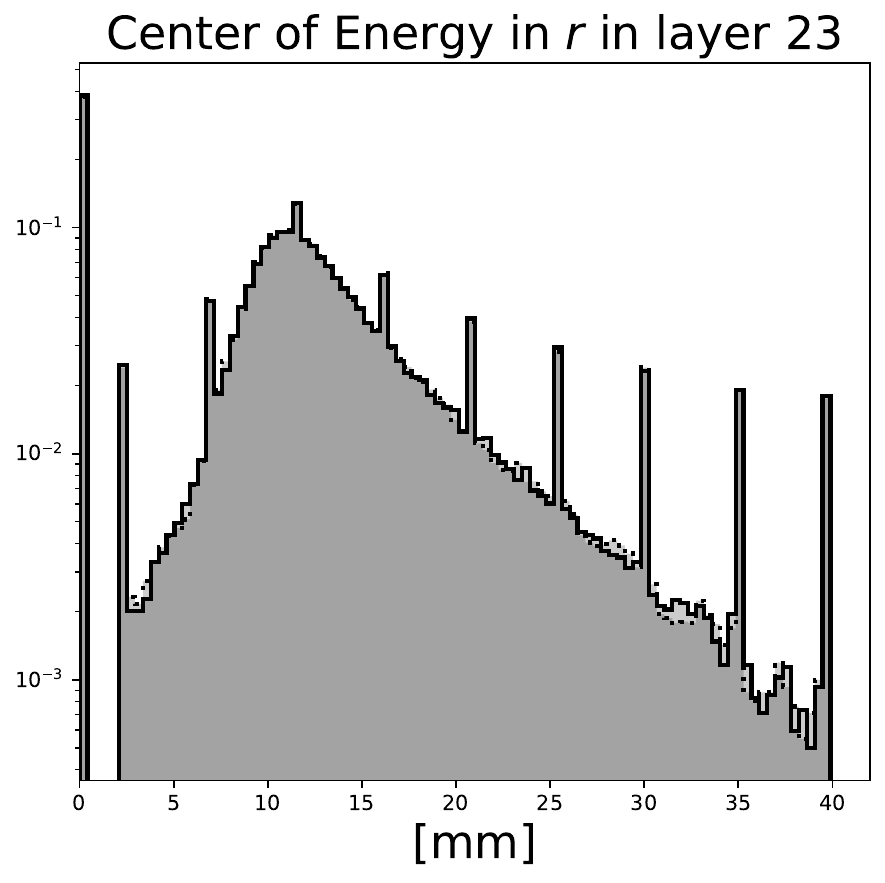} \hfill \includegraphics[height=0.1\textheight]{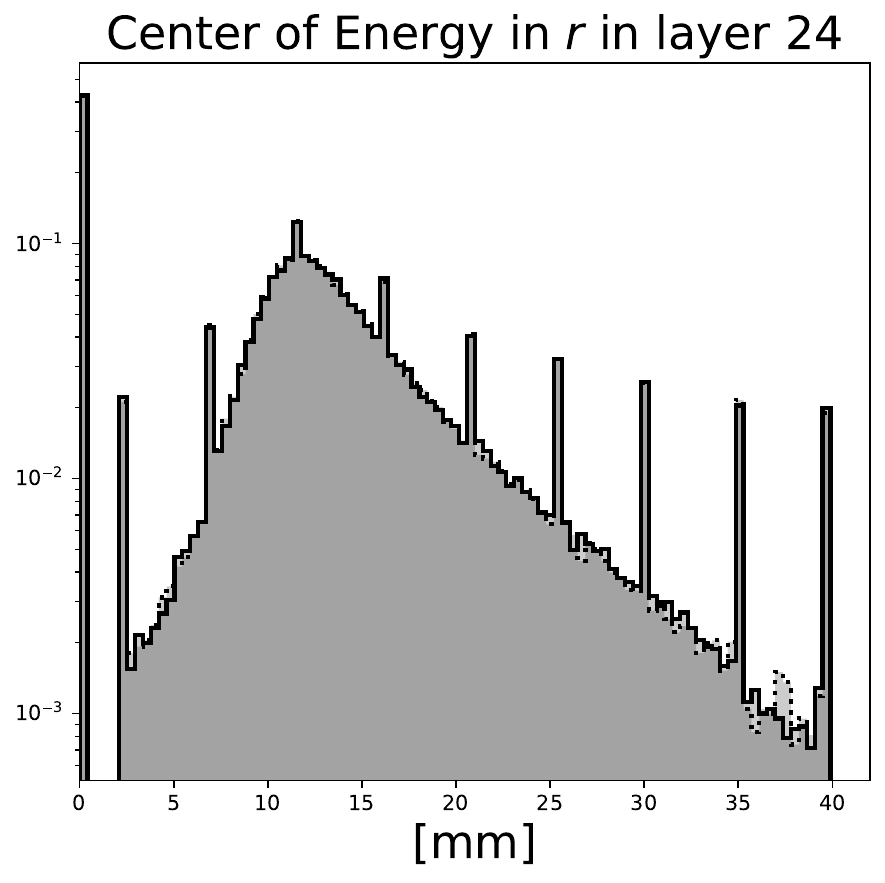}\\
    \includegraphics[height=0.1\textheight]{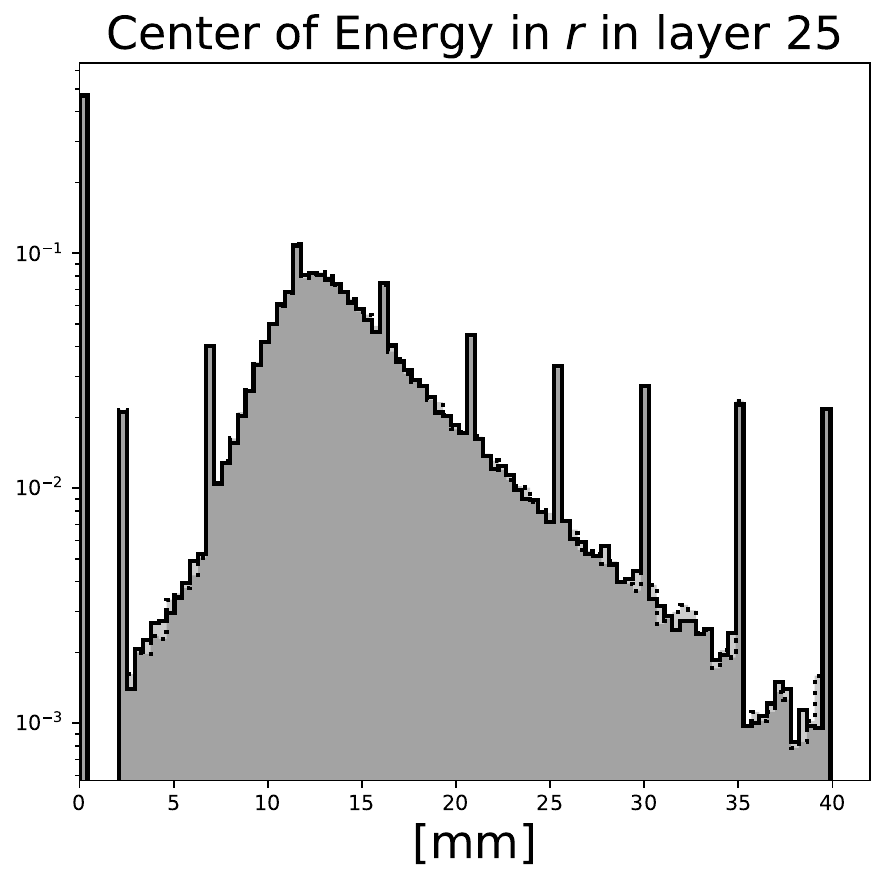} \hfill \includegraphics[height=0.1\textheight]{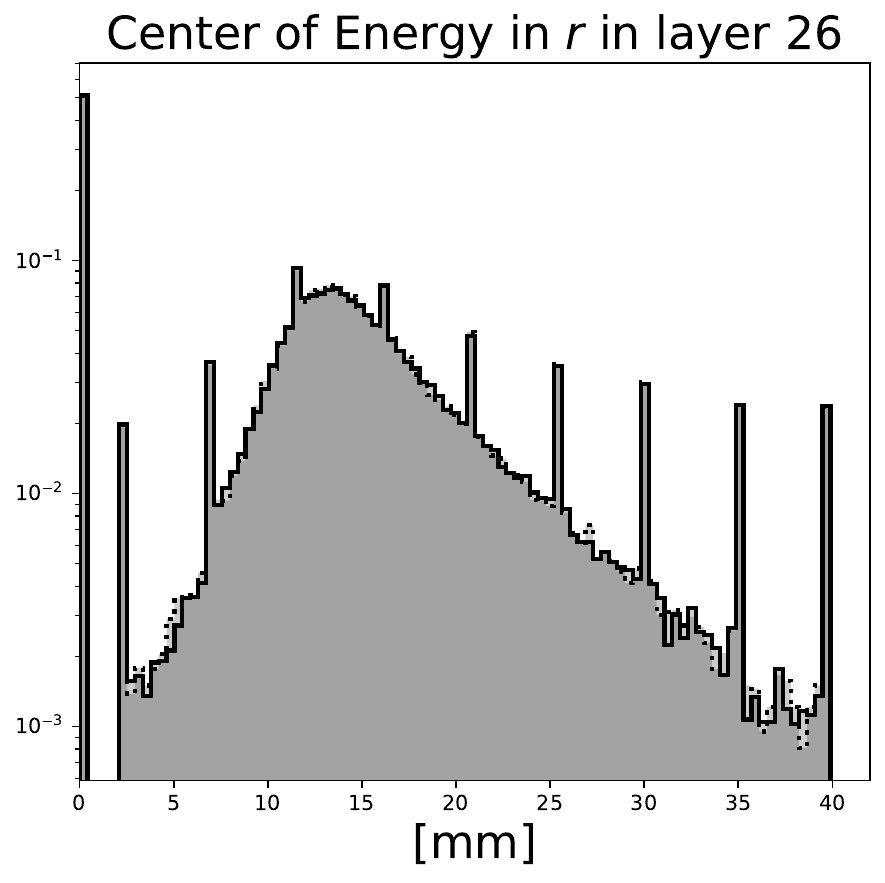} \hfill \includegraphics[height=0.1\textheight]{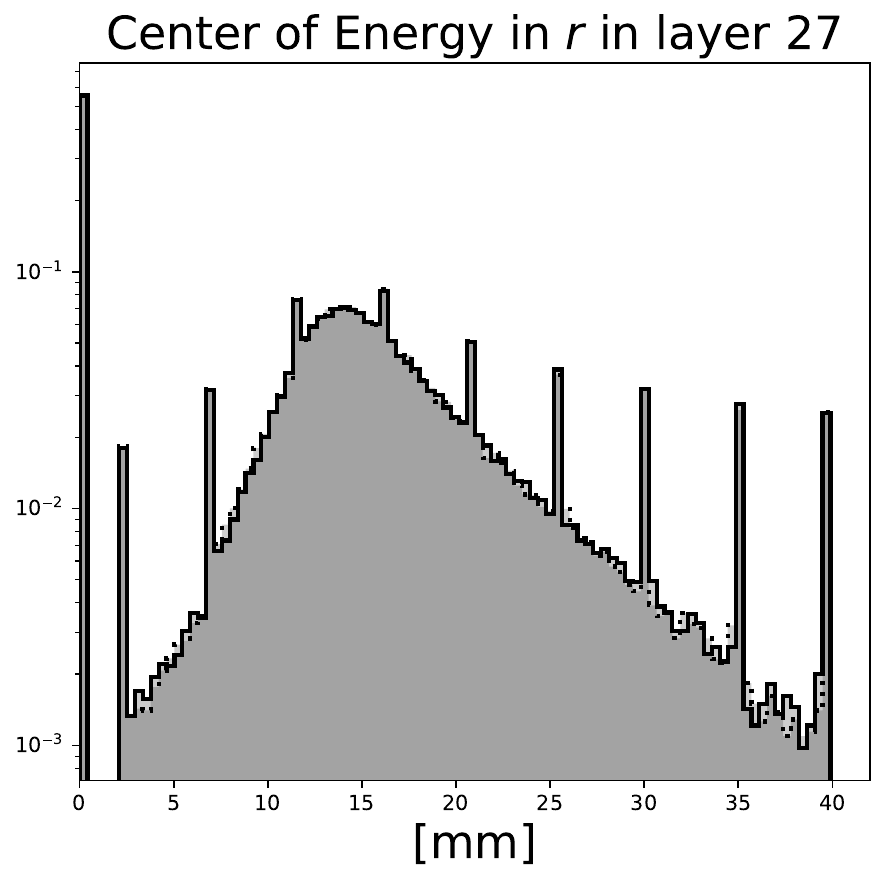} \hfill \includegraphics[height=0.1\textheight]{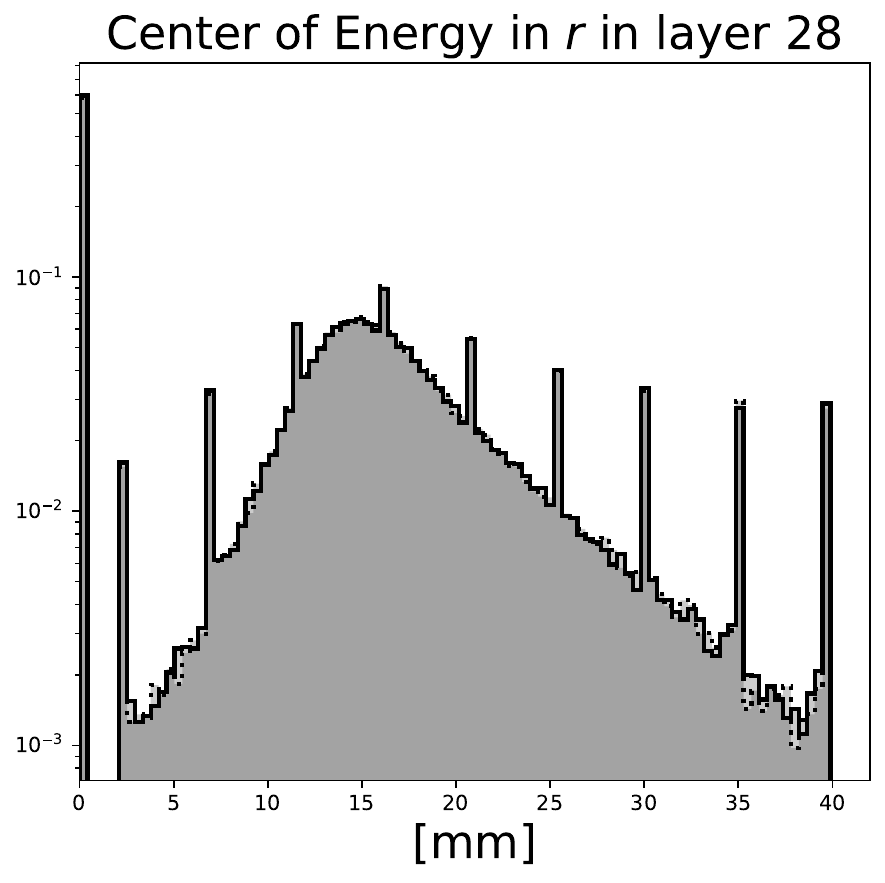} \hfill \includegraphics[height=0.1\textheight]{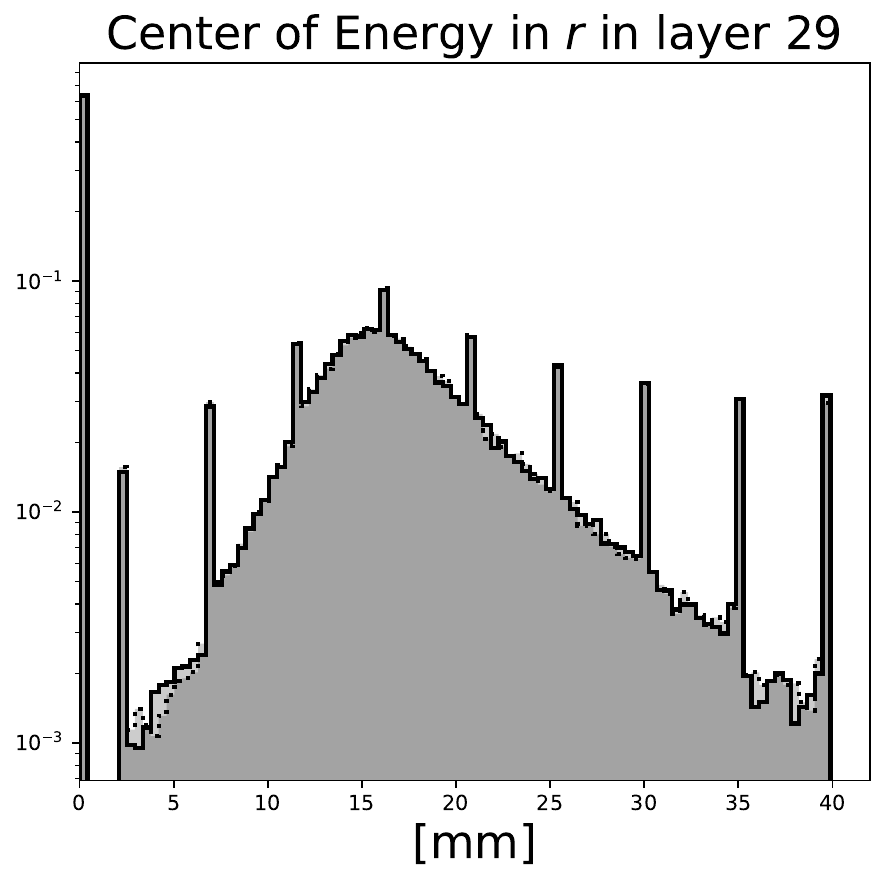}\\
    \includegraphics[height=0.1\textheight]{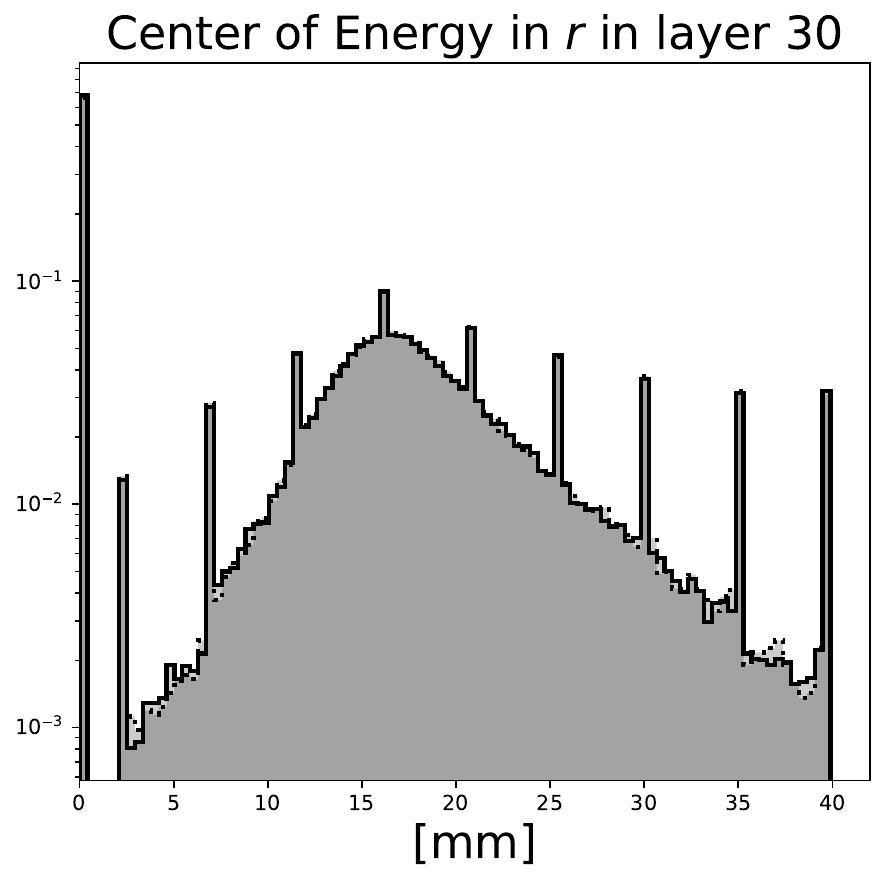} \hfill \includegraphics[height=0.1\textheight]{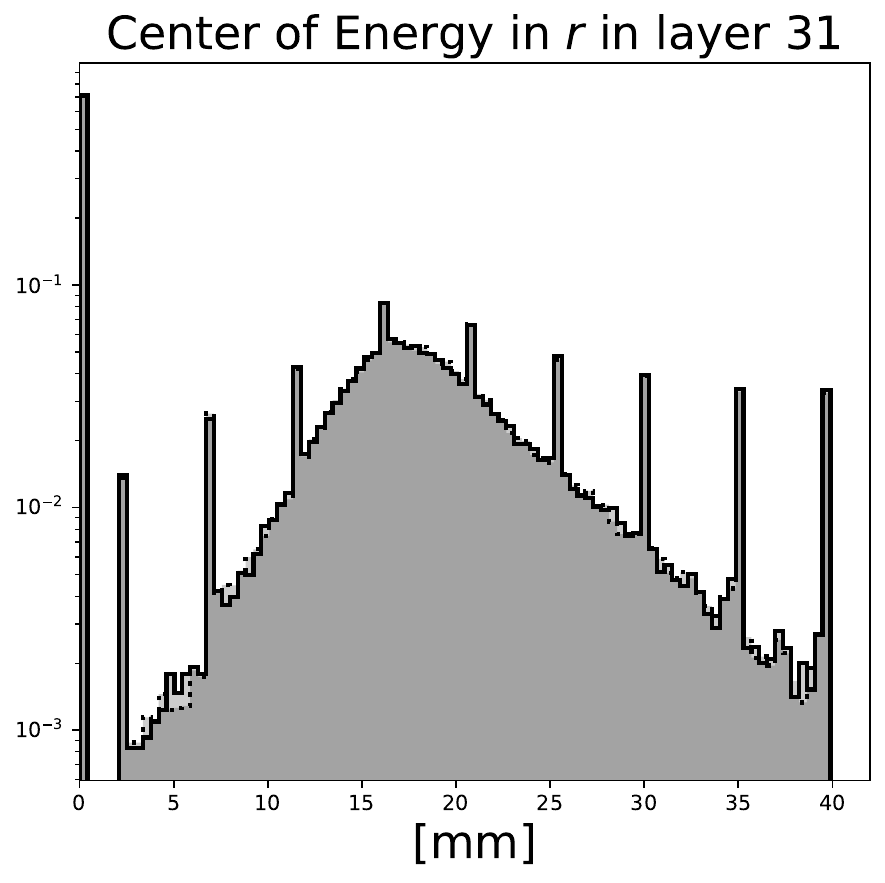} \hfill \includegraphics[height=0.1\textheight]{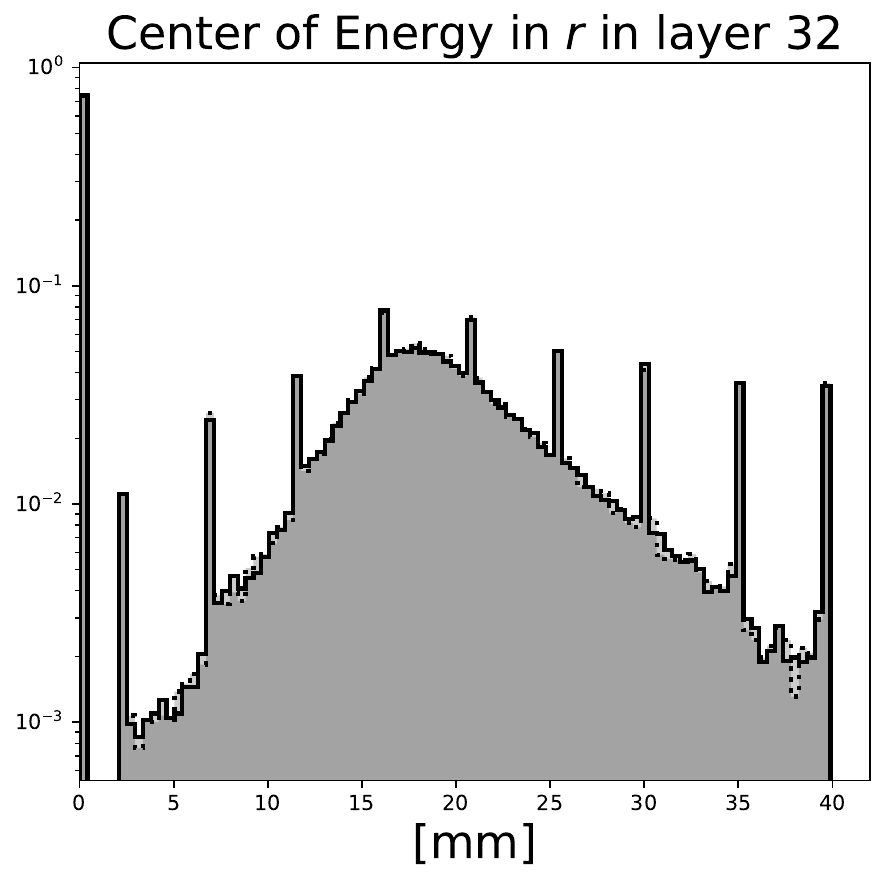} \hfill \includegraphics[height=0.1\textheight]{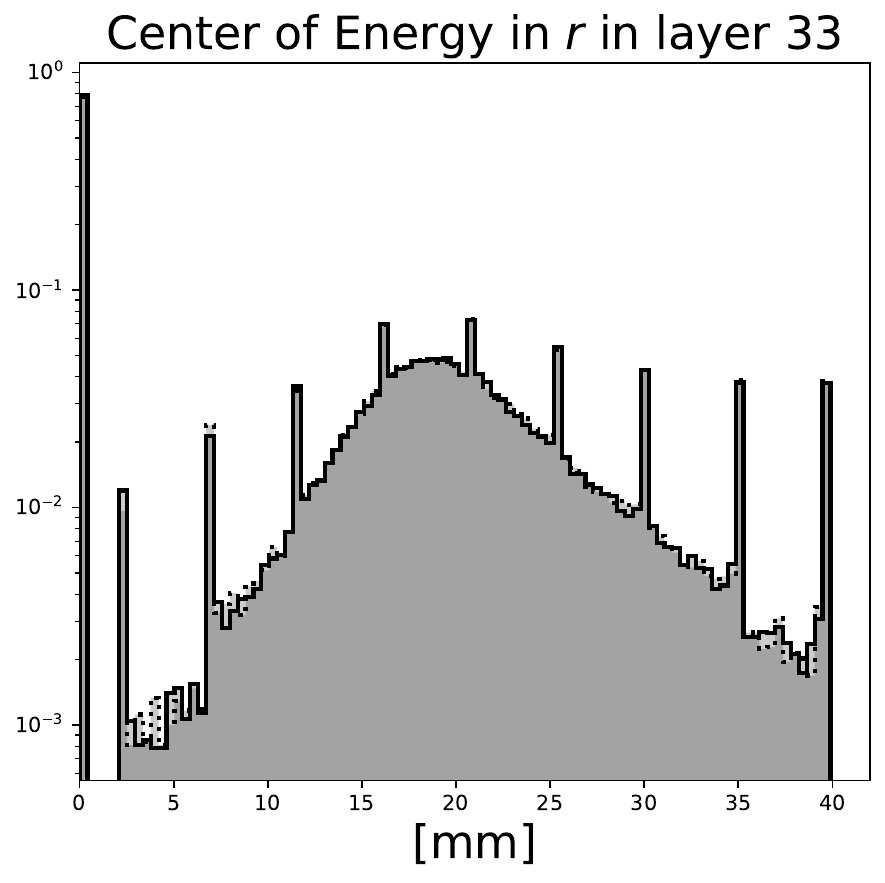} \hfill \includegraphics[height=0.1\textheight]{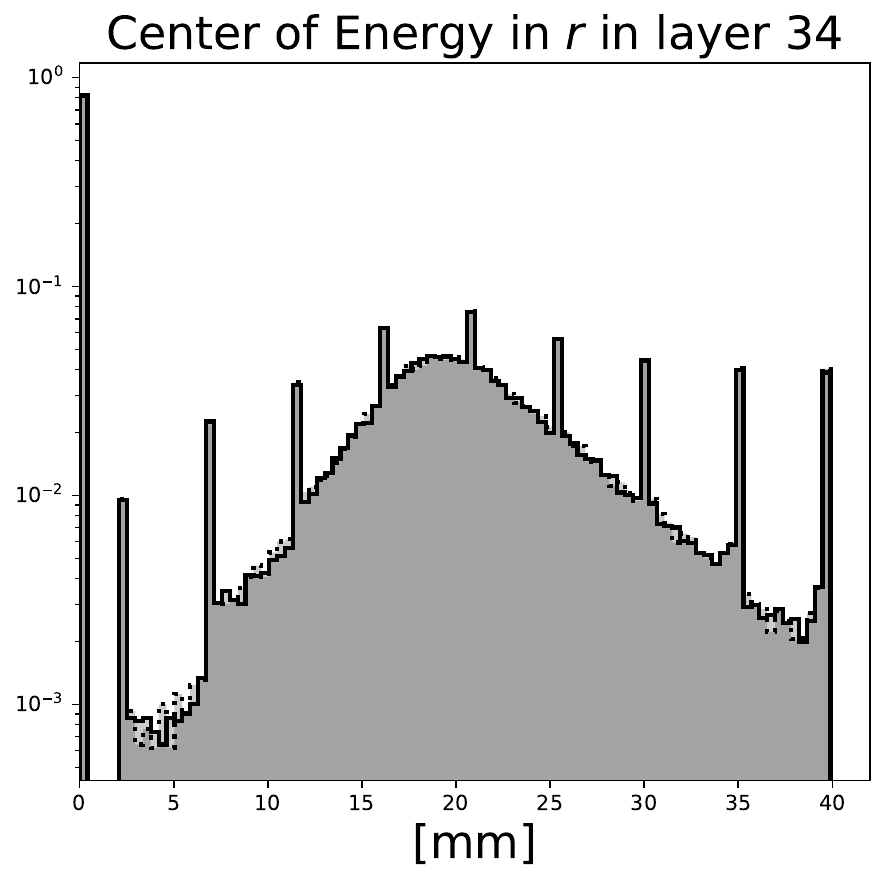}\\
    \includegraphics[height=0.1\textheight]{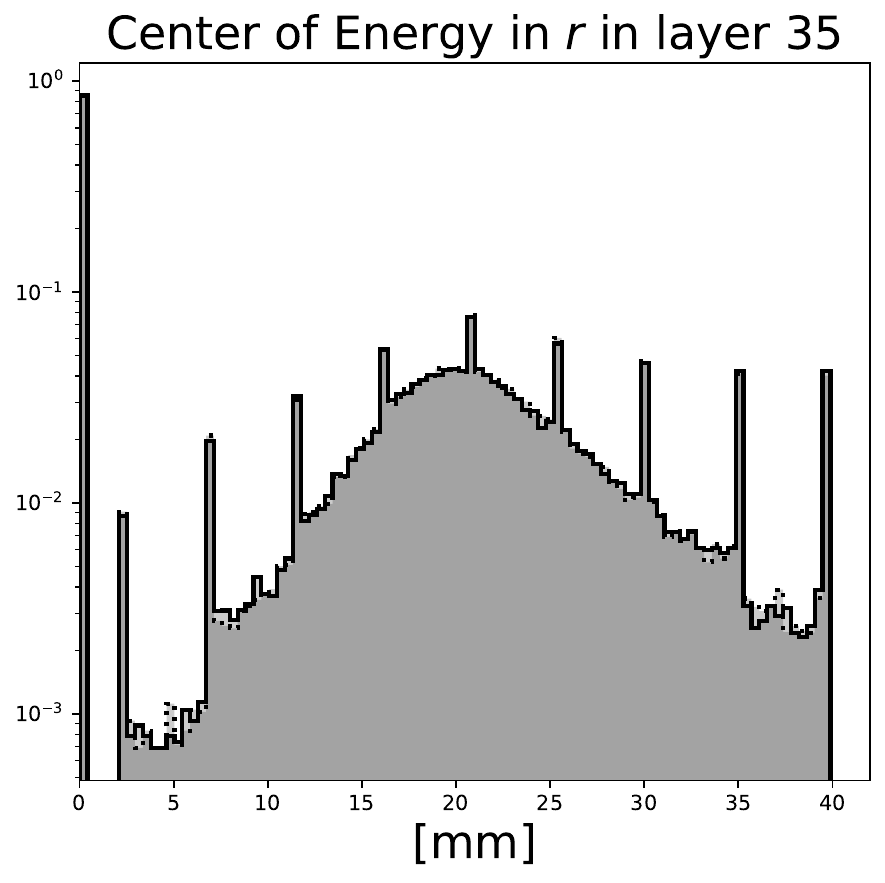} \hfill \includegraphics[height=0.1\textheight]{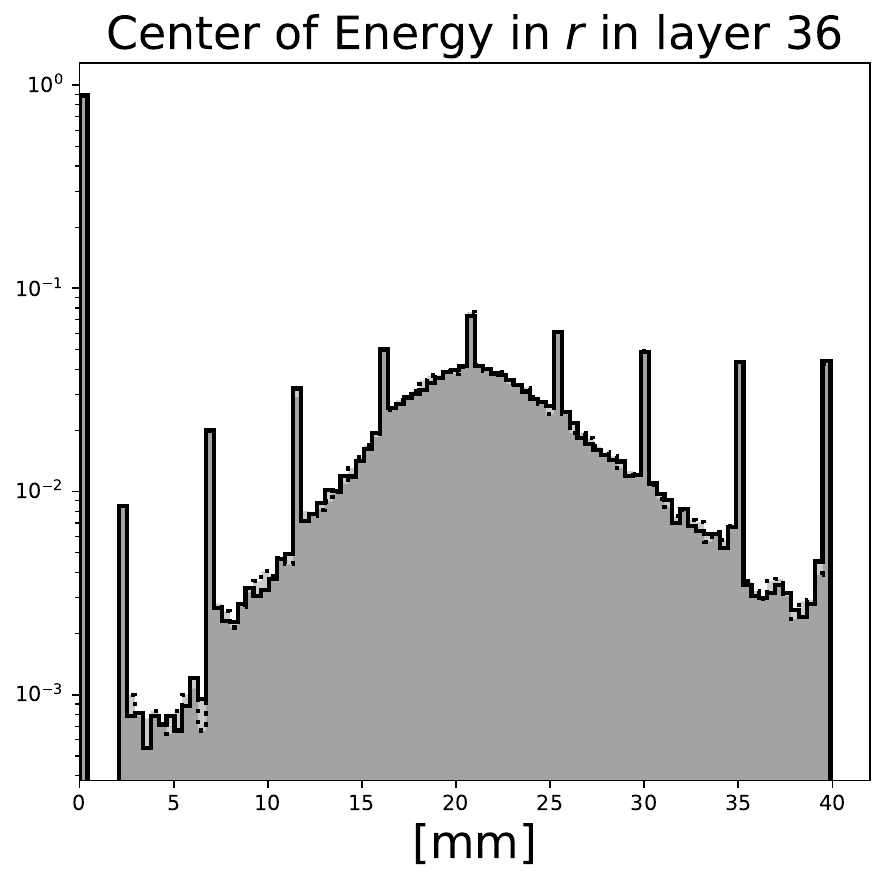} \hfill \includegraphics[height=0.1\textheight]{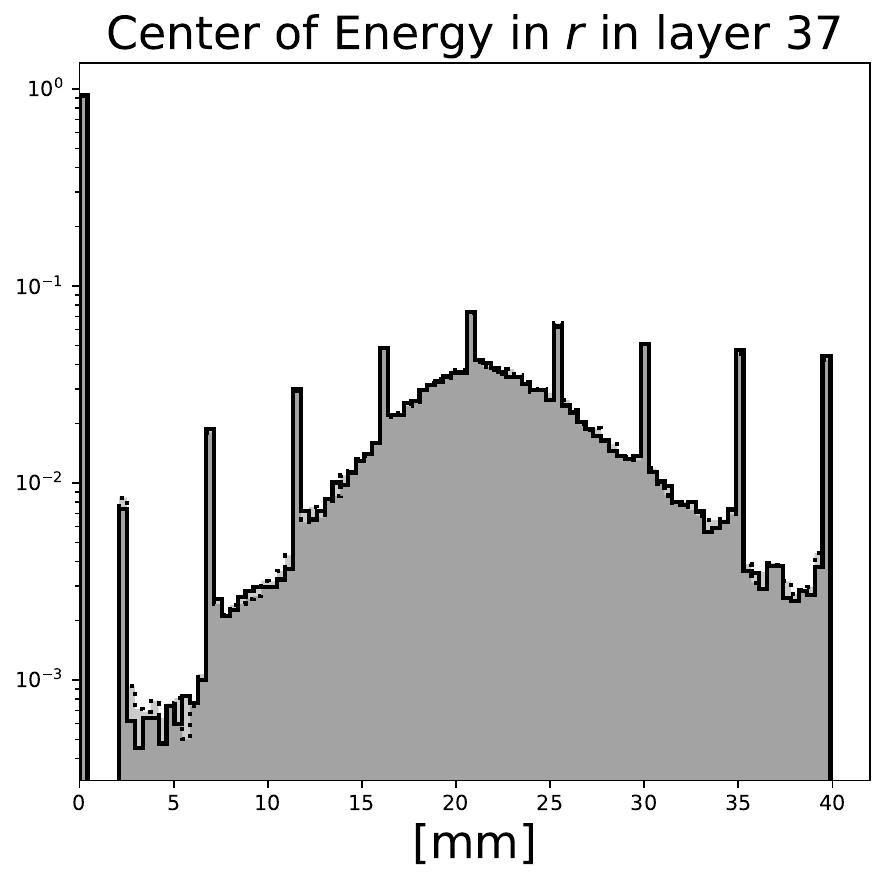} \hfill \includegraphics[height=0.1\textheight]{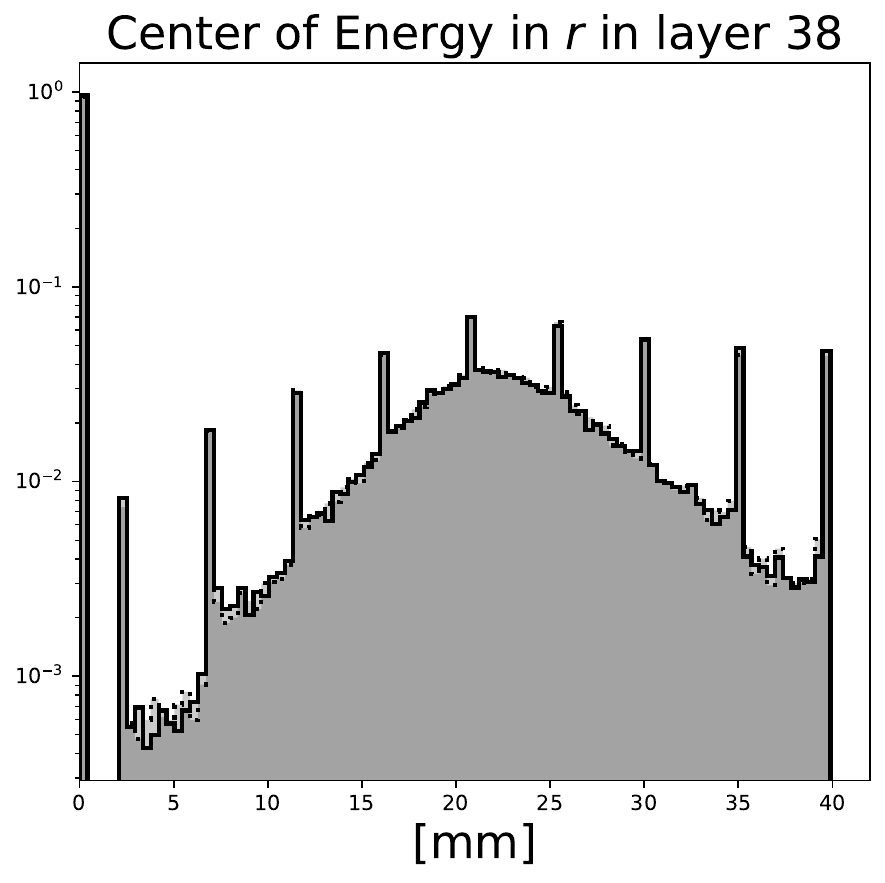} \hfill \includegraphics[height=0.1\textheight]{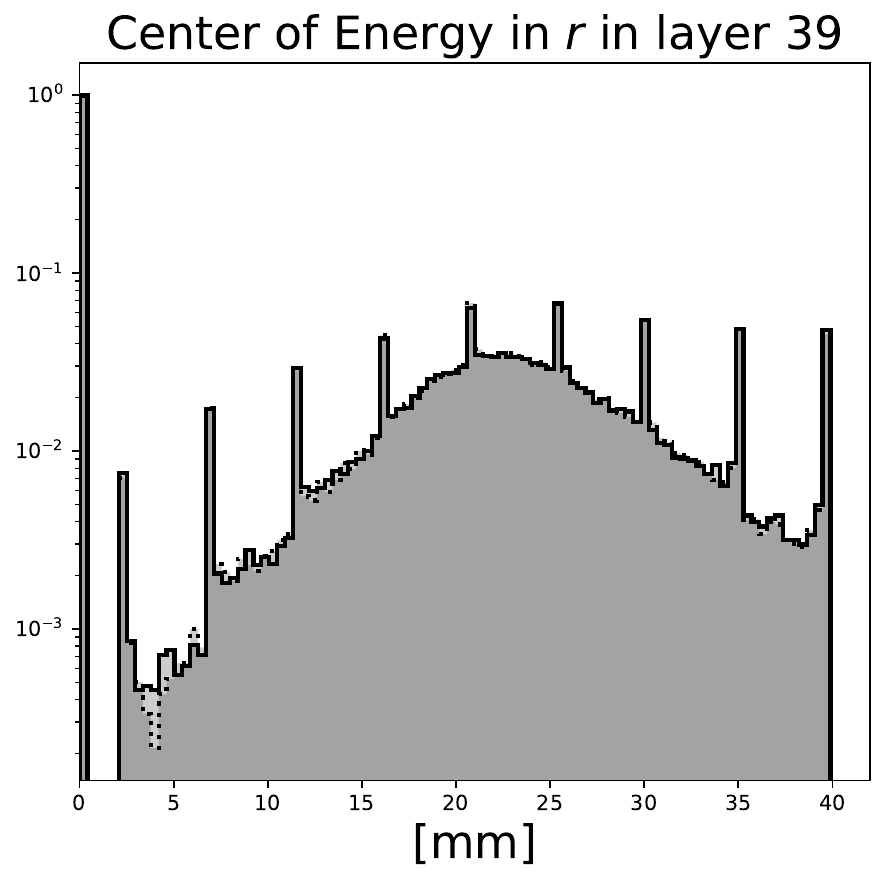}\\
    \includegraphics[height=0.1\textheight]{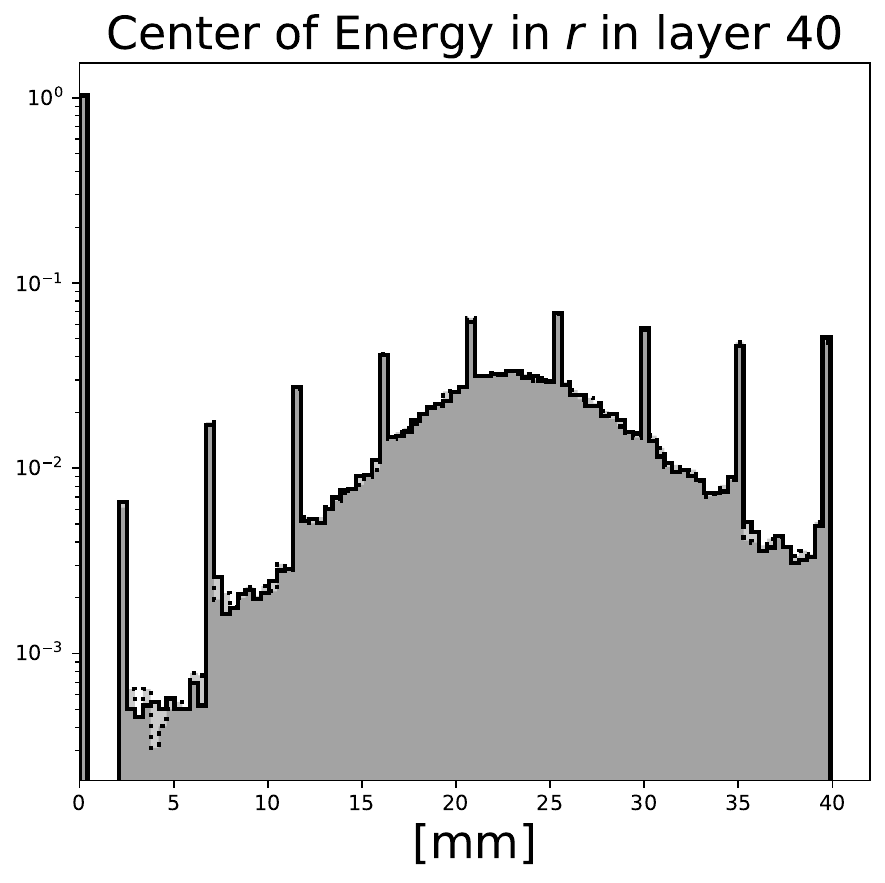} \hfill \includegraphics[height=0.1\textheight]{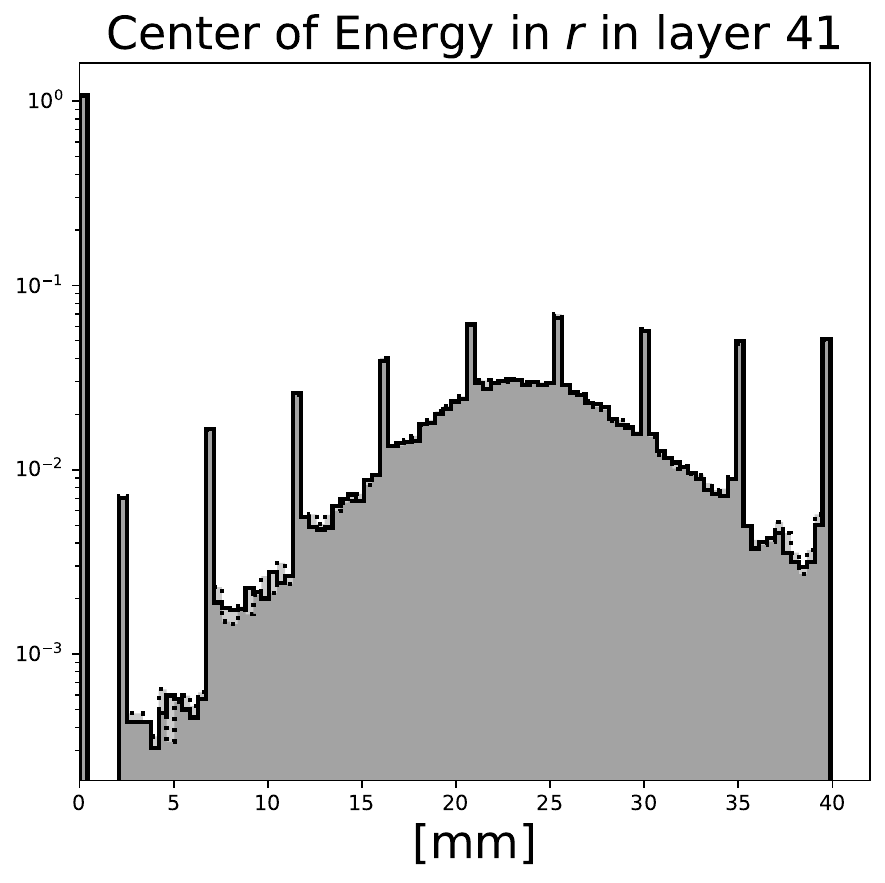} \hfill \includegraphics[height=0.1\textheight]{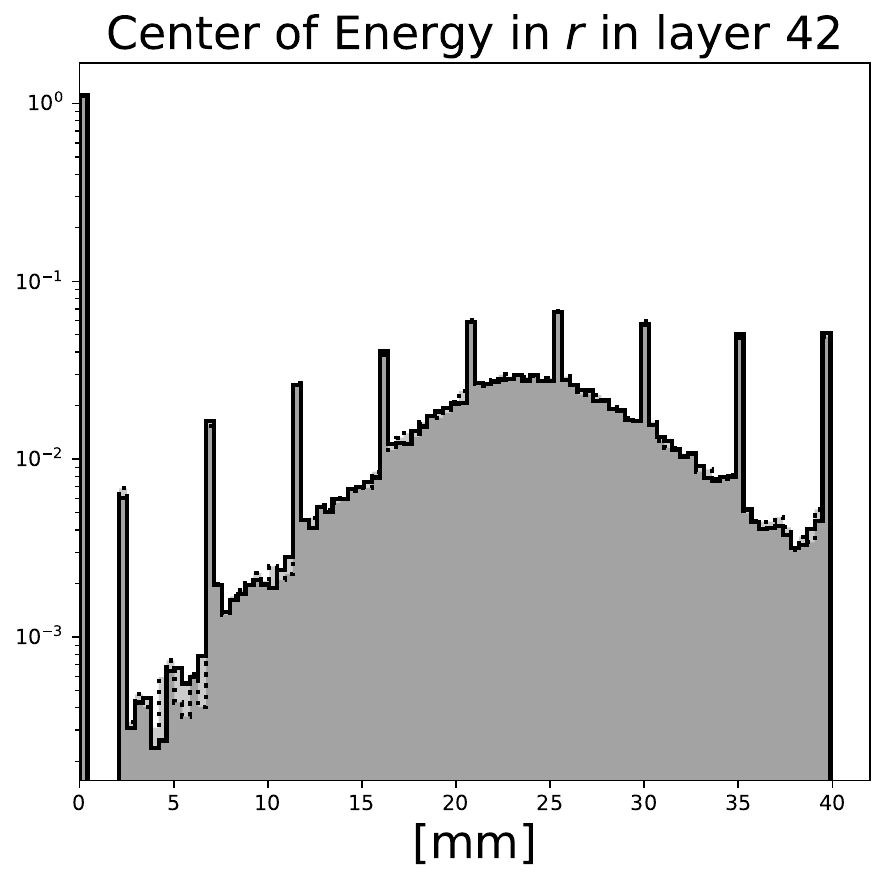} \hfill \includegraphics[height=0.1\textheight]{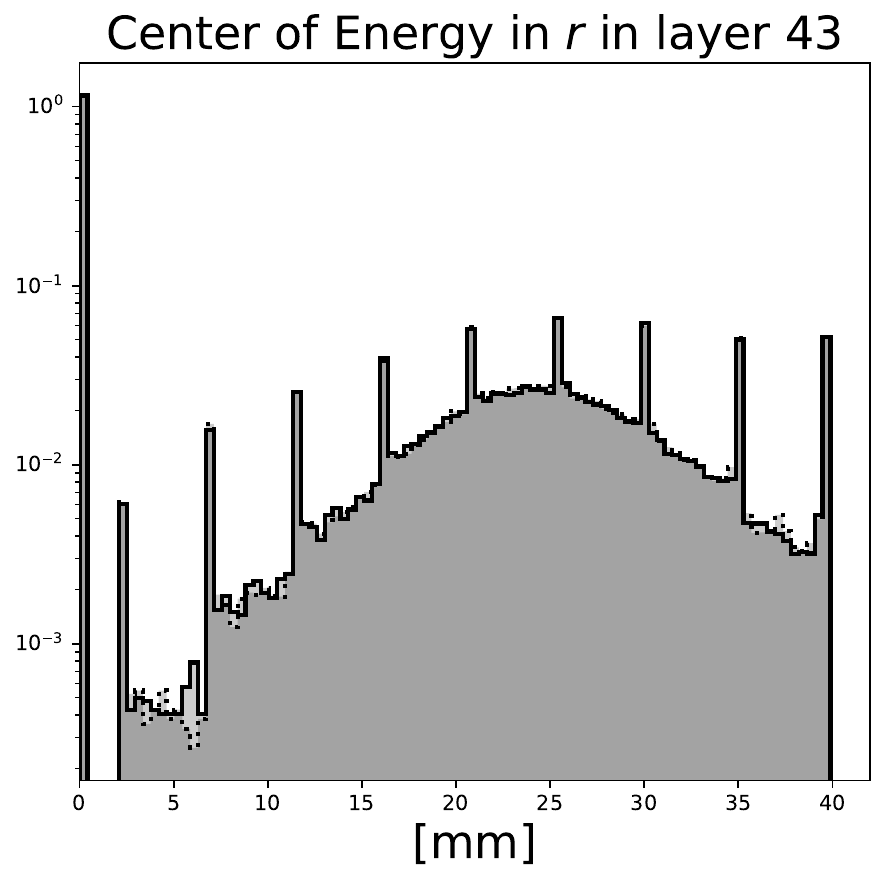} \hfill \includegraphics[height=0.1\textheight]{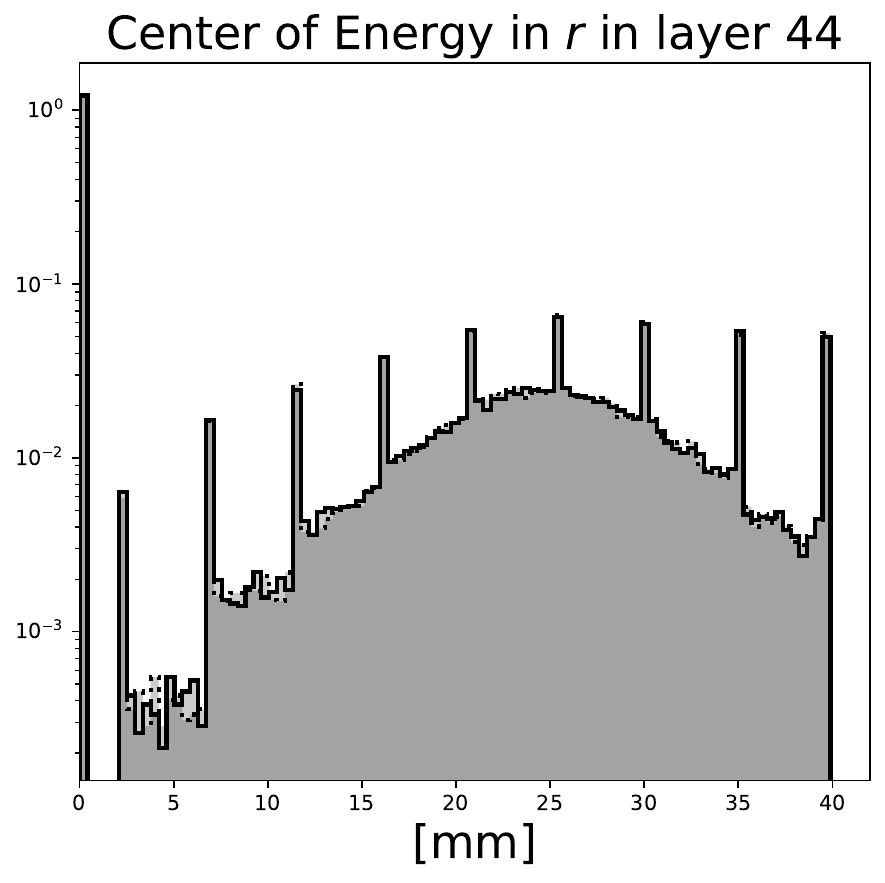}\\
    \includegraphics[width=0.5\textwidth]{figures/Appendix_reference/legend.pdf}
    \caption{Distribution of \geant training and evaluation data in centers of energy in $r$ direction for ds2. }
    \label{fig:app_ref.ds2.7}
\end{figure}

\begin{figure}[ht]
    \centering
    \includegraphics[height=0.1\textheight]{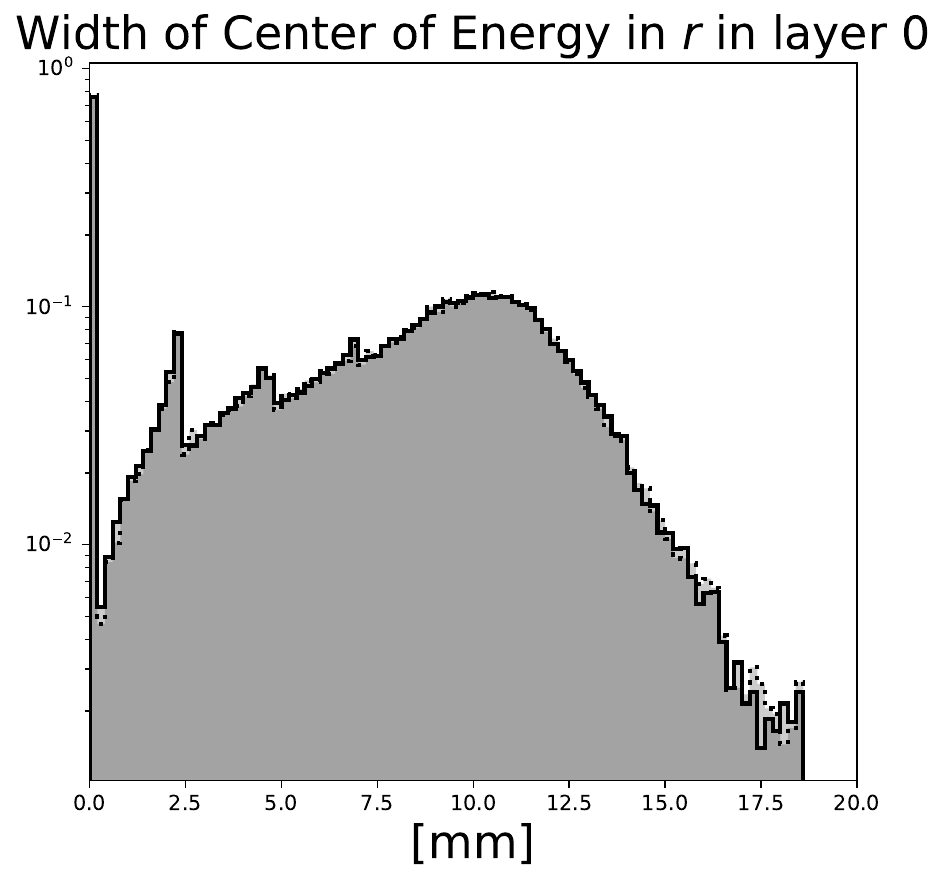} \hfill \includegraphics[height=0.1\textheight]{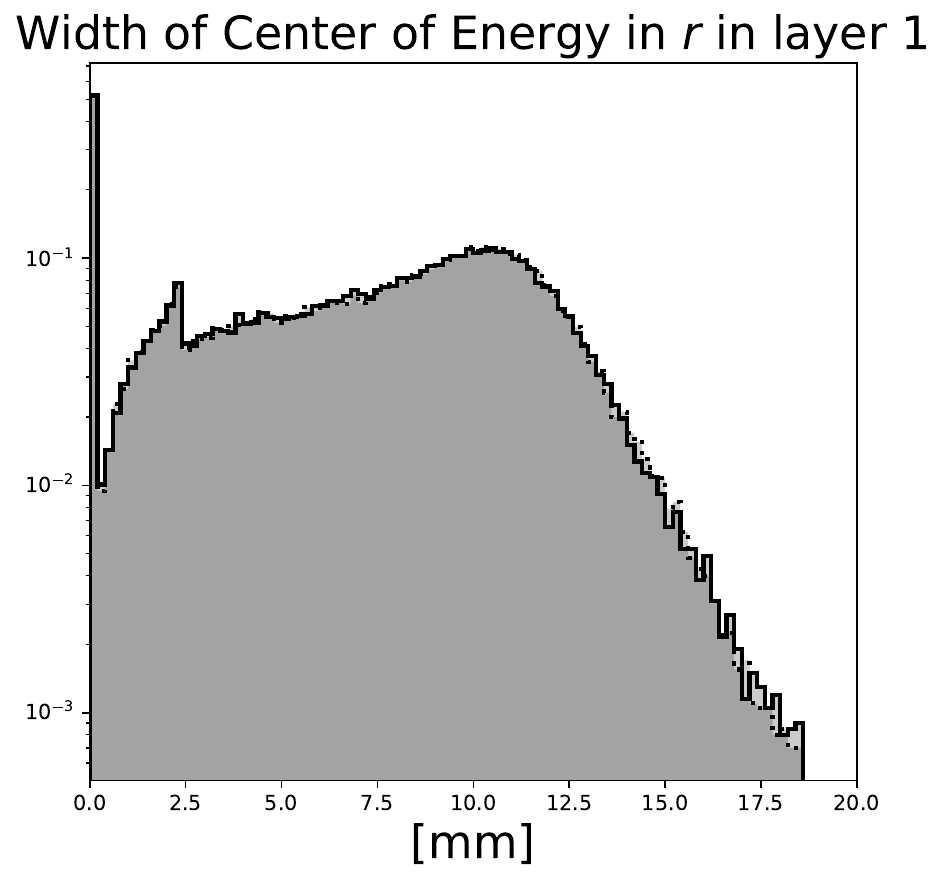} \hfill \includegraphics[height=0.1\textheight]{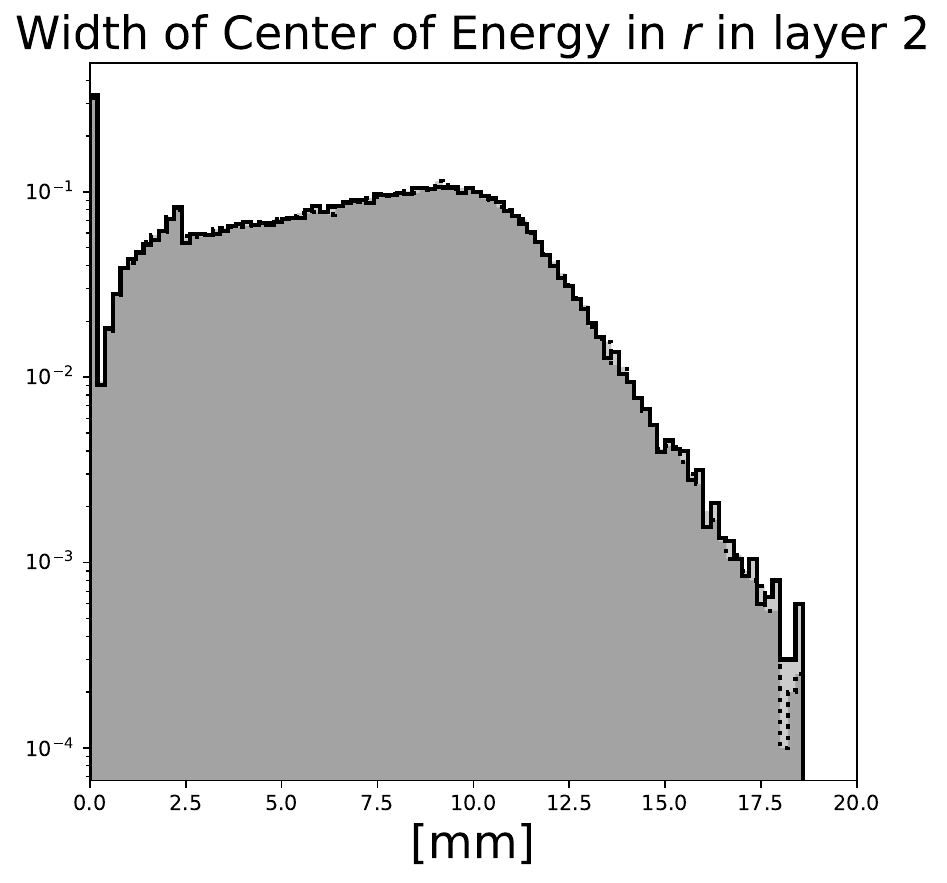} \hfill \includegraphics[height=0.1\textheight]{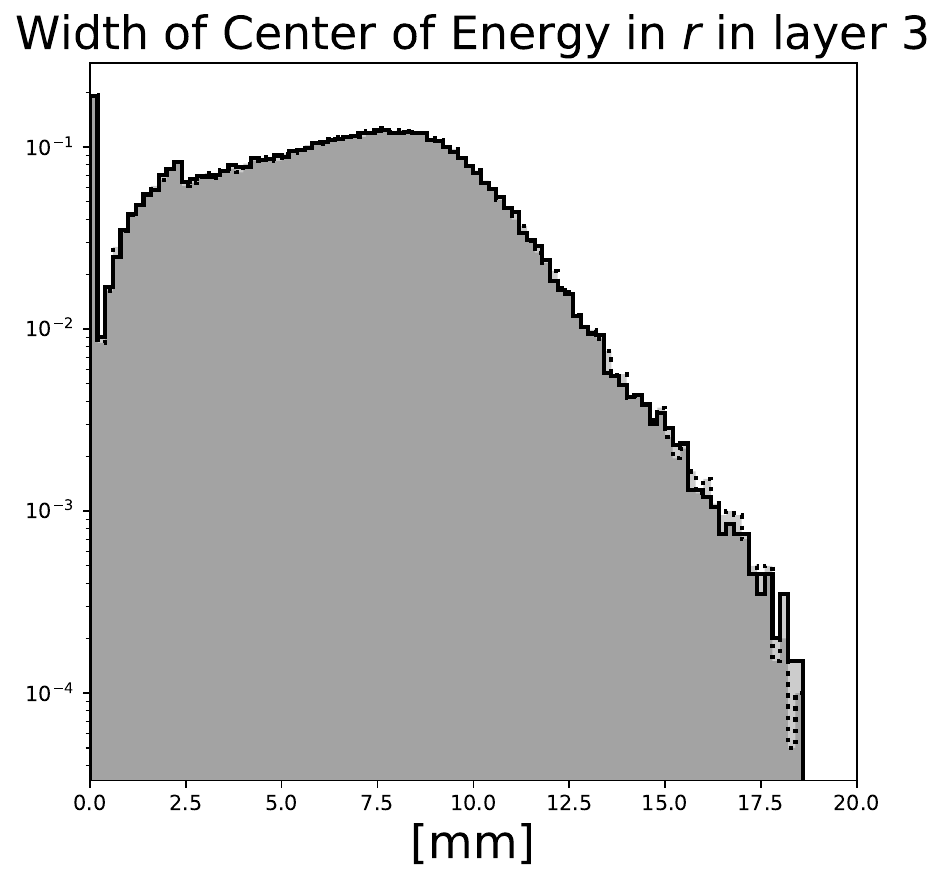} \hfill \includegraphics[height=0.1\textheight]{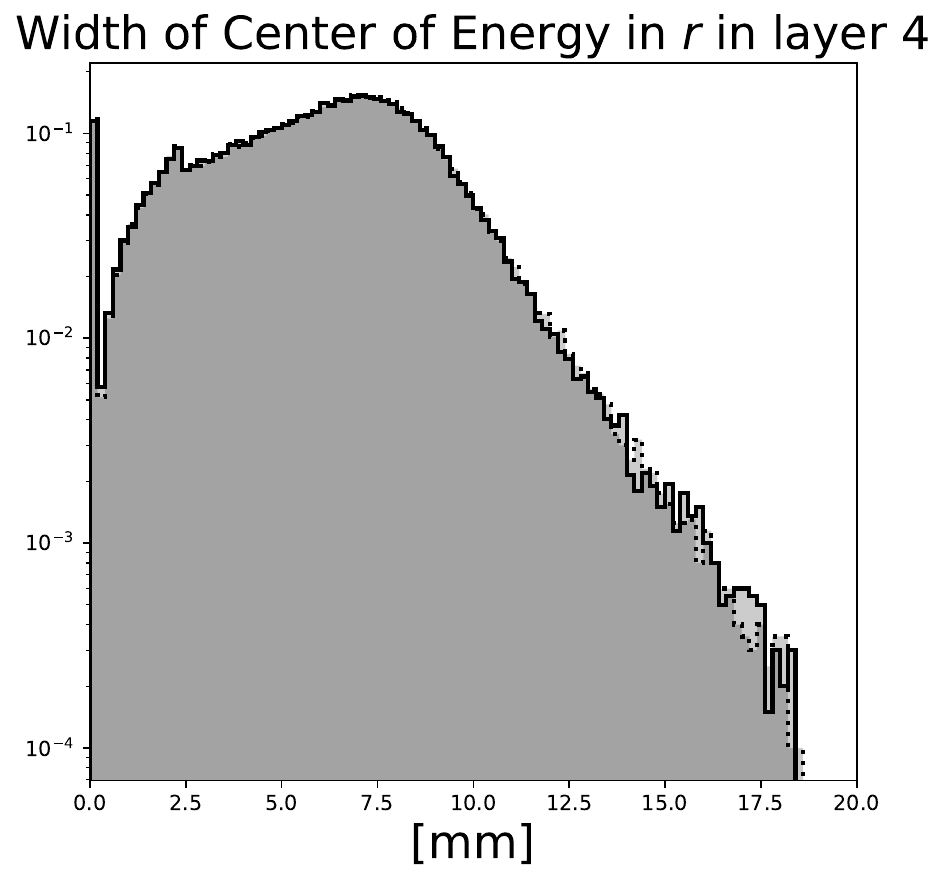}\\
    \includegraphics[height=0.1\textheight]{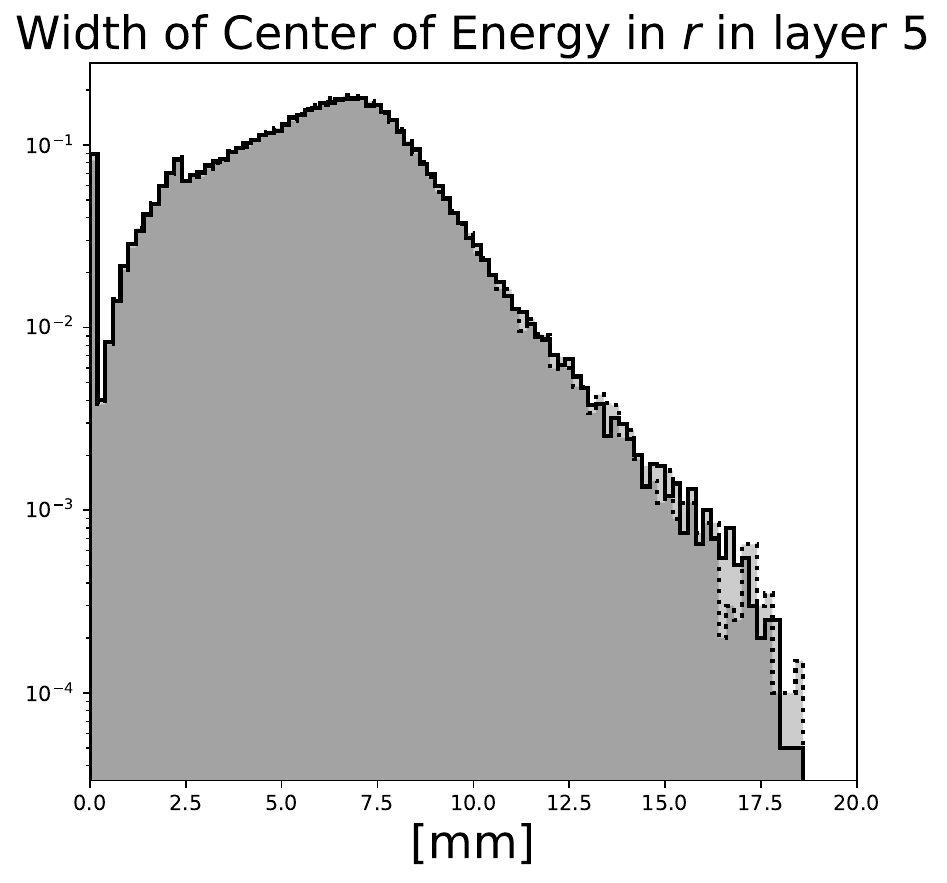} \hfill \includegraphics[height=0.1\textheight]{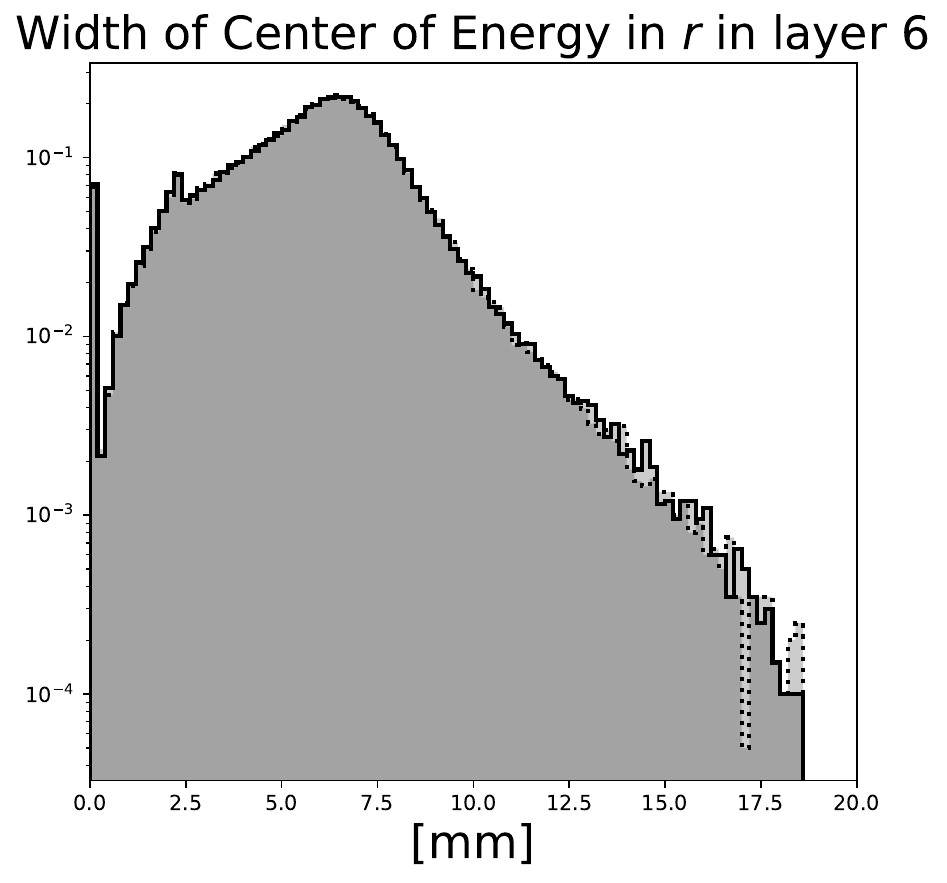} \hfill \includegraphics[height=0.1\textheight]{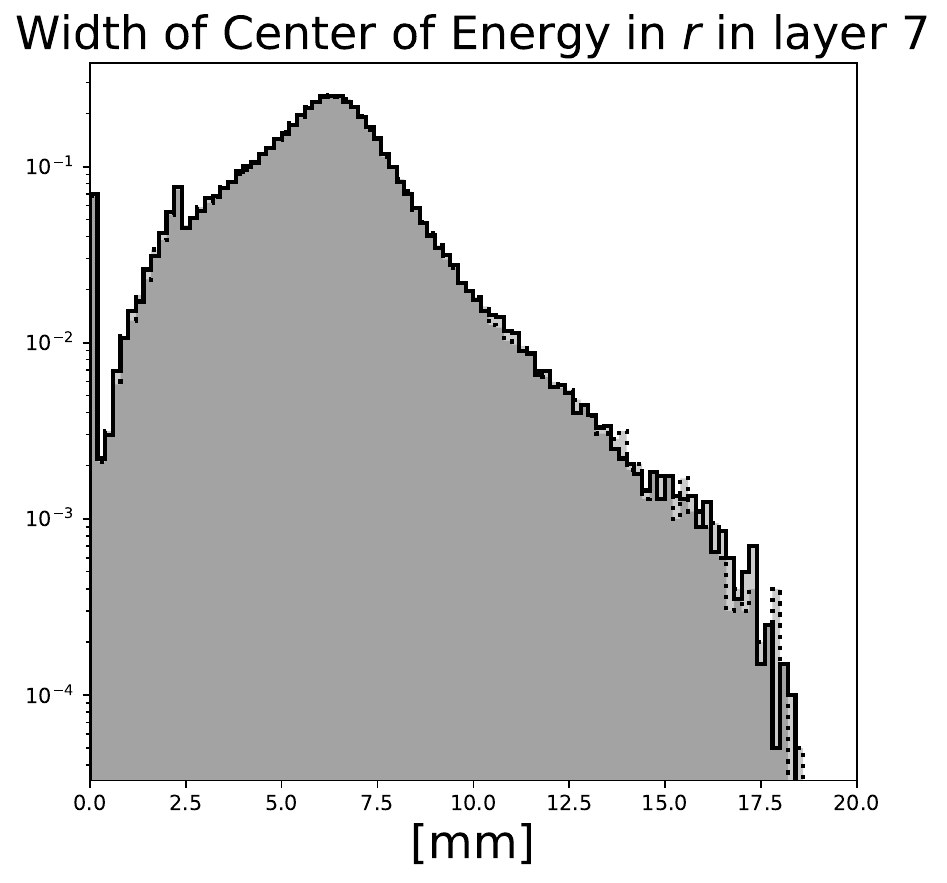} \hfill \includegraphics[height=0.1\textheight]{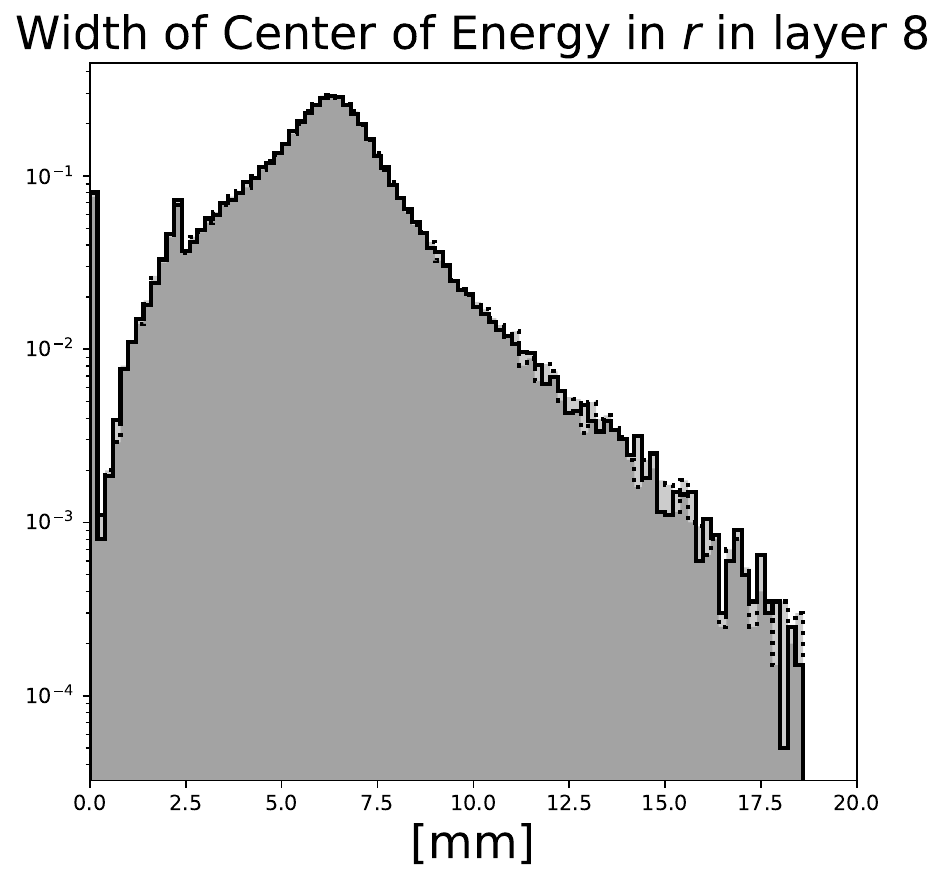} \hfill \includegraphics[height=0.1\textheight]{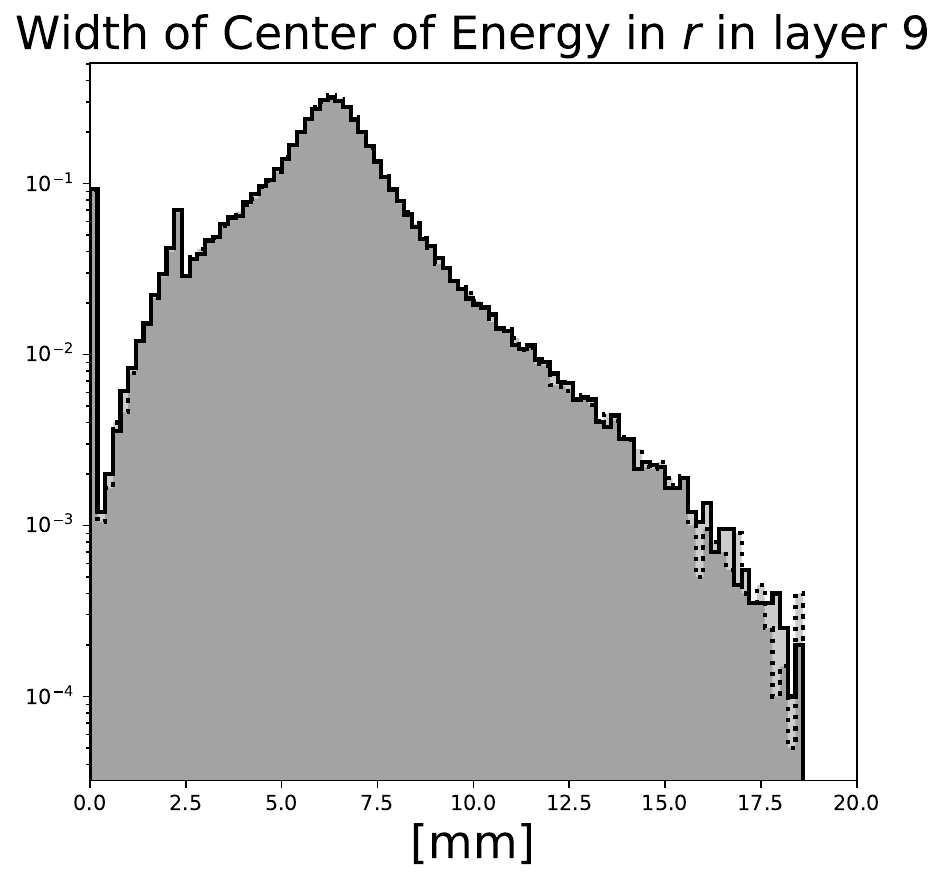}\\
    \includegraphics[height=0.1\textheight]{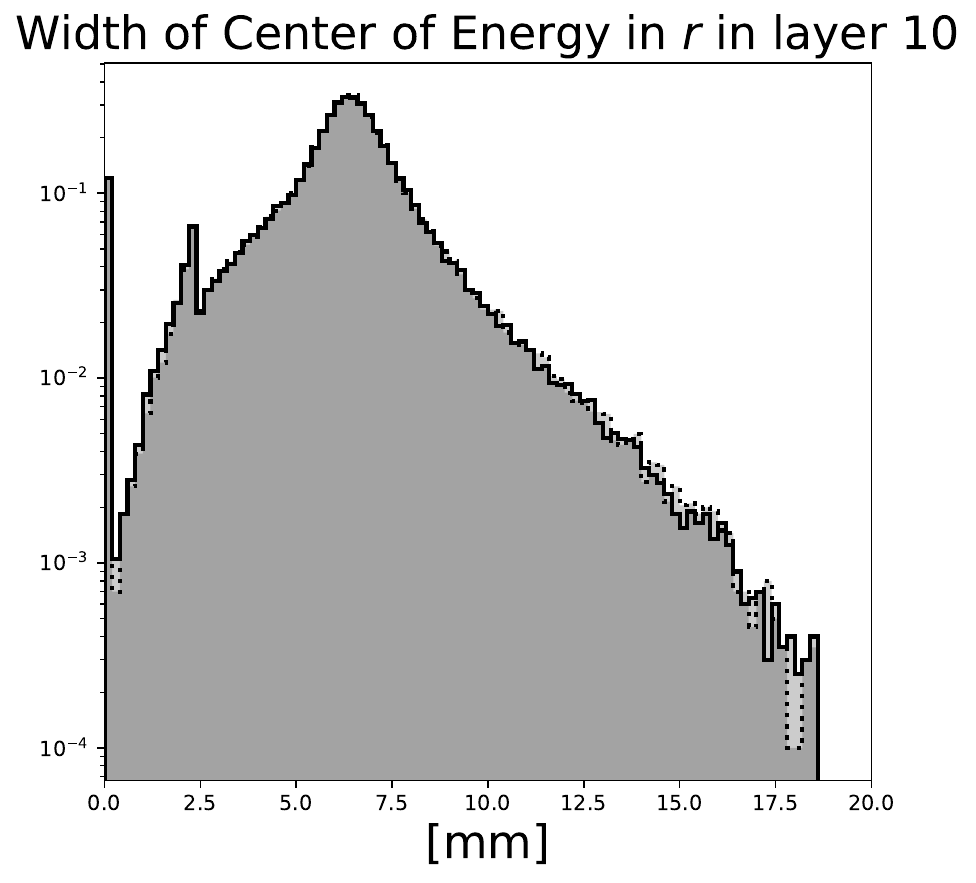} \hfill \includegraphics[height=0.1\textheight]{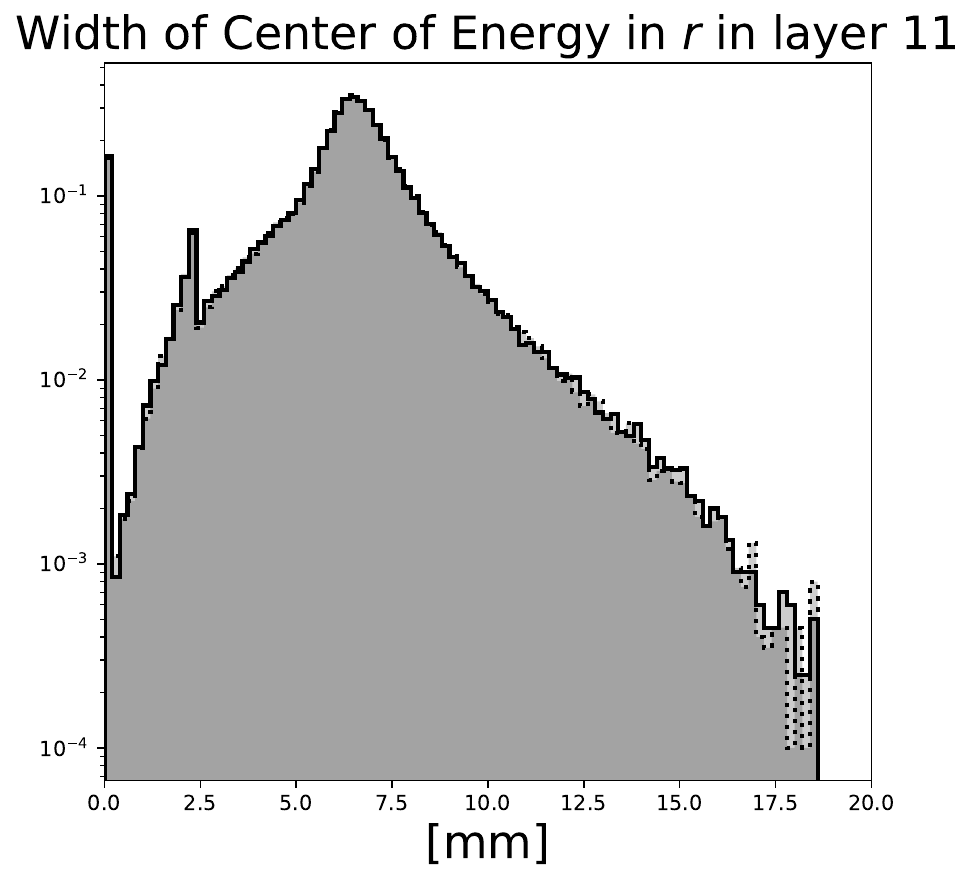} \hfill \includegraphics[height=0.1\textheight]{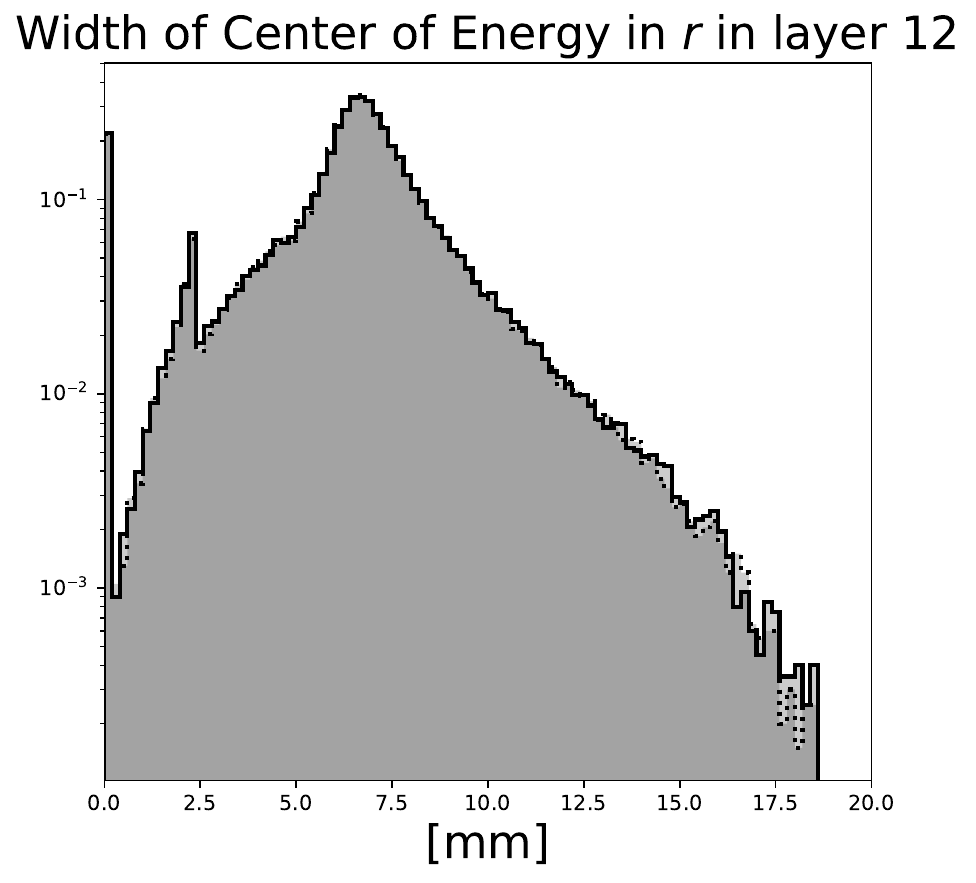} \hfill \includegraphics[height=0.1\textheight]{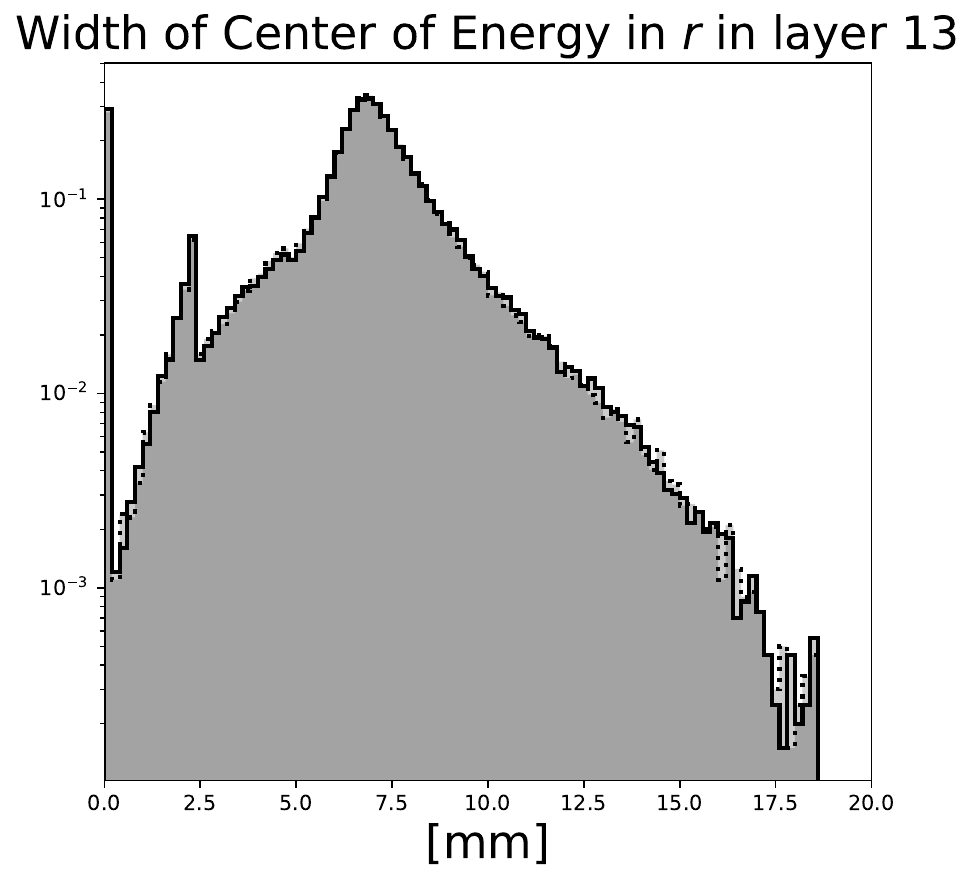} \hfill \includegraphics[height=0.1\textheight]{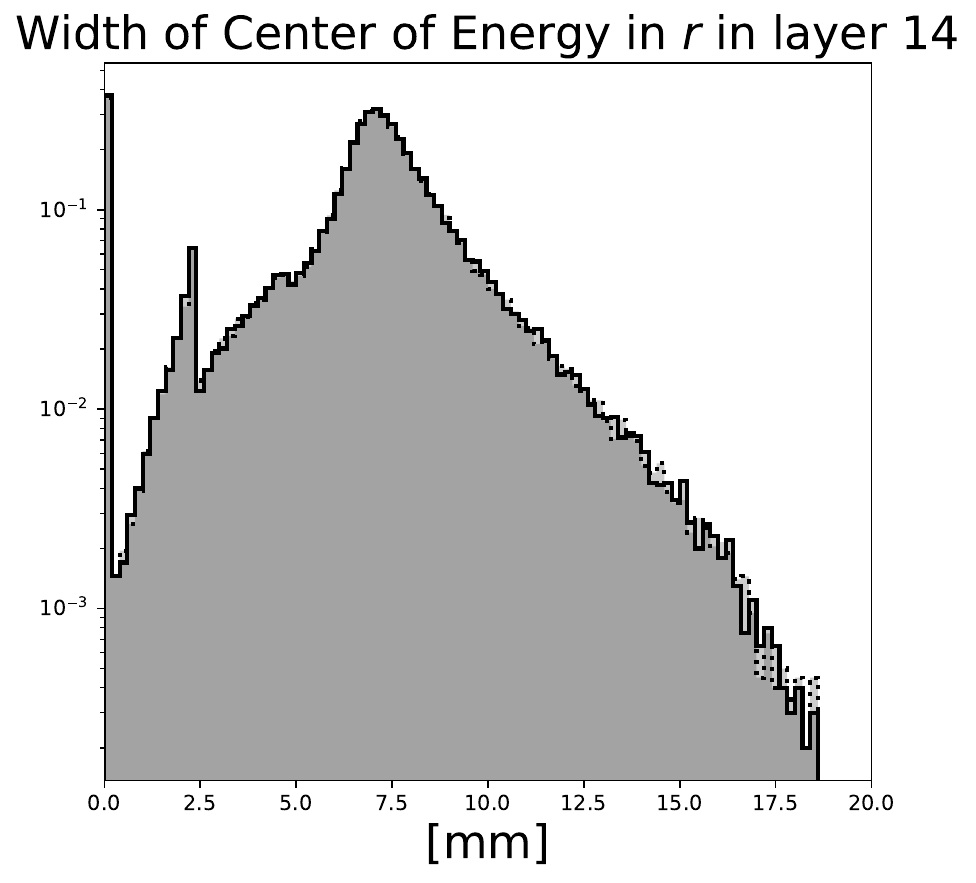}\\
    \includegraphics[height=0.1\textheight]{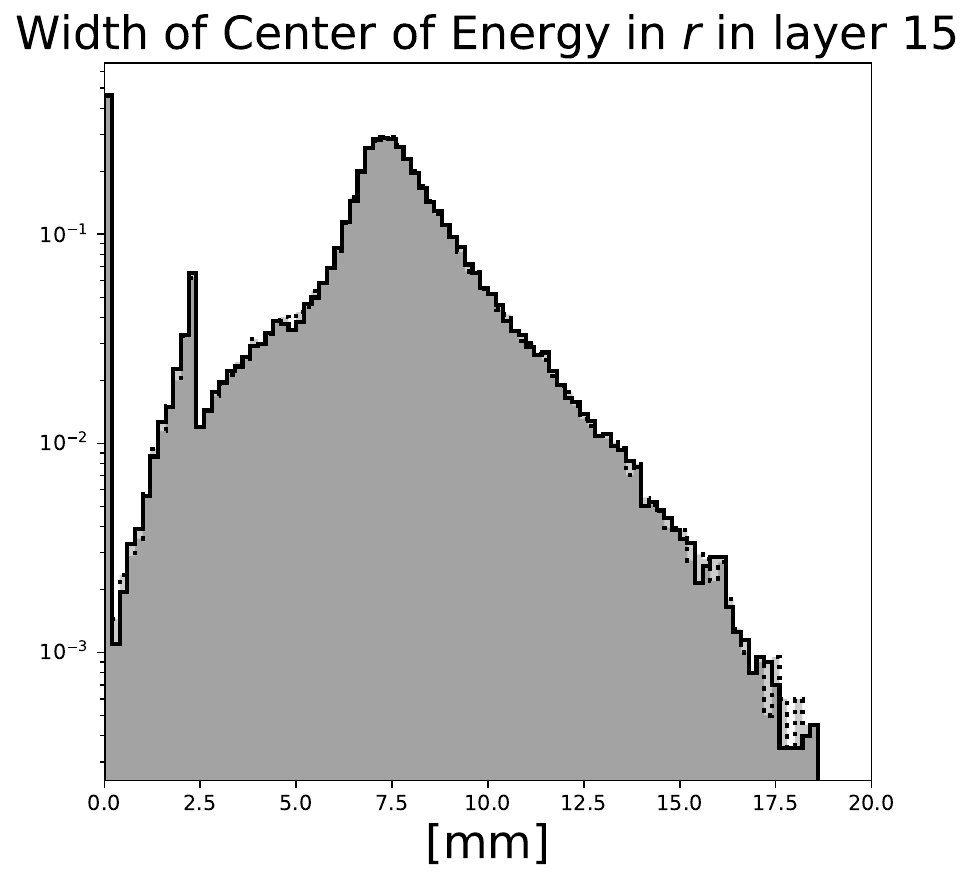} \hfill \includegraphics[height=0.1\textheight]{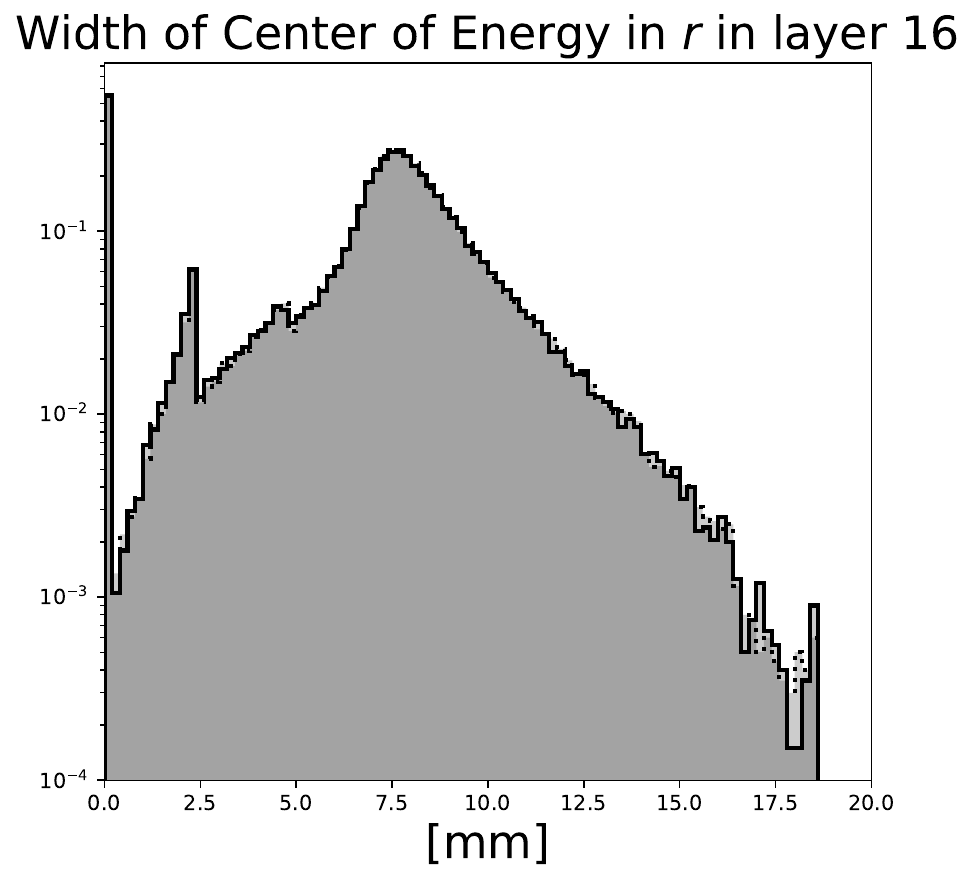} \hfill \includegraphics[height=0.1\textheight]{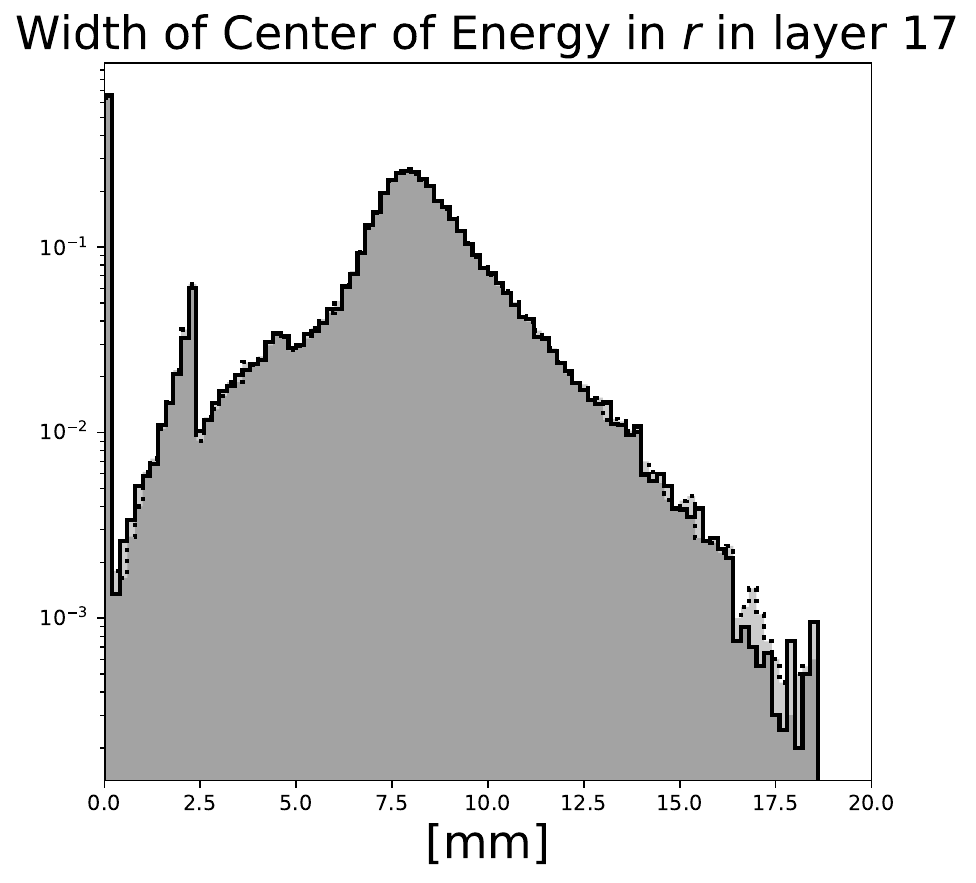} \hfill \includegraphics[height=0.1\textheight]{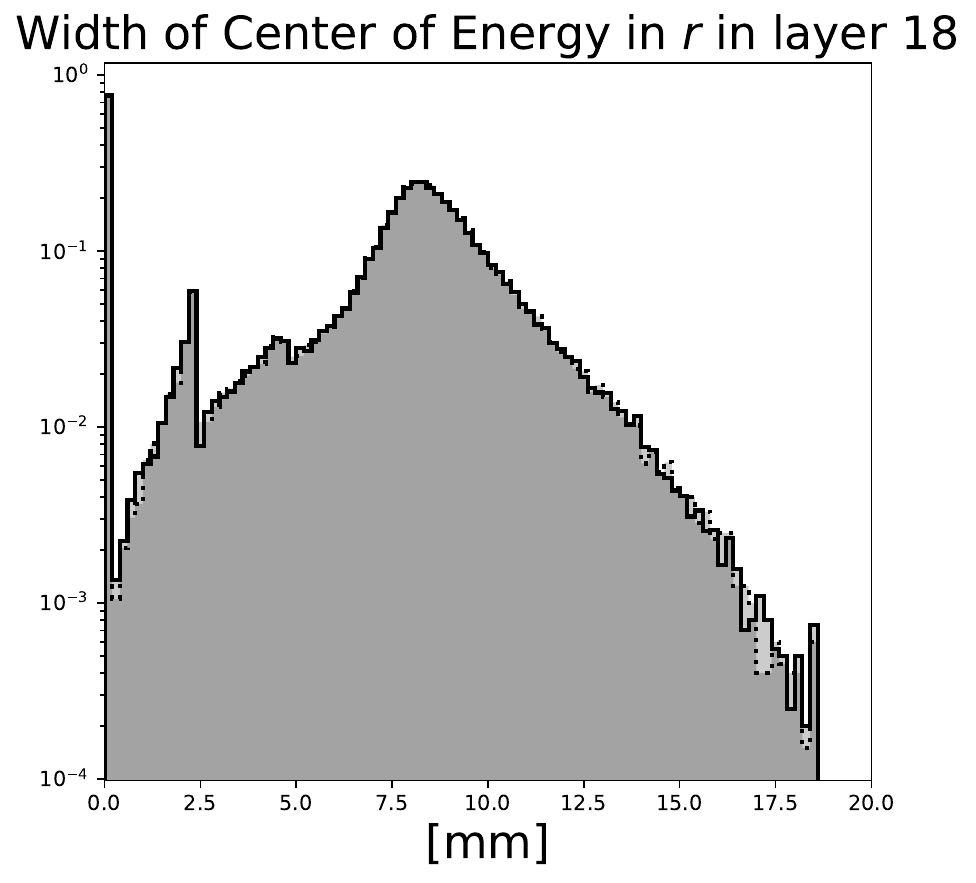} \hfill \includegraphics[height=0.1\textheight]{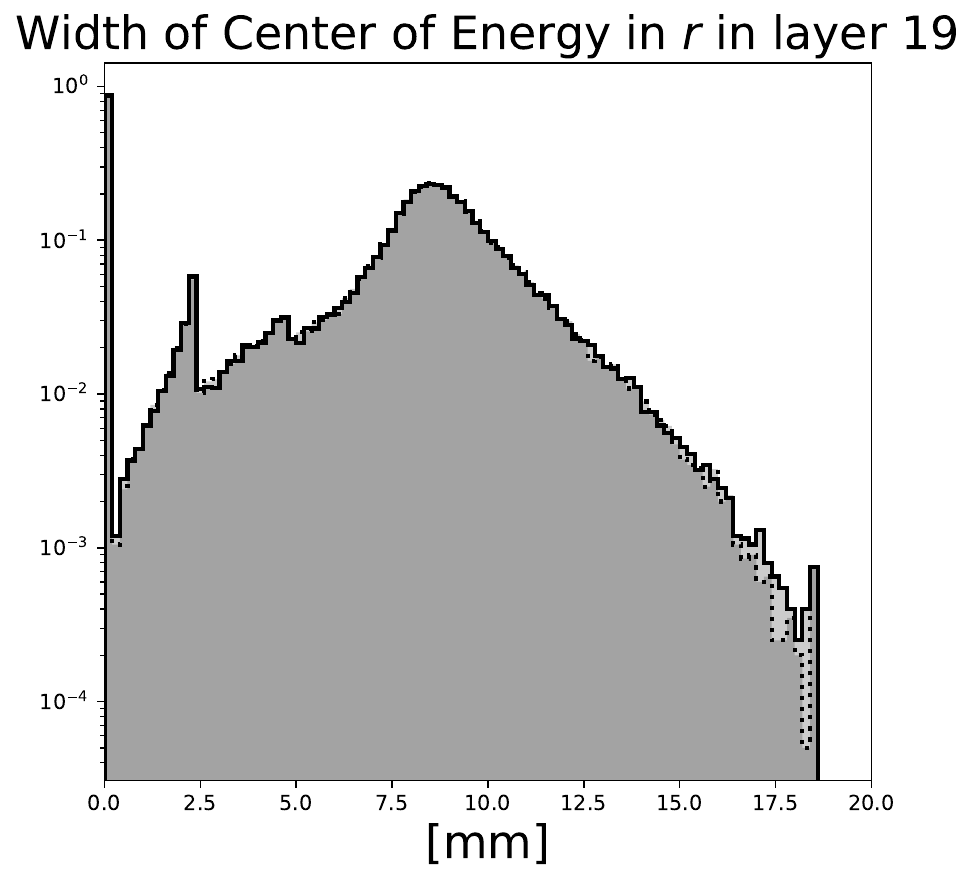}\\
    \includegraphics[height=0.1\textheight]{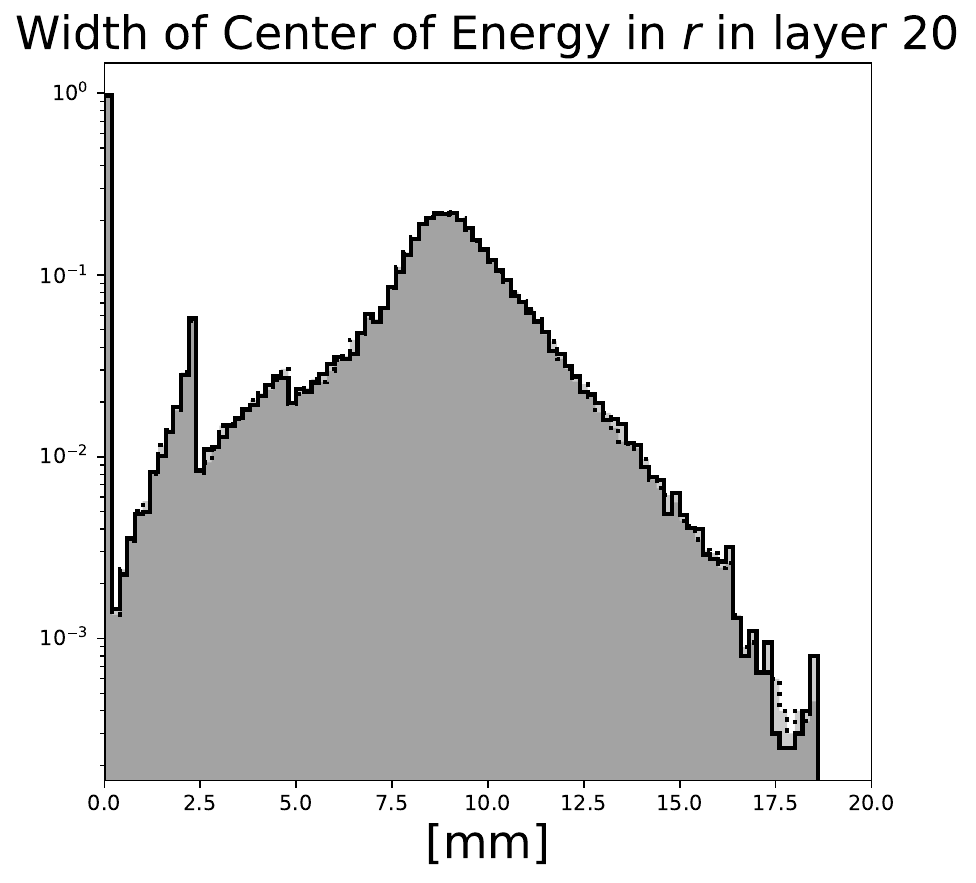} \hfill \includegraphics[height=0.1\textheight]{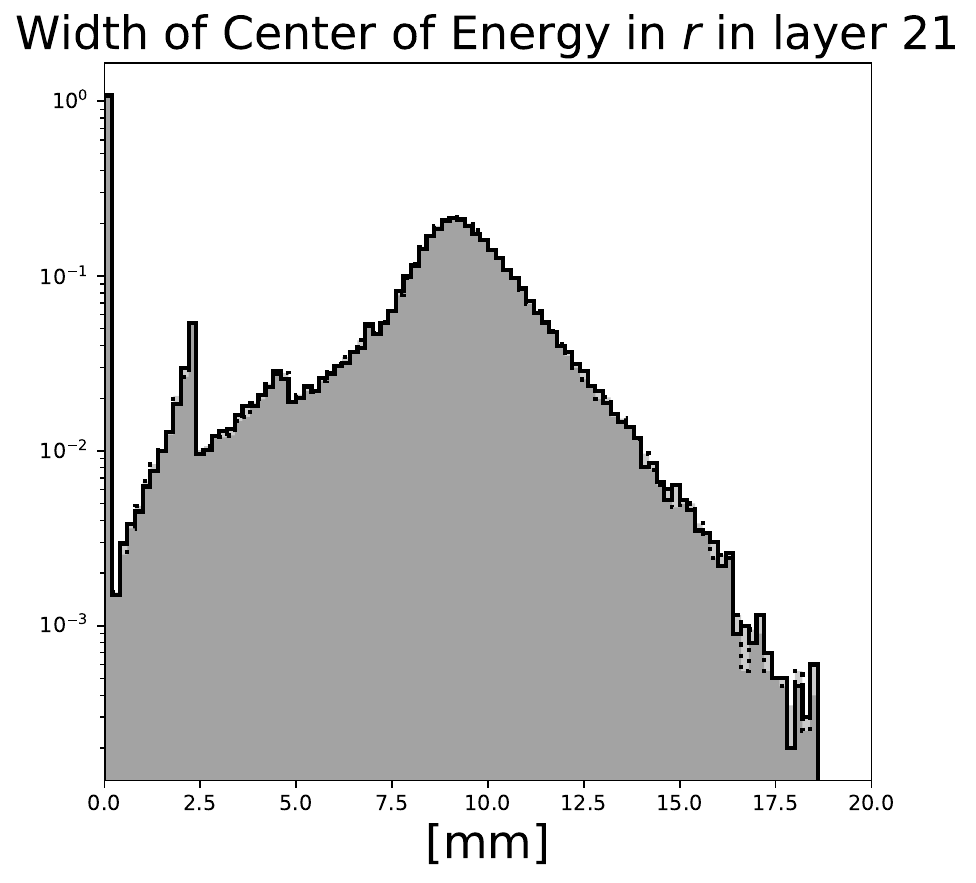} \hfill \includegraphics[height=0.1\textheight]{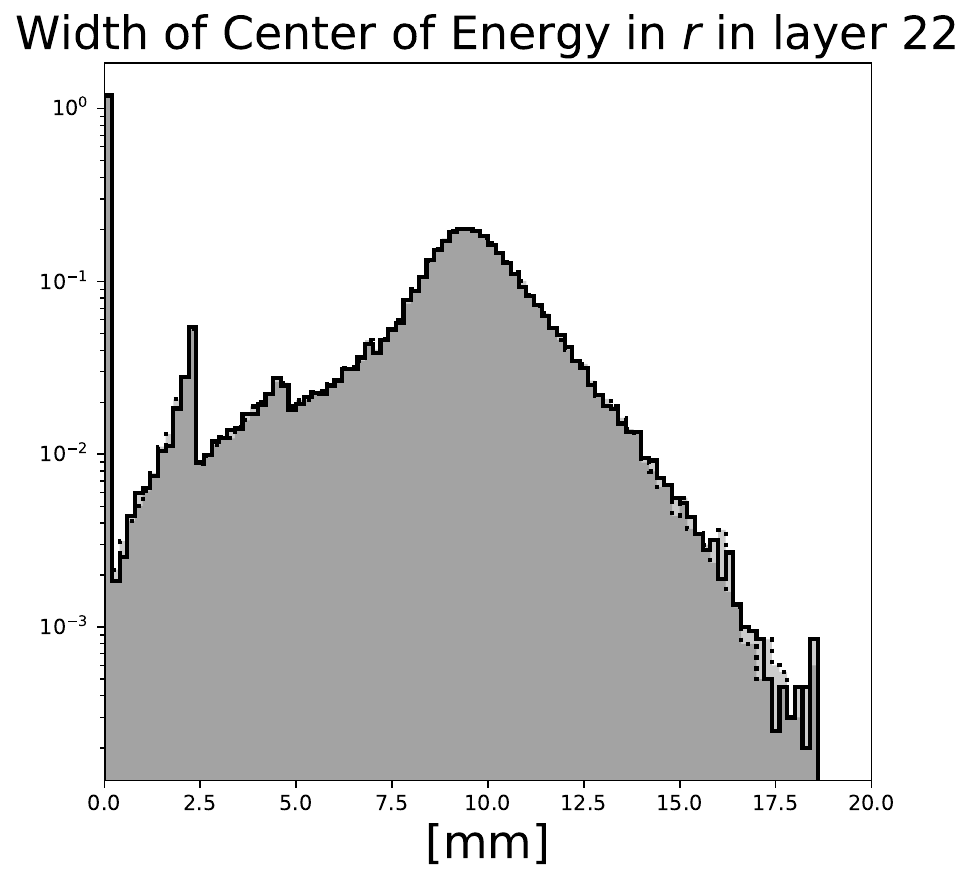} \hfill \includegraphics[height=0.1\textheight]{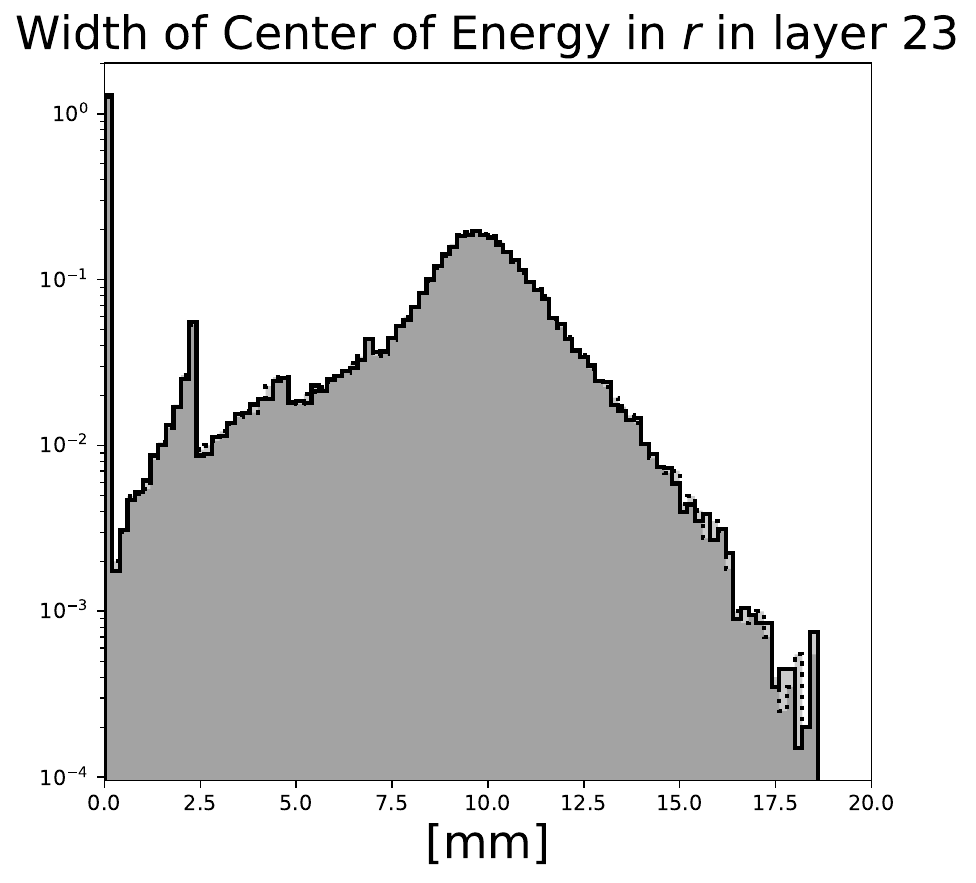} \hfill \includegraphics[height=0.1\textheight]{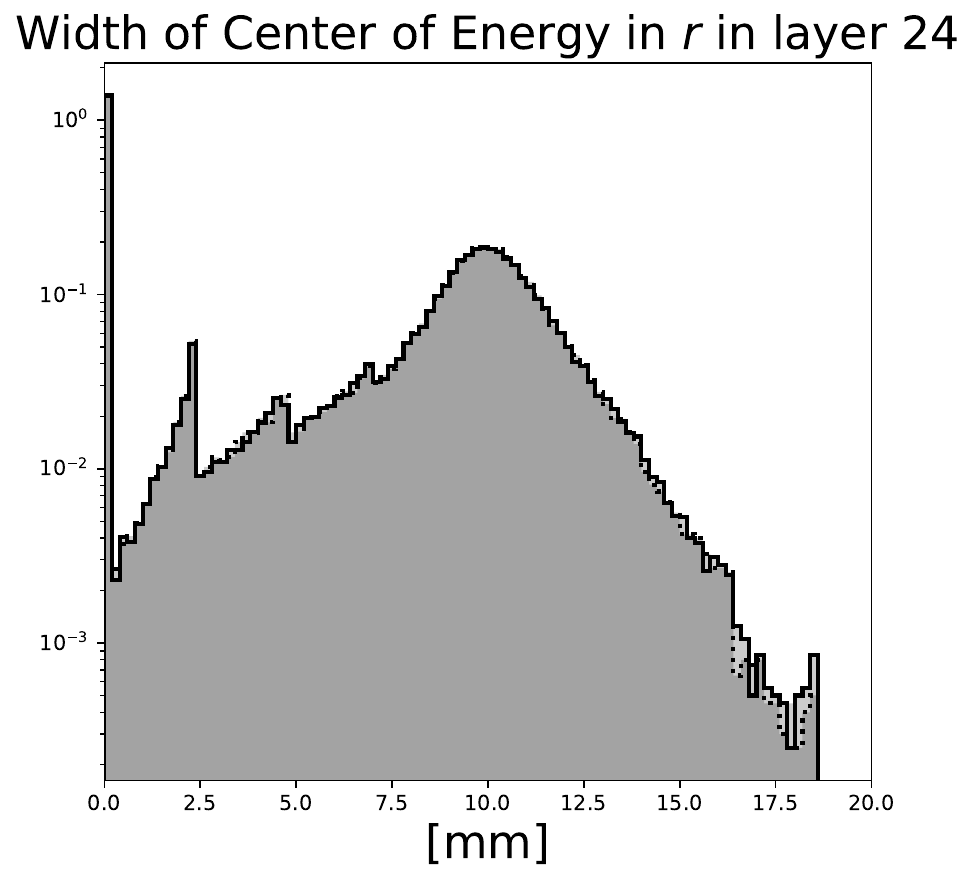}\\
    \includegraphics[height=0.1\textheight]{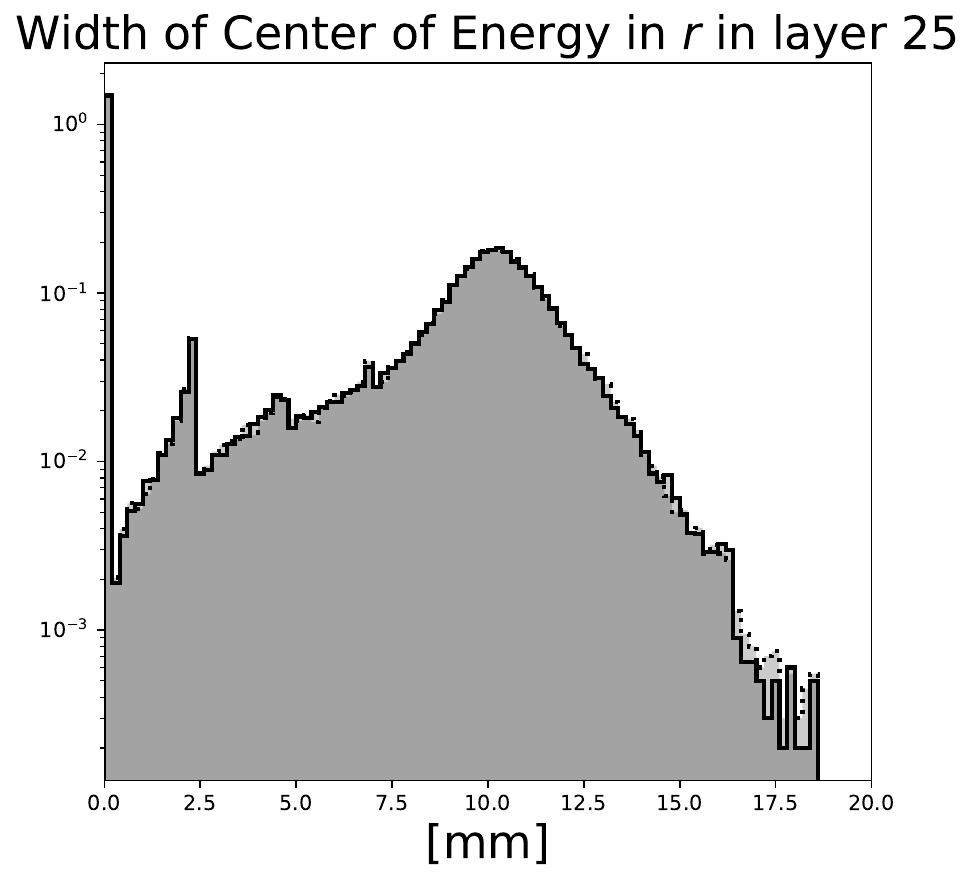} \hfill \includegraphics[height=0.1\textheight]{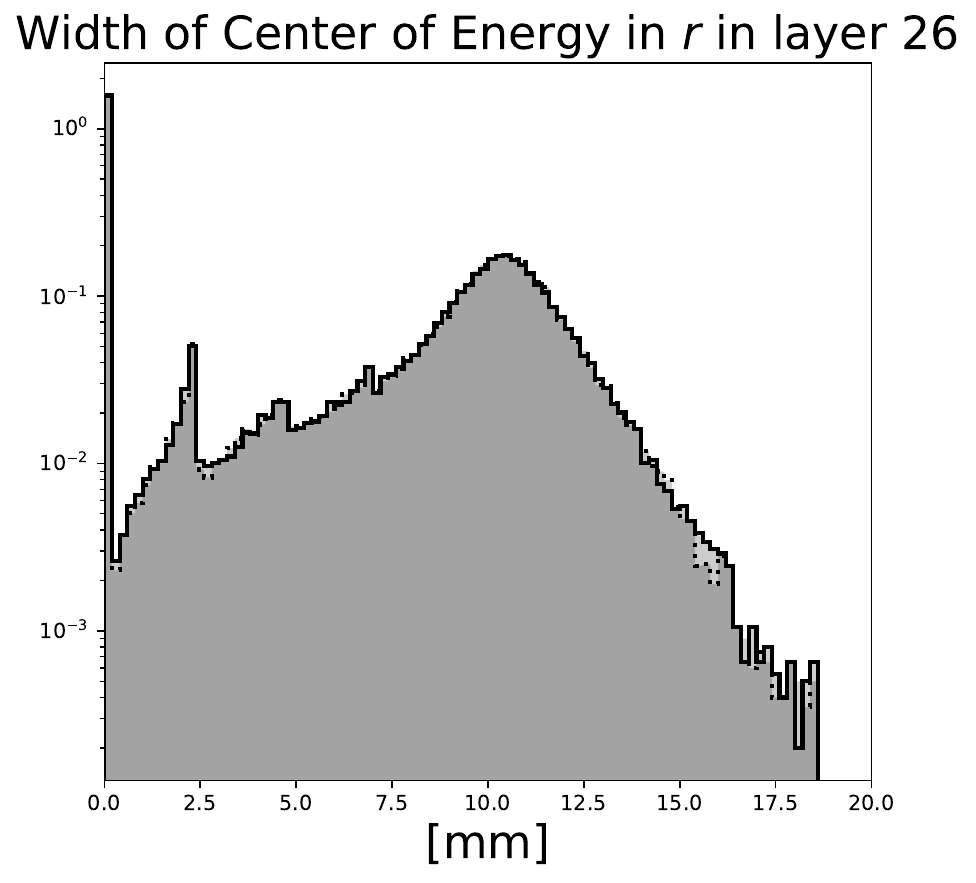} \hfill \includegraphics[height=0.1\textheight]{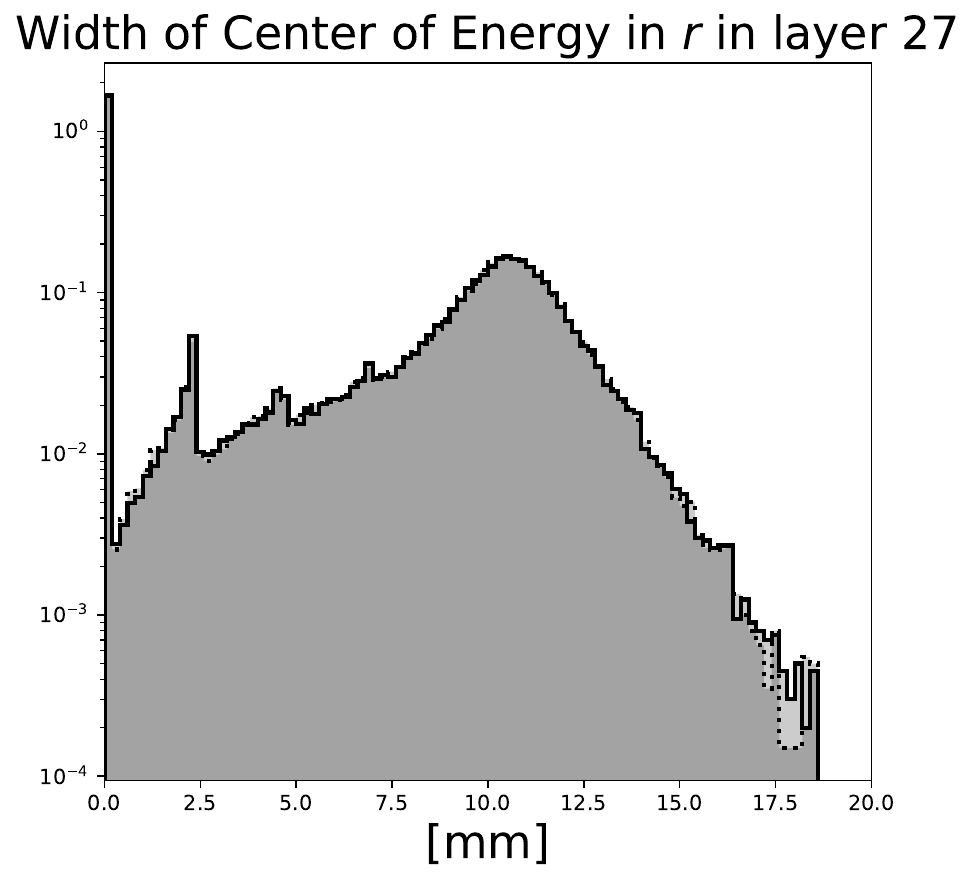} \hfill \includegraphics[height=0.1\textheight]{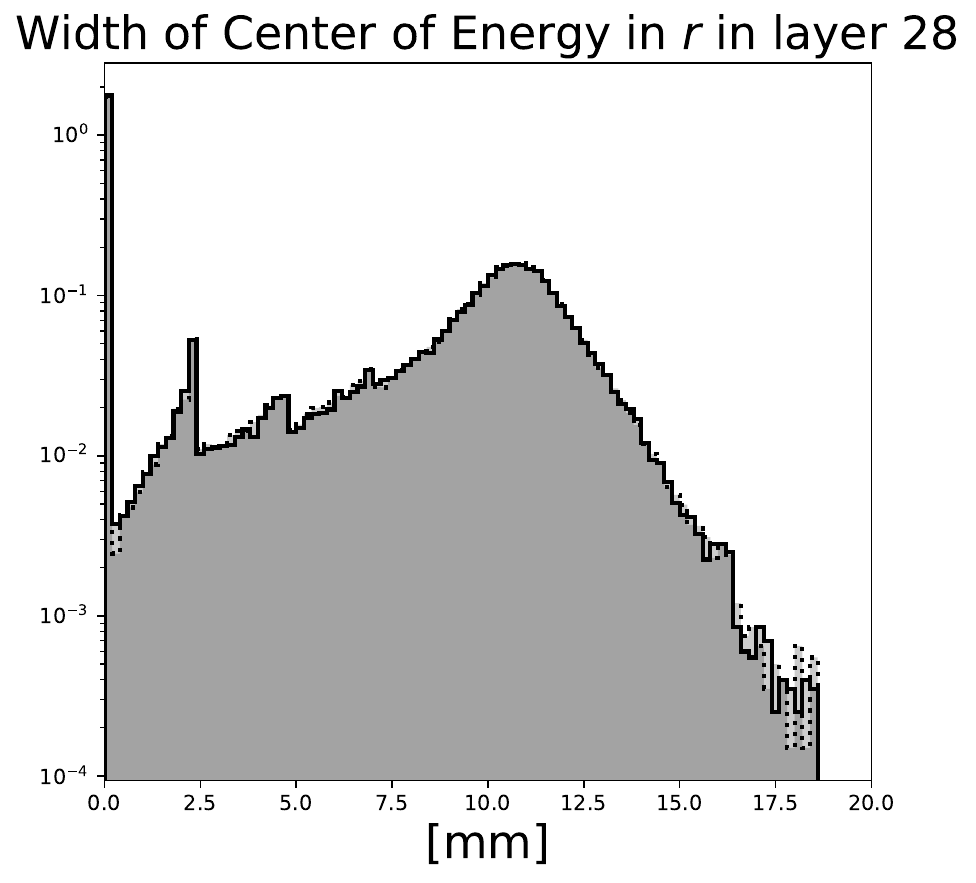} \hfill \includegraphics[height=0.1\textheight]{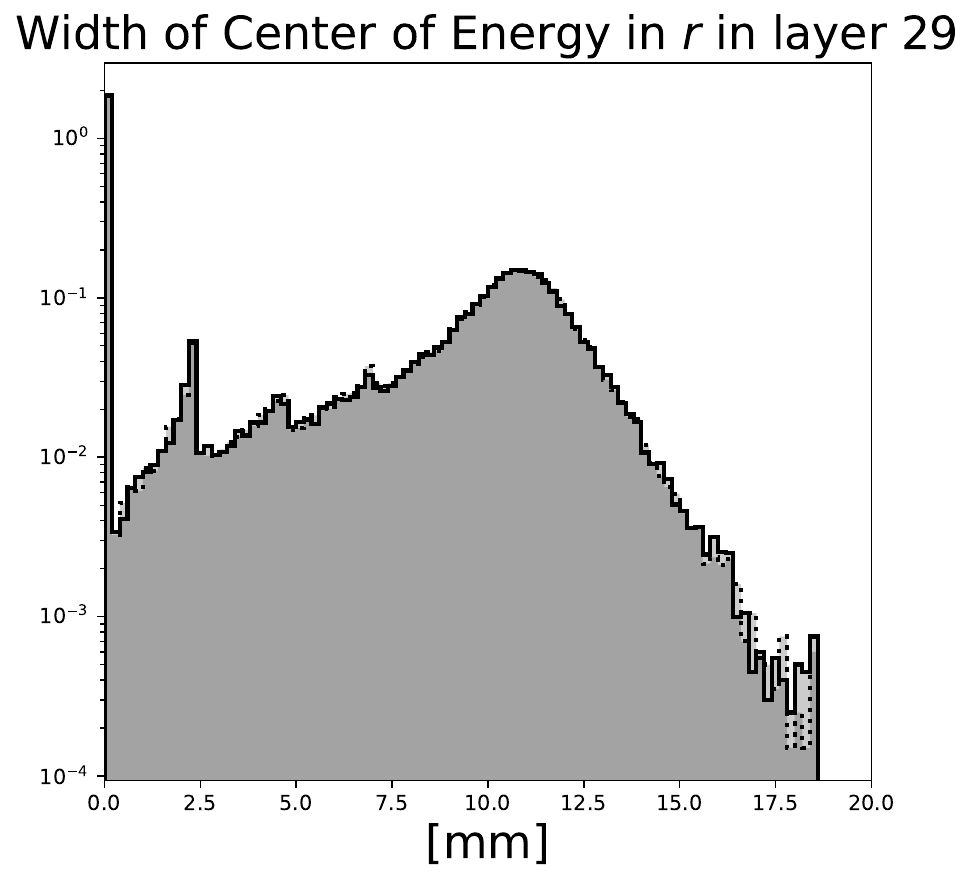}\\
    \includegraphics[height=0.1\textheight]{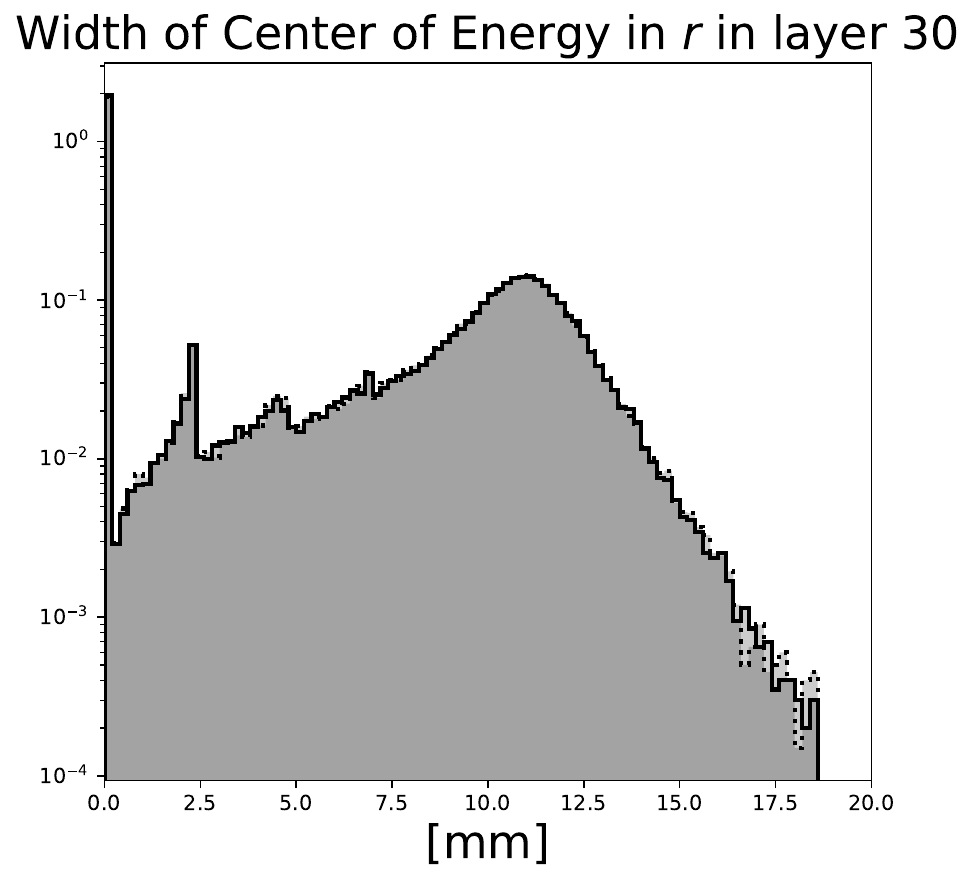} \hfill \includegraphics[height=0.1\textheight]{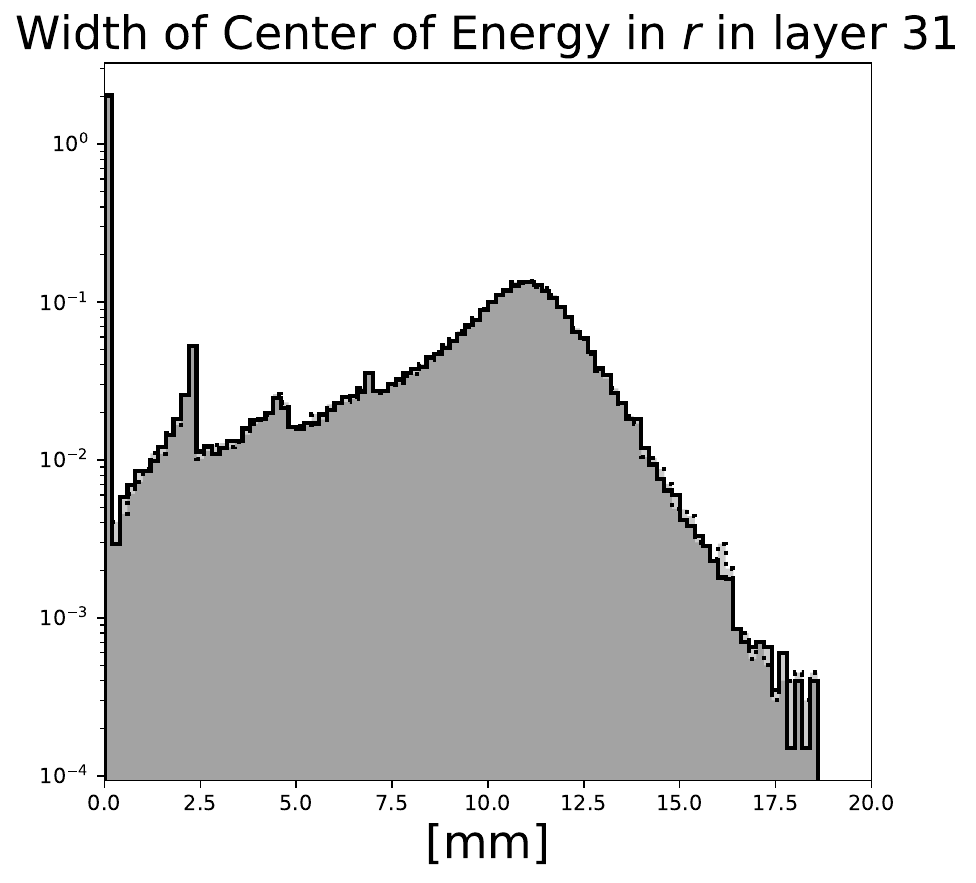} \hfill \includegraphics[height=0.1\textheight]{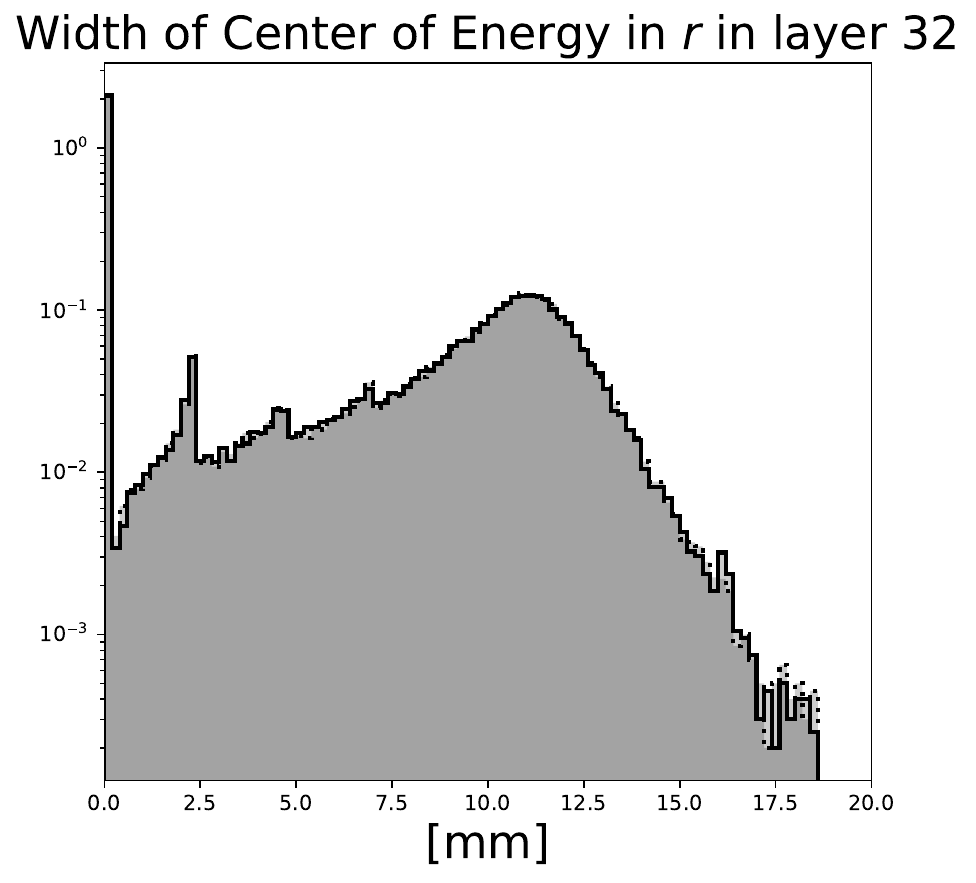} \hfill \includegraphics[height=0.1\textheight]{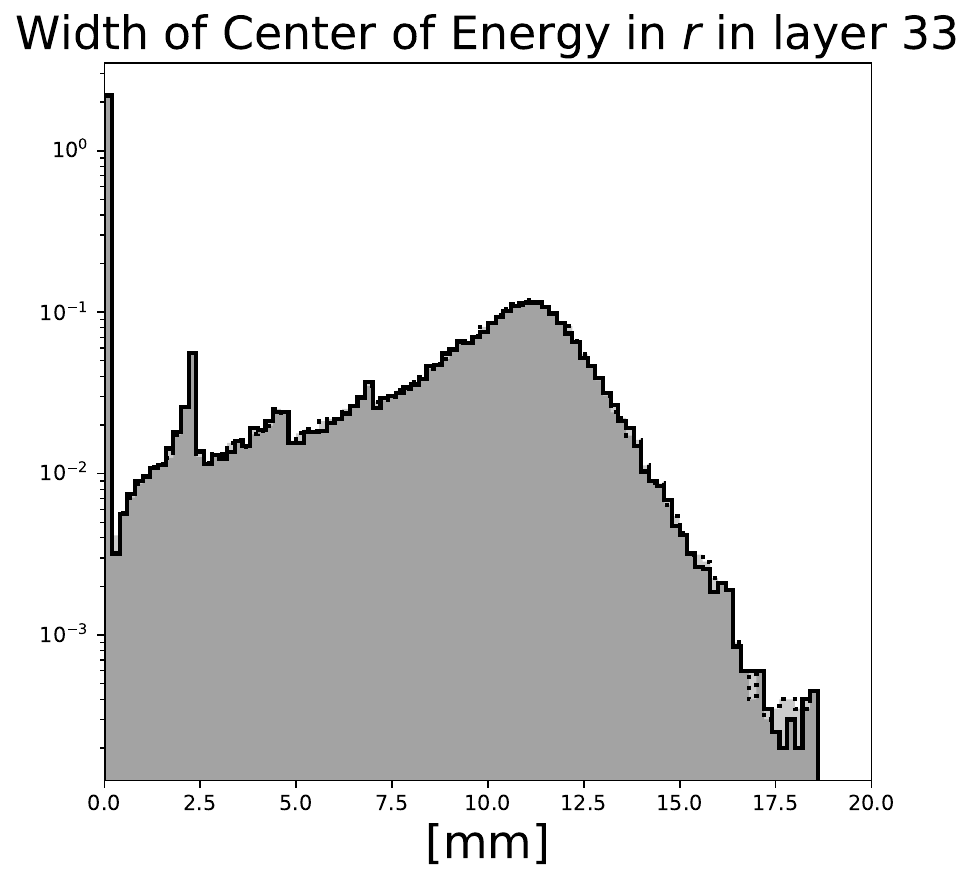} \hfill \includegraphics[height=0.1\textheight]{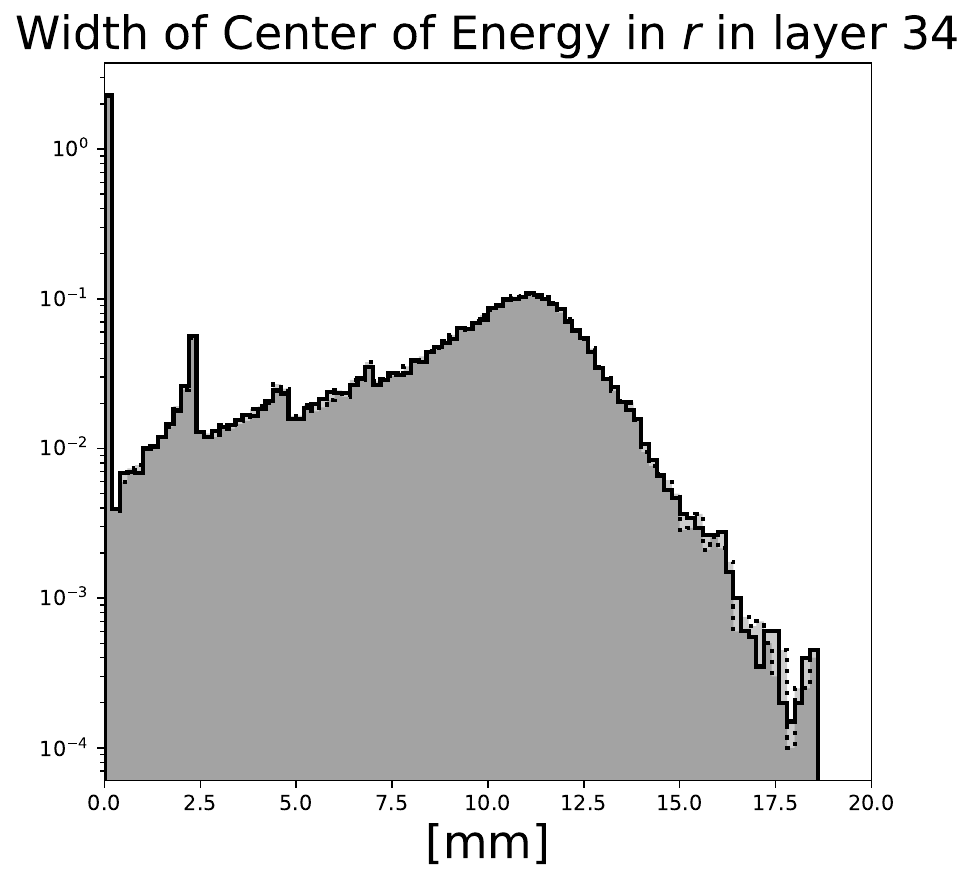}\\
    \includegraphics[height=0.1\textheight]{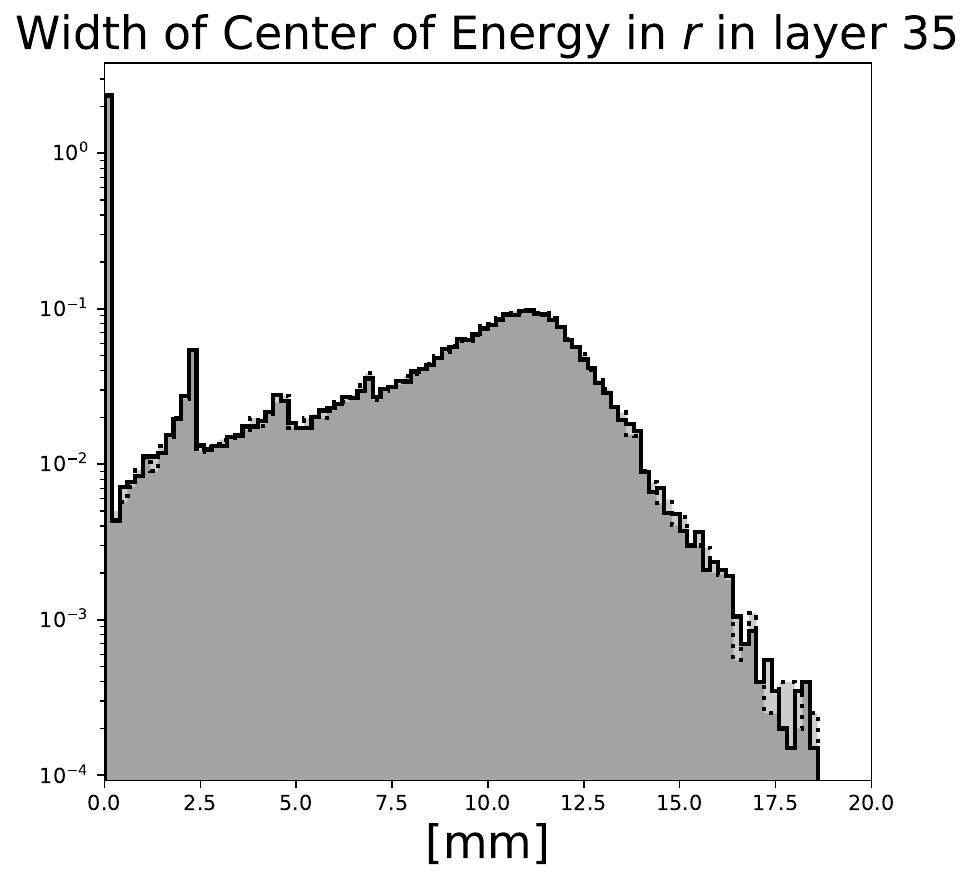} \hfill \includegraphics[height=0.1\textheight]{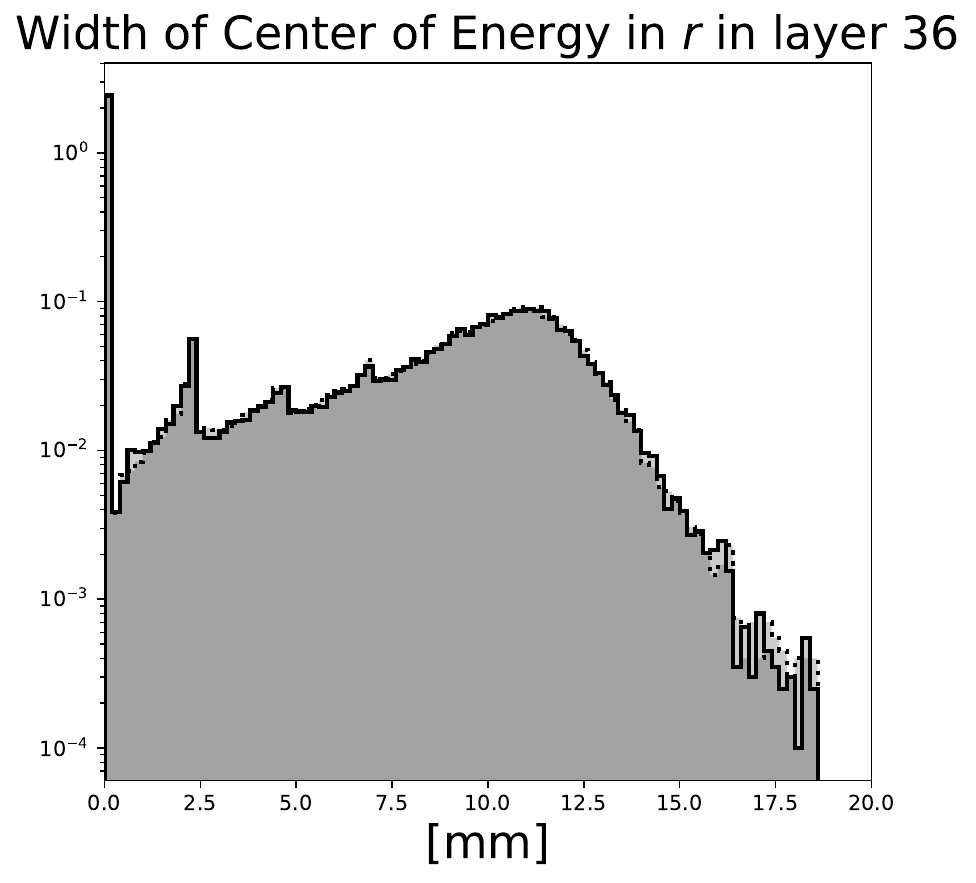} \hfill \includegraphics[height=0.1\textheight]{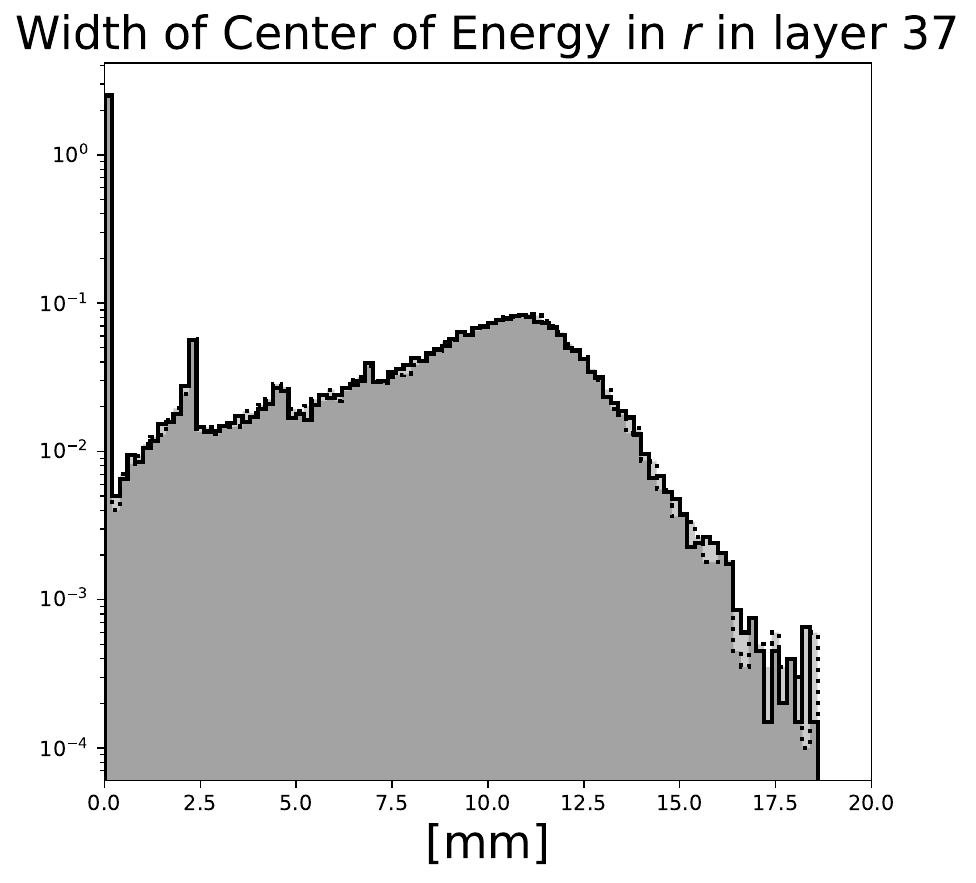} \hfill \includegraphics[height=0.1\textheight]{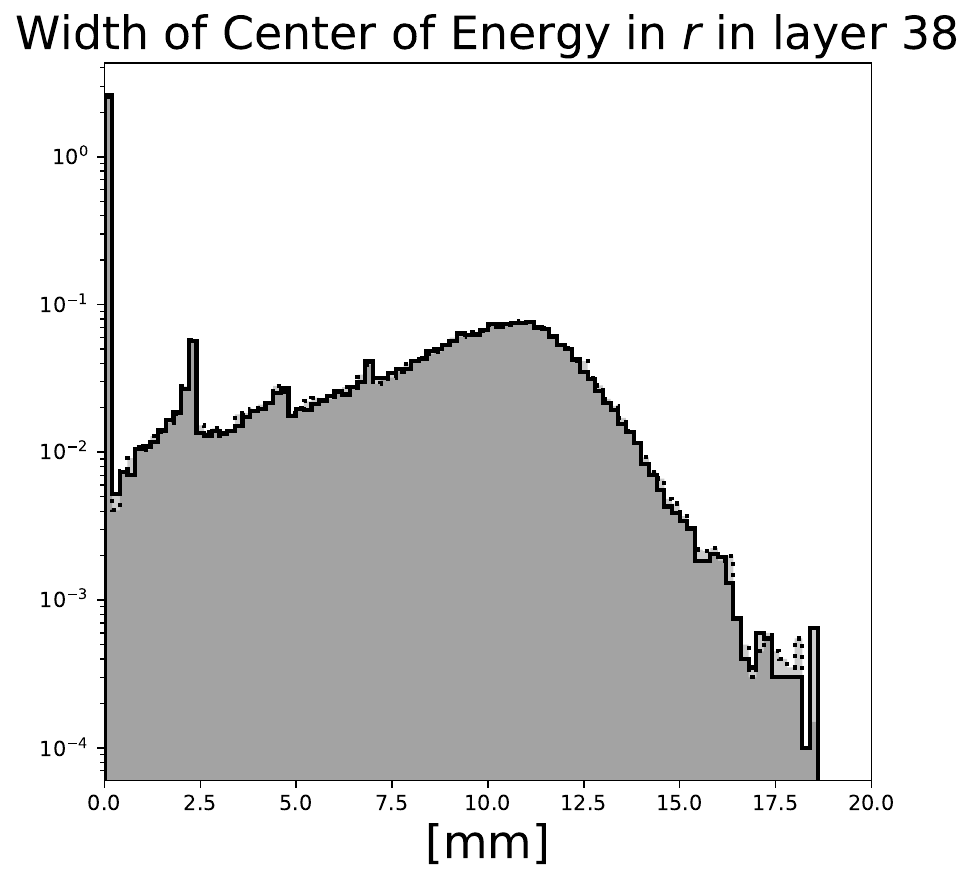} \hfill \includegraphics[height=0.1\textheight]{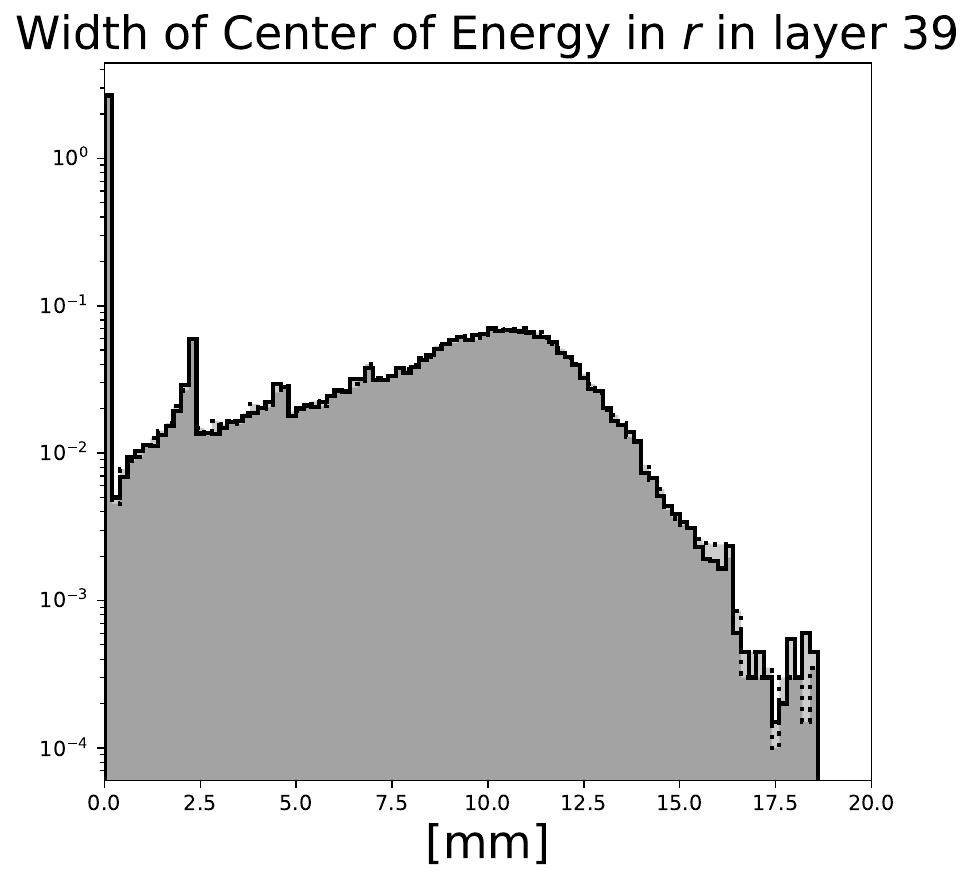}\\
    \includegraphics[height=0.1\textheight]{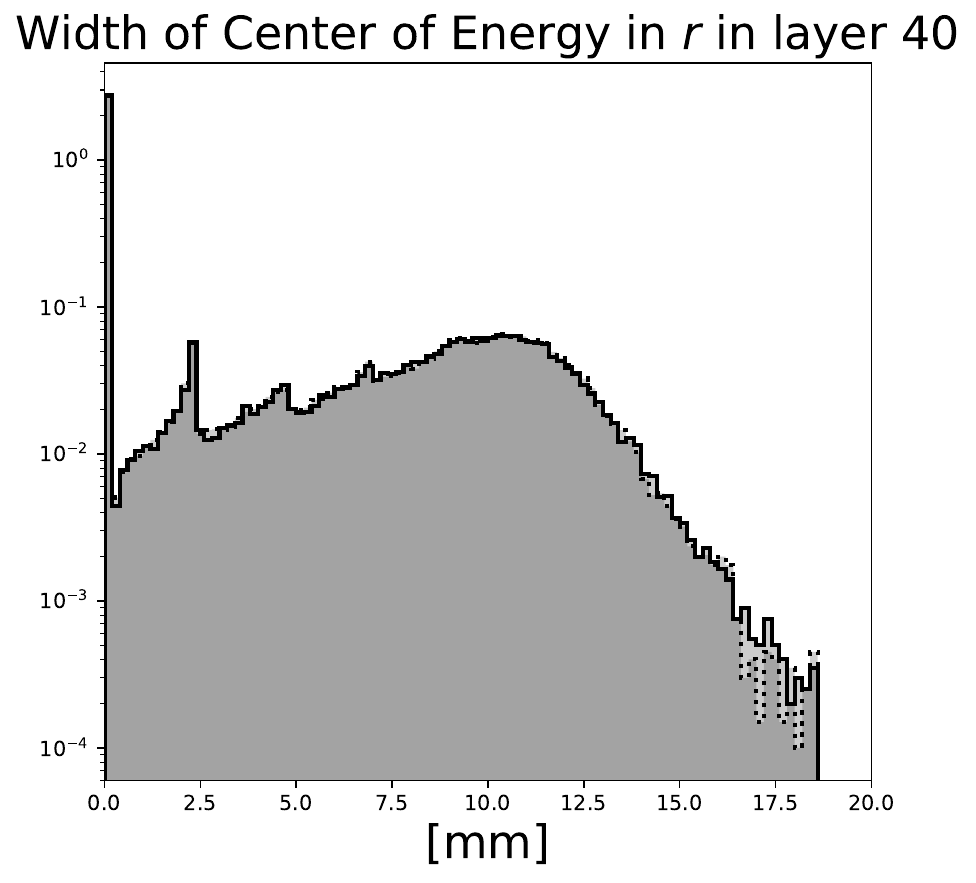} \hfill \includegraphics[height=0.1\textheight]{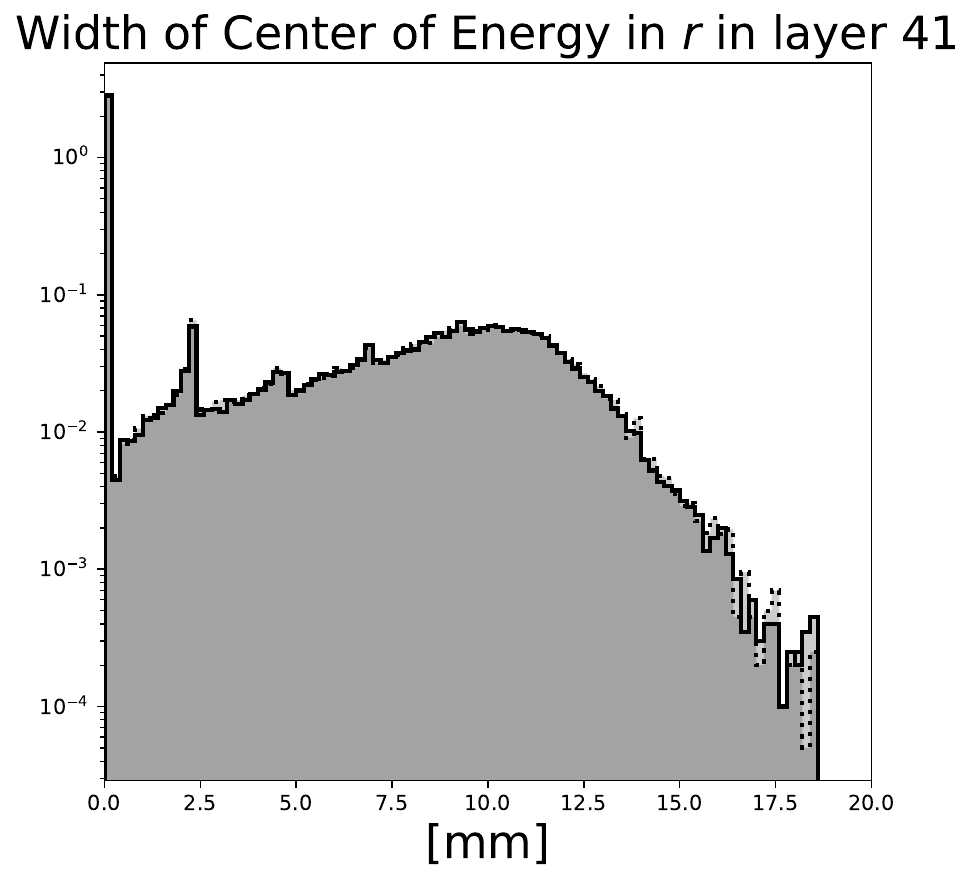} \hfill \includegraphics[height=0.1\textheight]{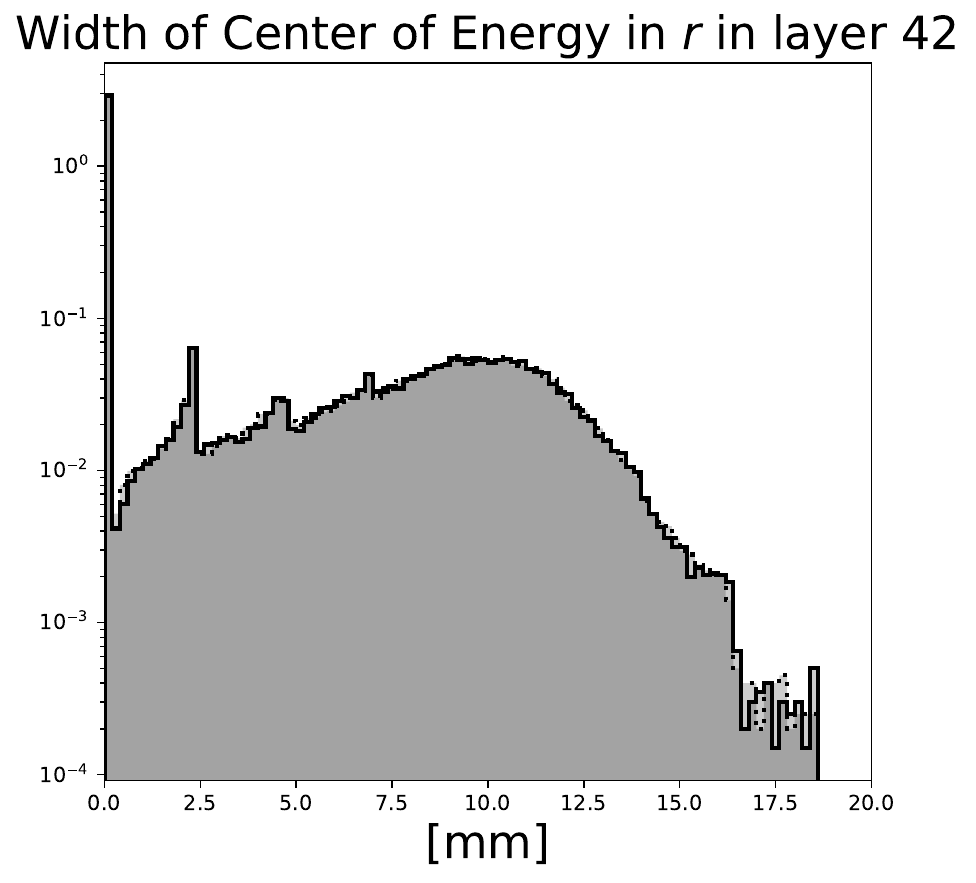} \hfill \includegraphics[height=0.1\textheight]{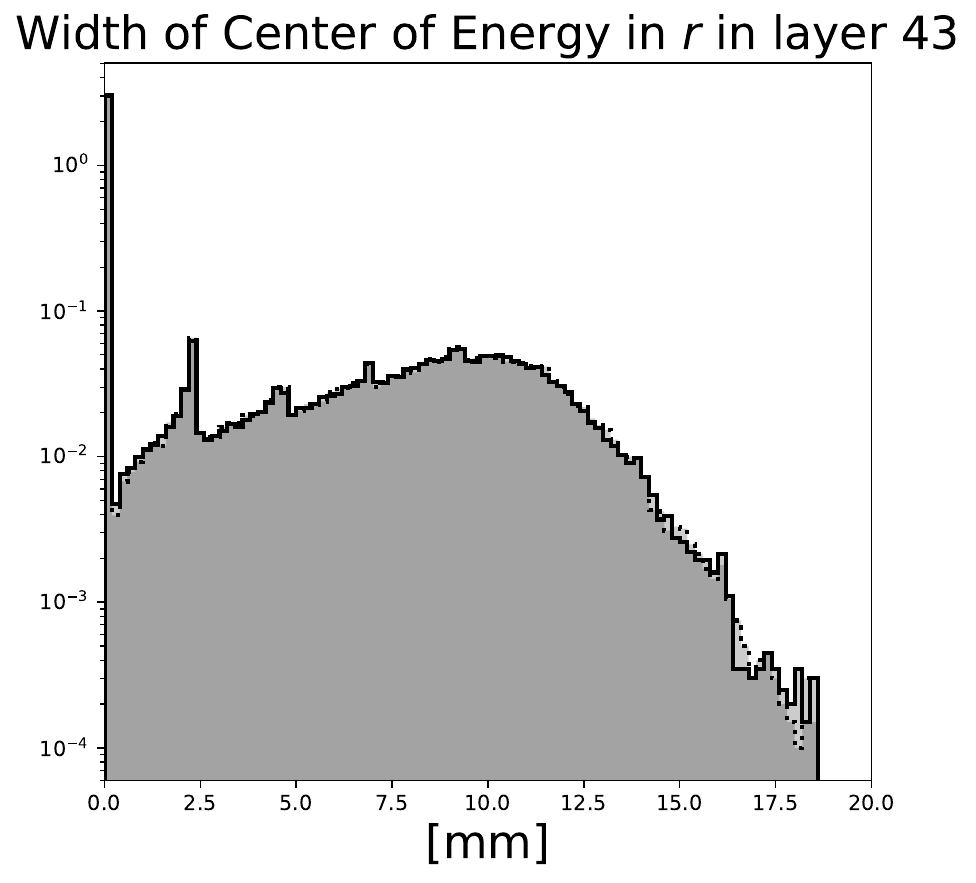} \hfill \includegraphics[height=0.1\textheight]{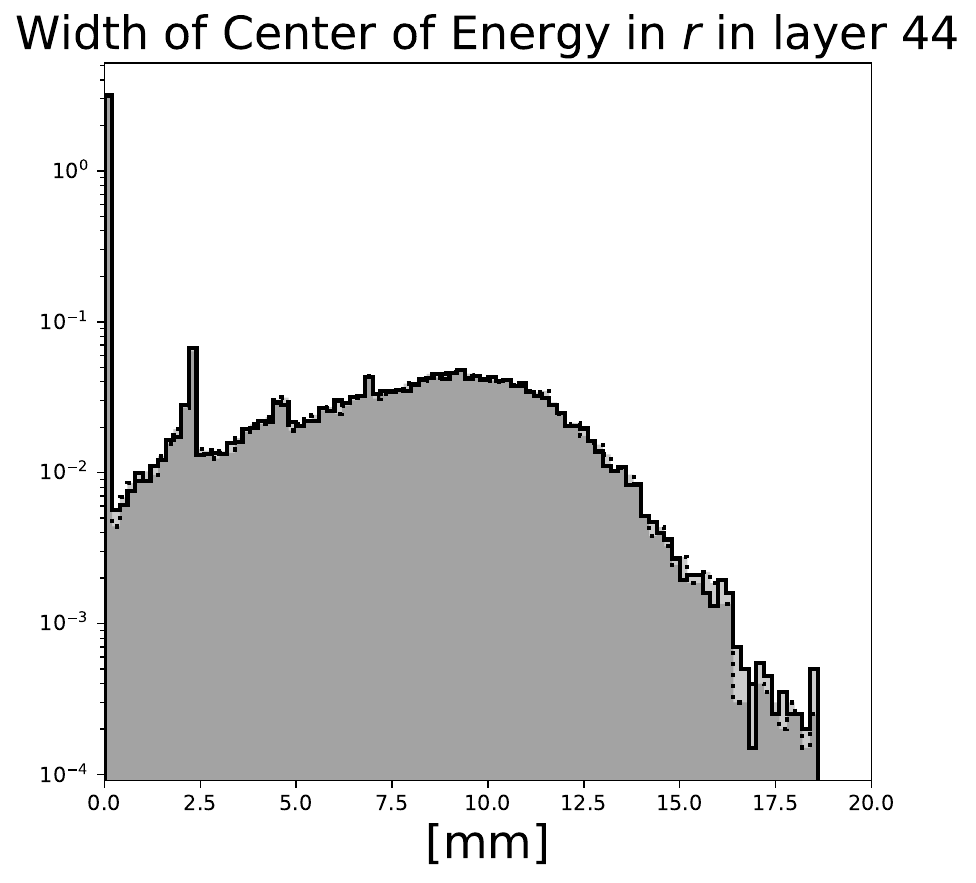}\\
    \includegraphics[width=0.5\textwidth]{figures/Appendix_reference/legend.pdf}
    \caption{Distribution of \geant training and evaluation data in width of the centers of energy in $r$ direction for ds2. }
    \label{fig:app_ref.ds2.8}
\end{figure}

\begin{figure}[ht]
    \centering
    \includegraphics[height=0.1\textheight]{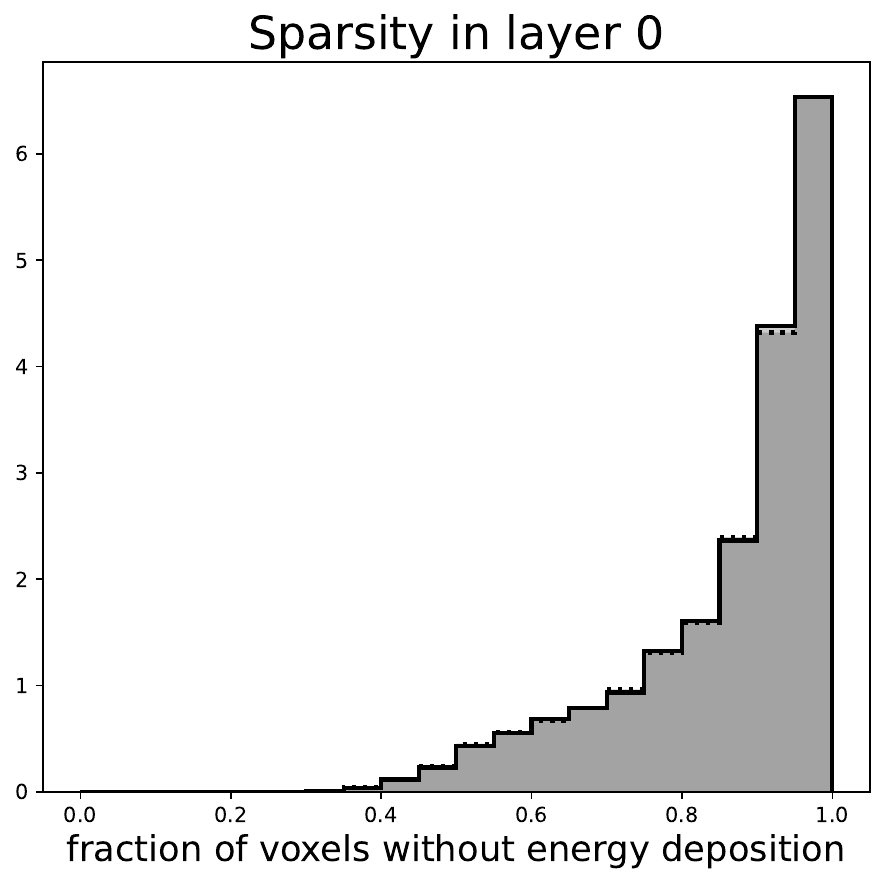} \hfill \includegraphics[height=0.1\textheight]{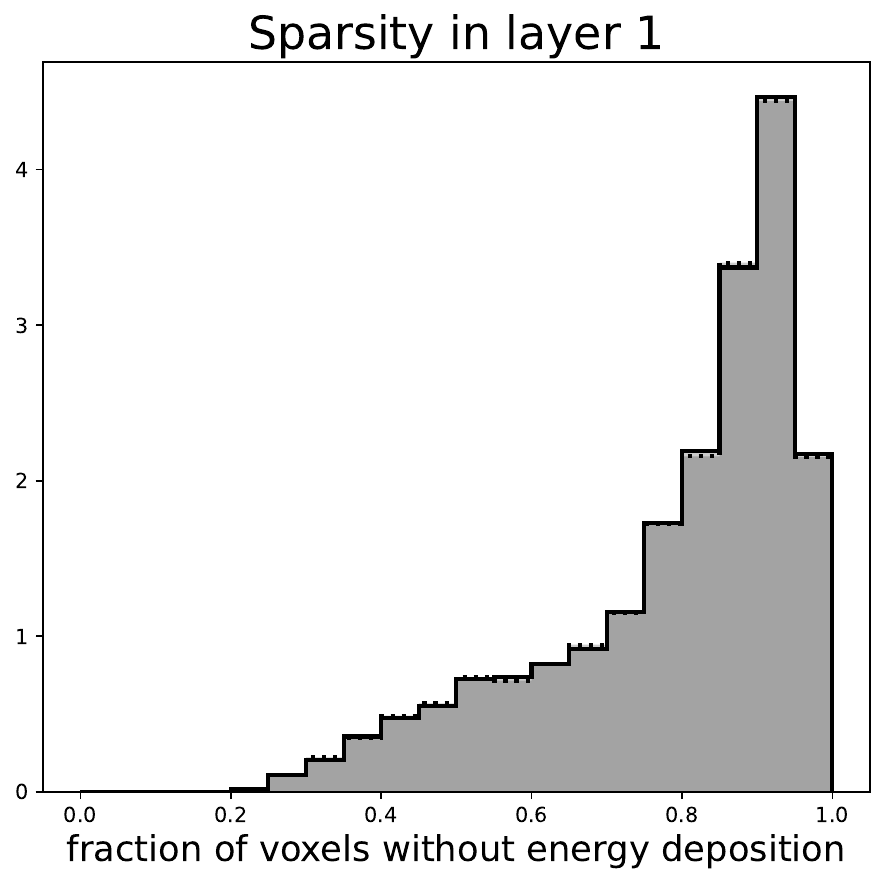} \hfill \includegraphics[height=0.1\textheight]{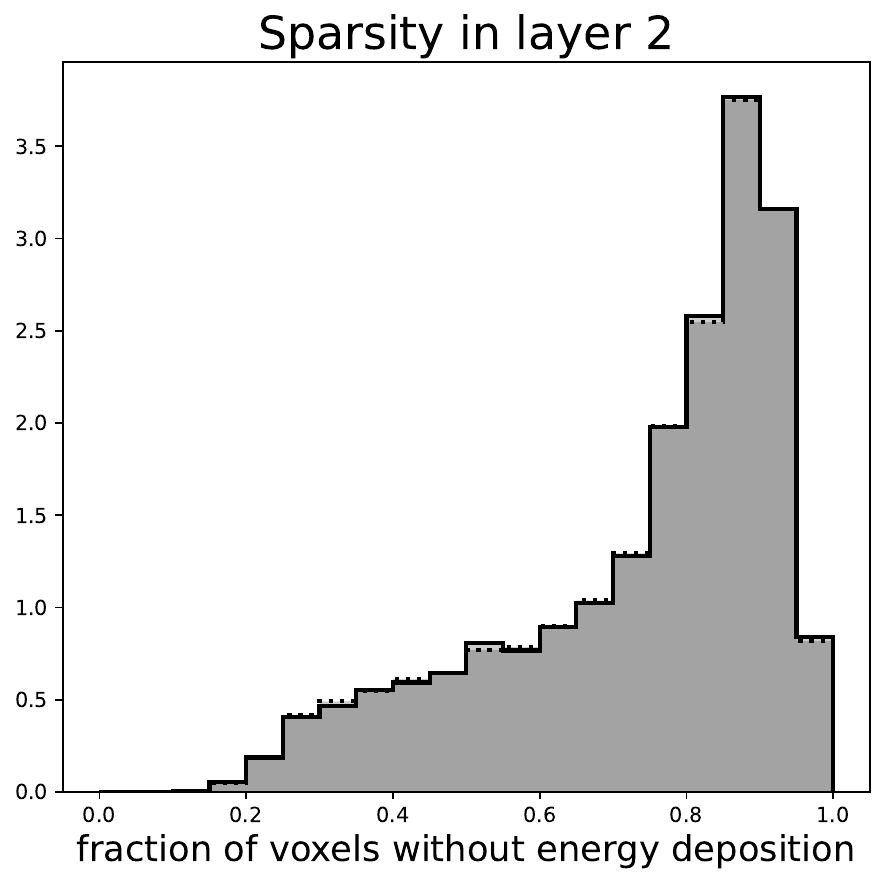} \hfill \includegraphics[height=0.1\textheight]{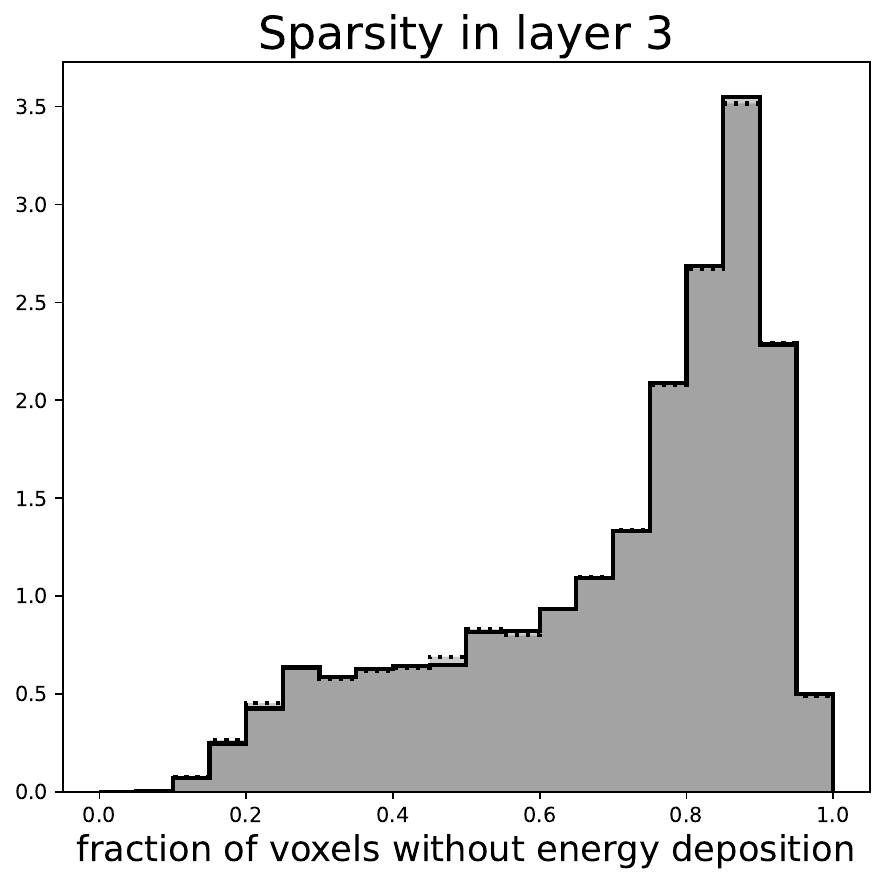} \hfill \includegraphics[height=0.1\textheight]{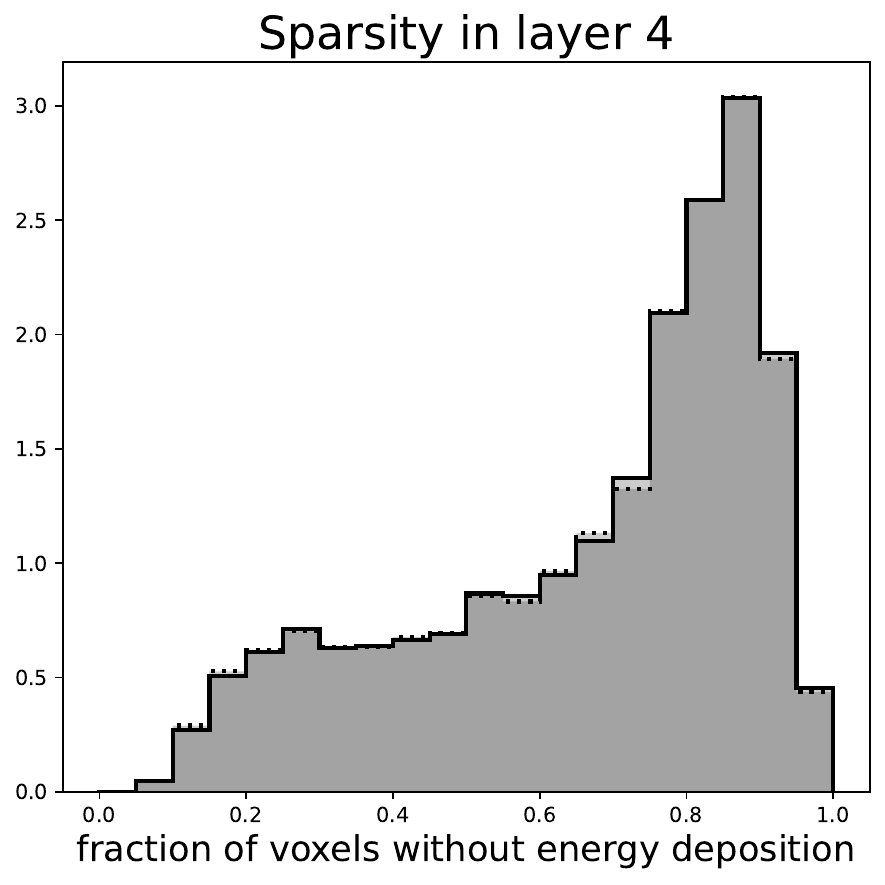}\\
    \includegraphics[height=0.1\textheight]{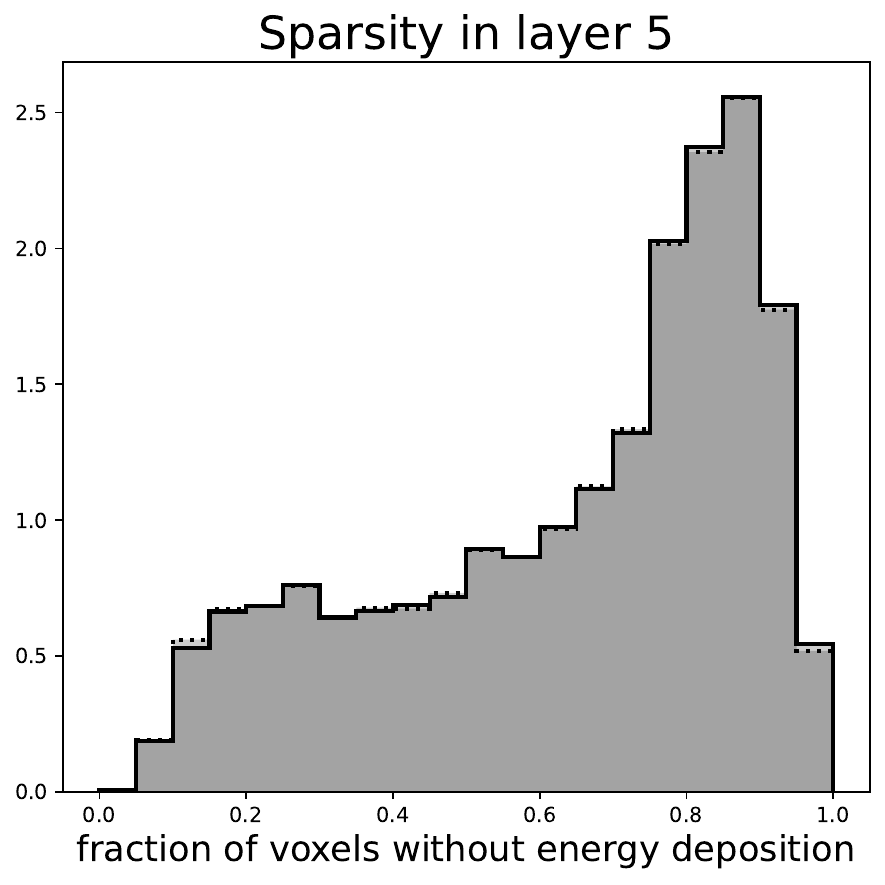} \hfill \includegraphics[height=0.1\textheight]{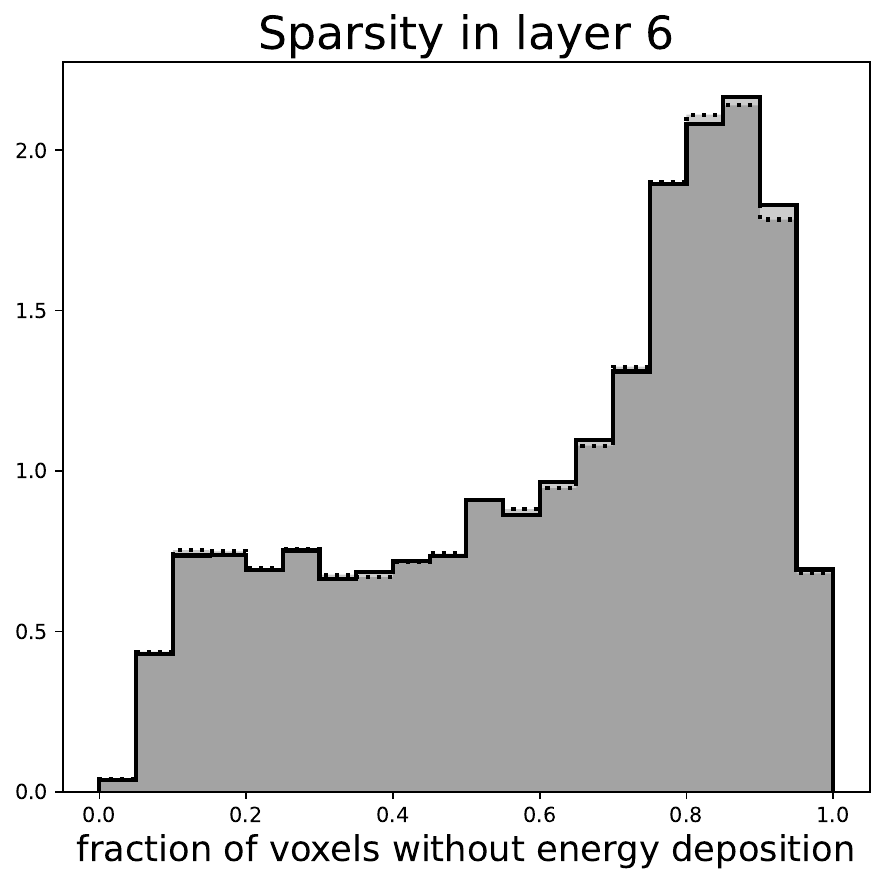} \hfill \includegraphics[height=0.1\textheight]{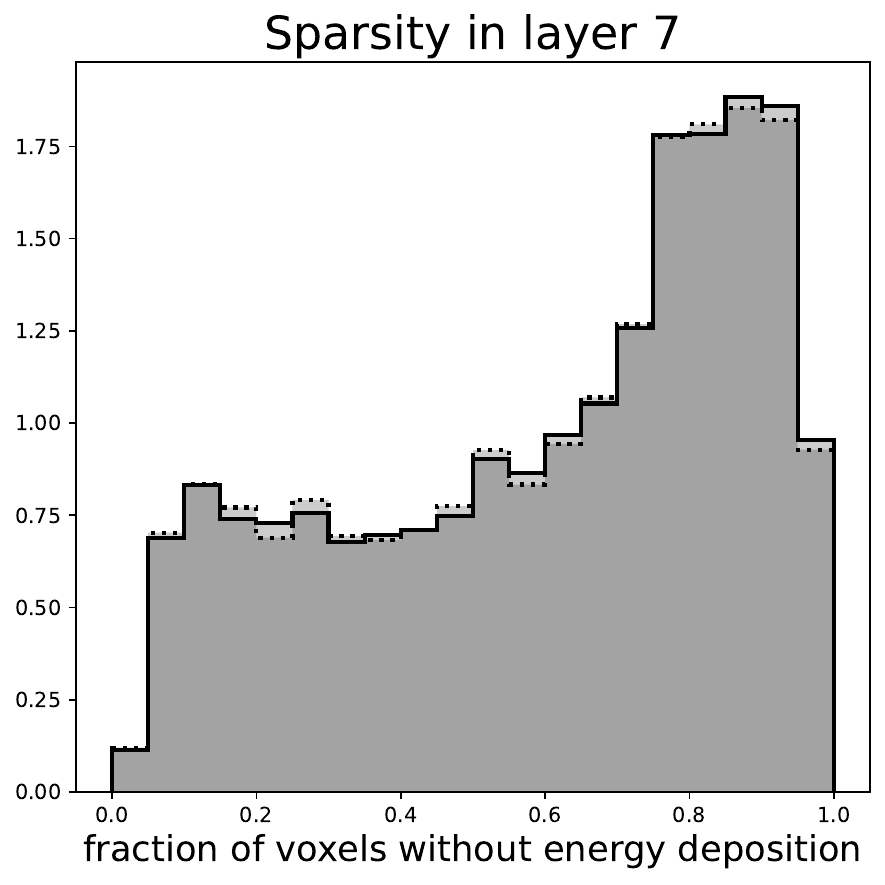} \hfill \includegraphics[height=0.1\textheight]{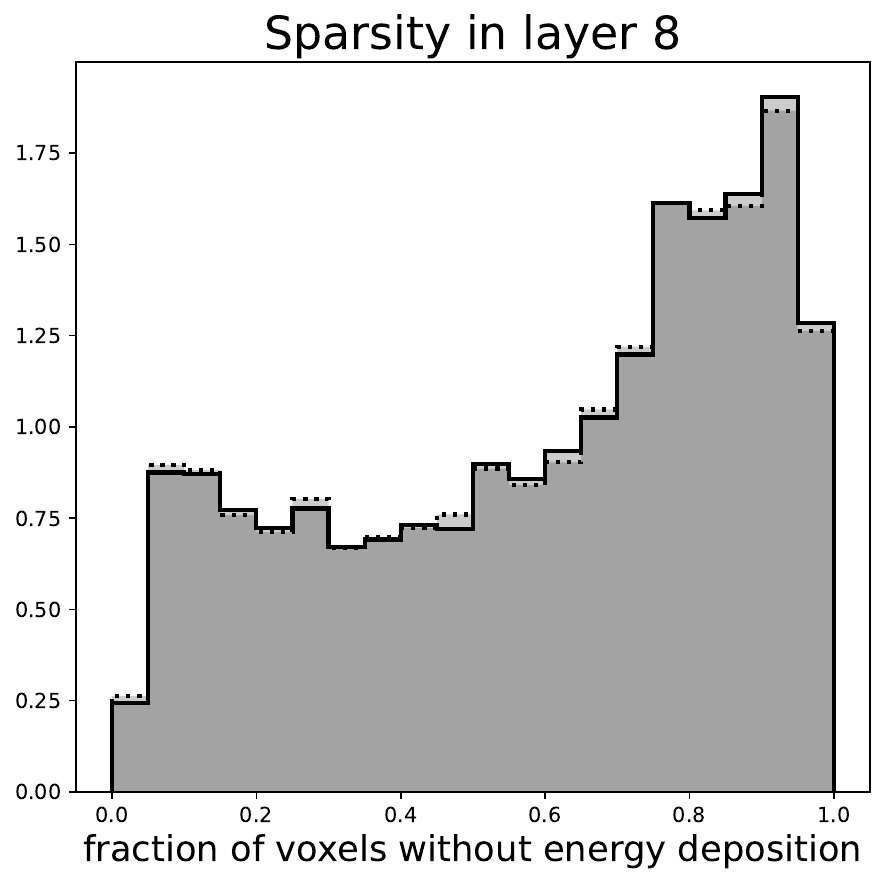} \hfill \includegraphics[height=0.1\textheight]{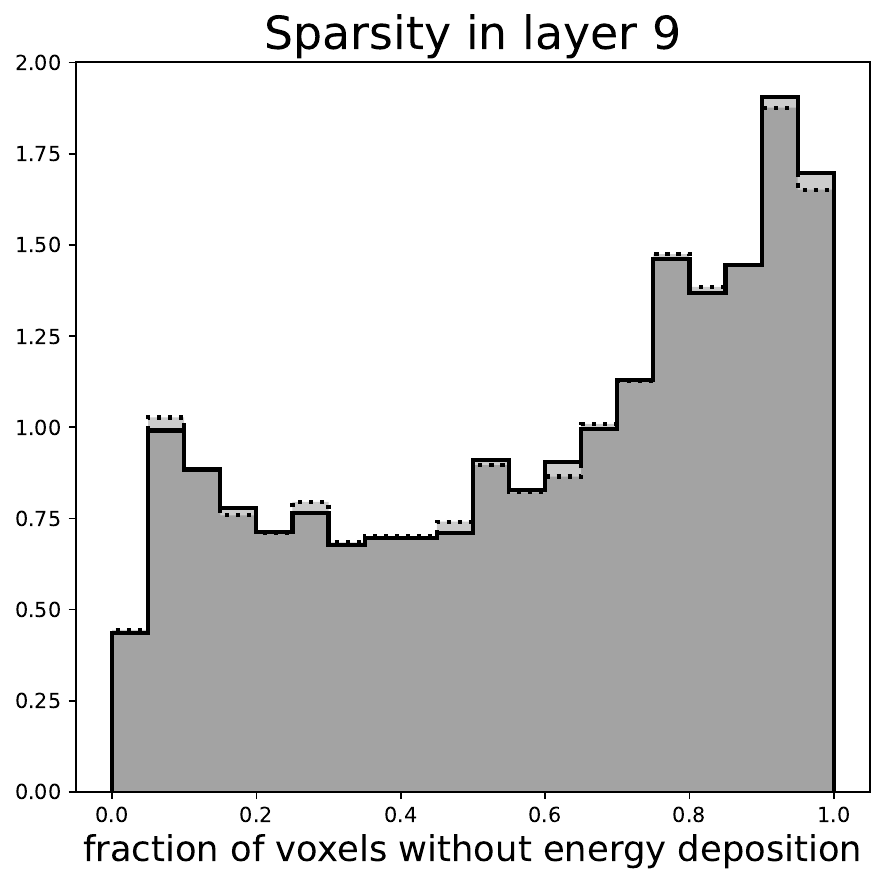}\\
    \includegraphics[height=0.1\textheight]{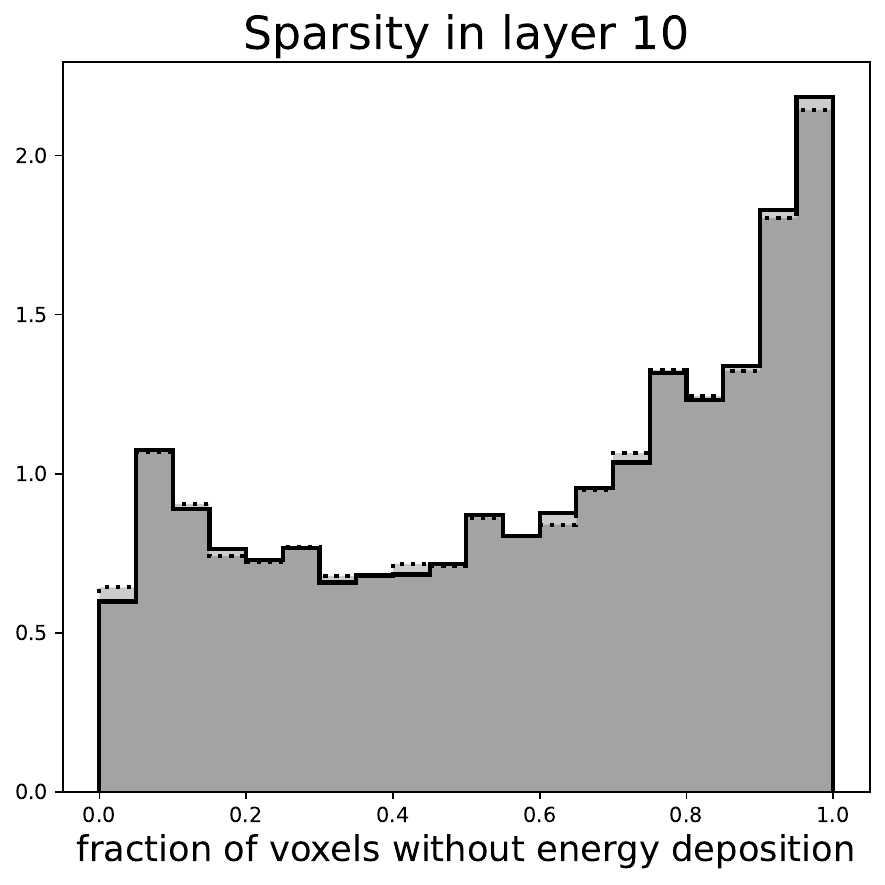} \hfill \includegraphics[height=0.1\textheight]{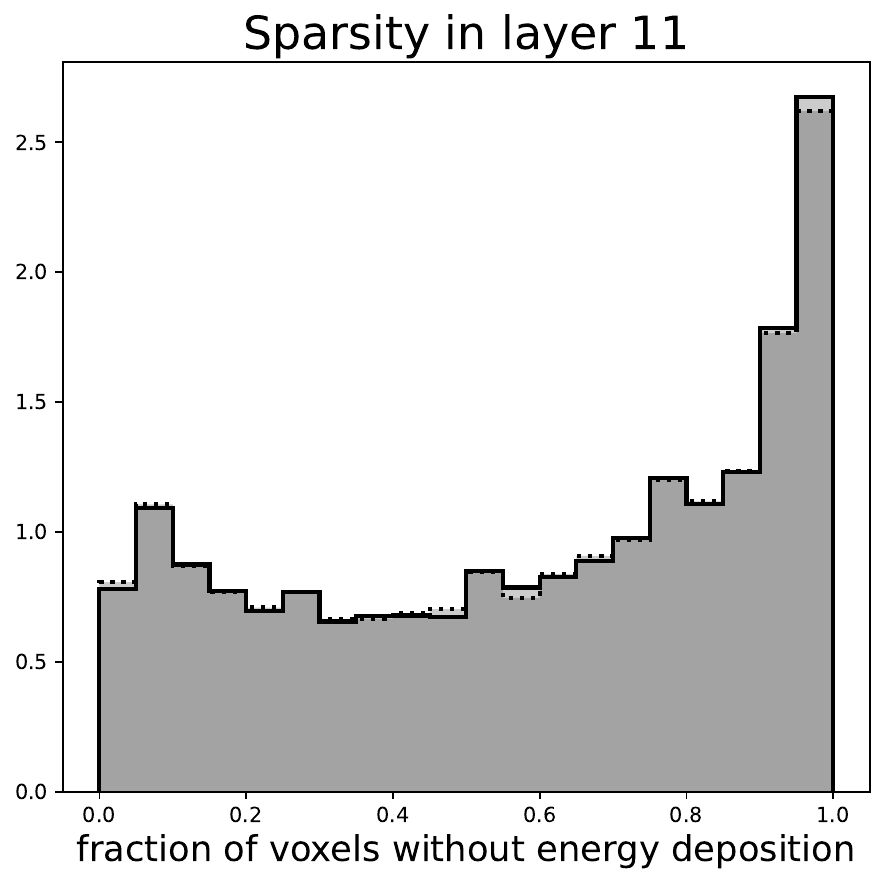} \hfill \includegraphics[height=0.1\textheight]{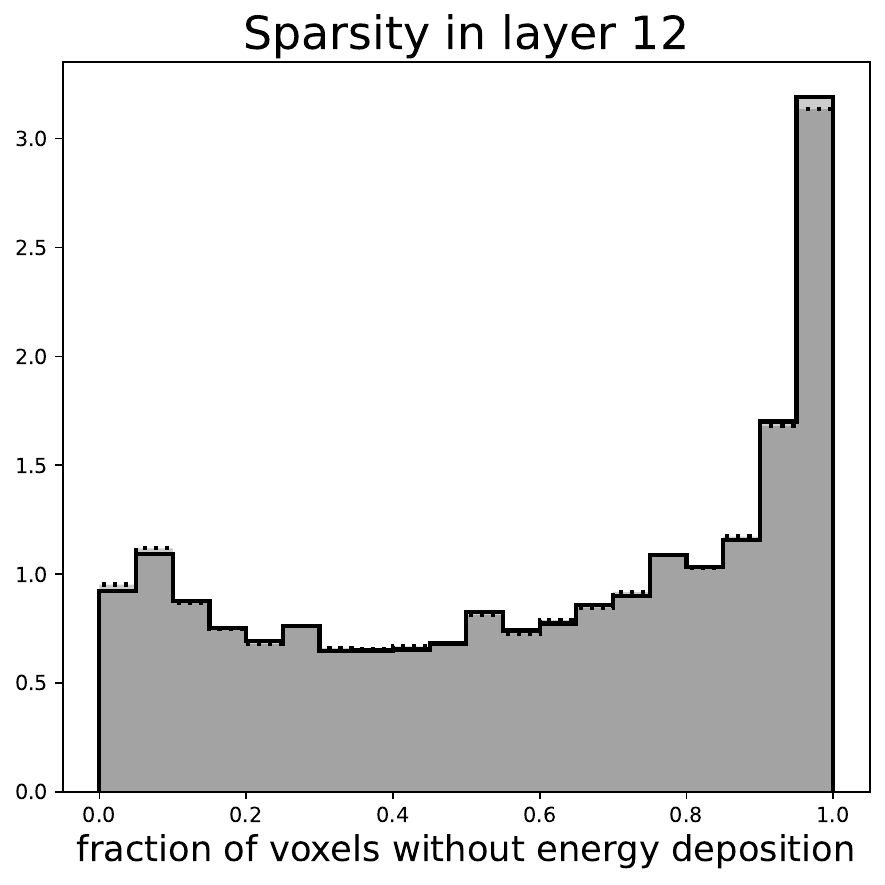} \hfill \includegraphics[height=0.1\textheight]{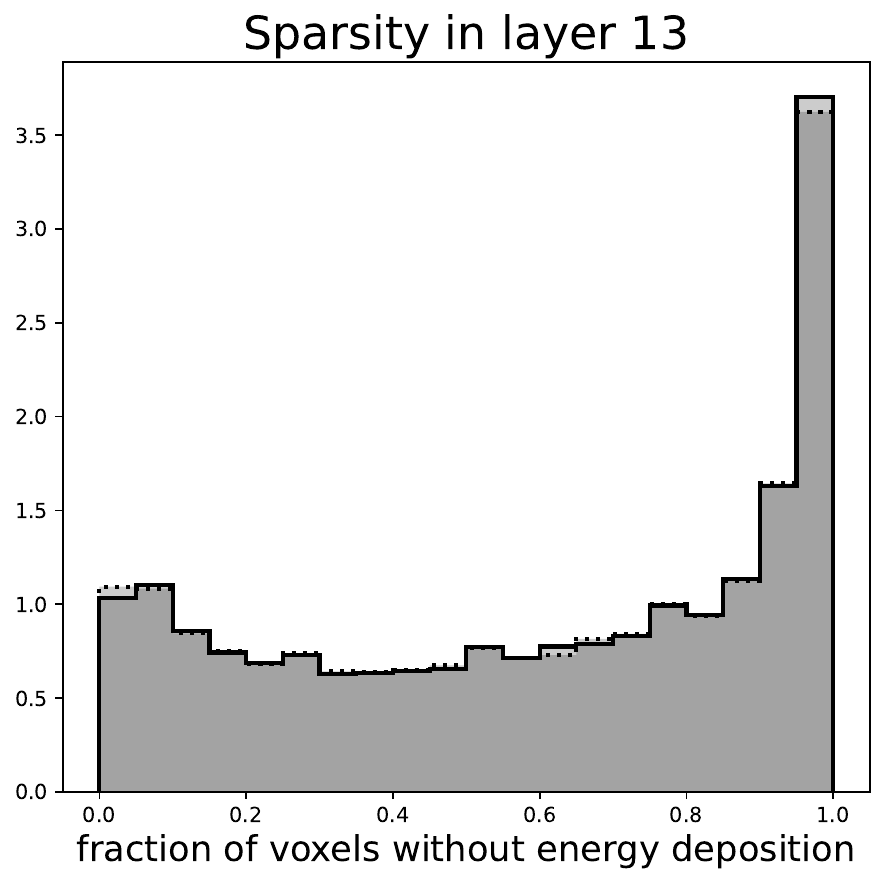} \hfill \includegraphics[height=0.1\textheight]{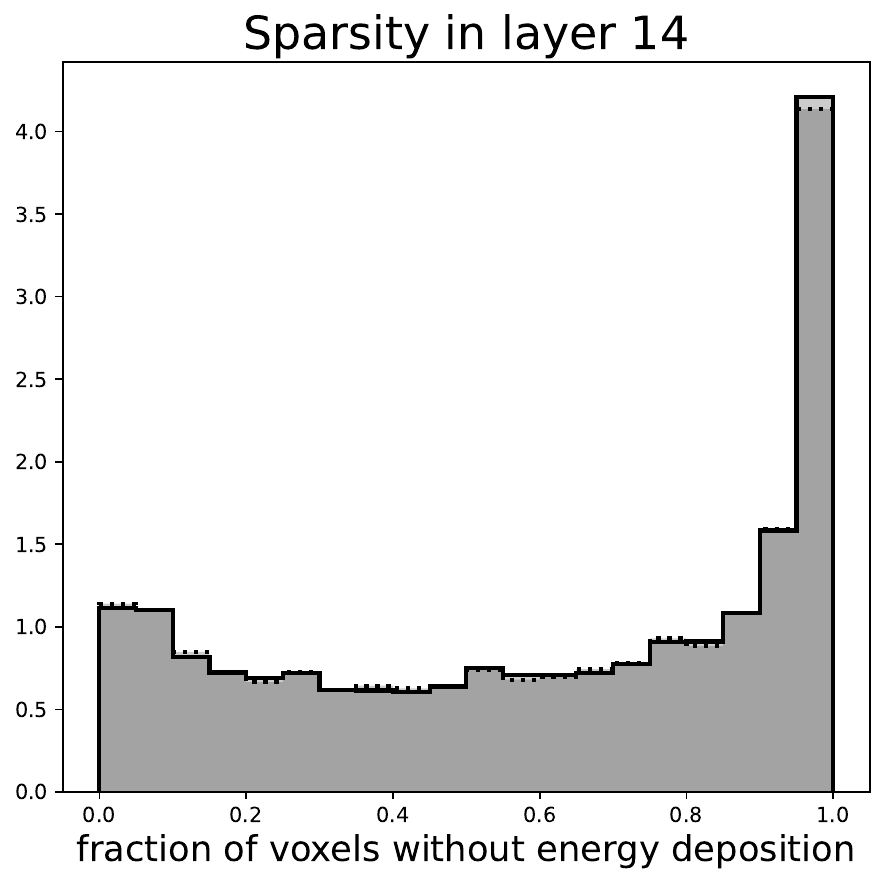}\\
    \includegraphics[height=0.1\textheight]{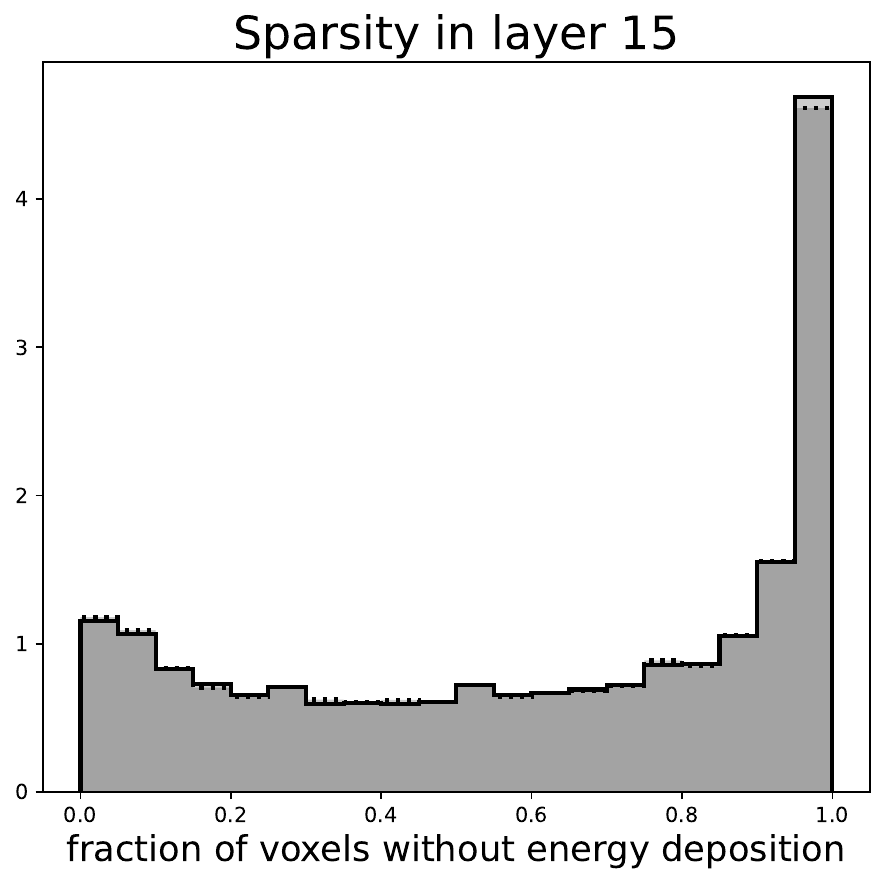} \hfill \includegraphics[height=0.1\textheight]{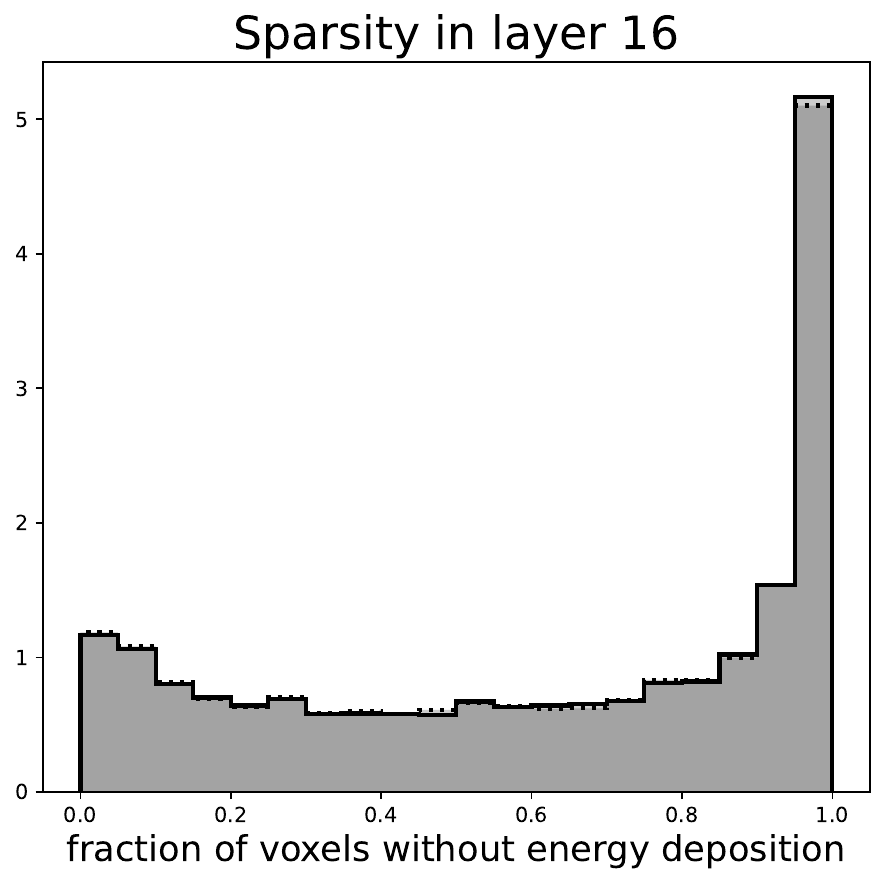} \hfill \includegraphics[height=0.1\textheight]{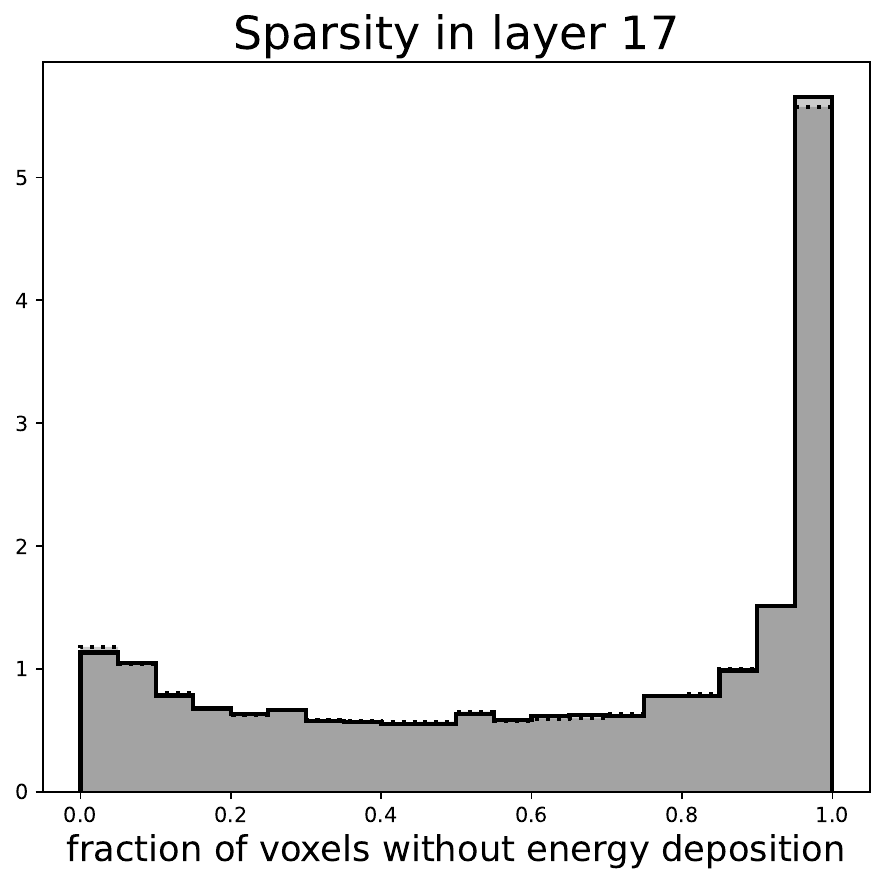} \hfill \includegraphics[height=0.1\textheight]{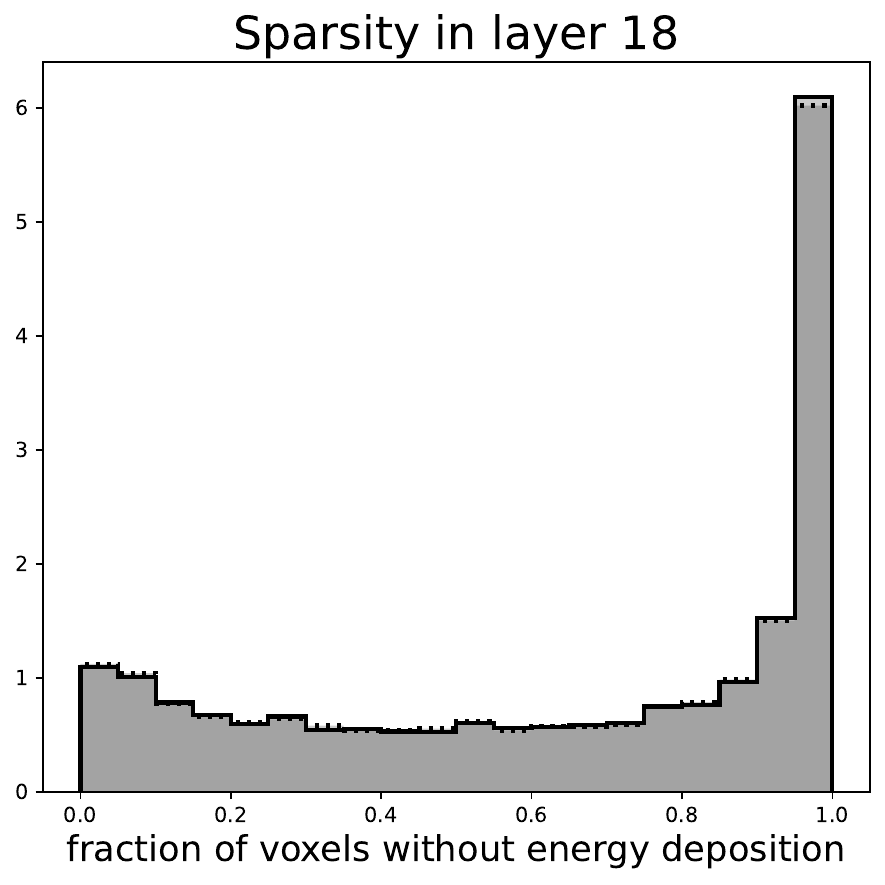} \hfill \includegraphics[height=0.1\textheight]{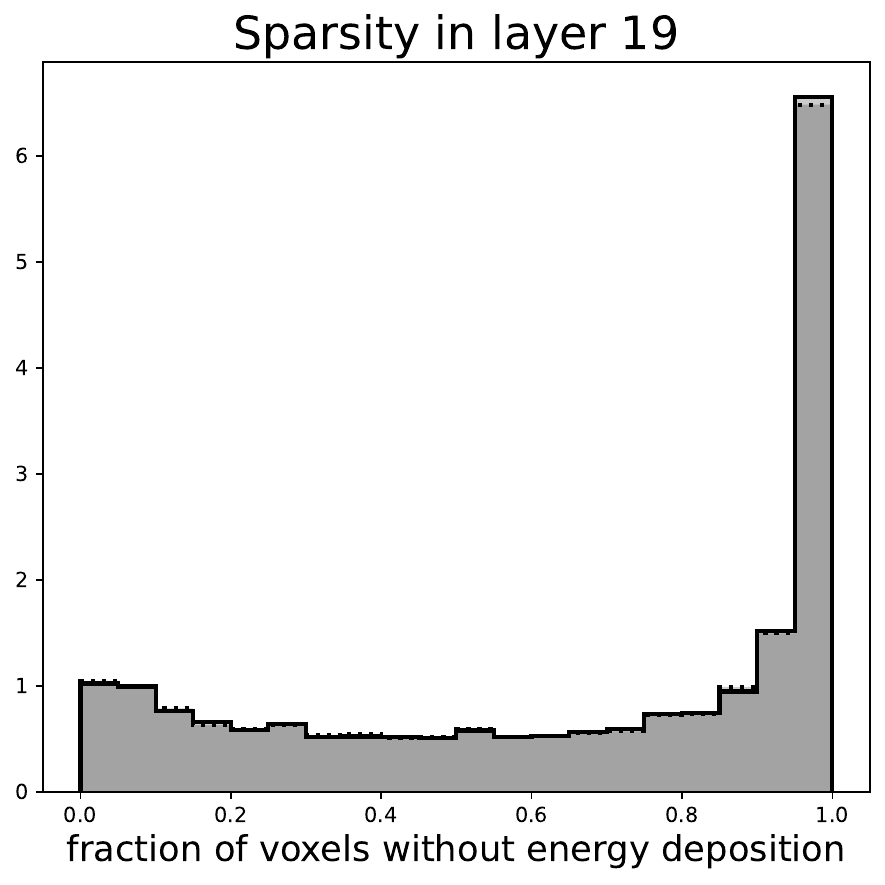}\\
    \includegraphics[height=0.1\textheight]{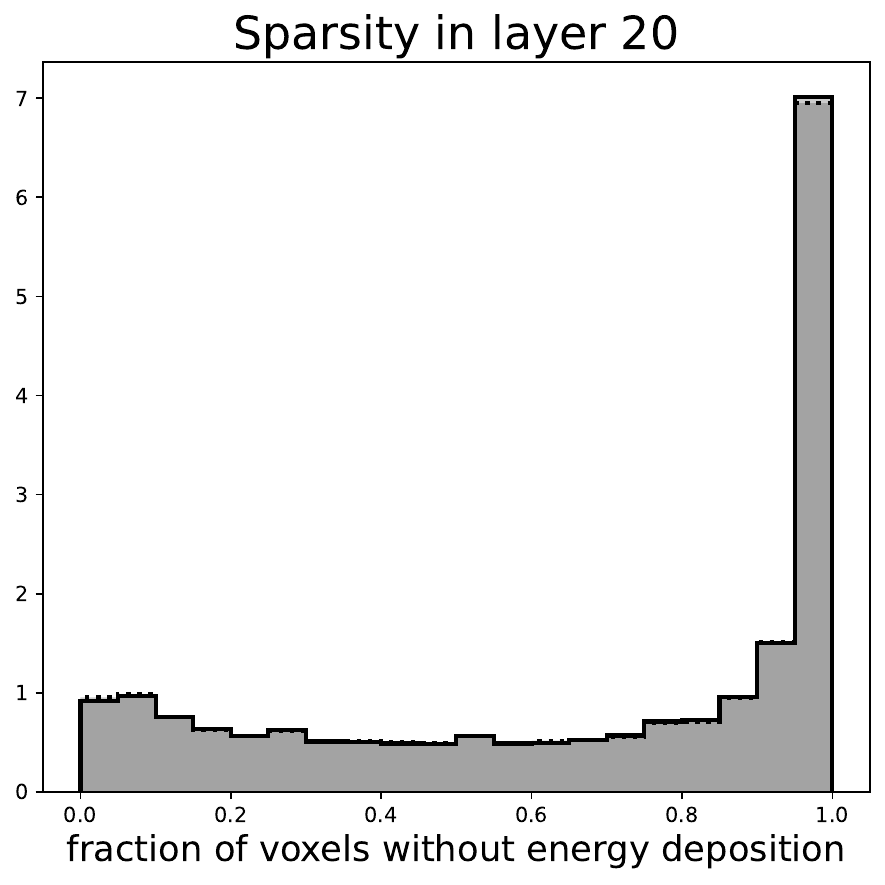} \hfill \includegraphics[height=0.1\textheight]{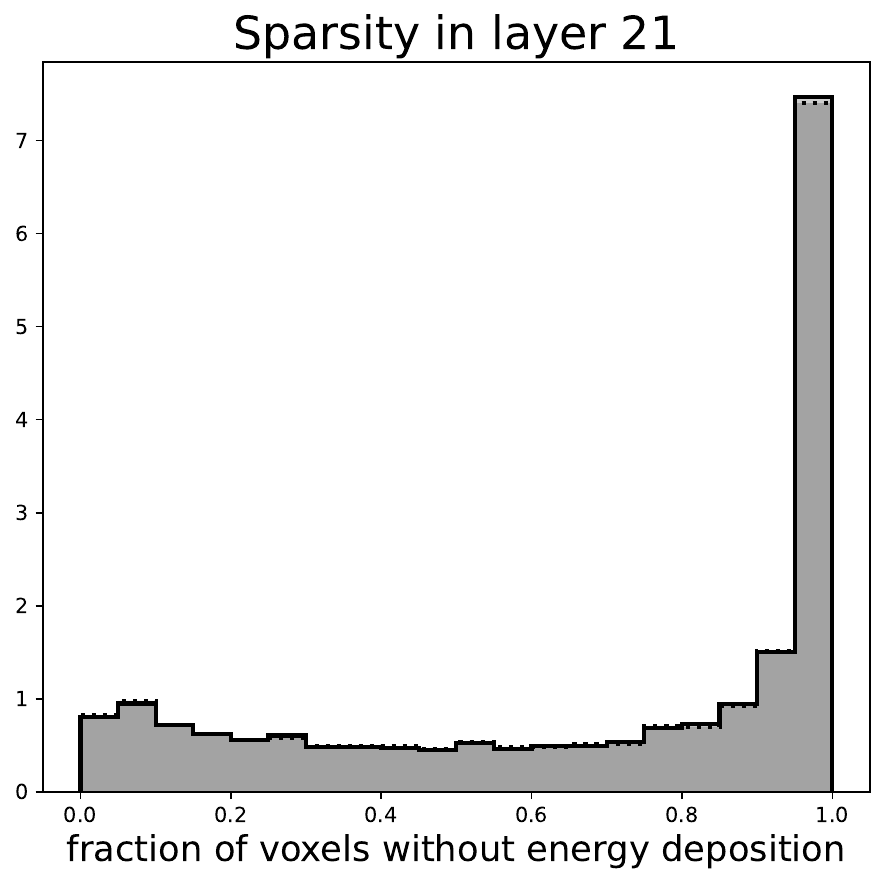} \hfill \includegraphics[height=0.1\textheight]{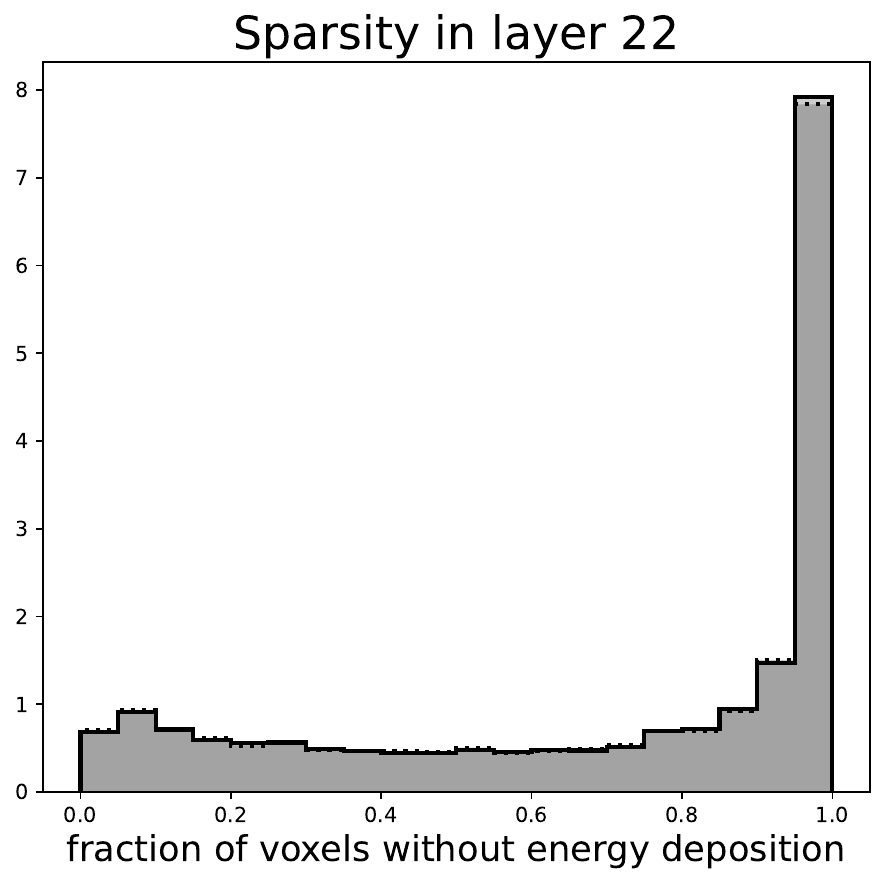} \hfill \includegraphics[height=0.1\textheight]{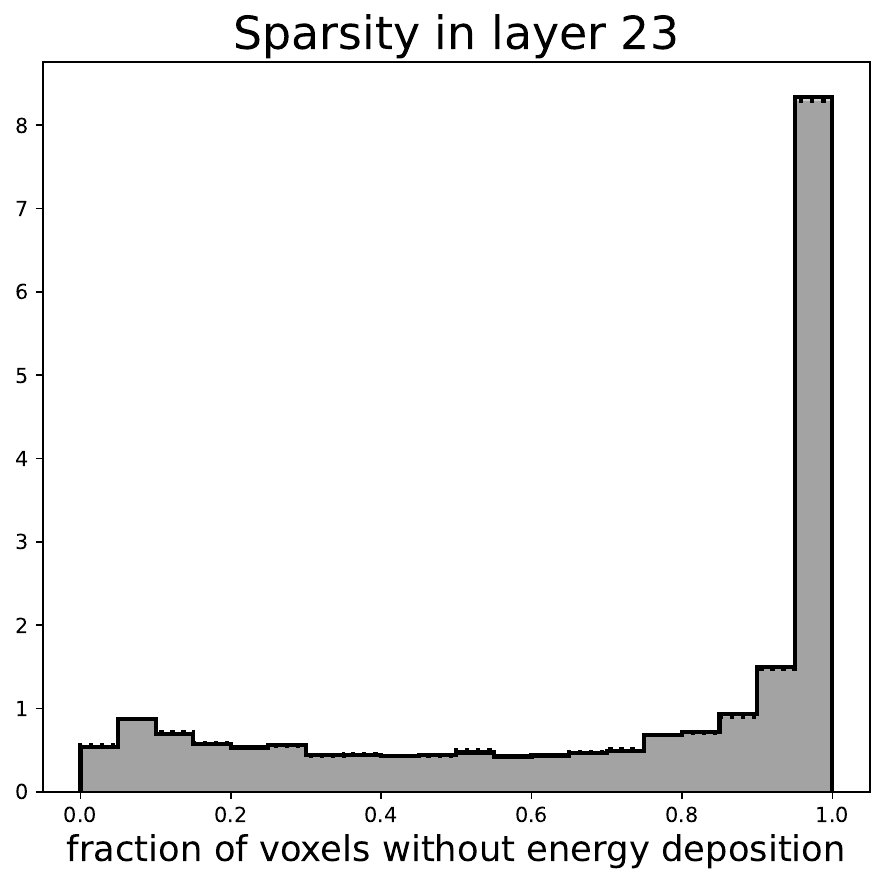} \hfill \includegraphics[height=0.1\textheight]{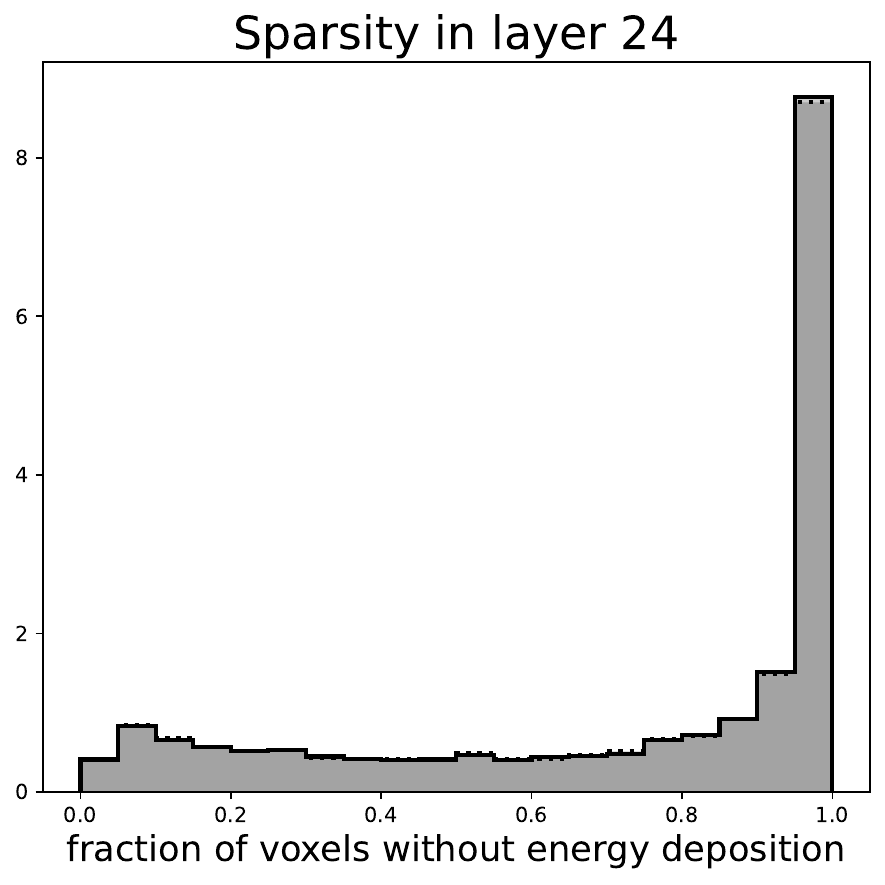}\\
    \includegraphics[height=0.1\textheight]{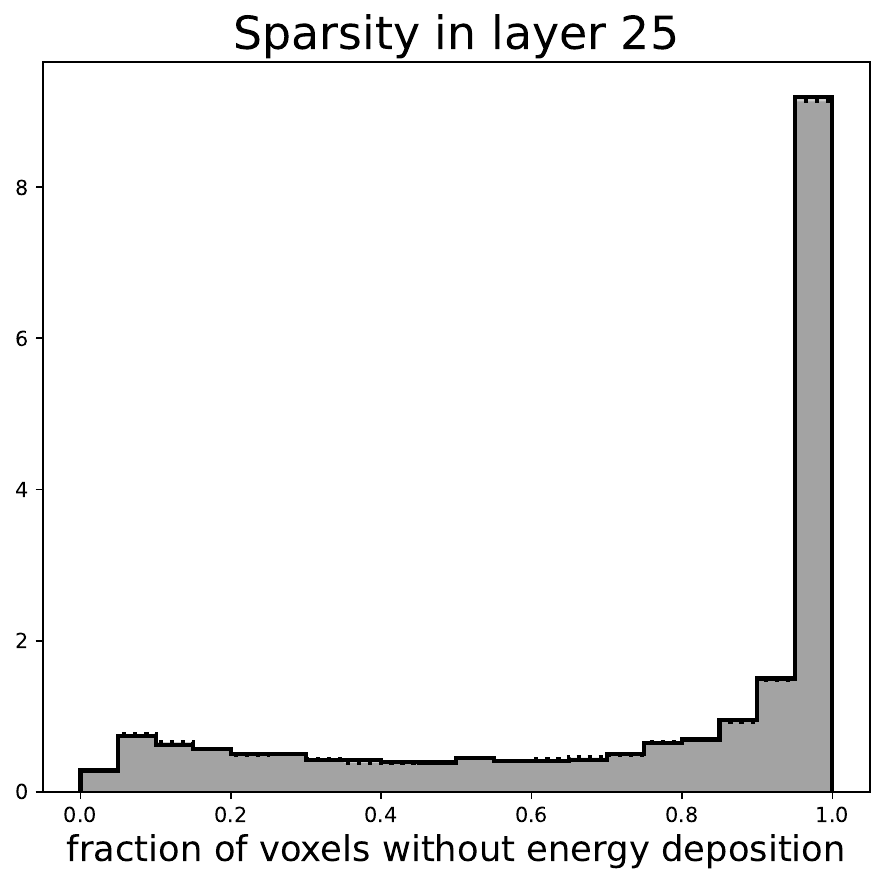} \hfill \includegraphics[height=0.1\textheight]{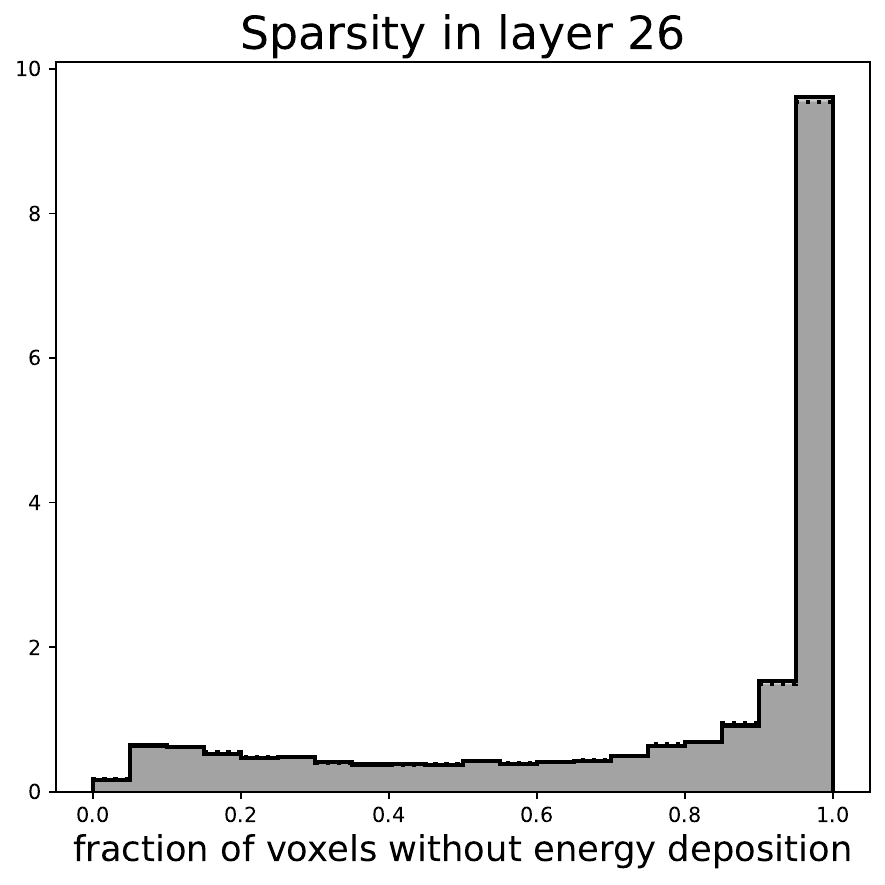} \hfill \includegraphics[height=0.1\textheight]{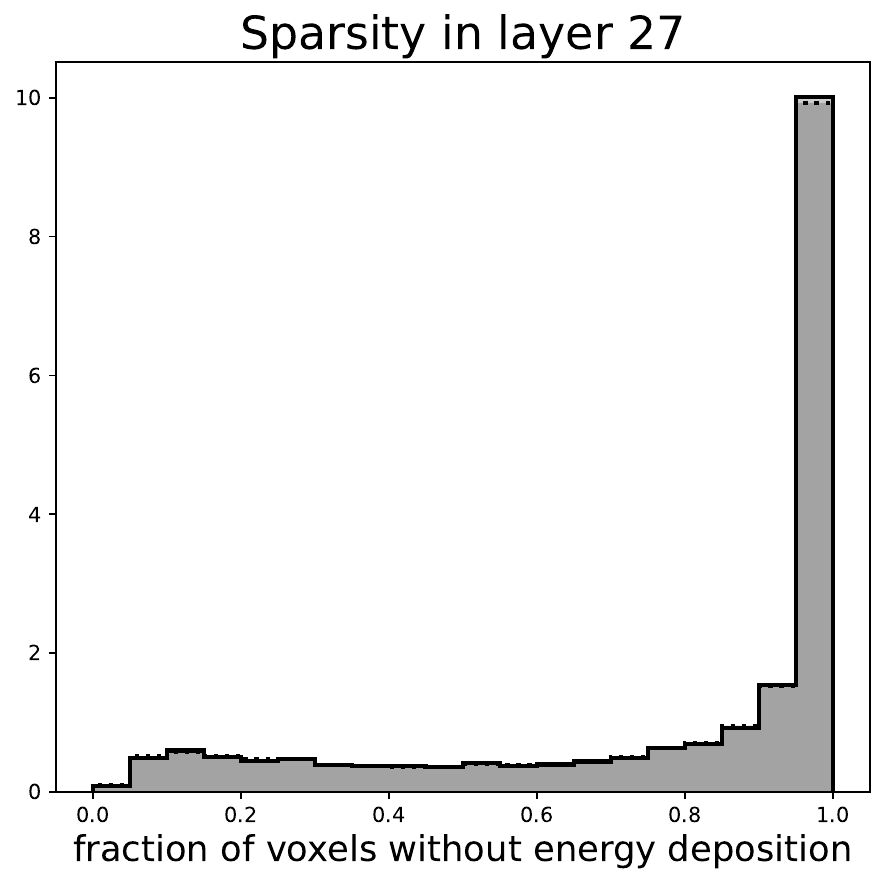} \hfill \includegraphics[height=0.1\textheight]{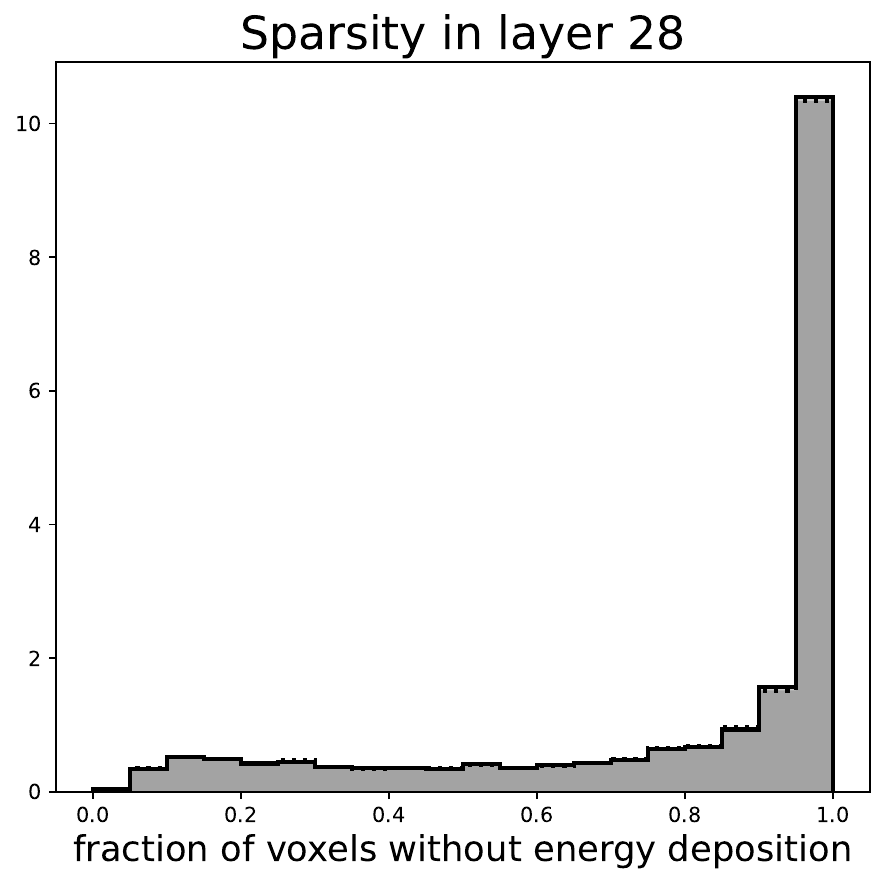} \hfill \includegraphics[height=0.1\textheight]{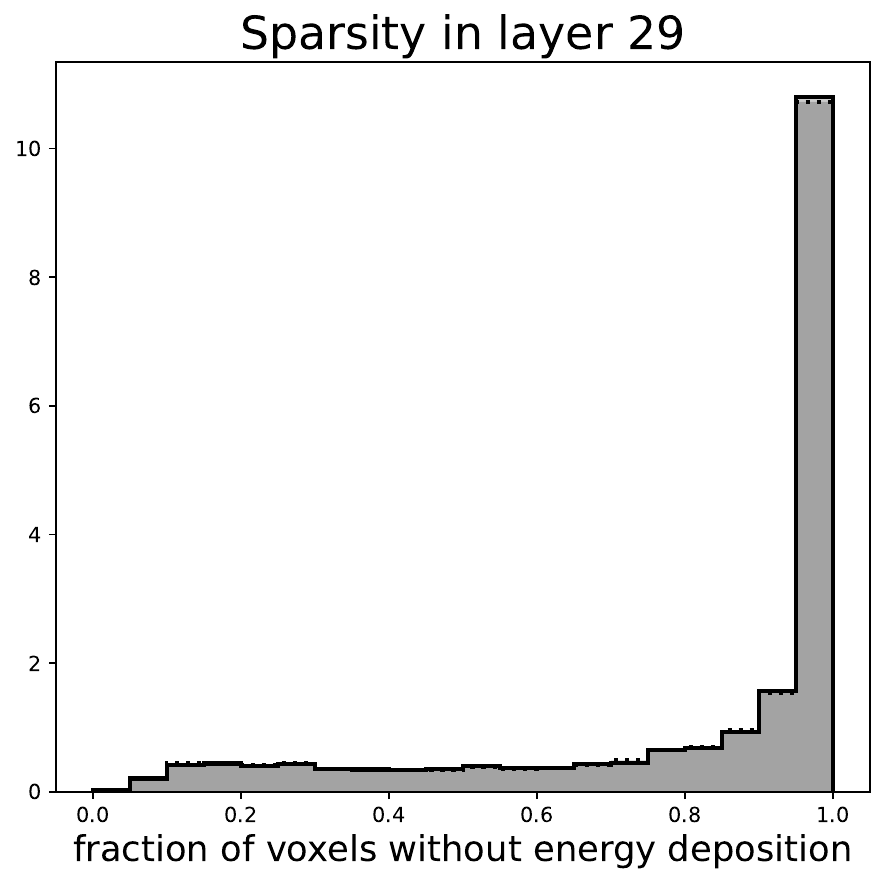}\\
    \includegraphics[height=0.1\textheight]{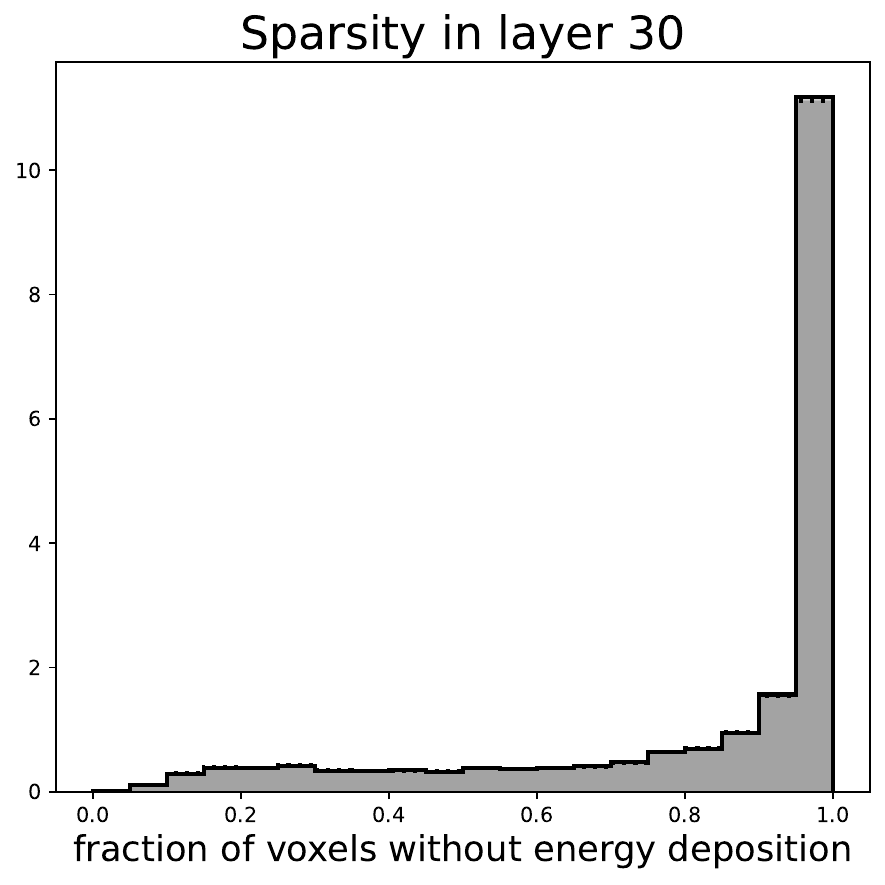} \hfill \includegraphics[height=0.1\textheight]{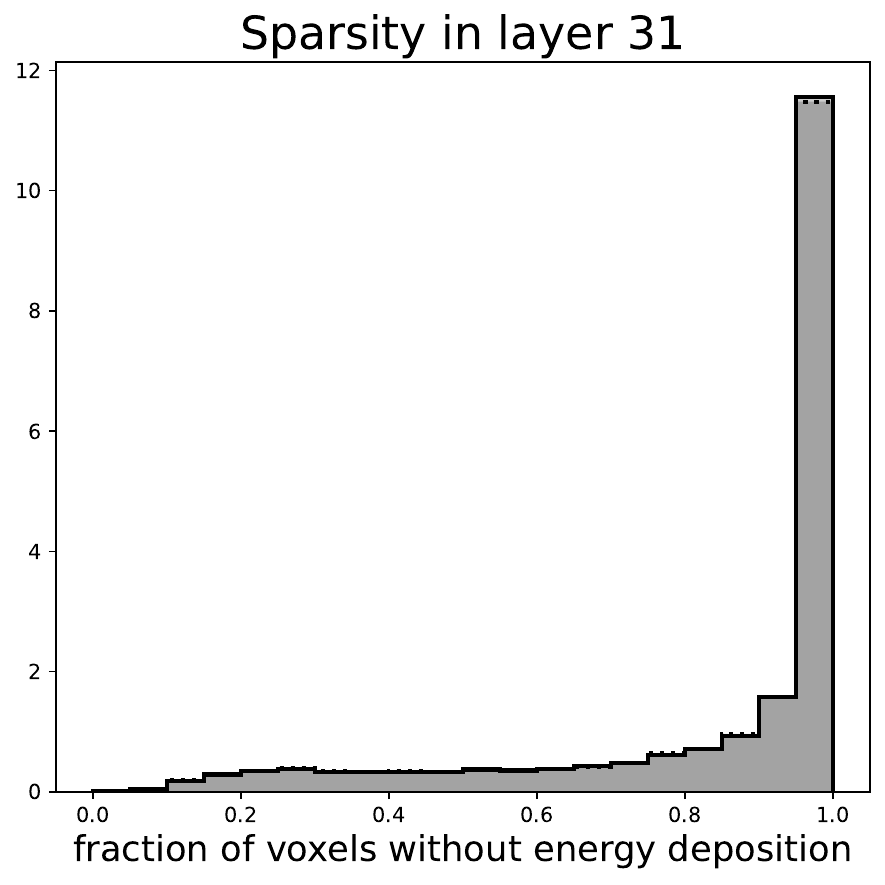} \hfill \includegraphics[height=0.1\textheight]{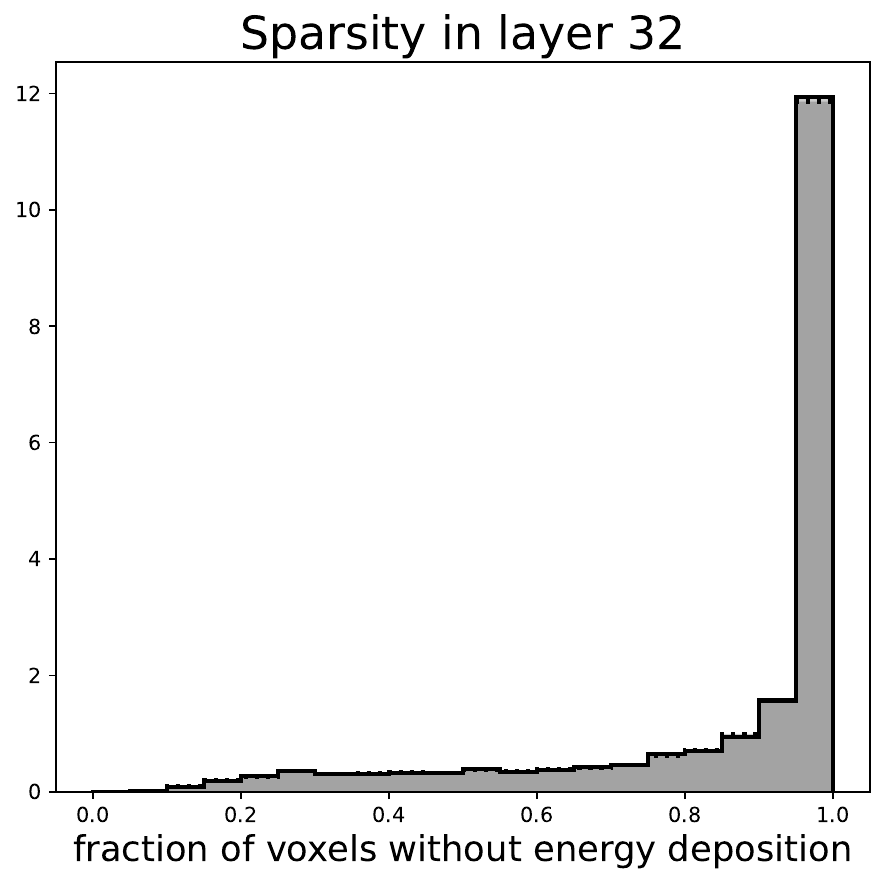} \hfill \includegraphics[height=0.1\textheight]{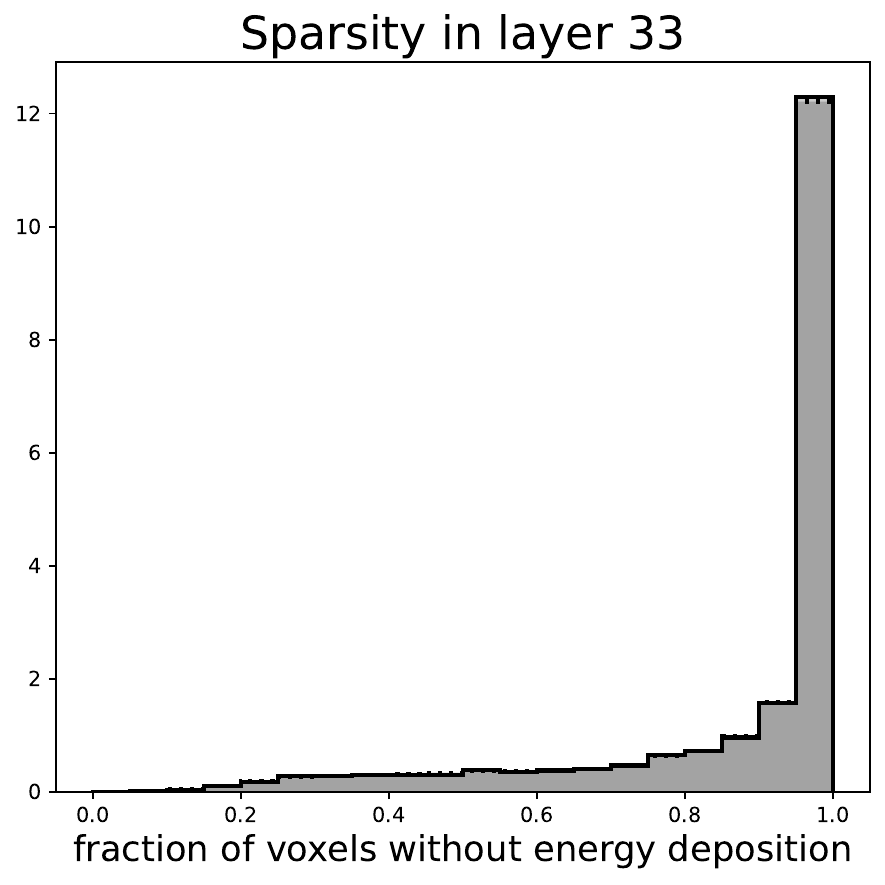} \hfill \includegraphics[height=0.1\textheight]{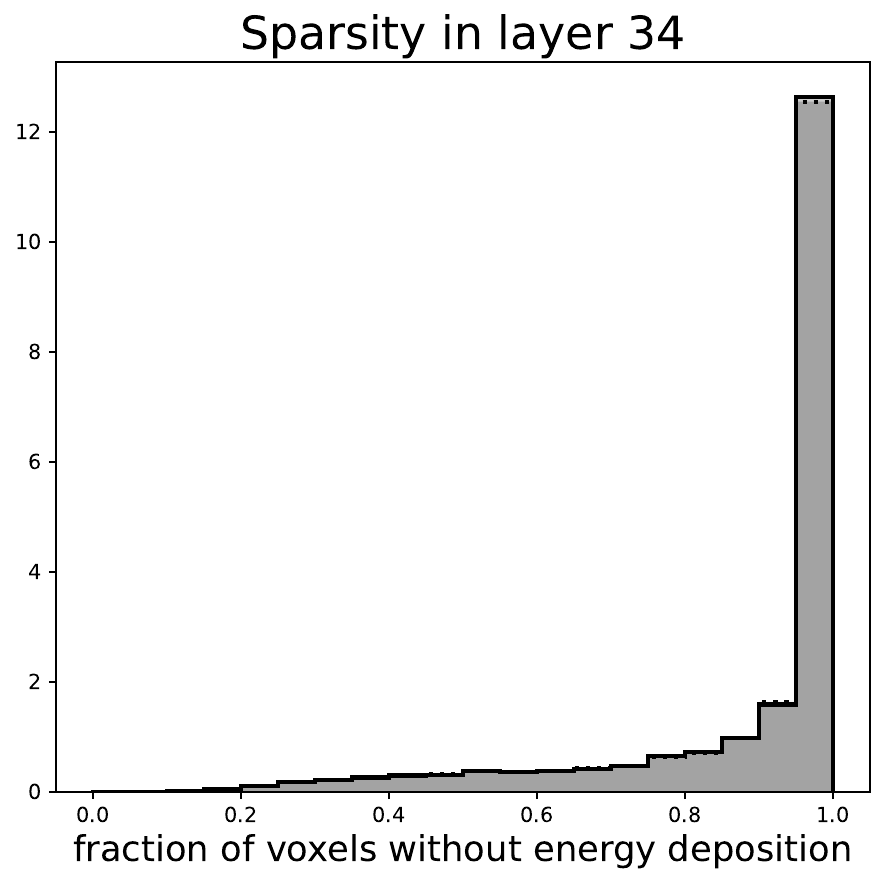}\\
    \includegraphics[height=0.1\textheight]{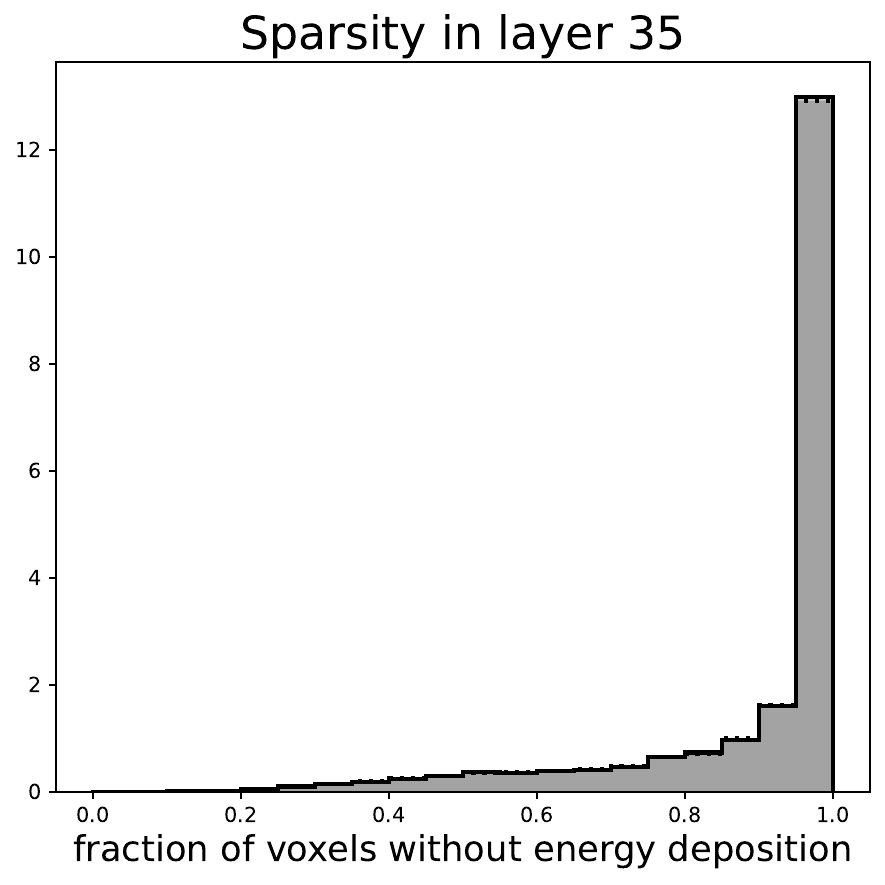} \hfill \includegraphics[height=0.1\textheight]{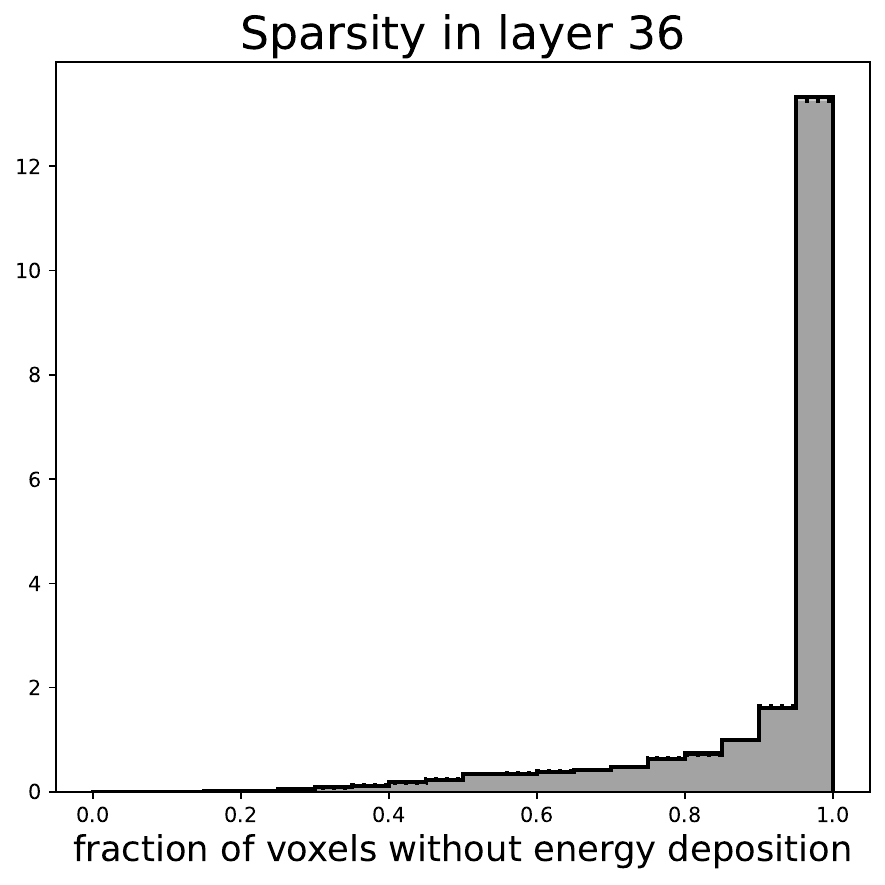} \hfill \includegraphics[height=0.1\textheight]{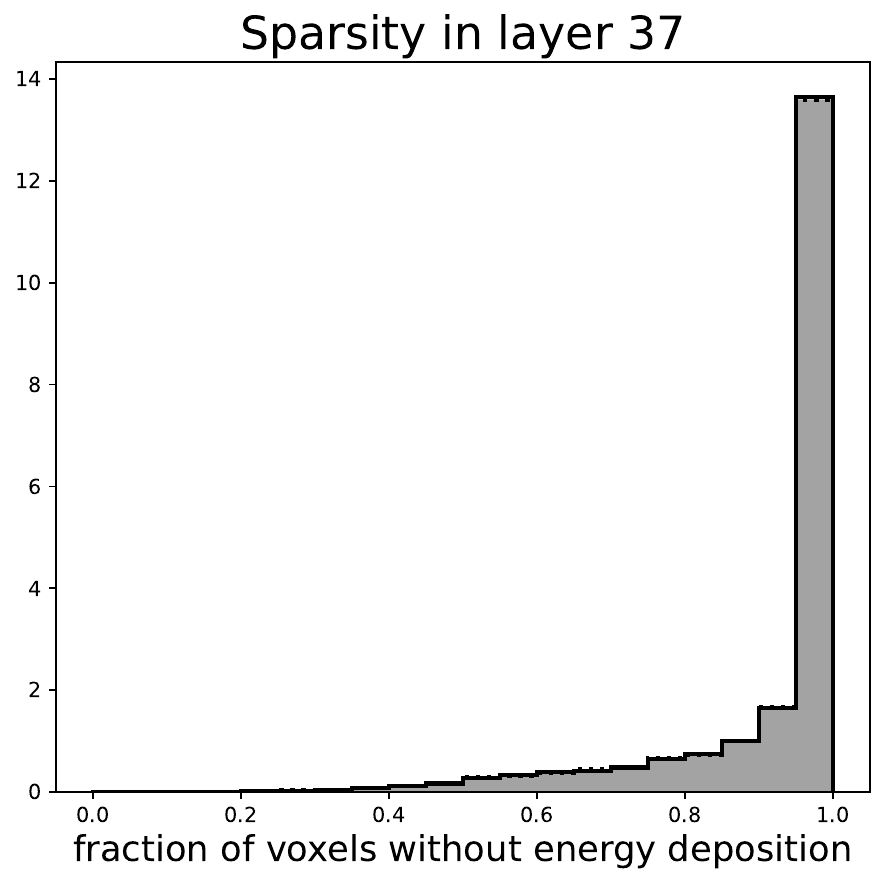} \hfill \includegraphics[height=0.1\textheight]{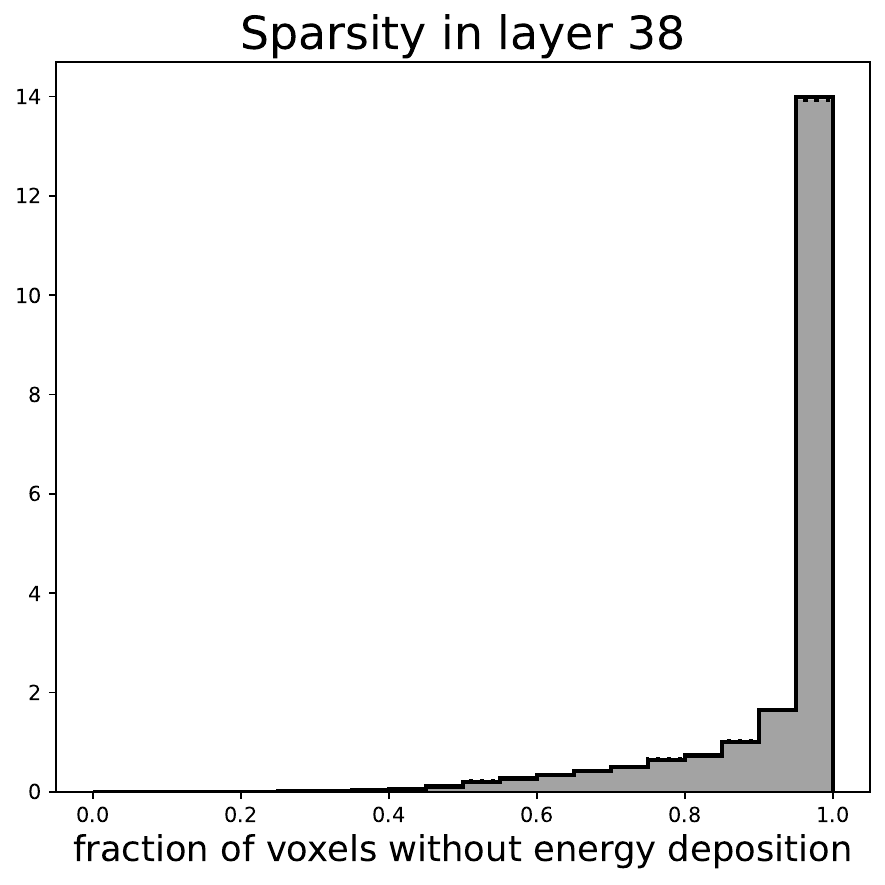} \hfill \includegraphics[height=0.1\textheight]{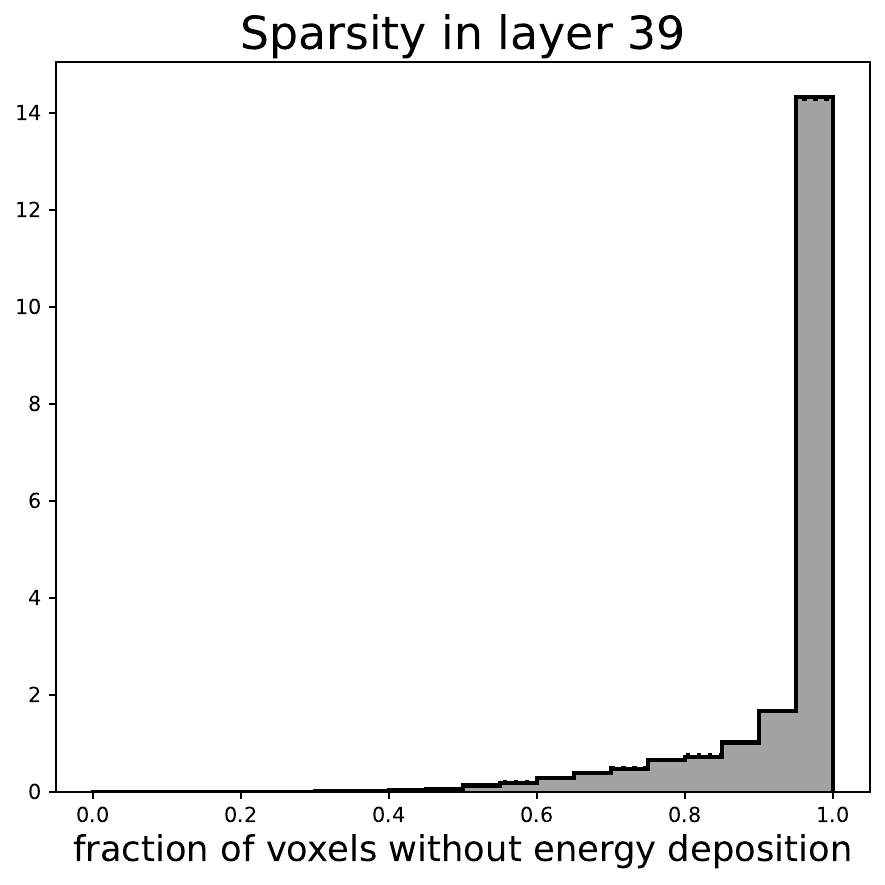}\\
    \includegraphics[height=0.1\textheight]{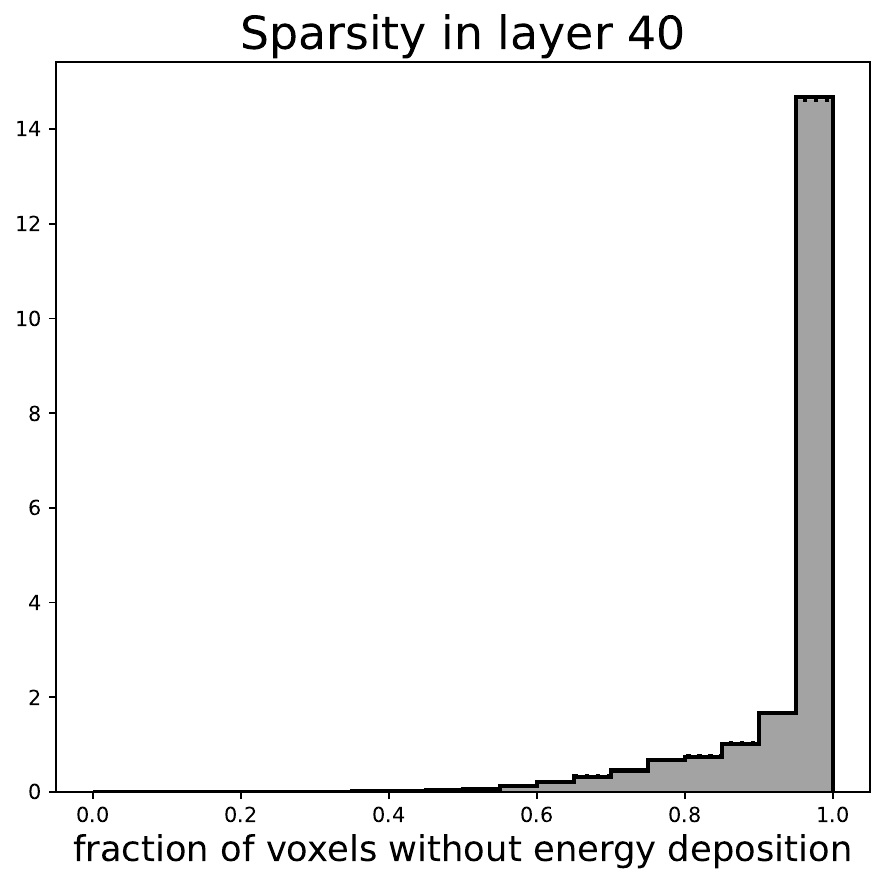} \hfill \includegraphics[height=0.1\textheight]{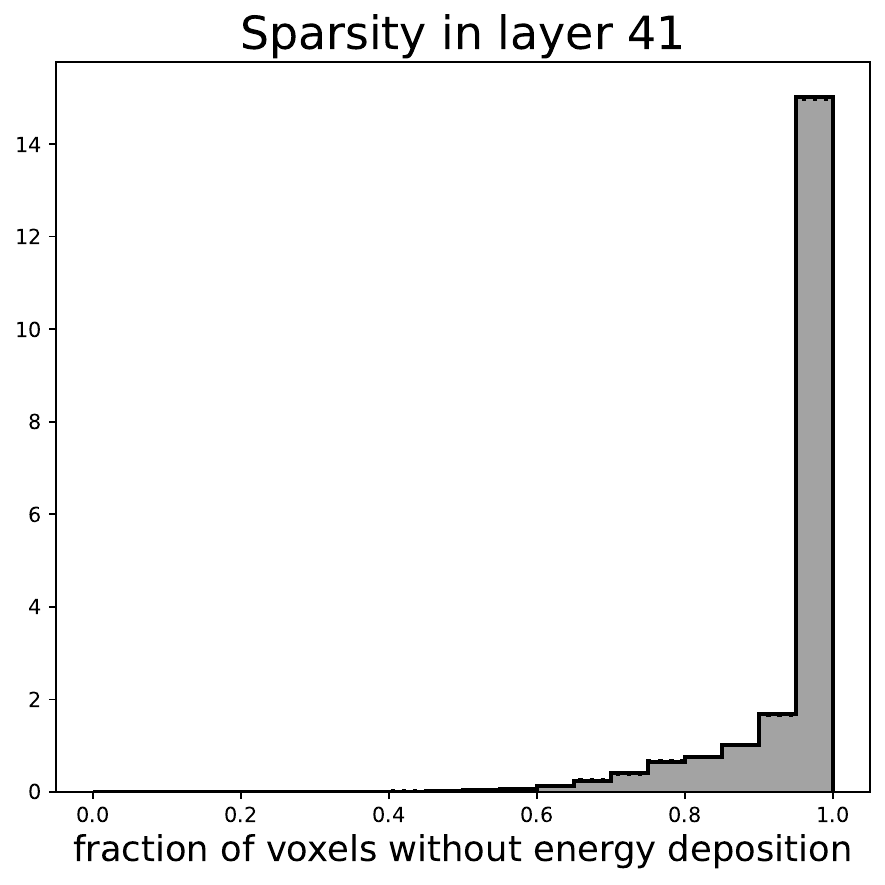} \hfill \includegraphics[height=0.1\textheight]{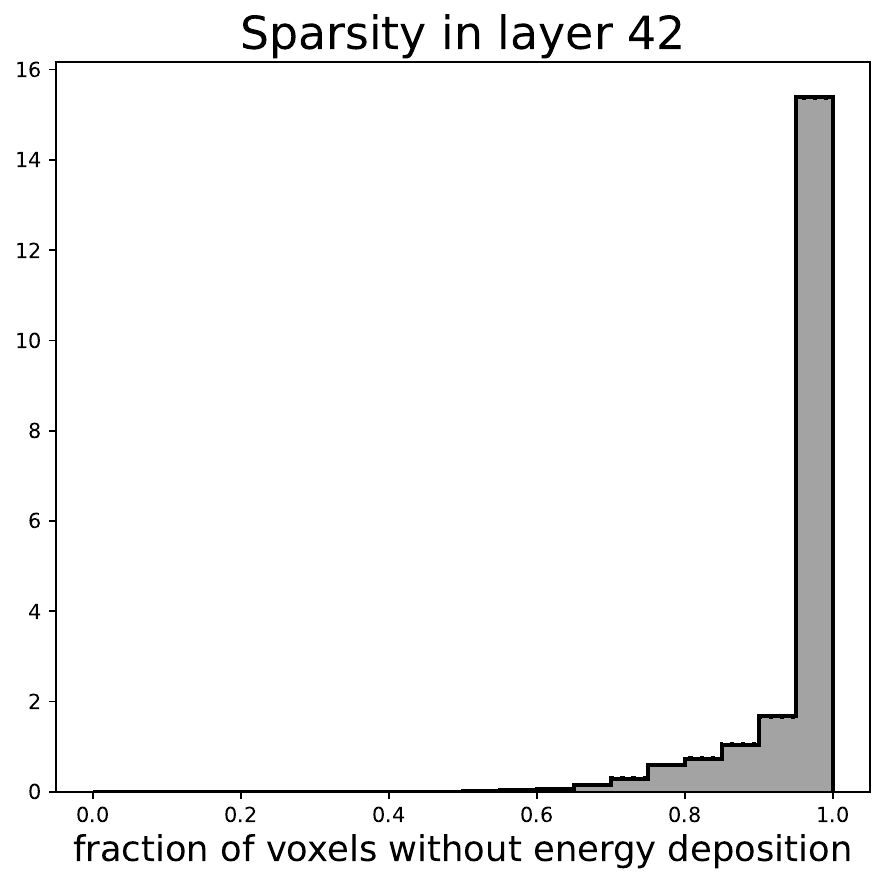} \hfill \includegraphics[height=0.1\textheight]{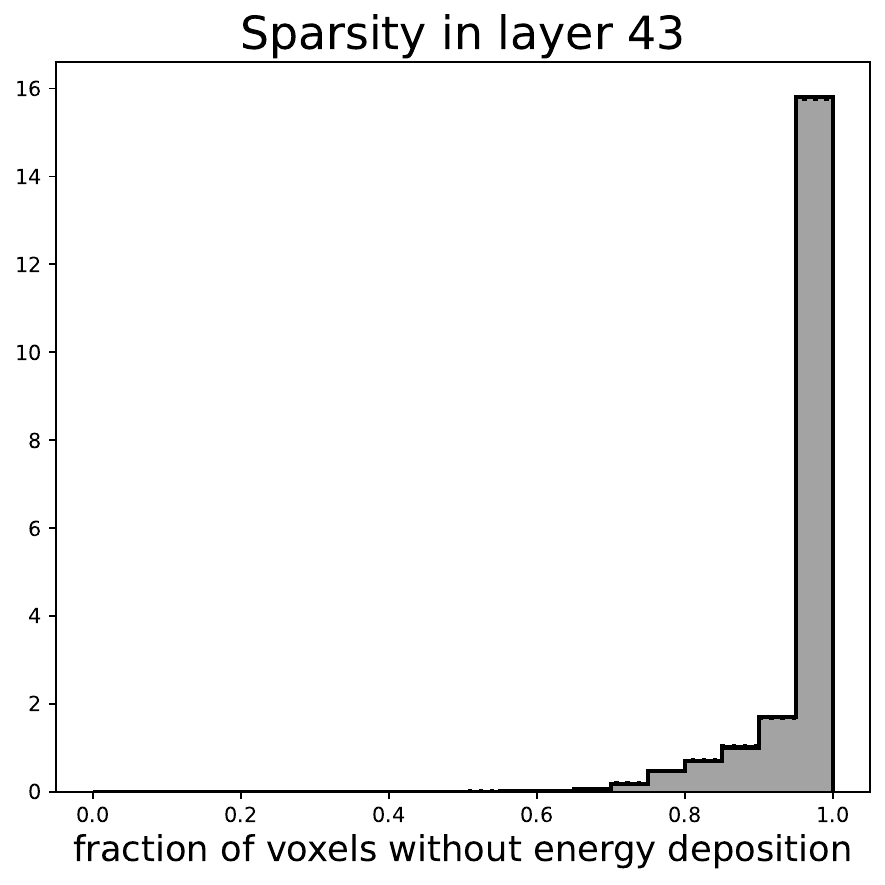} \hfill \includegraphics[height=0.1\textheight]{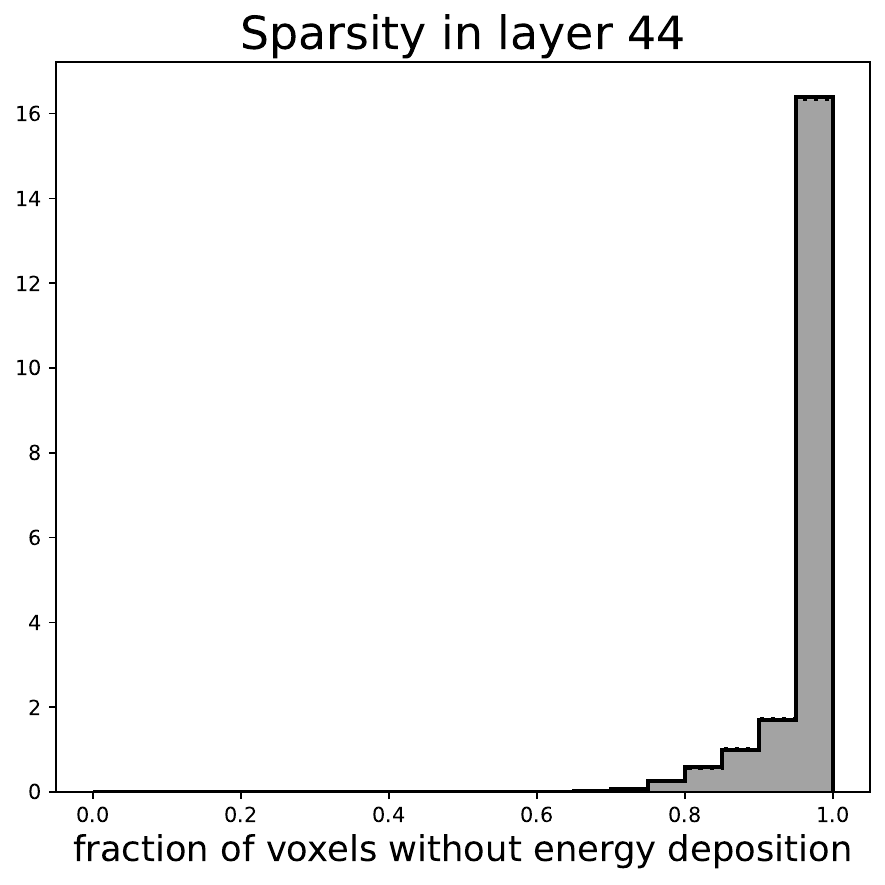}\\
    \includegraphics[width=0.5\textwidth]{figures/Appendix_reference/legend.pdf}
    \caption{Distribution of \geant training and evaluation data in sparsity for ds2. }
    \label{fig:app_ref.ds2.9}
\end{figure}

\subsection{\texorpdfstring{Dataset 3, Electrons (\dsIII)}{Dataset 3, Electrons}}
\begin{figure}[ht]
    \centering
    \hfill \includegraphics[height=0.2\textheight]{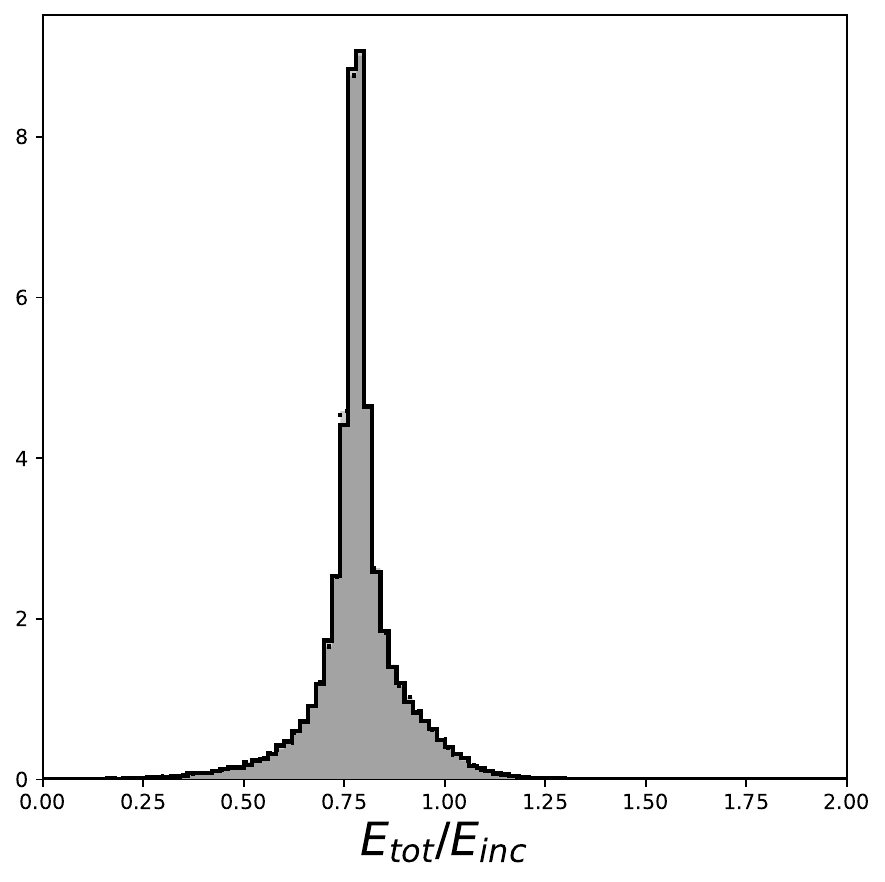} \hfill     \includegraphics[height=0.2\textheight]{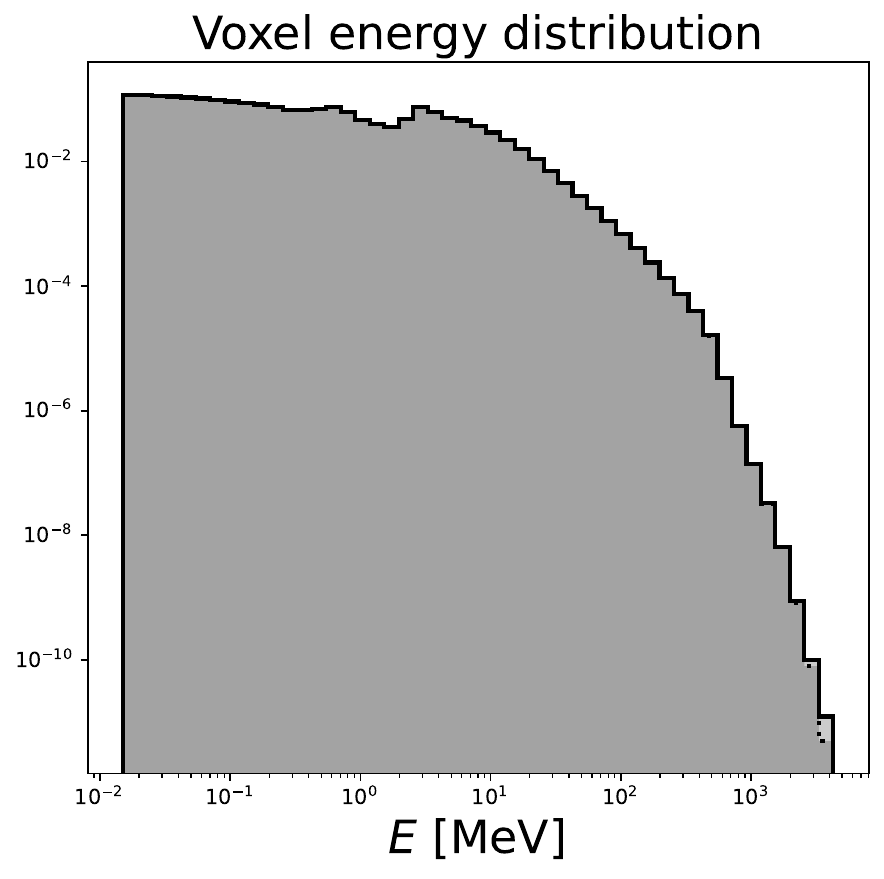} \hfill $ $\\
    \includegraphics[width=0.5\textwidth]{figures/Appendix_reference/legend.pdf}
    \caption{Distribution of \geant training and evaluation data in ratio of total deposited energy to incident energy and energy per voxel for ds3. }
    \label{fig:app_ref.ds3.1}
\end{figure}

\begin{figure}[ht]
    \centering
    \includegraphics[height=0.1\textheight]{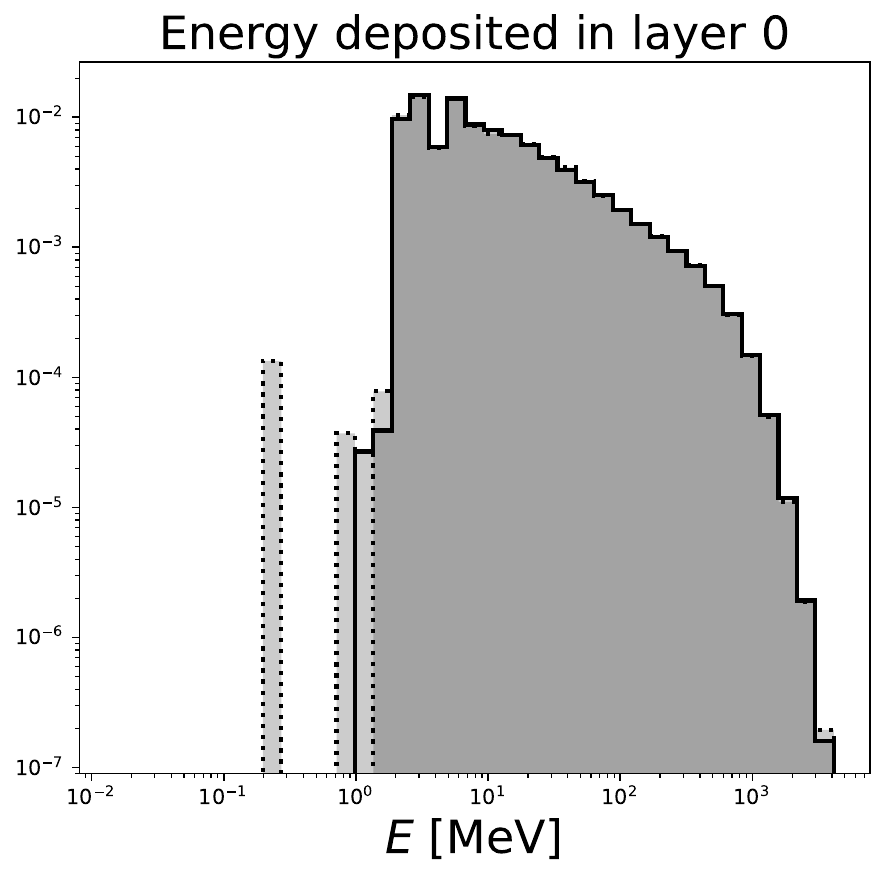} \hfill \includegraphics[height=0.1\textheight]{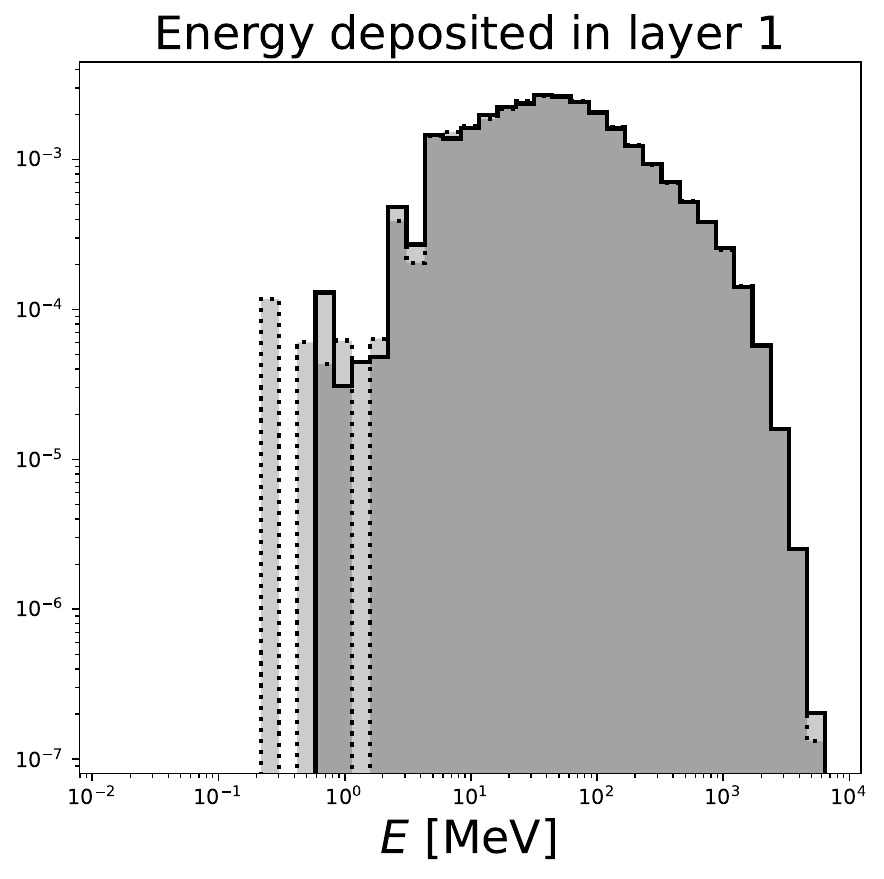} \hfill \includegraphics[height=0.1\textheight]{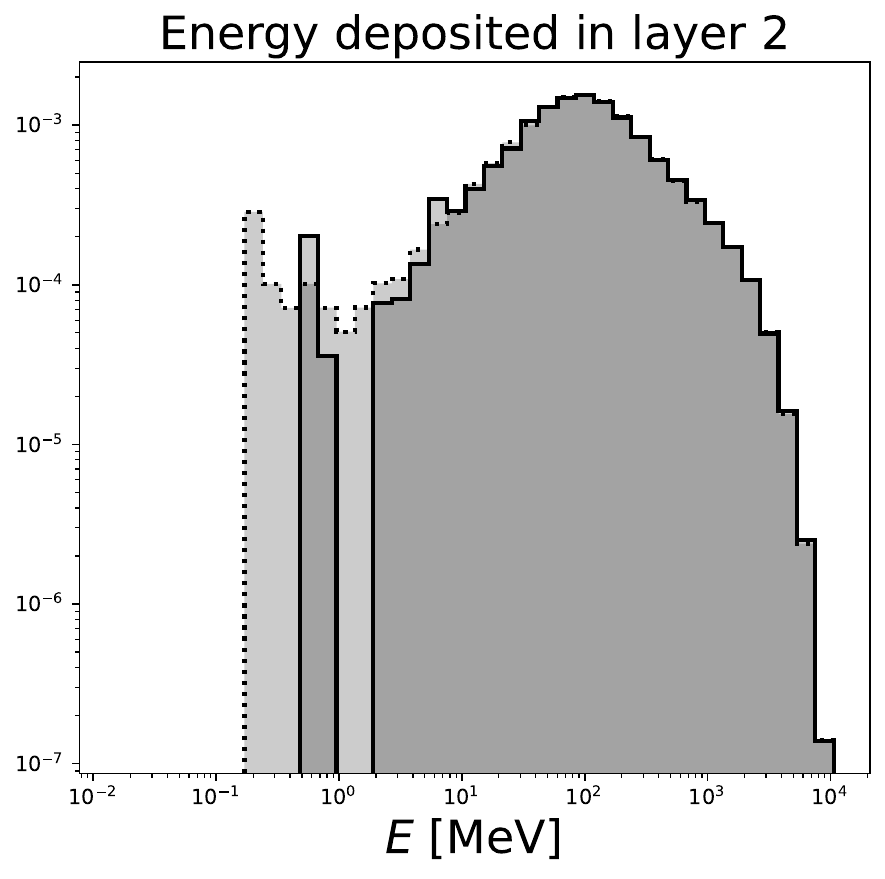} \hfill \includegraphics[height=0.1\textheight]{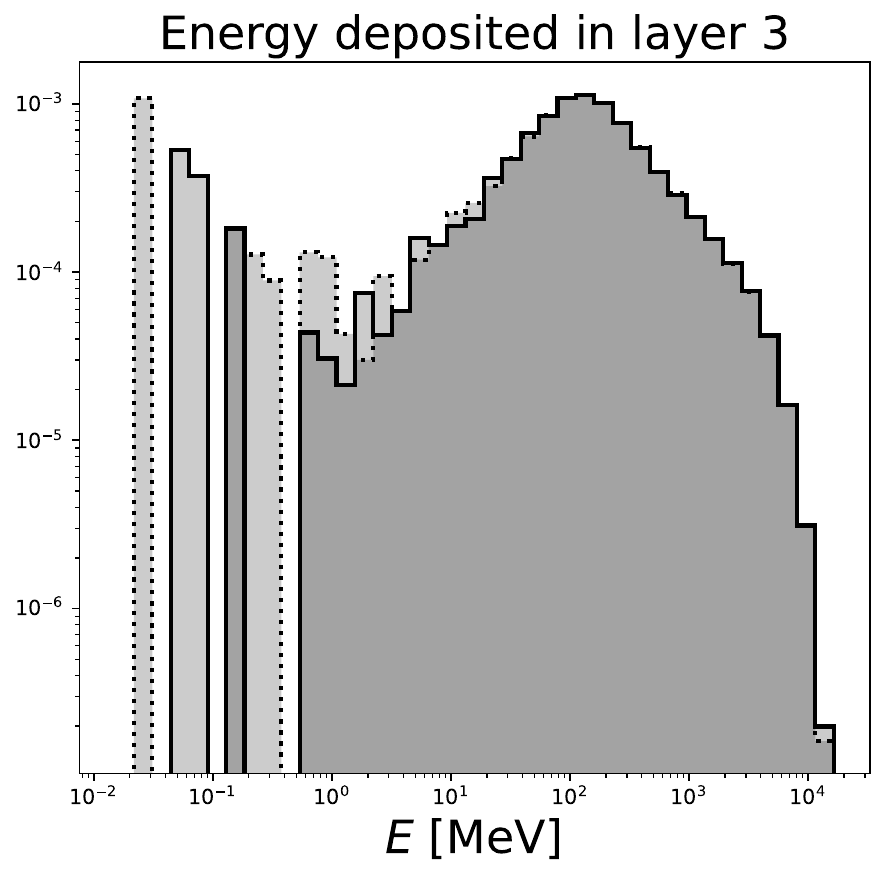} \hfill \includegraphics[height=0.1\textheight]{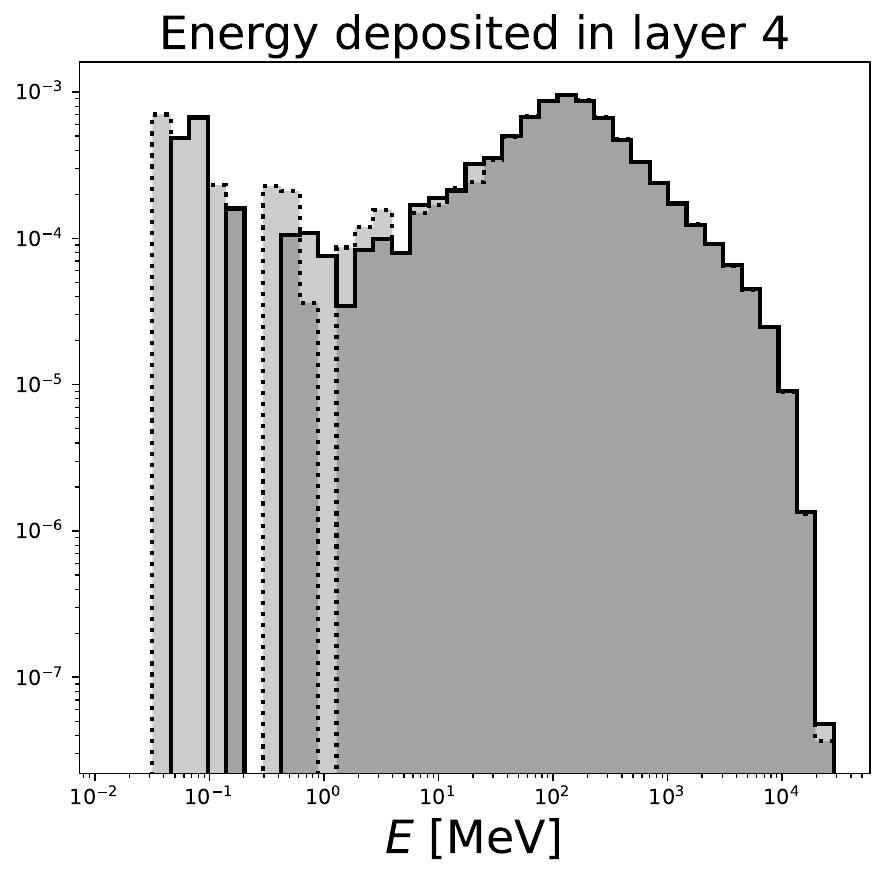}\\
    \includegraphics[height=0.1\textheight]{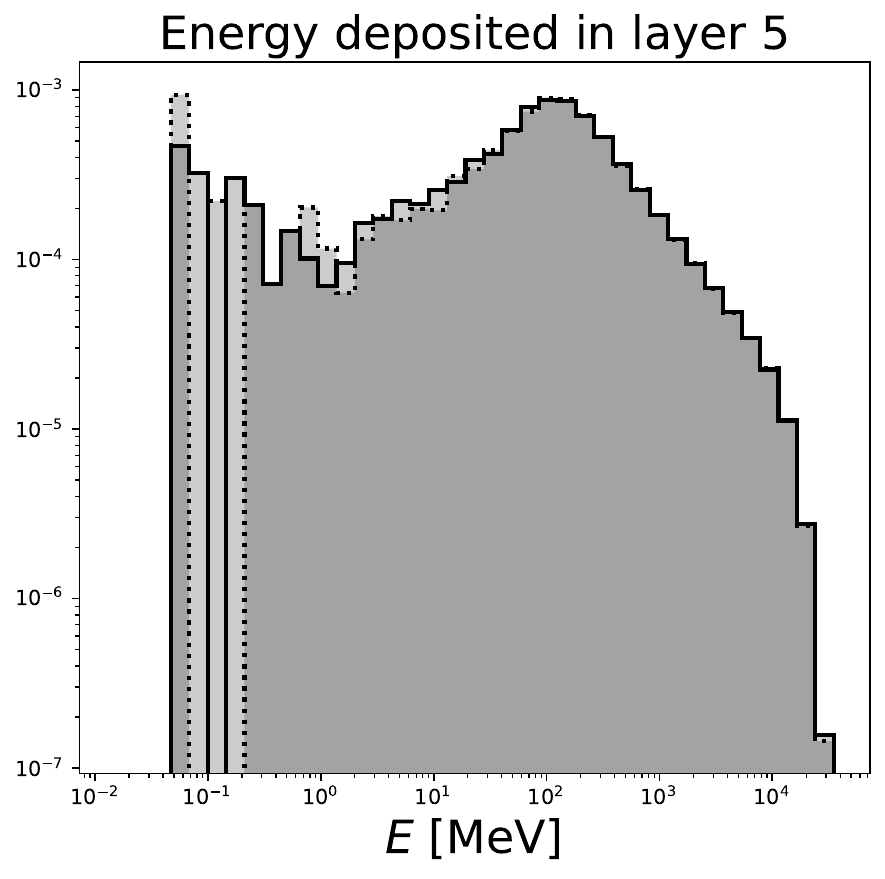} \hfill \includegraphics[height=0.1\textheight]{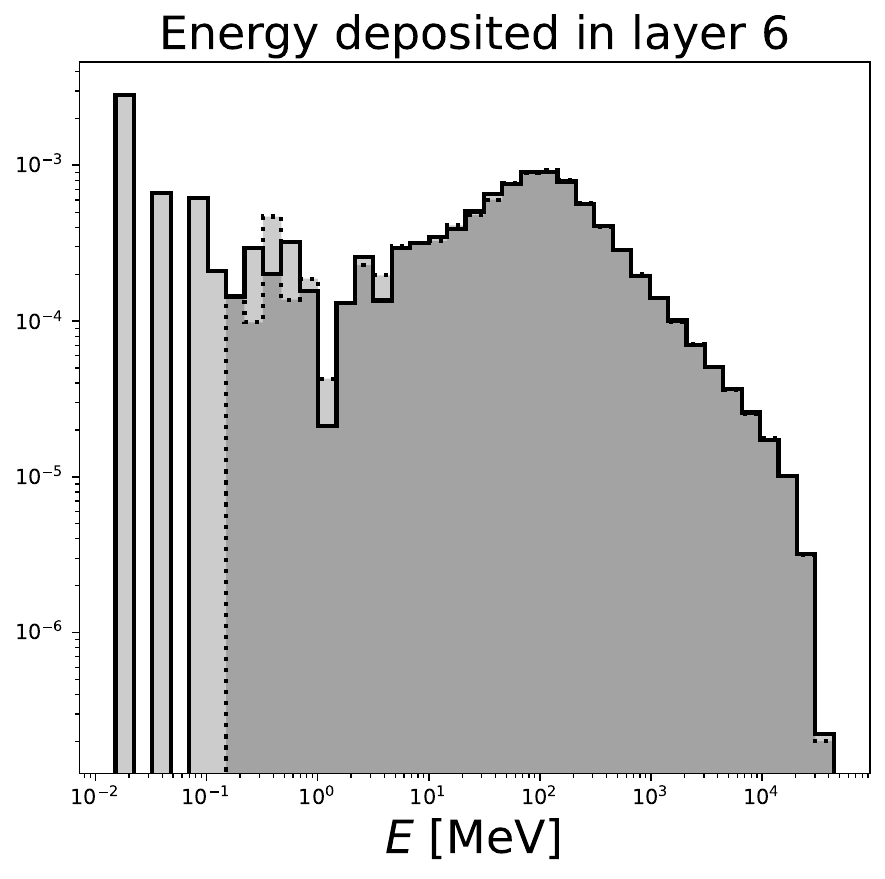} \hfill \includegraphics[height=0.1\textheight]{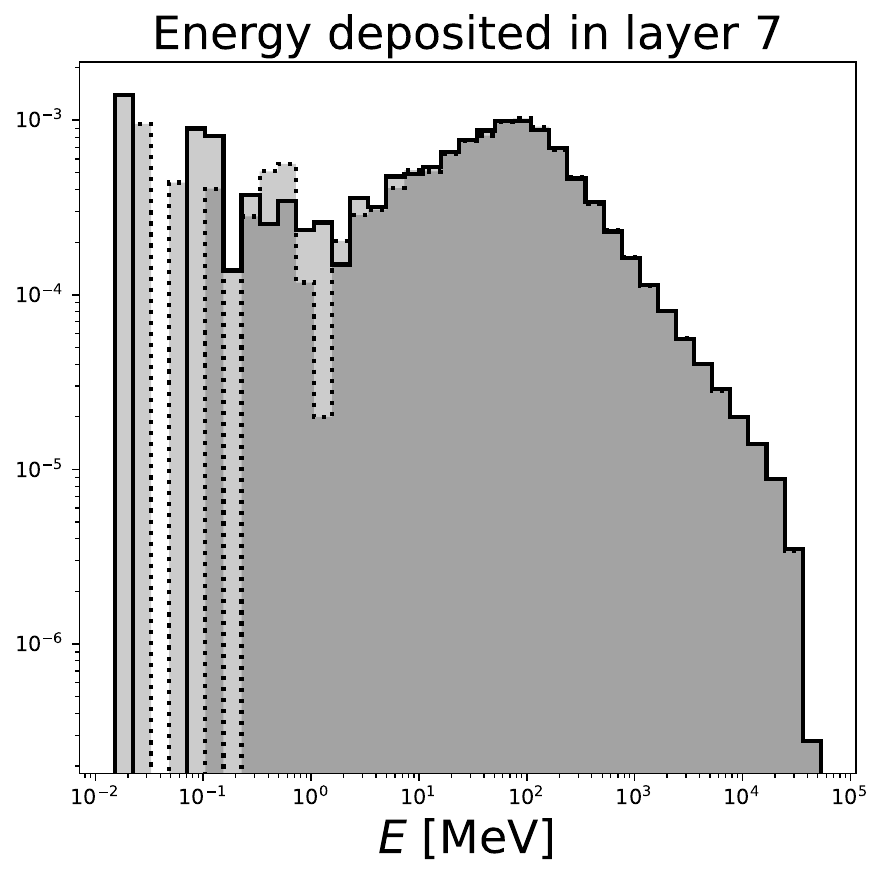} \hfill \includegraphics[height=0.1\textheight]{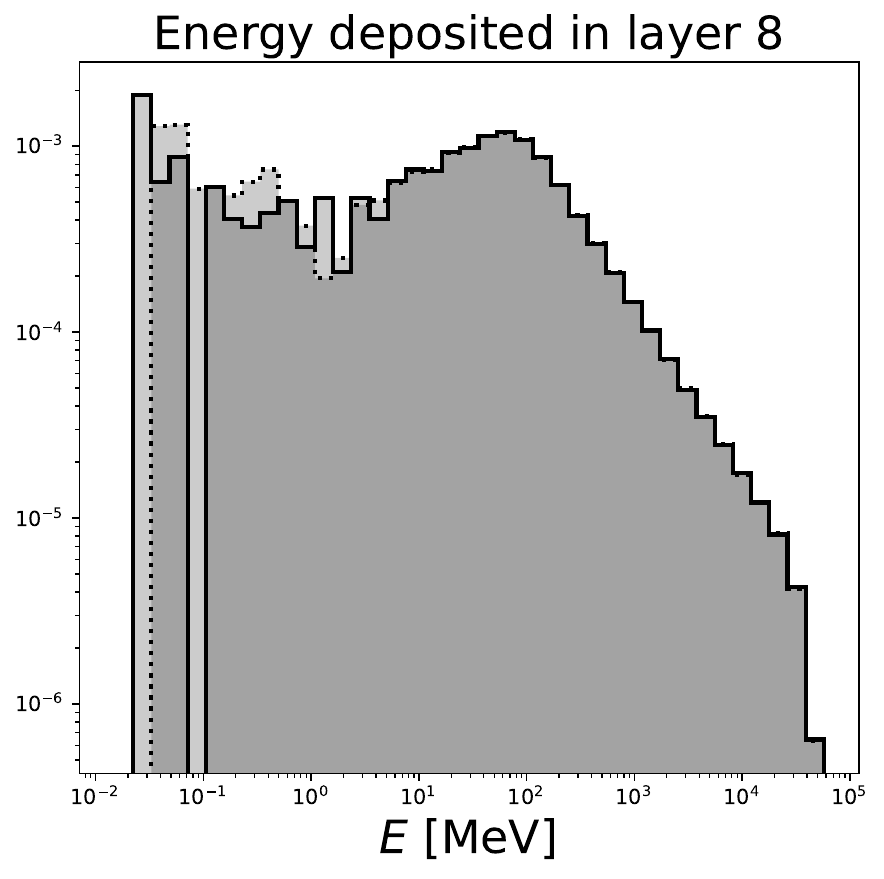} \hfill \includegraphics[height=0.1\textheight]{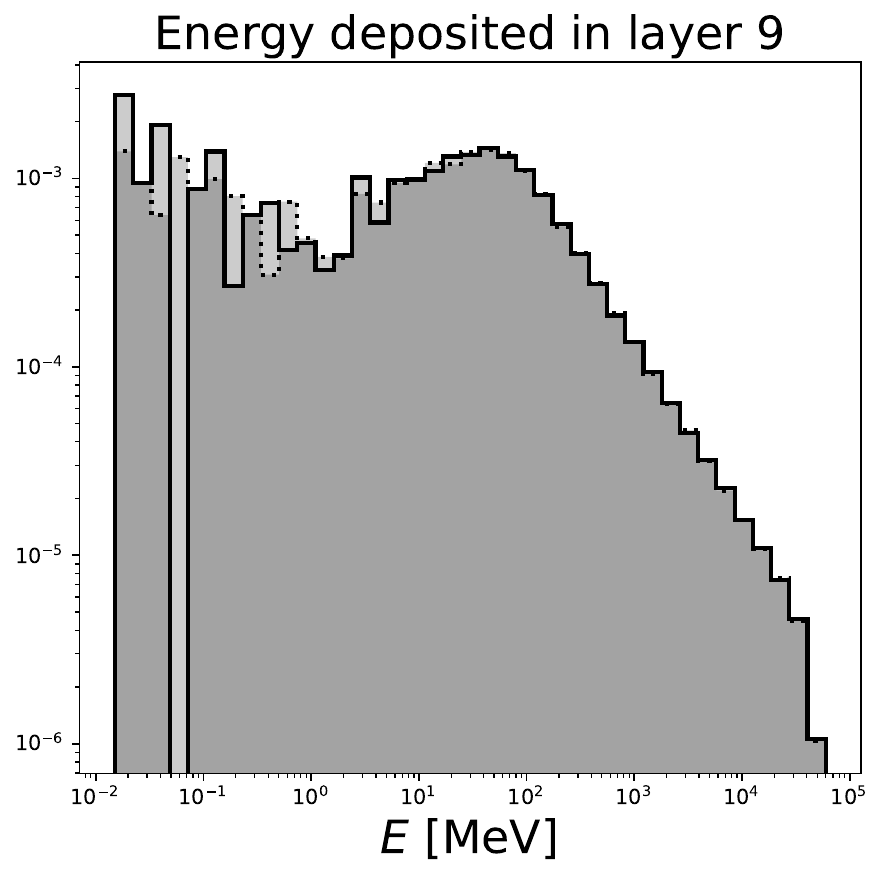}\\
    \includegraphics[height=0.1\textheight]{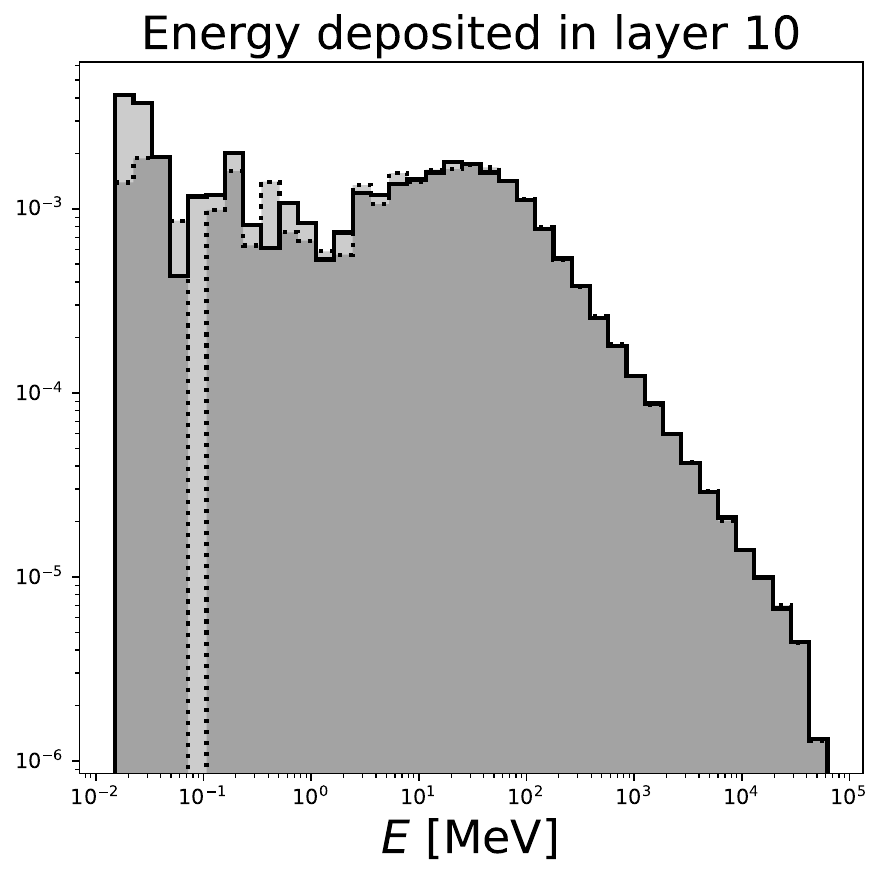} \hfill \includegraphics[height=0.1\textheight]{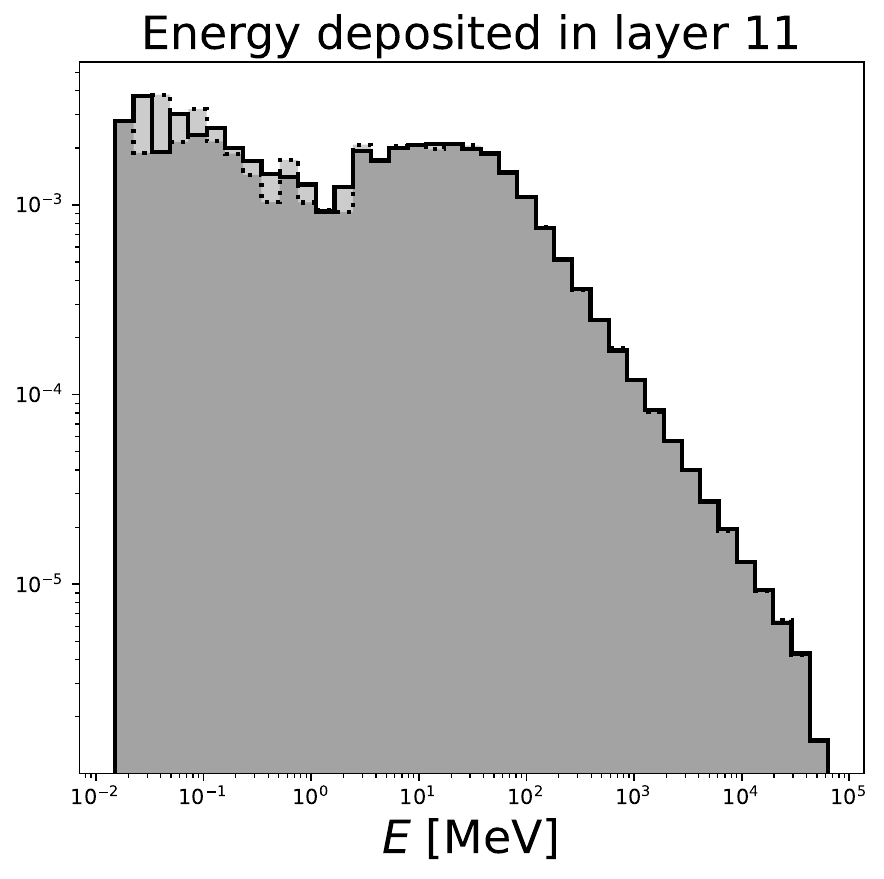} \hfill \includegraphics[height=0.1\textheight]{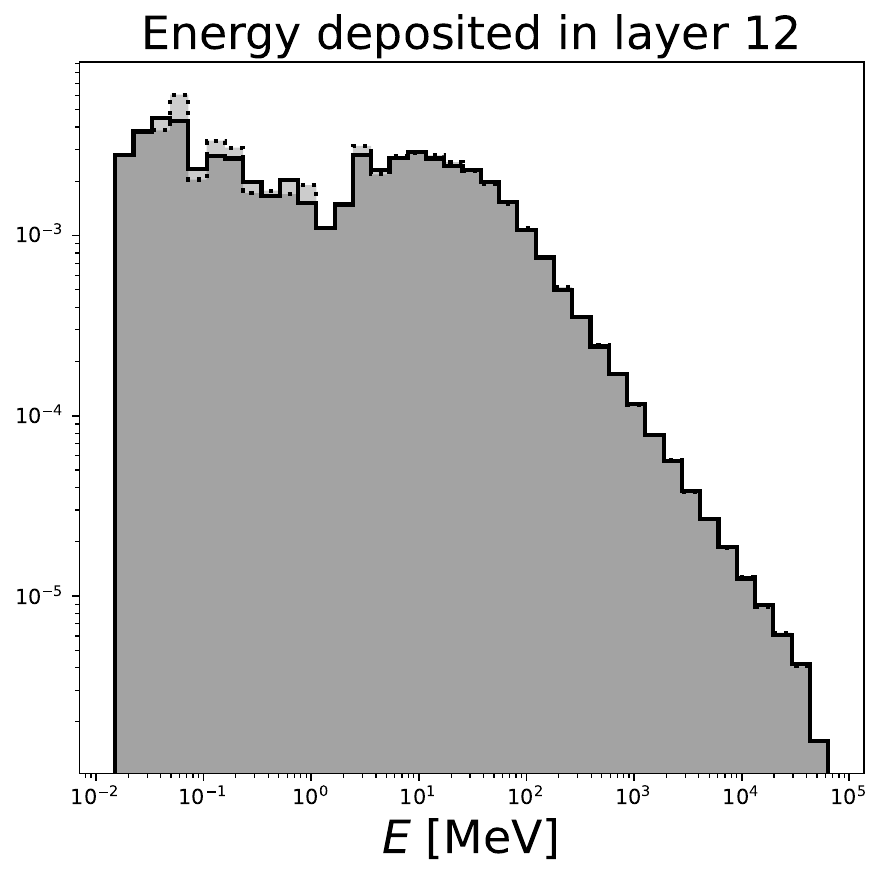} \hfill \includegraphics[height=0.1\textheight]{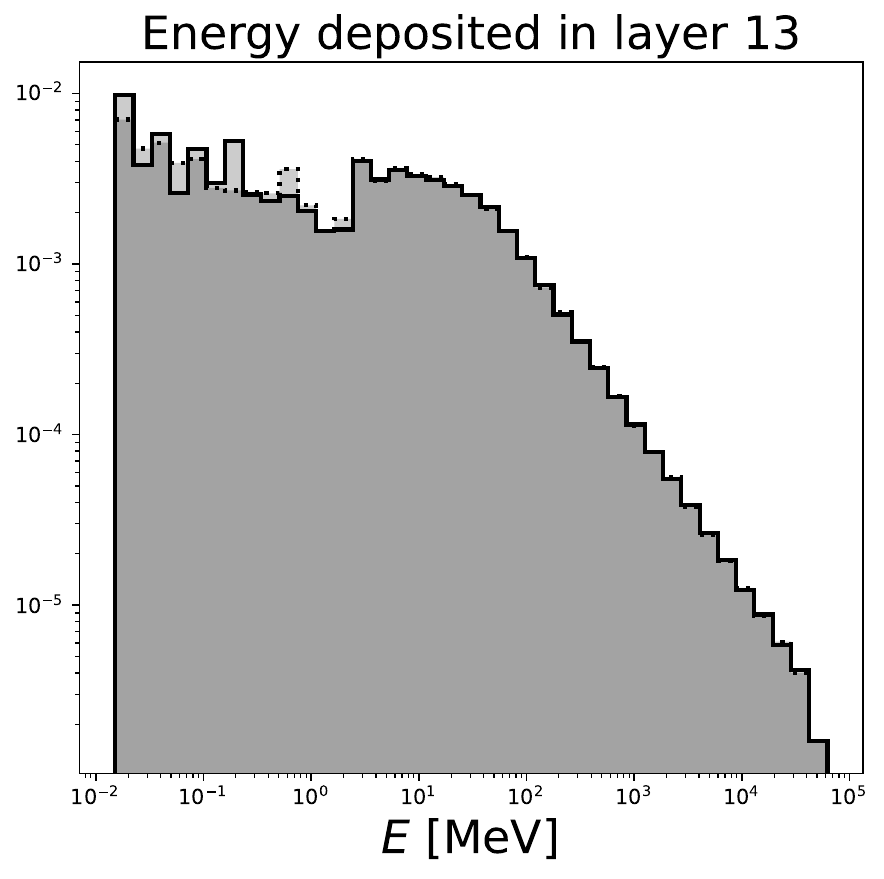} \hfill \includegraphics[height=0.1\textheight]{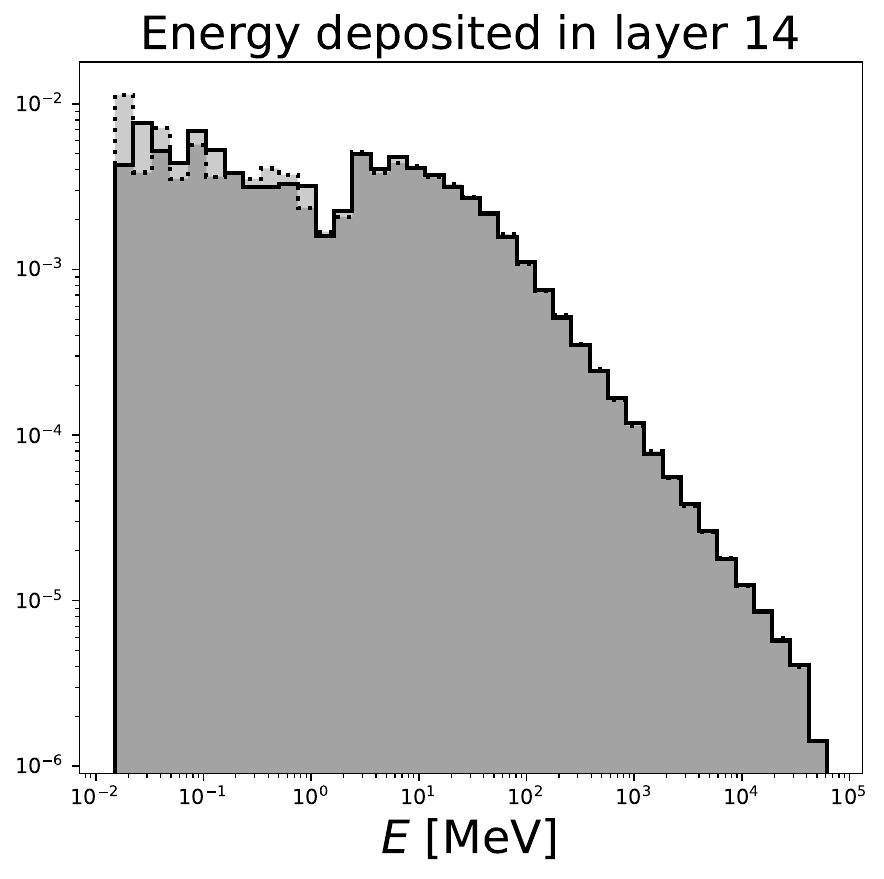}\\
    \includegraphics[height=0.1\textheight]{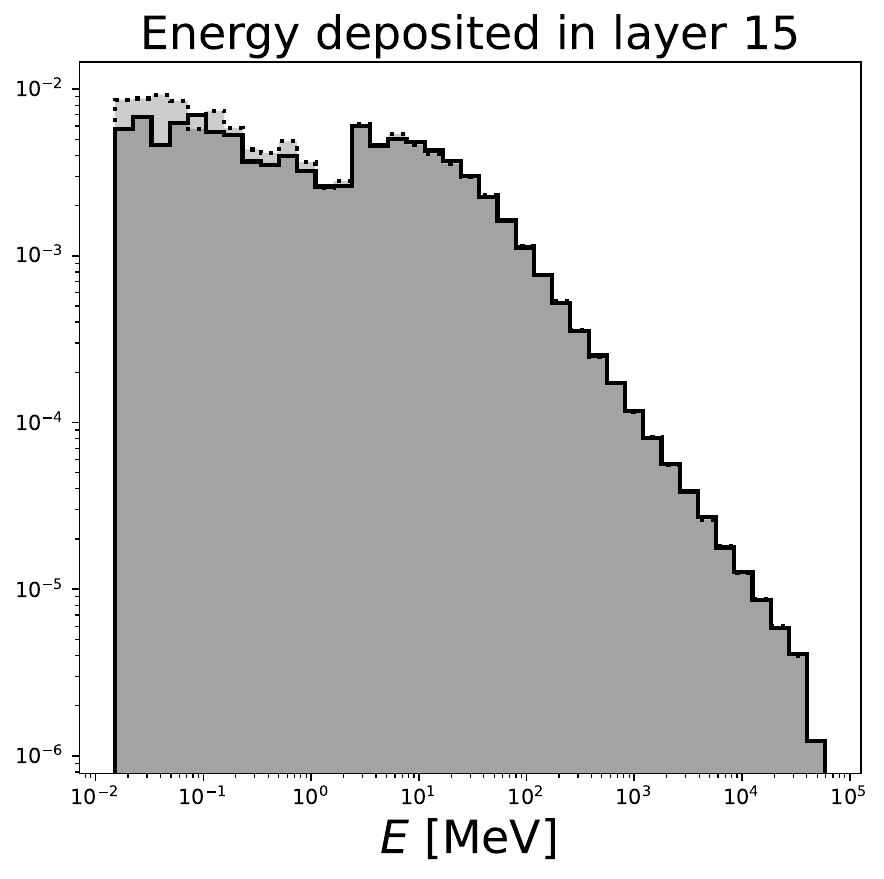} \hfill \includegraphics[height=0.1\textheight]{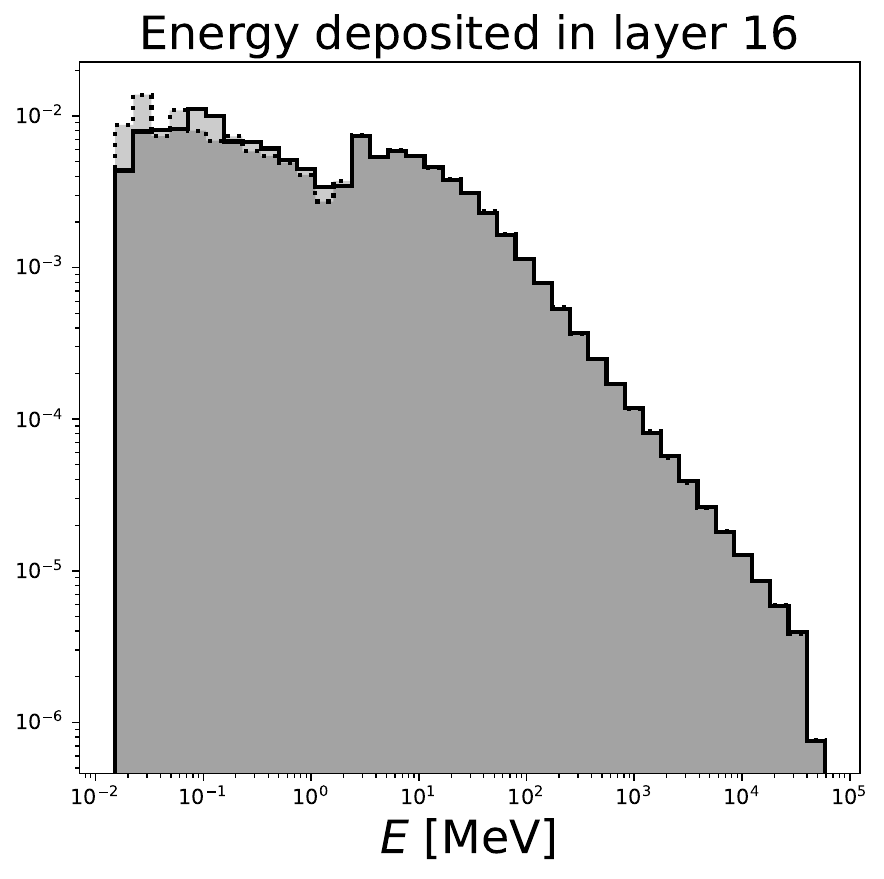} \hfill \includegraphics[height=0.1\textheight]{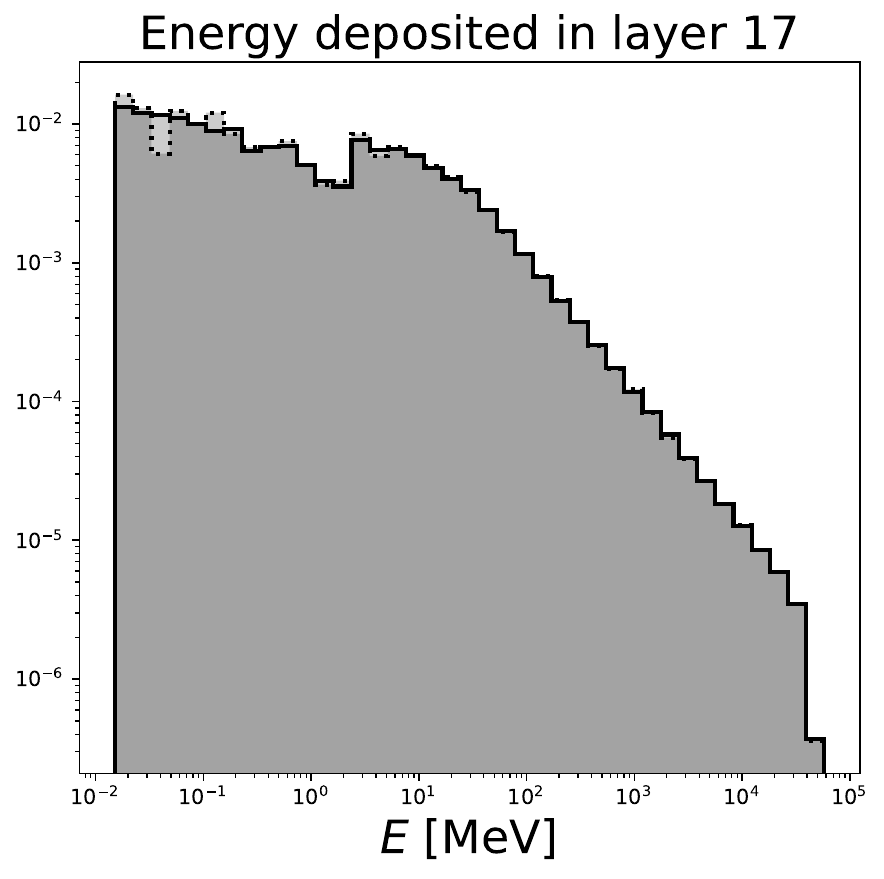} \hfill \includegraphics[height=0.1\textheight]{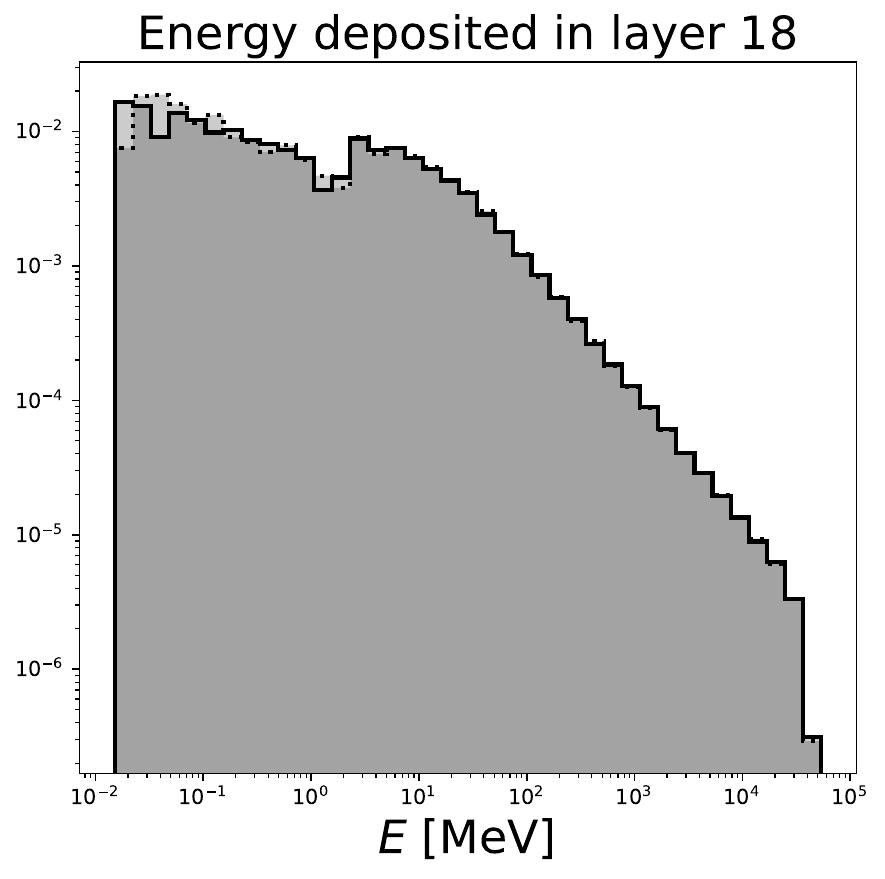} \hfill \includegraphics[height=0.1\textheight]{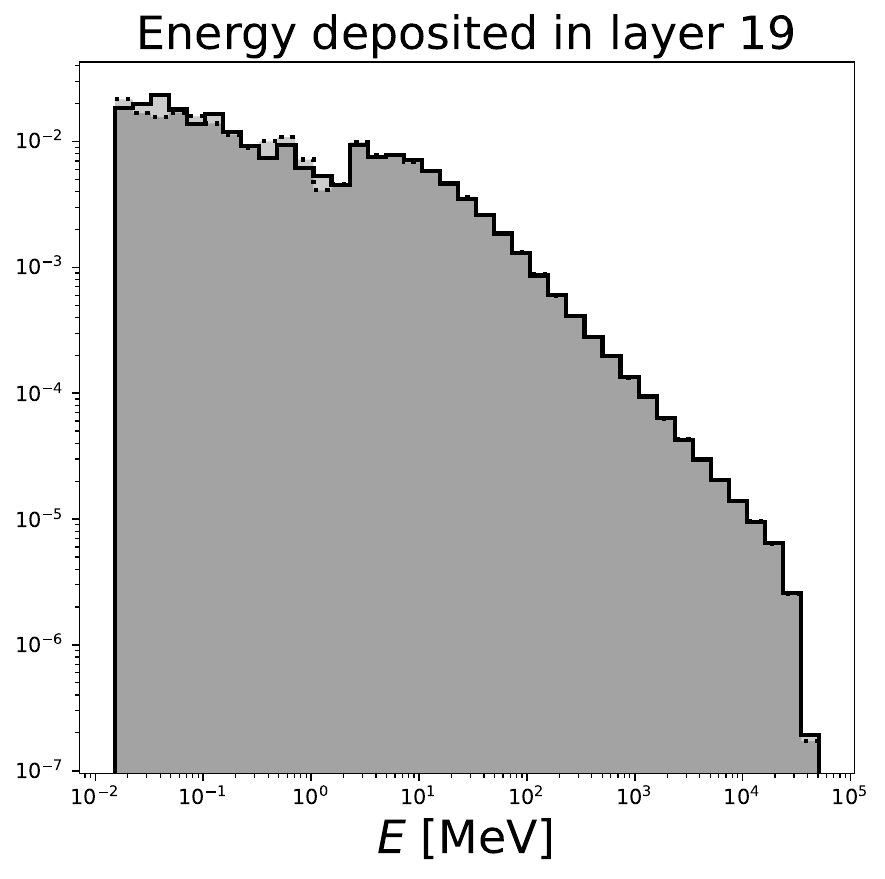}\\
    \includegraphics[height=0.1\textheight]{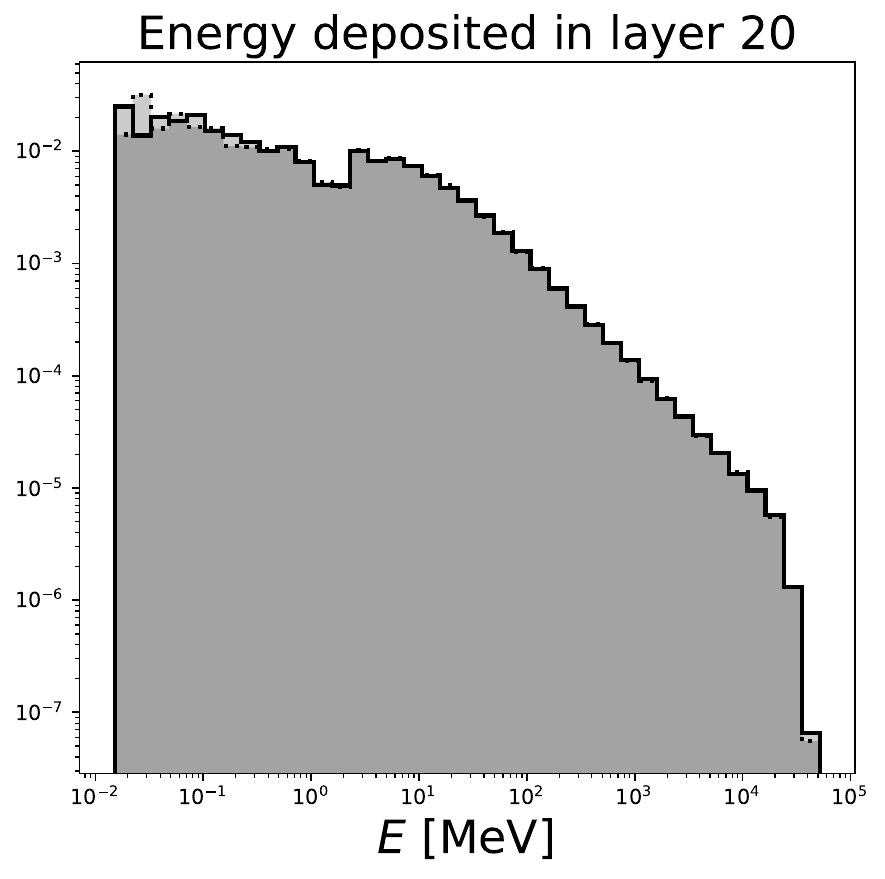} \hfill \includegraphics[height=0.1\textheight]{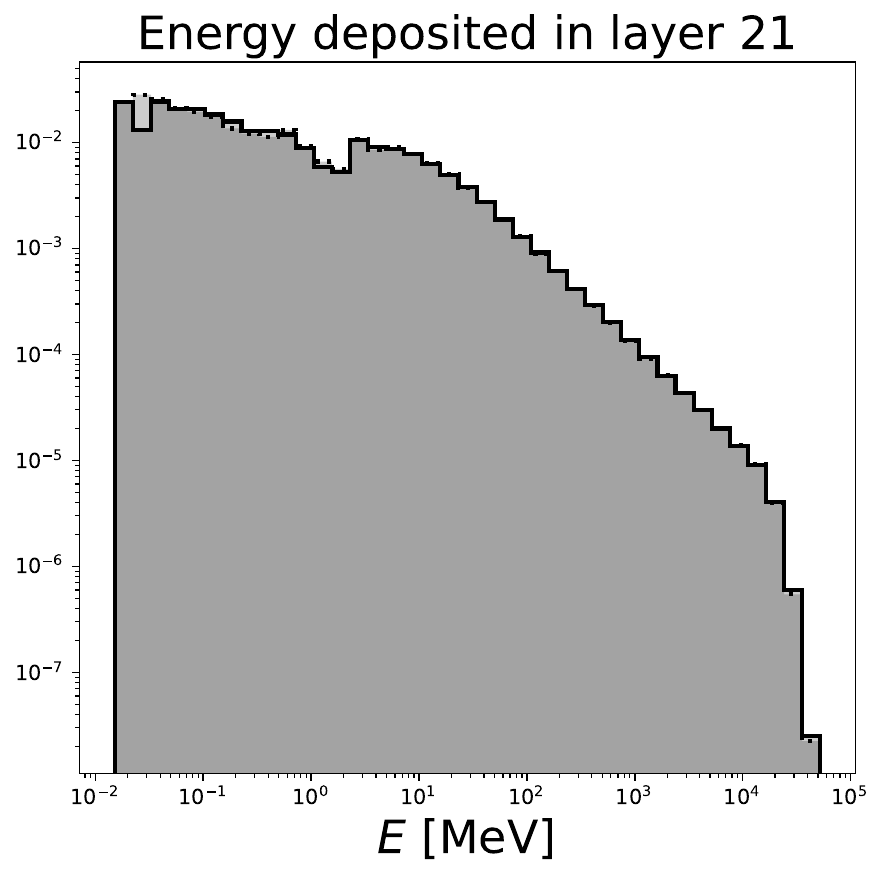} \hfill \includegraphics[height=0.1\textheight]{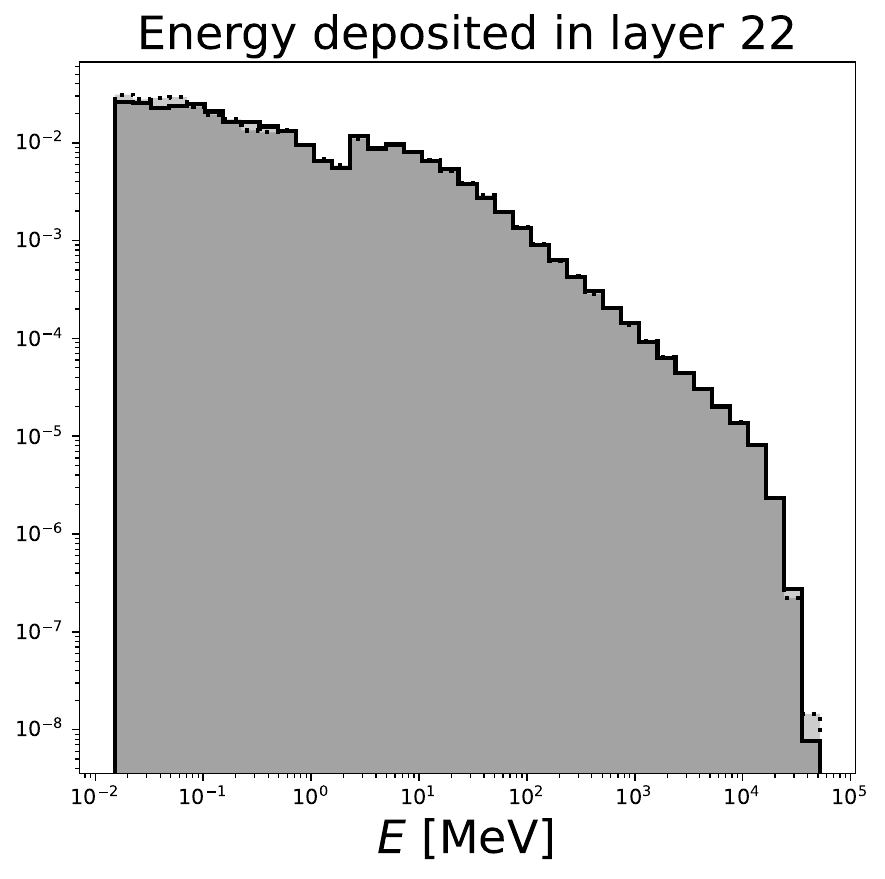} \hfill \includegraphics[height=0.1\textheight]{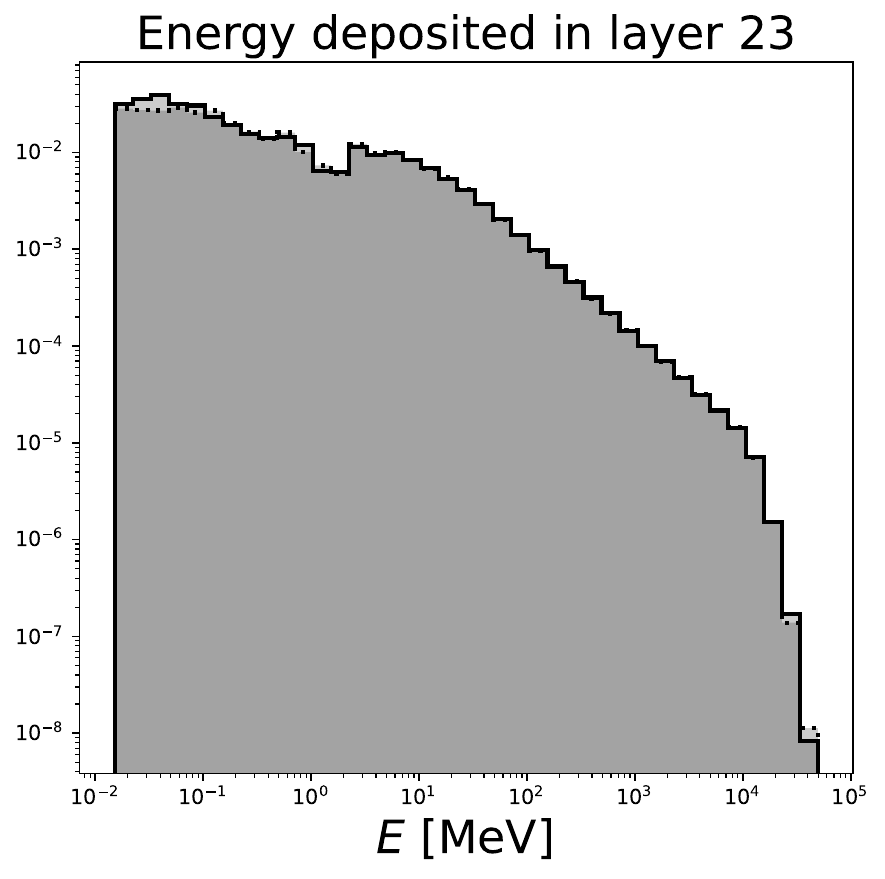} \hfill \includegraphics[height=0.1\textheight]{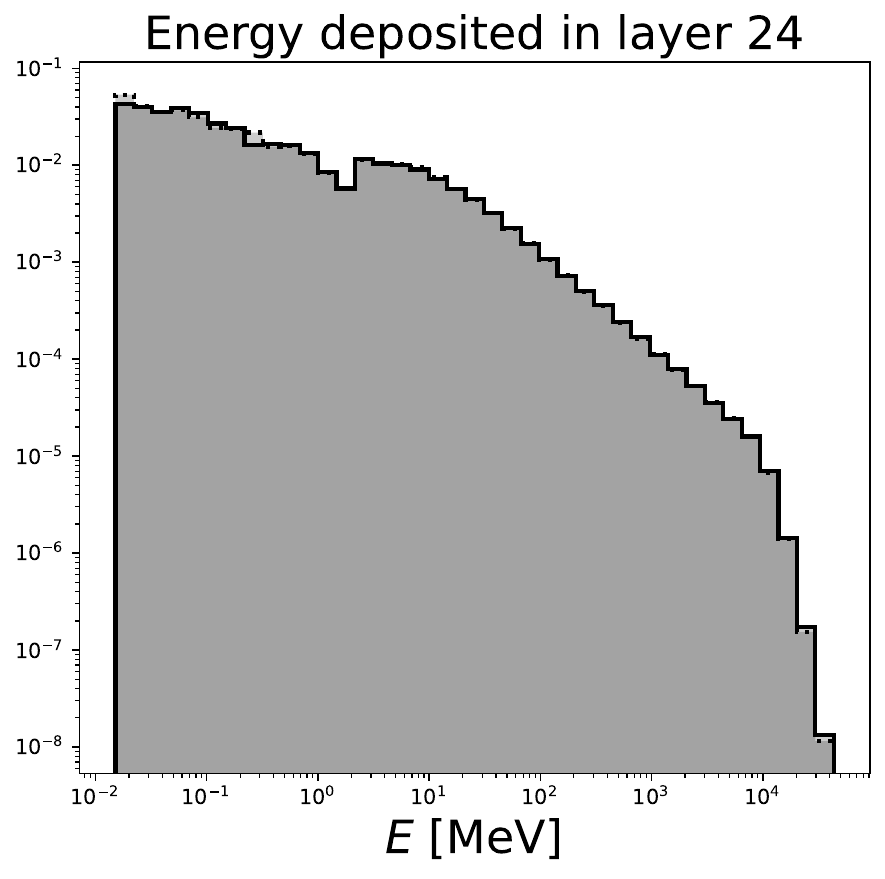}\\
    \includegraphics[height=0.1\textheight]{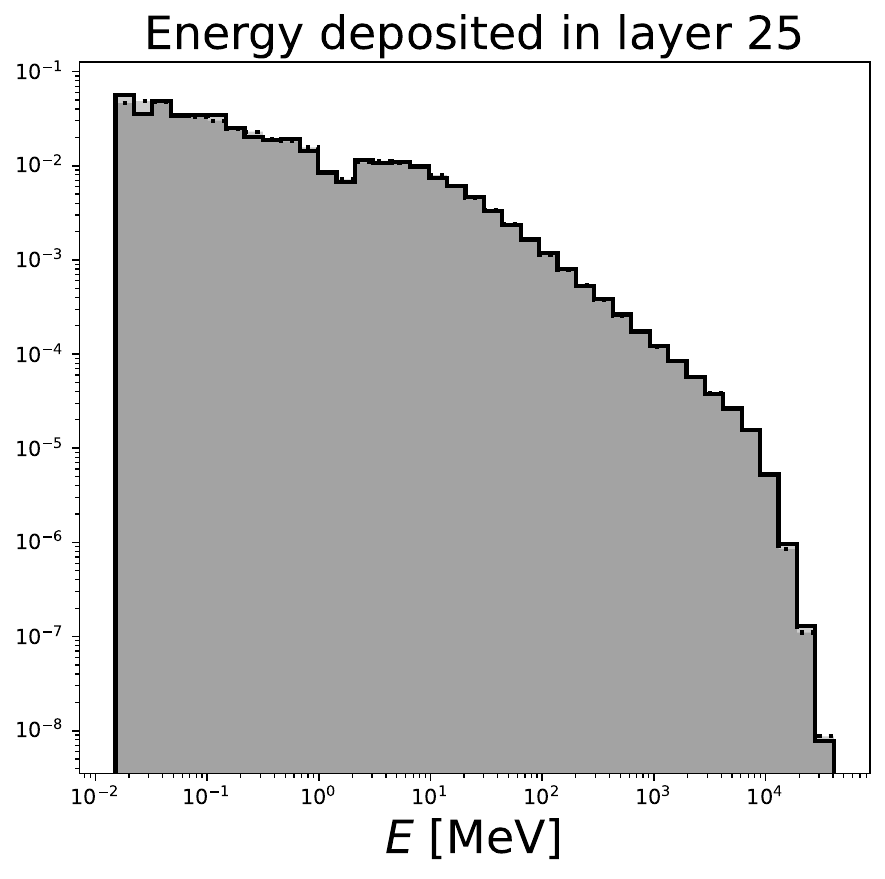} \hfill \includegraphics[height=0.1\textheight]{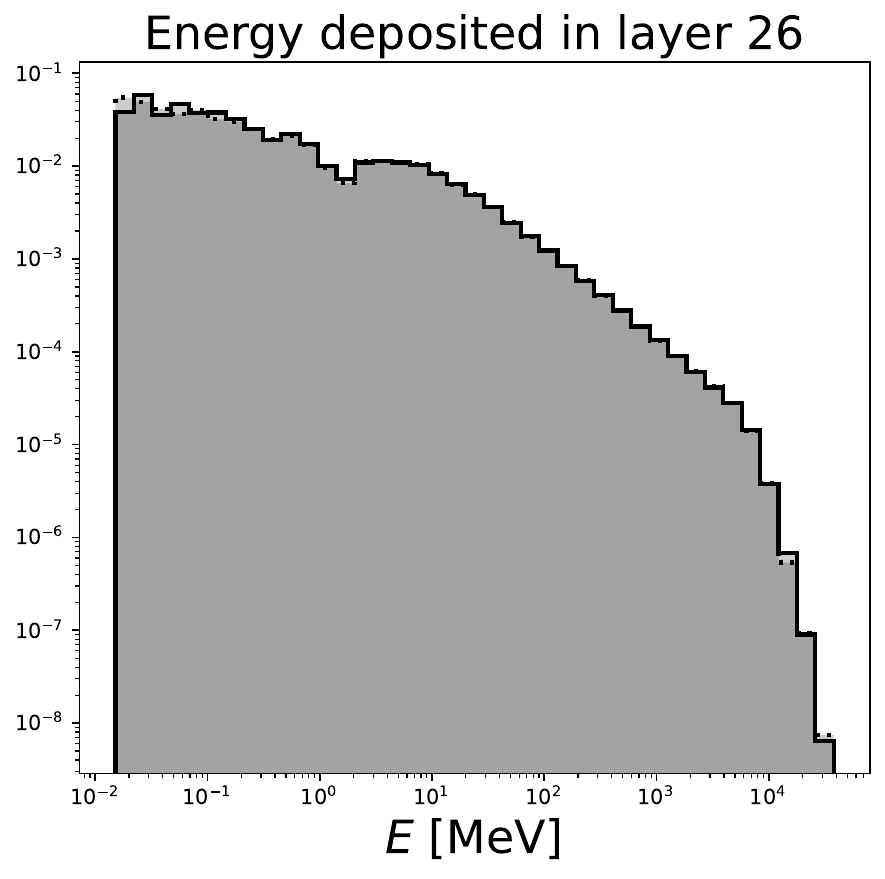} \hfill \includegraphics[height=0.1\textheight]{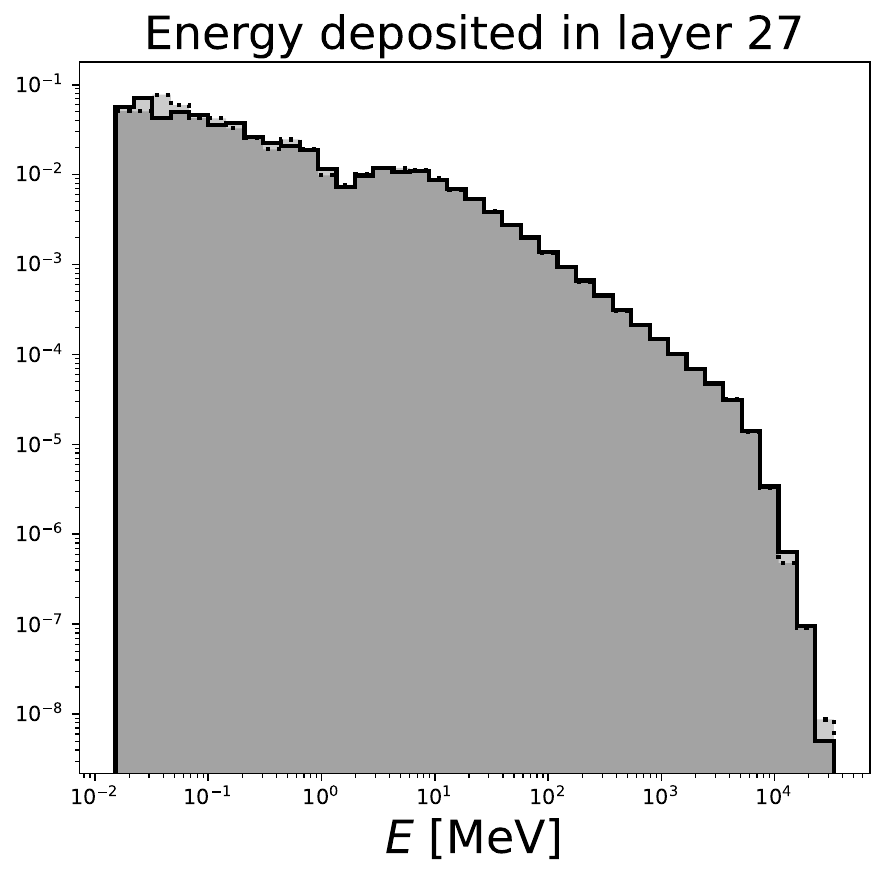} \hfill \includegraphics[height=0.1\textheight]{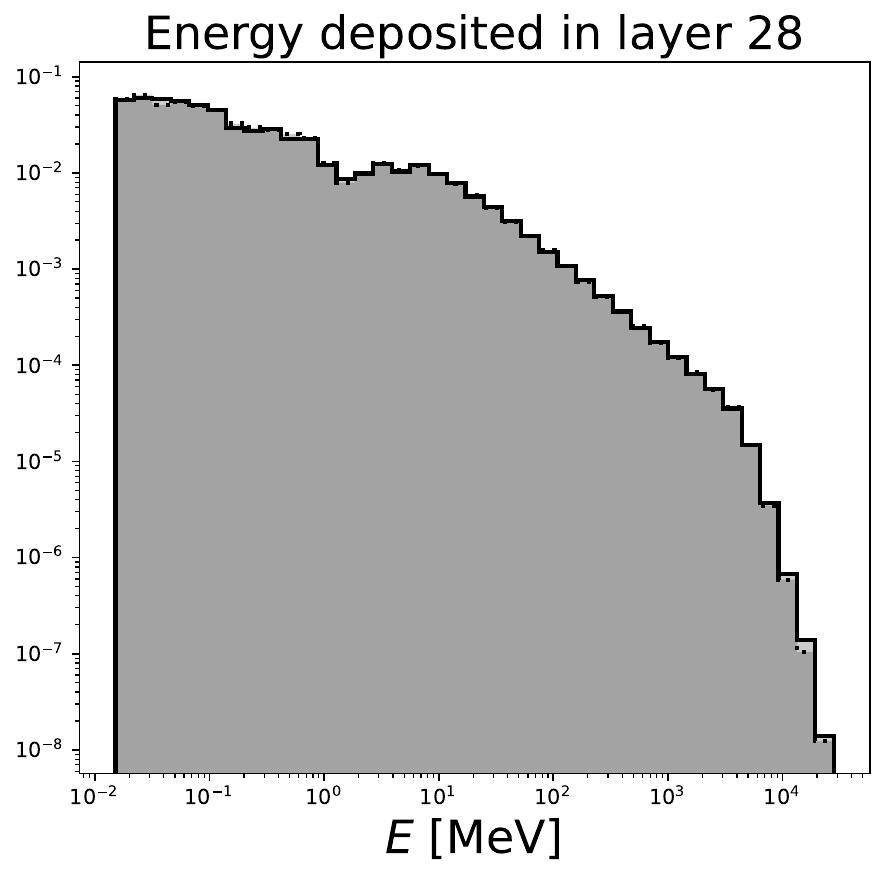} \hfill \includegraphics[height=0.1\textheight]{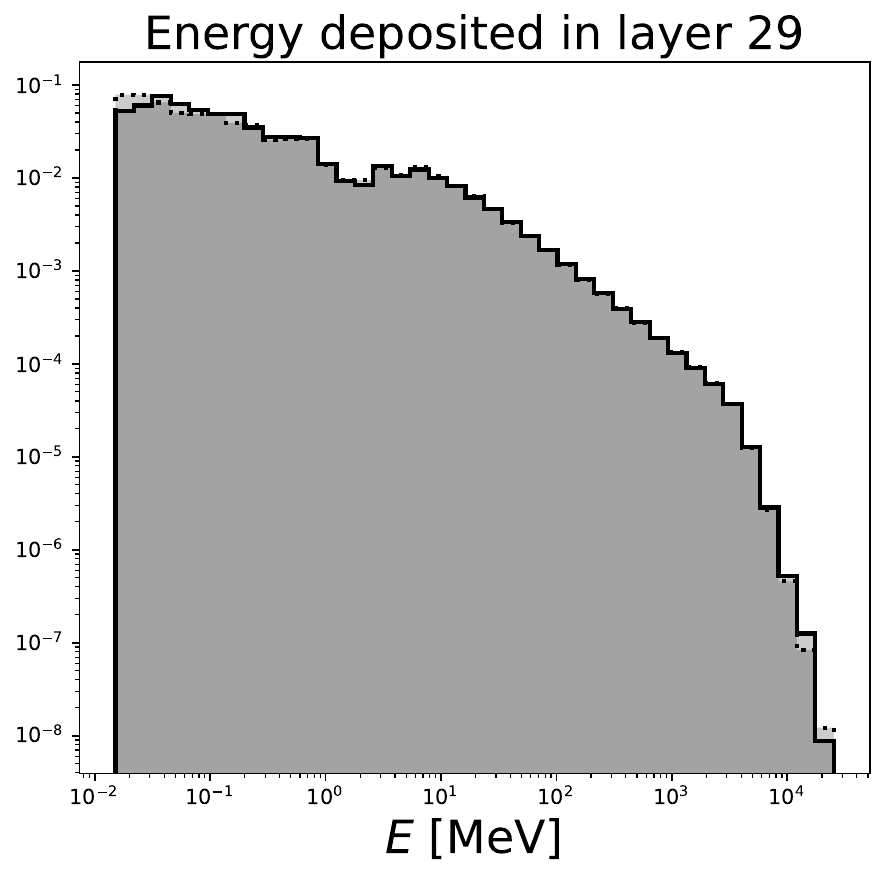}\\
    \includegraphics[height=0.1\textheight]{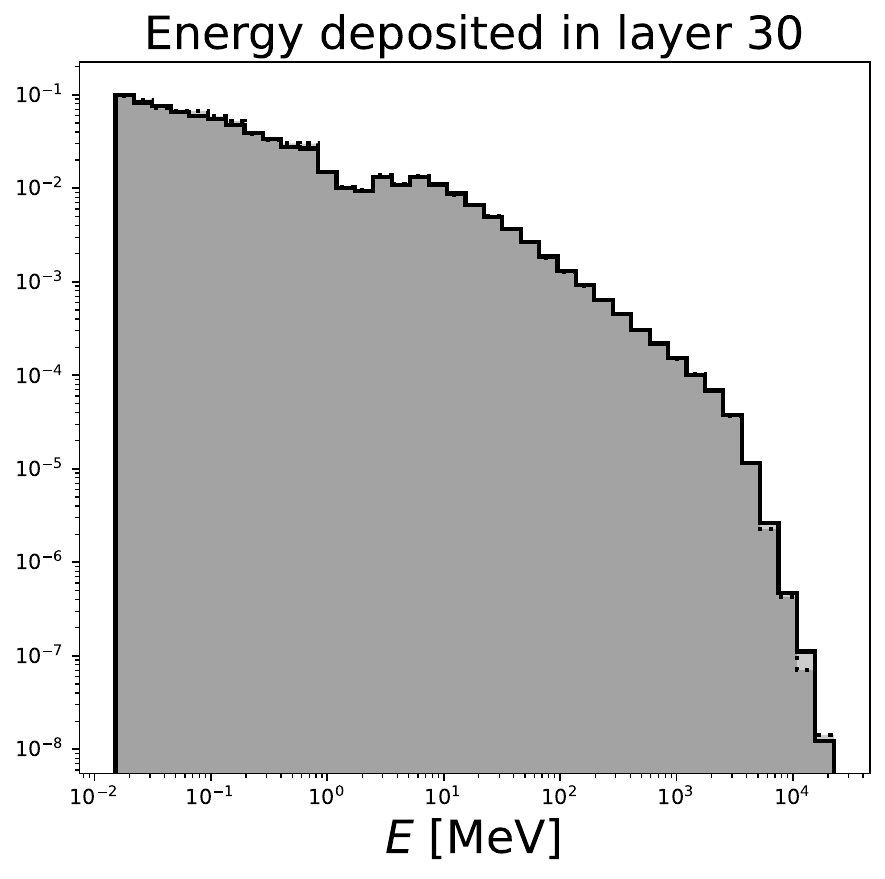} \hfill \includegraphics[height=0.1\textheight]{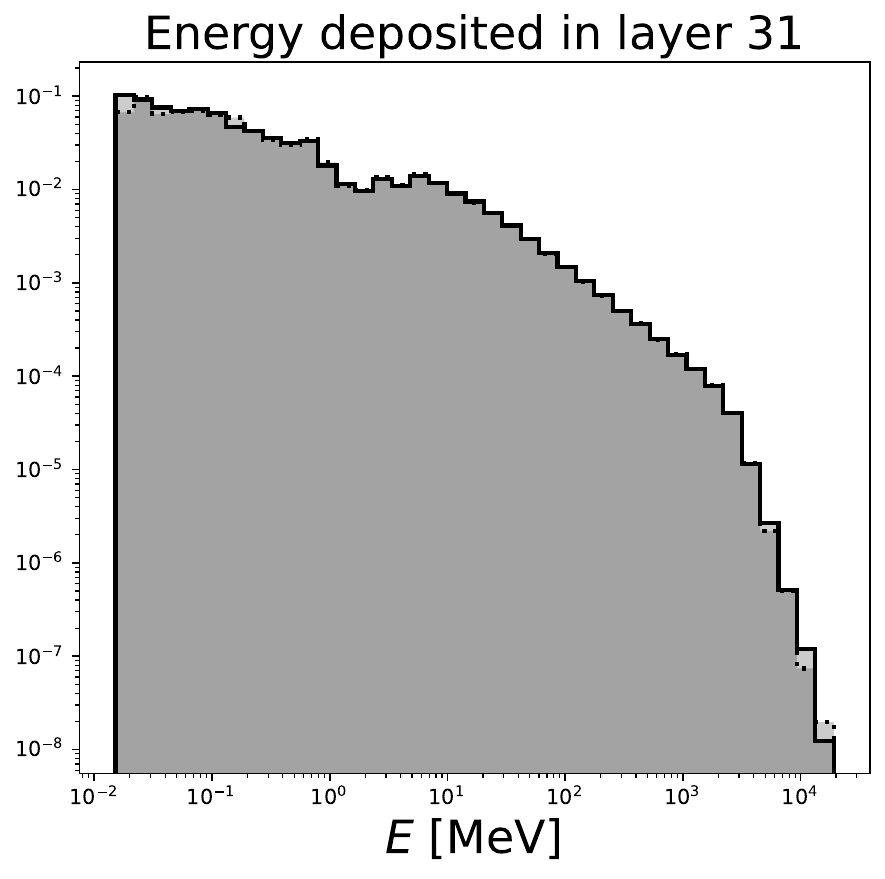} \hfill \includegraphics[height=0.1\textheight]{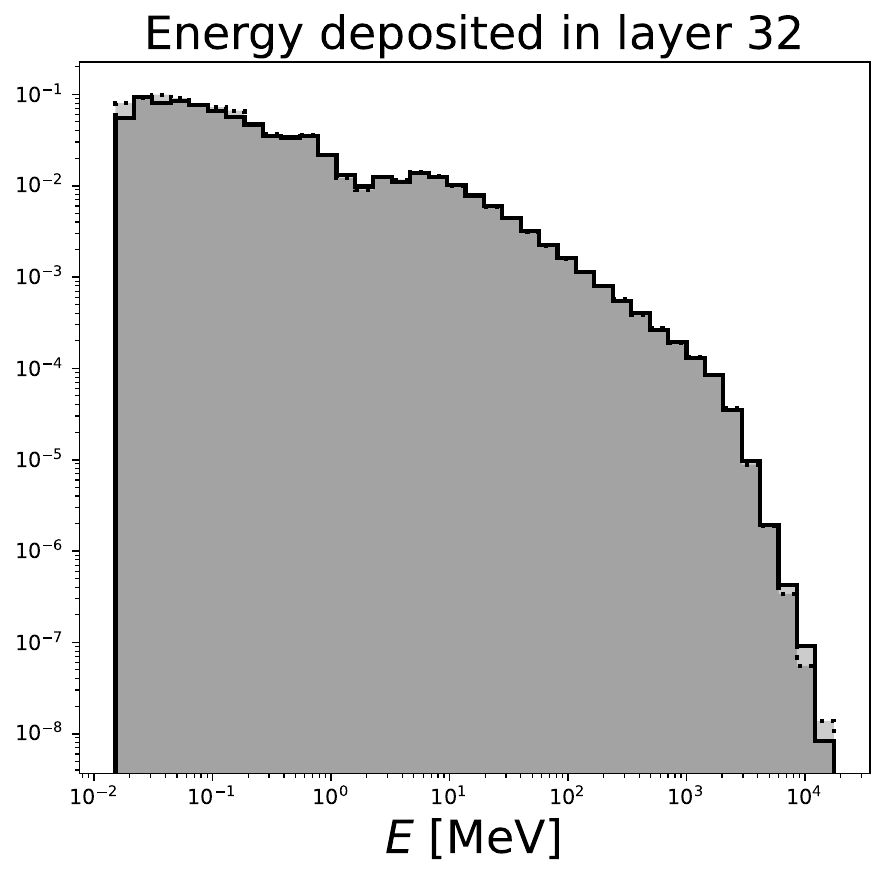} \hfill \includegraphics[height=0.1\textheight]{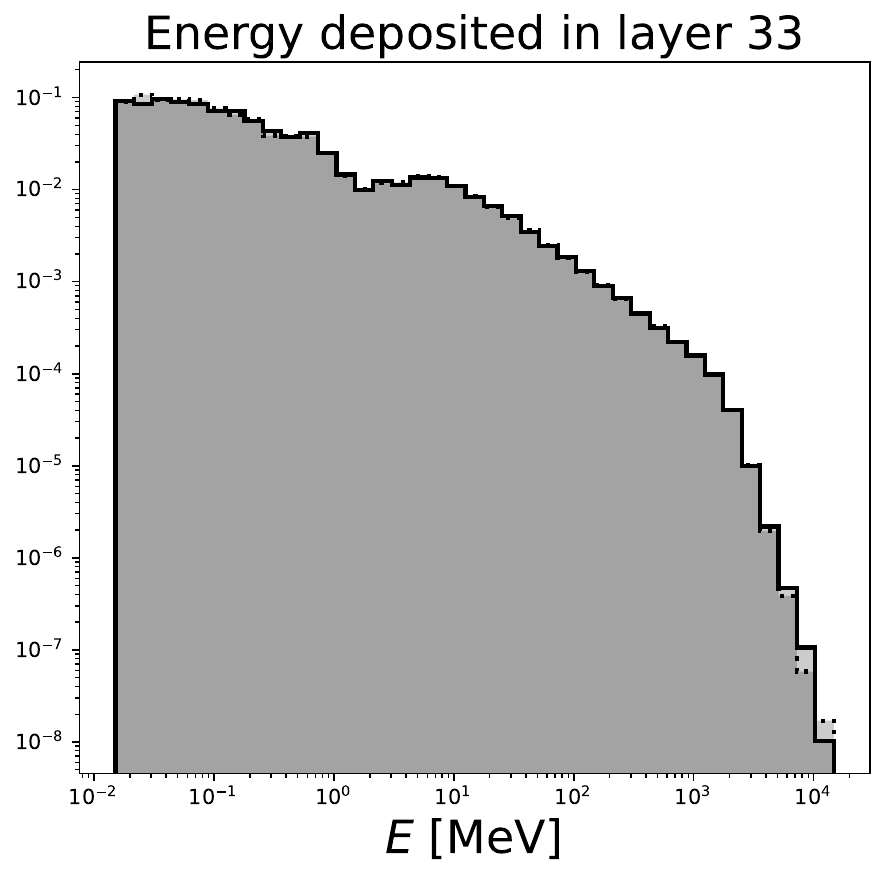} \hfill \includegraphics[height=0.1\textheight]{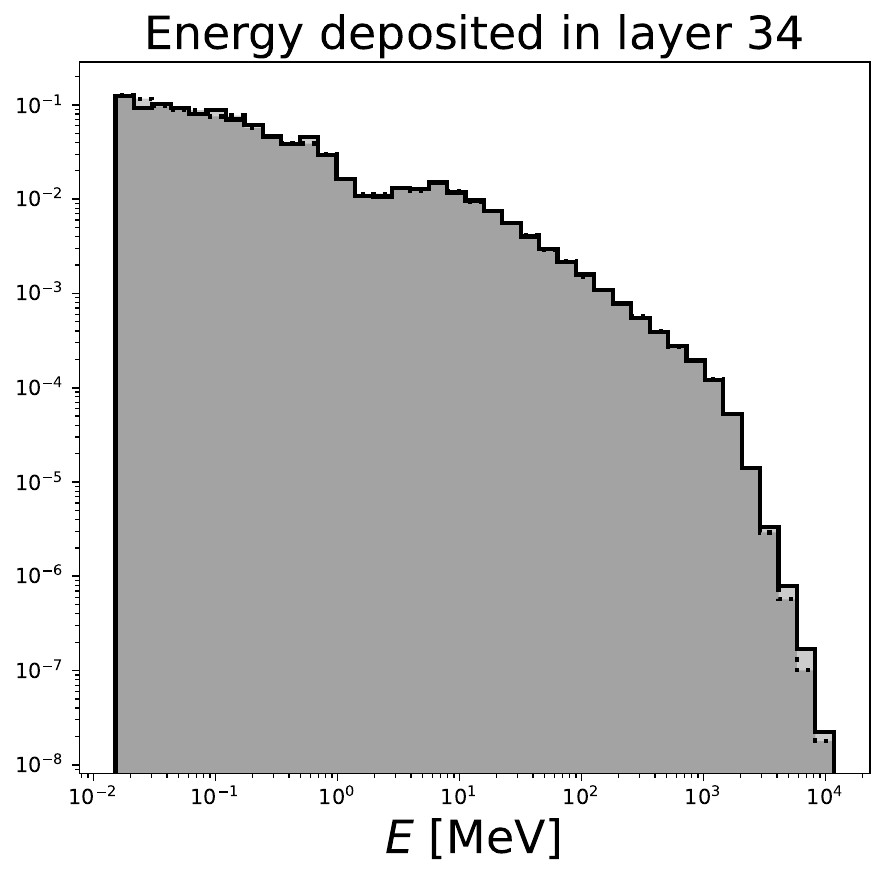}\\
    \includegraphics[height=0.1\textheight]{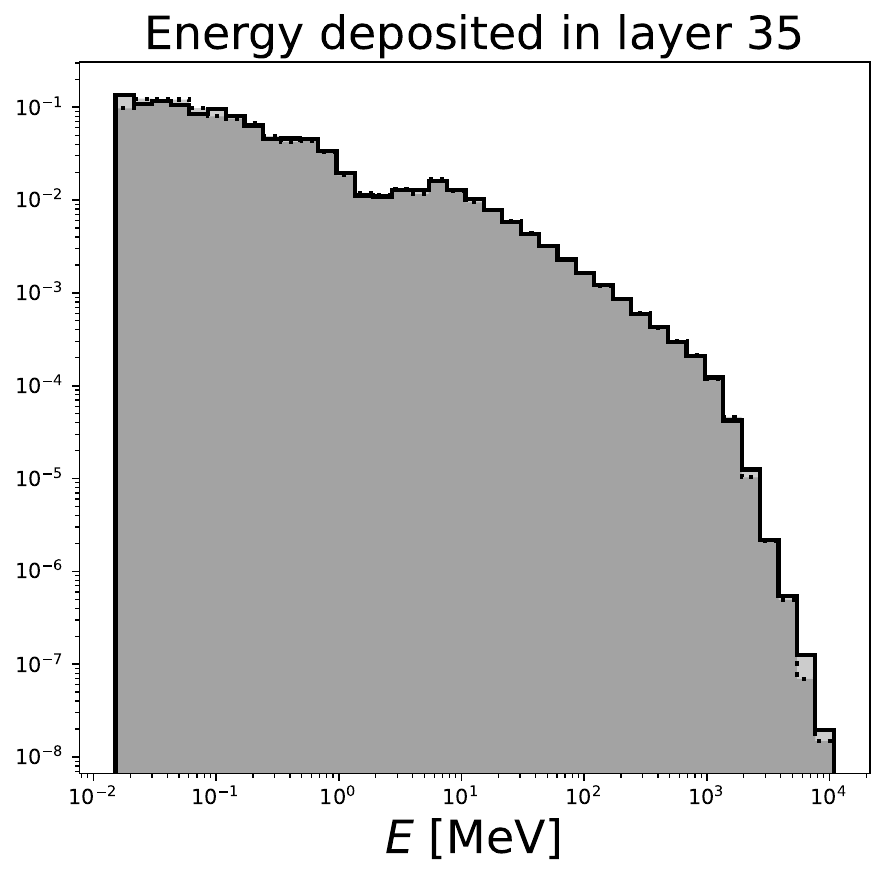} \hfill \includegraphics[height=0.1\textheight]{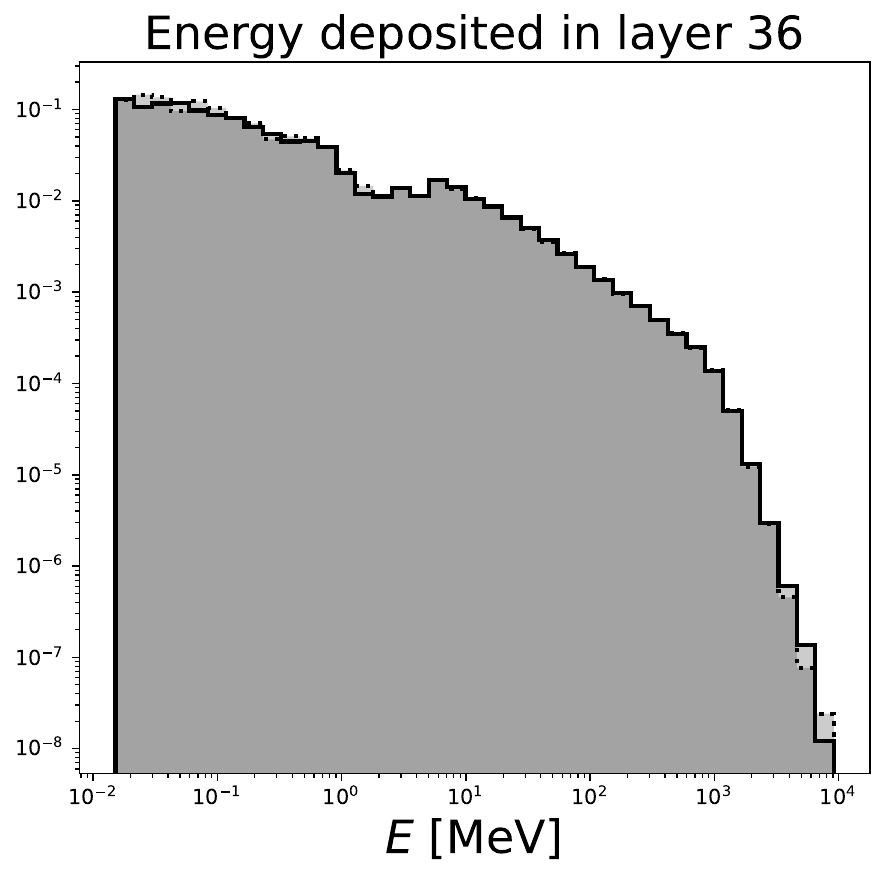} \hfill \includegraphics[height=0.1\textheight]{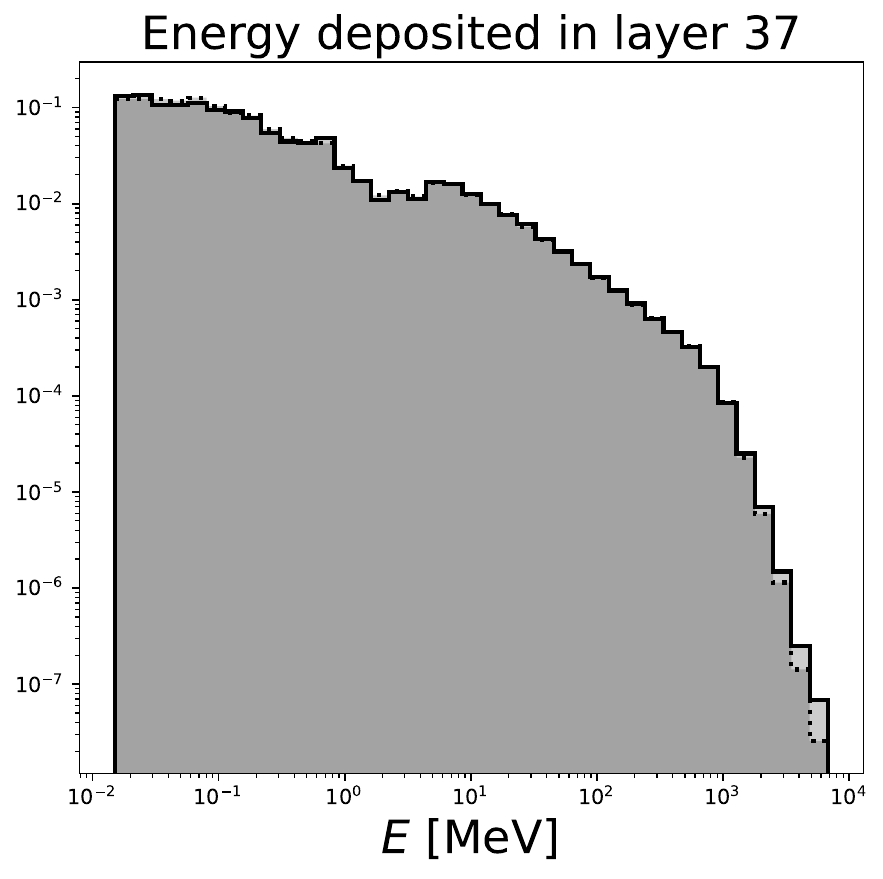} \hfill \includegraphics[height=0.1\textheight]{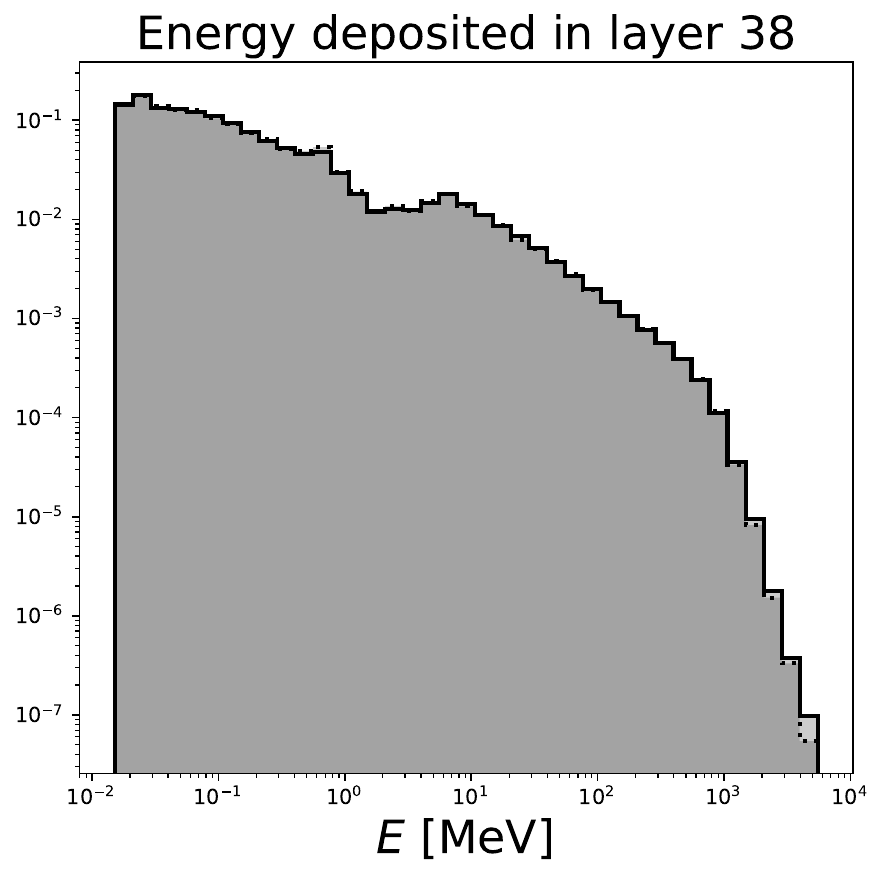} \hfill \includegraphics[height=0.1\textheight]{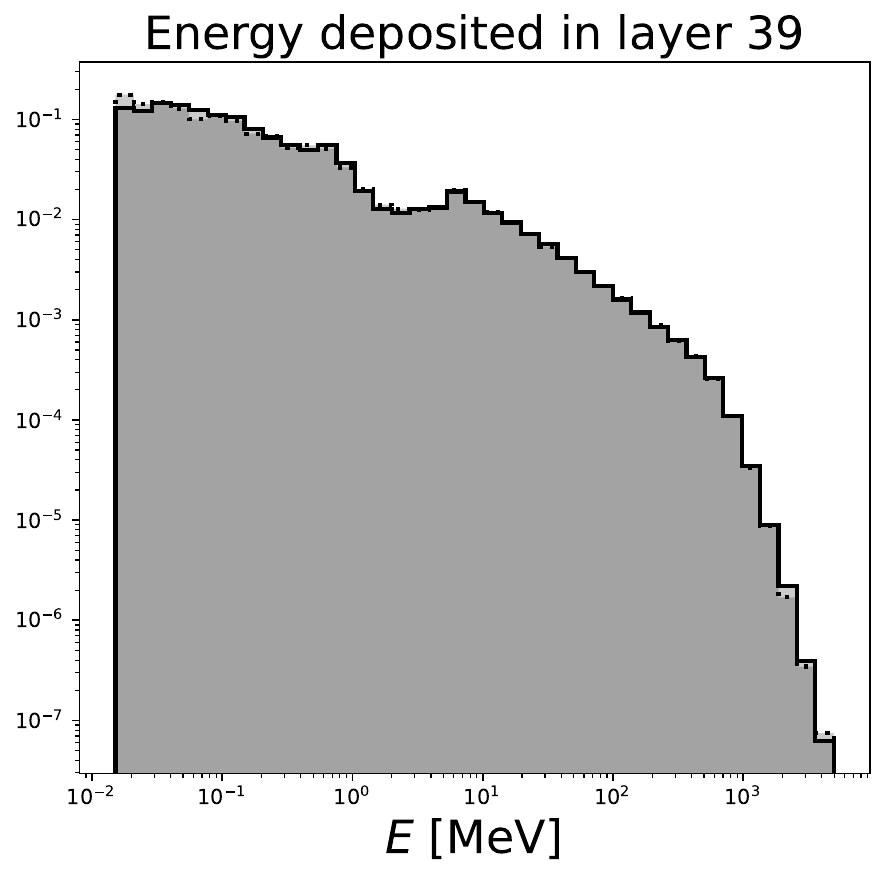}\\
    \includegraphics[height=0.1\textheight]{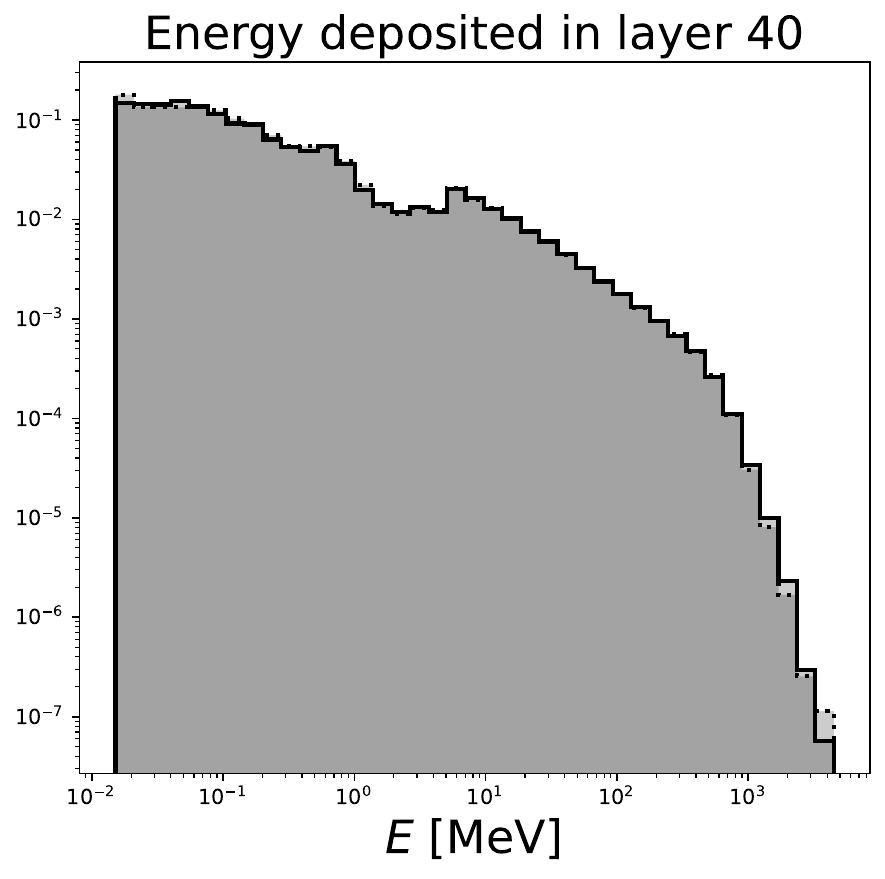} \hfill \includegraphics[height=0.1\textheight]{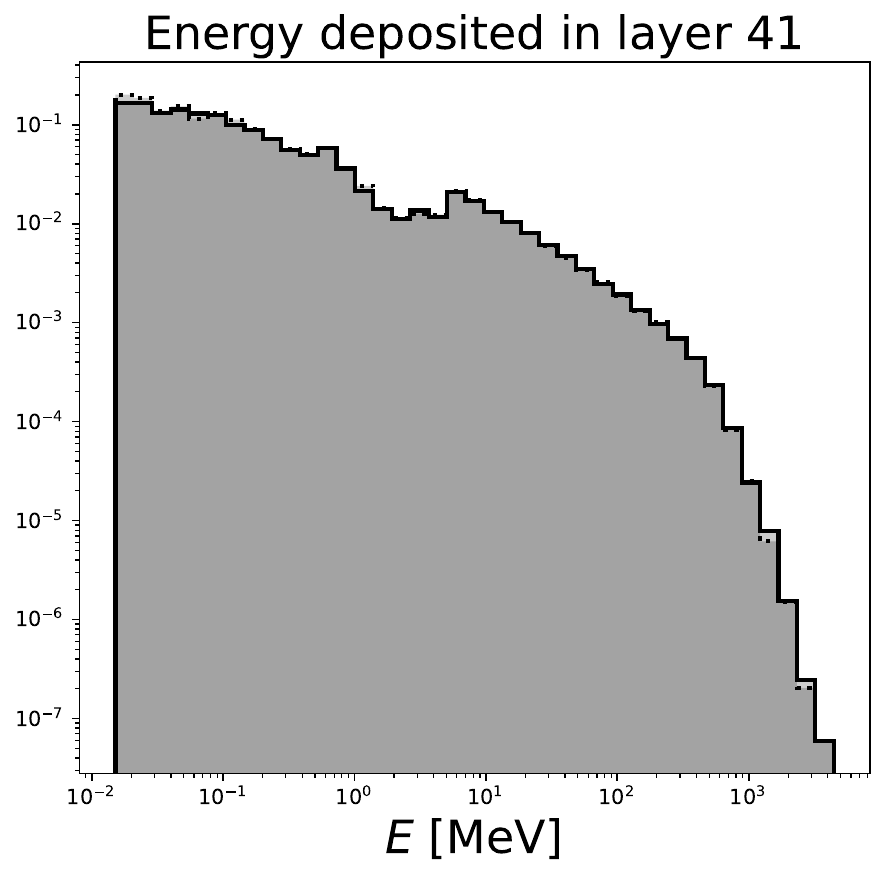} \hfill \includegraphics[height=0.1\textheight]{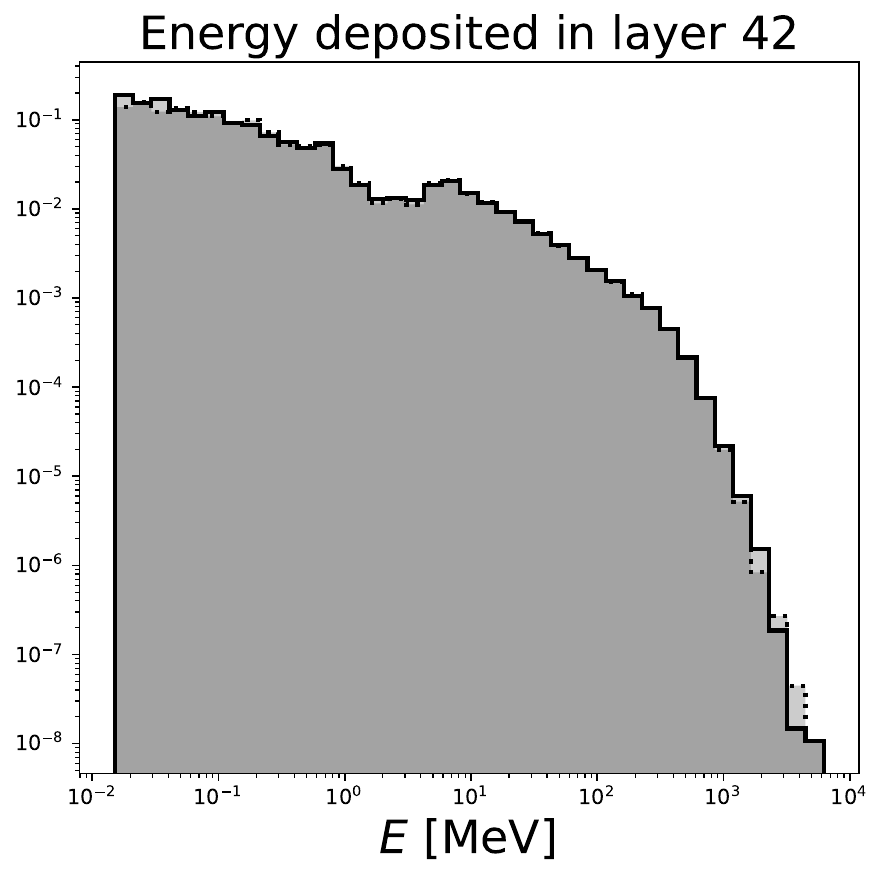} \hfill \includegraphics[height=0.1\textheight]{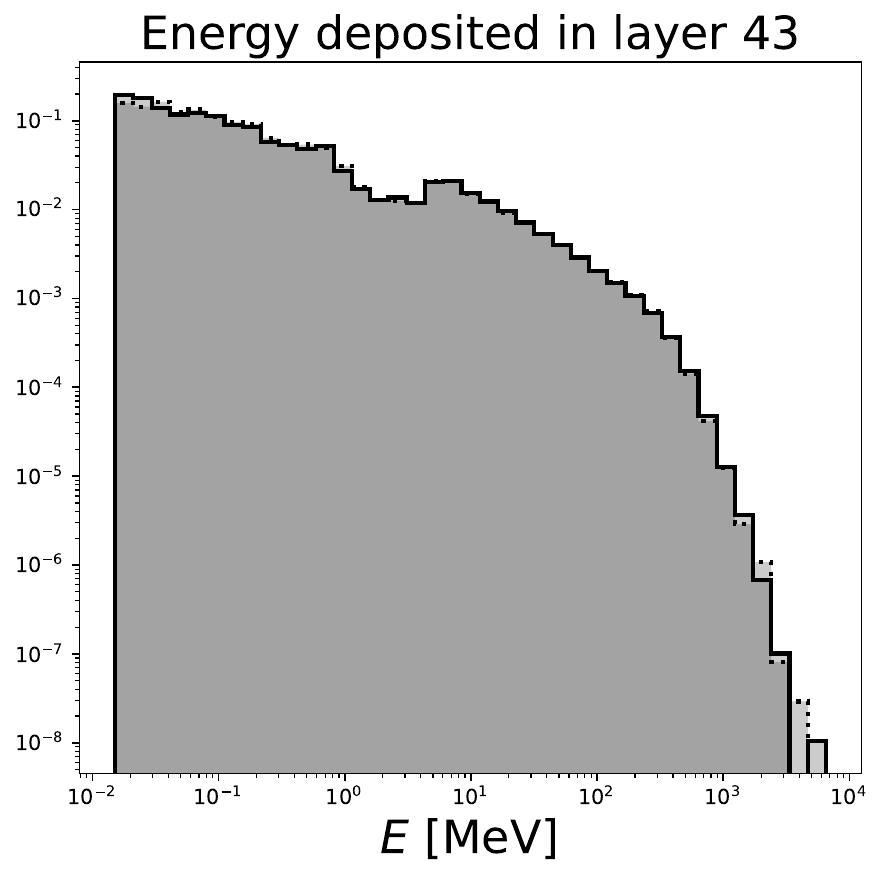} \hfill \includegraphics[height=0.1\textheight]{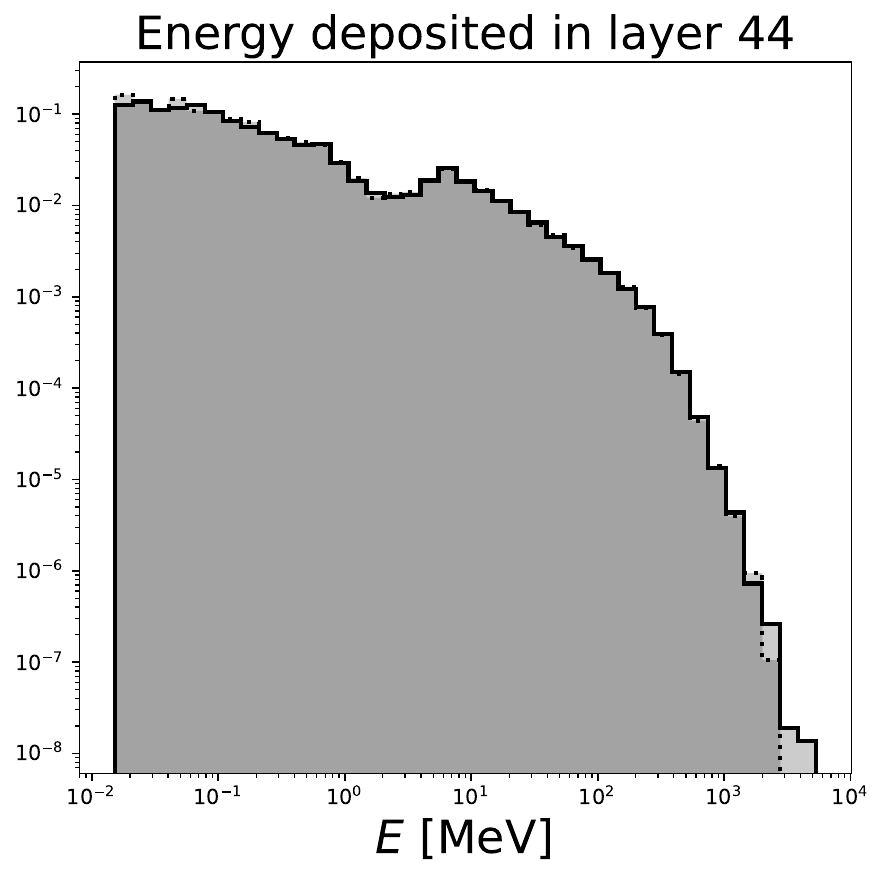}\\
    \includegraphics[width=0.5\textwidth]{figures/Appendix_reference/legend.pdf}
    \caption{Distribution of \geant training and evaluation data in layer energies $E_i$ for ds3. }
    \label{fig:app_ref.ds3.2}
\end{figure}

\begin{figure}[ht]
    \centering
    \includegraphics[height=0.1\textheight]{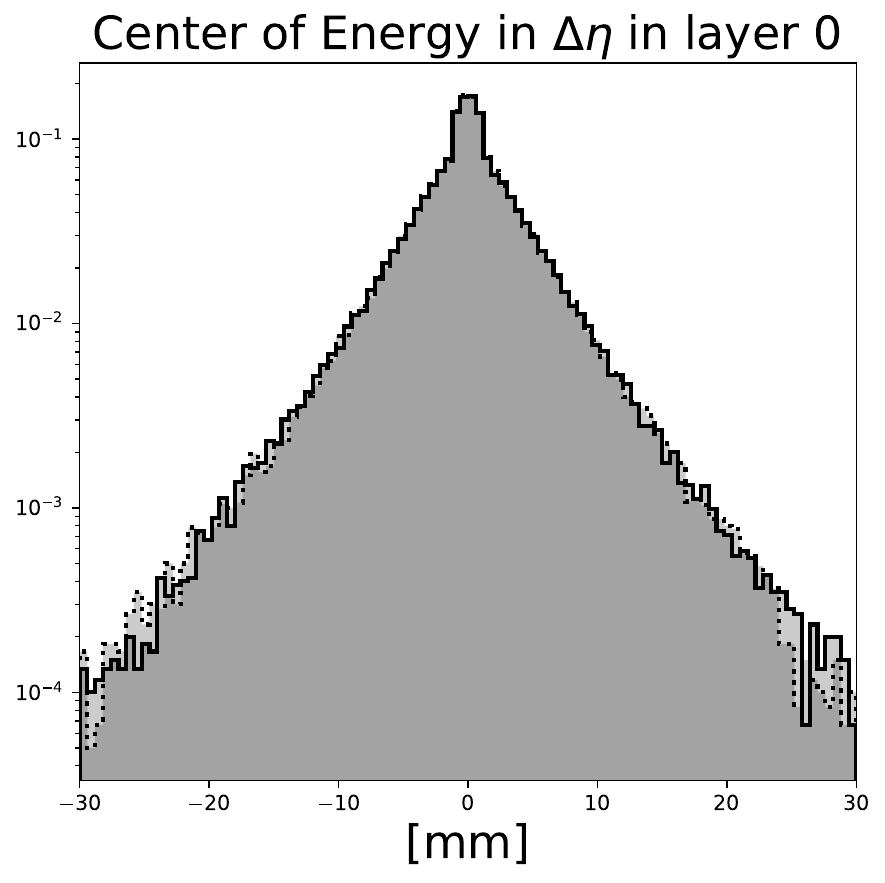} \hfill \includegraphics[height=0.1\textheight]{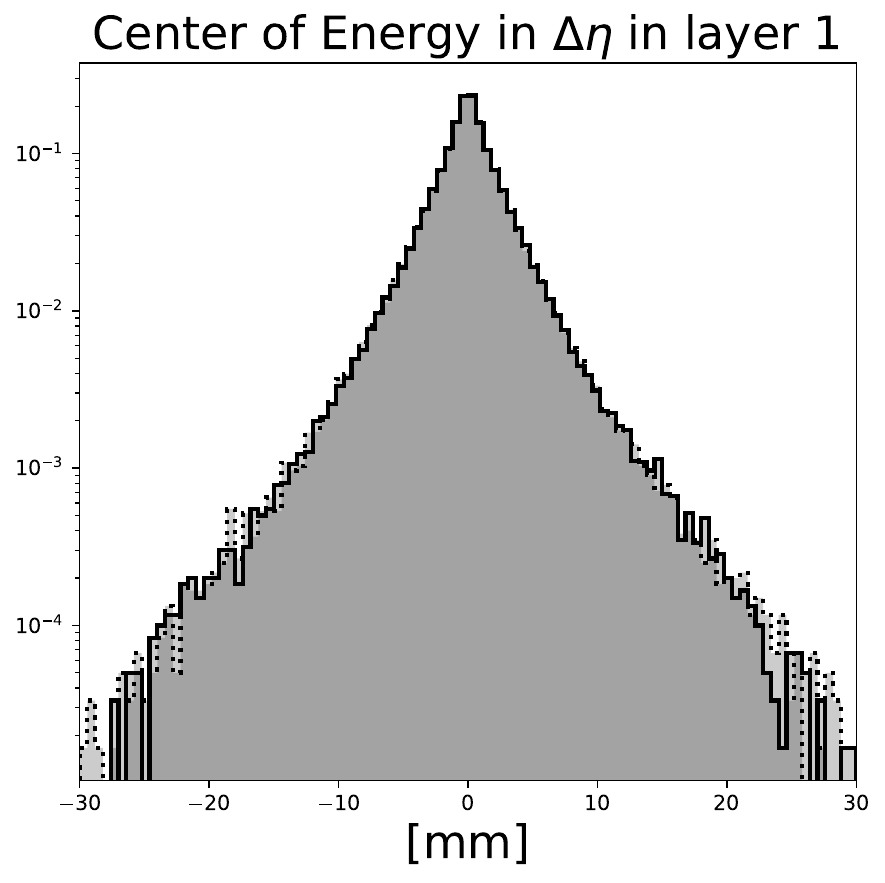} \hfill \includegraphics[height=0.1\textheight]{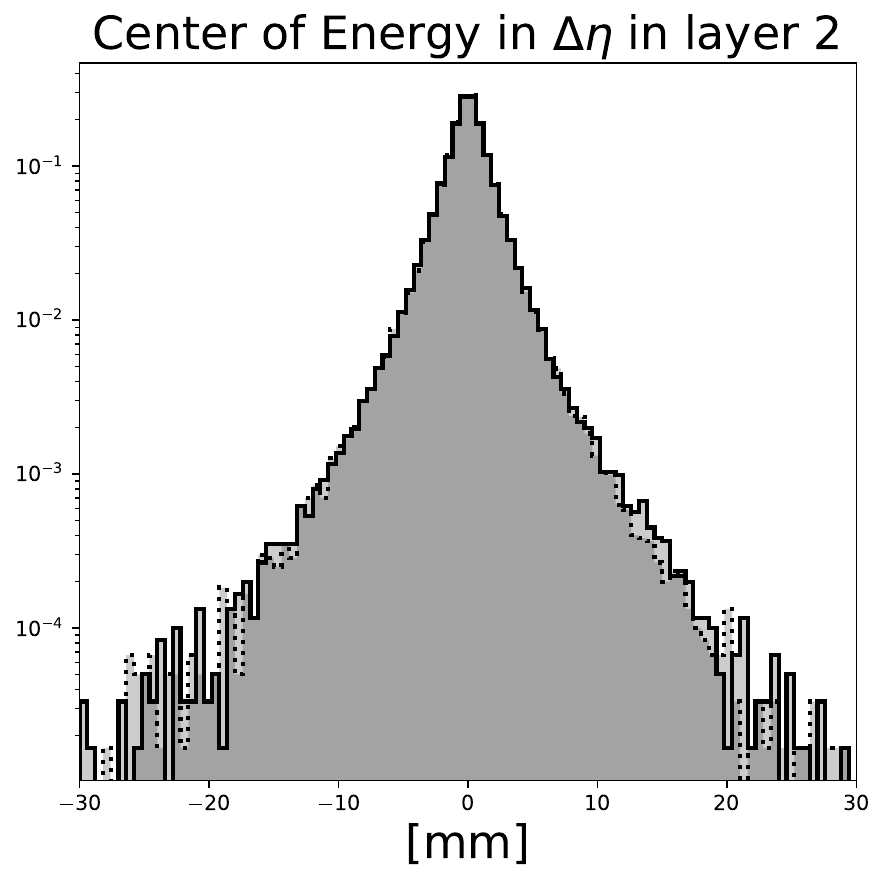} \hfill \includegraphics[height=0.1\textheight]{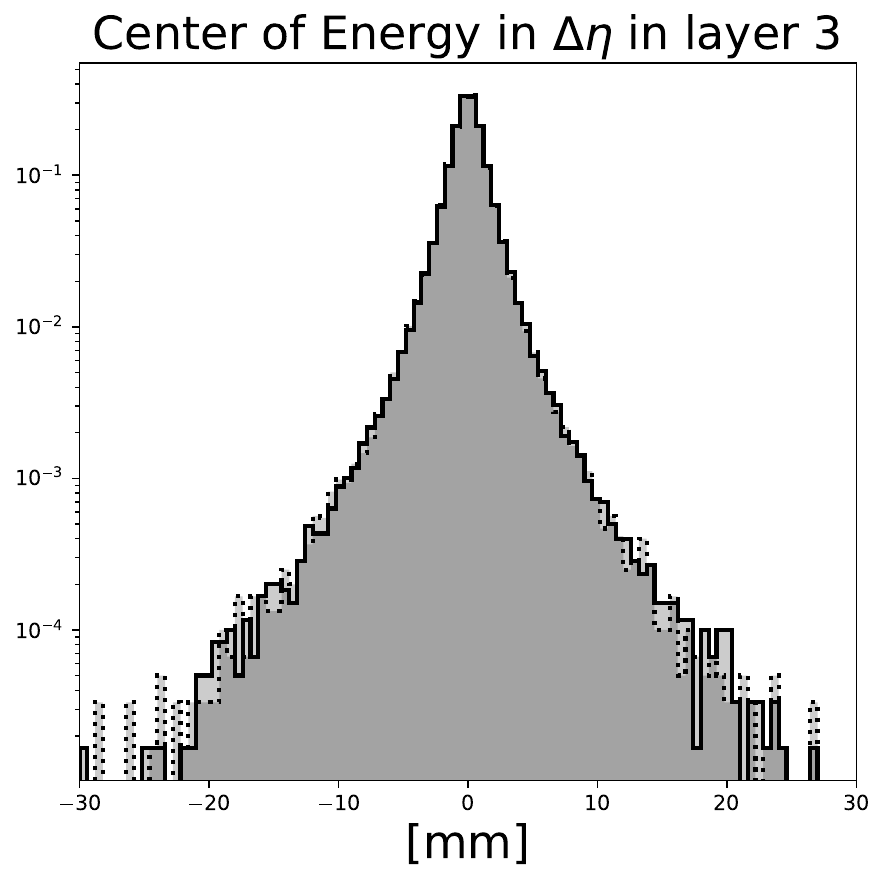} \hfill \includegraphics[height=0.1\textheight]{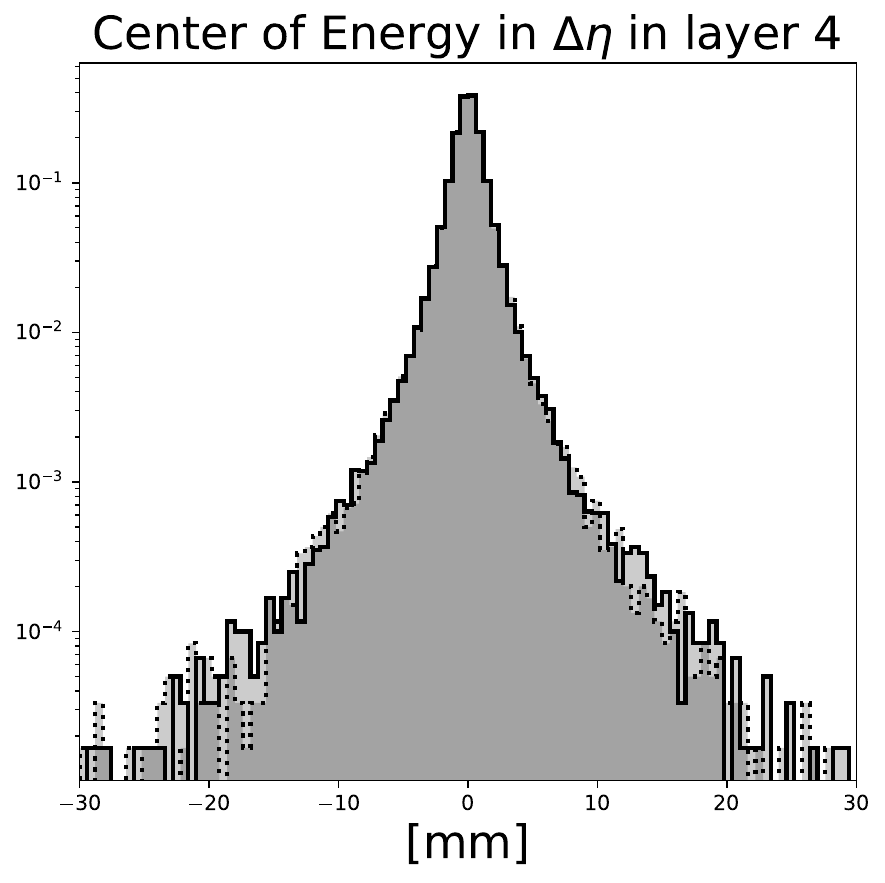}\\
    \includegraphics[height=0.1\textheight]{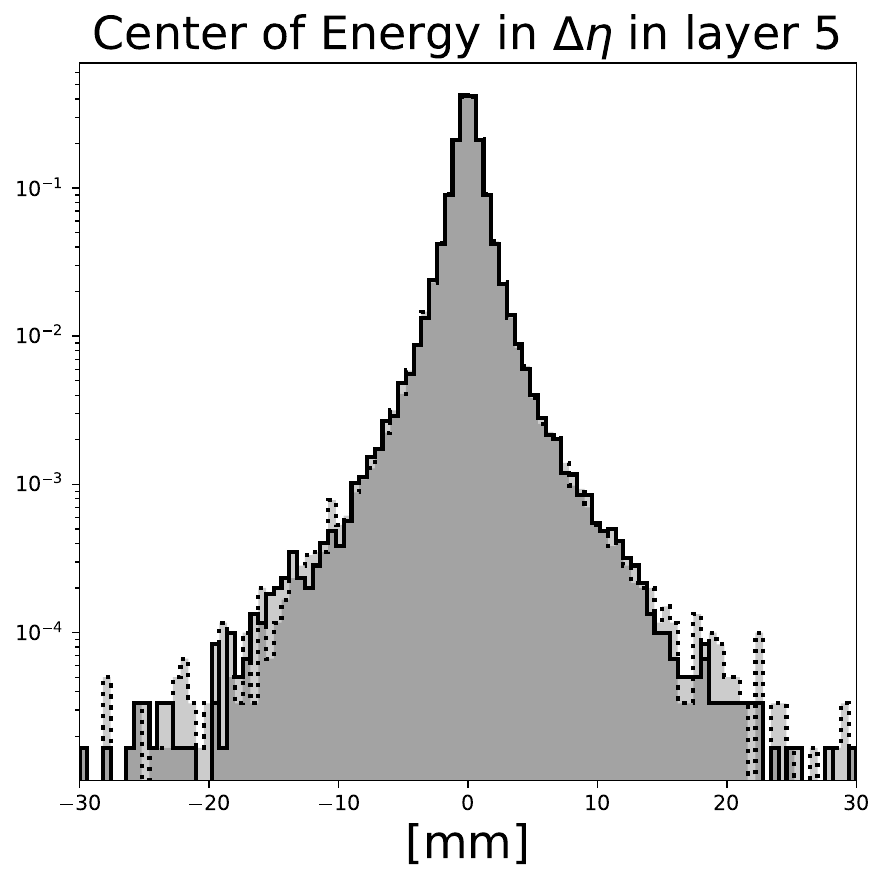} \hfill \includegraphics[height=0.1\textheight]{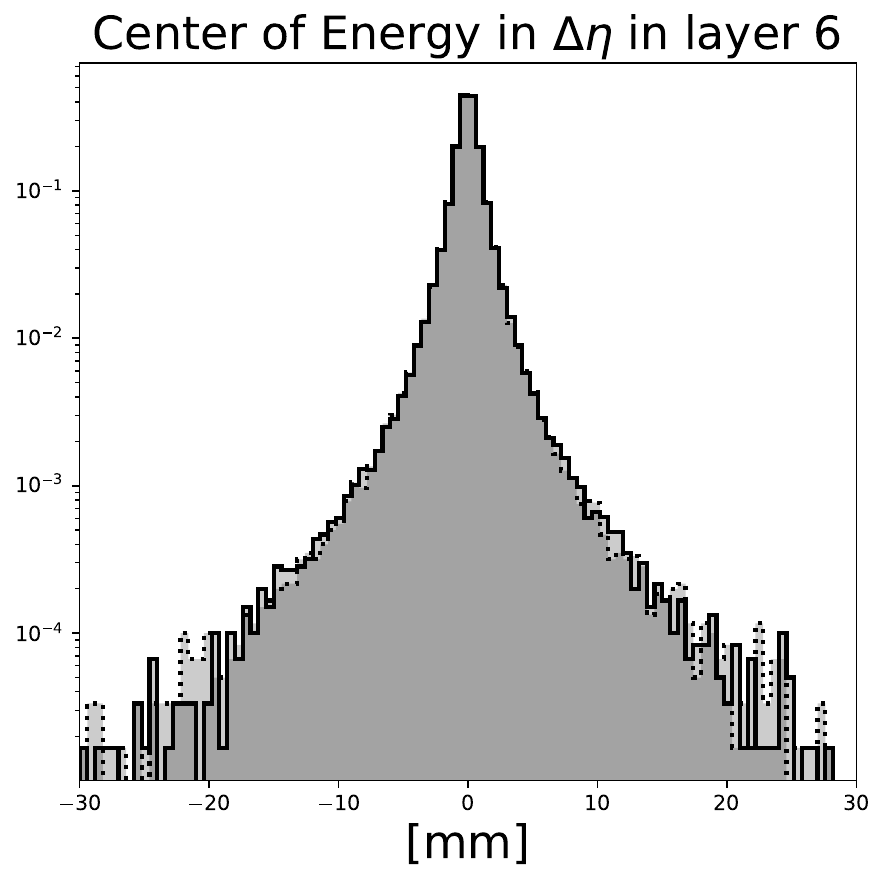} \hfill \includegraphics[height=0.1\textheight]{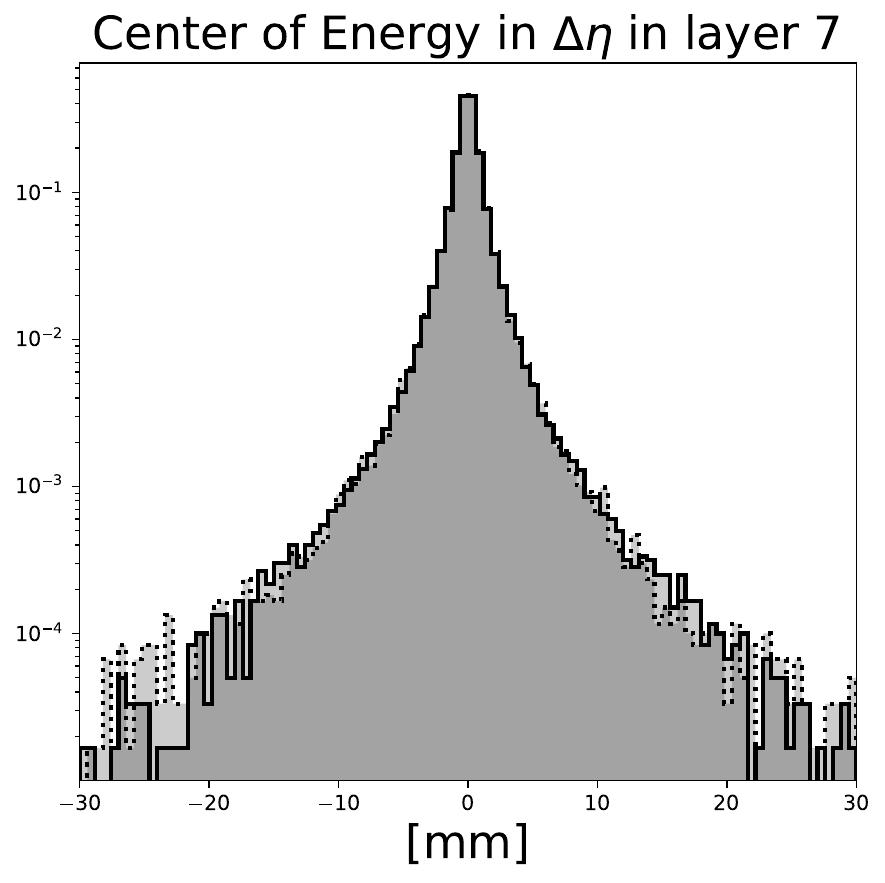} \hfill \includegraphics[height=0.1\textheight]{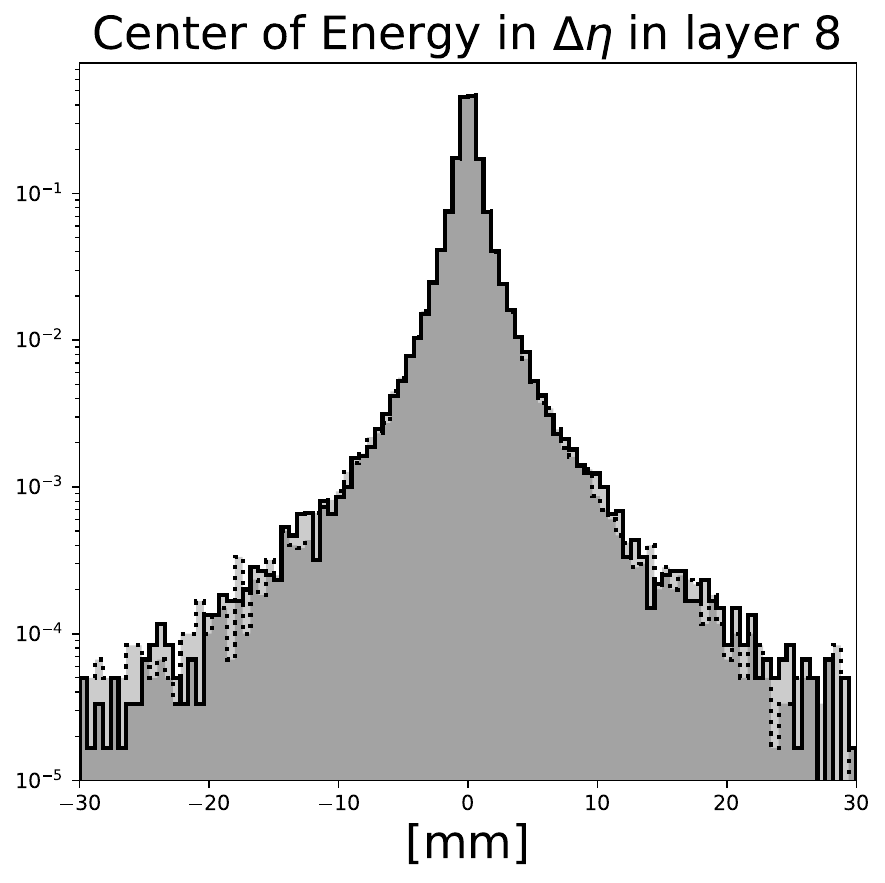} \hfill \includegraphics[height=0.1\textheight]{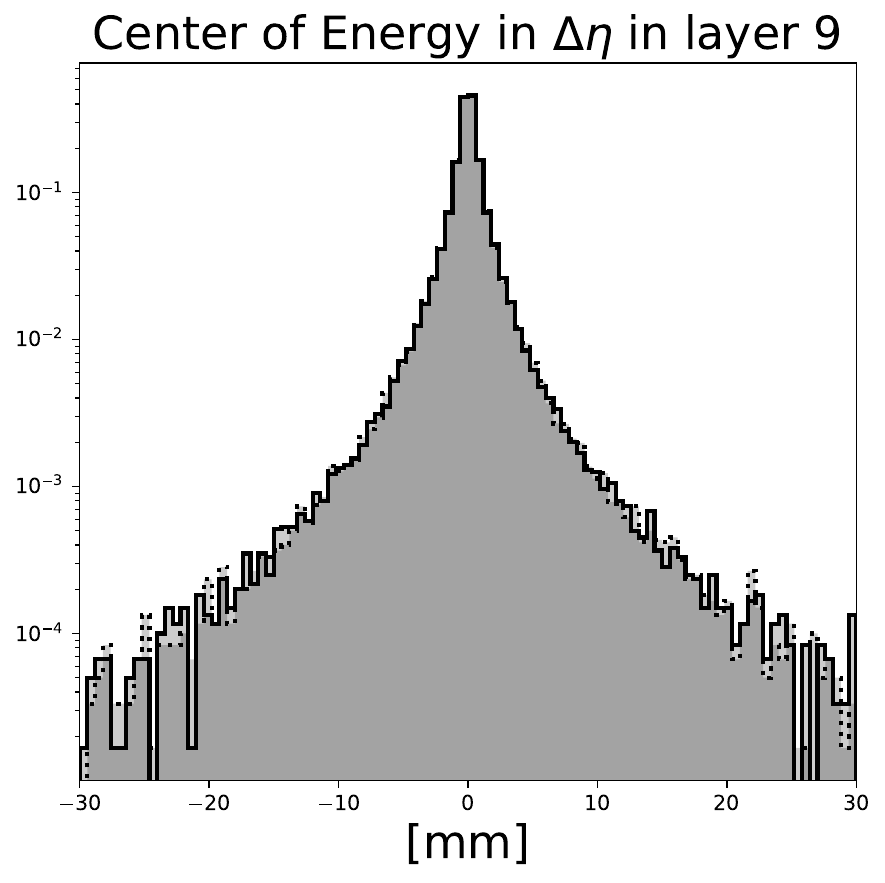}\\
    \includegraphics[height=0.1\textheight]{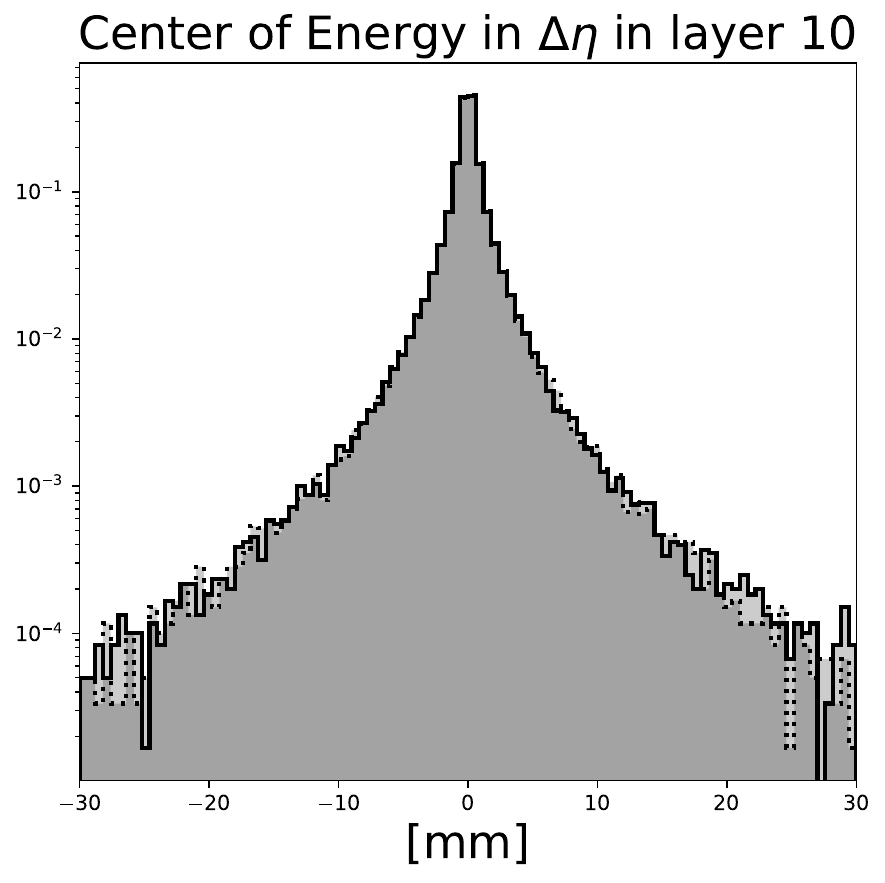} \hfill \includegraphics[height=0.1\textheight]{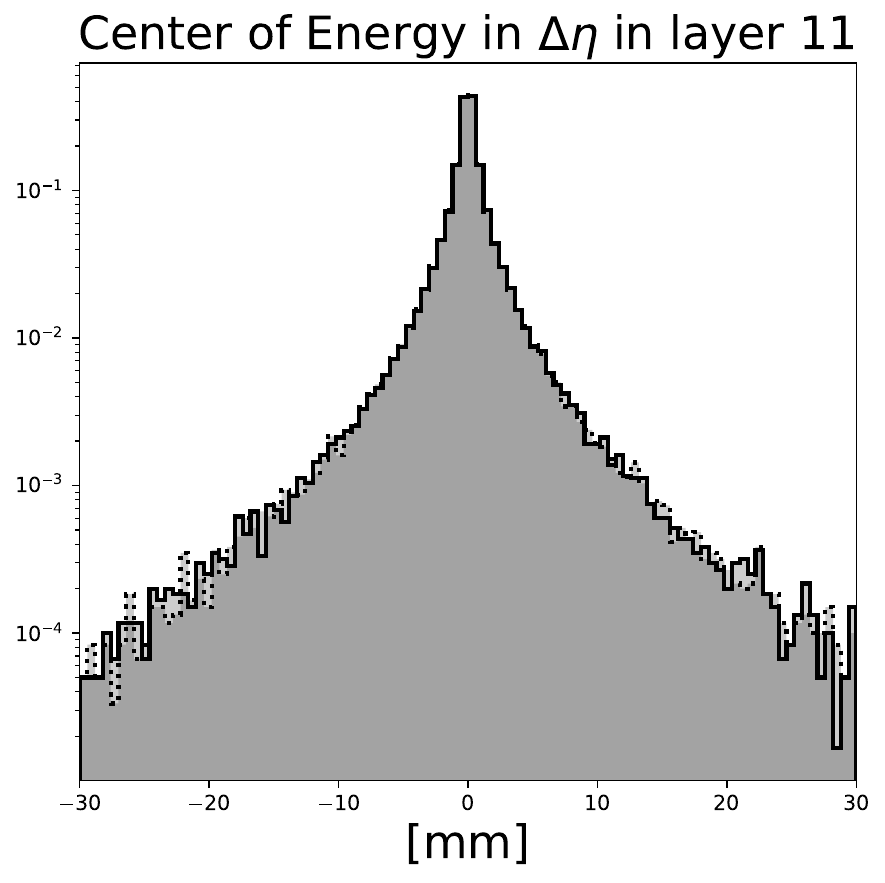} \hfill \includegraphics[height=0.1\textheight]{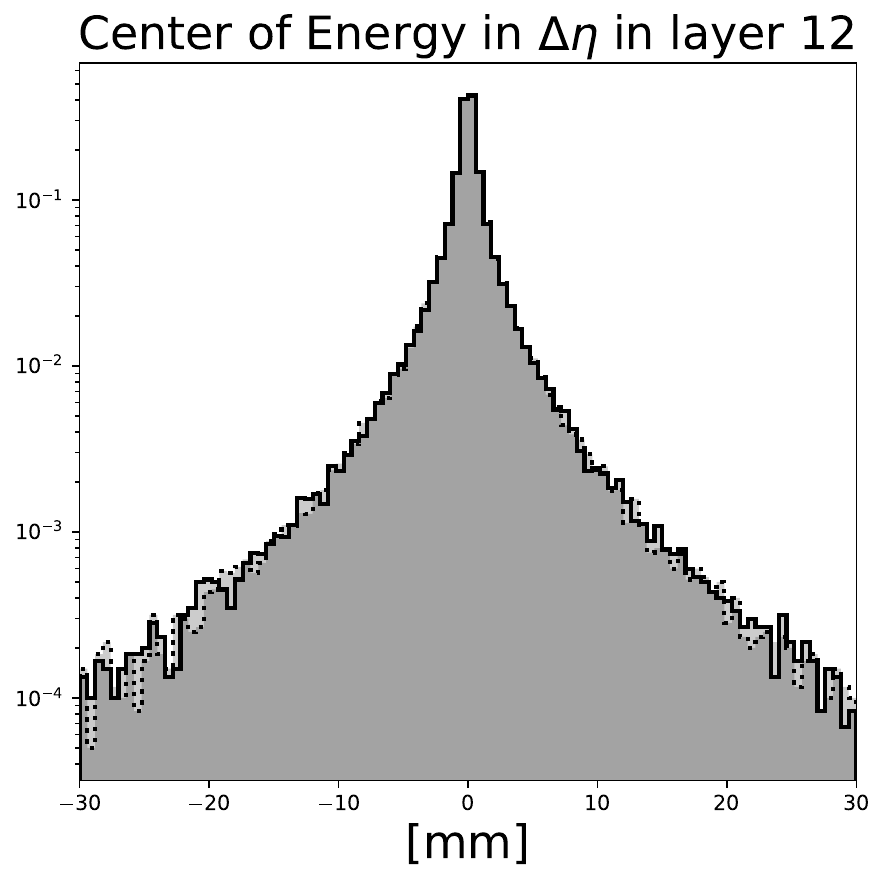} \hfill \includegraphics[height=0.1\textheight]{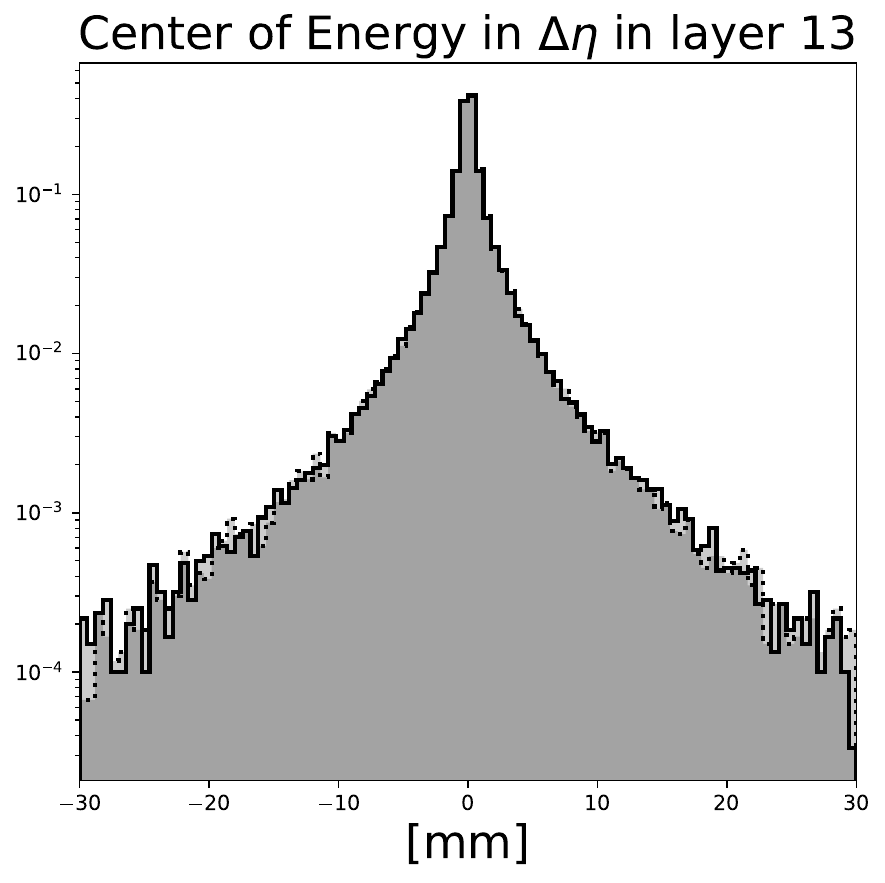} \hfill \includegraphics[height=0.1\textheight]{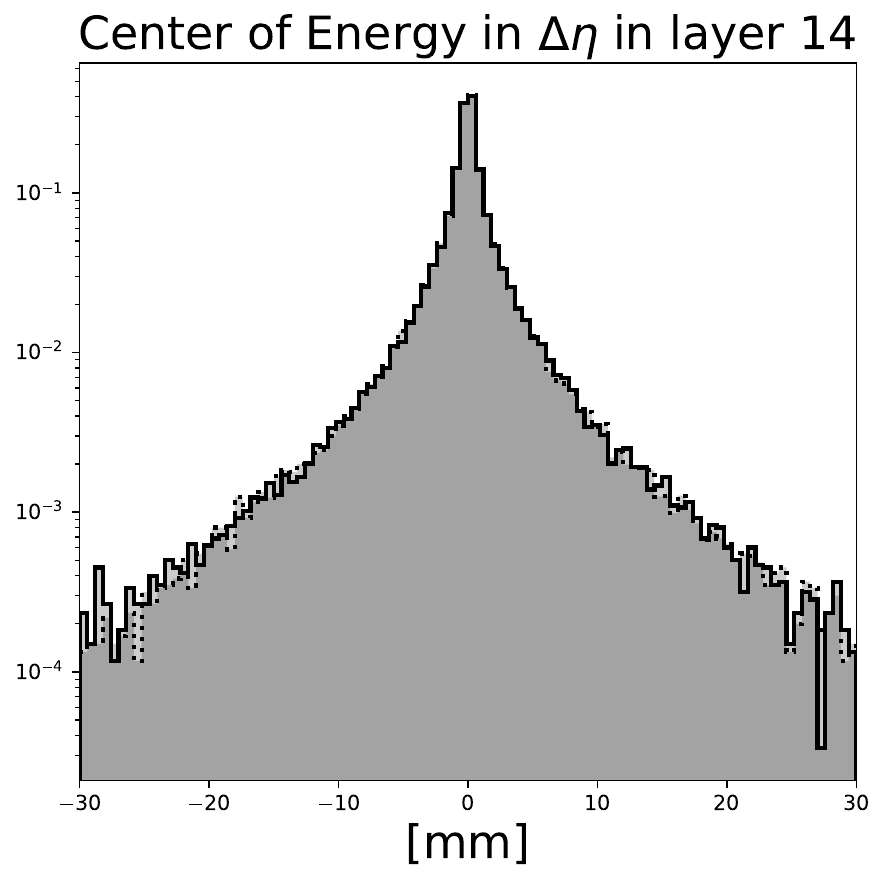}\\
    \includegraphics[height=0.1\textheight]{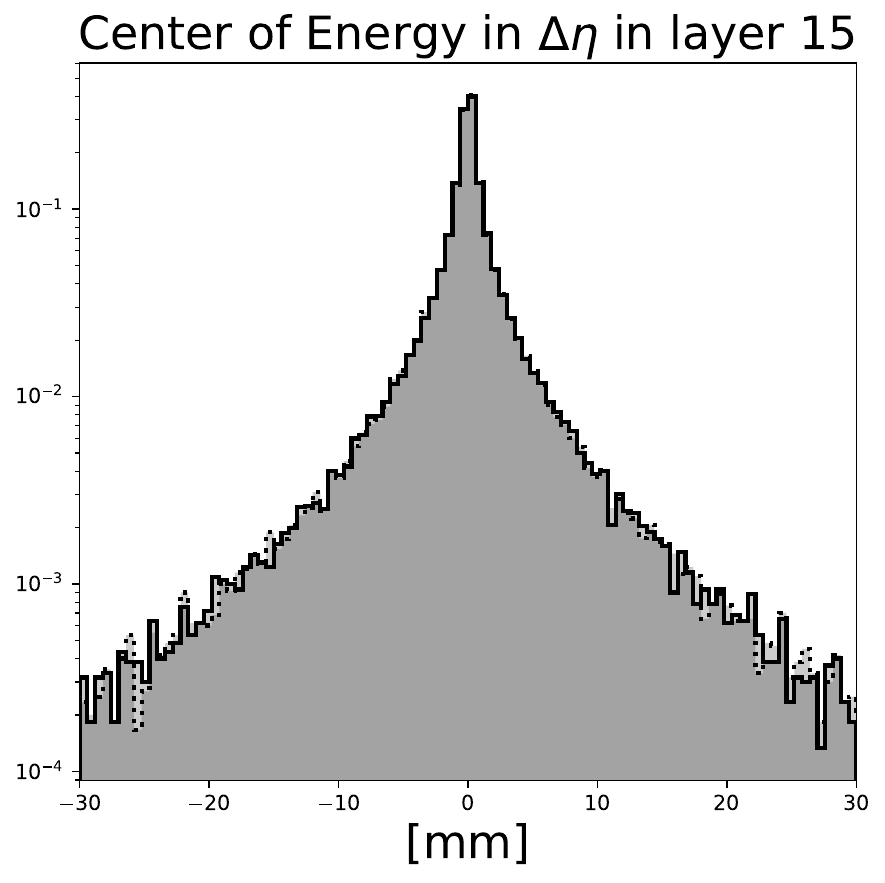} \hfill \includegraphics[height=0.1\textheight]{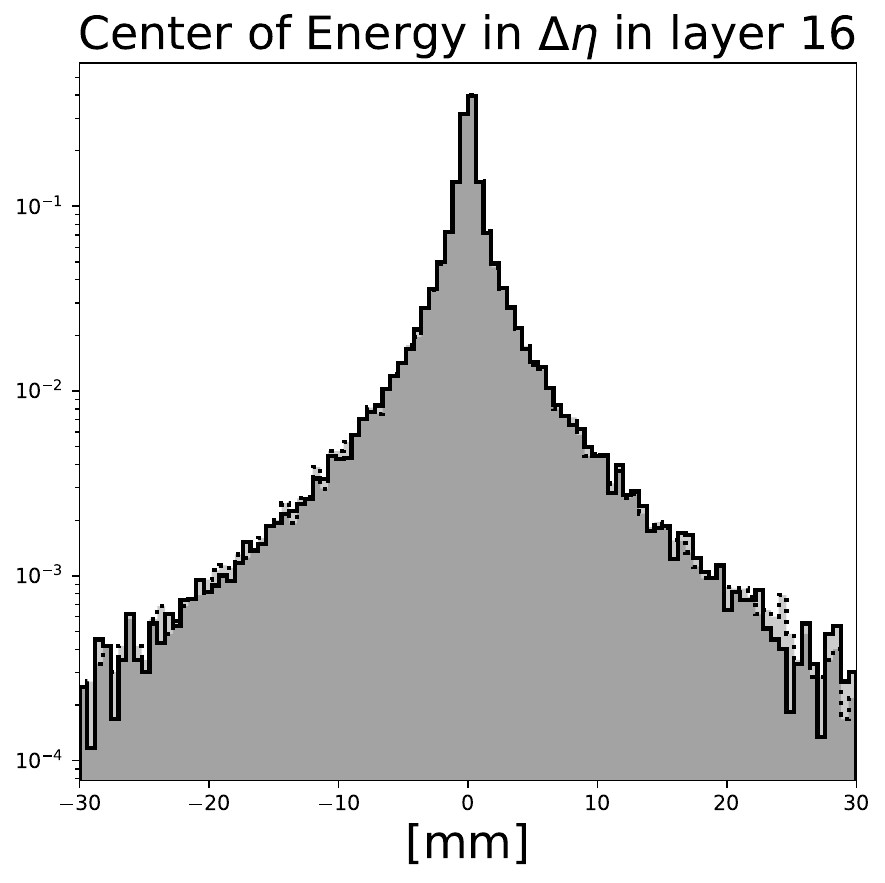} \hfill \includegraphics[height=0.1\textheight]{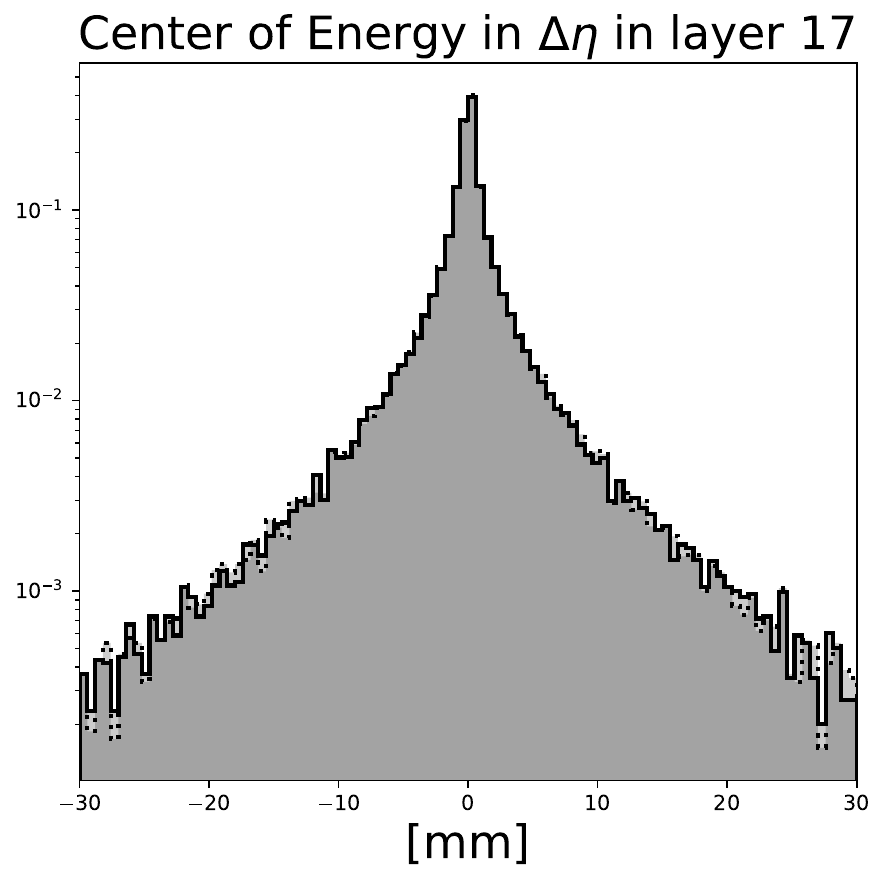} \hfill \includegraphics[height=0.1\textheight]{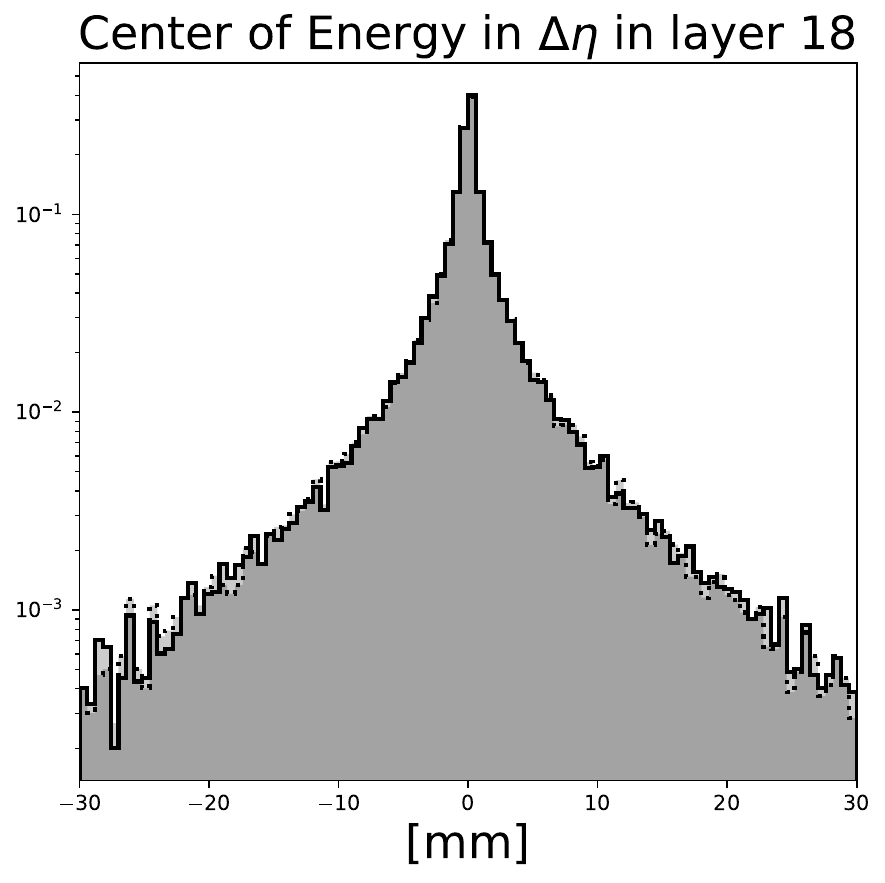} \hfill \includegraphics[height=0.1\textheight]{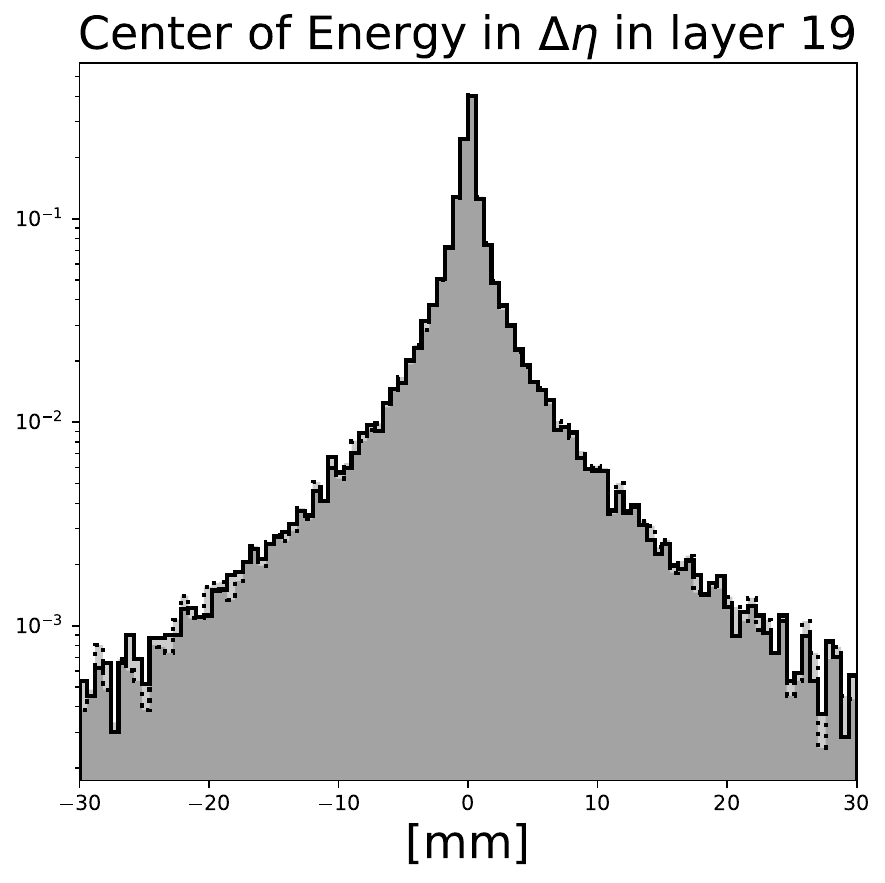}\\
    \includegraphics[height=0.1\textheight]{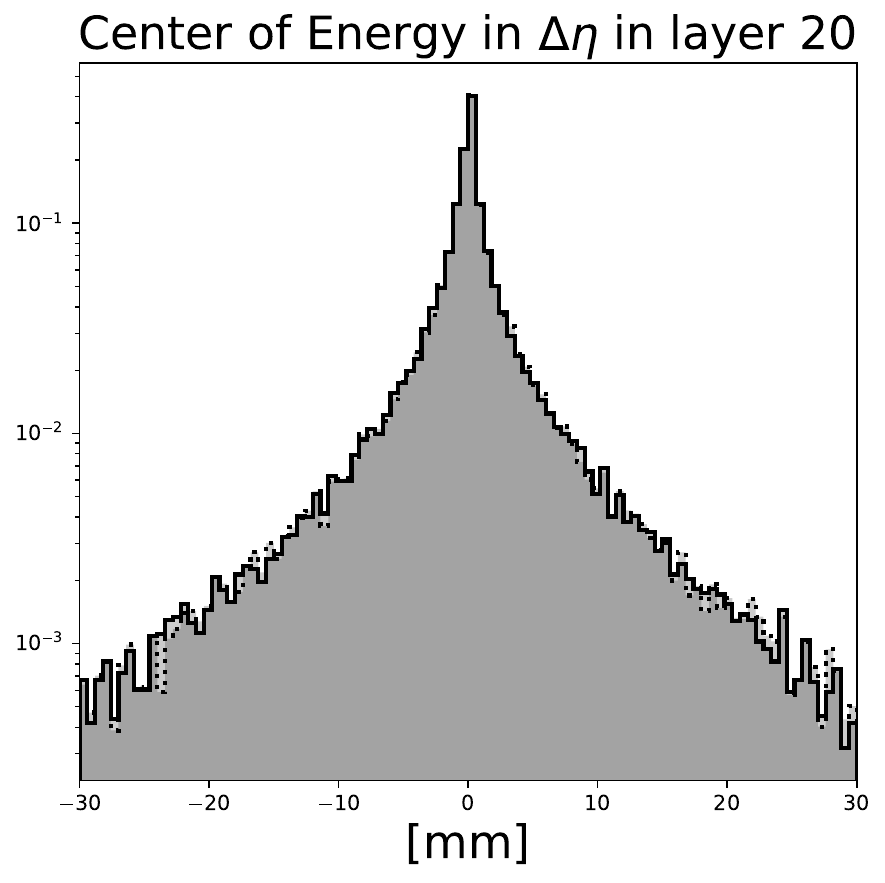} \hfill \includegraphics[height=0.1\textheight]{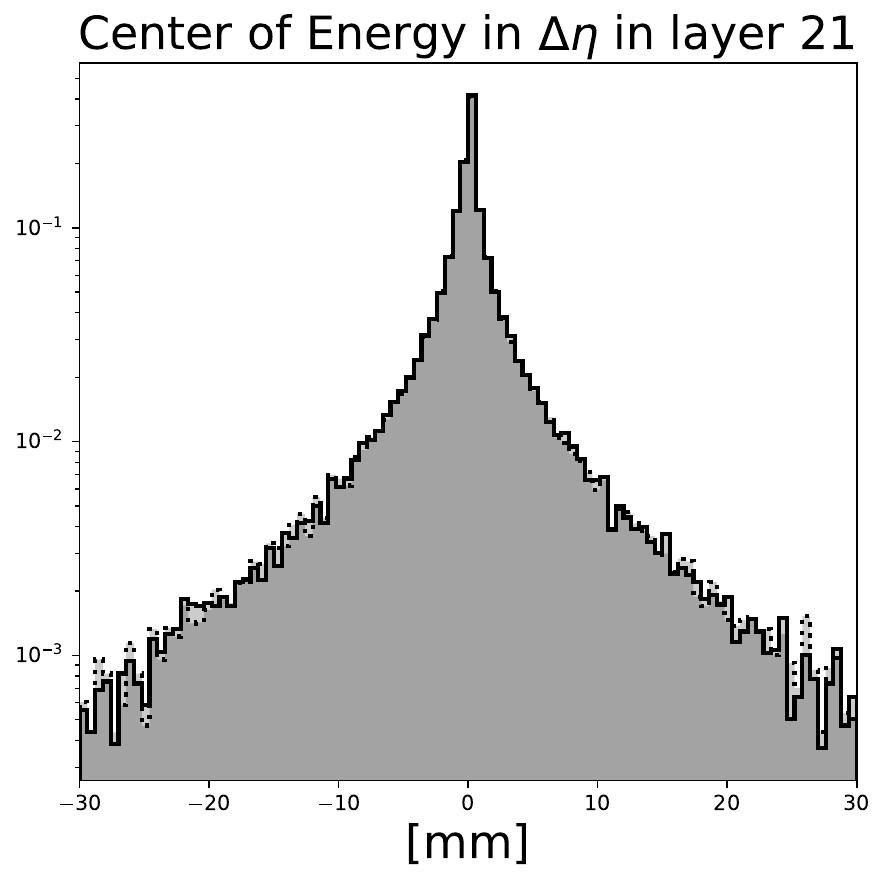} \hfill \includegraphics[height=0.1\textheight]{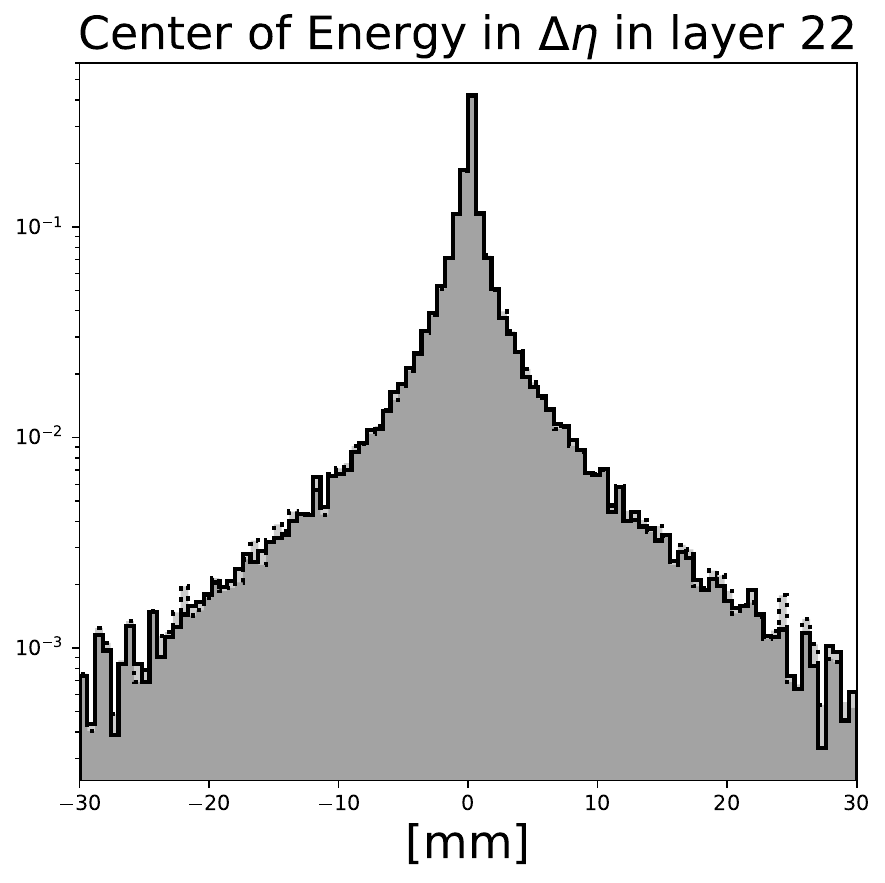} \hfill \includegraphics[height=0.1\textheight]{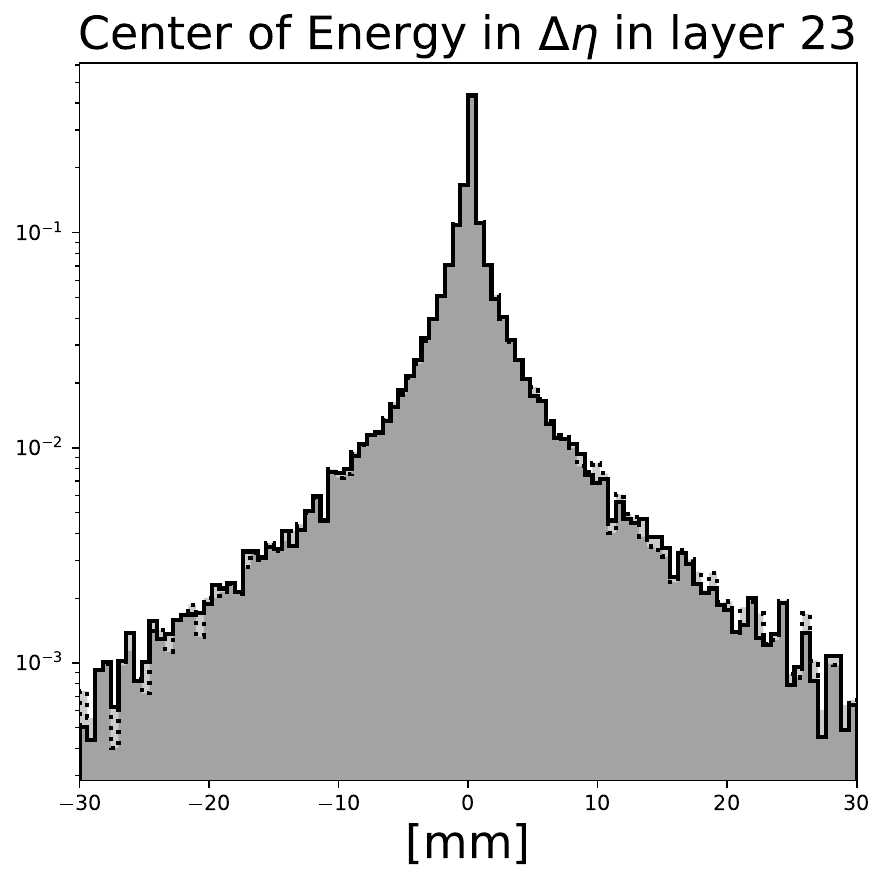} \hfill \includegraphics[height=0.1\textheight]{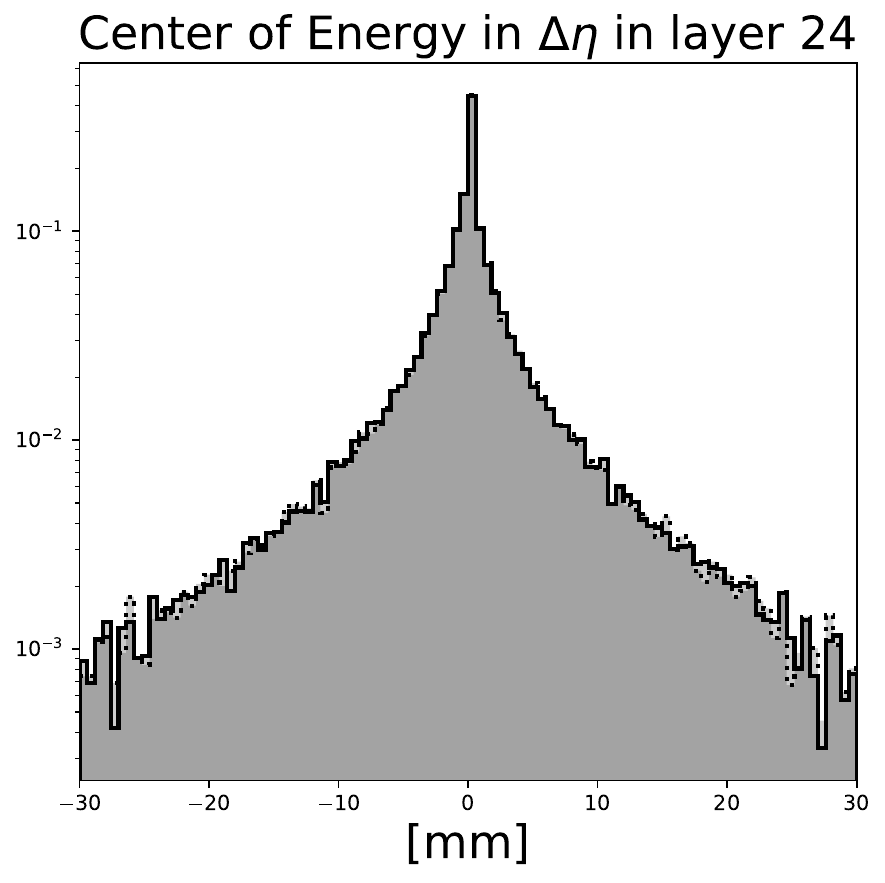}\\
    \includegraphics[height=0.1\textheight]{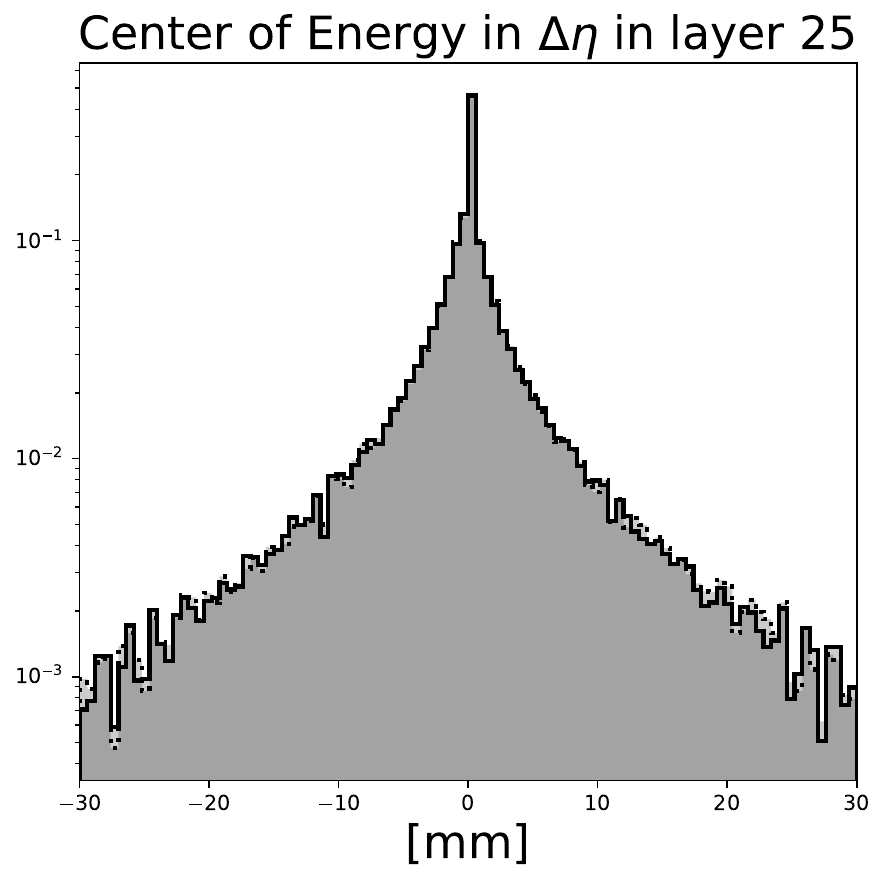} \hfill \includegraphics[height=0.1\textheight]{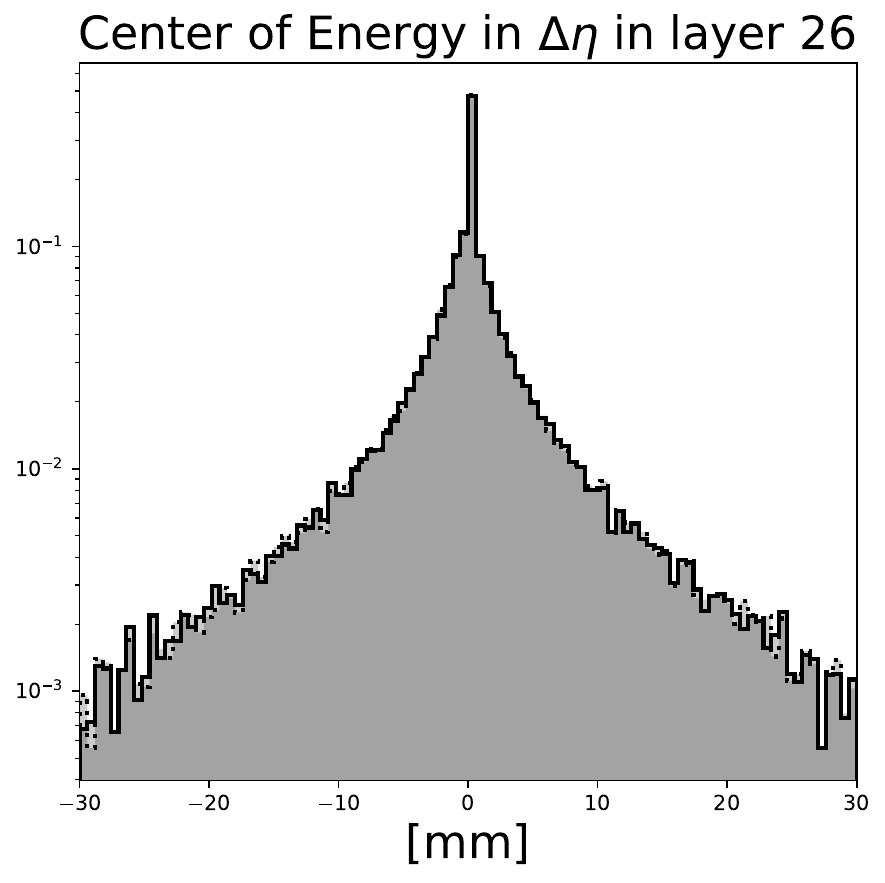} \hfill \includegraphics[height=0.1\textheight]{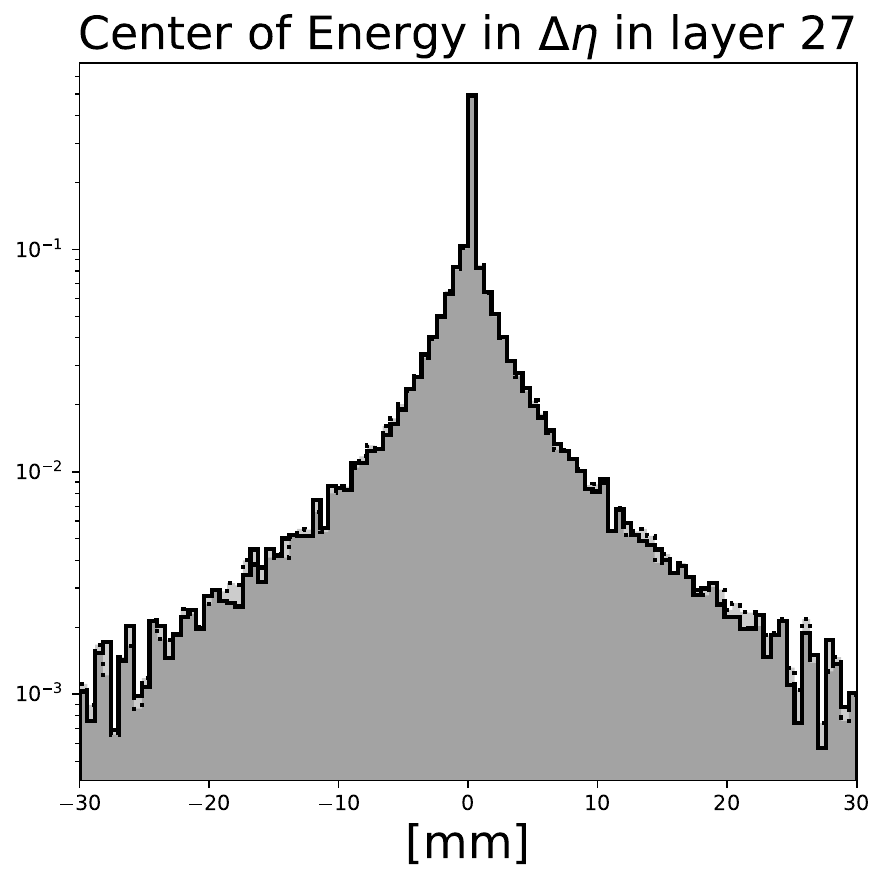} \hfill \includegraphics[height=0.1\textheight]{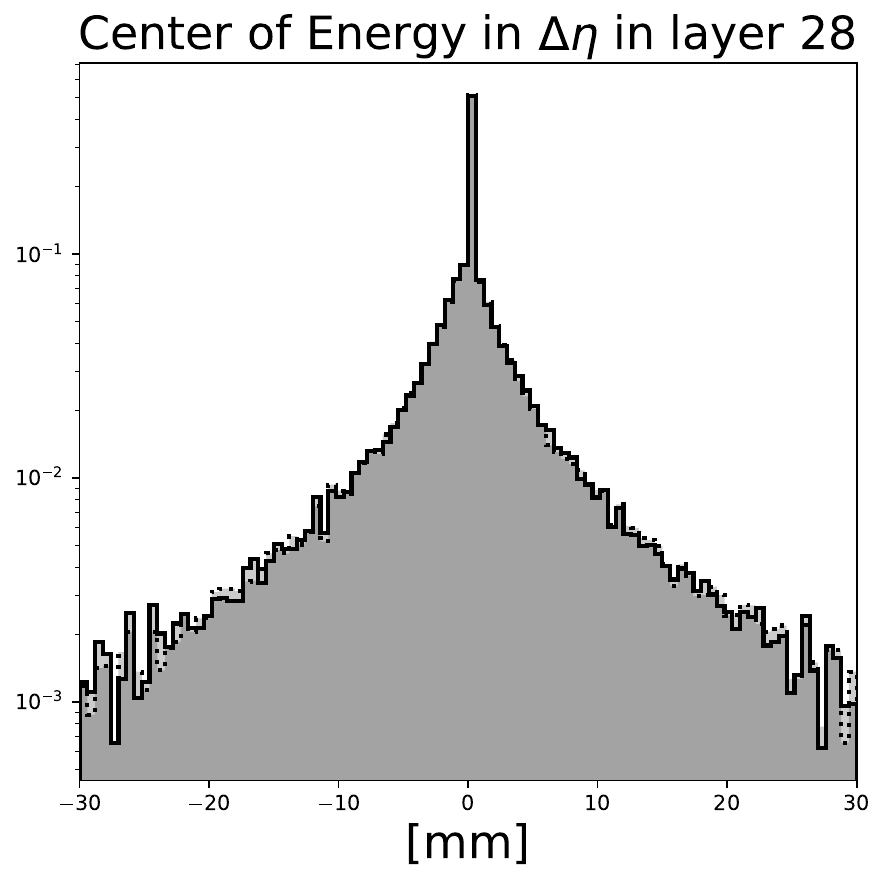} \hfill \includegraphics[height=0.1\textheight]{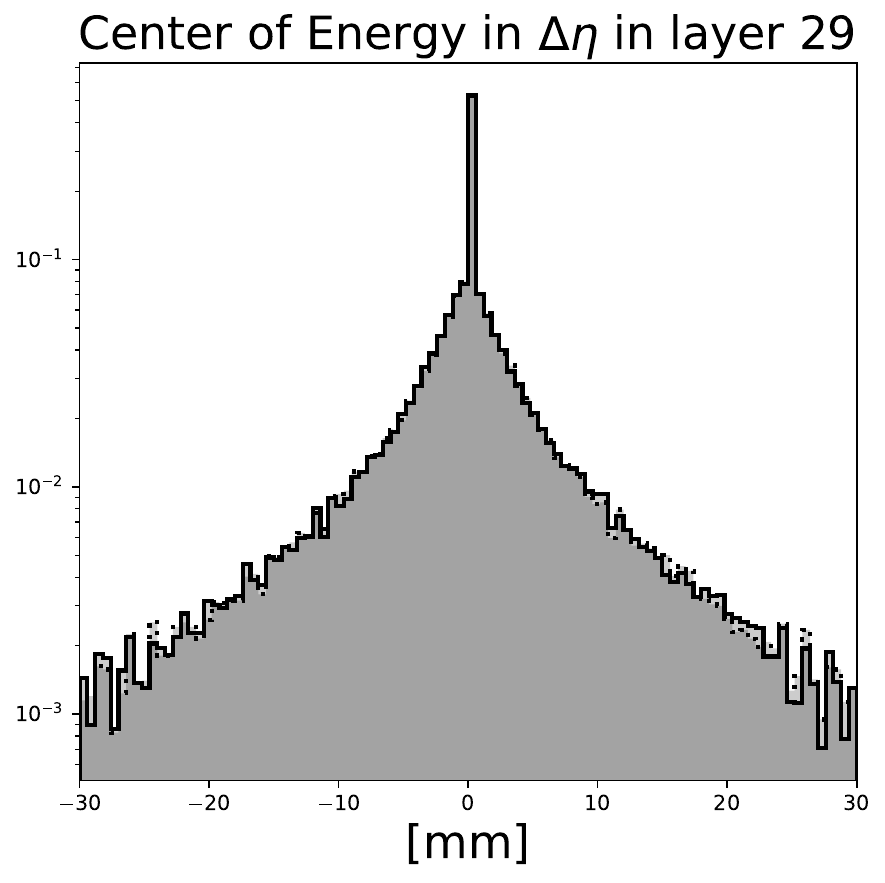}\\
    \includegraphics[height=0.1\textheight]{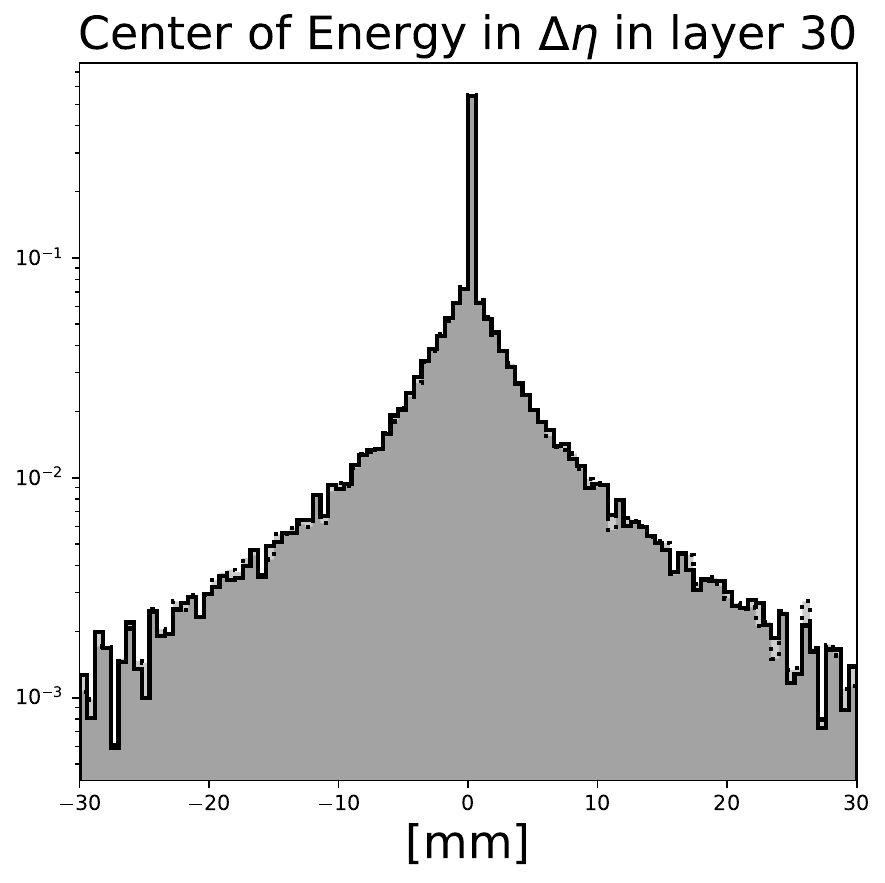} \hfill \includegraphics[height=0.1\textheight]{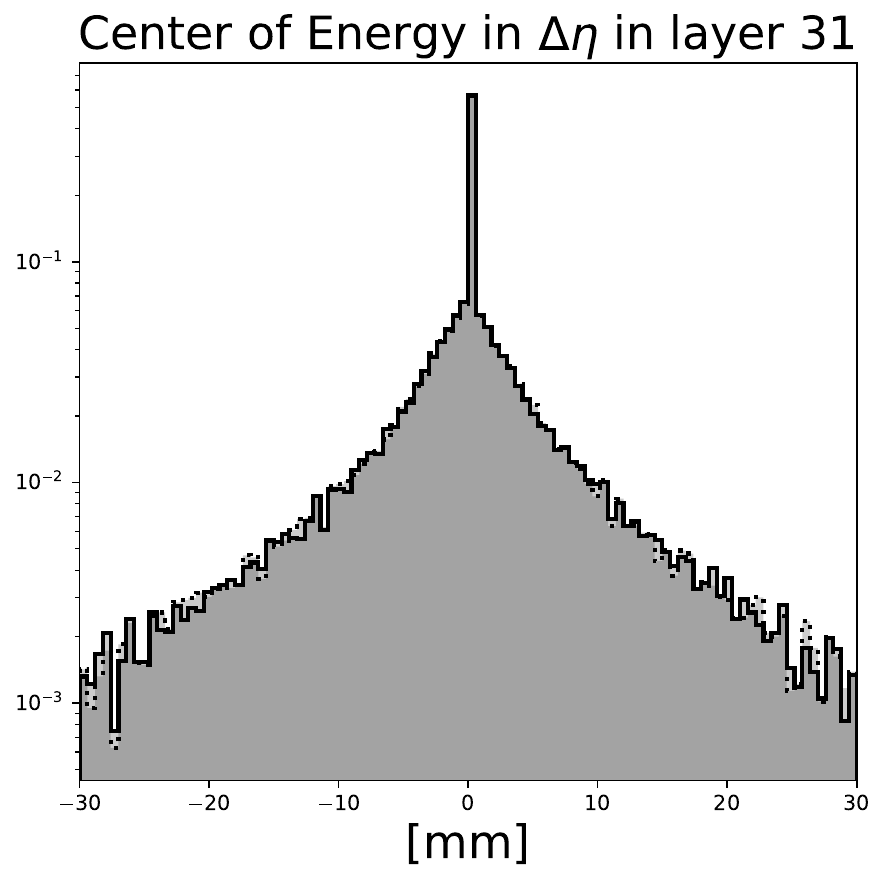} \hfill \includegraphics[height=0.1\textheight]{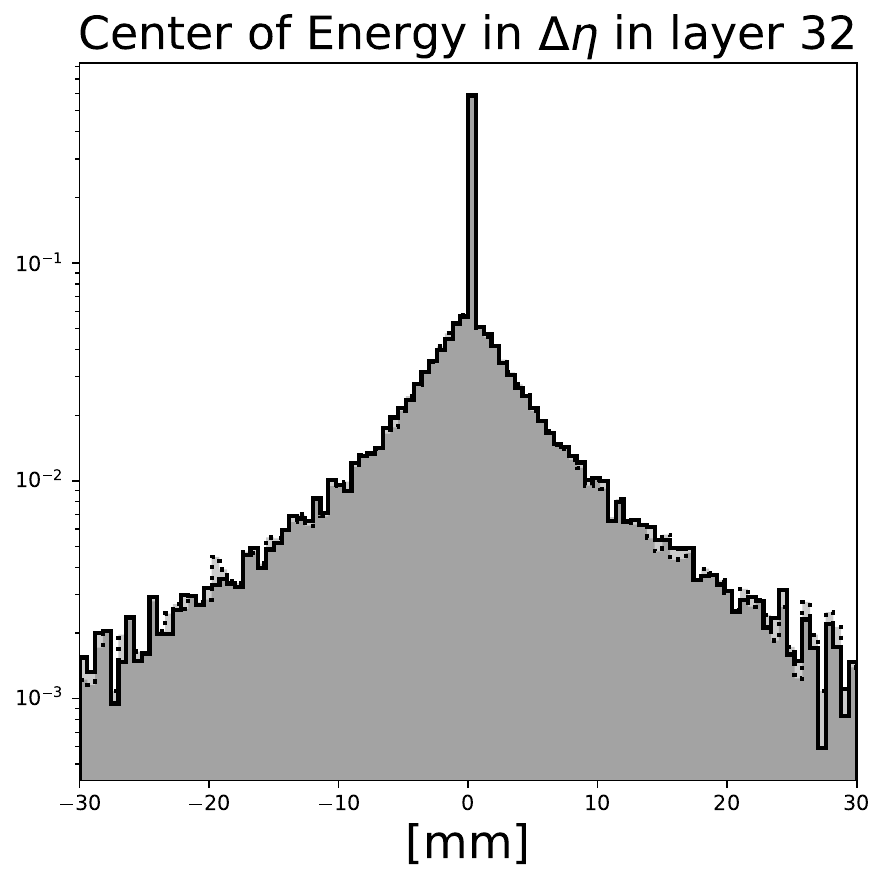} \hfill \includegraphics[height=0.1\textheight]{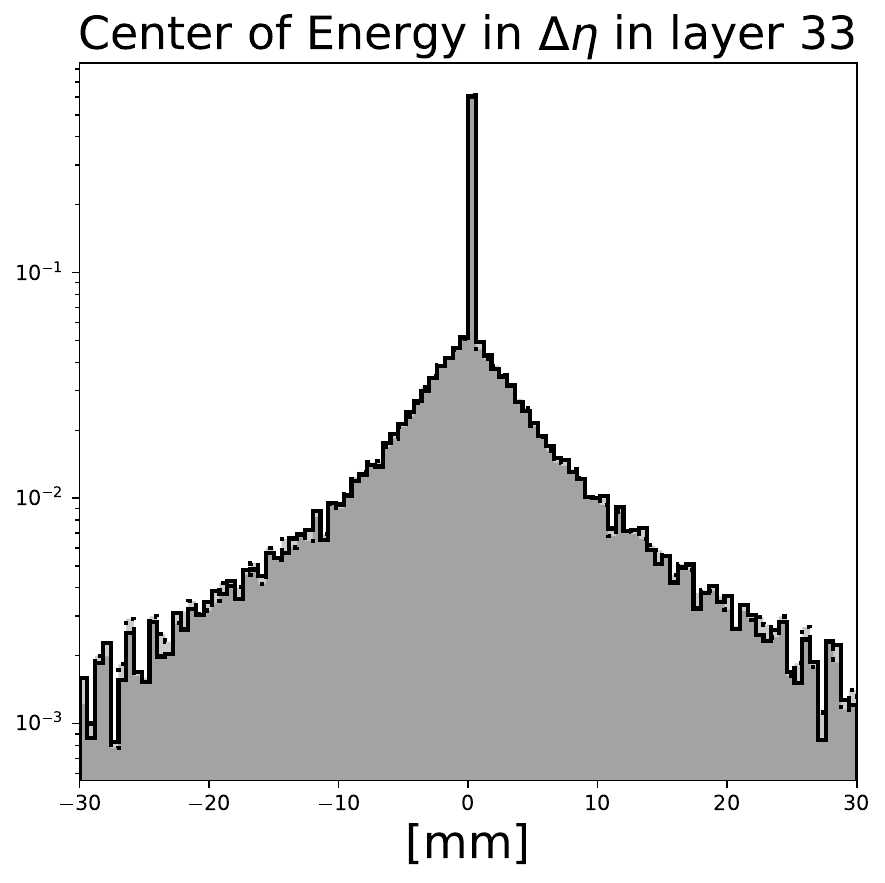} \hfill \includegraphics[height=0.1\textheight]{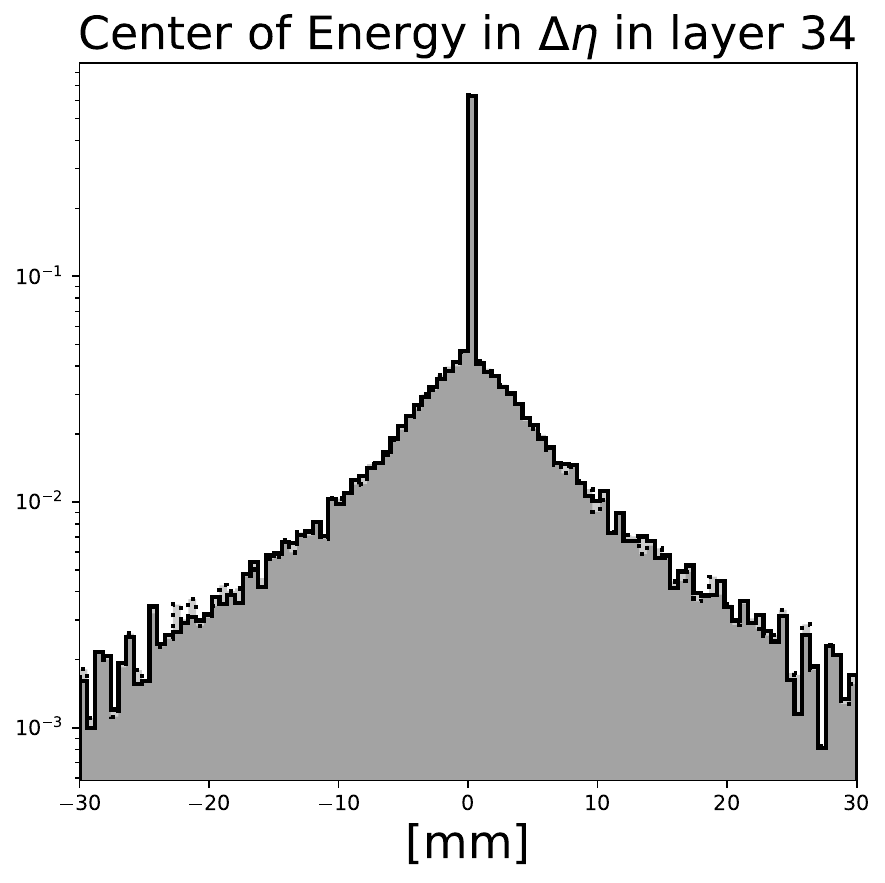}\\
    \includegraphics[height=0.1\textheight]{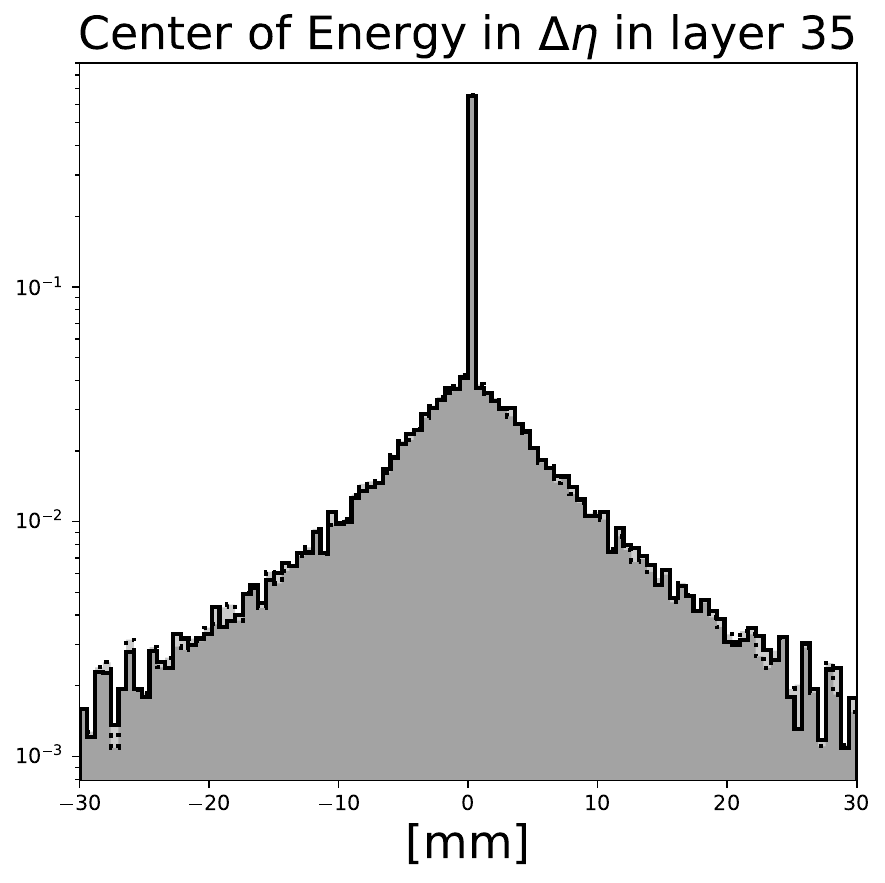} \hfill \includegraphics[height=0.1\textheight]{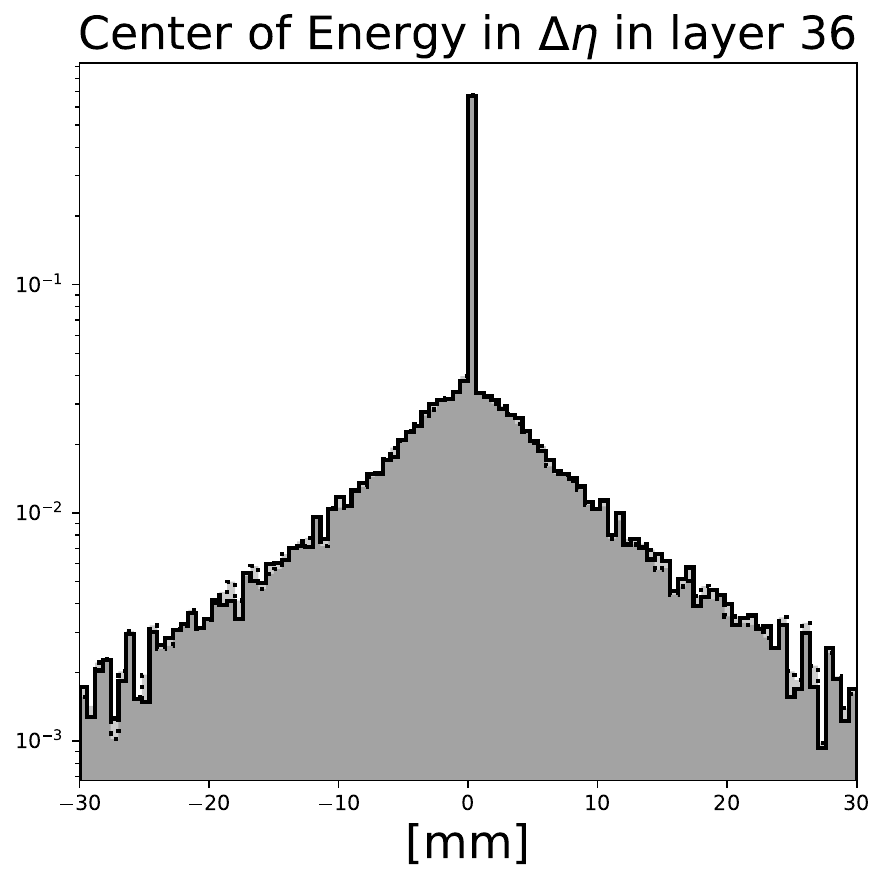} \hfill \includegraphics[height=0.1\textheight]{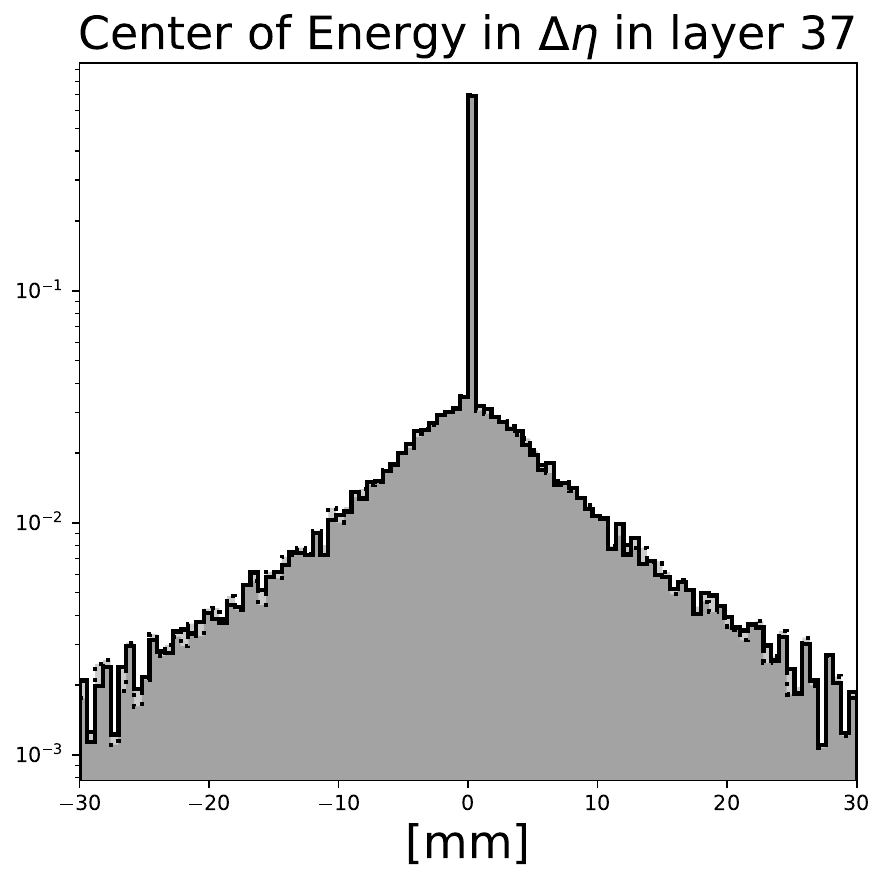} \hfill \includegraphics[height=0.1\textheight]{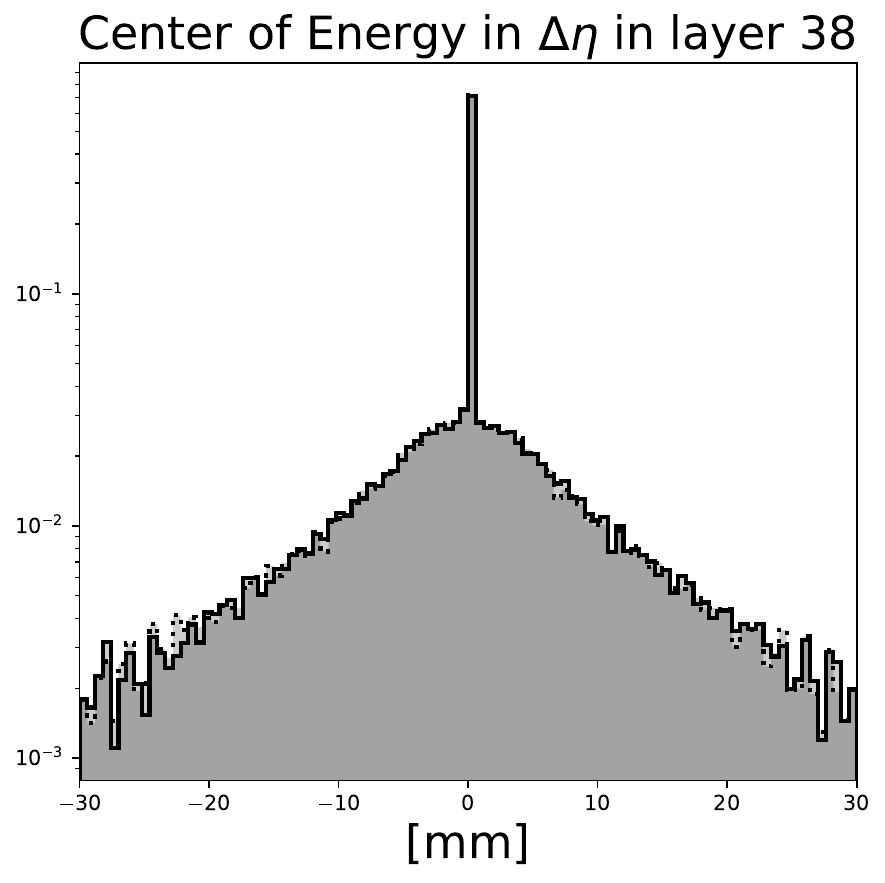} \hfill \includegraphics[height=0.1\textheight]{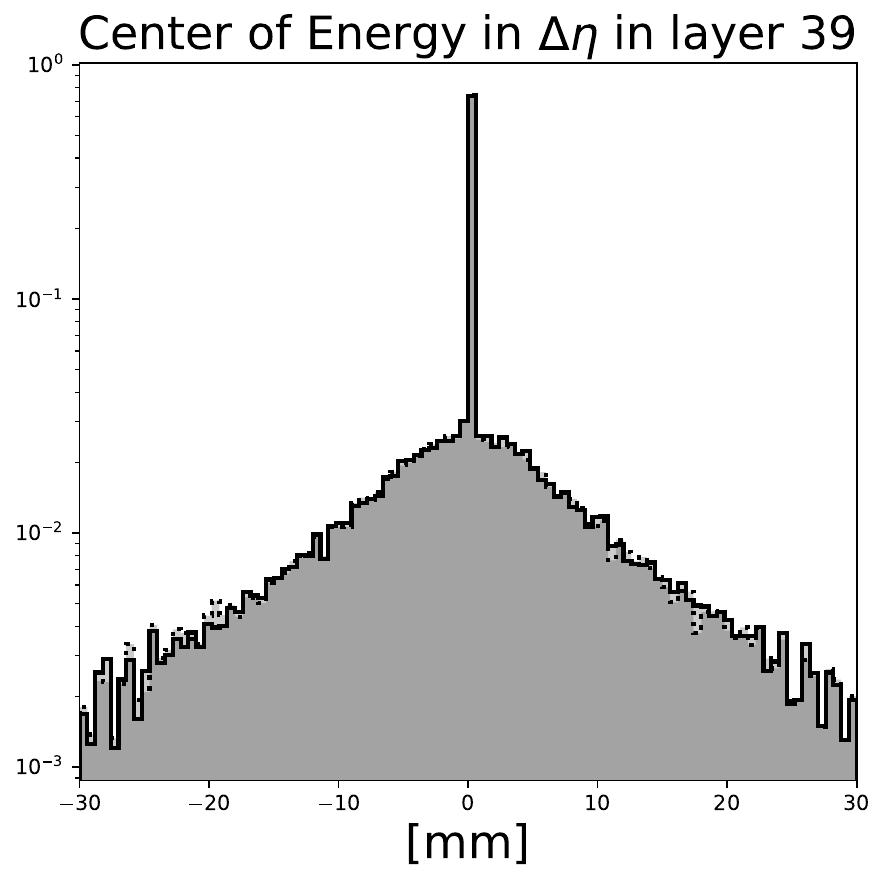}\\
    \includegraphics[height=0.1\textheight]{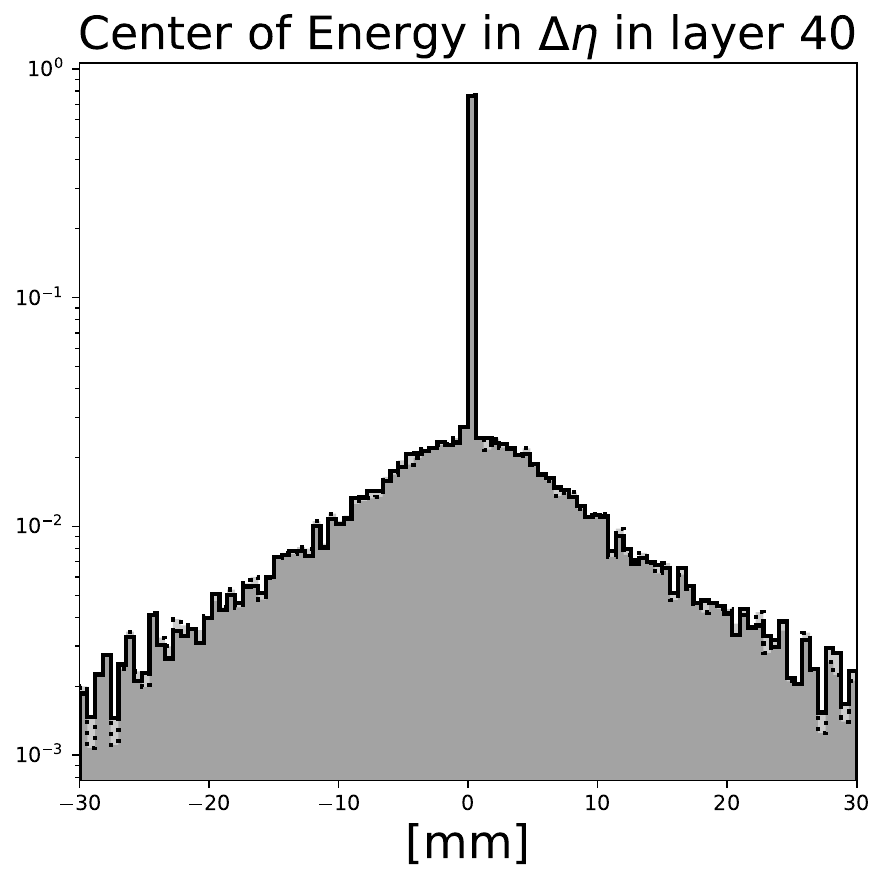} \hfill \includegraphics[height=0.1\textheight]{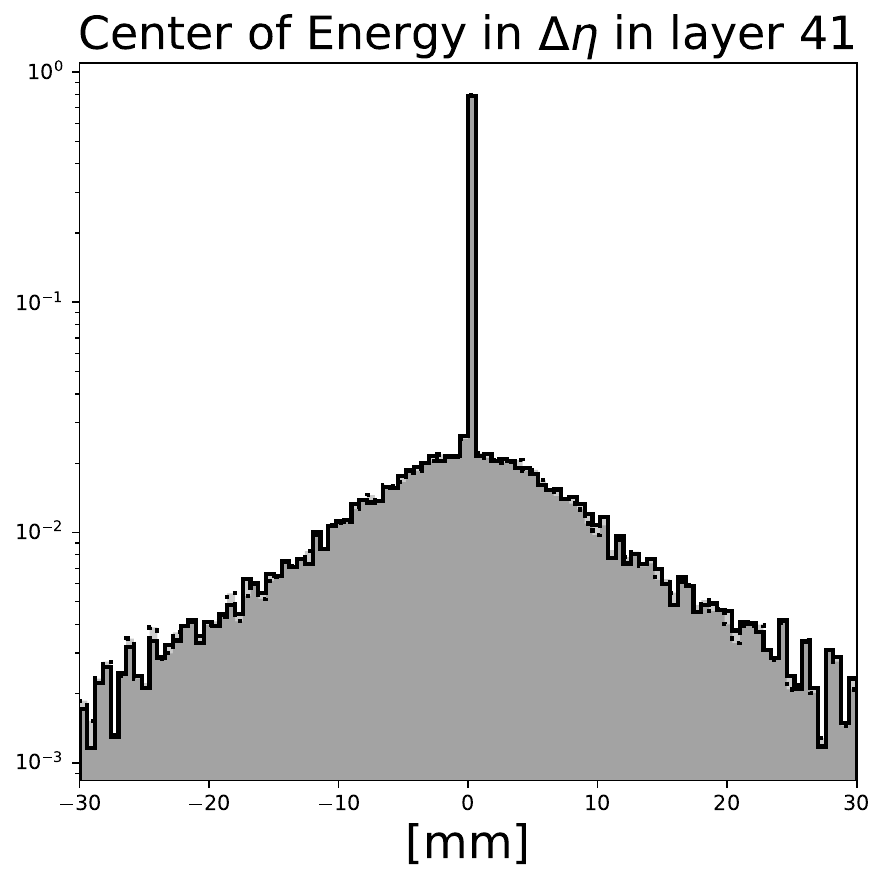} \hfill \includegraphics[height=0.1\textheight]{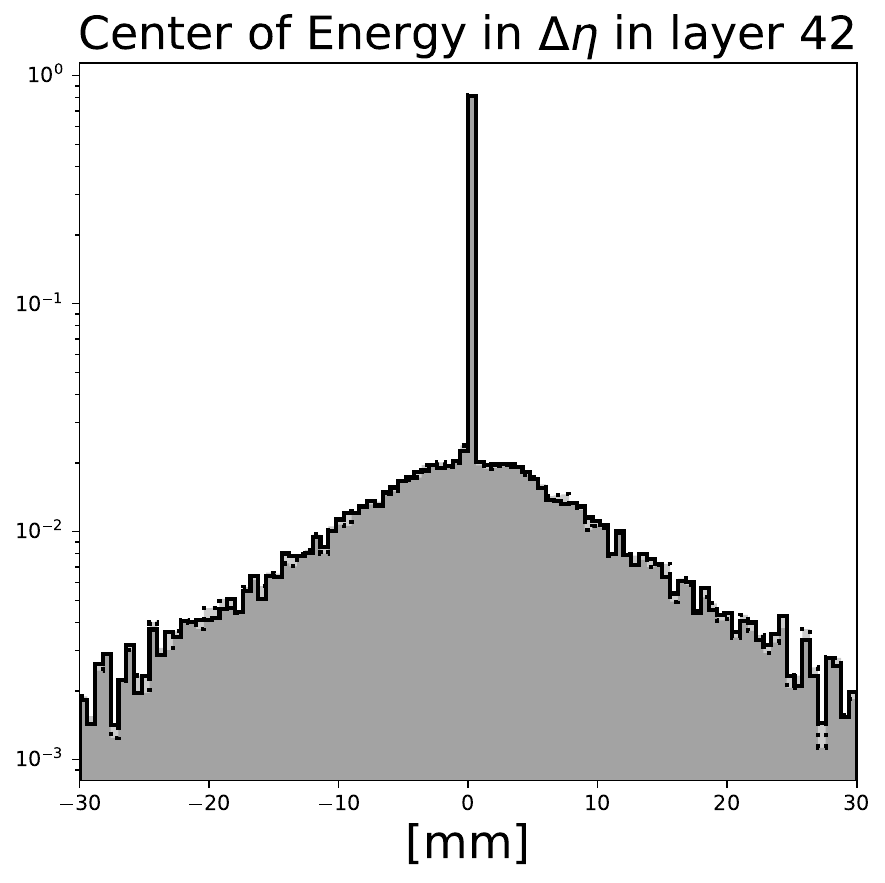} \hfill \includegraphics[height=0.1\textheight]{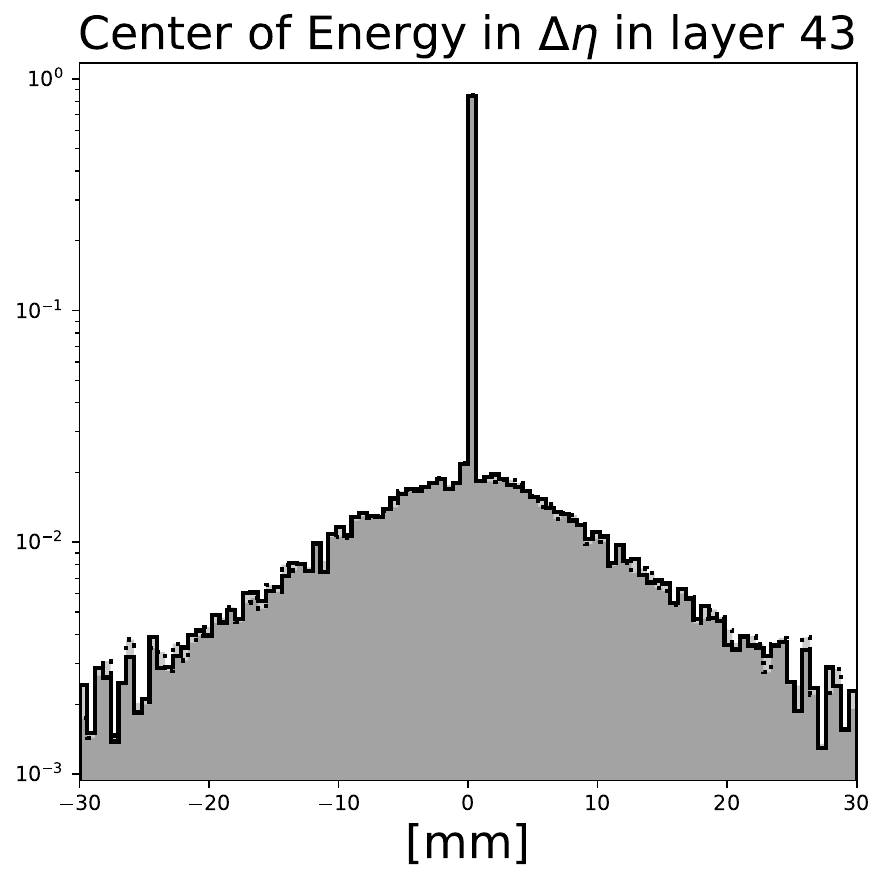} \hfill \includegraphics[height=0.1\textheight]{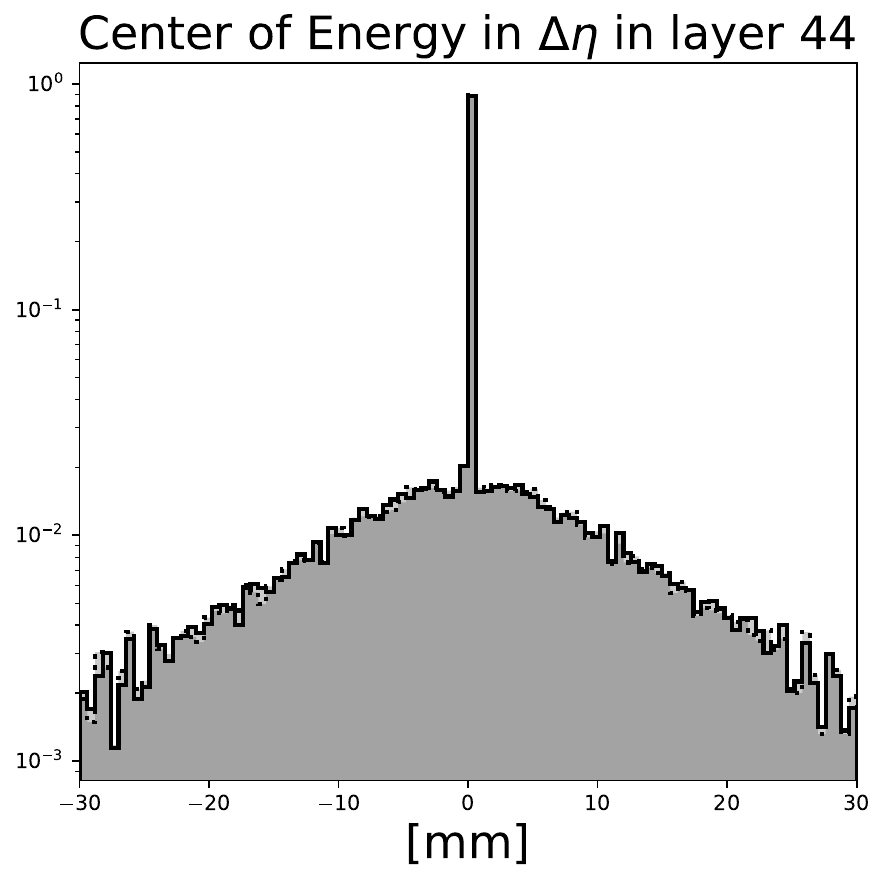}\\
    \includegraphics[width=0.5\textwidth]{figures/Appendix_reference/legend.pdf}
    \caption{Distribution of \geant training and evaluation data in centers of energy in $\eta$ direction for ds3. }
    \label{fig:app_ref.ds3.3}
\end{figure}

\begin{figure}[ht]
    \centering
    \includegraphics[height=0.1\textheight]{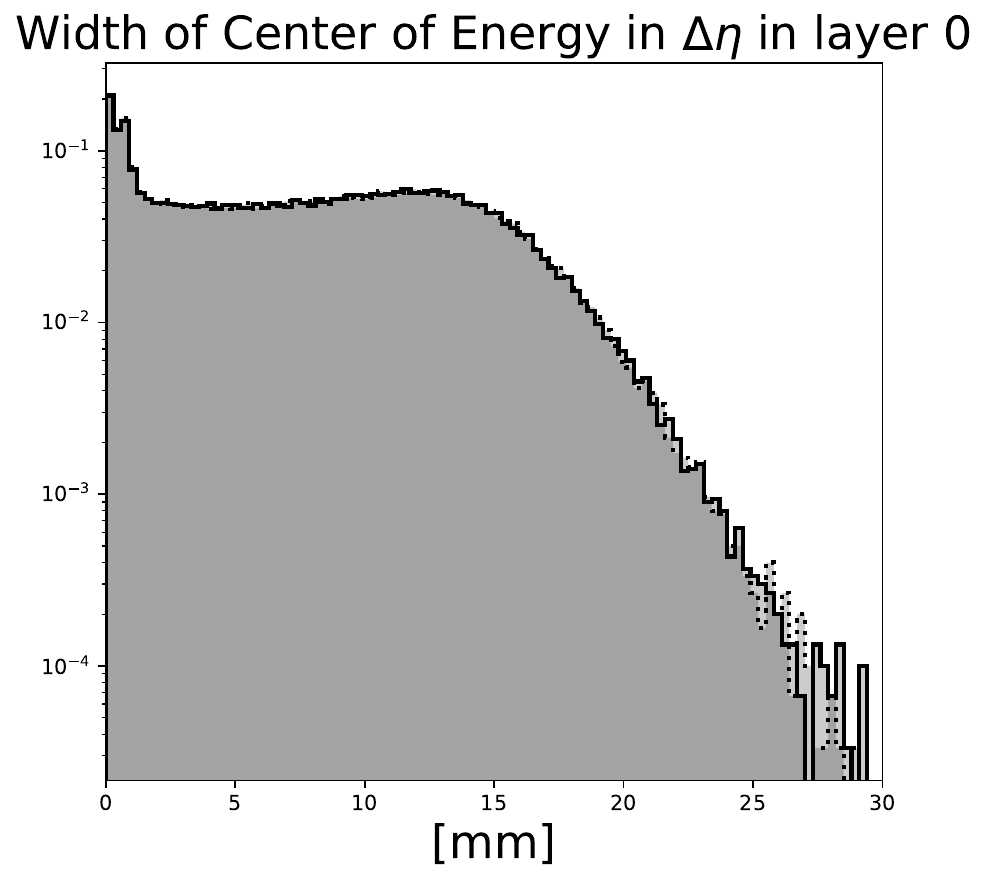} \hfill \includegraphics[height=0.1\textheight]{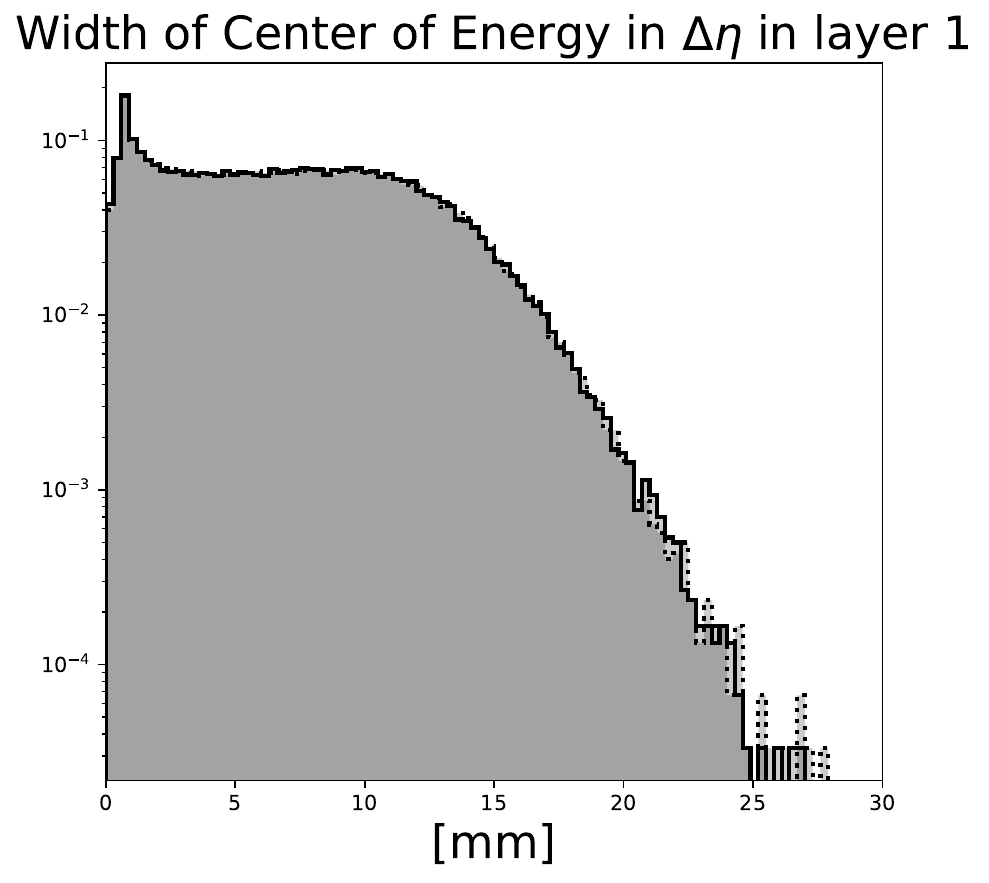} \hfill \includegraphics[height=0.1\textheight]{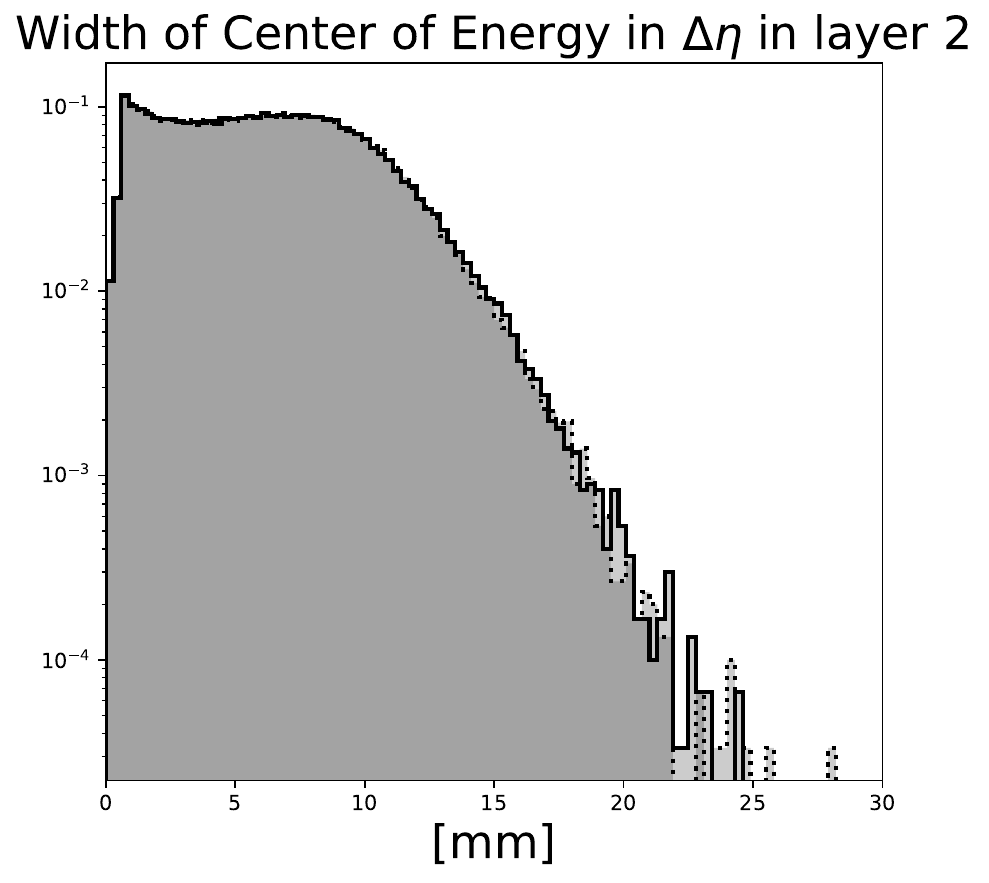} \hfill \includegraphics[height=0.1\textheight]{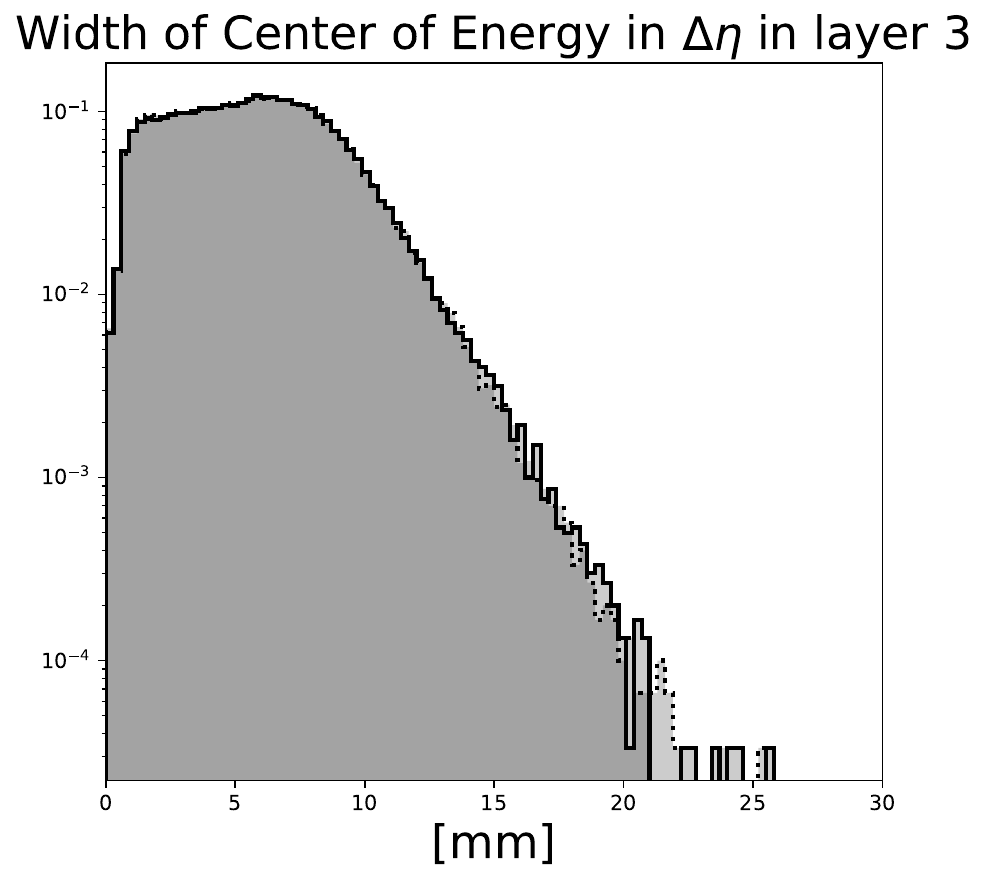} \hfill \includegraphics[height=0.1\textheight]{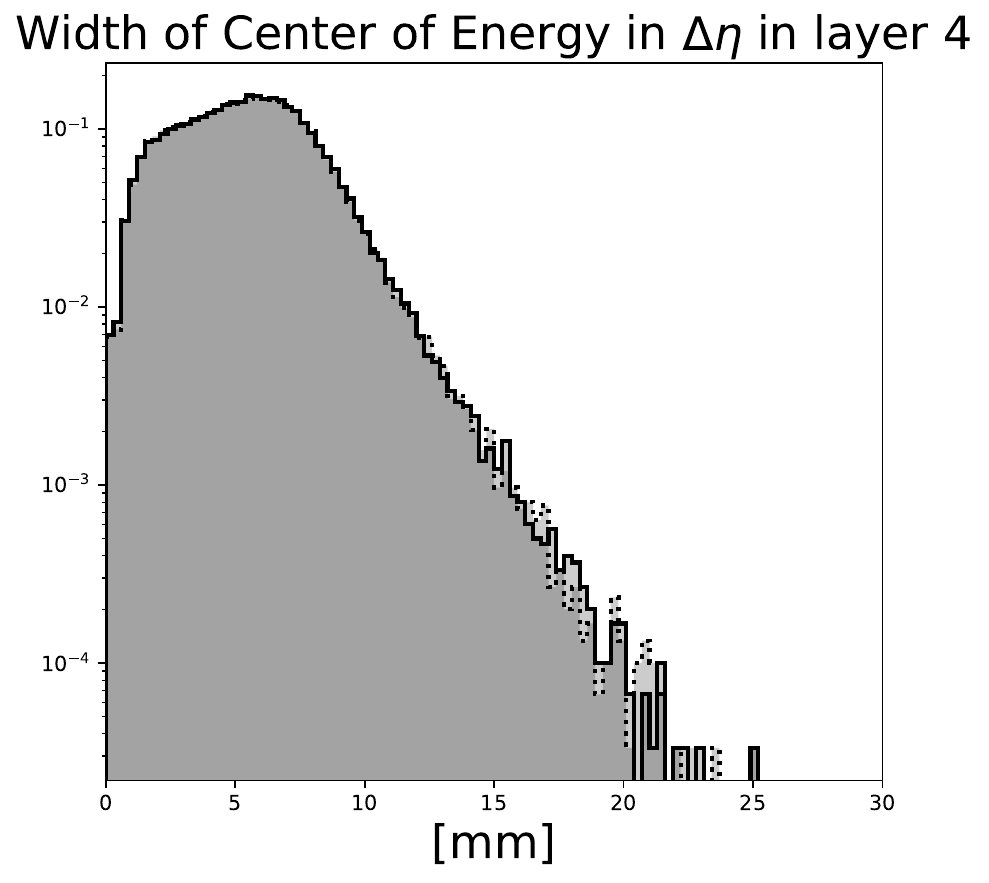}\\
    \includegraphics[height=0.1\textheight]{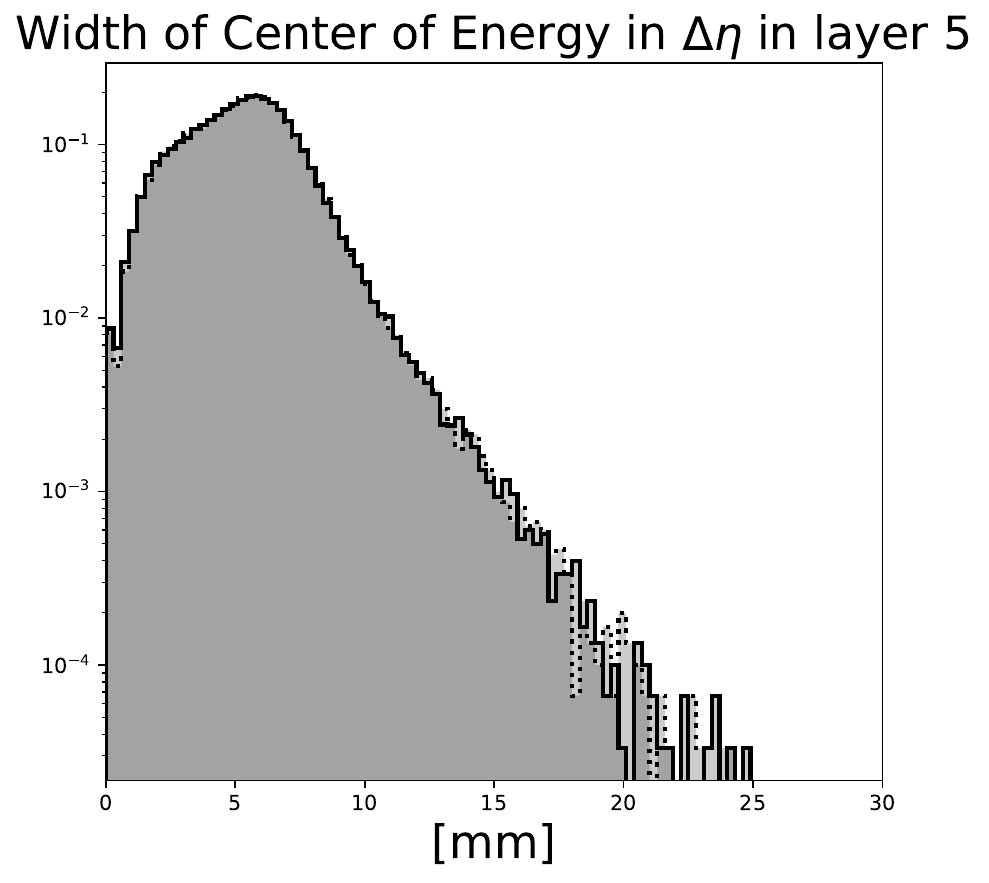} \hfill \includegraphics[height=0.1\textheight]{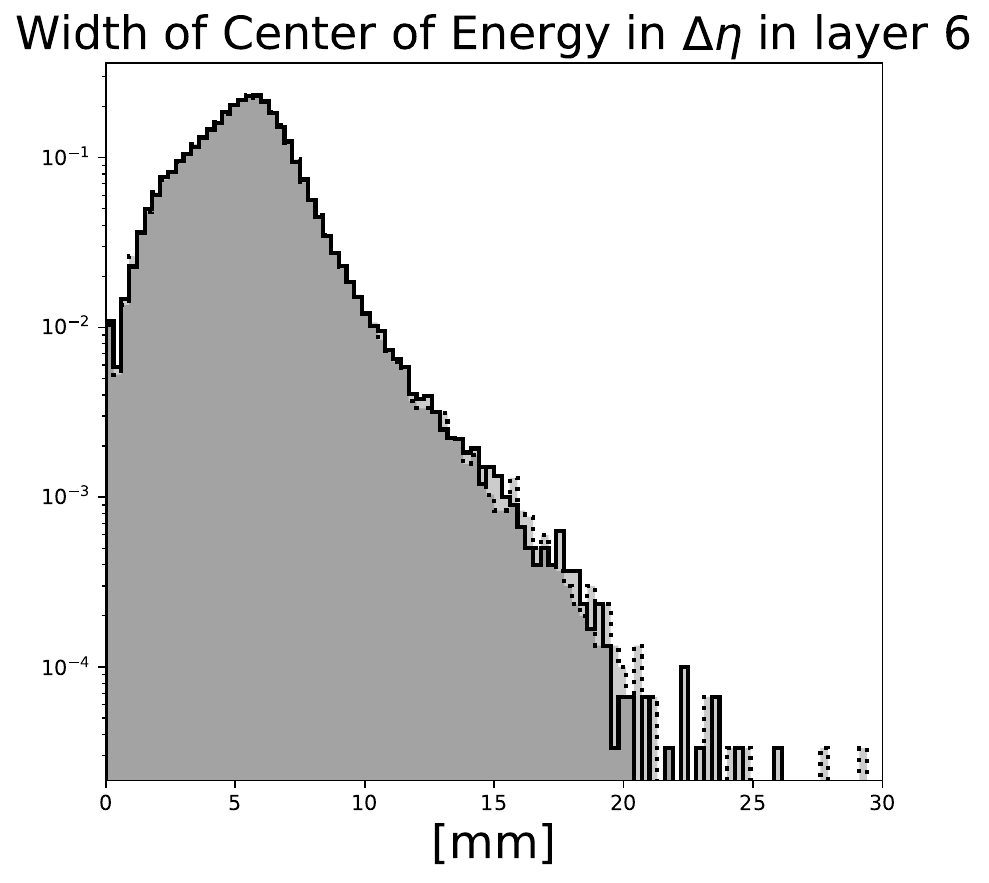} \hfill \includegraphics[height=0.1\textheight]{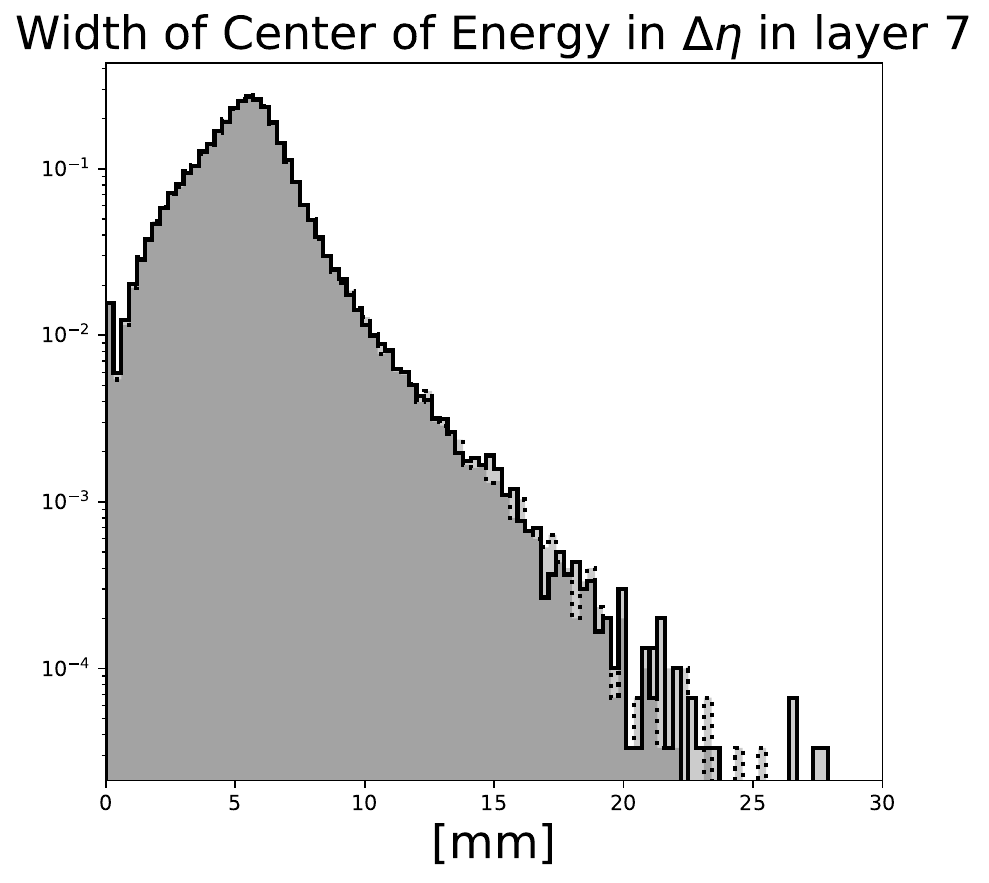} \hfill \includegraphics[height=0.1\textheight]{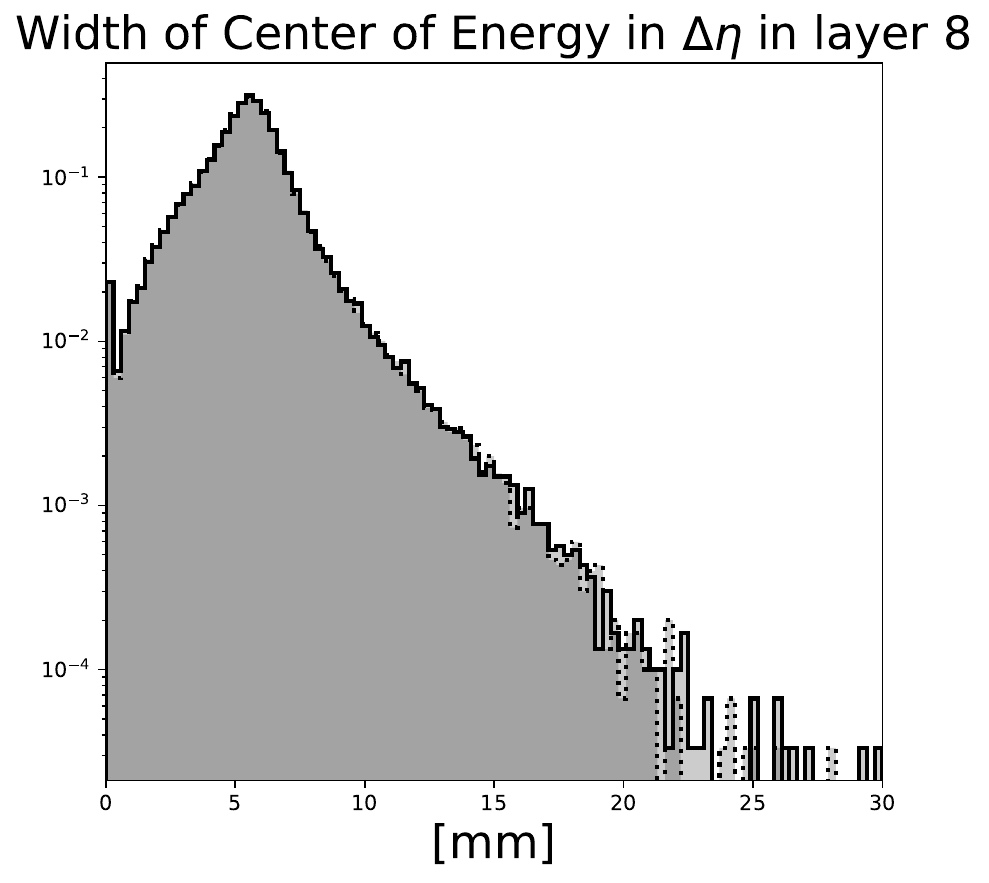} \hfill \includegraphics[height=0.1\textheight]{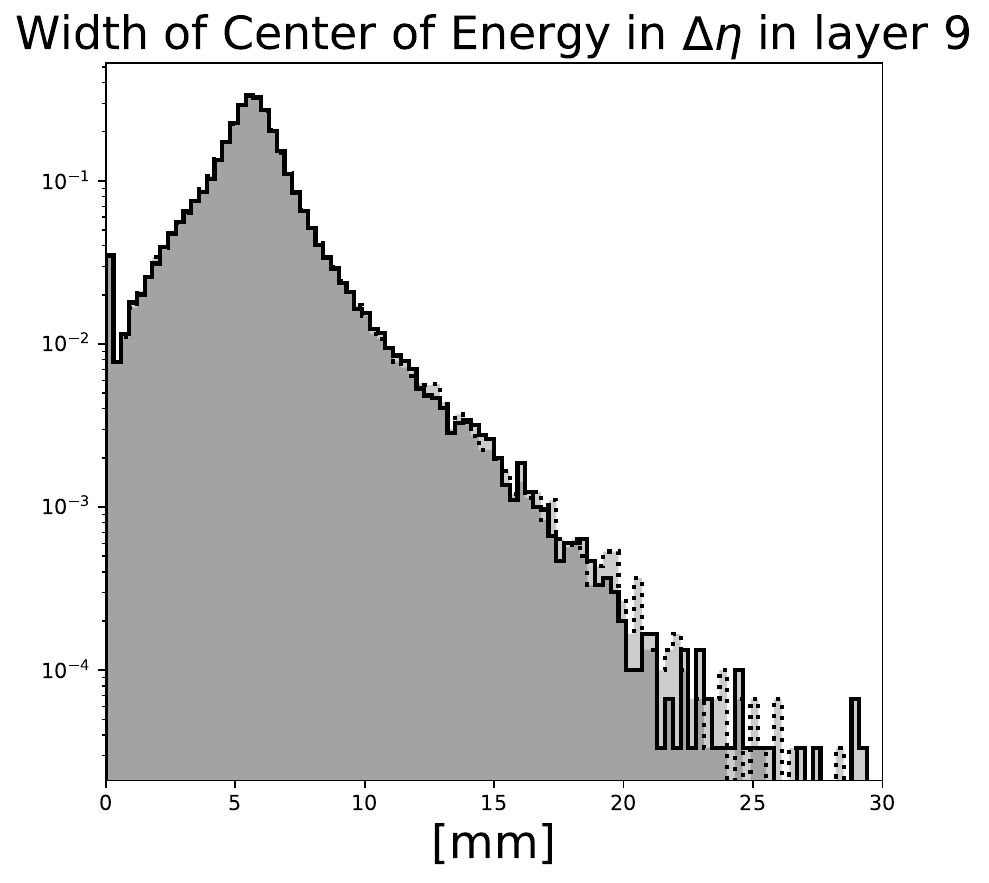}\\
    \includegraphics[height=0.1\textheight]{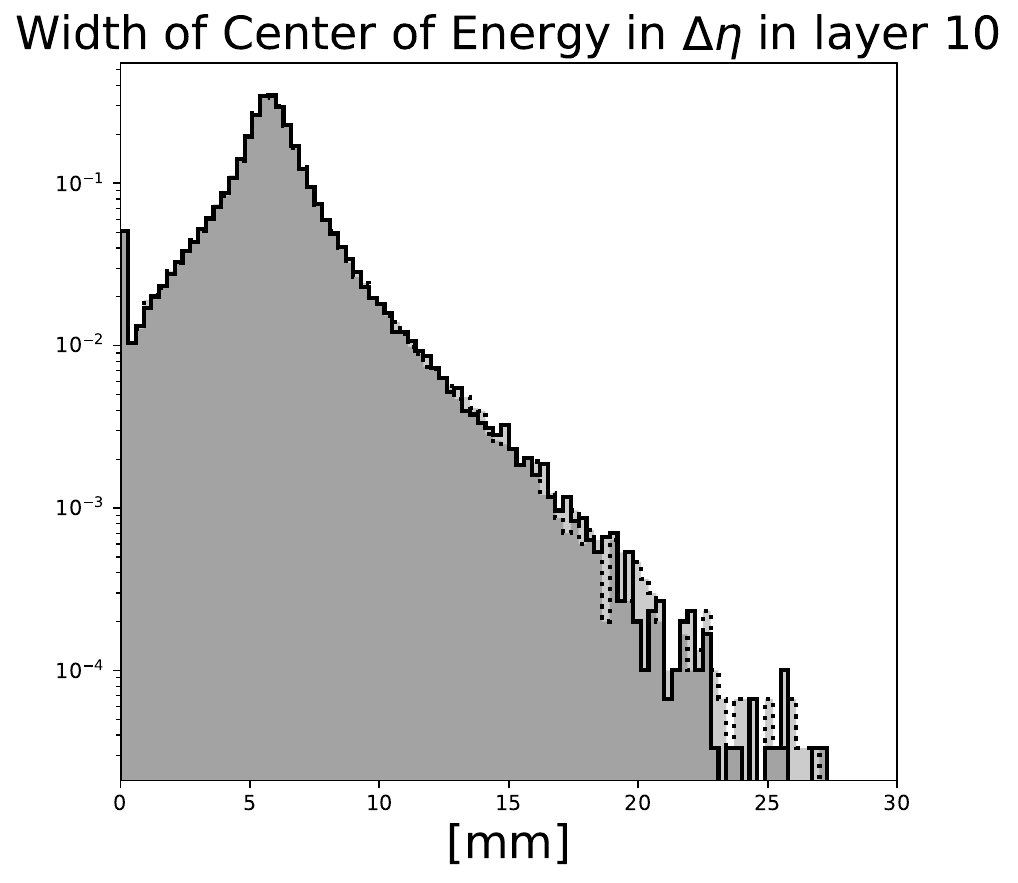} \hfill \includegraphics[height=0.1\textheight]{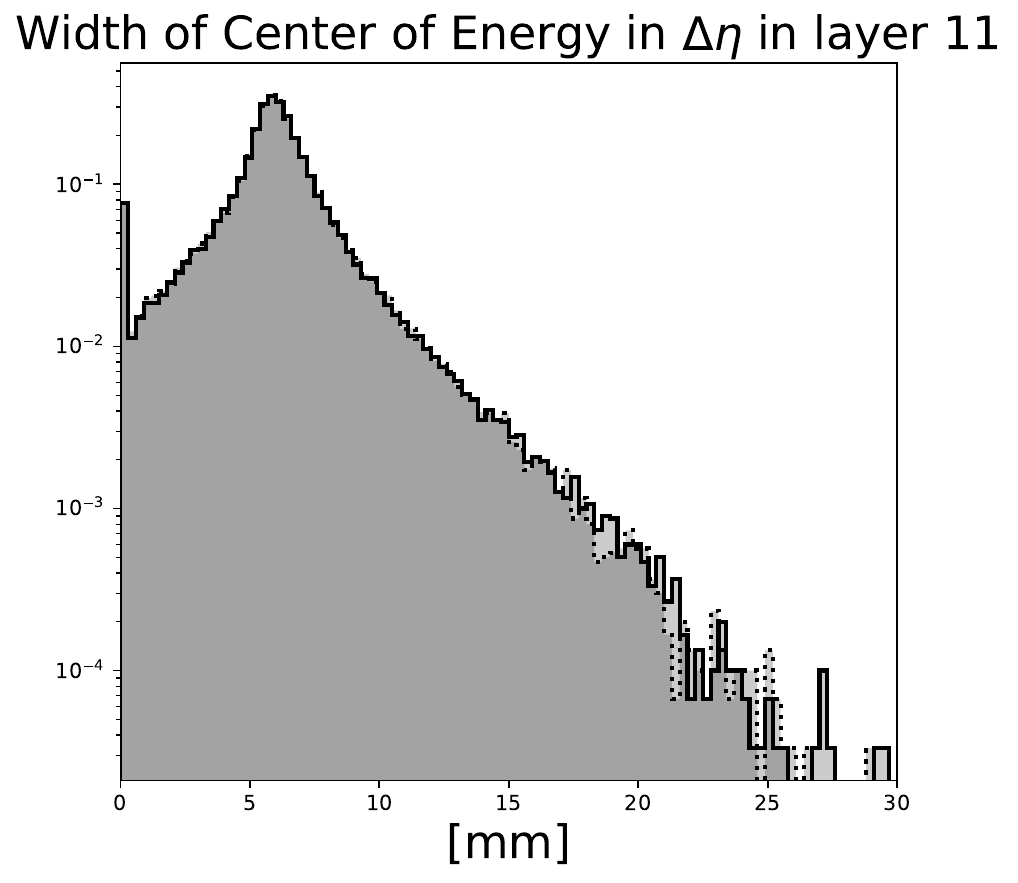} \hfill \includegraphics[height=0.1\textheight]{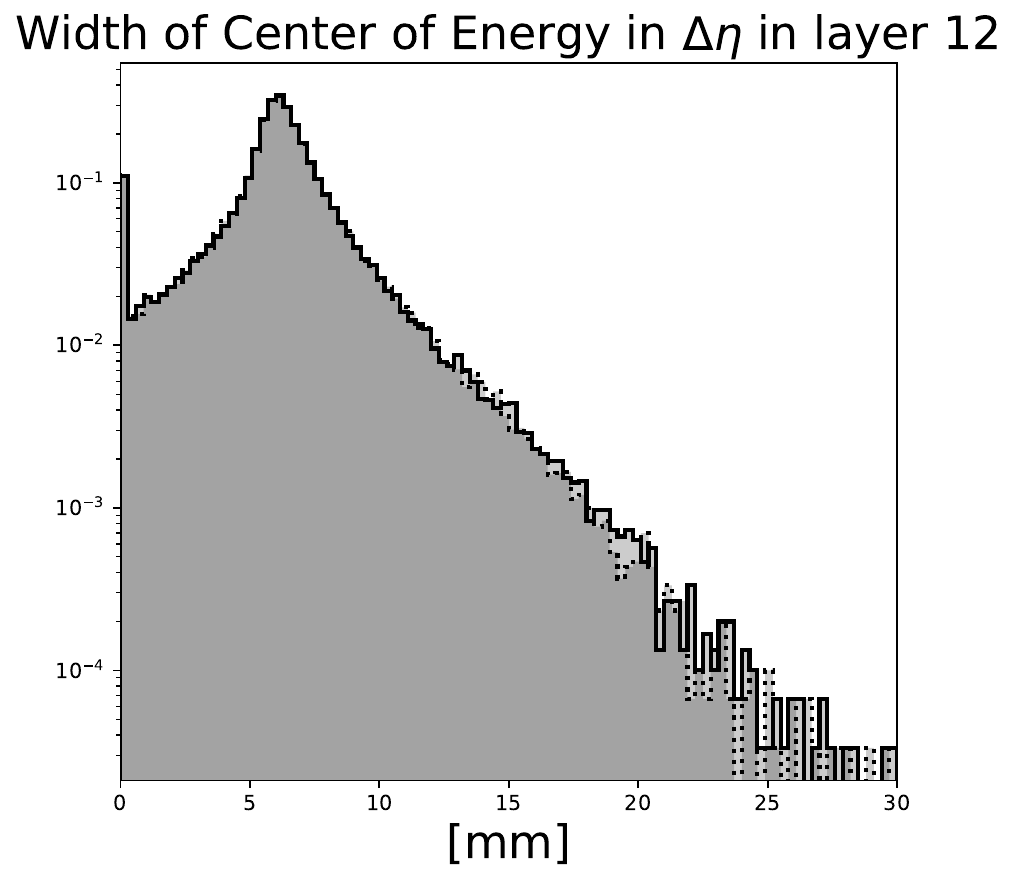} \hfill \includegraphics[height=0.1\textheight]{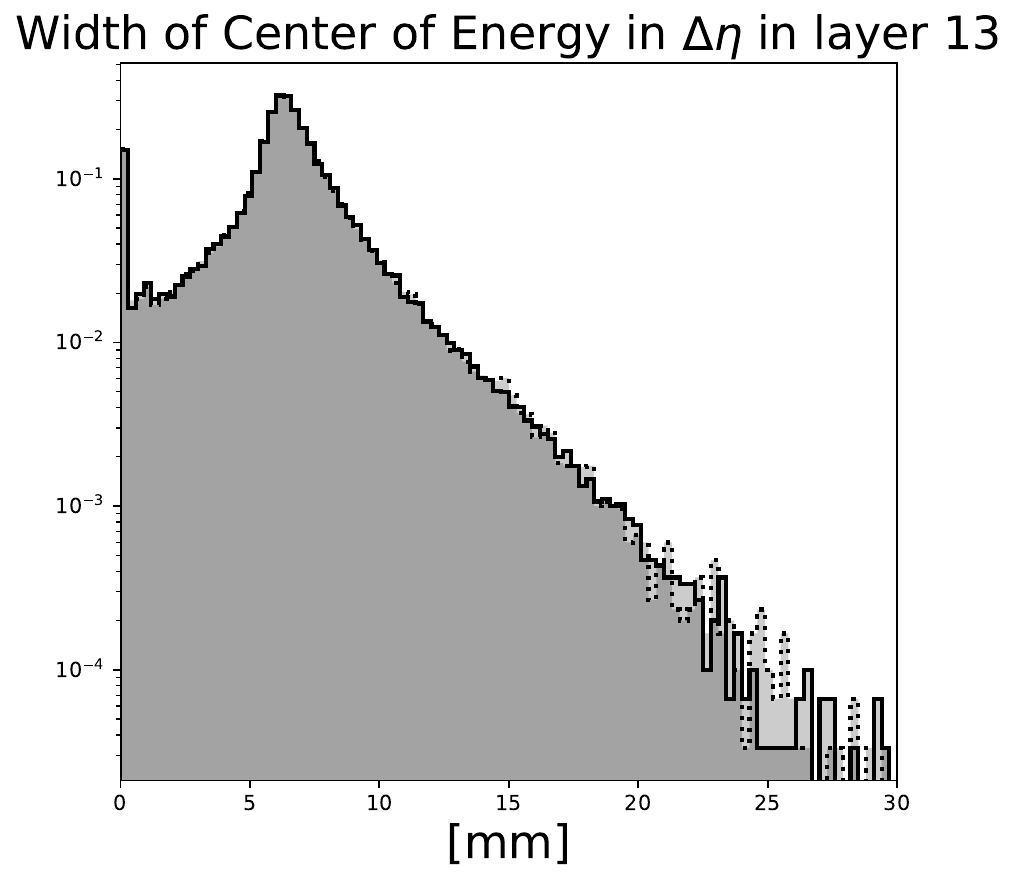} \hfill \includegraphics[height=0.1\textheight]{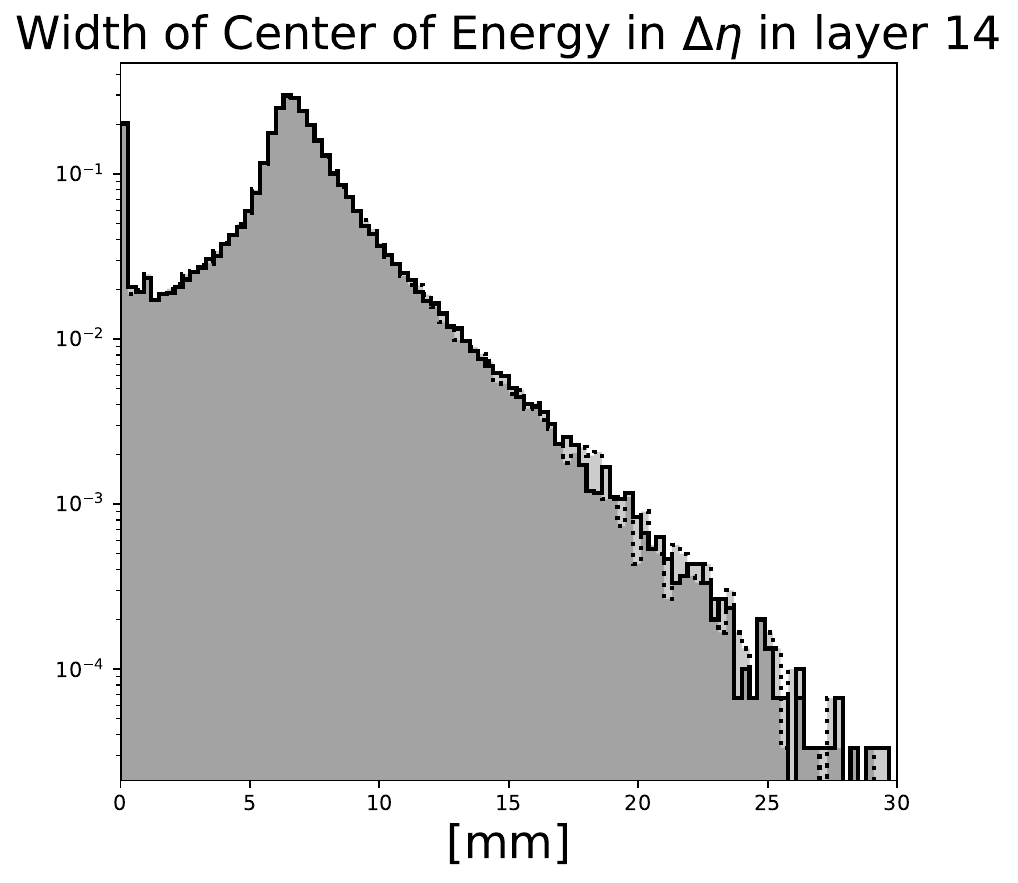}\\
    \includegraphics[height=0.1\textheight]{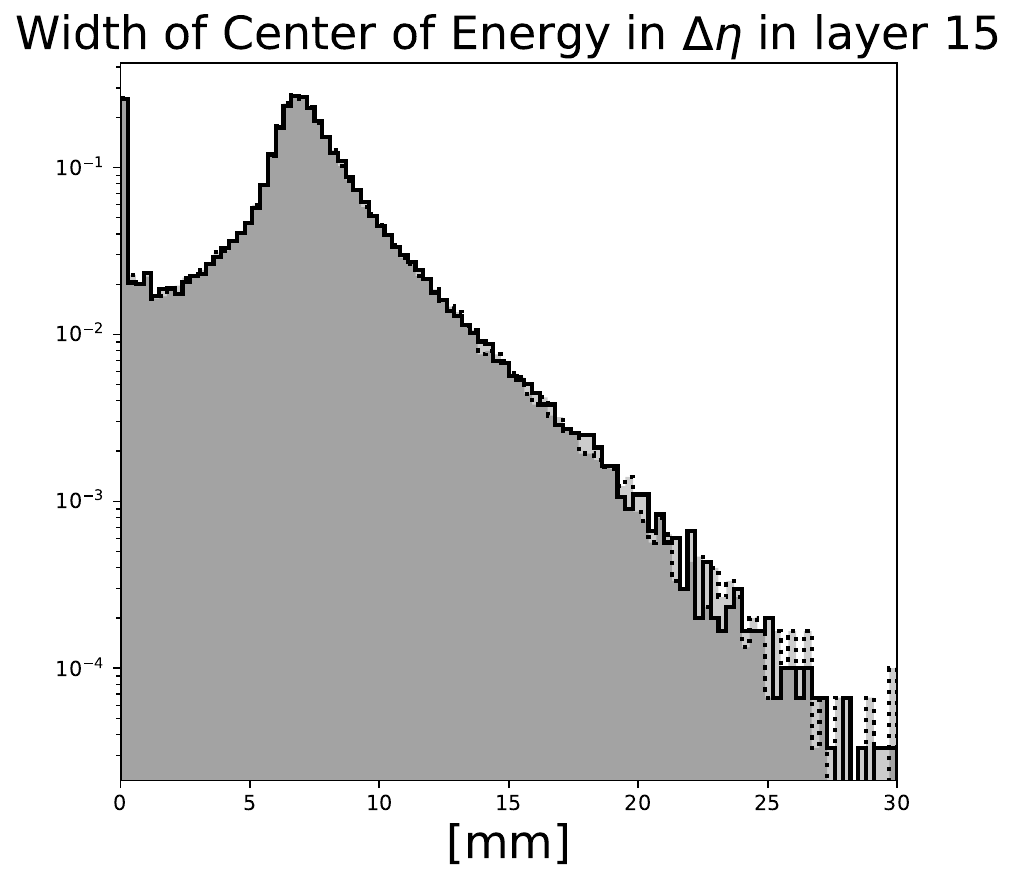} \hfill \includegraphics[height=0.1\textheight]{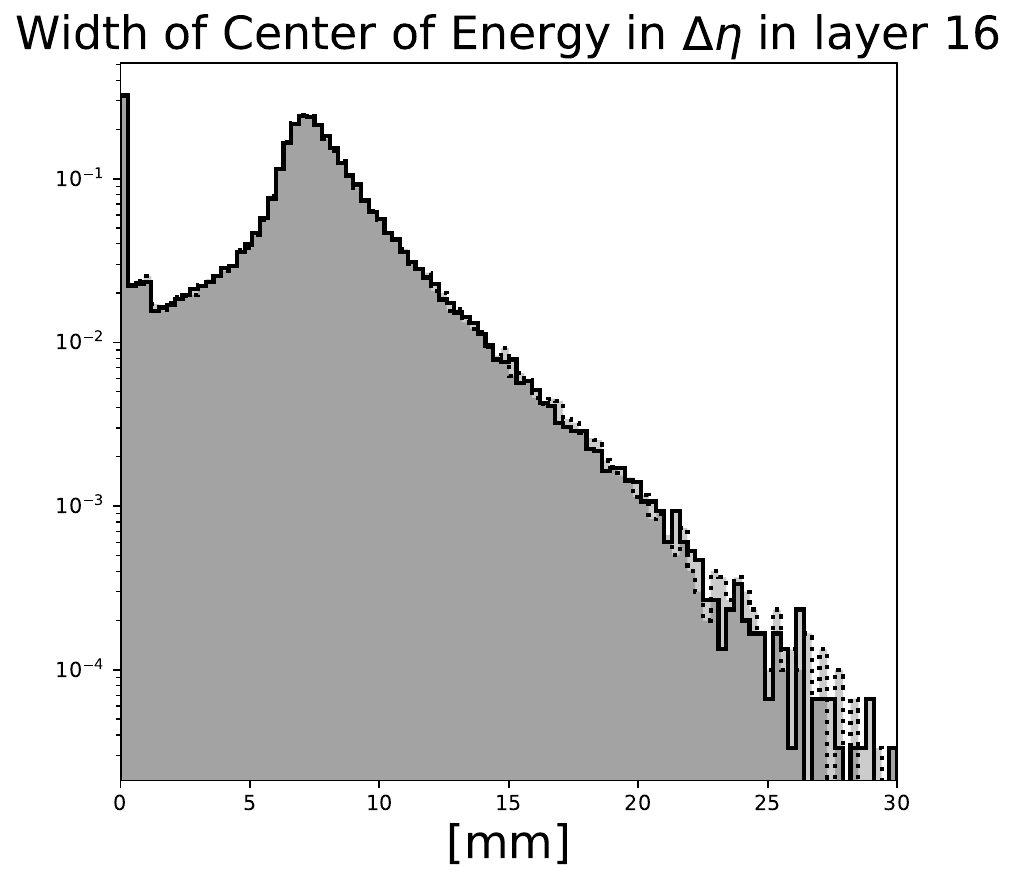} \hfill \includegraphics[height=0.1\textheight]{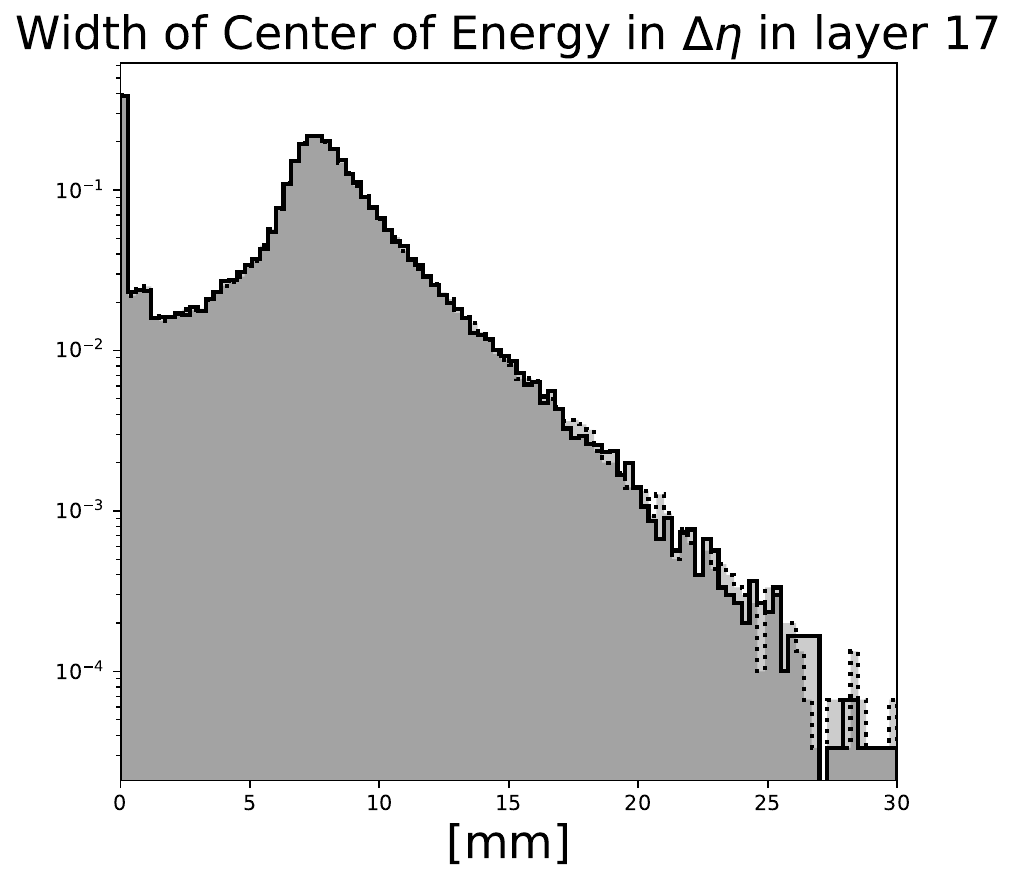} \hfill \includegraphics[height=0.1\textheight]{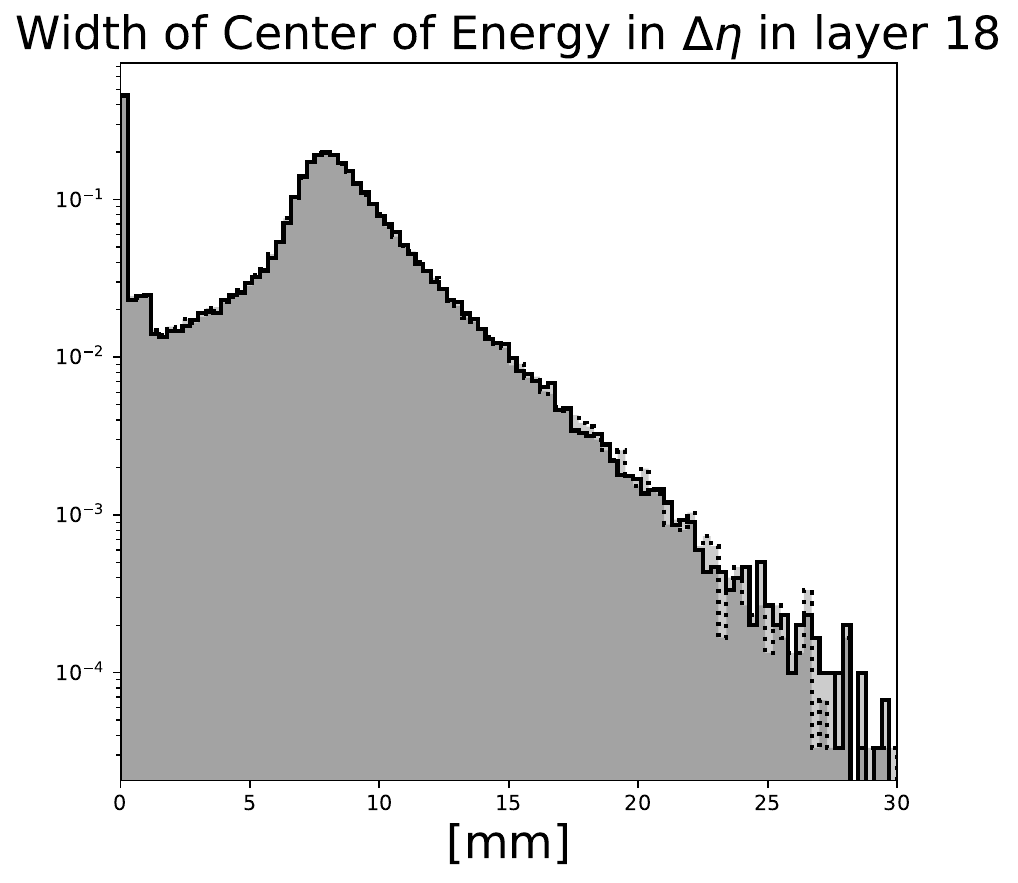} \hfill \includegraphics[height=0.1\textheight]{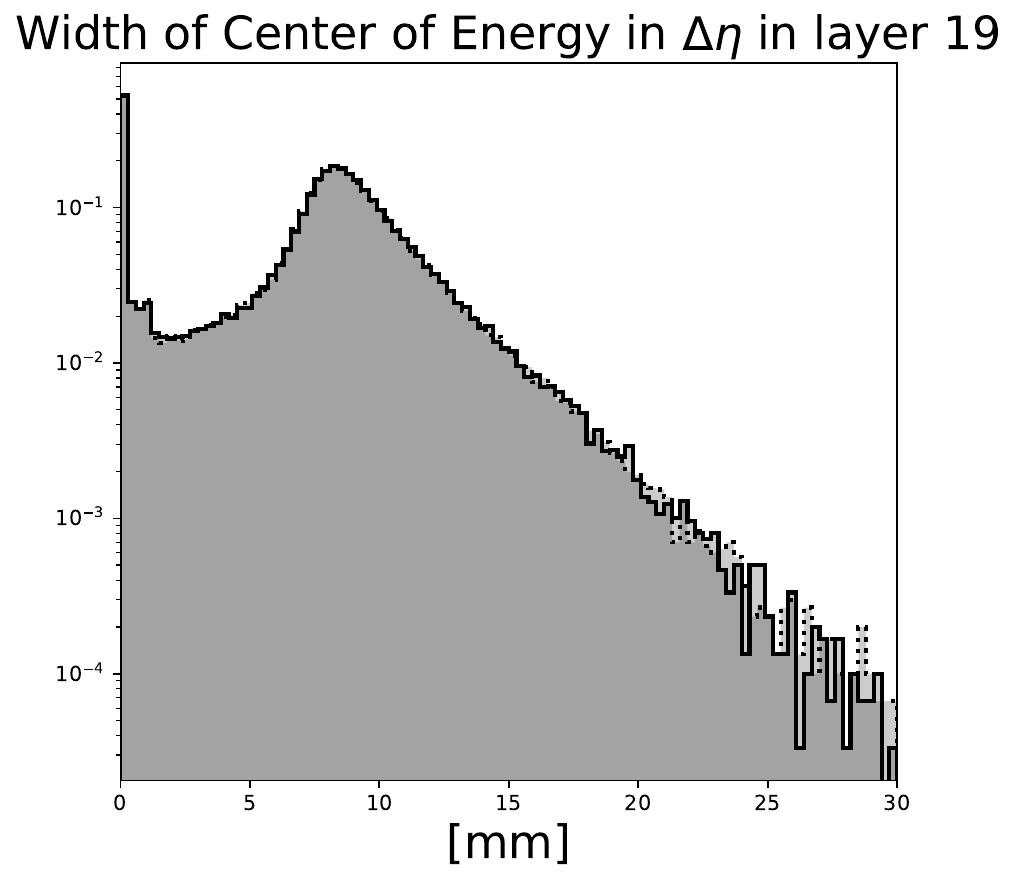}\\
    \includegraphics[height=0.1\textheight]{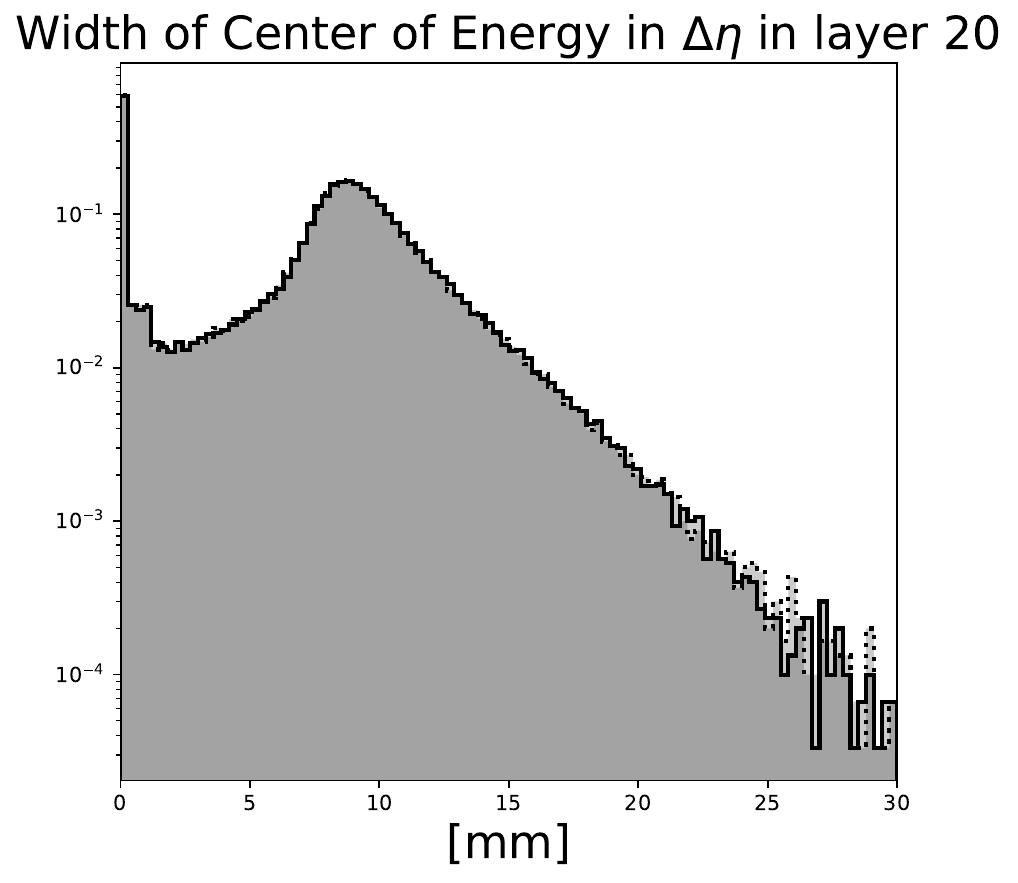} \hfill \includegraphics[height=0.1\textheight]{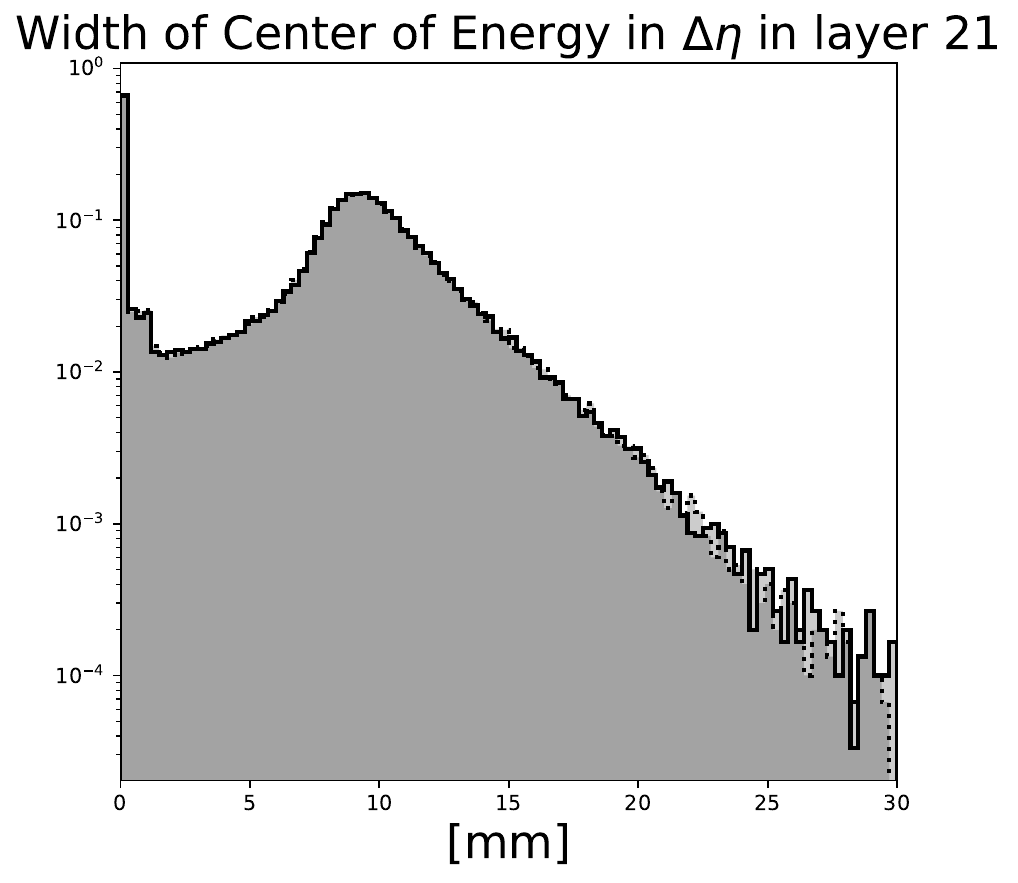} \hfill \includegraphics[height=0.1\textheight]{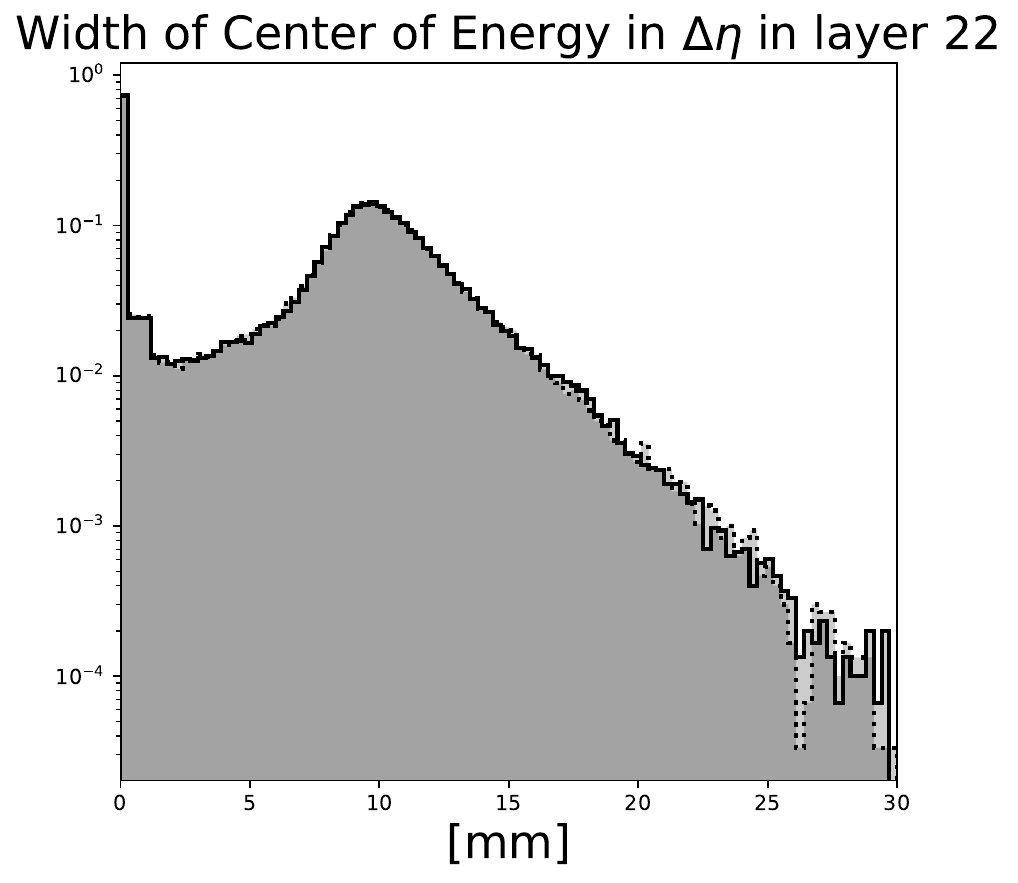} \hfill \includegraphics[height=0.1\textheight]{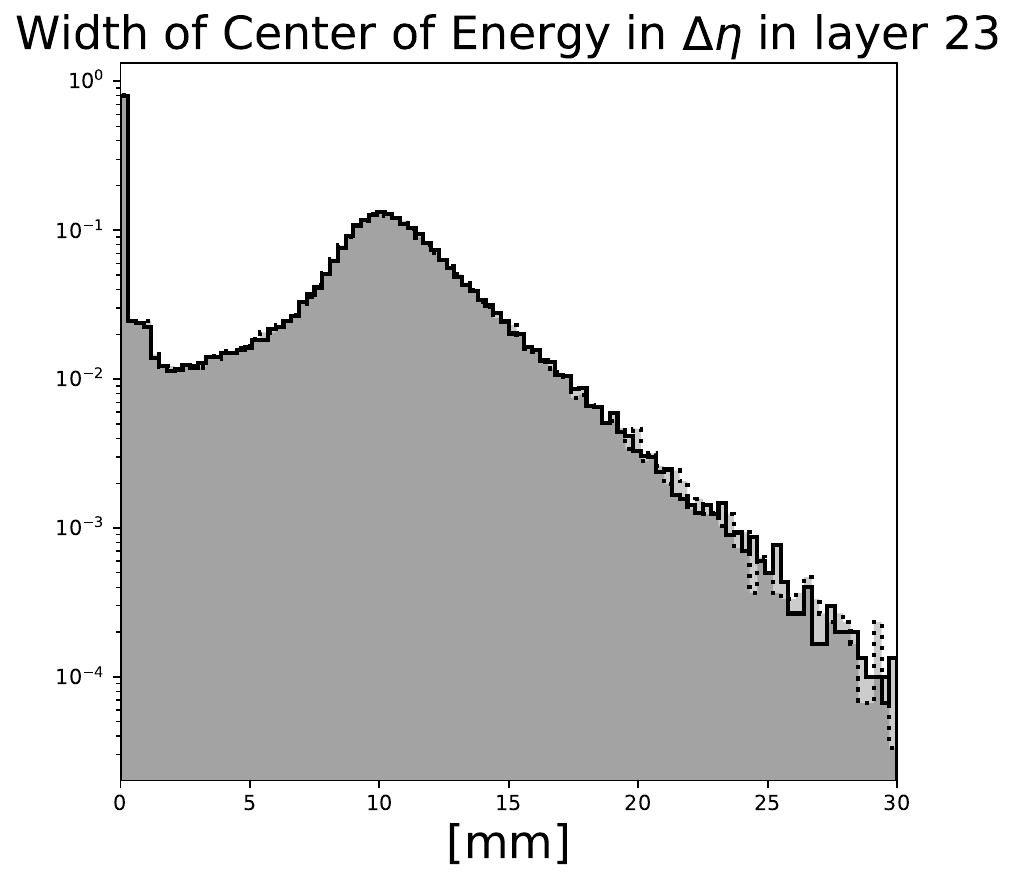} \hfill \includegraphics[height=0.1\textheight]{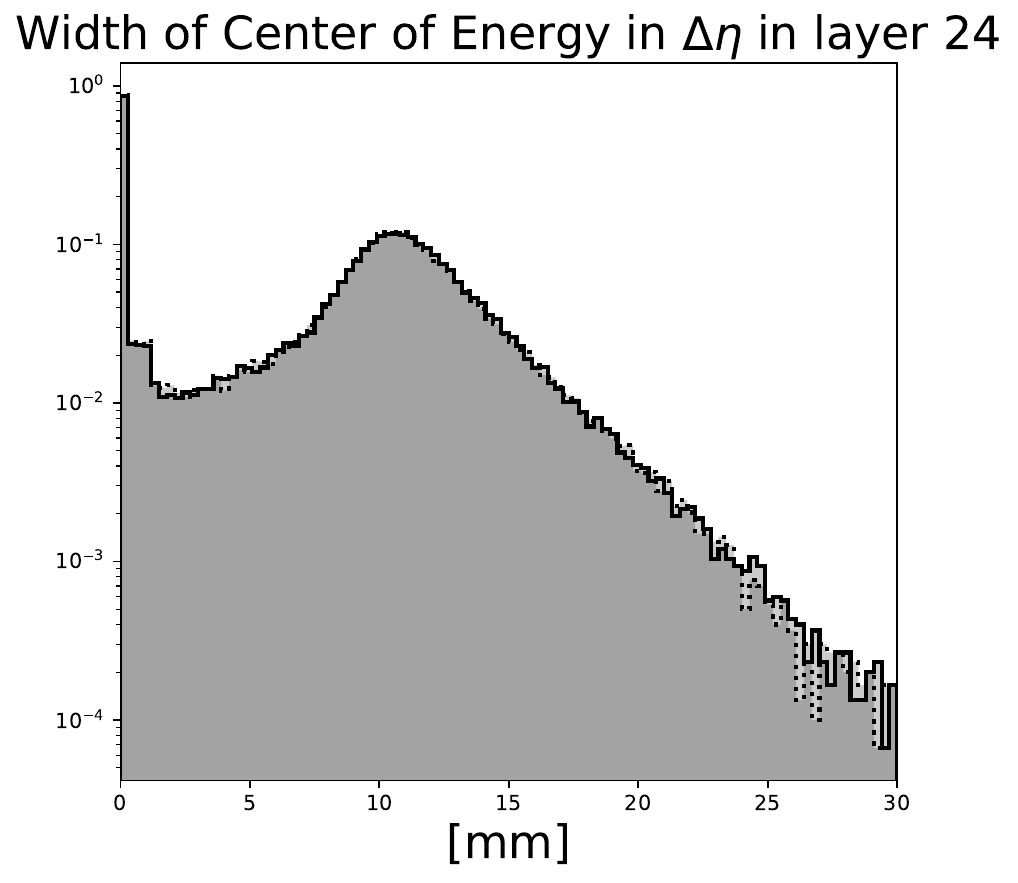}\\
    \includegraphics[height=0.1\textheight]{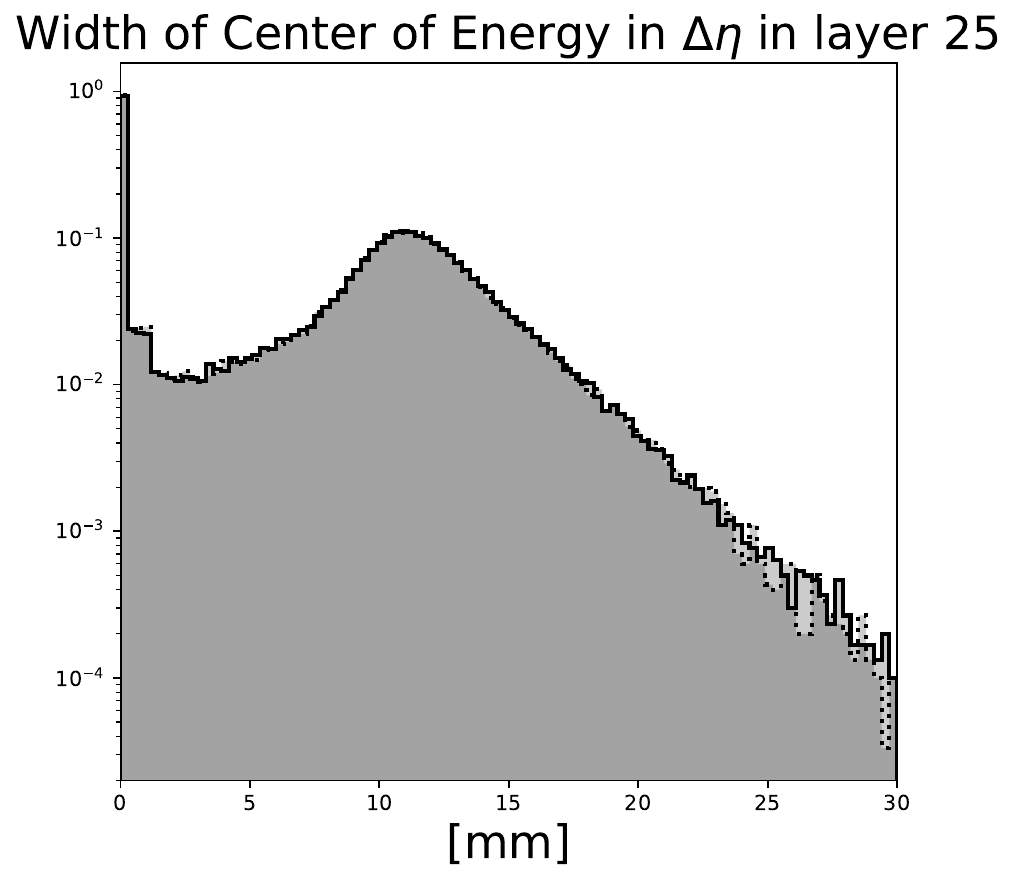} \hfill \includegraphics[height=0.1\textheight]{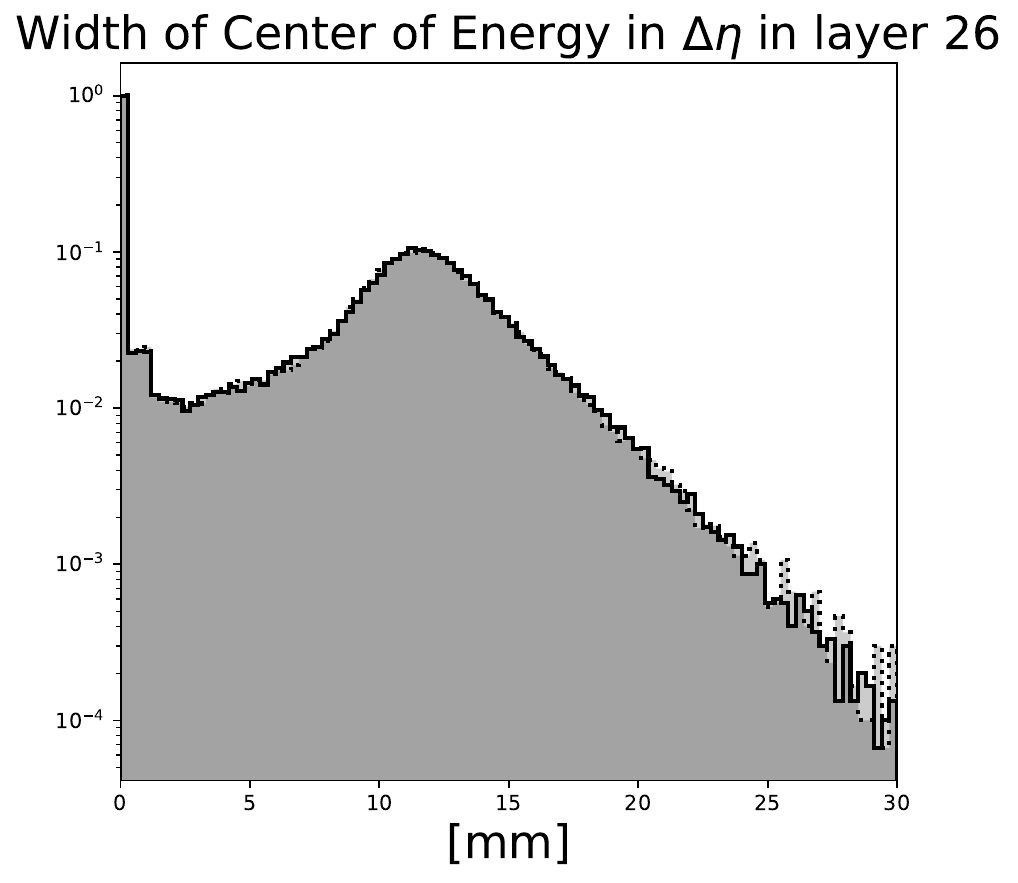} \hfill \includegraphics[height=0.1\textheight]{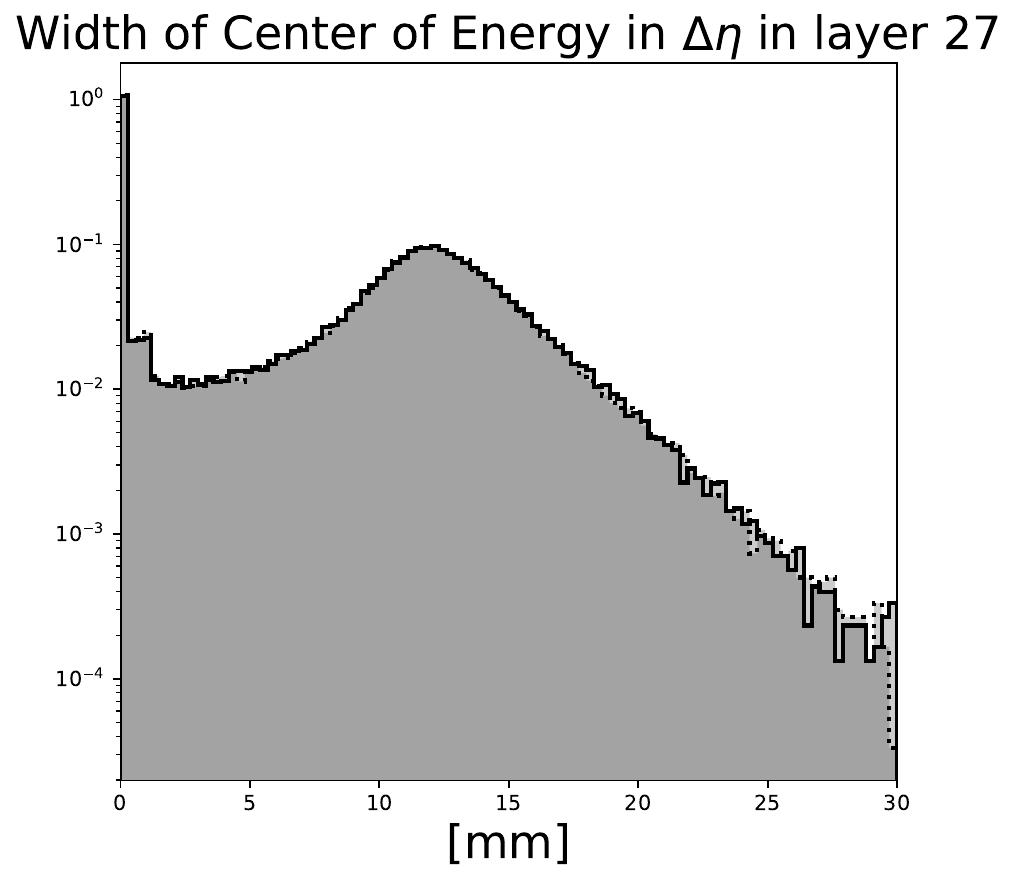} \hfill \includegraphics[height=0.1\textheight]{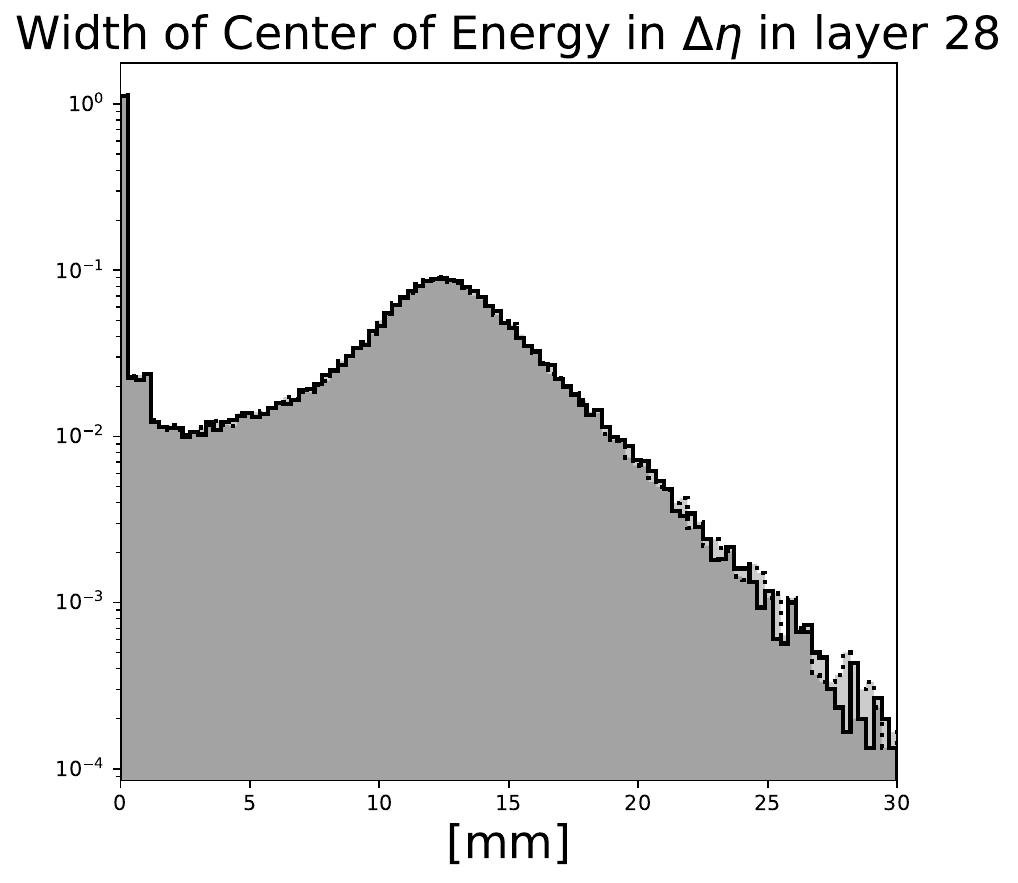} \hfill \includegraphics[height=0.1\textheight]{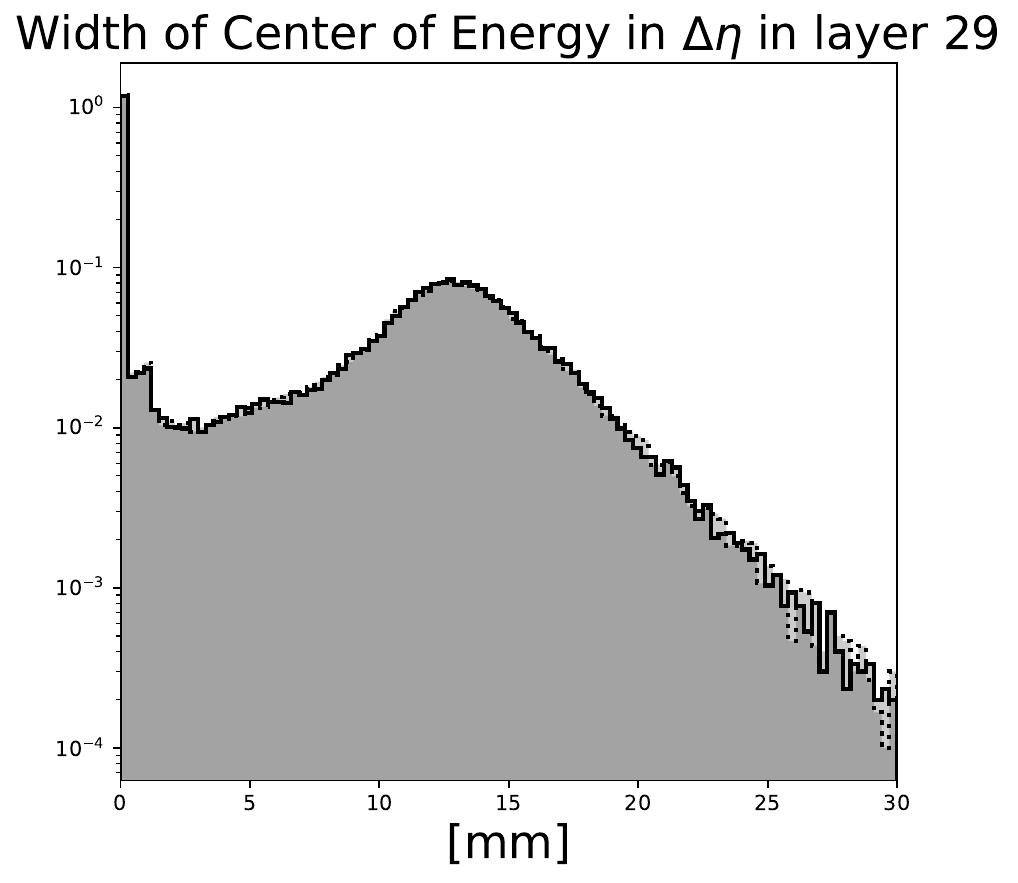}\\
    \includegraphics[height=0.1\textheight]{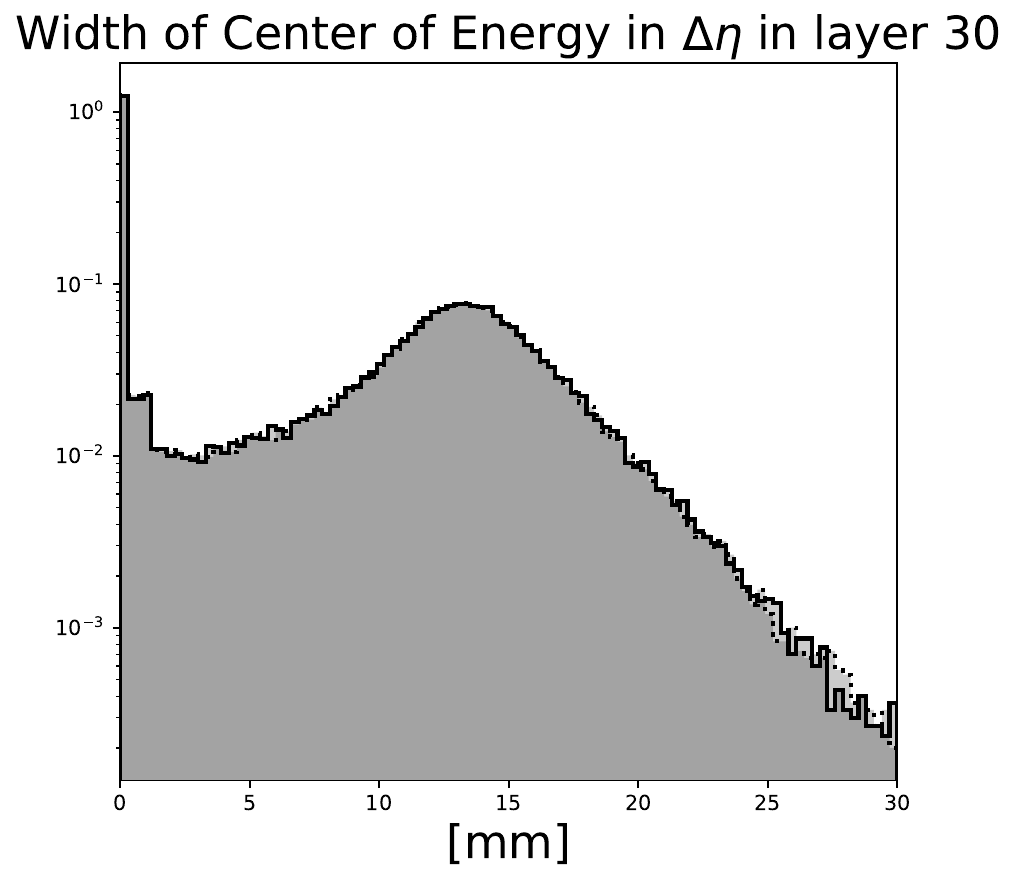} \hfill \includegraphics[height=0.1\textheight]{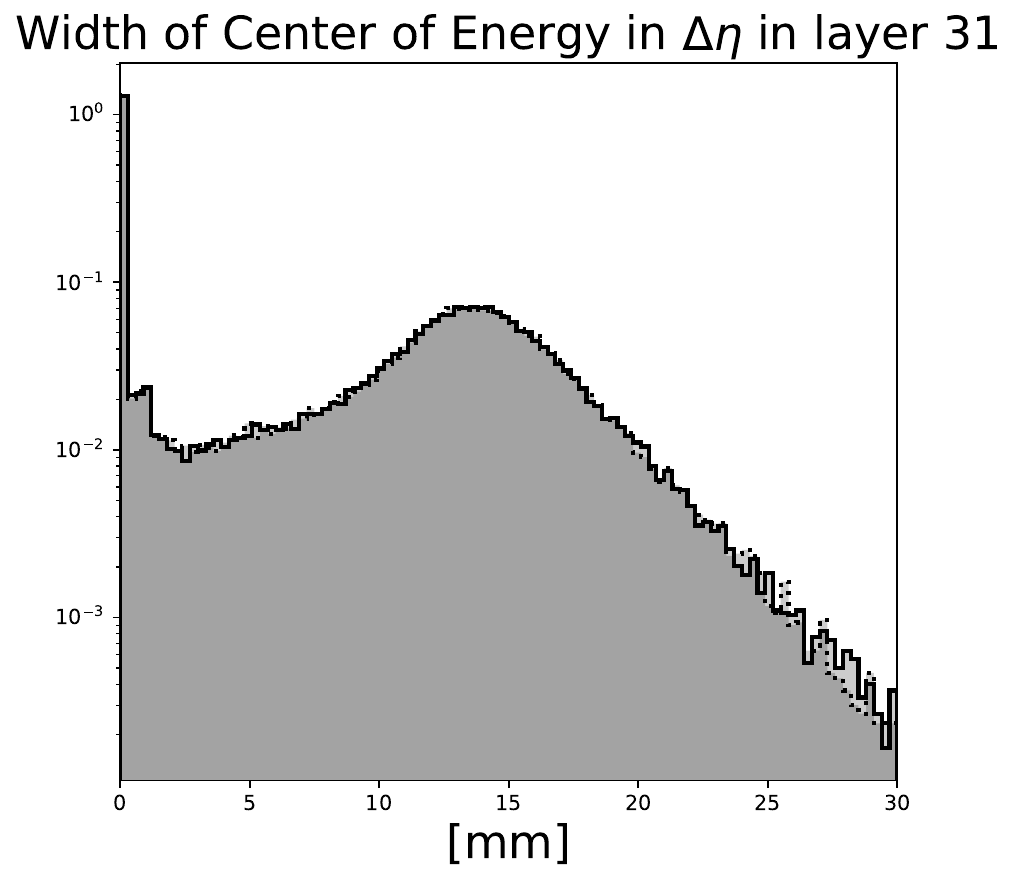} \hfill \includegraphics[height=0.1\textheight]{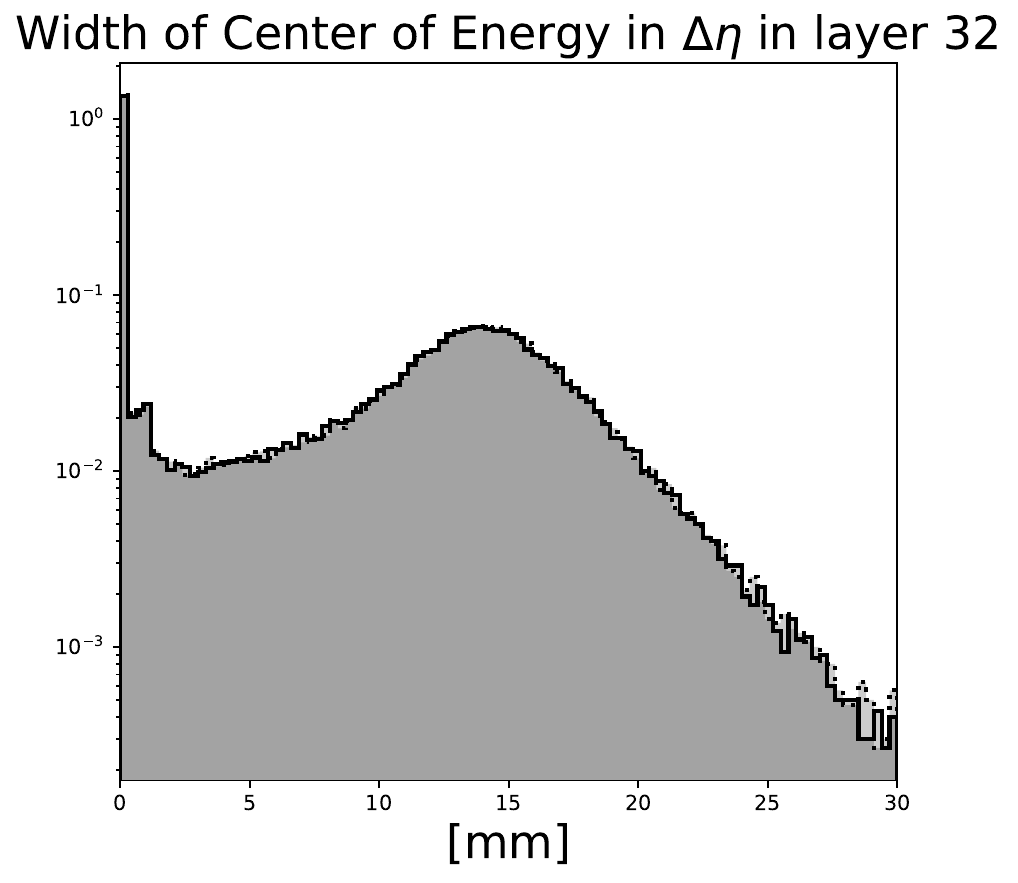} \hfill \includegraphics[height=0.1\textheight]{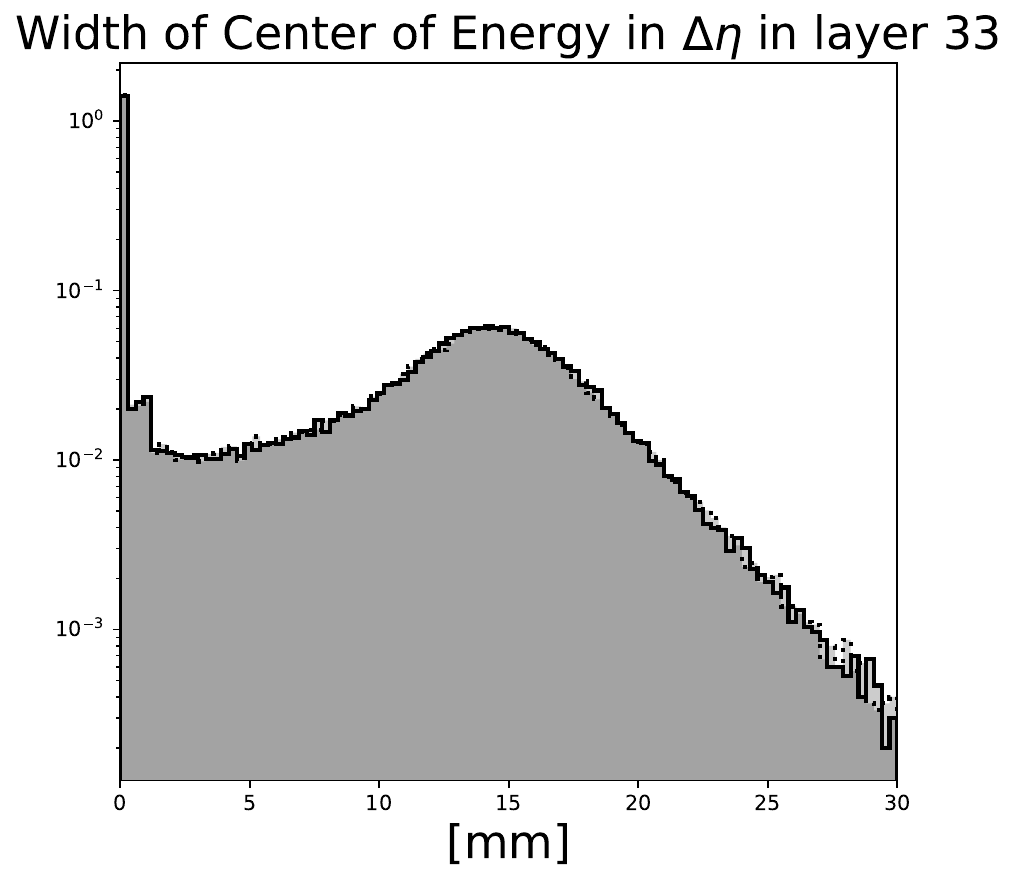} \hfill \includegraphics[height=0.1\textheight]{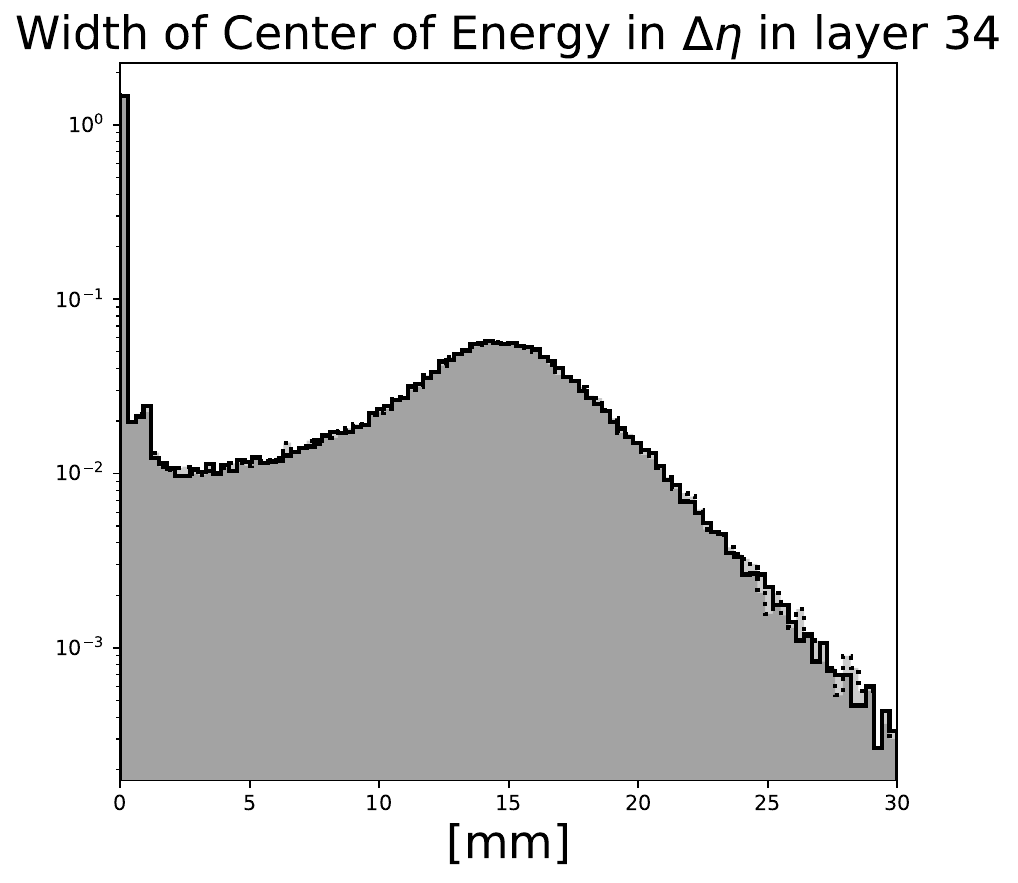}\\
    \includegraphics[height=0.1\textheight]{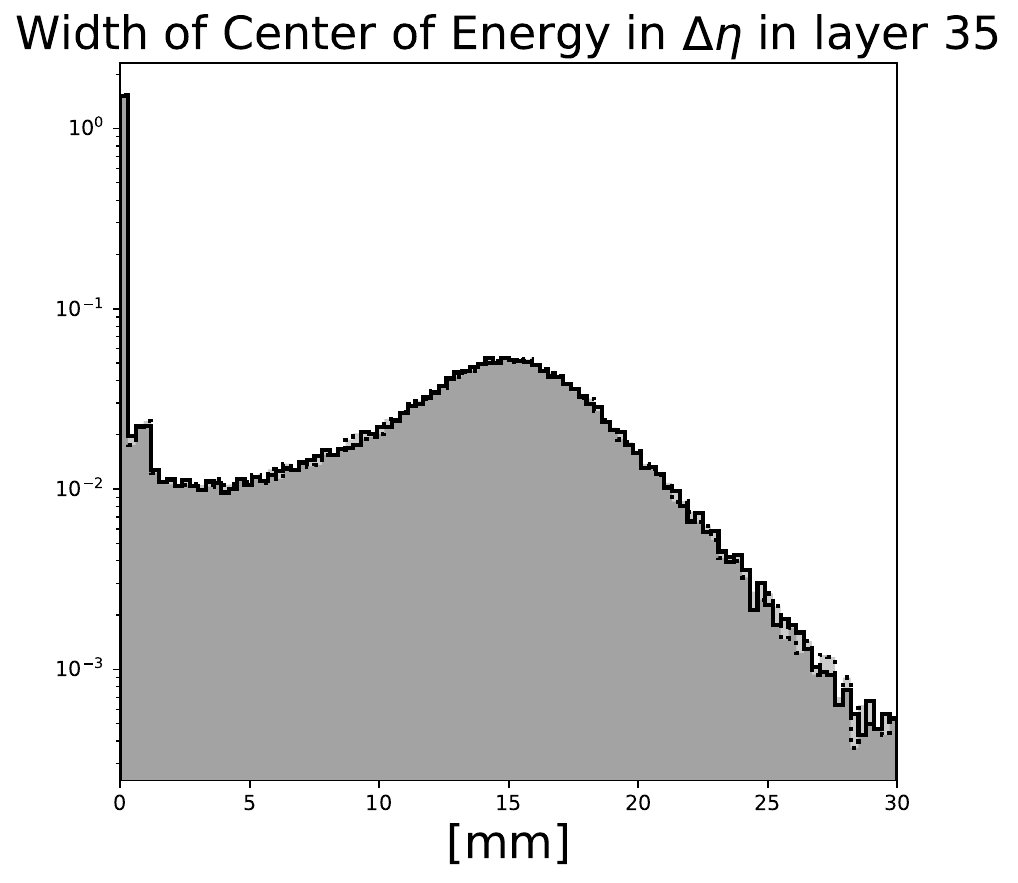} \hfill \includegraphics[height=0.1\textheight]{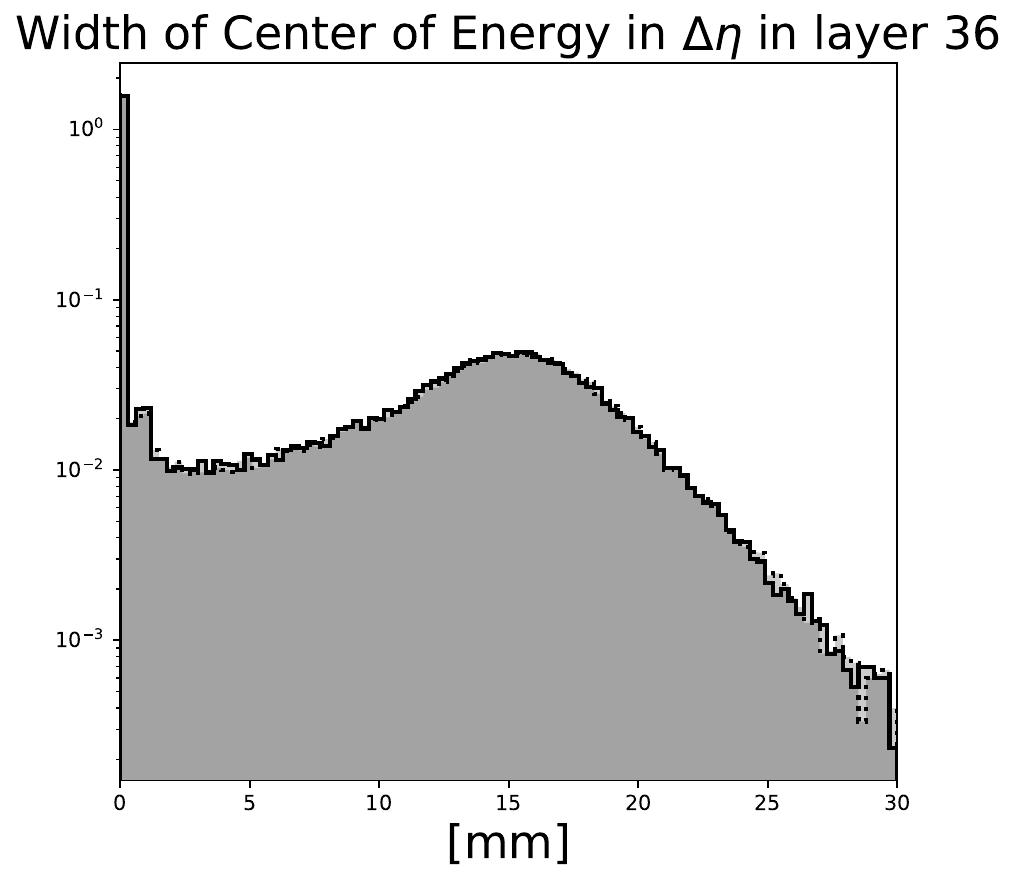} \hfill \includegraphics[height=0.1\textheight]{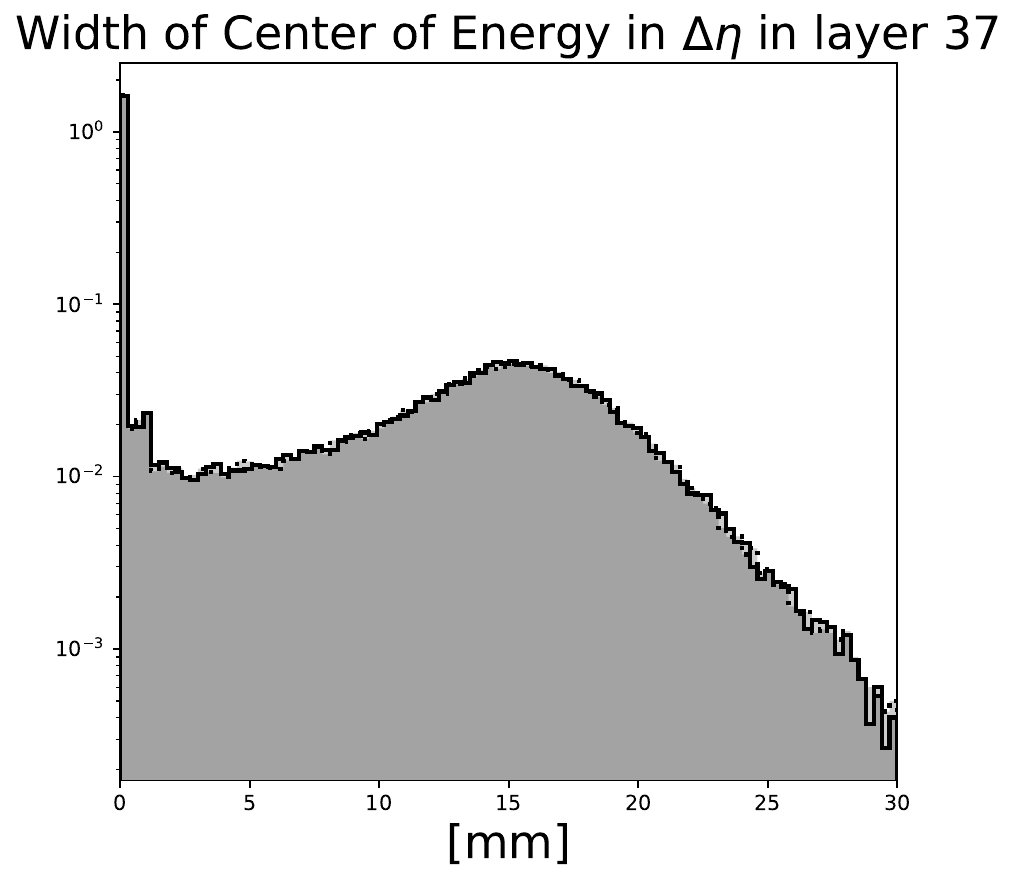} \hfill \includegraphics[height=0.1\textheight]{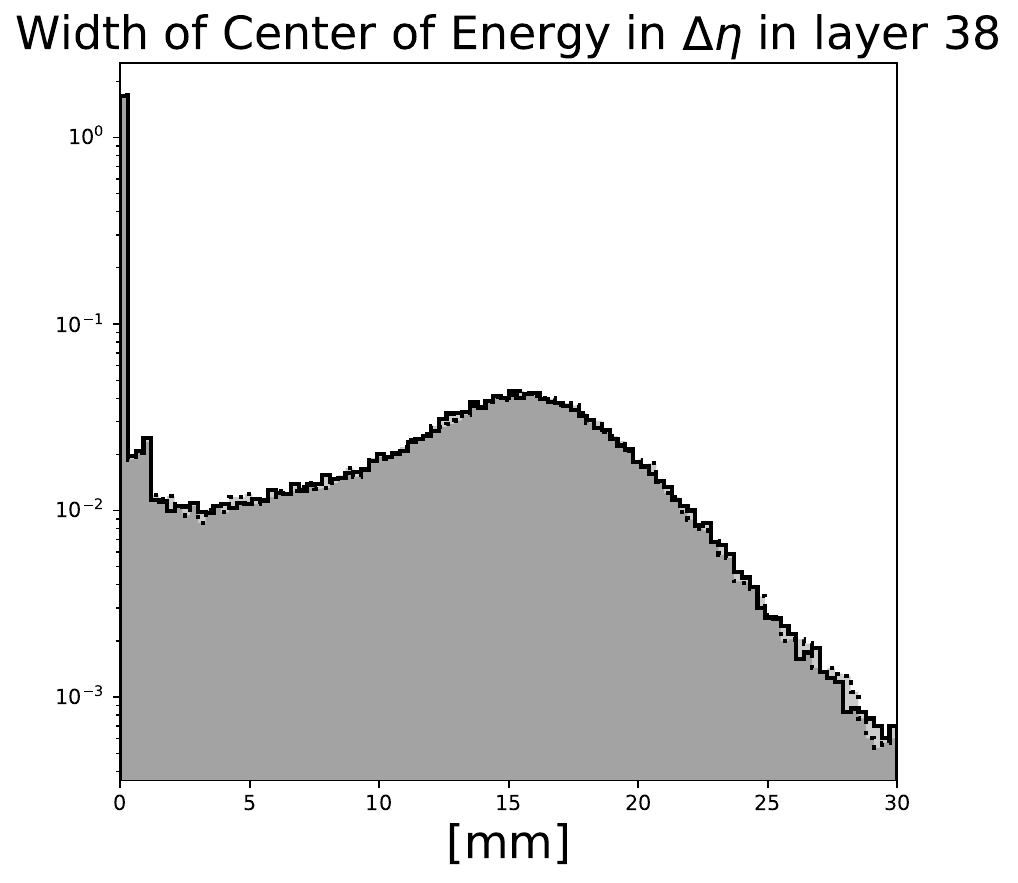} \hfill \includegraphics[height=0.1\textheight]{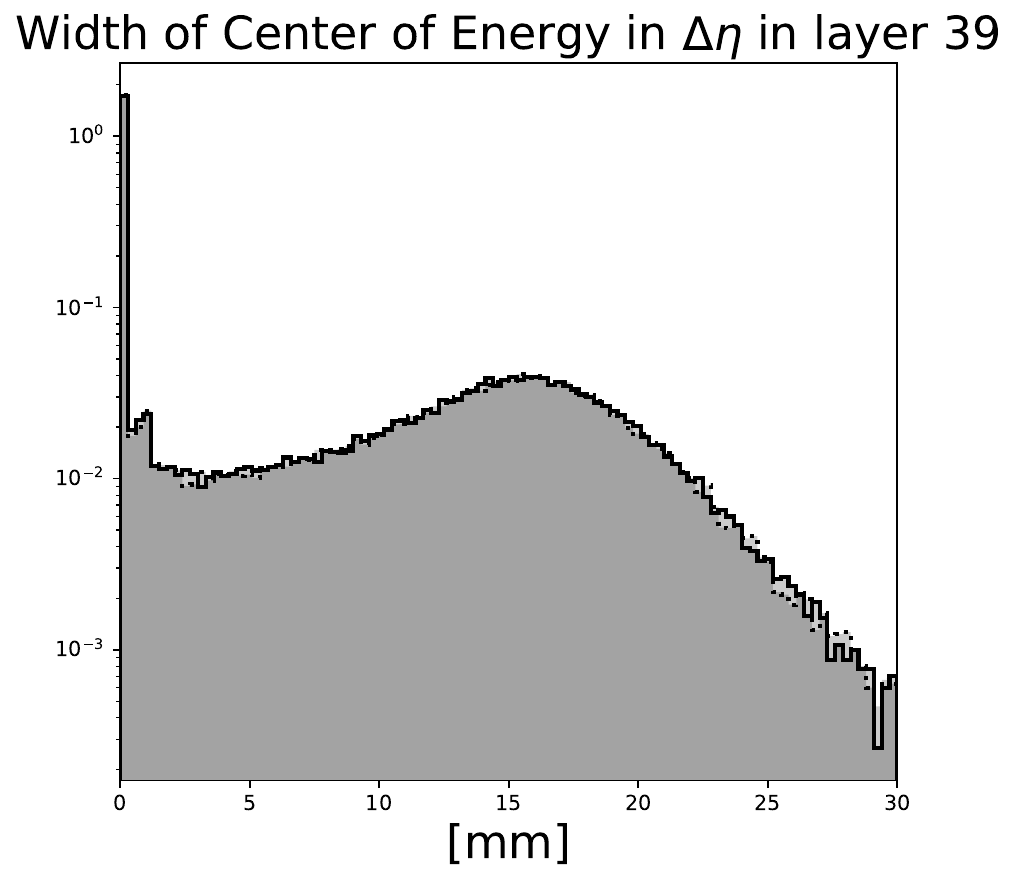}\\
    \includegraphics[height=0.1\textheight]{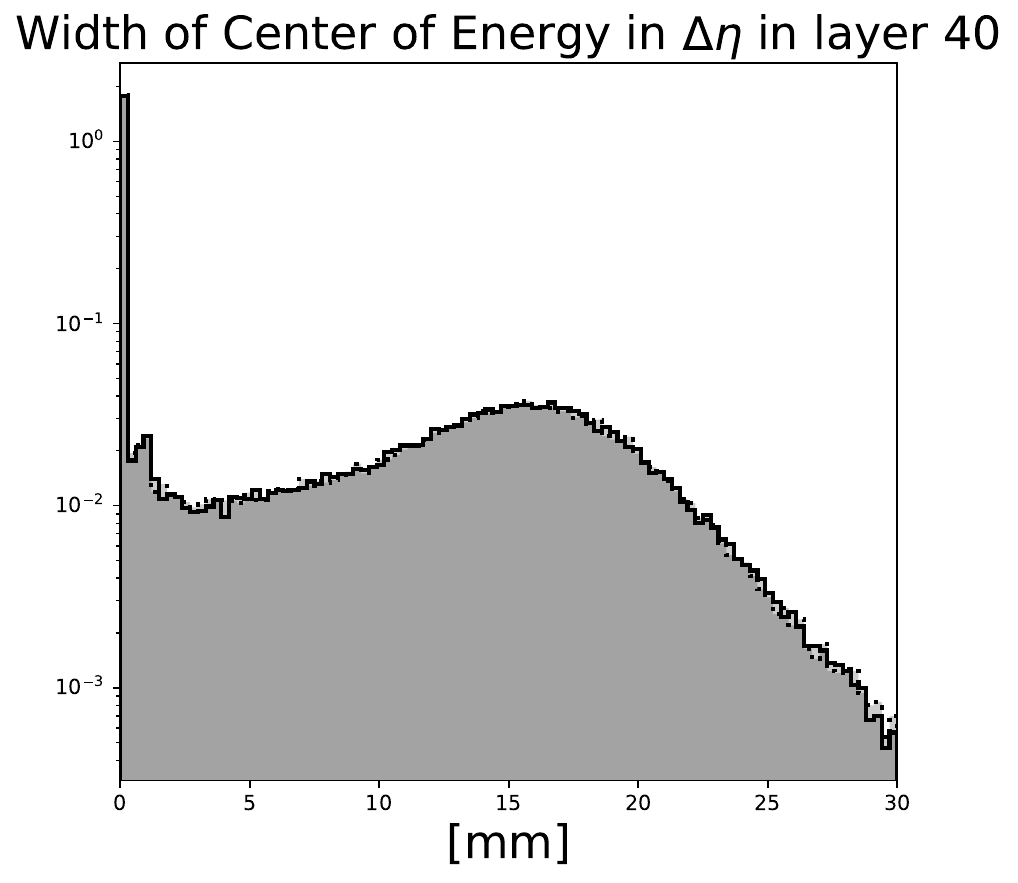} \hfill \includegraphics[height=0.1\textheight]{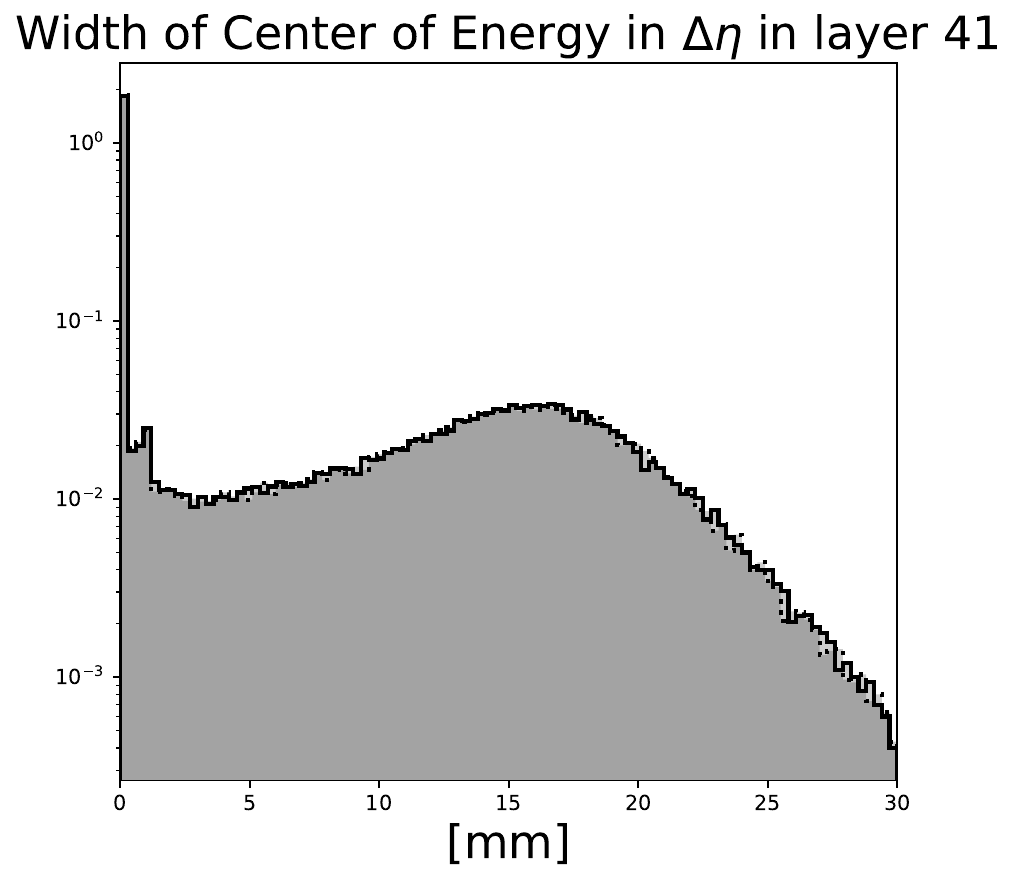} \hfill \includegraphics[height=0.1\textheight]{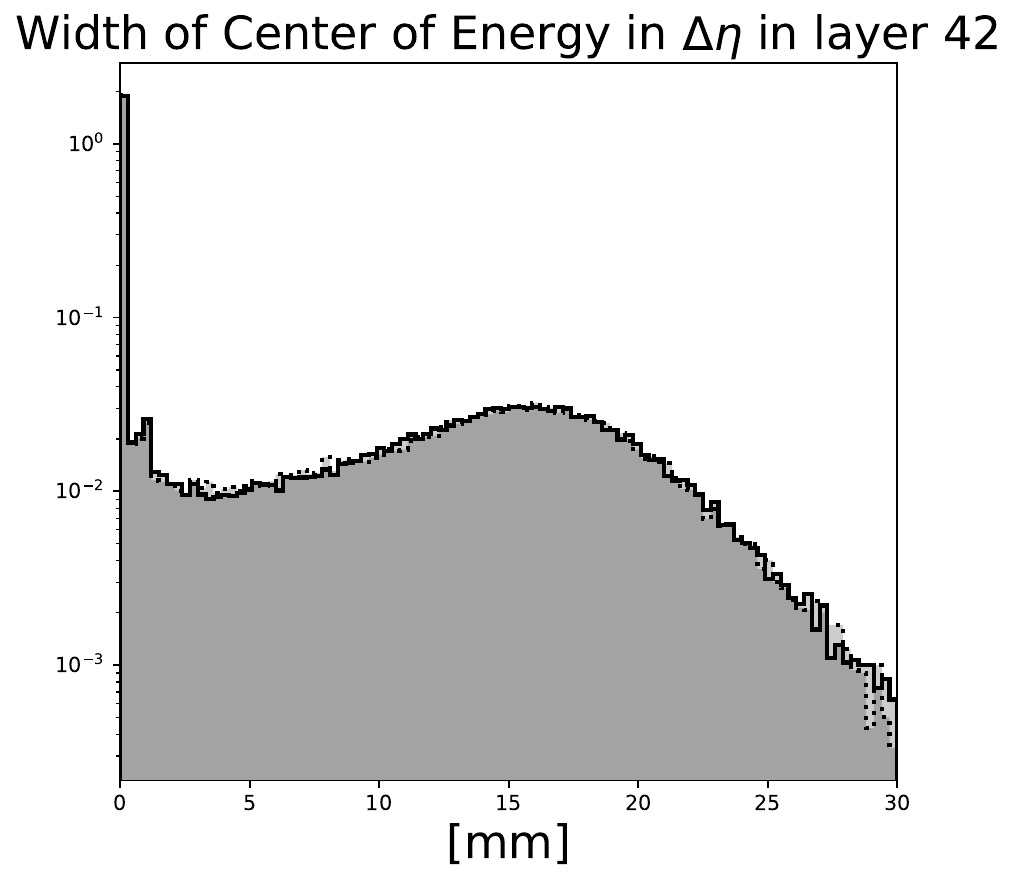} \hfill \includegraphics[height=0.1\textheight]{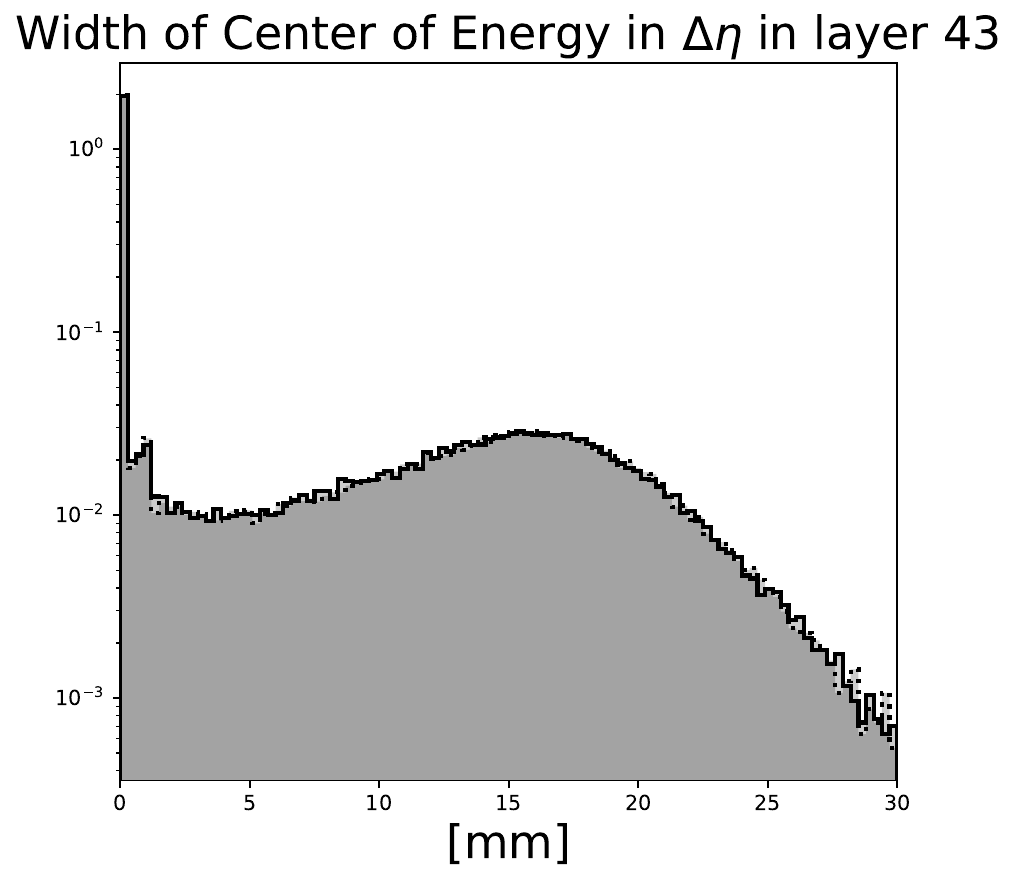} \hfill \includegraphics[height=0.1\textheight]{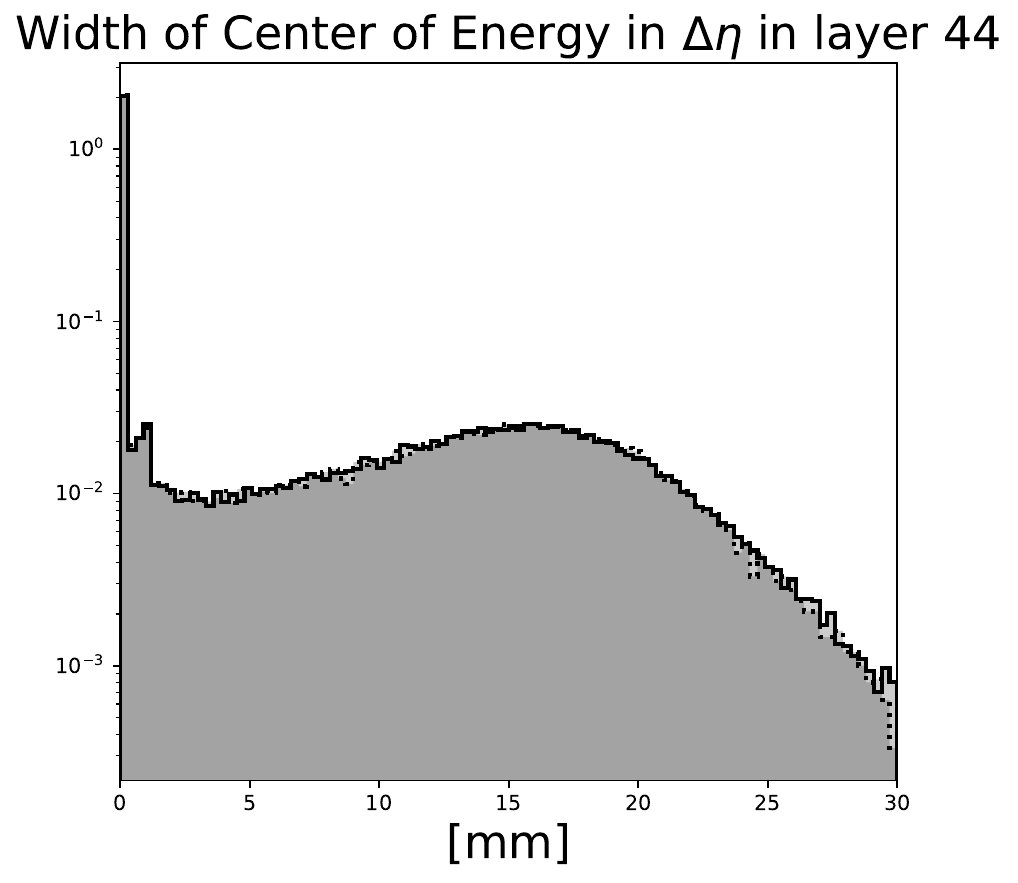}\\
    \includegraphics[width=0.5\textwidth]{figures/Appendix_reference/legend.pdf}
    \caption{Distribution of \geant training and evaluation data in width of the centers of energy in $\eta$ direction for ds3. }
    \label{fig:app_ref.ds3.4}
\end{figure}

\begin{figure}[ht]
    \centering
    \includegraphics[height=0.1\textheight]{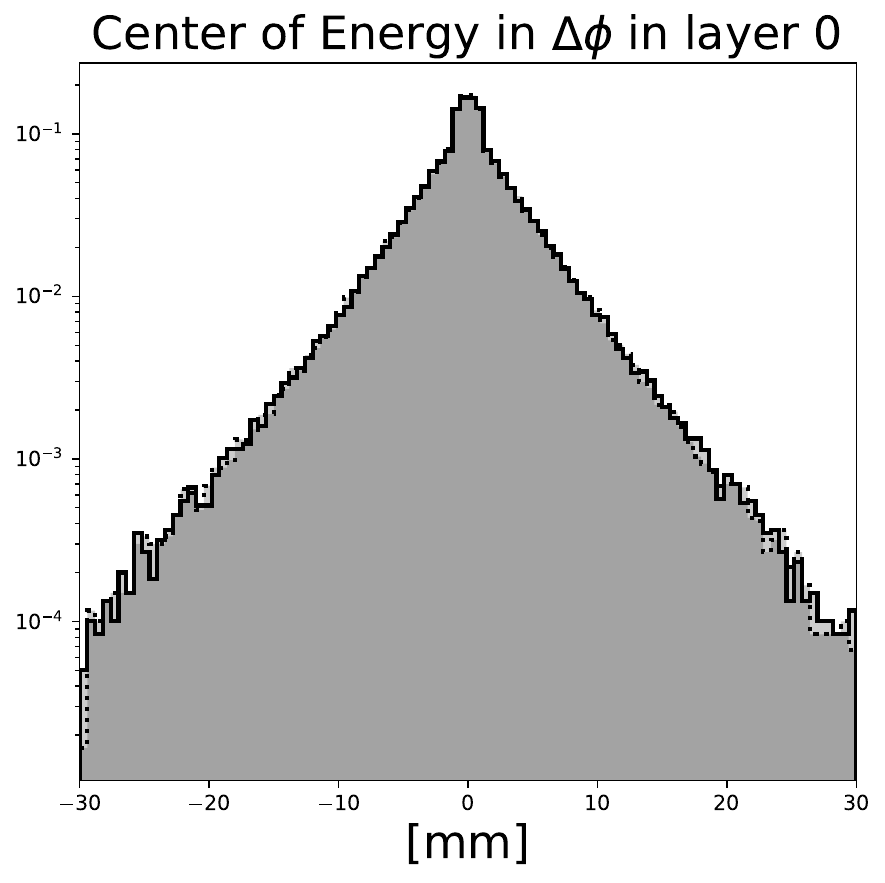} \hfill \includegraphics[height=0.1\textheight]{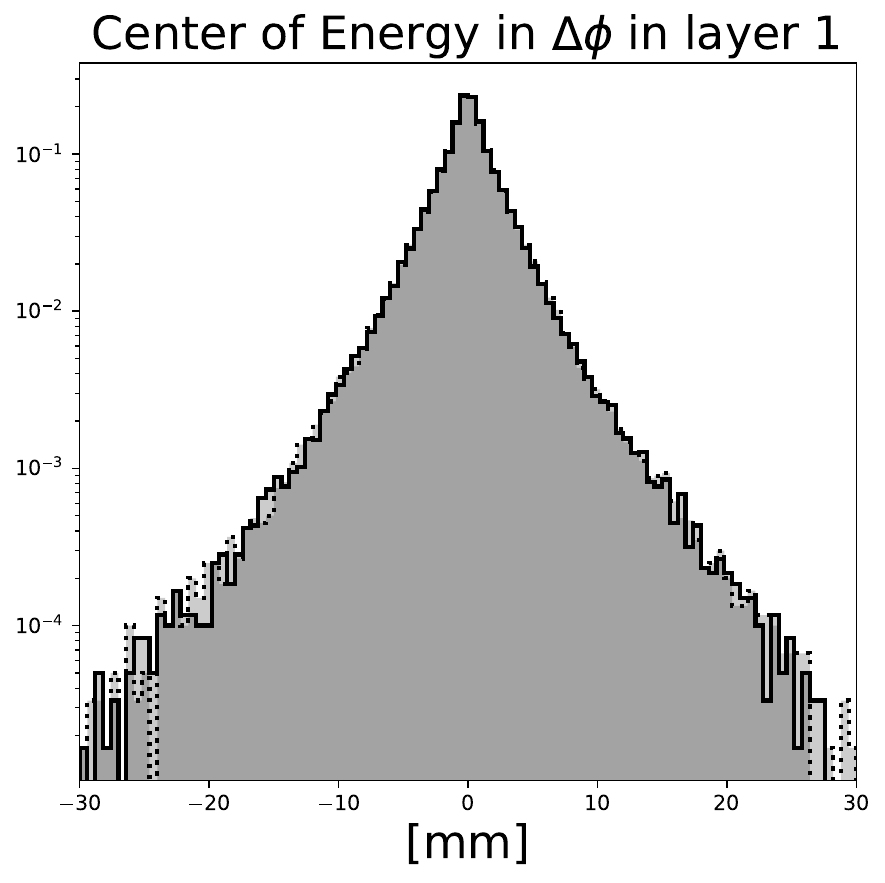} \hfill \includegraphics[height=0.1\textheight]{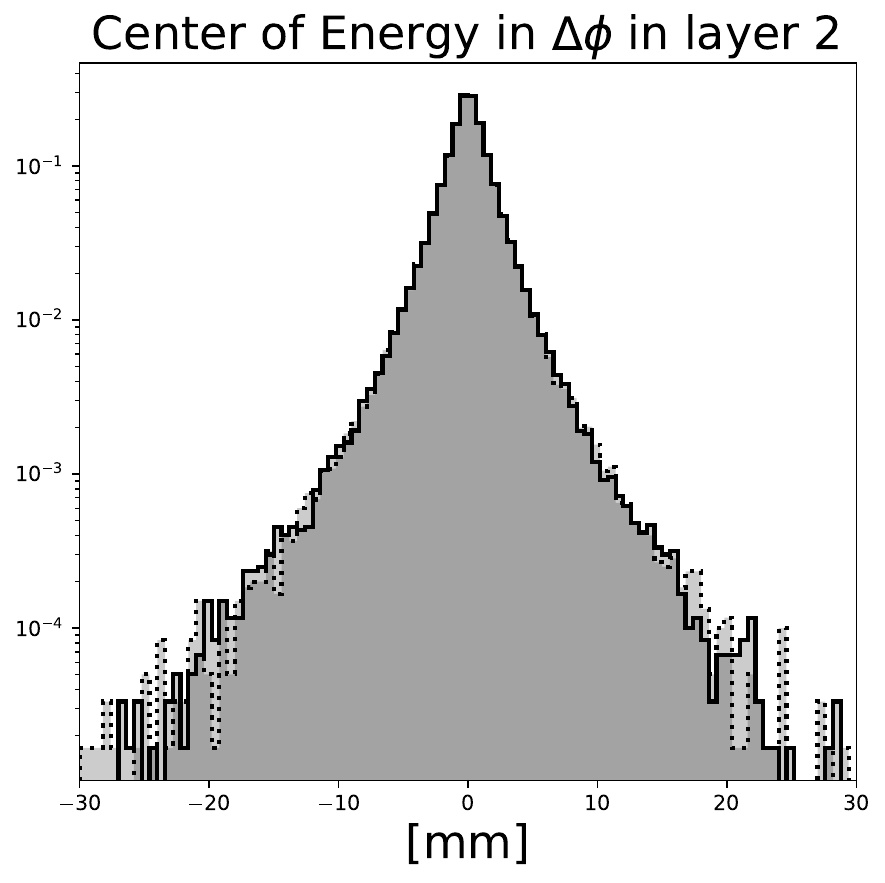} \hfill \includegraphics[height=0.1\textheight]{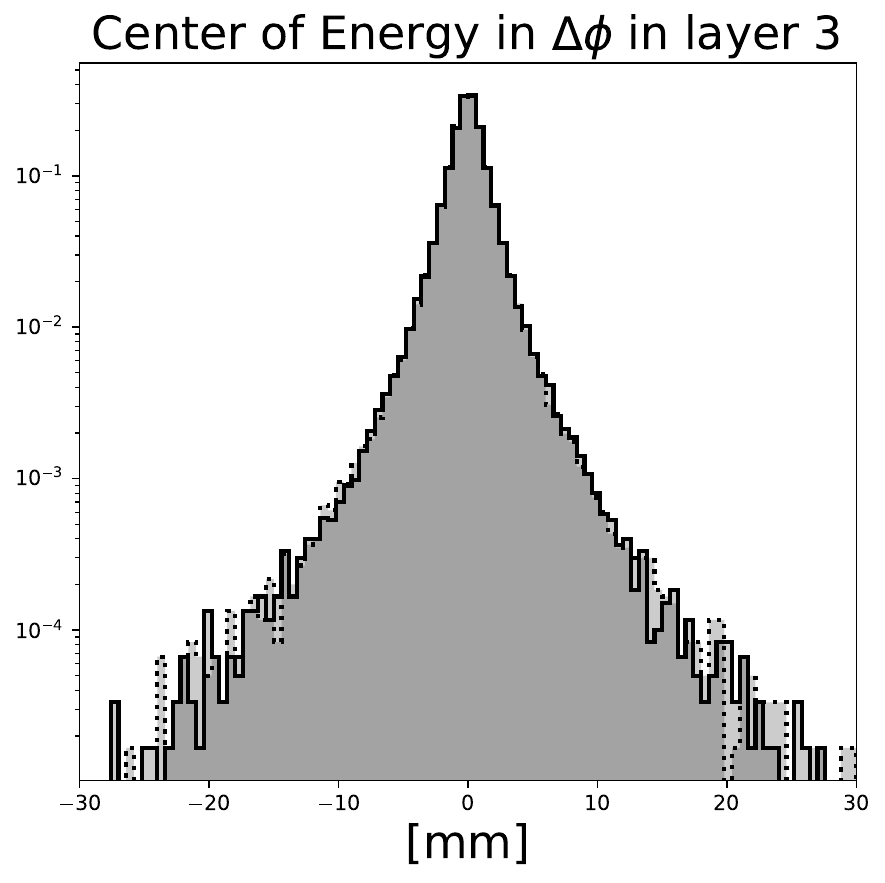} \hfill \includegraphics[height=0.1\textheight]{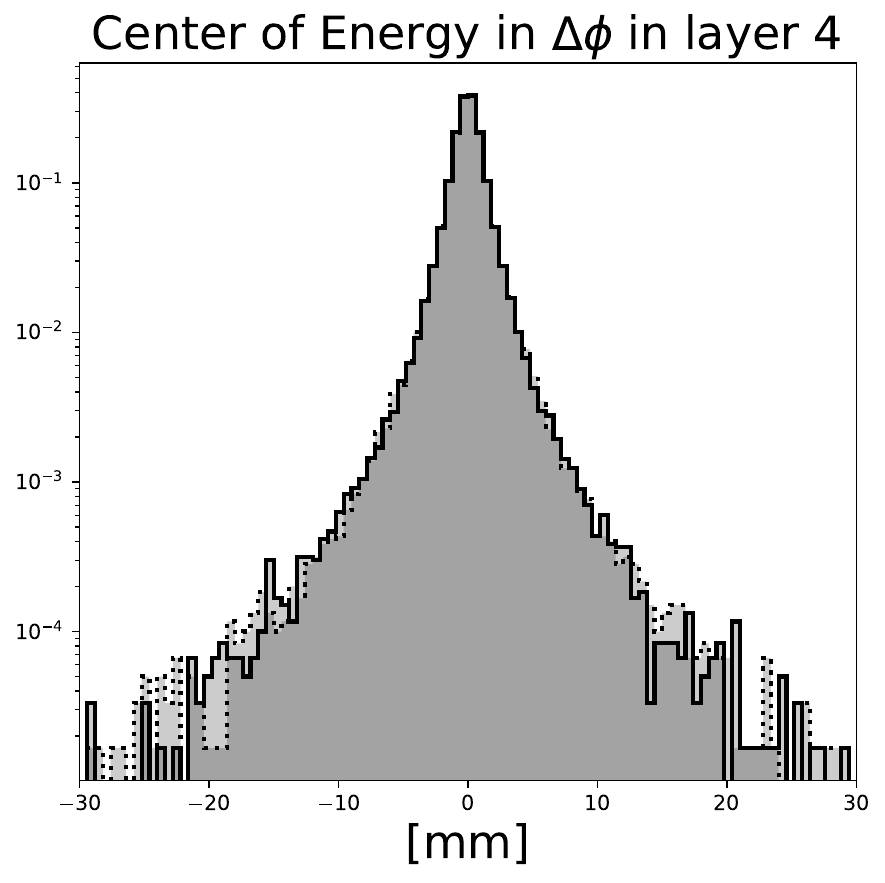}\\
    \includegraphics[height=0.1\textheight]{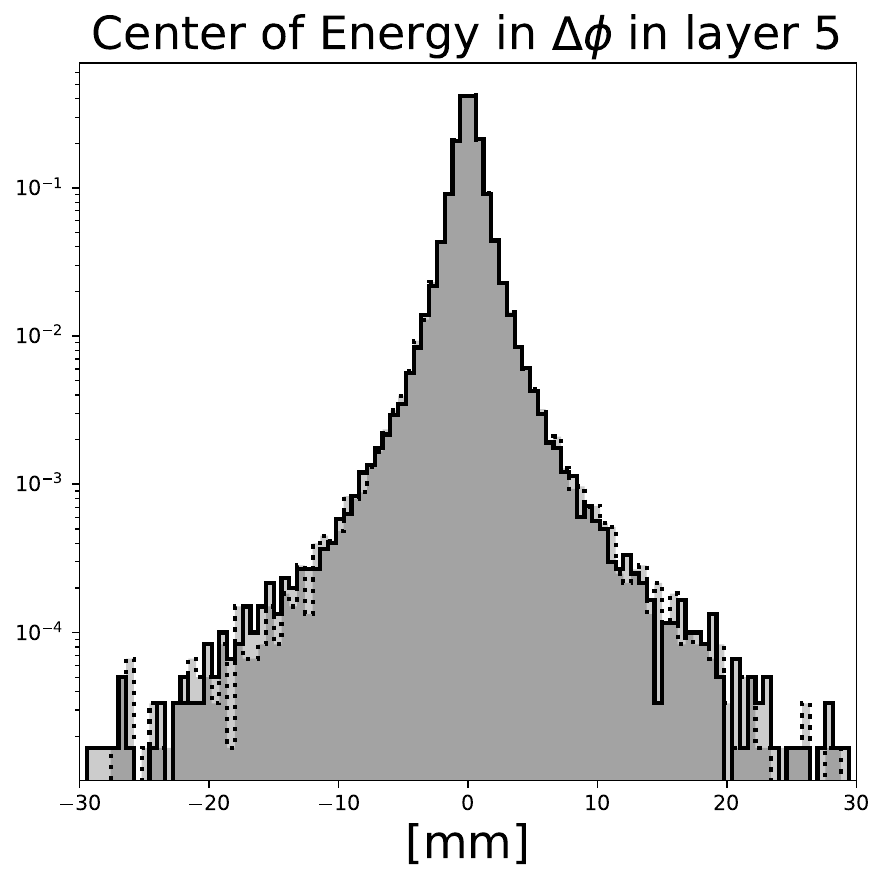} \hfill \includegraphics[height=0.1\textheight]{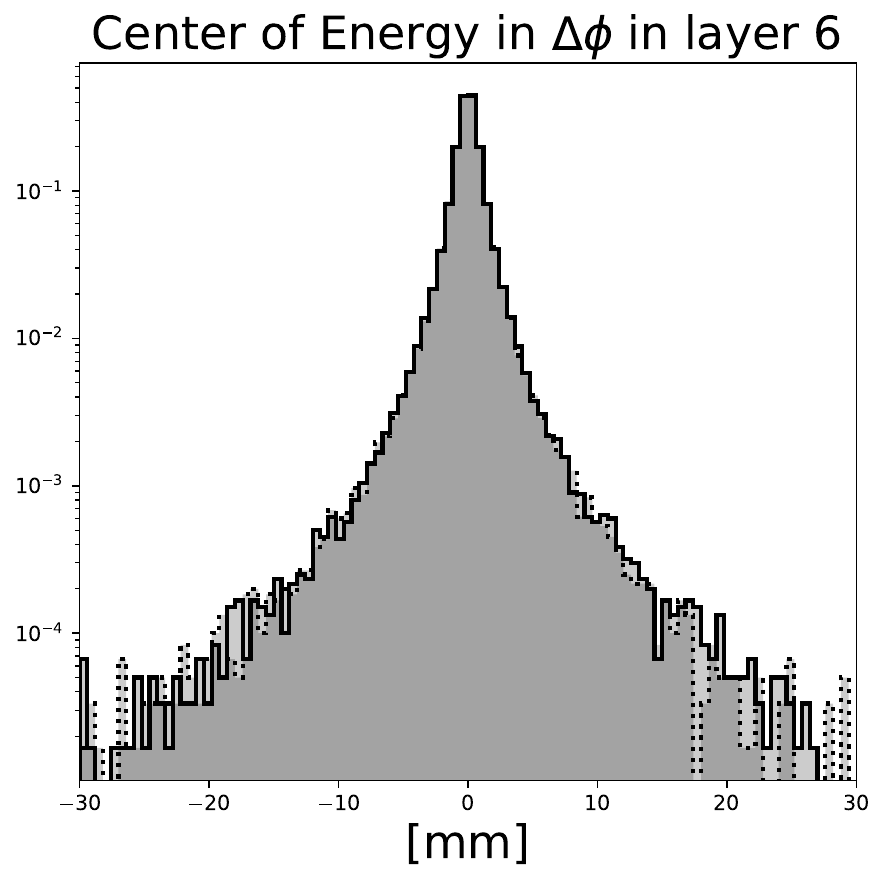} \hfill \includegraphics[height=0.1\textheight]{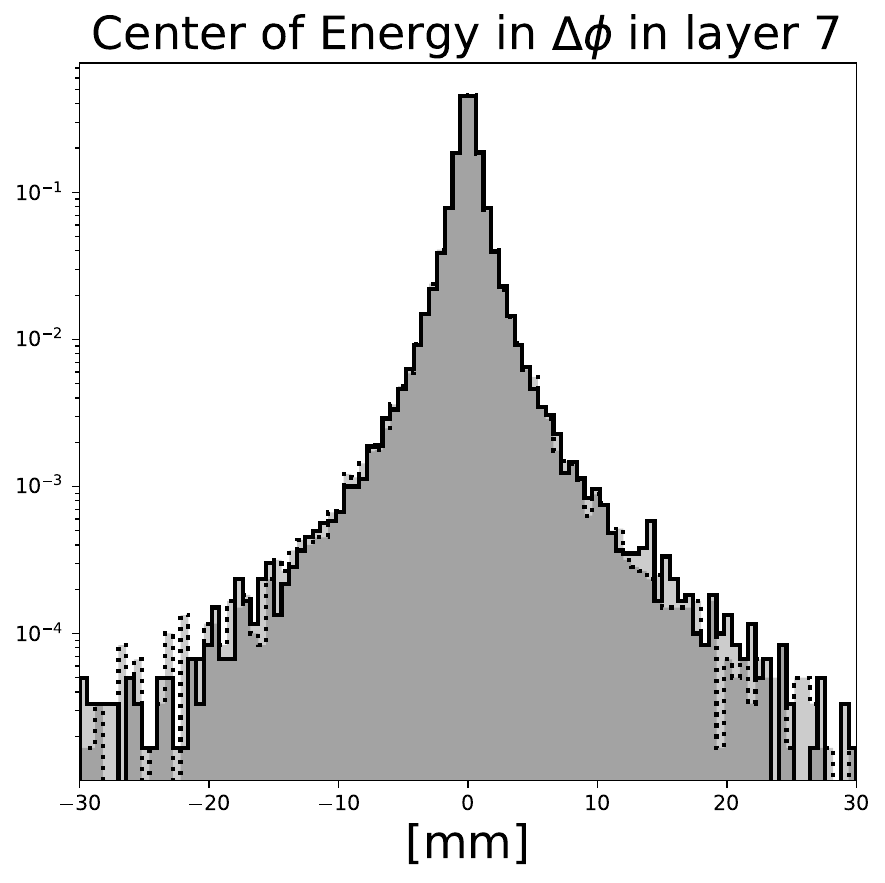} \hfill \includegraphics[height=0.1\textheight]{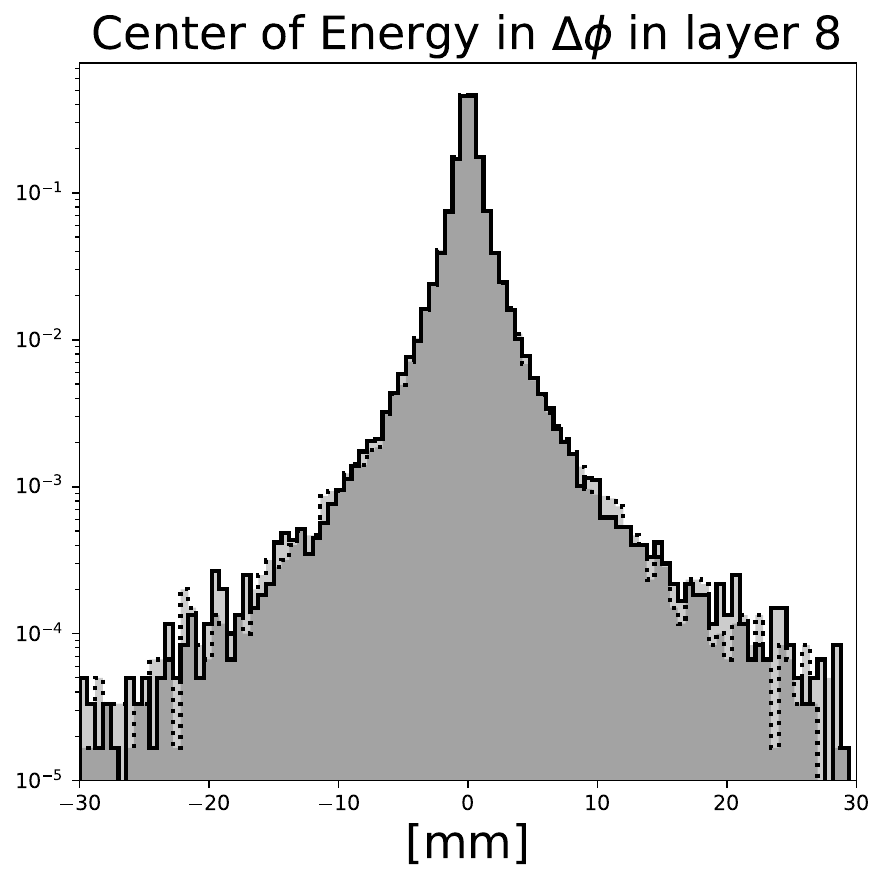} \hfill \includegraphics[height=0.1\textheight]{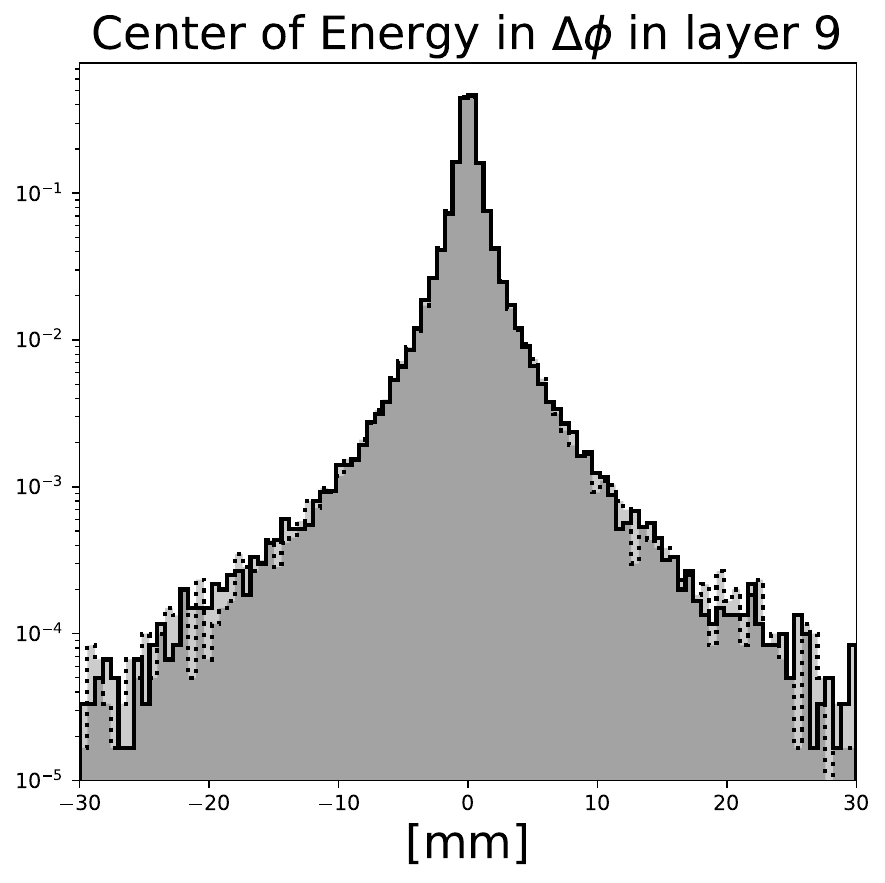}\\
    \includegraphics[height=0.1\textheight]{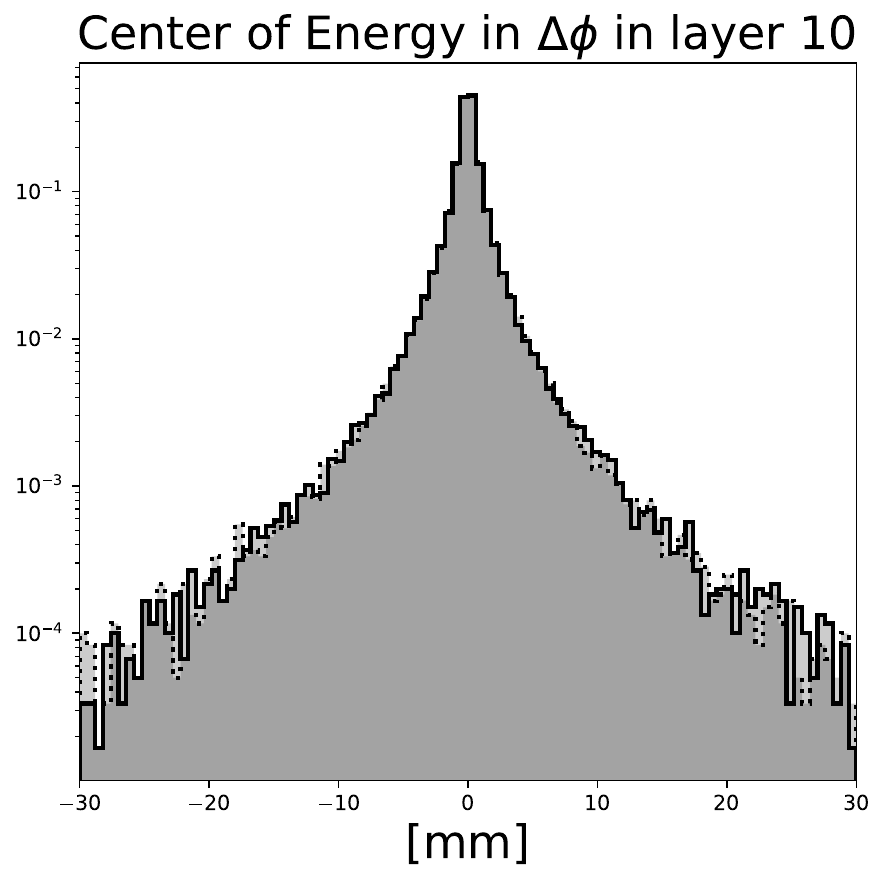} \hfill \includegraphics[height=0.1\textheight]{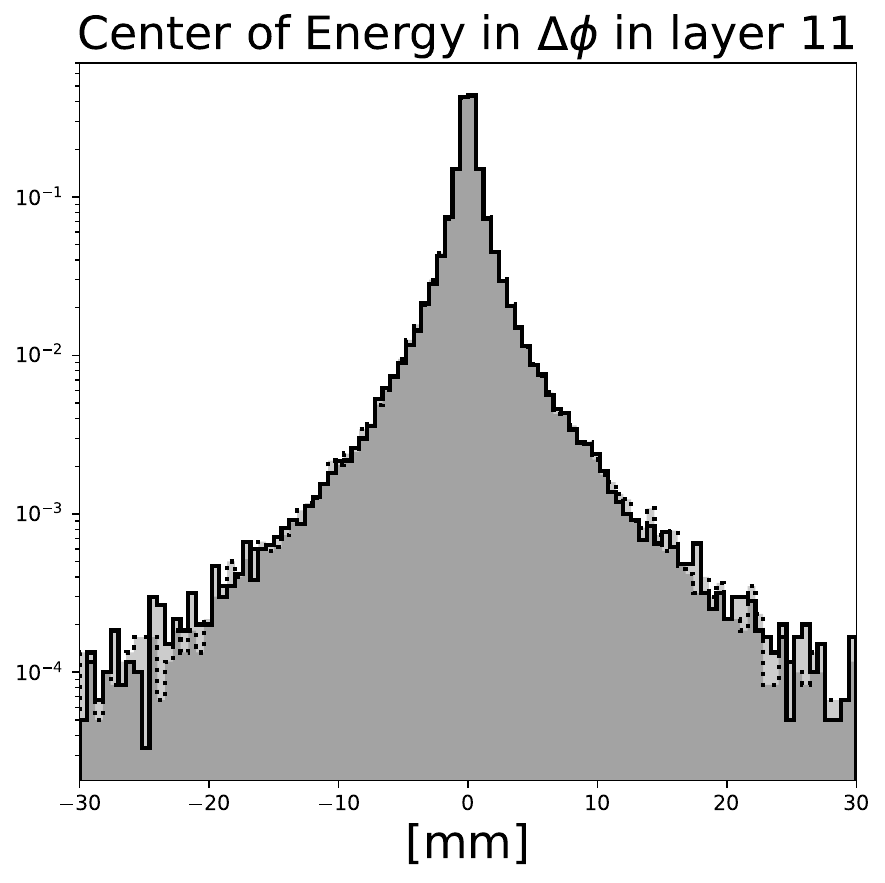} \hfill \includegraphics[height=0.1\textheight]{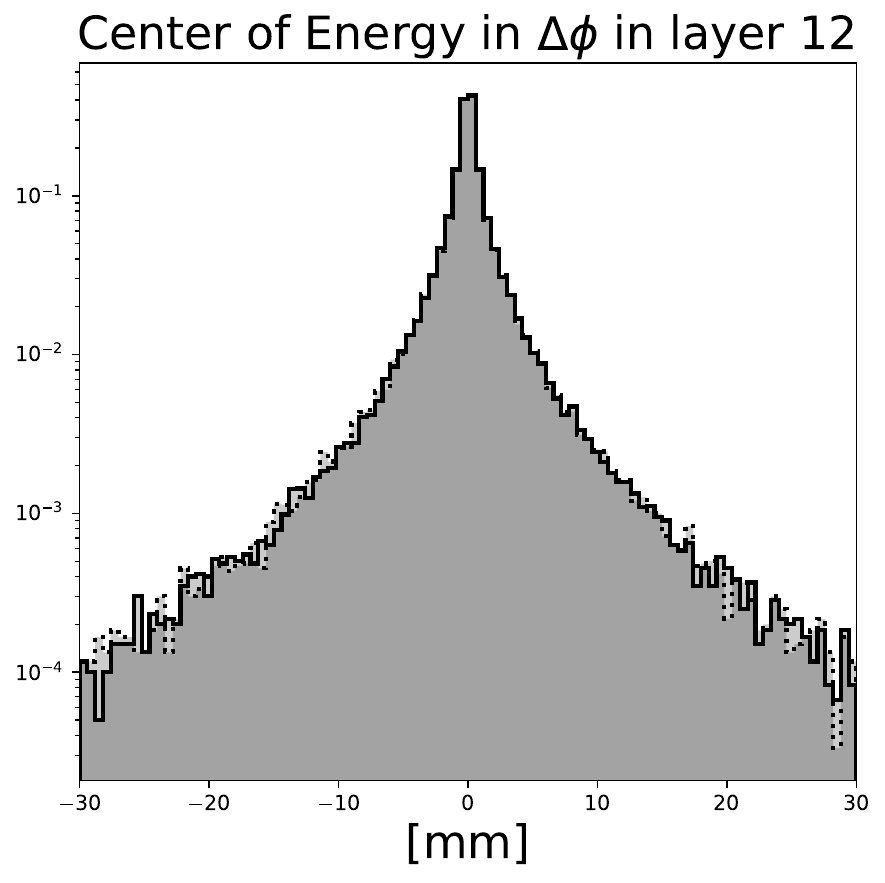} \hfill \includegraphics[height=0.1\textheight]{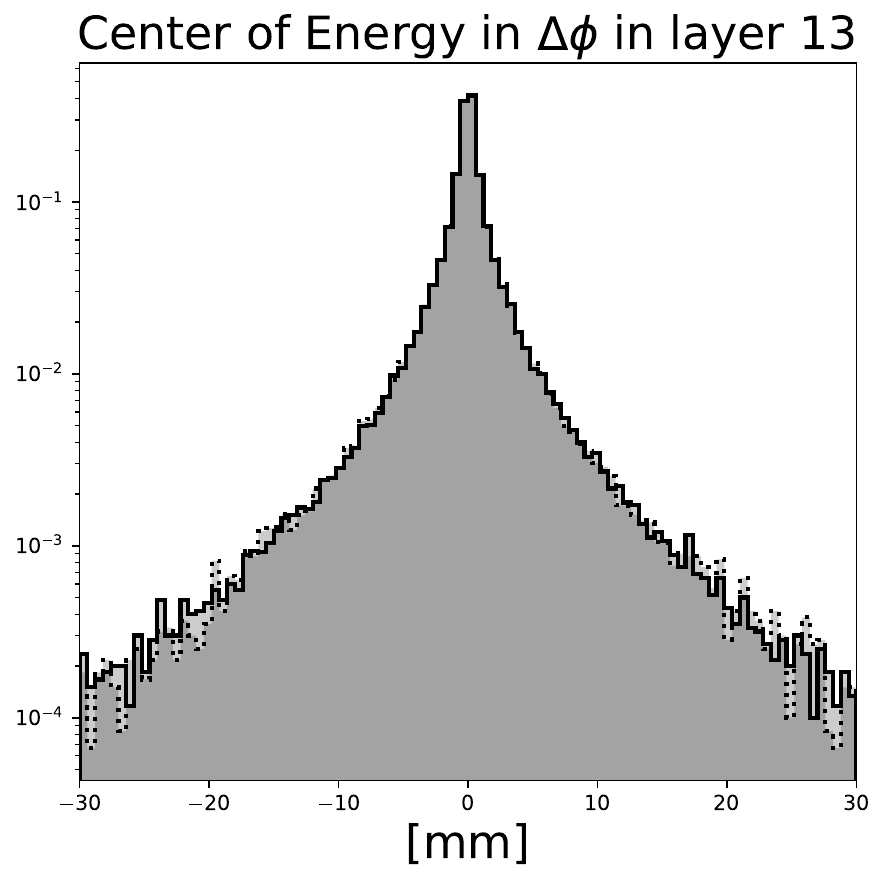} \hfill \includegraphics[height=0.1\textheight]{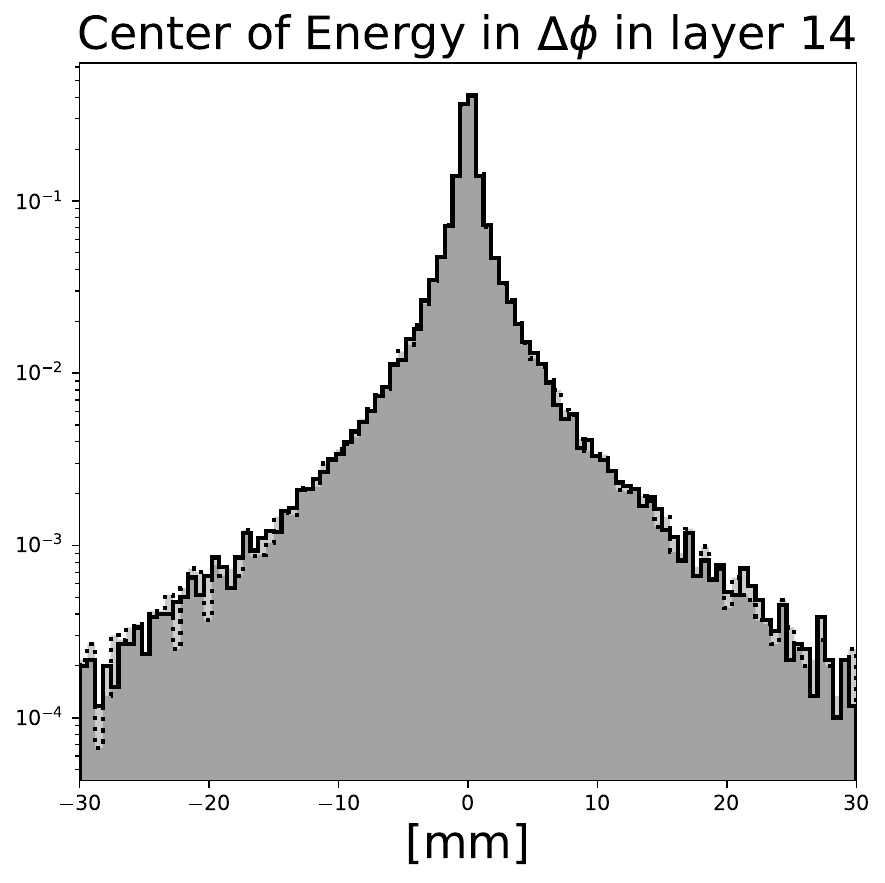}\\
    \includegraphics[height=0.1\textheight]{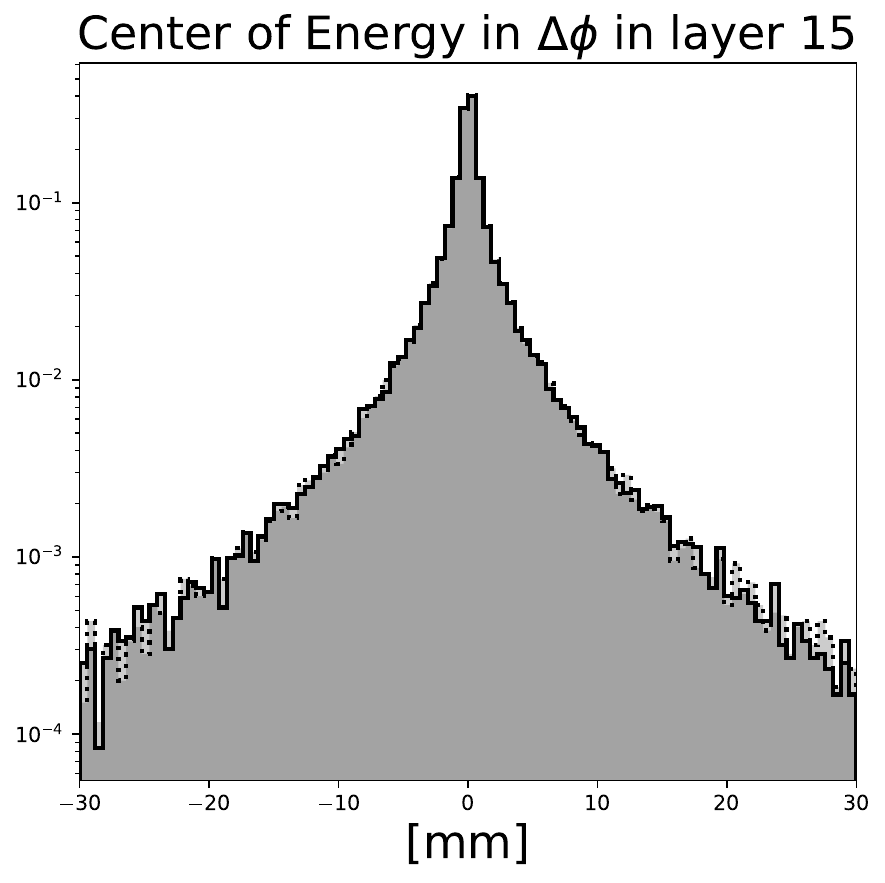} \hfill \includegraphics[height=0.1\textheight]{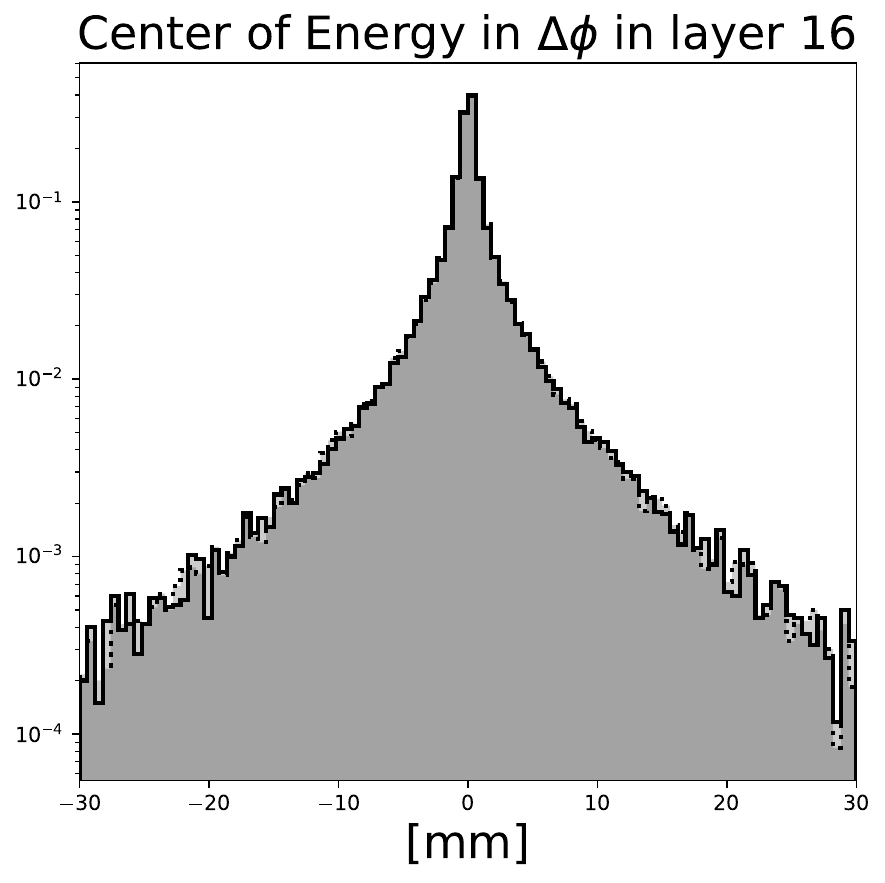} \hfill \includegraphics[height=0.1\textheight]{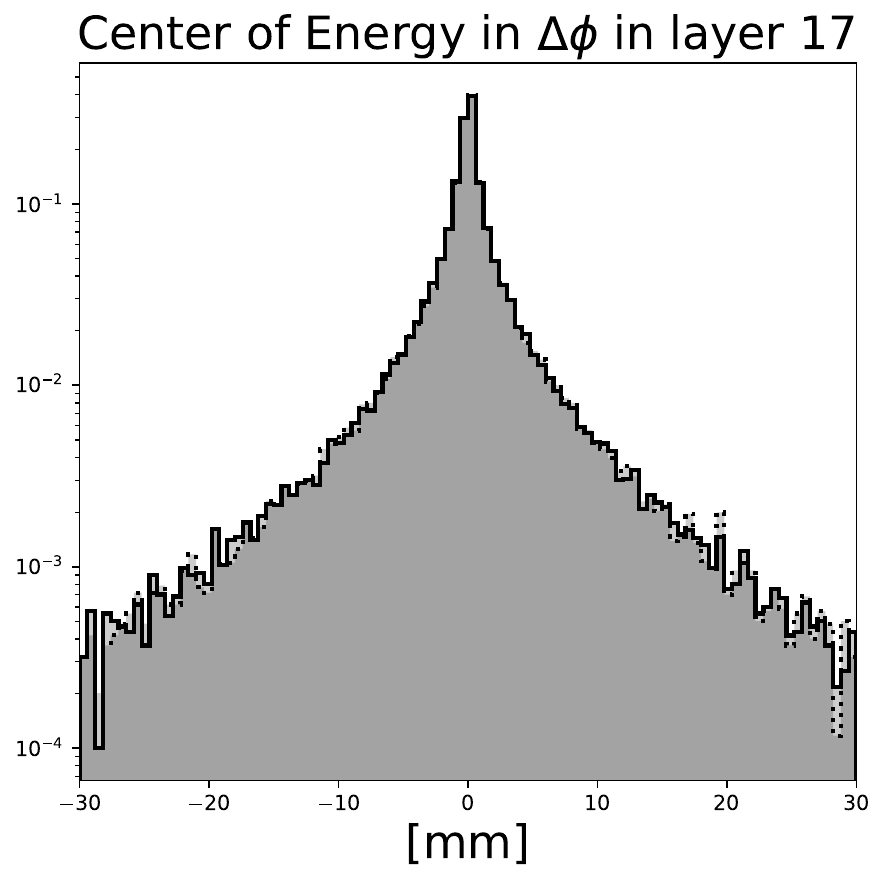} \hfill \includegraphics[height=0.1\textheight]{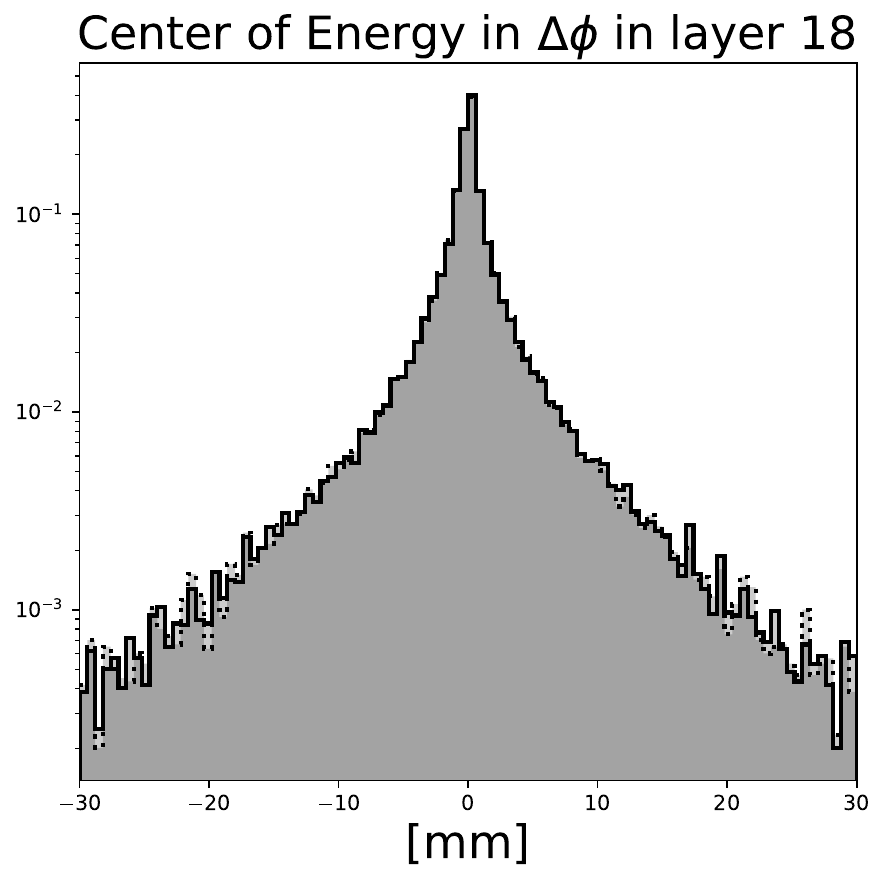} \hfill \includegraphics[height=0.1\textheight]{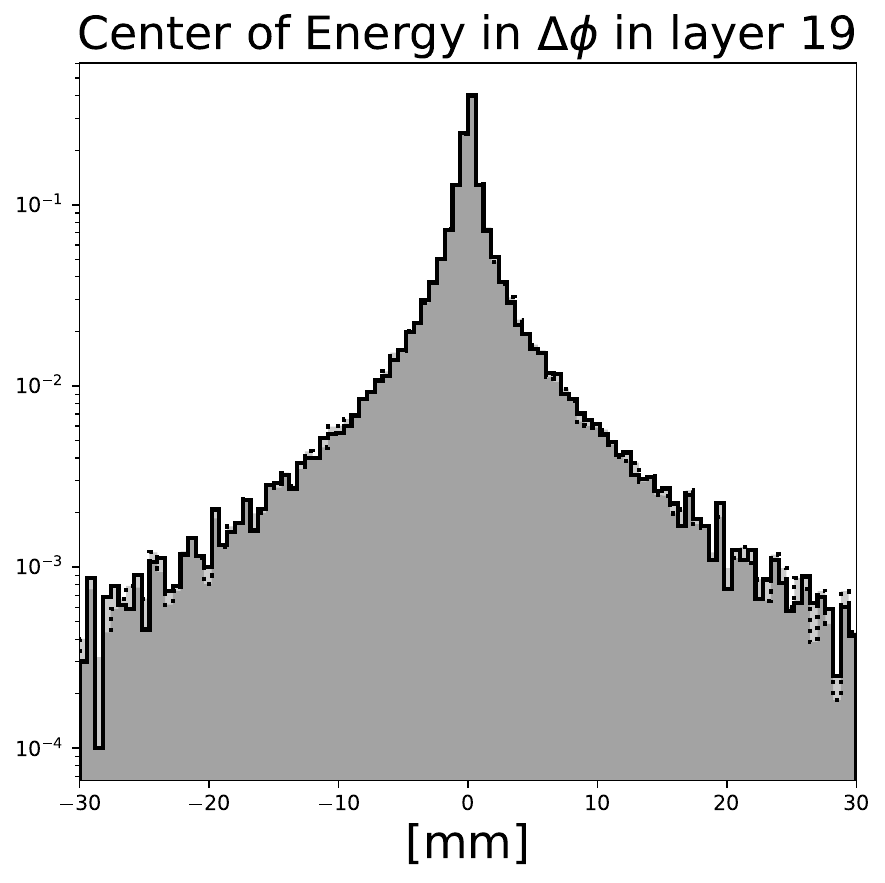}\\
    \includegraphics[height=0.1\textheight]{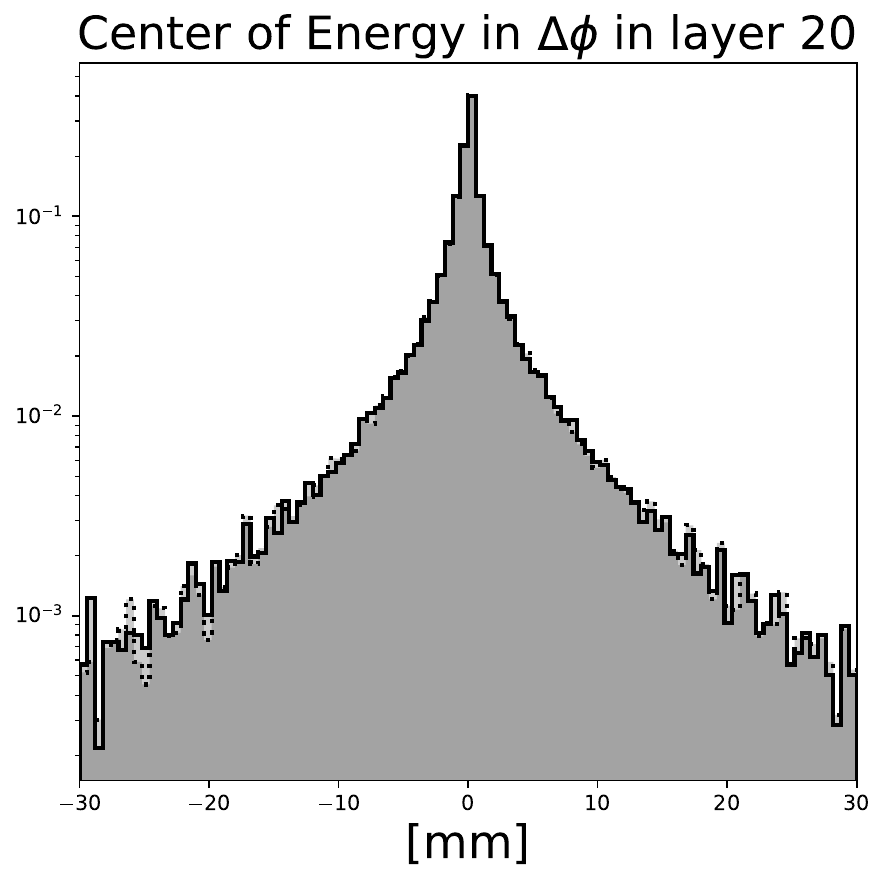} \hfill \includegraphics[height=0.1\textheight]{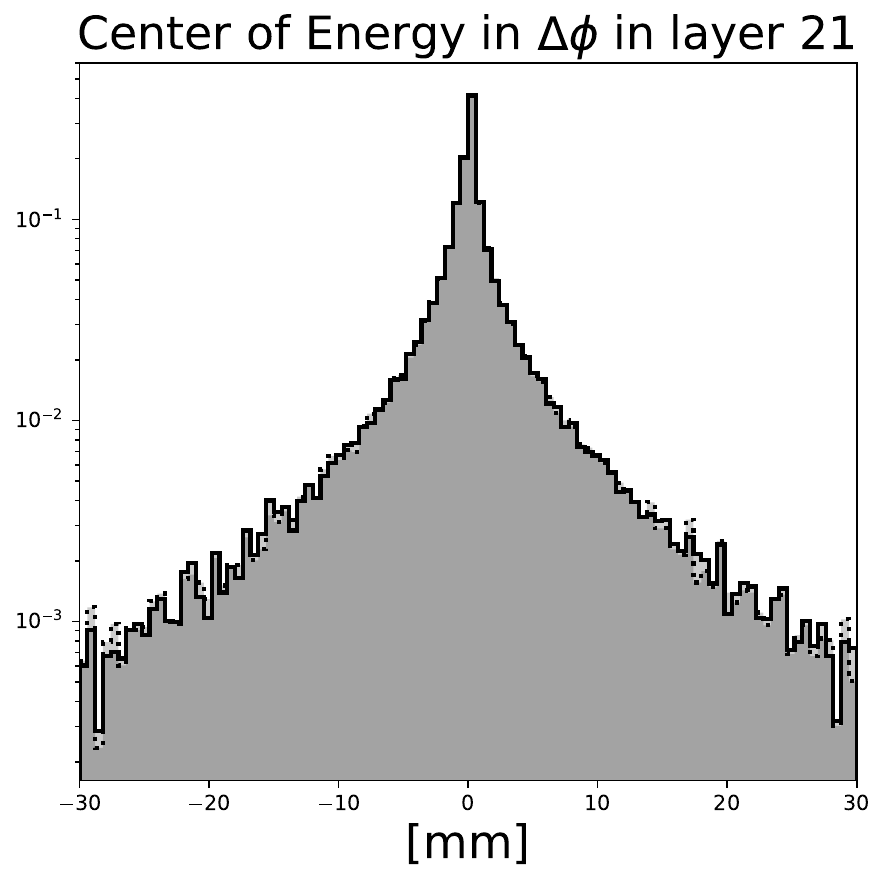} \hfill \includegraphics[height=0.1\textheight]{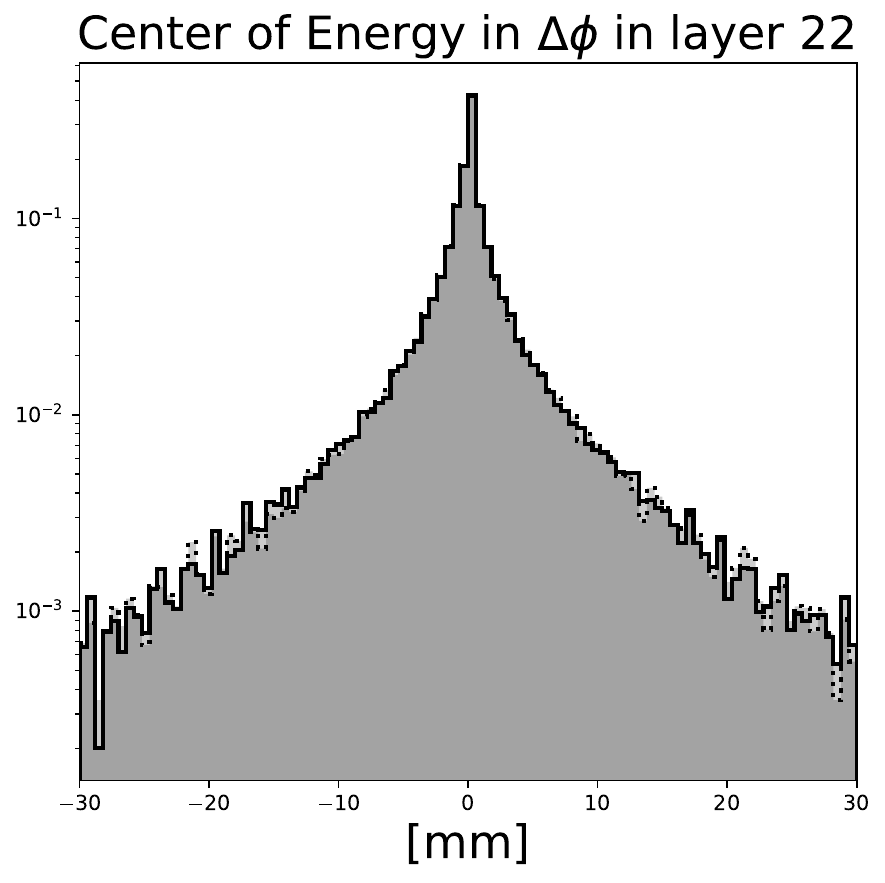} \hfill \includegraphics[height=0.1\textheight]{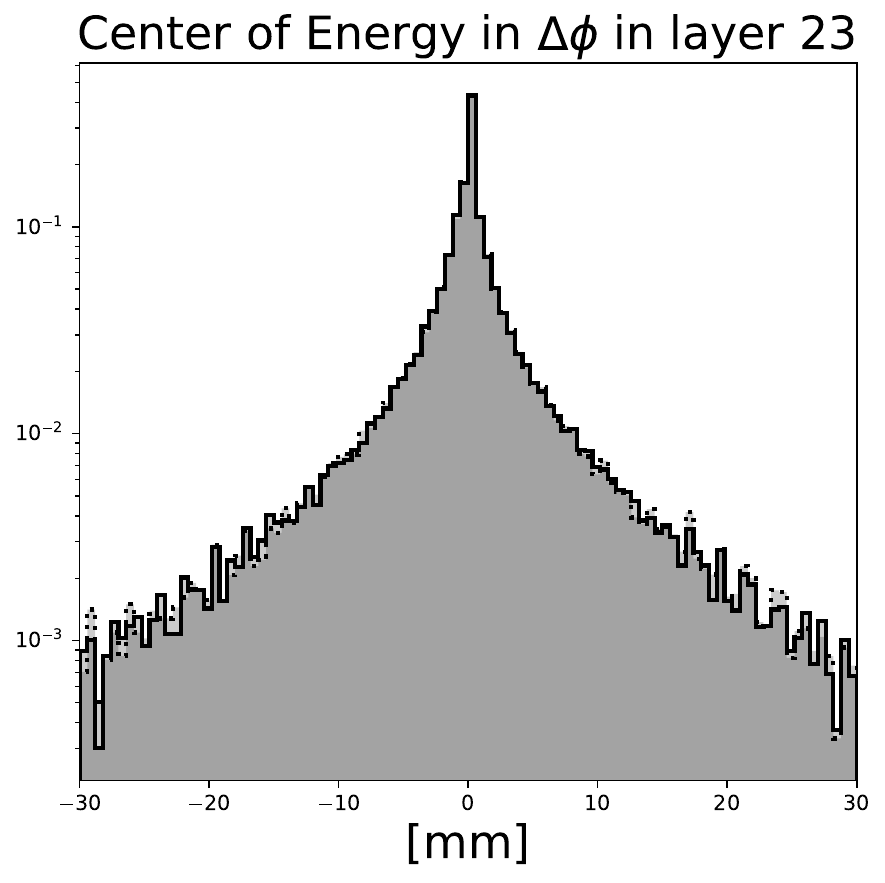} \hfill \includegraphics[height=0.1\textheight]{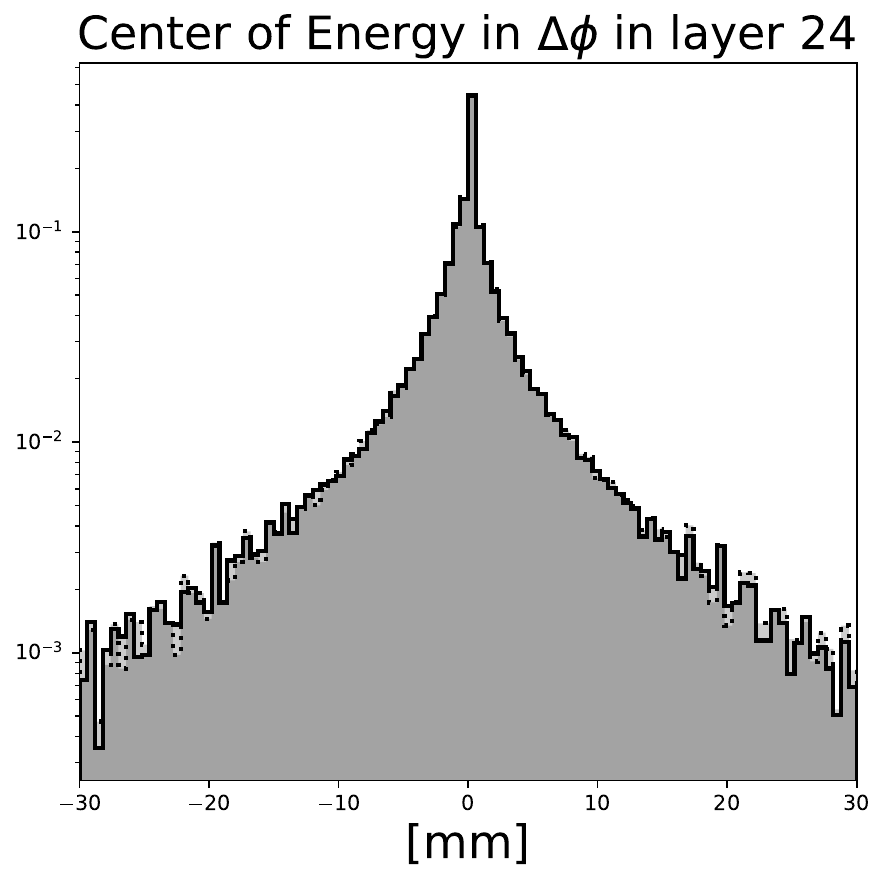}\\
    \includegraphics[height=0.1\textheight]{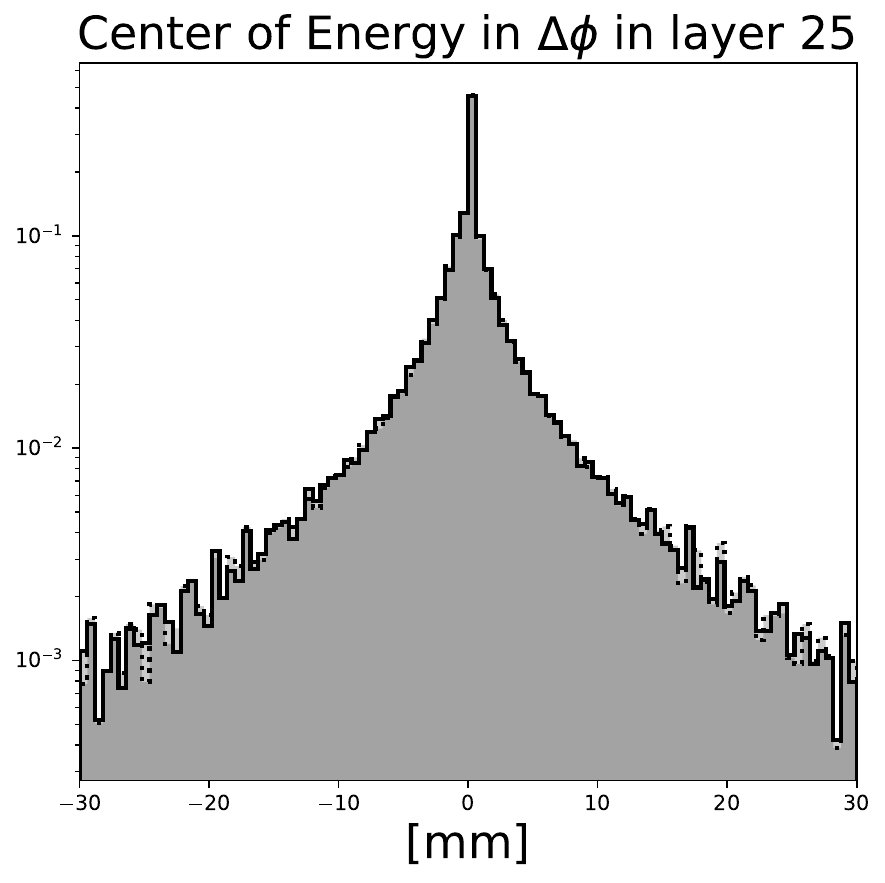} \hfill \includegraphics[height=0.1\textheight]{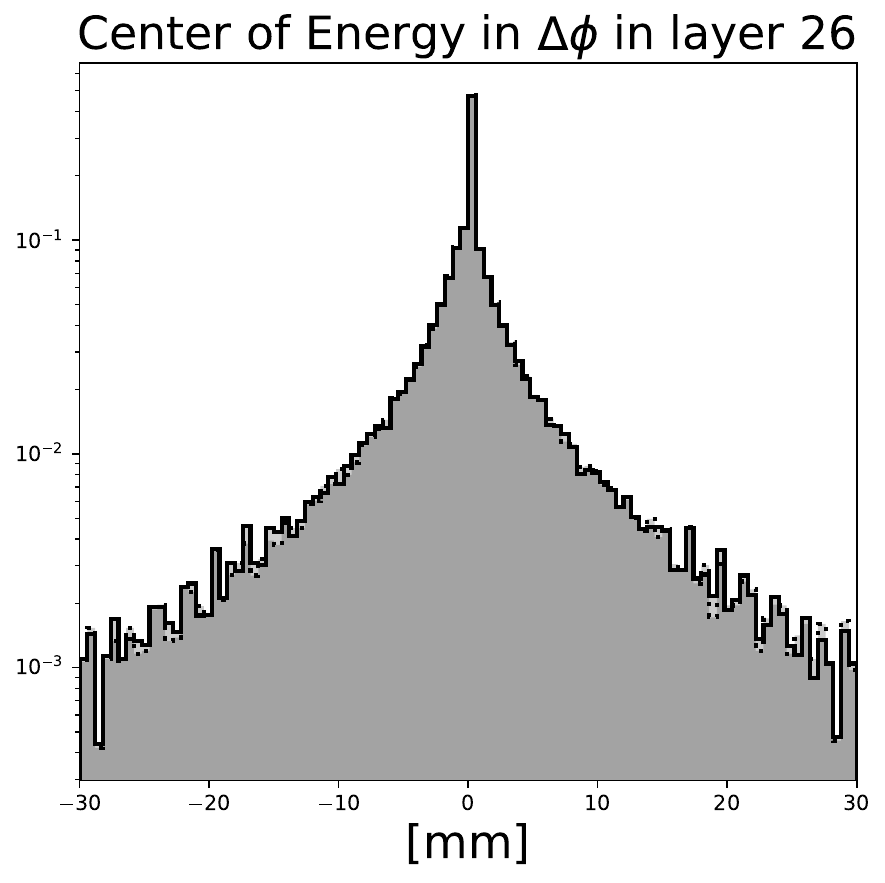} \hfill \includegraphics[height=0.1\textheight]{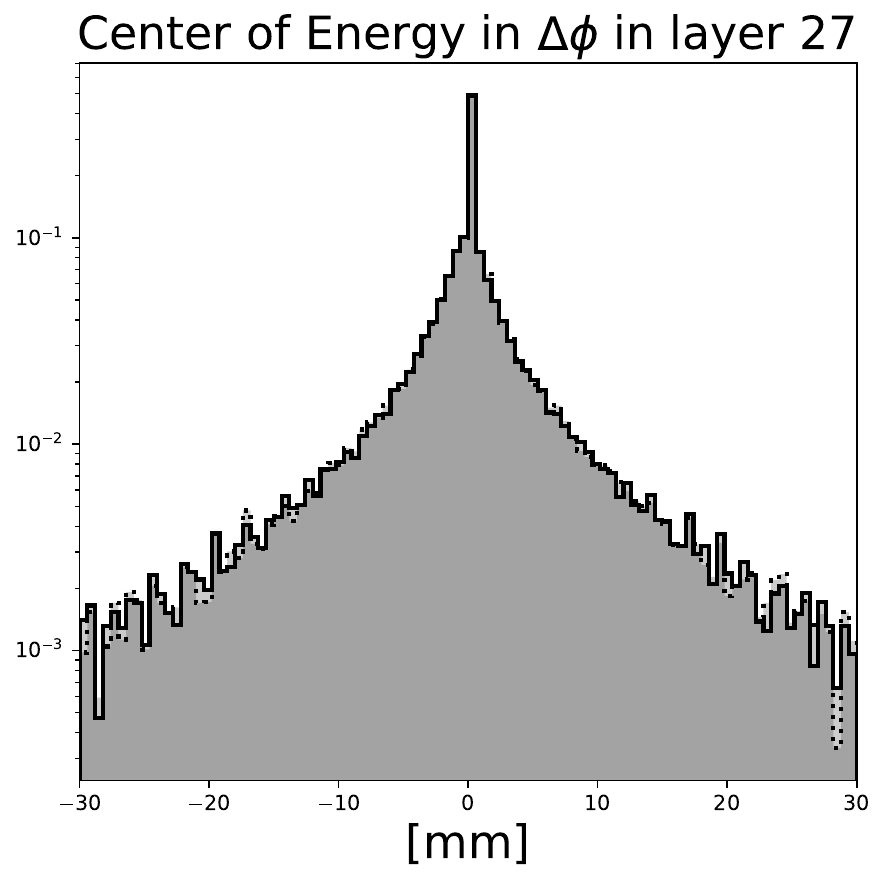} \hfill \includegraphics[height=0.1\textheight]{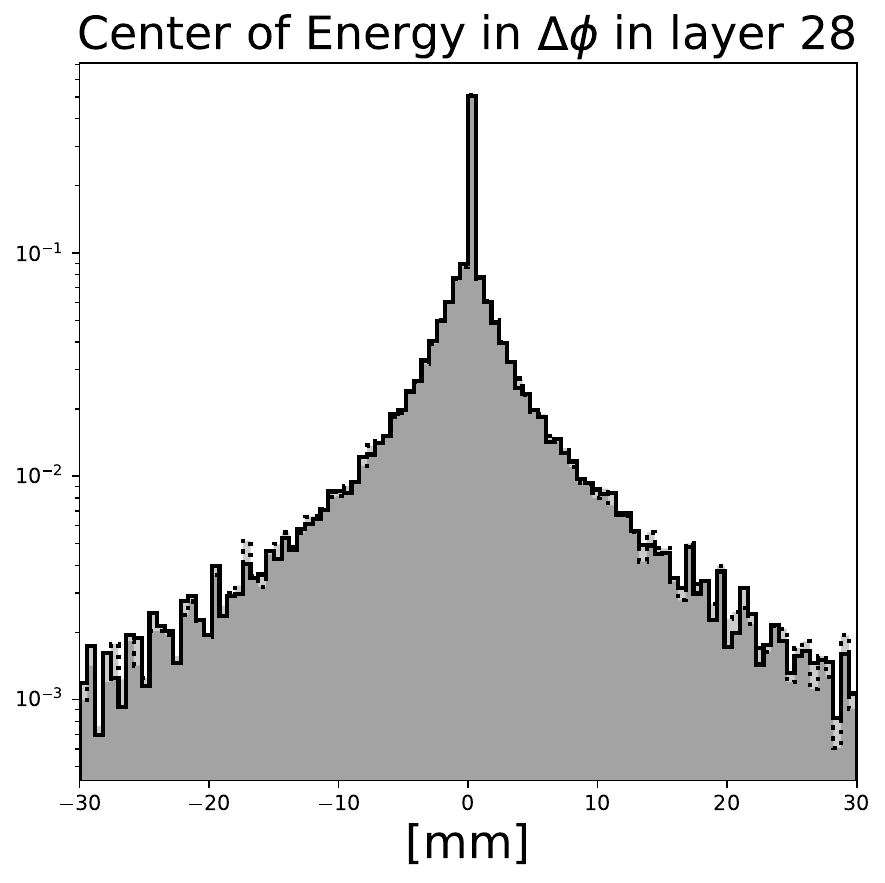} \hfill \includegraphics[height=0.1\textheight]{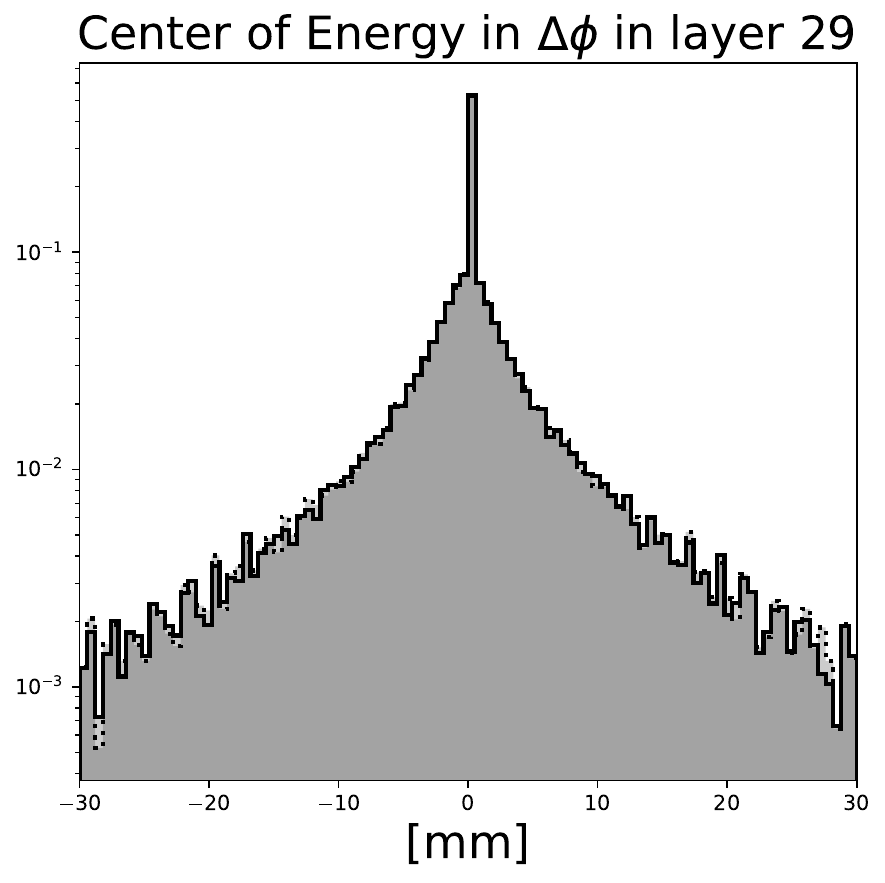}\\
    \includegraphics[height=0.1\textheight]{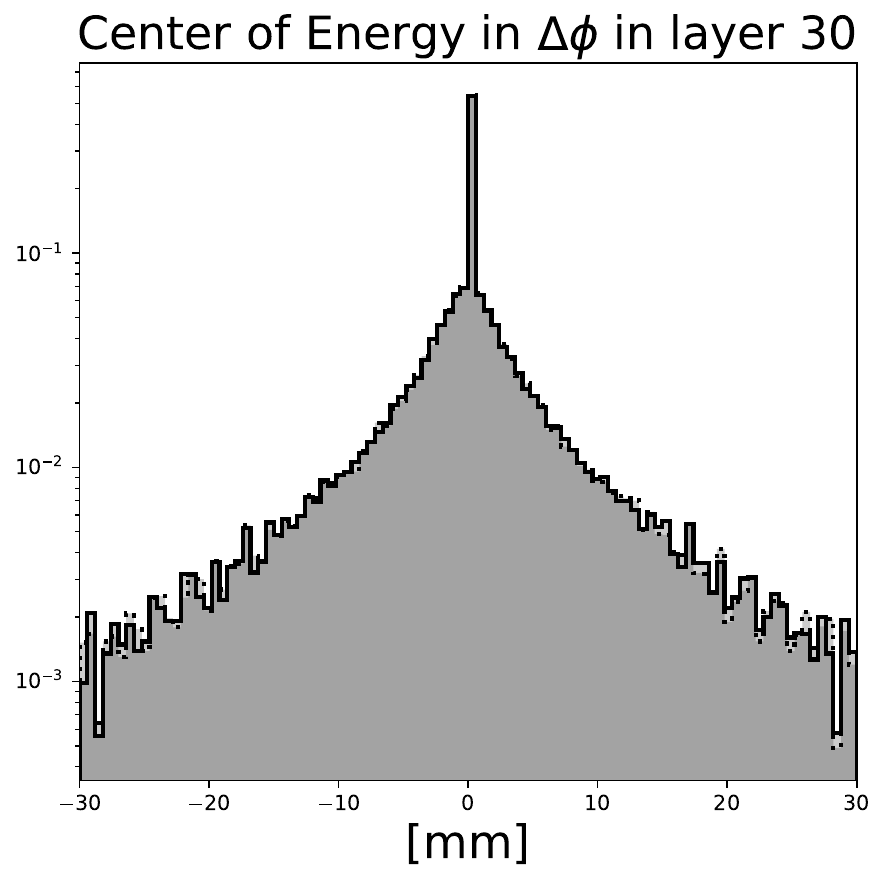} \hfill \includegraphics[height=0.1\textheight]{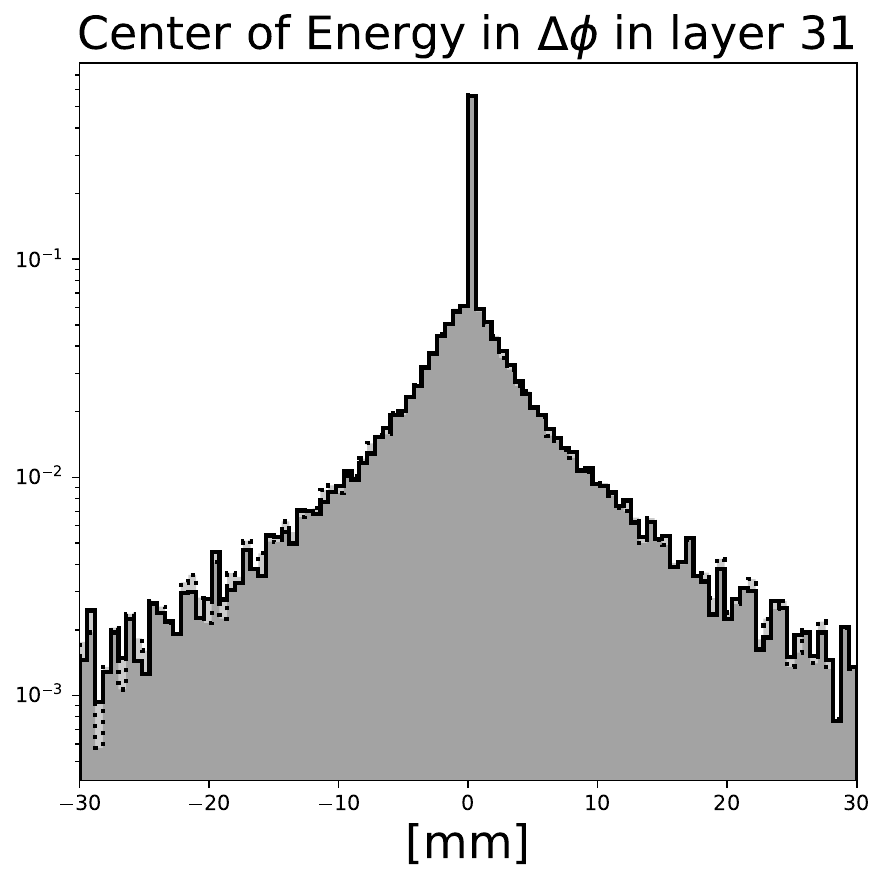} \hfill \includegraphics[height=0.1\textheight]{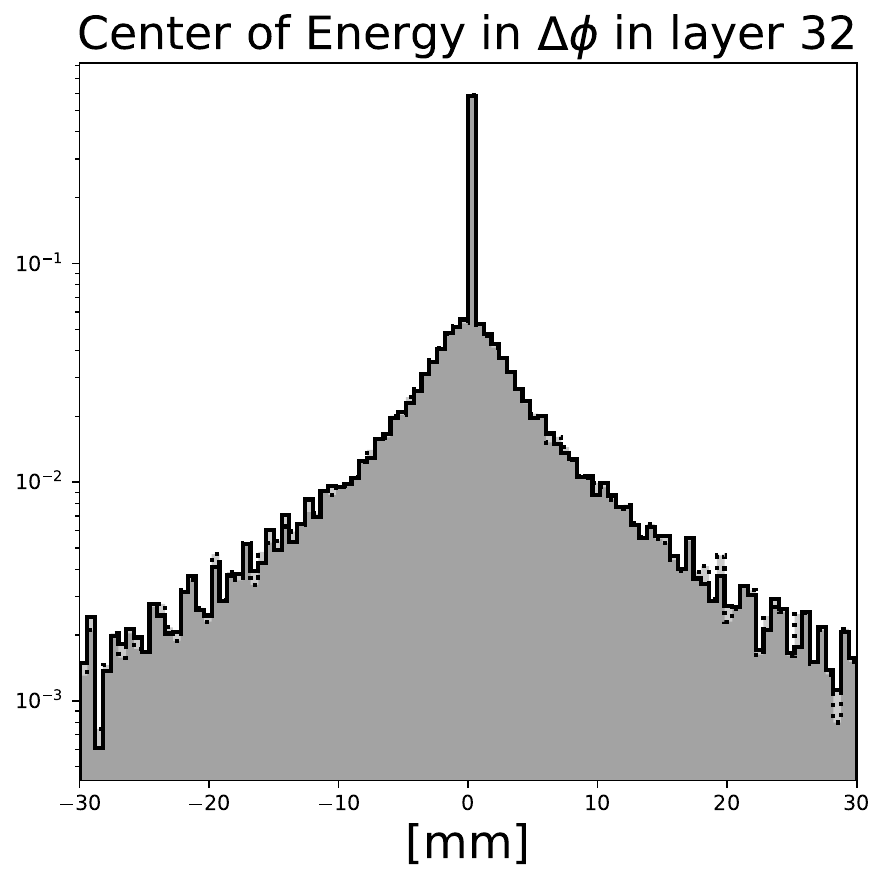} \hfill \includegraphics[height=0.1\textheight]{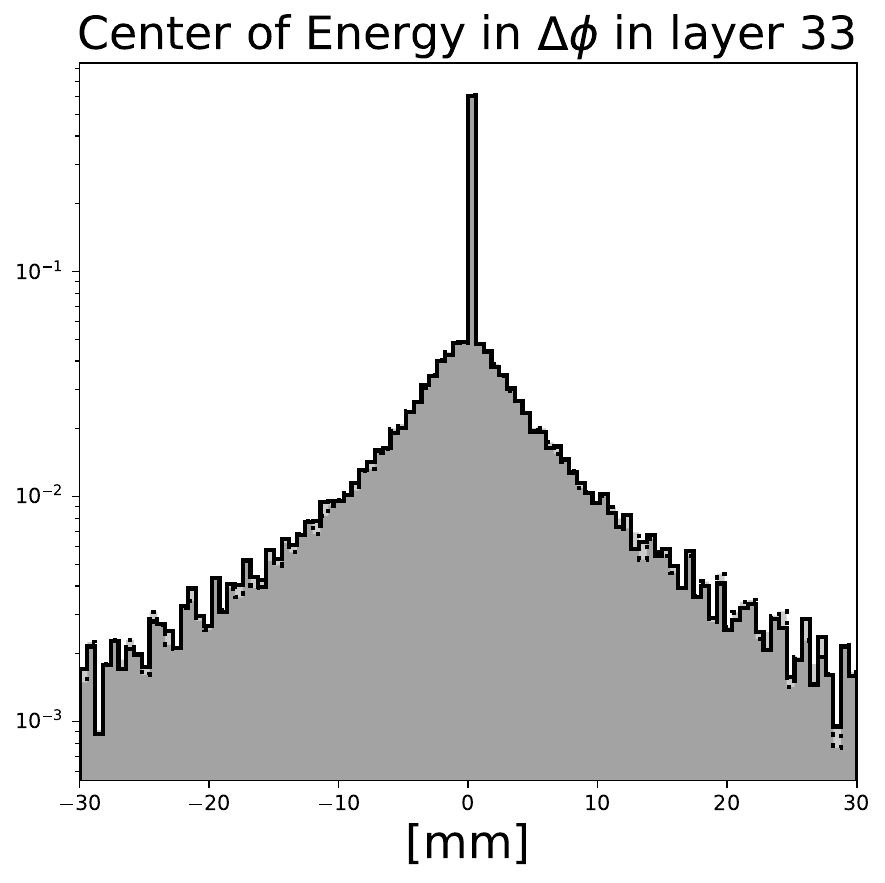} \hfill \includegraphics[height=0.1\textheight]{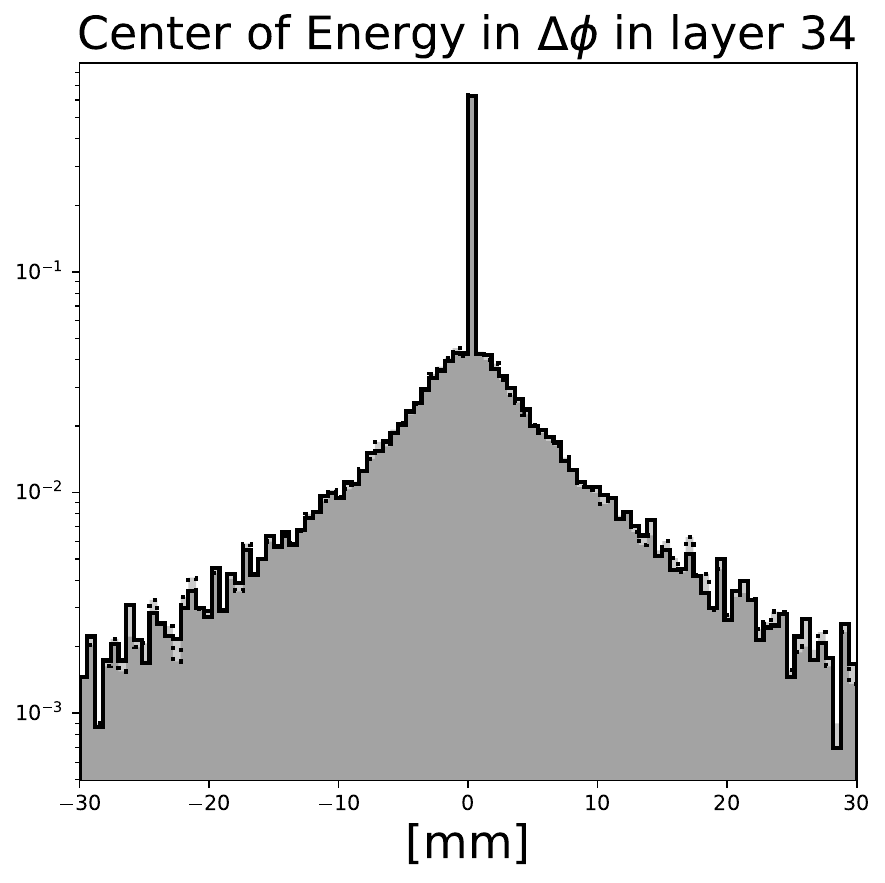}\\
    \includegraphics[height=0.1\textheight]{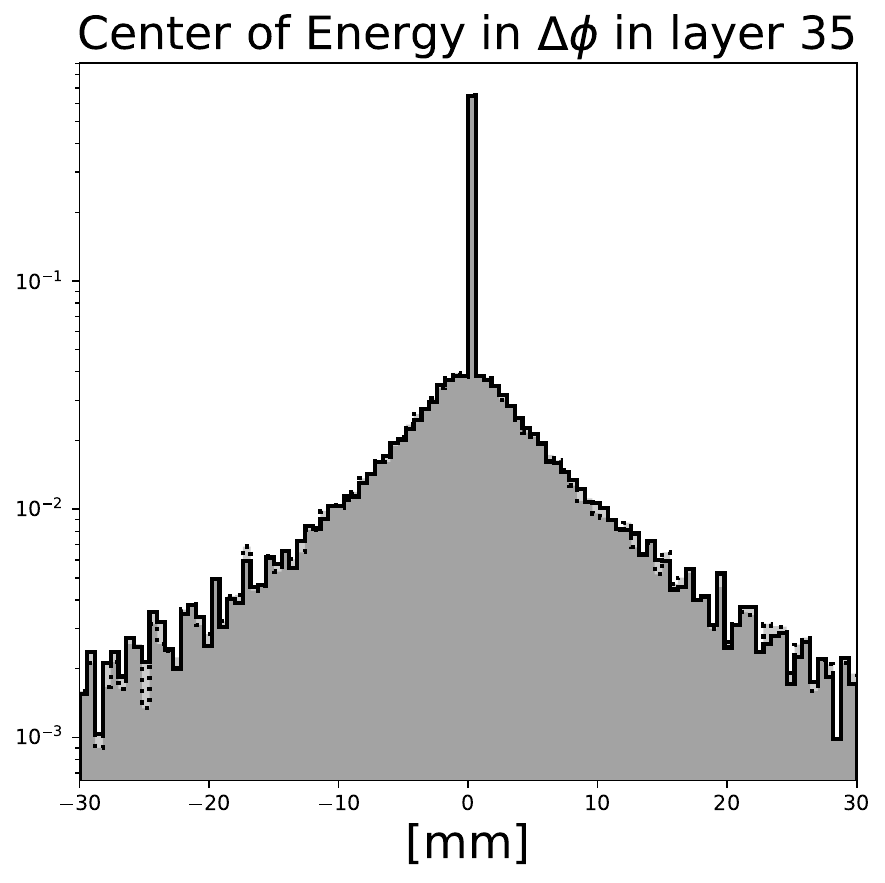} \hfill \includegraphics[height=0.1\textheight]{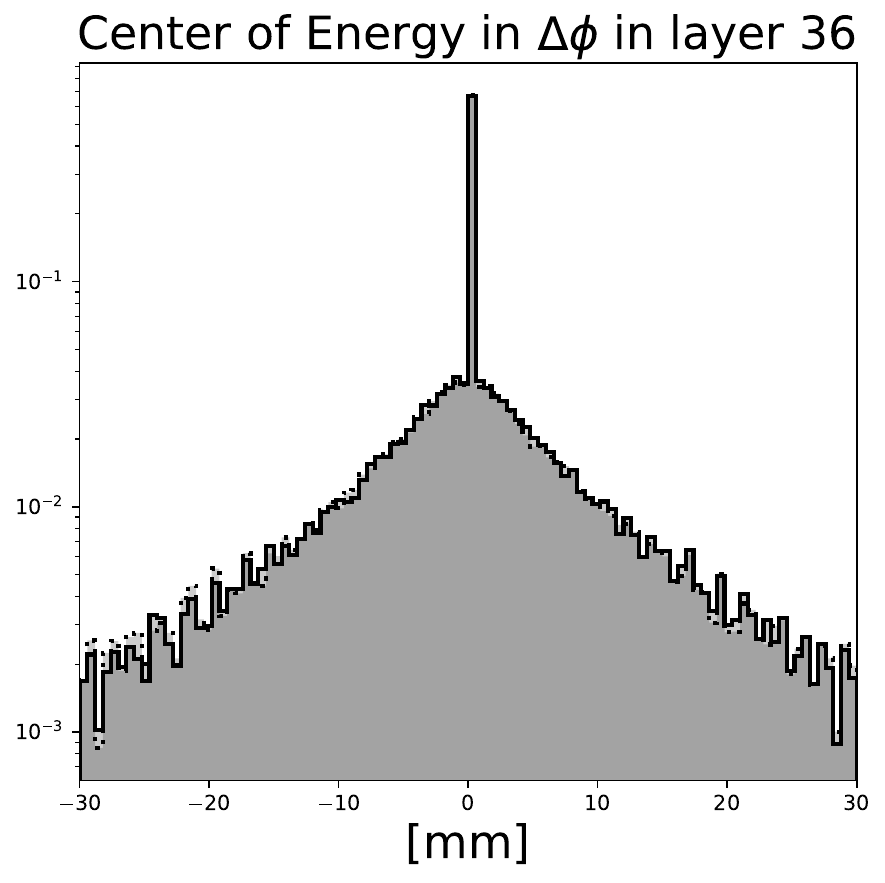} \hfill \includegraphics[height=0.1\textheight]{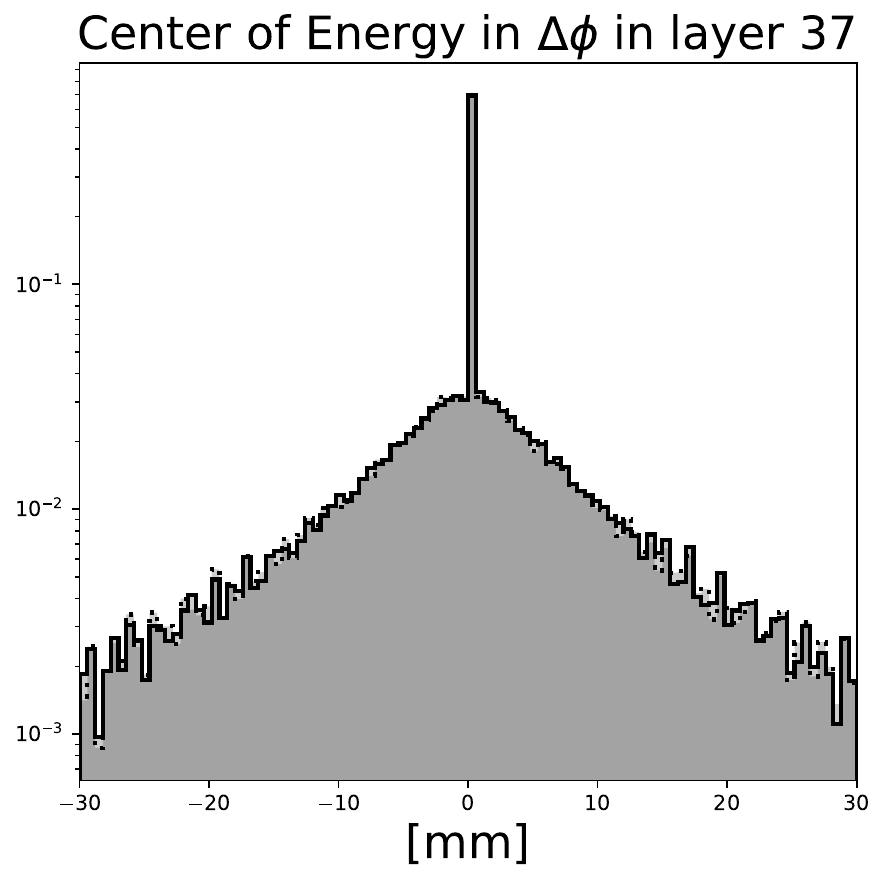} \hfill \includegraphics[height=0.1\textheight]{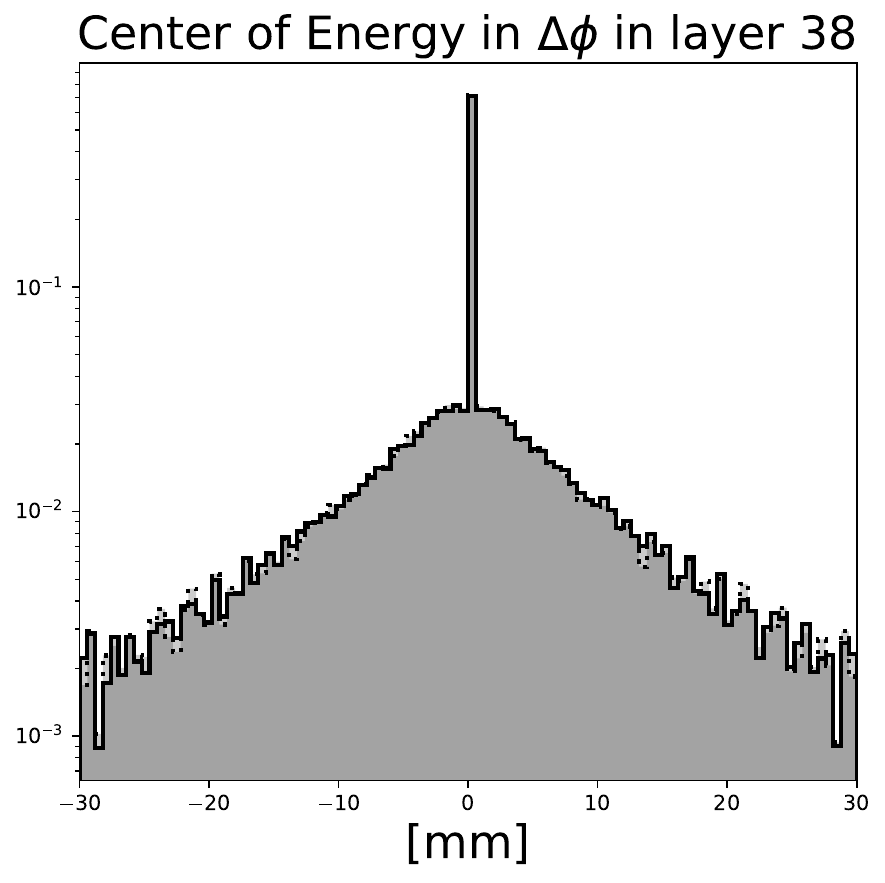} \hfill \includegraphics[height=0.1\textheight]{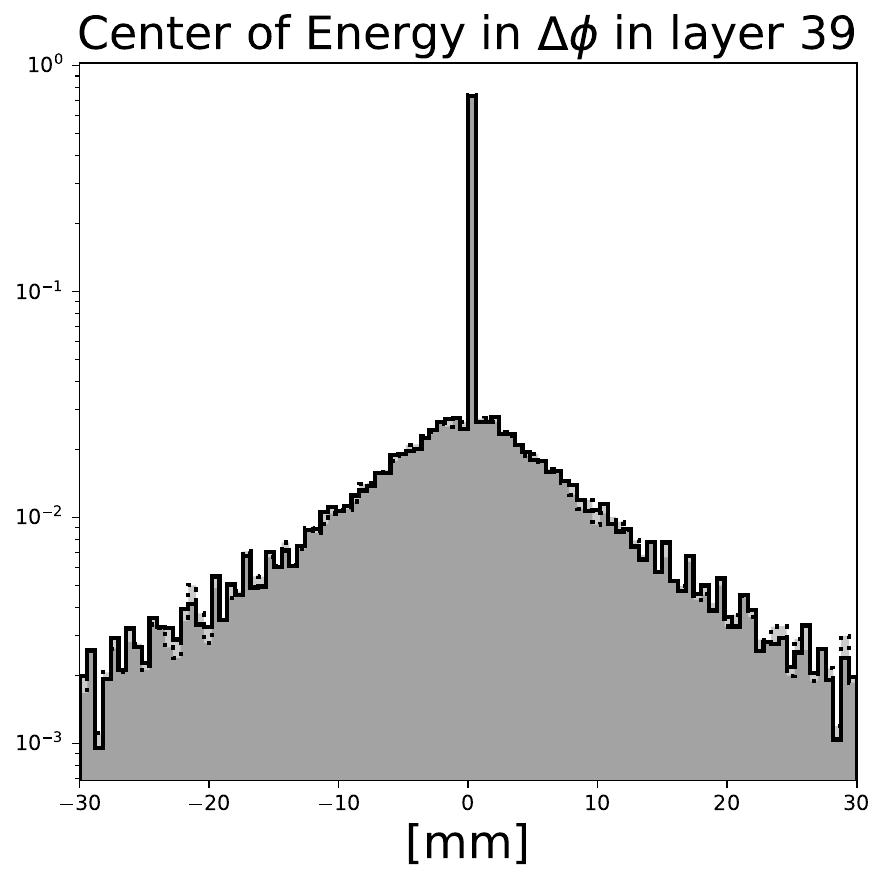}\\
    \includegraphics[height=0.1\textheight]{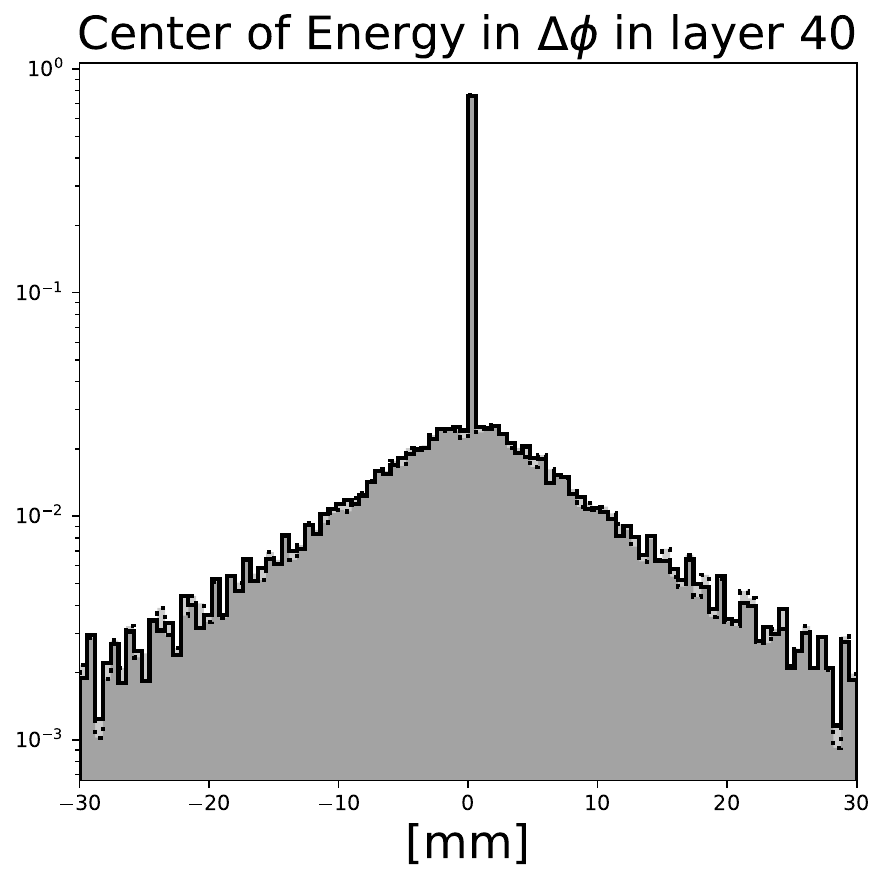} \hfill \includegraphics[height=0.1\textheight]{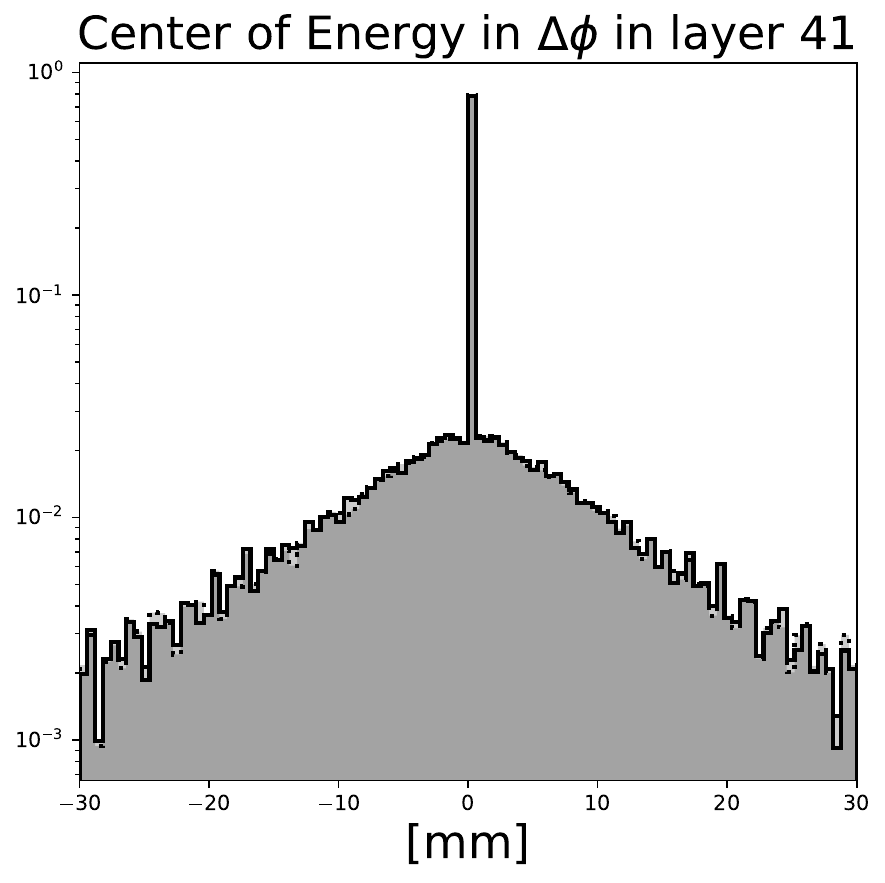} \hfill \includegraphics[height=0.1\textheight]{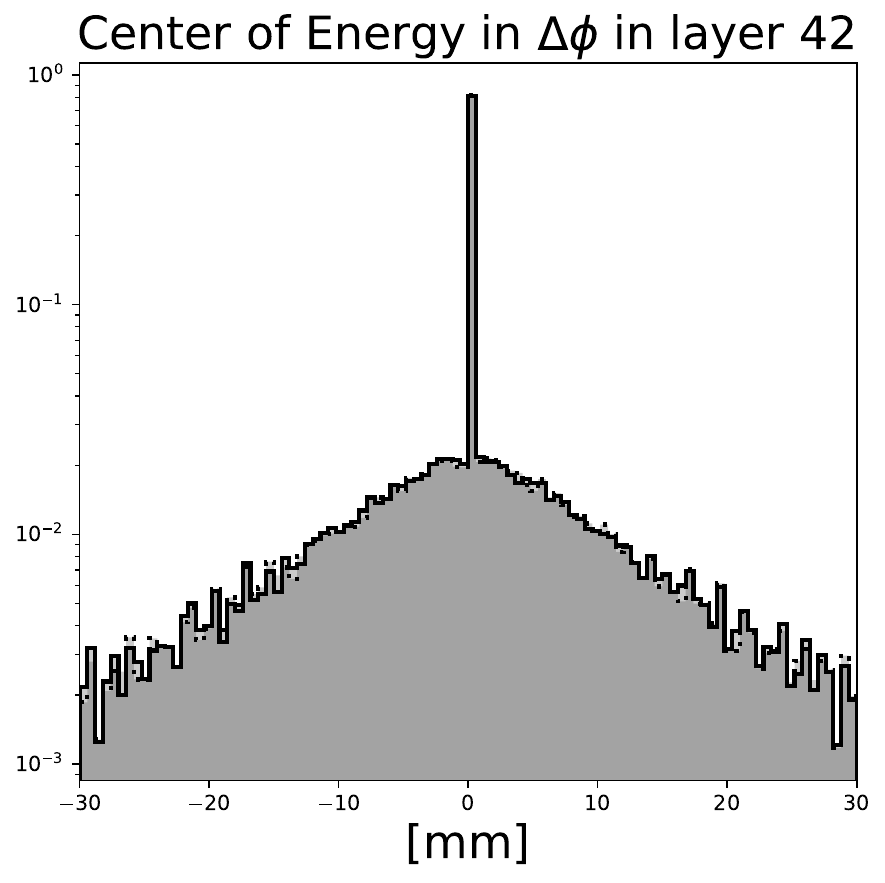} \hfill \includegraphics[height=0.1\textheight]{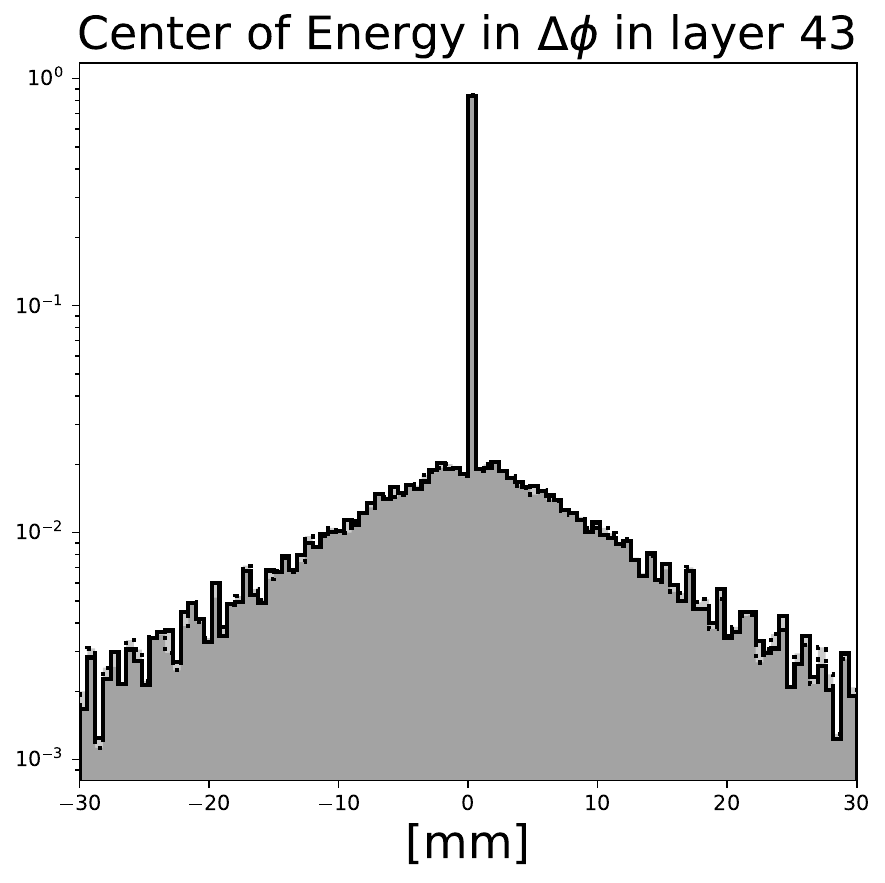} \hfill \includegraphics[height=0.1\textheight]{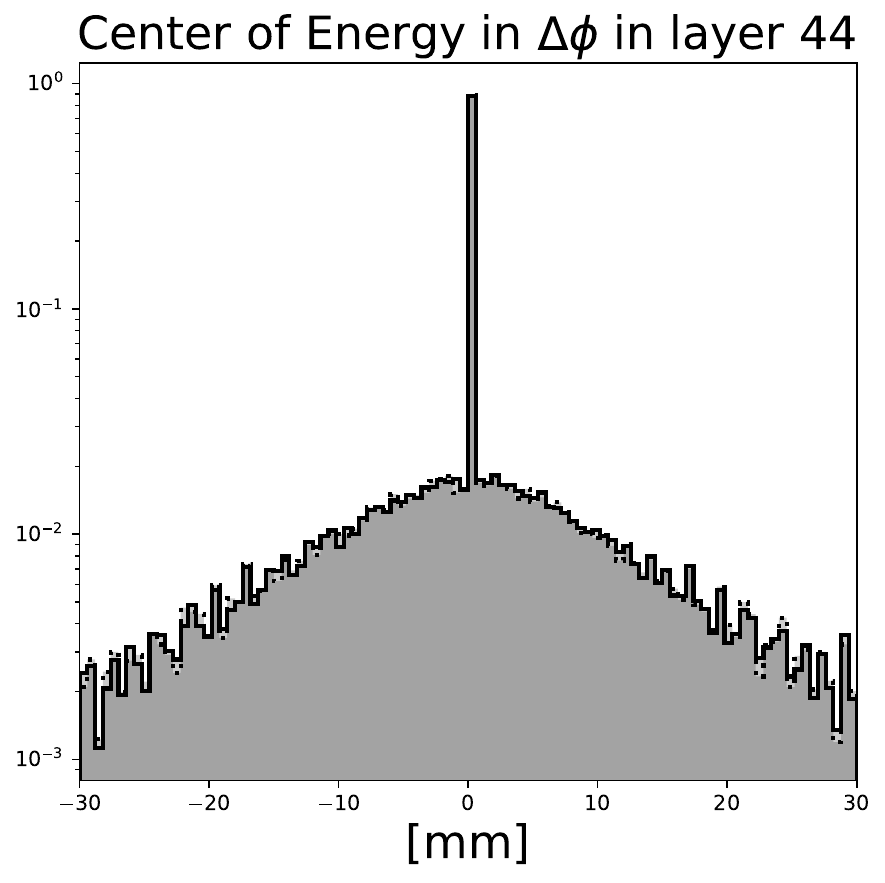}\\
    \includegraphics[width=0.5\textwidth]{figures/Appendix_reference/legend.pdf}
    \caption{Distribution of \geant training and evaluation data in centers of energy in $\phi$ direction for ds3. }
    \label{fig:app_ref.ds3.5}
\end{figure}

\begin{figure}[ht]
    \centering
    \includegraphics[height=0.1\textheight]{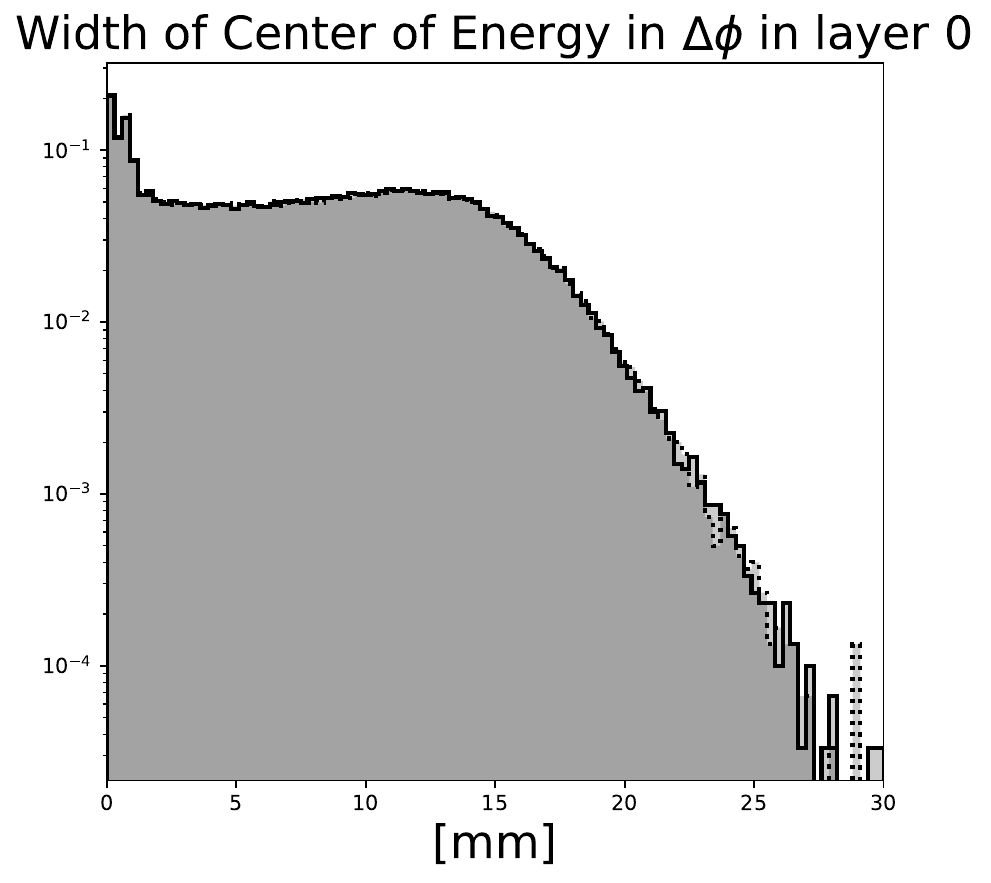} \hfill \includegraphics[height=0.1\textheight]{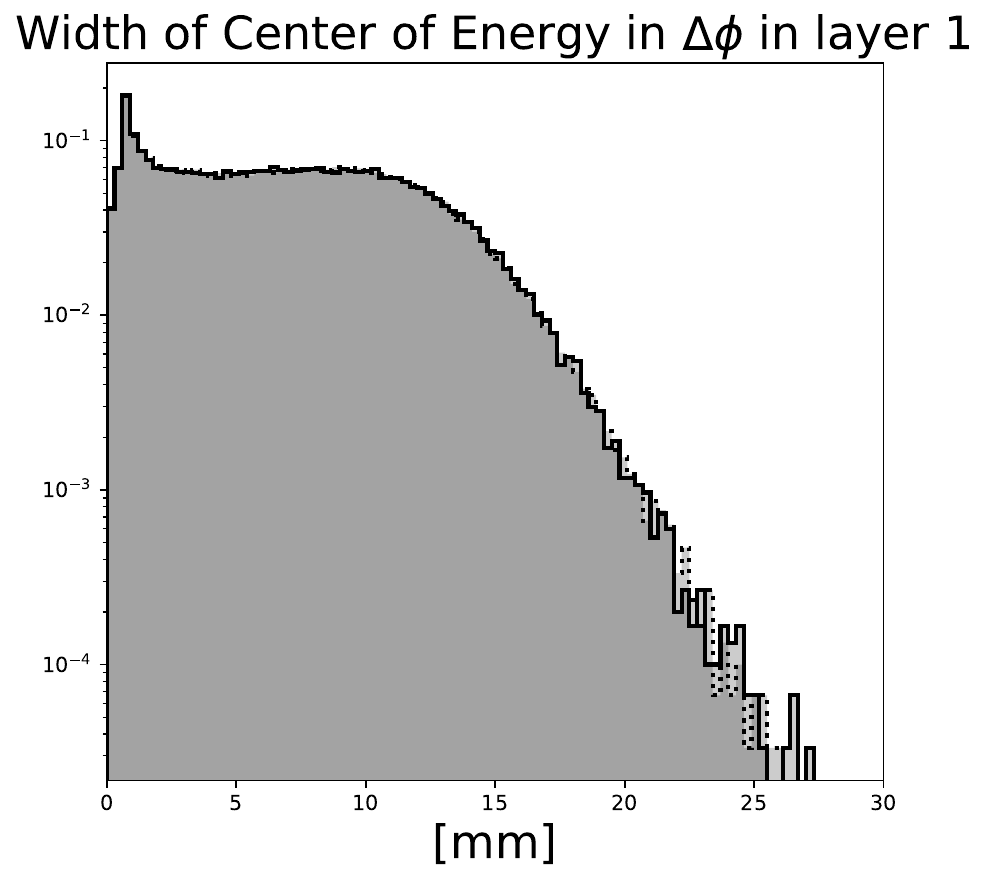} \hfill \includegraphics[height=0.1\textheight]{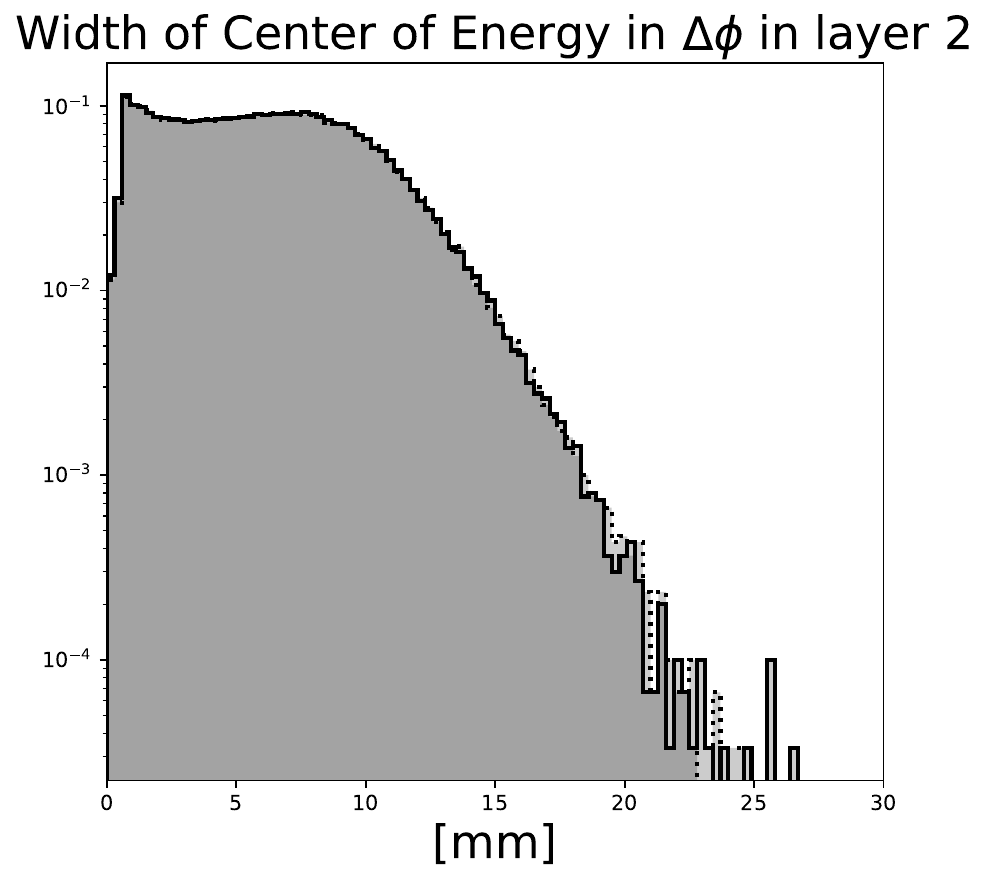} \hfill \includegraphics[height=0.1\textheight]{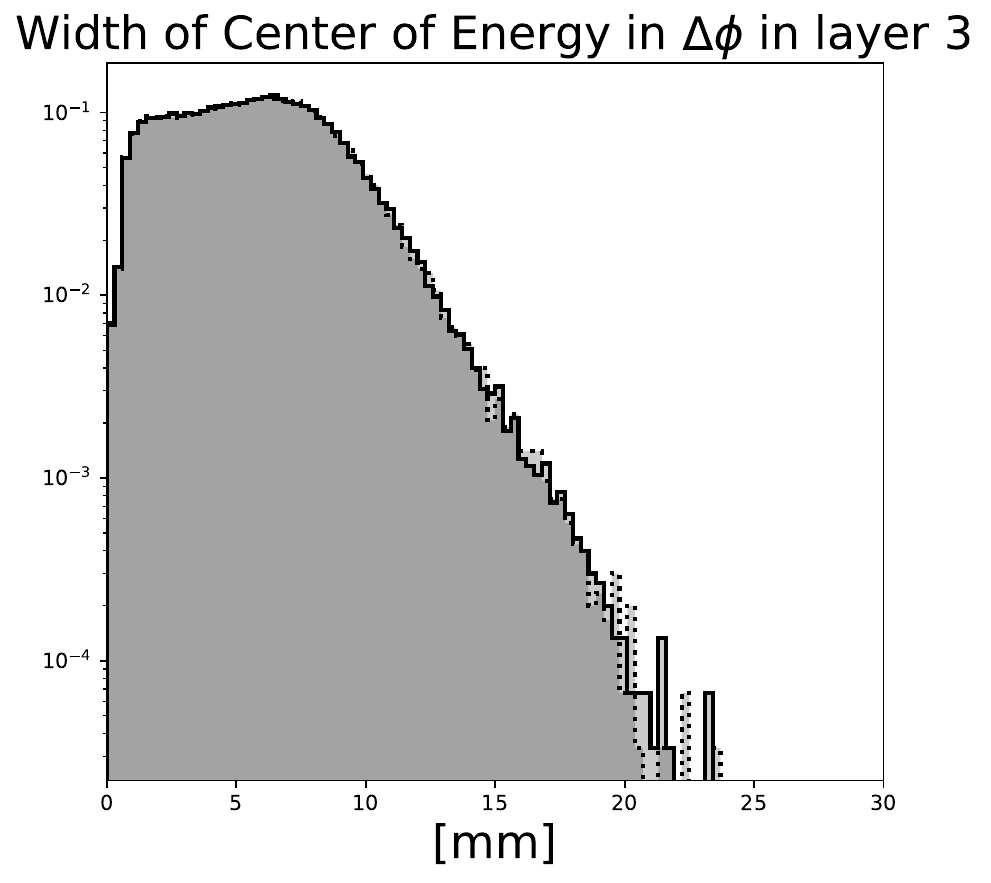} \hfill \includegraphics[height=0.1\textheight]{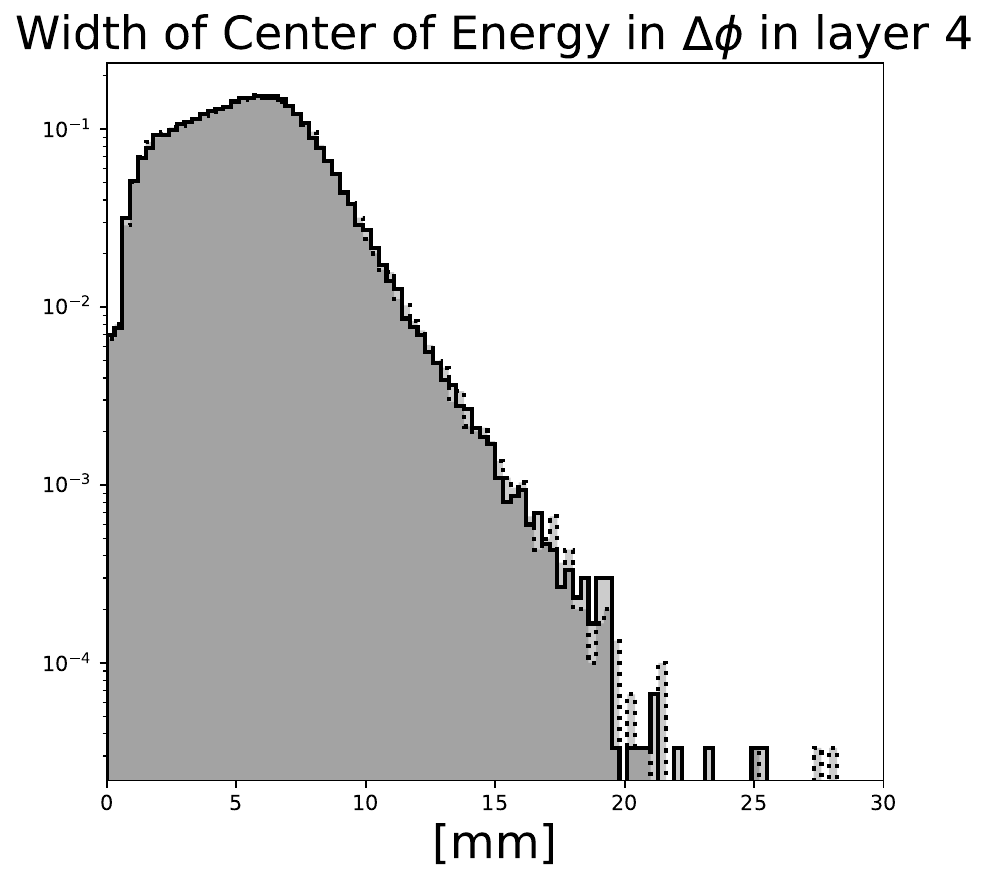}\\
    \includegraphics[height=0.1\textheight]{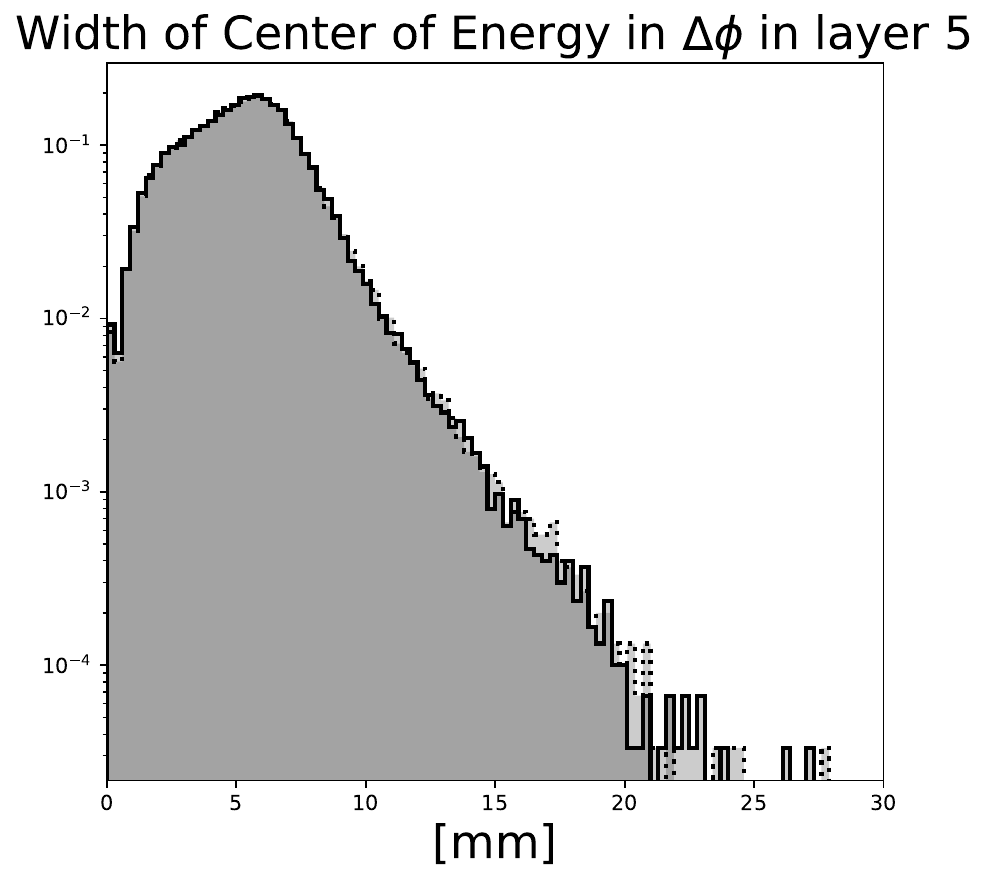} \hfill \includegraphics[height=0.1\textheight]{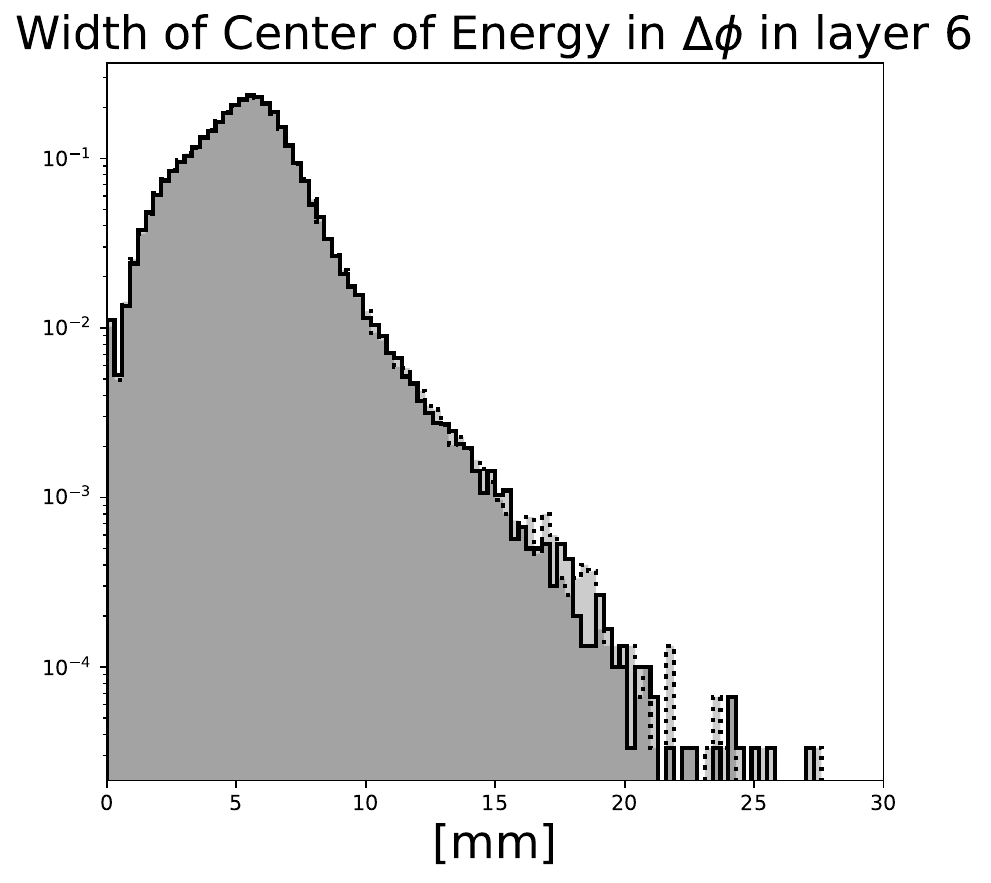} \hfill \includegraphics[height=0.1\textheight]{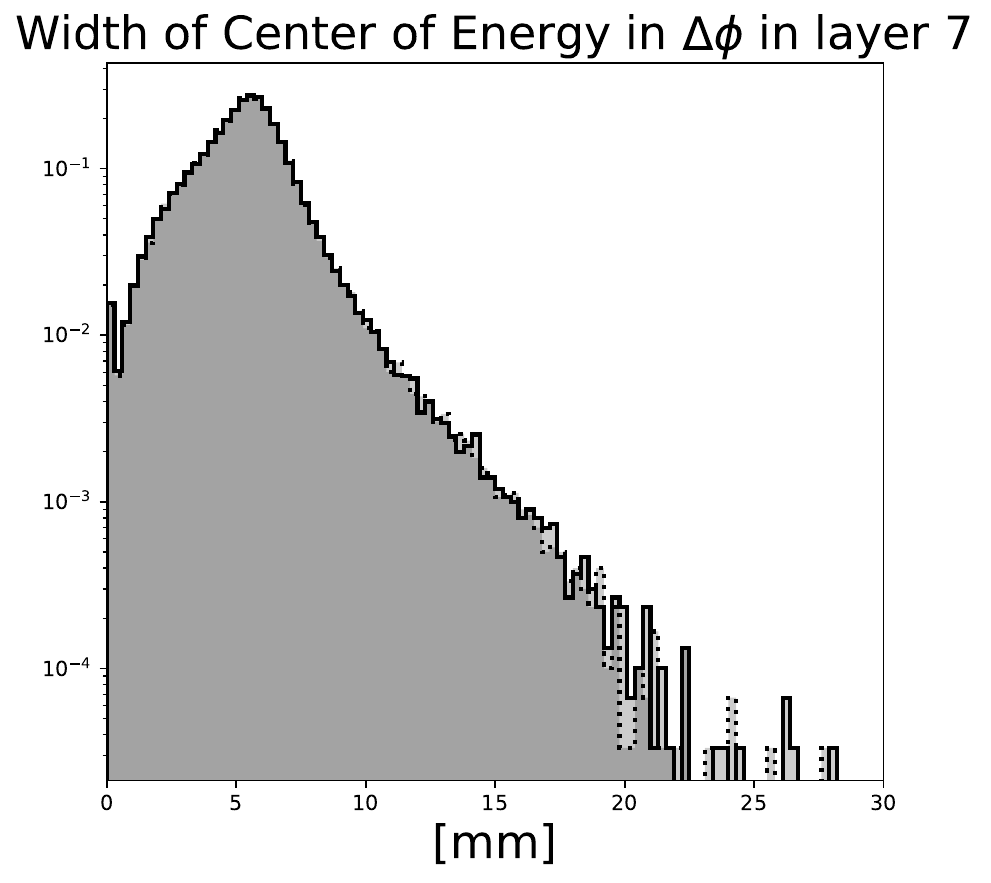} \hfill \includegraphics[height=0.1\textheight]{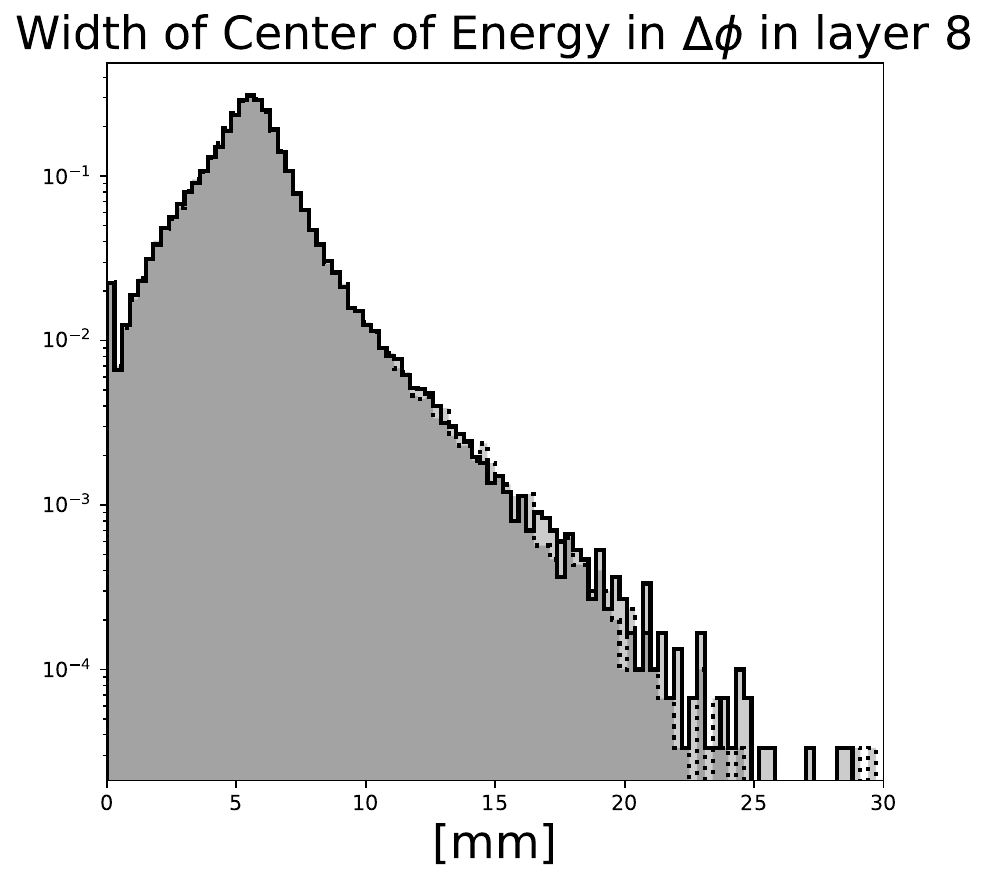} \hfill \includegraphics[height=0.1\textheight]{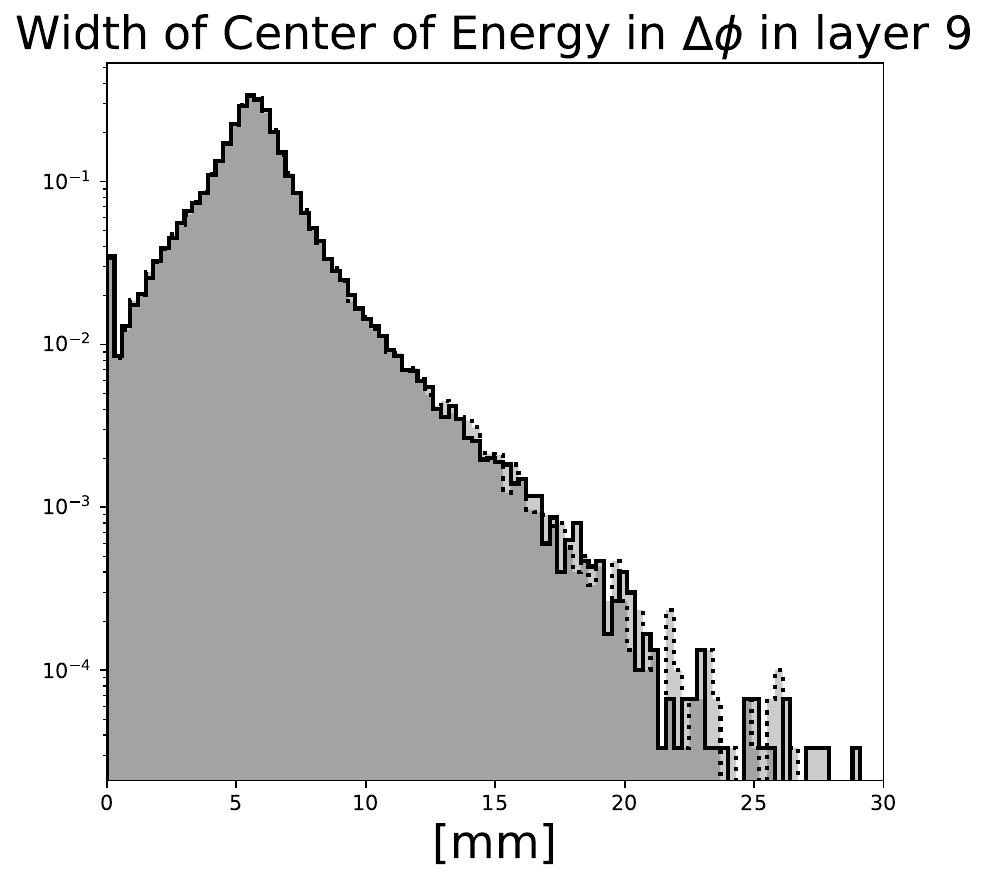}\\
    \includegraphics[height=0.1\textheight]{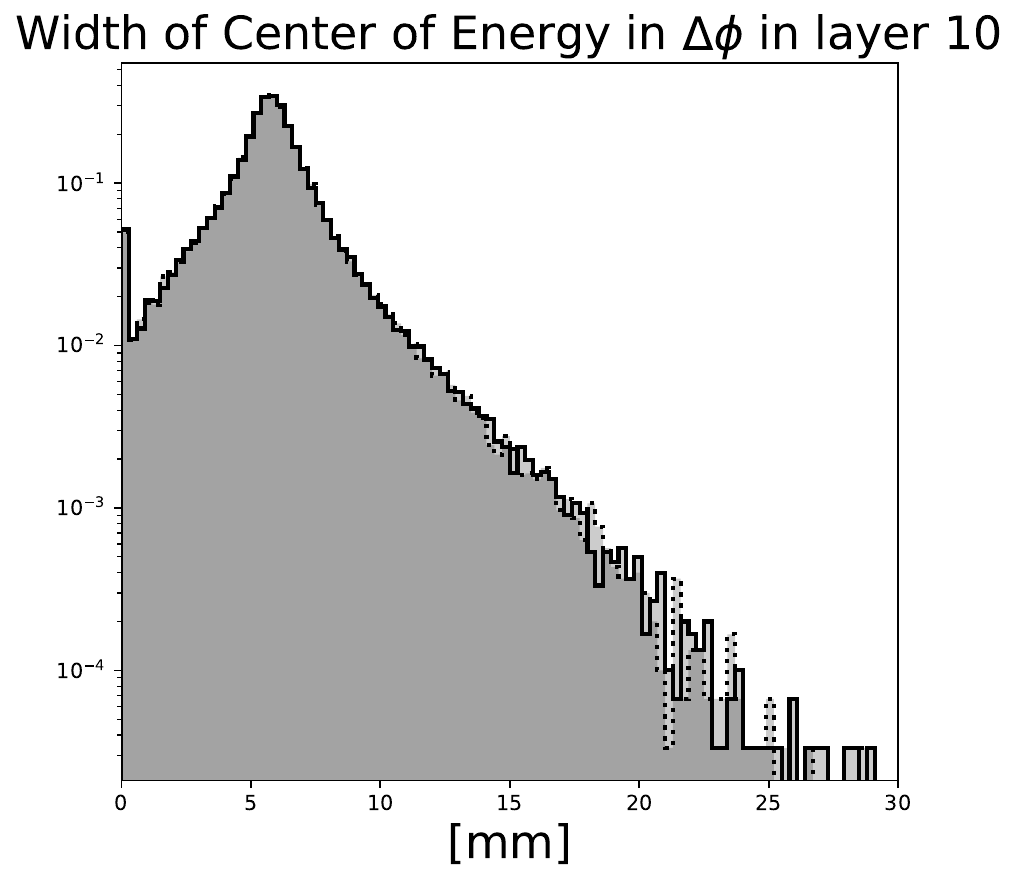} \hfill \includegraphics[height=0.1\textheight]{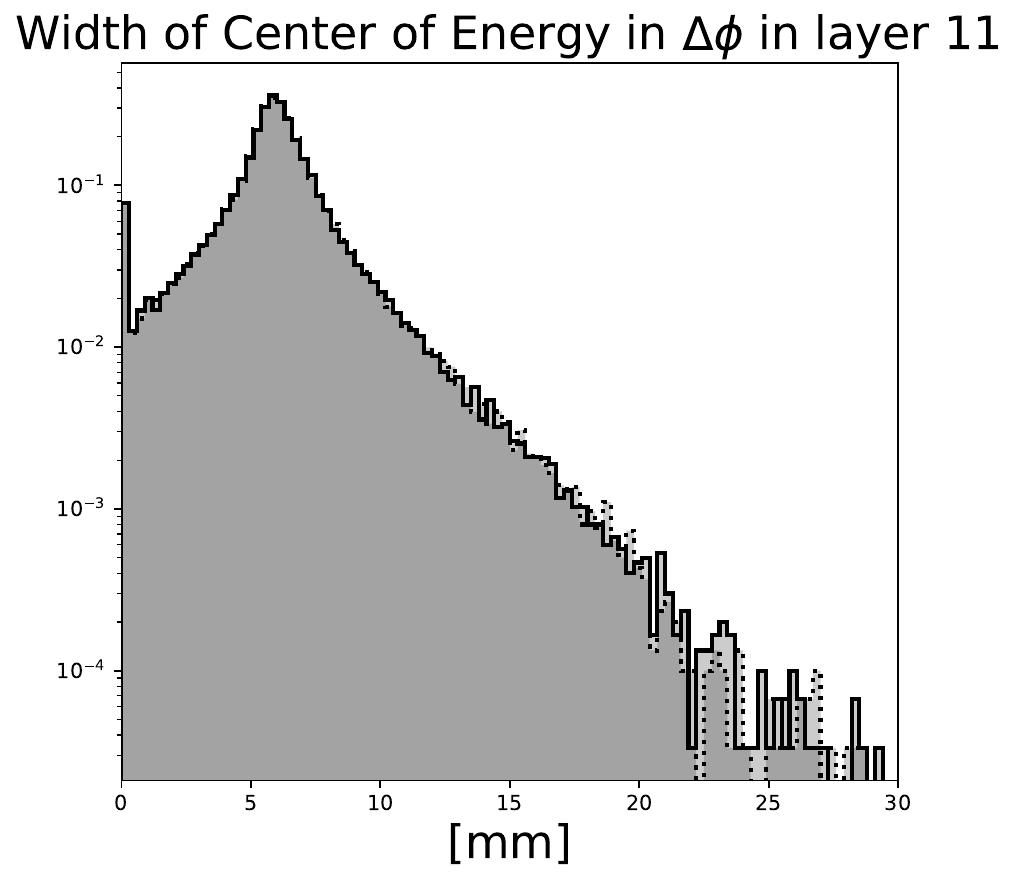} \hfill \includegraphics[height=0.1\textheight]{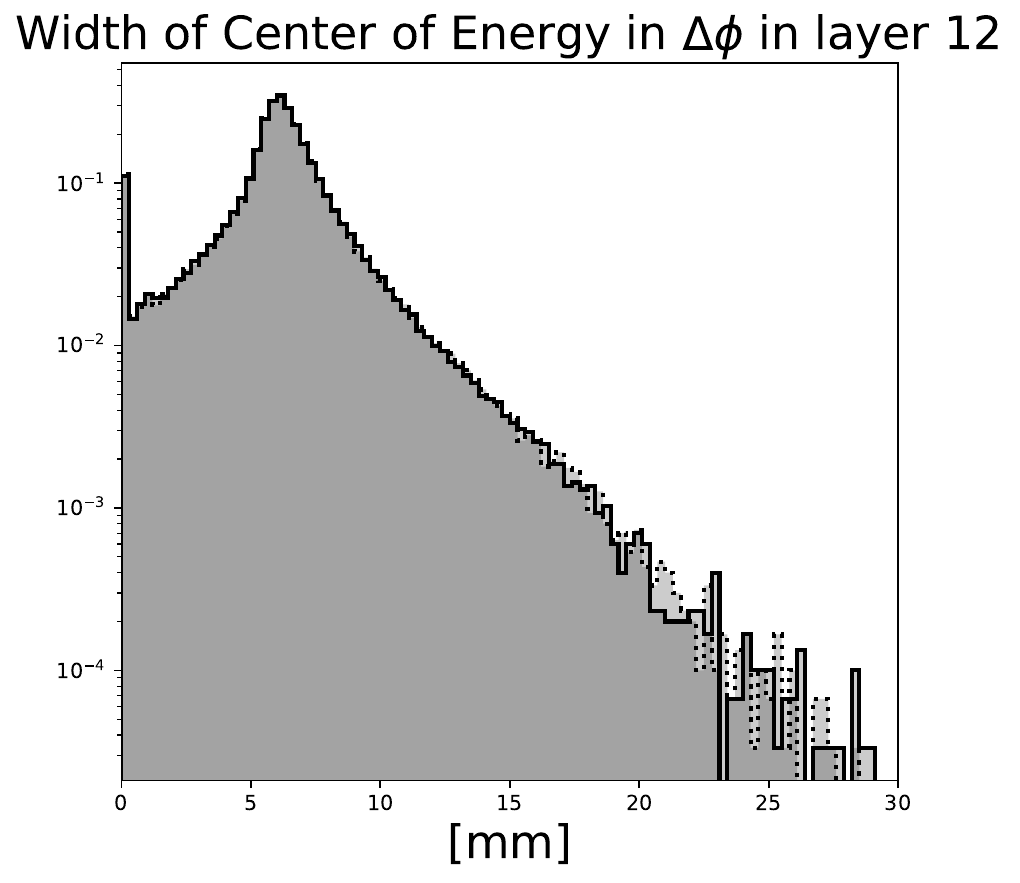} \hfill \includegraphics[height=0.1\textheight]{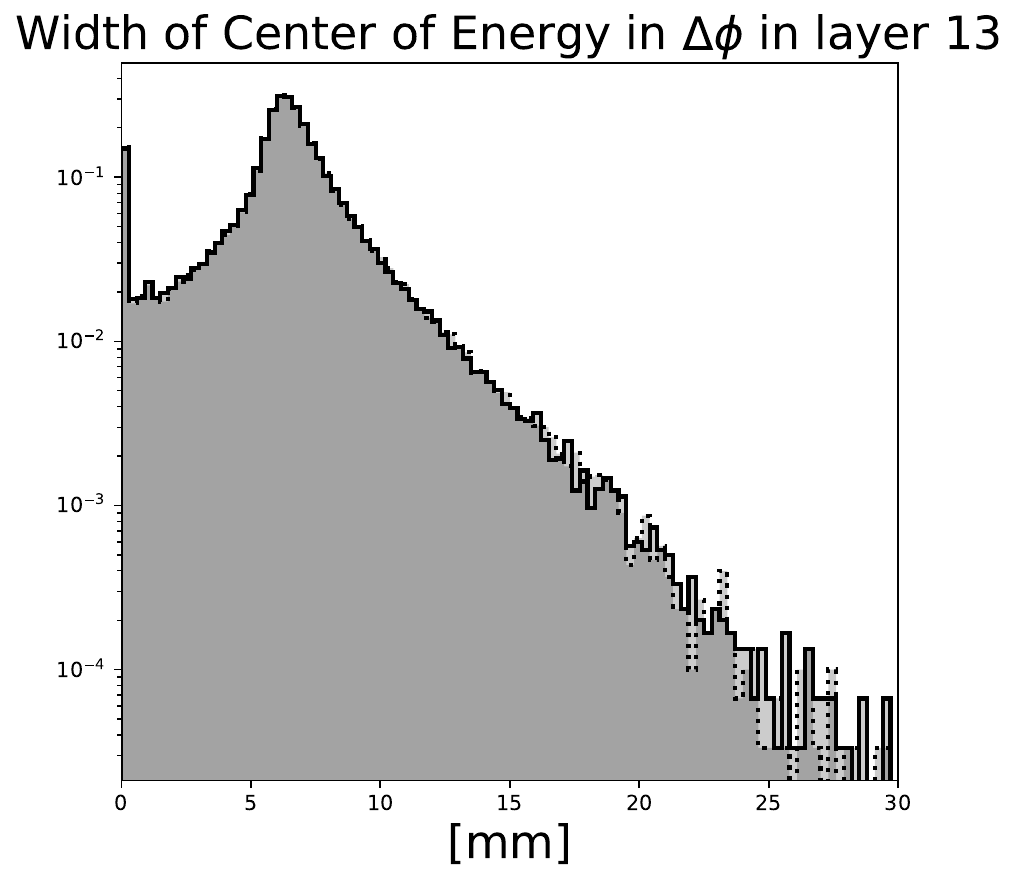} \hfill \includegraphics[height=0.1\textheight]{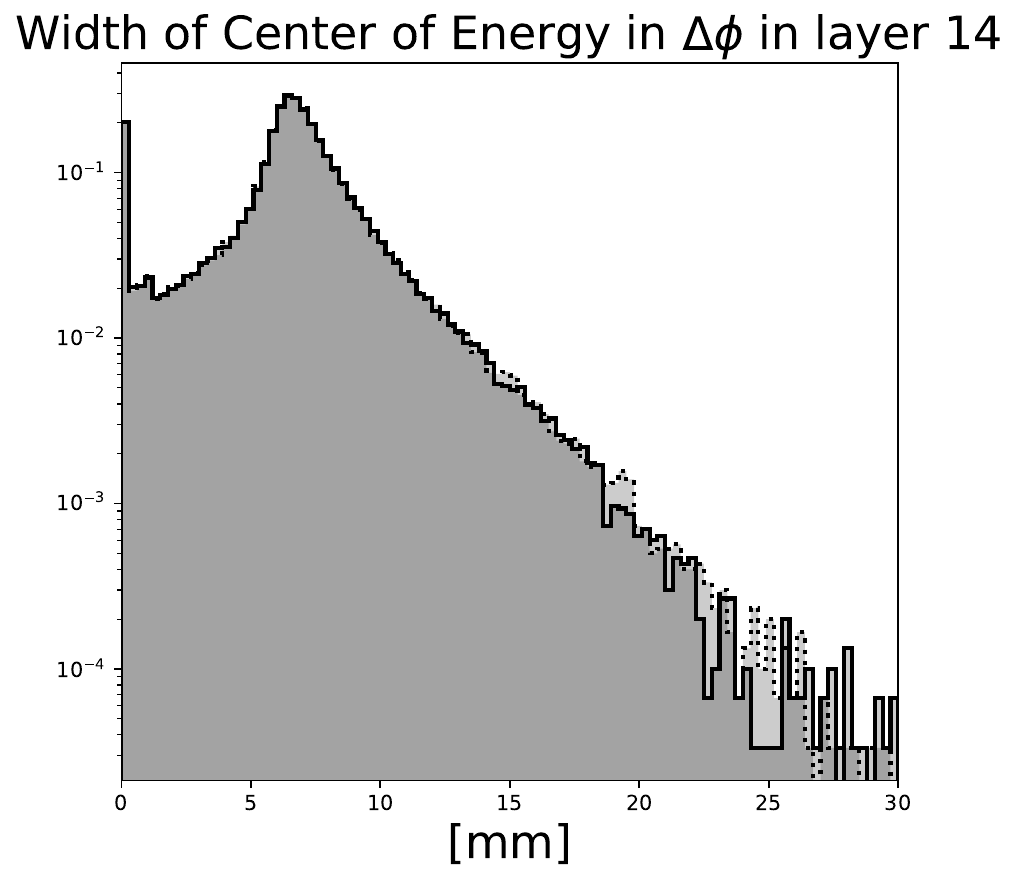}\\
    \includegraphics[height=0.1\textheight]{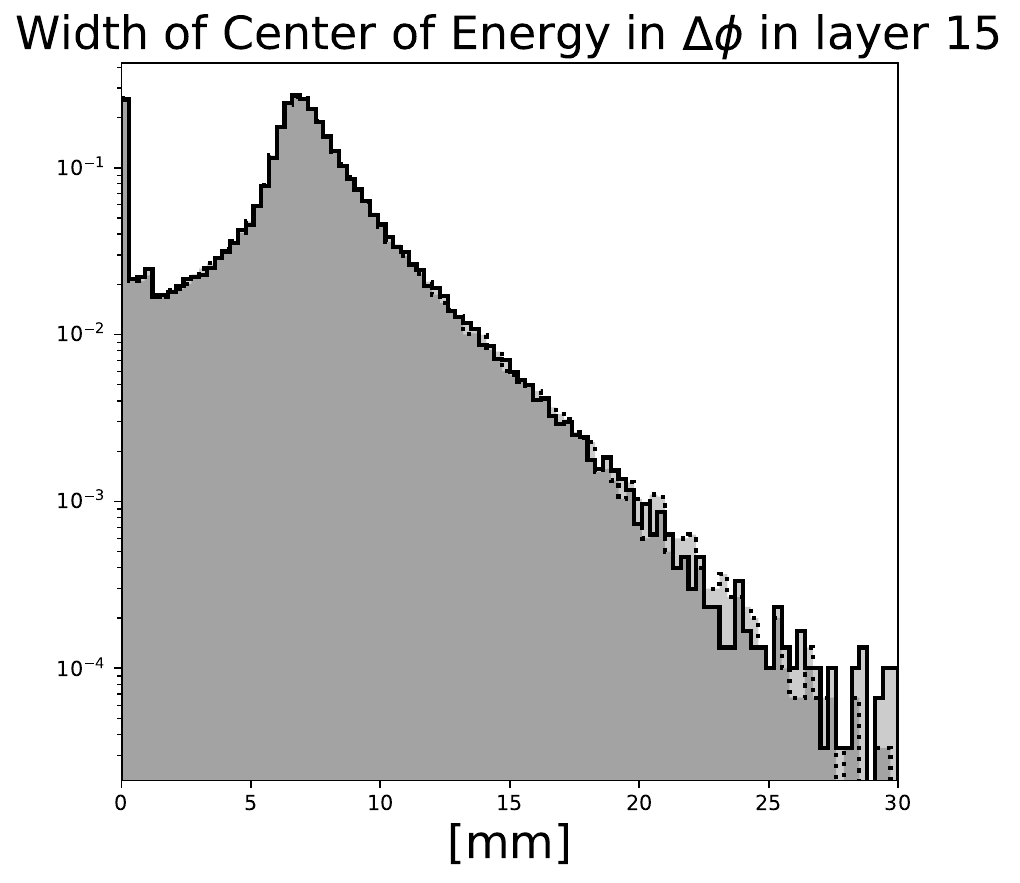} \hfill \includegraphics[height=0.1\textheight]{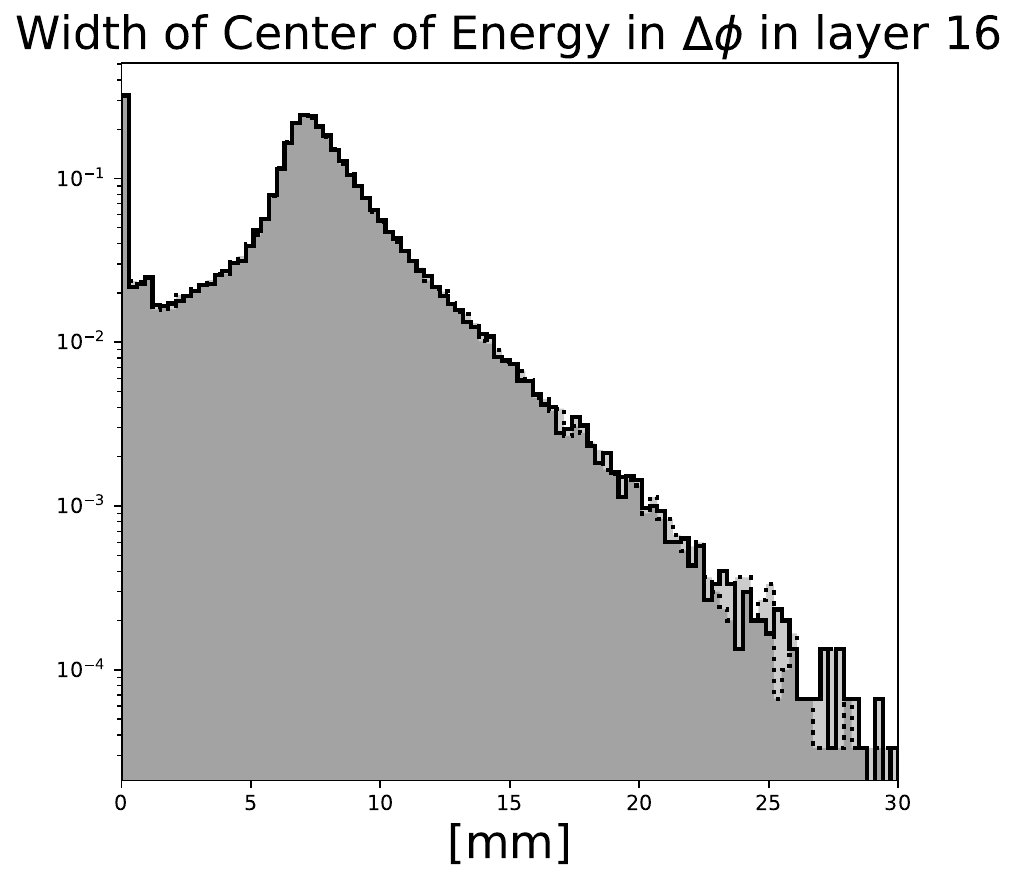} \hfill \includegraphics[height=0.1\textheight]{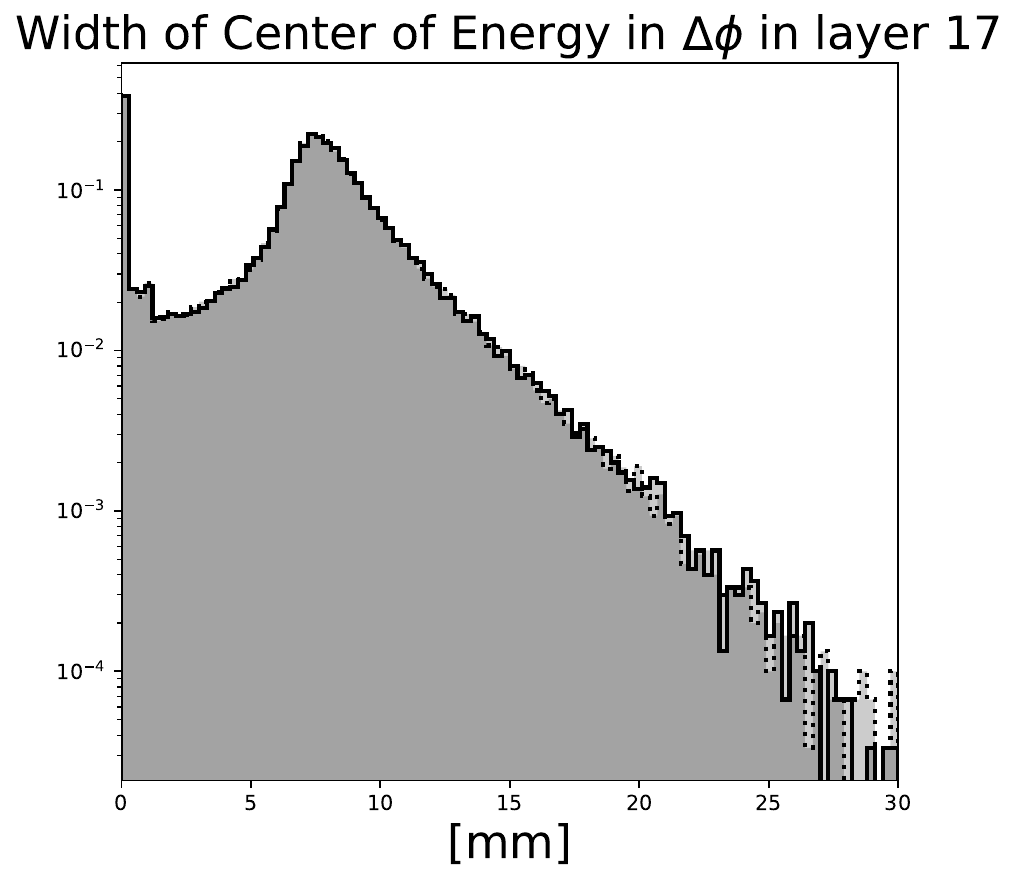} \hfill \includegraphics[height=0.1\textheight]{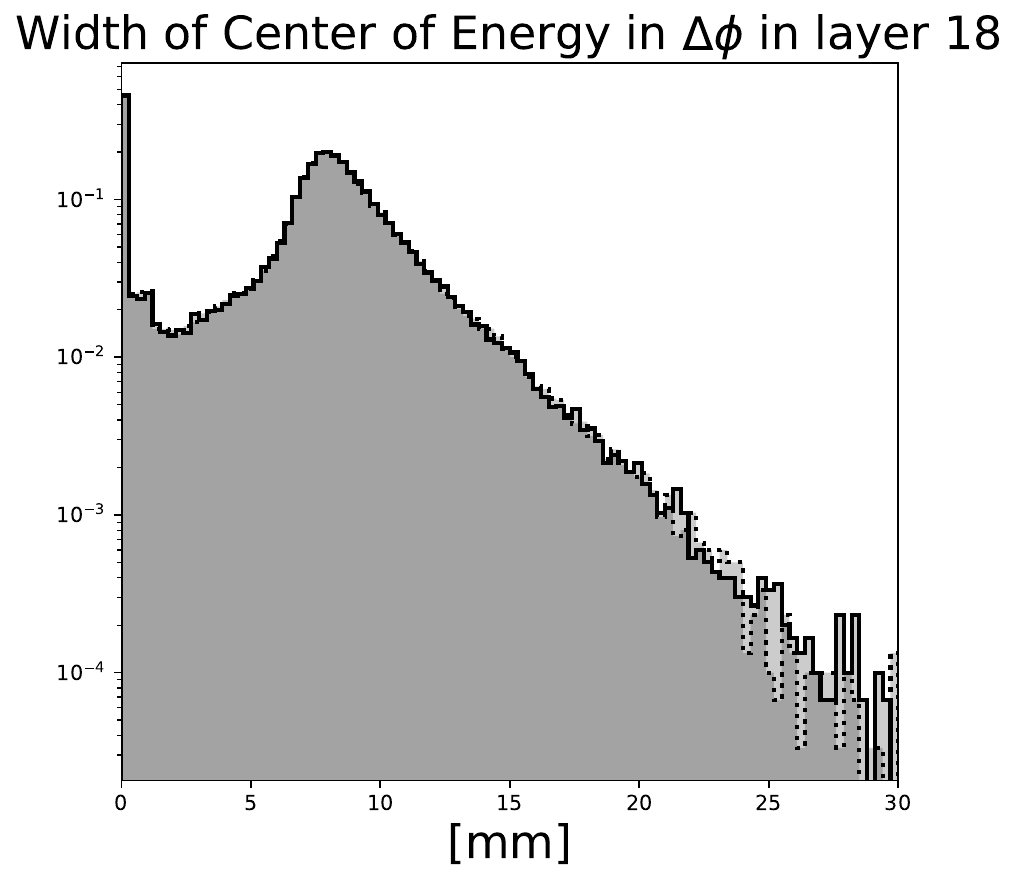} \hfill \includegraphics[height=0.1\textheight]{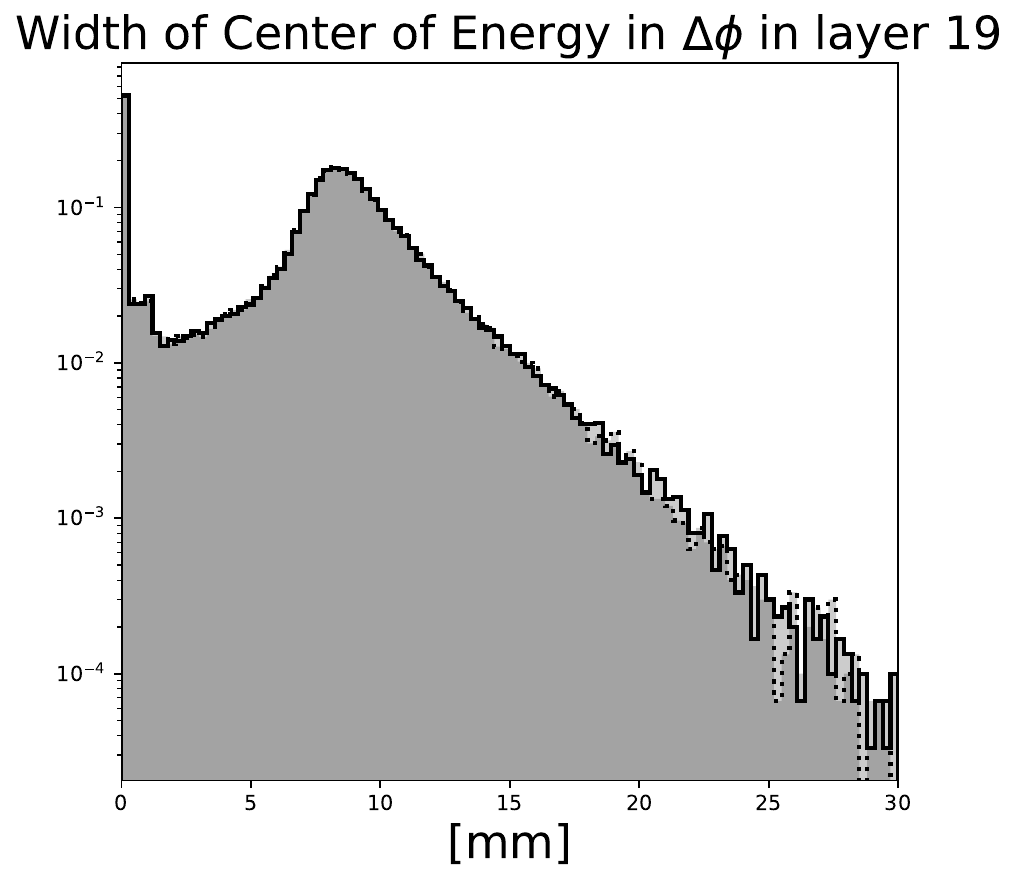}\\
    \includegraphics[height=0.1\textheight]{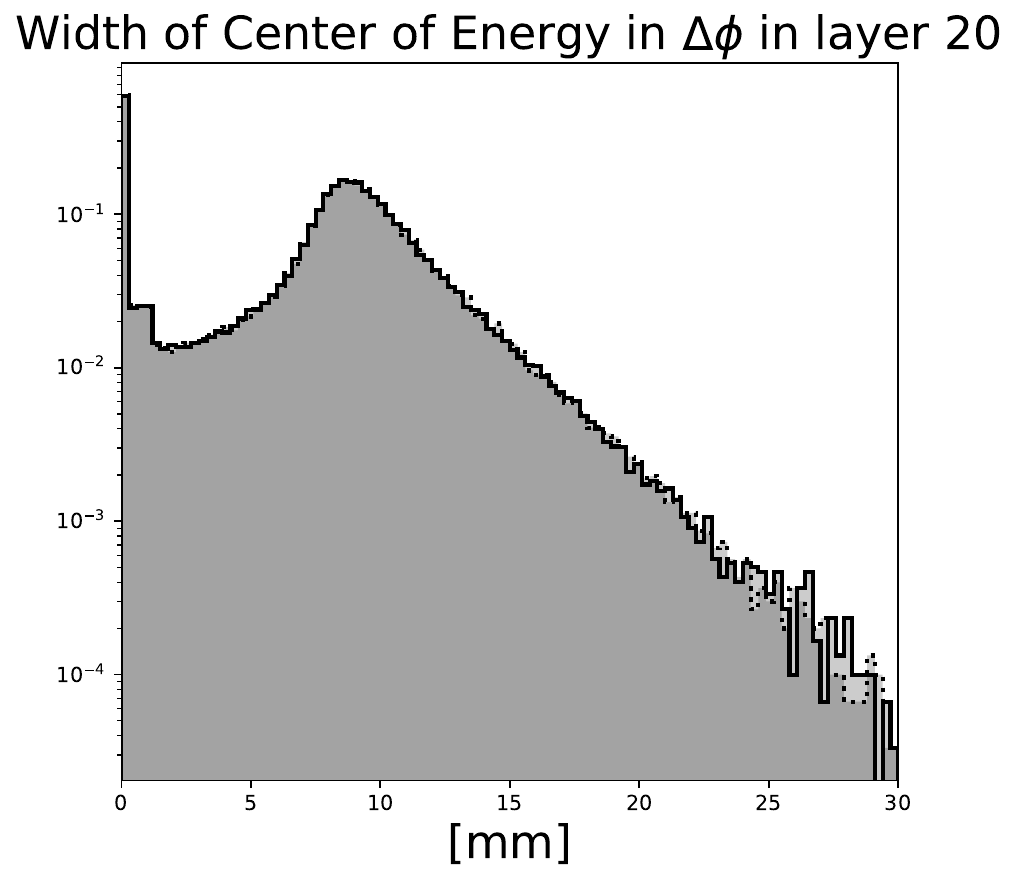} \hfill \includegraphics[height=0.1\textheight]{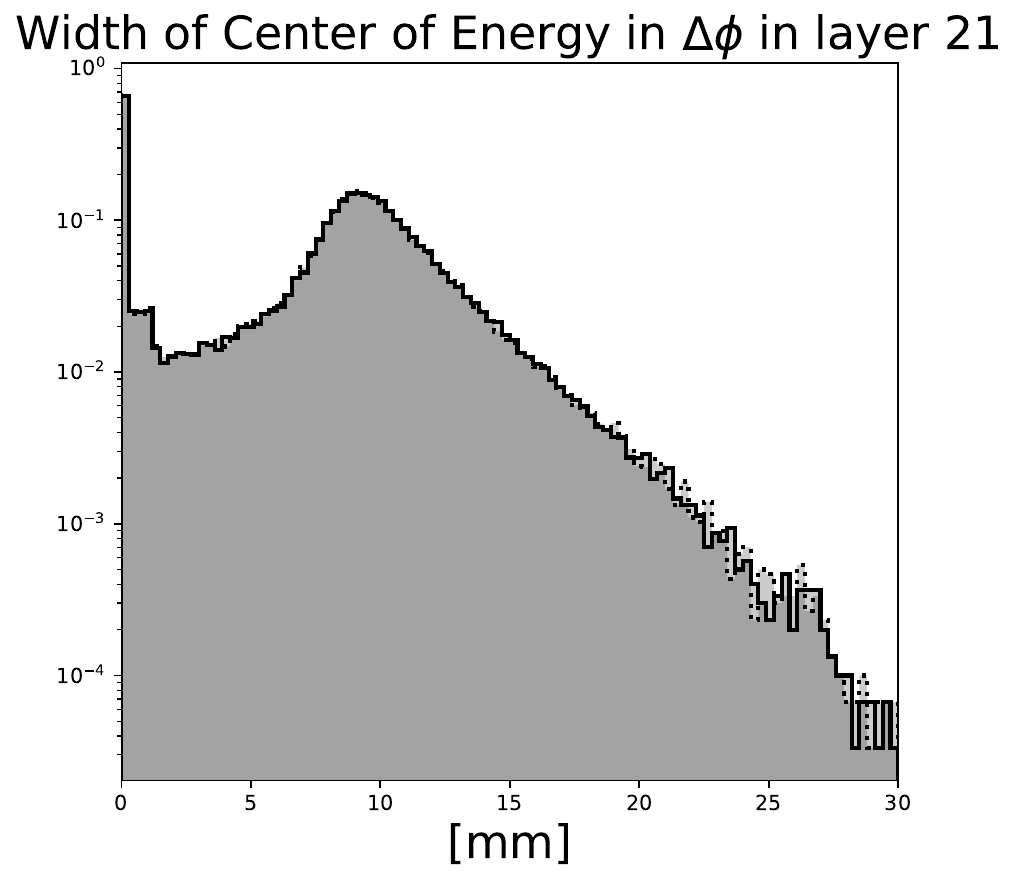} \hfill \includegraphics[height=0.1\textheight]{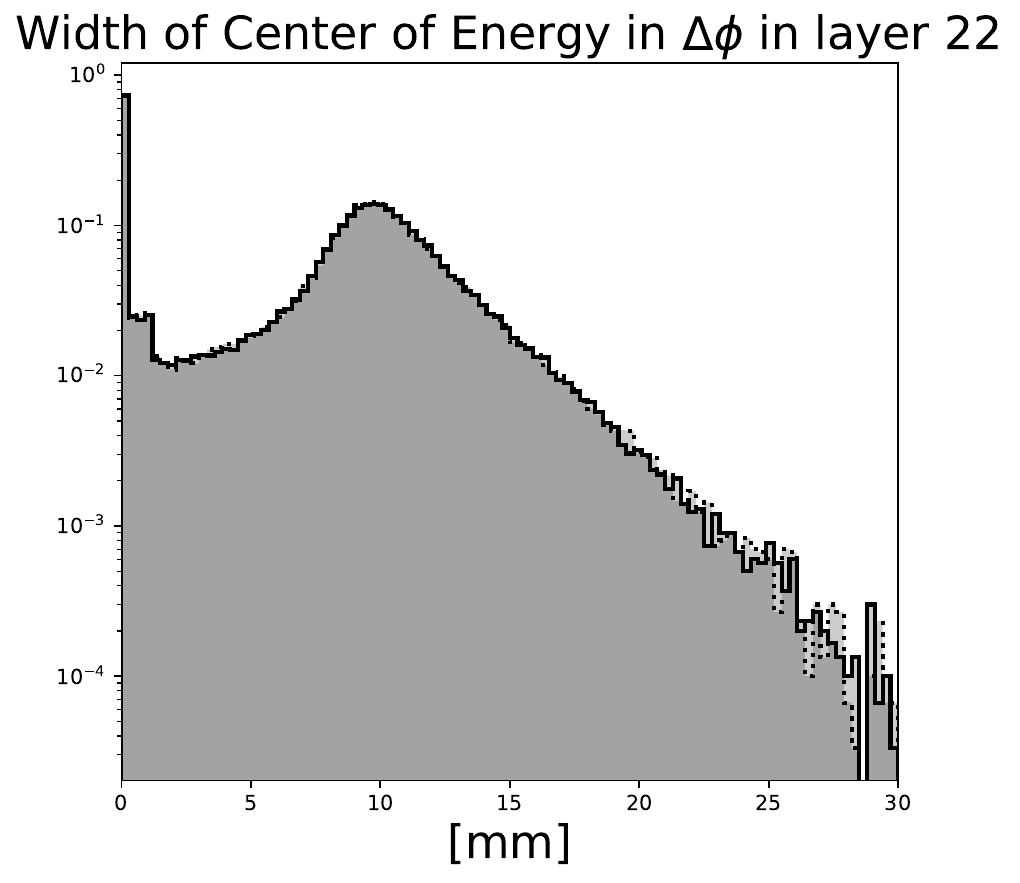} \hfill \includegraphics[height=0.1\textheight]{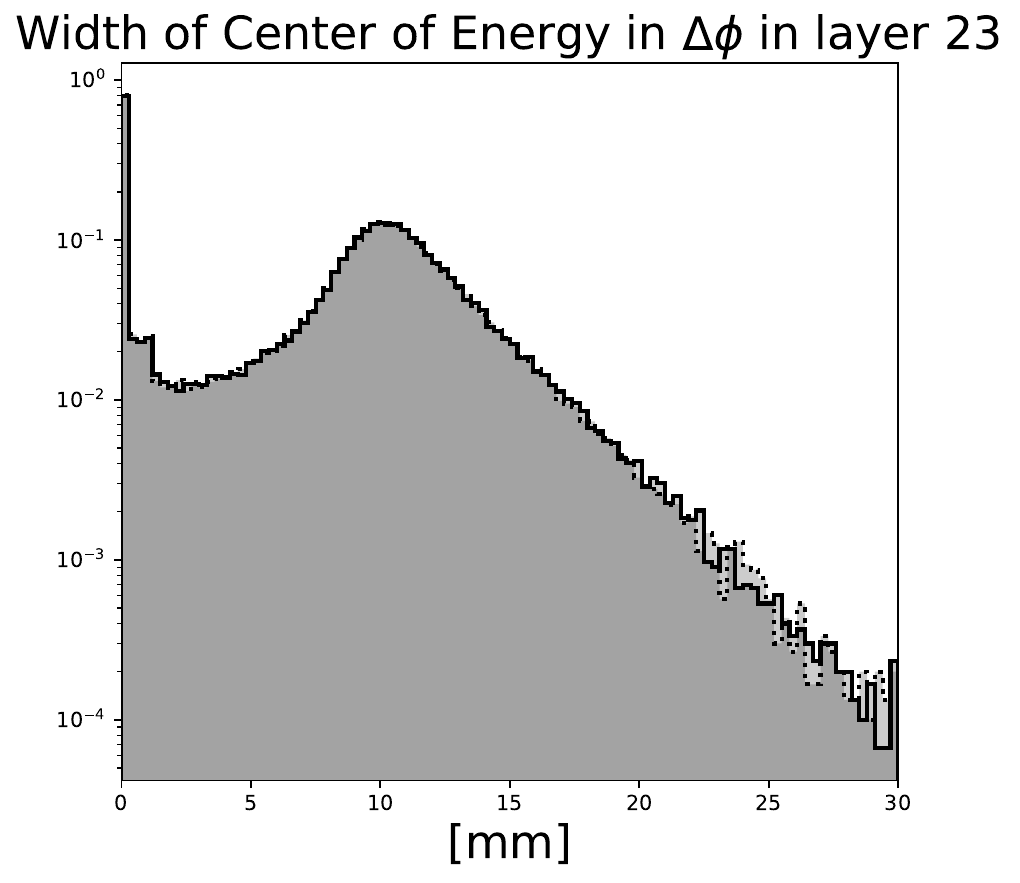} \hfill \includegraphics[height=0.1\textheight]{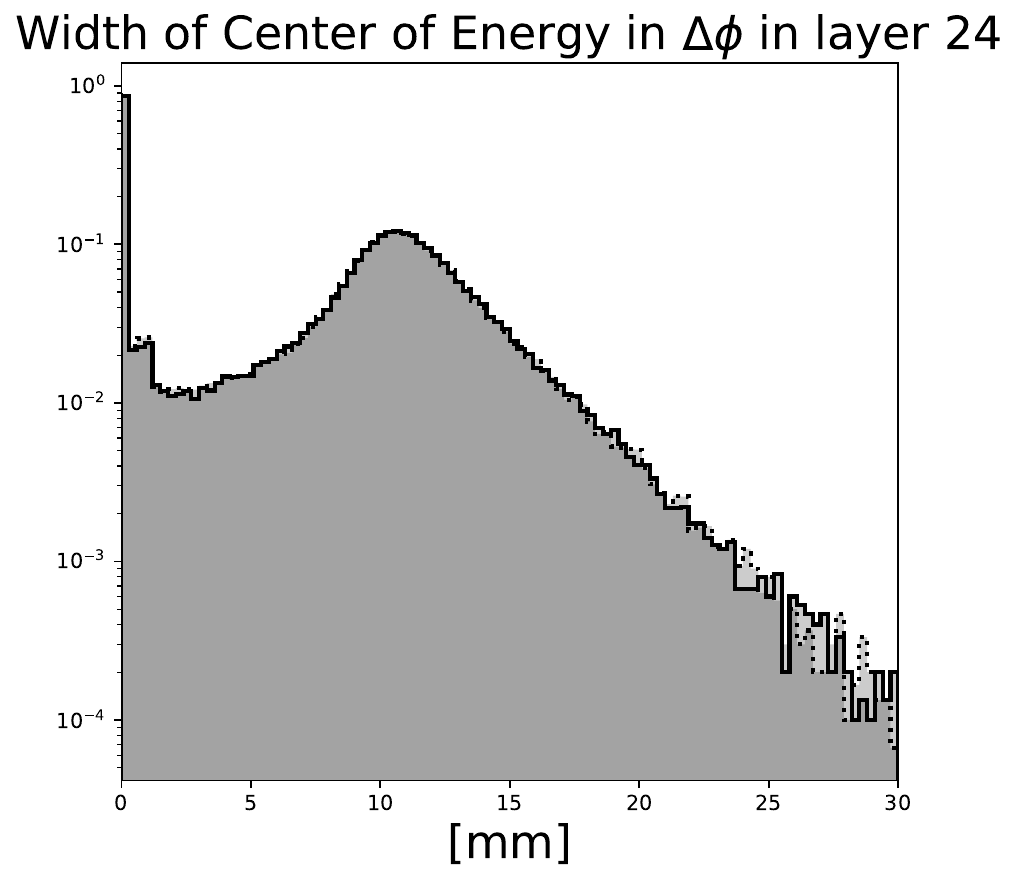}\\
    \includegraphics[height=0.1\textheight]{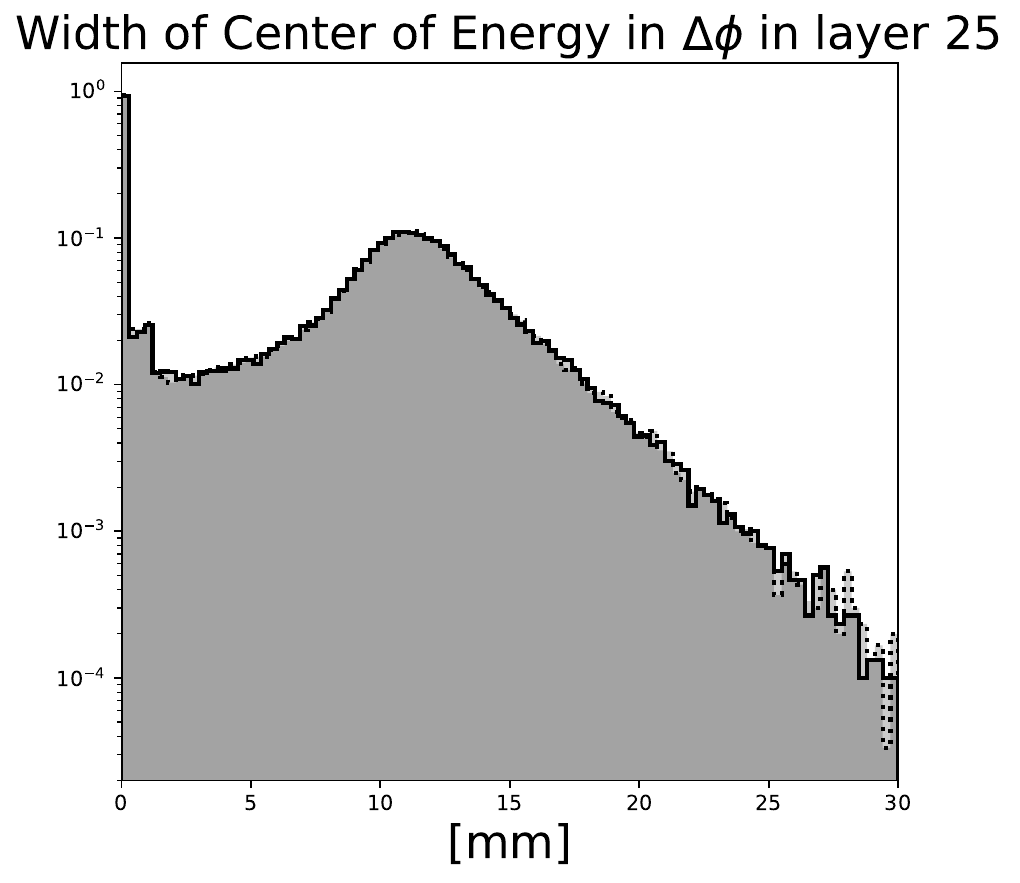} \hfill \includegraphics[height=0.1\textheight]{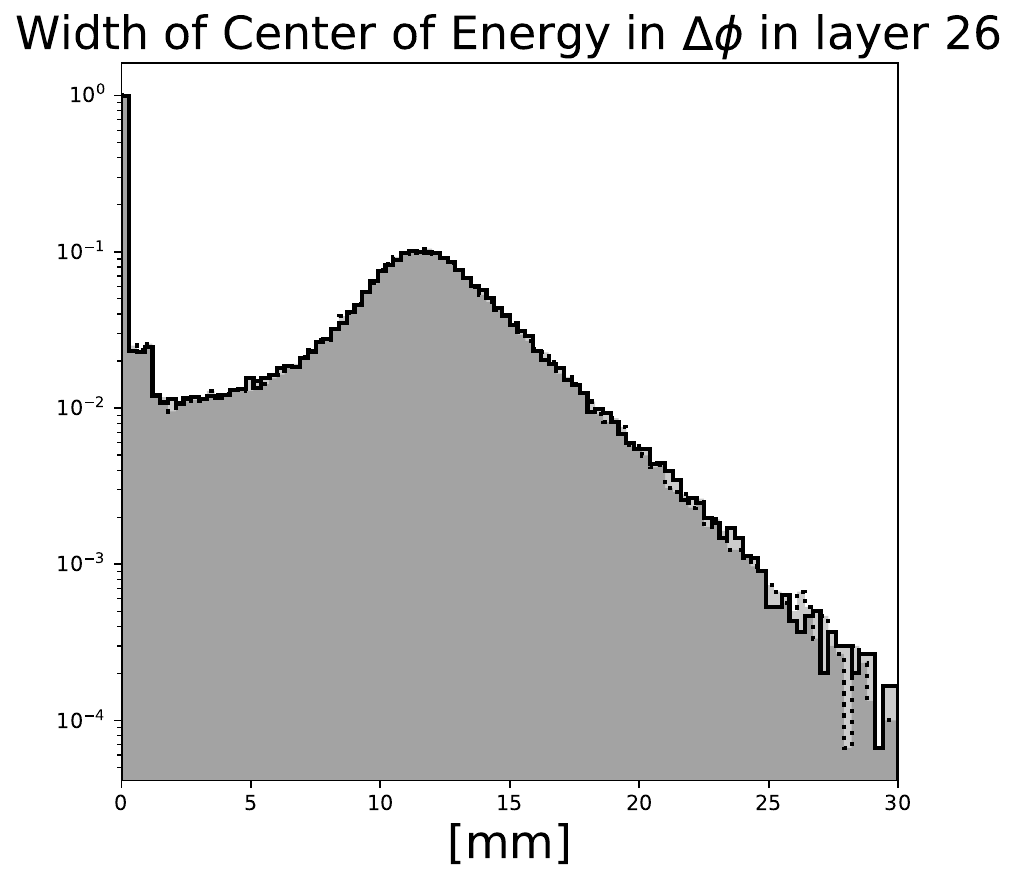} \hfill \includegraphics[height=0.1\textheight]{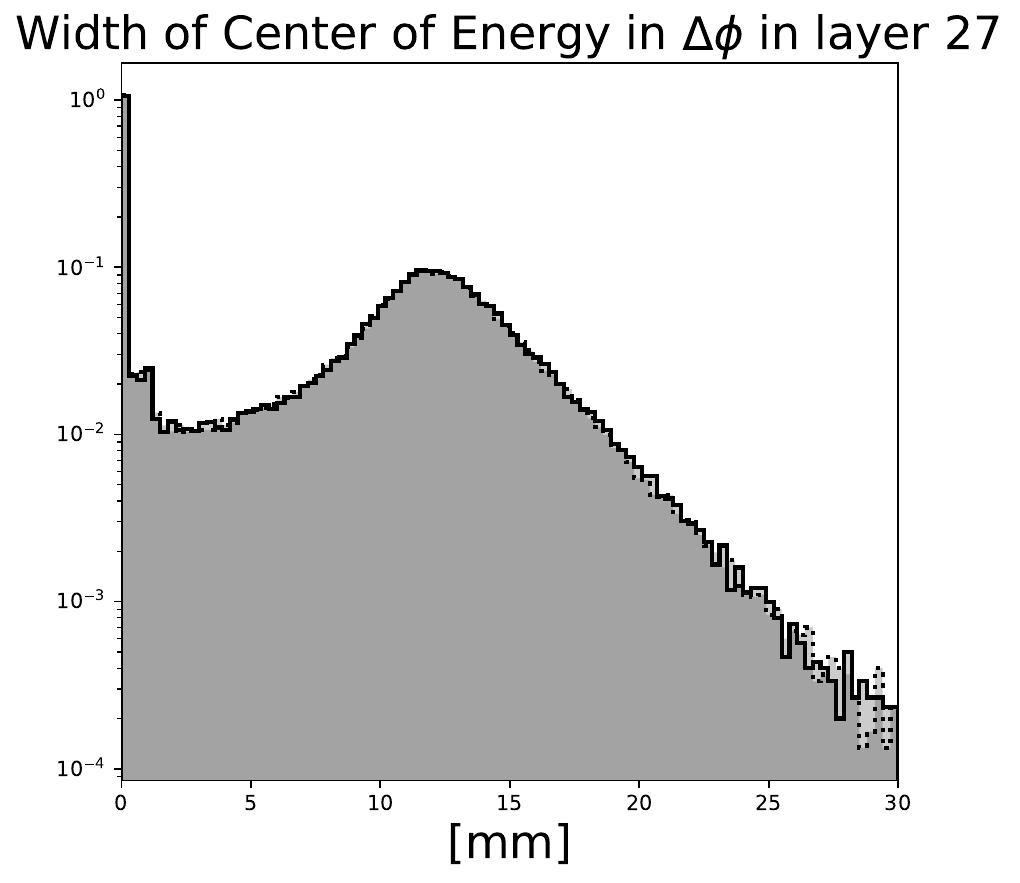} \hfill \includegraphics[height=0.1\textheight]{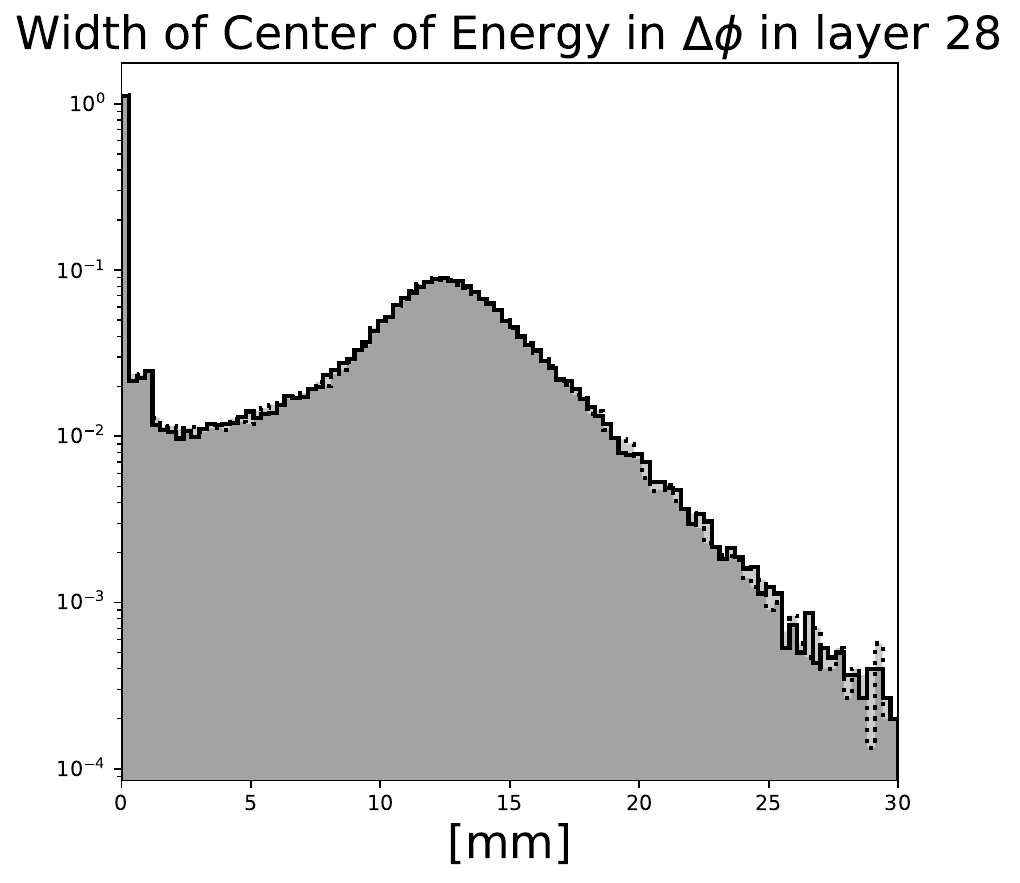} \hfill \includegraphics[height=0.1\textheight]{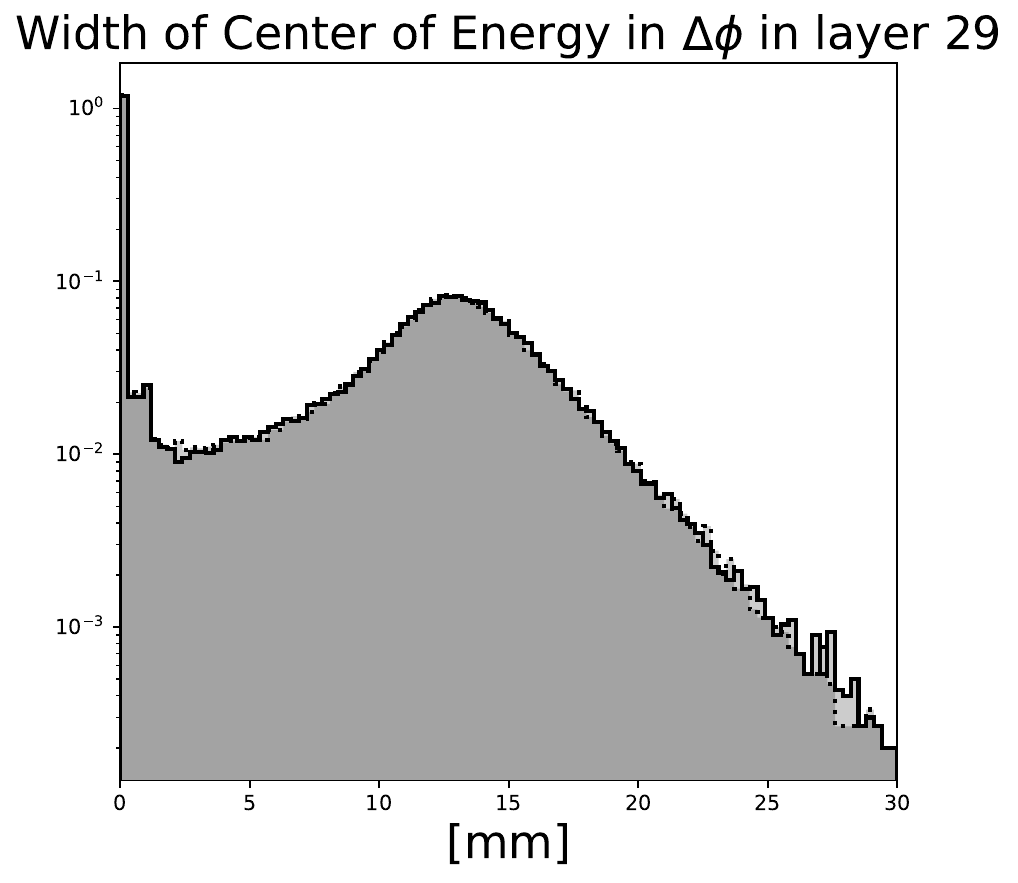}\\
    \includegraphics[height=0.1\textheight]{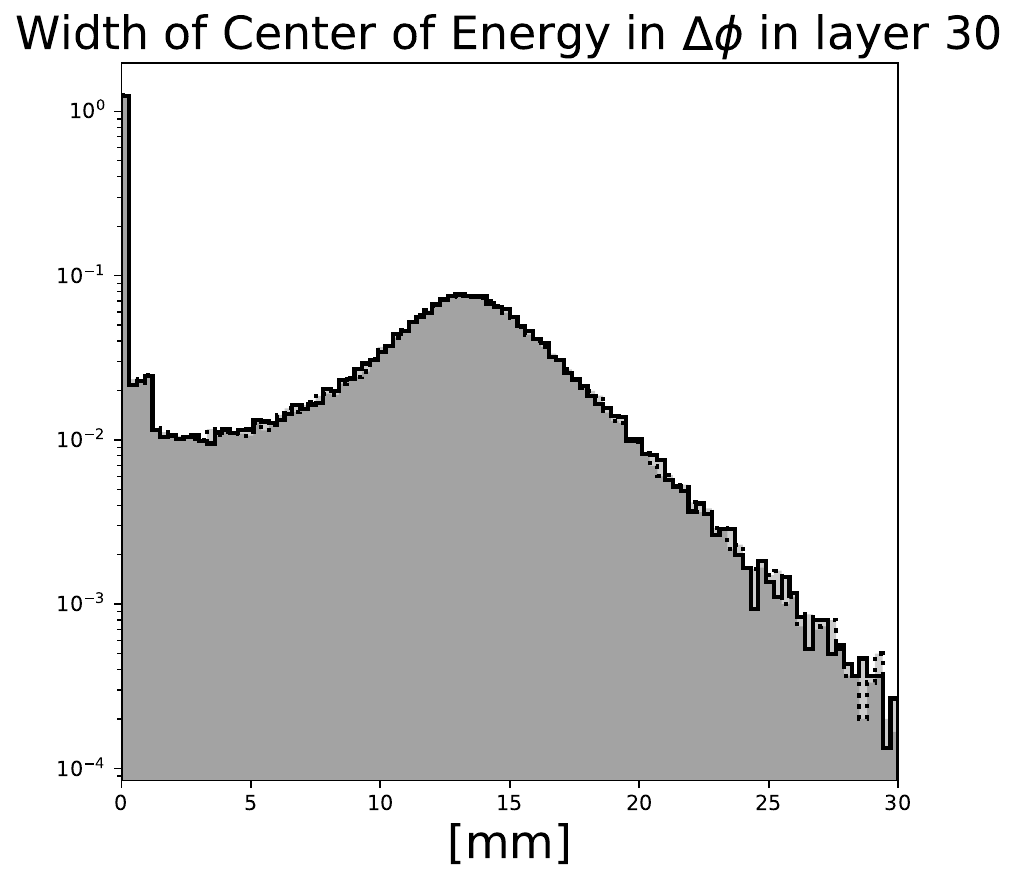} \hfill \includegraphics[height=0.1\textheight]{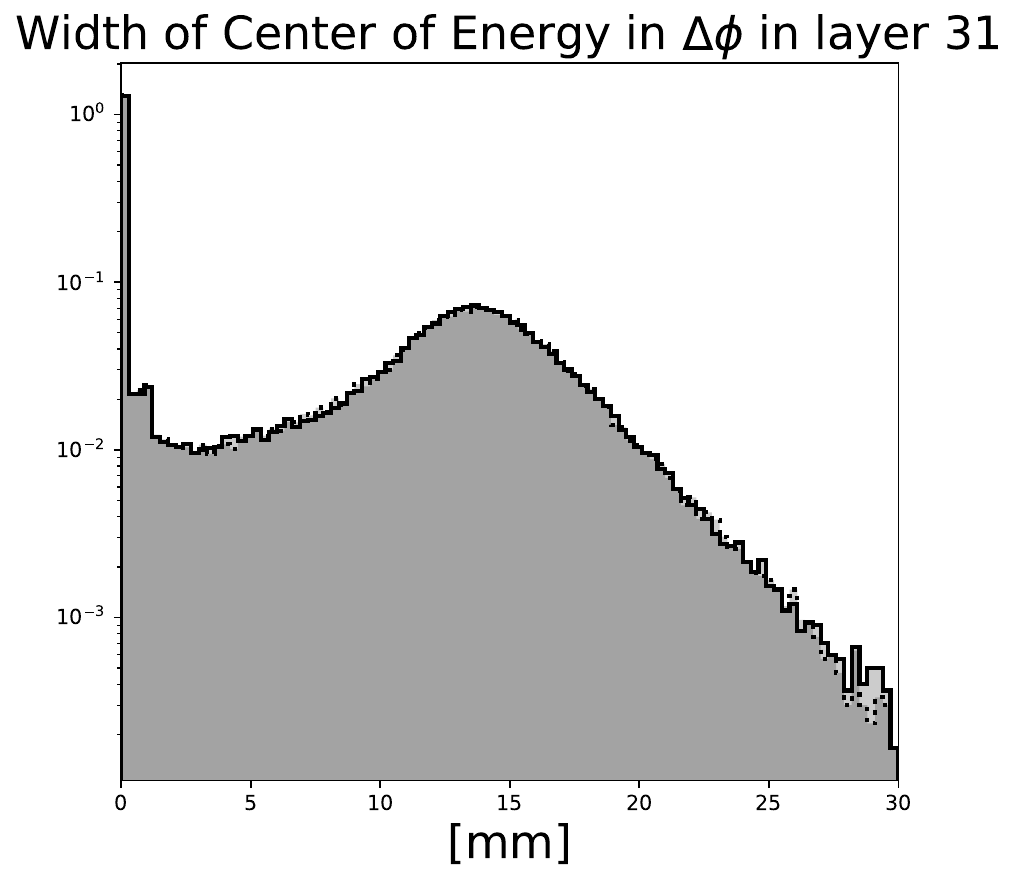} \hfill \includegraphics[height=0.1\textheight]{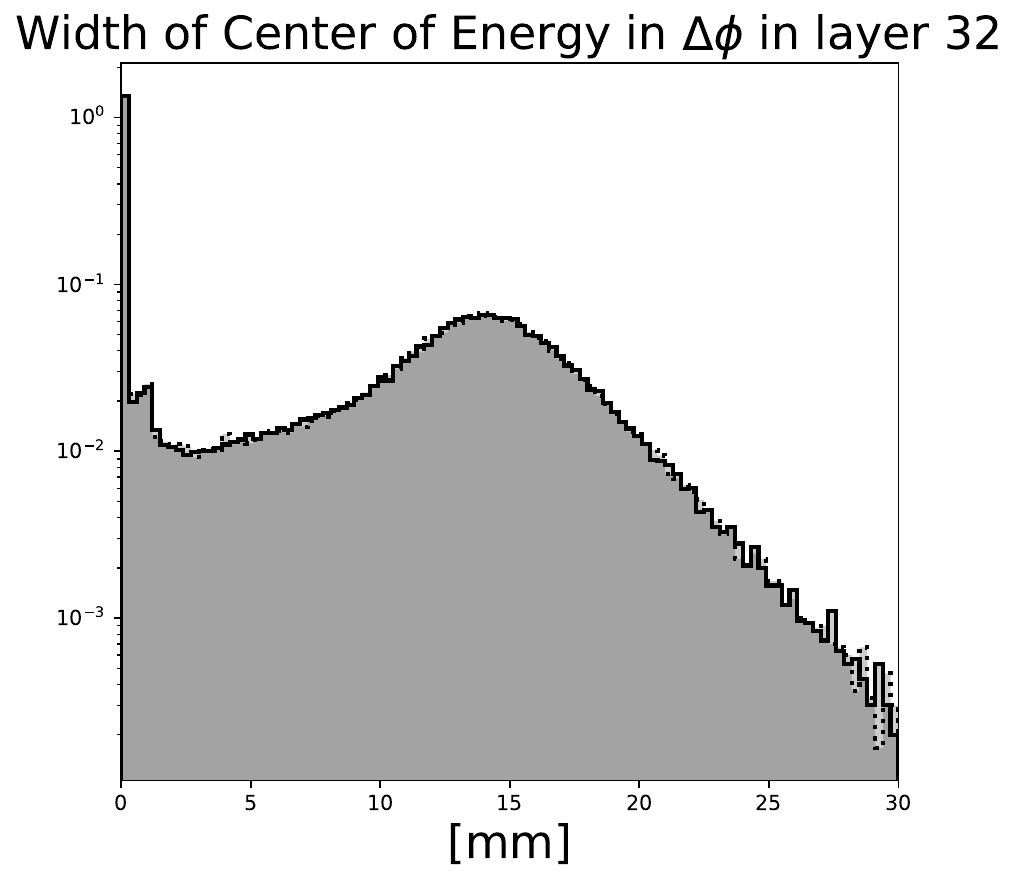} \hfill \includegraphics[height=0.1\textheight]{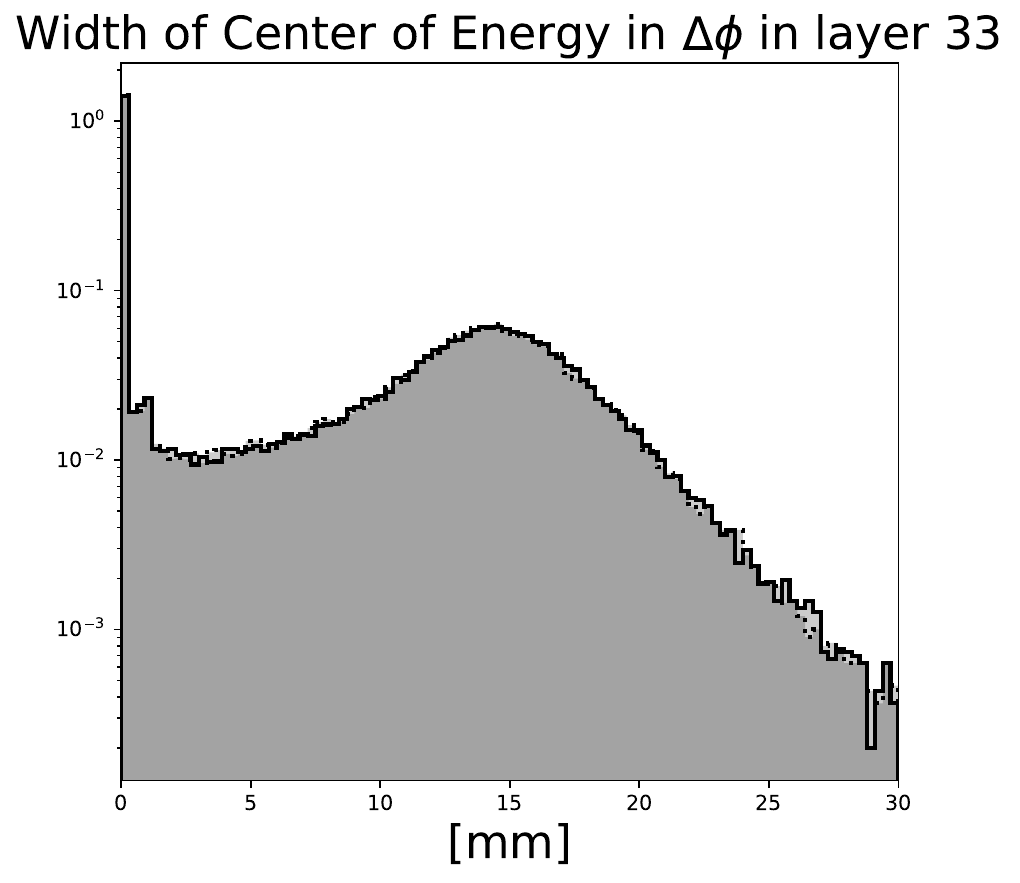} \hfill \includegraphics[height=0.1\textheight]{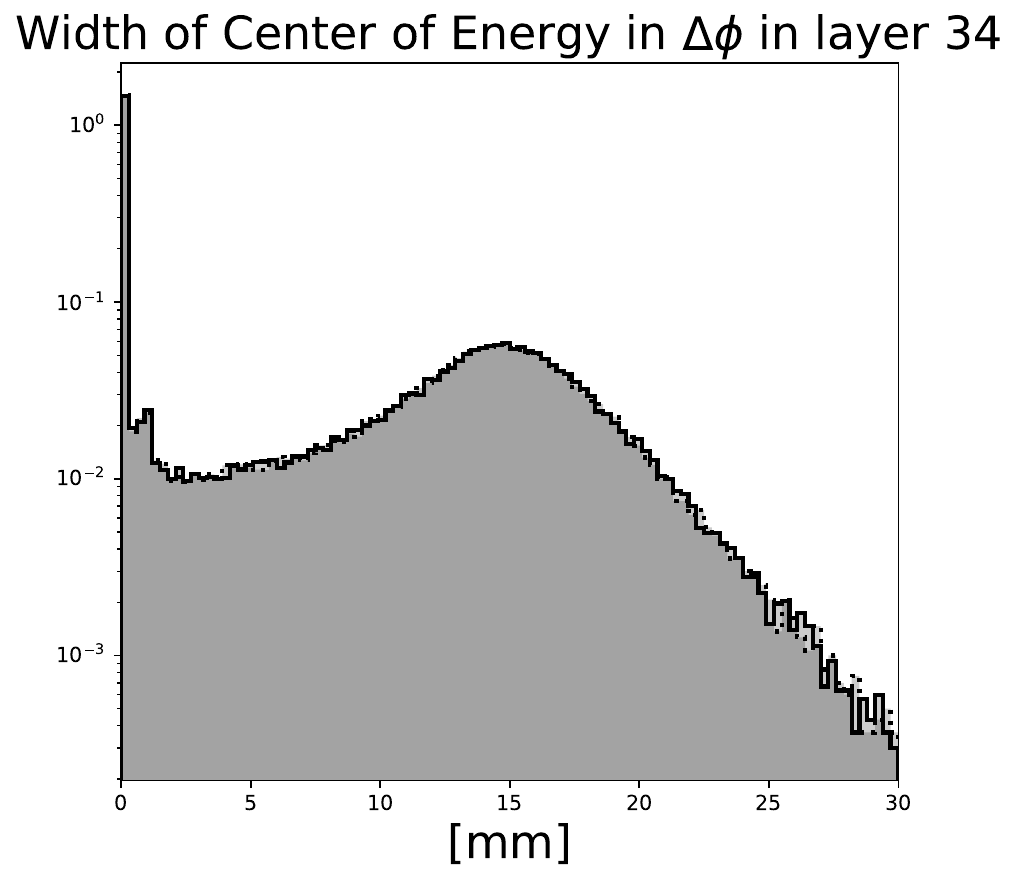}\\
    \includegraphics[height=0.1\textheight]{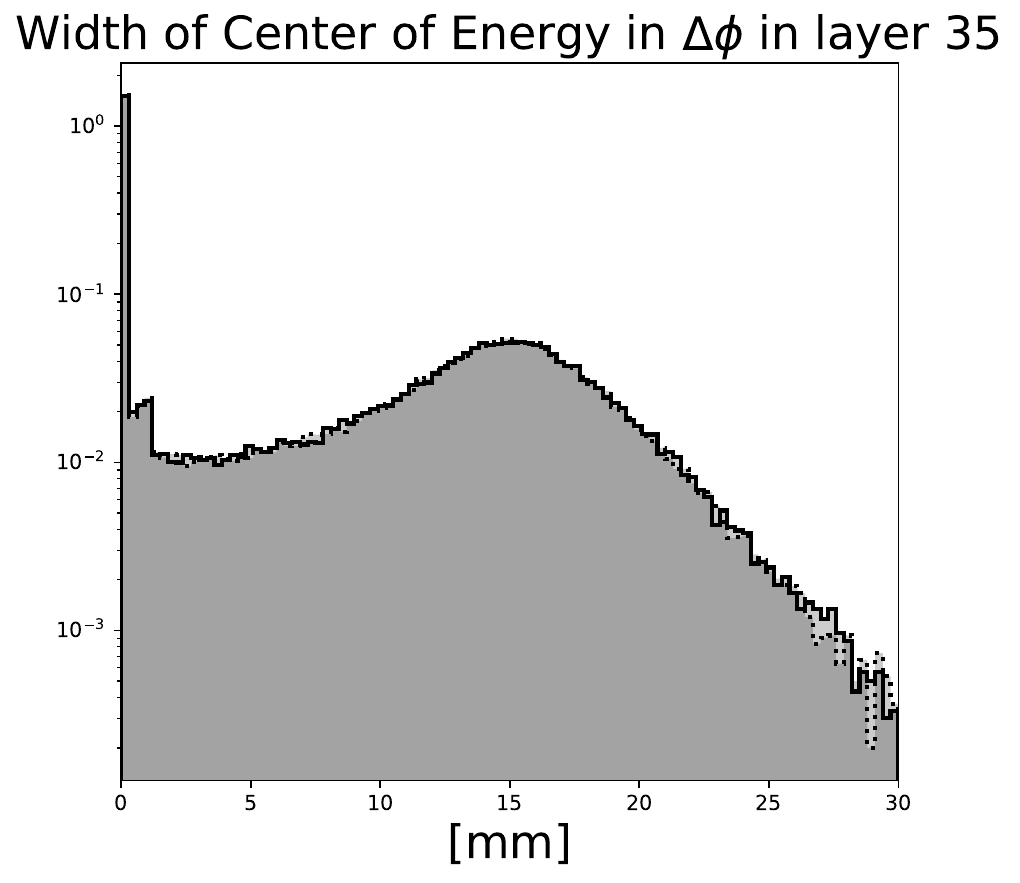} \hfill \includegraphics[height=0.1\textheight]{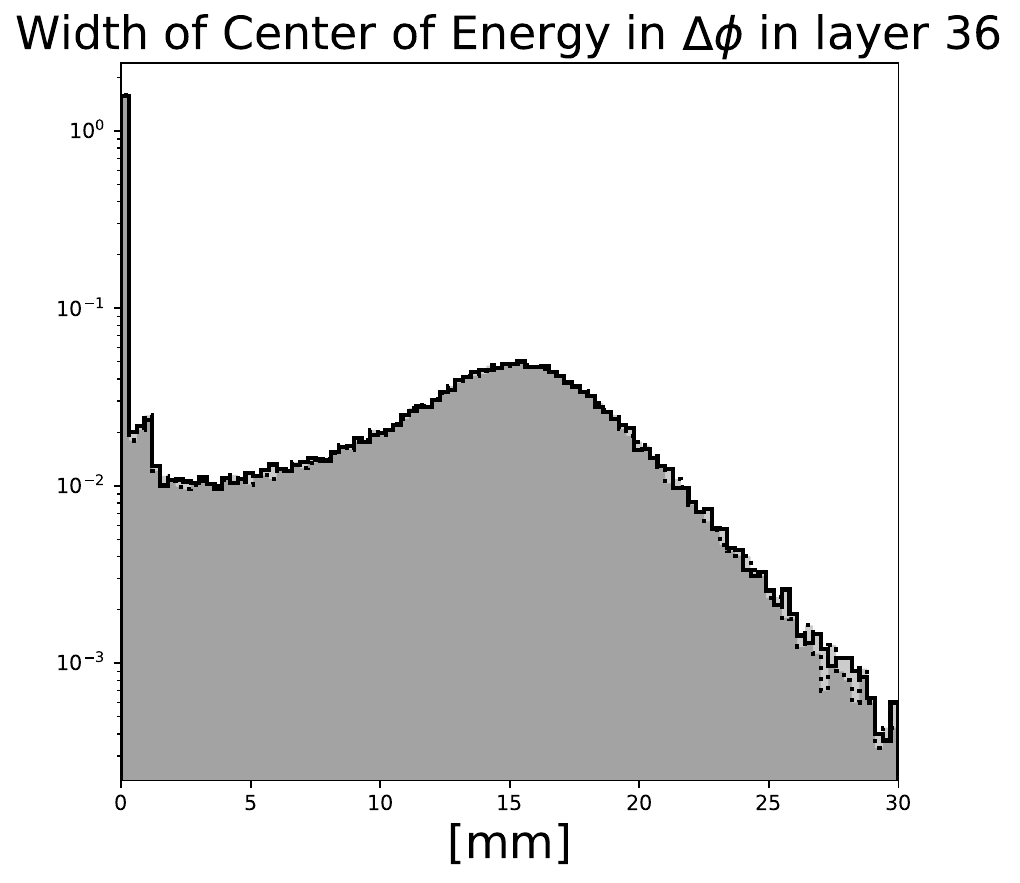} \hfill \includegraphics[height=0.1\textheight]{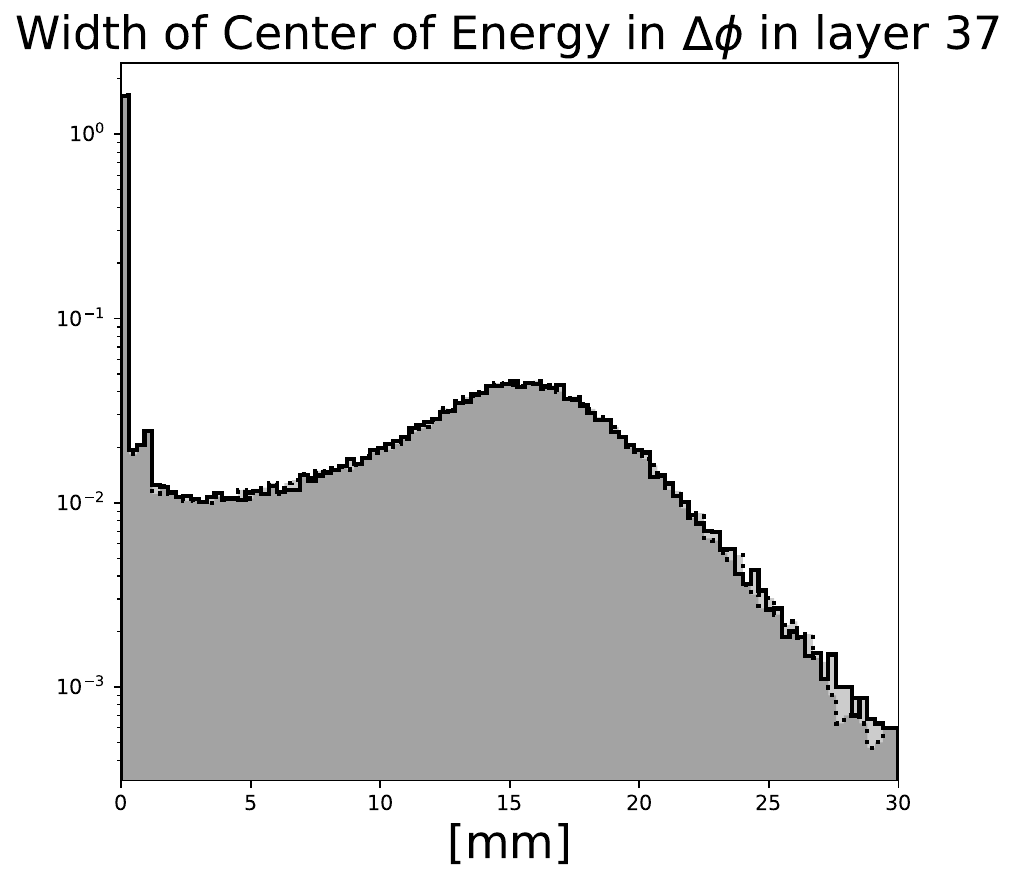} \hfill \includegraphics[height=0.1\textheight]{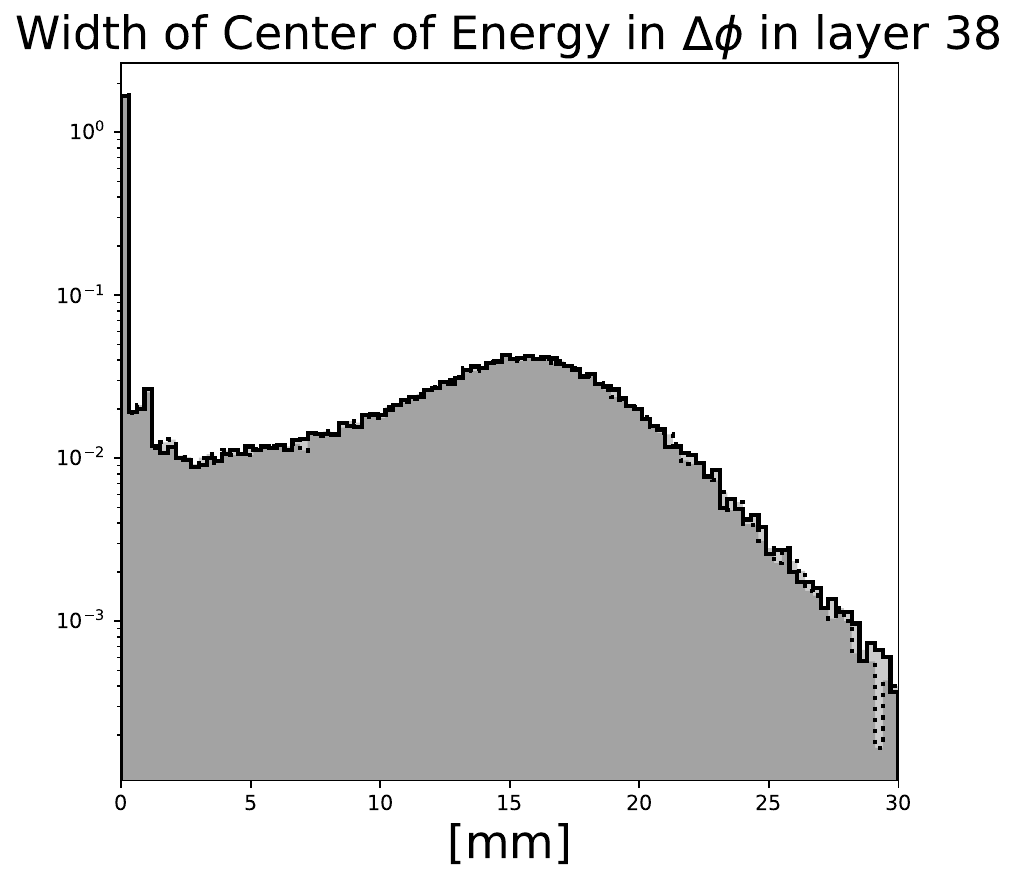} \hfill \includegraphics[height=0.1\textheight]{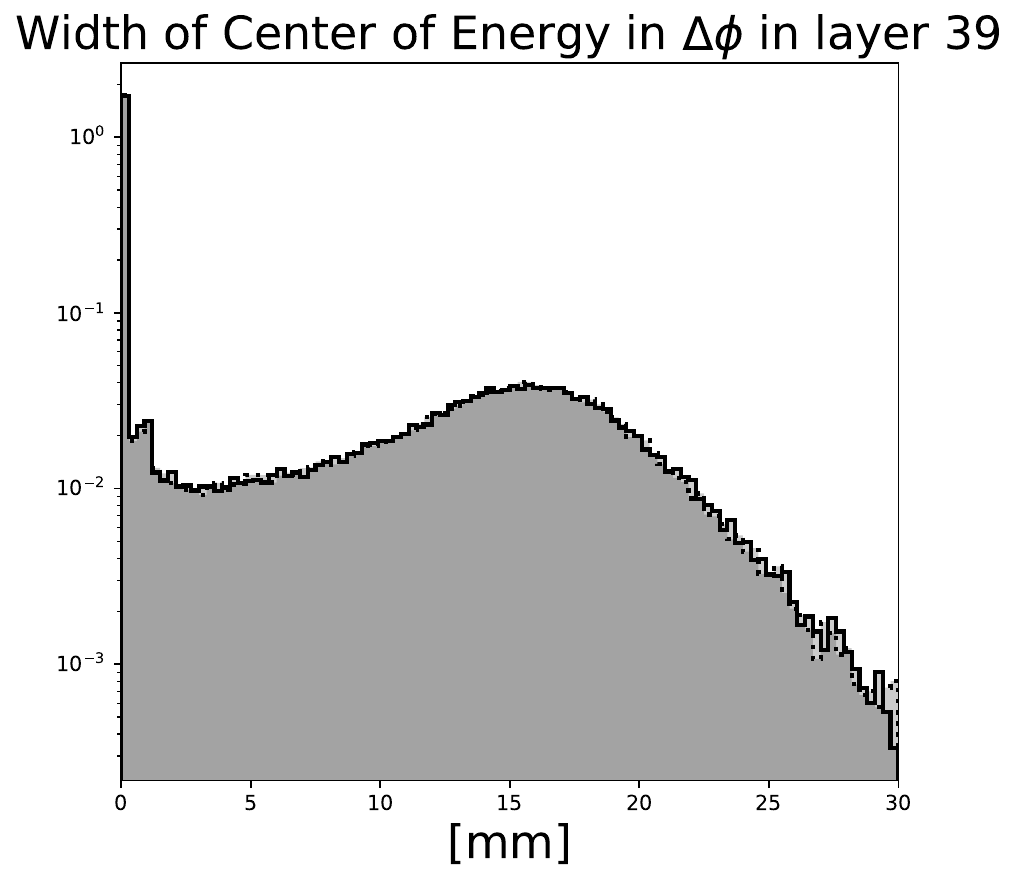}\\
    \includegraphics[height=0.1\textheight]{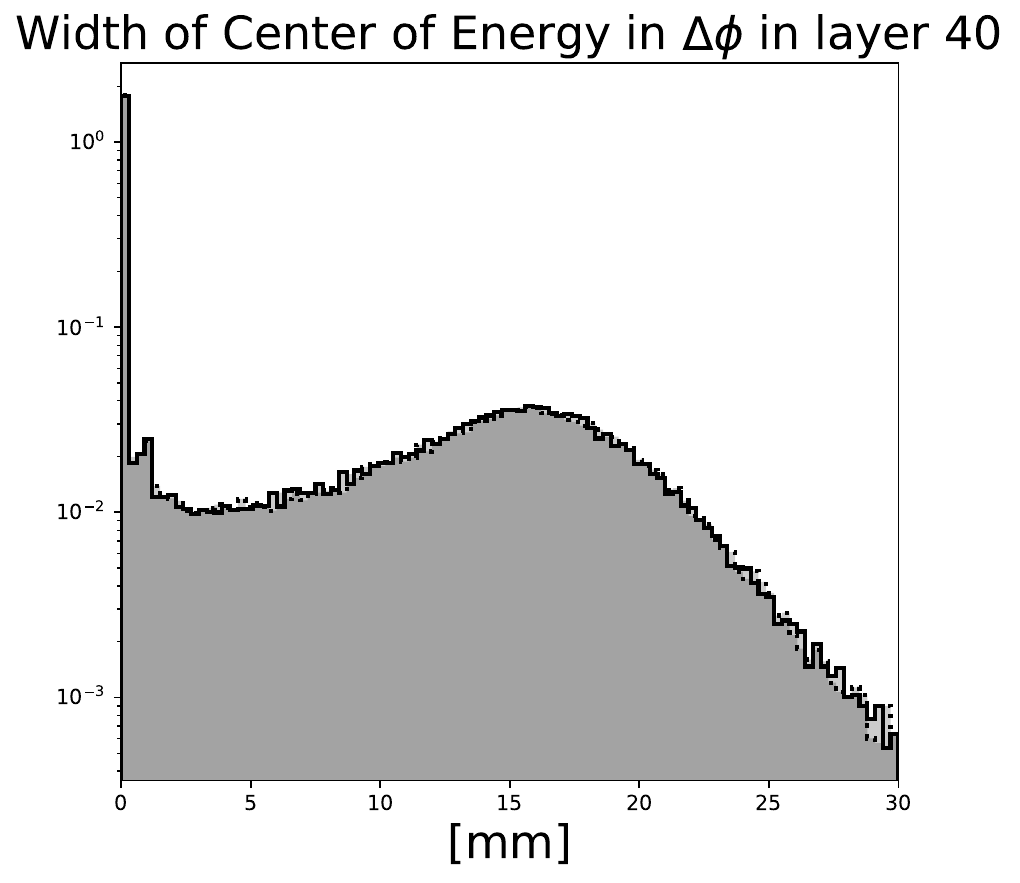} \hfill \includegraphics[height=0.1\textheight]{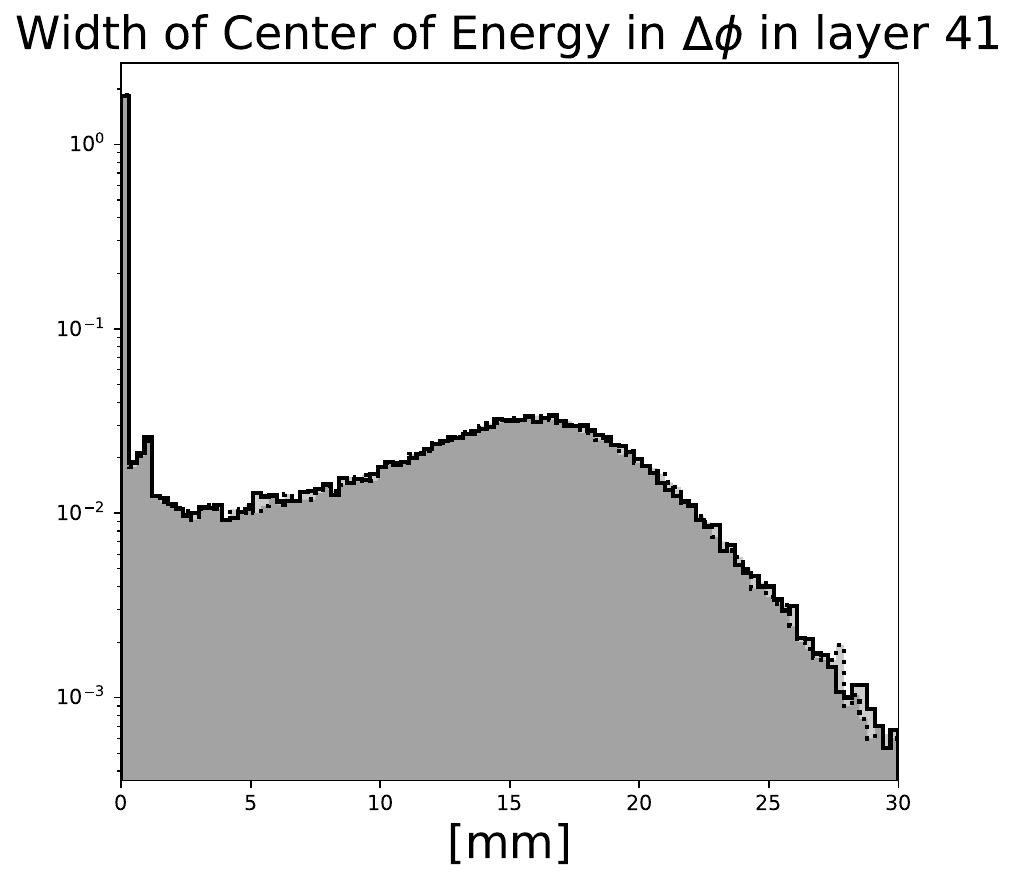} \hfill \includegraphics[height=0.1\textheight]{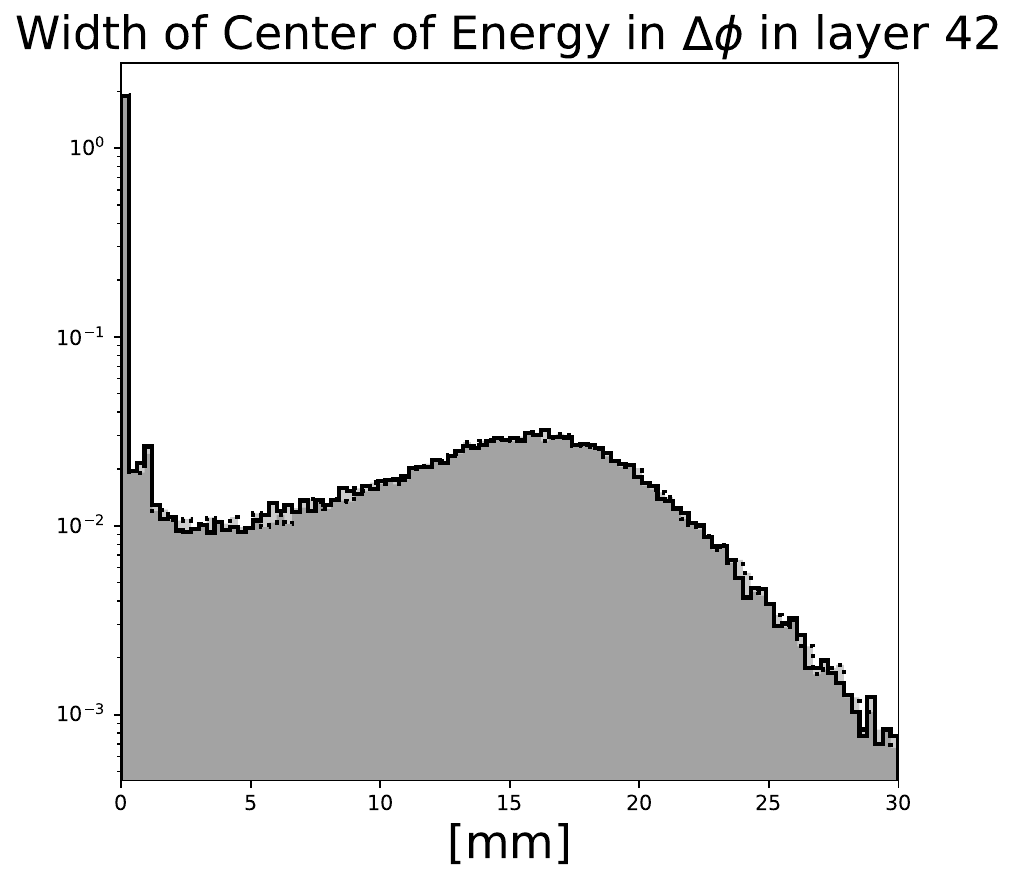} \hfill \includegraphics[height=0.1\textheight]{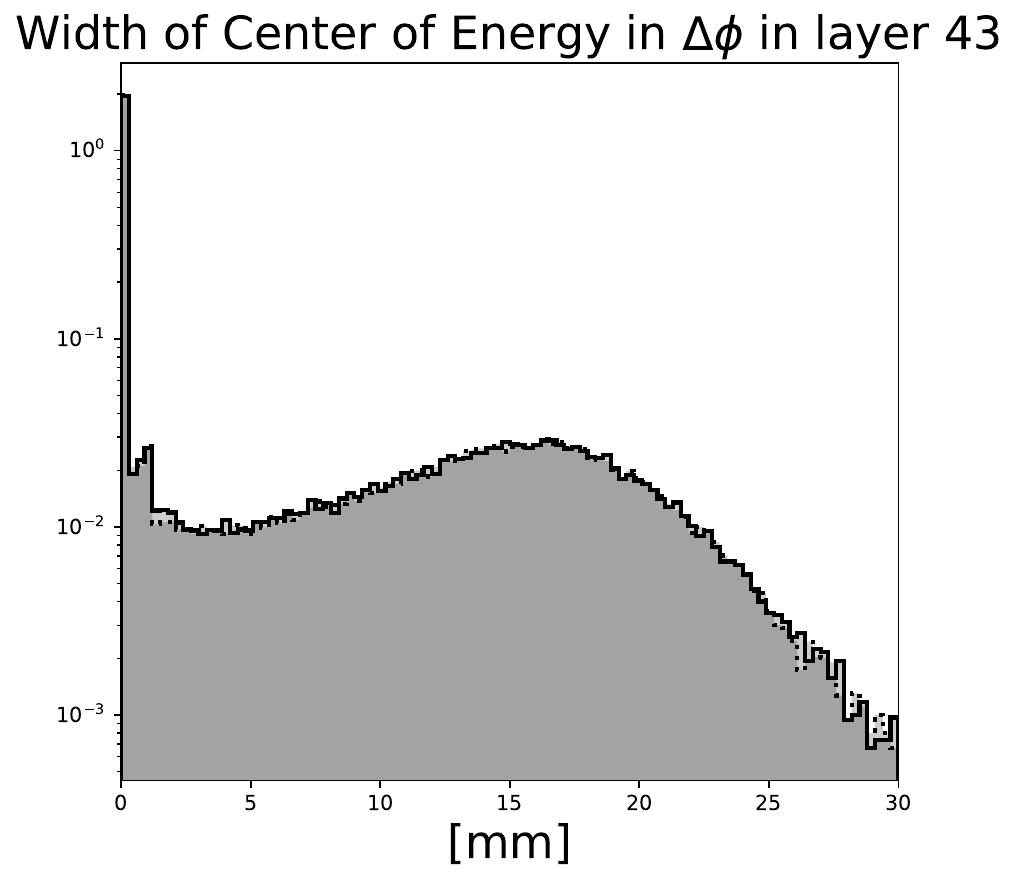} \hfill \includegraphics[height=0.1\textheight]{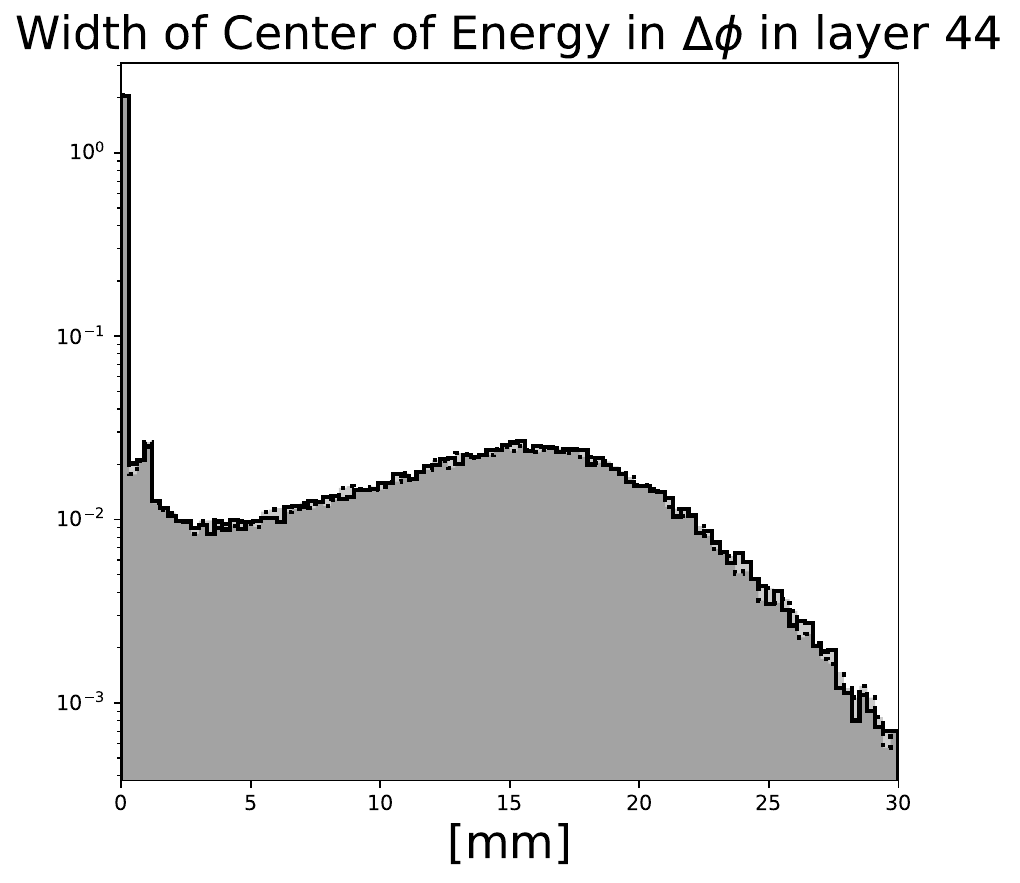}\\
    \includegraphics[width=0.5\textwidth]{figures/Appendix_reference/legend.pdf}
    \caption{Distribution of \geant training and evaluation data in width of the centers of energy in $\phi$ direction for ds3. }
    \label{fig:app_ref.ds3.6}
\end{figure}

\begin{figure}[ht]
    \centering
    \includegraphics[height=0.1\textheight]{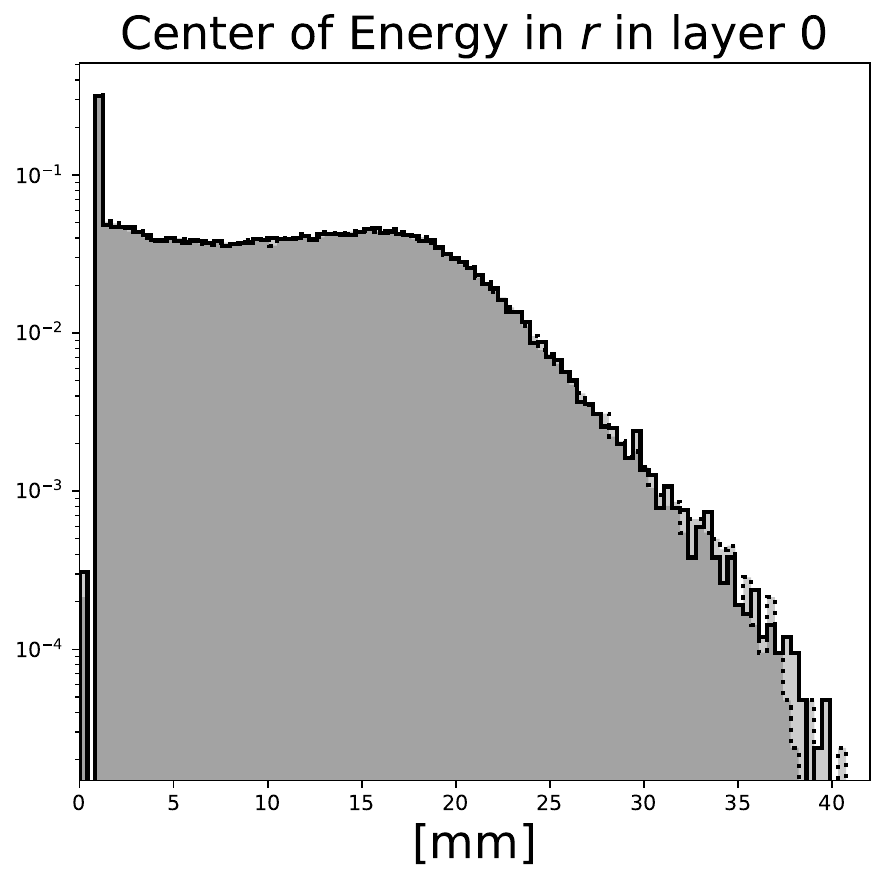} \hfill \includegraphics[height=0.1\textheight]{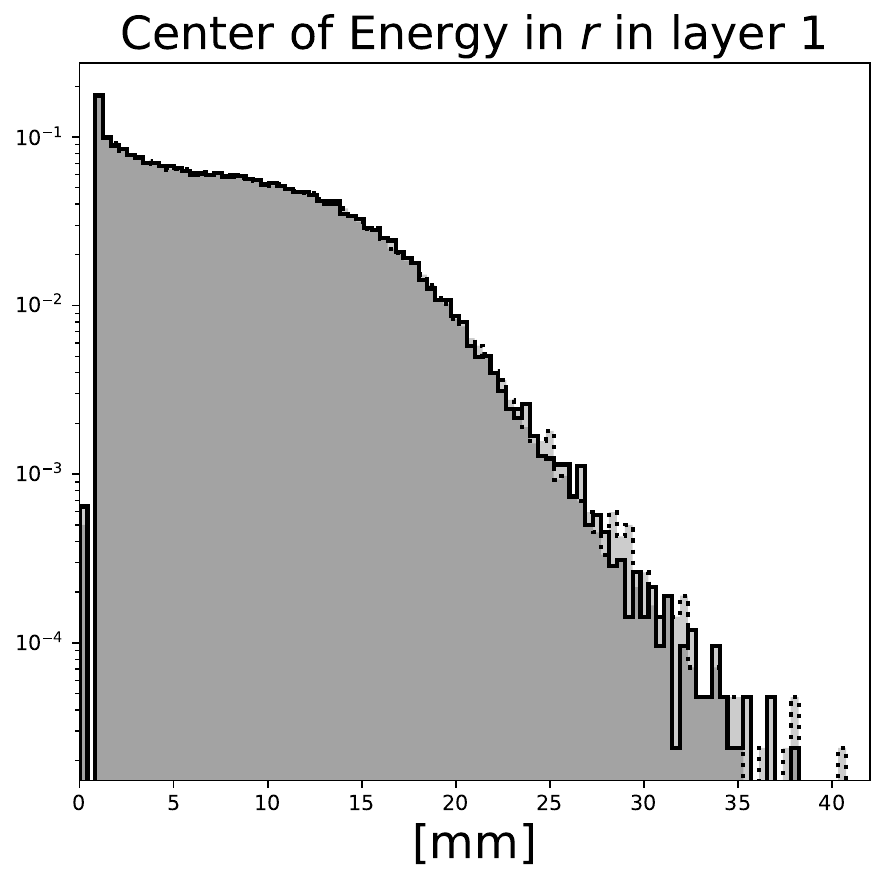} \hfill \includegraphics[height=0.1\textheight]{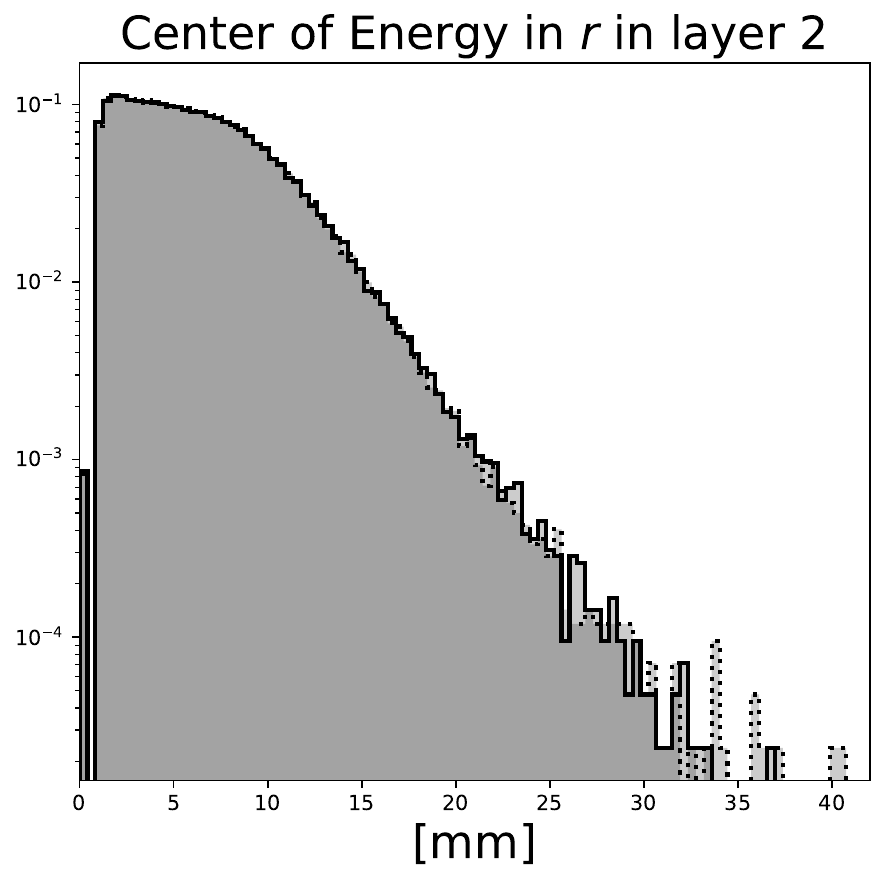} \hfill \includegraphics[height=0.1\textheight]{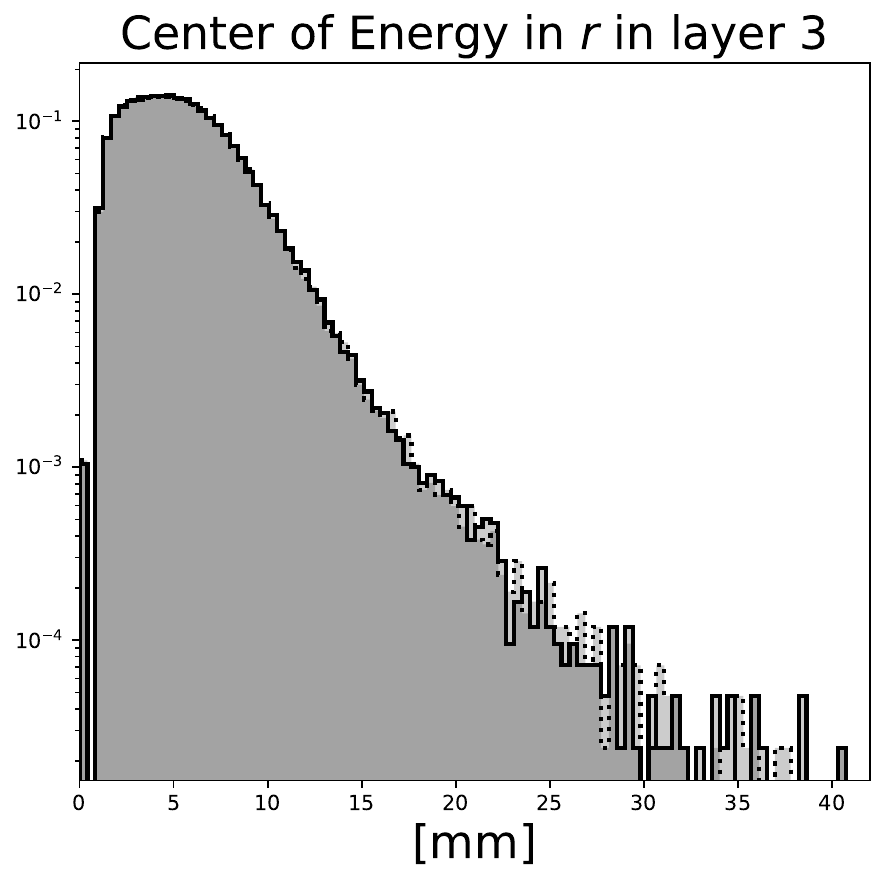} \hfill \includegraphics[height=0.1\textheight]{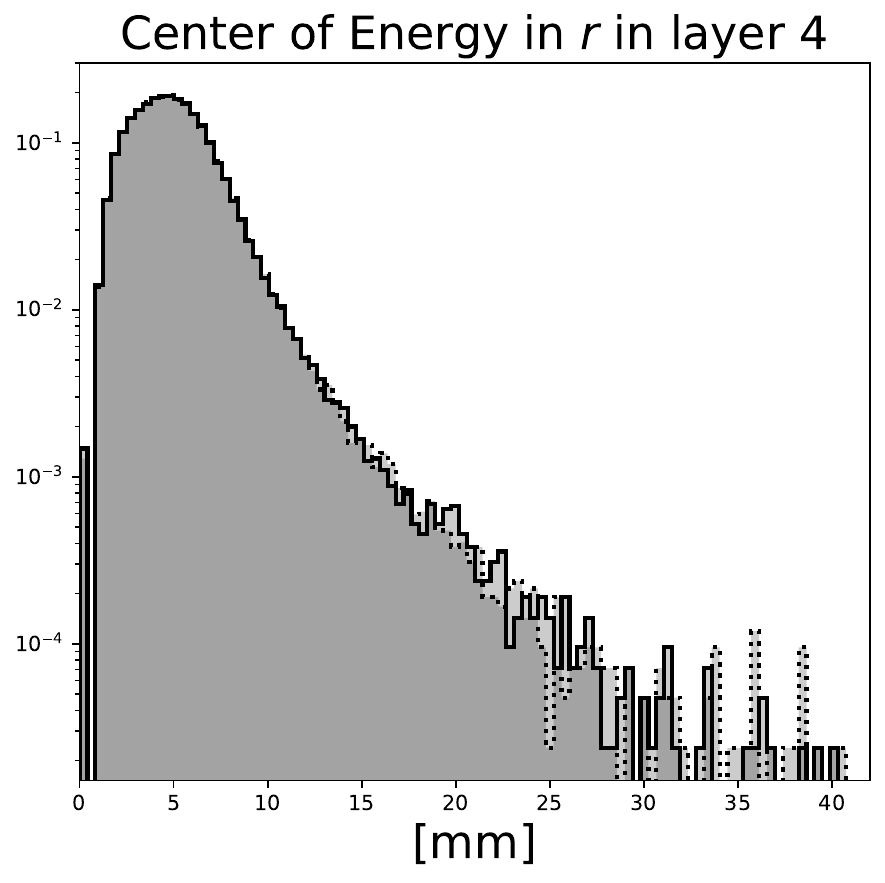}\\
    \includegraphics[height=0.1\textheight]{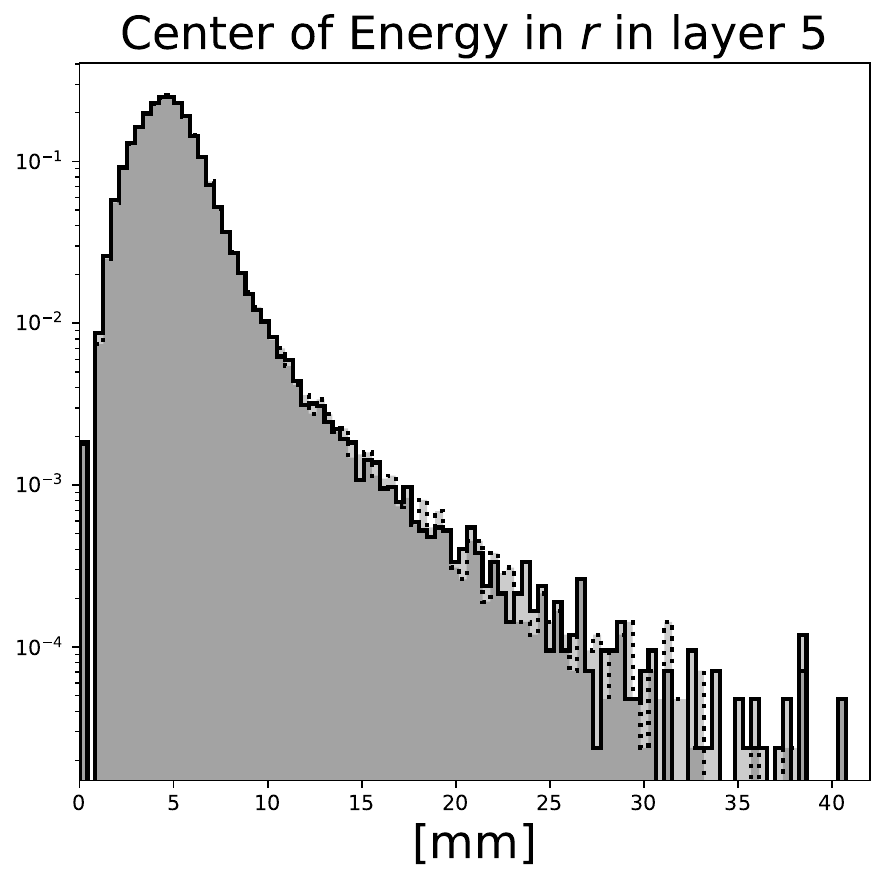} \hfill \includegraphics[height=0.1\textheight]{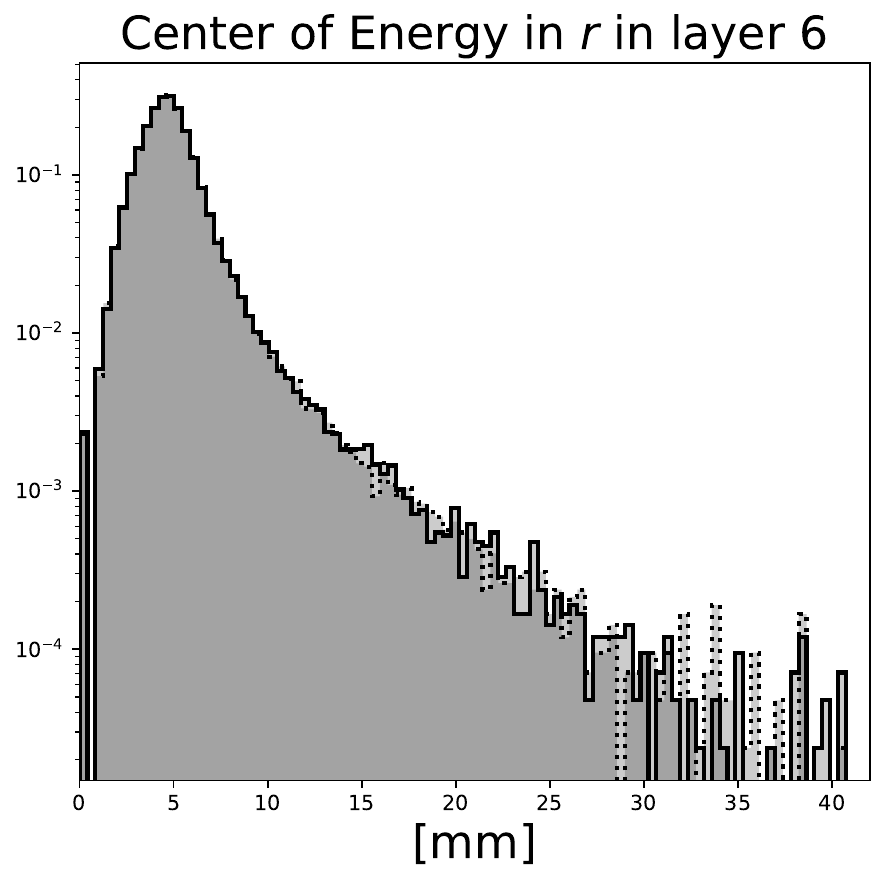} \hfill \includegraphics[height=0.1\textheight]{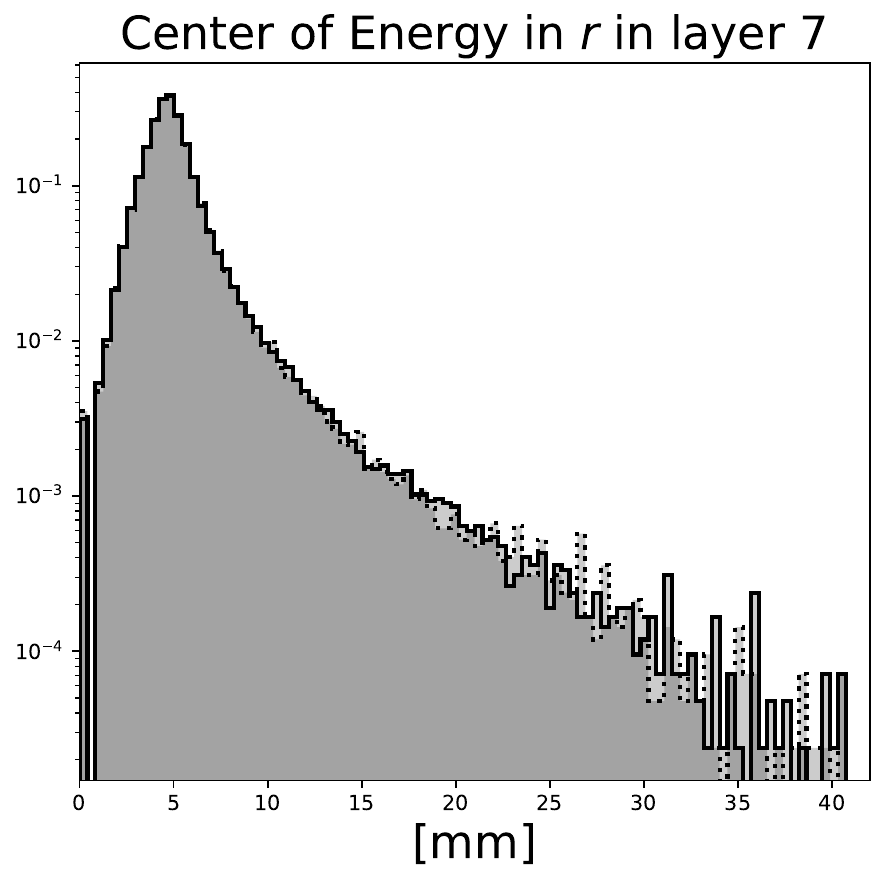} \hfill \includegraphics[height=0.1\textheight]{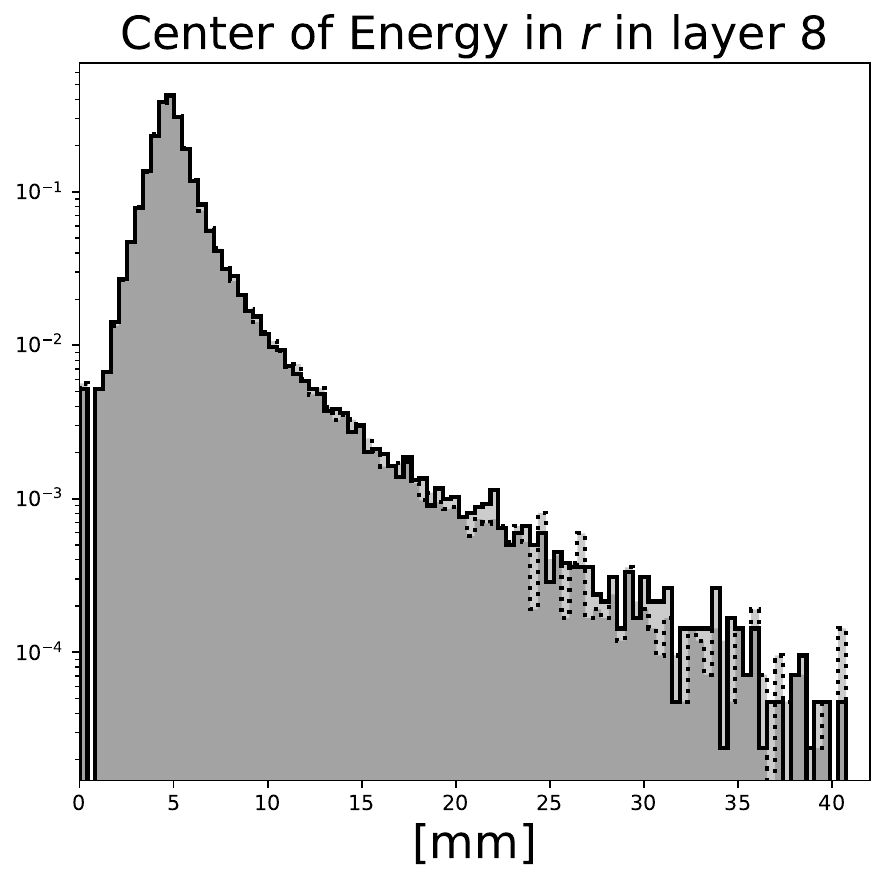} \hfill \includegraphics[height=0.1\textheight]{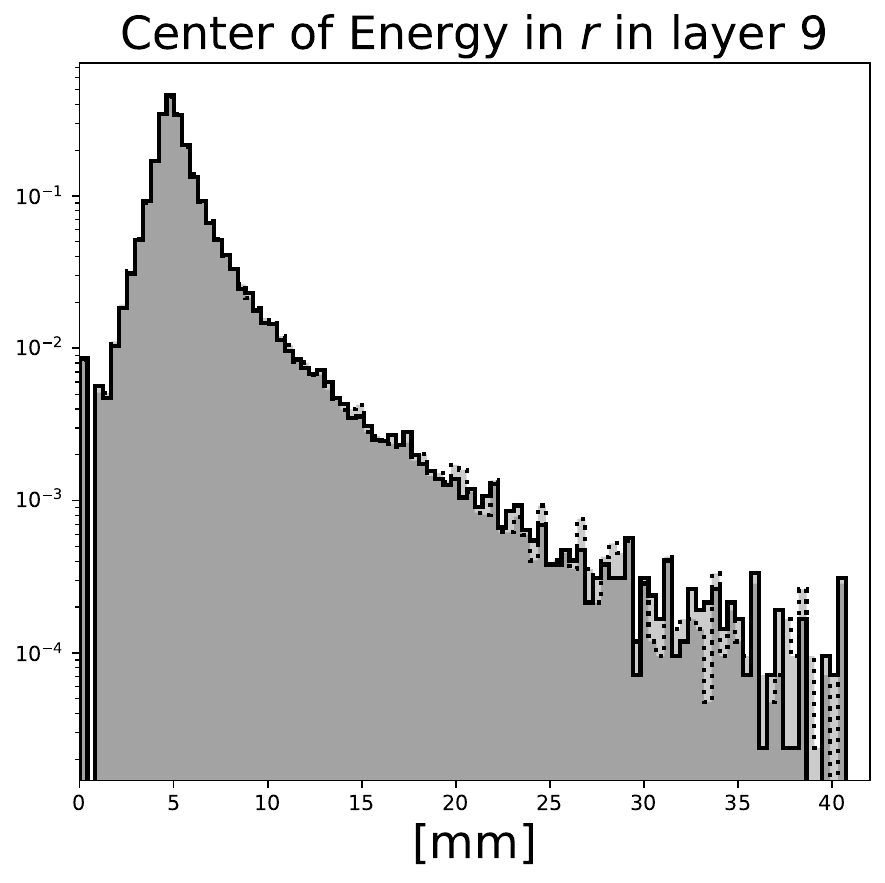}\\
    \includegraphics[height=0.1\textheight]{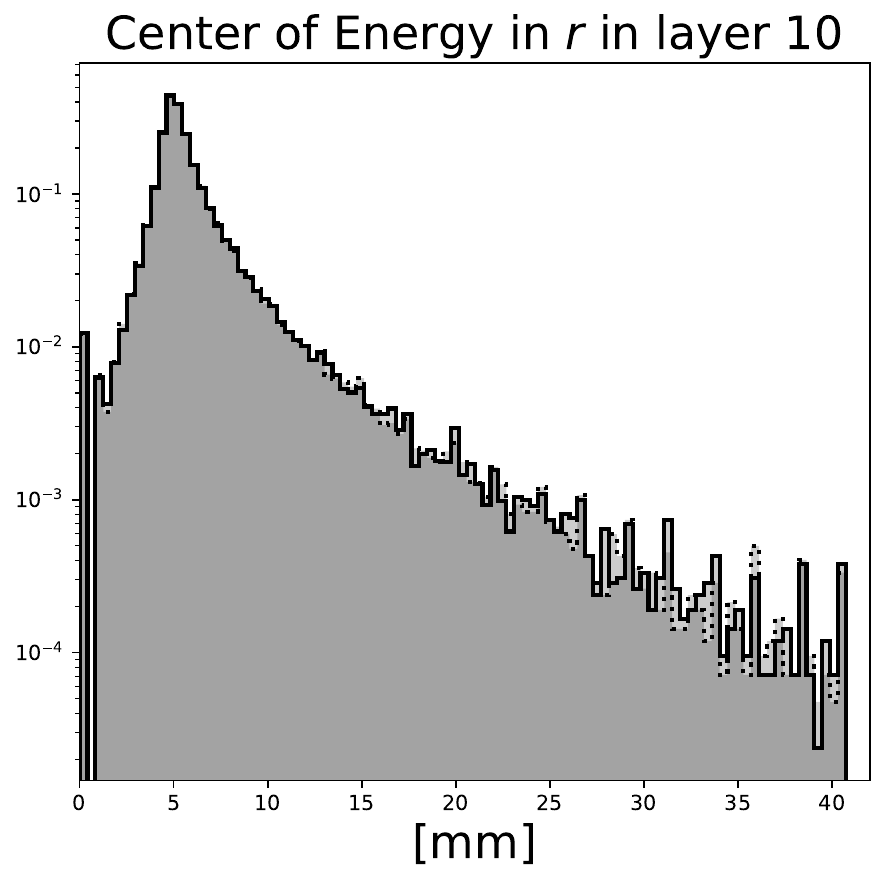} \hfill \includegraphics[height=0.1\textheight]{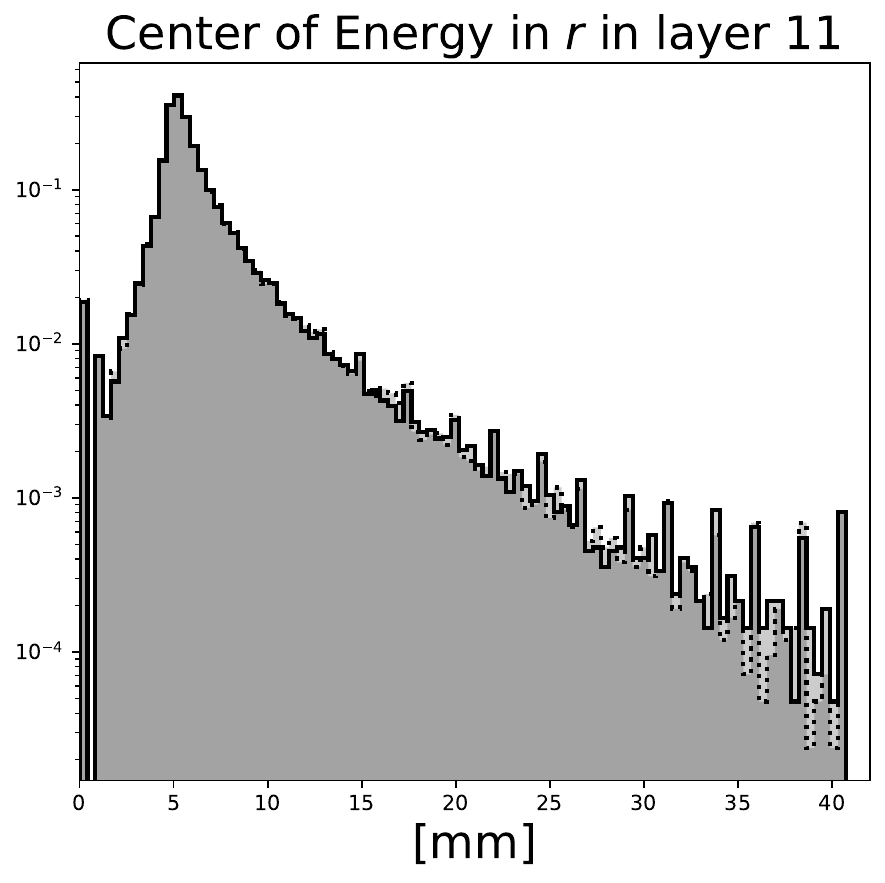} \hfill \includegraphics[height=0.1\textheight]{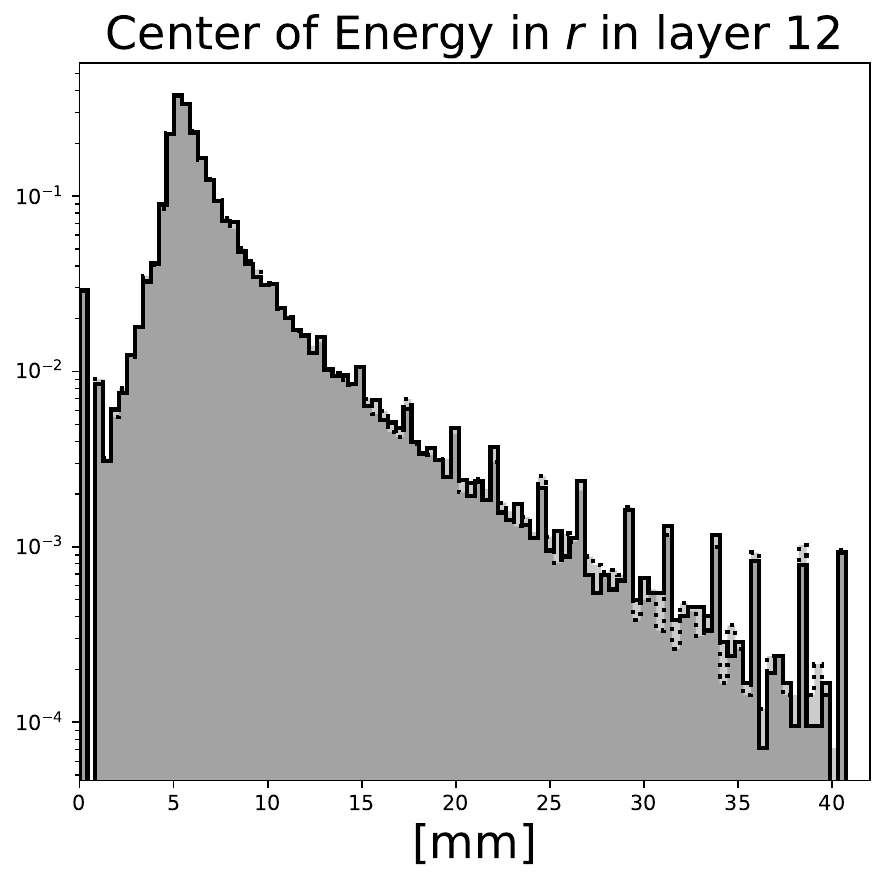} \hfill \includegraphics[height=0.1\textheight]{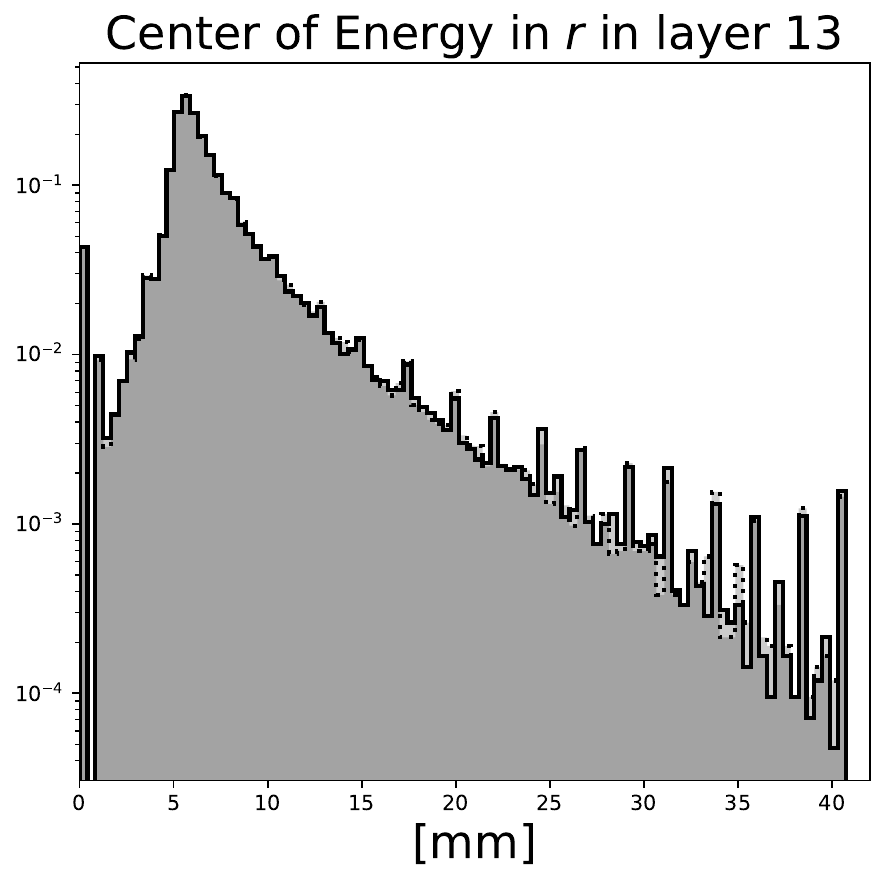} \hfill \includegraphics[height=0.1\textheight]{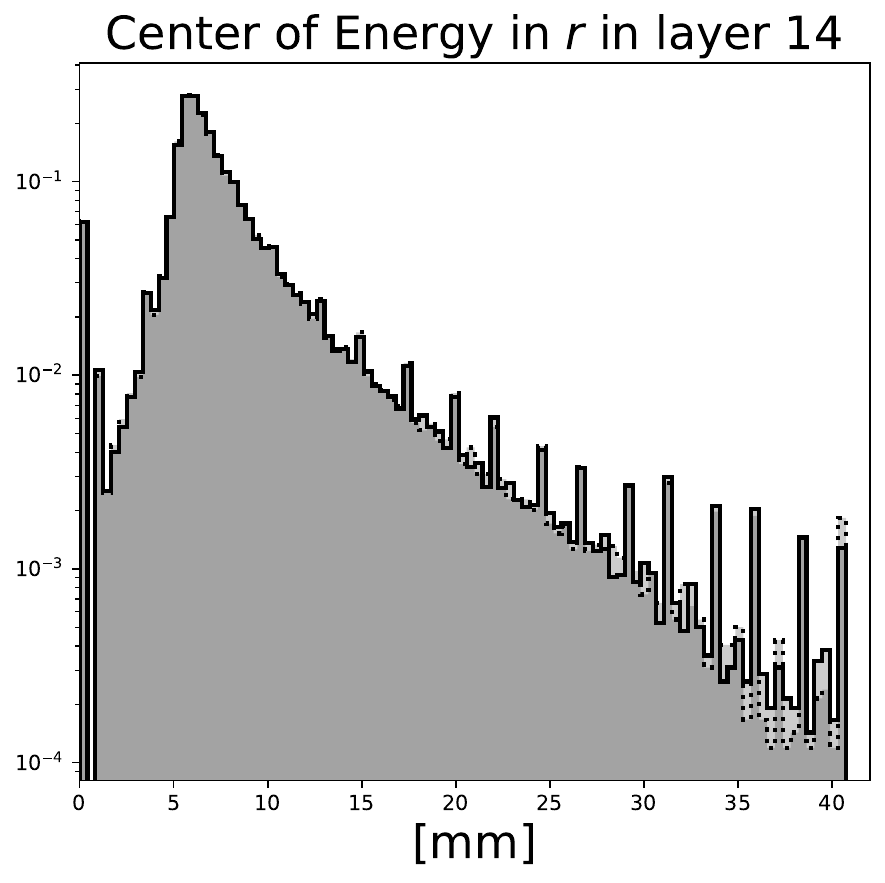}\\
    \includegraphics[height=0.1\textheight]{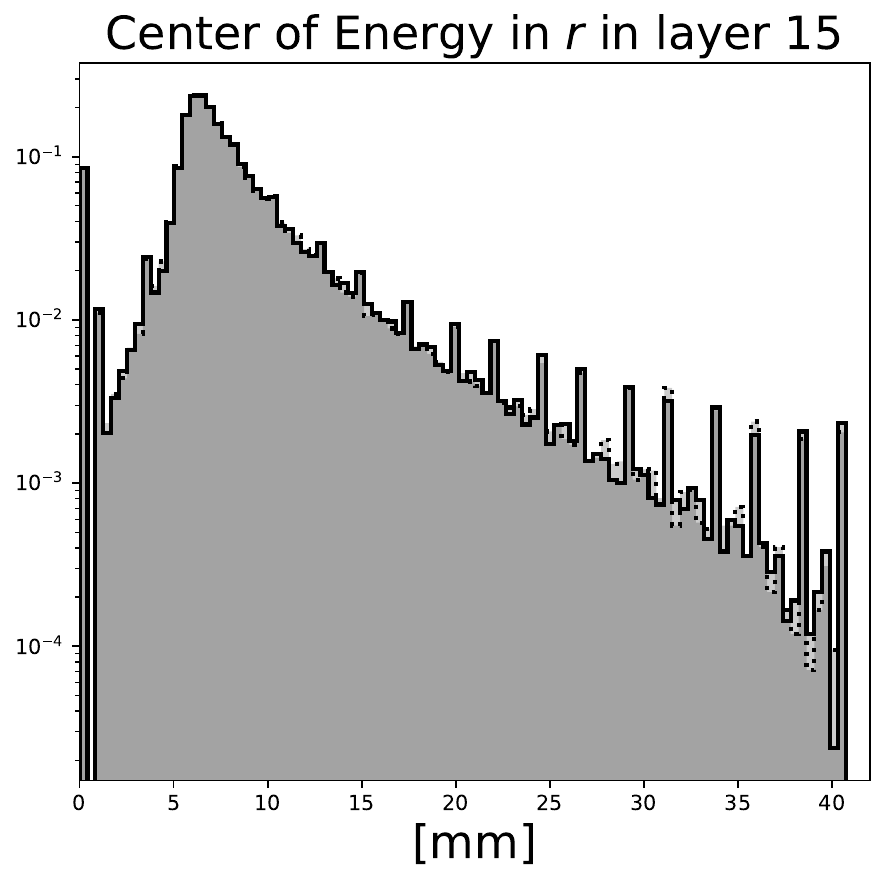} \hfill \includegraphics[height=0.1\textheight]{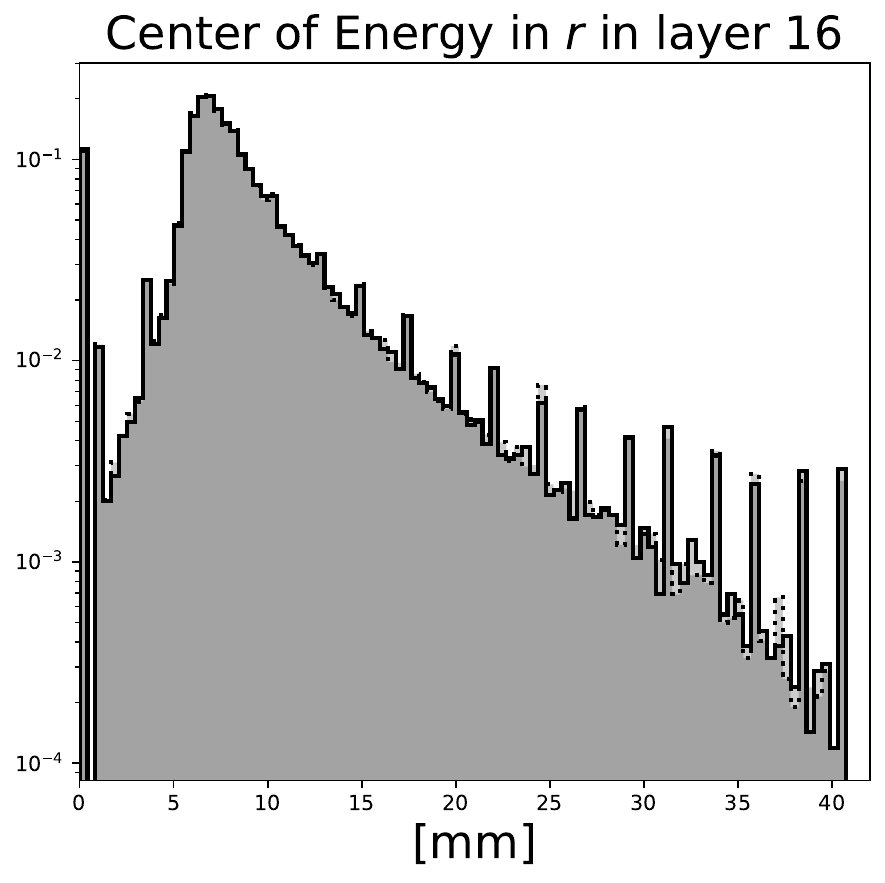} \hfill \includegraphics[height=0.1\textheight]{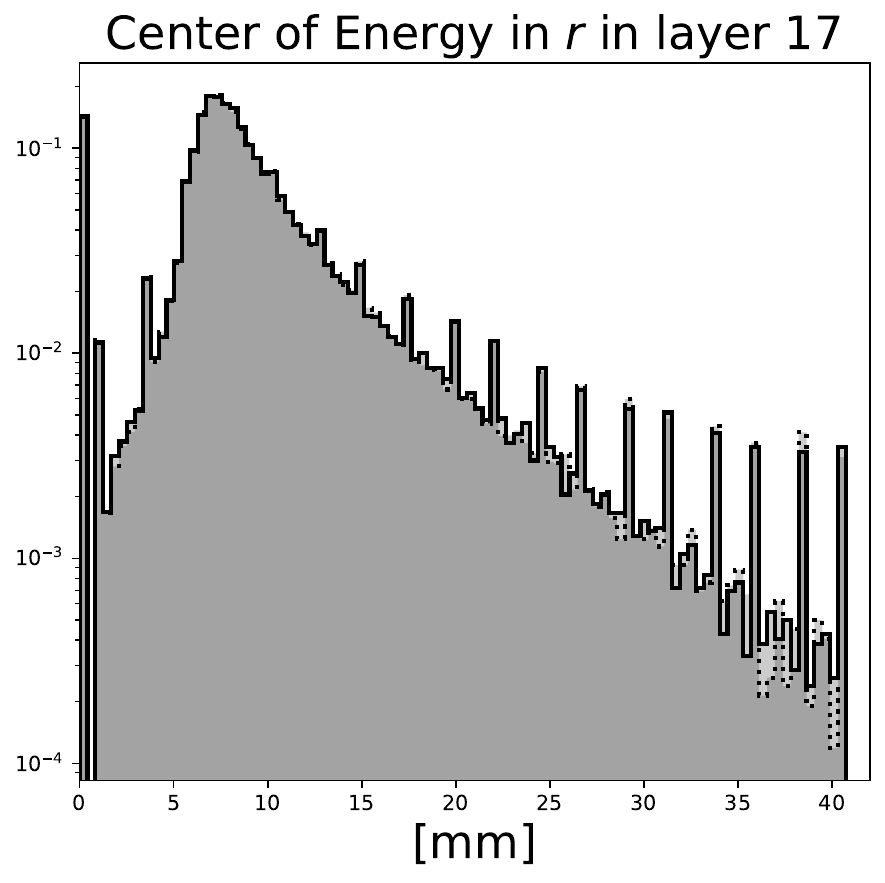} \hfill \includegraphics[height=0.1\textheight]{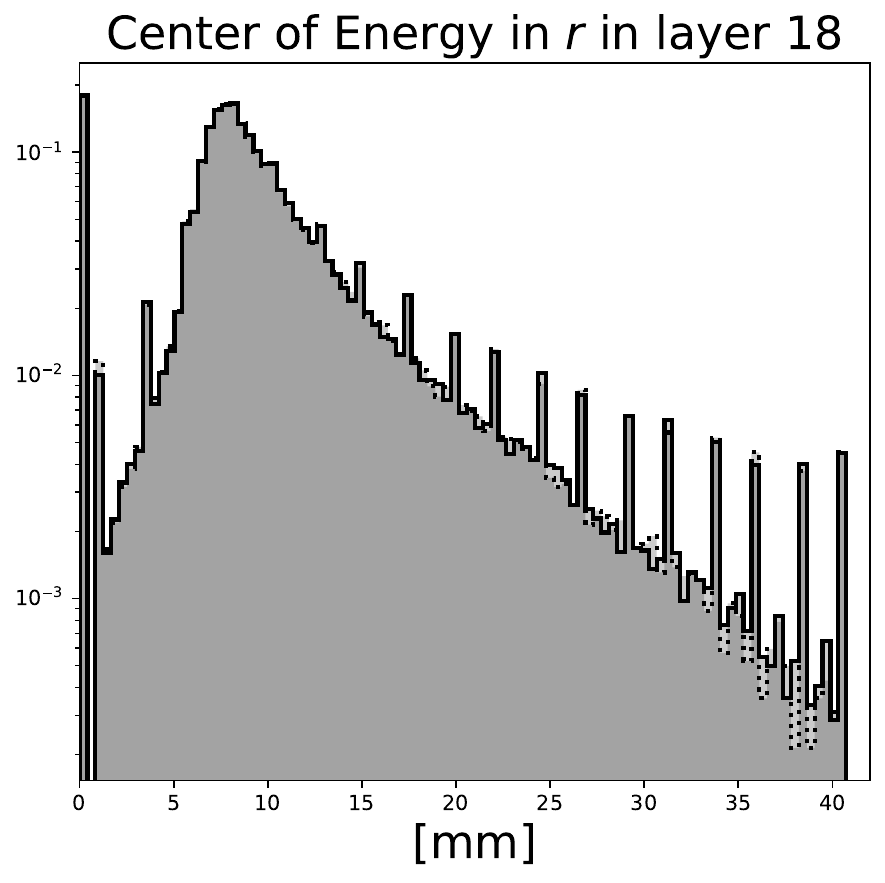} \hfill \includegraphics[height=0.1\textheight]{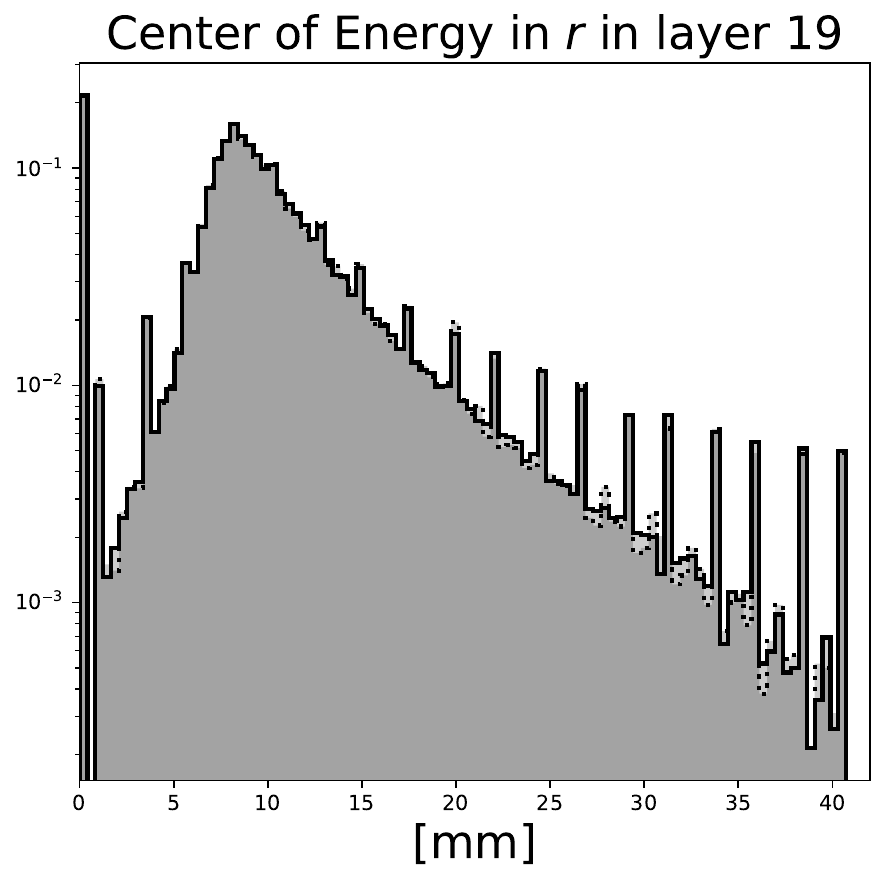}\\
    \includegraphics[height=0.1\textheight]{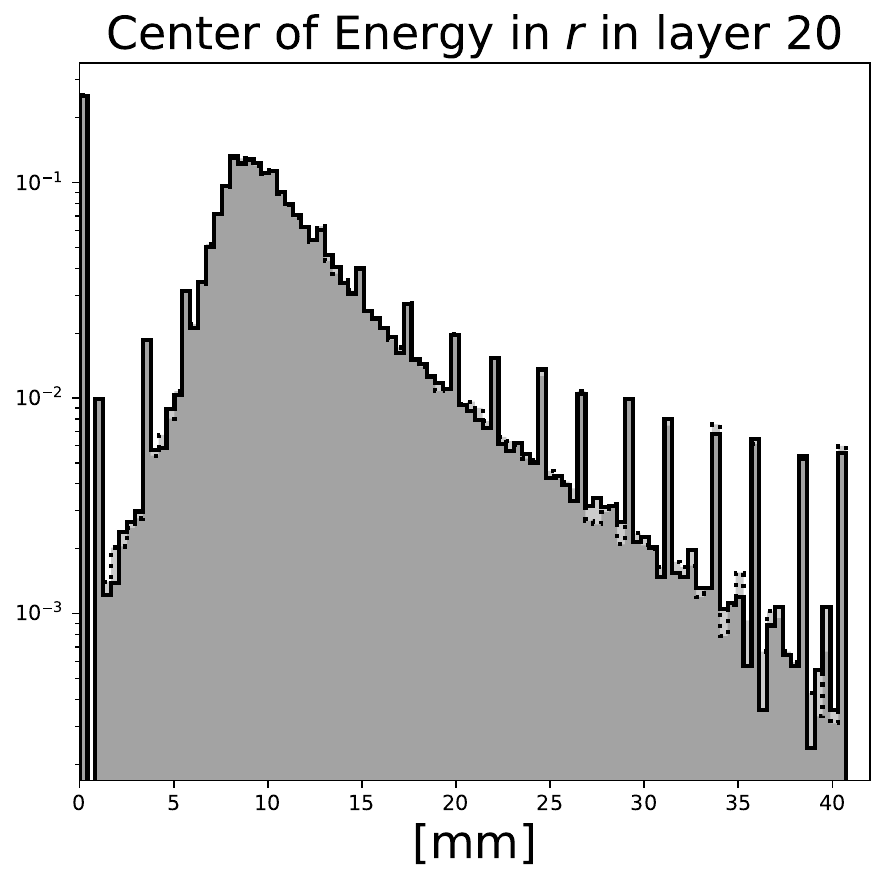} \hfill \includegraphics[height=0.1\textheight]{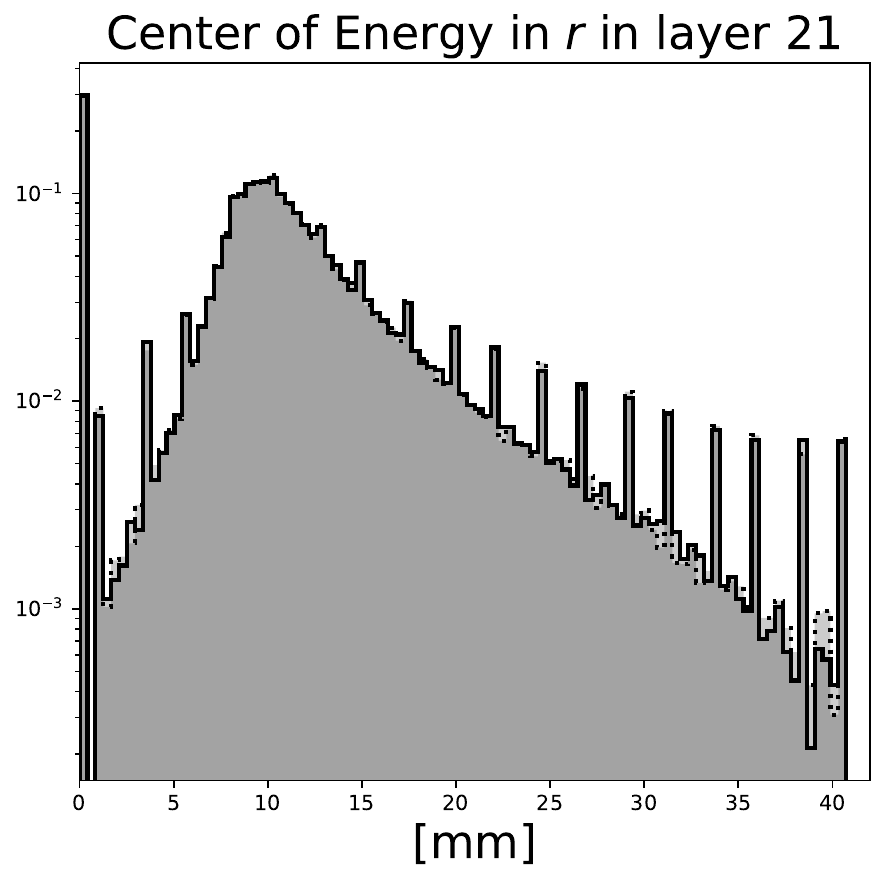} \hfill \includegraphics[height=0.1\textheight]{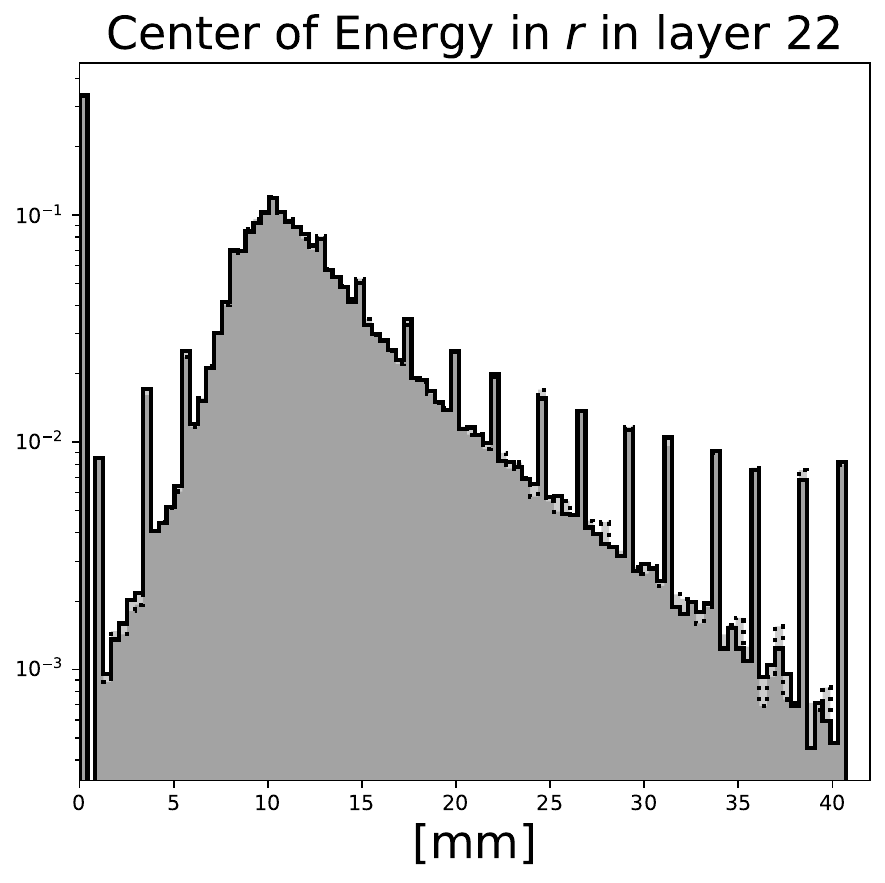} \hfill \includegraphics[height=0.1\textheight]{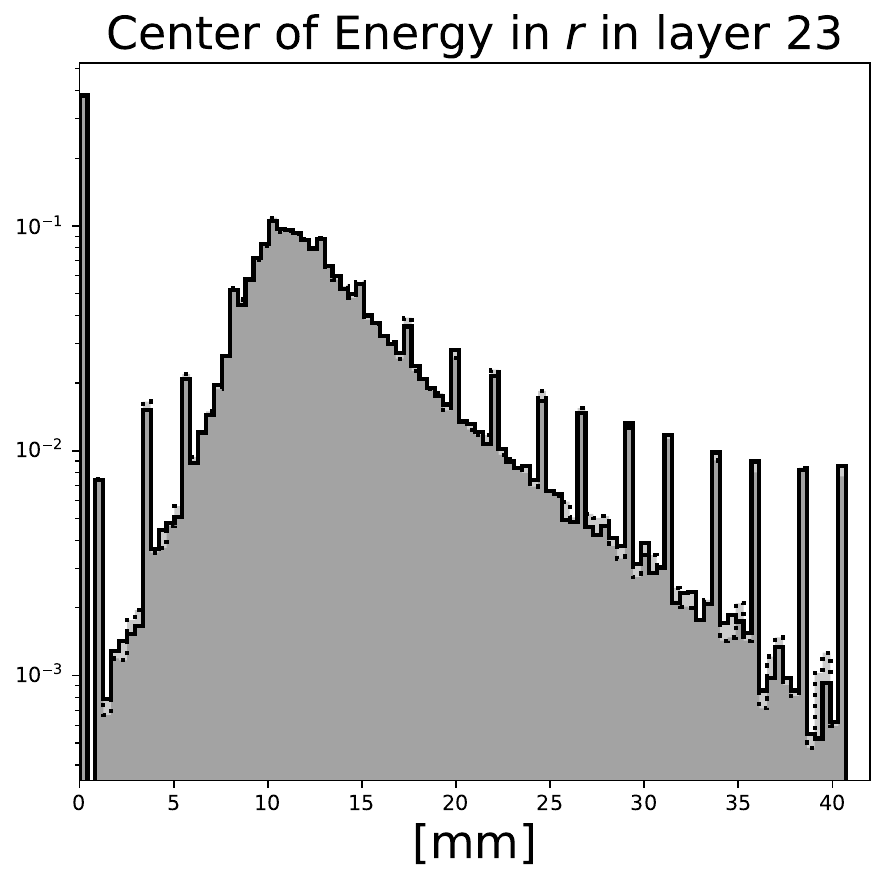} \hfill \includegraphics[height=0.1\textheight]{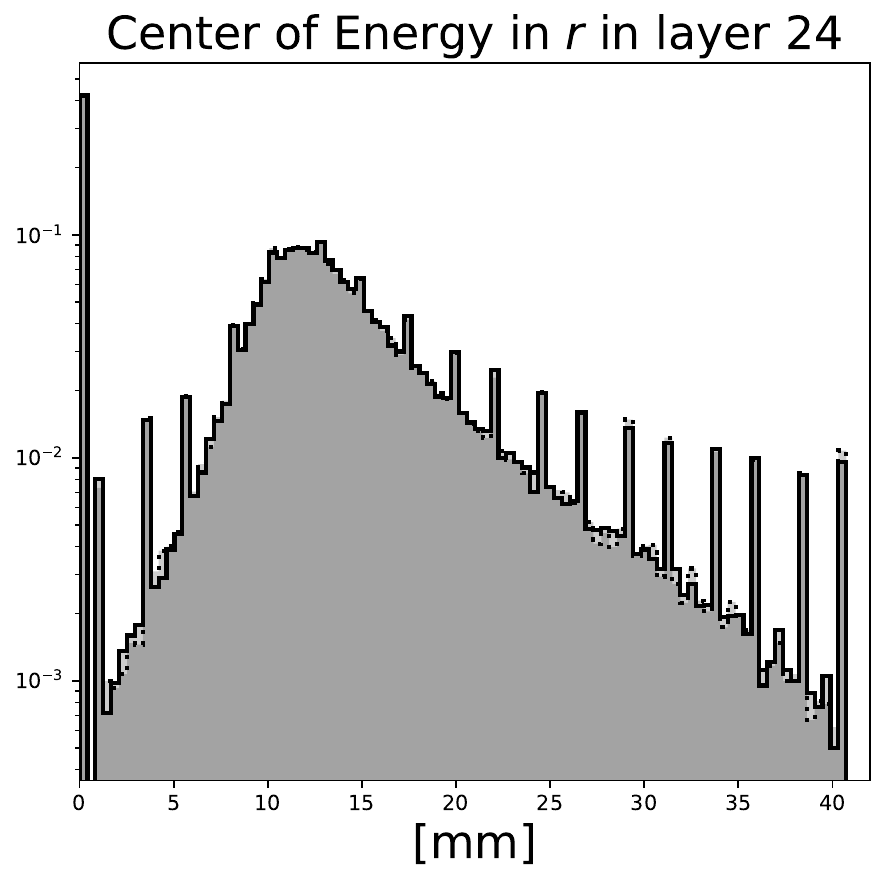}\\
    \includegraphics[height=0.1\textheight]{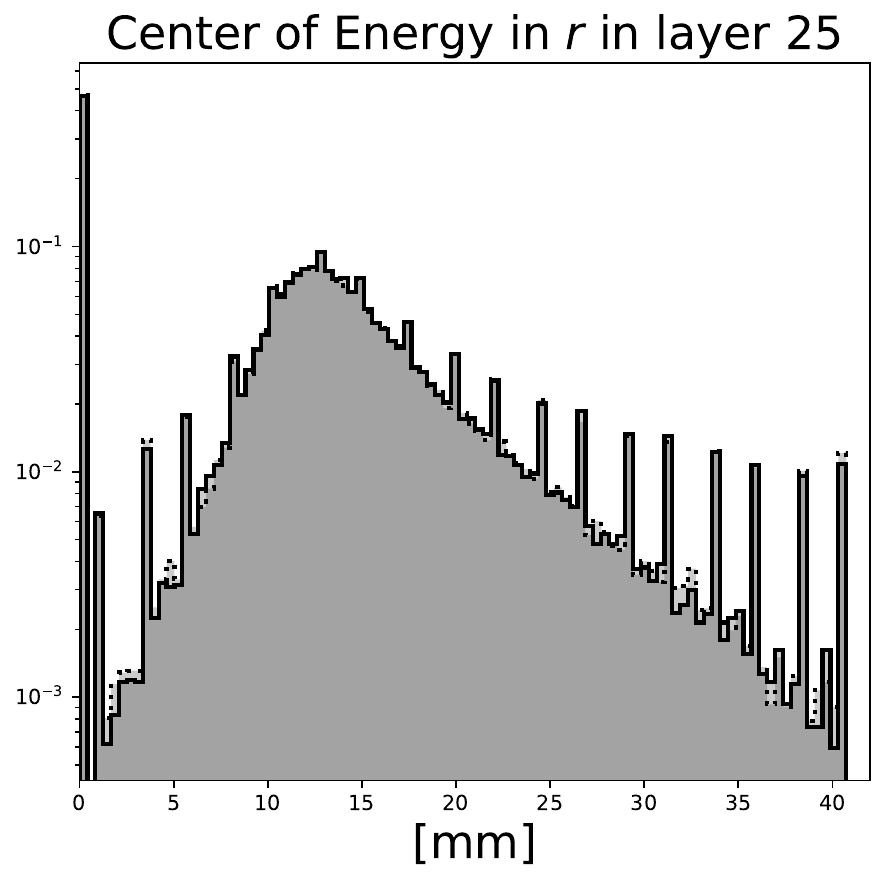} \hfill \includegraphics[height=0.1\textheight]{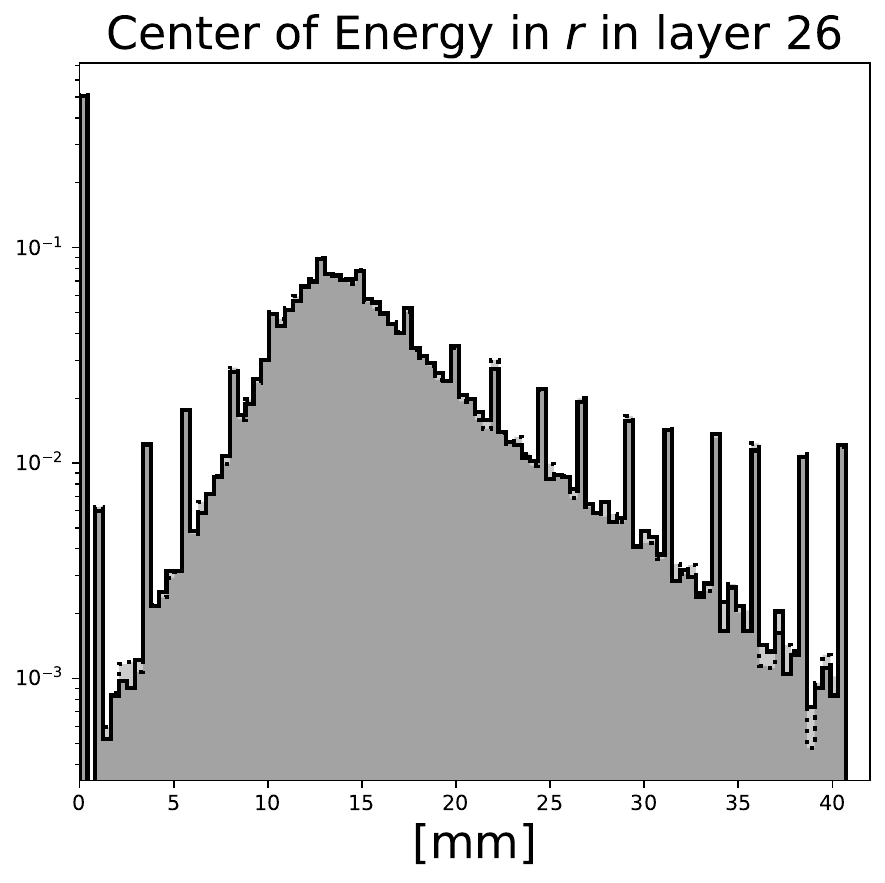} \hfill \includegraphics[height=0.1\textheight]{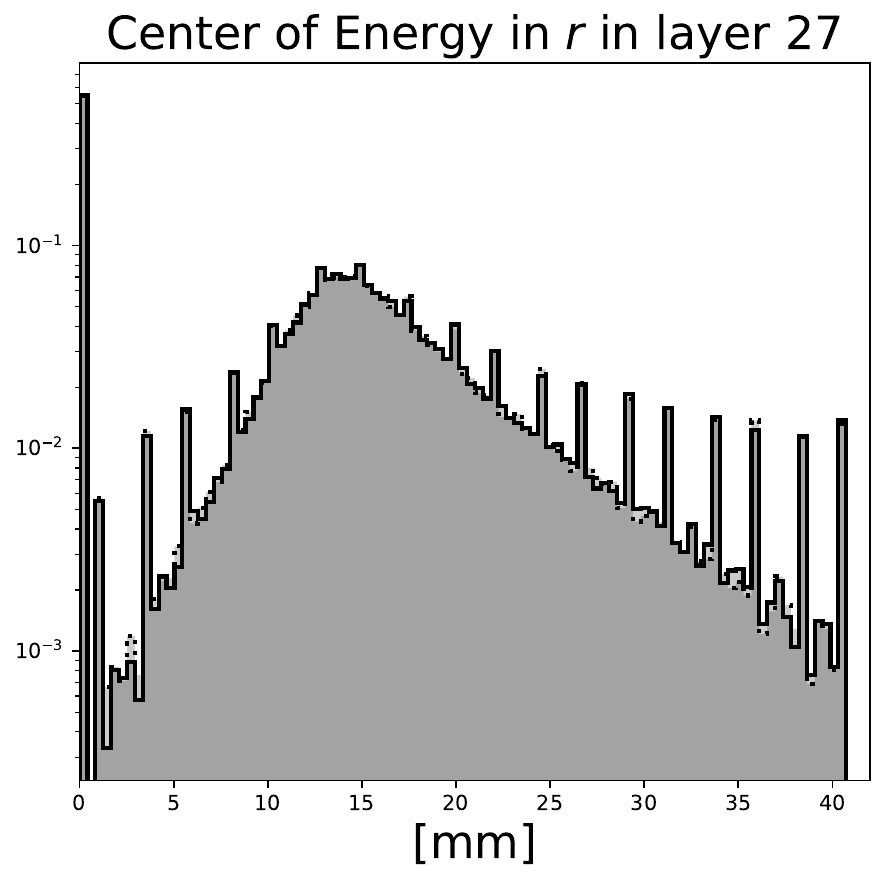} \hfill \includegraphics[height=0.1\textheight]{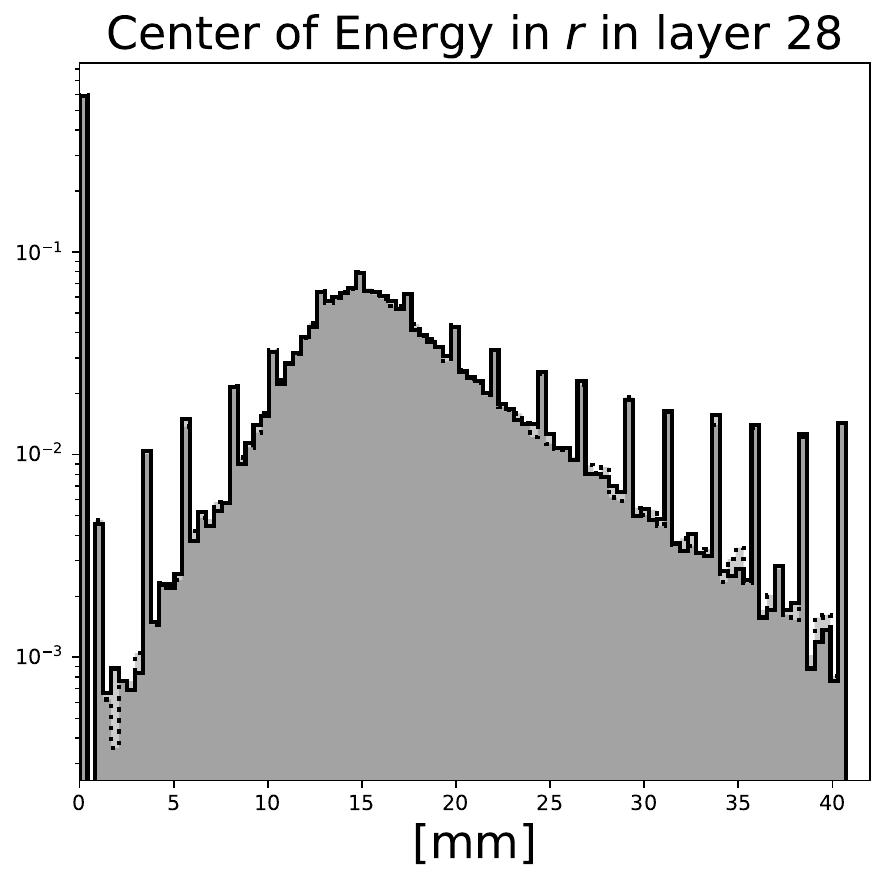} \hfill \includegraphics[height=0.1\textheight]{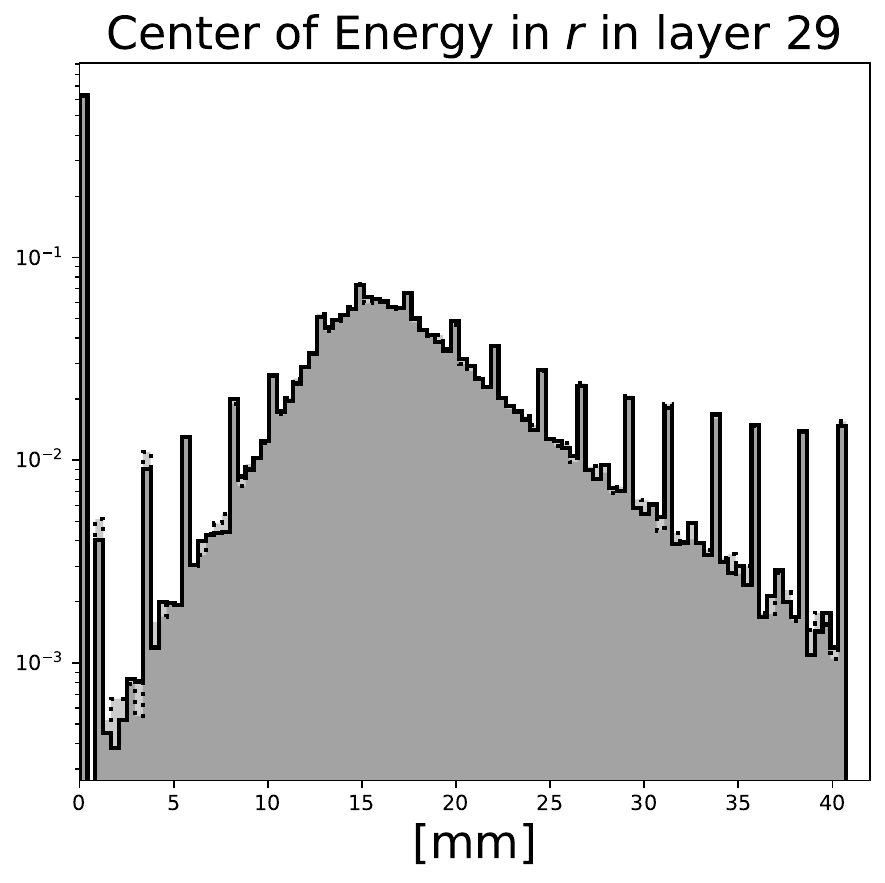}\\
    \includegraphics[height=0.1\textheight]{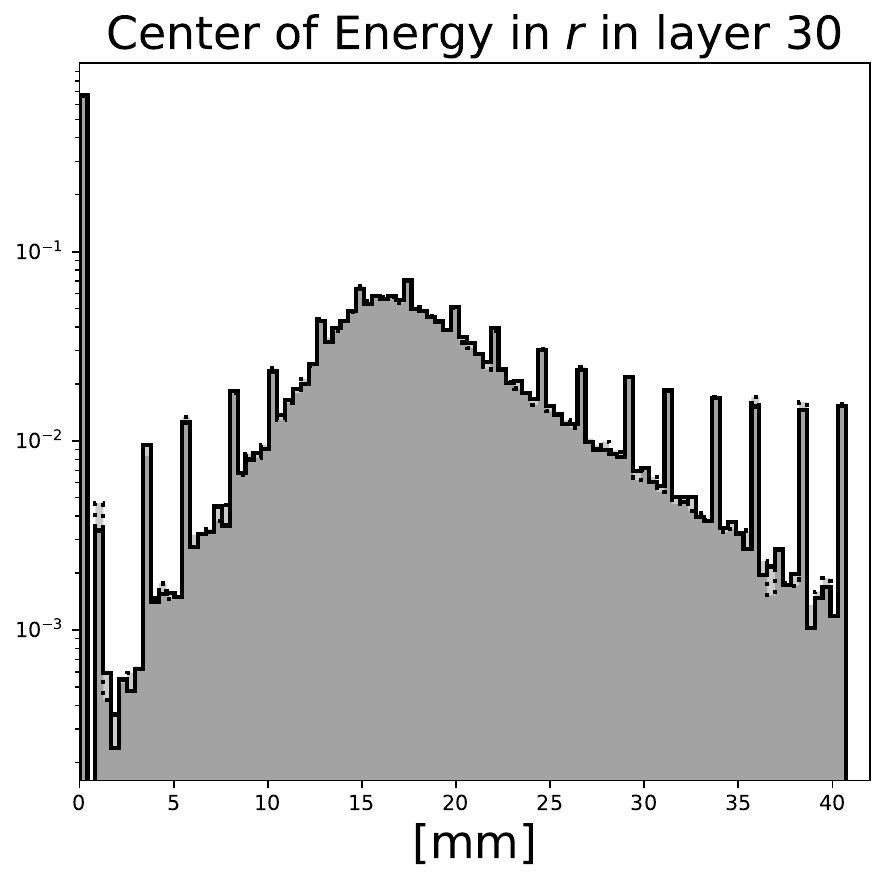} \hfill \includegraphics[height=0.1\textheight]{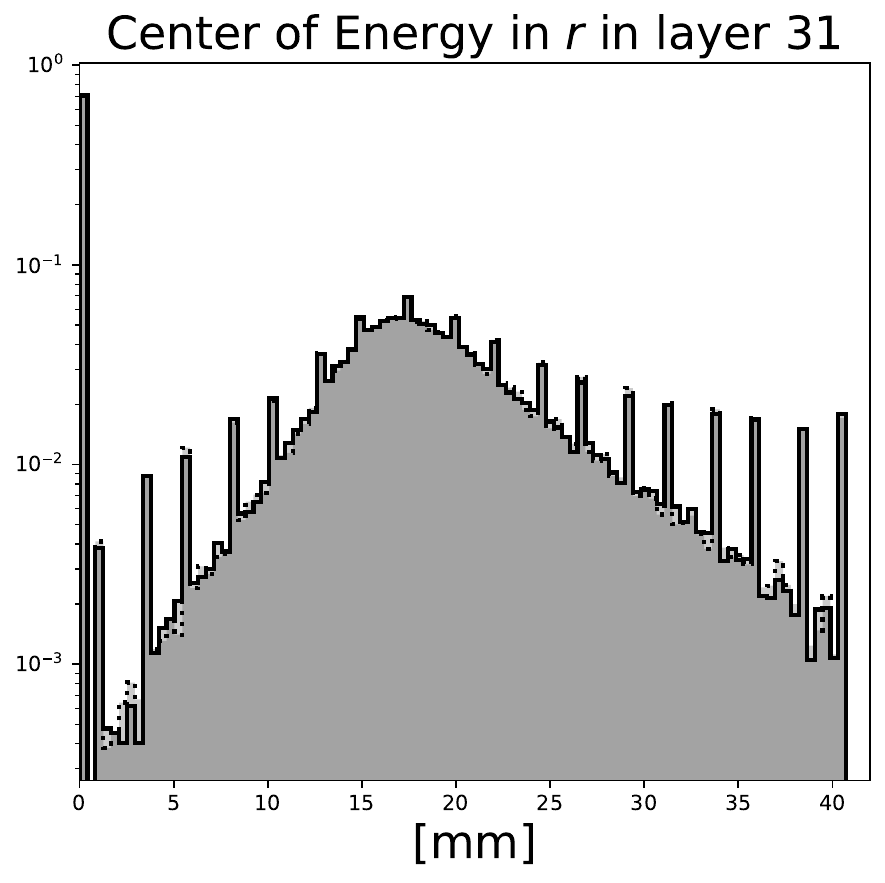} \hfill \includegraphics[height=0.1\textheight]{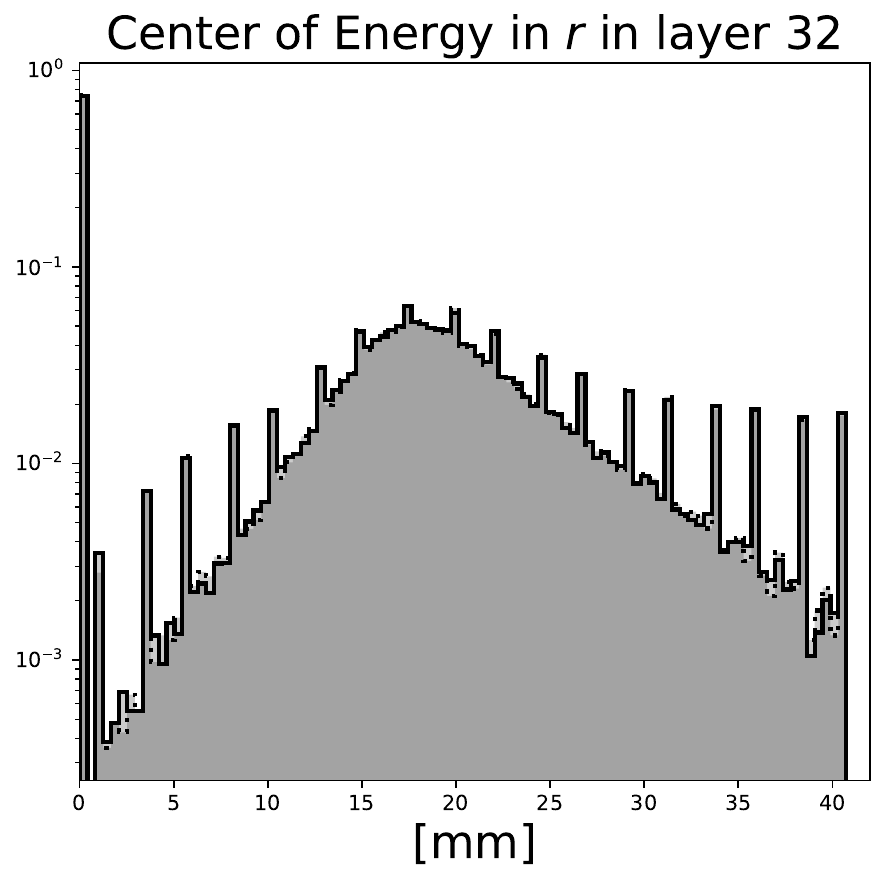} \hfill \includegraphics[height=0.1\textheight]{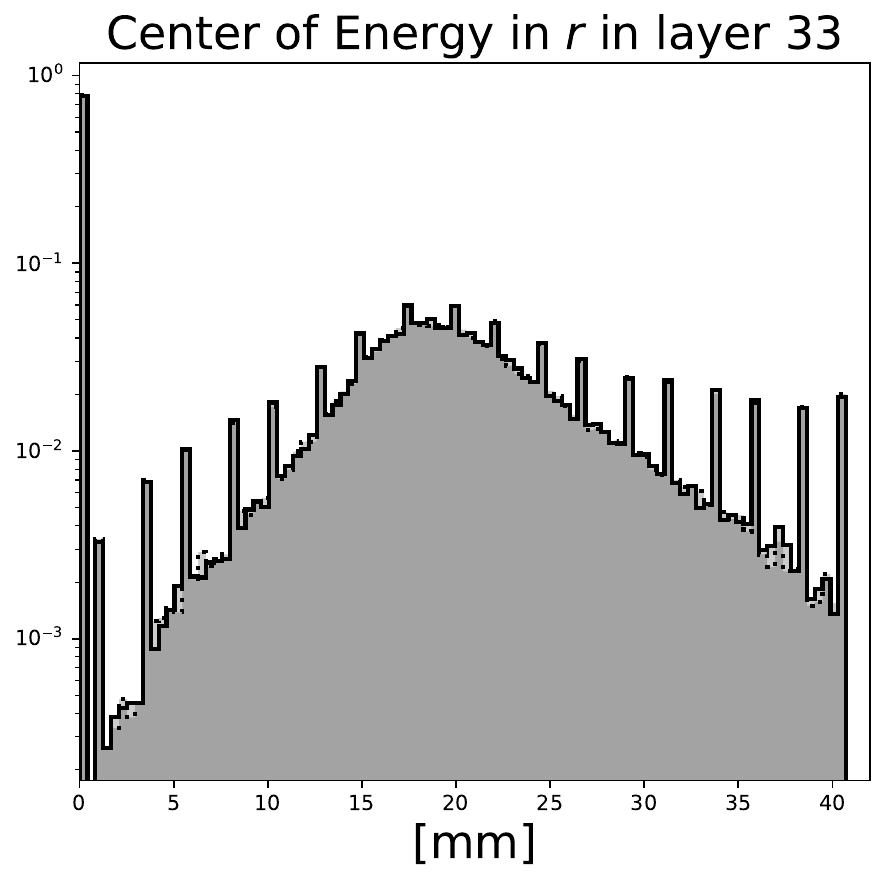} \hfill \includegraphics[height=0.1\textheight]{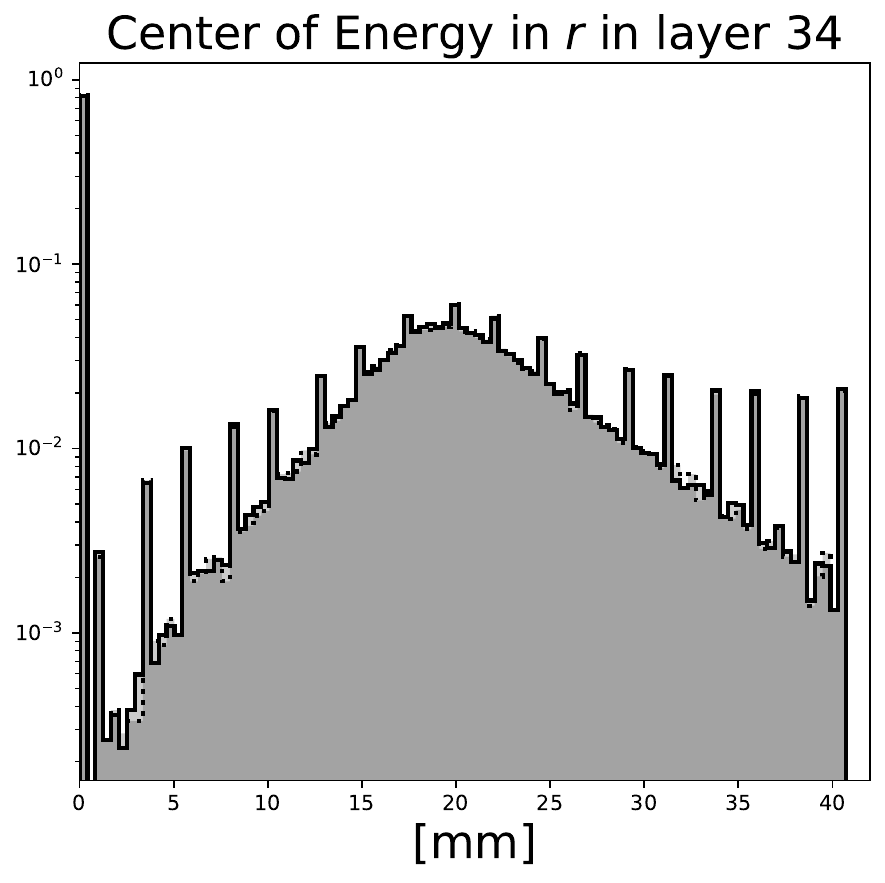}\\
    \includegraphics[height=0.1\textheight]{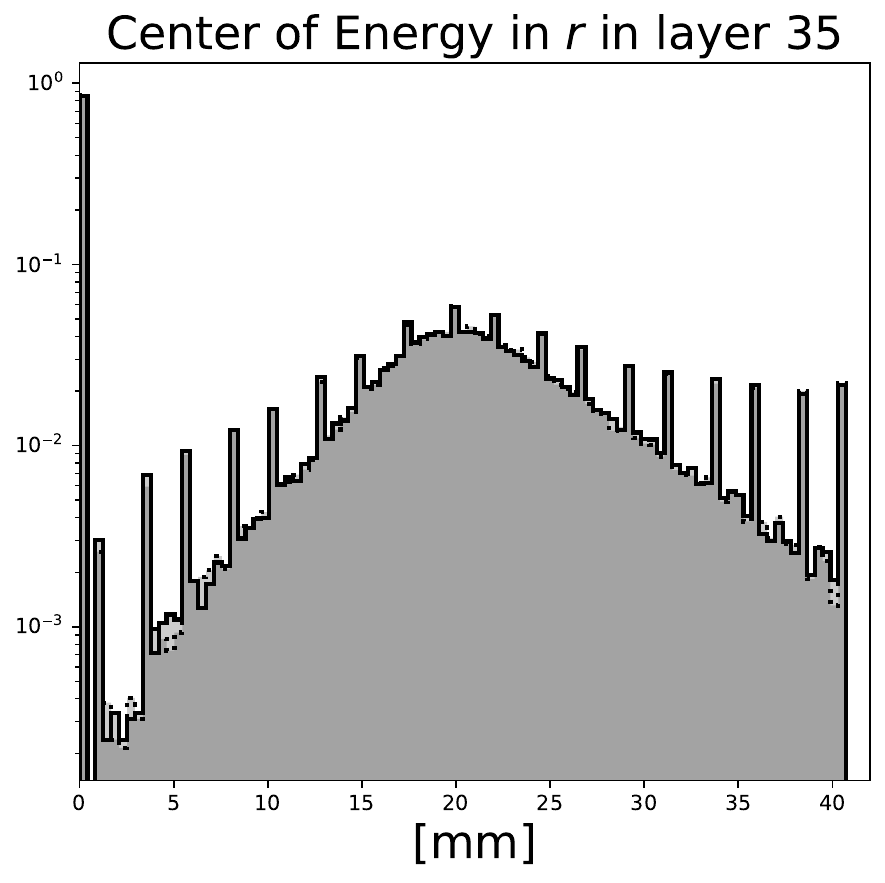} \hfill \includegraphics[height=0.1\textheight]{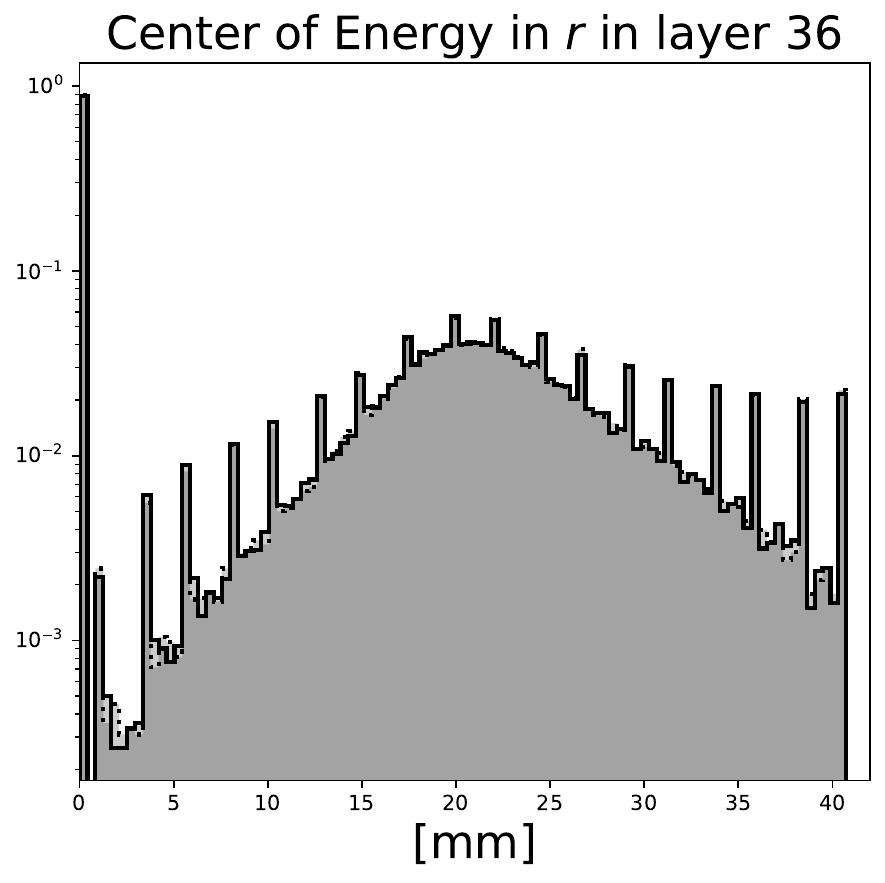} \hfill \includegraphics[height=0.1\textheight]{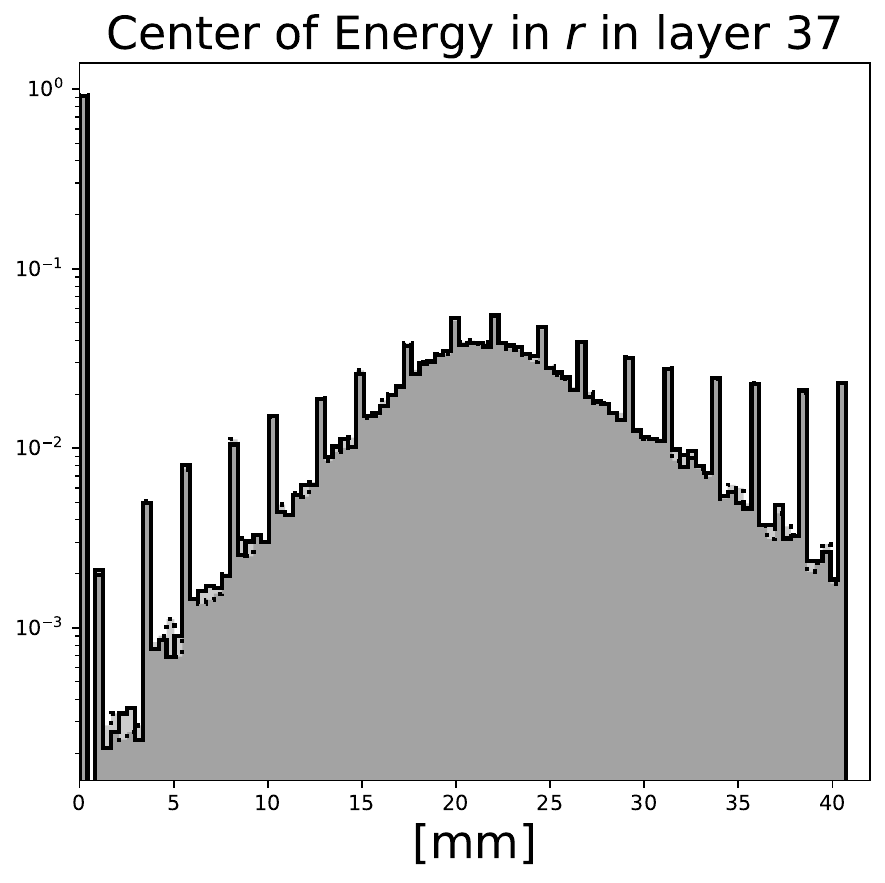} \hfill \includegraphics[height=0.1\textheight]{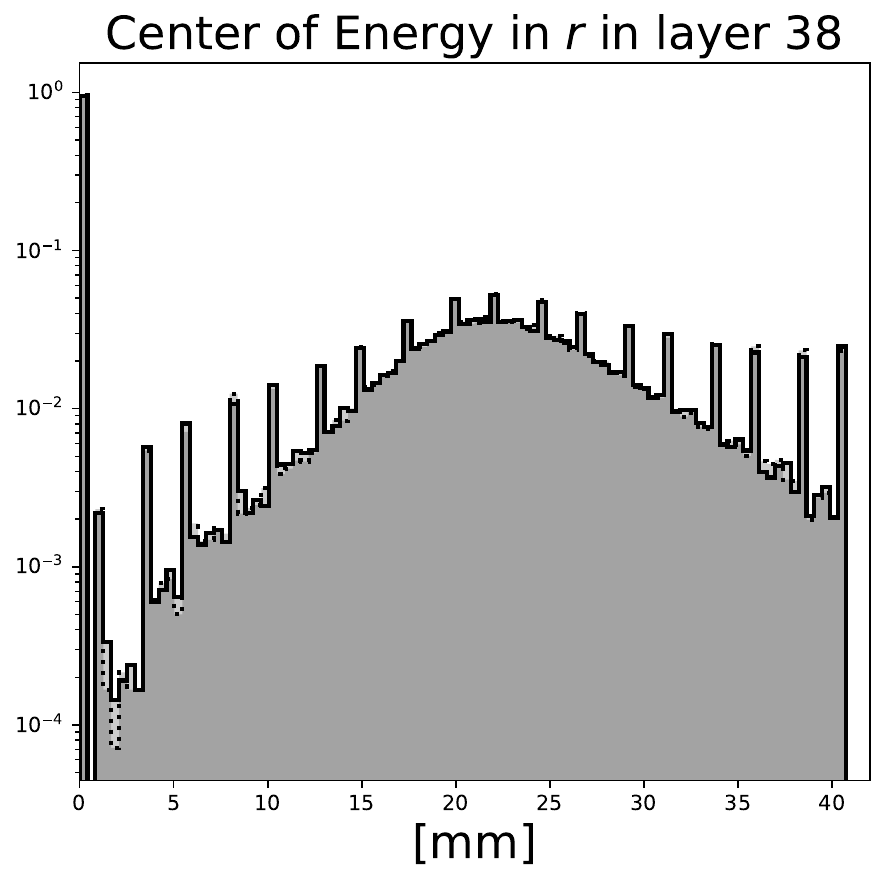} \hfill \includegraphics[height=0.1\textheight]{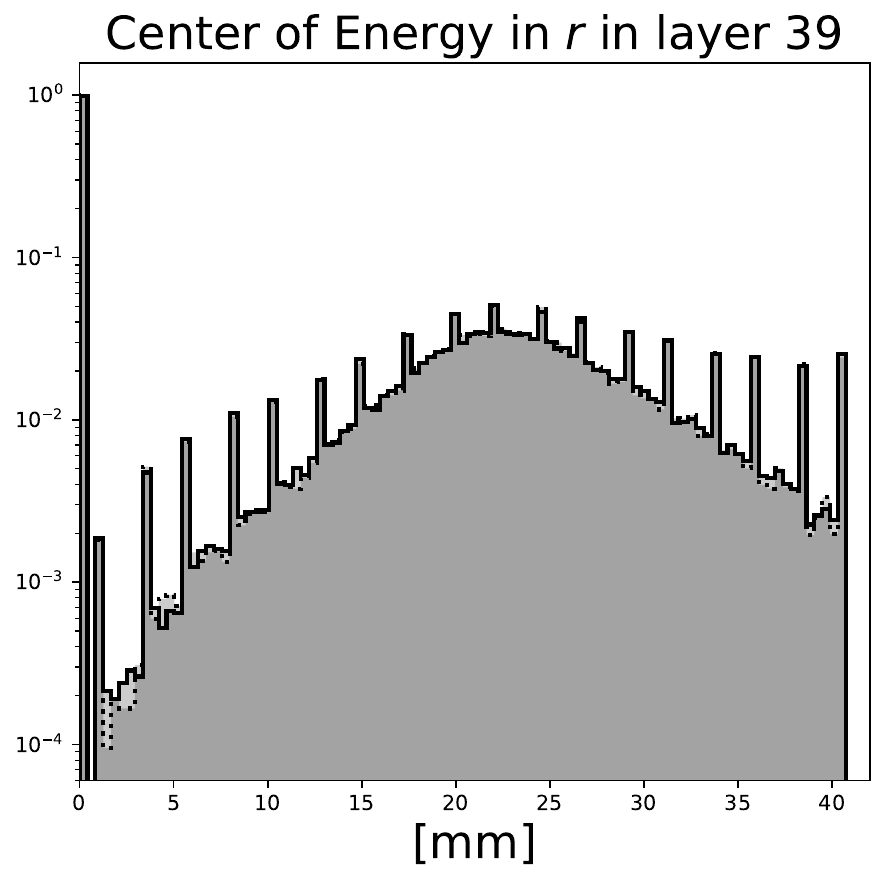}\\
    \includegraphics[height=0.1\textheight]{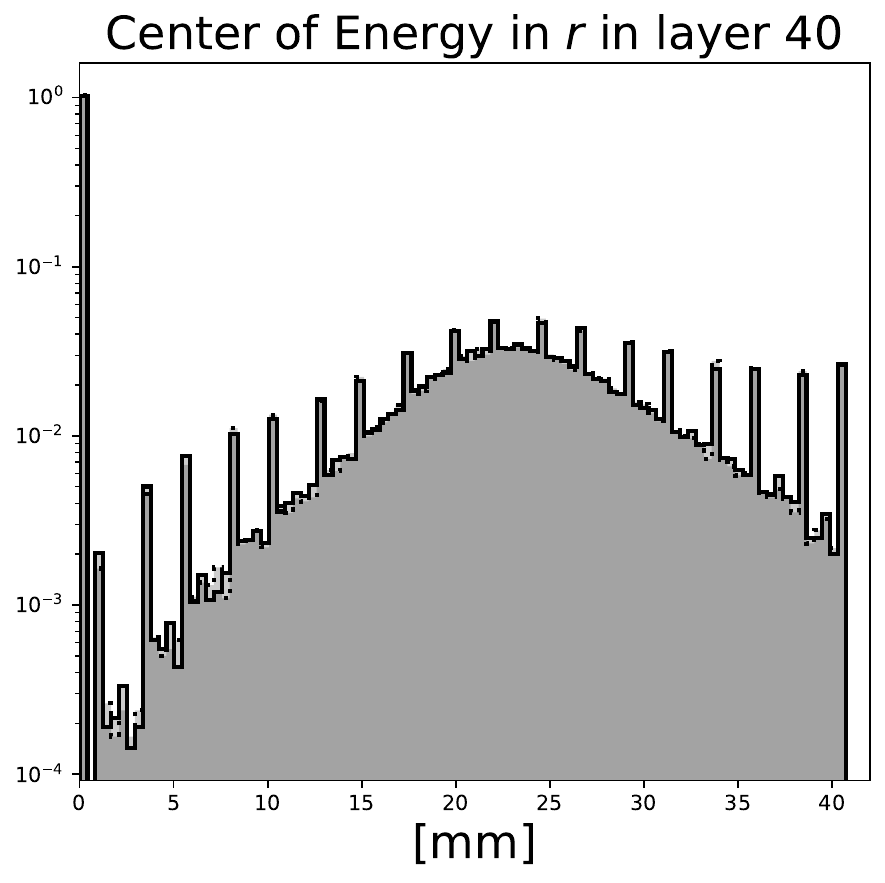} \hfill \includegraphics[height=0.1\textheight]{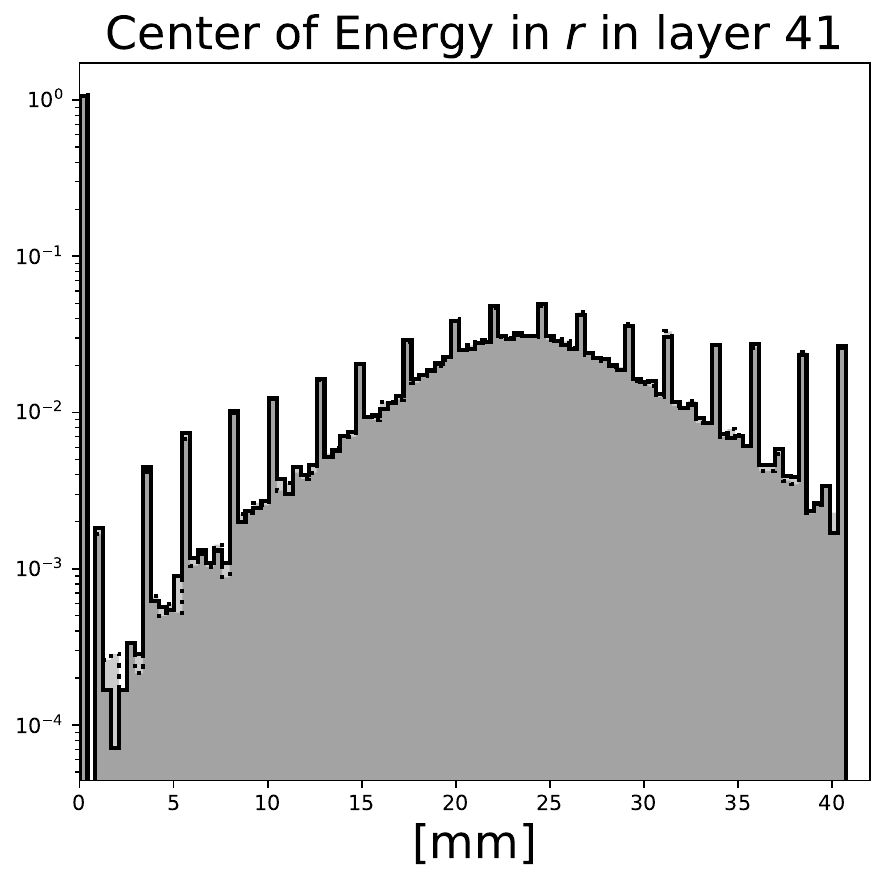} \hfill \includegraphics[height=0.1\textheight]{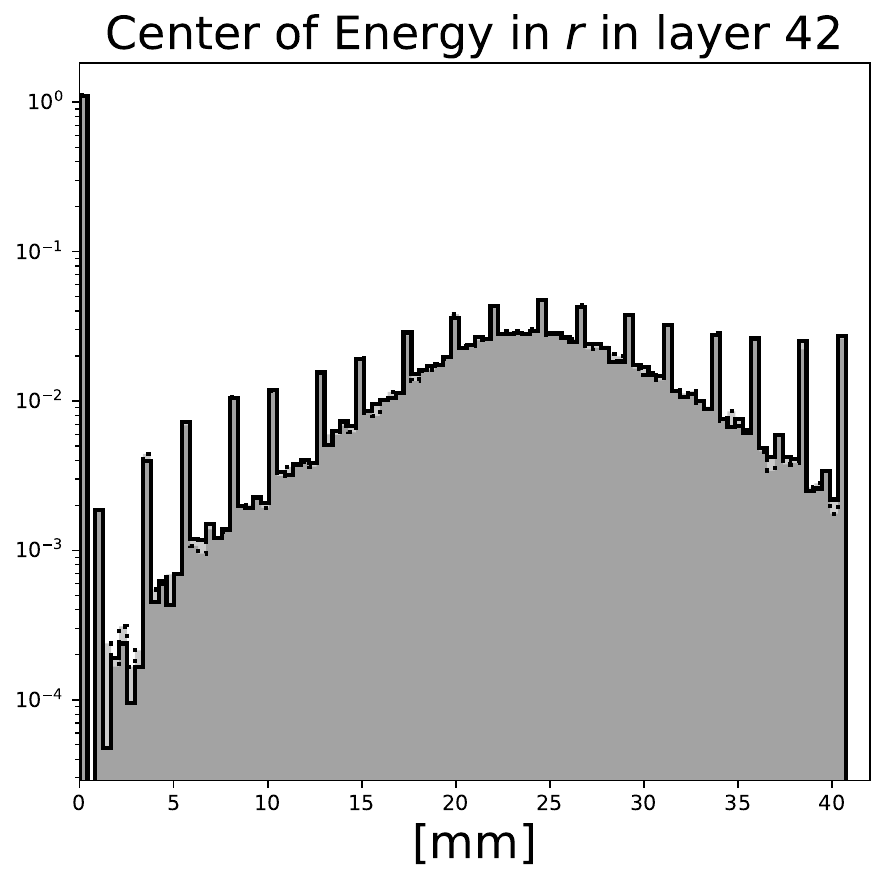} \hfill \includegraphics[height=0.1\textheight]{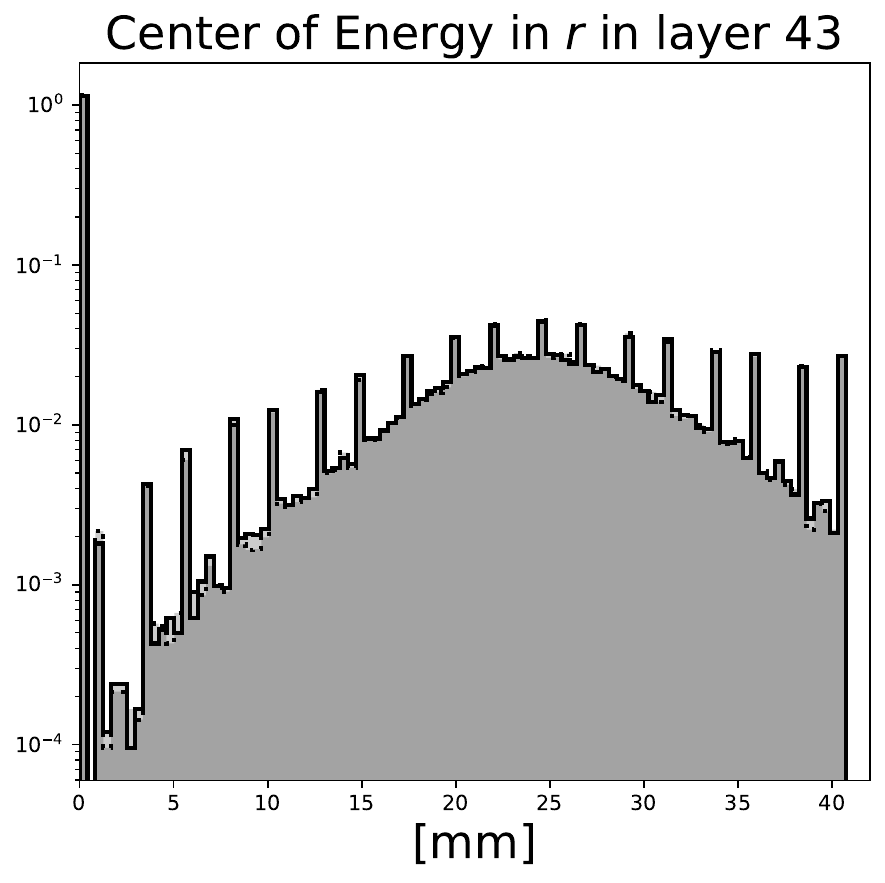} \hfill \includegraphics[height=0.1\textheight]{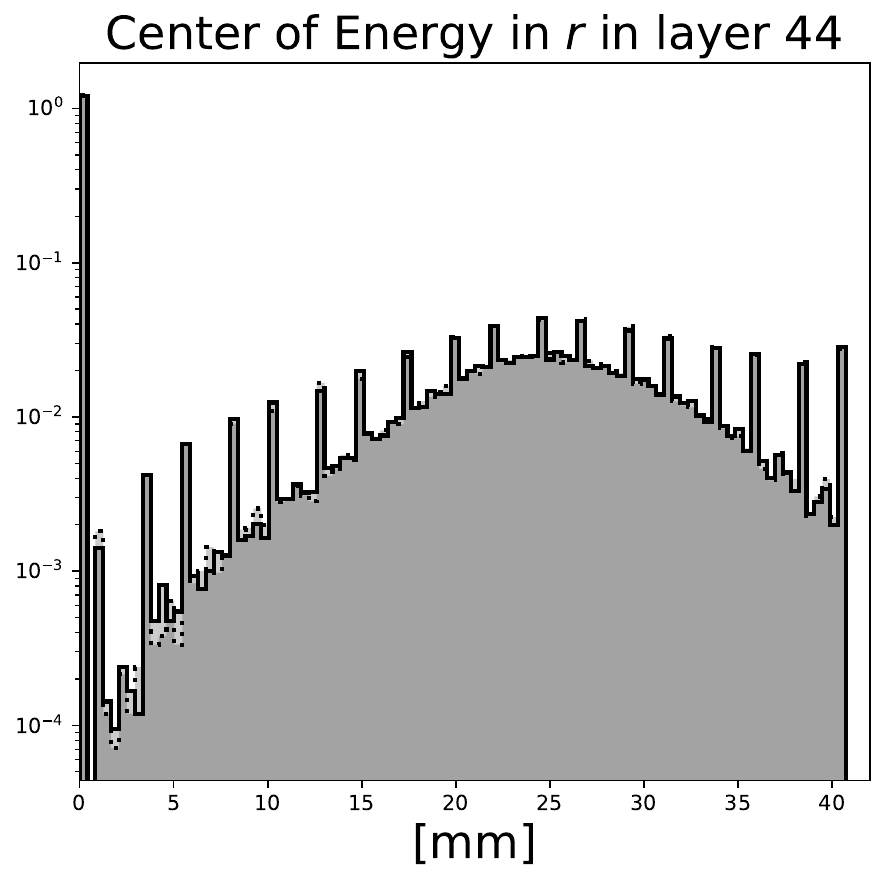}\\
    \includegraphics[width=0.5\textwidth]{figures/Appendix_reference/legend.pdf}
    \caption{Distribution of \geant training and evaluation data in centers of energy in $r$ direction for ds3. }
    \label{fig:app_ref.ds3.7}
\end{figure}

\begin{figure}[ht]
    \centering
    \includegraphics[height=0.1\textheight]{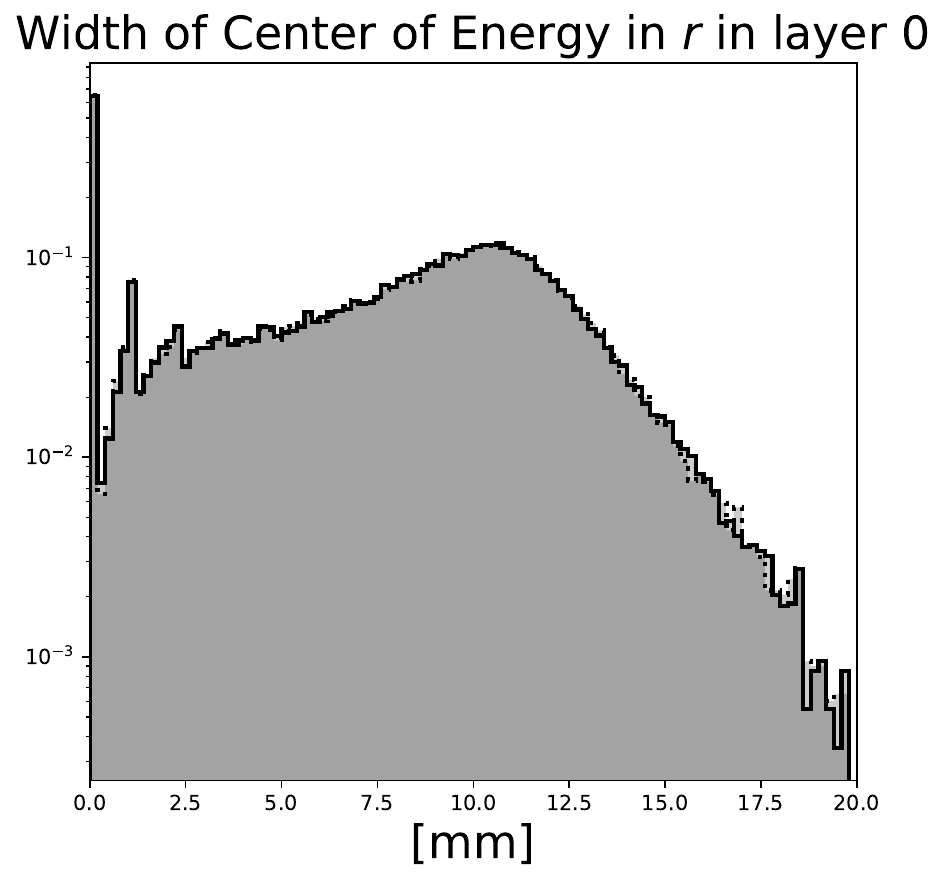} \hfill \includegraphics[height=0.1\textheight]{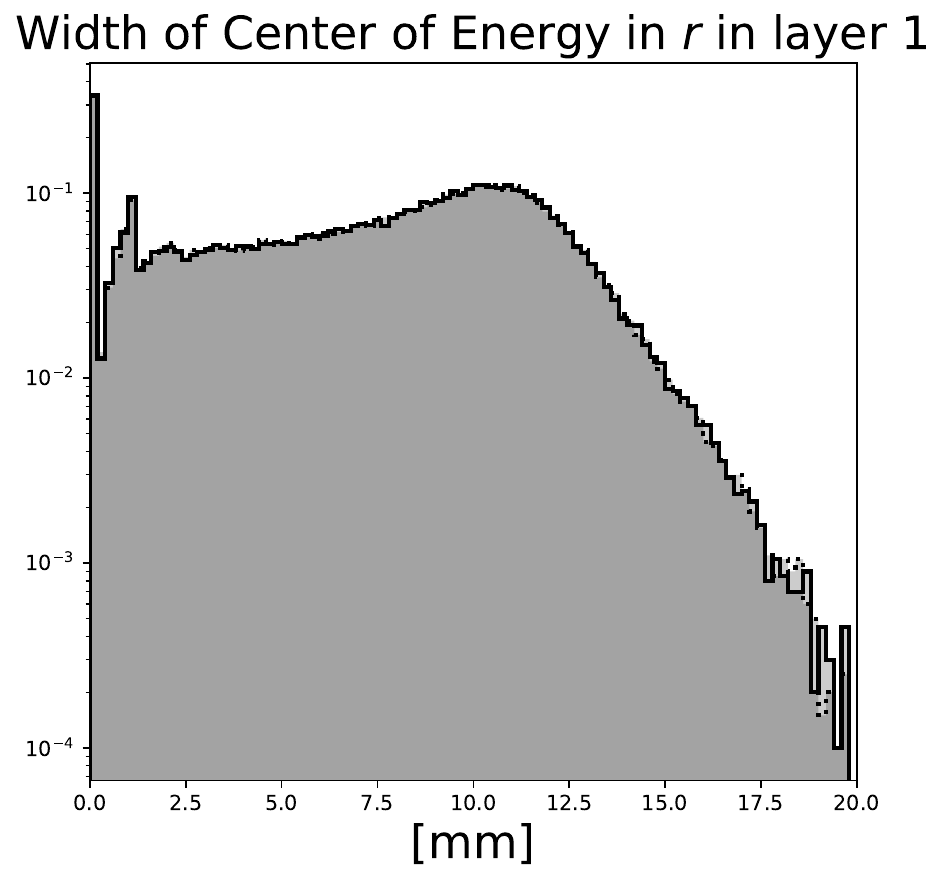} \hfill \includegraphics[height=0.1\textheight]{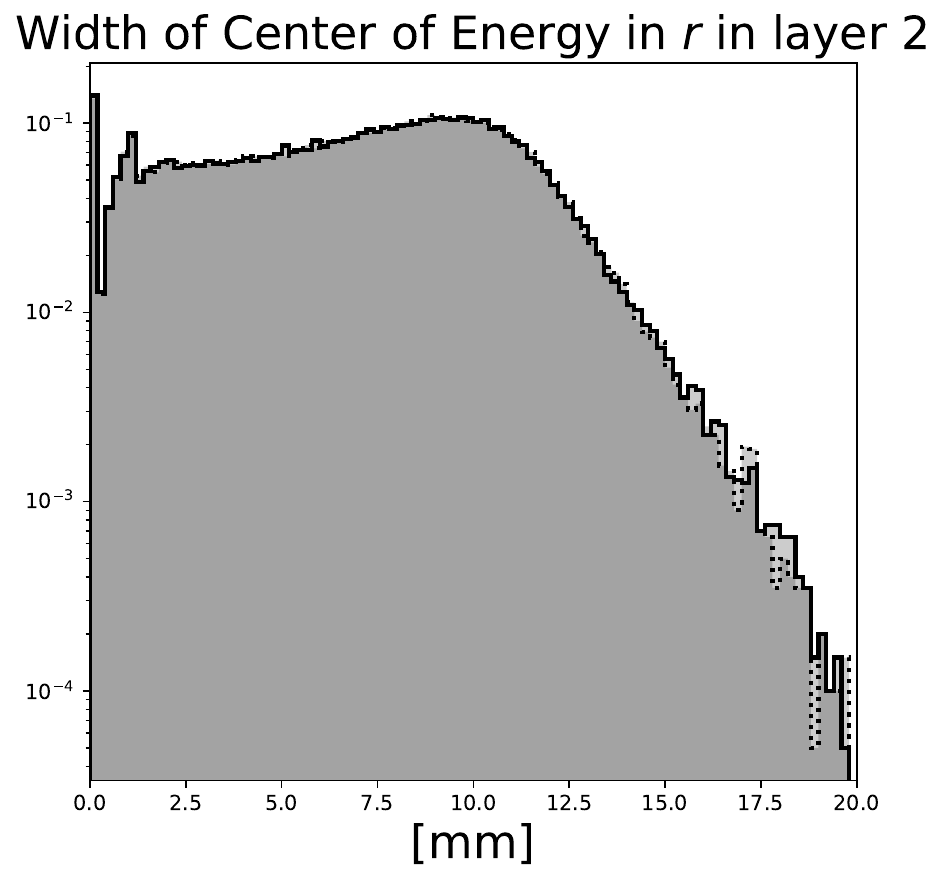} \hfill \includegraphics[height=0.1\textheight]{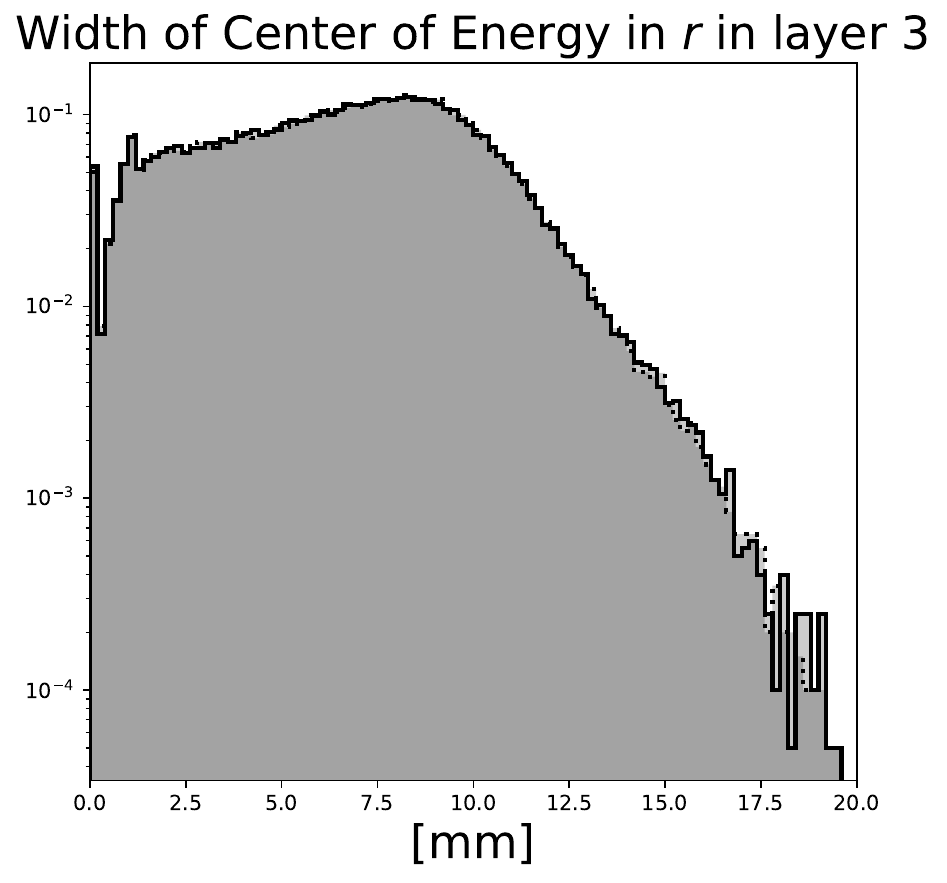} \hfill \includegraphics[height=0.1\textheight]{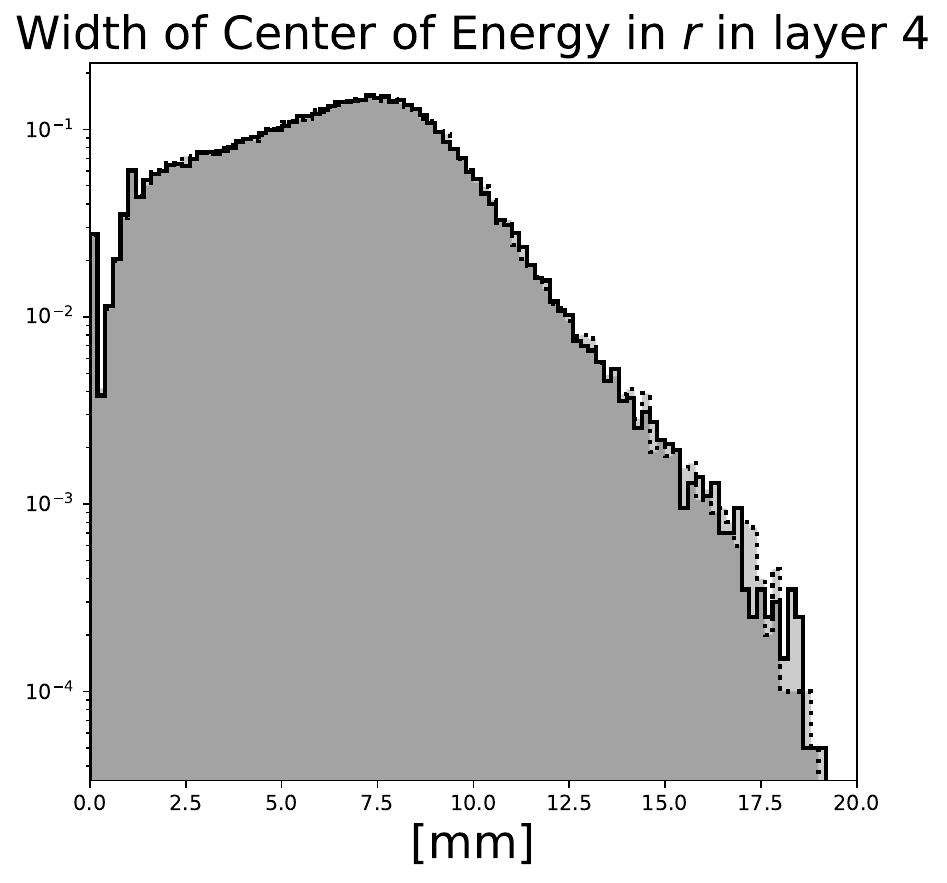}\\
    \includegraphics[height=0.1\textheight]{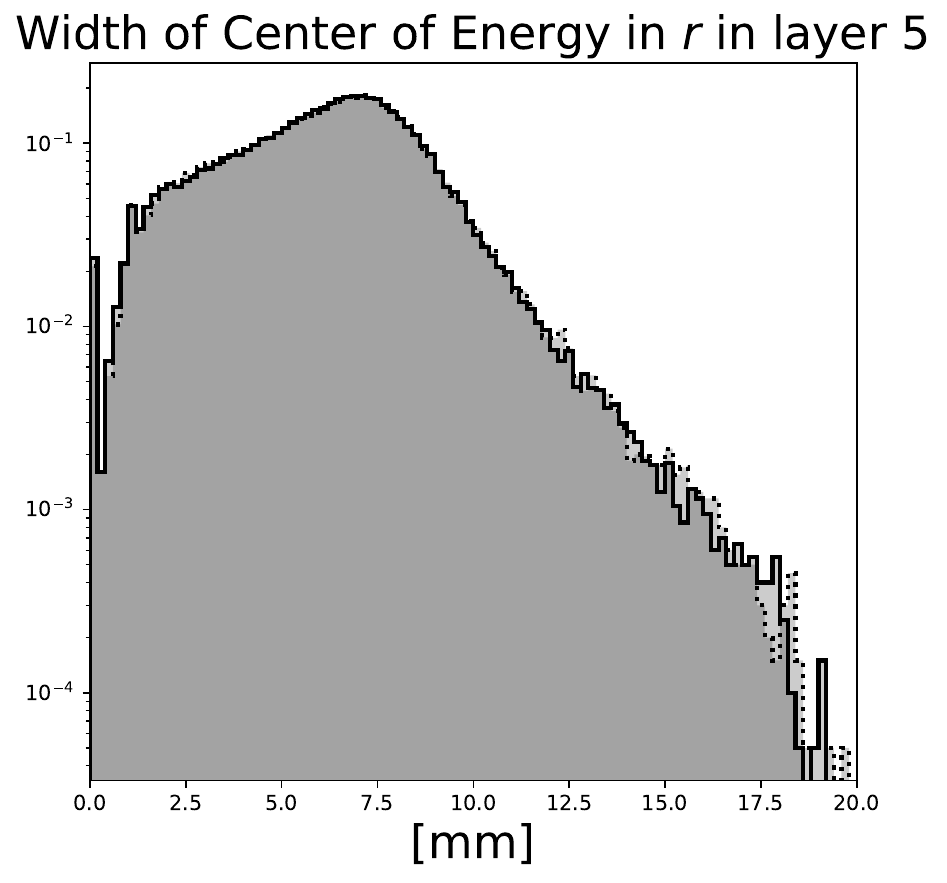} \hfill \includegraphics[height=0.1\textheight]{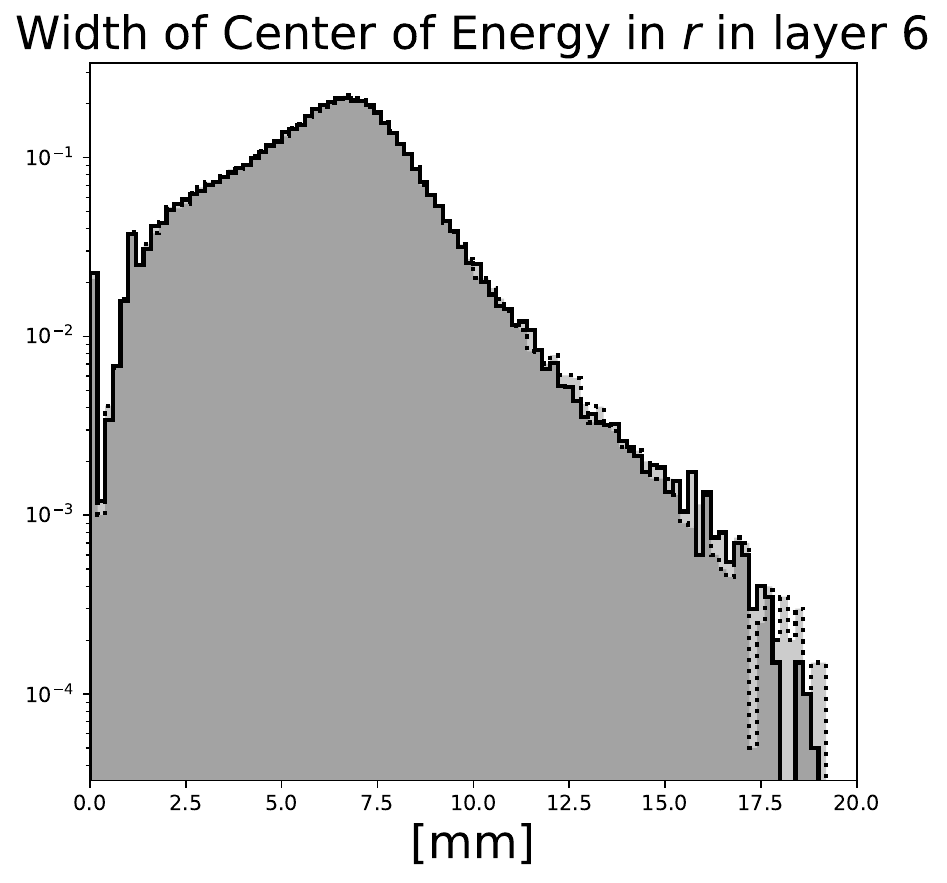} \hfill \includegraphics[height=0.1\textheight]{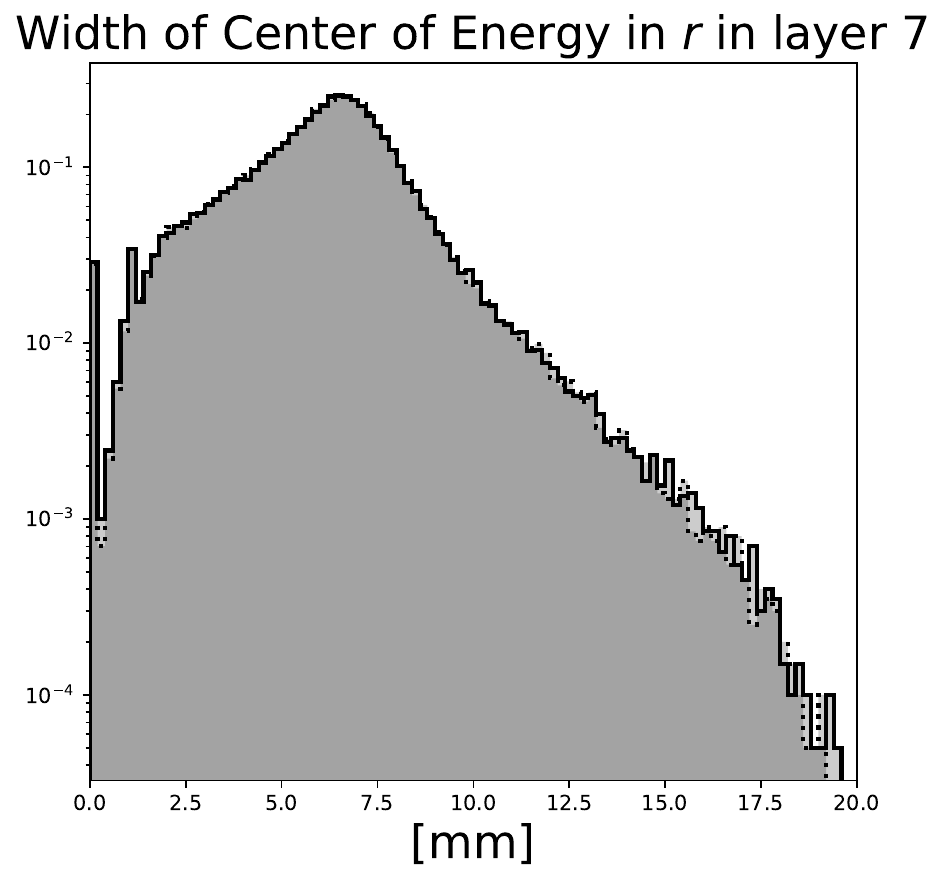} \hfill \includegraphics[height=0.1\textheight]{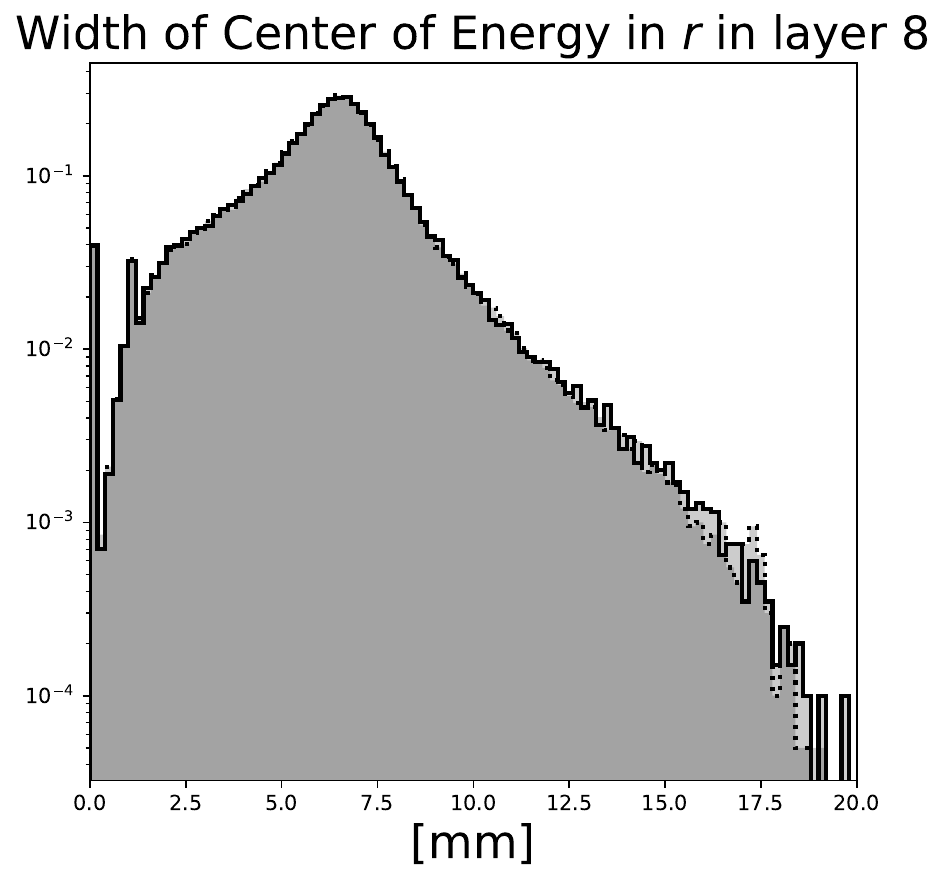} \hfill \includegraphics[height=0.1\textheight]{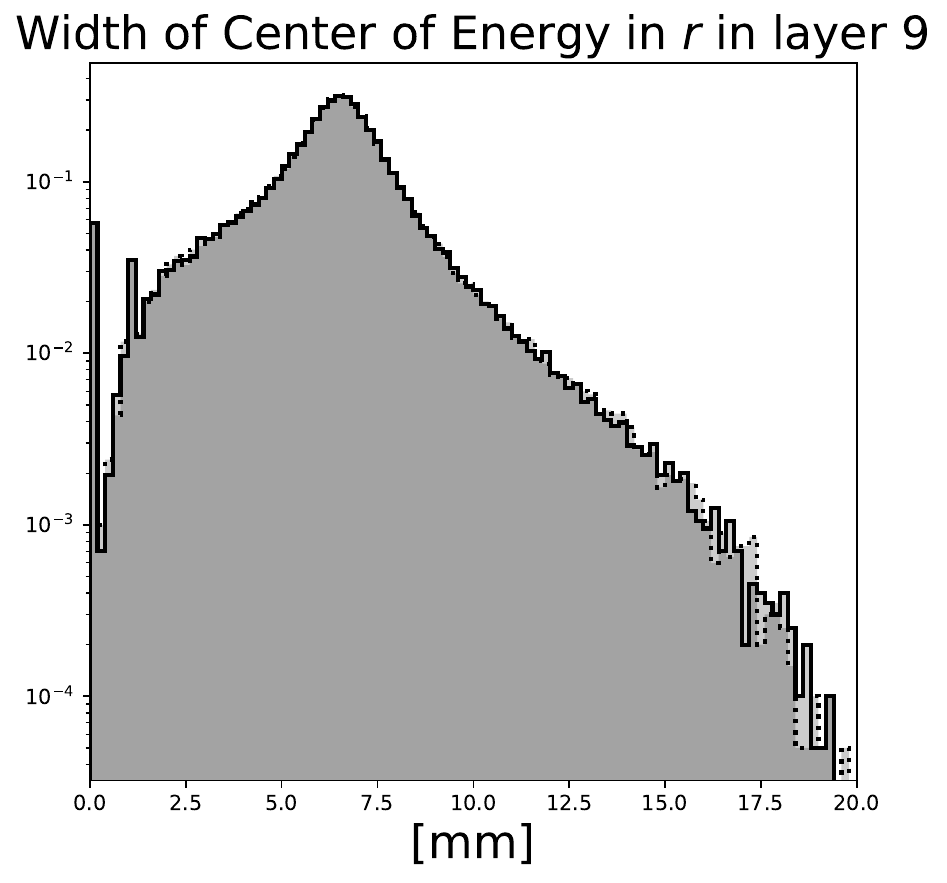}\\
    \includegraphics[height=0.1\textheight]{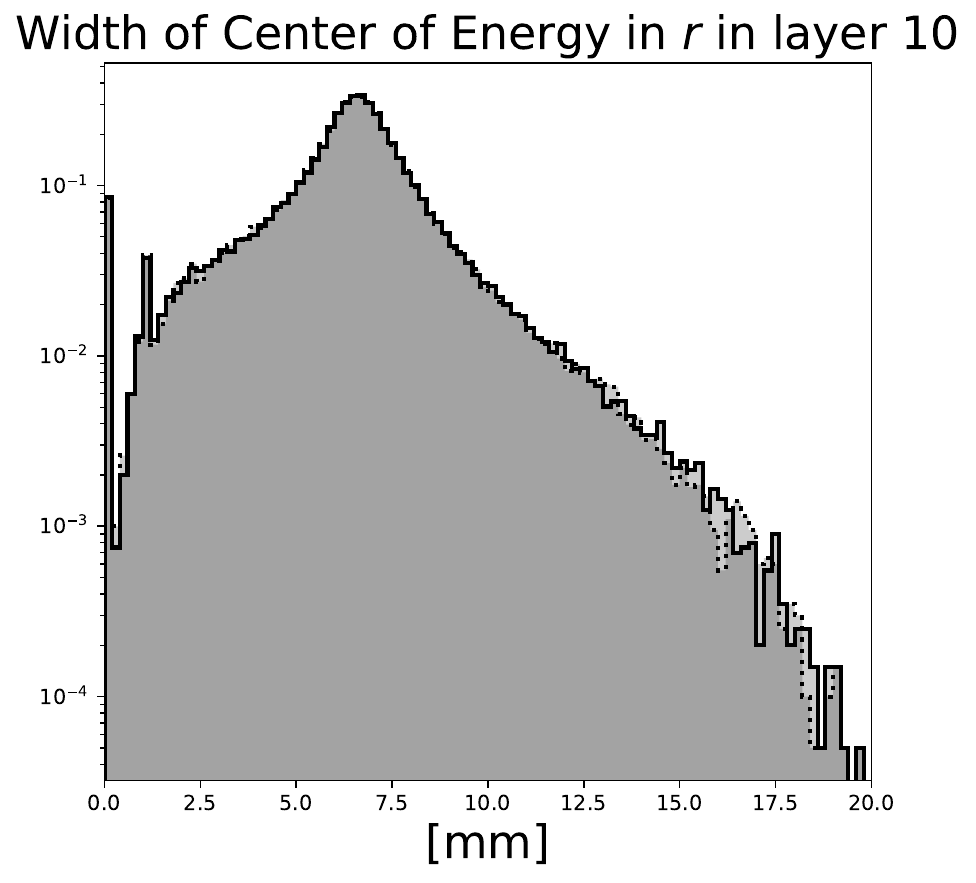} \hfill \includegraphics[height=0.1\textheight]{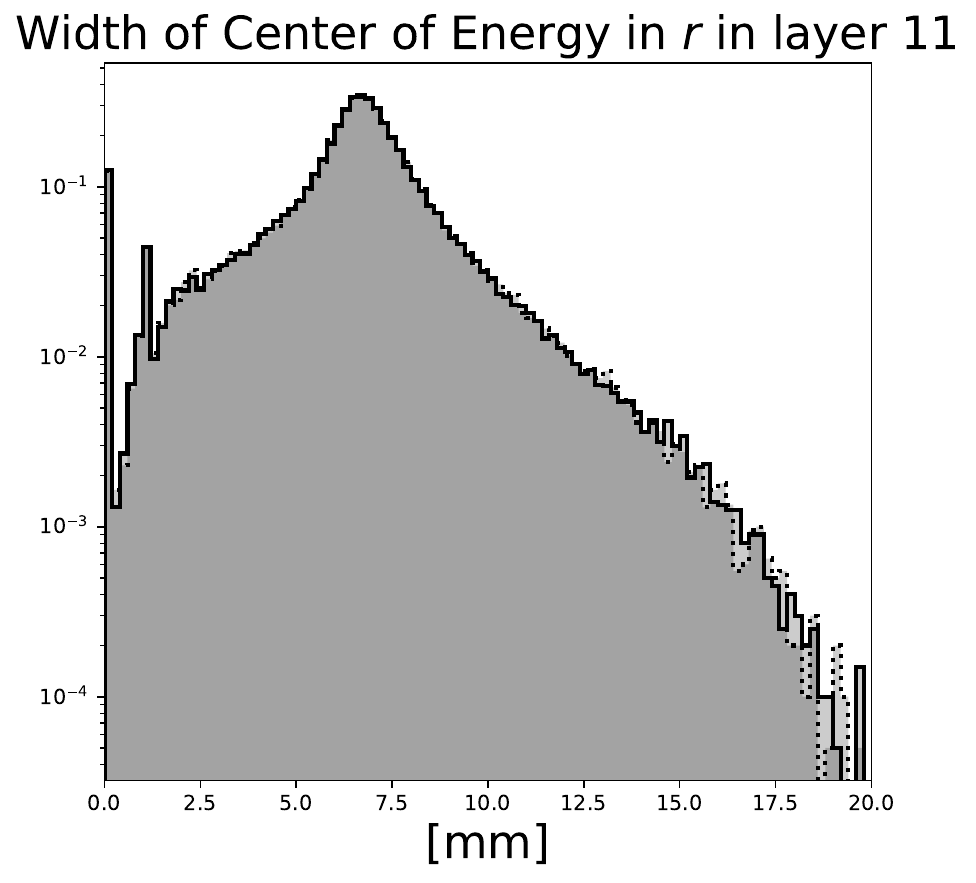} \hfill \includegraphics[height=0.1\textheight]{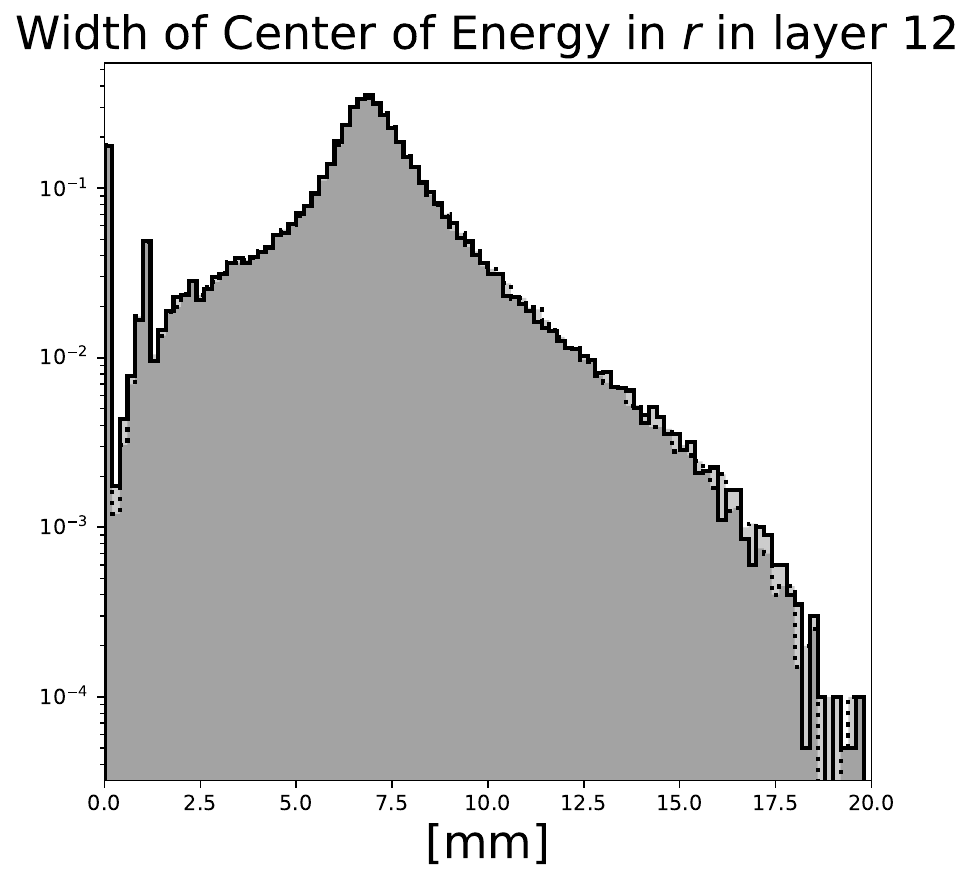} \hfill \includegraphics[height=0.1\textheight]{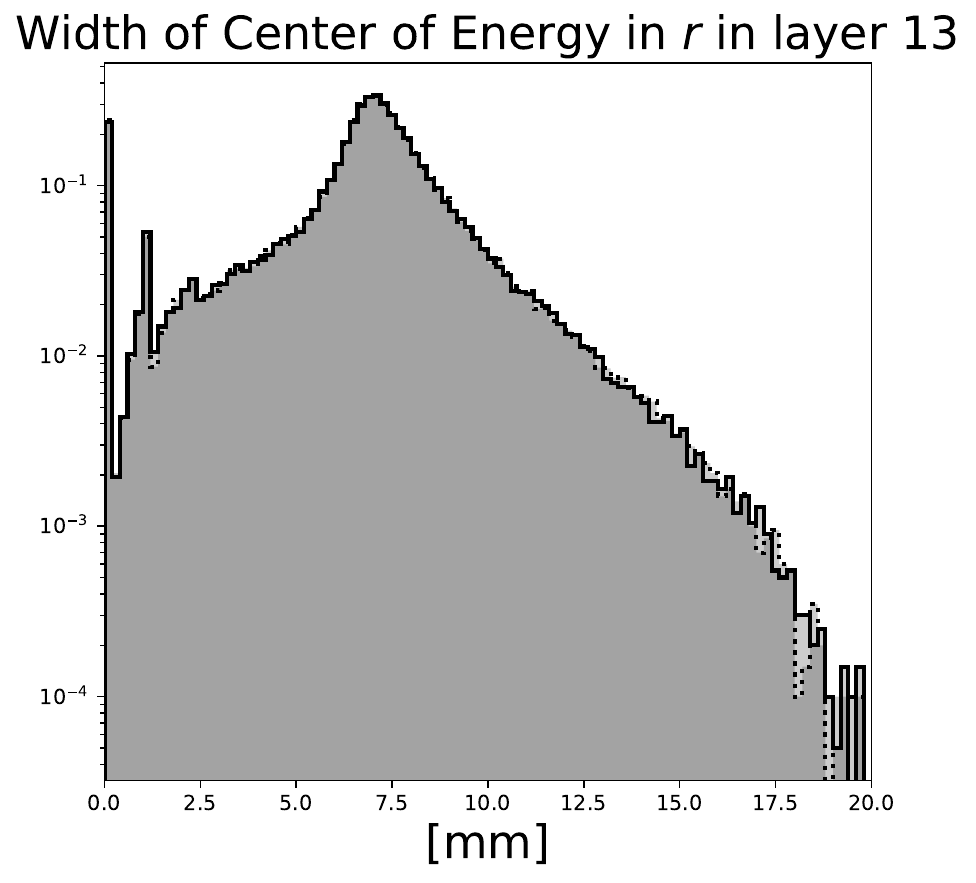} \hfill \includegraphics[height=0.1\textheight]{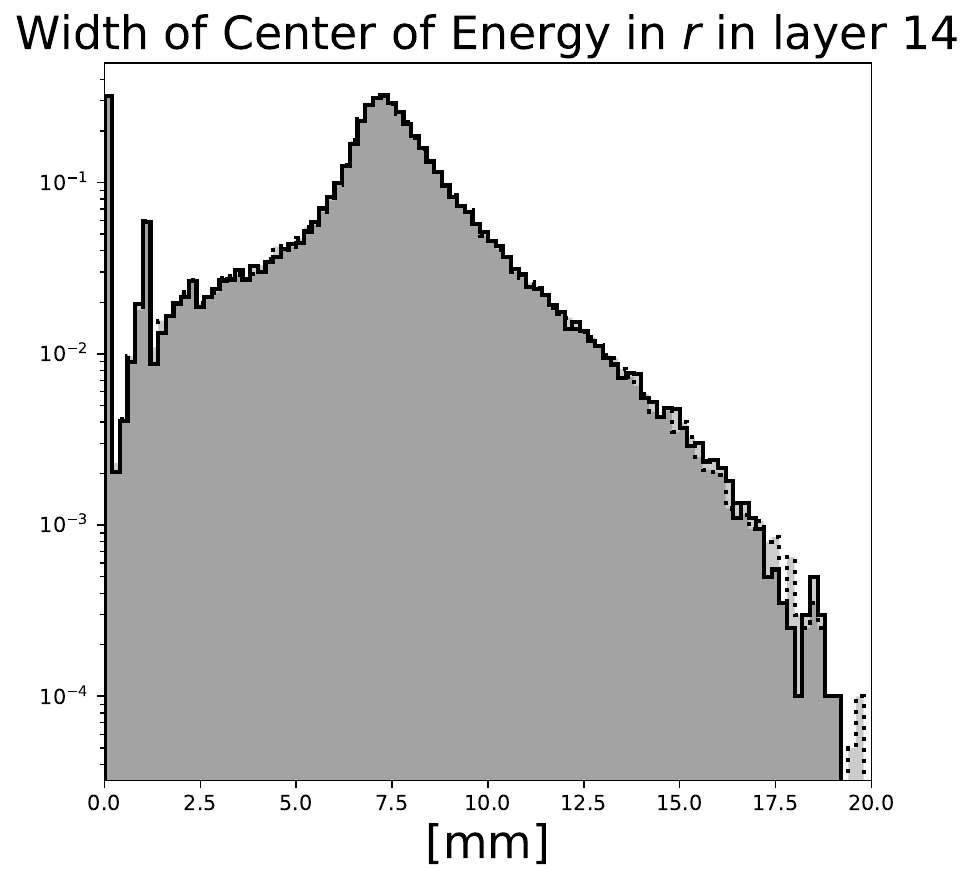}\\
    \includegraphics[height=0.1\textheight]{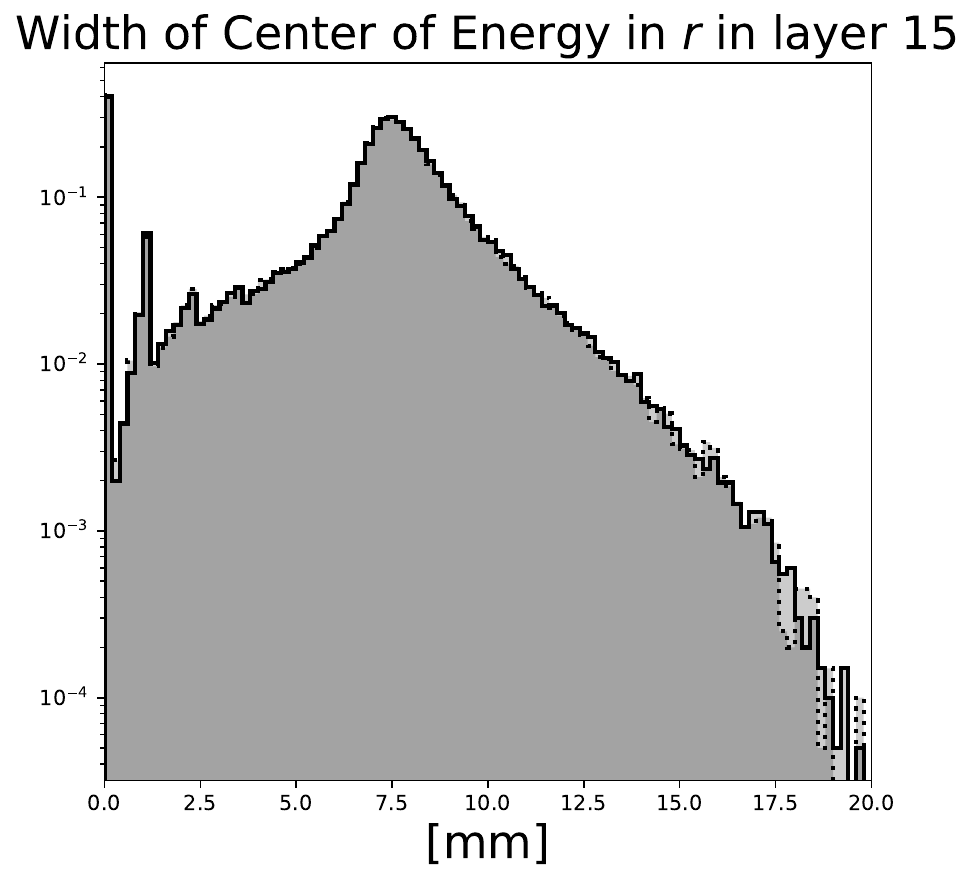} \hfill \includegraphics[height=0.1\textheight]{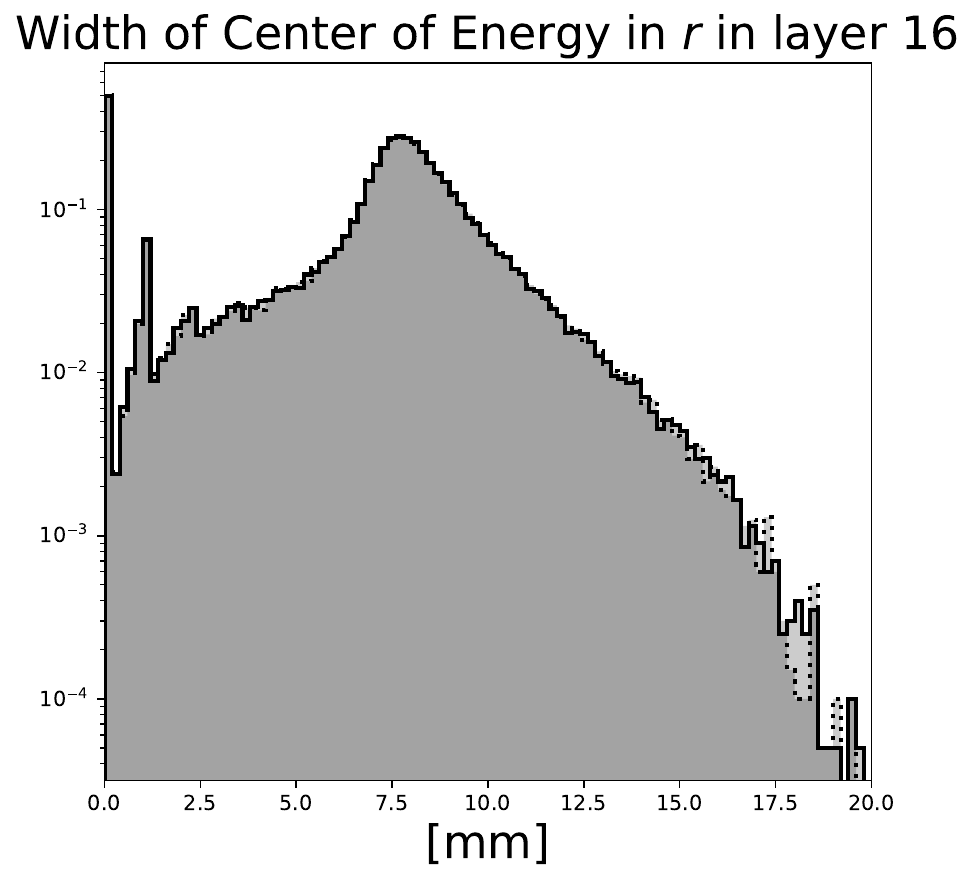} \hfill \includegraphics[height=0.1\textheight]{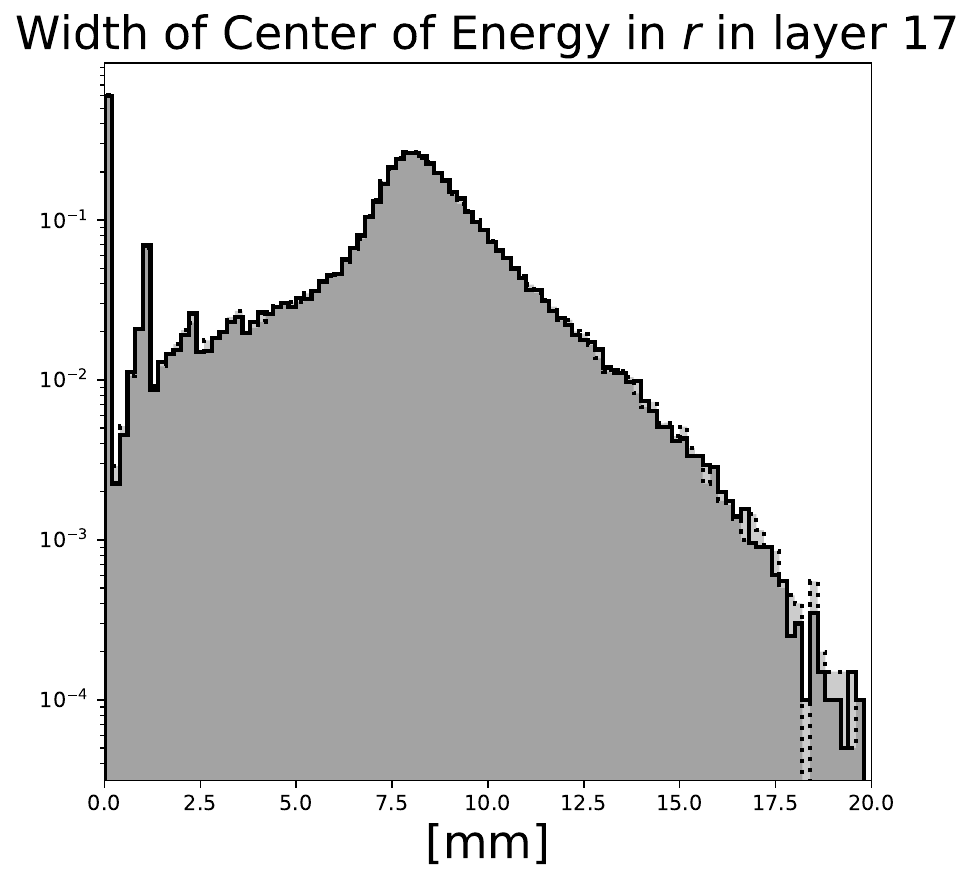} \hfill \includegraphics[height=0.1\textheight]{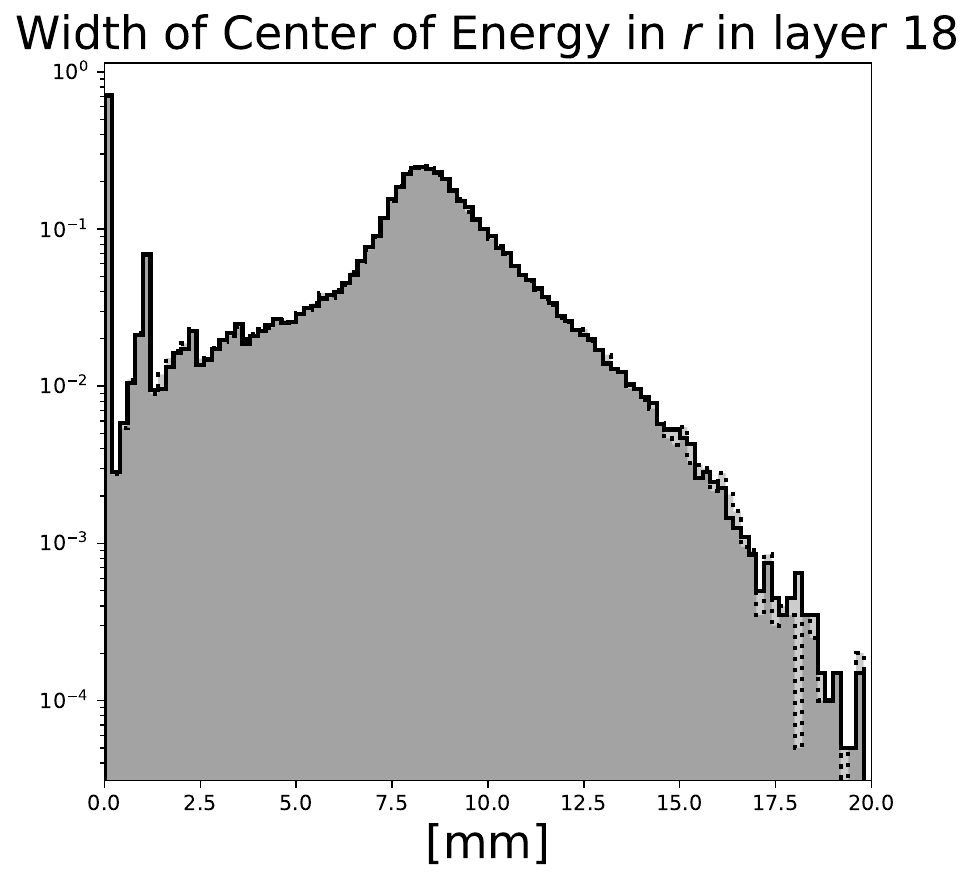} \hfill \includegraphics[height=0.1\textheight]{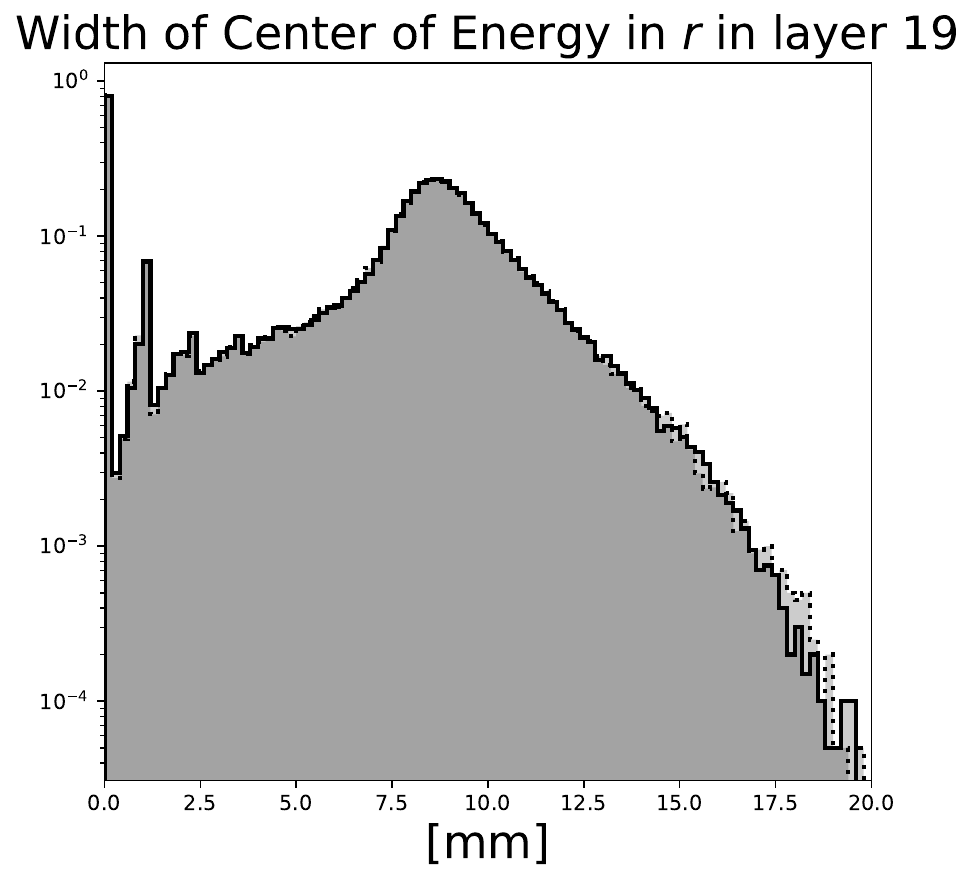}\\
    \includegraphics[height=0.1\textheight]{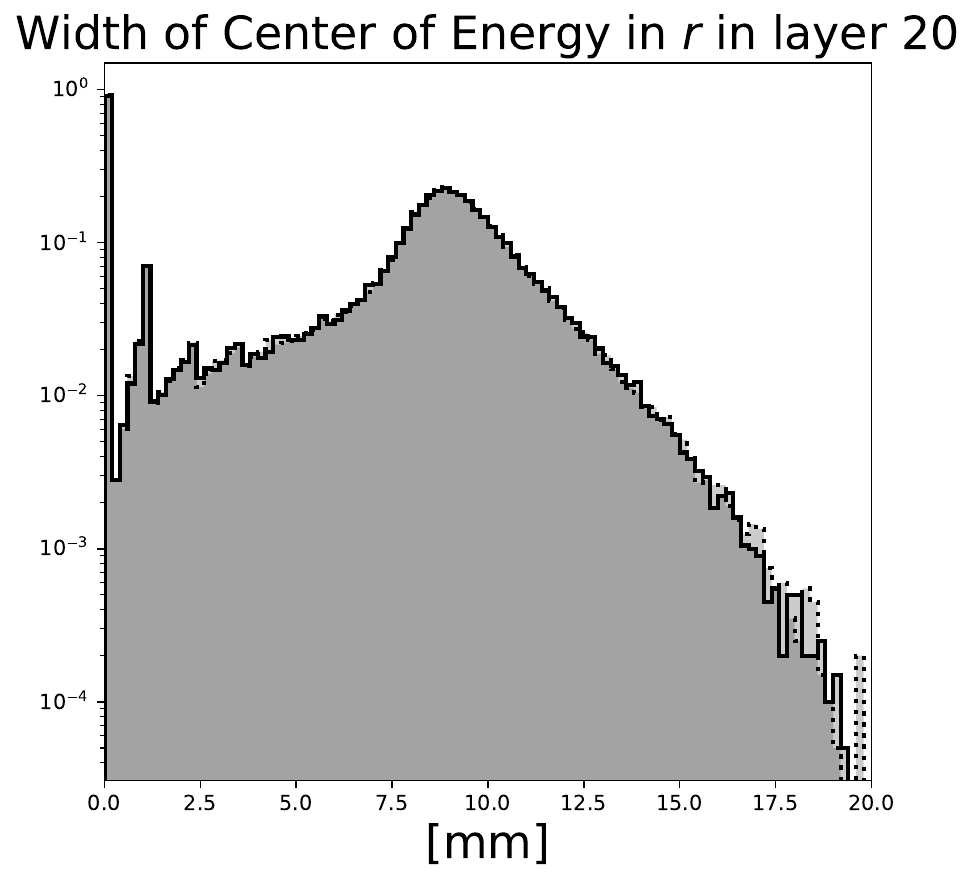} \hfill \includegraphics[height=0.1\textheight]{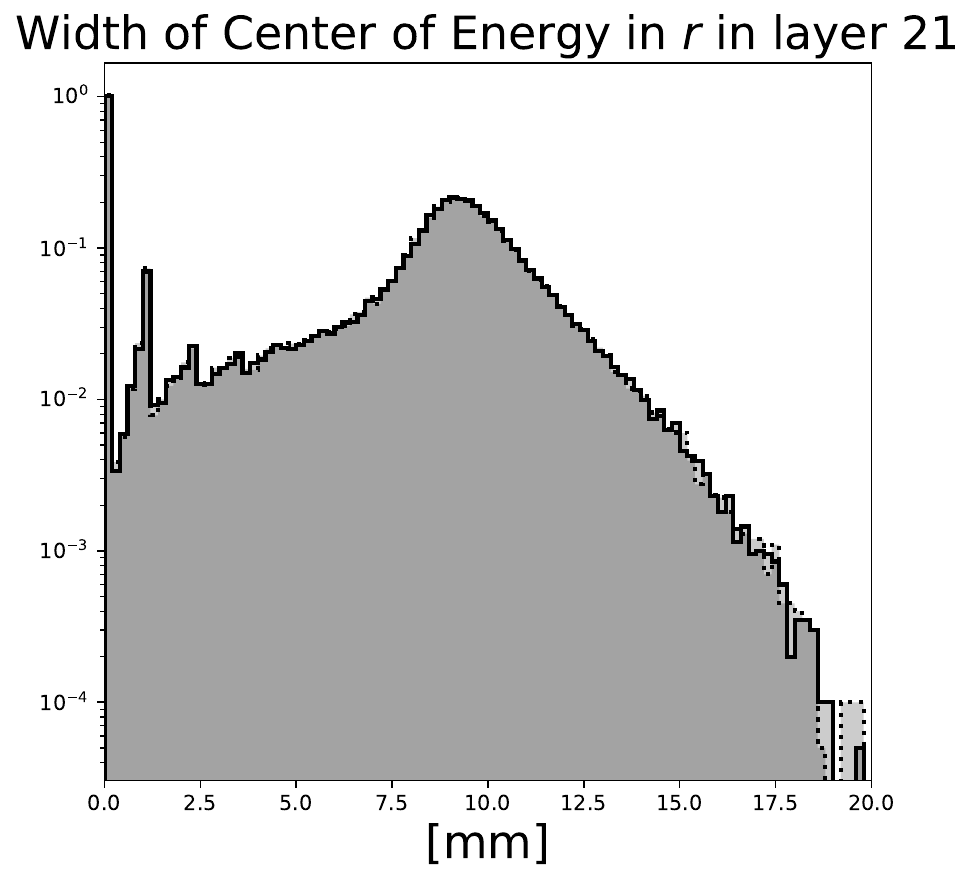} \hfill \includegraphics[height=0.1\textheight]{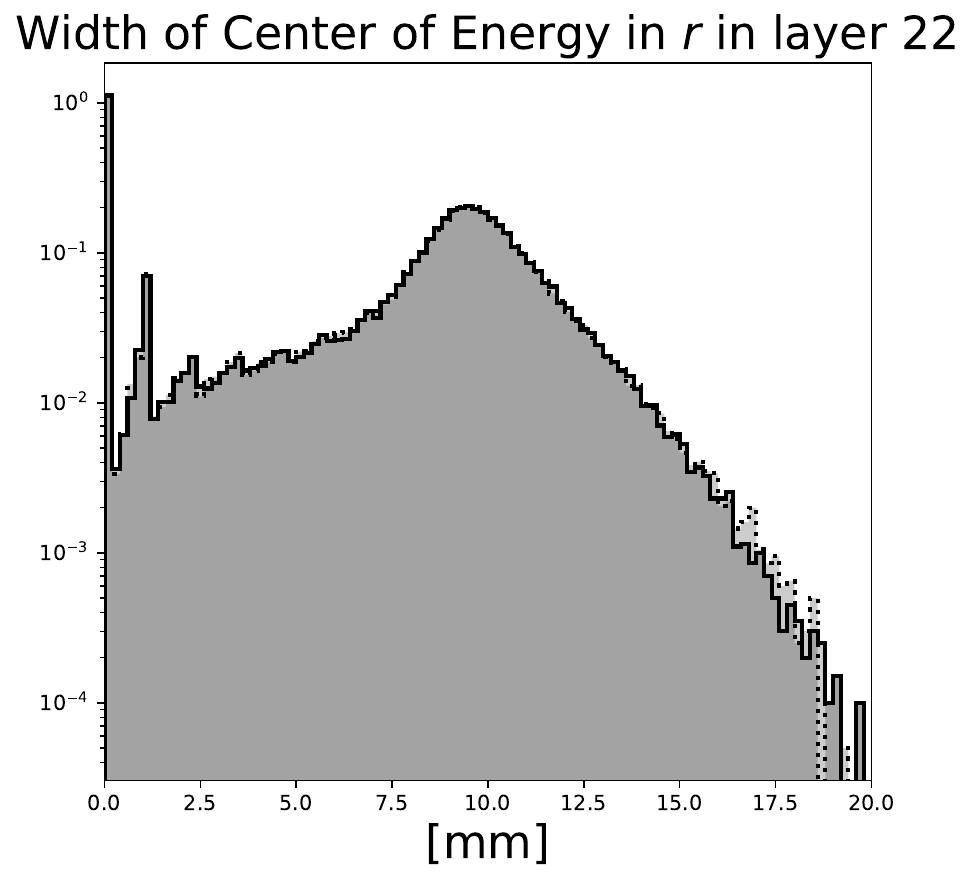} \hfill \includegraphics[height=0.1\textheight]{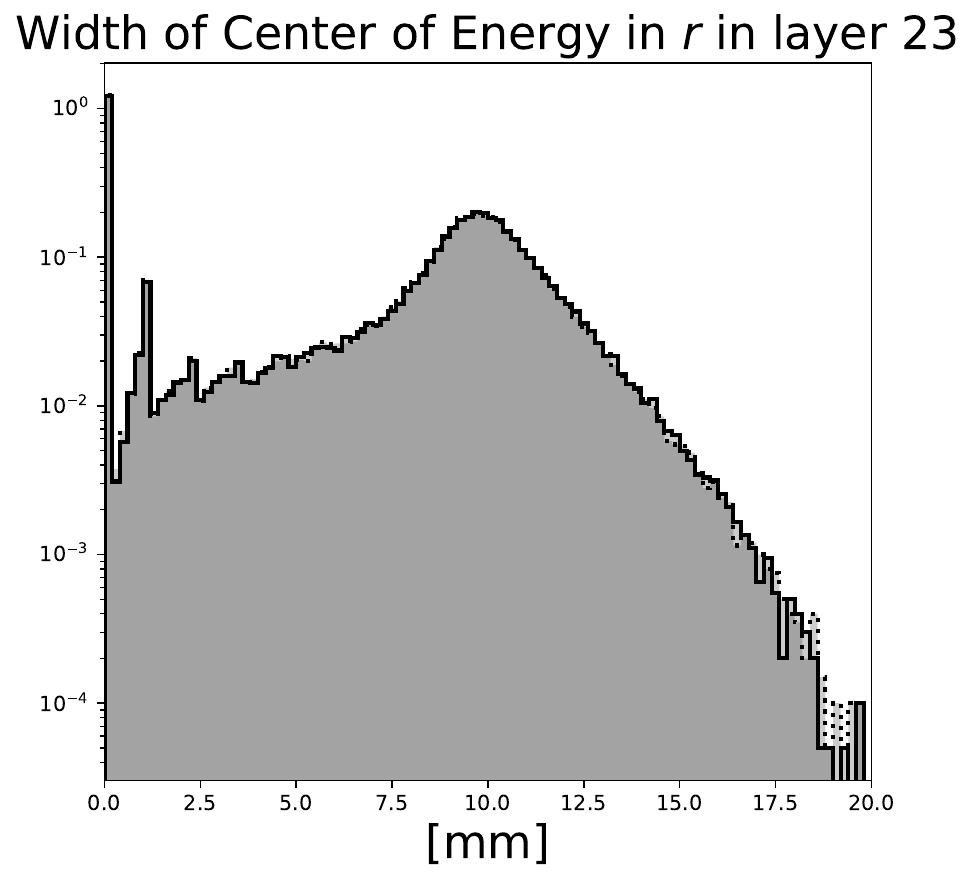} \hfill \includegraphics[height=0.1\textheight]{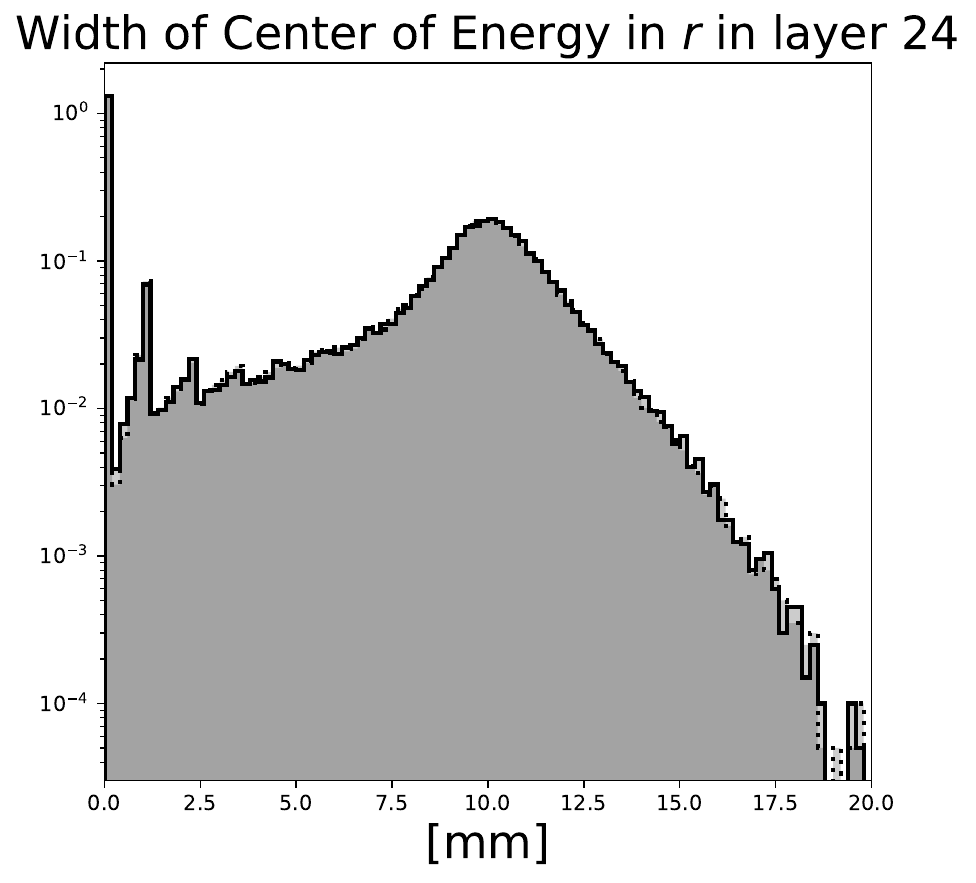}\\
    \includegraphics[height=0.1\textheight]{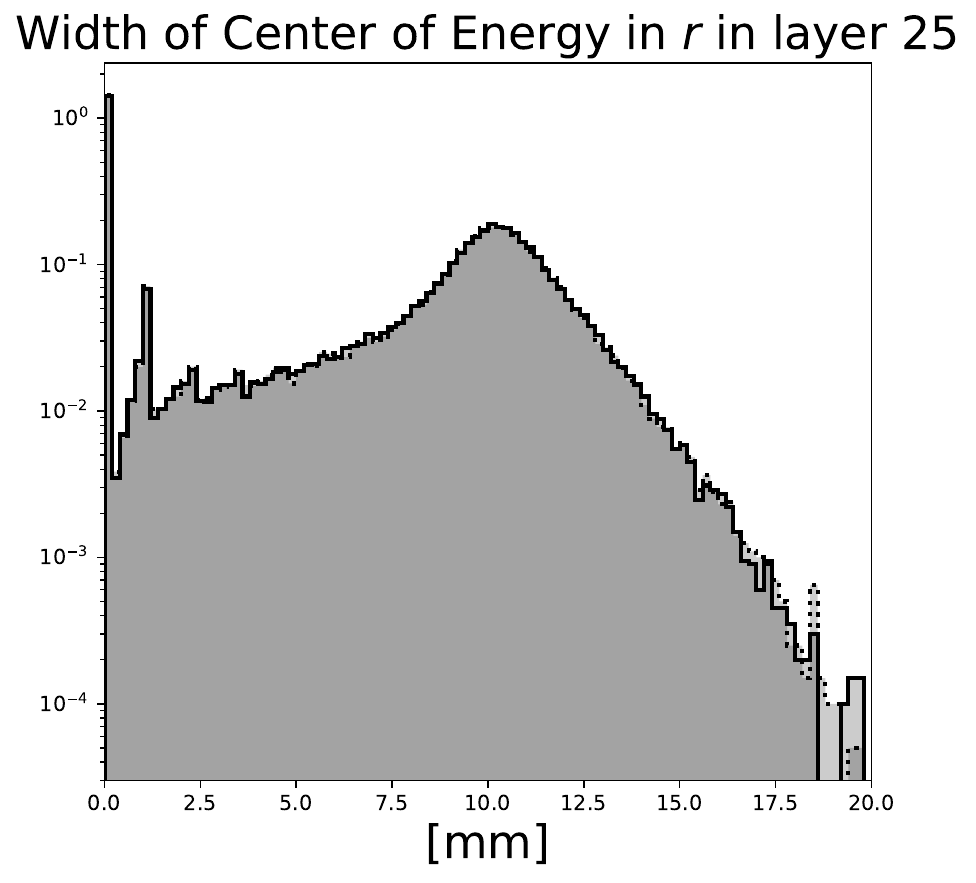} \hfill \includegraphics[height=0.1\textheight]{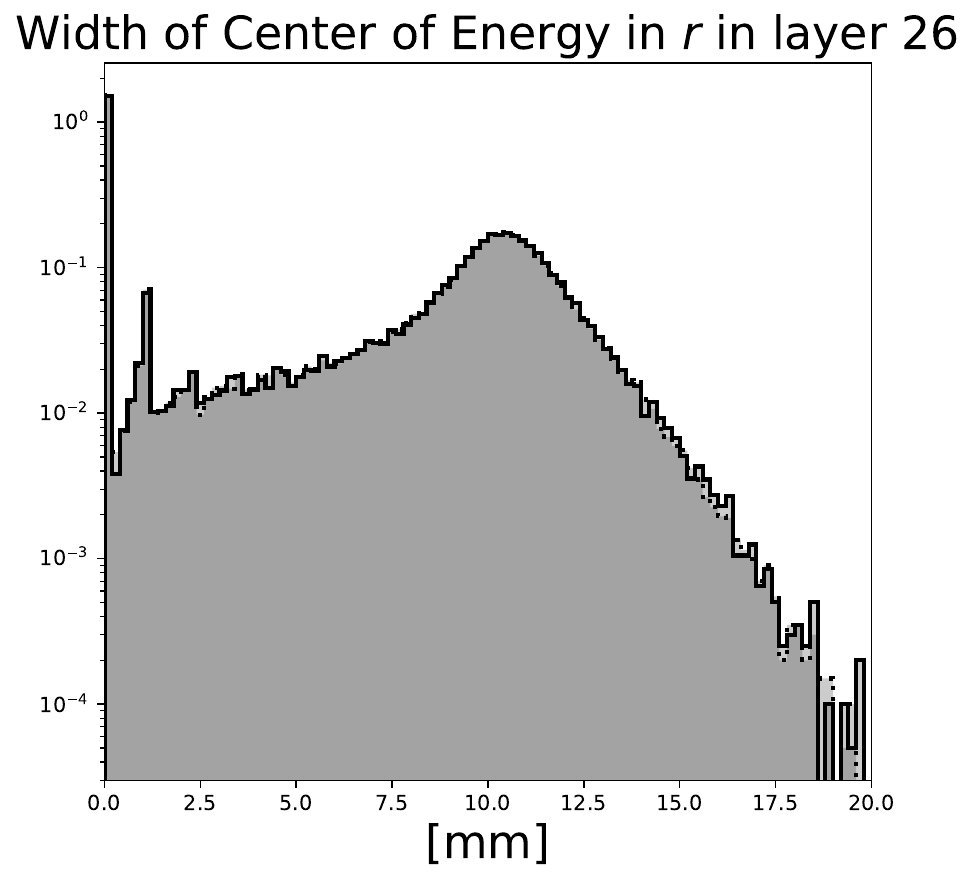} \hfill \includegraphics[height=0.1\textheight]{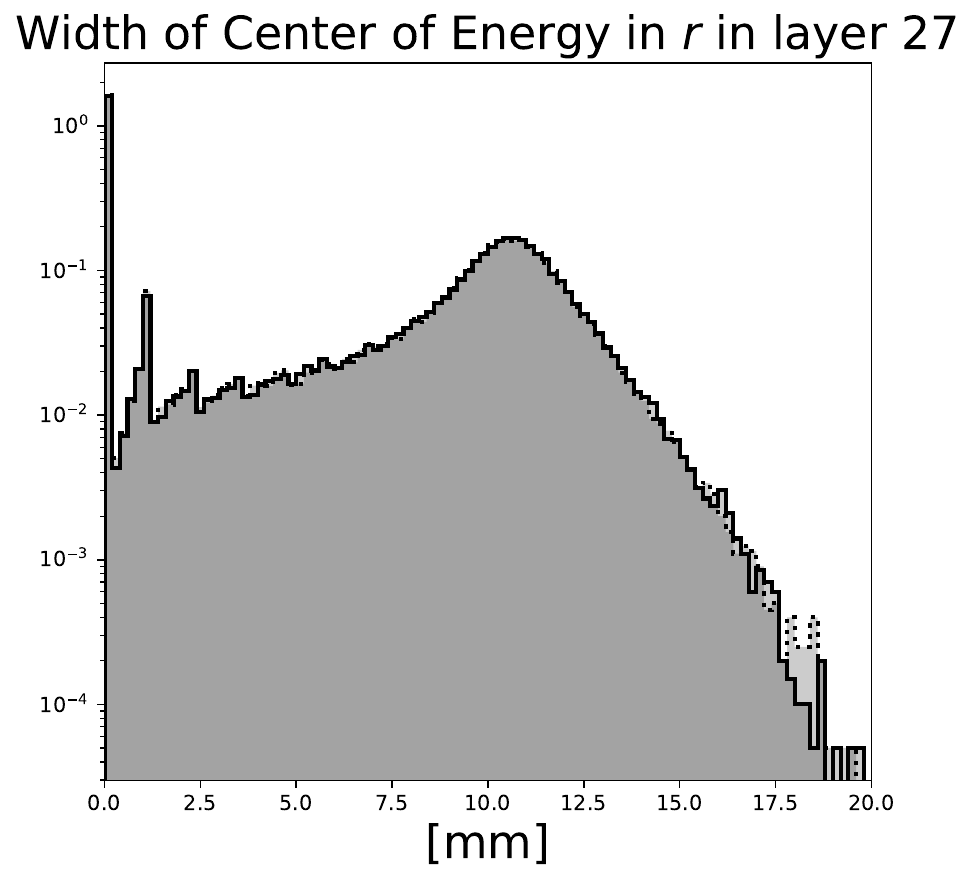} \hfill \includegraphics[height=0.1\textheight]{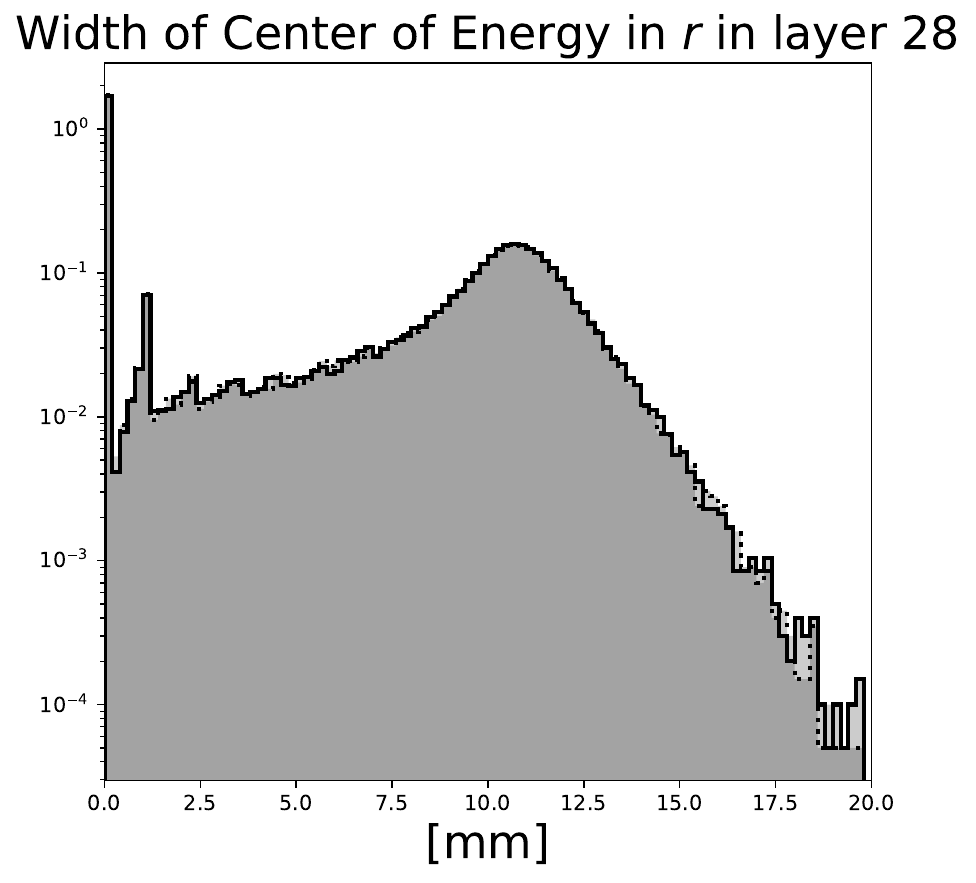} \hfill \includegraphics[height=0.1\textheight]{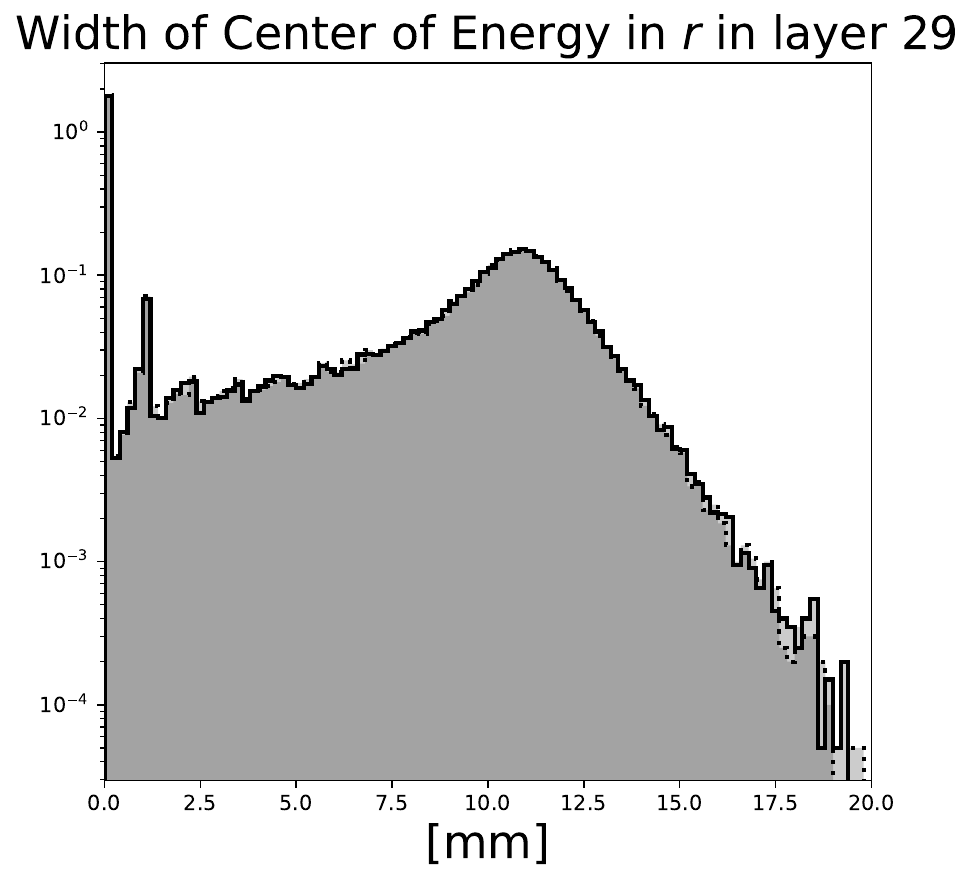}\\
    \includegraphics[height=0.1\textheight]{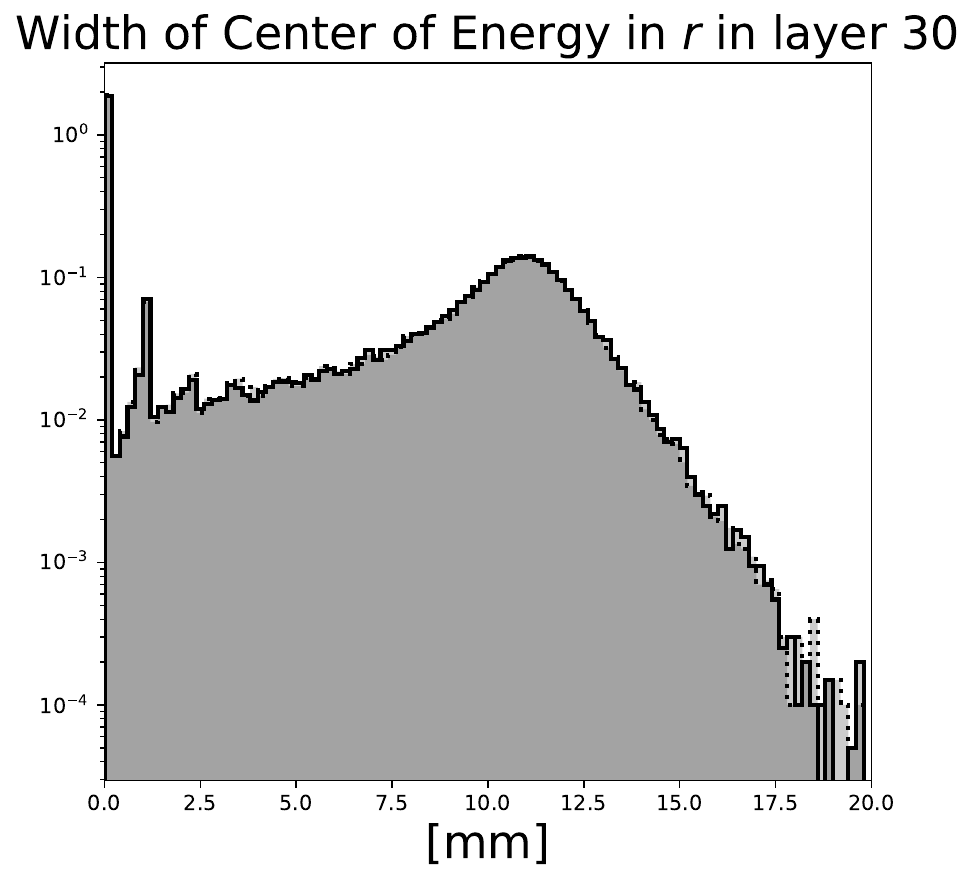} \hfill \includegraphics[height=0.1\textheight]{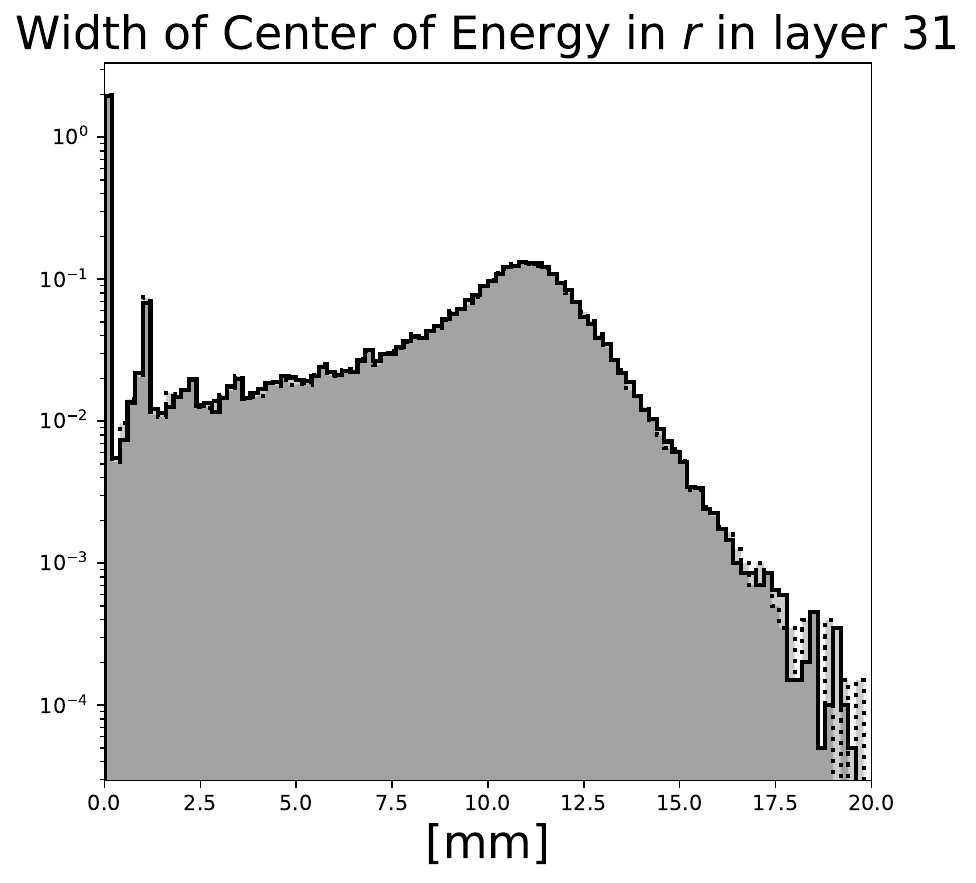} \hfill \includegraphics[height=0.1\textheight]{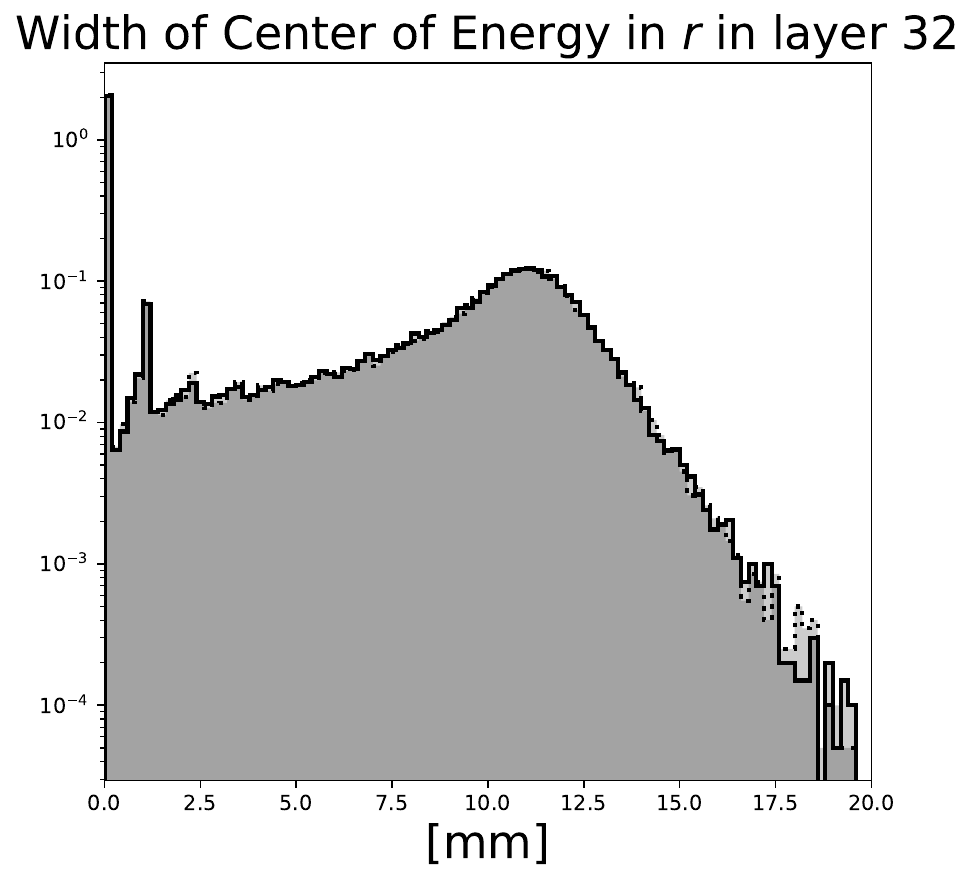} \hfill \includegraphics[height=0.1\textheight]{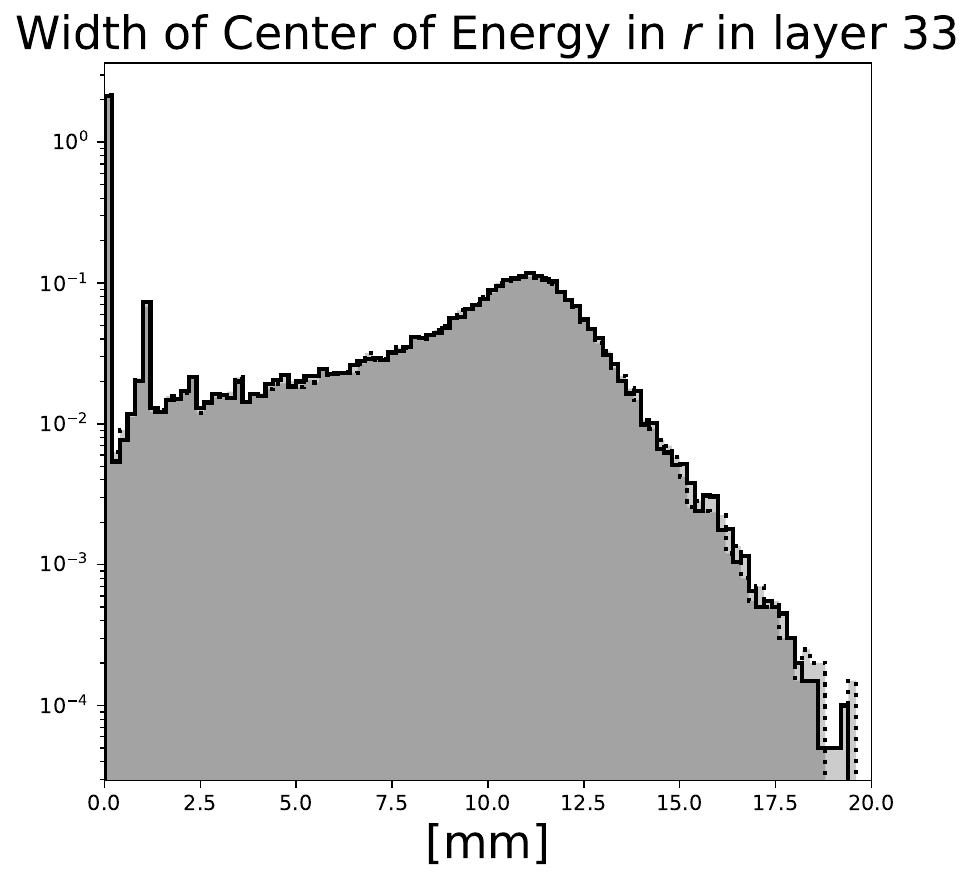} \hfill \includegraphics[height=0.1\textheight]{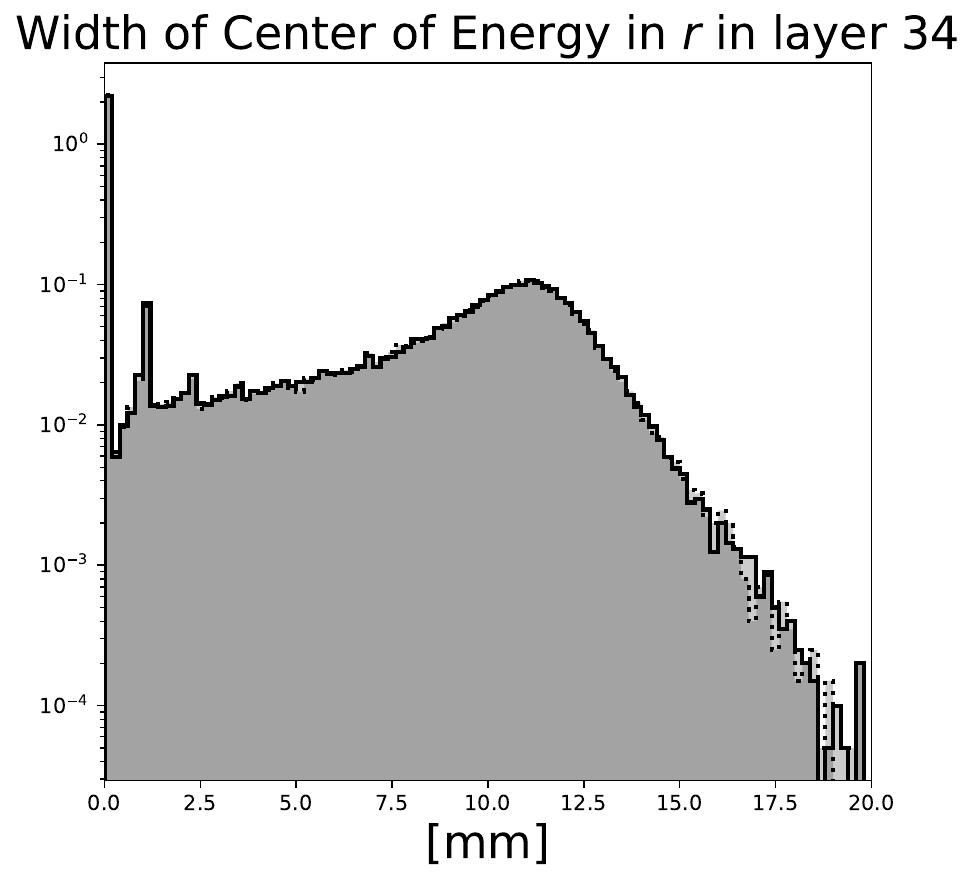}\\
    \includegraphics[height=0.1\textheight]{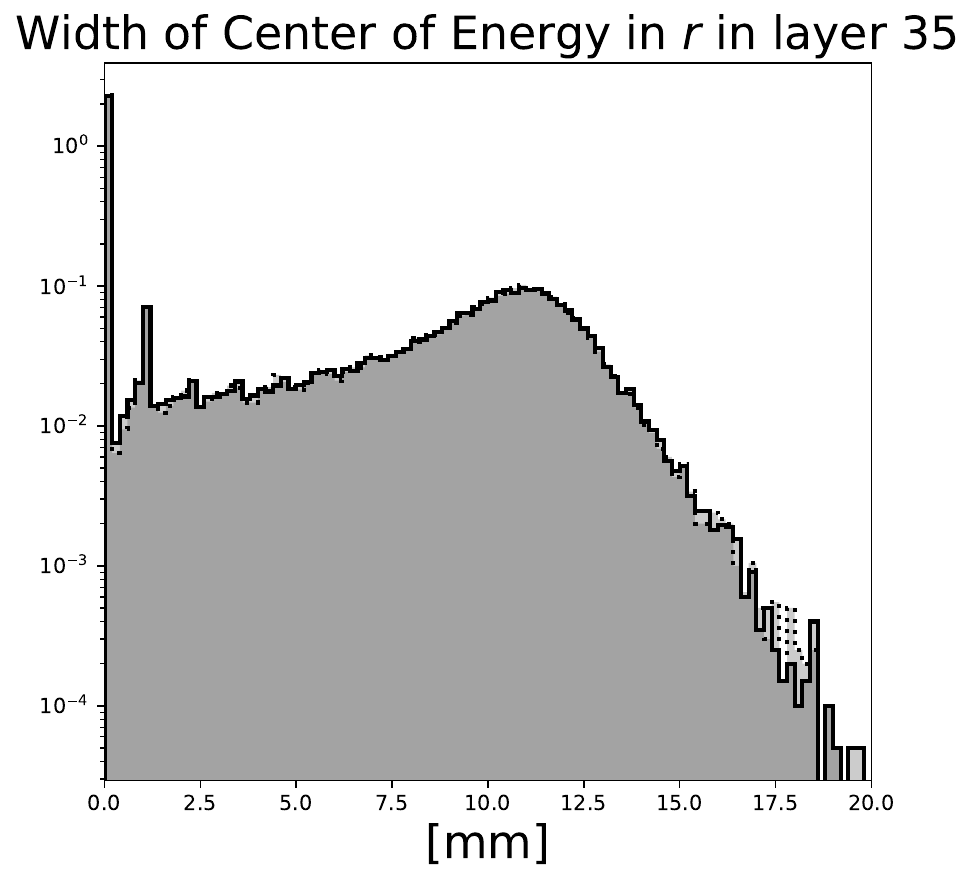} \hfill \includegraphics[height=0.1\textheight]{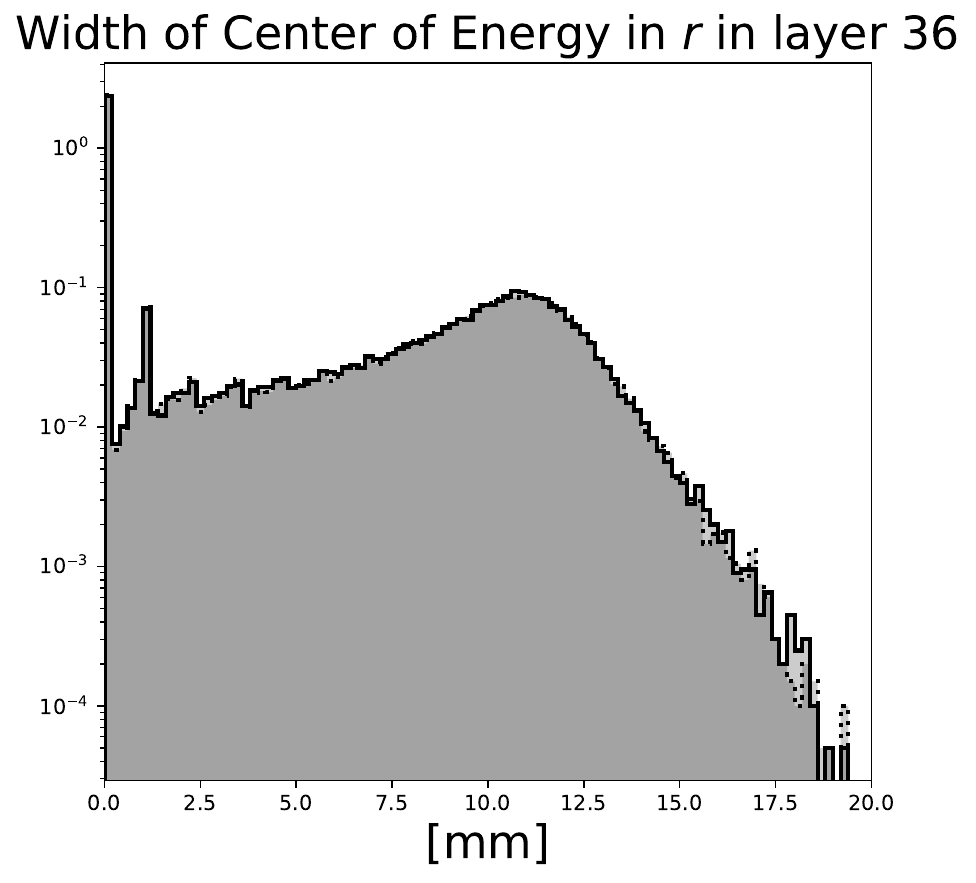} \hfill \includegraphics[height=0.1\textheight]{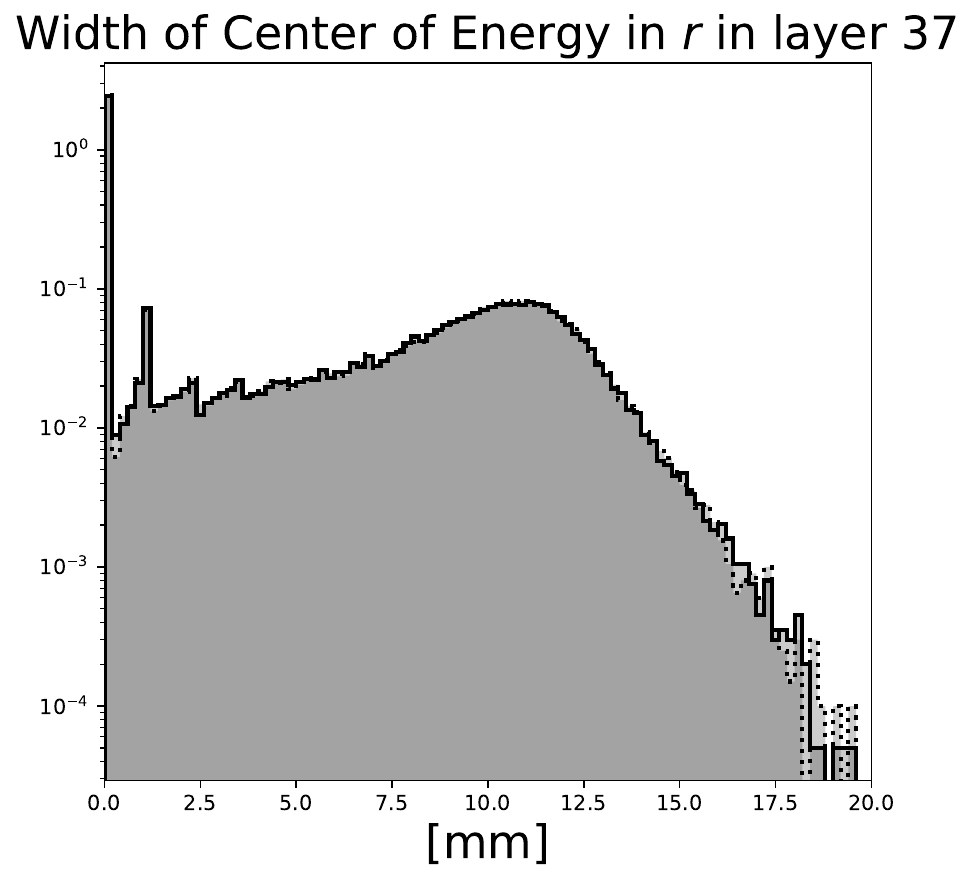} \hfill \includegraphics[height=0.1\textheight]{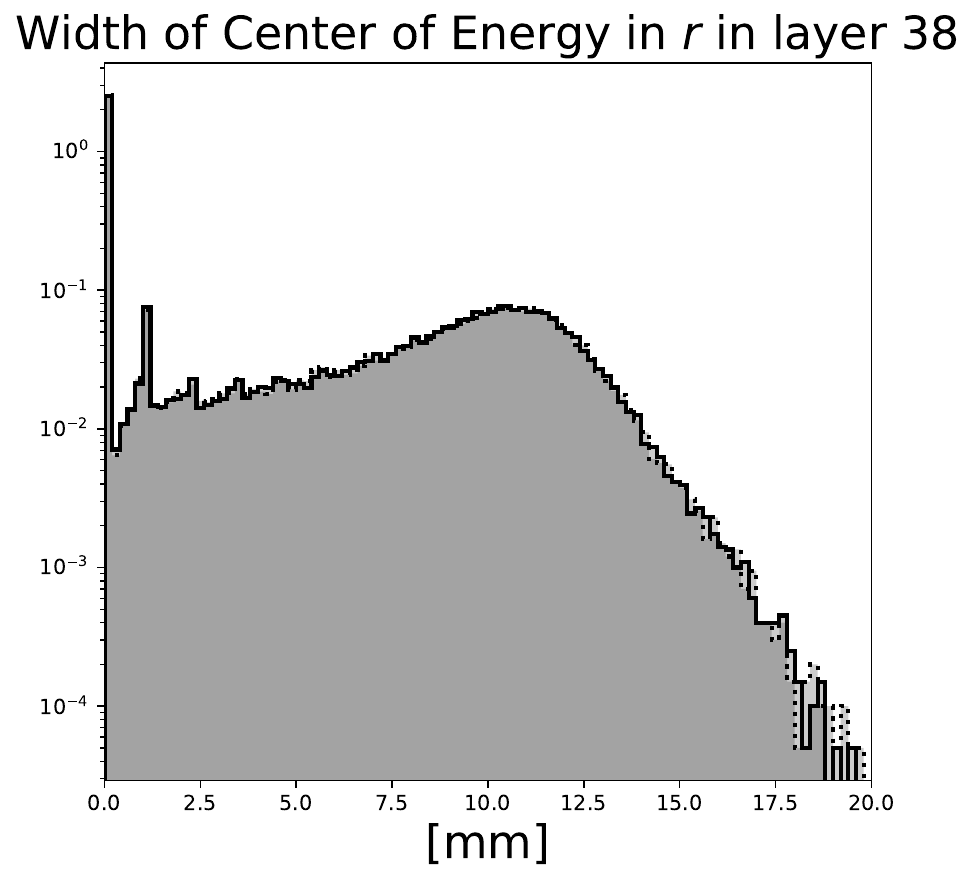} \hfill \includegraphics[height=0.1\textheight]{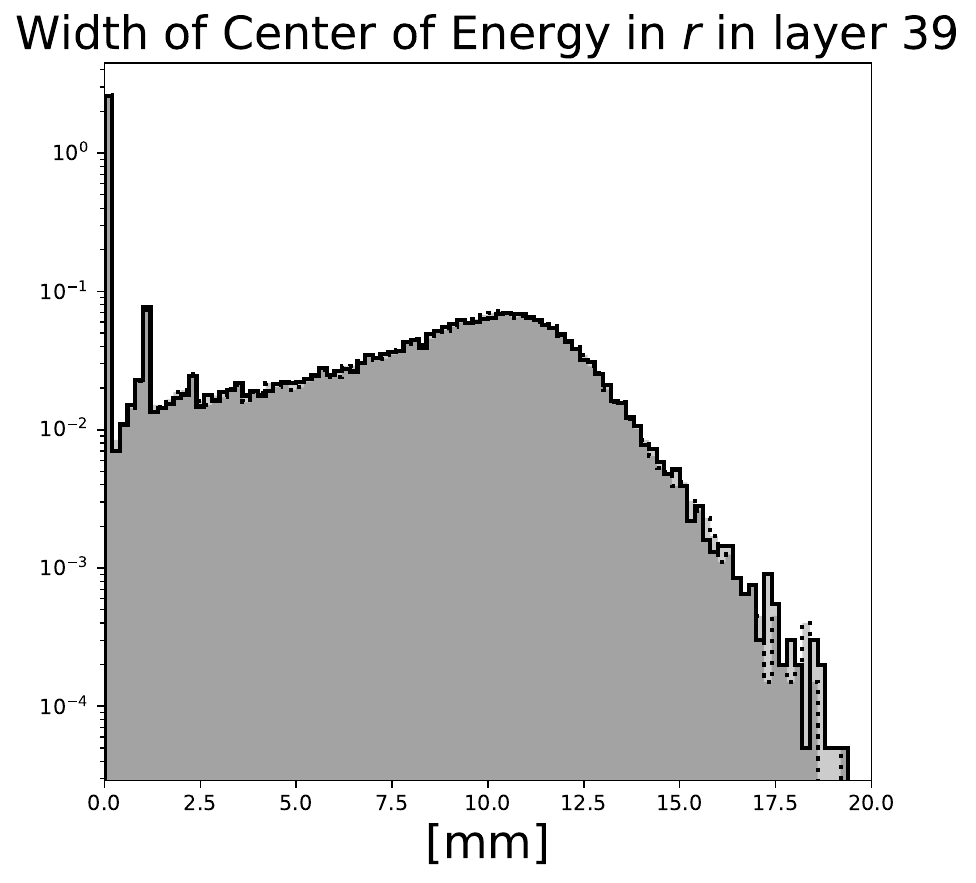}\\
    \includegraphics[height=0.1\textheight]{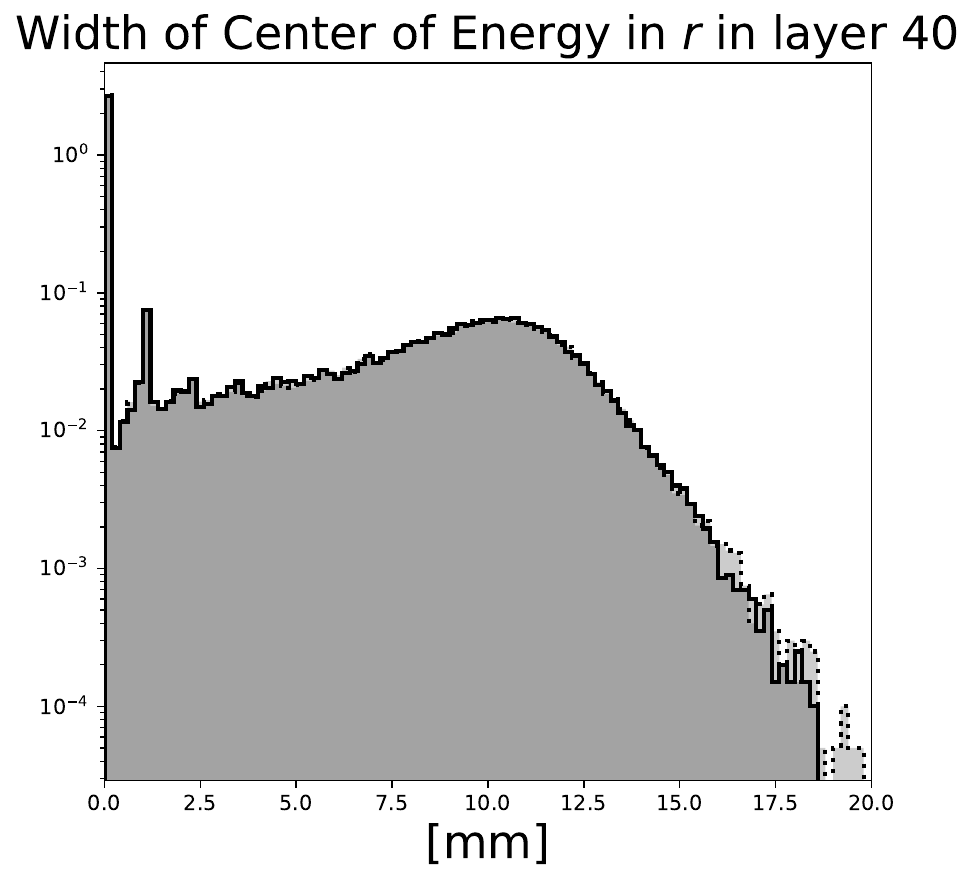} \hfill \includegraphics[height=0.1\textheight]{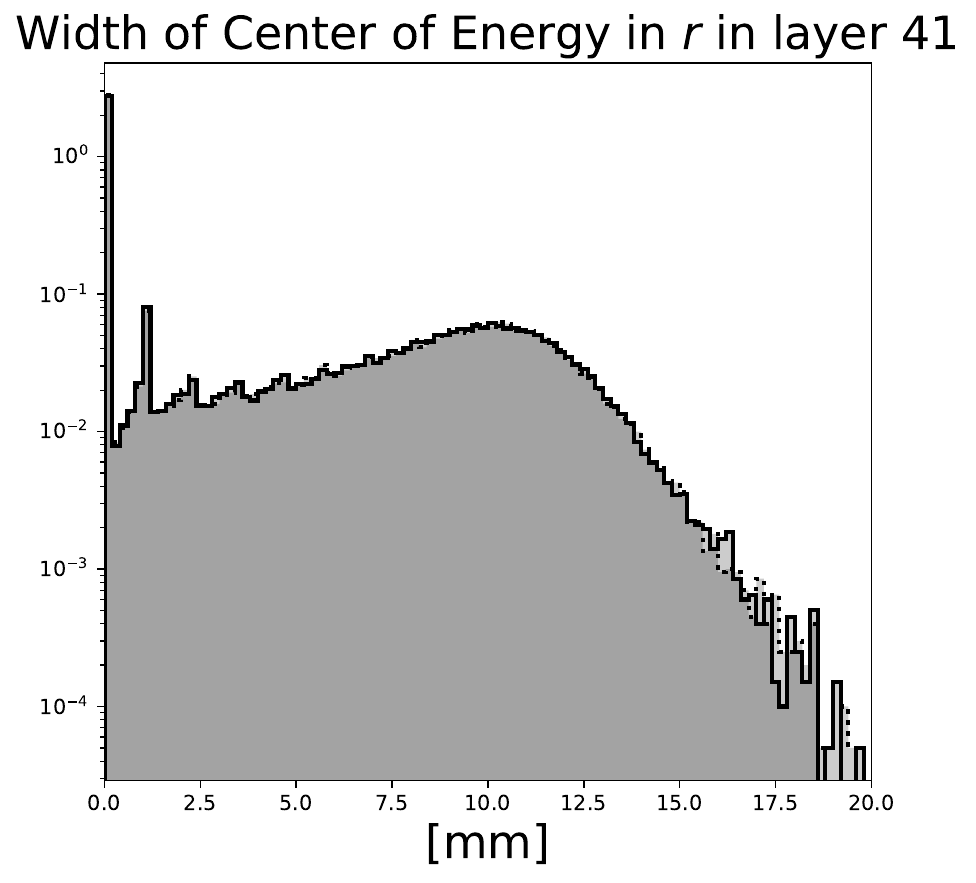} \hfill \includegraphics[height=0.1\textheight]{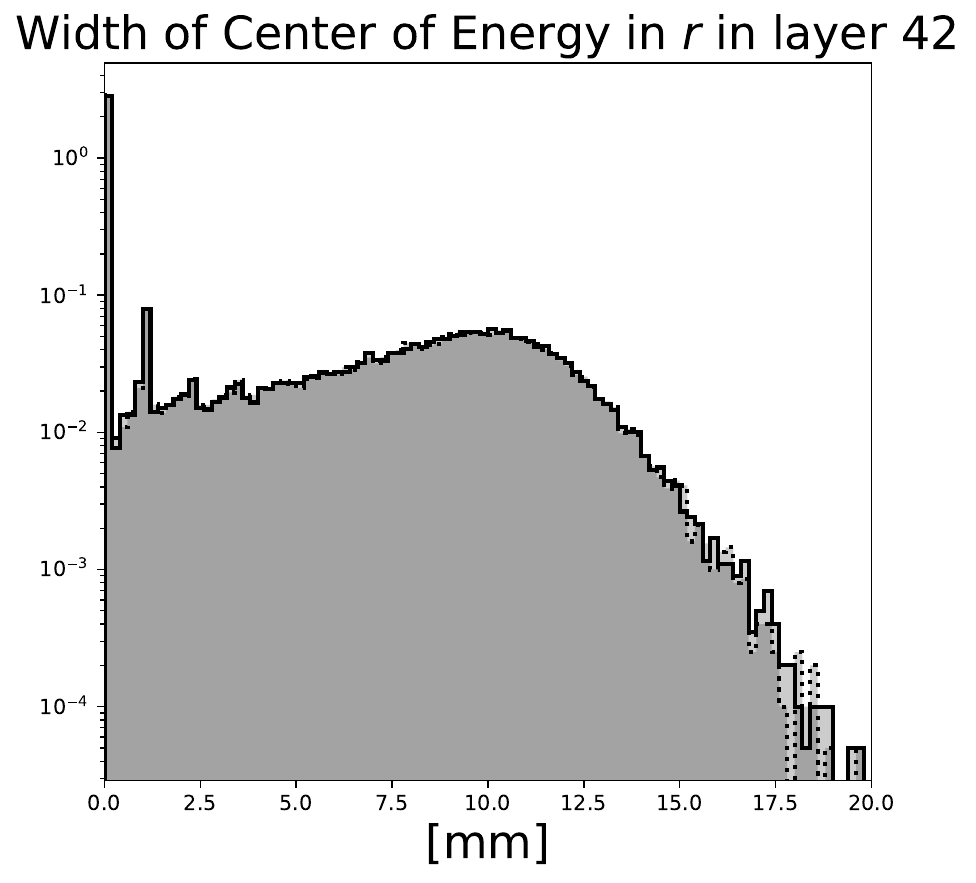} \hfill \includegraphics[height=0.1\textheight]{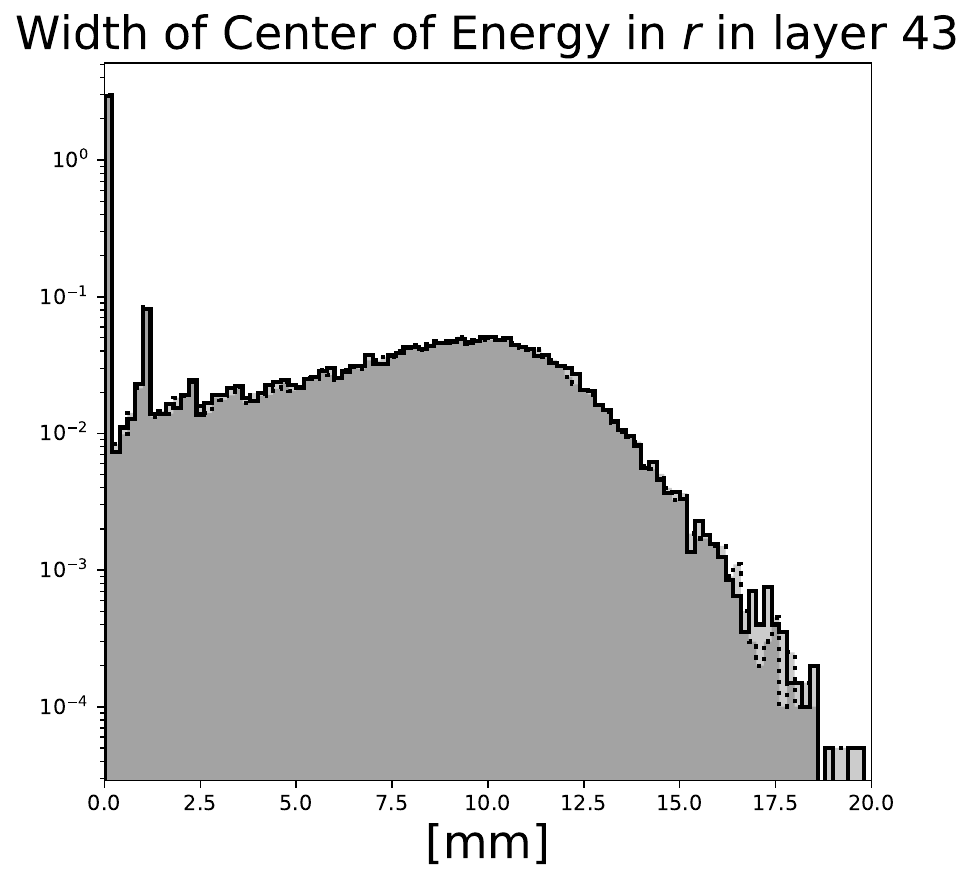} \hfill \includegraphics[height=0.1\textheight]{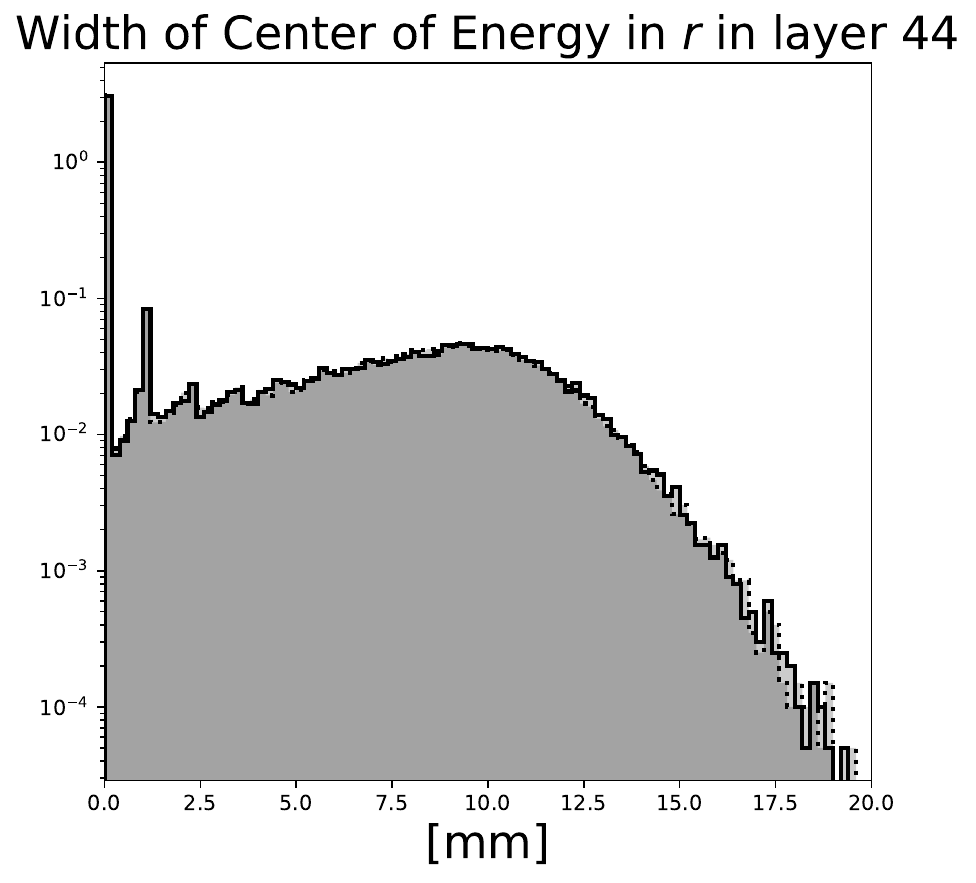}\\
    \includegraphics[width=0.5\textwidth]{figures/Appendix_reference/legend.pdf}
    \caption{Distribution of \geant training and evaluation data in width of the centers of energy in $r$ direction for ds3. }
    \label{fig:app_ref.ds3.8}
\end{figure}

\begin{figure}[ht]
    \centering
    \includegraphics[height=0.1\textheight]{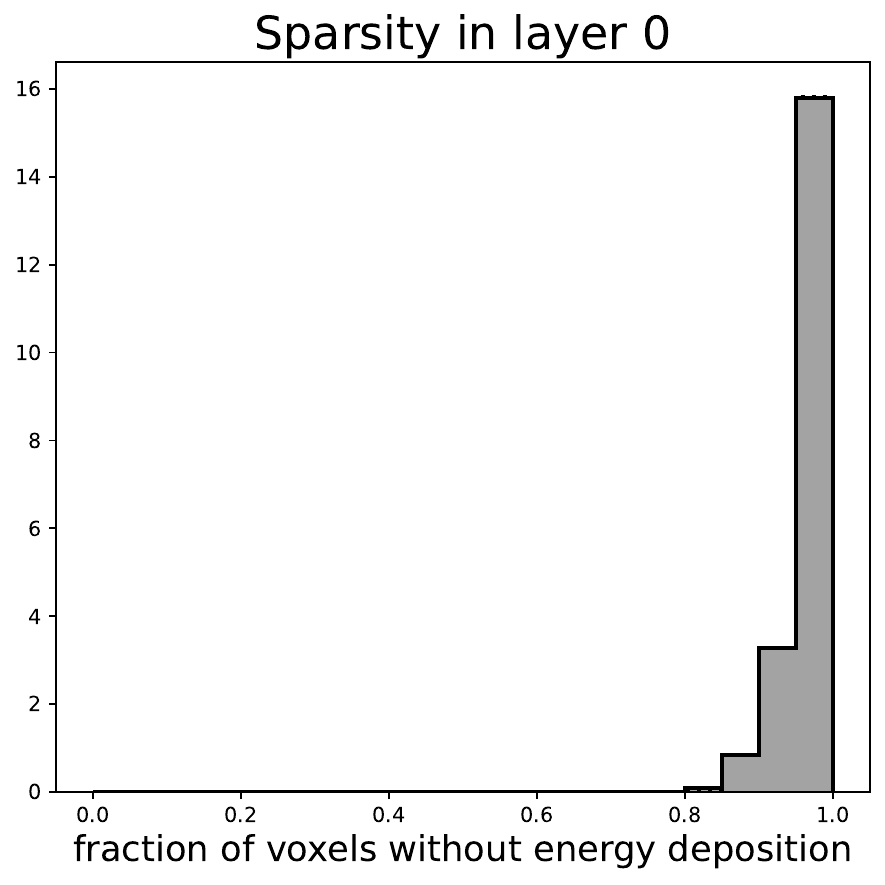} \hfill \includegraphics[height=0.1\textheight]{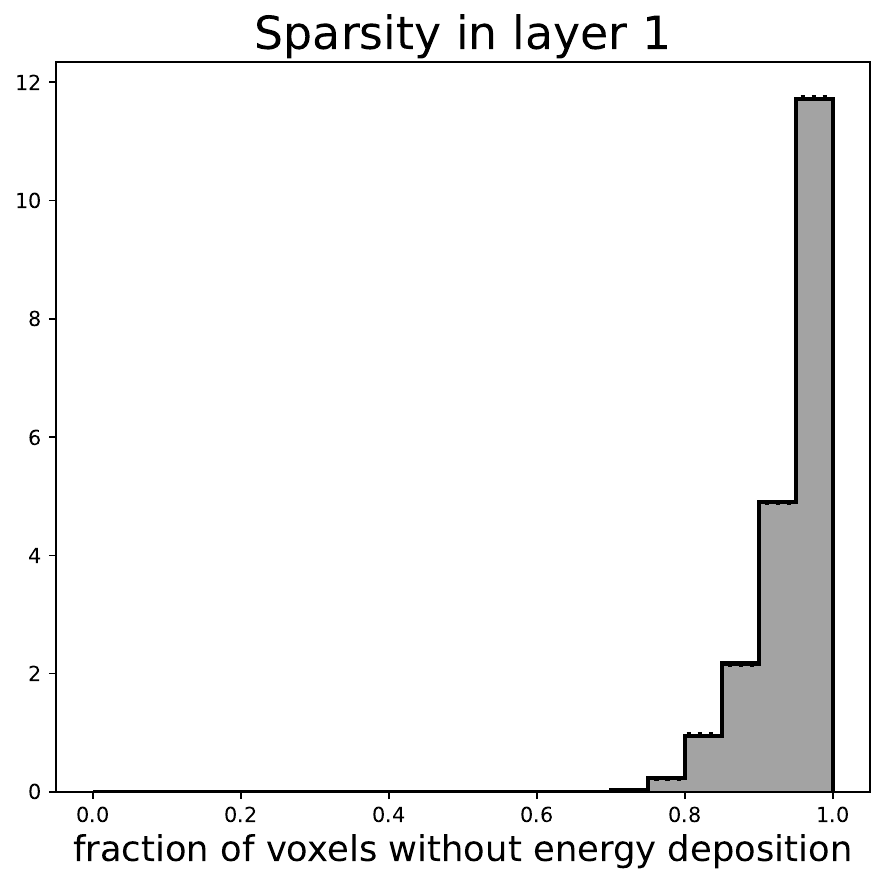} \hfill \includegraphics[height=0.1\textheight]{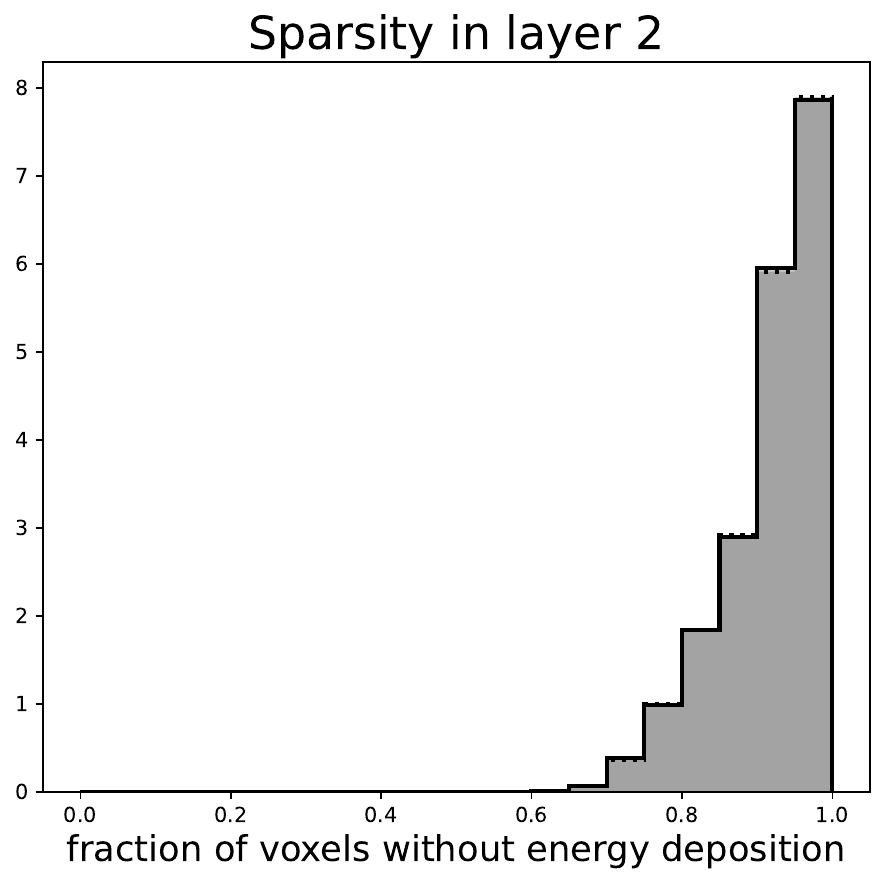} \hfill \includegraphics[height=0.1\textheight]{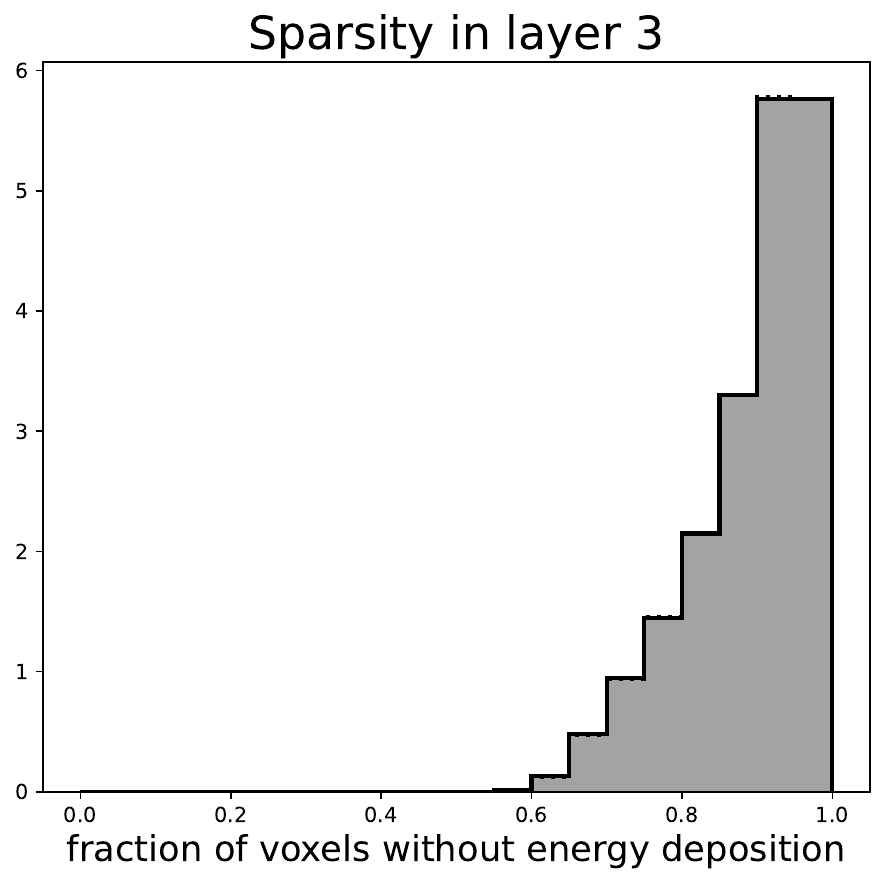} \hfill \includegraphics[height=0.1\textheight]{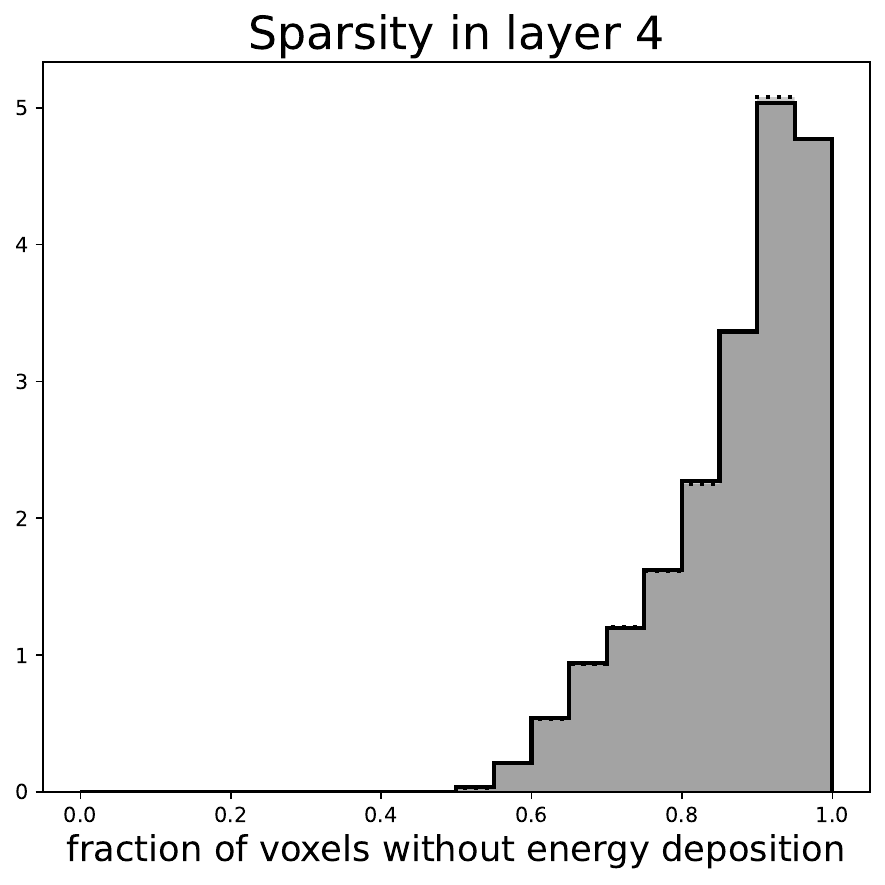}\\
    \includegraphics[height=0.1\textheight]{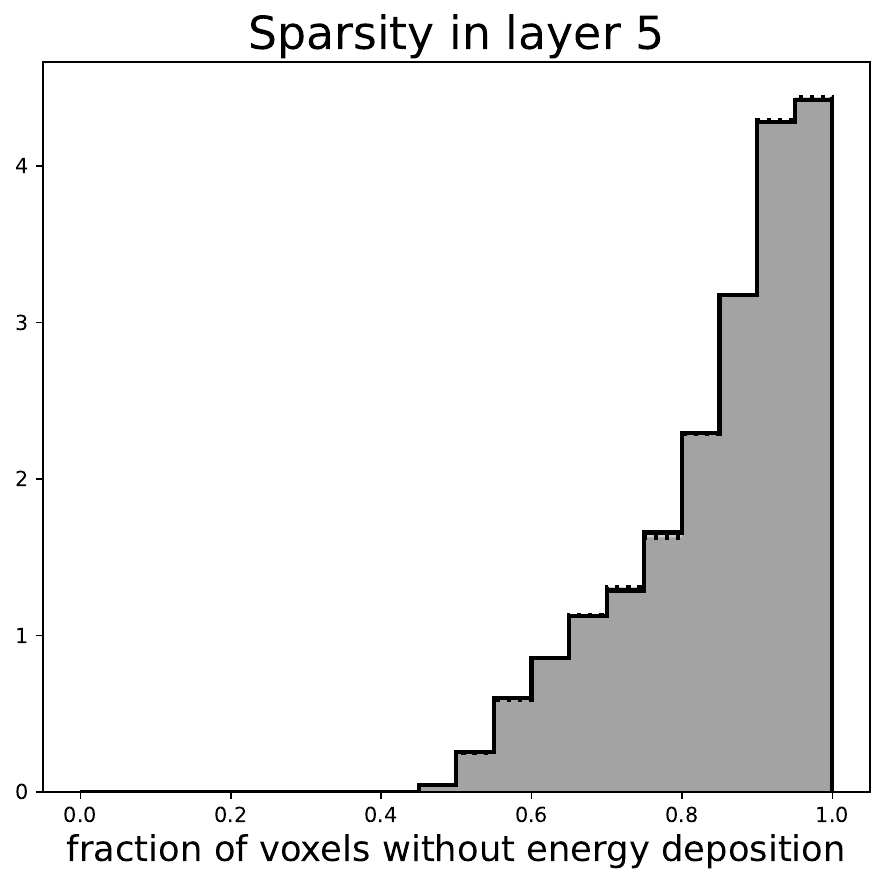} \hfill \includegraphics[height=0.1\textheight]{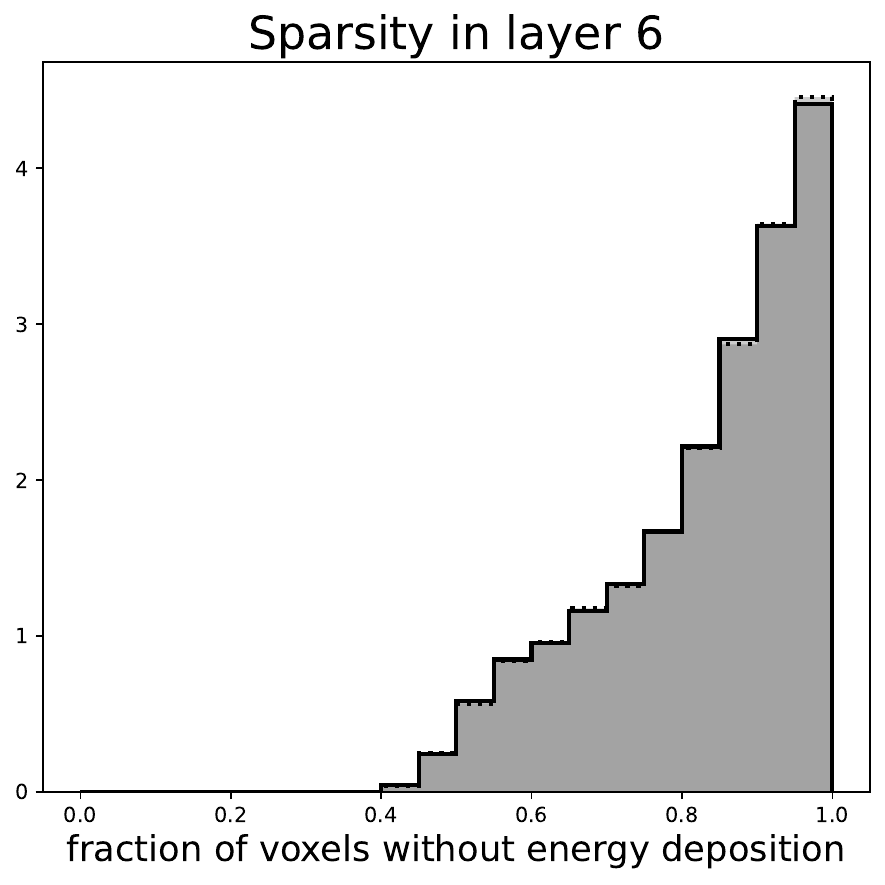} \hfill \includegraphics[height=0.1\textheight]{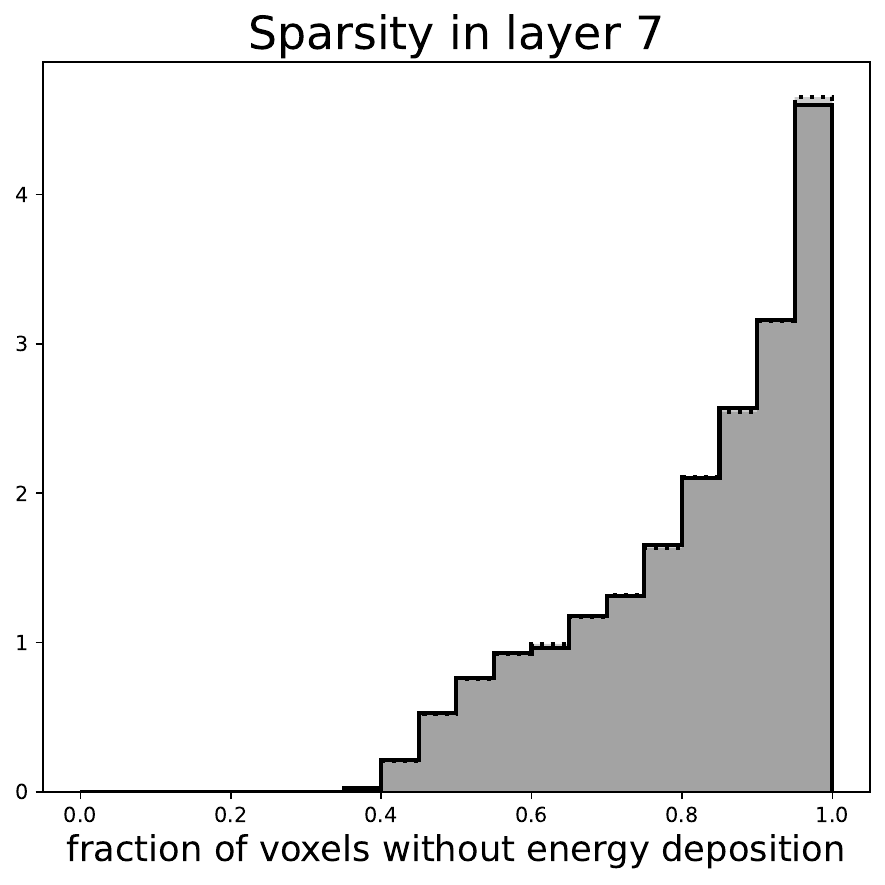} \hfill \includegraphics[height=0.1\textheight]{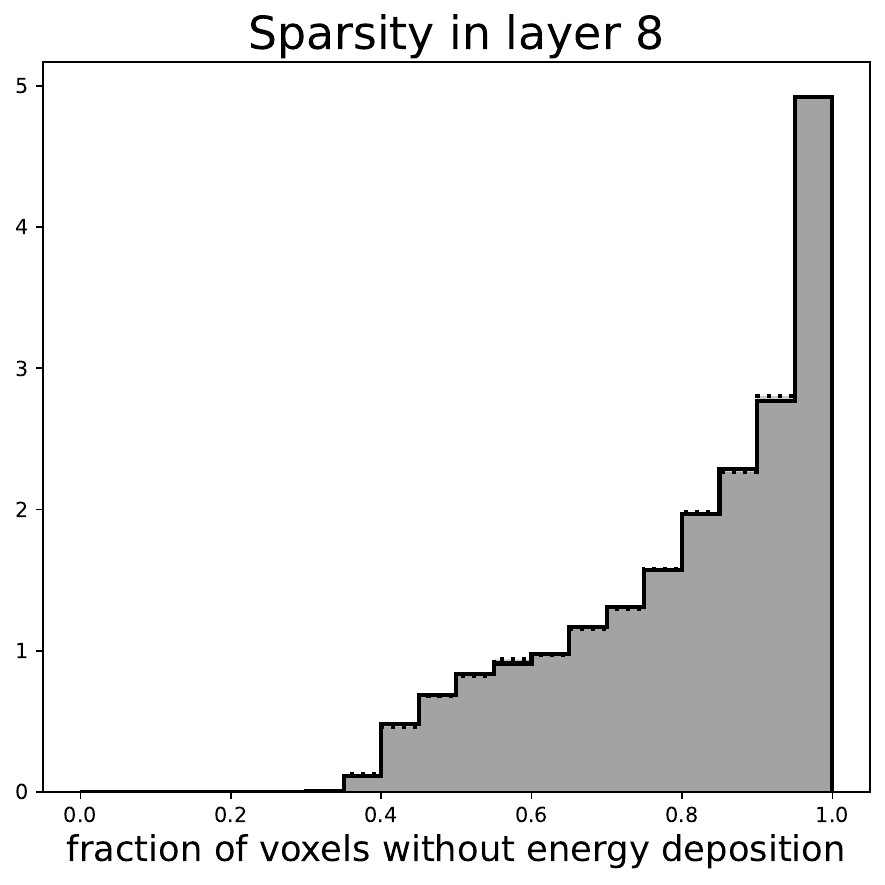} \hfill \includegraphics[height=0.1\textheight]{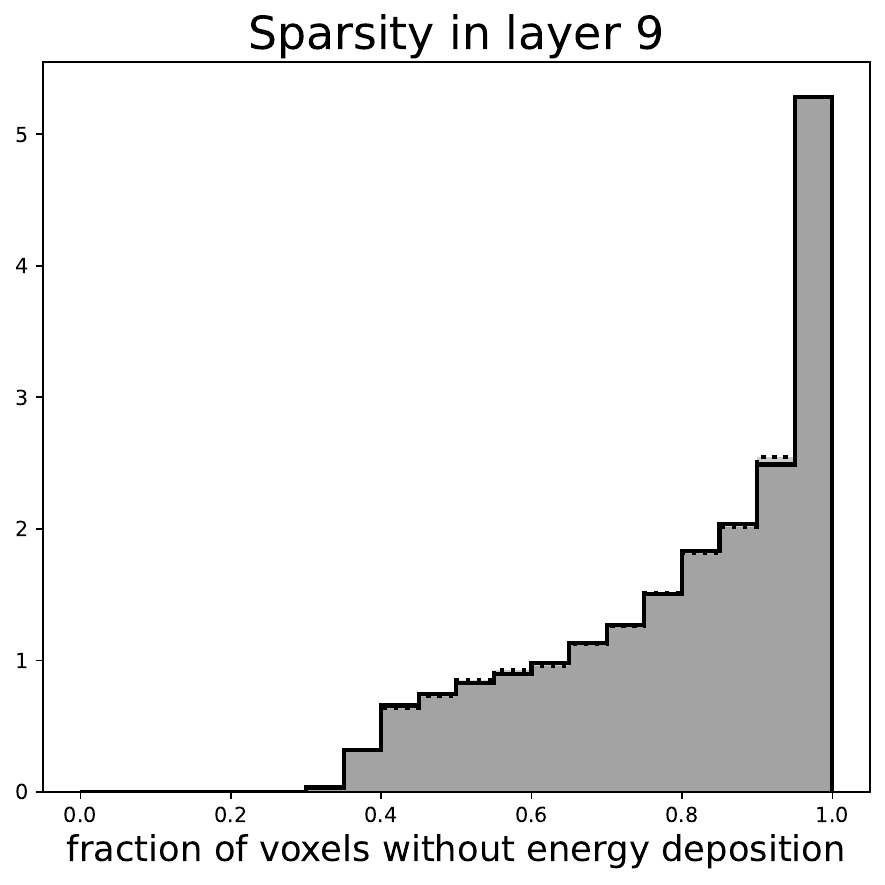}\\
    \includegraphics[height=0.1\textheight]{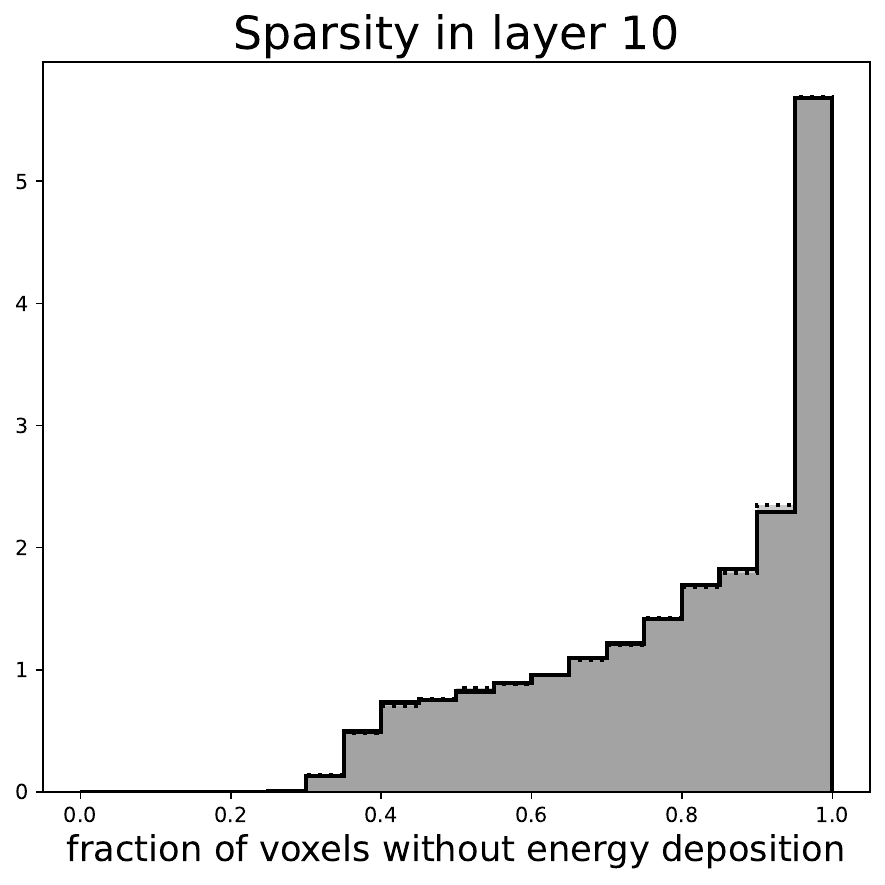} \hfill \includegraphics[height=0.1\textheight]{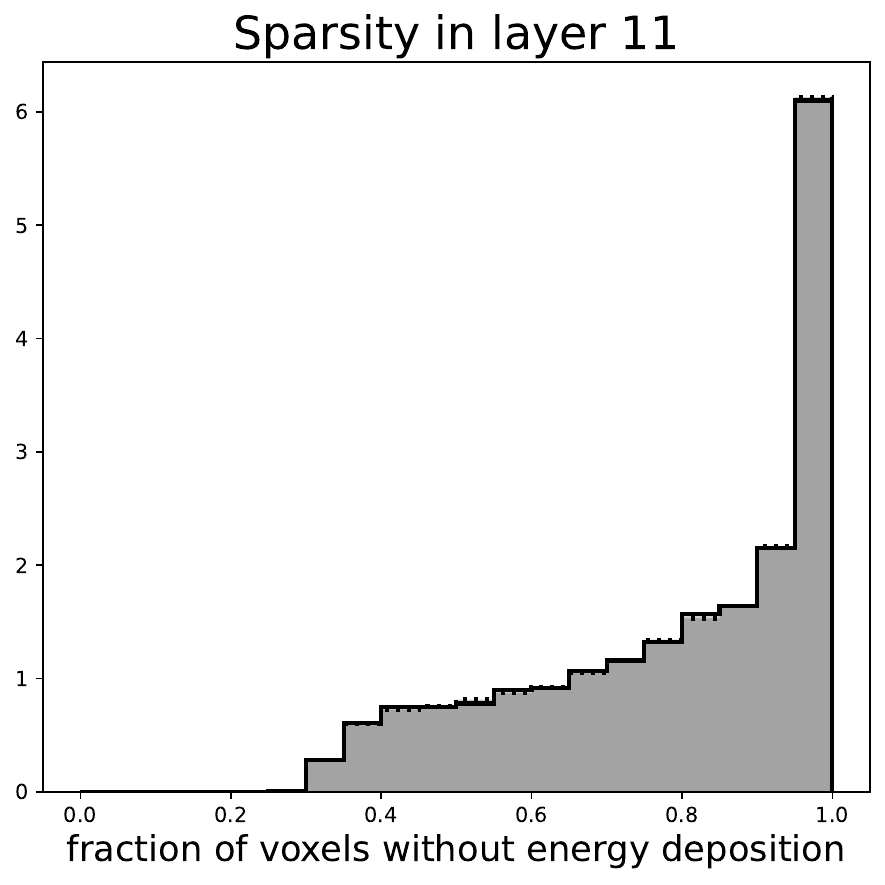} \hfill \includegraphics[height=0.1\textheight]{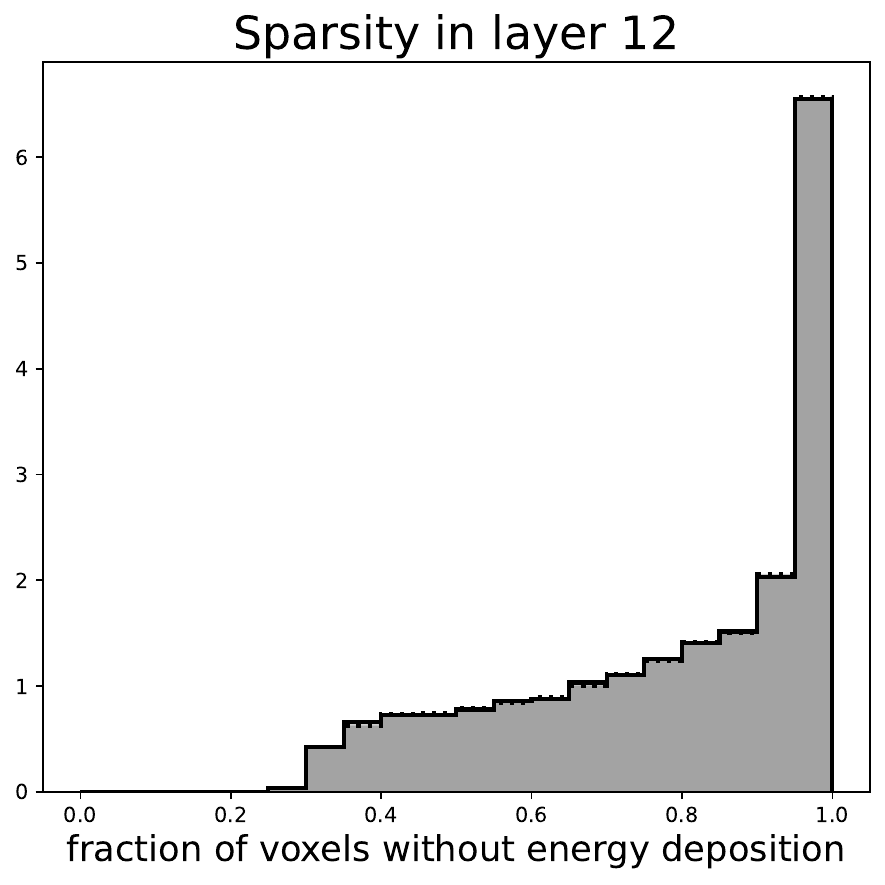} \hfill \includegraphics[height=0.1\textheight]{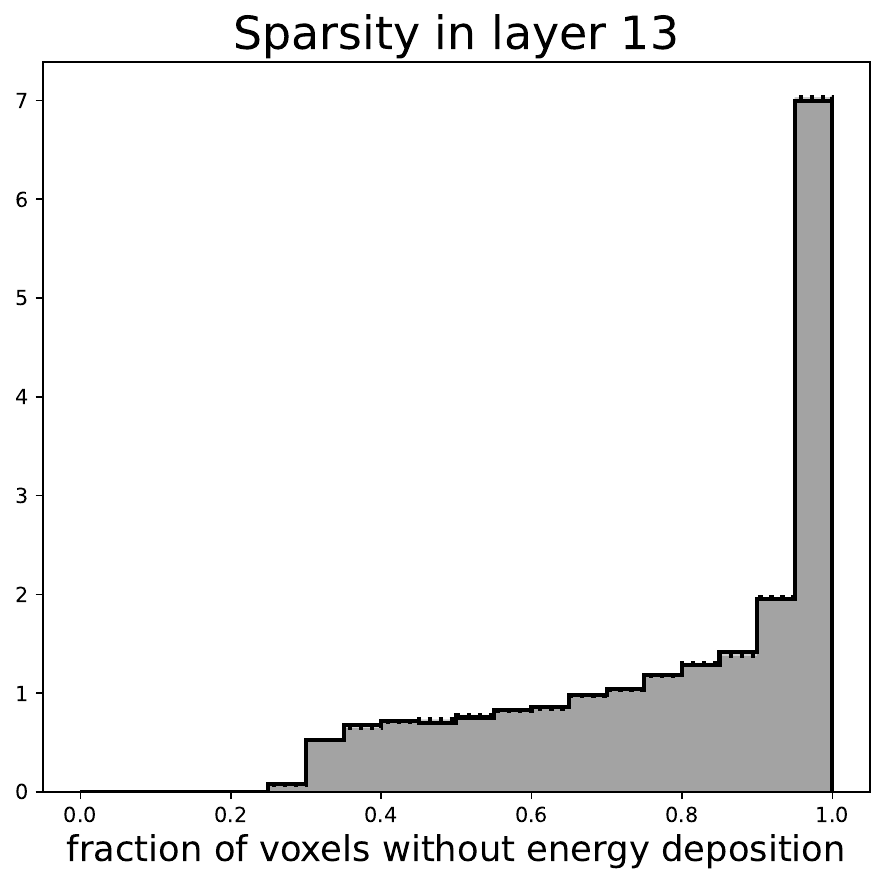} \hfill \includegraphics[height=0.1\textheight]{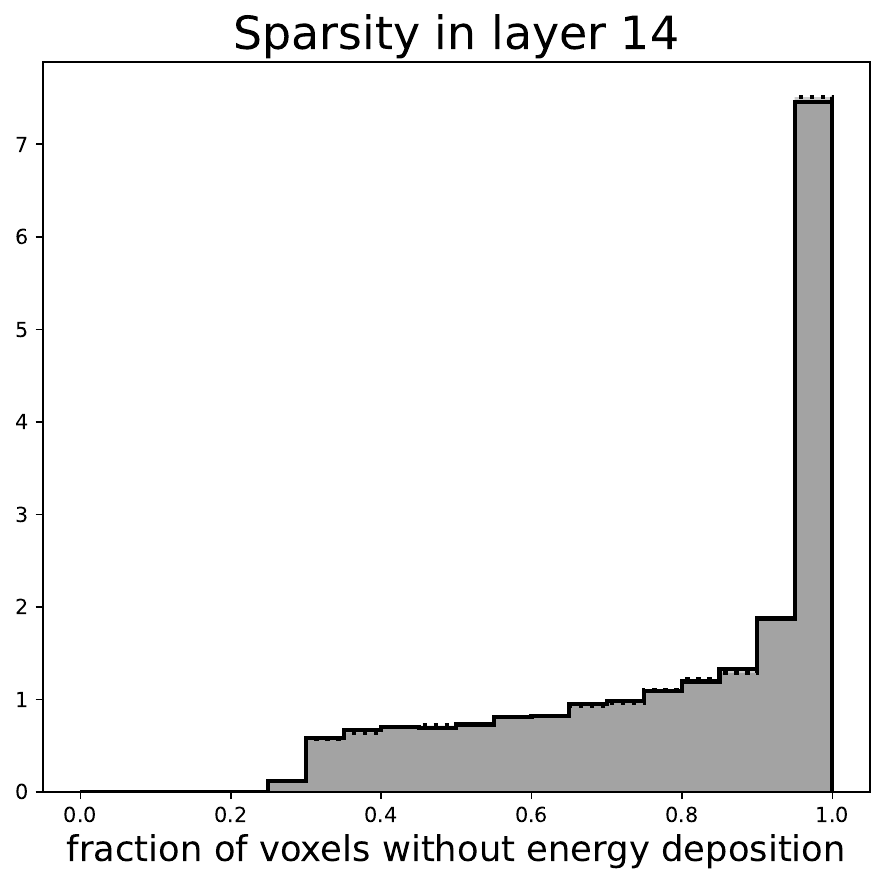}\\
    \includegraphics[height=0.1\textheight]{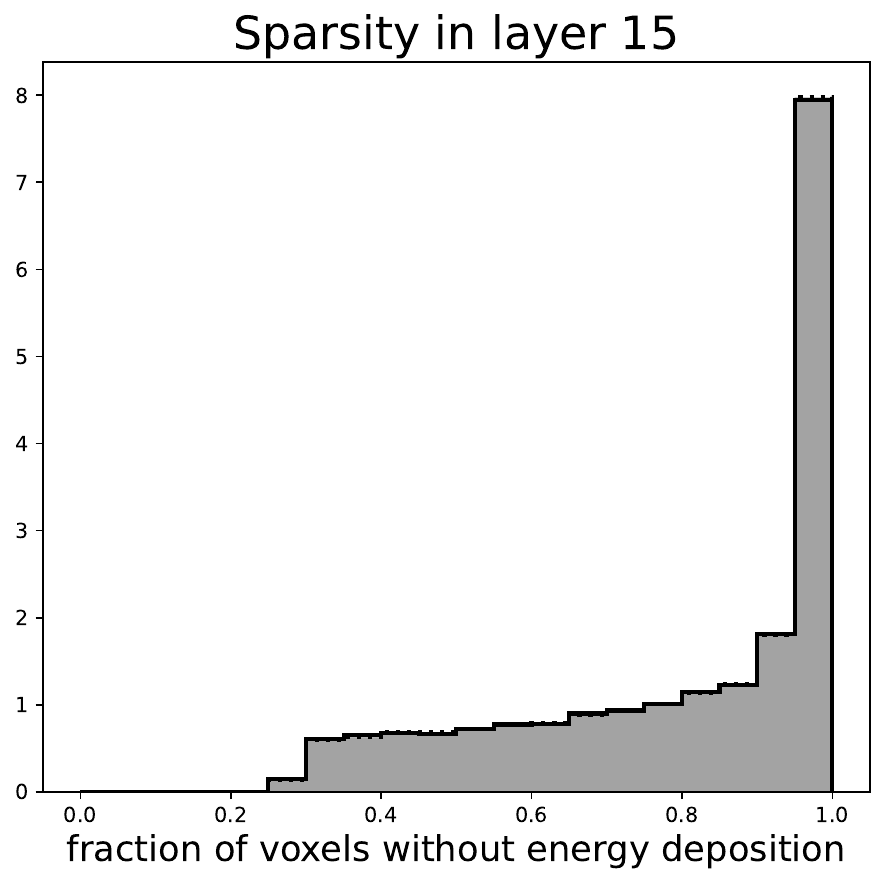} \hfill \includegraphics[height=0.1\textheight]{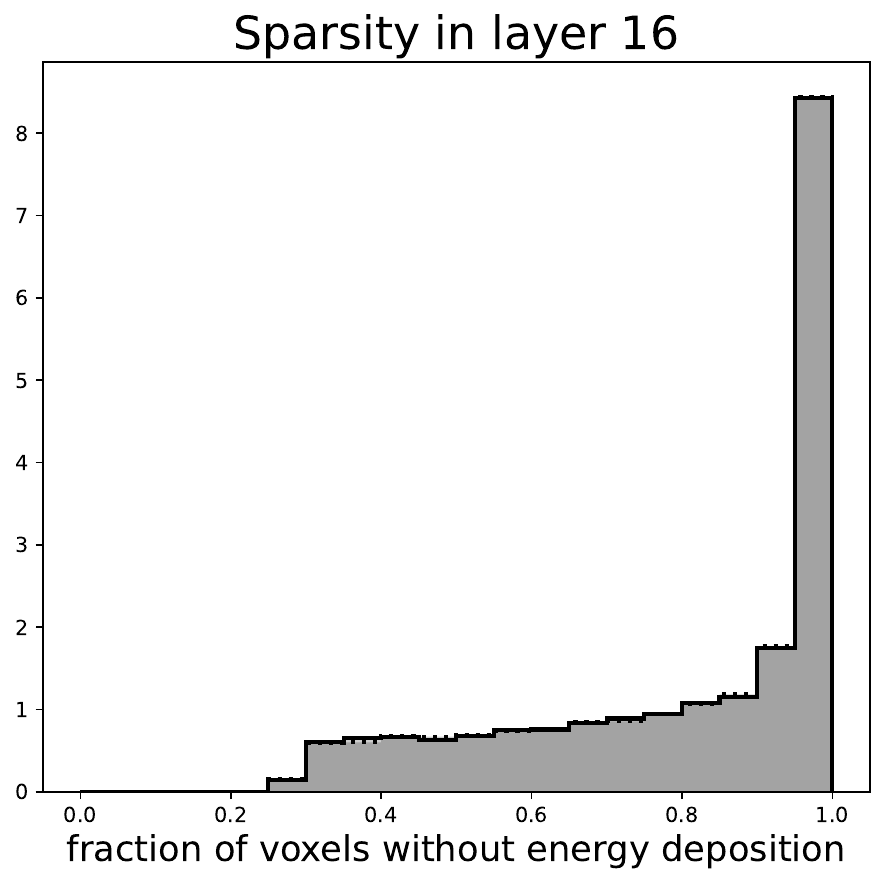} \hfill \includegraphics[height=0.1\textheight]{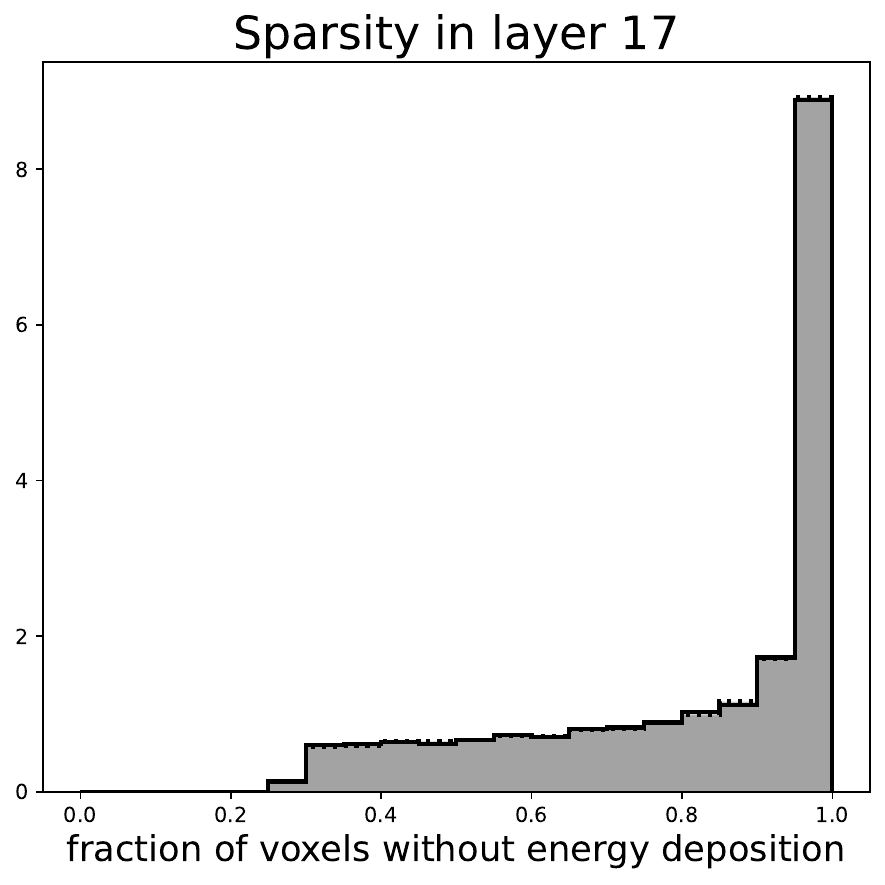} \hfill \includegraphics[height=0.1\textheight]{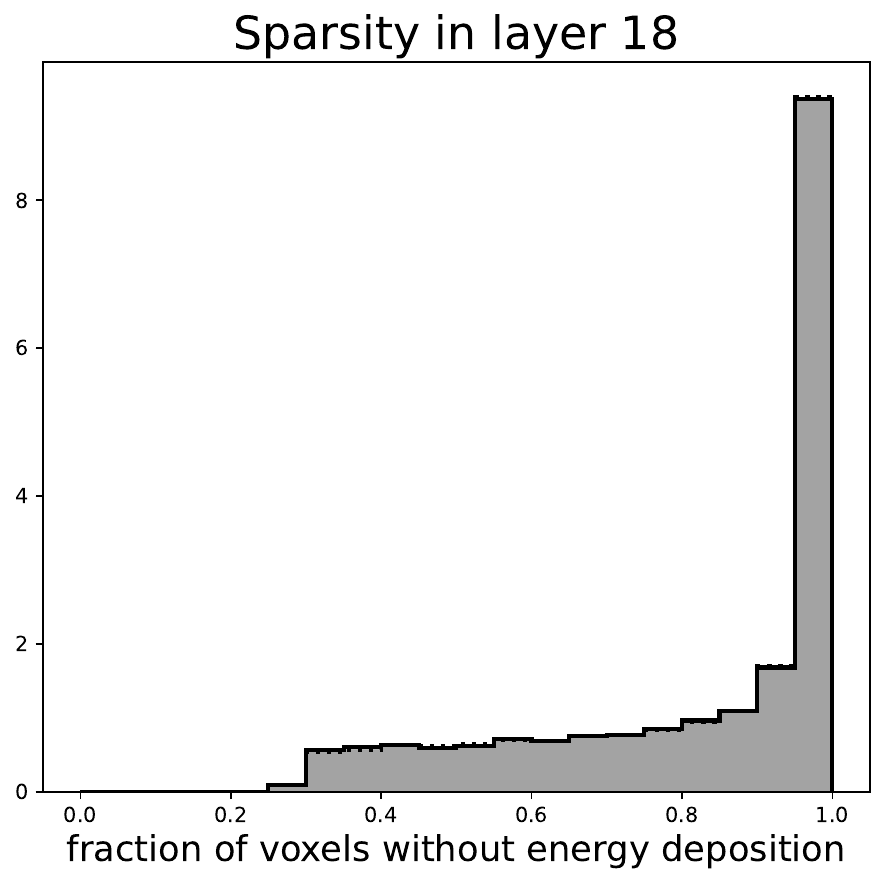} \hfill \includegraphics[height=0.1\textheight]{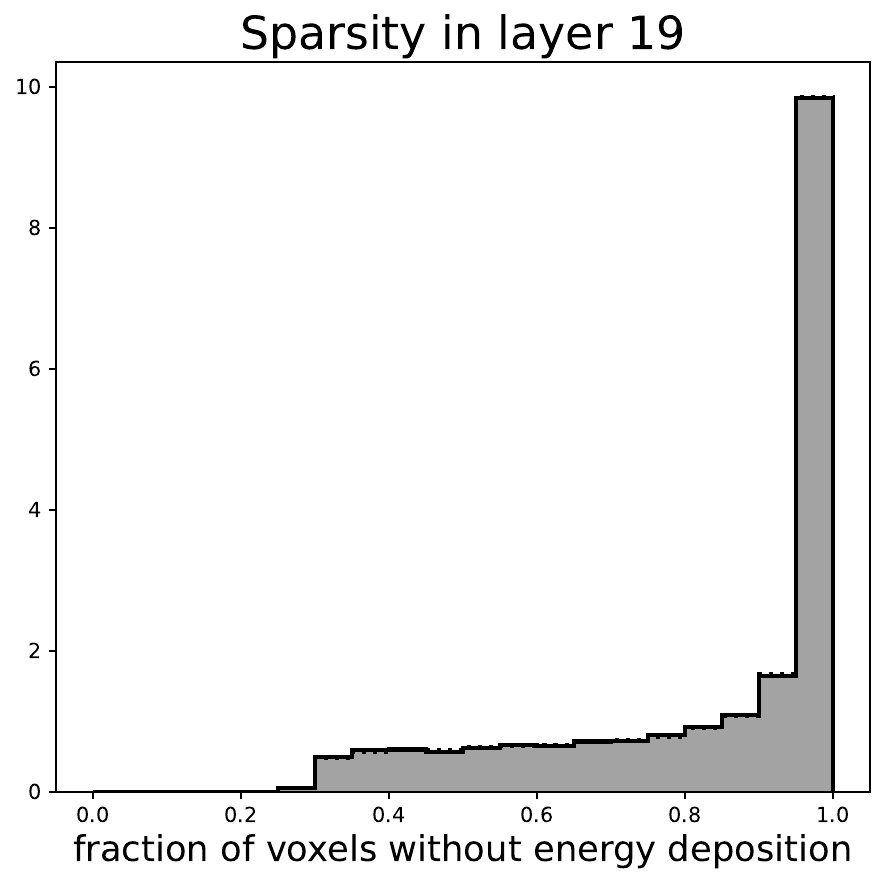}\\
    \includegraphics[height=0.1\textheight]{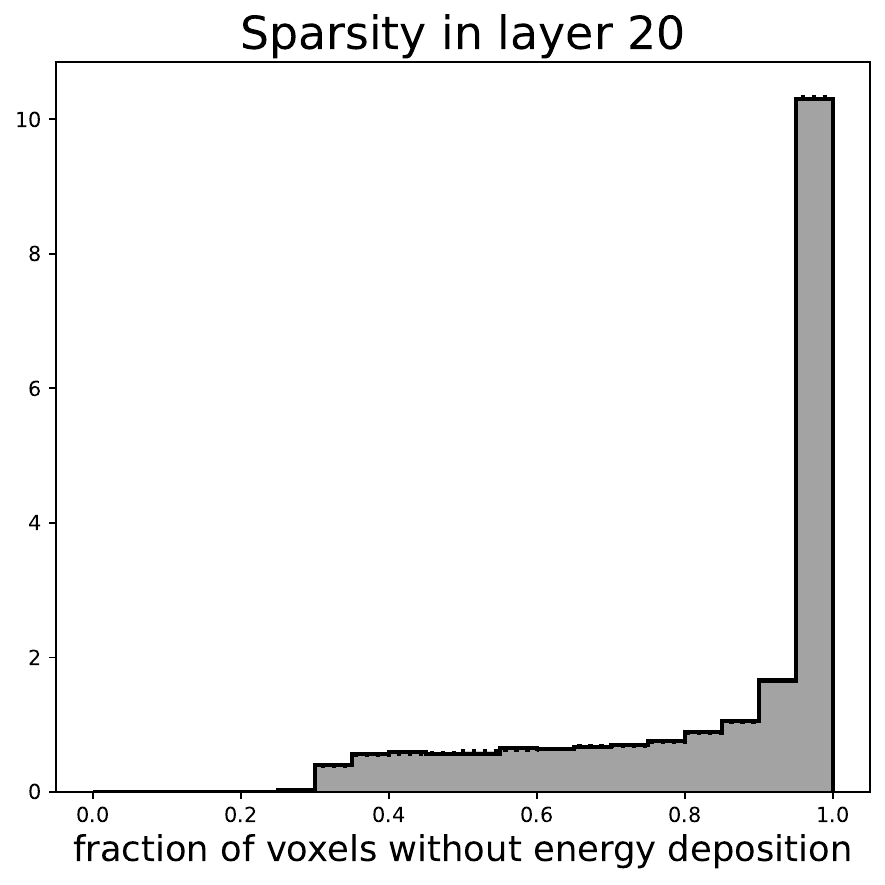} \hfill \includegraphics[height=0.1\textheight]{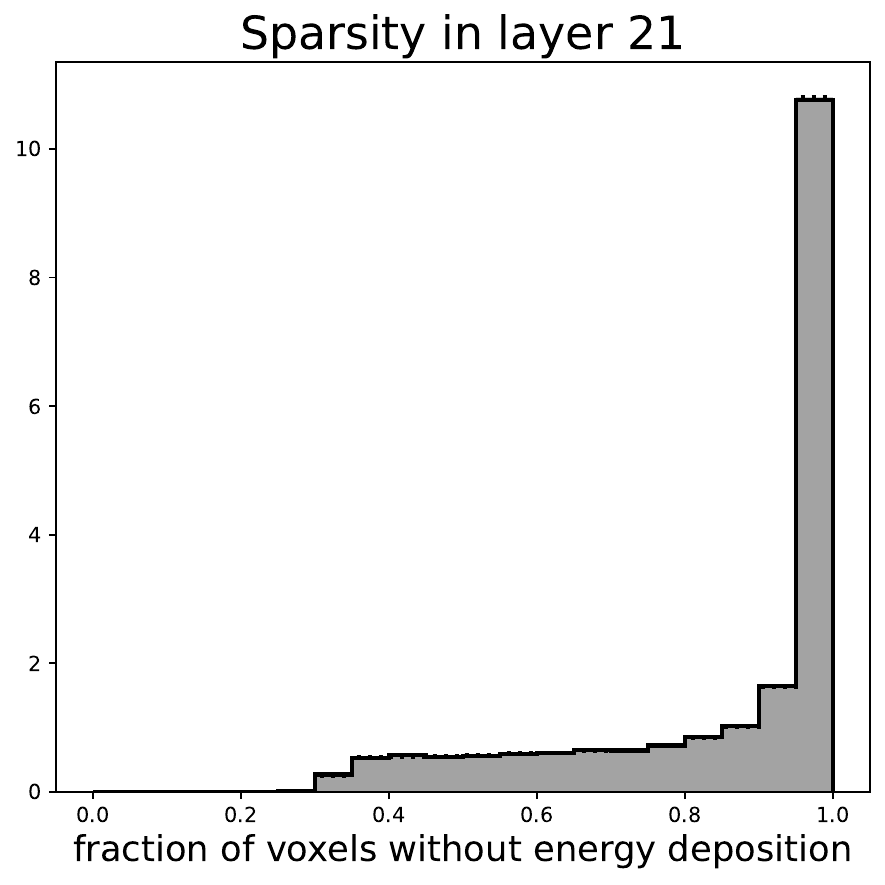} \hfill \includegraphics[height=0.1\textheight]{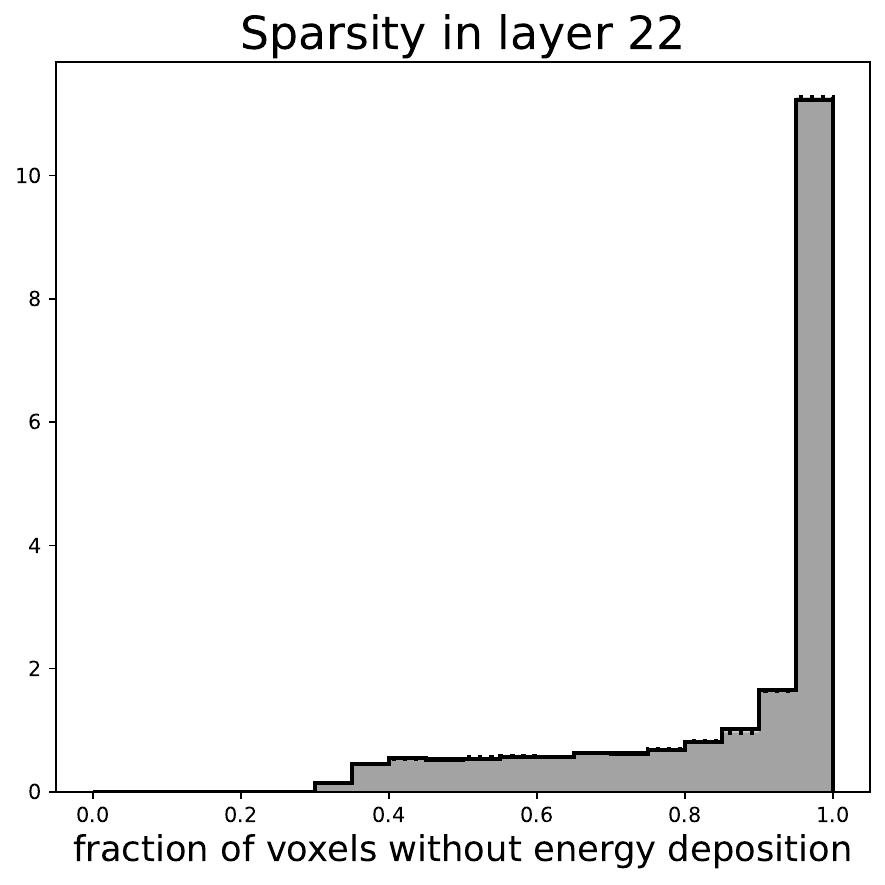} \hfill \includegraphics[height=0.1\textheight]{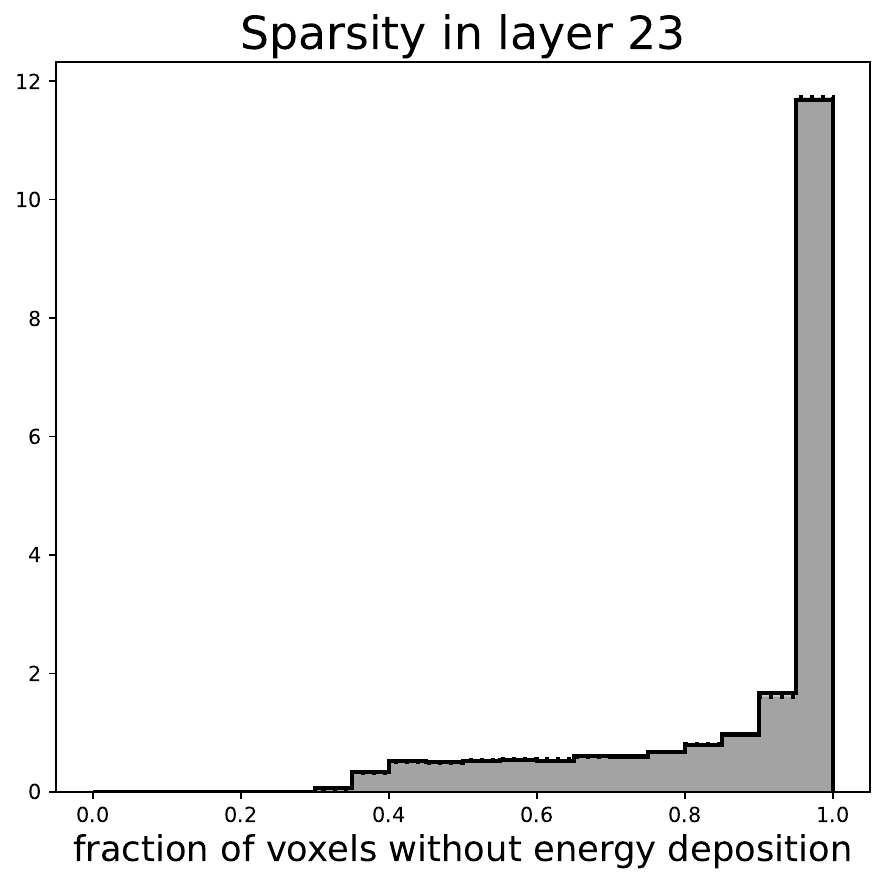} \hfill \includegraphics[height=0.1\textheight]{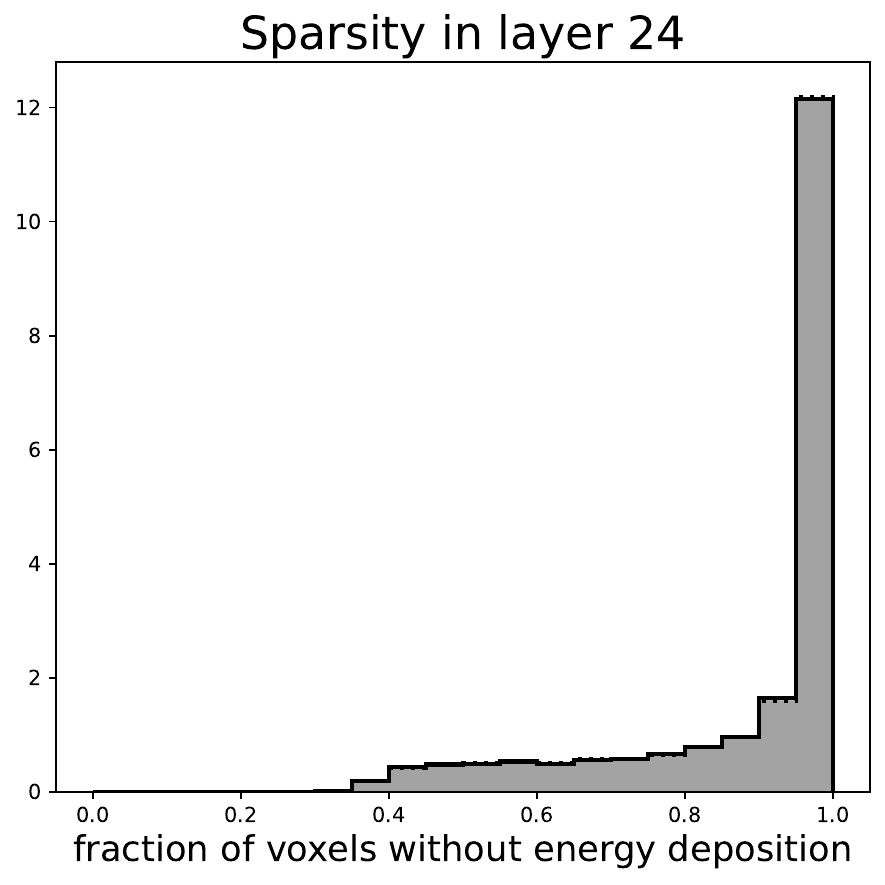}\\
    \includegraphics[height=0.1\textheight]{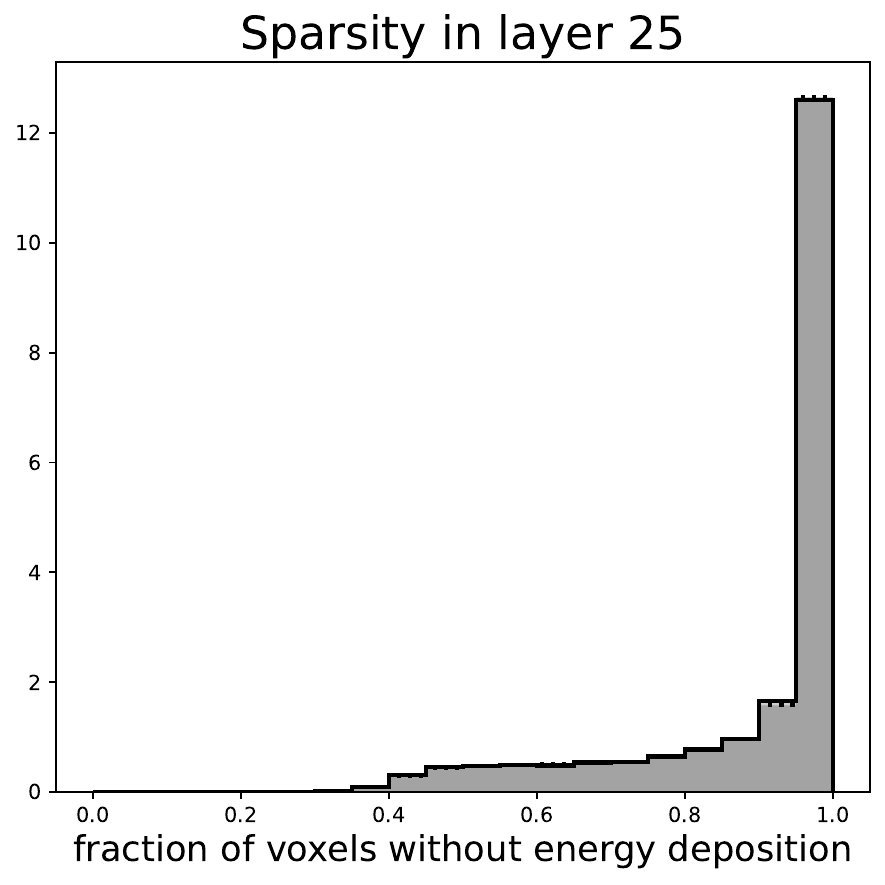} \hfill \includegraphics[height=0.1\textheight]{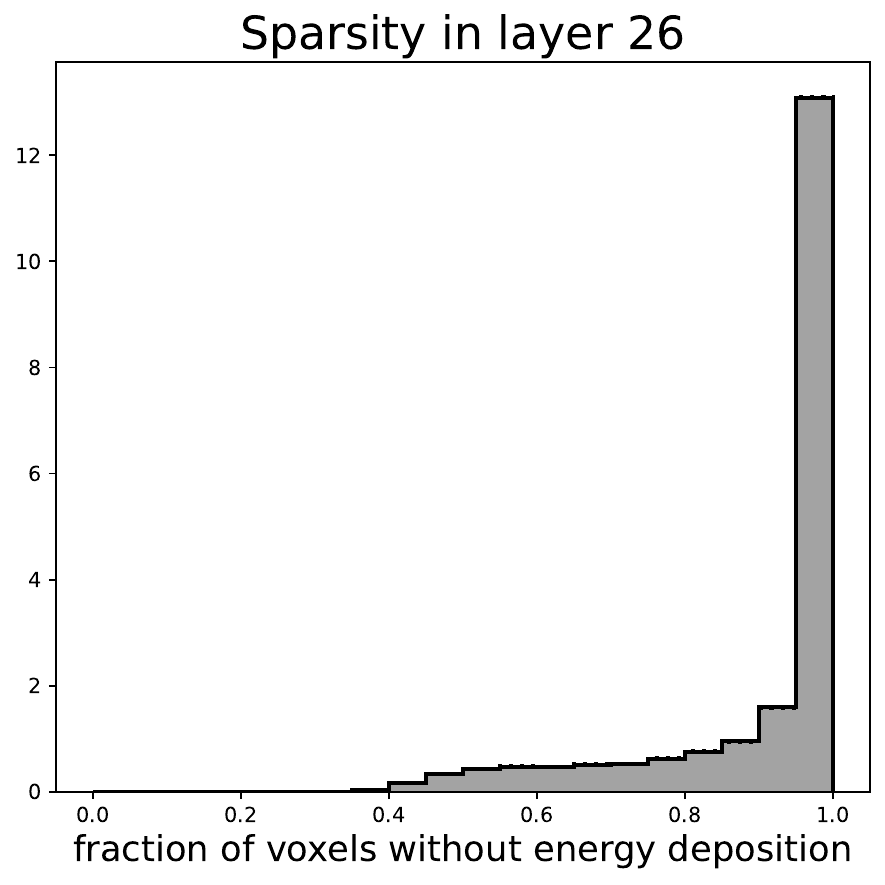} \hfill \includegraphics[height=0.1\textheight]{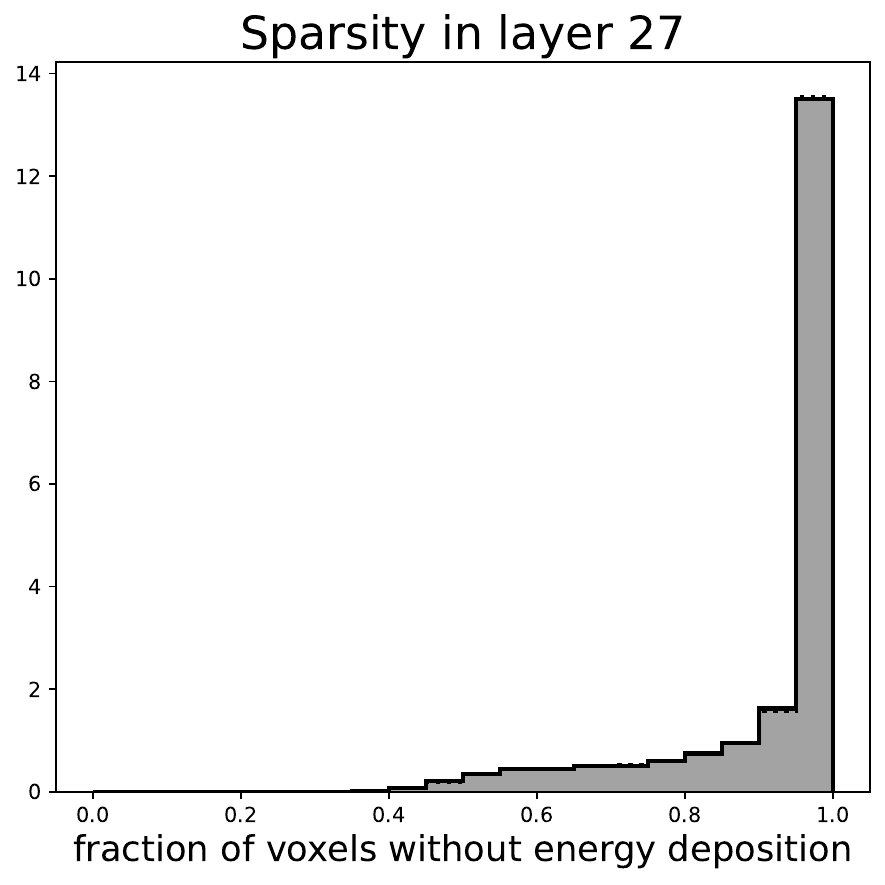} \hfill \includegraphics[height=0.1\textheight]{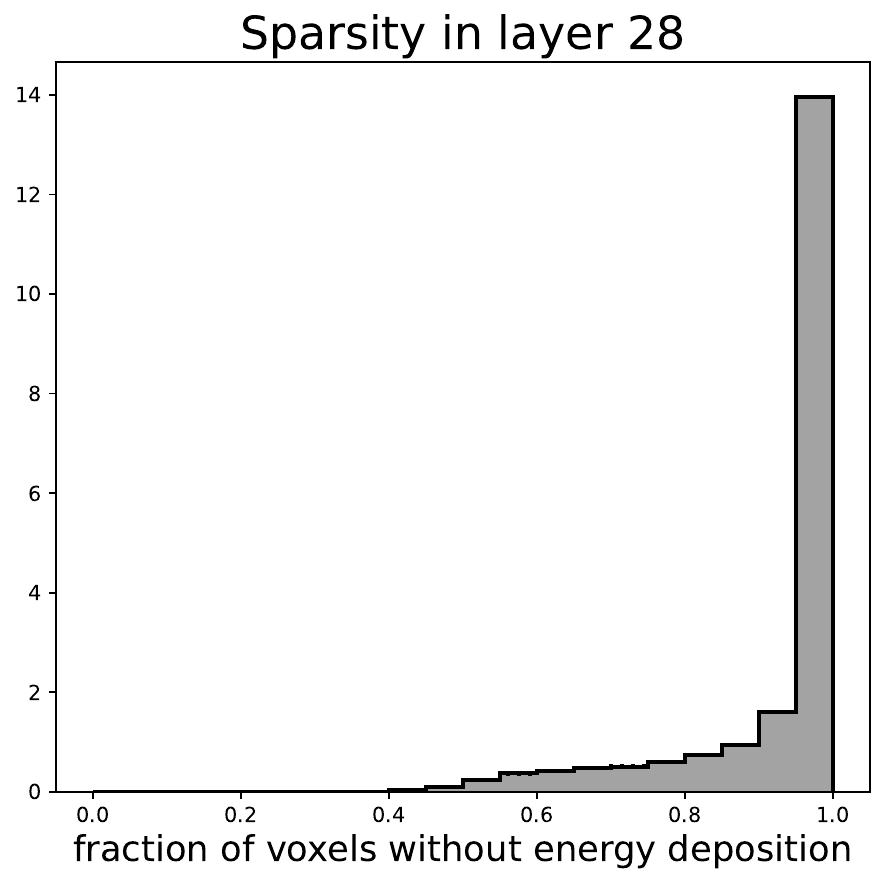} \hfill \includegraphics[height=0.1\textheight]{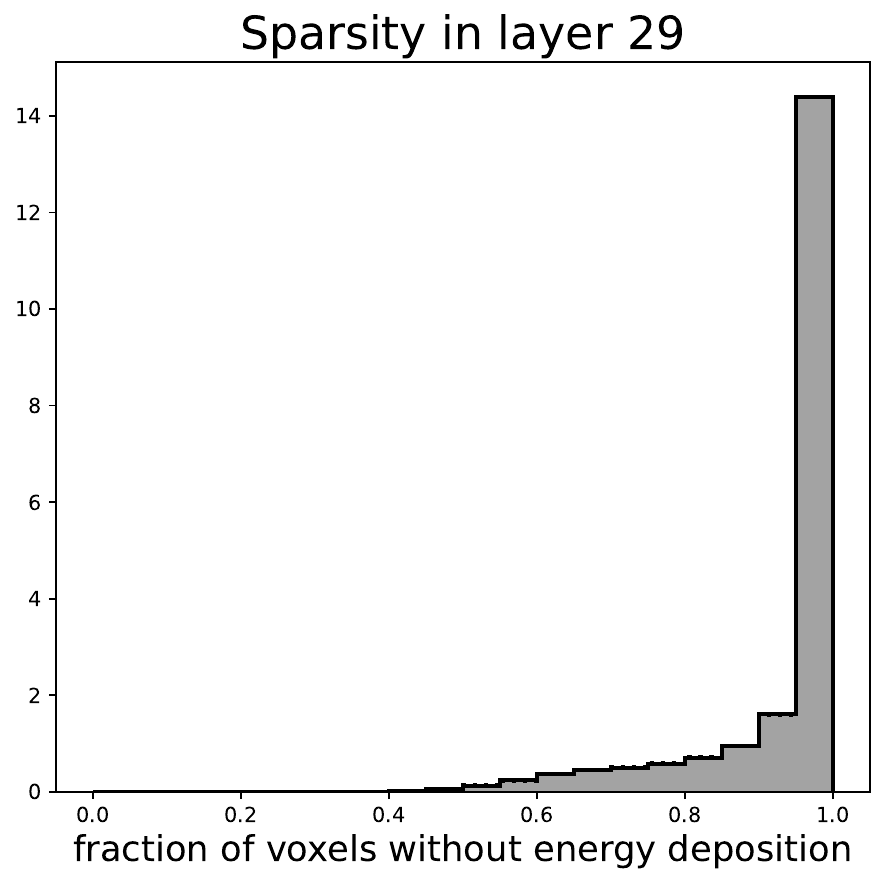}\\
    \includegraphics[height=0.1\textheight]{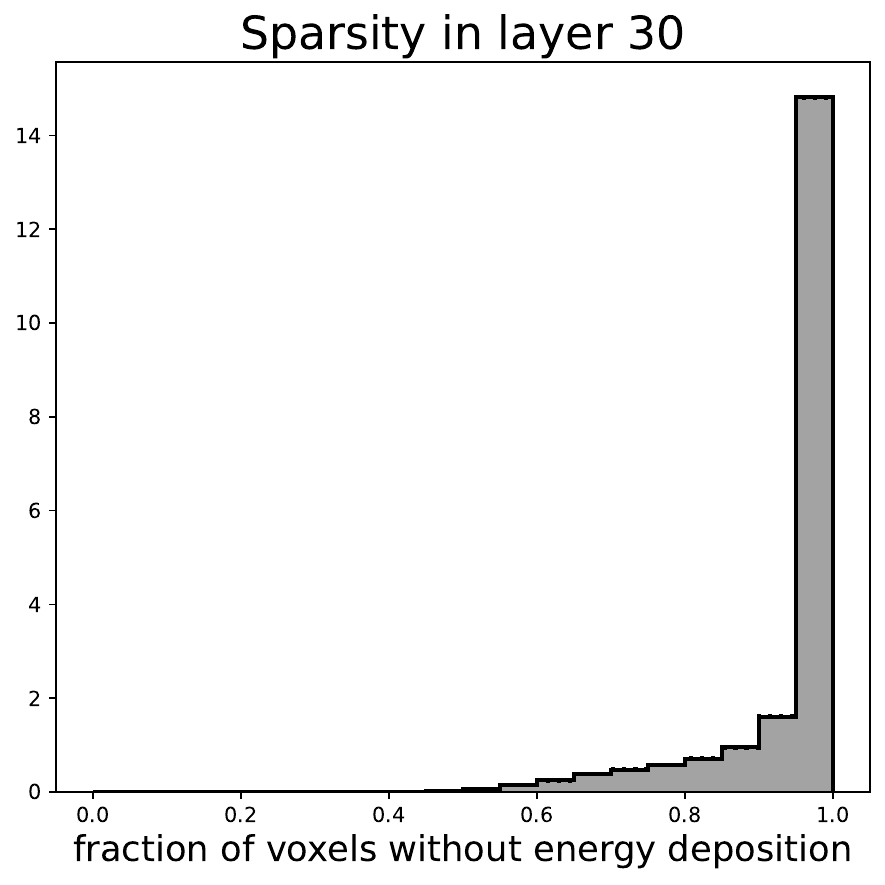} \hfill \includegraphics[height=0.1\textheight]{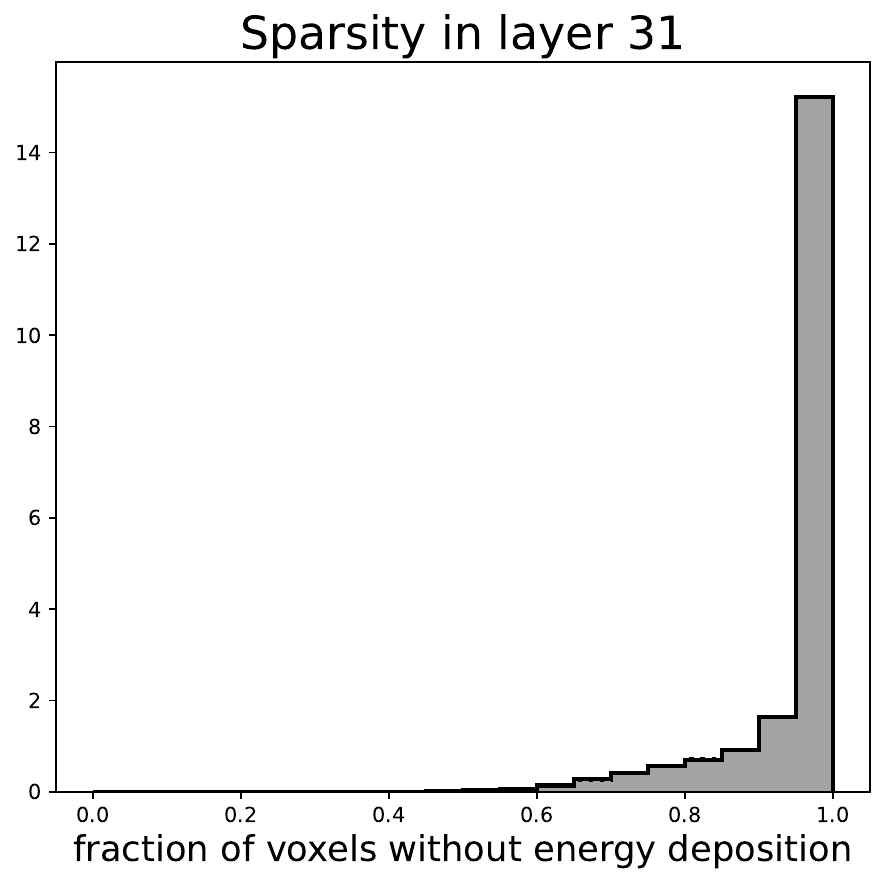} \hfill \includegraphics[height=0.1\textheight]{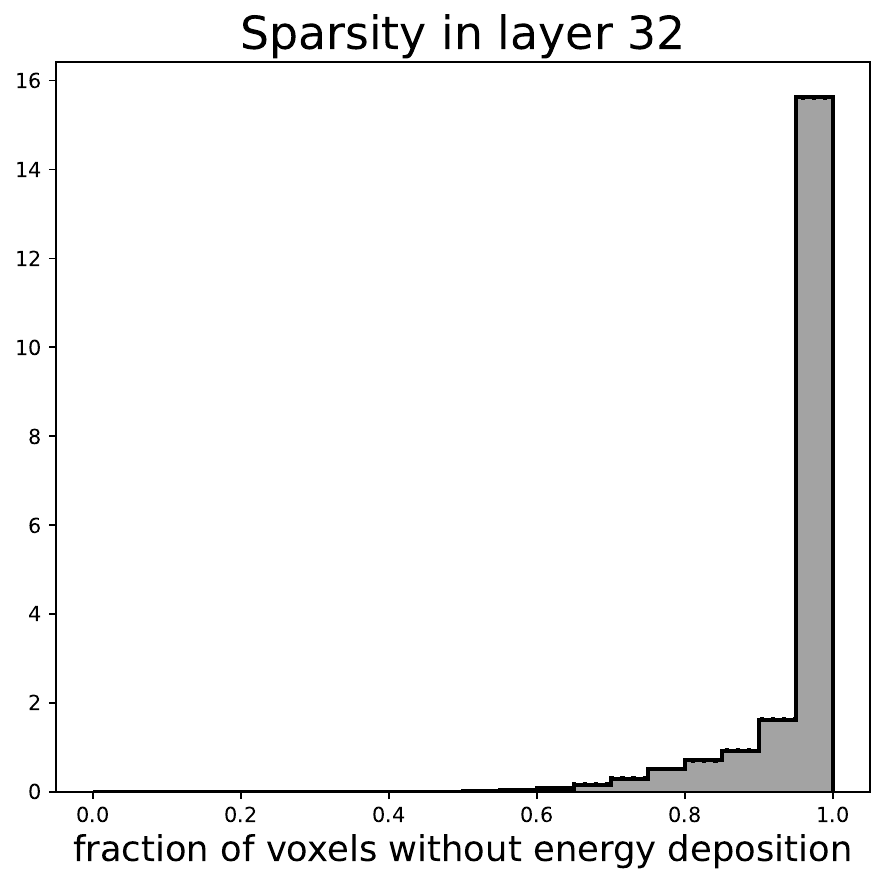} \hfill \includegraphics[height=0.1\textheight]{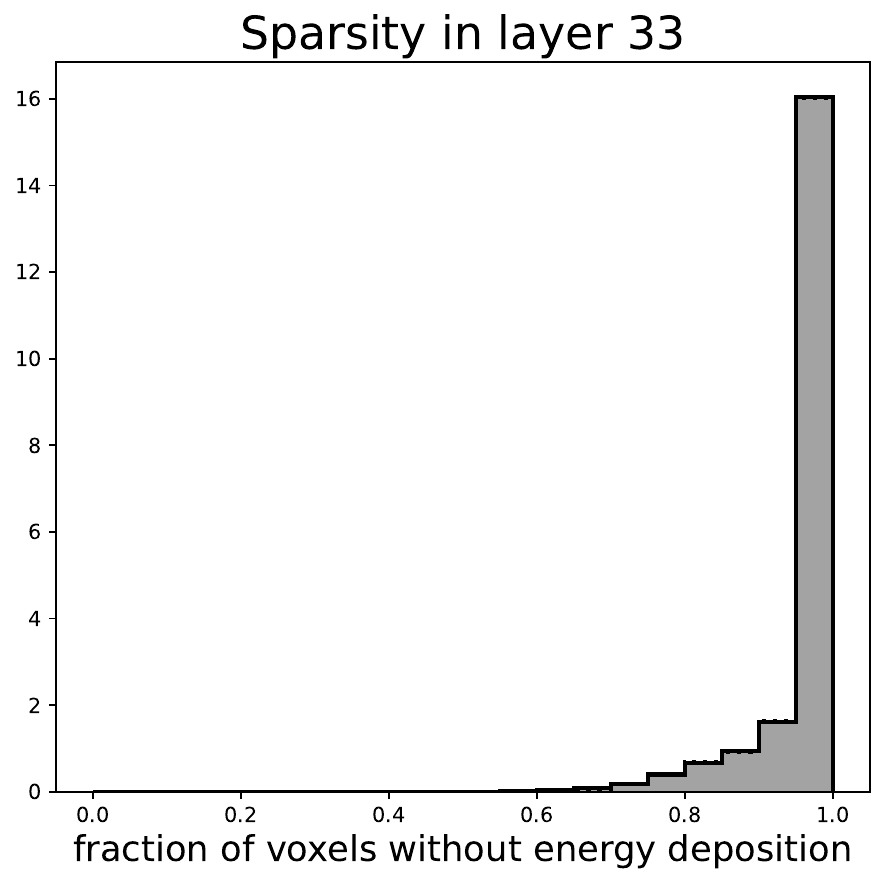} \hfill \includegraphics[height=0.1\textheight]{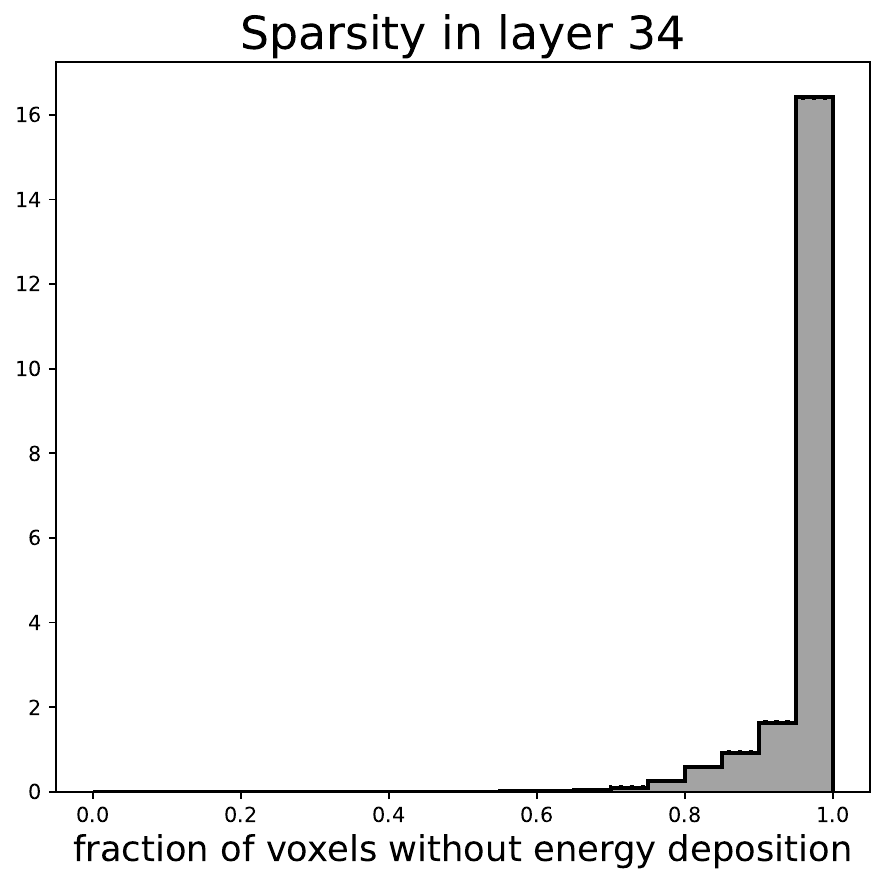}\\
    \includegraphics[height=0.1\textheight]{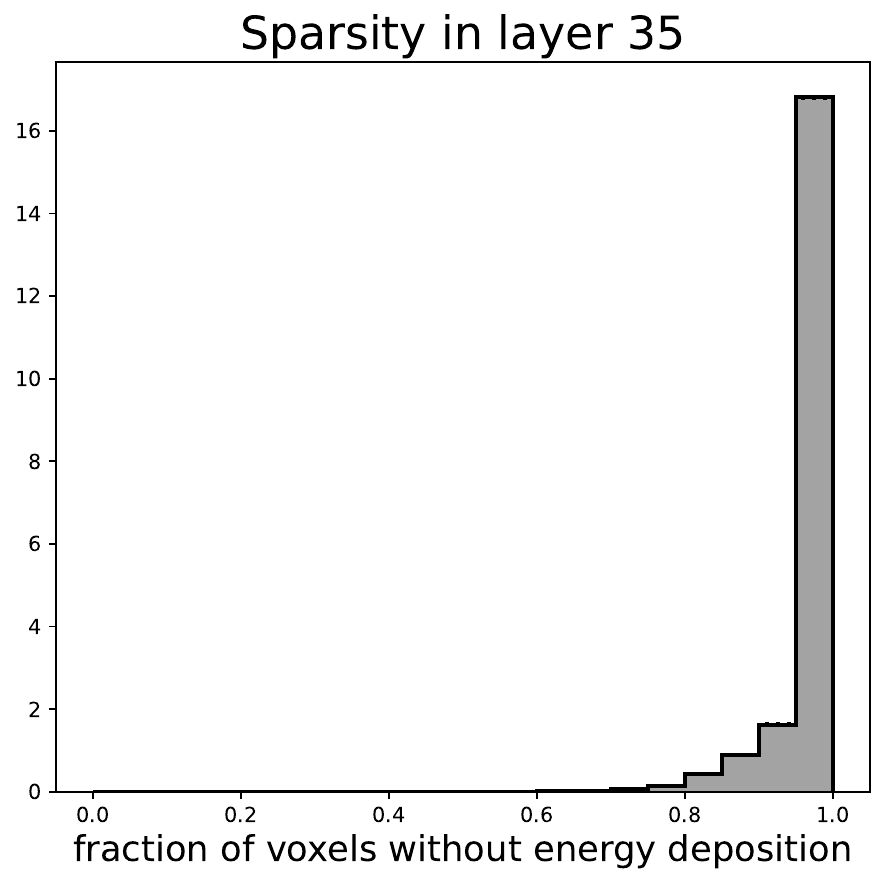} \hfill \includegraphics[height=0.1\textheight]{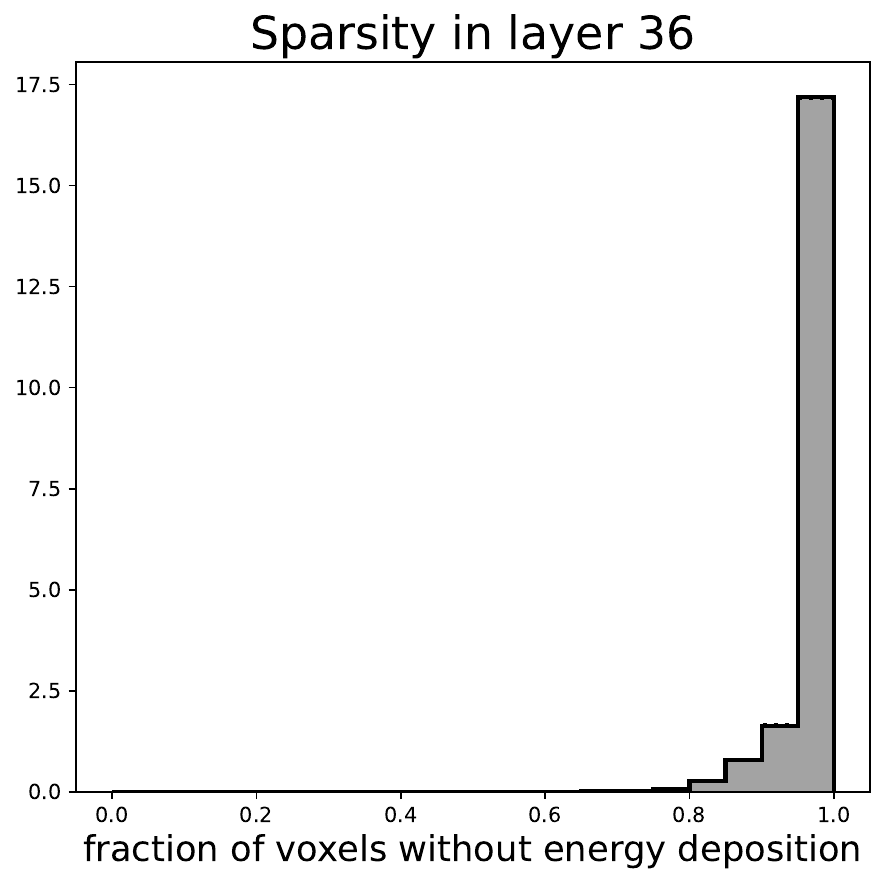} \hfill \includegraphics[height=0.1\textheight]{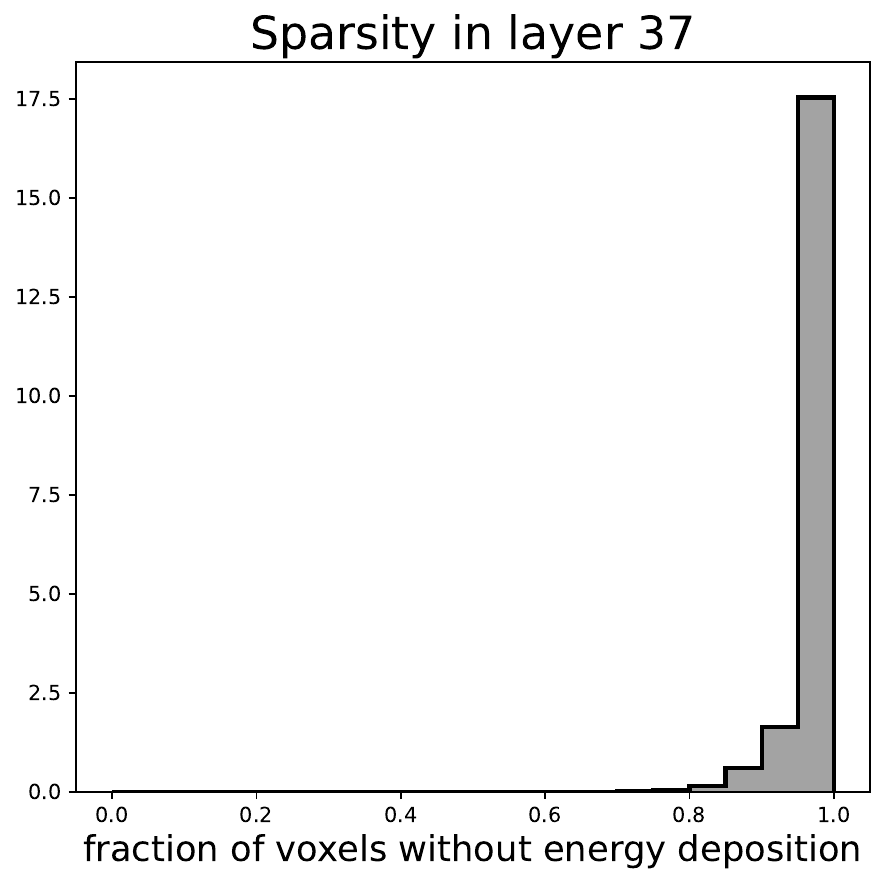} \hfill \includegraphics[height=0.1\textheight]{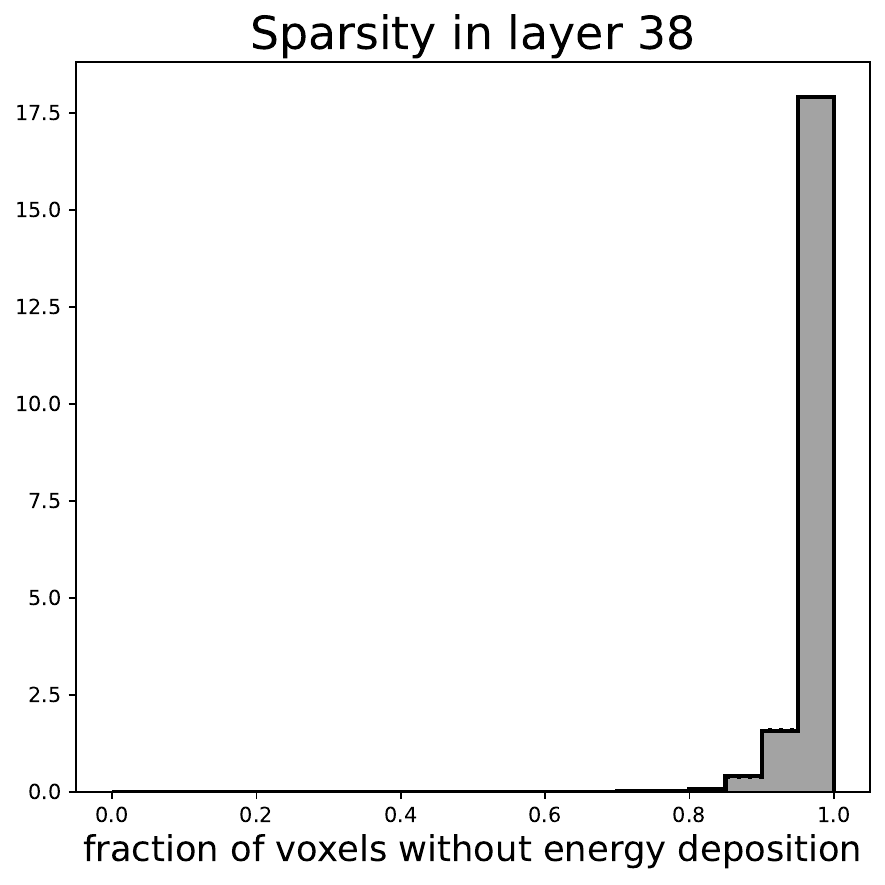} \hfill \includegraphics[height=0.1\textheight]{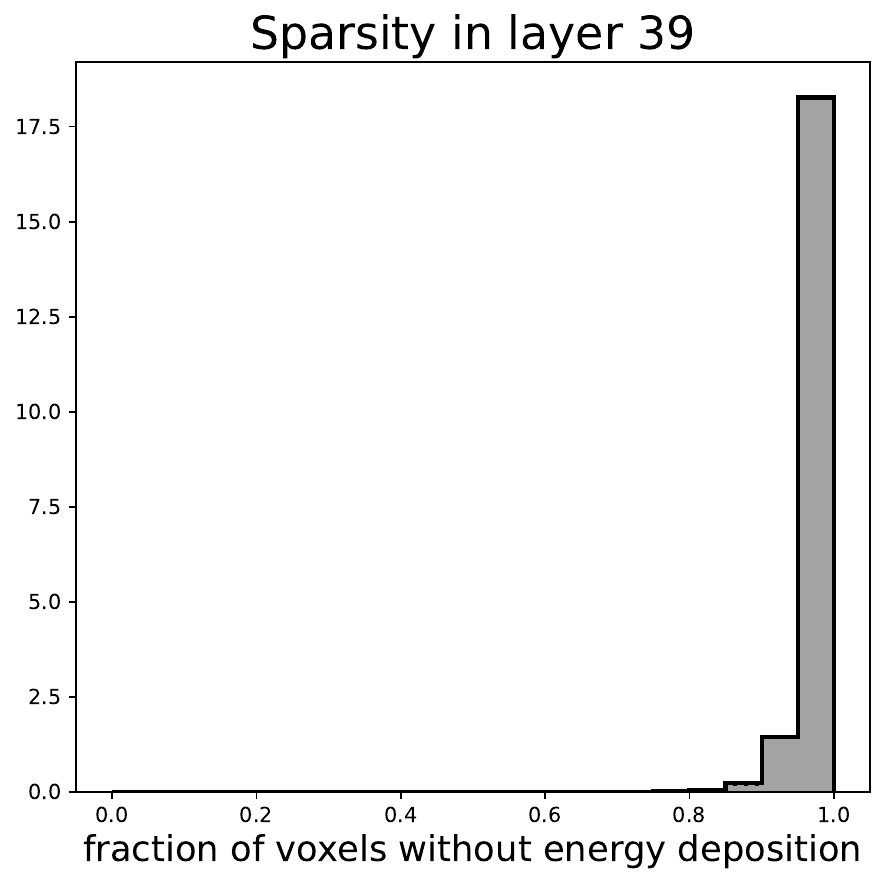}\\
    \includegraphics[height=0.1\textheight]{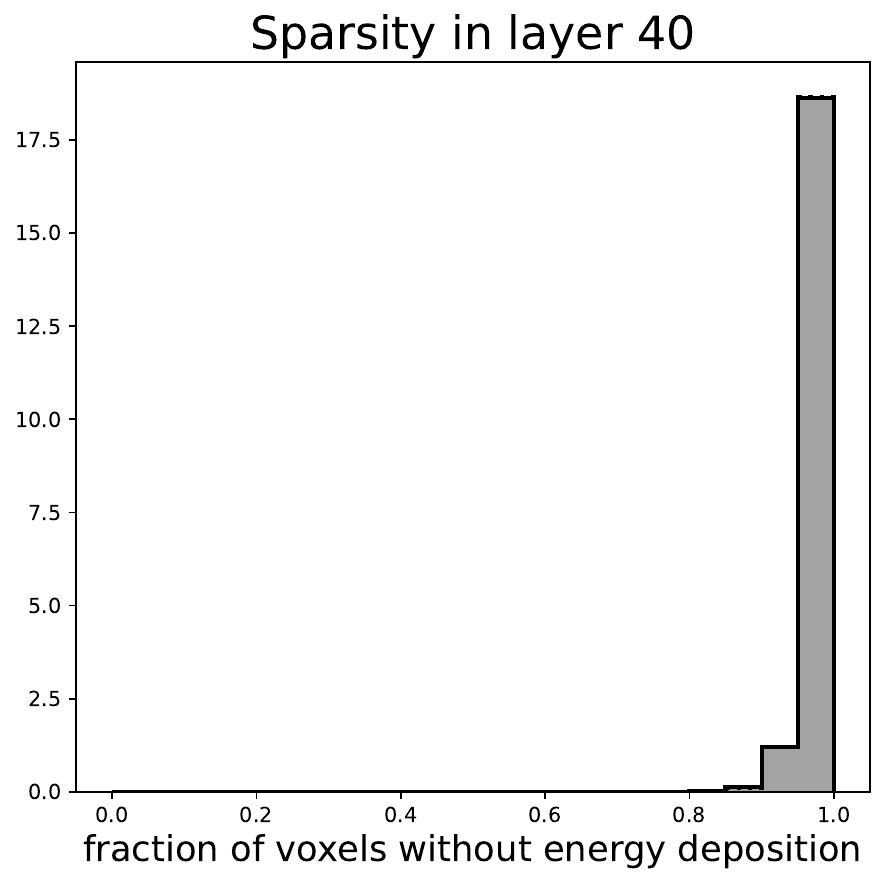} \hfill \includegraphics[height=0.1\textheight]{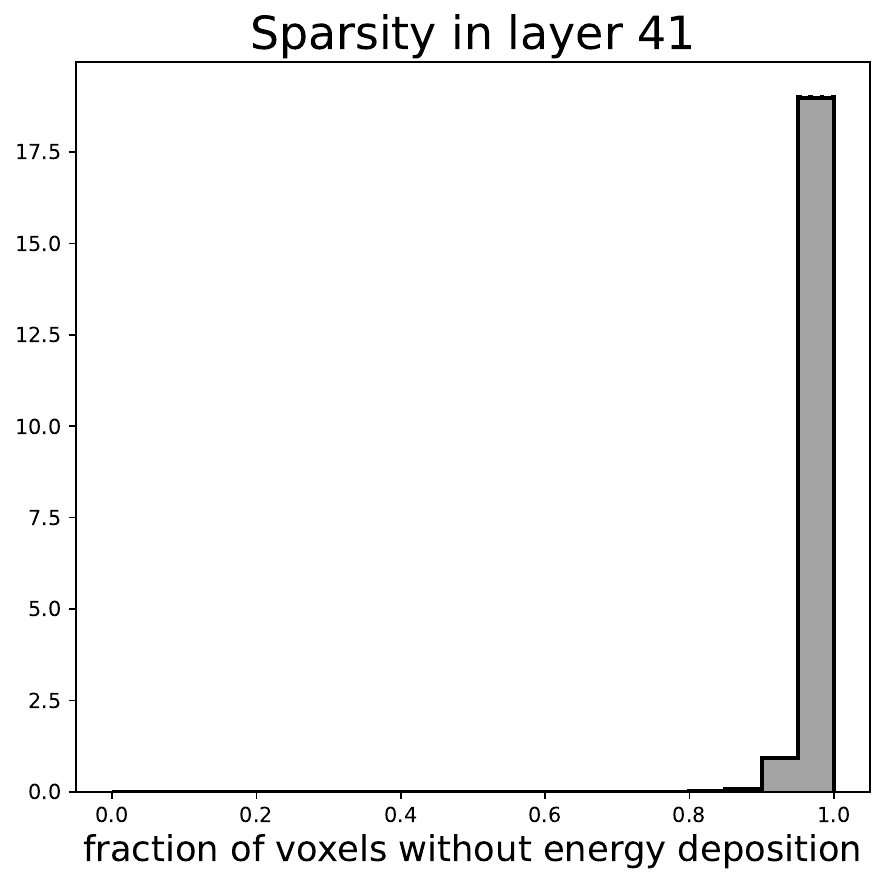} \hfill \includegraphics[height=0.1\textheight]{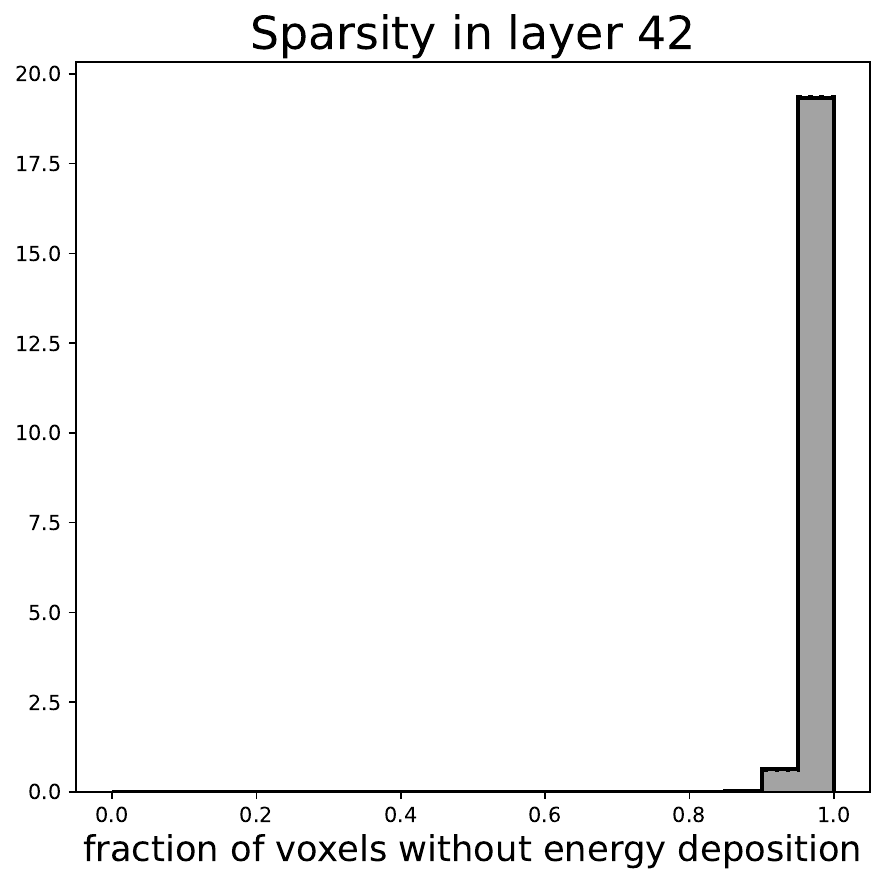} \hfill \includegraphics[height=0.1\textheight]{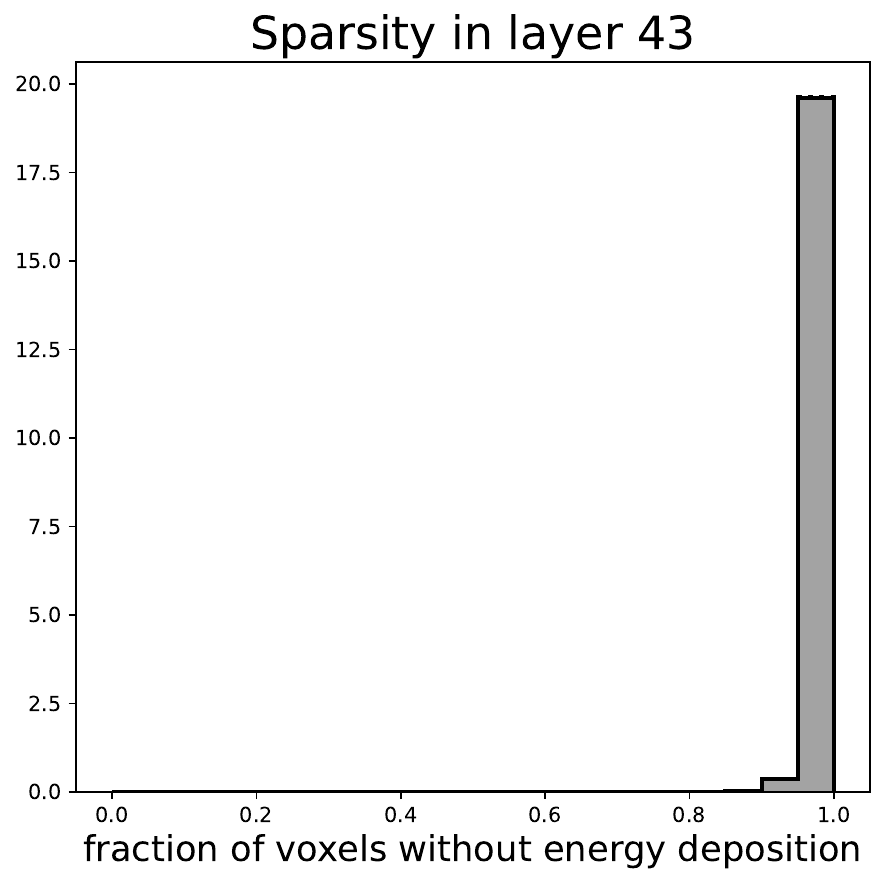} \hfill \includegraphics[height=0.1\textheight]{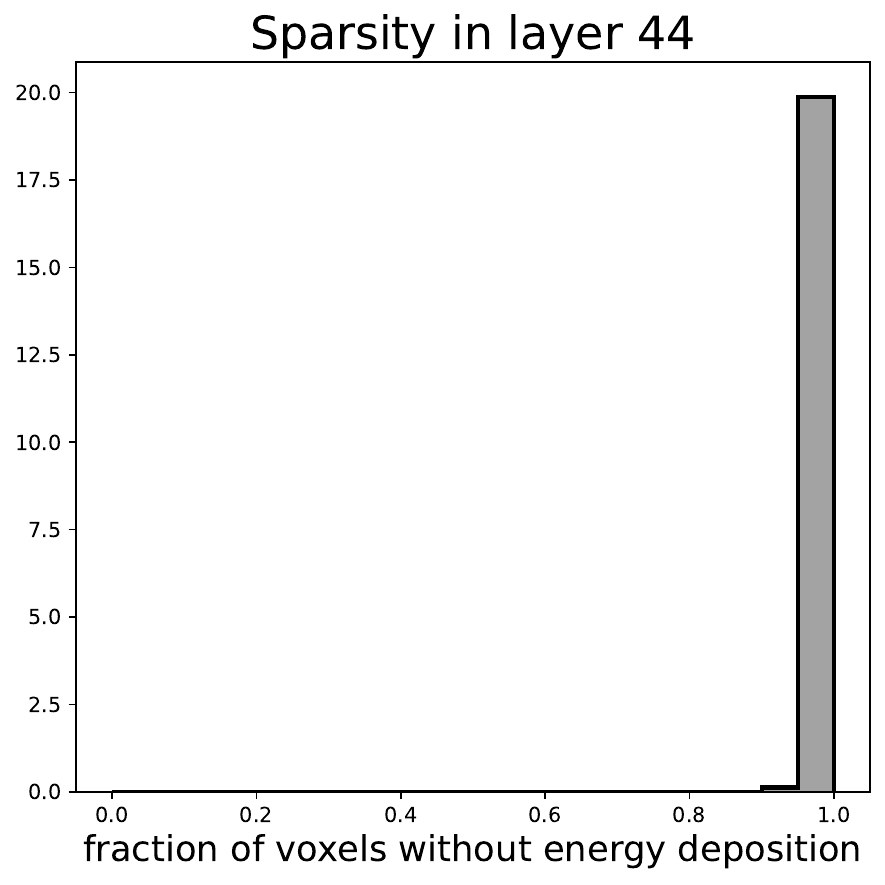}\\
    \includegraphics[width=0.5\textwidth]{figures/Appendix_reference/legend.pdf}
    \caption{Distribution of \geant training and evaluation data in sparsity for ds3. }
    \label{fig:app_ref.ds3.9}
\end{figure}

\section{Consistency check of the multiclass classifier}
\label{app:crosscheck}
\begin{figure}[ht]
    \centering
    \includegraphics[width=\textwidth]{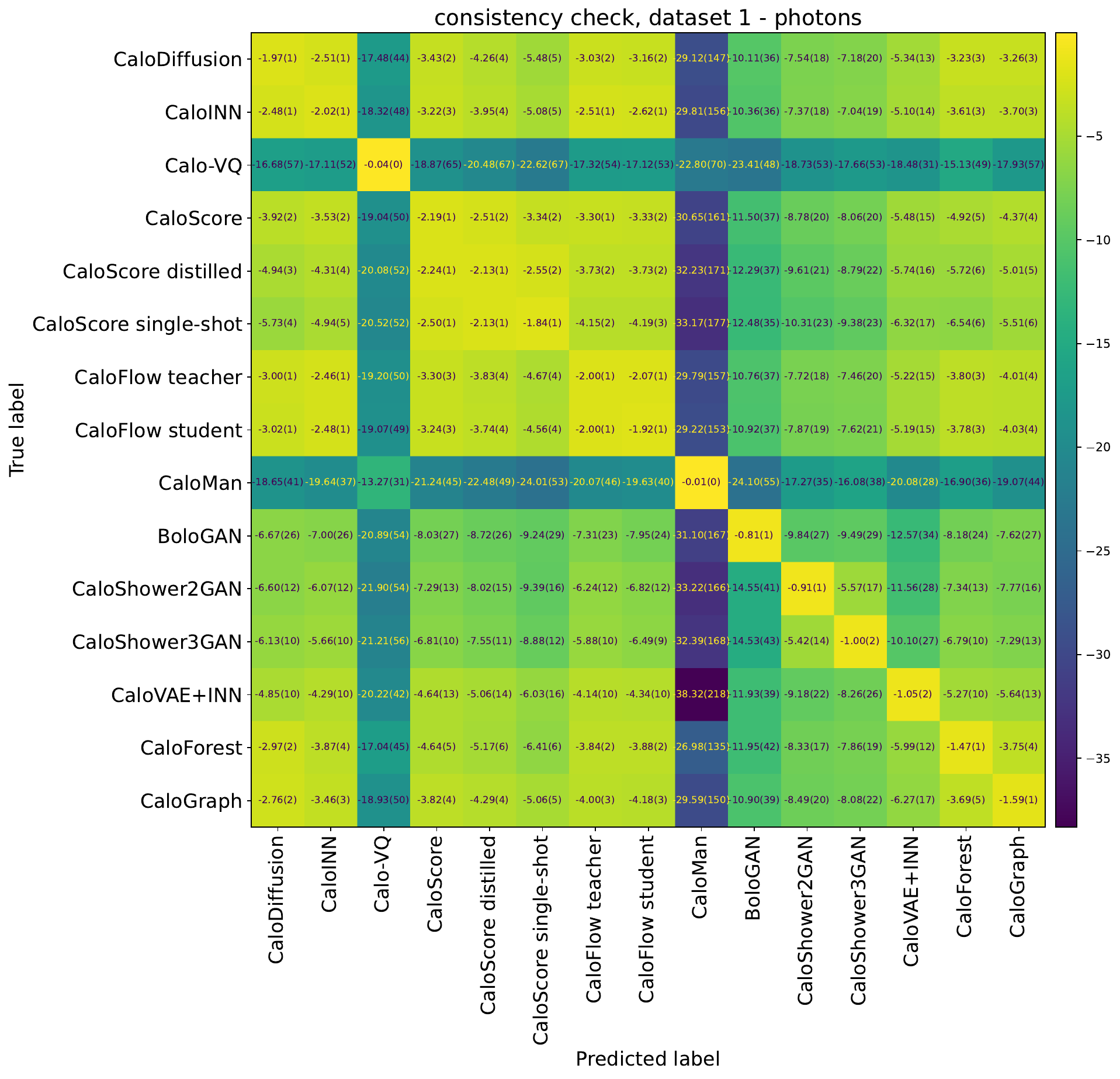}
    \caption{Log posteriors for evaluating the DNN multiclass classifier on submission test sets for \dsIph. }
    \label{fig:consistency.ds1photons}
\end{figure}

\begin{figure}[ht]
    \centering
    \includegraphics[width=\textwidth]{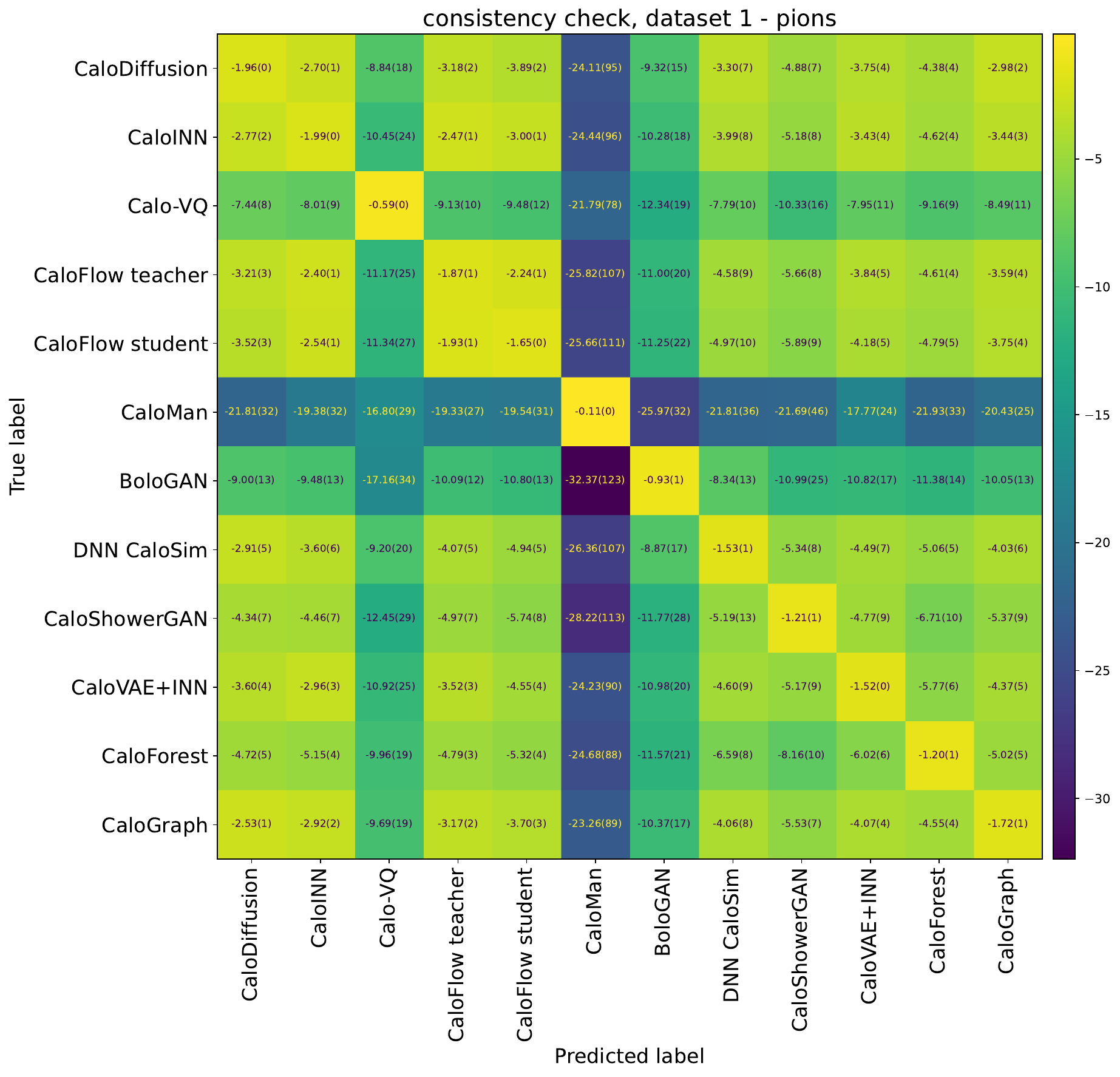}
    \caption{Log posteriors for evaluating the DNN multiclass classifier on submission test sets for \dsIpi. }
    \label{fig:consistency.ds1pions}
\end{figure}

\begin{figure}[ht]
    \centering
    \includegraphics[width=\textwidth]{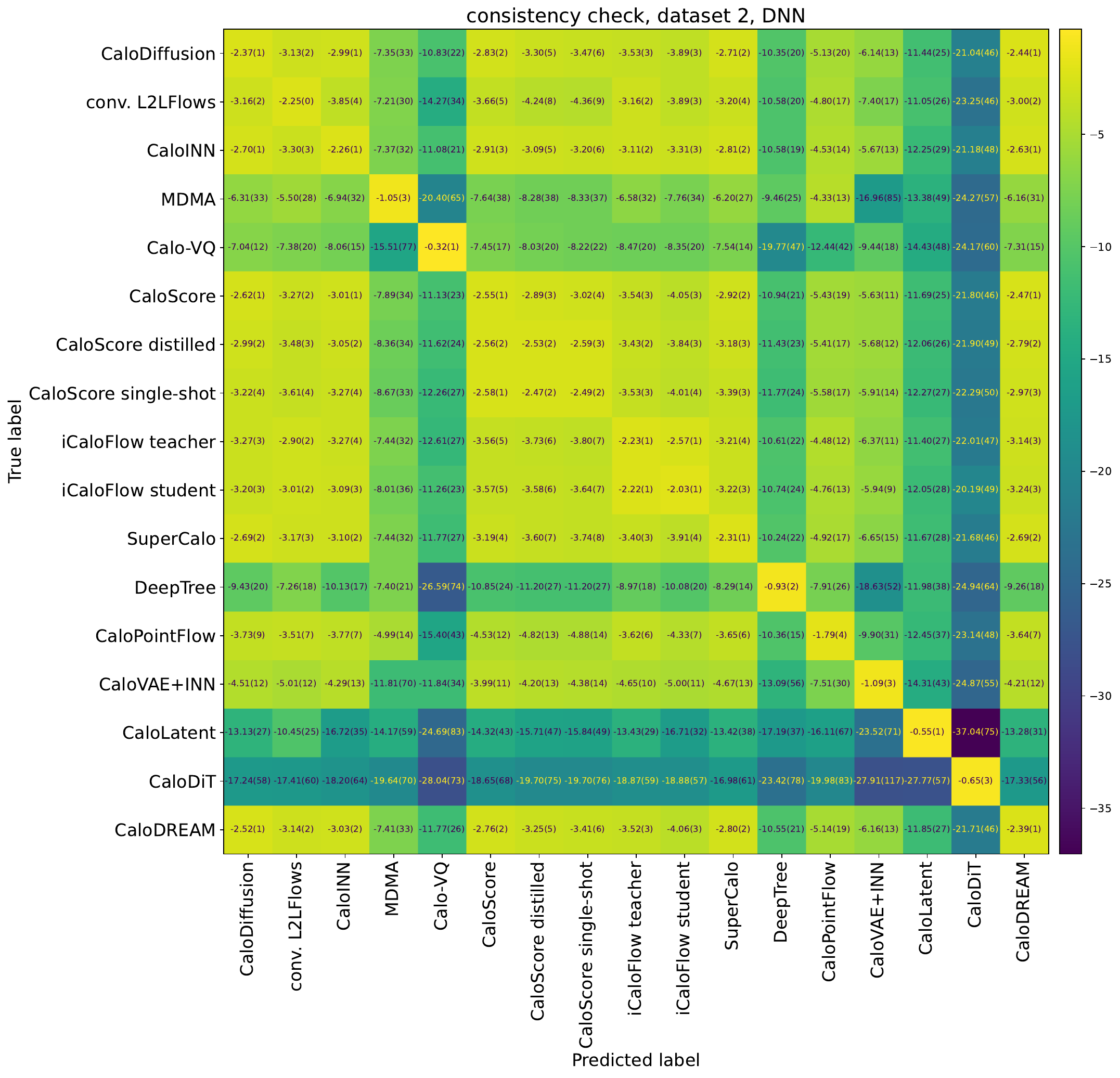}
    \caption{Log posteriors for evaluating the DNN multiclass classifier on submission test sets for \dsII. }
    \label{fig:consistency.ds2}
\end{figure}

\begin{figure}[ht]
    \centering
    \includegraphics[width=\textwidth]{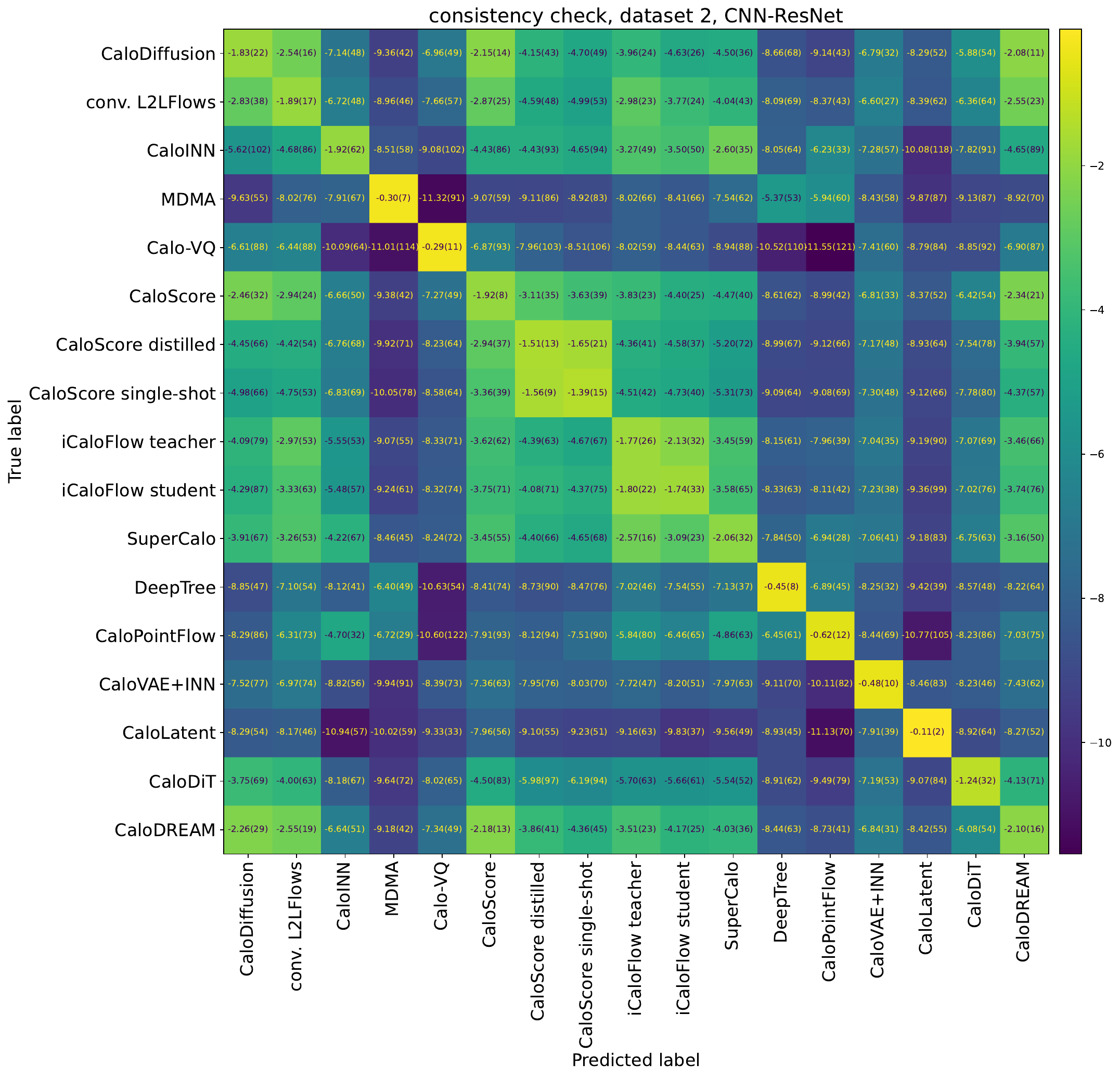}
    \caption{Log posteriors for evaluating the CNN ResNet multiclass classifier on submission test sets for \dsII. }
    \label{fig:consistency.ds2.ResNet}
\end{figure}

\begin{figure}[ht]
    \centering
    \includegraphics[width=\textwidth]{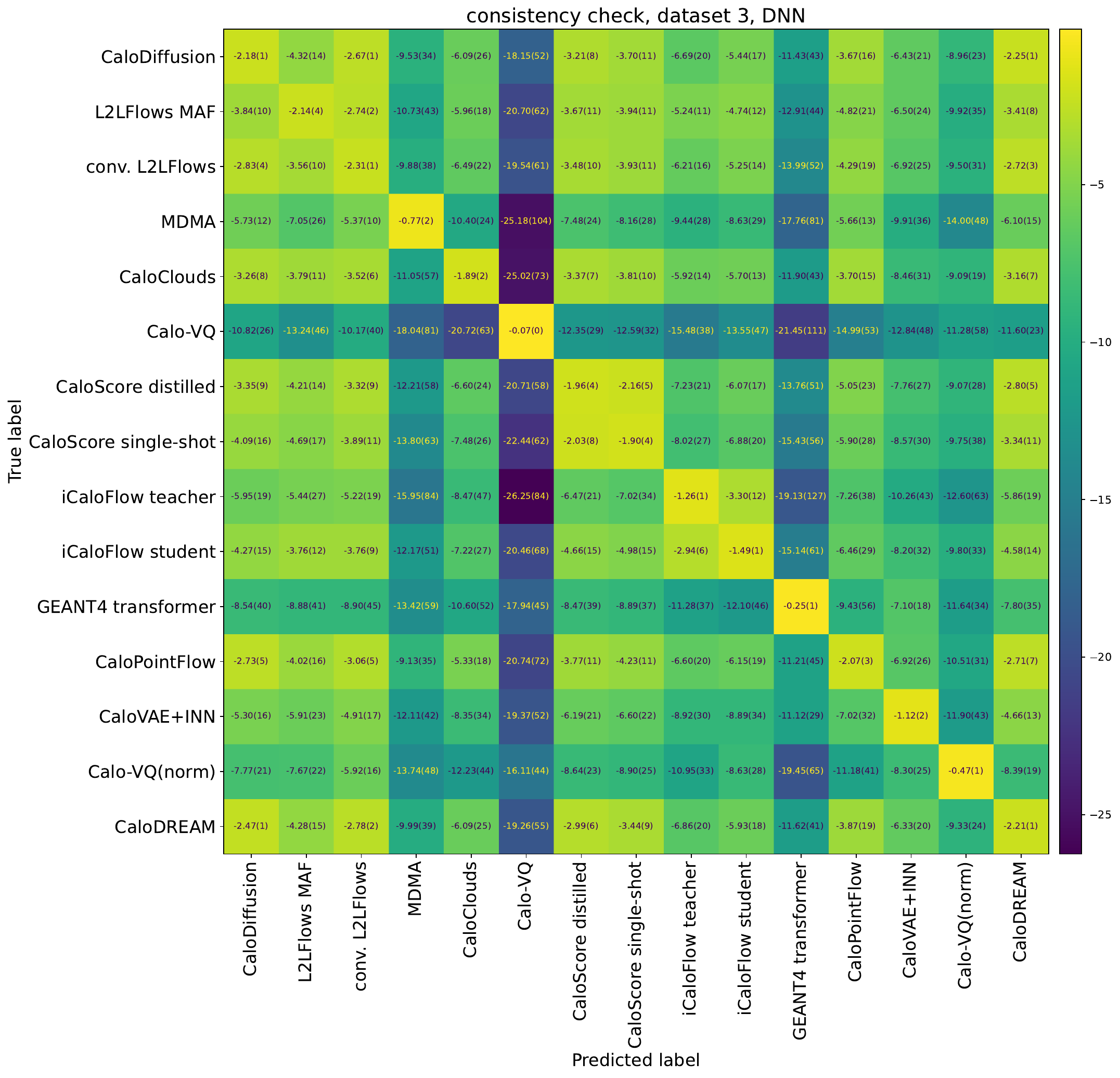}
    \caption{Log posteriors for evaluating the DNN multiclass classifier on submission test sets for \dsIII. }
    \label{fig:consistency.ds3}
\end{figure}

\begin{figure}[ht]
    \centering
    \includegraphics[width=\textwidth]{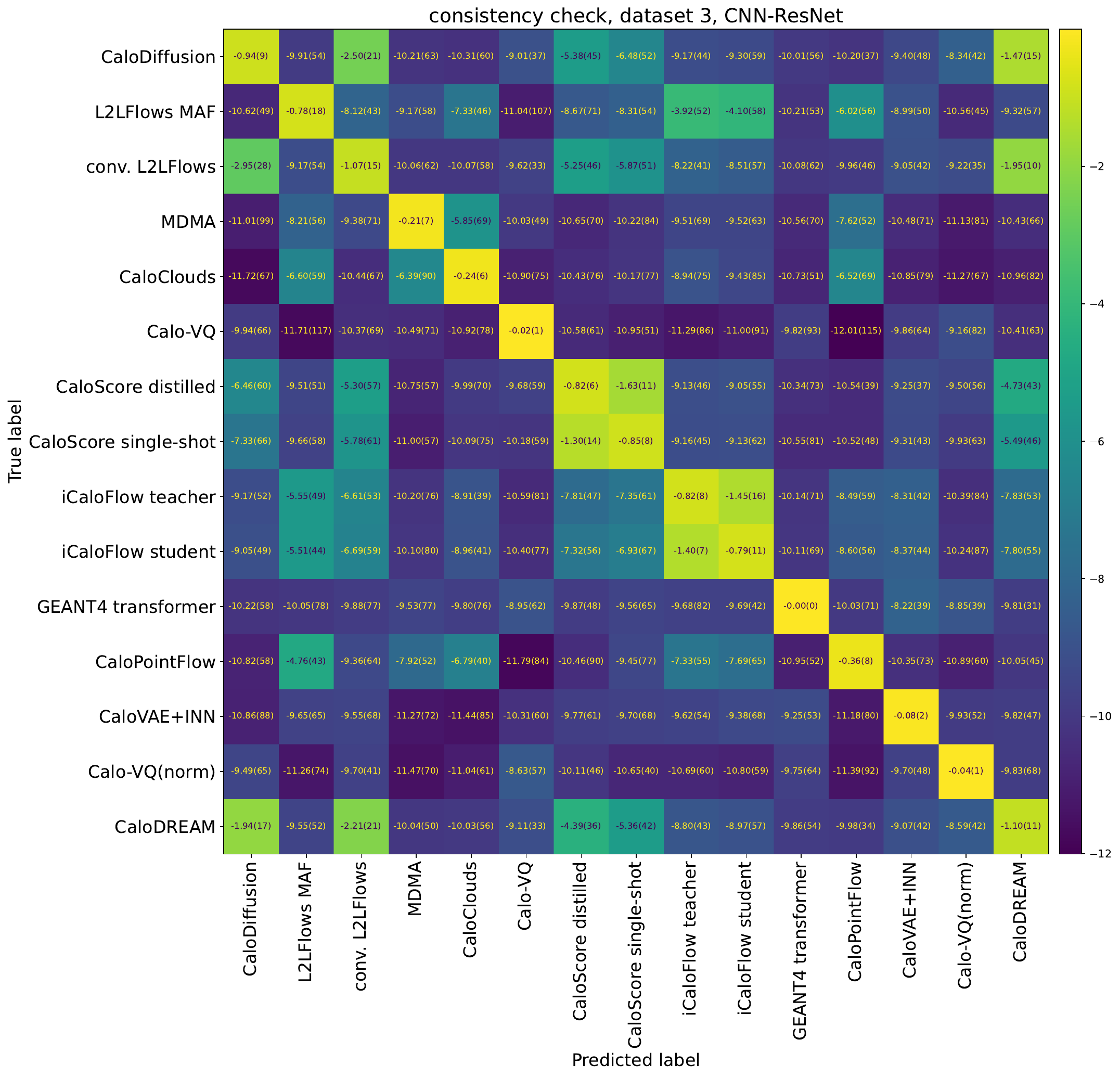}
    \caption{Log posteriors for evaluating the CNN ResNet multiclass classifier on submission test sets for \dsIII. }
    \label{fig:consistency.ds3.ResNet}
\end{figure}

A well-trained multiclass classifier identifies samples from each submission correctly. We show this test here in terms of the log posterior of \eref{eq:log.post}. We show the mean and standard deviation of ten independent trainings and subsequent determination of the log posterior. In \fref{fig:consistency.ds1photons} to \fref{fig:consistency.ds3.ResNet} we show all log posteriors in terms of confusion matrices. The consistency condition of \eref{eq:log.post.crosscheck} can be read line by line in them: In each line, the largest entry is in the diagonal position. This holds for all tests, except for the DNN classifier of dataset 2, where a \submMikuni submission was confused as being \submPalacios, which could bias the test with the \geant dataset towards \submPalacios. In addition, there are a few cases where a class confusion is within error bars, but these concern mostly models that are distilled versions from each other, most notably between \submMikuniDist and \submMikuniSingle for both, datasets 2 and 3. 

\section{Numerical Results in Tables}
\label{app:tables}
In this appendix we show tables with all the results that went into the figures of \sref{sec:results}. Bold-faced entries indicate the best submission(s) within the given errorbars. 

\subsection{\texorpdfstring{Dataset 1, Photons (\dsIph)}{Dataset 1, Photons}}
\begin{table}
\caption{\label{tab:ds1-photons.aucs}Low-level and high-level AUCs for evaluating \geant\ vs.~submission of \dsIph, averaged over 10 independent evaluation runs. For visualization, see \Fref{fig:ds1-photons.aucs}.}
\centering
\begin{tabular}{@{}l|rr}
\br
Submission & low-level AUC $\downarrow$ & high-level AUC $\downarrow$ \\
\mr
\geant &$0.499\pm 0.002$&$0.499\pm 0.003$\\ 
\mr
\submAmramCite        & $0.635 \pm 0.003 $ & $\mathbf{0.536 \pm 0.003}$\\
\submFavaroCite       & $\mathbf{0.626 \pm 0.004}$ & $ 0.638 \pm 0.003 $\\
\submLiuCite          & $0.998 \pm 0.000 $ & $ 0.989 \pm 0.001 $\\
\submMikuniCite       & $0.751 \pm 0.002 $ & $ 0.552 \pm 0.005 $\\
\submMikuniDistCite   & $0.816 \pm 0.004 $ & $ 0.641 \pm 0.003 $\\
\submMikuniSingleCite & $0.866 \pm 0.003 $ & $ 0.726 \pm 0.004 $\\
\submPangTCite        & $0.733 \pm 0.003 $ & $ 0.636 \pm 0.002 $\\
\submPangSCite        & $0.761 \pm 0.002 $ & $ 0.667 \pm 0.004 $\\
\submReyesCite        & $1.000 \pm 0.000 $ & $ 0.999 \pm 0.000 $\\
\submRinaldiCite      & $0.927 \pm 0.003 $ & $ 0.966 \pm 0.001 $\\
\submZhangTwoCite     & $0.938 \pm 0.004 $ & $ 0.942 \pm 0.002 $\\
\submZhangThreeCite   & $0.928 \pm 0.004 $ & $ 0.947 \pm 0.002 $\\
\submErnstCite        & $0.889 \pm 0.003 $ & $ 0.966 \pm 0.001 $\\
\submCresswellCite    & $0.845 \pm 0.002 $ & $ 0.924 \pm 0.002 $\\
\submKobylyanskyCite  & $0.820 \pm 0.002 $ & $ 0.672 \pm 0.004 $\\
\br
\end{tabular}
\end{table}

\begin{table}
\caption{\label{tab:ds1-photons.kpd}KPD and FPD for evaluating \geant\ vs.~submission of \dsIph. For visualization, see \Fref{fig:ds1-photons.kpd}.}
\centering
\begin{tabular}{@{}l|rr}
\br
Submission & KPD $\cdot 10^{3}$  $\downarrow$ & FPD $\cdot 10^{3} $$\downarrow$ \\
\mr
\geant & $0.0279 \pm 0.0592$  &  $0.1192 \pm 0.0534$\\ 
\mr
\submAmramCite        & $\mathbf{-0.0100 \pm 0.0515}$  & $  \mathbf{0.6497 \pm 0.1308}$ \\
\submFavaroCite       & $\mathbf{ 0.2327 \pm 0.2165}$  & $  3.8974 \pm 0.1707$ \\
\submLiuCite          & $ 2.4053 \pm 0.1363$  & $ 35.9531 \pm 0.3235$ \\
\submMikuniCite       & $\mathbf{ 0.0008 \pm 0.0579}$  & $  \mathbf{0.4381 \pm 0.1296}$ \\
\submMikuniDistCite   & $\mathbf{ 0.0319 \pm 0.0504}$  & $  1.1200 \pm 0.1310$ \\
\submMikuniSingleCite & $ 0.2214 \pm 0.0537$  & $  3.0866 \pm 0.1348$ \\
\submPangTCite        & $\mathbf{ 0.0824 \pm 0.0679}$  & $  3.1124 \pm 0.0938$ \\
\submPangSCite        & $\mathbf{ 0.0902 \pm 0.0723}$  & $  3.1512 \pm 0.1080$ \\
\submReyesCite        & $11.1063 \pm 0.5458$  & $141.3752 \pm 0.4676$ \\
\submRinaldiCite      & $11.6268 \pm 1.1852$  & $142.4424 \pm 1.3268$ \\
\submZhangTwoCite     & $ 2.8025 \pm 0.4946$  & $ 52.0972 \pm 0.3547$ \\
\submZhangThreeCite   & $ 1.1561 \pm 0.2029$  & $ 20.5146 \pm 0.3082$ \\
\submErnstCite        & $ 1.2138 \pm 0.0542$  & $ 14.5877 \pm 0.1710$ \\
\submCresswellCite    & $ 2.2297 \pm 0.3132$  & $ 33.5196 \pm 0.5523$ \\
\submKobylyanskyCite  & $ 1.1558 \pm 0.2367$  & $ 15.7884 \pm 0.2592$ \\
\br
\end{tabular}
\end{table}
\begin{table}
\caption{\label{tab:ds1-photons.multi}Log-posterior scores for \dsIph \geant test data, averaged over 10 independent classifier trainings. For visualization, see \Fref{fig:ds1-photons.multi}.}
\centering
\begin{tabular}{@{}l|r}
\br
Submission & Log-posterior $\uparrow$ \\
\mr
\submAmramCite        & $\mathbf{ -2.1893 \pm 0.0053}$\\
\submFavaroCite       & $ -2.2046 \pm 0.0083$\\
\submLiuCite          & $-17.9096 \pm 0.4605$\\
\submMikuniCite       & $ -3.3126 \pm 0.0254$\\
\submMikuniDistCite   & $ -4.2047 \pm 0.0408$\\
\submMikuniSingleCite & $ -5.4383 \pm 0.0507$\\
\submPangTCite        & $ -2.7946 \pm 0.0133$\\
\submPangSCite        & $ -2.9796 \pm 0.0143$\\
\submReyesCite        & $-30.3461 \pm 1.5767$\\
\submRinaldiCite      & $ -9.7551 \pm 0.3416$\\
\submZhangTwoCite     & $ -7.1072 \pm 0.1705$\\
\submZhangThreeCite   & $ -6.7926 \pm 0.1889$\\
\submErnstCite        & $ -5.2225 \pm 0.1405$\\
\submCresswellCite    & $ -3.6188 \pm 0.0307$\\
\submKobylyanskyCite  & $ -3.5833 \pm 0.0310$\\
\br
\end{tabular}
\end{table}

\begin{table}
\caption{\label{tab:ds1-photons.prdc}Precision, density, recall, and coverage for \dsIph submissions. A visualization is shown in \fref{fig:ds1-photons.prdc}.}
\centering
\begin{tabular}{@{}l|rrrr}
\br
Submission & Precision $\uparrow$ & Density $\uparrow$ & Recall $\uparrow$ & Coverage $\uparrow$ \\
\mr
\geant &  0.704  & 0.992 &  0.699  & 0.964 \\
\mr
\submAmramCite        &  0.730  & 1.152 &  0.665  & 0.974 \\
\submFavaroCite       &	 0.607  & 0.831 &  0.716  & 0.911 \\
\submLiuCite          &	 0.957  & 115.883 & 0.003 & 0.952 \\
\submMikuniCite       &	 0.662  & 0.891 &  0.708  & 0.939 \\
\submMikuniDistCite   &	 0.613  & 0.781 &  0.722  & 0.907 \\
\submMikuniSingleCite &	 0.482  & 0.441 &  0.807  & 0.782 \\
\submPangTCite        &	 0.595  & 0.799 &  0.718  & 0.913 \\
\submPangSCite        &	 0.617  & 0.859 &  0.701  & 0.912 \\
\submReyesCite        &	 0.888  & 612.387  & 0.010 & 0.888 \\
\submRinaldiCite      &	 0.207  & 0.315 &  0.194  & 0.411\\
\submZhangTwoCite     &	 0.761  & 72.665  & 0.036  & 0.832 \\
\submZhangThreeCite   &	 0.734  & 65.729  & 0.039  & 0.795 \\
\submErnstCite        &	 0.873  & 73.873  & 0.139  &  0.973 \\
\submCresswellCite    &	 0.906  & 17.494 & 0.186  & 0.957 \\
\submKobylyanskyCite  &	 0.657  & 0.982 &  0.670  & 0.933 \\
\br
\end{tabular}
\end{table}
\begin{table}
\caption{\label{tab:ds1-photons.numparam}Number of trainable parameters in training and for generation for \dsIph submissions. A visualization is shown in \fref{fig:ds1-photons.numparam}. }
\centering
\begin{tabular}{@{}l|rr}
\br
& \multicolumn{2}{c}{number of parameters $\downarrow$}  \\
Submission & total & generator only  \\
\mr
\submAmramCite        &$\mathbf{521\,581}$ &       $521\,581$  \\
\submFavaroCite       &    $18\,821\,350$ &    $18\,821\,350$  \\
\submLiuCite          &     $4\,060\,878$ &     $2\,152\,637$  \\
\submMikuniCite       &     $2\,447\,366$ &     $2\,447\,366$  \\
\submMikuniDistCite   &     $4\,894\,732$ &     $2\,447\,366$  \\
\submMikuniSingleCite &     $4\,894\,732$ &     $2\,447\,366$  \\
\submPangTCite        &    $28\,043\,810$ &    $28\,043\,810$  \\
\submPangSCite        &    $84\,500\,898$ &    $56\,554\,930$  \\
\submReyesCite        &    $19\,276\,658$ &    $17\,061\,148$  \\
\submRinaldiCite      &     $1\,185\,520$ &       $368\,558$  \\
\submZhangTwoCite     &     $1\,183\,606$ &$\mathbf{367\,380}$  \\
\submZhangThreeCite   &     $1\,696\,459$ &       $472\,120$  \\
\submErnstCite        &    $15\,747\,908$ &     $8\,321\,308$  \\
\submCresswellCite    & $3\,837\,598\,845$ & $3\,837\,598\,845$  \\
\submKobylyanskyCite  &       $823\,617$ &       $823\,617$  \\
\br
\end{tabular}
\end{table}
\begin{table}
\caption{\label{tab:ds1-photons.timing.CPU}Timing of \dsIph submissions on a CPU. The symbols ${}^\ast$,  ${}^\dagger$, ${}^\ddag$, and ${}^\diamond$ indicate that only 1000 / 10\,000 / 20\,000 / 50\,000 events were generated in timing the submission. A visualization of these timings is shown in \fref{fig:ds1-photons.timing}.}
\begin{tabular}{@{}l|r@{}c@{}lr@{}c@{}lr@{}c@{}l}
\br
\multirow{2}{*}{Submission} & \multicolumn{9}{c}{CPU [ms per shower]} \\
& \multicolumn{3}{c}{batch size 1} & \multicolumn{3}{c}{batch size 100} & \multicolumn{3}{c}{batch size 10\,000} \\
\mr
\submAmramCite        & $15\,677 $&$\,\pm\,$&$633^\dagger$ & $4266 $&$\,\pm\, $&$106^\dagger$  & $4500 $&$\,\pm\,$&$ 357^\ddag$ \\
\submFavaroCite       & $\mathbf{37.9} $&$\,\pm\,$&$ \mathbf{3.2}$ & $2.78 $&$\,\pm\,$&$ 0.34$ & $2.88 $&$\,\pm\,$&$ 0.31$ \\
\submLiuCite          & $93.6 $&$\,\pm\,$&$ 5.6$ & $10.4 $&$\,\pm\,$&$ 1.8$ & $14.1 $&$\,\pm\,$&$ 3.1$ \\
\submMikuniCite       & $19\,324 $&$\,\pm\,$&$ 729^\ast$ & $9425 $&$\,\pm\,$&$ 322^\ast$ & $12\,871 $&$\,\pm\,$&$ 1197^\dagger$ \\
\submMikuniDistCite   & $2456 $&$\,\pm\,$&$ 178^\dagger$ & $1078 $&$\,\pm\,$&$ 54^\dagger$ & $1276 $&$\,\pm\,$&$ 190^\ddag$ \\
\submMikuniSingleCite & $223.3 $&$\,\pm\,$&$ 10.1$ & $20.5 $&$\,\pm\,$&$ 2.0$ & $19.8 $&$\,\pm\,$&$ 3.5$ \\
\end{tabular}
\end{table}

\begin{table}
\caption{\label{tab:ds1-photons.timing.GPU}Timing of \dsIph submissions on a GPU. The symbols ${}^\ast $ and $ {}^\ddag$ indicate that only 5000 or 10\,000 events were generated in timing the submission; a -- indicates a model that does not run on a GPU; and ``CUDA o.o.m'' ran out of VRAM on the GPU. A visualization of these timings is shown in \fref{fig:ds1-photons.timing}.}
\begin{tabular}{@{}l|r@{}c@{}lr@{}c@{}lr@{}c@{}l}
\br
\multirow{2}{*}{Submission} & \multicolumn{9}{c}{GPU [ms per shower]} \\
& \multicolumn{3}{c}{batch size 1} & \multicolumn{3}{c}{batch size 100} & \multicolumn{3}{c}{batch size 10\,000} \\
\mr
\submAmramCite        & $5593$ & $\,\pm\,$ & $64^\ast$ & $75.2$ & $\,\pm\,$ & $2.1$  & \quad $24.4$ & $\,\pm\,$ & $0.1$ \\
\submFavaroCite       & $\mathbf{24.6}$ & $\,\pm\,$ & $\mathbf{1.6}$ & $\mathbf{0.51}$ & $\,\pm\,$ & $\mathbf{0.03}$ & $0.19$ & $\,\pm\,$ & $0.01$ \\
\submLiuCite          & $48.2$ & $\,\pm\,$ & $0.6$ & $0.81$ & $\,\pm\,$ & $0.08$ & $\mathbf{0.16}$ & $\,\pm\,$ & $\mathbf{0.01}$ \\
\submMikuniCite       & $4706$ & $\,\pm\,$ & $171^\ast$ & $60.4$ & $\,\pm\,$ & $4.1$  & $36.00$ & $\,\pm\,$ & $0.01$ \\
\submMikuniDistCite   & $756.6$ & $\,\pm\,$ & $22.8$ & $8.8$ & $\,\pm\,$ & $0.2$   & $4.96$ & $\,\pm\,$ & $0.02$ \\
\submMikuniSingleCite & $189.4$ & $\,\pm\,$ & $12.5$ & $2.0$ & $\,\pm\,$ & $0.0$   & $0.56$ & $\,\pm\,$ & $0.02$ \\
\end{tabular}
\end{table}
\subsection{\texorpdfstring{Dataset 1, Pions (\dsIpi)}{Dataset 1, Pions}}
\begin{table}
\caption{\label{tab:ds1-pions.aucs}Low-level and high-level AUCs for evaluating \geant\ vs.~submission of \dsIpi, averaged over 10 independent evaluation runs. For visualization, see \Fref{fig:ds1-pions.aucs}.}
\centering
\begin{tabular}{@{}l|rr}
\br
Submission & low-level AUC $\downarrow$ & high-level AUC $\downarrow$  \\
\mr
\geant &$0.609\pm 0.004$& $0.558\pm 0.002$\\ 
\mr
\submAmramCite       &$ \mathbf{0.680 \pm 0.002} $& $ \mathbf{0.652 \pm 0.006} $\\
\submFavaroCite      &$ 0.784 \pm 0.002 $& $ 0.732 \pm 0.002 $\\
\submLiuCite         &$ 0.958 \pm 0.002 $& $ 0.976 \pm 0.001 $\\
\submPangTCite       &$ 0.845 \pm 0.002 $& $ 0.797 \pm 0.002 $\\
\submPangSCite       &$ 0.884 \pm 0.002 $& $ 0.827 \pm 0.004 $\\
\submReyesCite       &$ 0.999 \pm 0.000 $& $ 0.999 \pm 0.000 $\\
\submRinaldiCite     &$ 0.913 \pm 0.002 $& $ 0.969 \pm 0.001 $\\
\submSalamaniDNNCite &$ \mathbf{0.676 \pm 0.004} $& $ 0.819 \pm 0.002 $\\
\submZhangCite       &$ 0.889 \pm 0.002 $& $ 0.922 \pm 0.001 $\\
\submErnstCite       &$ 0.853 \pm 0.003 $& $ 0.921 \pm 0.002 $\\
\submCresswellCite   &$ 0.909 \pm 0.002 $& $ 0.965 \pm 0.001 $\\
\submKobylyanskyCite &$ 0.811 \pm 0.005 $& $ 0.763 \pm 0.002 $\\
\br
\end{tabular}
\end{table}

\begin{table}
\caption{\label{tab:ds1-pions.kpd}KPD and FPD for evaluating \geant\ vs.~submission of \dsIpi. For visualization, see \Fref{fig:ds1-pions.kpd}.}
\centering
\begin{tabular}{@{}l|rr}
\br
Submission & KPD $\cdot 10^{3}$  $\downarrow$ & FPD $\cdot 10^{3} $$\downarrow$ \\
\mr
\geant & $-0.0075 \pm 0.0362$ &  $0.5110 \pm 0.0730$  \\ 
\mr
\submAmramCite       &  $ \mathbf{0.0893 \pm 0.0572}$  & $  \mathbf{2.7746 \pm 0.0922}$ \\
\submFavaroCite      &  $ 1.4781 \pm 0.2448$  & $ 29.2598 \pm 0.2133$ \\
\submLiuCite         &  $ 6.2679 \pm 0.2652$  & $126.9924 \pm 0.6750$ \\
\submPangTCite       &  $ 0.8083 \pm 0.0923$  & $ 25.6634 \pm 0.3002$ \\
\submPangSCite       &  $ 0.9937 \pm 0.0846$  & $ 25.6868 \pm 0.2292$ \\
\submReyesCite       &  $31.1636 \pm 0.9840$  & $524.4263 \pm 0.9067$ \\
\submRinaldiCite     &  $25.2457 \pm 0.9748$  & $498.6887 \pm 2.9504$ \\
\submSalamaniDNNCite &  $20.0149 \pm 1.0029$  & $464.6882 \pm 2.2500$ \\
\submZhangCite       &  $ 3.0518 \pm 0.1067$  & $113.2271 \pm 0.6673$ \\
\submErnstCite       &  $ 4.3241 \pm 0.2868$  & $ 82.9985 \pm 0.8746$ \\
\submCresswellCite   &  $ 9.5250 \pm 0.9264$  & $204.7435 \pm 1.6359$ \\
\submKobylyanskyCite &  $ 0.8500 \pm 0.2796$  & $ 22.3235 \pm 0.4428$ \\
\br
\end{tabular}
\end{table}

\begin{table}
\caption{\label{tab:ds1-pions.multi}Log-posterior scores for \dsIpi \geant test data, averaged over 10 independent classifier trainings. For visualization, see \Fref{fig:ds1-pions.multi}.}
\centering
\begin{tabular}{@{}l|r}
\br
Submission & Log-posterior $\uparrow$ \\
\mr
\submAmramCite       & $ -2.3189 \pm 0.0135$\\
\submFavaroCite      & $ -3.0949 \pm 0.0234$\\
\submLiuCite         & $ -8.5998 \pm 0.1748$\\
\submPangTCite       & $ -3.7110 \pm 0.0316$\\
\submPangSCite       & $ -4.5623 \pm 0.0365$\\
\submReyesCite       & $-25.9528 \pm 1.0659$\\
\submRinaldiCite     & $ -8.0202 \pm 0.1331$\\
\submSalamaniDNNCite & $ \mathbf{-1.9262 \pm 0.0239}$\\
\submZhangCite       & $ -4.7157 \pm 0.0807$\\
\submErnstCite       & $ -4.0618 \pm 0.0511$\\
\submCresswellCite   & $ -4.8811 \pm 0.0406$\\
\submKobylyanskyCite & $ -3.4762 \pm 0.0360$\\
\br
\end{tabular}
\end{table}

\begin{table}
\caption{\label{tab:ds1-pions.prdc}Precision, density, recall, and coverage for \dsIpi submissions. A visualization is shown in \fref{fig:ds1-pions.prdc}.}
\centering
\begin{tabular}{@{}l|rrrr}
\br
Submission & Precision $\uparrow$ & Density $\uparrow$ & Recall $\uparrow$ & Coverage $\uparrow$ \\
\mr
\geant & 0.643  & 0.782  &  0.739  & 0.913 \\
\mr
\submAmramCite       &  0.632 &  0.750  & 0.743 & 0.895 \\
\submFavaroCite      &  0.474 &  0.474  & 0.789 & 0.734 \\
\submLiuCite         &  0.954 & 54.862  & 0.134 & 0.947 \\
\submPangTCite       &  0.394 &  0.390  & 0.799 & 0.621 \\
\submPangSCite       &  0.414 &  0.445  & 0.772 & 0.625 \\
\submReyesCite       &  0.669 & 27.709  & 0.133 & 0.584 \\
\submRinaldiCite     &  0.268 &  0.487  & 0.335 & 0.386 \\
\submSalamaniDNNCite &  0.945 &  0.788  & 1.000 & 0.931 \\
\submZhangCite       &  0.710 &  2.803  & 0.185 & 0.855 \\
\submErnstCite       &  0.709 &  5.857  & 0.416 & 0.808 \\
\submCresswellCite   &  0.643 &  1.625  & 0.490 & 0.661 \\
\submKobylyanskyCite &  0.626 &  0.800  & 0.687 & 0.827  \\
\br
\end{tabular}
\end{table}
\begin{table}
\caption{\label{tab:ds1-pions.numparam}Number of trainable parameters in training and for generation for \dsIpi submissions. A visualization is shown in \fref{fig:ds1-pions.numparam}. }
\centering
\begin{tabular}{@{}l|rr}
\br
& \multicolumn{2}{c}{number of parameters $\downarrow$} \\
Submission & total & generator only \\
\mr
\submAmramCite       &$\mathbf{525\,901}$ &$\mathbf{525\,901}$  \\
\submFavaroCite      &    $26\,592\,624$ &    $26\,592\,624$  \\
\submLiuCite         &     $4\,314\,739$ &     $2\,237\,538$  \\
\submPangTCite       &    $57\,079\,326$ &    $57\,079\,326$  \\
\submPangSCite       &   $110\,389\,398$ &    $53\,426\,622$  \\
\submReyesCite       &    $18\,452\,248$ &    $16\,032\,327$  \\
\submRinaldiCite     &     $1\,678\,334$ &       $848\,733$  \\
\submSalamaniDNNCite &     $6\,052\,063$ &     $3\,169\,663$  \\
\submZhangCite       &     $1\,715\,742$ &       $880\,541$  \\
\submErnstCite       &    $17\,426\,875$ &     $9\,165\,275$  \\
\submCresswellCite   & $5\,297\,822\,388$ & $5\,297\,822\,388$  \\
\submKobylyanskyCite &       $823\,617$ &       $823\,617$  \\
\br
\end{tabular}
\end{table}
\begin{table}
\caption{\label{tab:ds1-pions.timing.CPU}Timing of \dsIpi submissions on a CPU. The symbols ${}^\diamond$, ${}^\ddag$, ${}^\ast$, and ${}^\dagger$ indicate that only 100 / 1000 / 10\,000 / 20\,000 events were generated in timing the submission. A visualization of these timings is shown in \fref{fig:ds1-pions.timing}.}
\begin{tabular}{@{}l|r@{}c@{}lr@{}c@{}lr@{}c@{}l}
\br
\multirow{2}{*}{Submission} & \multicolumn{9}{c}{CPU [ms per shower]} \\
& \multicolumn{3}{c}{batch size 1} & \multicolumn{3}{c}{batch size 100} & \multicolumn{3}{c}{batch size 10\,000} \\
\mr
\submAmramCite       & $15\,144$ & $\,\pm\,$ & $1126^{\ddag}$ & $4571$ & $\,\pm\,$ & $333^\ast$ & $4501$ & $\,\pm\,$ & $181^\dagger$ \\
\submFavaroCite      & $42.8$ & $\,\pm\,$ & $2.9$ & $3.92$ & $\,\pm\,$ & $0.36$ & $4.57$ & $\,\pm\,$ & $0.3$ \\
\submLiuCite         & $108.2$ & $\,\pm\,$ & $10.5$ & $12.7$ & $\,\pm\,$ & $1.4$ & $16.5$ & $\,\pm\,$ & $4.3$ \\
\submPangTCite       & $197\,570$ & $\,\pm\,$ & $34\,423^\diamond$ & $5430$ & $\,\pm\,$ & $489^\ast$ & $3509$ & $\,\pm\,$ & $177^\dagger$ \\
\submPangSCite       & $620.1$ & $\,\pm\,$ & $18.4$ & $14.2$ & $\,\pm\,$ & $2.2$ & $10.2$ & $\,\pm\,$ & $0.6$ \\
\submReyesCite       & $605.5$ & $\,\pm\,$ & $38.3$ & $12.1$ & $\,\pm\,$ & $1.3$ & $7.38$ & $\,\pm\,$ & $0.54$ \\
\submRinaldiCite     & $79.6$ & $\,\pm\,$ & $3.1$ & $1.38$ & $\,\pm\,$ & $0.05$ & $0.49$ & $\,\pm\,$ & $0.03$ \\
\submSalamaniDNNCite & $\mathbf{3.85}$ & $\,\pm\,$ & $\mathbf{0.96}$ & $\mathbf{0.47}$ & $\,\pm\,$ & $\mathbf{0.04}$ & $\mathbf{0.39}$ & $\,\pm\,$ & $\mathbf{0.03}$ \\
\submZhangCite       & $1163$ & $\,\pm\,$ & $301^\ast$ & $70.7$ & $\,\pm\,$ & $6.3$ & $1.60$ & $\,\pm\,$ & $0.07$ \\
\submErnstCite       & $40.8$ & $\,\pm\,$ & $1.1$ & $1.72$ & $\,\pm\,$ & $0.14$ & $1.20$ & $\,\pm\,$ & $0.14$ \\
\submCresswellCite   & $37\,876$ & $\,\pm\,$ & $1905^\ast$ & $432.8$ & $\,\pm\,$ & $22.9$ & $41.2$ & $\,\pm\,$ & $4.2$ \\
\submKobylyanskyCite & $3168$ & $\,\pm\,$ & $135^\ast$ & $1263$ & $\,\pm\,$ & $112^\ast$ & $2419$ & $\,\pm\,$ & $153^\dagger$ \\
\br
\end{tabular}
\end{table}

\begin{table}
\caption{\label{tab:ds1-pions.timing.GPU}Timing of \dsIpi submissions on a GPU. The symbols ${}^\ast, {}^\ddag$, and ${}^\dagger$ indicate that only 4000 / 5000 / 10\,000 events were generated in timing the submission; a -- indicates a model that does not run on a GPU; and ``CUDA o.o.m'' ran out of VRAM on the GPU. A visualization of these timings is shown in \fref{fig:ds1-pions.timing}.}
\begin{tabular}{@{}l|r@{}c@{}lr@{}c@{}lr@{}c@{}l}
\br
\multirow{2}{*}{Submission} & \multicolumn{9}{c}{GPU [ms per shower]} \\
& \multicolumn{3}{c}{batch size 1} & \multicolumn{3}{c}{batch size 100} & \multicolumn{3}{c}{batch size 10\,000} \\
\mr
\submAmramCite       & $5673$ & $\,\pm\,$ & $49^\ddag$ & $76.9$ & $\,\pm\,$ & $2.2$ & \quad$27.0$ & $\,\pm\,$ & $0.2$ \\
\submFavaroCite      & $24.7$ & $\,\pm\,$ & $2.0$ & $0.44$ & $\,\pm\,$ & $0.01$ & $\mathbf{0.20}$ & $\,\pm\,$ & $\mathbf{0.01}$ \\
\submLiuCite         & $53.3$ & $\,\pm\,$ & $0.8$ & $0.83$ & $\,\pm\,$ & $0.05$ & $\mathbf{0.18}$ & $\,\pm\,$ & $\mathbf{0.01}$ \\
\submPangTCite       & $6166$ & $\,\pm\,$ & $112^\ast$ & $70.1$ & $\,\pm\,$ & $1.0$ & $17.7$ & $\,\pm\,$ & $0.0$ \\
\submPangSCite       & $77.4$ & $\,\pm\,$ & $3.6$ & $1.00$ & $\,\pm\,$ & $0.02$ & $0.25$ & $\,\pm\,$ & $0.08$ \\
\submReyesCite       & $181.4$ & $\,\pm\,$ & $3.6$ & $2.07$ & $\,\pm\,$ & $0.02$ & $0.27$ & $\,\pm\,$ & $0.01$ \\
\submRinaldiCite     & $209.5$ & $\,\pm\,$ & $12.9$ & $1.59$ & $\,\pm\,$ & $0.13$ & $0.48$ & $\,\pm\,$ & $0.03$ \\
\submSalamaniDNNCite & $\mathbf{2.34}$ & $\,\pm\,$ & $\mathbf{0.17}$ & $\mathbf{0.32}$ & $\,\pm\,$ & $\mathbf{0.01}$ & $0.29$ & $\,\pm\,$ & $0.01$ \\
\submZhangCite       & $1119$ & $\,\pm\,$ & $56^\dagger$ & $122.6$ & $\,\pm\,$ & $1.5$ & $2.06$ & $\,\pm\,$ & $0.06$ \\
\submErnstCite       & $34.2$ & $\,\pm\,$ & $0.5$ & $0.68$ & $\,\pm\,$ & $0.02$ & $0.27$ & $\,\pm\,$ & $0.01$ \\
\submCresswellCite   & &--& & &--& & &--& \\
\submKobylyanskyCite & $1475$ & $\,\pm\,$ & $22^\dagger$ & $28.4$ & $\,\pm\,$ & $0.02$ & \multicolumn{3}{c}{CUDA o.o.m.} \\
\br
\end{tabular}
\end{table}
\subsection{\texorpdfstring{Dataset 2, Electrons (\dsII)}{Dataset 2, Electrons}}
\begin{table}
\caption{\label{tab:ds2.aucs}Low-level and high-level AUCs for evaluating \geant\ vs.~submission of \dsII, averaged over 10 independent evaluation runs. For visualization, see \Fref{fig:ds2.aucs}.}
\begin{tabular}{@{}l|rrr}
\br
\multirow{2}{*}{Submission} & \multicolumn{3}{c}{AUC $\downarrow$} \\
& low-level & high-level  &ResNet \\ 
\mr
\geant                & $ 0.500 \pm 0.002 $ & $ 0.499 \pm 0.002 $ & $ 0.500 \pm 0.004 $\\ 
\mr
\submAmramCite        & $ 0.577 \pm 0.004 $ & $ 0.591 \pm 0.009 $ & $ \mathbf{0.680 \pm 0.006} $\\
\submBussConvCite     & $ 0.708 \pm 0.004 $ & $ 0.737 \pm 0.002 $ & $ 0.941 \pm 0.003 $\\
\submFavaroCite       & $ 0.743 \pm 0.002 $ & $ 0.865 \pm 0.003 $ & $ 0.994 \pm 0.000 $\\
\submKaechCite        & $ 0.942 \pm 0.005 $ & $ 0.987 \pm 0.001 $ & $ 1.000 \pm 0.000 $\\
\submLiuCite          & $ 0.986 \pm 0.001 $ & $ 0.994 \pm 0.000 $ & $ 0.999 \pm 0.000 $\\
\submMikuniCite       & $ 0.595 \pm 0.003 $ & $ 0.666 \pm 0.002 $ & $ 0.795 \pm 0.011 $\\
\submMikuniDistCite   & $ 0.710 \pm 0.002 $ & $ 0.891 \pm 0.003 $ & $ 0.965 \pm 0.002 $\\
\submMikuniSingleCite & $ 0.747 \pm 0.003 $ & $ 0.902 \pm 0.002 $ & $ 0.973 \pm 0.002 $\\
\submPangITCite       & $ 0.763 \pm 0.004 $ & $ 0.837 \pm 0.005 $ & $ 0.970 \pm 0.002 $\\
\submPangISCite       & $ 0.819 \pm 0.004 $ & $ 0.886 \pm 0.003 $ & $ 0.975 \pm 0.002 $\\
\submPangSuperCite    & $ 0.694 \pm 0.006 $ & $ 0.757 \pm 0.004 $ & $ 0.986 \pm 0.001 $\\
\submSchamCite        & $ 0.963 \pm 0.002 $ & $ 0.927 \pm 0.002 $ & $ 0.999 \pm 0.000 $\\
\submSchnakeCite      & $ 0.863 \pm 0.005 $ & $ 0.908 \pm 0.004 $ & $ 0.999 \pm 0.000 $\\
\submErnstCite        & $ 0.907 \pm 0.004 $ & $ 1.000 \pm 0.000 $ & $ 0.993 \pm 0.001 $\\
\submMadulaCite       & $ 0.983 \pm 0.001 $ & $ 0.995 \pm 0.001 $ & $ 1.000 \pm 0.000 $\\
\submCardosoCite      & $ 0.984 \pm 0.001 $ & $ 0.912 \pm 0.002 $ & $ 0.988 \pm 0.001 $\\
\submPalaciosCite     & $ \mathbf{0.531 \pm 0.003} $ & $ \mathbf{0.521 \pm 0.002} $ & $ \mathbf{0.681 \pm 0.015} $\\
\br
\end{tabular}
\end{table}

\begin{table}
\caption{\label{tab:ds2.kpd}KPD and FPD for evaluating \geant\ vs.~submission of \dsII. For visualization, see \Fref{fig:ds2.kpd}.}
\centering
\begin{tabular}{@{}l|rr}
\br
Submission & KPD $\cdot 10^{3}$  $\downarrow$ & FPD $\cdot 10^{3} $ $\downarrow$ \\
\mr
\geant                &$-0.0276 \pm 0.0215$  &  $  10.7760 \pm  0.7901$ \\ 
\mr
\submAmramCite        &$ 0.1741 \pm 0.0422$  &  $ 146.9334 \pm  0.8703$ \\
\submBussConvCite     &$ 0.2705 \pm 0.0897$  &  $ 157.4047 \pm  0.9684$ \\
\submFavaroCite       &$ 2.8210 \pm 0.4194$  &  $ 732.8274 \pm  5.3303$ \\
\submKaechCite        &$ 4.9624 \pm 0.2728$  &  $ 864.9781 \pm  5.1452$ \\
\submLiuCite          &$ 8.5212 \pm 0.5043$  &  $1315.7233 \pm  7.0344$ \\
\submMikuniCite       &$ 0.1486 \pm 0.0568$  &  $ 112.4790 \pm  0.9080$ \\
\submMikuniDistCite   &$ 1.0129 \pm 0.0738$  &  $ 638.8525 \pm  1.5996$ \\
\submMikuniSingleCite &$ 0.9294 \pm 0.0684$  &  $ 546.2661 \pm  1.9396$ \\
\submPangITCite       &$ 0.5679 \pm 0.1375$  &  $ 377.0613 \pm  1.8961$ \\
\submPangISCite       &$ 1.0406 \pm 0.2190$  &  $ 449.2585 \pm  3.2844$ \\
\submPangSuperCite    &$ 0.5564 \pm 0.1900$  &  $ 300.8183 \pm  2.7275$ \\
\submSchamCite        &$ 0.6803 \pm 0.1285$  &  $ 292.6319 \pm  2.9330$ \\
\submSchnakeCite      &$ 0.3241 \pm 0.0392$  &  $ 494.0547 \pm  1.7906$ \\
\submErnstCite        &$45.6091 \pm 0.8315$  &  $5443.4295 \pm 27.2305$ \\
\submMadulaCite       &$ 2.8791 \pm 0.1998$  &  $ 962.9750 \pm  2.4089$ \\
\submCardosoCite      &$11.0322 \pm 0.4274$  &  $1690.9873 \pm  6.7650$ \\
\submPalaciosCite     &$ \mathbf{0.0231 \pm 0.0364}$  &  $  \mathbf{24.6488 \pm  1.0350}$ \\
\br
\end{tabular}
\end{table}

\begin{table}
\caption{\label{tab:ds2.multi.dnn}Log-posterior scores for \dsII \geant test data, averaged over 10 independent DNN classifier trainings. For visualization, see \Fref{fig:ds2.multi.dnn}.}
\centering
\begin{tabular}{@{}l|r}
\br
Submission & Log-posterior $\uparrow$ \\
\mr
\submAmramCite        & $ -2.5226 \pm 0.0094$\\
\submBussConvCite     & $ -3.1295 \pm 0.0219$\\
\submFavaroCite       & $ -3.2032 \pm 0.0153$\\
\submKaechCite        & $ -7.4399 \pm 0.3234$\\
\submLiuCite          & $-11.8863 \pm 0.2627$\\
\submMikuniCite       & $ -2.8415 \pm 0.0238$\\
\submMikuniDistCite   & $ -3.4226 \pm 0.0556$\\
\submMikuniSingleCite & $ -3.5974 \pm 0.0685$\\
\submPalaciosCite     & $ \mathbf{-2.4102 \pm 0.0105}$\\
\submPangITCite       & $ -3.6423 \pm 0.0286$\\
\submPangISCite       & $ -4.2617 \pm 0.0360$\\
\submPangSuperCite    & $ -2.8204 \pm 0.0237$\\
\submCardosoCite      & $-22.1206 \pm 0.4549$\\
\submSchamCite        & $-10.5062 \pm 0.2094$\\
\submSchnakeCite      & $ -5.2306 \pm 0.1996$\\
\submErnstCite        & $ -6.4103 \pm 0.1439$\\
\submMadulaCite       & $-11.6683 \pm 0.2679$\\
\br
\end{tabular}
\end{table}

\begin{table}
\caption{\label{tab:ds2.multi.resnet}Log-posterior scores for \dsII \geant test data, averaged over 10 independent CNN ResNet classifier trainings. For visualization, see \Fref{fig:ds2.multi.resnet}.}
\centering
\begin{tabular}{@{}l|r}
\br
Submission & Log-posterior $\uparrow$ \\
\mr
\submAmramCite        & $ \mathbf{-1.9901 \pm 0.2358}$\\
\submBussConvCite     & $ -2.4500 \pm 0.1546$\\
\submFavaroCite       & $ -7.2706 \pm 0.4780$\\
\submKaechCite        & $ -9.3476 \pm 0.4435$\\
\submLiuCite          & $ -7.0740 \pm 0.4898$\\
\submMikuniCite       & $ \mathbf{-2.1544 \pm 0.1303}$\\
\submMikuniDistCite   & $ -4.1027 \pm 0.4292$\\
\submMikuniSingleCite & $ -4.6509 \pm 0.4837$\\
\submPalaciosCite     & $ \mathbf{-1.9761 \pm 0.1203}$\\
\submPangITCite       & $ -3.9376 \pm 0.2385$\\
\submPangISCite       & $ -4.6476 \pm 0.2627$\\
\submPangSuperCite    & $ -4.4702 \pm 0.4064$\\
\submCardosoCite      & $ -5.8461 \pm 0.5629$\\
\submSchamCite        & $ -8.5889 \pm 0.6987$\\
\submSchnakeCite      & $ -9.0910 \pm 0.4704$\\
\submErnstCite        & $ -6.9001 \pm 0.3468$\\
\submMadulaCite       & $ -8.2169 \pm 0.5325$\\

\br
\end{tabular}
\end{table}

\begin{table}
\caption{\label{tab:ds2.prdc}Precision, density, recall, and coverage for \dsII submissions. A visualization is shown in \fref{fig:ds2.prdc}.}
\centering
\begin{tabular}{@{}l|rrrr}
\br
Submission & Precision $\uparrow$ & Density $\uparrow$ & Recall $\uparrow$ & Coverage $\uparrow$ \\
\mr
\geant & 0.239  &  1.021   & 0.241 & 0.971 \\
\mr
\submAmramCite        &   0.239  &    1.236  &  0.235 & 0.933  \\
\submBussConvCite     &	  0.231  &    1.656  &  0.177 & 0.969  \\
\submFavaroCite       &	  0.193  &    4.573  &  0.090 & 0.957  \\
\submKaechCite        &	  0.003  &    0.009  &  0.937 & 0.033  \\
\submLiuCite          &	  0.345  &  254.397  &  0.217 & 0.868  \\
\submMikuniCite       &	  0.228  &    1.013  &  0.228 & 0.933  \\
\submMikuniDistCite   &	  0.197  &    1.407  &  0.181 & 0.880  \\
\submMikuniSingleCite &	  0.171  &    1.056  &  0.208 & 0.852  \\
\submPangITCite       &	  0.152  &    0.809  &  0.253 & 0.817  \\
\submPangISCite       &	  0.155  &    1.354  &  0.253 & 0.827  \\
\submPangSuperCite    &	  0.120  &    0.347  &  0.310 & 0.692  \\
\submSchamCite        &	  0.003  &    0.013  &  0.834 & 0.045  \\
\submSchnakeCite      &	  0.016  &    0.575  &  0.335 & 0.487  \\
\submErnstCite        &	  0.739  & 1793.855  &  0.026 & 0.961  \\
\submMadulaCite       &	  0.016  &    0.176  &  0.622 & 0.235  \\
\submCardosoCite      &	  0.500  &   10.228  &  0.060 & 0.924  \\
\submPalaciosCite     &	  0.253  &    1.146  &  0.220 & 0.976  \\
\br
\end{tabular}
\end{table}
\begin{table}
\caption{\label{tab:ds2.numparam}Number of trainable parameters in training and for generation for \dsII submissions. A visualization is shown in \fref{fig:ds2.numparam}. }
\centering
\begin{tabular}{@{}l|rr}
\br
& \multicolumn{2}{c}{number of parameters $\downarrow$} \\
Submission & total & generator only \\
\mr
\submAmramCite        &     $517\,969$ &     $517\,969$  \\
\submBussConvCite     & $158\,017\,226$ & $158\,017\,226$  \\
\submFavaroCite       & $270\,999\,370$ & $270\,999\,370$  \\
\submKaechCite        &$\mathbf{108\,656}$ &$\mathbf{66\,416}$  \\
\submLiuCite          &   $3\,317\,546$ &   $1\,231\,433$  \\
\submMikuniCite       &  $14\,436\,206$ &  $14\,436\,206$  \\
\submMikuniDistCite   &  $28\,872\,412$ &  $14\,436\,206$  \\
\submMikuniSingleCite &  $28\,872\,412$ &  $14\,436\,206$  \\
\submPangITCite       &  $19\,470\,168$ &  $19\,470\,168$  \\
\submPangISCite       &  $41\,237\,080$ &  $24\,519\,512$  \\
\submPangSuperCite    &  $87\,465\,608$ &  $87\,465\,608$  \\
\submSchamCite        &   $2\,240\,496$ &     $527\,676$  \\
\submSchnakeCite      &  $14\,215\,334$ &  $14\,215\,334$  \\
\submErnstCite        &  $96\,356\,674$ &  $48\,393\,824$  \\
\submMadulaCite       &  $10\,707\,408$ &   $1\,942\,402$  \\
\submCardosoCite      &   $1\,221\,544$ &   $1\,221\,544$  \\
\submPalaciosCite     &  $28\,427\,393$ &  $28\,427\,393$  \\
\br
\end{tabular}
\end{table}
\begin{table}
\caption{\label{tab:ds2.timing.CPU}Timing of \dsII submissions on a CPU. Superscripts indicate if fewer than 100\,000 events were generated in timing the submission. ``LLVM: o.o.m.'' crashes with an LLVM out of memory error, ``o.o.m.'' crashes with an memory allocation error, ``$>17\,280$'' translates to $>$48h/batch. A visualization of these timings is shown in \fref{fig:ds2.timing}.}
\begin{tabular}{@{}l|r@{}c@{}lr@{}c@{}lr@{}c@{}l}
\br
\multirow{2}{*}{Submission} & \multicolumn{9}{c}{CPU [ms per shower]} \\ 
& \multicolumn{3}{c}{batch size 1} & \multicolumn{3}{c}{batch size 100} & \multicolumn{3}{c}{batch size 10\,000} \\
\mr
\submAmramCite        & $36\,502$ & $\,\pm\,$ & $2127^{(1000)}$ & $23\,317$ & $\,\pm\,$ & $3343^{(1000)}$ & \multicolumn{3}{c}{$>17\,280$} \\
\submBussConvCite     & $1969$ & $\,\pm\,$ & $234^{(10\,000)}$ & $121.2$ & $\,\pm\,$ & $4.4$ & \quad $89.2$ & $\,\pm\,$ & $13.2$ \\
\submFavaroCite       & $387.7$ & $\,\pm\,$ & $31.6$ & $59.7$ & $\,\pm\,$ & $10.1$ & $46.2$ & $\,\pm\,$ & $0.7$ \\
\submKaechCite        & $\mathbf{14.3}$ & $\,\pm\,$ & $\mathbf{0.6}$ & $19.9$ & $\,\pm\,$ & $5.3$ & $32.4$ & $\,\pm\,$ & $4.5$ \\
\submLiuCite          & $168.4$ & $\,\pm\,$ & $7.3$ & $26.1$ & $\,\pm\,$ & $1.2$ & $36.2$ & $\,\pm\,$ & $9.6$ \\
\submMikuniCite       & $133\,549$ & $\,\pm\,$ & $5286^{(1000)}$ & $147\,694$ & $\,\pm\,$ & $12\,981^{(500)}$ & \multicolumn{3}{c}{LLVM: o.o.m.} \\
\end{tabular}
\end{table}

\begin{table}
\caption{\label{tab:ds2.timing.GPU}Timing of \dsII submissions on a GPU. Superscripts indicate if fewer than 100\,000 events were generated in timing the submission, ``CUDA o.o.m.'' ran out of VRAM on the GPU; and ``array o.o.m.'' crashed because created arrays were too large. A visualization of these timings is shown in \fref{fig:ds2.timing}.}
\begin{tabular}{@{}l|r@{}c@{}lr@{}c@{}lr@{}c@{}l}
\br
\multirow{2}{*}{Submission} & \multicolumn{9}{c}{GPU [ms per shower]} \\
& \multicolumn{3}{c}{batch size 1} & \multicolumn{3}{c}{batch size 100} & \multicolumn{3}{c}{batch size 10\,000} \\
\mr
\submAmramCite        & $5291$ & $\,\pm\,$ & $88^{(5000)}$ & $99.5$ & $\,\pm\,$ & $1.8$ & \multicolumn{3}{c}{CUDA o.o.m.} \\
\submBussConvCite     & $1409$ & $\,\pm\,$ & $29^{(10\,000)}$ & $14.2$ & $\,\pm\,$ & $0.1$ & \quad $1.64$ & $\,\pm\,$ & $0.01$ \\
\submFavaroCite       & $53.4$ & $\,\pm\,$ & $0.9$ & $\mathbf{1.18}$ & $\,\pm\,$ & $\mathbf{0.03}$ & $0.65$ & $\,\pm\,$ & $0.03$ \\
\submKaechCite        & $\mathbf{6.4}$ & $\,\pm\,$ & $\mathbf{0.1}$ & $\mathbf{1.2}$ & $\,\pm\,$ & $\mathbf{0.1}$ & \multicolumn{3}{c}{CUDA o.o.m.} \\
\submLiuCite          & $72.8$ & $\,\pm\,$ & $1.0$ & $\mathbf{1.1}$ & $\,\pm\,$ & $\mathbf{0.1}$ & $\mathbf{0.36}$ & $\,\pm\,$ & $\mathbf{0.01}$ \\
\submMikuniCite       & $2389$ & $\,\pm\,$ & $92^{(5000)}$ & $241.3$ & $\,\pm\,$ & $0.6$ & \multicolumn{3}{c}{array o.o.m.} \\
\end{tabular}
\end{table}
\subsection{\texorpdfstring{Dataset 3, Electrons (\dsIII)}{Dataset 3, Electrons}}
\begin{table}
\caption{\label{tab:ds3.aucs}Low-level and high-level AUCs for evaluating \geant\ vs.~submission of \dsIII, averaged over 10 independent evaluation runs. For visualization, see \Fref{fig:ds3.aucs}.}
\centering
\begin{tabular}{@{}l|rrr}
\br
\multirow{2}{*}{Submission} & \multicolumn{3}{c}{AUC $\downarrow$} \\
& low-level & high-level  &ResNet \\
\mr
\geant                 &$ 0.498 \pm 0.002 $&$ 0.500 \pm 0.003 $& $ 0.499 \pm 0.002 $\\ 
\mr
\submAmramCite         &$ \mathbf{0.561 \pm 0.003} $&$ 0.607 \pm 0.005 $& $ \mathbf{0.656 \pm 0.015} $\\
\submBussMAFCite       &$ 0.720 \pm 0.016 $&$ 0.946 \pm 0.002 $& $ 1.000 \pm 0.000 $\\
\submBussConvCite      &$ 0.588 \pm 0.004 $&$ 0.733 \pm 0.006 $& $ 0.919 \pm 0.003 $\\
\submKaechCite         &$ 0.944 \pm 0.002 $&$ 0.987 \pm 0.001 $& $ 1.000 \pm 0.000 $\\
\submKorolCite         &$ 0.865 \pm 0.005 $&$ 0.980 \pm 0.001 $& $ 1.000 \pm 0.000 $\\
\submLiuCite           &$ 0.996 \pm 0.001 $&$ 0.998 \pm 0.000 $& $ 1.000 \pm 0.000 $\\
\submLiuNormCite       &$ 0.975 \pm 0.003 $&$ 0.994 \pm 0.000 $& $ 1.000 \pm 0.000 $\\
\submMikuniDistCite    &$ 0.776 \pm 0.005 $&$ 0.924 \pm 0.002 $& $ 0.994 \pm 0.001 $\\
\submMikuniSingleCite  &$ 0.807 \pm 0.005 $&$ 0.939 \pm 0.001 $& $ 0.995 \pm 0.002 $\\
\submPangITCite        &$ 0.911 \pm 0.003 $&$ 0.962 \pm 0.001 $& $ 1.000 \pm 0.000 $\\
\submPangISCite        &$ 0.891 \pm 0.003 $&$ 0.971 \pm 0.001 $& $ 1.000 \pm 0.000 $\\
\submSalamaniTransCite &$ 0.886 \pm 0.011 $&$ 1.000 \pm 0.000 $& $ 1.000 \pm 0.000 $\\
\submSchnakeCite       &$ 0.720 \pm 0.012 $&$ 0.945 \pm 0.002 $& $ 1.000 \pm 0.000 $\\
\submErnstCite         &$ 0.881 \pm 0.005 $&$ 1.000 \pm 0.000 $& $ 1.000 \pm 0.000 $\\
\submPalaciosCite      &$ 0.630 \pm 0.005 $&$ \mathbf{0.524 \pm 0.004} $& $ 0.802 \pm 0.014 $\\
\br
\end{tabular}
\end{table}

\begin{table}
\caption{\label{tab:ds3.kpd}KPD and FPD for evaluating \geant\ vs.~submission of \dsIII. For visualization, see \Fref{fig:ds3.kpd}.}
\centering
\begin{tabular}{@{}l|rr}
\br
Submission & KPD $\cdot 10^{3}$  $\downarrow$ & FPD $\cdot 10^{3} $$\downarrow$ \\
\mr
\geant                 &$ -0.0091 \pm 0.0466$  & $    8.7578 \pm  0.5587$ \\ 
\mr
\submAmramCite         &$  0.2278 \pm 0.0978$  & $   71.2380 \pm  1.9208$ \\
\submBussMAFCite       &$  1.5398 \pm 0.1831$  & $  665.4975 \pm  1.6930$ \\
\submBussConvCite      &$  0.3245 \pm 0.1521$  & $  171.6365 \pm  1.7965$ \\
\submKaechCite         &$  1.6705 \pm 0.1370$  & $  588.6035 \pm  2.5358$ \\
\submKorolCite         &$  5.1826 \pm 0.7016$  & $  948.2275 \pm  4.6265$ \\
\submLiuCite           &$  5.6838 \pm 0.2075$  & $ 1193.9149 \pm  2.8258$ \\
\submLiuNormCite       &$  3.9937 \pm 0.3564$  & $  930.8472 \pm  3.4598$ \\
\submMikuniDistCite    &$  1.7304 \pm 0.2490$  & $  610.8560 \pm  4.0175$ \\
\submMikuniSingleCite  &$  1.5934 \pm 0.1380$  & $  584.0234 \pm  2.9294$ \\
\submPangITCite        &$  2.8602 \pm 0.3240$  & $  897.5908 \pm  5.2608$ \\
\submPangISCite        &$  2.5991 \pm 0.2299$  & $  841.1136 \pm  5.1413$ \\
\submSalamaniTransCite &$241.0380 \pm 2.6919$  & $22947.3168 \pm 23.3703$ \\
\submSchnakeCite       &$  2.0229 \pm 0.4123$  & $  670.7538 \pm  3.3806$ \\
\submErnstCite         &$ 83.0692 \pm 0.7260$  & $11060.7266 \pm 13.9947$ \\
\submPalaciosCite      &$  \mathbf{0.0098 \pm 0.0145}$  & $   \mathbf{20.7469 \pm  1.0767}$ \\
\br
\end{tabular}
\end{table}
\begin{table}
\caption{\label{tab:ds3.multi.dnn}Log-posterior scores for \dsIII \geant test data, averaged over 10 independent DNN classifier trainings. For visualization, see \Fref{fig:ds3.multi.dnn}.}
\centering
\begin{tabular}{@{}l|r}
\br
Submission & Log-posterior $\uparrow$ \\
\mr
\submAmramCite         & $ -2.3860 \pm 0.0063$\\
\submBussMAFCite       & $ -4.3836 \pm 0.1634$\\
\submBussConvCite      & $ -2.6992 \pm 0.0171$\\
\submKaechCite         & $ -9.8424 \pm 0.3800$\\
\submKorolCite         & $ -5.9925 \pm 0.2542$\\
\submLiuCite           & $-19.8196 \pm 0.6060$\\
\submLiuNormCite       & $ -9.6010 \pm 0.2645$\\
\submMikuniDistCite    & $ -3.2759 \pm 0.0844$\\
\submMikuniSingleCite  & $ -3.8002 \pm 0.1135$\\
\submPalaciosCite      & $ \mathbf{-2.2337 \pm 0.0132}$\\
\submPangITCite        & $ -6.7583 \pm 0.2071$\\
\submPangISCite        & $ -5.8949 \pm 0.1835$\\
\submSalamaniTransCite & $-12.0696 \pm 0.4366$\\
\submSchnakeCite       & $ -3.6853 \pm 0.1748$\\
\submErnstCite         & $ -6.4805 \pm 0.2134$\\
\br
\end{tabular}
\end{table}

\begin{table}
\caption{\label{tab:ds3.multi.resnet}Log-posterior scores for \dsIII \geant test data, averaged over 10 independent CNN ResNet classifier trainings. For visualization, see \Fref{fig:ds3.multi.resnet}.}
\centering
\begin{tabular}{@{}l|r}
\br
Submission & Log-posterior $\uparrow$ \\
\mr
\submAmramCite         & $ \mathbf{-1.1254 \pm 0.1035}$\\
\submBussMAFCite       & $ -9.8665 \pm 0.5211$\\
\submBussConvCite      & $ -2.3183 \pm 0.1994$\\
\submKaechCite         & $-10.1538 \pm 0.6095$\\
\submKorolCite         & $-10.2566 \pm 0.5892$\\
\submLiuCite           & $ -8.9864 \pm 0.3448$\\
\submLiuNormCite       & $ -8.3526 \pm 0.4014$\\
\submMikuniDistCite    & $ -5.3080 \pm 0.4710$\\
\submMikuniSingleCite  & $ -6.3708 \pm 0.5287$\\
\submPalaciosCite      & $ -1.4174 \pm 0.1267$\\
\submPangITCite        & $ -9.1767 \pm 0.4420$\\
\submPangISCite        & $ -9.3465 \pm 0.5997$\\
\submSalamaniTransCite & $ -9.9694 \pm 0.5631$\\
\submSchnakeCite       & $-10.1480 \pm 0.3690$\\
\submErnstCite         & $ -9.3558 \pm 0.4615$\\
\br
\end{tabular}
\end{table}

\begin{table}
\caption{\label{tab:ds3.prdc}Precision, density, recall, and coverage for \dsIII submissions. A visualization is shown in \fref{fig:ds3.prdc}.}
\centering
\begin{tabular}{@{}l|rrrr}
\br
Submission & Precision $\uparrow$ & Density $\uparrow$ & Recall $\uparrow$ & Coverage $\uparrow$ \\
\mr
\geant & 0.098  &  1.145  & 0.095  &  0.980  \\
\mr
\submAmramCite         &   0.122  &    1.118   & 0.090 & 0.973 \\
\submBussMAFCite       &   0.079  &    6.126   & 0.068 & 0.870 \\
\submBussConvCite      &   0.109  &    1.667   & 0.065 & 0.887 \\
\submKaechCite         &   0.000  &    0.097   & 0.790 & 0.233 \\
\submKorolCite         &   0.001  &    0.000   & 0.841 & 0.002 \\
\submLiuCite           &   0.313  &  199.747   & 0.000 & 0.948 \\
\submLiuNormCite       &   0.561  &  227.616   & 0.002 & 1.000 \\
\submMikuniDistCite    &   0.078  &    1.020   & 0.085 & 0.708 \\
\submMikuniSingleCite  &   0.038  &    0.262   & 0.153 & 0.536 \\
\submPangITCite        &   0.070  &    0.989   & 0.123 & 0.786 \\
\submPangISCite        &   0.079  &    1.463   & 0.115 & 0.793 \\
\submSalamaniTransCite &   0.623  & 1570.369   & 0.074 & 0.478 \\
\submSchnakeCite       &   0.004  &    0.045   & 0.284 & 0.171 \\
\submErnstCite         &   0.879  & 1990.146   & 0.009 & 0.787 \\
\submPalaciosCite      &   0.114  &    1.637   & 0.079 & 0.989 \\ 
\br
\end{tabular}
\end{table}
\begin{table}
\caption{\label{tab:ds3.numparam}Number of trainable parameters in training and for generation for \dsIII submissions. A visualization is shown in \fref{fig:ds3.numparam}. }
\centering
\begin{tabular}{@{}l|rr}
\br
& \multicolumn{2}{c}{number of parameters $\downarrow$} \\
Submission & total & generator only  \\
\mr
\submAmramCite         &     $1\,221\,153$ &     $1\,221\,153$  \\
\submBussMAFCite       &   $556\,526\,578$ &   $556\,526\,578$  \\
\submBussConvCite      &   $194\,964\,482$ &   $194\,964\,482$  \\
\submKaechCite         &$\mathbf{108\,656}$ &$\mathbf{66\,416}$  \\
\submKorolCite         &    $77\,475\,856$ &    $77\,475\,856$  \\
\submLiuCite           &     $2\,155\,763$ &       $876\,050$  \\
\submLiuNormCite       &     $2\,767\,443$ &     $1\,471\,282$  \\
\submMikuniDistCite    &    $28\,872\,412$ &    $14\,436\,206$  \\
\submMikuniSingleCite  &    $28\,872\,412$ &    $14\,436\,206$  \\
\submPangITCite        &    $95\,088\,152$ &    $95\,088\,152$  \\
\submPangISCite        &   $187\,423\,704$ &    $95\,088\,152$  \\
\submSalamaniTransCite & $1\,262\,921\,306$ & $1\,262\,921\,306$  \\
\submSchnakeCite       &    $14\,215\,334$ &    $14\,215\,334$  \\
\submErnstCite         &   $204\,609\,270$ &    $93\,935\,070$  \\
\submPalaciosCite      &     $8\,253\,575$ &     $8\,253\,575$  \\
\br
\end{tabular}
\end{table}
\begin{table}
\caption{\label{tab:ds3.timing.CPU}Timing of \dsIII submissions on a CPU. Superscripts indicate if fewer than 100\,000 events were generated in timing the submission, ``LLVM: o.o.m.'' crashes with an LLVM out of memory error, ``o.o.m.'' crashes with an memory allocation error, ``$>17\,280$'' translates to $>$48h/batch. A visualization of these timings is shown in \fref{fig:ds3.timing}.}
\hspace*{-3em}
\begin{tabular}{@{}l|r@{}c@{}lr@{}c@{}lr@{}c@{}l}
\br
\multirow{2}{*}{Submission} & \multicolumn{9}{c}{CPU [ms per shower]} \\
& \multicolumn{3}{c}{batch size 1} & \multicolumn{3}{c}{batch size 100} & \multicolumn{3}{c}{batch size 10\,000} \\
\mr
\submAmramCite         & $205\,235$ & $\,\pm\,$ & $6154^{(100)}$ & $269\,230$ & $\,\pm\,$ & $50\,007^{(100)}$ & \multicolumn{3}{c}{o.o.m.} \\
\submBussMAFCite       & $880\,131$ & $\,\pm\,$ & $61\,699^{(100)}$ & $141\,317$ & $\,\pm\,$ & $8151^{(1000)}$ & \multicolumn{3}{c}{$>17\,280$} \\
\submBussConvCite      & $2357$ & $\,\pm\,$ & $283^{(10\,000)}$ & $340.2$ & $\,\pm\,$ & $15.2^{(10\,000)}$ & \quad $428.7$ & $\,\pm\,$ & $25.7$ \\
\submKaechCite         & $\mathbf{32.1}$ & $\,\pm\,$ & $\mathbf{1.4}$ & $193.7$ & $\,\pm\,$ & $12.7$ & \multicolumn{3}{c}{o.o.m.} \\
\submKorolCite         & $2404$ & $\,\pm\,$ & $96^{(10\,000)}$ & $3924$ & $\,\pm\,$ & $701^{(10\,000)}$ & \multicolumn{3}{c}{o.o.m.} \\
\submLiuCite           & $260.0$ & $\,\pm\,$ & $16.3$ & $68.1$ & $\,\pm\,$ & $4.7$ & $58.2$ & $\,\pm\,$ & $8.4$ \\
\end{tabular}
\end{table}

\begin{table}
\caption{\label{tab:ds3.timing.GPU}Timing of \dsIII submissions on a GPU. Superscripts indicate if fewer than 100\,000 events were generated in timing the submission, ``CUDA o.o.m.'' ran out of VRAM on the GPU. A visualization of these timings is shown in \fref{fig:ds3.timing}.}
\begin{tabular}{@{}l|r@{}c@{}lr@{}c@{}lr@{}c@{}l}
\br
\multirow{2}{*}{Submission} & \multicolumn{9}{c}{GPU [ms per shower]} \\
& \multicolumn{3}{c}{batch size 1} & \multicolumn{3}{c}{batch size 100} & \multicolumn{3}{c}{batch size 10\,000} \\
\mr
\submAmramCite         & $6171$ & $\,\pm\,$ & $45^{(2500)}$ & $810.2$ & $\,\pm\,$ & $2.4^{(10\,000)}$ & \multicolumn{3}{c}{CUDA o.o.m.} \\
\submBussMAFCite       & $235\,537$ & $\,\pm\,$ & $4910^{(100)}$ & $2454$ & $\,\pm\,$ & $120^{(10\,000)}$ & \quad $3112$ & $\,\pm\,$ & $1^{(10\,000)}$ \\
\submBussConvCite      & $1430$ & $\,\pm\,$ & $16^{(10\,000)}$ & $16.0$ & $\,\pm\,$ & $0.2$ & $6.8$ & $\,\pm\,$ & $0.1$ \\
\submKaechCite         & $\mathbf{9.51}$ & $\,\pm\,$ & $\mathbf{0.13}$ & $5.47$ & $\,\pm\,$ & $0.30$ & \multicolumn{3}{c}{CUDA o.o.m.} \\
\submKorolCite         & $94.9$ & $\,\pm\,$ & $1.1$ & $25.3$ & $\,\pm\,$ & $0.2$ & \multicolumn{3}{c}{CUDA o.o.m.} \\
\submLiuCite           & $127.4$ & $\,\pm\,$ & $1.9$ & $\mathbf{1.8}$ & $\,\pm\,$ & $\mathbf{0.1}$ & $0.98$ & $\,\pm\,$ & $0.02$ \\
\end{tabular}
\end{table}

\section{Generation time vs. number of parameters}
\label{app:time.num_param}
\begin{figure}[ht]
\centering
\includegraphics[width=\textwidth]{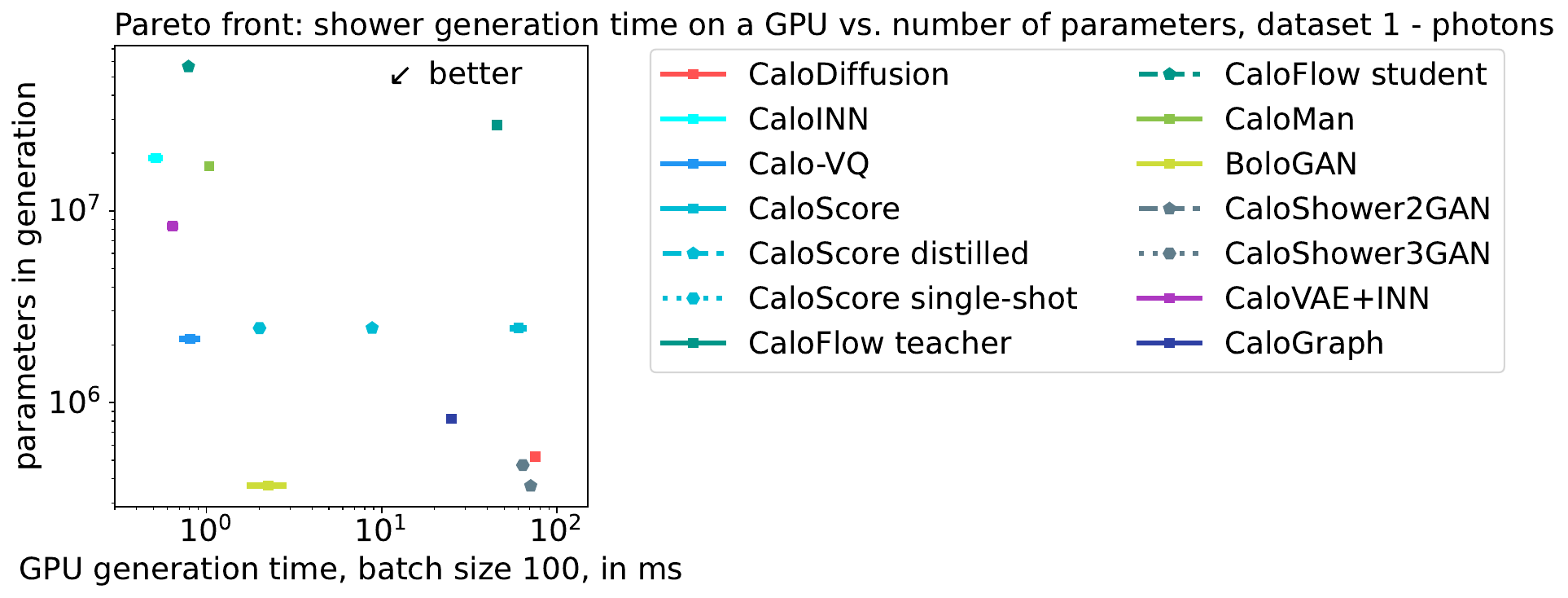}
\caption{Pareto front in number of trainable parameters in generation (from \fref{fig:ds1-photons.numparam} and \tref{tab:ds1-photons.numparam}) and generation speed (from \fref{fig:ds1-photons.timing} and \tref{tab:ds1-photons.timing.GPU}).}
\label{fig:ds1-photons.pareto.speed.size}
\end{figure}

\begin{figure}[ht]
\centering
\includegraphics[width=\textwidth]{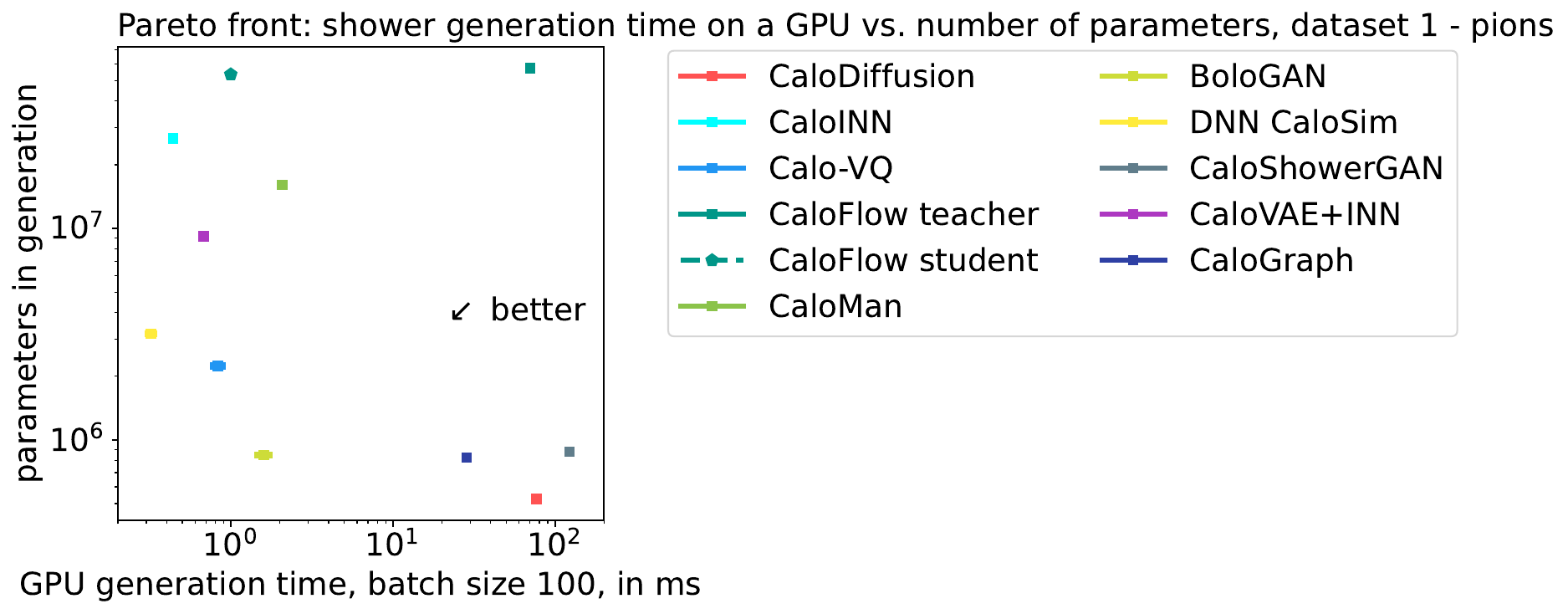}
\caption{Pareto front in number of trainable parameters in generation (from \fref{fig:ds1-pions.numparam} and \tref{tab:ds1-pions.numparam}) and generation speed (from \fref{fig:ds1-pions.timing} and \tref{tab:ds1-pions.timing.GPU}).}
\label{fig:ds1-pions.pareto.speed.size}
\end{figure}

\begin{figure}[ht]
\centering
\includegraphics[width=\textwidth]{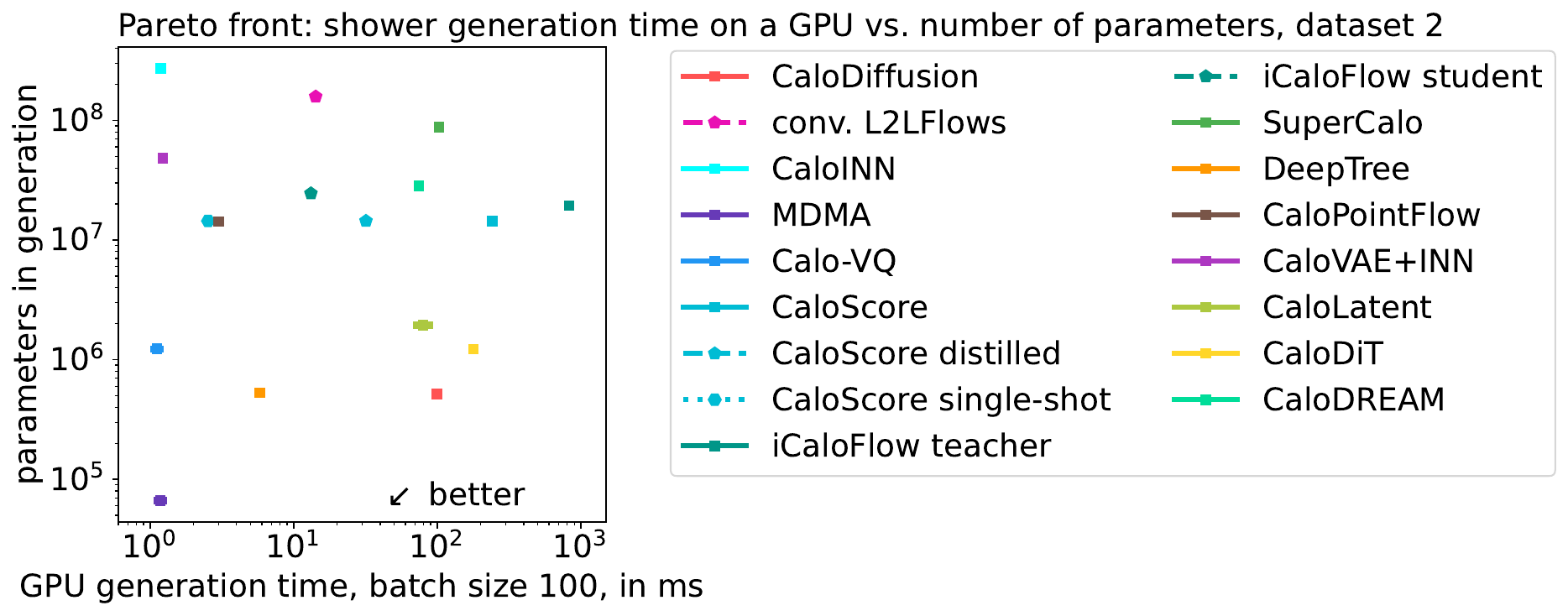}
\caption{Pareto front in number of trainable parameters in generation (from \fref{fig:ds2.numparam} and \tref{tab:ds2.numparam}) and generation speed (from \fref{fig:ds2.timing} and \tref{tab:ds2.timing.GPU}).}
\label{fig:ds2.pareto.speed.size}
\end{figure}

\begin{figure}[ht]
\centering
\includegraphics[width=\textwidth]{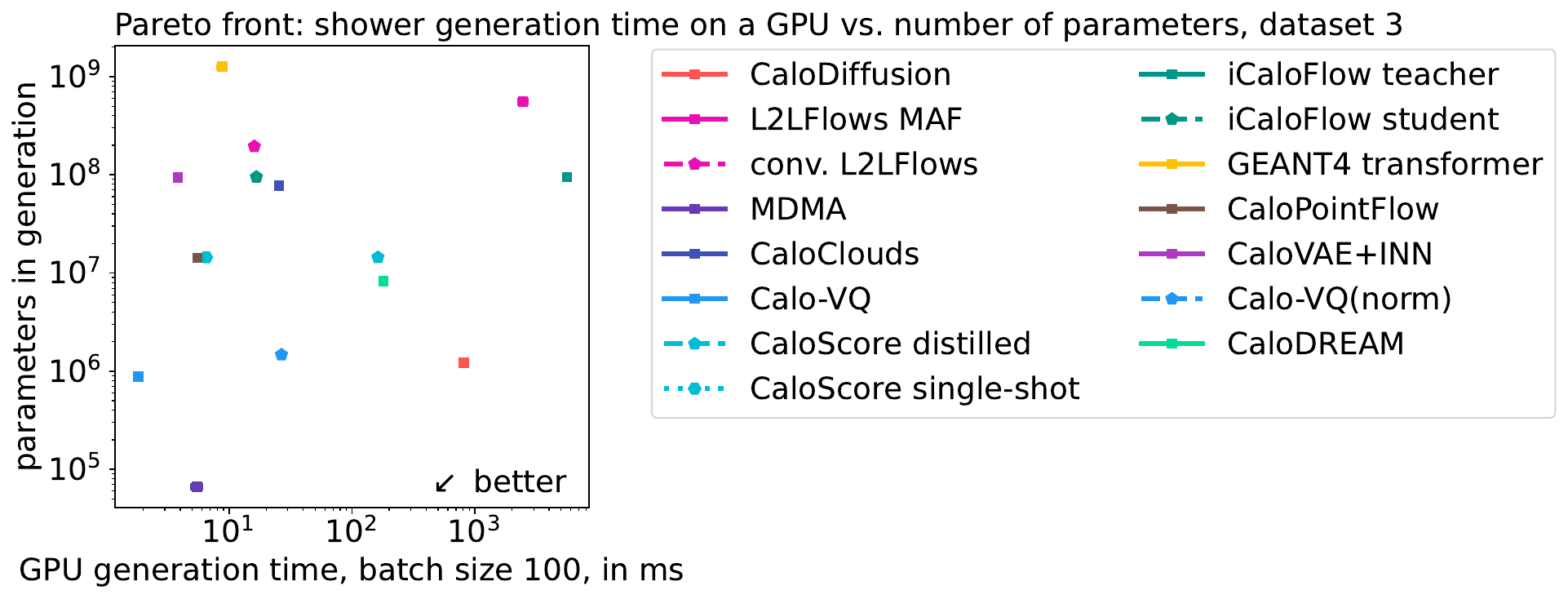}
\caption{Pareto front in number of trainable parameters in generation (from \fref{fig:ds3.numparam} and \tref{tab:ds3.numparam}) and generation speed (from \fref{fig:ds3.timing} and \tref{tab:ds3.timing.GPU}).}
\label{fig:ds3.pareto.speed.size}
\end{figure}
When looking at the Pareto front in generation speed (here taken as the time it takes to generate in batches of 100 on a GPU) and the model size (in terms of number of trainable parameters in generation), we barely see an actual front emerging. The generation time strongly depends on the model architecture and not so much on the actual size of the submissions, as can be seen for example by the \cf examples in \dsIph in \fref{fig:ds1-photons.pareto.speed.size}: \submPangS has more parameters than \submPangT, but is almost 2 orders of magnitude faster in sampling. The diffusion model \submAmram even has the fewest number of parameters in \fref{fig:ds1-photons.pareto.speed.size}, but is one of the slowest in sampling. Datasets 2 (in \fref{fig:ds2.pareto.speed.size}) and 3 (in \fref{fig:ds3.pareto.speed.size}) show even less of a front, but with \submKaech a submission clearly in the sweet spot at few parameters and fast generation.

\clearpage

\section*{References}
\markboth{\uppercase{References}}{}
\addcontentsline{toc}{section}{References}
\bibliographystyle{iopart-num}
\bibliography{literature}

\end{document}